\documentclass[prx,twocolumn,superscriptaddress]{revtex4}
\usepackage{amsmath}
\usepackage{amssymb}
\usepackage{amsthm}
\usepackage{amsfonts}
\usepackage{listings}
\usepackage{longtable}
\usepackage{enumerate}
\usepackage{latexsym}
\usepackage{color}
\usepackage{setspace}
\usepackage{blindtext}
\usepackage{xcolor}
\usepackage{float}

\usepackage{bm}
\usepackage{hyperref}
\usepackage{subfigure}

\usepackage{booktabs}

\usepackage{array,etoolbox}
\preto\tabular{\setcounter{magicrownumbers}{0}}
\newcounter{magicrownumbers}

\hypersetup{
 pdfnewwindow=true, colorlinks=true,
 linkcolor=blue, anchorcolor=blue,
 citecolor=blue, filecolor=blue,
 menucolor=blue, urlcolor=blue}

\newcommand{\beginsupplement}{%
        \setcounter{table}{0}
        \renewcommand{\thetable}{S\arabic{table}}%
        \setcounter{figure}{0}
        \renewcommand{\thefigure}{S\arabic{figure}}%
     }

\usepackage{psfrag}

\usepackage{graphicx}

\def\ie{{\it i.e.},\ }
\def\eg{{\it e.g.}\ }

\input{epsf}
\newcommand{\webposbr}{\href{https://github.com/zjwang11/UnconvMat/blob/master/src_pos2aBR.tar.gz}{\ttfamily pos2aBR}}
\newcommand{\webatomly}{\href{https://atomly.net}{atomly}}
\newcommand{\webirvsp}{\href{https://github.com/zjwang11/irvsp/blob/master/src_irvsp_v2.tar.gz}{\ttfamily irvsp}}

\begin{document}

\tolerance 10000

\newcommand{\vk}{{\bf k}}

\draft

\title{Unconventional Materials: the mismatch between electronic charge centers and atomic positions}
\author{Jiacheng Gao}
\thanks{These authors contributed equally to this work.}
\affiliation{Beijing National Laboratory for Condensed Matter Physics,
and Institute of Physics, Chinese Academy of Sciences, Beijing 100190, China}
\affiliation{University of Chinese Academy of Sciences, Beijing 100049, China}

\author{Yuting Qian}
\thanks{These authors contributed equally to this work.}
\affiliation{Beijing National Laboratory for Condensed Matter Physics,
and Institute of Physics, Chinese Academy of Sciences, Beijing 100190, China}
\affiliation{University of Chinese Academy of Sciences, Beijing 100049, China}

\author{Huaxian Jia}
\thanks{These authors contributed equally to this work.}
\affiliation{Beijing National Laboratory for Condensed Matter Physics,
and Institute of Physics, Chinese Academy of Sciences, Beijing 100190, China}
\affiliation{University of Chinese Academy of Sciences, Beijing 100049, China}

\author{Zhaopeng Guo}
\affiliation{Beijing National Laboratory for Condensed Matter Physics,
and Institute of Physics, Chinese Academy of Sciences, Beijing 100190, China}
\affiliation{University of Chinese Academy of Sciences, Beijing 100049, China}

\author{Zhong Fang}
\affiliation{Beijing National Laboratory for Condensed Matter Physics,
and Institute of Physics, Chinese Academy of Sciences, Beijing 100190, China}
\affiliation{University of Chinese Academy of Sciences, Beijing 100049, China}

\author{Miao Liu}
\email{mliu@iphy.ac.cn}
\affiliation{Beijing National Laboratory for Condensed Matter Physics,
and Institute of Physics, Chinese Academy of Sciences, Beijing 100190, China}
\affiliation{University of Chinese Academy of Sciences, Beijing 100049, China}

\author{Hongming Weng}
\email{hmweng@iphy.ac.cn}
\affiliation{Beijing National Laboratory for Condensed Matter Physics,
and Institute of Physics, Chinese Academy of Sciences, Beijing 100190, China}
\affiliation{University of Chinese Academy of Sciences, Beijing 100049, China}

\author{Zhijun Wang}
\email{wzj@iphy.ac.cn}
\affiliation{Beijing National Laboratory for Condensed Matter Physics,
and Institute of Physics, Chinese Academy of Sciences, Beijing 100190, China}
\affiliation{University of Chinese Academy of Sciences, Beijing 100049, China}

{
\begin{abstract}
The complete band representations (BRs) have been constructed in the work of topological quantum chemistry. Each BR is expressed by either a localized orbital at a Wyckoff site in real space, or by a set of irreducible representations in momentum space. In this work, we define unconventional materials with a common feature of the mismatch between average electronic centers and atomic positions. They can be effectively diagnosed as whose occupied bands can be expressed as a sum of elementary BRs (eBRs), but not a sum of atomic-orbital-induced BRs (aBRs). The existence of an essential BR at an empty site is described by nonzero real-space invariants (RSIs). The `valence' states can be derived by the aBR decomposition, and unconventional materials are supposed to have an \emph{uncompensated} total `valence' state. The high-throughput screening for unconventional materials has been performed through the first-principles calculations. We have discovered 423 unconventional compounds, including thermoelectronic materials, higher-order topological insulators, electrides, hydrogen storage materials, hydrogen evolution reaction electrocatalysts, electrodes, and superconductors. The diversity of these interesting properties and applications would be widely studied in the future.
\newline
\textbf{Keywords:} Unconventional materials, Band representations, Real-space invariants, Thermoelectric materials.
\end{abstract}

\maketitle 

\section{Introduction}
For the past decade, topological insulators (TIs) and semimetals have been intensively studied~\cite{bernevig2006quantum,hasan2010colloquium,qi2011topological,fu2011topological,yan2017topological,wang2016hourglass,ma2017experimental,Nie2021PRB,song2019all,zhang2009topological,wang2012dirac,wang2013three,weng2015weyl,hsieh2012topological,liu2020symmetry,gao2021high}. Many exotic physical properties are proposed in the topological materials, such as spin-momentum-locking Dirac-cone surface states, quantum anomalous Hall effect, Fermi-arc states, negative magnetoresistivity, and chiral anomaly, which have attracted broad interest in condensed matter physics~\cite{wang2013three,weng2015weyl,hsieh2012topological,liu2020symmetry,gao2021high,chang2013experimental}. Recently, topological quantum chemistry (TQC)~\cite{bradlyn2017topological,vergniory2019complete} and related theories~\cite{po2017symmetry,song2017} provided a general framework to diagnose whether the band structure of a material is topological from irreducible representations (irreps) at several high-symmetry $k$-points (HSKPs). If the irreps of a band structure are the same as those of a BR in TQC, which is a space group representation formed by exponentially decayed symmetric orbitals in real space, representing a trivial (atomic) insulator, then the band structure is consistent with topologically trivial state; otherwise, it must be topological (Fig.~\ref{fig:flowchart}). However, among topologically trivial compounds, we have overlooked a large amount of {\it{unconventional materials}}, whose occupied bands can be decomposed as a sum of eBRs (\ie generators of BRs), but not a sum of aBRs. They possess the unconventional feature of the mismatch between the average electronic centers and the atomic positions (previously known as obstructed atomic limits~\cite{bradlyn2017topological,PhysRevB.97.035139,Nie2021PRB}). In fact, many interesting properties, such as low work function, strong hydrogen affinity, electrocatalysis, etc., can be expected in these unconventional materials.

\begin{figure}[t]
    \includegraphics[width=8 cm]{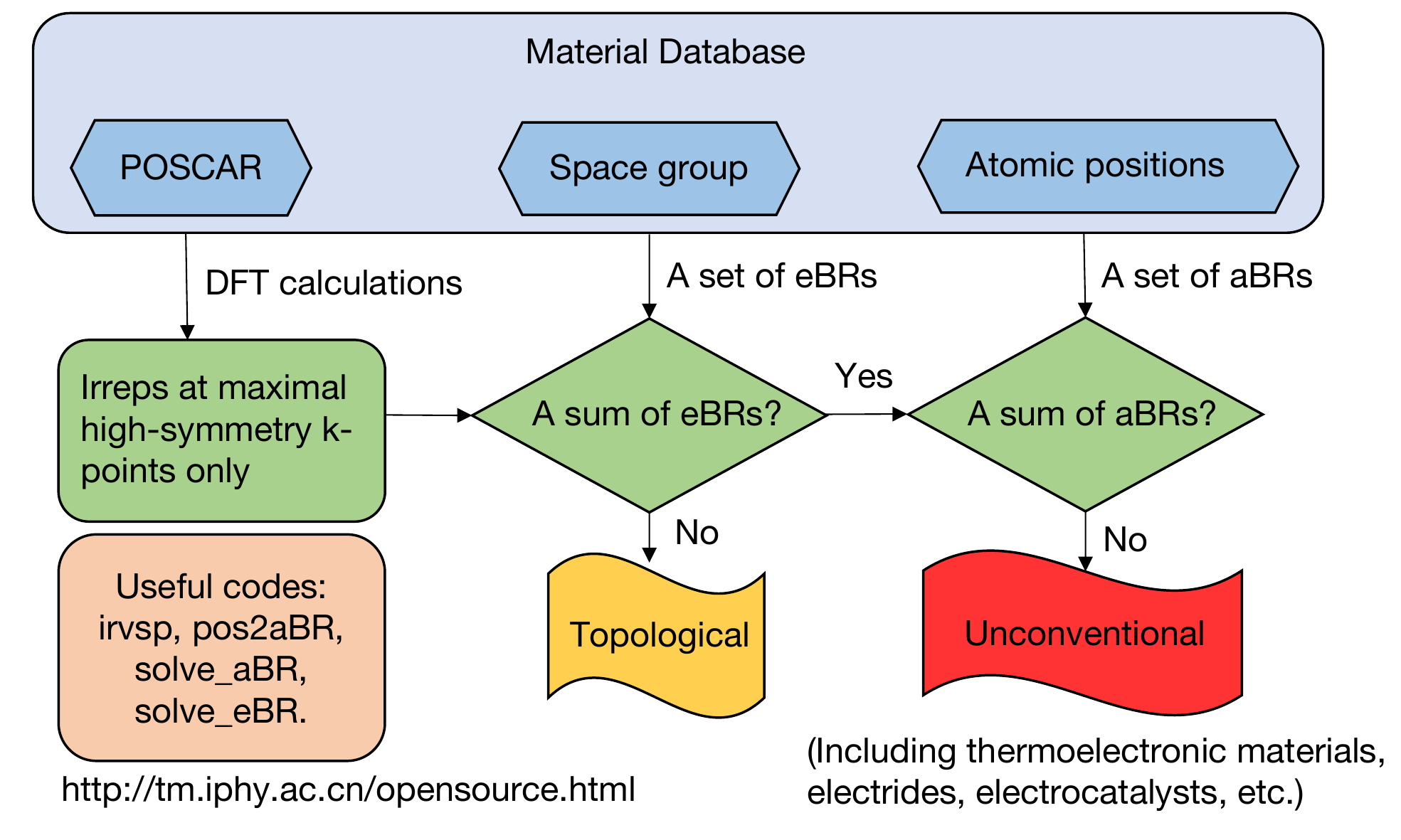}
    \caption{(Color online) The workflow of high-throughput screening for unconventional materials by solving eBR/aBR decomposition in the theory of TQC. Some useful codes are released \href{http://tm.iphy.ac.cn/UnconvMat.html}{online}.
    Unconventional materials possess the common feature of the mismatch between average electronic centers and atomic positions, where a diversity of interesting properties are expected, such as low work function, strong hydrogen affinity, electrocatalysis, etc.}
    \label{fig:flowchart}
\end{figure}

In this work, we have performed the high-throughput screening for unconventional materials in the materials database. We have computed irreps at maximal HSKPs in density-functional theory (DFT) calculations. The aBRs are generated from atomic configurations and positions in a crystal. Based on compatibility relations (CRs) and BRs of the TQC, the BR decomposition of an unconventional material is solved to be a sum of eBRs, but not a sum of aBRs, \eg ``aBRs + an essential BR” (Fig.~\ref{fig:abr2ebr}).
The essentiality of the BR is described by nonzero RSIs on an empty site.
One can derive the `valence' states from the solved aBRs. The unconventional materials are supposed to be a group of materials with an \emph{uncompensated} total `valence' state. 
We find 423 unconventional compounds and tabulate their detailed information in the supplementary materials (SM). A diversity of interesting properties have been discovered in these compounds, including thermoelectric materials, higher-order TIs, electrides, solid-state hydrogen storage materials, hydrogen evolution reaction (HER) electrocatalysts, electrodes, and superconductors.

\begin{figure}[!t]
    \centering
    \includegraphics[scale=0.26]{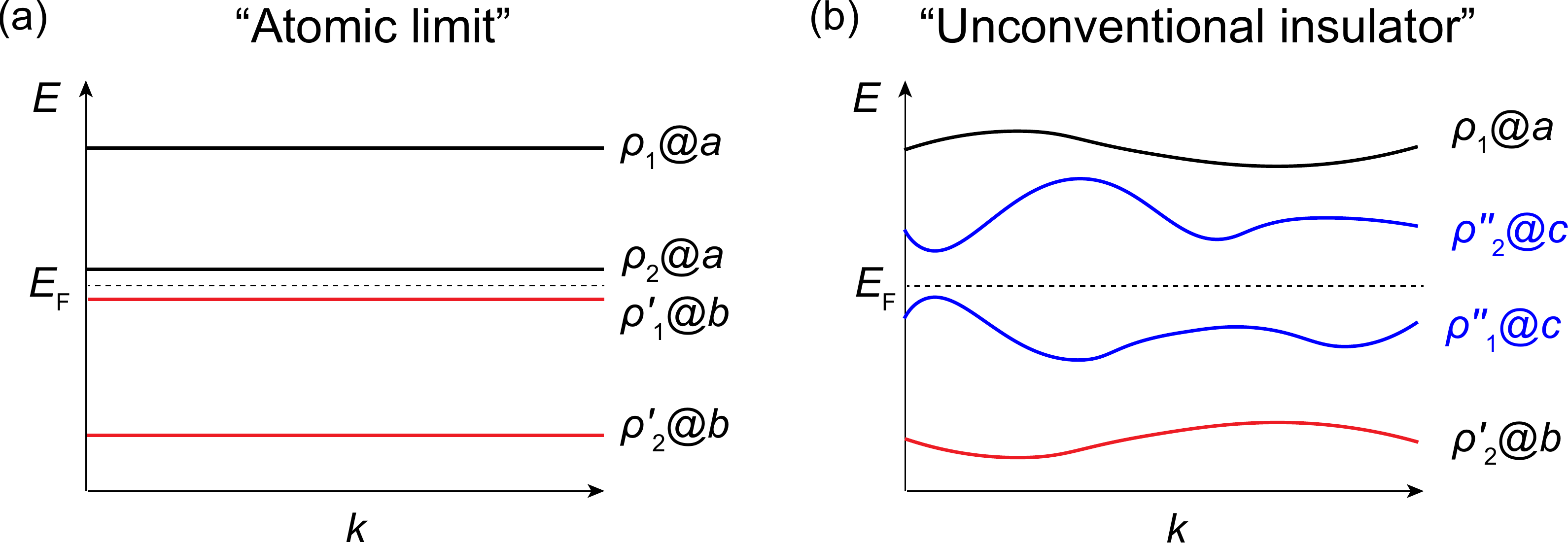}
    \caption{(Color online) Schematic diagram of an unconventional compound $AB$. The elements $A$ and $B$ sit at the $a$ and $b$ Wyckoff sites, respectively. (a) Energy `bands' for the hypothetical structure with large (infinite) lattice constants (\ie the atomic limit). Each set of flat bands are aBRs formed by  atomic orbitals, \eg $A$-$s$, $A$-$p$, $B$-$s$, $B$-$p$, etc. (b) Energy bands for the synthesized crystal (with experimental lattice constants). The aBR decomposition for occupied bands is solved to be $\rho'_2@ b + {\color{blue} \rho''_1@c}$ (which is empty/hollow in the crystal).}
    \label{fig:abr2ebr}
\end{figure}

\section{calculation method}
We swept through materials with the Inorganic Crystal Structure Database (ICSD) numbers on the \webatomly ~website. The Vienna \emph{ab-initio} simulation package (VASP)~\cite{KRESSE199615,vasp} with the projector augmented wave method~\cite{paw1,paw2} based on density functional theory was employed for the first-principles calculations. The generalized gradient approximation of Perdew-Burke-Ernzerhof type~\cite{pbe} was adopt for the exchange-correlation potential. The cutoff energy of plane wave basis set was set to be 125\% ENMAX value in the pseudopotential file. A $\Gamma$-centered grid with 30 $k$-points per 1/\AA~was used for self-consistent calculations. For simplicity, we did not consider any magnetic configurations in the calculations. 
Electron-phonon coupling calculations were performed in the framework of density functional perturbation theory, as implemented in the QUANTUM ESPRESSO package~\cite{QE2009}.

The general workflow of our high-throughput screening is given in Fig.~\ref{fig:flowchart}. First, we obtain the crystal structures of synthesized compounds in the materials database (\ie atomic elements, Wyckoff positions, and space group number). In TQC theory, a set of eBRs are well defined for a certain space group, while the list of aBRs are defined by atomic elements and positions in a crystal, generated by a homemade program \webposbr~\cite{Nie2021PRB}. Then, the electronic states at maximal HSKPs are obtained in DFT calculations and their irreps  are assigned by the program \webirvsp~\cite{gao2021irvsp}. Since spin-orbit coupling is not included, the obtained irreps are single-valued.  Next, we check if these irreps can be decomposed into a sum of eBRs (\href{http://tm.iphy.ac.cn/UnconvMat.html}{\ttfamily eBR decomposition}). If yes, we further check if they are a sum of aBRs (\href{http://tm.iphy.ac.cn/UnconvMat.html}{\ttfamily aBR decomposition}).  When they are a sum of eBRs but a not a sum of aBRs, we come across an unconventional material. The full list of unconventional materials in our searching are tabulated in the SM.

\begin{figure}[!b]
    \includegraphics[scale=0.16]{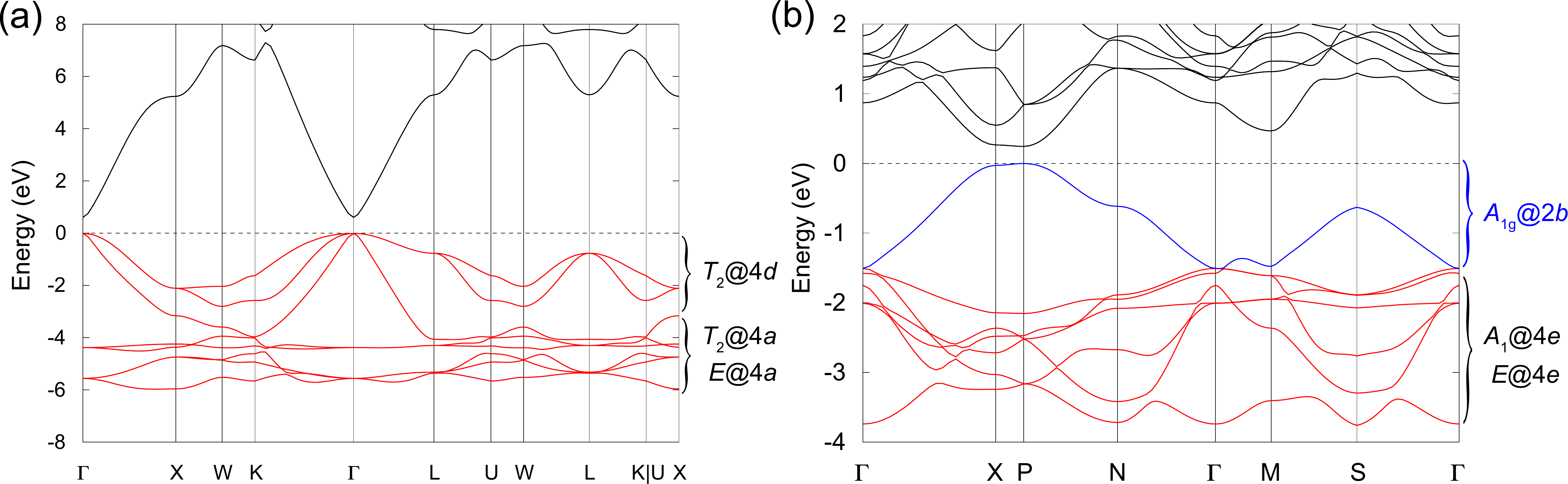}
    \caption{(Color online) Band structures of (a) conventional ZnO and (b) unconventional Ca$_2$As compounds. The aBR decompositions for occupied bands are presented in the figure.}
    \label{fig:bandstate}
\end{figure}

\section{Results and discussion}
\subsection{Basic concepts in the TQC}
The TQC theory tabulates the CRs for 230 space groups, and constructs a complete list of BRs. The CRs suggest that the symmetry eigenvalues of a band structure rely on the irreps only at maximal HSKPs. For a given space group, a certain orbital (irrep $\rho$; labelled by the site-symmetry group) at a Wyckoff site ($q$) can form a bundle of energy bands in momentum space (labelled by a set of irreps of $k$-little groups).  The set of irreps is usually regarded as a BR of $\rho @ q$ in the space group. 
The topologically trivial band structure is a sum of eBRs. On the other hand, by matching the irreps of a band structure with the BRs, one can infer that the band structure belongs to a certain (elementary) BR (\ie $\rho_0 @ q_0$), which tells the average charge center ($q_0$) and the site-symmetry character ($\rho_0$). The BR analysis/decomposition can be widely used in materials computation.

\begin{table}[!h]
\tiny
\caption{
The aBRs, BR decompositions and valence states are obtained for $F\bar43m$ zinc-blende ZnO and $I4/mmm$ Ca$_2$As. 
}\label{table:ZnO}
\begin{tabular}{ccccccccc}
\multicolumn{2}{c}{$F\bar43m$ ZnO}  \\
\toprule[0.5pt]
 Atom &WKS($q$) &  Symm.&States &Conf. &\multicolumn{2}{c}{Irreps($\rho$)}& aBRs($\rho@q$)& Occ.\\
 \hline
 Zn & $4a $&$ \bar{4}3m$& Zn$^{2+}$& $ 4s^2 3d^{10} $  & $ s $  &$: A_1 $ &$ A_1@ 4a$& \\
 \cline{6-9}
     &     &       & &            & $ d_{z^2,x^2} $  &$: E   $ &$   E@ 4a$& Yes\\
    &     &       & &            & $ d_{xy,yz,xz} $  &$: T_2 $ &$ T_2@ 4a$& Yes\\
 \hline
 O & $4d $&$ \bar{4}3m$& O$^{2-}$&$ 2p^4 $  & $ p_{x,y,z} $  &$: T_2 $ &$ T_2@ 4d$& Yes\\
\bottomrule[0.5pt]
\multicolumn{2}{c}{$~$}  \\
\multicolumn{2}{c}{$I4/mmm$ Ca$_2$As}  \\
\toprule[0.5pt]
 Atom &WKS($q$) &  Symm.&States &Conf. &\multicolumn{2}{c}{Irreps($\rho$)}& aBRs($\rho@q$)& Occ.\\
\hline
As&$4e$& $4mm$  & As$^{3-}$ & $4p^3$ & $p_z$&$:A_1$   &$A_1@4e$ & Yes\\
    &        &  &  &    &$p_{x,y}$&$:E$& $E@4e$& Yes\\
  \hline
Ca(1)& $4e$& $4mm$  & Ca$^{2+}$ &$4s^2$&$s$&$:A_1$    & $A_1@4e$&\\
  \hline
Ca(2)& $4c$& $mmm$  & Ca$^{2+}$ &$4s^2$ & $s$&$:A_{g}$   & $A_{g}@4c$&\\
  \hline
 \multicolumn{7}{r}{$~$} & {\color{blue} $A_{1g}@2b$}& Yes \\
\bottomrule[0.5pt]
\end{tabular}
\end{table}

\subsubsection{The aBR decomposition}
The electronic states of a compound originate from the BRs induced by its atomic orbitals. In the hypothetical structure [Fig.~\ref{fig:abr2ebr}(a)], the `flat' bands are the original aBRs. After hybridization, the reconstructed occupied bands in unconventional materials can be a sum of eBRs, but not a sum of aBRs.
After solving the aBR decomposition (Fig.~\ref{fig:bandstate}), the occupied bands for topologically trivial insulators can be generally classified into two cases: i) solved to be a sum of aBRs, \eg zinc-blende structure ZnO (conventional) ; or ii) not a sum of aBRs, such as some aBRs + an essential BR for Ca$_2$As (unconventional). In the latter case, the essential BR tells the average charge center (\ie an empty site) and the site-symmetry character of the `loose' electrons. 

\subsubsection{`Valence' states from the aBR decomposition}
In both cases, one can derive atomic `valence' states from the solved aBRs of the crystal. Hereafter, valence states in this work are referred to the TQC `valence' states based on the aBR decomposition. The irreps for the occupied bands in ZnO are obtained at four HSKPs in SM. The aBR decomposition is solved to be $E@4a$ + $T_2@4a$ + $T_2@4d$ (denoted by `Yes' in the last column of Table~\ref{table:ZnO}), corresponding to Zn $d$ and O $p_{x,y,z}$ orbitals, respectively. Considering the atomic configurations of Zn and O, one can conclude that their valence states are Zn$^{2+}$ and O$^{2-}$. Thus, the insulator ZnO has a compensated total valence state. However, in Ca$_2$As, the aBR decomposition for its occupied bands is $A_1@4e+E@4e+$ {\color{blue} $A_{1g}@2b$}. Hereafter, the BR colored in blue indicates the essential BR at an empty site.
The $A_1@4e$ and $E@4e$ aBRs correspond to As $p_z$ and $ p_{x,y}$ orbitals. The valence states are derived to be Ca$^{2+}$ and As$^{3-}$. The uncompensated total valence state (+1) implies that it is unconventional, which is consistent with the electride nature~\cite{Nie2021PRB,hirayama2018electrides}.

\begin{figure*}[!ht]
    \includegraphics[scale=0.2]{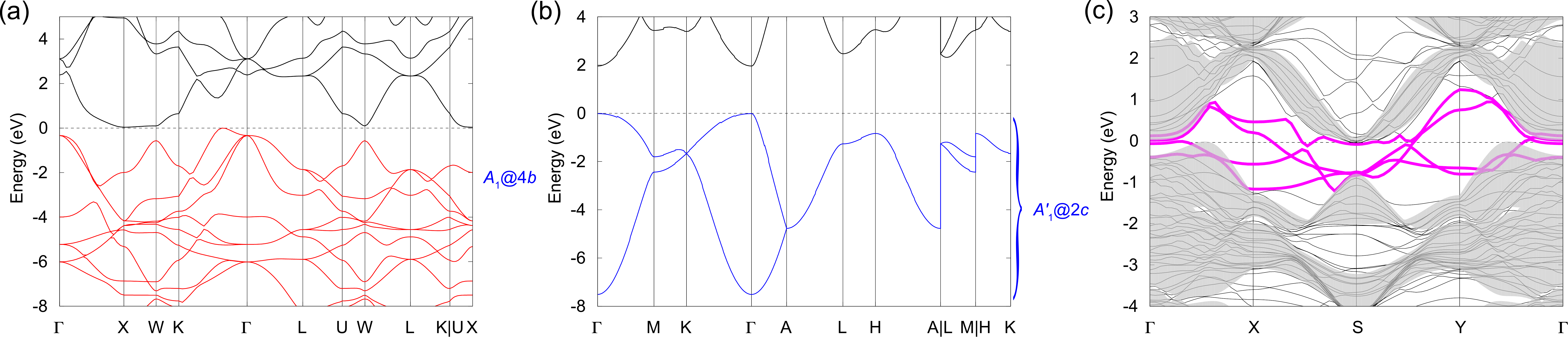}
      \caption{(Color online) Band structures of intermetallic semiconductors (a) Be$_5$Pt and (b)Na-hP4, where the essential BRs are indicated. (c) Ingap surface states in the slab calculation of Be$_5$Pt. The projected bulk band structure is shown in gray (shaded area).}
    \label{fig:Al2Ru}
\end{figure*}

\subsubsection{The essential BR and real-space indicators}
The essential BR of an empty site can be described by the RSIs, which are local quantum numbers at a Wyckoff site protected by its site-symmetry group (isomorphic to point-group symmetry). The RSIs were introduced to characterize the topological fragile phases and determine the number of gap closings under the specific twisted boundary conditions in 2D~\cite{song2020twisted}. Here we generalize the concept of the RSIs to all Wyckoff sites in 230 space groups, especially for maximum Wyckoff sites. The essentiality of the BR for an unconventional insulator is described by the non-zero RSIs on an empty site.

In the atomic limit of an unconventional insulator [hypothetical structure; Fig.~\ref{fig:abr2ebr}(a)], the RSI is zero on the empty site,  while it becomes nonzero in the synthesized crystal [Fig.~\ref{fig:abr2ebr}(b)].
The charge centers of occupied electronic bands can not move away from the empty site symmetrically without closing the band gap. The essential BR at the empty site implies the disagreement between average electronic centers and atomic positions.

\begin{table}[!b]
\tiny
\caption{
The aBRs, BR decompositions and valence states for the unconventional intermetallic semiconductors.
}\label{table:Be5Pt_orb}
\begin{tabular}{ccccccccc}
\multicolumn{2}{c}{$F\bar43m$ Be$_5$Pt}  \\
\toprule[0.5pt]
 Atom &WKS($q$) &  Symm.& States& Conf. &\multicolumn{2}{c}{Irreps($\rho$)}& aBRs($\rho@q$)& Occ.\\ 
\hline
 Pt   &$ 4a$& $ \bar{4}3m$& Pt$^{0}$&$6s^15d^9$   & $  s $ &: $ A_1 $ &   $A_1@4a $ \\
\cline{6-9}
      &$   $& $     $& & &     $ d_{z^2,x^2}   $ &:  $ E   $ &   $  E@4a  $  & Yes\\
      &$   $& $     $& &       & $ d_{xy,xz,yz} $ &  $: T_2 $ & $ T_2@4a  $& Yes \\
\hline
 Be(1)&$ 4d$& $ \bar{4}3m$& Be$^{2+}$&$2s^2$   & $ s  $ &: $ A_1 $ &   $A_1@4d $ & \\
 \hline
 Be(2)&$16e$& $  3m $& Be$^{0}$&$2s^2$   & $ s  $ &: $ A_1$ &   $A_1@16e $ & Yes\\
\hline
 \multicolumn{7}{r}{$~$} &{\color{blue} $A_1@4b$} & Yes \\
\bottomrule[0.5pt]
\multicolumn{2}{c}{$~$}  \\
\multicolumn{2}{c}{$P6_3/mmc$ Na-hP4}  \\
\toprule[0.5pt]
 Atom &WKS($q$) &  Symm.& States& Conf. &\multicolumn{2}{c}{Irreps($\rho$)}& aBRs($\rho@q$)& Occ.\\ 
\hline
 Na(1)&$ 2a$& $ \bar{3}m$& Na$^{1+}$ & $3s^1$   & $ s  $ &: $ A_{1g} $ &   $A_{1g}@2a $ & \\
\hline
 Na(2)& $2d$& $\bar{6}2m$& Na$^{1+}$ & $3s^1$   & $ s  $ &: $ A_1'$ &   $A_1'@2d $ & \\
\hline
 \multicolumn{7}{r}{$~$} & {\color{blue} $A_1'@2c$} & Yes \\
\bottomrule[0.5pt]
\end{tabular}
\end{table}

\subsection{Unconventional insulators}
\subsubsection{Intermetallic semiconductors}
Firstly, in the searching results, there are many unconventional semiconductors with only metallic elements, which are known as intermetallic semiconductors in literatures~\cite{amon2019interplay,ma2009transparent,volkov1994semiconducting}. 
Semiconducting substances form one of the most important families of functional materials. However, semiconductors containing only metals are very rare. The chemical mechanisms behind their ground-state properties are not fully understood. Our investigations for unconventional materials can reveal the semi-conduction behaviour in the intermetallic compounds and provide an effective way to search for them by computing the irreps at several HSKPs only. 

We take {Be$_5$Pt} and Na-hP4 as two representatives. Their band structures are shown in Fig.~\ref{fig:Al2Ru}(a,b), where we can find there are clear band gaps at E$_F$. The list of aBRs are given in Table~\ref{table:Be5Pt_orb}. With the computed irreps, the aBRs decomposition of Be$_5$Pt are solved (\ie $E@4a+T_2@4a+A_1@16e$+{\color{blue}$A_1@4b$}) and given in the last column (denoted by `yes'), suggesting that one Be(2)-$s$ and five Pt-$d$ orbitals are occupied. Thus, one can derive the valence state for each Wyckoff atom, which is presented in the 4th column of Table~\ref{table:Be5Pt_orb}.
The results for Na-hP4 are presented in Table~\ref{table:Be5Pt_orb} as well.
The uncompensated valence state in total reveals that there is an essential BR, \ie {\color{blue}$A_1@4b$} for Be$_5$Pt and {\color{blue}$A_1'@2c$} for Na-hP4. In addition, some thermoelectric materials (Al$_2$Ru, AlVFe$_2$, etc.) are also unconventional intermetallic semiconductors.

\begin{figure}[!b]
    \includegraphics[scale=0.16]{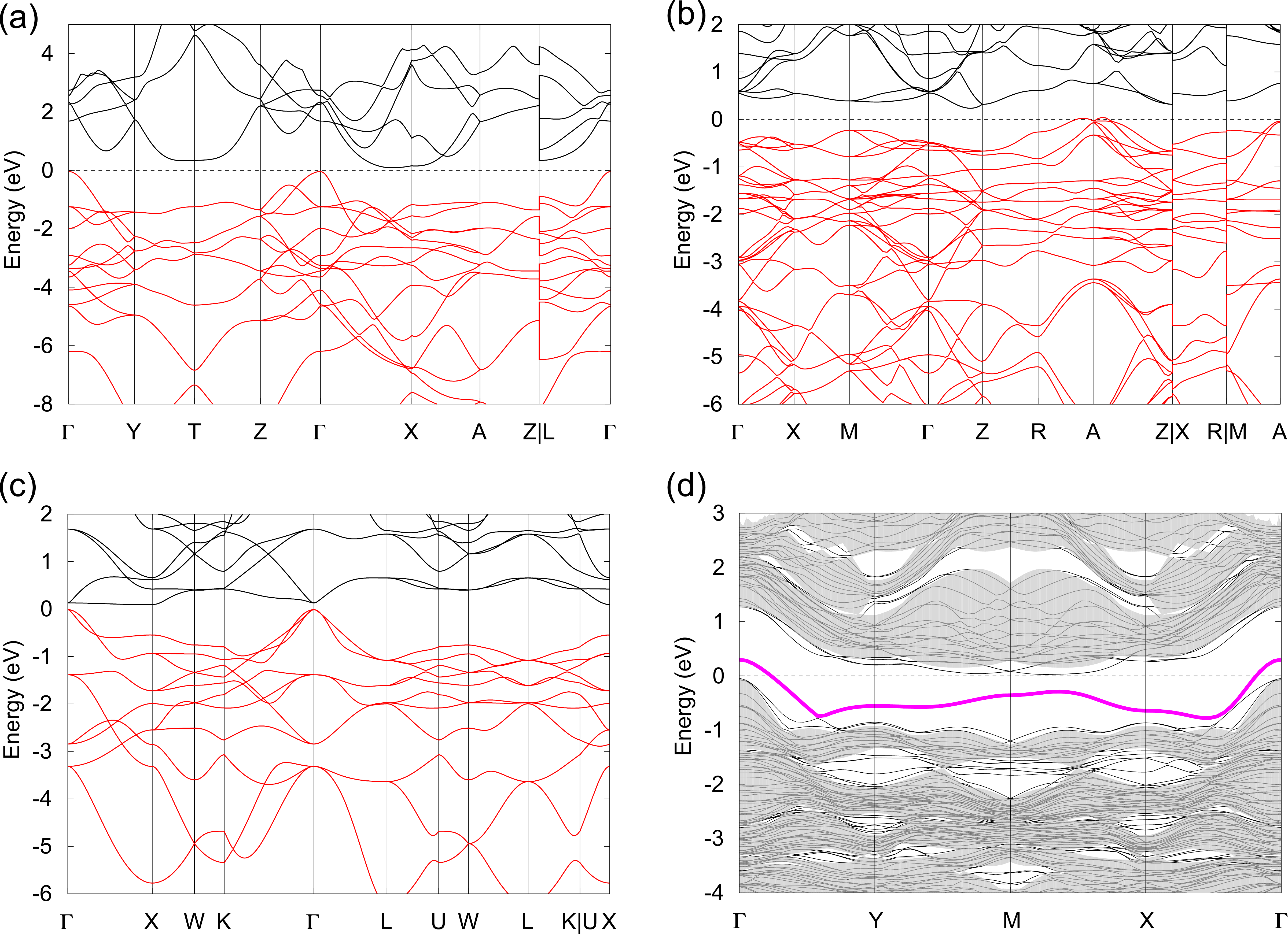}
    \caption{(Color online) Band structures of thermoelectric materials. (a)Al$_2$Ru has the  essential BR of $A_g@16d$. (b) In$_3$Ru has the  essential BR of $A_g@2b$  (c) TiFe$_2$Sn  has the essential BR of $A_g@24d$. (d) Surface states in the slab calculation of Al$_2$Ru. The projected bulk band structure is shown in gray (shaded area).}
    \label{fig:thermo}
\end{figure}

The hard X-ray photoelectron spectroscopy experiment~\cite{amon2019interplay} of Be$_5$Pt shows a complete filling of Pt-$5d$ levels. The higher binding energy of Be(1)-$s$ than Be(2)-$s$ reflects the difference of valence states between Be(1) and Be(2). These facts are consistent with the aBR analysis. 
The electrical resistivity measured on bulk samples shows a metal-like temperature dependence, while for the microscale samples, it decreases monotonically with increasing temperature, corresponding to a semiconducting behaviour inferred from the band-structure calculations~\cite{amon2019interplay}. 
In Fig.~\ref{fig:Al2Ru}(c), the in-gap surface states emerge in the slab calculation of Be$_5$Pt, where the (001) termination cuts through the empty site of the essential BR.
We conjecture that its flatness at low temperatures in microscale samples and the semimetallic behaviour in bulk samples are contributed to the existence of surface states on the grain boundaries.

\subsubsection{Thermoelectric materials}
Secondly, many unconventional semiconductors are formed by transition metals and Group 13-15 ($p$-block) elements, such as Al$_2$Ru/Ga$_2$Ru, In$_3$Ru/Ga$_3$Ir, and TiFe$_2$Sn/AlVFe$_2$/VGaFe$_2$ (\ie Heusler compounds XY$_2$Z with 24 valence electrons),  which were previously known as thermoelectronic materials~\cite{volkov1994semiconducting,takagiwa2010thermoelectric,takagiwa2012thermoelectric,alvarez2018semiclassical,taranova2019influence,rai2017electronic,wagner2011ruin3}. The band structures of Al$_2$Ru , In$_3$Ru and TiFe$_2$Sn are shown in Fig.~\ref{fig:thermo}. The aBR decompositions for occupied bands of these materials are solved. The results in Table.~\ref{tab:thermo} indicate that the essential BR is {\color{blue}$A_g@16d$} for Al$_2$Ru/Ga$_2$Ru, {\color{blue}$A_g@2b$} for In$_3$Ru, and {\color{blue}$A_g@24d$} for TiFe$_2$Sn/AlVFe$_2$/VGaFe$_2$. Thermoelectric properties, superparamagnetism and negative giant magnetoresistance have been widely studied in these Heusler compounds XY$_2$Z with 24 valence electrons~\cite{rai2017electronic,buffon2017thermoelectric,lue2001superparamagnetism,endo1998anomalous}.

As we know, the thermoelectric efficiency of materials at temperature $T$ is characterized through the figure of merit given by the relation $zT = S^2\sigma T/\kappa$, where $S$ is the thermoelectric or Seebeck coefficient, $\sigma$ is the electronic conductivity and $\kappa$ is the thermal conductivity. The essential BRs of the unconventional materials suggest that surface states can emerge and conduct electricity when the surface cuts through the empty sites. The in-gap surface states are obtained in the (001)-slab calculation of Al$_2$Ru in Fig.~\ref{fig:thermo}(d). The insulating bulk states suggest that thermal conductivity due to bulk carries is weak. But the electronic conductivity due to the grain boundaries is substantial. These facts of these unconventional semiconductors are good for thermoelectric efficiency.

\begin{table}[!htb]
\tiny
\caption{
The aBRs, BR decompositions and valence states for $Fddd$ Al$_2$Ru, $P4_2/mnm$ In$_3$Ru, $Fm\bar3m$ TiFe$_2$Sn/AlVFe$_2$.
}\label{tab:thermo}
\begin{tabular}{ccccccccc}
\multicolumn{2}{c}{$Fddd$ Al$_2$Ru}  \\
\toprule[0.5pt]
 Atom &WKS($q$) &  Symm.& States& Conf. &\multicolumn{2}{c}{Irreps($\rho$)}& aBRs($\rho@q$)& Occ.\\ 
\hline
Al & $16f$   & 2         & Al$^{1+}$& $3s^2 3p^1$ &$s $          &:$A  $  &$A  @16f$ & Yes   \\
 \cline{6-9}
  &         &           &  &      &$p_z $        &:$A  $  &$A  @16f$ &  \\
  &         &           &  &      &$p_x/p_y $    &:$B  $  &$B  @16f$ &  \\
 \hline
Ru & $8b $   &222        & Ru$^{2+}$& $5s^1 4d^7$ &$s $          &:$A  $  &$A  @8b$ &  \\
 \cline{6-9}
  &         &           &  &      &$d_{z^2}/d_{x^2}$ &:$A  $  &$A  @8b$ & Yes   \\
  &         &           &  &      &$d_{xy}$      &:$B_1$  &$B_1@8b$ & Yes  \\
  &         &           &  &      &$d_{yz}$      &:$B_3$  &$B_3@8b$ & Yes  \\
  &         &           &  &      &$d_{xz}$      &:$B_2$  &$B_2@8b$   \\
 \hline
 \multicolumn{7}{r}{$~$} & {\color{blue} $A_g@16d$} & Yes \\
\bottomrule[0.5pt]
\multicolumn{2}{c}{$~$}  \\
\multicolumn{2}{c}{$P4_2/mnm$ In$_3$Ru}  \\
\toprule[0.5pt]
 Atom &WKS($q$) &  Symm.&States &Conf. &\multicolumn{2}{c}{Irreps($\rho$)}& aBRs($\rho@q$)& Occ.\\
 \hline    
 Ru & $4f $&$  mm2$& Ru$^0$ &$ 5s^14d^7  $  & $ s $  &$: A_{1} $ &$ A_{1}@ 4f$&\\
 \cline{6-9}
                    & & & &  & $ d_{x^2} $  &$: A_{1} $ &$ A_{1}@ 4f$&\\
                      && & & & $ d_{z^2} $  &$: A_{1} $ &$ A_{1}@ 4f$& Yes\\
                      && & & & $ d_{xy} $  &$: A_{2} $ &$ A_{2}@ 4f$& Yes\\
                      && & & & $ d_{yz} $  &$: B_{2} $ &$ B_{2}@ 4f$& Yes\\
                      && & & & $ d_{xz} $  &$: B_{1} $ &$ B_{1}@ 4f$& Yes\\
 \hline                      
 In & $4c $&$  2/m$& In$^{1-}$&$ 5s^25p^1 $  & $ s $  &$: A_{g} $ &$ A_{g}@ 4c$& Yes\\
 \cline{6-9}
                     & & & &  & $ p_z $  &$: A_{u} $ &$ A_{u}@ 4c$&\\
                      && & & & $ p_x $  &$: B_{u} $ &$ B_{u}@ 4c$& Yes\\
                      && & & & $ p_y $  &$: B_{u} $ &$ B_{u}@ 4c$&\\
  \hline                     
 In & $8j $&$    m$& In$^{1+}$&$ 5s^25p^1 $  & $ s $  &$: A' $ &$ A'@ 8j$& Yes\\
 \cline{6-9}
                    & & & &  & $ p_x $  &$: A' $ &$ A'@ 8j$&\\
                      && & & & $ p_y $  &$: A' $ &$ A'@ 8j$&\\
                      && & & & $ p_z $  &$: A'' $ &$ A''@ 8j$&\\    
 \hline
 \multicolumn{7}{r}{$~$} & {\color{blue} $A_{g}@2b$}& Yes \\                      
\bottomrule[0.5pt]
\multicolumn{2}{c}{$~$}  \\
\multicolumn{4}{c}{$Fm\bar3m$ TiFe$_2$Sn (AlVFe$_2$)}  \\
\toprule[0.5pt]
 Atom &WKS($q$) &  Symm.&States &Conf. &\multicolumn{2}{c}{Irreps($\rho$)}& aBRs($\rho@q$)& Occ.\\
\hline
 Fe & $8c $&$ \bar{4}3m$& Fe$^{2+}$&$ 3d^74s^1 $  & $ s $  &$: A_{1} $ &$ A_{1}@ 8c$&\\
 \cline{6-9}
                     & & & & & $ d_{z^2,x^2} $  &$: E $ &$ E@ 8c$&\\
                     & & & & & $ d_{xy,xz,yz} $  &$: T_{2} $ &$ T_{2}@ 8c$& Yes\\
 \hline                
 Sn & $4b $&$ m\bar{3}m$& Sn$^{4+}$&$ 5s^25p^2 $  & $ s $  &$: A_{1g} $ &$ A_{1g}@ 4b$&\\
 \cline{6-9}
  (Al)                  &         & &(Al$^{3+}$) & ($3s^23p^1$)& $ p_{x,y,z} $  &$: T_{1u} $ &$ T_{1u}@ 4b$&\\
 \hline               
 Ti & $4a $&$ m\bar{3}m$& Ti$^{4+}$&$ 4s^13d^3 $  & $ s $  &$: A_{1g} $ &$ A_{1g}@ 4a$&\\
 \cline{6-9}
                    (V) & & & (V$^{5+}$)& ($4s^1 3d^4$)& $ d_{z^2,x^2} $  &$: E_{g} $ &$ E_{g}@ 4a$&\\
                     & & & & & $ d_{xy,xz,yz} $  &$: T_{2g} $ &$ T_{2g}@ 4a$&\\
 \hline
 \multicolumn{7}{r}{$~$} & {\color{blue} $A_{g}@24d$}& Yes \\
\bottomrule[0.5pt]
\end{tabular}
\end{table}

\subsubsection{Higher-order topological insulators}
Thirdly, the nonmetal materials used to be known as covalent compounds. As the common covalent state has the average charge center located on $X-X$ bonds, the covalent compounds would fit the definition of unconventional materials as well. Since they are well-known and studied, we exclude these compounds without any metal elements in our high-throughput screening for simplicity.

\begin{figure}[!t]
    \includegraphics[scale=0.12]{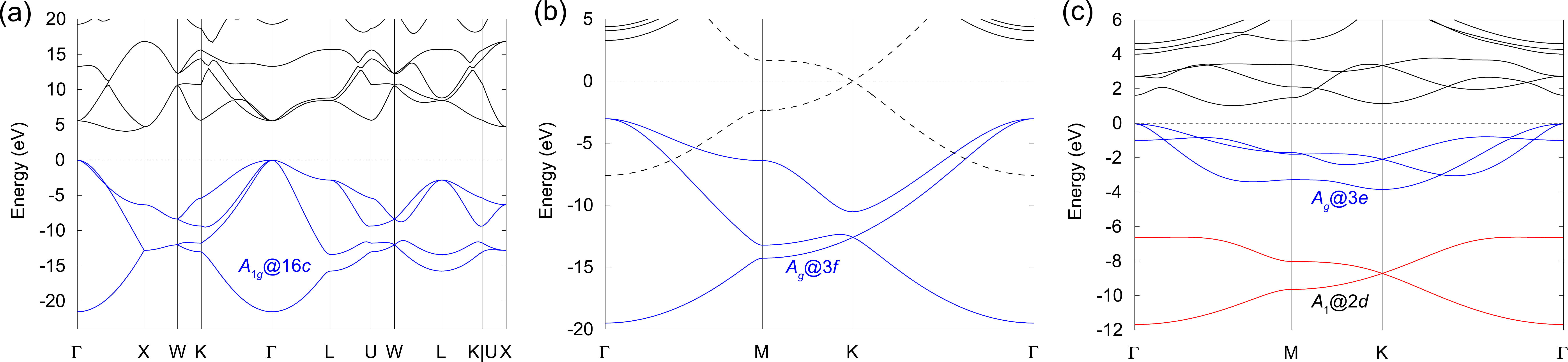}
    \caption{(Color online) Band structures of (a) diamond, (b) graphene and (c) $\beta$-antimonene. The essential BRs are shown by blue bands. In graphene, the m$_z$-even (-odd) bands are plotted in solid (dashed) lines.}
    \label{fig:HOTI}
\end{figure}

\begin{figure*}[!ht]
    \includegraphics[scale=0.24]{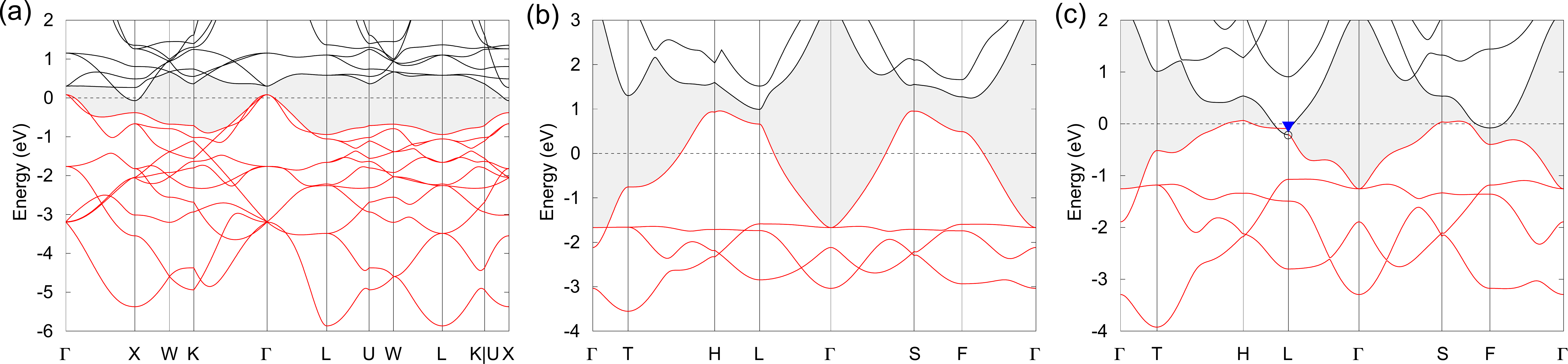}
    \caption{(Color online) Three classes of unconventional  `neat' metals, where there is almost a direct band gap (indicated by shaded area) about/above $E_F$. (a) VGaFe$_2$ of class I: There is a direct energy gap in the band structure, while there is no global gap for the entire Brillouin zone. (b) Ca$_2$N of class II: It is metallic due to the odd total number of electrons. However there would be a gap with one more electron. (c) Y$_2$C of class III: There is an overall band gap except some $k$-point; namely, there can be a band inversion, denoted by a circle and a triangle.}
    \label{fig:VGaFe2}
\end{figure*}

Here, some semiconducting covalent elements can be regarded as higher-order TIs in literature\cite{PhysRevB.104.085205, lee2020two, PhysRevResearch.3.023121}. Strictly speaking, The 3rd-order TIs in 3D and 2nd-order TIs in 2D are topologically trivial (witout particle-hole symmetry). However, they are proved to present a filling anomaly: a mismatch between the number of electrons in a symmetric geometry and the number of electrons required for charge neutrality~\cite{PhysRevB.99.245151}.
For example, the 3D crystals of diamond (or silicon) have $sp^3$ hybridization and the occupied bands belong to the BR of $16c$ sites, corresponding to four $C-C$ bond centers [Fig.~\ref{fig:HOTI}(a)]. In this sense, they can be regarded as a 3rd-order TI in 3D, while the two-dimensional graphene can be regarded as a 2nd-order TI in the $m_z$-even subspace (\ie $s,p_x,p_y$ orbitals). The $sp^2$ hybridization in graphene can be understood by the BR of $3f$ sites, corresponding to three $C-C$ bond centers in a unit cell. Note that $p_z$ orbital belongs to the $m_z$-odd subspace, shown as dashed bands in Fig.~\ref{fig:HOTI}(b). 
In a broad sense, the $7\times 7$ reconstruction on Si-(111) surface could be related to the unconventional nature of silicon due to the presence of substantial charge on the terminations.

Recently, the graphene-like buckled structure ($\beta$-phase) of Sb was found to have the best stability~\cite{zhang2016semiconducting}. The Sb monolayer ($\beta$-antimonene) has been successfully exfoliated using micromechanical technology~\cite{ares2016mechanical}, and has been synthesized on various substrates via van der Waals epitaxial growth~\cite{ji2016two}.
The band structure of $\beta$-antimonene is obtained in Fig.~\ref{fig:HOTI}(c). The aBR decomposition is solved to be $A_1@2d$ + {\color{blue}$A_g@3e$} (Table~\ref{table:beta-Sb}).  The Sb atoms are located at $2d$ site of space group $P\bar{3}m1$ (No. 164), while the $3e$ sites of the essential BR are the centers of Sb$-$Sb bonds. In terms of charge fractionalization and polarization, it corresponds to the $h^{(6)}_{3c}$ primitive generator class in Benalcazar et al.’s notation in Ref.~\cite{PhysRevResearch.1.033074,PhysRevB.99.245151,PhysRevB.102.115104}. In a strict way, the primitive generator would be $h^{(\bar{3})}_{3c}$ after generalizing their notations to $\bar 3$ layer group.
It implies that there is no net dipole in the plane and the corner charge fractionalization will be $e/2$ in each $\pi/6$ sector in the spinless case and $e$ in the spinful case. We note that even though it is the inversion that protects the edge states, hosting corner states at all corners require the presence of the S$_6   (\equiv IC_3$) symmetry operation and a hexagon-shaped island.

\begin{table}[!b]
\tiny
\caption{
The aBRs, BR decomposition and valence states for $P\bar{3}m1$ $\beta$-antimonene.
}\label{table:beta-Sb}
\begin{tabular}{ccccccccc}
\toprule[0.5pt]
 Atom &WKS($q$) &  Symm.& States& Conf. &\multicolumn{2}{c}{Irreps($\rho$)}& aBRs($\rho@q$)& Occ.\\ 
\hline
 Sb &$ 2d$& $  3m$& Sb$^{3+}$ & $5s^25p^3$   & $ s  $ &: $ A_{1} $ &   $A_{1}@2d $ & Yes\\
 \cline{6-9}
    &     &       &           &              & $ p_z    $ &: $ A_1$ &   $A_1@2d $ & \\
    &     &       &           &              & $ p_x,p_y$ &: $ E  $ &   $E@2d $ & \\
 \hline
\multicolumn{7}{r}{$~$} & {\color{blue} $A_g@3e$} & Yes \\
\bottomrule[0.5pt]
\end{tabular}
\end{table}

\subsection{Unconventional metals}
Then, we can generalize the concept of unconventional insulators into unconventional `neat' metals. 
A `neat' metallic compound is supposed to have an overall band gap shadowed about/above $E_F$ in Fig.~\ref{fig:VGaFe2}. We classify them into three classes. In class I, \eg VGaFe$_2$ in Fig.~\ref{fig:VGaFe2}(a), there is a direct energy gap in the band structure, while there is no global gap in the entire Brillouin zone. In class II, \eg Ca$_2$N in Fig.~\ref{fig:VGaFe2}(b), it is metallic due to the odd total number of electrons. However there would be a gap with one more electron. In class III, \eg Y$_2$C in Fig.~\ref{fig:VGaFe2}(c), there is an overall band gap except some $k$-point; namely, there can be a band inversion.

Unconventional metals are well defined as long as a set of `occupied' bands are defined in the metallic compounds. 
The occupied states are defined by simply counting energy bands at HSKPs in class I and class II. For complicated metals of class III, one has to work a little bit to get the set of `occupied' bands at HSKPs by solving CRs. We need to eliminate the band inversion via switching irreps by hand. The results of unconventional metals contain many functional materials, such as electrides, solid-state hydrogen storage materials, HER electrocatalysts, \emph{etc}.

\subsubsection{Electrides}
An electride is usually defined an ionic crystal with excess electrons confined in particular vacancies, which is an excellent example of unconventional metals~\cite{Nie2021PRB,nie2021sixfold}. To achieve an electride, it is empirically known that three criteria should be satisfied: excess electrons, lattice vacancies, and suitable electronegativity of the elements. By the definition, the electrides are consistent with unconventional materials. They have an essential BR, which is not induced by any atomic orbitals in the crystal, but formed by the electrons at the vacancies. Therefore, there are many electride candidates picked out by our high-through screening for unconventional materials, like Ca$_2$N and Ca$_2$As. Additionally, as we did in Refs.~\cite{Nie2021PRB,nie2021sixfold}, the electrides (\ie Y$_2$C, LaCl, Li$_{12}$Mg$_3$Si$_4$ and C12A7) with relatively complicated band structures are also diagnosed by the aBR decomposition. The calculated band structures and partial electron density (PED) are shown in Fig.~\ref{fig:ca2n}.

\begin{figure}[!t]
    \includegraphics[scale=0.1]{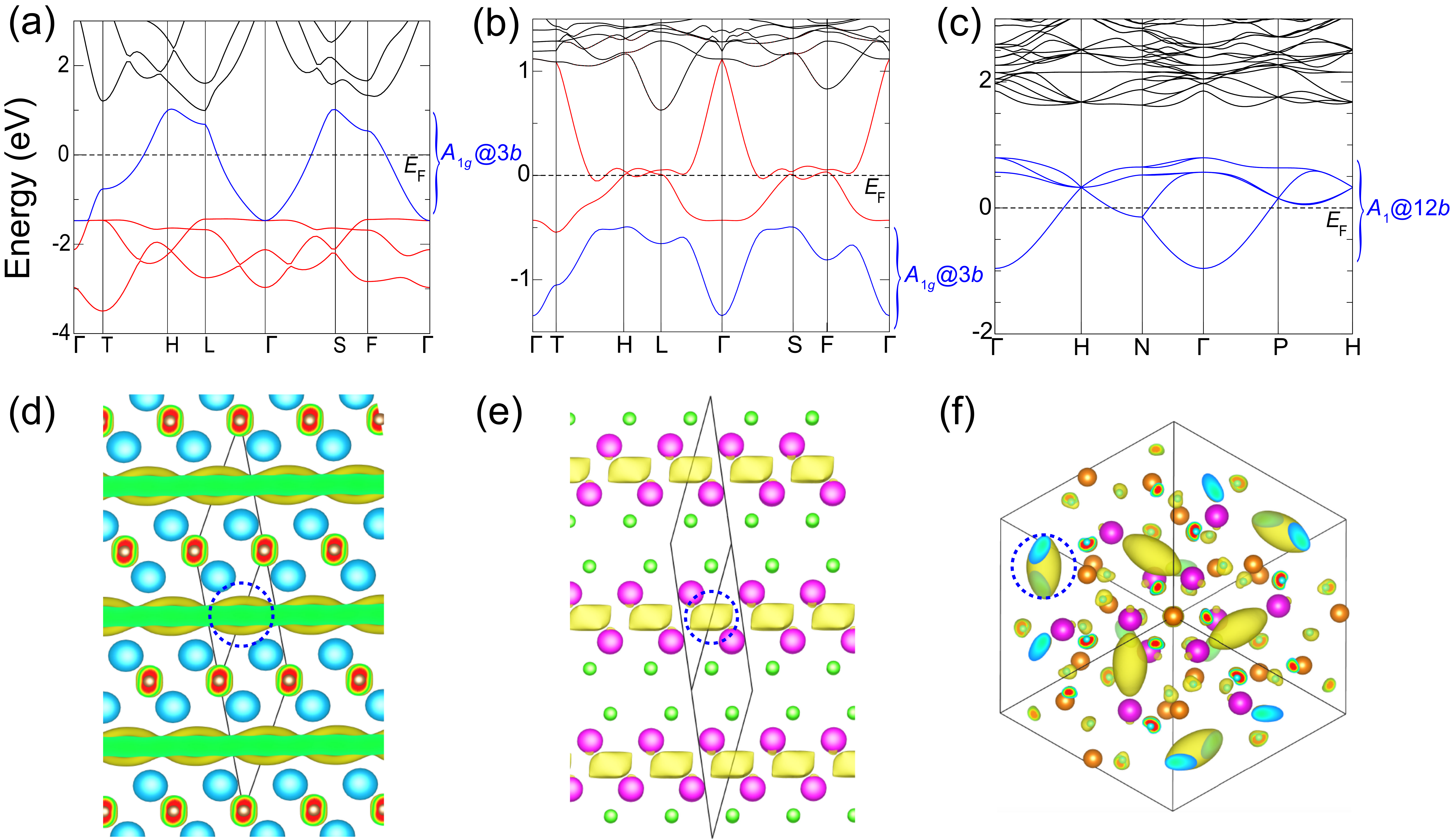}
    \caption{(Color online) (a-c) Band structures of the electrides: Ca$_2$N, LaCl, and C12A7, respectively. (d-f) The calculated PED of the blue-colored bands for Ca$_2$N, LaCl, and C12A7, respectively, with the dashed blue circles marked the essential site (adapted from Ref.~\cite{Nie2021PRB}).
}
    \label{fig:ca2n}
\end{figure}

The previous search for electrides is done mainly by analyzing charge density around E$_F$, where electron localization function (ELF) analysis has proved to be effective. 
However, the symmetry analysis is lacking. By analyzing the symmetry eigenvalues (or irreps) alone at several HSKPs in first-principles calculations, the BR analysis of TQC theory leads to the clear understanding of three characteristics of electrides as discussed in Ref.~\cite{Nie2021PRB}. First, the floating bands are induced from the BRs of vacancies, indicating that their average charge densities are located at the vacancies in real space. Second, due to the loose confinement, the floating bands are usually close to the E$_F$, which is very likely to induce the band inversion and nontrivial band topology. Third, the excess anionic electrons in vacancies present a strong hydrogen affinity. The absorption of hydrogen usually moves those floating bands far below E$_F$ and lowers the total energy (stabilizing the lattice). A significant amount of hydrogen is found experimentally in the crystals of Lanthanum monochloride~\cite{araujo1981lanthanum} and Ca$_5$P$_3$ crystals~\cite{xie2015new}. 
Note that an unconventional material is necessary but not sufficient for an electride.
For safety, one may need to compute the charge distribution to confirm electride nature in the selected unconventional materials.

\begin{figure}[!t]
    \includegraphics[scale=0.18]{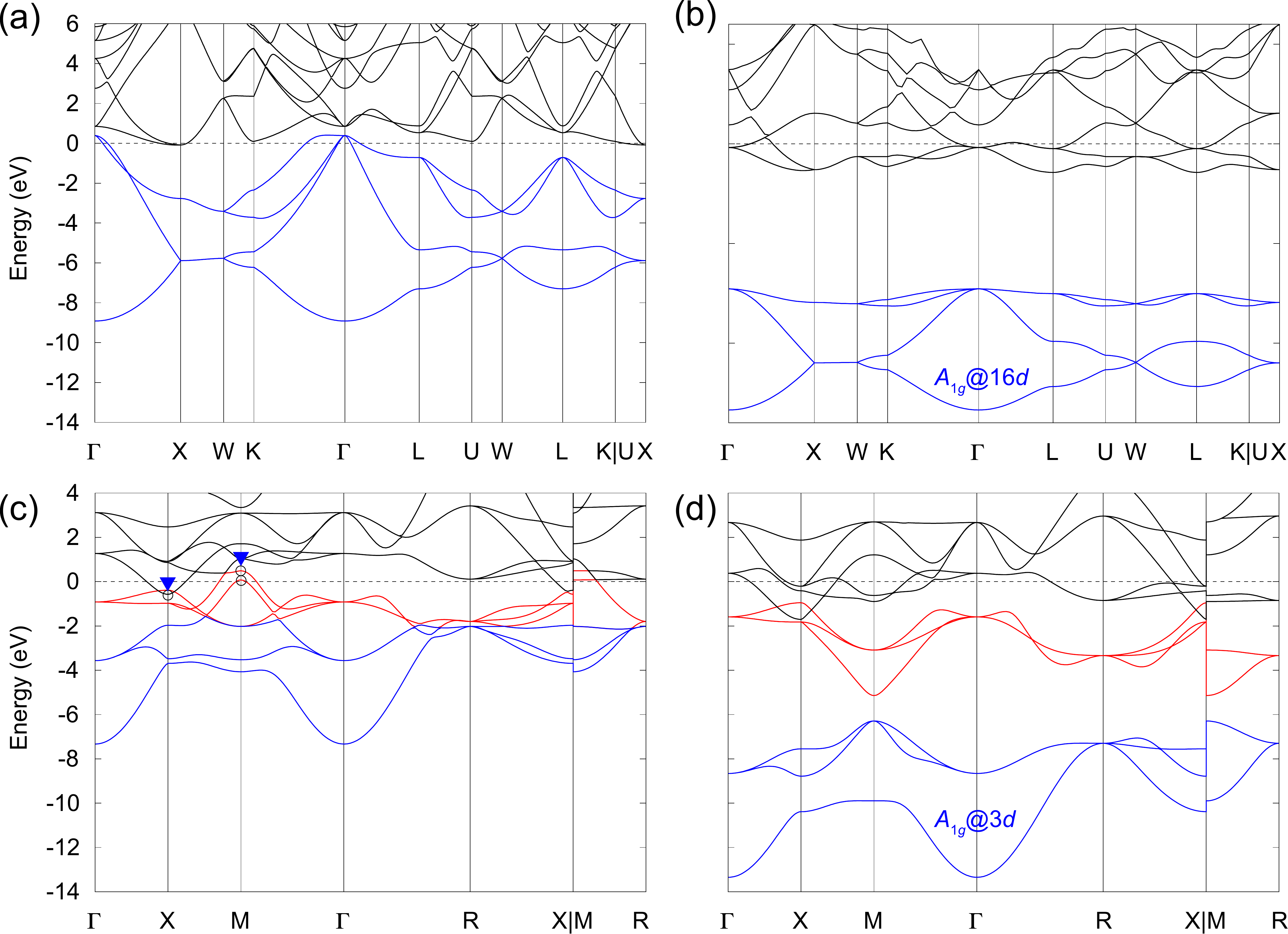}
    \caption{(Color online) Band structures of (a)LiAl and (c)TiFe. Panels (b) and (d) present the band structures after absorbing hydrogen atoms at the empty sites.}
    \label{fig:TiFe}
\end{figure}

\subsubsection{Solid-state hydrogen storage materials}
We found some unconventional metal alloys, LiAl/LiB ~\cite{kang2004hydrogen,choi2011reaction}, which are well-known solid-state hydrogen storage materials. When the hydrogen molecule comes in contact with the surface of solid-state hydrogen storage materials, it dissociates into two hydrogen atoms which diffuse in the solid and form a chemical bond with the solid material (\ie metal hydrides). The crystals of LiAl have the structure of $Fd\bar{3}m$ (No. 227). The eight valence electrons (two formulas per unit cell) form the bands of the BR $A_{1g}@16d$ (which is empty). 
Hydrogen forms metal hydrides with some metals and alloys, leading to solid state storage under moderate temperature and pressure that gives them important safety advantages over the gaseous and liquid storage methods. 
Hence, metal hydride storage is a safe, volume-efficient storage method for on-board vehicle applications.

Then, we also check the unconventional metal TiFe of class III by hand~\cite{sujan2020overview,haliciouglu2012experimental}. In its band structure of Fig.~\ref{fig:TiFe}(c), there is an overall band gap except two band inverted HSKPs, \ie X and M. After removing the band inversions (exchanging the irreps denoted by triangles and circles), the `occupied' bands are solved to be $A_{1g}@1b$ + $T_{2g}@1a$ + {\color{blue} $A_{1g}@3d$}. 
It can be seen clearly that the bands of the essential BR appears around E$_F$. By absorbing the hydrogen atoms on $3d$ site, the energy bands of the essential BR decrease dramatically as shown in Fig.~\ref{fig:TiFe}(d).

\begin{figure}[!b]
    \includegraphics[scale=0.13]{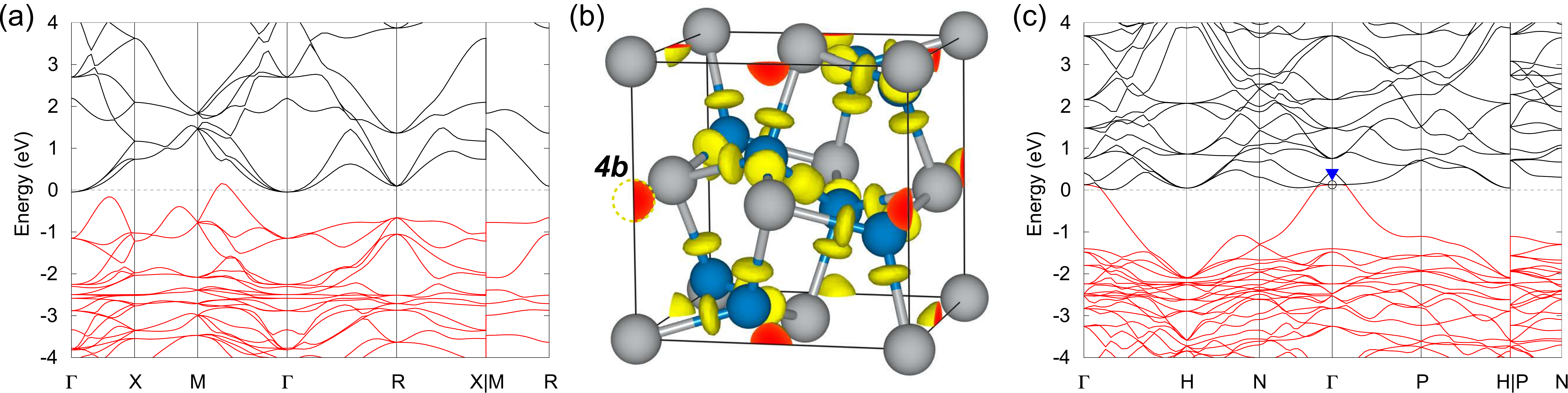}
    \caption{(Color online) Band structures of (a)NiP$_2$ and (c)CoP$_3$. (b) The ELF plot of NiP$_2$.}
    \label{fig:NiP2}
\end{figure}

\begin{figure*}[htbp]
	\includegraphics[width=6.8in]{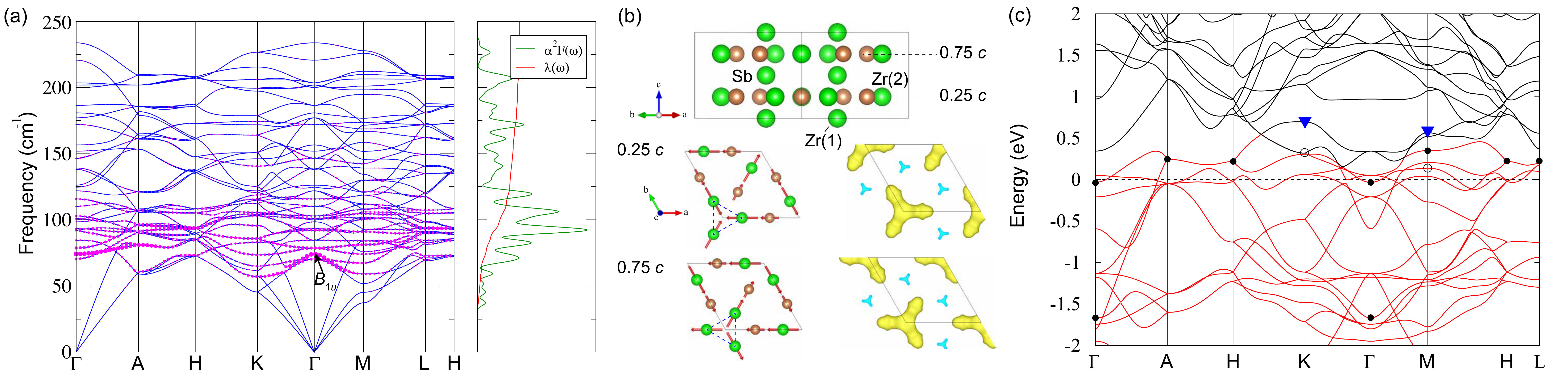}
	\caption{(Color online)
		(a) Phonon spectrum, Eliashberg spectral functions $\alpha^{2}F(\omega)$, and the frequency-dependent coupling $\lambda(\omega)$. The electron-phonon couplings $\lambda_{\textbf{q}v}$ are represented by magenta circles. (b) Side view of the crystal structure of Zr$_{5}$Sb$_{3}$. The planes of $z$ = 0.25 $c$ and $z$ = 0.75 $c$ are marked by dash lines. The phonon vibration mode of B$_{1u}$ is indicated by the arrows, \ie compression of the triangle in $z$ = 0.25 $c$ plane and expansion of the triangle in $z$ = 0.75 $c$plane. The calculated PED of bands of the energy range 0.0 eV to 0.3 eV is shown and coincides with the B$_{1u}$ phonon mode. (c) The band structure of Zr$_{5}$Sb$_{3}$.The BR decomposition is solved for the lower 38 energy bands in red. The essential BR is $A_1'@2a$.
	}\label{fig:Zr5Sb3}
\end{figure*}

\subsubsection{Electrocatalysts and electrodes}
The orthogonal NiP and cubic NiP$_2$ compounds are found to be unconventional metals. The essential BR is solved to be {\color{blue}$A_{g}@4a$} for NiP and {\color{blue}$A_{1g}@4b$} for NiP$_2$. The band structure of cubic NiP$_2$ is presented in Fig.~\ref{fig:NiP2}. To show that the electronic centers are not located at the atoms, the ELF are plotted for NiP$_2$ in Fig.~\ref{fig:NiP2}(b). 
On the other hand, the HER, which generates molecular hydrogen through the electrochemical reduction of water, underpins many clean-energy technologies~\cite{2001hydrogen,wang2021first}. 
The pyrite structure-type transition metal dichalcogenides ($MX_2$, where $M=$ Fe, Co, or Ni and $X=$ S or Se) have emerged as an interesting family of low-cost materials with high catalyticactivity toward the HER~\cite{owens2019nip2}.
We believe that these maximum electron distributions offset from the atomic positions benefit to catalyze electrochemical reaction, such as HERs and negative electrodes for Li-ion batteries~\cite{gillot2005electrochemical,hayashi2009electrochemical}. 

Additionally, the aBR decomposition for the skutterudite-type CoP$_3$ compound suggests that it is unconventional with band inversion. From the plotted band structure in Fig.~\ref{fig:NiP2}(c), one can find that the band inversion happens between low-energy states at $\Gamma$, denoted by a circle and a triangle, respectively. The electrochemical reaction of lithium with the CoP$_3$ compound has been studied in Ref.~\cite{alcantara2002electrochemical}. Co$_{1-x}$Ni$_x$P$_3$ exhibits much better electronic properties for obtaining high energy density supercapacitors~\cite{jiang2020novel} and NiP$_3$ is proved to be a promising negative electrode for Li- and Na-ion batteries~\cite{fullenwarth2014nip}.

\subsubsection{Superconductors}
The compound Zr$_5$Sb$_3$ is experimentally found to be the first superconductor in the large family of compounds with Mn$_5$Si$_3$-type structure (No. 193)~\cite{Zr5Sb32013}, which is believed to be superconducting due to the electron-phonon coupling. 
The superconducting transition temperature ($T_{c}$) is estimated using Allen-Dynes modified McMillian equation~\cite{McMillan1968,Allen1975},
\begin{equation}\label{eq:Tc}
	\begin{split}
		T_{c}=\frac{\omega_{log}}{1.2k_{B}}\exp[\frac{-1.04(1+\lambda)}{\lambda(1-0.62\mu^{*})-\mu^{*}}]
	\end{split}
\end{equation}
where $k_{B}$ is the Boltzmann constant, $\mu^{*}$ is the effective screened Coulomb repulsion constant, typically $\sim$ 0.1, $\lambda=\Sigma_{\textbf{q}v}\lambda_{\textbf{q}v}=2\int_{0}^{\infty}d\omega\frac{\alpha ^{2}F(\omega)}{\omega}$ is electron-phonon coupling constant, and $\omega_{log}$ is logarithmic average phonon frequency. 
With $\mu^{*}$ = 0.10 and $\lambda$ = 0.53, T$_{c}$ of Zr$_{5}$Sb$_{3}$ is estimated to be 2.2 K, which is consistent to the experimental value ($\sim$ 2.3 K)~\cite{Zr5Sb32013}. The phonon spectrum of Zr$_{5}$Sb$_{3}$ is shown in Fig.~\ref{fig:Zr5Sb3}(a). The contributions of $\lambda$ mainly come from the phonon modes of 50 cm$^{-1}<\omega<$ 150 cm$^{-1}$. Among these phonon modes, we find that the low-frequency B$_{1u}$ phonon mode at $\Gamma$ has higher $\lambda_{\textbf{q}v}$ than others. As shown in Fig.~\ref{fig:Zr5Sb3}(b), the B$_{1u}$ phonon mode is an in-plane vibration mode of Zr(2) atoms, which form triangles in $z = 0.25c$ and $z = 0.75c$ planes.

\begin{table}[!b]
\tiny
\caption{
 Atomic positions, valence states and aBRs of 38 energy bands in $P6_3/mcm$ Zr$_5$Sb$_3$.
}\label{table:Zr5Sb3_orb}
\begin{tabular}{ccccccccc}
\toprule[0.5pt]
 Atom &WKS($q$) &  Symm.& States& Conf. &\multicolumn{2}{c}{Irreps($\rho$)}& aBRs($\rho@q$)& Occ.\\ 
\hline
Zr(1) & $4d$   & 32        & Zr$^{2-}$& $5s^1 4d^3$ &$s $          &:$A_1$  &$A_1@4d$ & Yes   \\
 \cline{6-9}
  &         &           &  &      &$d_{z^2} $        &:$A_1  $  &$A_1@4d$ &  \\
  &         &           &  &      &$d_{xz},d_{yz}$   &:$E$ & $E@4d$ & Yes \\
    &         &           &  &    &$d_{x^2},d_{xy}$  &:$E$ & $E@4d$ &  \\
 \hline
Zr(2) & $6g$   & mm2       & Zr$^{4+}$  & $5s^1 4d^3$ &$s $          &:$A_1$  &$A_1@4d$ & \\
 \cline{6-9}
  &         &           &  &      &$d_{z^2} $        &:$A_1$ &$A_1@6g$ &  \\
  &         &           &  &      &$d_{x^2} $        &:$A_1$ &$A_1@6g$ &  \\
  &         &           &  &      &$d_{xy}  $        &:$A_2$ &$A_2@6g$ &  \\
  &         &           &  &      &$d_{yz}  $        &:$B_2$ &$B_2@6g$ &  \\
  &         &           &  &      &$d_{xz}  $        &:$B_1$ &$B_1@6g$ &  \\
 \hline
Sb & $6g $   &mm2        & Sb$^{3-}$& $5s^2 5p^3$ &$s $  &:$A_1$  &$A_1@6g$ & Yes \\
 \cline{6-9}
  &         &           &  &      &$p_{z}$      &:$A_1$  &$A_1@6g$ & Yes  \\
  &         &           &  &      &$p_{y}$      &:$B_2$  &$B_2@6g$ & Yes  \\
  &         &           &  &      &$p_{x}$      &:$B_1$  &$B_1@6g$ & Yes  \\
 \hline
 \multicolumn{7}{r}{$~$} & {\color{blue} $A_1'@2a$} & Yes \\
\bottomrule[0.5pt]
\end{tabular}
\end{table}

In our calculations, the aBR decomposition of Zr$_5$Sb$_3$ superconductor suggests that it is a complicated unconventional metallic material.
Its band structure is presented in Fig.~\ref{fig:Zr5Sb3}(c). One can find that there is an overall band gap between the red-colored bands (corresponding to 76 electrons) and higher energy bands, while two band inversions happen between the bands denoted by circles and triangles at K and M points, respectively. The aBR decomposition is solved to be $A_1@4d+E@4d+2A_1@6g+B_1@6g+B_2@6g$+{\color{blue}${A'}_1@2a$} (Table \ref{table:Zr5Sb3_orb}). The center of the essential BR is $2a$ site, being the center of the Zr(2) triangle. The energy bands of the essential BR are denoted by black dots, which are located energetically near E$_F$. The PED of the bands near E$_F$ is shown in Fig.~\ref{fig:Zr5Sb3}(b), which is consistent with the essential BR.
Surprisingly, the electronic distribution of the low-energy states (\ie the essential BR) coincides with the B$_{1u}$ phonon mode. It gives rise to strong interaction (higher $\lambda_{\textbf{q}v}$), which contributes mainly to the superconductivity in Zr$_{5}$Sb$_{3}$. 

Additionally, the authors of Ref.~\cite{gao2020electronic} introduce a BR sitting on an empty $1b$ site (an essential BR) in the minimum tight-binding model in the parent compound NdNiO$_2$ of Ni-based superconductors, prepared by removing an apical O atom in NdNiO$_3$. The $1b$ site was sit by the apical O atom in the precursor. On the other hand, using DFT calculations, the stoichiometric NdNiO$_2$ is found to be significantly unstable. Instead, they argue that the H impurity can be expected to be present in NdNiO$_2$ in Ref.~\cite{malyi2021bulk}. 
A detailed aBR decomposition for energy bands in NdNiO$_2$ is given in SM. It clearly shows that the essential BR of $A_{1g}@1b$ is around E$_F$ and could be shifted downwards significantly by absorbing H atom at $1b$ site. Like the elctrides~\cite{araujo1981lanthanum,xie2015new}, some H impurities should benefit to stabilize the crystal structure, and it would be hard to remove them completely during the preparation.

\section{Discussion}
Due to the limitation of length of this paper, we only discuss on some examples in the main text. The list unconventional materials are tabulated in the SM. The detailed informations, such as ICSD number, chemical formula, number of atoms, number of electrons, direct band gap, indirect gap, space group number,
atomic Wyckoff sites, and the essential BR, are presented in the table. We believe that there are more interesting functional materials among them.

In conclusion, we demonstrate that the analysis of irreps and BRs in TQC theory provides an effective way to identify the origin of the energy bands from their
symmetry eigenvalues (or irreps) alone. 
The aBR decomposition is efficient to diagnose the mismatch between the average electronic charge centers and the atomic positions in unconventional materials. The essential BR of an empty symmetry site can be described by the nonzero RSIs. It's worth noting that not all unconventional materials can be found be this symmetry-based method (aBR decomposition), especially for those with electronic charge centered on the non-maximal Wyckoff positions. Like topological insulators with zero symmetry-based indicators, there are still some unconventional materials with zero RSIs (a sum of aBRs), which need to explore in future. The presence of the electronic distribution away from nuclei gives rise to a diversity of interesting properties and applications, such as thermoelectronic materials, higher-order TIs, electrides, hydrogen storage materials, HER electrocatalysts, \emph{etc}. These interesting properties in unconventional materials would draw broad interest in the future.

\noindent \textbf{Note added}
During the refereeing stages of this work, we found some works with similar topic appear~\cite{xu2021fillingenforced,xu2021threedimensional,li2021obstructed}.

\noindent \textbf{Author contributions}
Zhijun Wang, Hongming Weng and Miao Liu proposed and supervised the project. Jiacheng Gao, Yuting Qian and Huaxian Jia carried out the high-throughput calculations. All authors contributed to writing of the manuscript.

\ \\
\noindent \textbf{Acknowledgments}
This work was supported by the National Natural Science Foundation of China (Grants No. 11974395, No. 12188101), the Strategic Priority Research Program of Chinese Academy of Sciences (Grant No. XDB33000000), and the Center for Materials Genome. H.W. acknowledges support from the Ministry of Science and Technology of China under grant numbers 2016YFA0300600 and 2018YFA0305700, the Chinese Academy of Sciences under grant number XDB28000000, the Science Challenge Project (No. TZ2016004), the K. C. Wong Education Foundation (GJTD-2018-01), Beijing Municipal Science \& Technology Commission (Z181100004218001) and Beijing Natural Science Foundation (Z180008).

\bibliography{main}

} 

\clearpage
\begin{widetext}
\beginsupplement{}
\setcounter{section}{0}
\renewcommand{\thesubsection}{\arabic{subsection}}
\renewcommand{\thesubsubsection}{\alph{subsubsection}}

\section*{SUPPLEMENTARY MATERIALS}

\subsection*{A. The programs used for solving the aBR decomposition} 
\label{programs}
To decompose occupied bands into aBRs, we first need to get a full list of aBRs in a material. For example in ZnO, O sits on the $4d$ Wyckoff site and its electron configuration is $2s^22p^4$. Its $s$ orbital  induces the $A_1@4d$ aBR, while the $p_x, p_y, p_z$ orbitals on $4d$ induce the $T_2@4d$ aBR. On the other hand, the s/d orbitals of Zn atoms ($4a$) induce the $A_1@4a/E@4a,T_2@4a$ aBR. The full list of aBRs in ZnO contains only these five aBRs. We develop the code \webposbr\ to extract the aBRs of a material. The code is released on the \href{https://github.com/zjwang11/UnconvMat/blob/master/src_pos2aBR.tar.gz}{https://github.com/zjwang11/UnconvMat/blob/master/src\_pos2aBR.tar.gz}. One can download and compile the code by the following command.
\lstset{language=bash, keywordstyle=\color{blue!70}, basicstyle=\ttfamily, frame=shadowbox}
\begin{lstlisting}
$ ./configure.sh
$ source ~/.bashrc
$ make
\end{lstlisting}

The program \webposbr\ converts \texttt{PPOSCAR} to \texttt{POSCAR\_std}. The file \texttt{PPOSCAR} is the primitive cell generated by \href{https://phonopy.github.io/phonopy/}{Phonopy}. The standardized structure \texttt{POSCAR\_std} is for the first-principles calculations in VASP, in order to be compatible with the programs \webirvsp,
\href{http://tm.iphy.ac.cn/UnconvMat.html}{\ttfamily solve\_BR},
and \href{http://tm.iphy.ac.cn/UnconvMat.html}{\ttfamily solve\_CR} in the framework of TQC.
To standardize a crystal structure and generate a full list of its aBRs,
one can either paste \texttt{PPOSCAR} onto the website \href{http://tm.iphy.ac.cn/UnconvMat.html}{\ttfamily http://tm.iphy.ac.cn/UnconvMat.html} (The outputs of the website are pasted in Fig.~\ref{fig:webfigure}), or run \webposbr\ locally by the following command.
\lstset{language=bash, keywordstyle=\color{blue!70}, basicstyle=\ttfamily, frame=shadowbox}
\begin{lstlisting}
$ phonopy --symmetry --tolerance 0.01 -c POSCAR
$ pos2aBR 
\end{lstlisting}

\indent After finishing the calculations in VASP, one can get the irreps of occupied states via \webirvsp. In ZnO, the space group number is 216 and there are 18 electrons in one primitive cell (depending on the pseudopotential files). Thus the command is,
\lstset{language=bash, keywordstyle=\color{blue!70}, basicstyle=\ttfamily, frame=shadowbox}
\begin{lstlisting}
$ irvsp -sg 216 -nb 1 18
\end{lstlisting}
\indent Meanwhile, the files \texttt{tqc.txt} and \texttt{tqc.data} (Fig.~\ref{fig:webfigure}) are generated for BR decomposition in the TQC. 
By pasting \texttt{tqc.data} onto the website \href{http://tm.iphy.ac.cn/UnconvMat.html}{\ttfamily UnconvMat.html}, one can solve the aBR/eBR decomposition and CR conditions conveniently.
\begin{figure}[hb]
    \centering 
    \includegraphics[width=500pt]{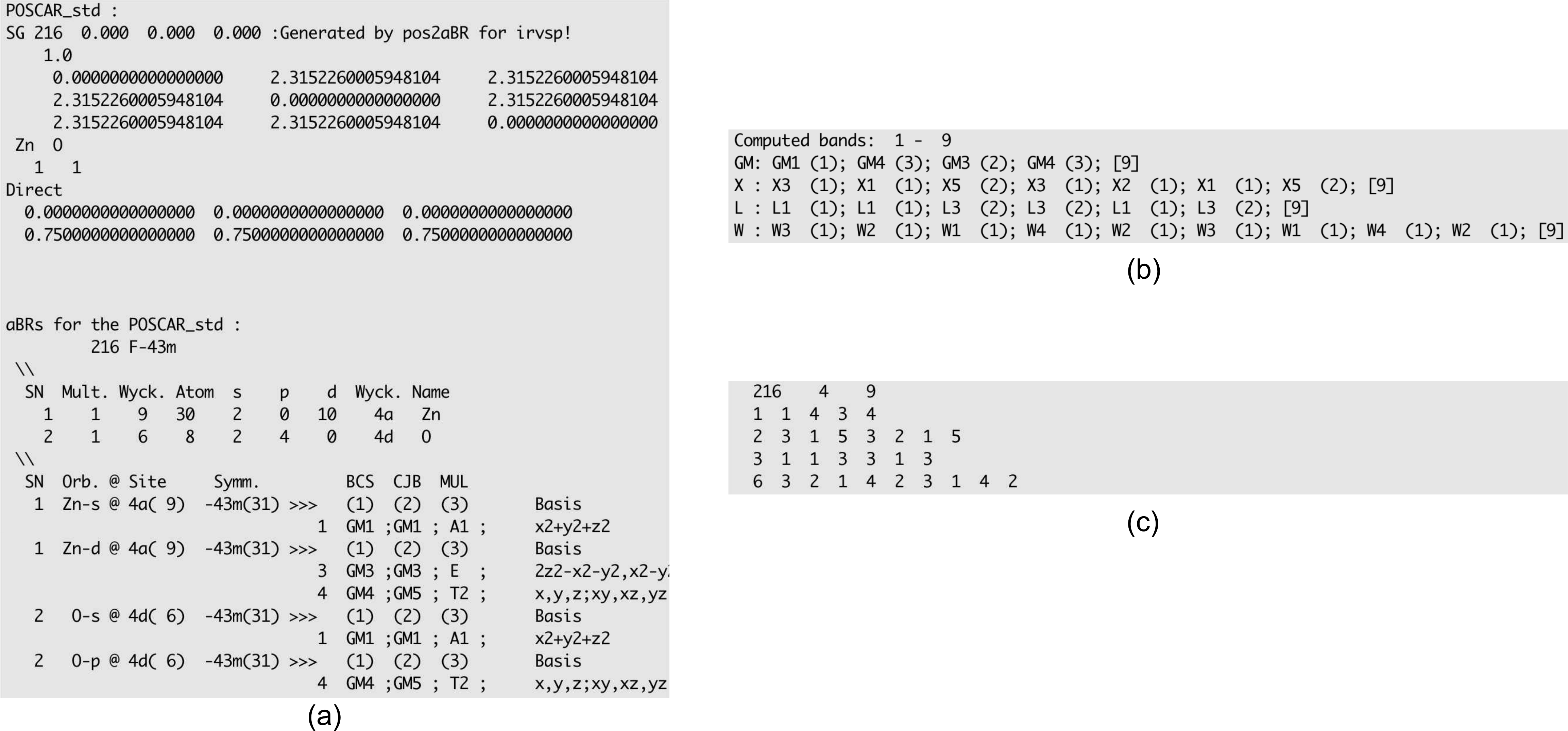}
    \caption{(a) The output of \webposbr\ of ZnO containing the standardized \texttt{POSCAR} and the list of aBRs (screenshot of the web).  \texttt{tqc.txt}(b) and \texttt{tqc.data}(c) files of ZnO, generated by \webirvsp.}
    \label{fig:webfigure}
\end{figure}

\clearpage
\subsection*{B. Band representations of ZnO and Ca$_2$As}
The irreps for occupied bands of ZnO and Ca$_2$As on high-symmetry $k$-points are listed in Table~\ref{table:irrep}.
\begin{table*}[!htb]
\tiny 
\caption{
Irreps are computed for the occupied bands in conventional ZnO and unconventional Ca$_2$As compounds (\ie \texttt{tqc.txt} generated by the program \webirvsp). The corresponding BRs in the TQC are presented as well.
}\label{table:irrep}
\begin{tabular}{p{1cm}|p{1.2cm}|p{1cm}|p{1cm}|p{1cm}p{0.4cm}r|p{1.5cm}|p{1.5cm}|p{1cm}|p{1.2cm}|p{1.2cm}}
\hline
\hline
 \#216& $\Gamma$(GM) & X & L &W  && \#139 & $\Gamma$(GM) & M&P&X&N\\
\cline{1-5}\cline{7-12}
ZnO   & GM4(3) GM3(2) GM4(3) & X1(1) X5(2) X3(1) X2(1) X1(1) X5(2)    & L1(1)  L3(2) L3(2) L1(1) L3(2)     & W2(1)  W1(1)  W4(1) W2(1) W3(1) W1(1) W4(1) W2(1) && Ca$_2$As & GM1+(1) GM5+(2) GM1+(1) GM3$-$(1) GM5$-$(2)& M1+(1) M3--(1) M5+(2) M5--(2) M1+(1)&P5(2) P5(2) P1(1) P3(1) P3(1)&X4--(1) X3$-(1)$ X4+(1) X1+(1) X3+(1) X2--(1) X1+(1)&N2--(1) N1+(1) N1+(1) N2+(1) N2--(1) N1--(1) N2--(1) \\
\cline{1-5}\cline{7-12}
\cline{1-5}\cline{7-12}
\multicolumn{5}{l}{BRs}  \\
\cline{1-5}\cline{7-12}
$E@4a$ & GM3 & X1  X2 & L3 & W1  W2  && $A_1@4e$&GM3-- GM1+&M3-- M1+&P3 P1&X2-- X1+ &N2-- N1+\\
\cline{1-5}\cline{7-12}
$T_2@4a$ & GM4 & X3  X5 & L1  L3 & W2 W3 W4 && $E@4e$&GM5-- GM5+ &M5-- M5+ &P5 P5&X3-- X3+ X4-- X4+&N1-- N1+ N2-- N2+\\
\cline{1-5}\cline{7-12}
$T_2@4d$ & GM4 & X1 X5 & L1 L3 & W1 W2 W4 &&  & & & & & \\
\cline{1-5}\cline{7-12}
 &  &  &  &  && {\color{blue}$A_{1g}@2b$}& GM1+ & M1+ &P3&X1+&N2-\\
\hline
\hline
\end{tabular}
\end{table*}

\subsection*{C. The aBR decomposition of NdNiO$_2$} 
\label{nibased}
Around the E$_F$ in NdNiO$_2$, there are 17 energy bands (blue colored) in the energy range (from -10 eV to 7 eV). By solving the CR conditions, we find that there are are two band inversions between them and other higher energy bands, at M and Z points (Fig.~\ref{fig:NdNiO2}). After removing these band inversions by hand, the aBR decomposition for these 17 bands are solved to be $\{B_{1u},B_{2u},B_{3u}\}@2f$ (O-$p$) + 
$\{A_{1g},B_{1g},B_{2g},E_{g}\}@1a$ (Ni-$d$) +
$\{A_{1g},B_{1g},B_{2g},E_{g}\}@1d$ (Nd-$d$) +{\color{blue} $A_{1g}@1b$}, as listed in Table~\ref{table:NdNiO2}.

\begin{figure*}[htbp]
	\includegraphics[width=5.5in]{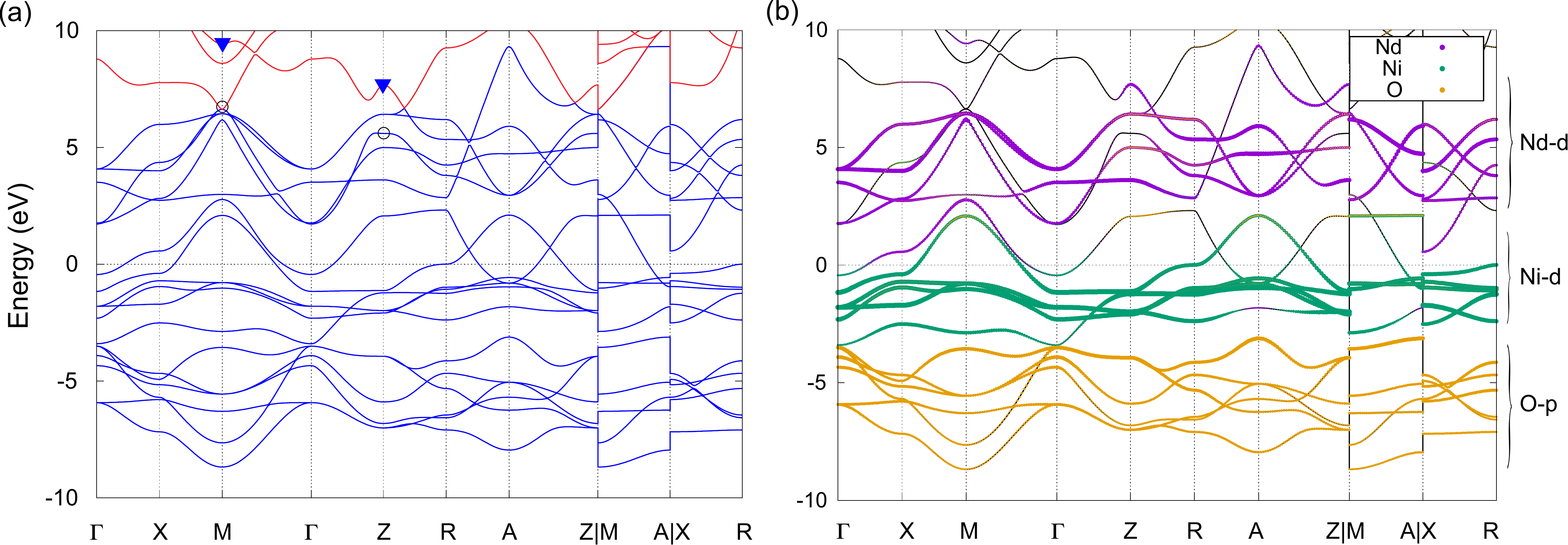}
	\caption{(color online).
		The band structure of NdNiO$_2$ without considering the 4$f$ electrons of Nd. 
	}\label{fig:NdNiO2}
\end{figure*}

\begin{table}[!h]
\caption{
Atomic positions and valence states of $P4/mmm$ NdNiO$_2$. 
}\label{table:NdNiO2}
\begin{tabular}{c|c|c|cc|c}
\hline
\hline
 Atom &WKS($q$) &  Symm.&\multicolumn{2}{c|}{Irreps($\rho$)}& aBRs($\rho@q$)\\
 \hline
 Nd & $1d $&$4/mmm$&  $ d_{z^2} $  &$: A_{1g} $ &$ A_{1g}@ 1d$\\
    &      &       &  $ d_{x^2-y^2} $  &$: B_{1g} $ &$ B_{1g}@ 1d$\\
    &      &       &  $ d_{xy} $  &$: B_{2g} $ &$ B_{2g}@ 1d$\\
    &      &       &  $ d_{xz,yz} $  &$: E_{g} $ &$ E_{g}@ 1d$\\
 \hline
 Ni & $1a $&$4/mmm$&  $ d_{z^2} $  &$: A_{1g} $ &$ A_{1g}@ 1a$\\
    &      &       &  $ d_{x^2-y^2} $  &$: B_{1g} $ &$ B_{1g}@ 1a$\\
    &      &       &  $ d_{xy} $  &$: B_{2g} $ &$ B_{2g}@ 1a$\\
    &      &       &  $ d_{xz,yz} $  &$: E_{g} $ &$ E_{g}@ 1a$\\
 \hline
 O & $2f $&$mmm$   &  $ p_z $  &$: B_{1u} $ &$ B_{1u}@ 2f$\\
    &     &        &  $ p_y $  &$: B_{2u} $ &$ B_{2u}@ 2f$\\
    &     &        &  $ p_x $  &$: B_{3u} $ &$ B_{3u}@ 2f$\\
 \hline
\hline
\end{tabular}
\end{table}

\clearpage
\subsection*{D. Table of searching results} 
\label{tab:electride}
The results of the automatic high-throughput screening are listed in Table~3: 1) We remove the compounds without any metal elements to exclude covalent materials; 2) We add one more electron for the metals with odd number of valence electrons; 3) We remove large-gap compounds ($E_{ind} > 0.8$eV) for simplicity.
Only unconventional `neat' metals of class I and class II can be captured in the process.
The first and second columns are the ICSD number and the chemical formula. The `NoA' represents the number of atoms in a primitive cell. The `NoE' represents the number of valence electrons (determined by pseudopotential files). The `E$_g$' and `E$_{ind}$' represent the direct band gap and indirect band gap, respectively. The `SG' represents the space group number. In the column of `Wyckoff positions', we listed the atomic Wyckoff positions of a crystal. For materials with more than one same type non-metal atoms in the primitive cell, we listed the shortest length of these bonds under `Length of bonds'. The Wyckoff positions of the center of the shortest bonds are listed under `Center'. The band structures are presented in Fig.~S3-S55.

In addition, for the compounds including nonmetals ($X$) in the table of unconventional materials, we also provide the length of the nearest $X-X$ bonds (`Length of bonds') and the corresponding Wyckoff sites of bond centers (`Center'), which can be used to further check whether the presence of the essential BR is due to the covalent bonds of the $X$ dimerization.
For materials with only one essential BR, they are listed in the last column.
\begin{longtable}[!h]{|c|c|c|c|c|c|c|p{2.5cm}<{\centering}|p{1.5cm}|p{1cm}|c|}

\caption{The results of the automatic high-throughput screening for unconventional materials. Click ICSD numbers and formulas to obtain irreps (\texttt{tqc.txt} files) and electronic band structures respectively.}\label{tab:search}
\\ \hline
 ICSD & Formula & NoA & NoE & E$_{g}$(eV) & E$_{ind}$(eV) & SG & Wyckoff positions & Length of bonds & Center & The essential BR\\
\hline
\hyperref[sec:tqc167876]{167876} & \hyperref[subfig:167876]{TaN} & 2 & 10 &  0.000 & -0.977 & 216 & 4c,4a & &&A1@4d\\
\hline
\hyperref[sec:tqc167870]{167870} & \hyperref[subfig:167870]{TcN} & 2 & 12 &  0.243 & -0.333 & 216 & 4c,4a & &&A1@4d\\
\hline
\hyperref[sec:tqc187182]{187182} & \hyperref[subfig:187182]{MoN} & 2 & 11 & $\varnothing$ & $\varnothing$ & 216 & 4a,4d & &&A1@4c\\
\hline
\hyperref[sec:tqc167875]{167875} & \hyperref[subfig:167875]{HfN} & 2 & 9 & $\varnothing$ & $\varnothing$ & 216 & 4c,4a & &&A1@4d\\
\hline
\hyperref[sec:tqc186876]{186876} & \hyperref[subfig:186876]{MoP} & 2 & 11 & $\varnothing$ & $\varnothing$ & 216 & 4c,4a & &&A1@4d\\
\hline
\hyperref[sec:tqc186243]{186243} & \hyperref[subfig:186243]{IrN} & 2 & 14 &  0.000 & -2.506 & 216 & 4c,4a & &&A1@4d\\
\hline
\end{longtable}
\begin{figure}[htp]
 \centering
\subfigure[TaN SG216 NoA=2 NoE=10]{
\label{subfig:167876}
\includegraphics[scale=0.32,angle=270]{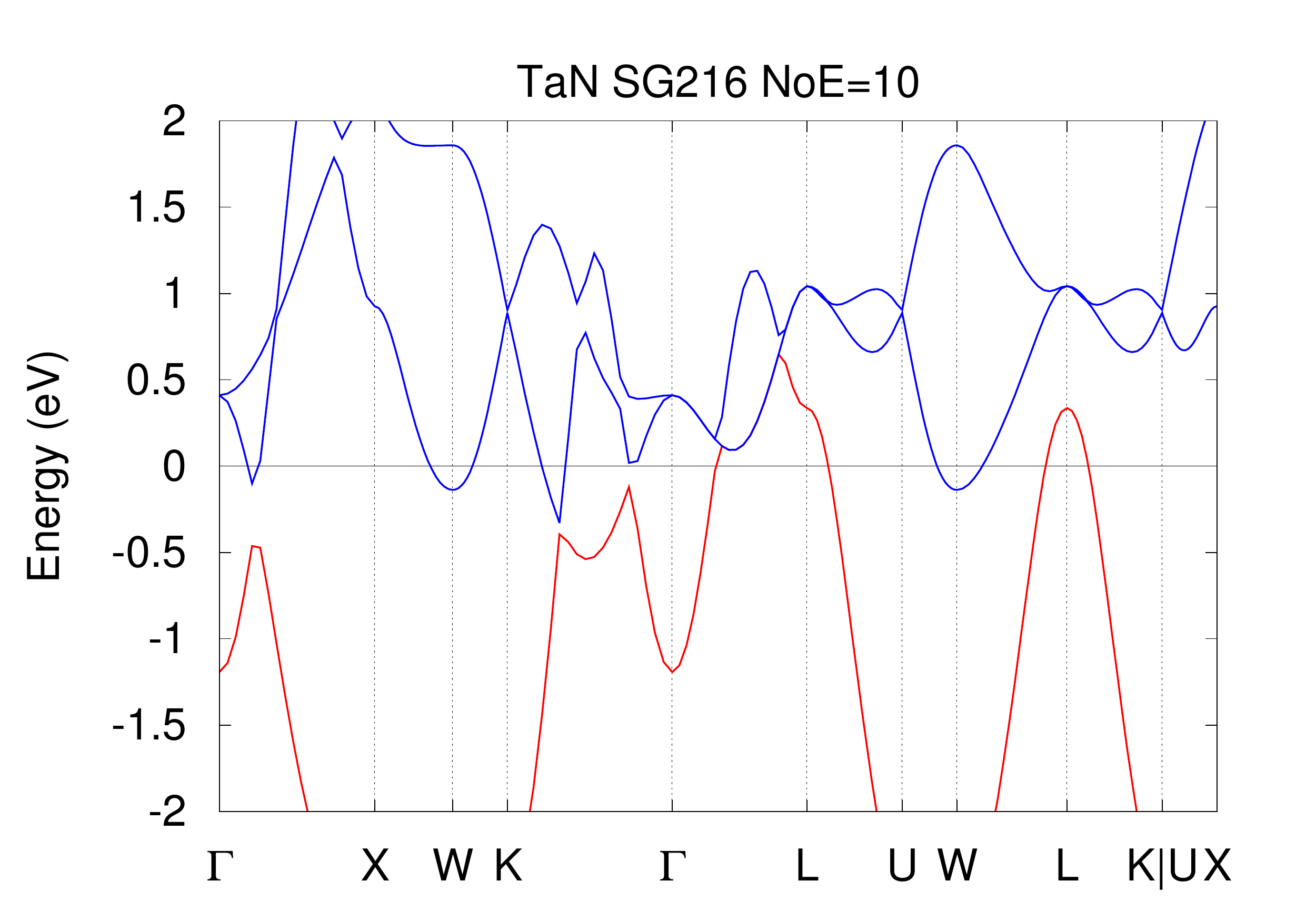}
}
\subfigure[TcN SG216 NoA=2 NoE=12]{
\label{subfig:167870}
\includegraphics[scale=0.32,angle=270]{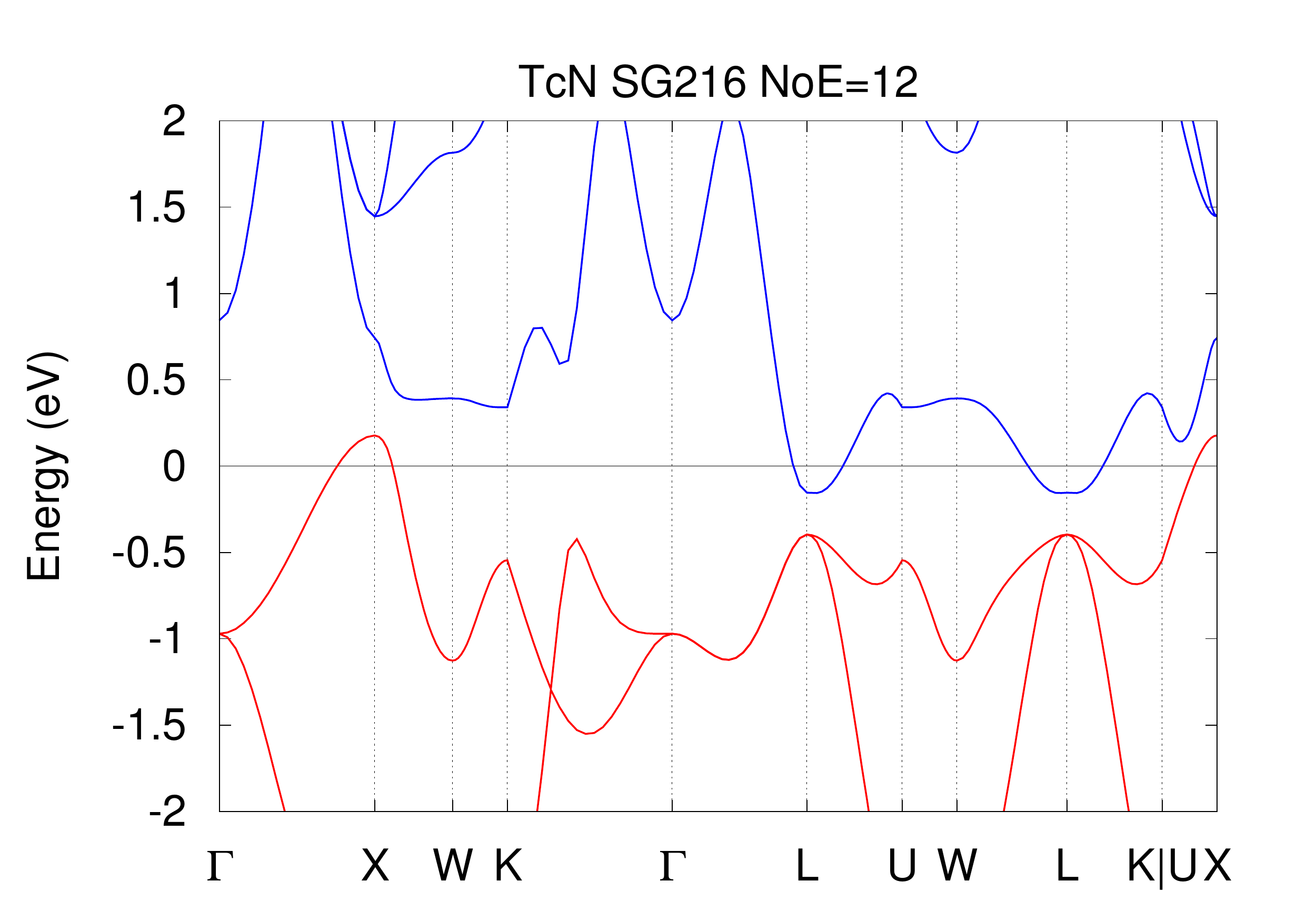}
}
\subfigure[MoN SG216 NoA=2 NoE=11]{
\label{subfig:187182}
\includegraphics[scale=0.32,angle=270]{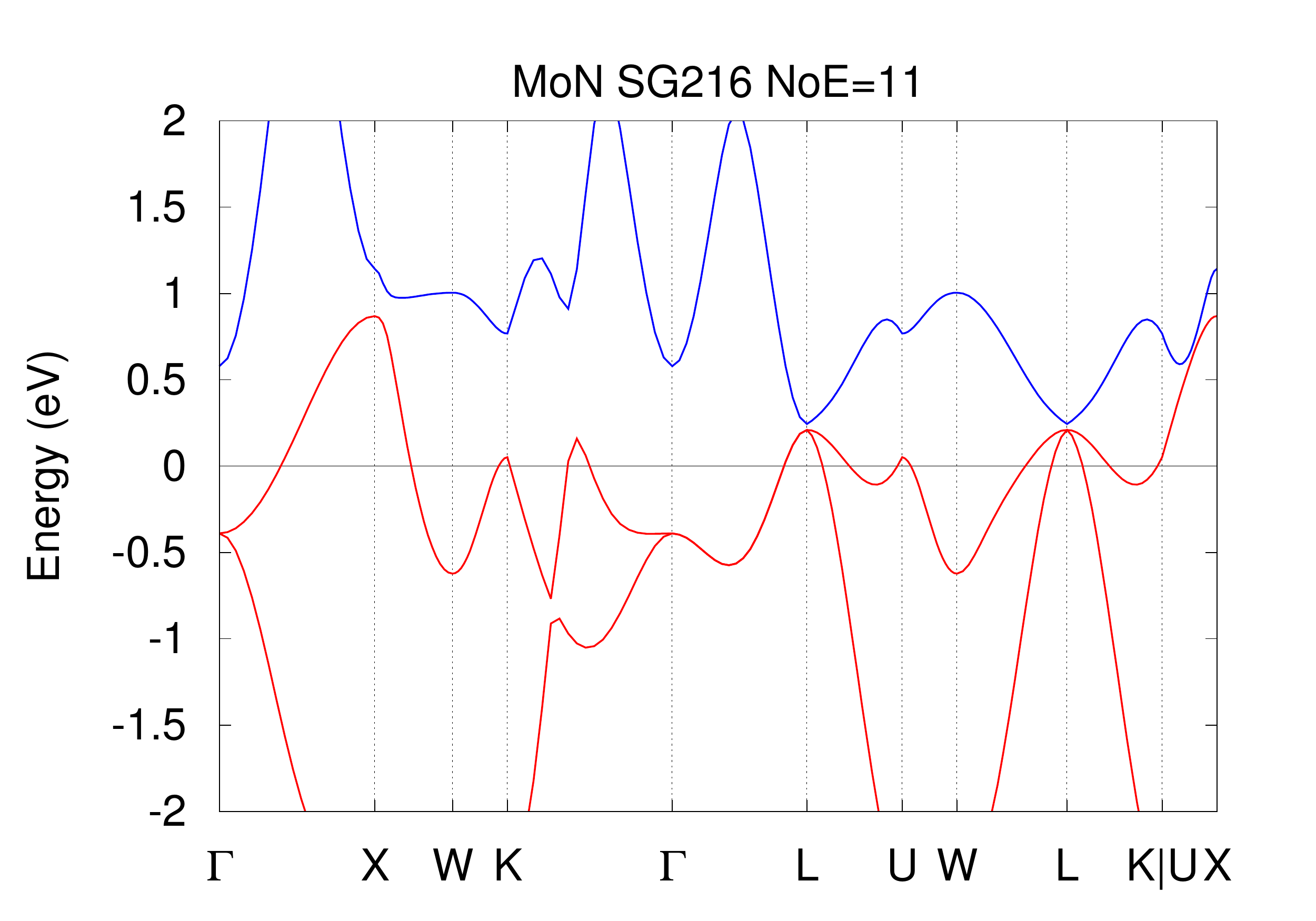}
}
\subfigure[HfN SG216 NoA=2 NoE=9]{
\label{subfig:167875}
\includegraphics[scale=0.32,angle=270]{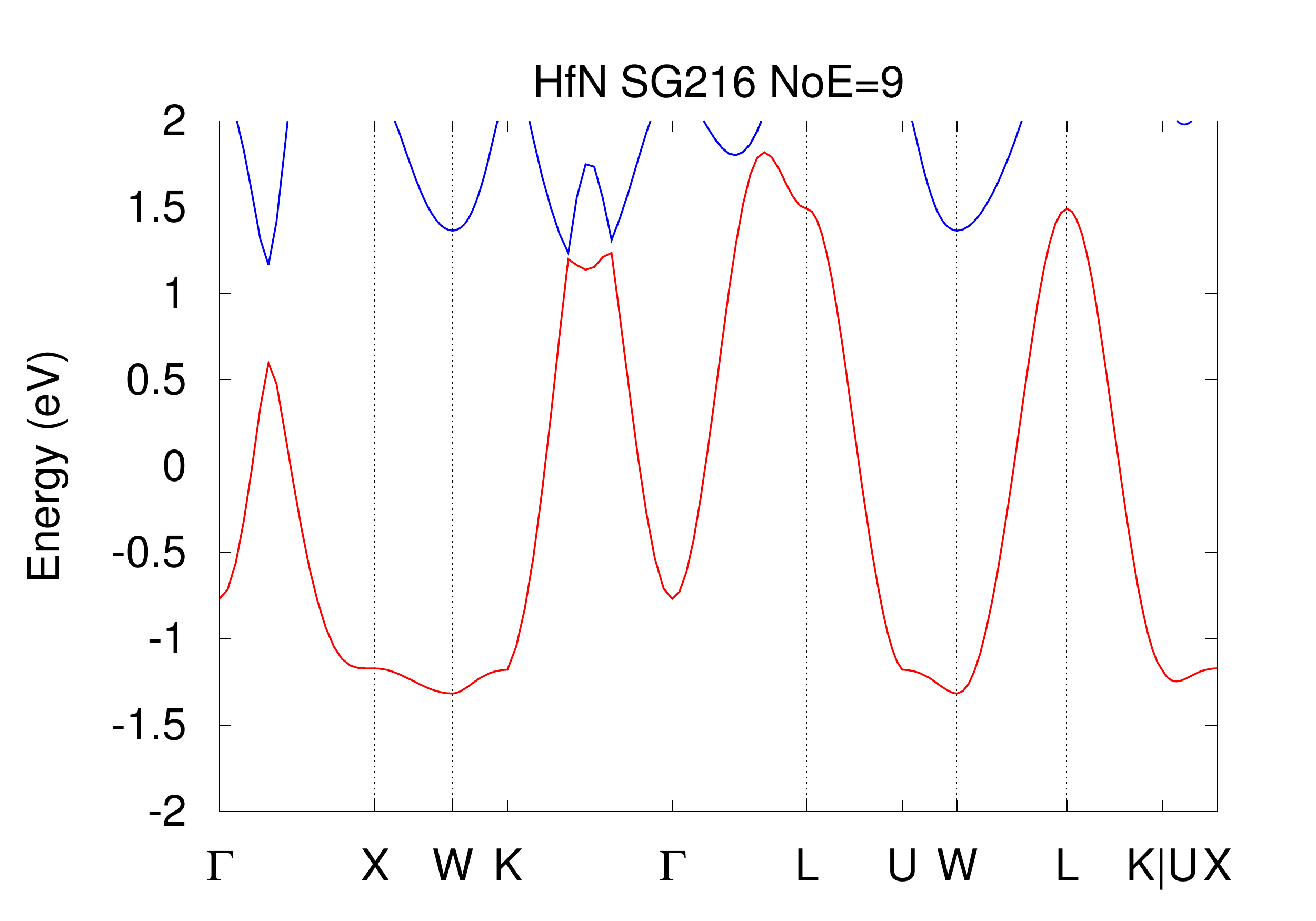}
}
\subfigure[MoP SG216 NoA=2 NoE=11]{
\label{subfig:186876}
\includegraphics[scale=0.32,angle=270]{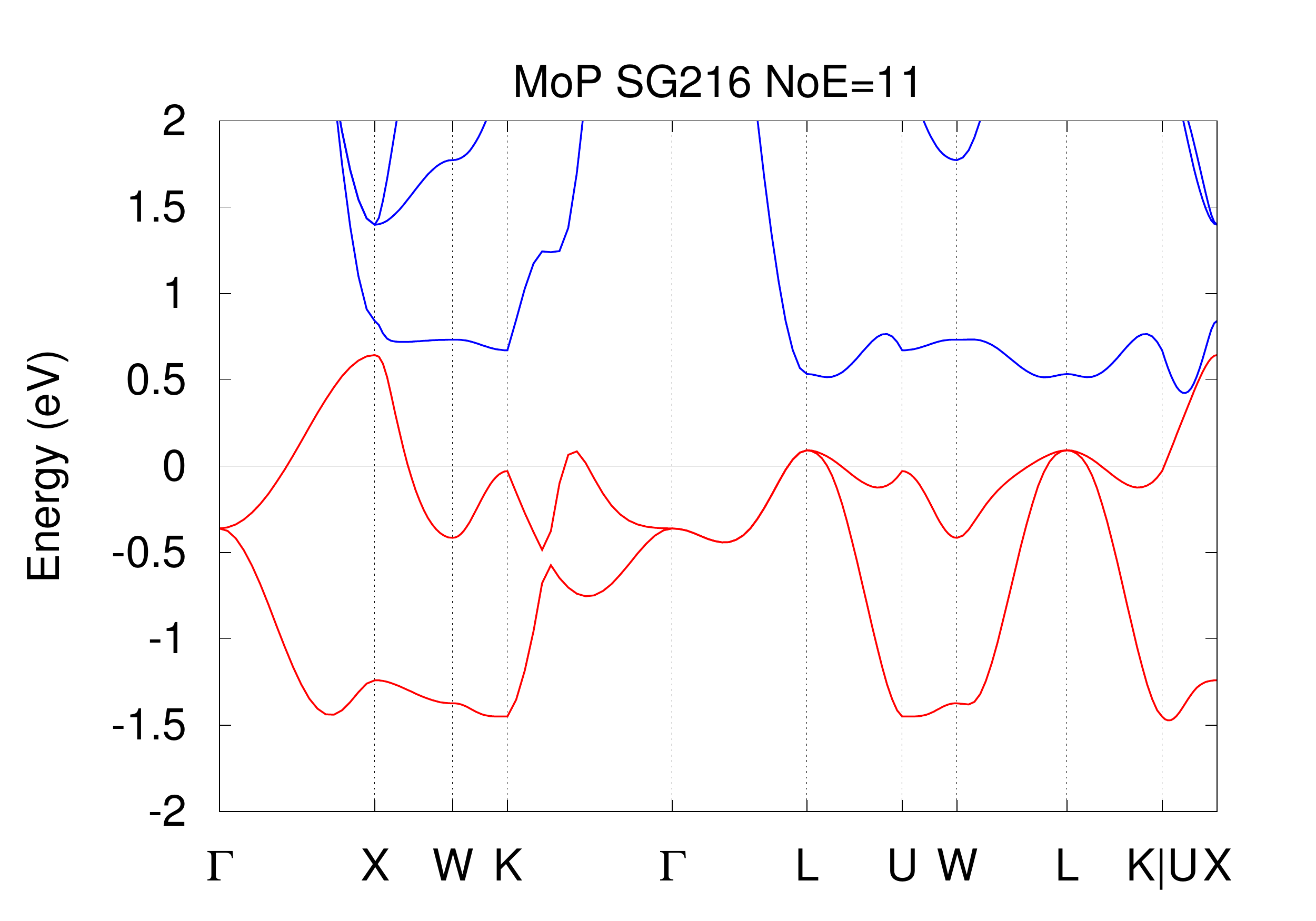}
}
\subfigure[IrN SG216 NoA=2 NoE=14]{
\label{subfig:186243}
\includegraphics[scale=0.32,angle=270]{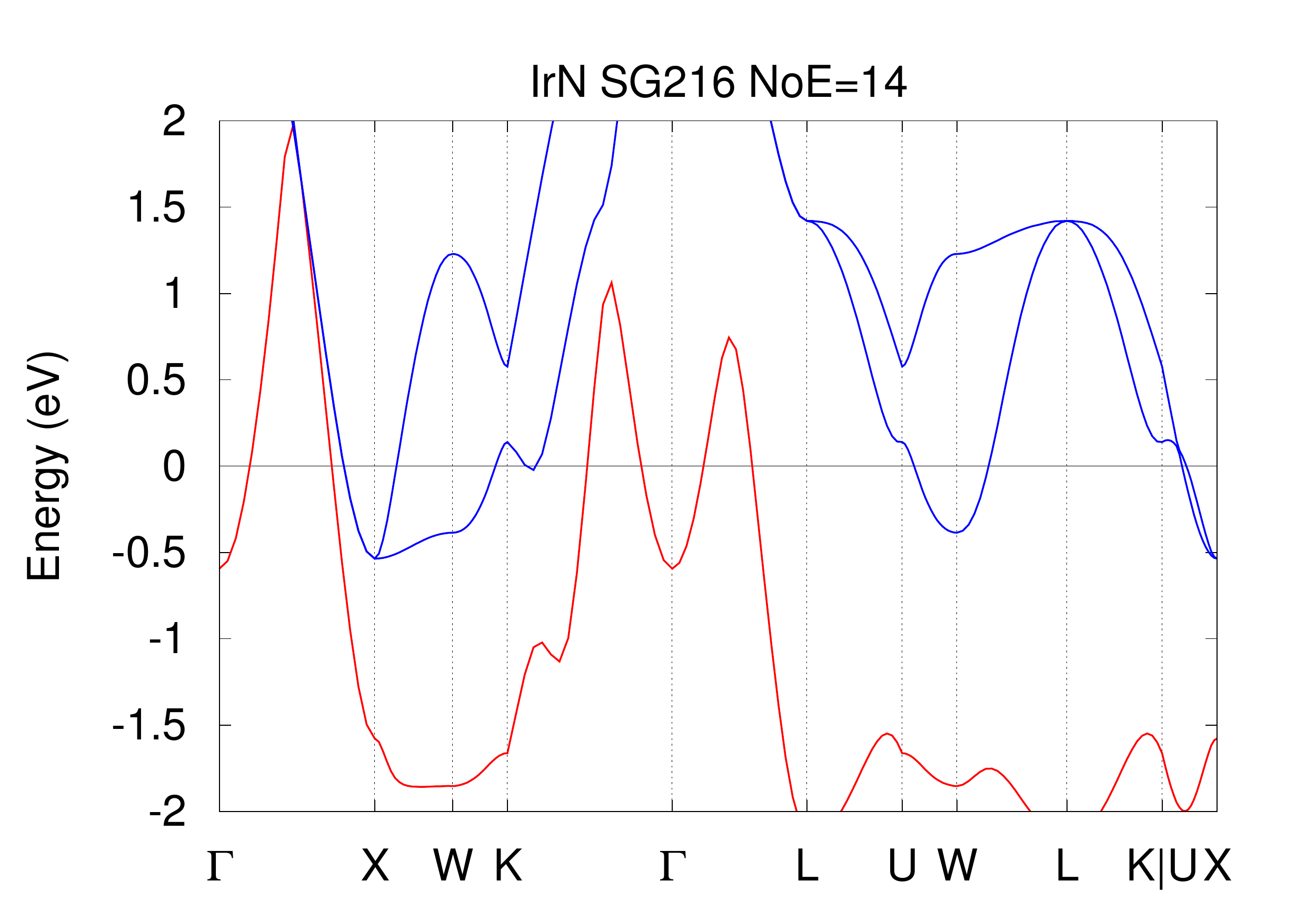}
}
\subfigure[RhN SG216 NoA=2 NoE=14]{
\label{subfig:183192}
\includegraphics[scale=0.32,angle=270]{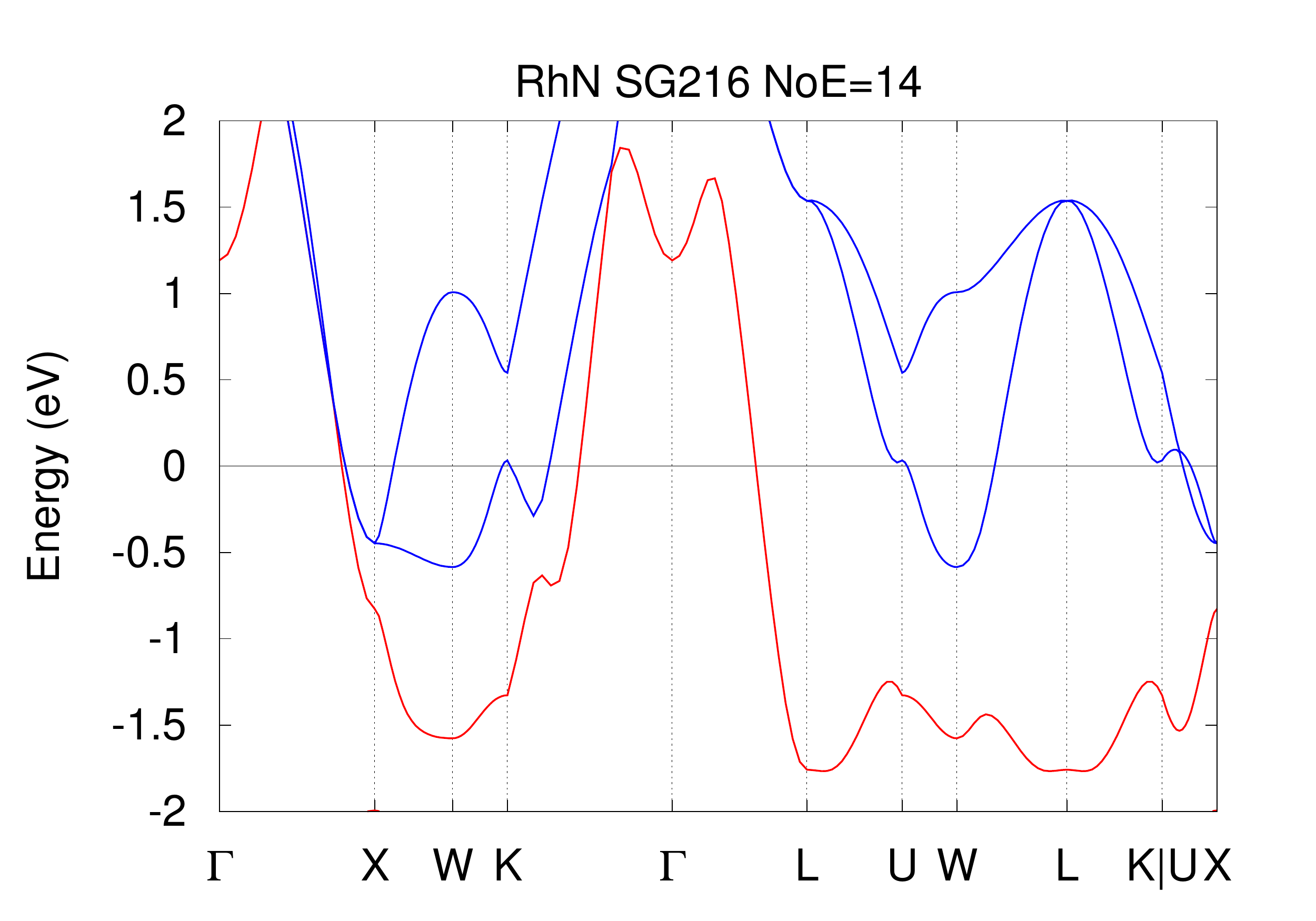}
}
\subfigure[Cu SG12 NoA=2 NoE=22]{
\label{subfig:150682}
\includegraphics[scale=0.32,angle=270]{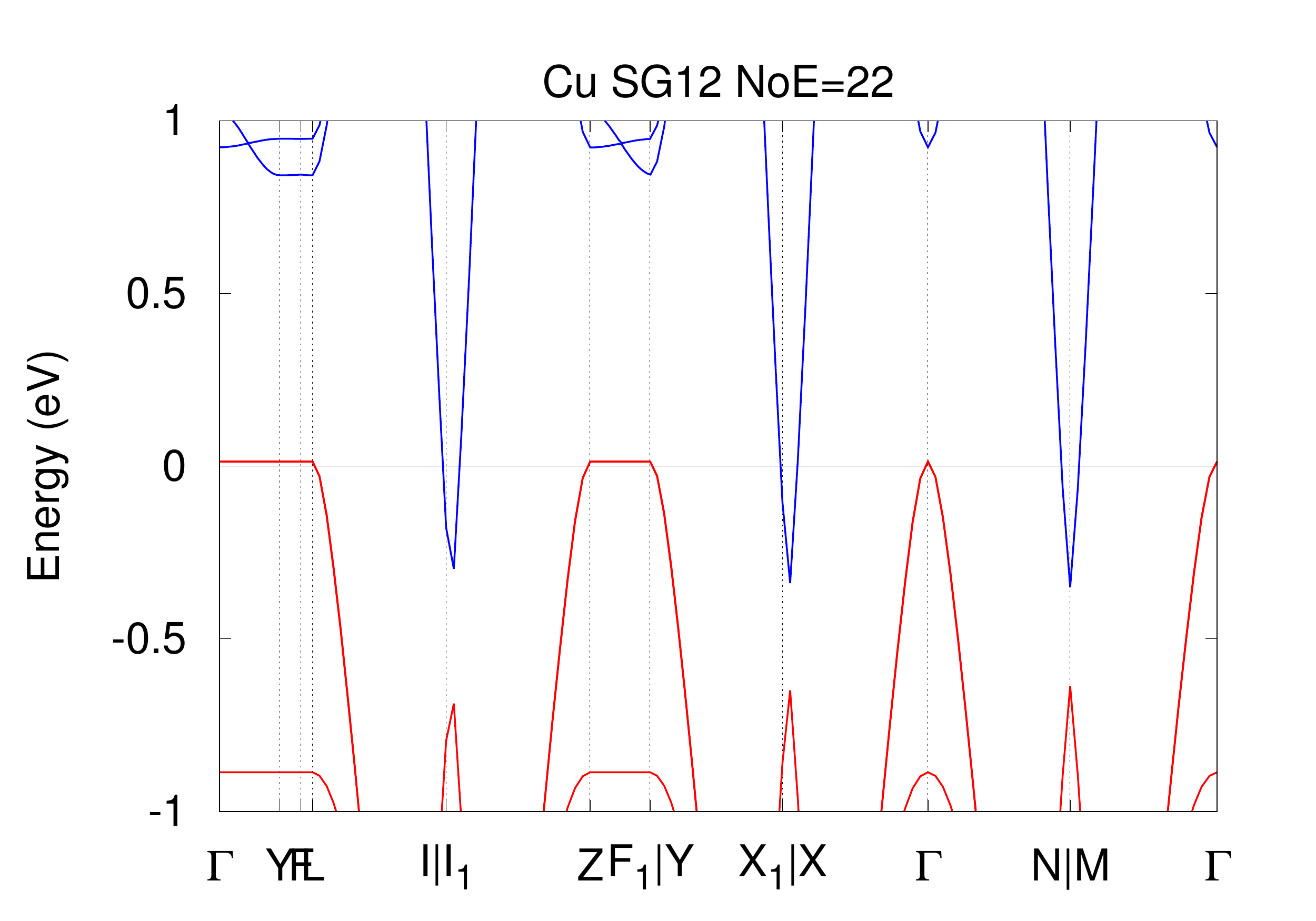}
}
\caption{\hyperref[tab:electride]{back to the table}}
\end{figure}

\begin{figure}[htp]
 \centering
\subfigure[GdO SG216 NoA=2 NoE=15]{
\label{subfig:24981}
\includegraphics[scale=0.32,angle=270]{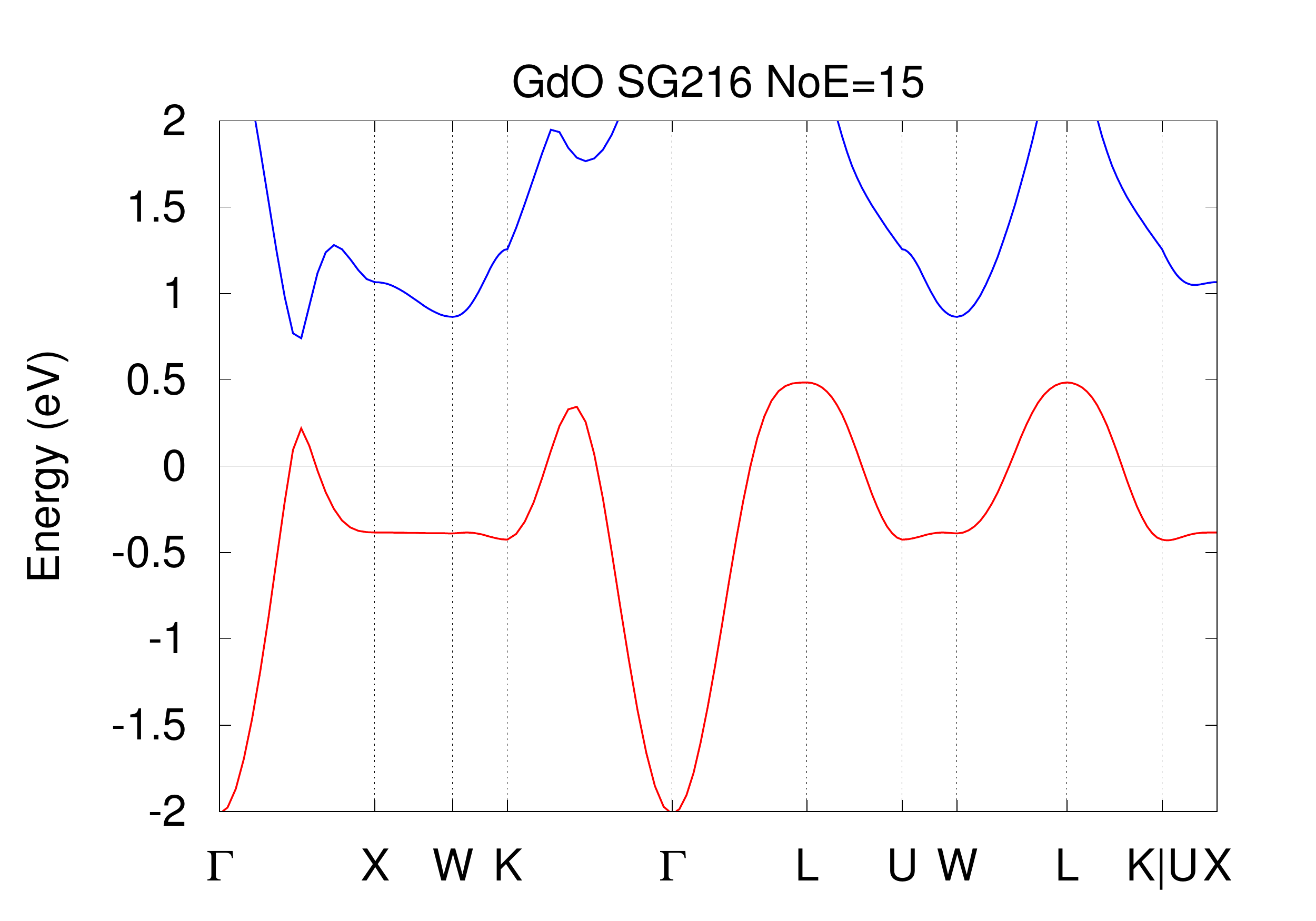}
}
\subfigure[RuN SG216 NoA=2 NoE=13]{
\label{subfig:183190}
\includegraphics[scale=0.32,angle=270]{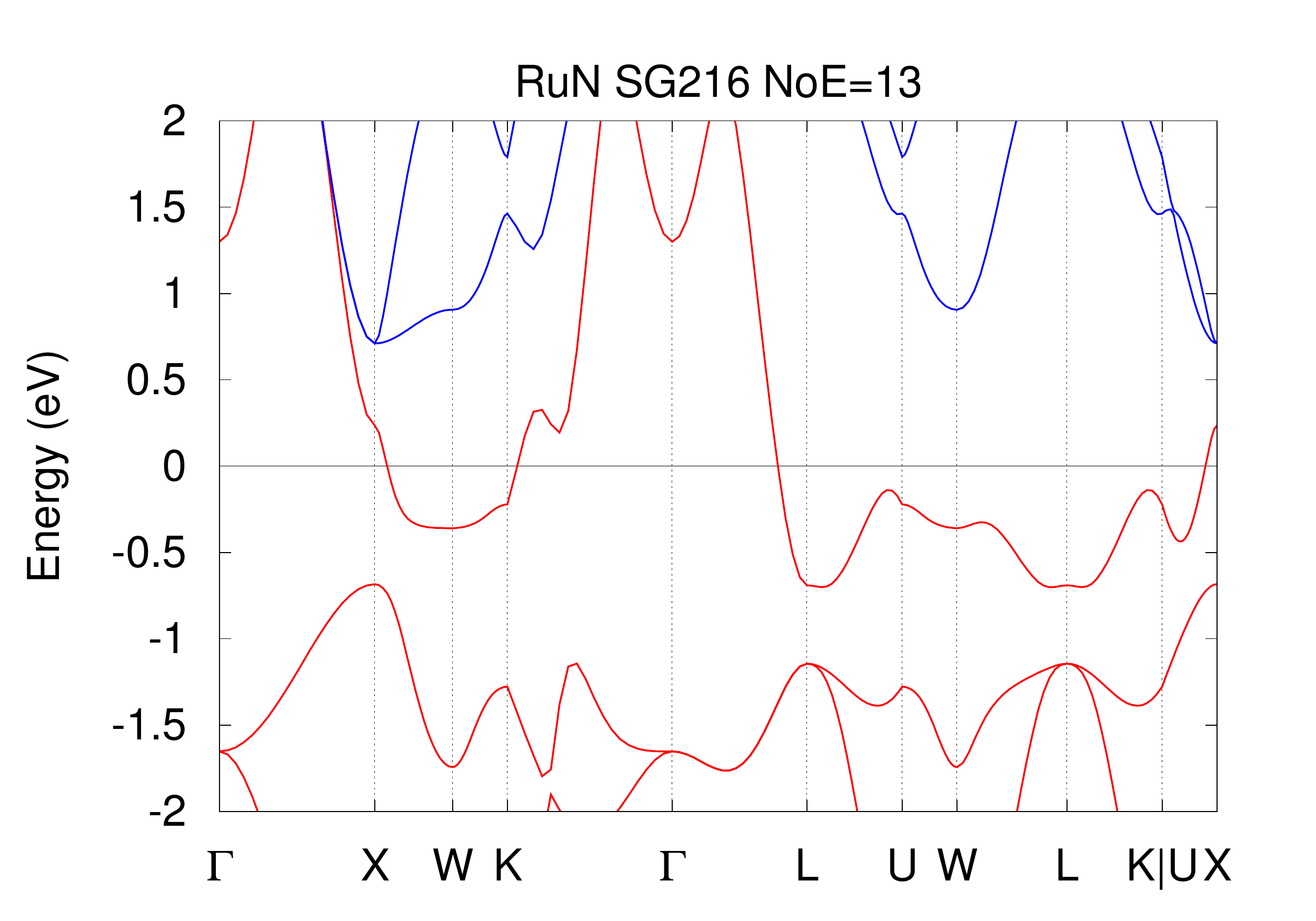}
}
\subfigure[MnSn SG216 NoA=2 NoE=11]{
\label{subfig:191171}
\includegraphics[scale=0.32,angle=270]{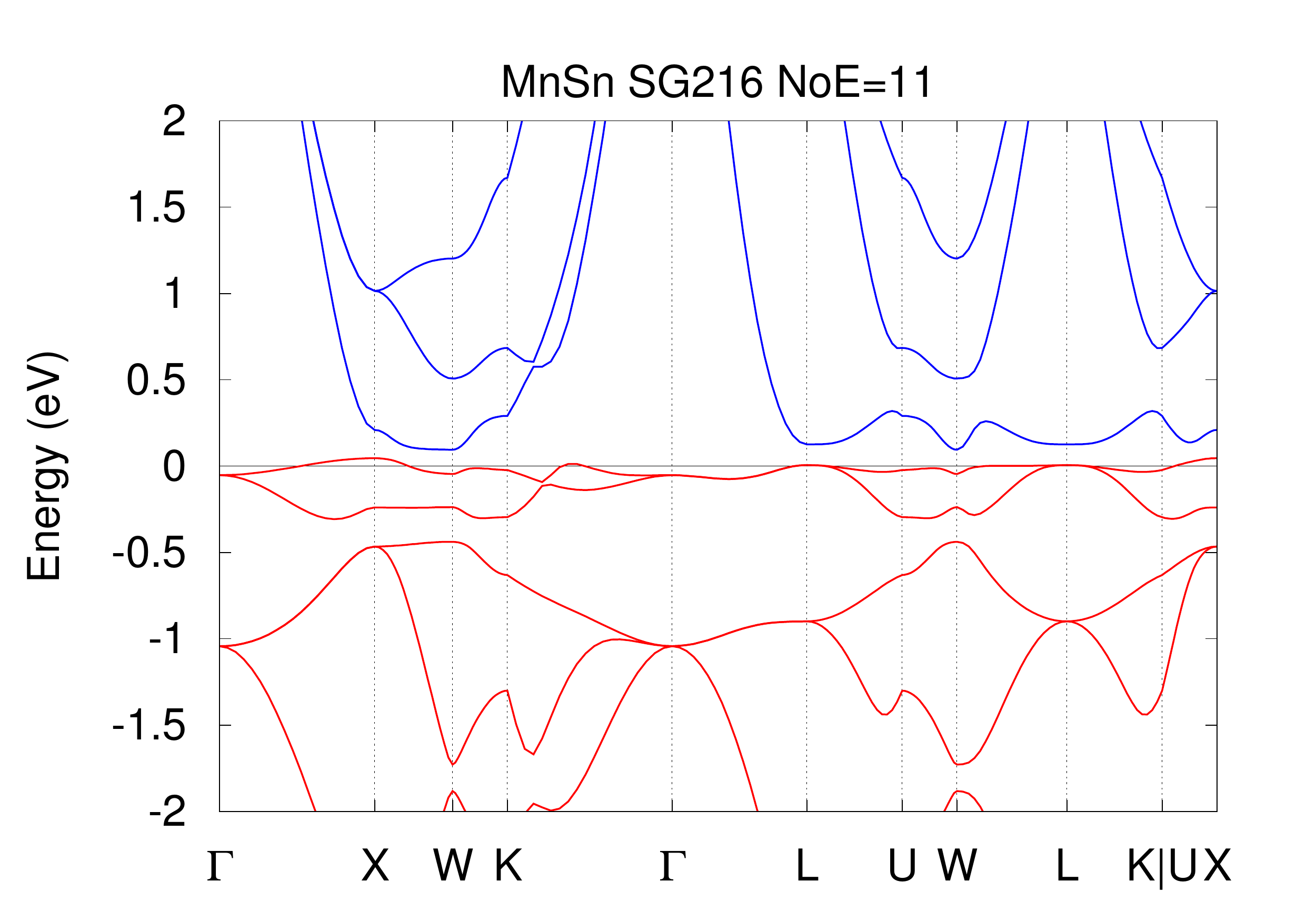}
}
\subfigure[NbN SG216 NoA=2 NoE=16]{
\label{subfig:183184}
\includegraphics[scale=0.32,angle=270]{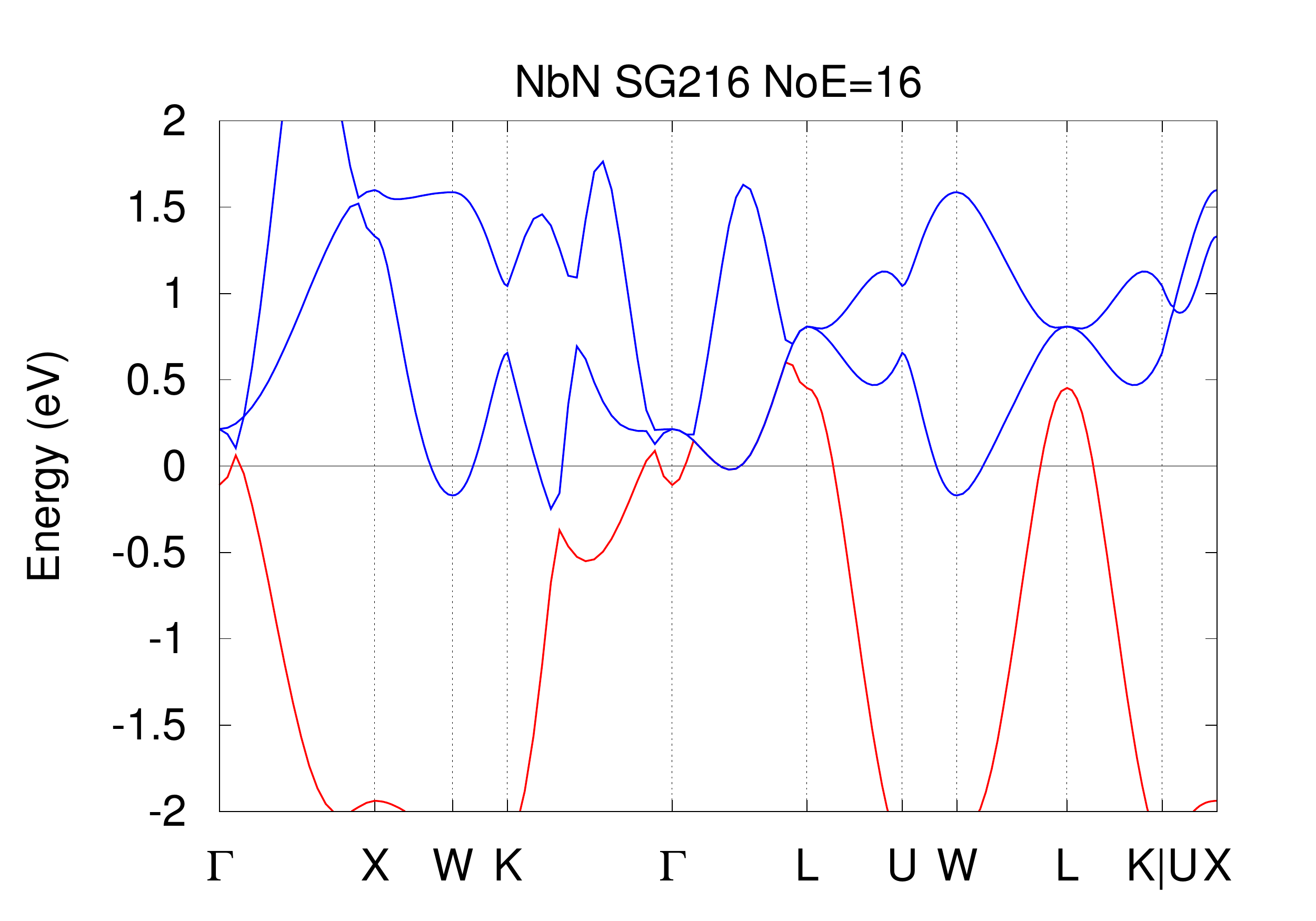}
}
\subfigure[ZrN SG216 NoA=2 NoE=17]{
\label{subfig:183182}
\includegraphics[scale=0.32,angle=270]{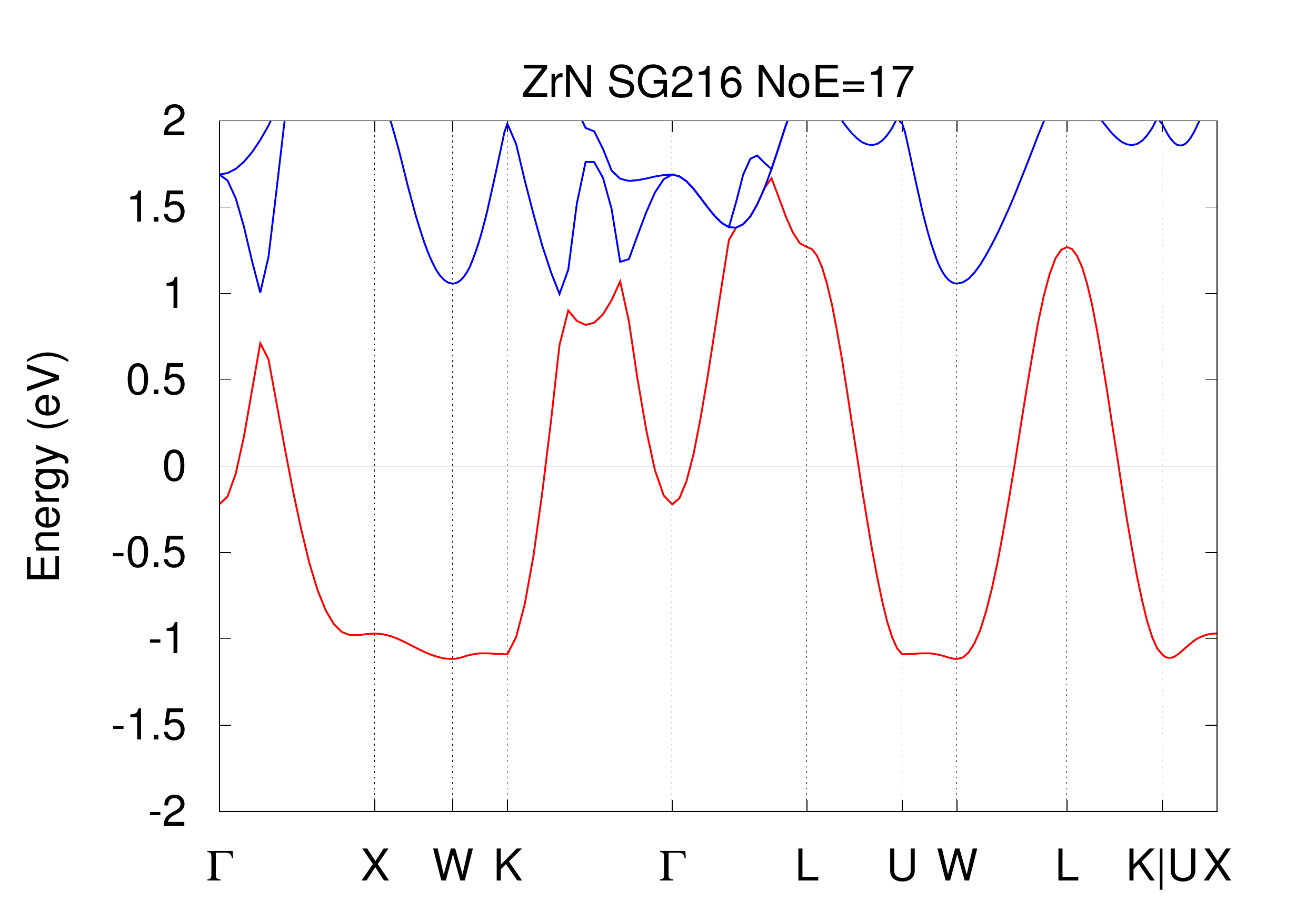}
}
\subfigure[FeN SG216 NoA=2 NoE=13]{
\label{subfig:41258}
\includegraphics[scale=0.32,angle=270]{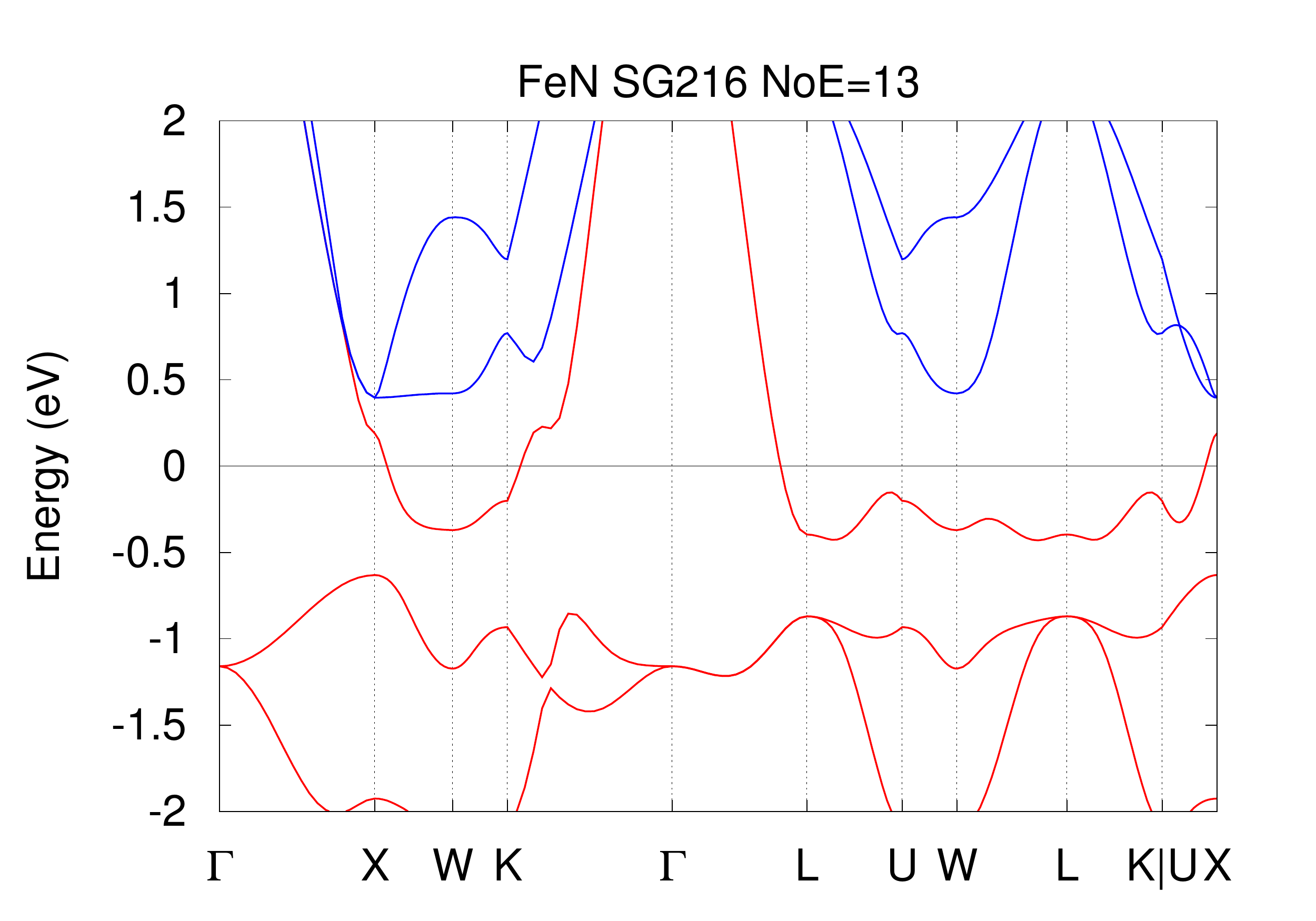}
}
\subfigure[Si SG227 NoA=2 NoE=8]{
\label{subfig:60389}
\includegraphics[scale=0.32,angle=270]{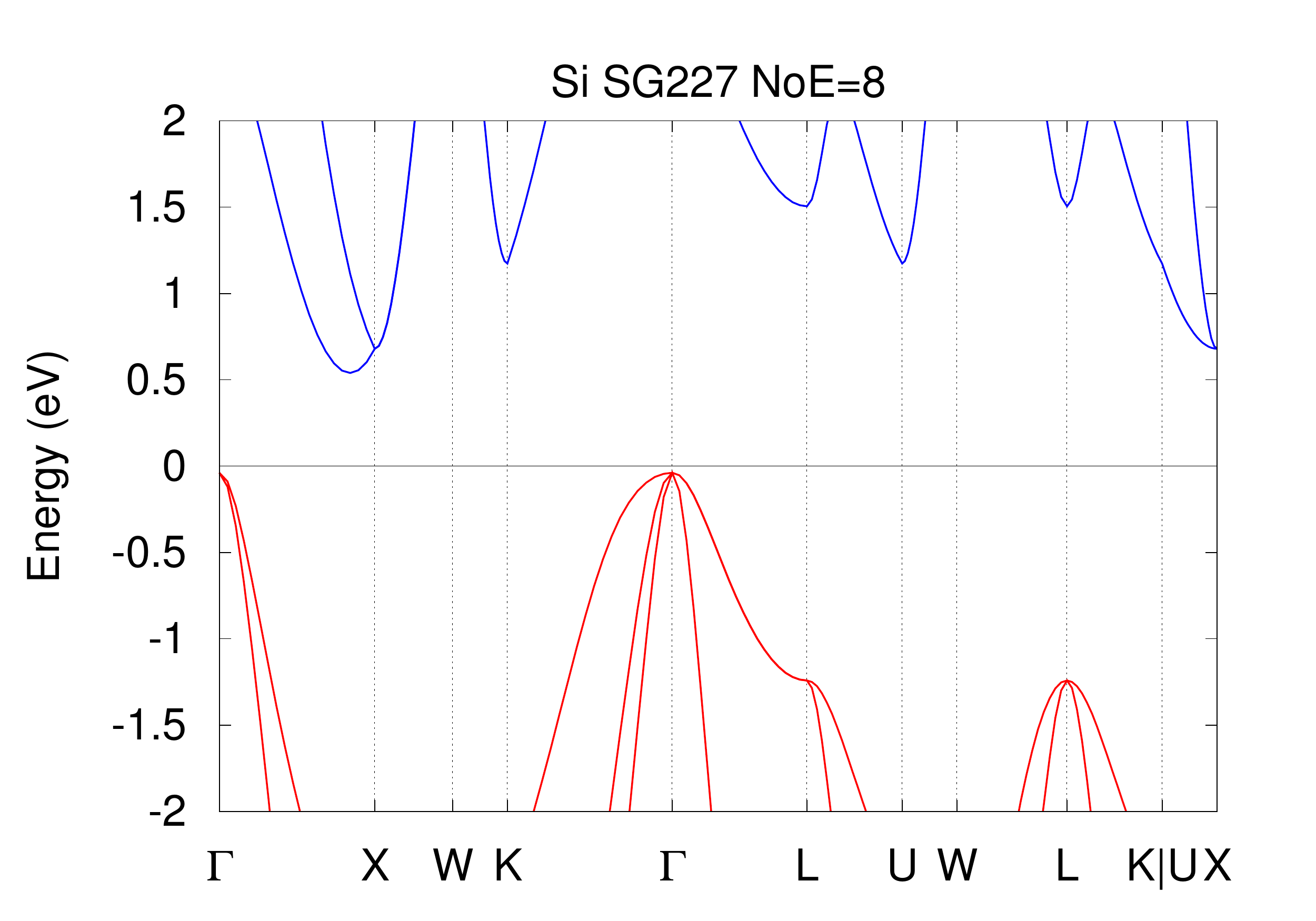}
}
\subfigure[SiP SG216 NoA=2 NoE=9]{
\label{subfig:30334}
\includegraphics[scale=0.32,angle=270]{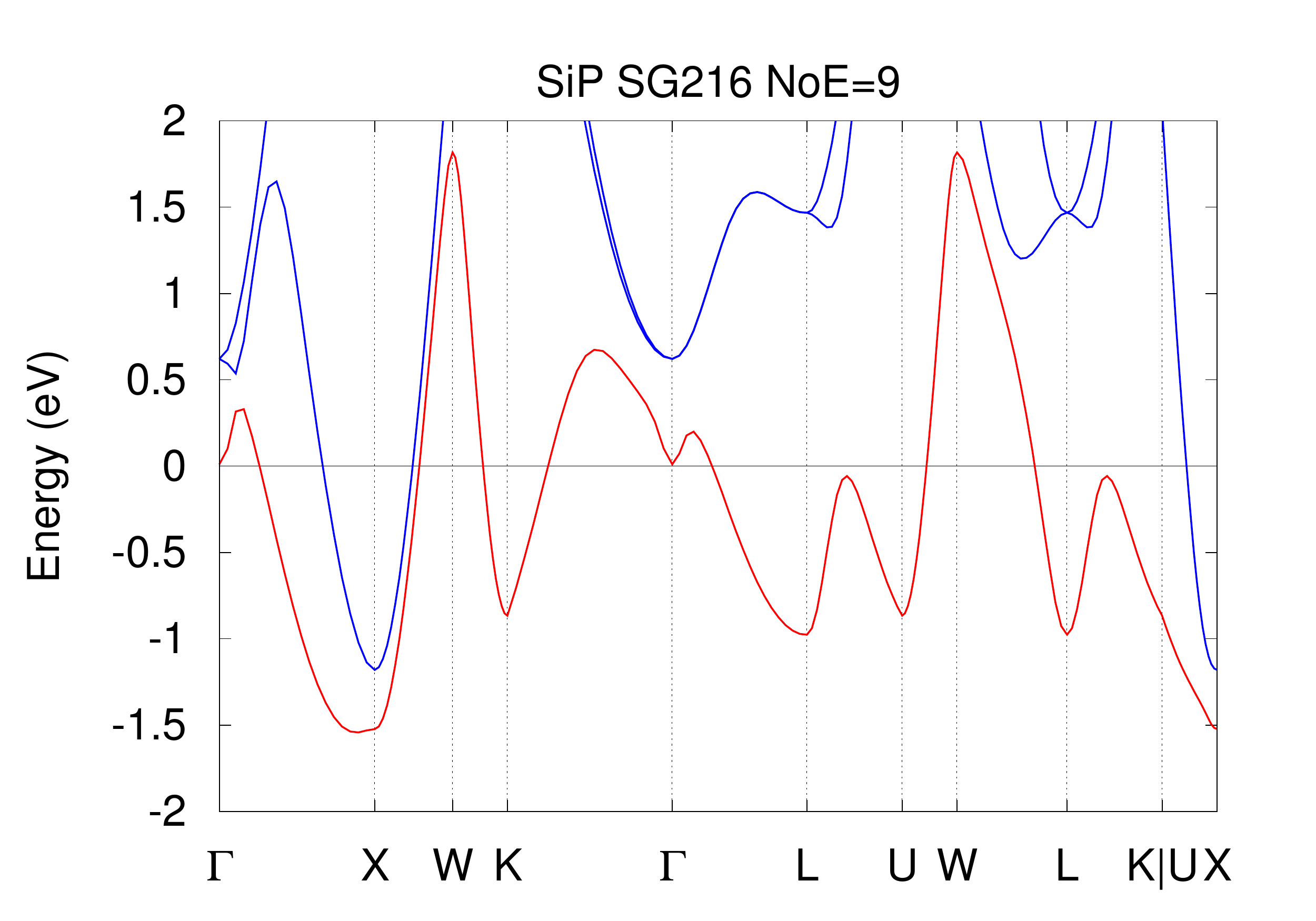}
}
\caption{\hyperref[tab:electride]{back to the table}}
\end{figure}

\begin{figure}[htp]
 \centering
\subfigure[CrN SG216 NoA=2 NoE=11]{
\label{subfig:181079}
\includegraphics[scale=0.32,angle=270]{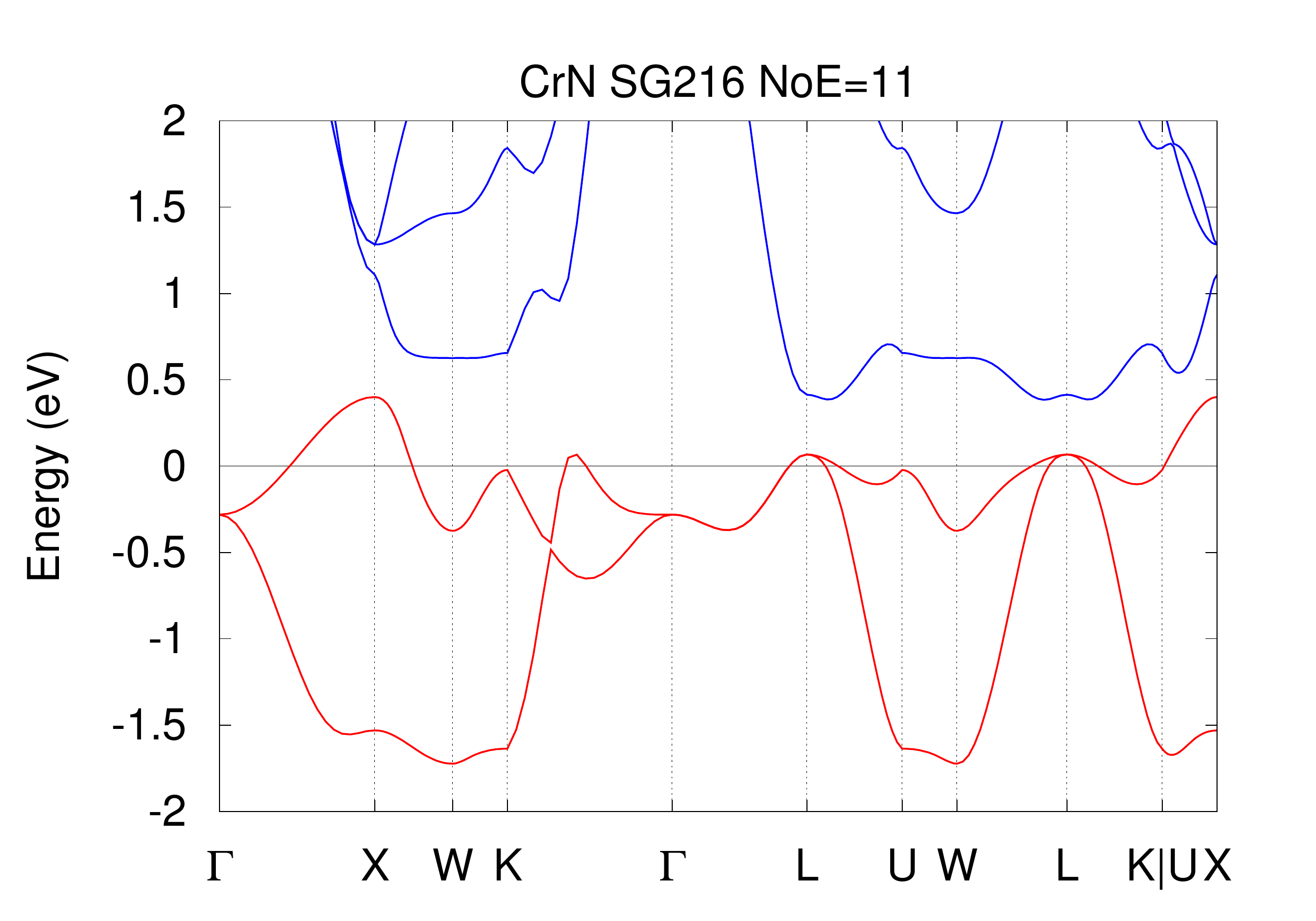}
}
\subfigure[MnN SG216 NoA=2 NoE=12]{
\label{subfig:236788}
\includegraphics[scale=0.32,angle=270]{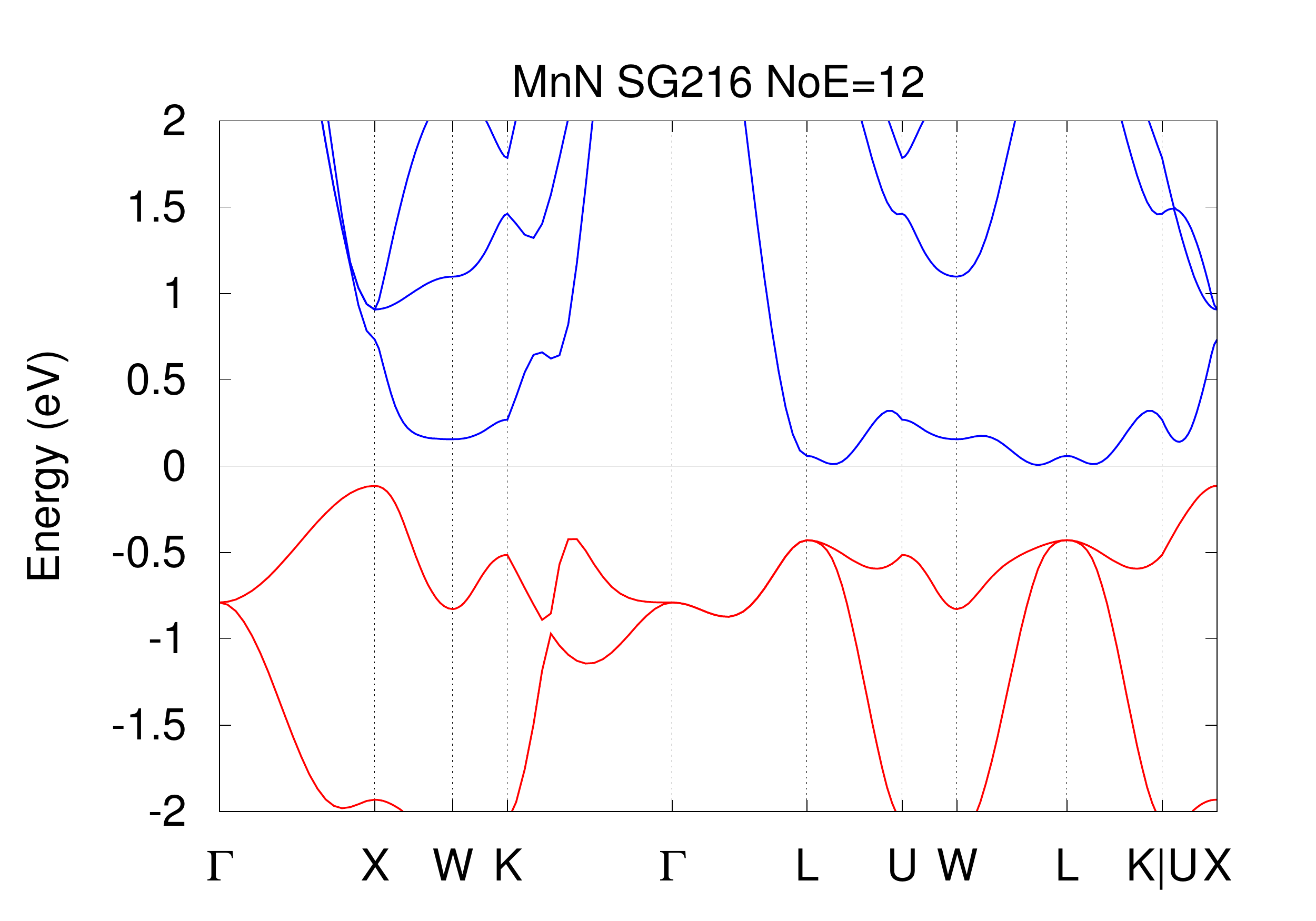}
}
\subfigure[OsN SG216 NoA=2 NoE=13]{
\label{subfig:167879}
\includegraphics[scale=0.32,angle=270]{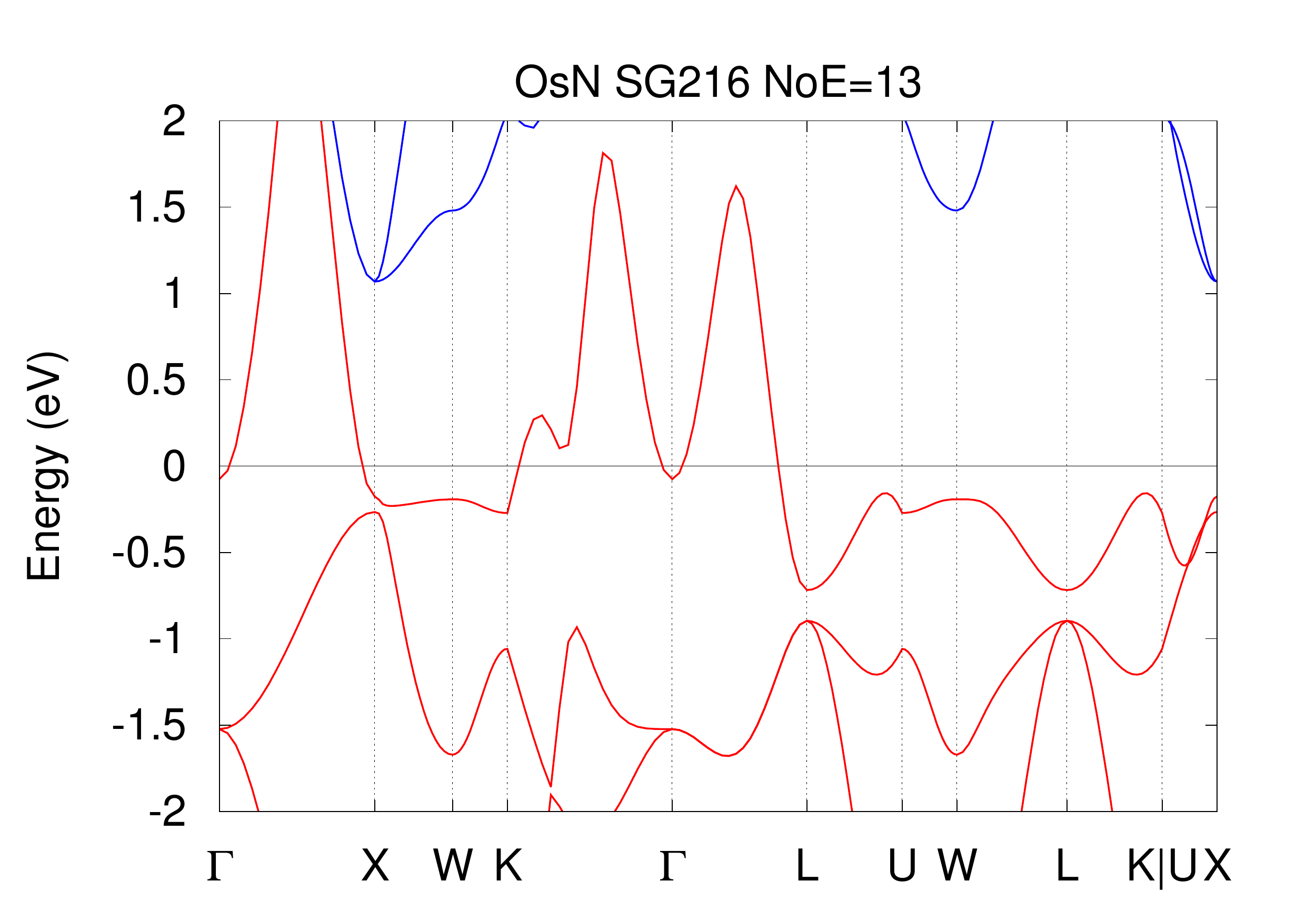}
}
\subfigure[MnP SG216 NoA=2 NoE=12]{
\label{subfig:191788}
\includegraphics[scale=0.32,angle=270]{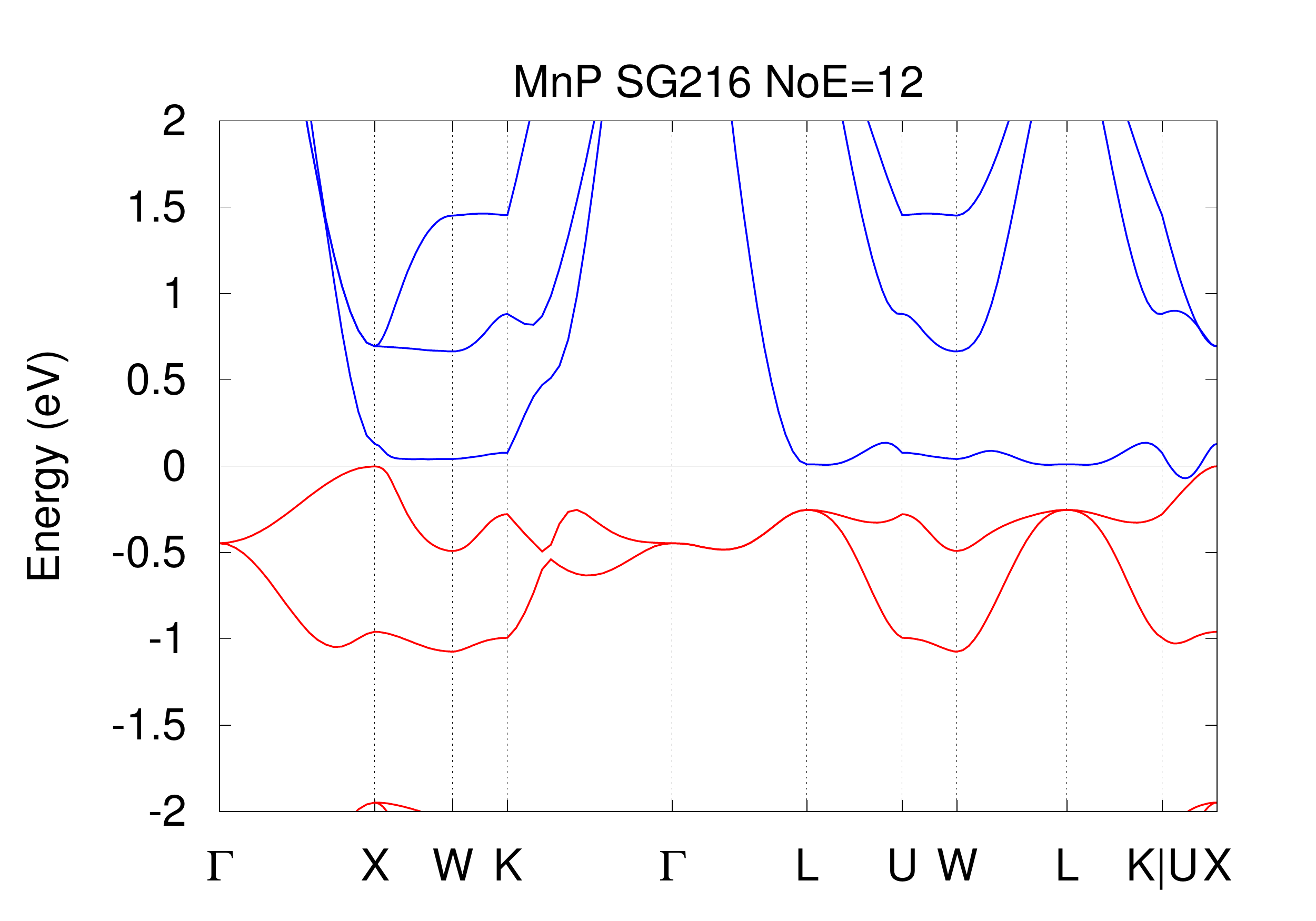}
}
\subfigure[Sr$_{2}$N SG166 NoA=3 NoE=25]{
\label{subfig:414330}
\includegraphics[scale=0.32,angle=270]{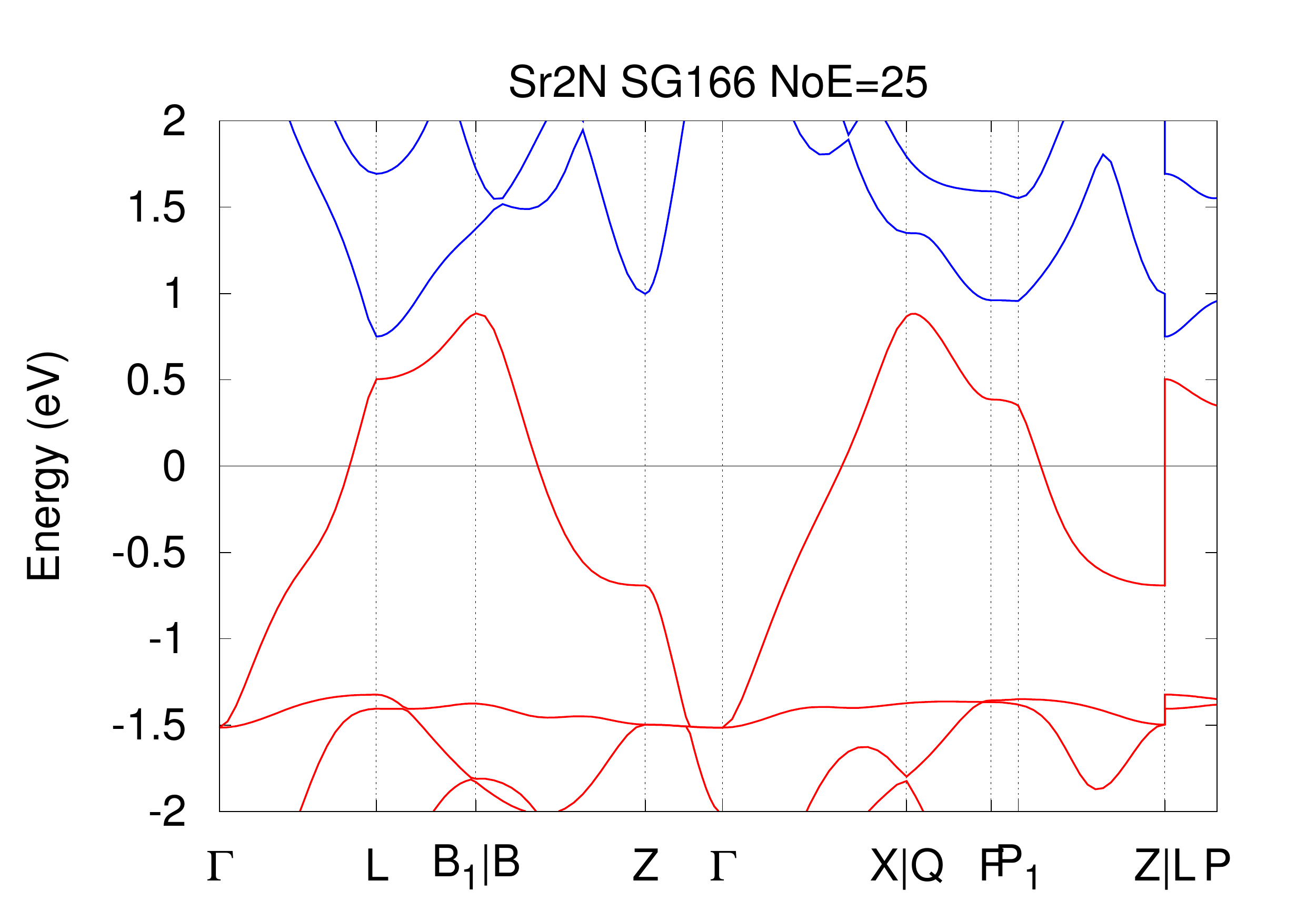}
}
\subfigure[LiFeP SG107 NoA=3 NoE=14]{
\label{subfig:187135}
\includegraphics[scale=0.32,angle=270]{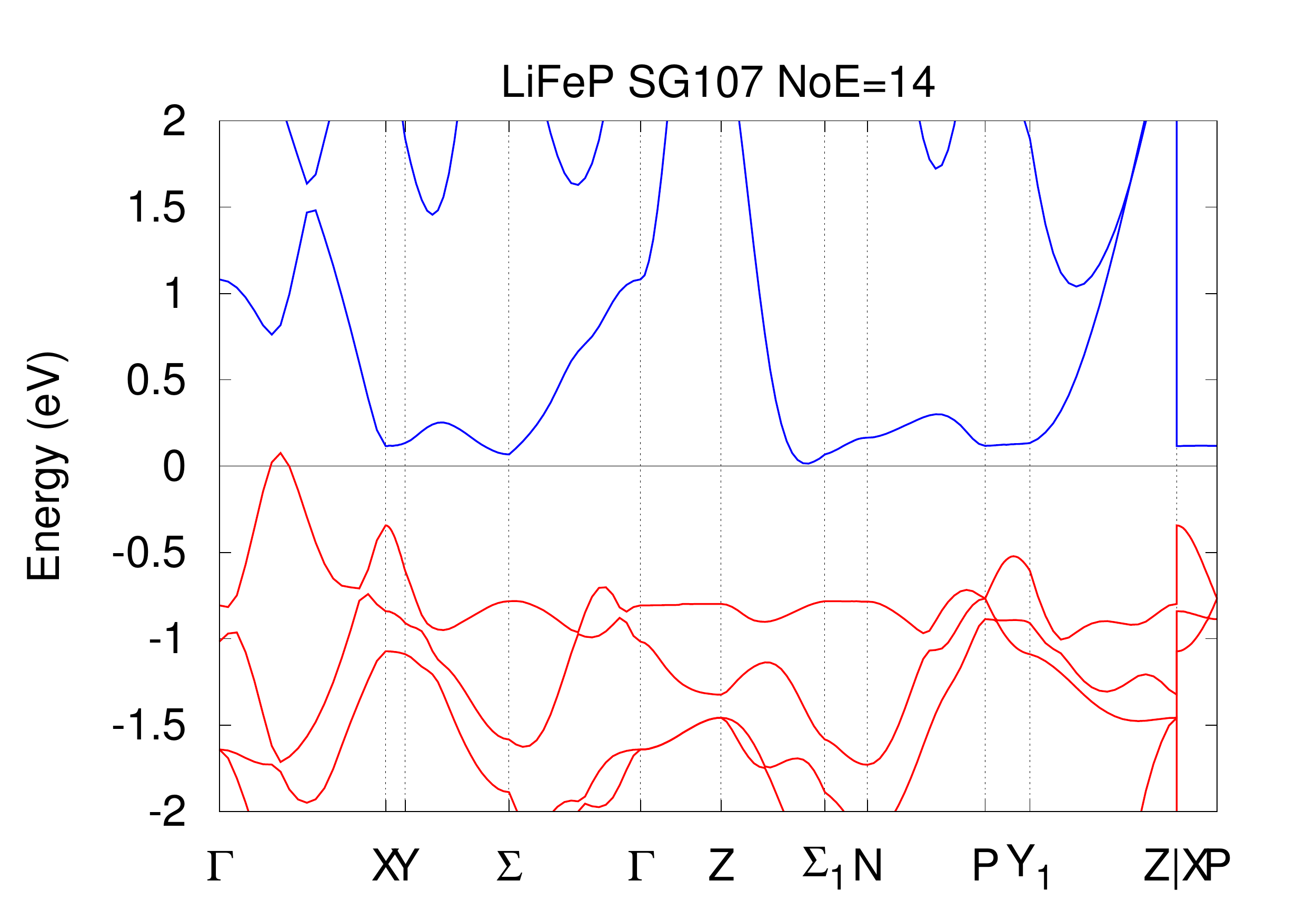}
}
\subfigure[MnNiSb SG216 NoA=3 NoE=22]{
\label{subfig:54255}
\includegraphics[scale=0.32,angle=270]{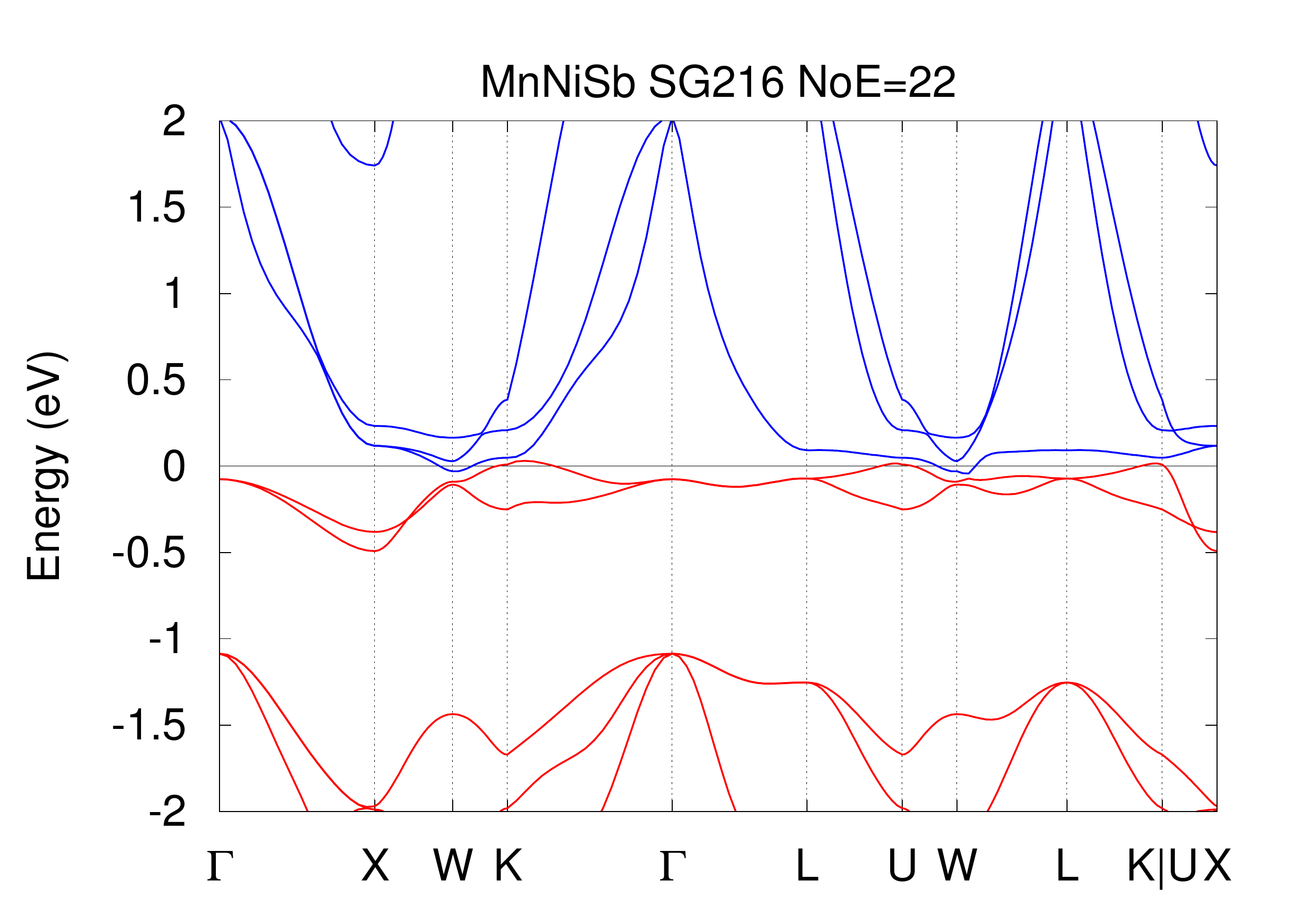}
}
\subfigure[Ca$_{2}$N SG166 NoA=3 NoE=25]{
\label{subfig:22231}
\includegraphics[scale=0.32,angle=270]{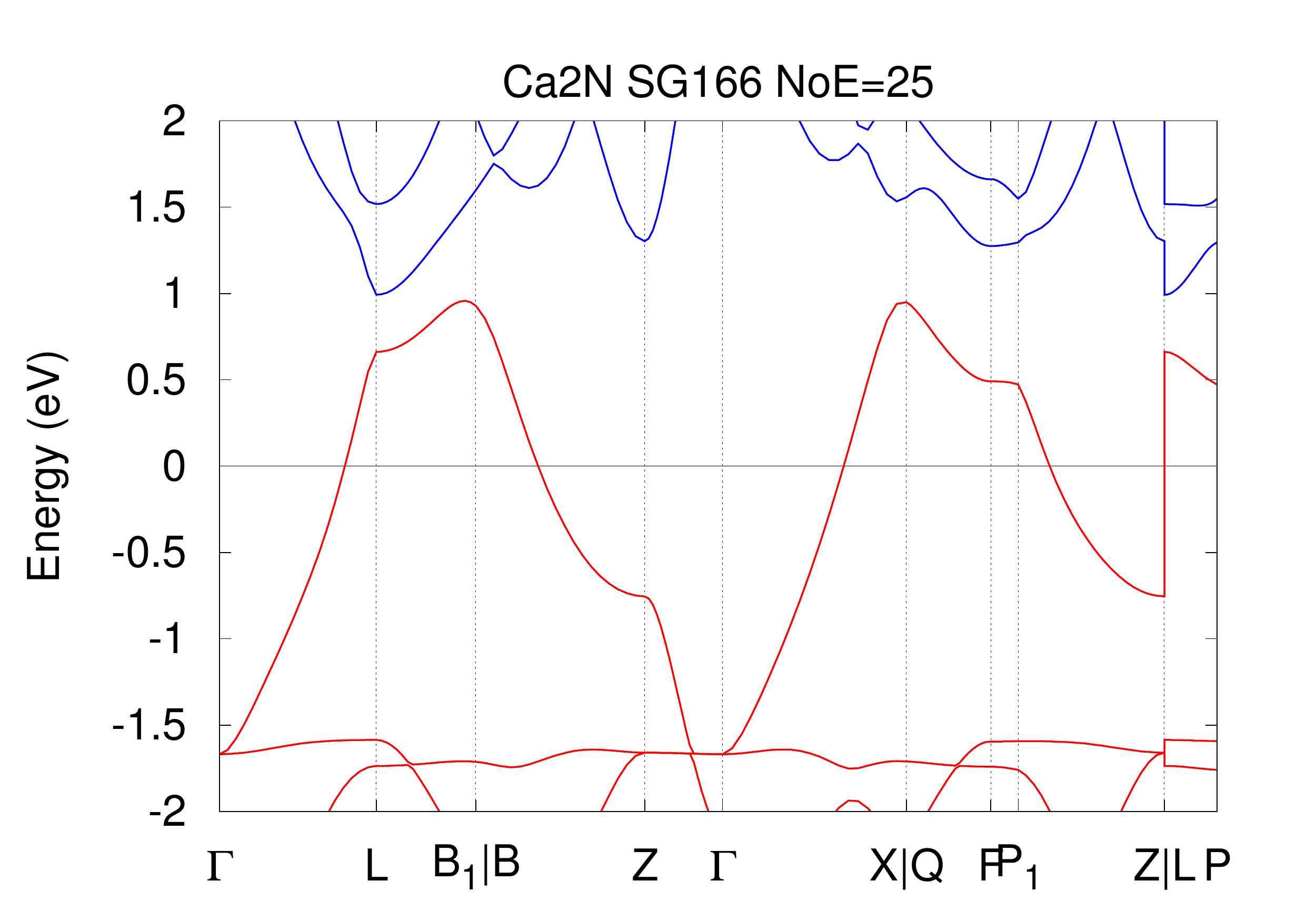}
}
\caption{\hyperref[tab:electride]{back to the table}}
\end{figure}

\begin{figure}[htp]
 \centering
\subfigure[RbO$_{2}$ SG139 NoA=3 NoE=21]{
\label{subfig:647338}
\includegraphics[scale=0.32,angle=270]{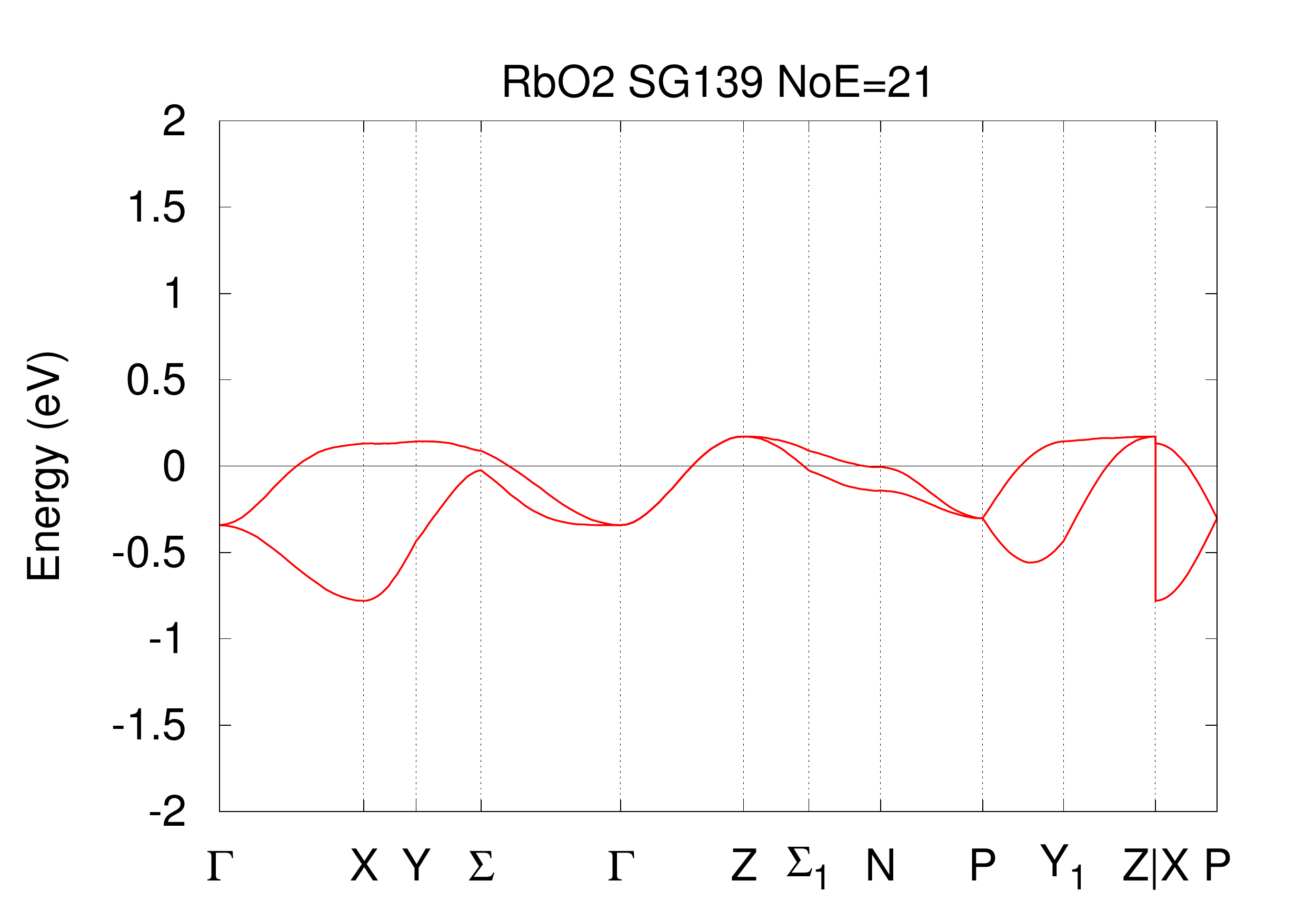}
}
\subfigure[TaN$_{2}$ SG187 NoA=3 NoE=15]{
\label{subfig:290430}
\includegraphics[scale=0.32,angle=270]{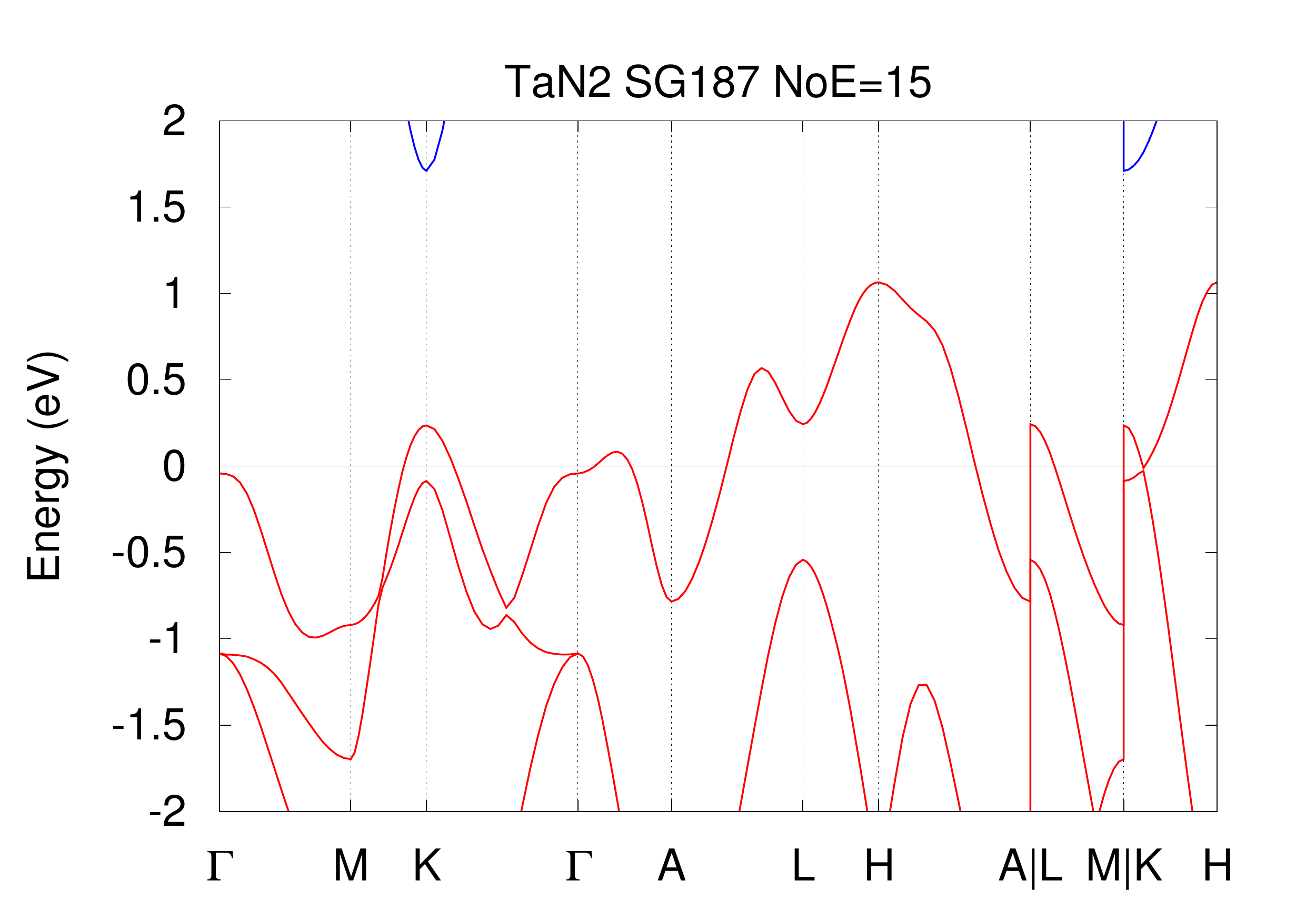}
}
\subfigure[MnSnAu SG216 NoA=3 NoE=22]{
\label{subfig:54465}
\includegraphics[scale=0.32,angle=270]{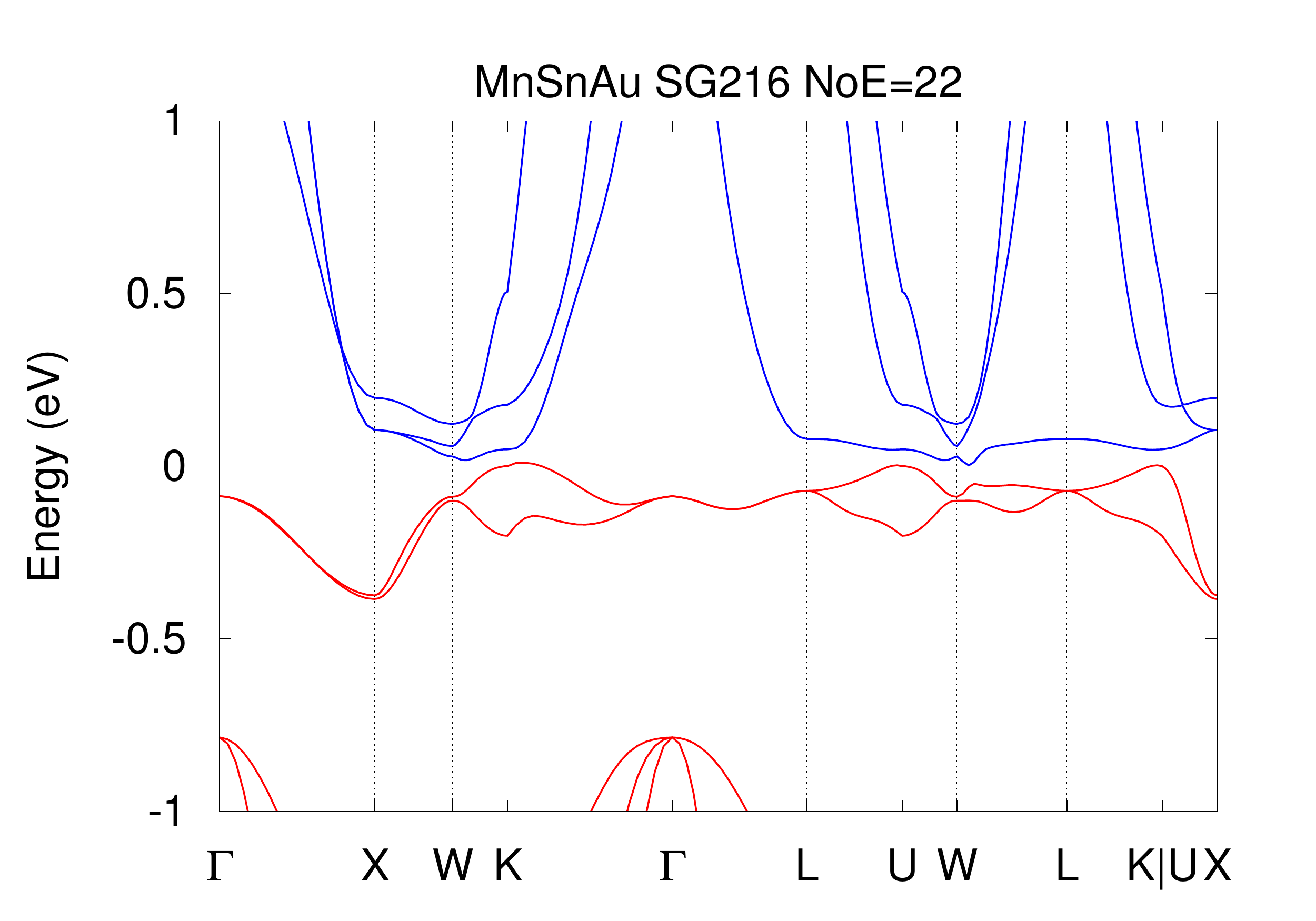}
}
\subfigure[NbS$_{2}$ SG187 NoA=3 NoE=23]{
\label{subfig:67443}
\includegraphics[scale=0.32,angle=270]{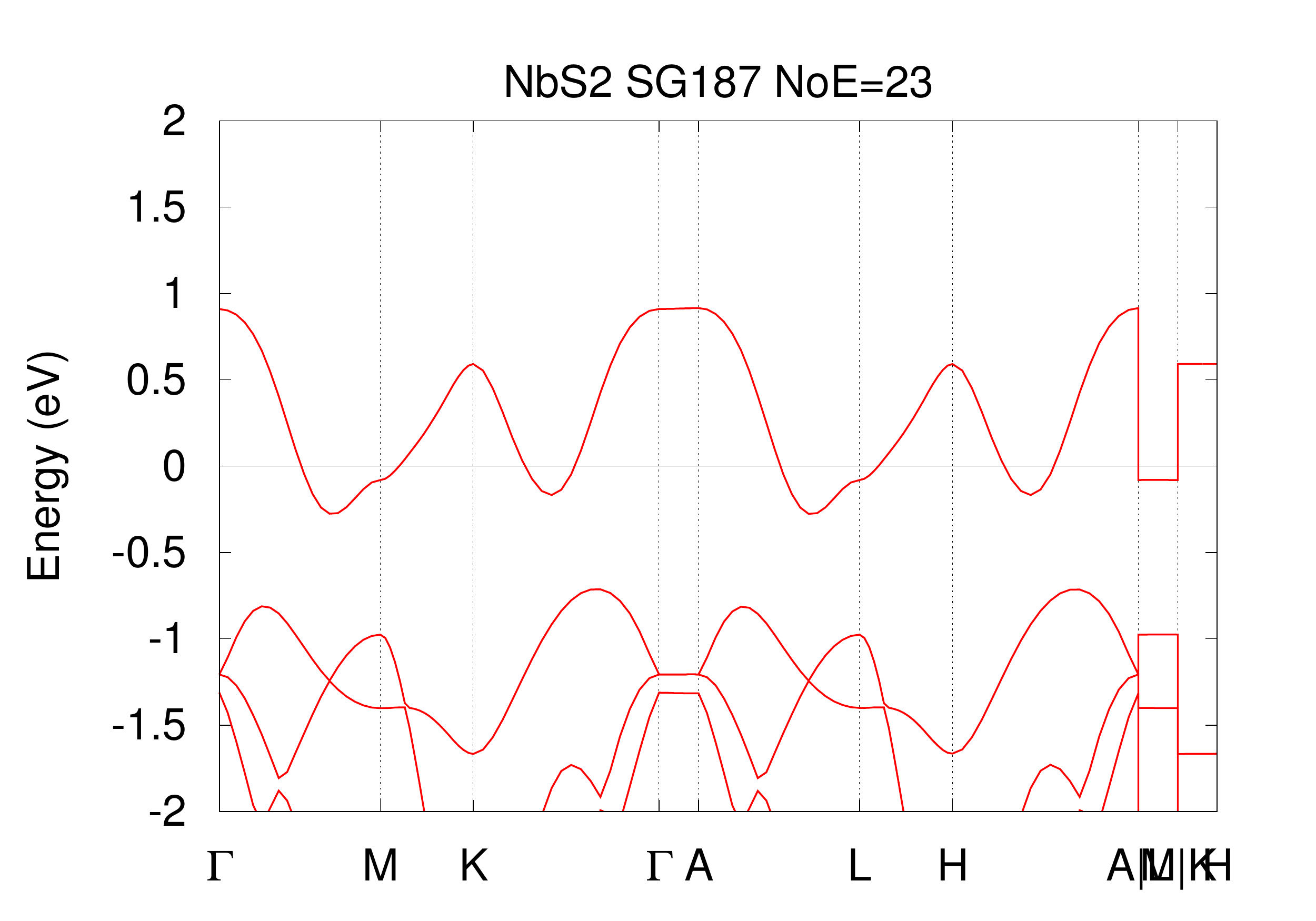}
}
\subfigure[Al$_{2}$Os SG139 NoA=3 NoE=14]{
\label{subfig:58108}
\includegraphics[scale=0.32,angle=270]{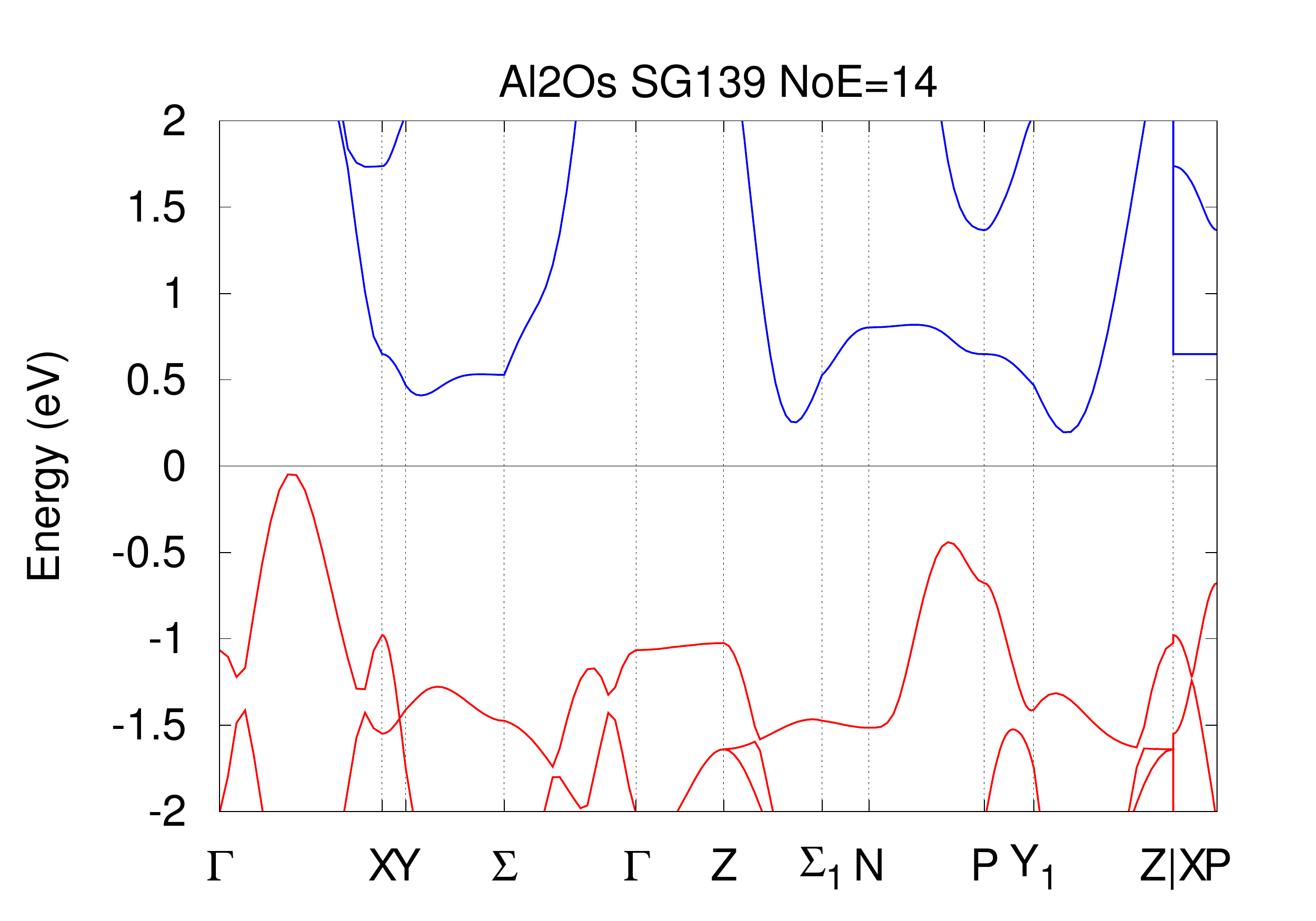}
}
\subfigure[KO$_{2}$ SG139 NoA=3 NoE=21]{
\label{subfig:38245}
\includegraphics[scale=0.32,angle=270]{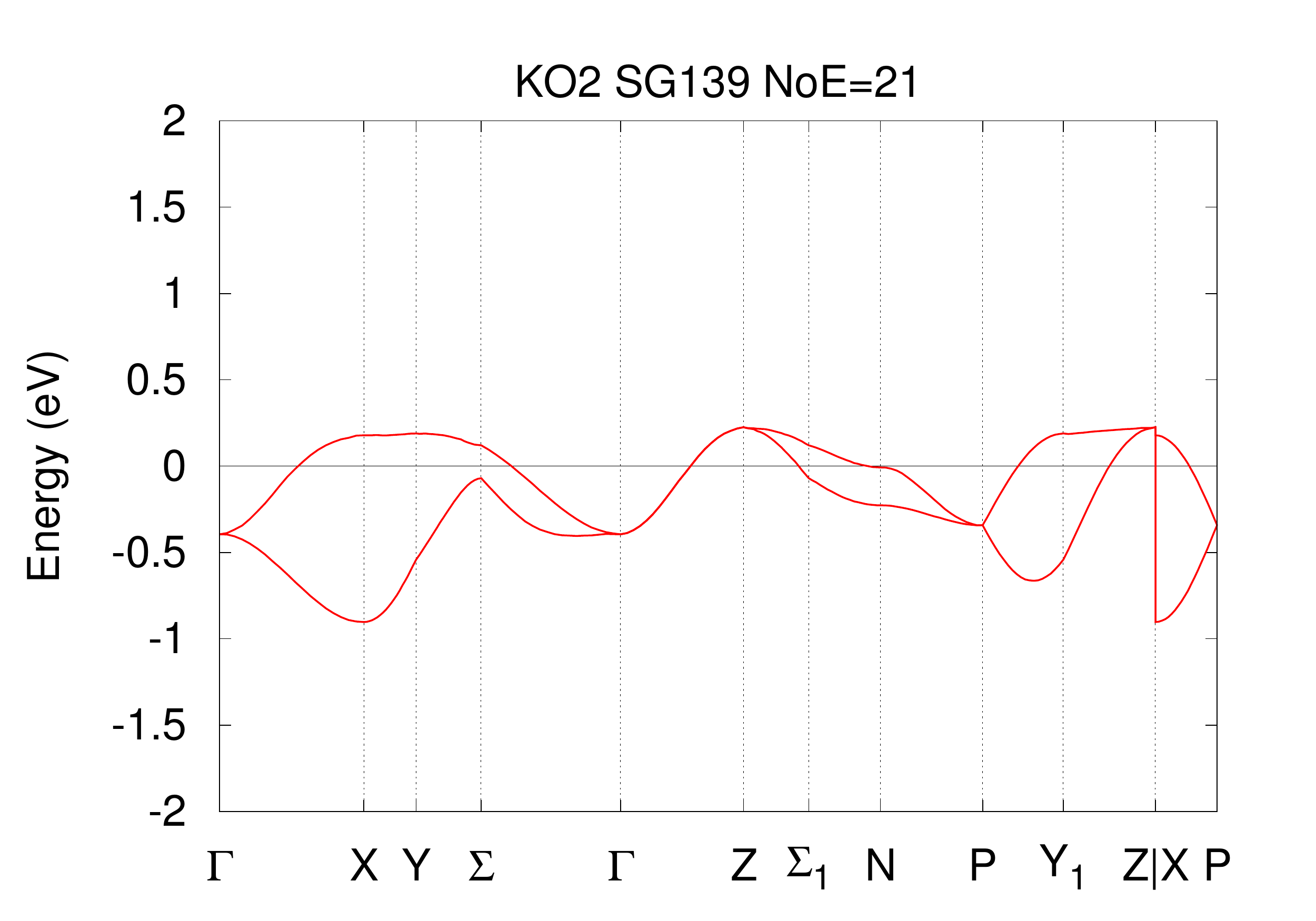}
}
\subfigure[HfN$_{2}$ SG187 NoA=3 NoE=14]{
\label{subfig:290427}
\includegraphics[scale=0.32,angle=270]{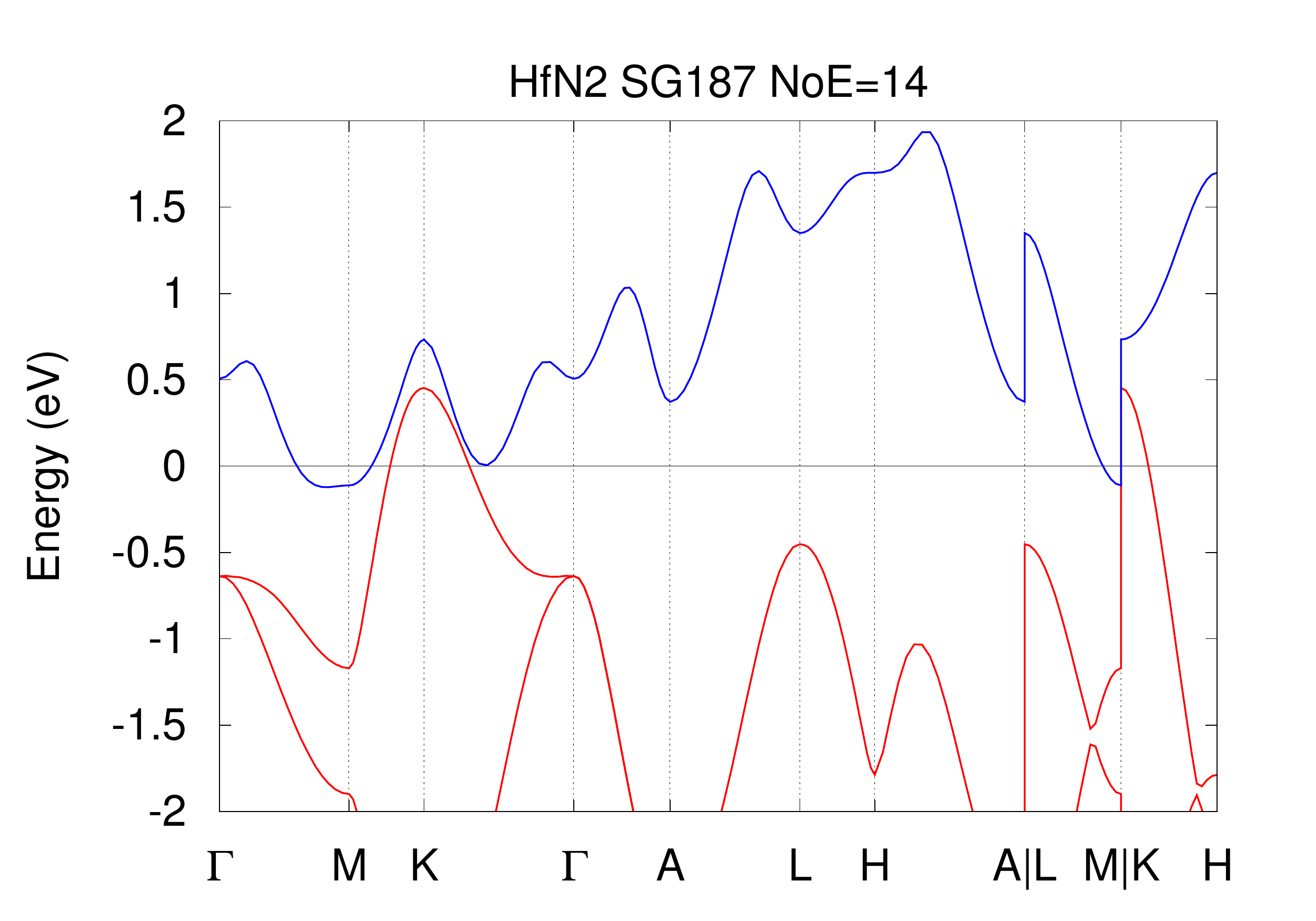}
}
\subfigure[HgO$_{2}$ SG12 NoA=3 NoE=24]{
\label{subfig:48214}
\includegraphics[scale=0.32,angle=270]{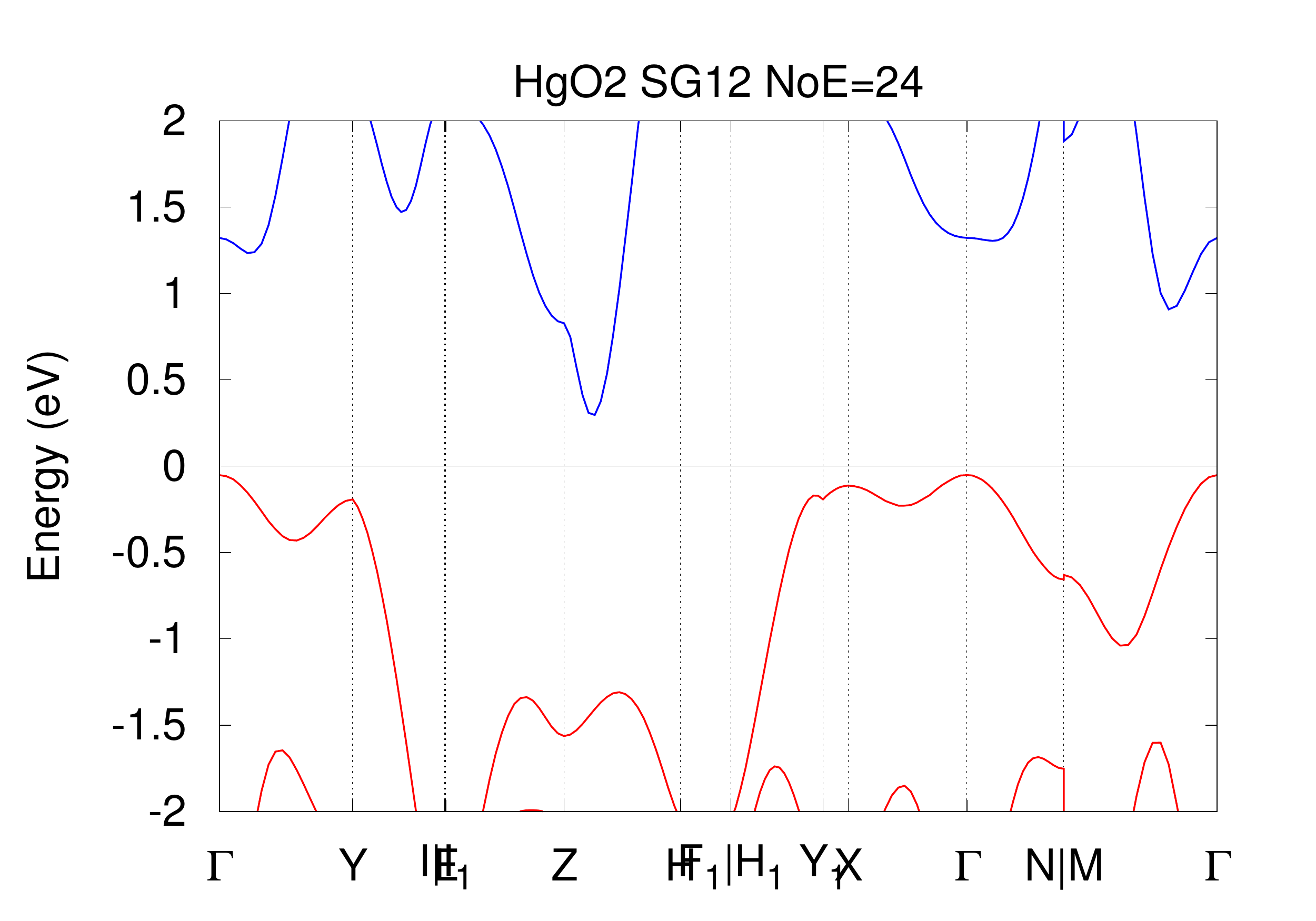}
}
\caption{\hyperref[tab:electride]{back to the table}}
\end{figure}

\begin{figure}[htp]
 \centering
\subfigure[CrSe$_{2}$ SG12 NoA=3 NoE=18]{
\label{subfig:251718}
\includegraphics[scale=0.32,angle=270]{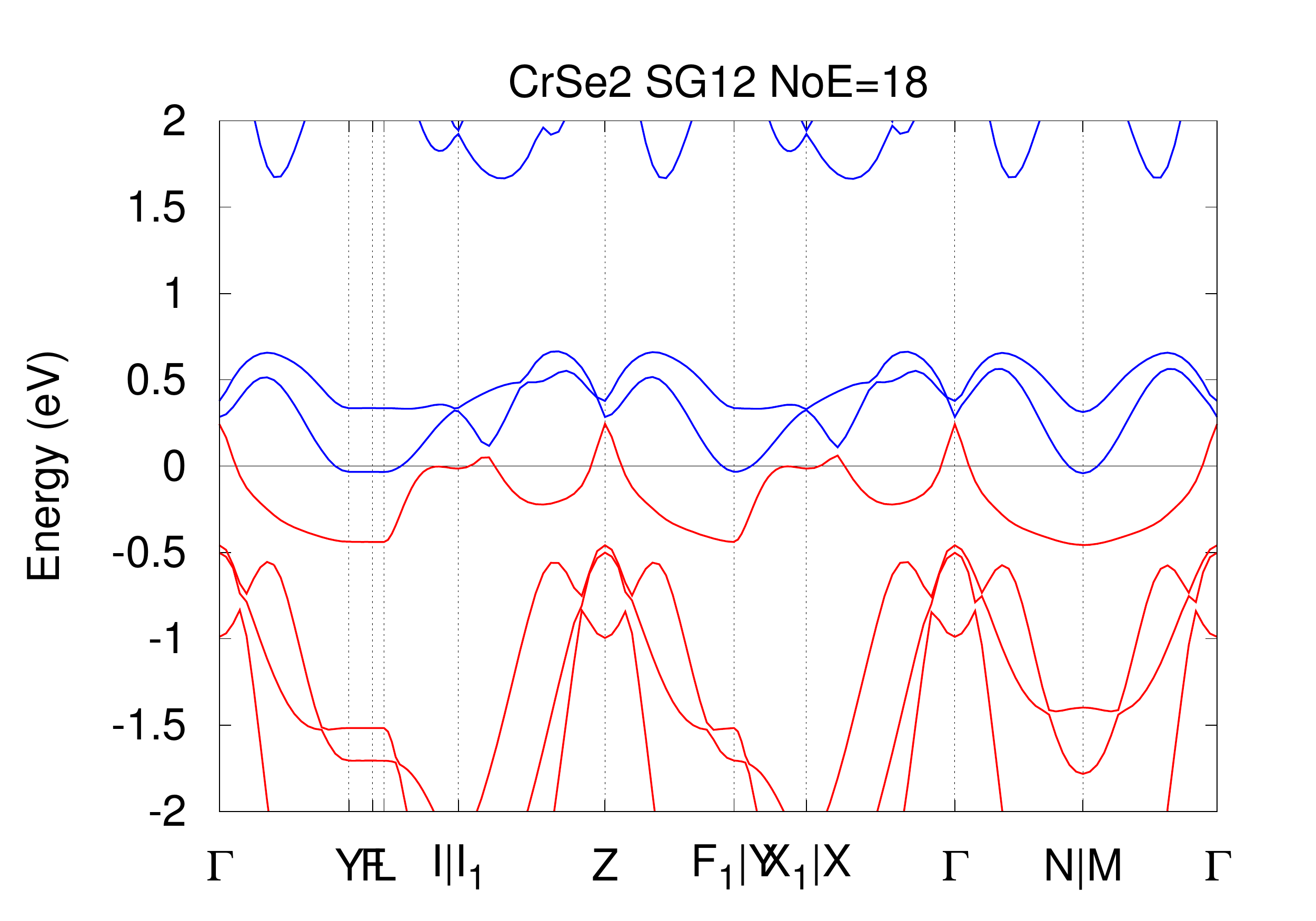}
}
\subfigure[VGaFe$_{2}$ SG225 NoA=4 NoE=24]{
\label{subfig:631850}
\includegraphics[scale=0.32,angle=270]{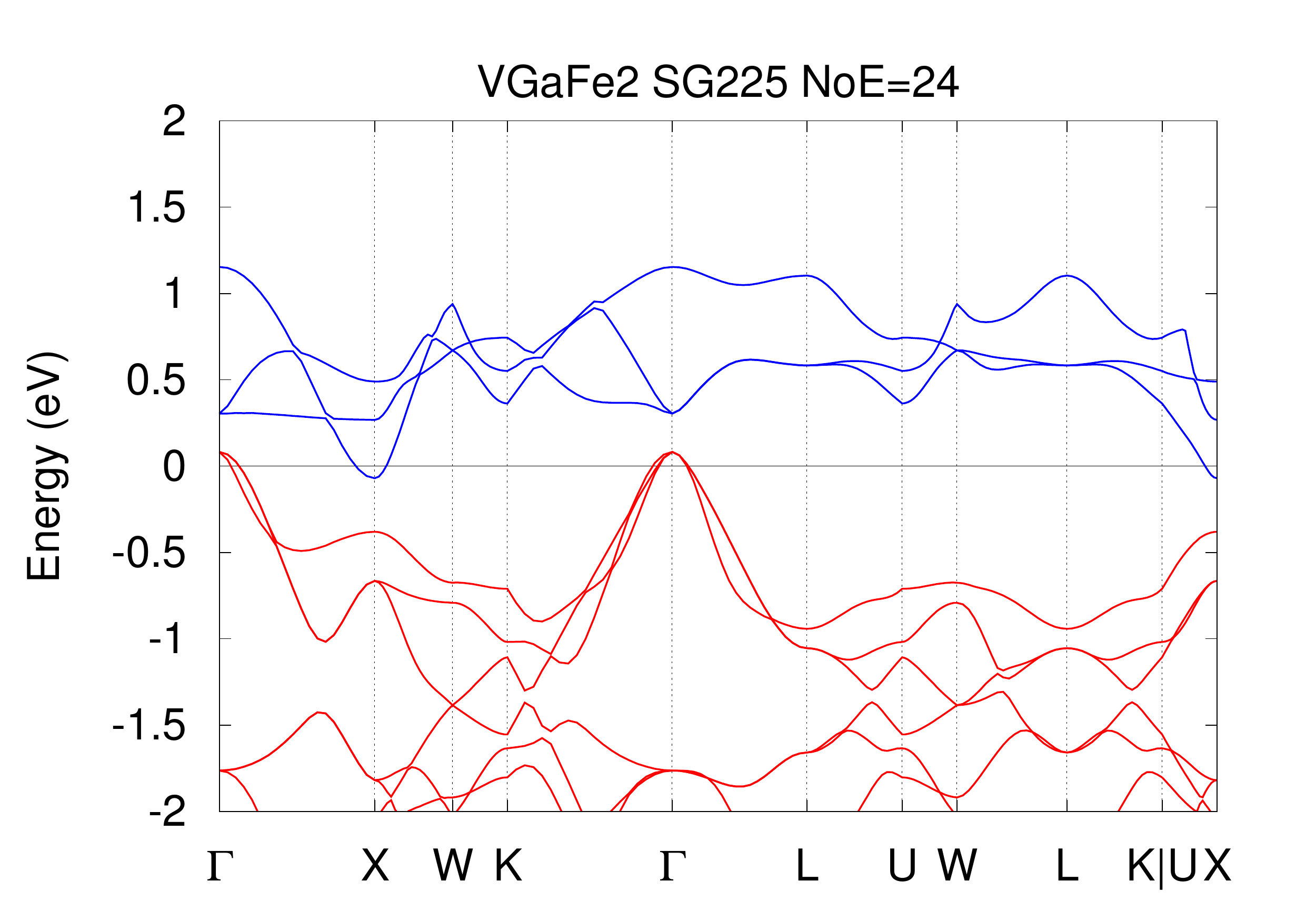}
}
\subfigure[Si SG194 NoA=4 NoE=16]{
\label{subfig:30101}
\includegraphics[scale=0.32,angle=270]{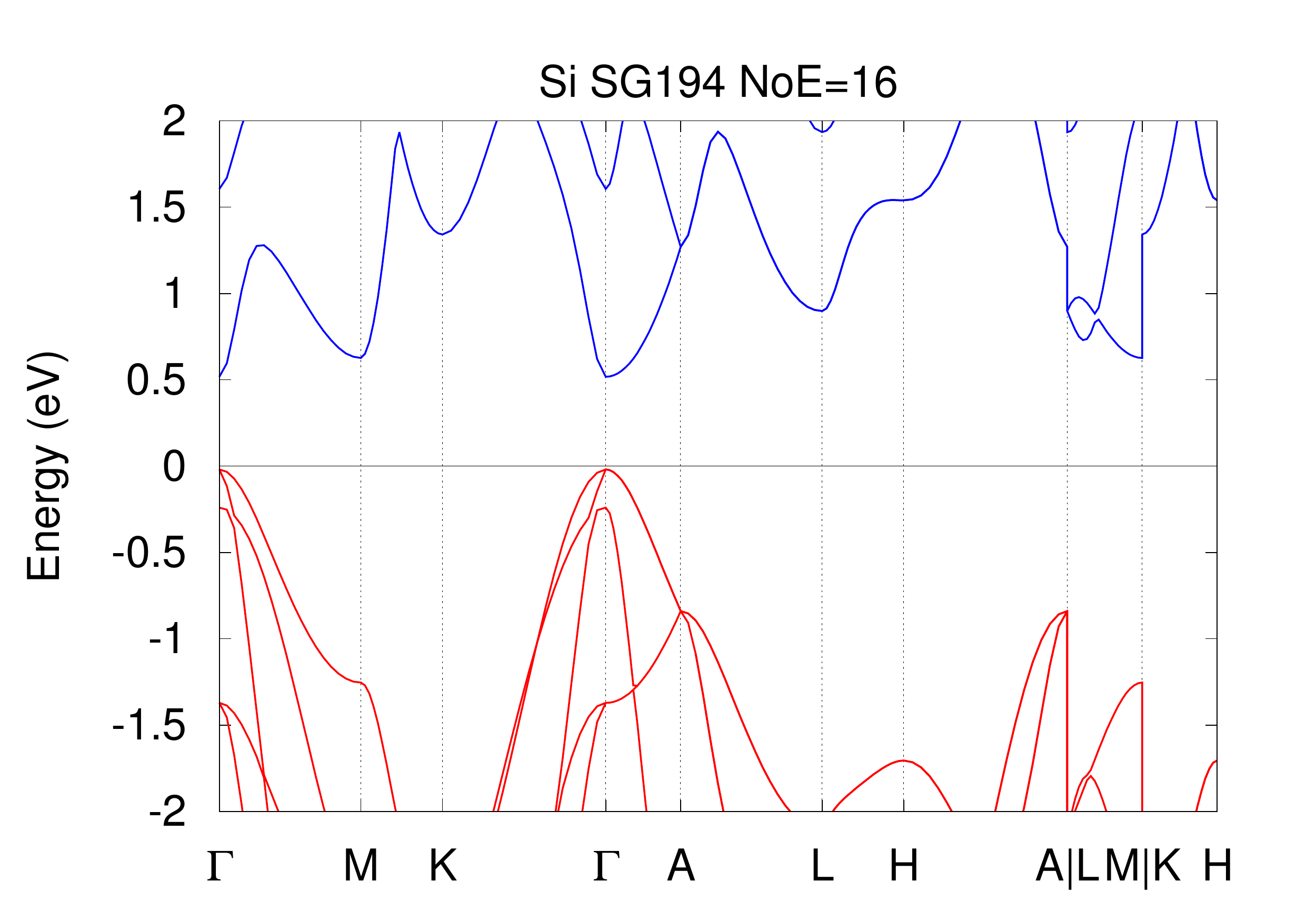}
}
\subfigure[TiFe$_{2}$Sn SG225 NoA=4 NoE=24]{
\label{subfig:633766}
\includegraphics[scale=0.32,angle=270]{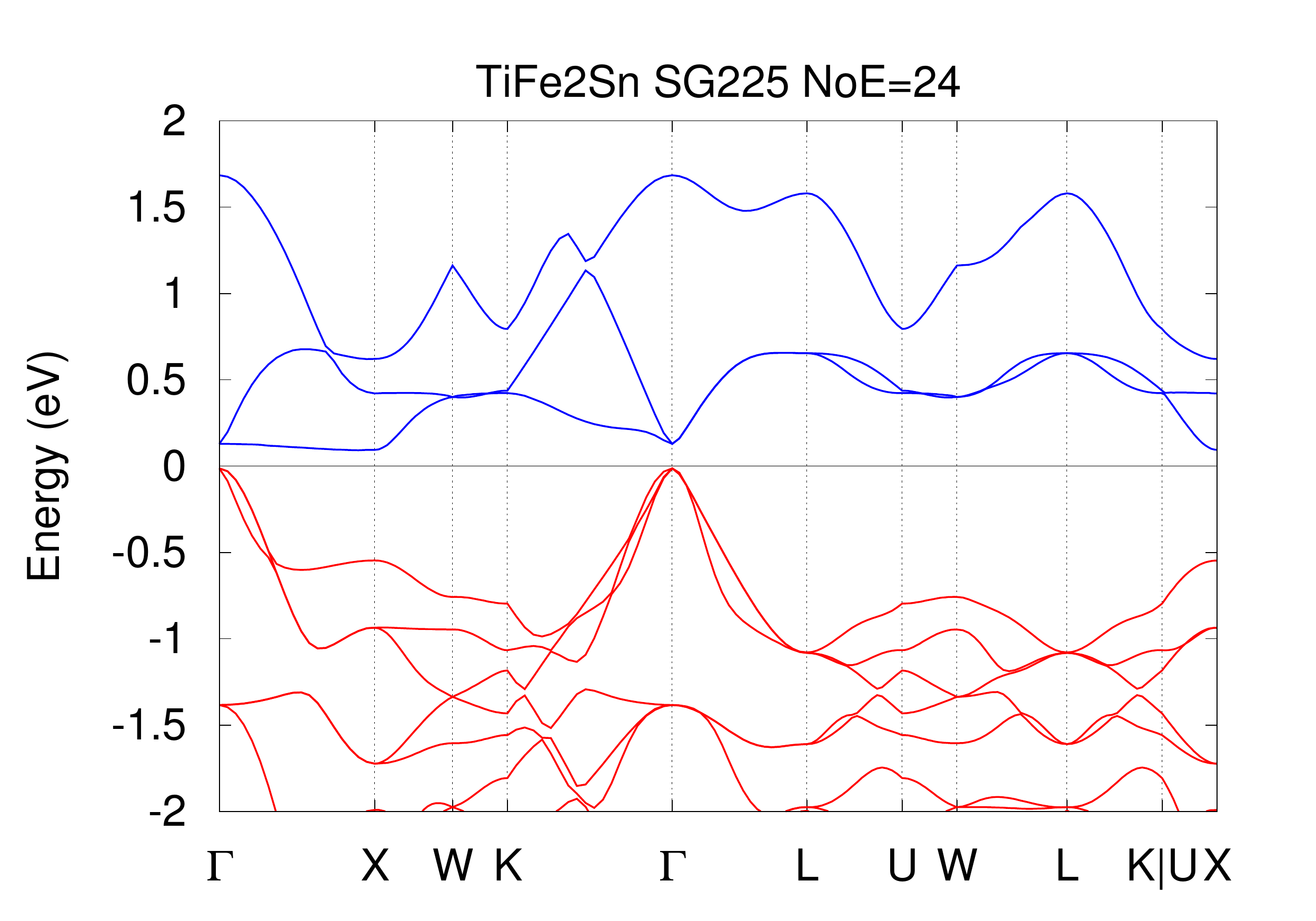}
}
\subfigure[TiAlFe$_{2}$ SG225 NoA=4 NoE=23]{
\label{subfig:57827}
\includegraphics[scale=0.32,angle=270]{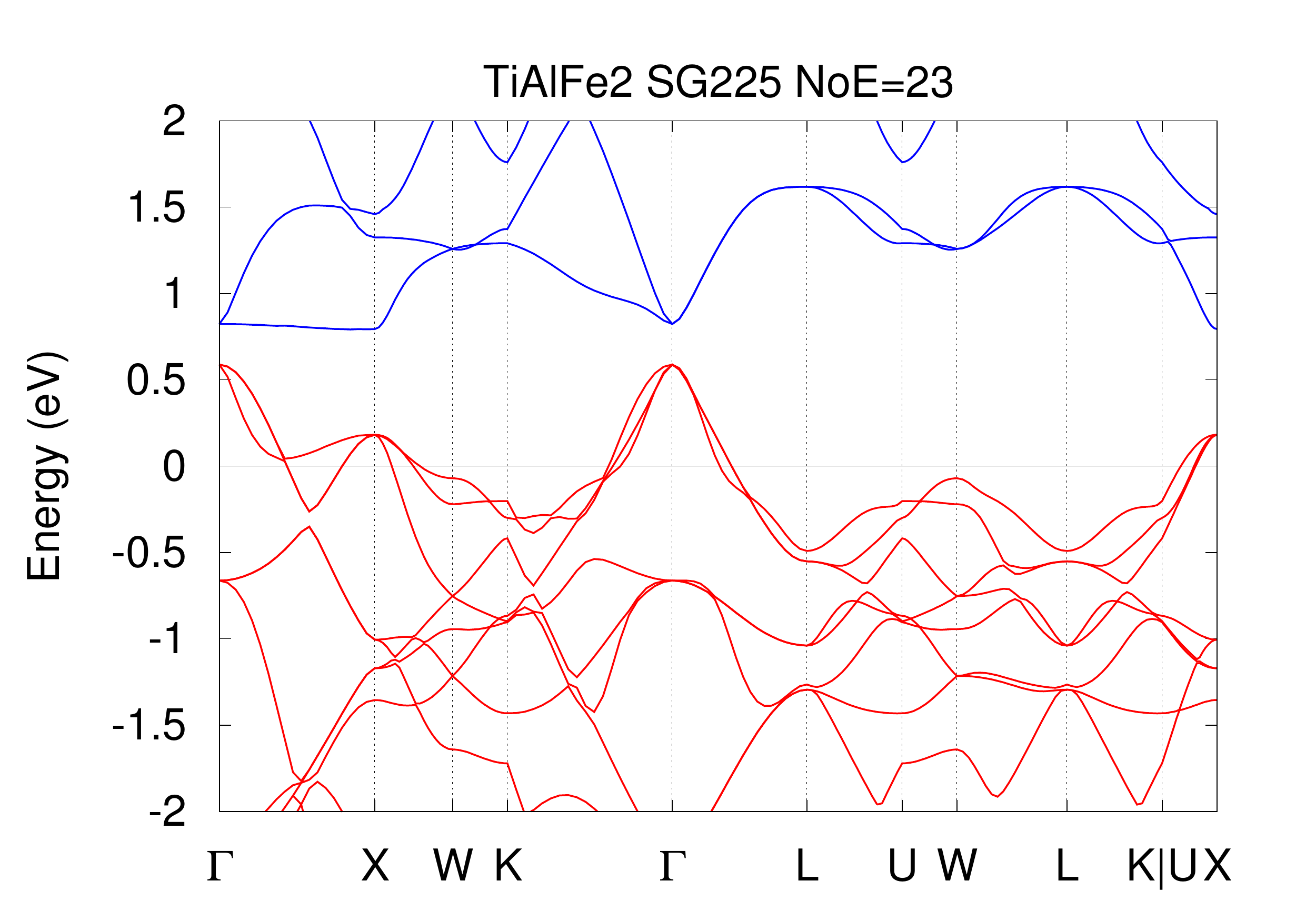}
}
\subfigure[ZrCl SG164 NoA=4 NoE=38]{
\label{subfig:35701}
\includegraphics[scale=0.32,angle=270]{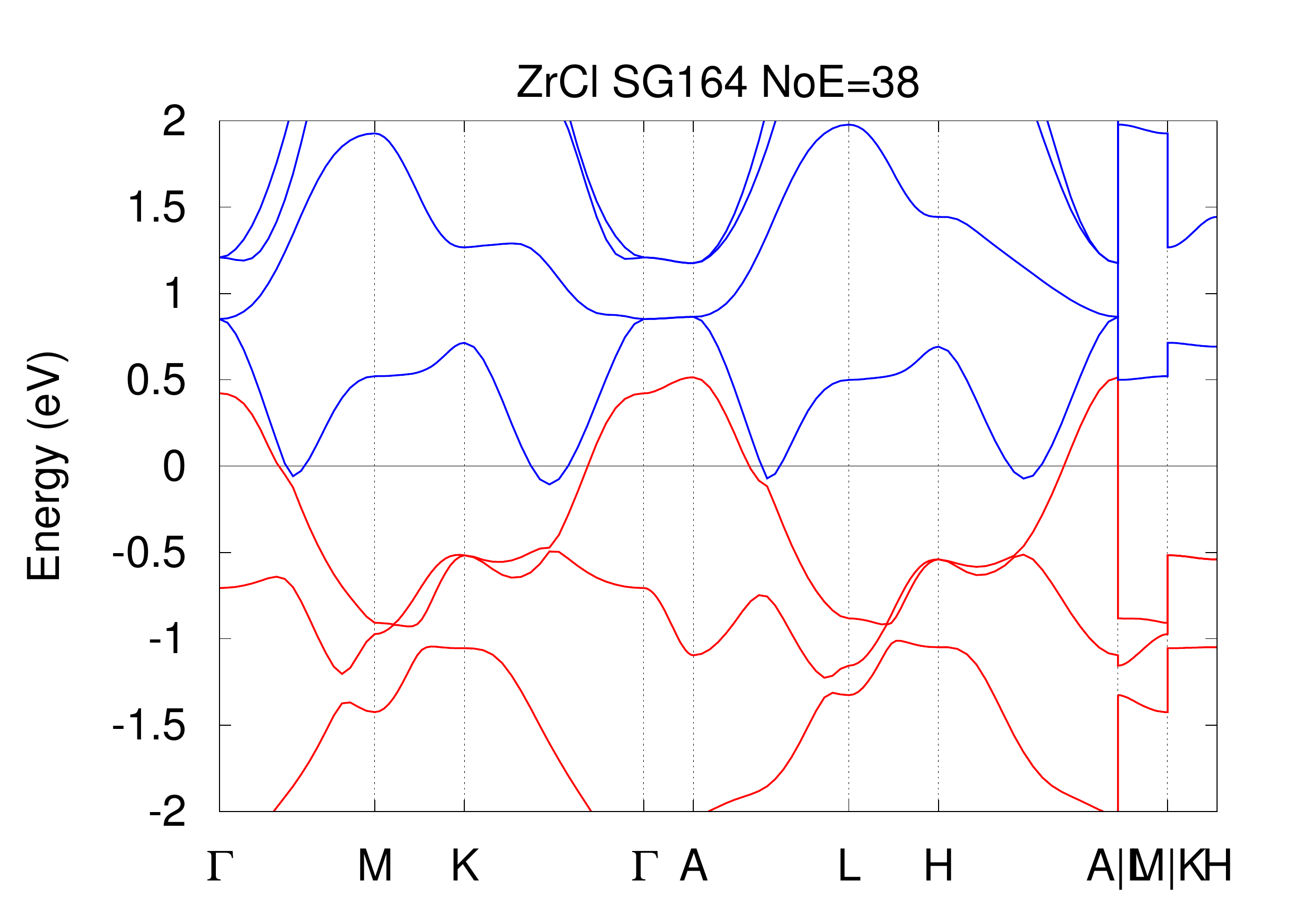}
}
\subfigure[FeSiRu$_{2}$ SG225 NoA=4 NoE=28]{
\label{subfig:633246}
\includegraphics[scale=0.32,angle=270]{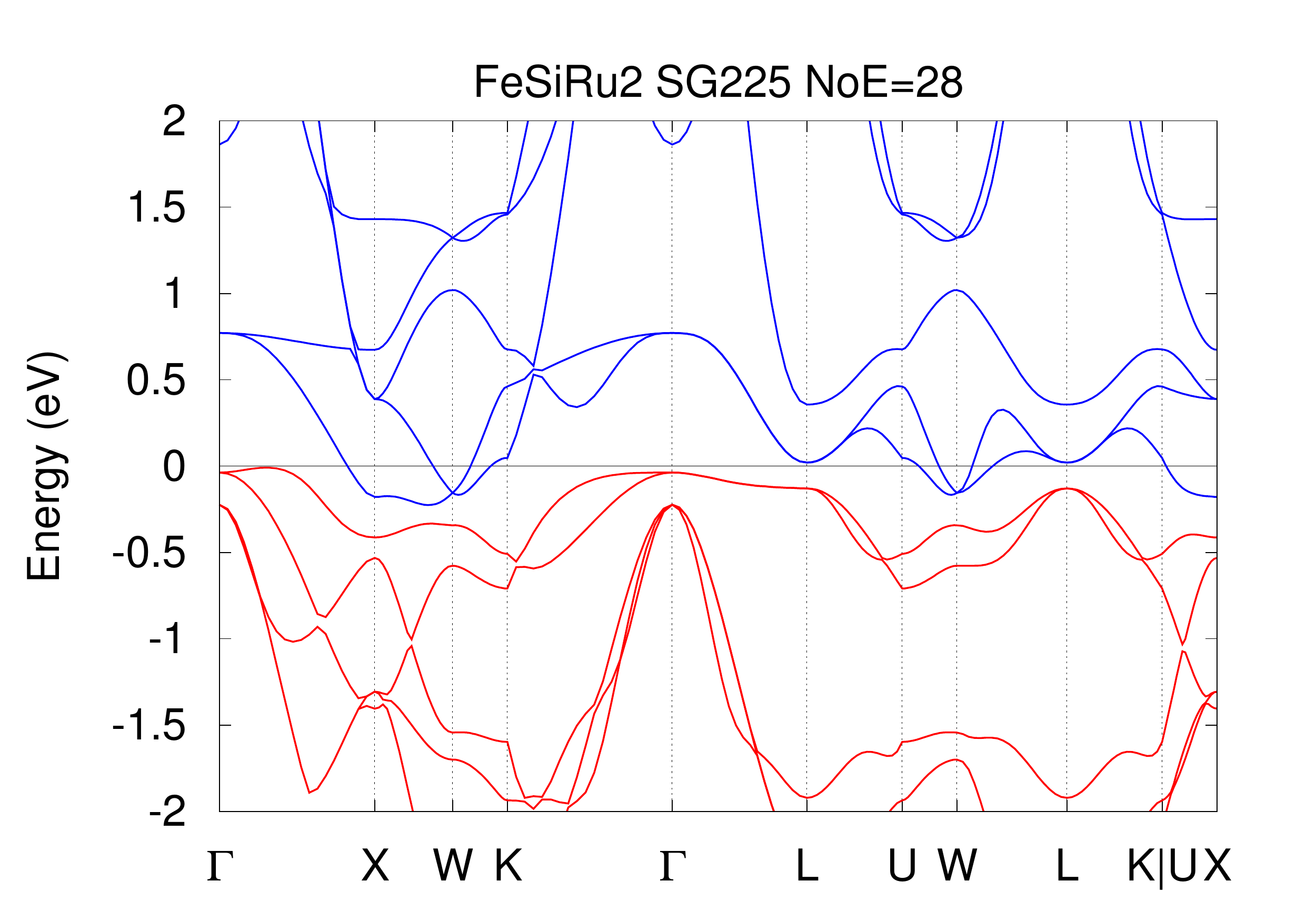}
}
\subfigure[AlVFe$_{2}$ SG225 NoA=4 NoE=24]{
\label{subfig:57832}
\includegraphics[scale=0.32,angle=270]{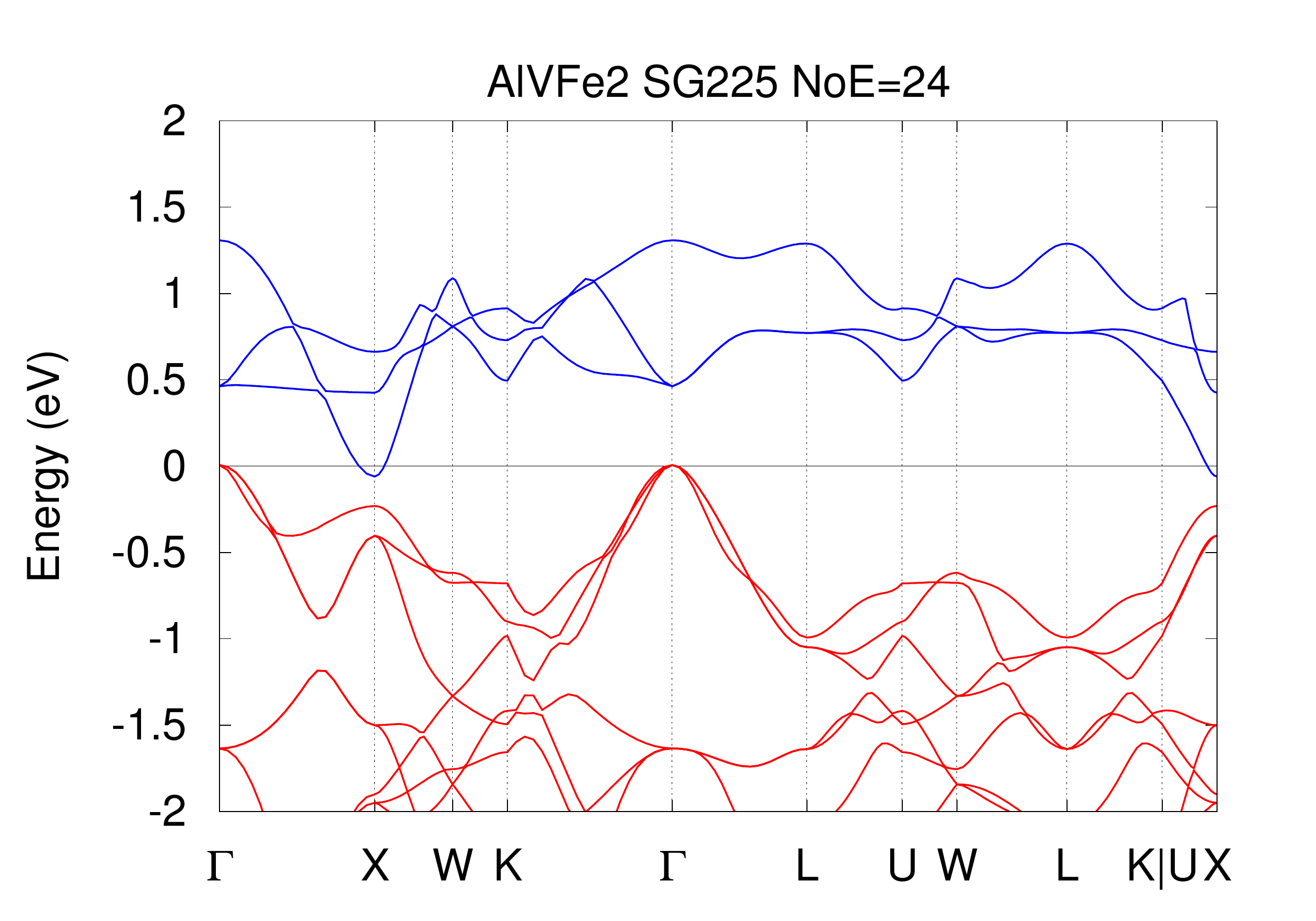}
}
\caption{\hyperref[tab:electride]{back to the table}}
\end{figure}

\begin{figure}[htp]
 \centering
\subfigure[LiAl SG227 NoA=4 NoE=8]{
\label{subfig:240110}
\includegraphics[scale=0.32,angle=270]{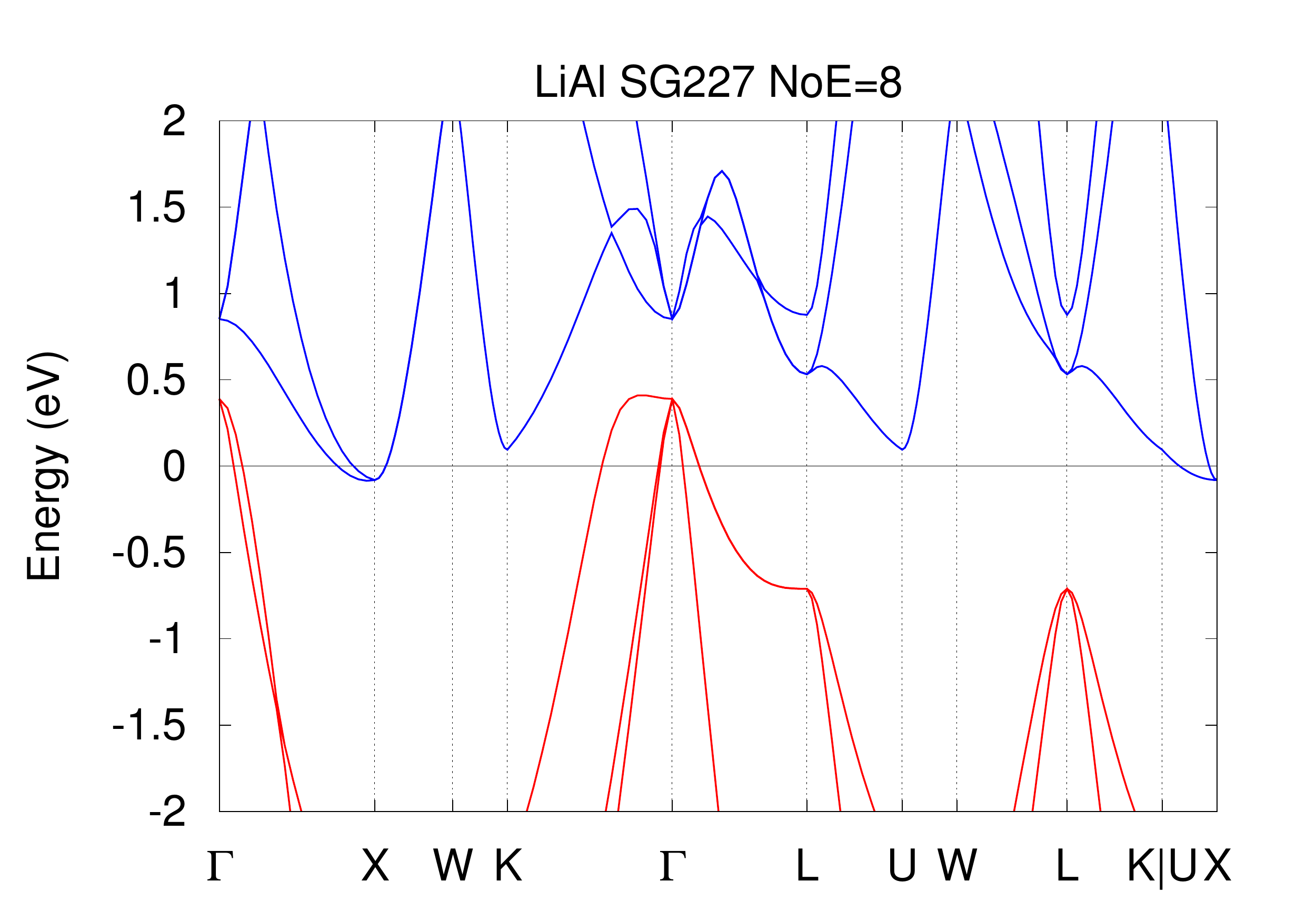}
}
\subfigure[Mg$_{3}$In SG221 NoA=4 NoE=9]{
\label{subfig:51975}
\includegraphics[scale=0.32,angle=270]{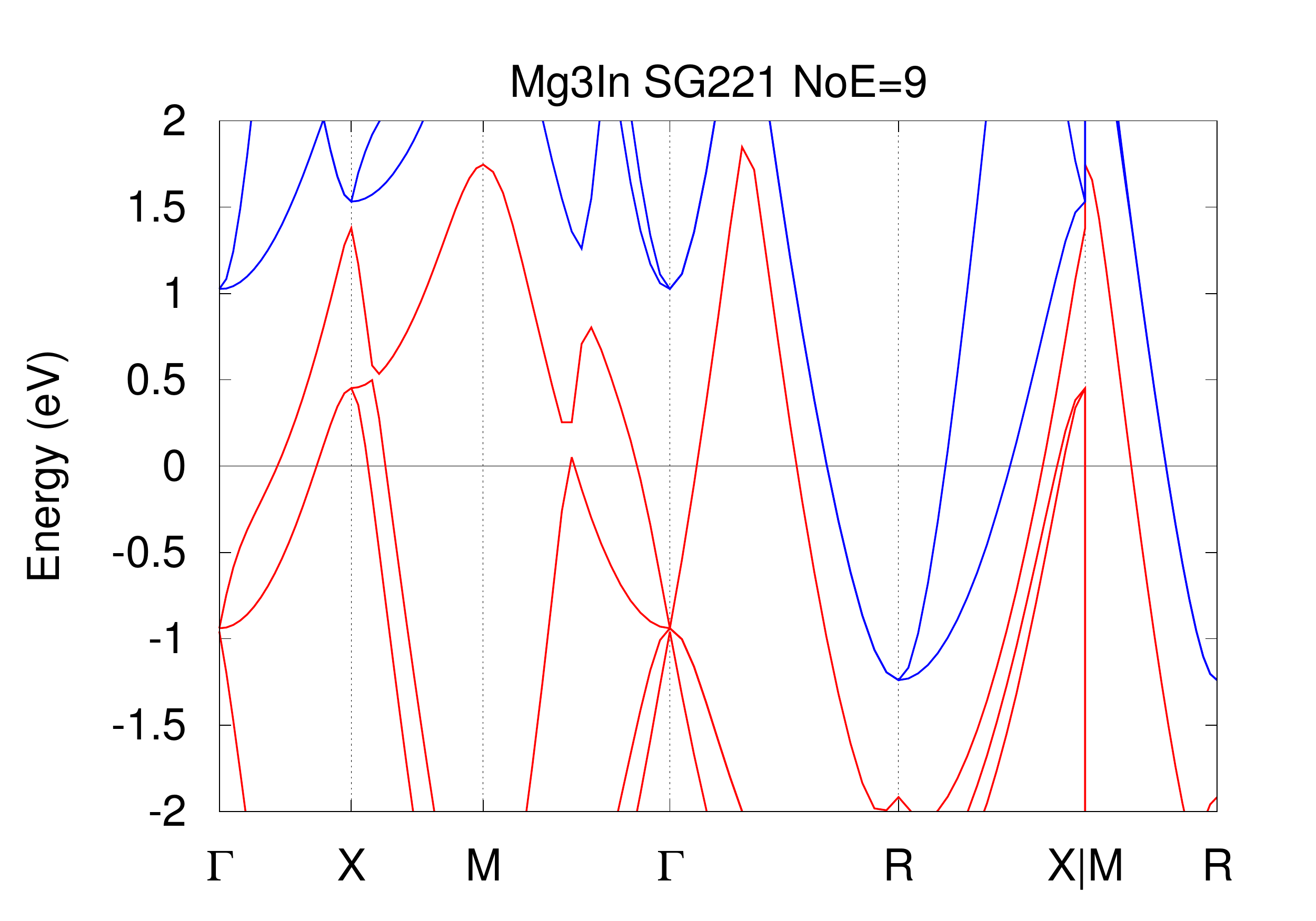}
}
\subfigure[ZrCd SG129 NoA=4 NoE=48]{
\label{subfig:620612}
\includegraphics[scale=0.32,angle=270]{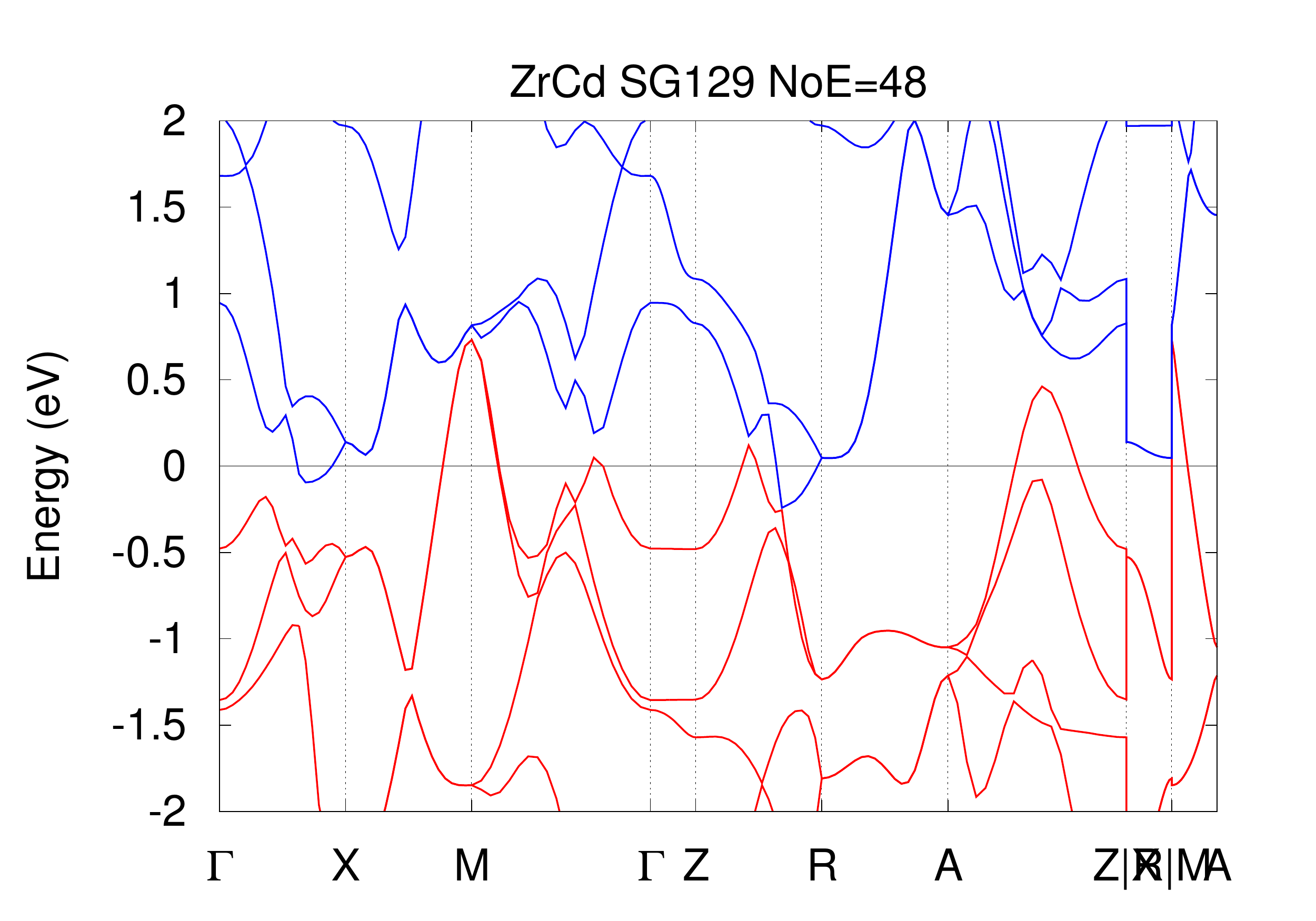}
}
\subfigure[HfBe SG63 NoA=4 NoE=12]{
\label{subfig:616286}
\includegraphics[scale=0.32,angle=270]{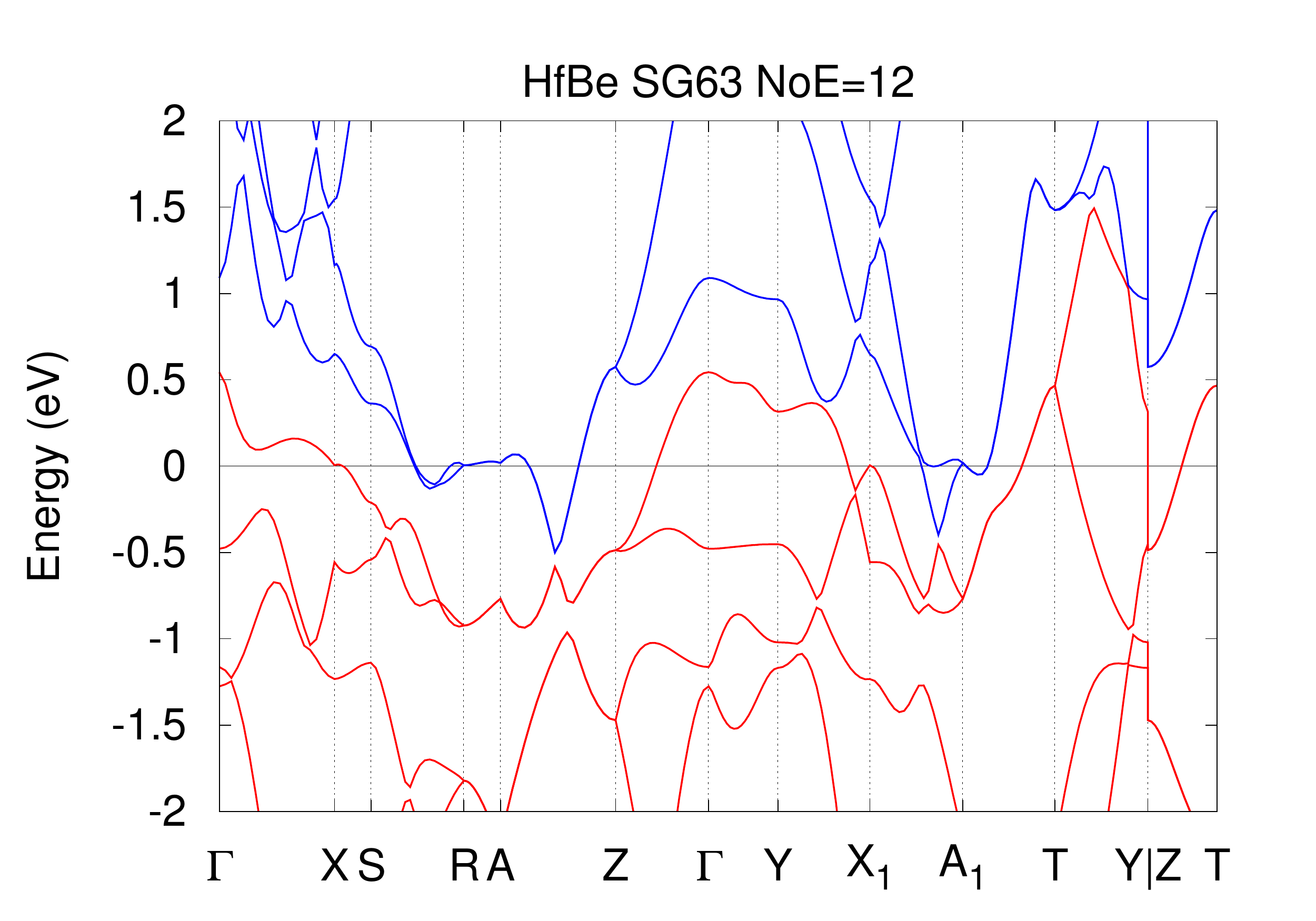}
}
\subfigure[SrMgIn$_{3}$ SG119 NoA=5 NoE=21]{
\label{subfig:249592}
\includegraphics[scale=0.32,angle=270]{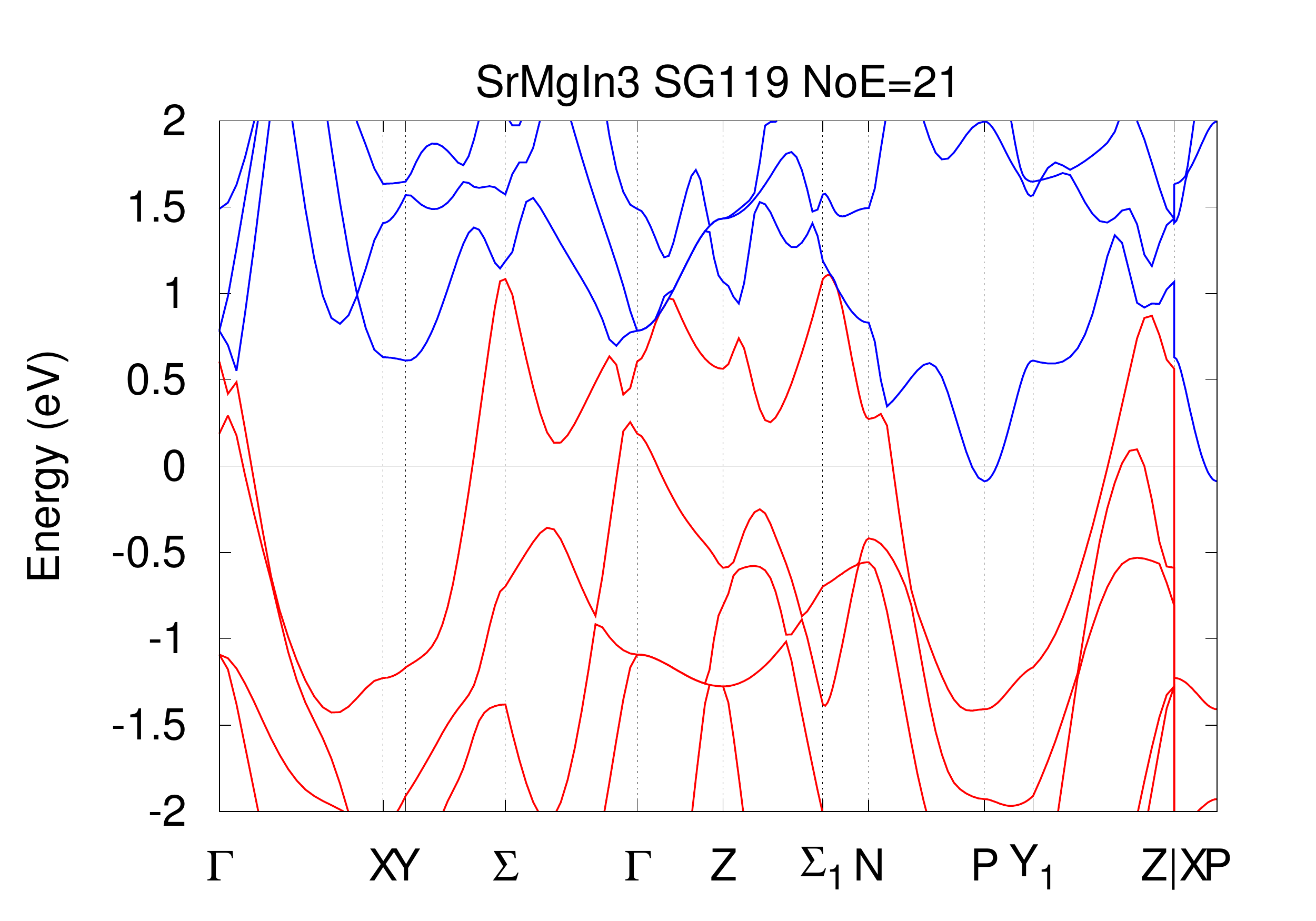}
}
\subfigure[ScInCu$_{4}$ SG216 NoA=6 NoE=50]{
\label{subfig:416528}
\includegraphics[scale=0.32,angle=270]{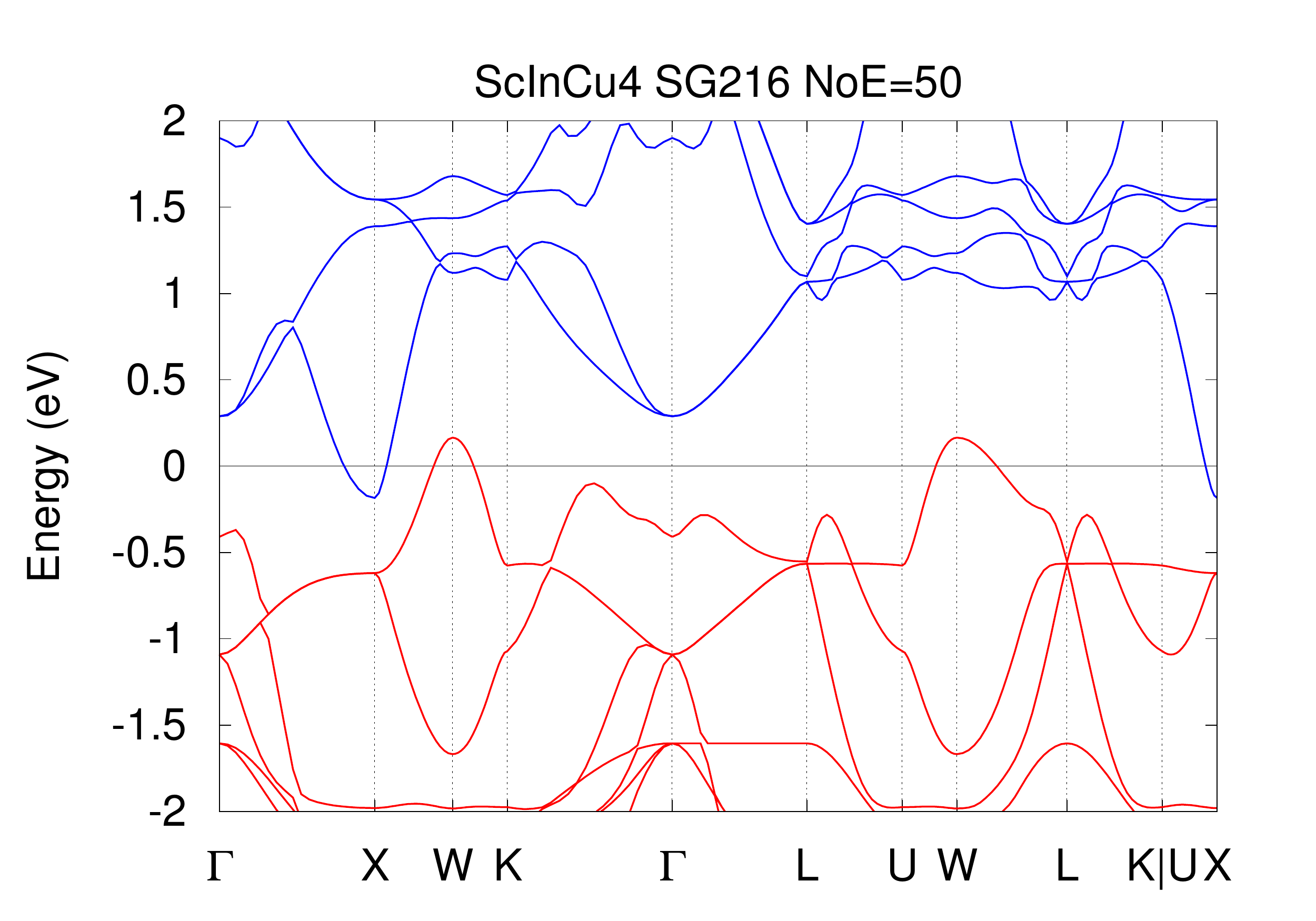}
}
\subfigure[MgCu$_{4}$Sn SG216 NoA=6 NoE=50]{
\label{subfig:103055}
\includegraphics[scale=0.32,angle=270]{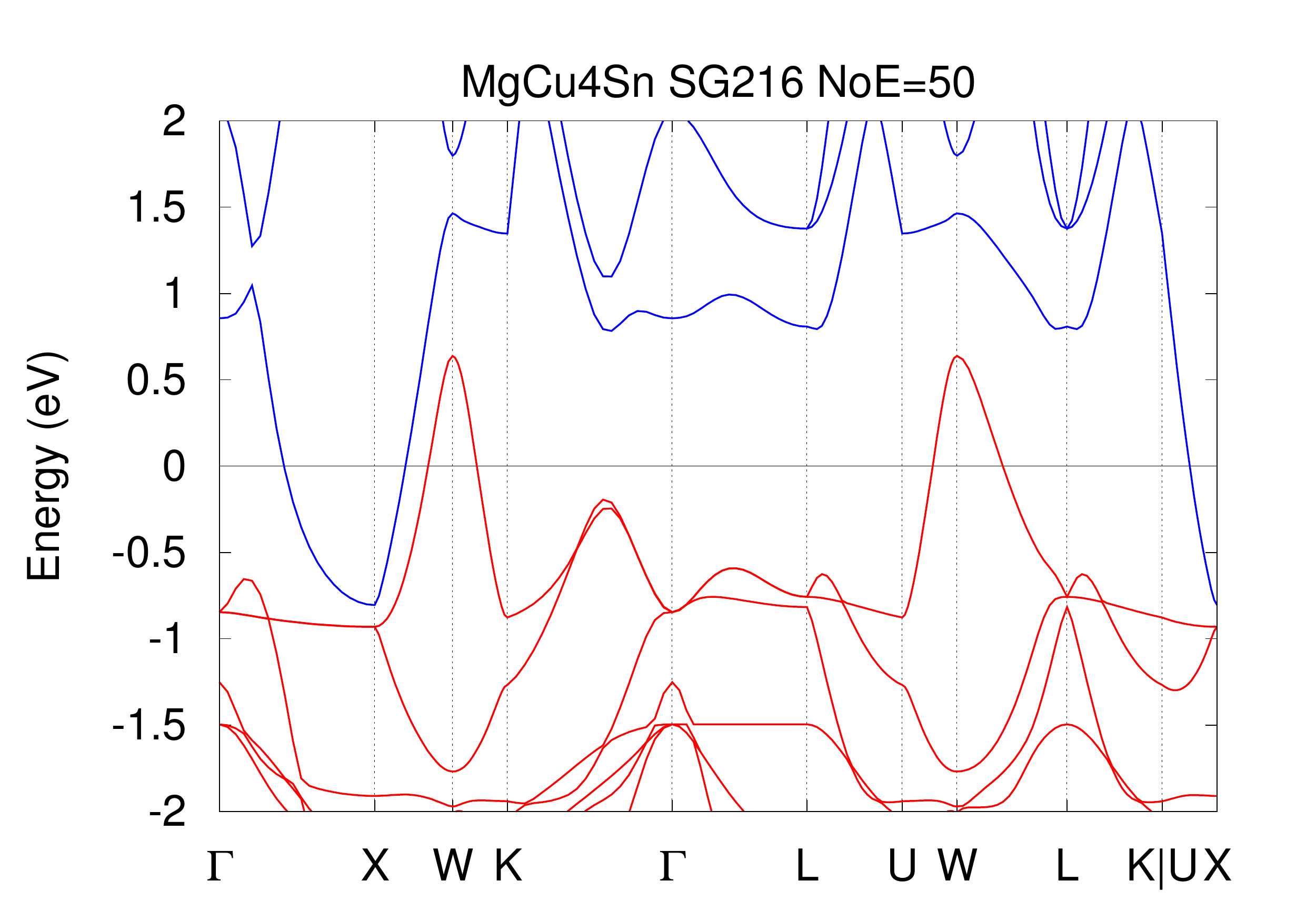}
}
\subfigure[P$_{2}$Ru SG58 NoA=6 NoE=36]{
\label{subfig:42607}
\includegraphics[scale=0.32,angle=270]{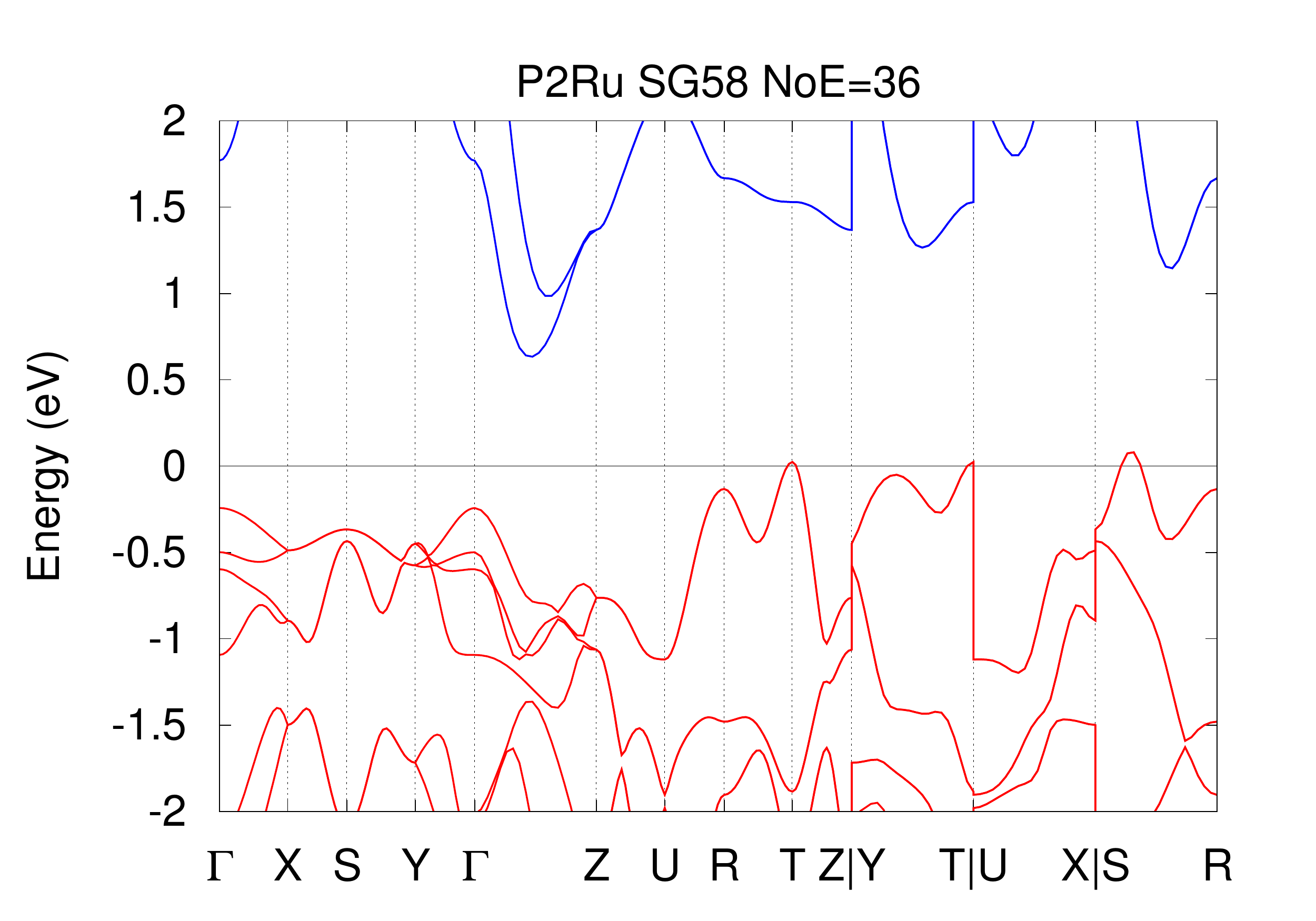}
}
\caption{\hyperref[tab:electride]{back to the table}}
\end{figure}

\begin{figure}[htp]
 \centering
\subfigure[Sb$_{2}$Ru SG58 NoA=6 NoE=36]{
\label{subfig:43652}
\includegraphics[scale=0.32,angle=270]{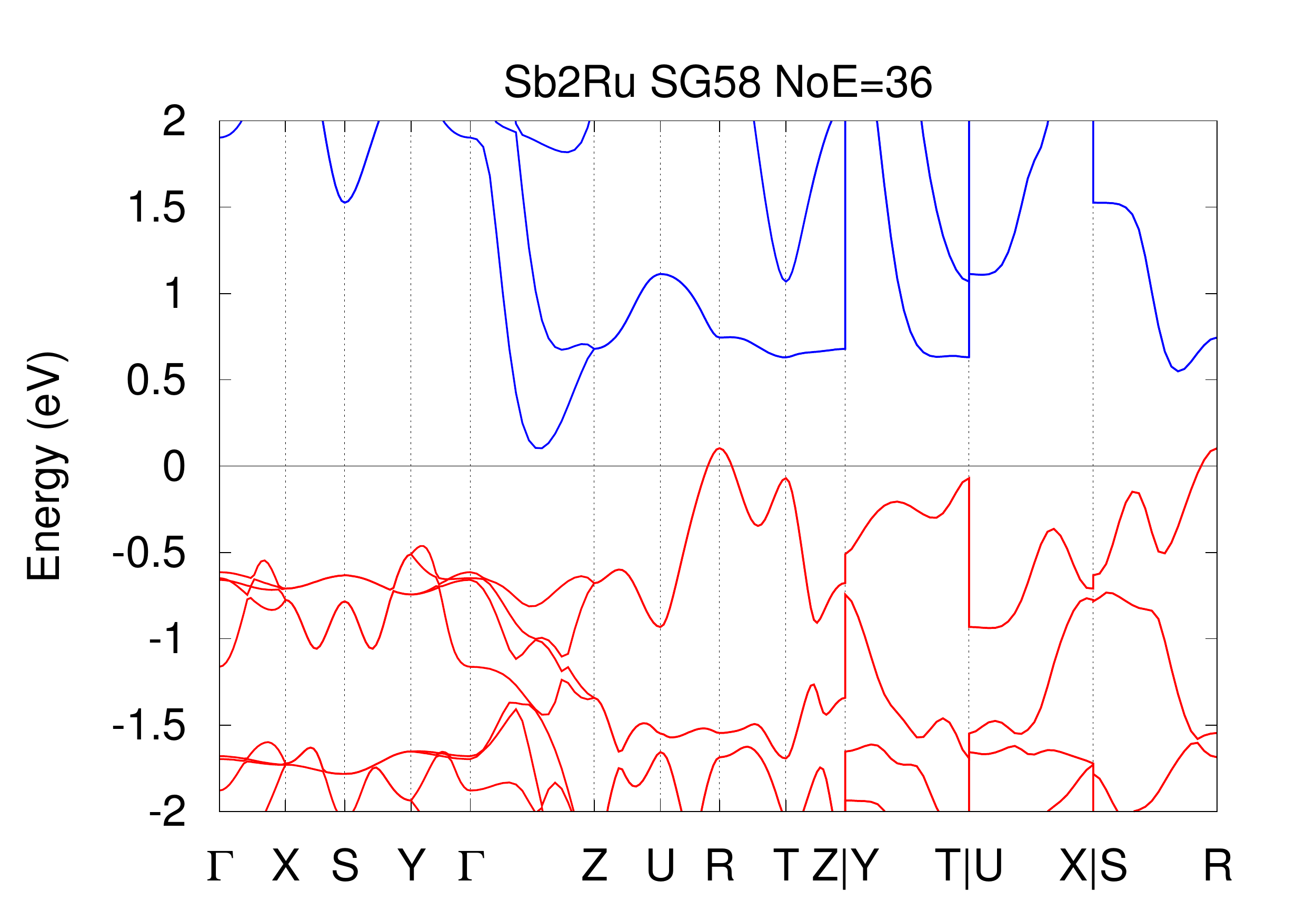}
}
\subfigure[YMgCu$_{4}$ SG216 NoA=6 NoE=57]{
\label{subfig:163696}
\includegraphics[scale=0.32,angle=270]{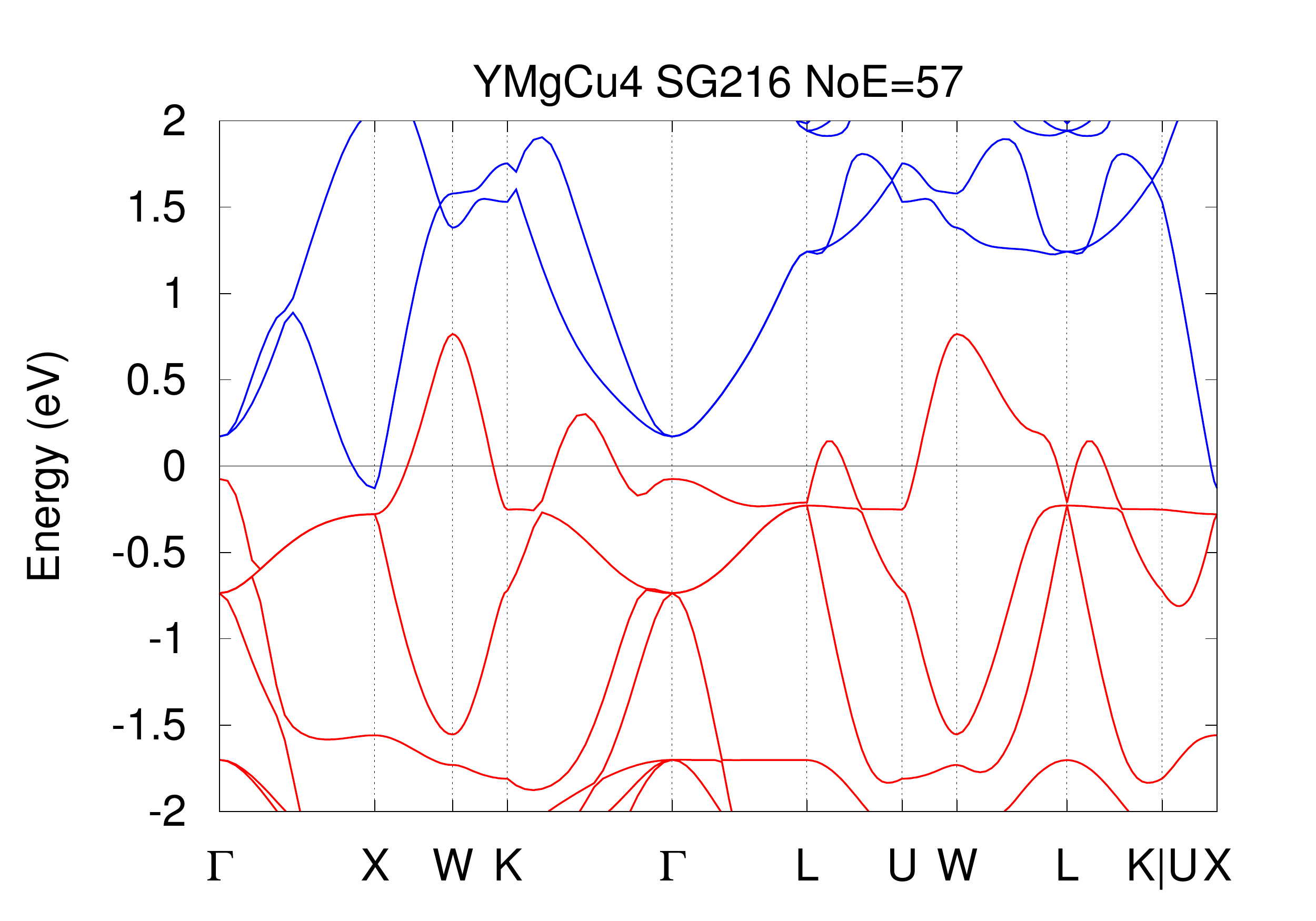}
}
\subfigure[FeSb$_{2}$ SG58 NoA=6 NoE=36]{
\label{subfig:186627}
\includegraphics[scale=0.32,angle=270]{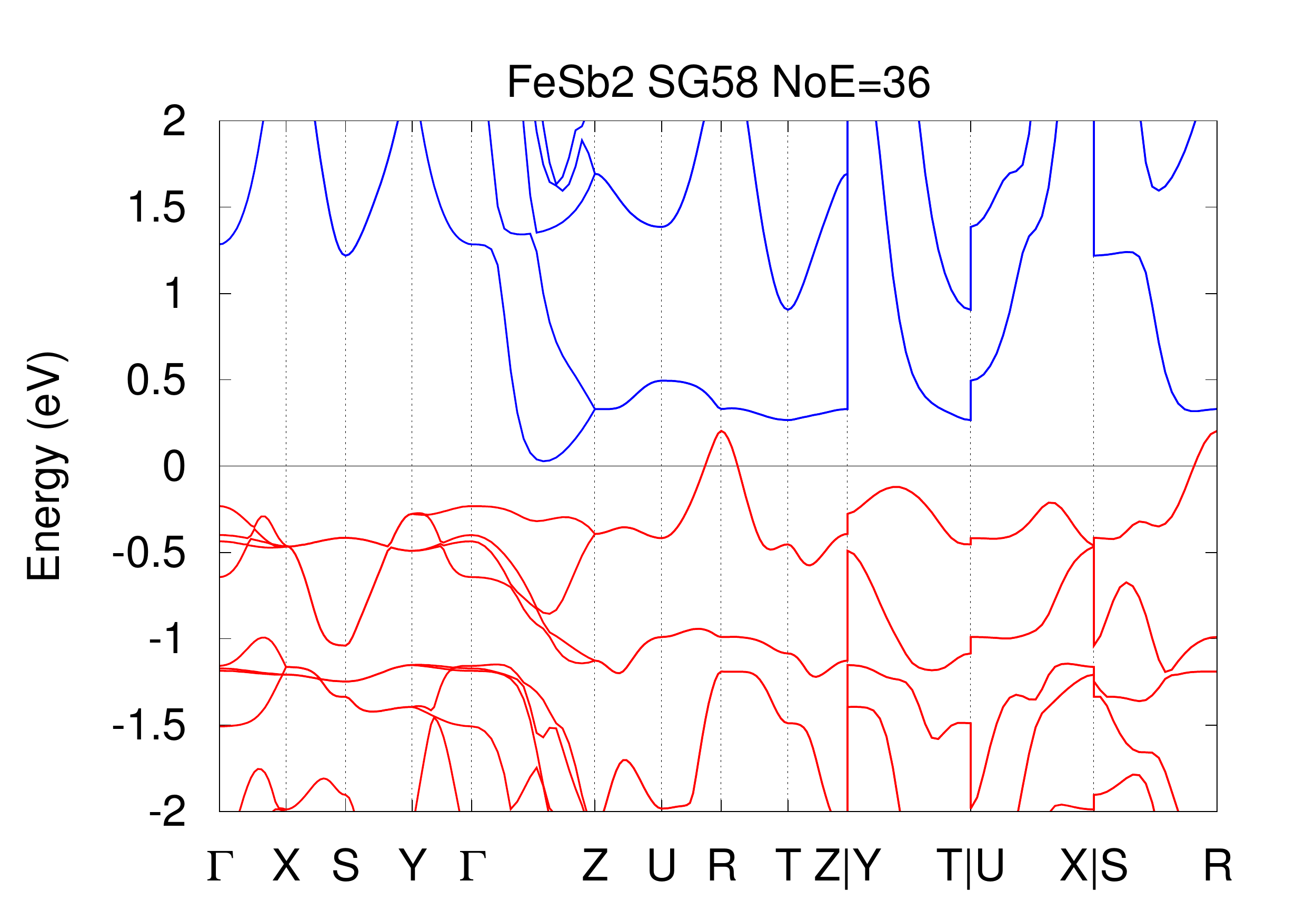}
}
\subfigure[Be$_{5}$Co SG216 NoA=6 NoE=19]{
\label{subfig:616209}
\includegraphics[scale=0.32,angle=270]{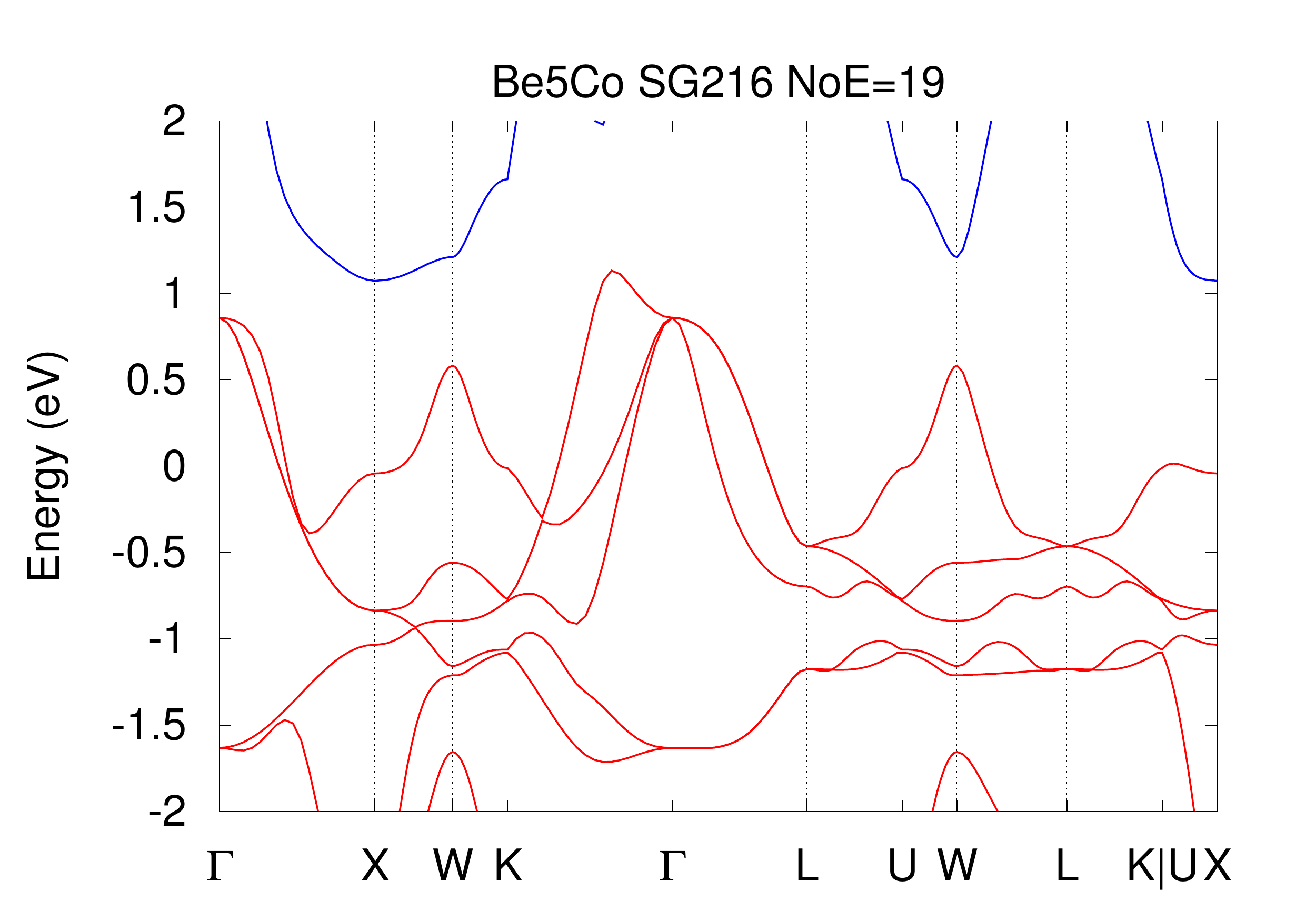}
}
\subfigure[Be$_{5}$Pt SG216 NoA=6 NoE=20]{
\label{subfig:616395}
\includegraphics[scale=0.32,angle=270]{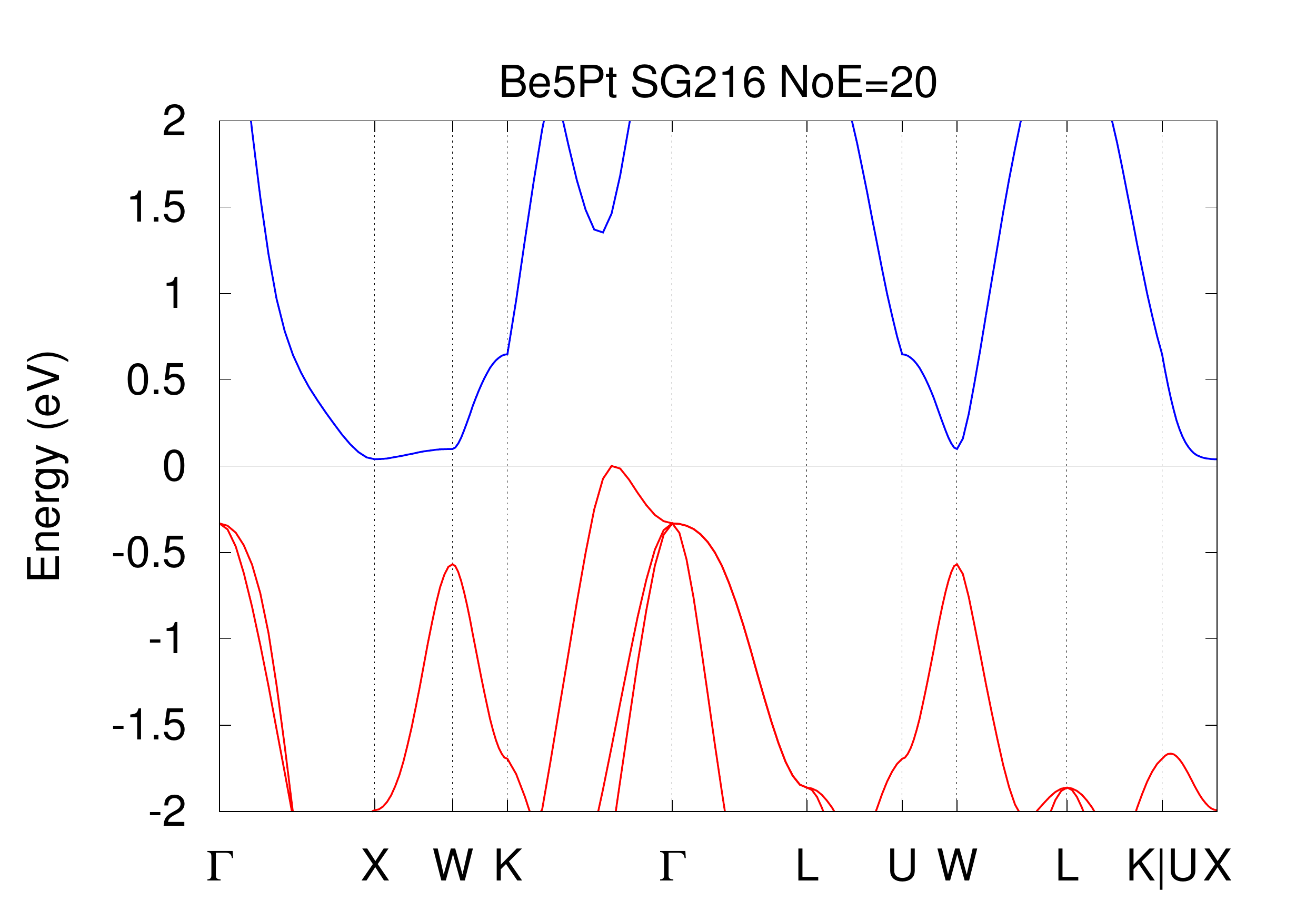}
}
\subfigure[HfGaAu SG187 NoA=6 NoE=36]{
\label{subfig:156265}
\includegraphics[scale=0.32,angle=270]{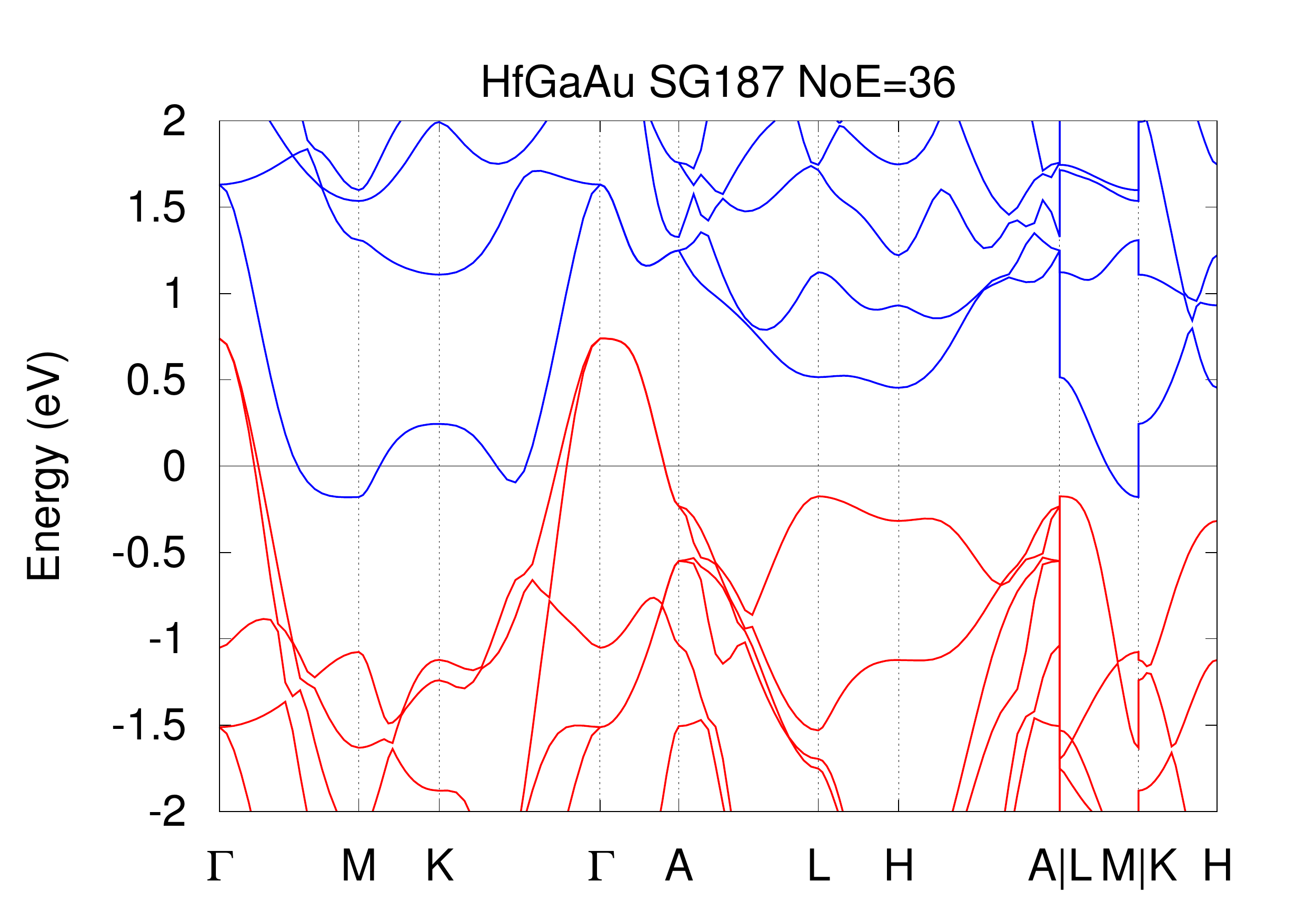}
}
\subfigure[P$_{2}$W SG12 NoA=6 NoE=32]{
\label{subfig:648286}
\includegraphics[scale=0.32,angle=270]{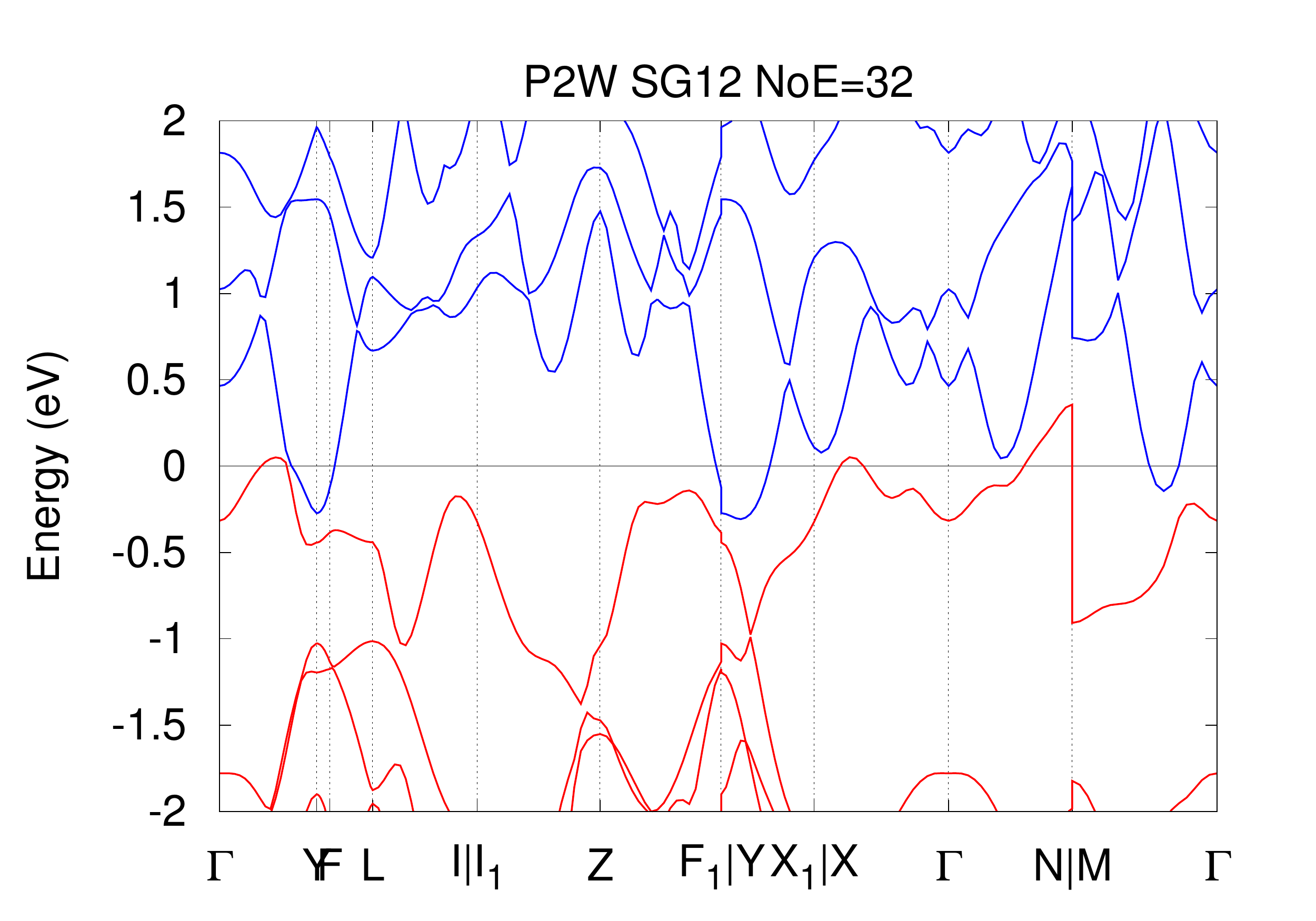}
}
\subfigure[Al$_{2}$Ru SG70 NoA=6 NoE=28]{
\label{subfig:58156}
\includegraphics[scale=0.32,angle=270]{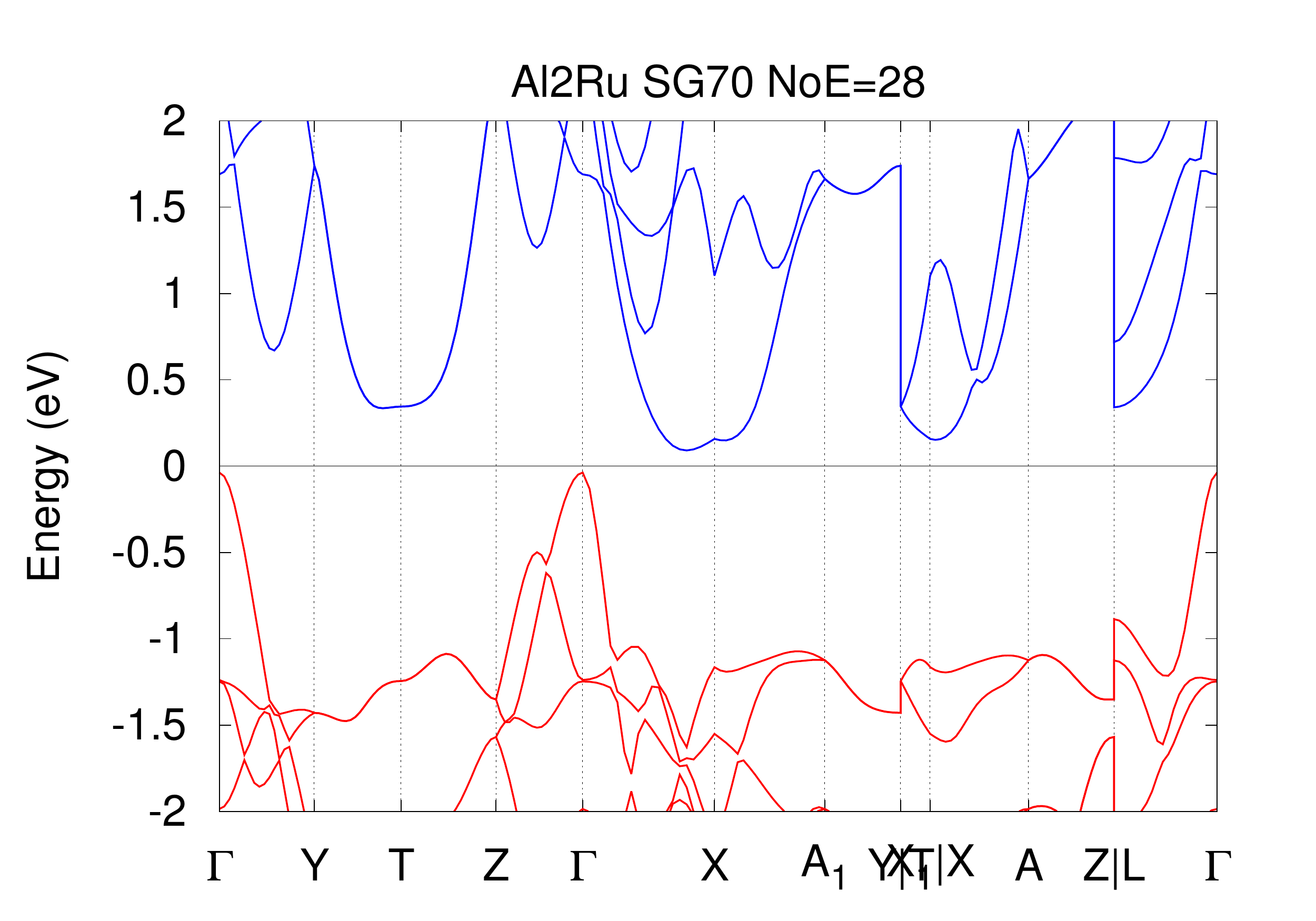}
}
\caption{\hyperref[tab:electride]{back to the table}}
\end{figure}

\begin{figure}[htp]
 \centering
\subfigure[HoCdCu$_{4}$ SG216 NoA=6 NoE=65]{
\label{subfig:415195}
\includegraphics[scale=0.32,angle=270]{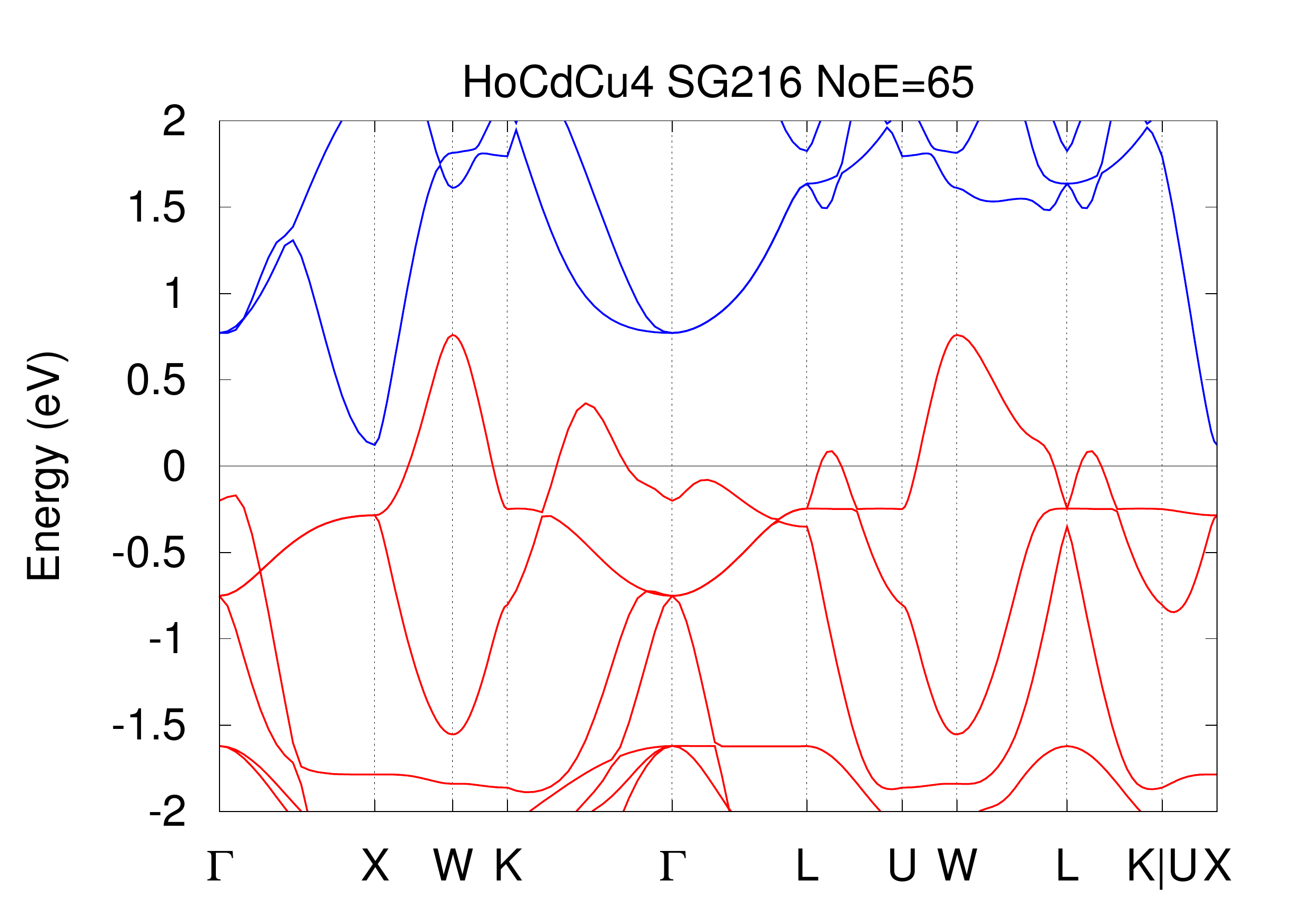}
}
\subfigure[YInCu$_{4}$ SG216 NoA=6 NoE=58]{
\label{subfig:628179}
\includegraphics[scale=0.32,angle=270]{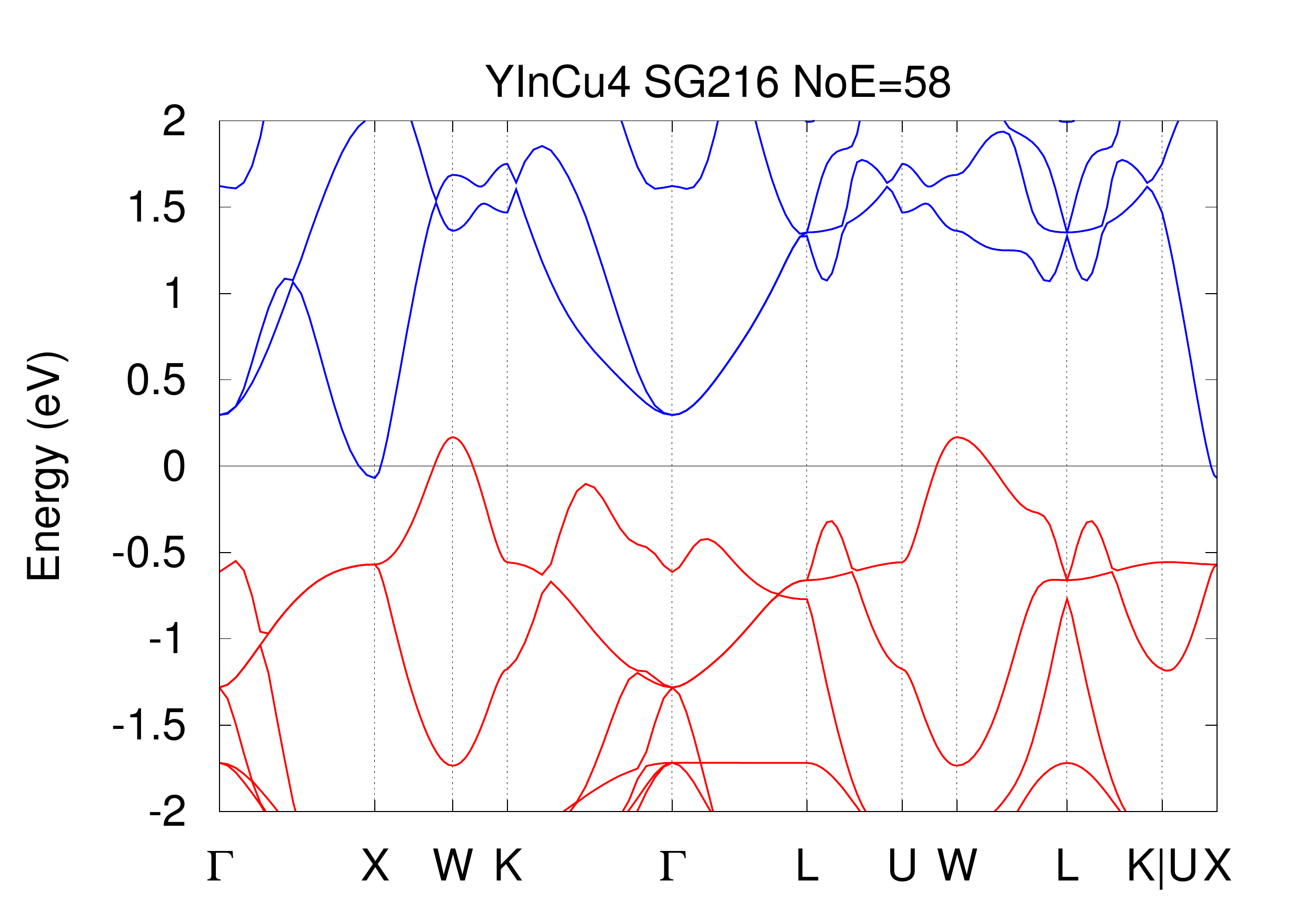}
}
\subfigure[ZrGaAu SG187 NoA=6 NoE=52]{
\label{subfig:156264}
\includegraphics[scale=0.32,angle=270]{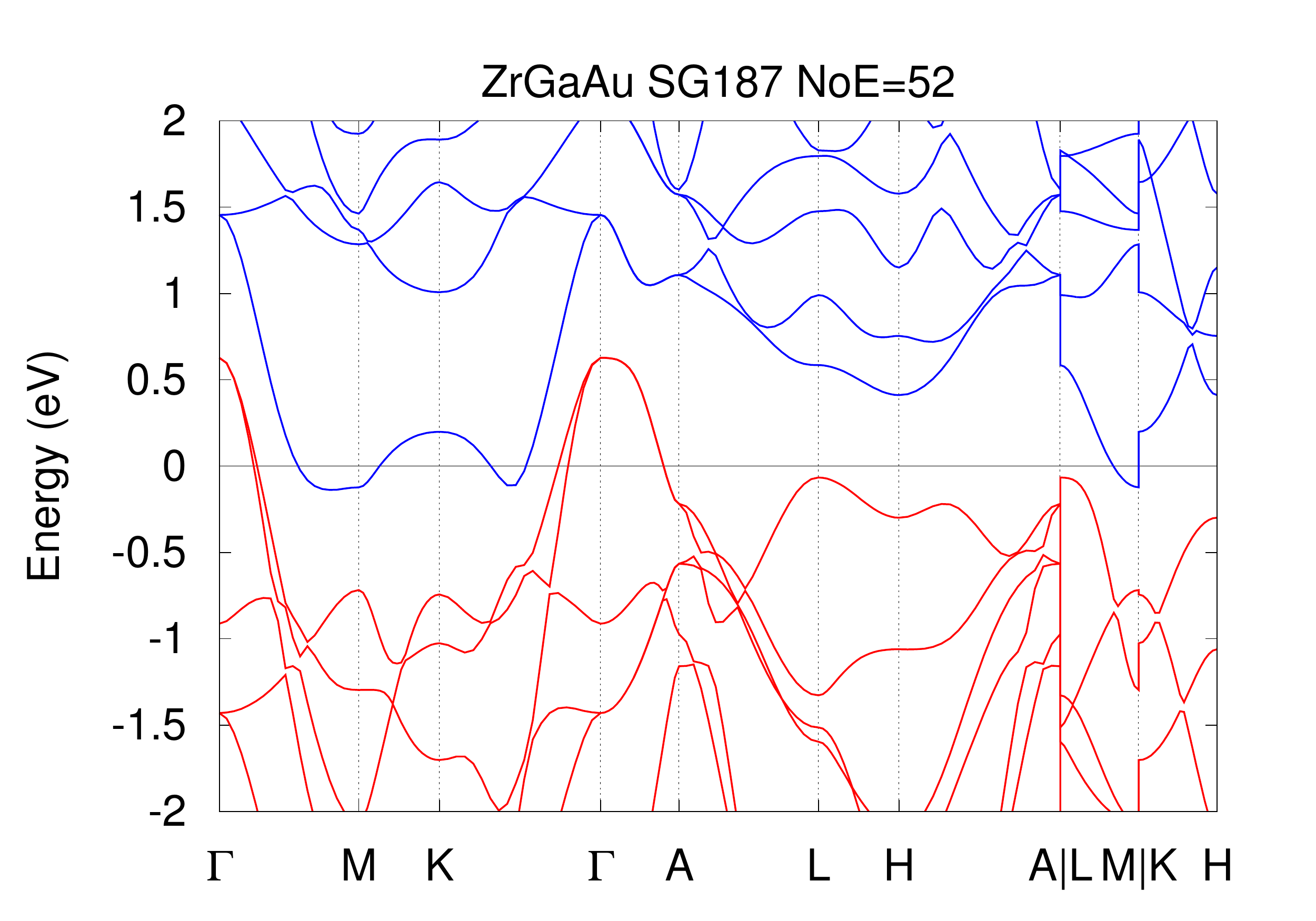}
}
\subfigure[SrAl$_{2}$ SG74 NoA=6 NoE=32]{
\label{subfig:609407}
\includegraphics[scale=0.32,angle=270]{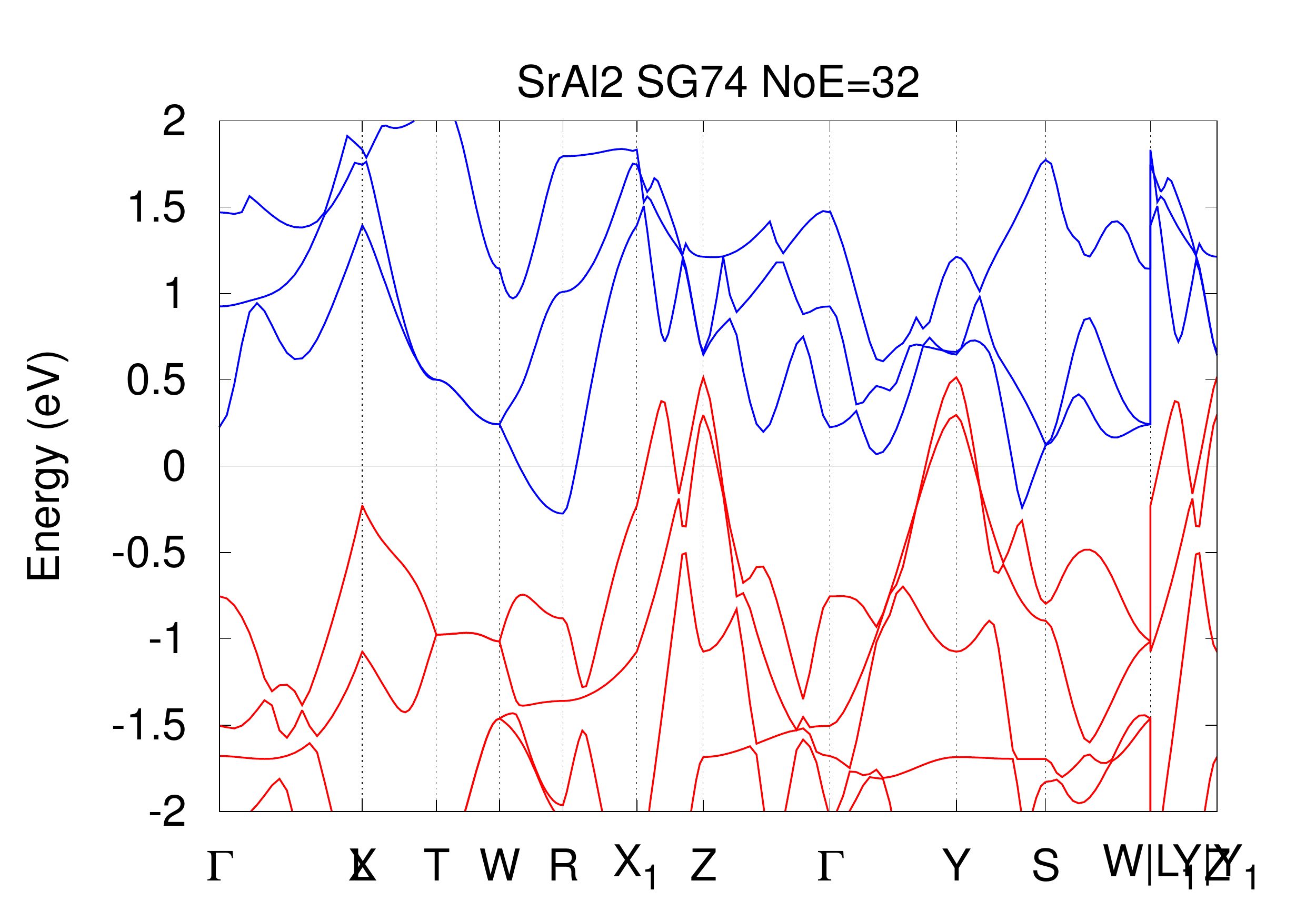}
}
\subfigure[HfN$_{2}$ SG194 NoA=6 NoE=28]{
\label{subfig:290428}
\includegraphics[scale=0.32,angle=270]{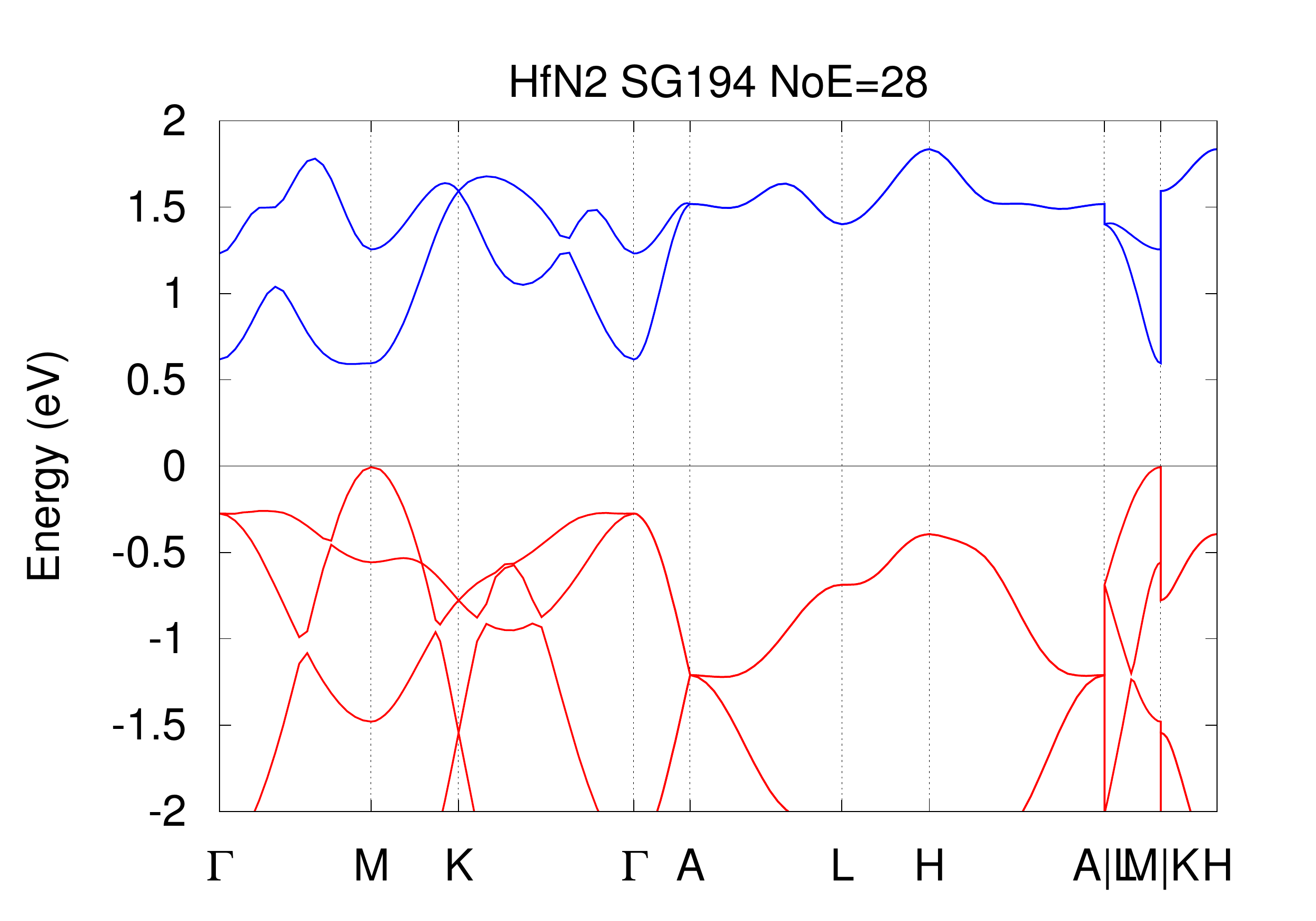}
}
\subfigure[KSb$_{2}$ SG12 NoA=6 NoE=38]{
\label{subfig:80945}
\includegraphics[scale=0.32,angle=270]{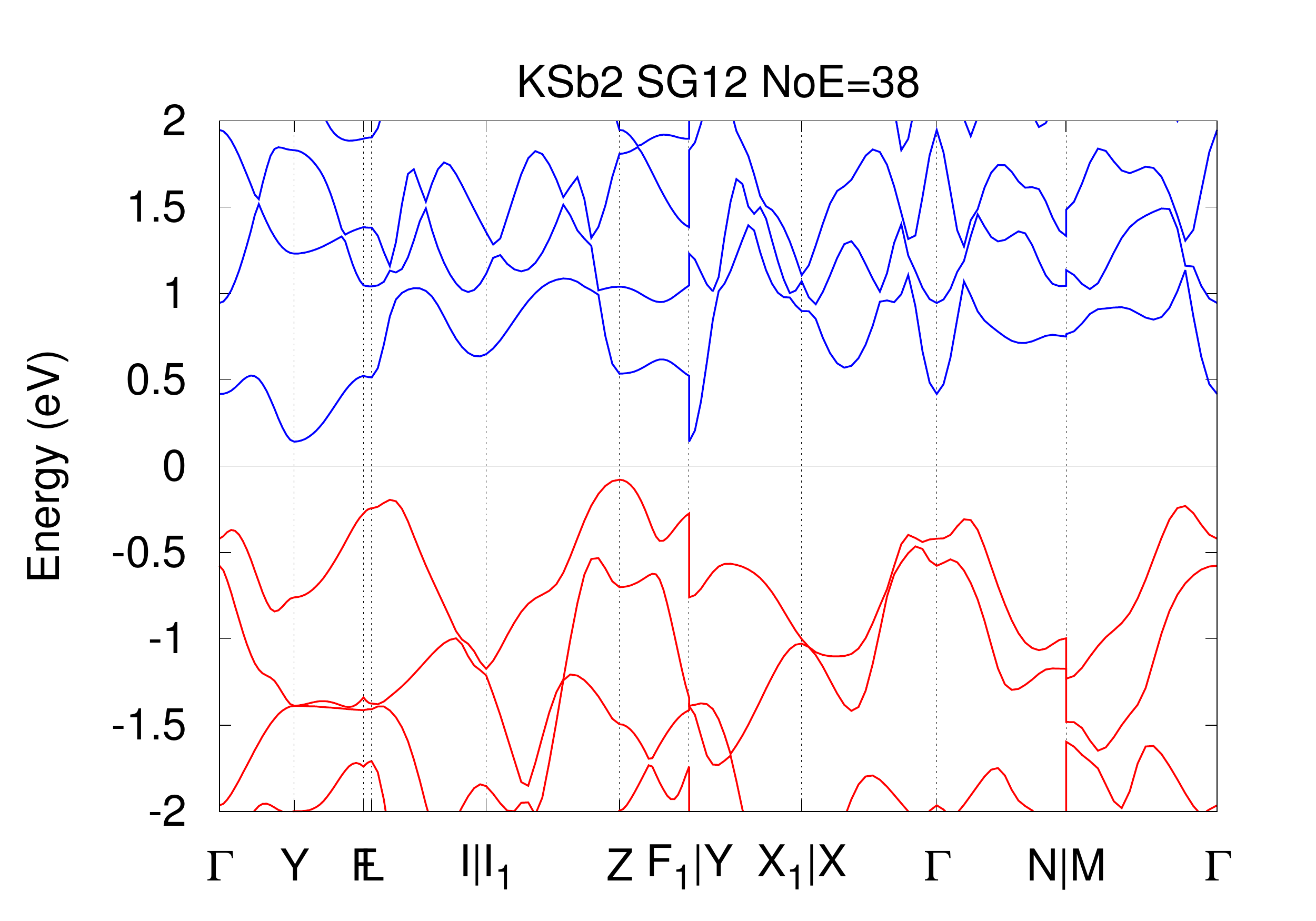}
}
\subfigure[ErCdCu$_{4}$ SG216 NoA=6 NoE=65]{
\label{subfig:415196}
\includegraphics[scale=0.32,angle=270]{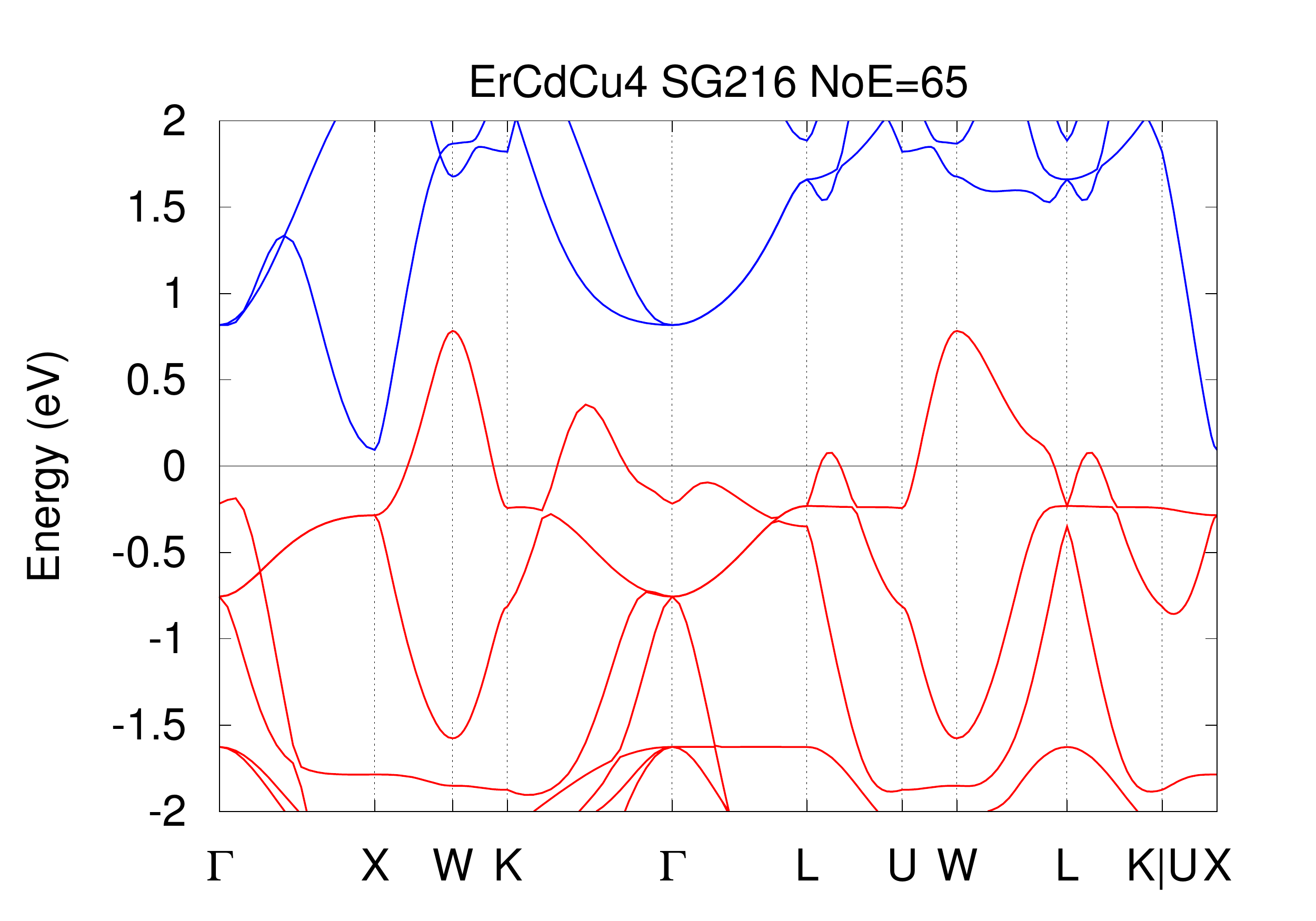}
}
\subfigure[ScSiAu SG187 NoA=6 NoE=36]{
\label{subfig:71998}
\includegraphics[scale=0.32,angle=270]{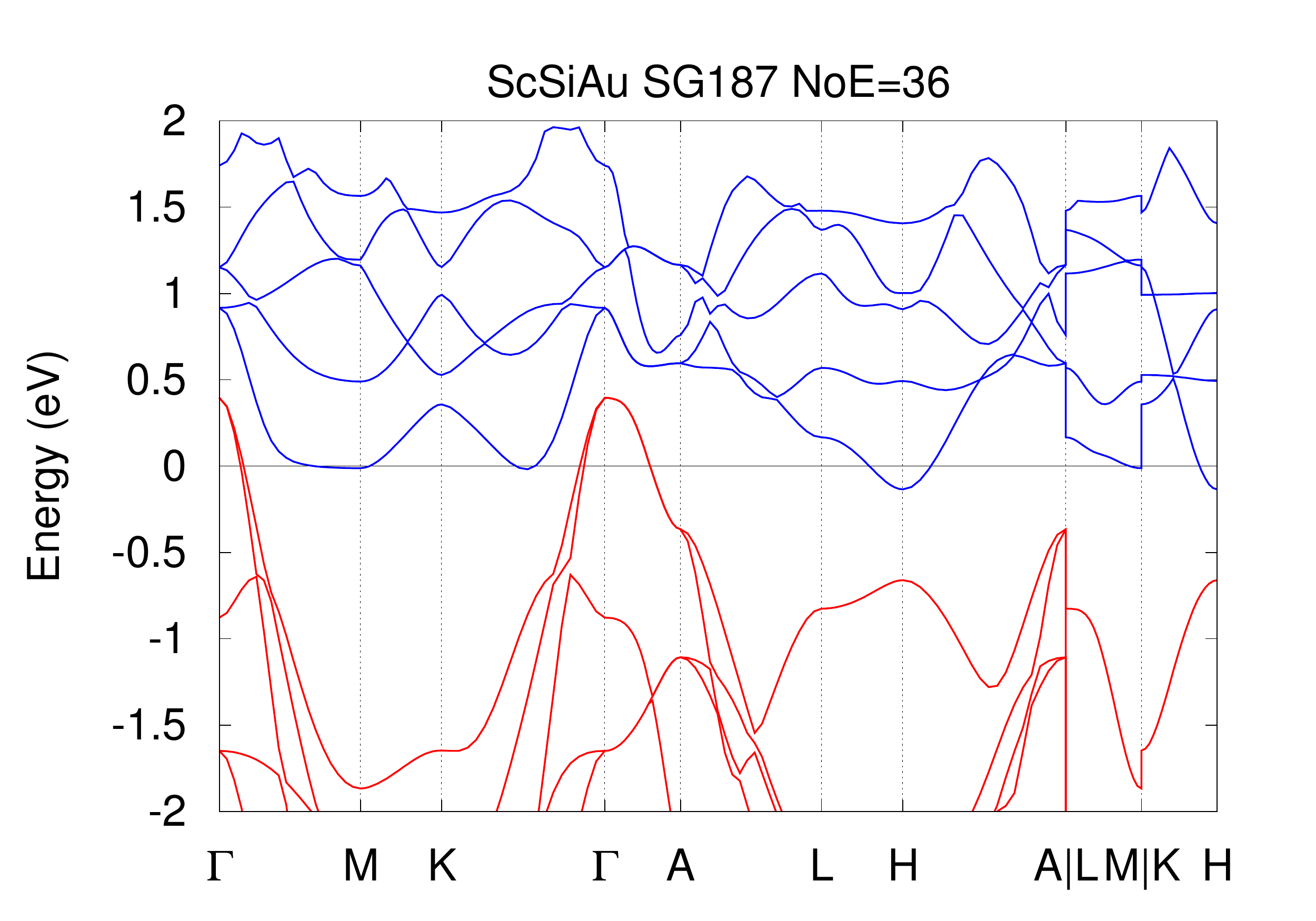}
}
\caption{\hyperref[tab:electride]{back to the table}}
\end{figure}

\begin{figure}[htp]
 \centering
\subfigure[BaSe$_{2}$ SG15 NoA=6 NoE=44]{
\label{subfig:16358}
\includegraphics[scale=0.32,angle=270]{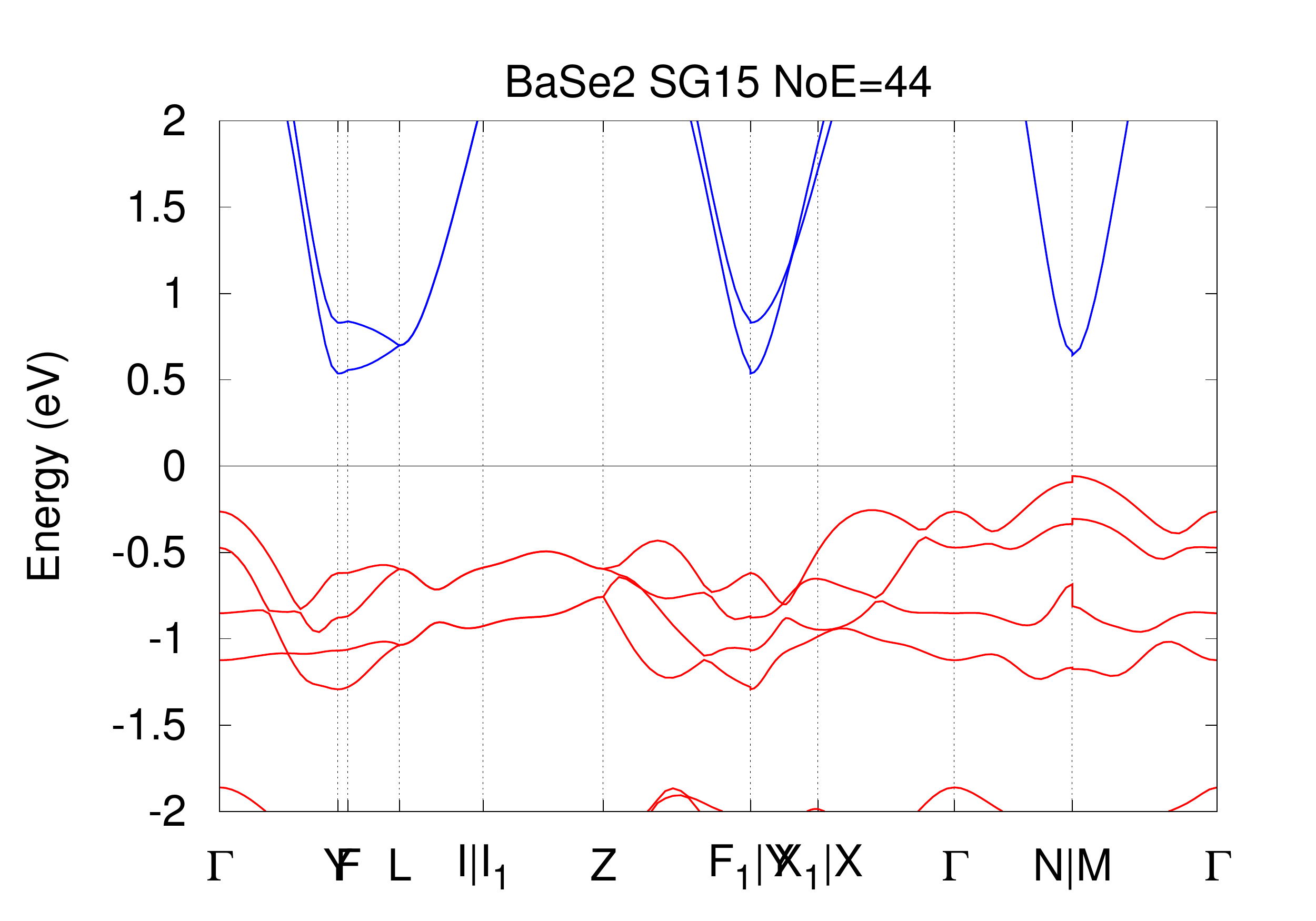}
}
\subfigure[FeSe$_{2}$ SG58 NoA=6 NoE=40]{
\label{subfig:44751}
\includegraphics[scale=0.32,angle=270]{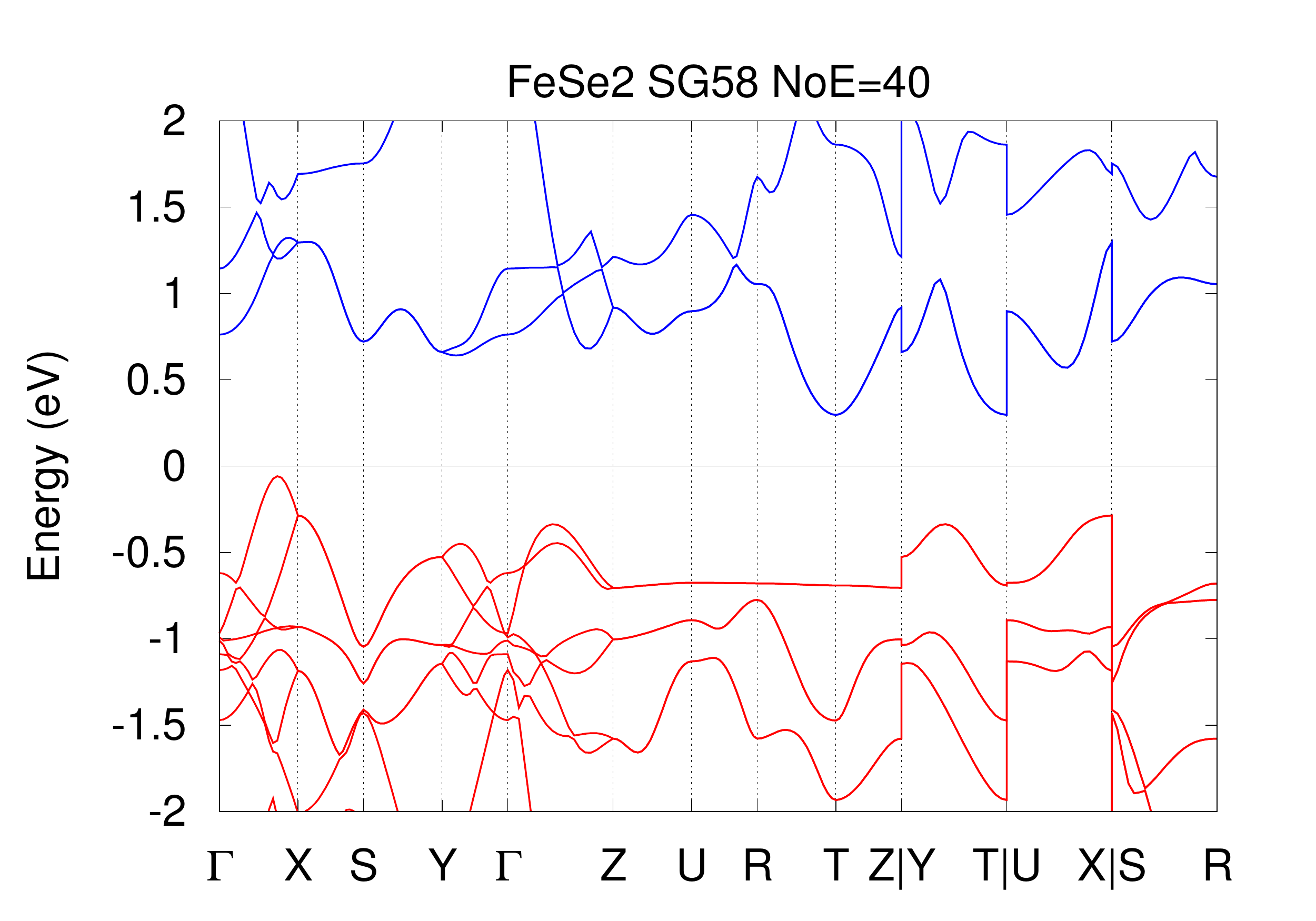}
}
\subfigure[RbSb$_{2}$ SG12 NoA=6 NoE=38]{
\label{subfig:419402}
\includegraphics[scale=0.32,angle=270]{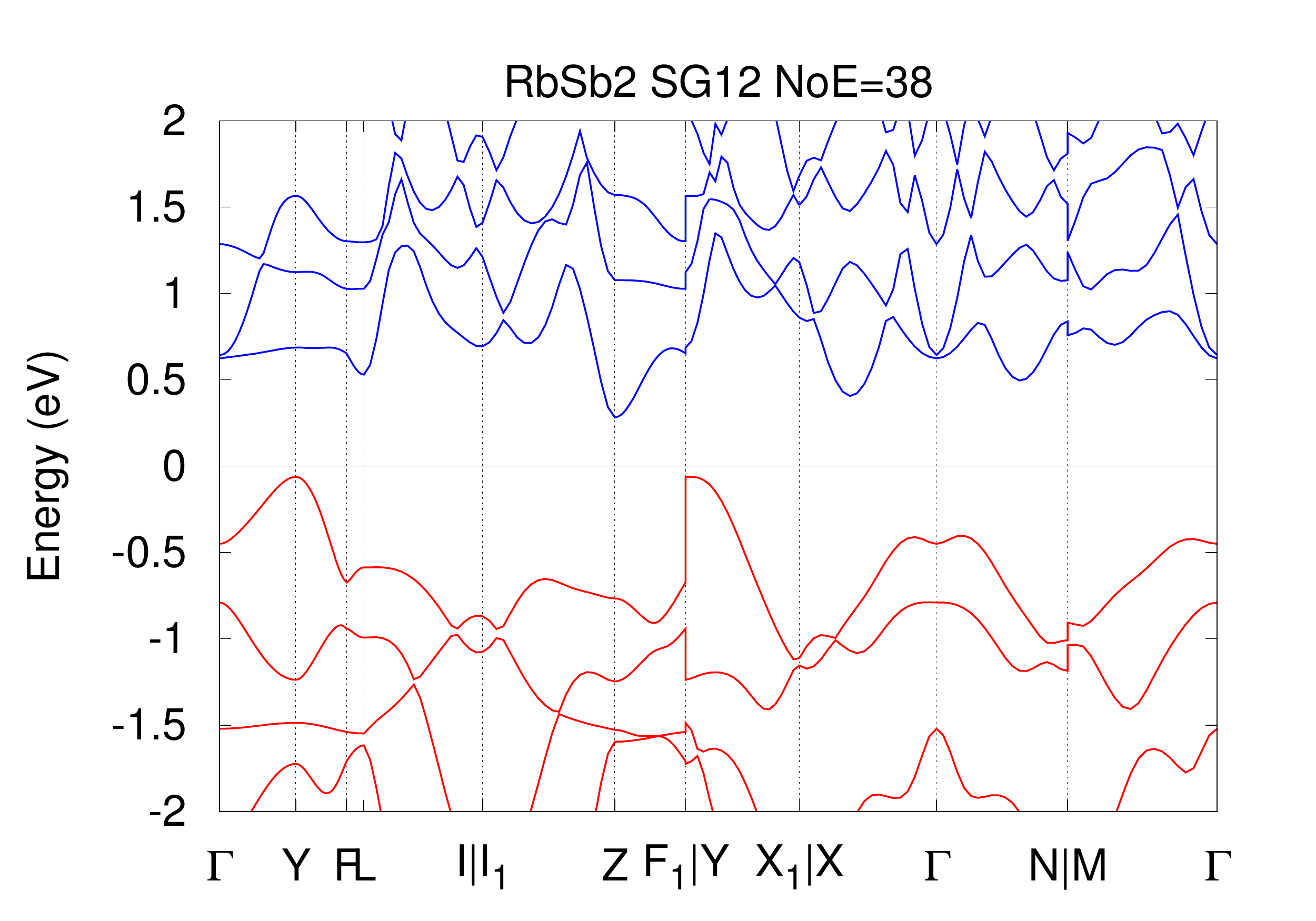}
}
\subfigure[ZrCu$_{4}$Ag SG216 NoA=6 NoE=67]{
\label{subfig:196260}
\includegraphics[scale=0.32,angle=270]{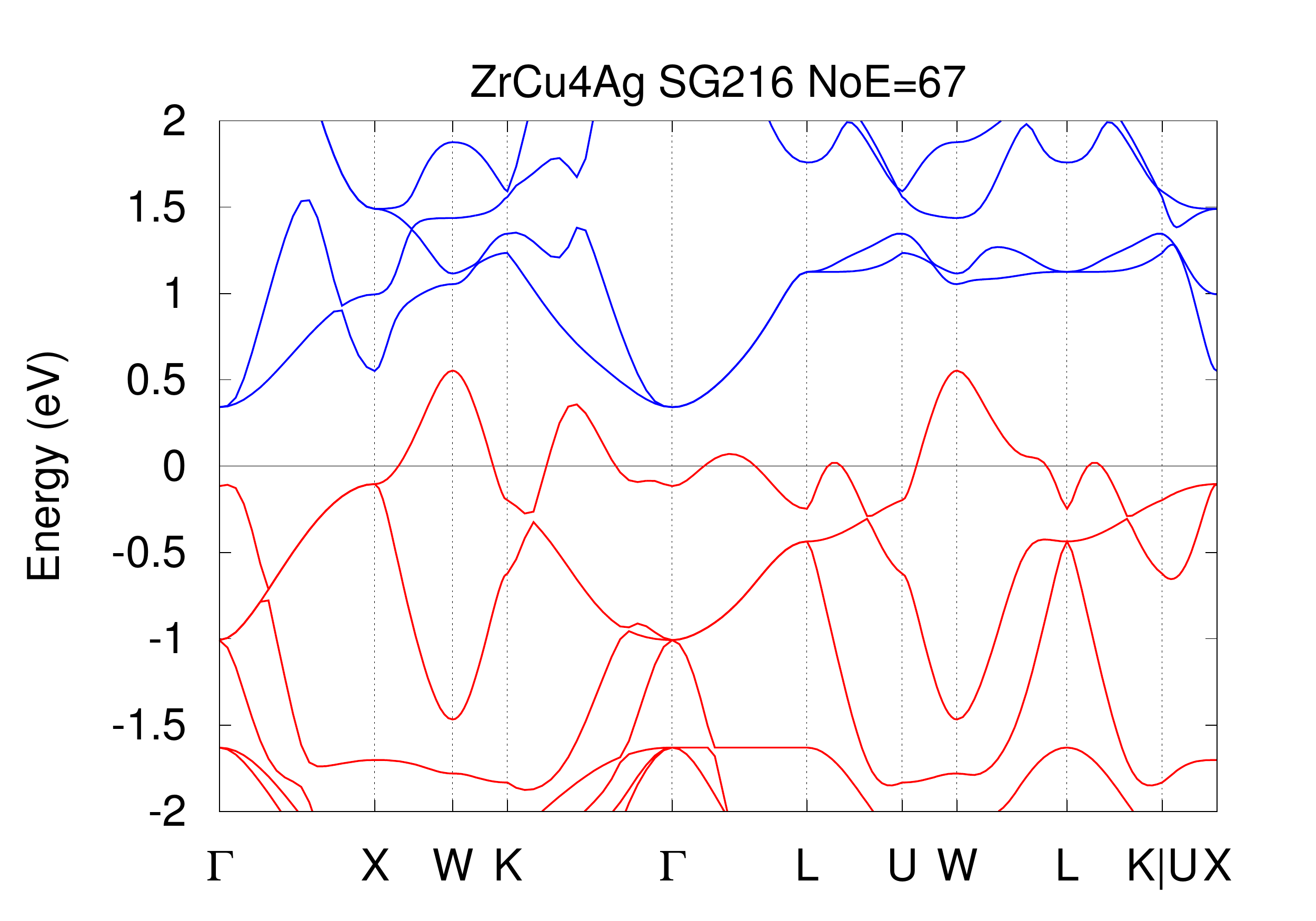}
}
\subfigure[PtN$_{2}$ SG58 NoA=6 NoE=40]{
\label{subfig:166463}
\includegraphics[scale=0.32,angle=270]{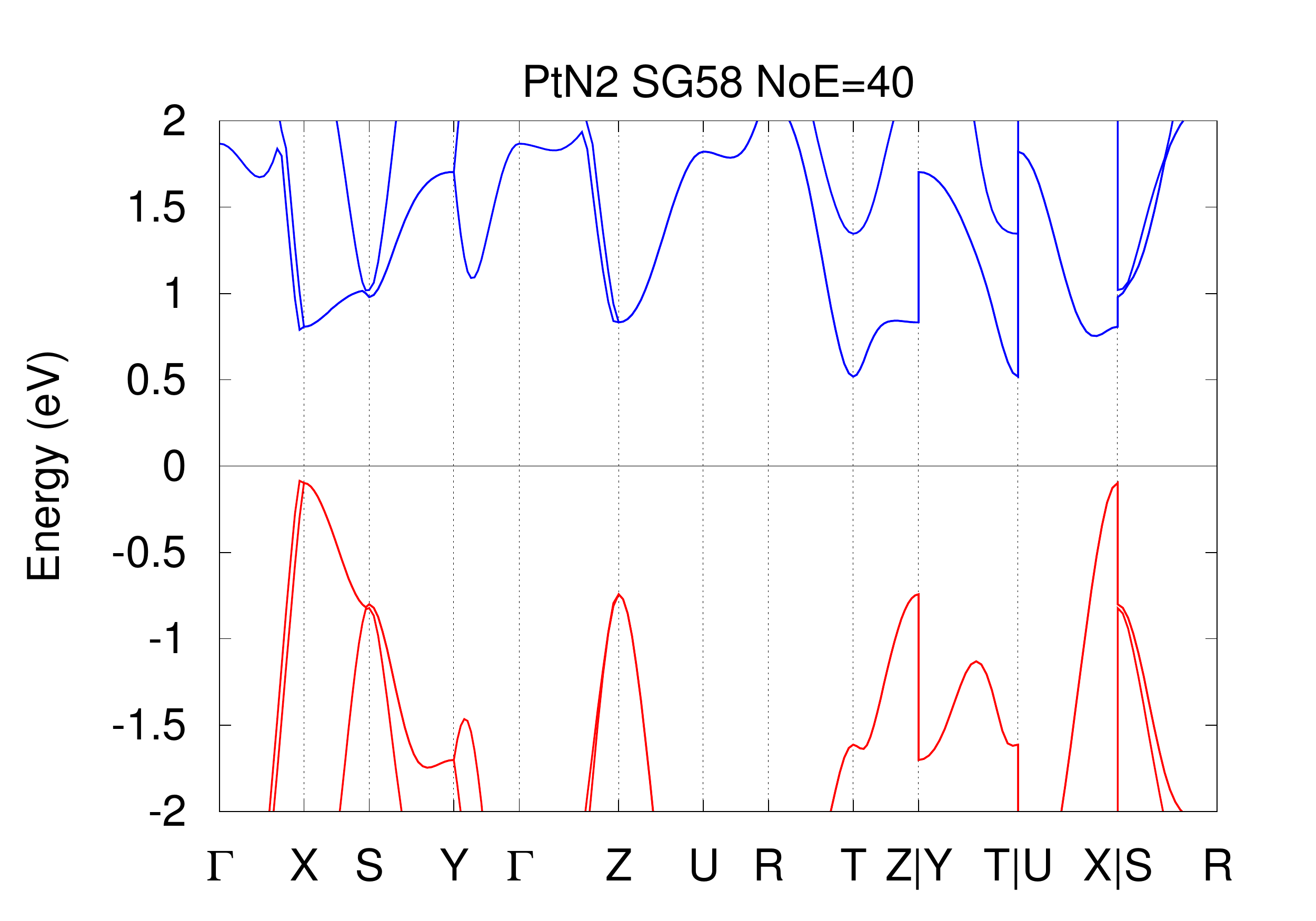}
}
\subfigure[Sb$_{2}$Os SG58 NoA=6 NoE=36]{
\label{subfig:238254}
\includegraphics[scale=0.32,angle=270]{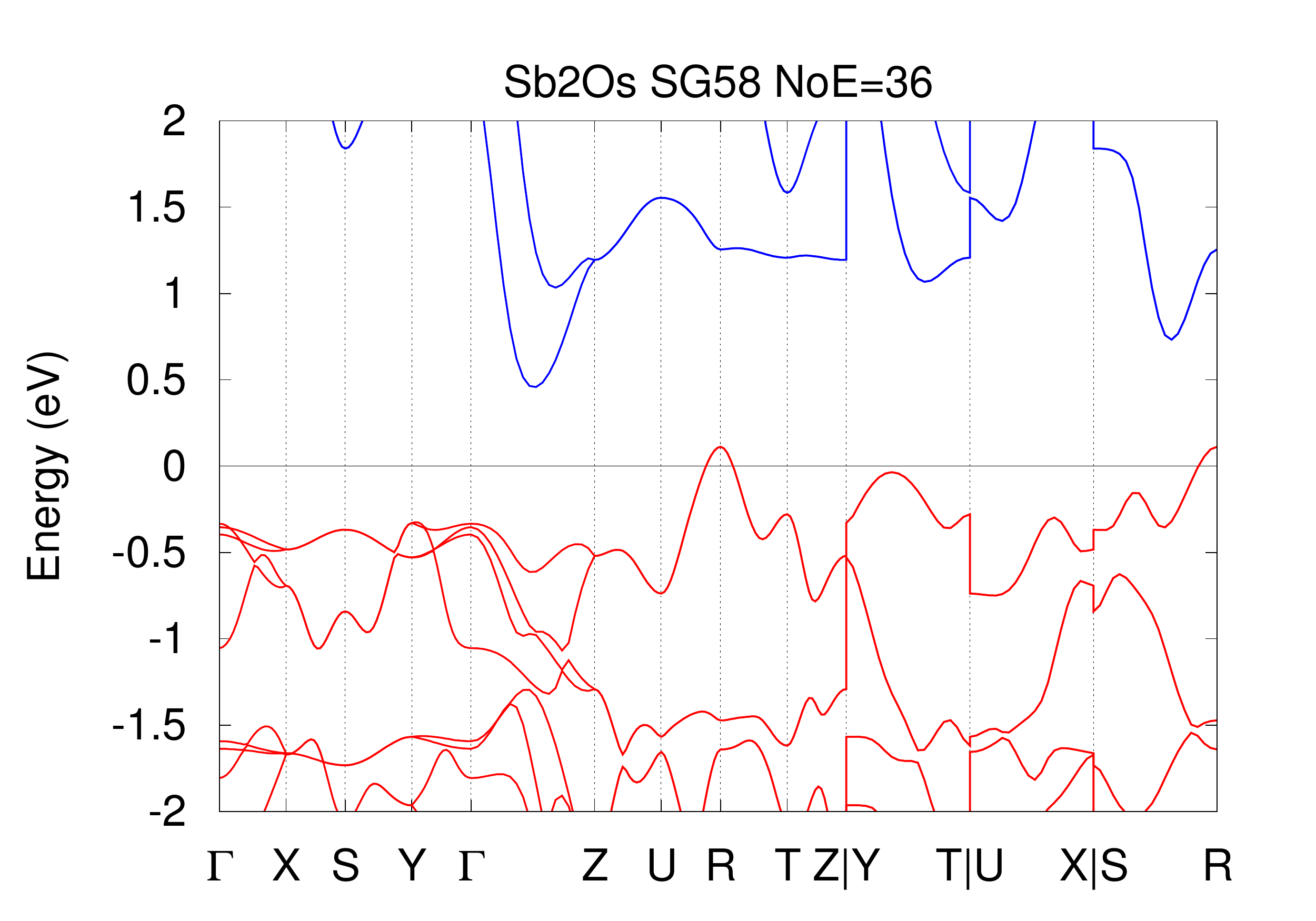}
}
\subfigure[YbNi$_{4}$Au SG216 NoA=6 NoE=59]{
\label{subfig:612234}
\includegraphics[scale=0.32,angle=270]{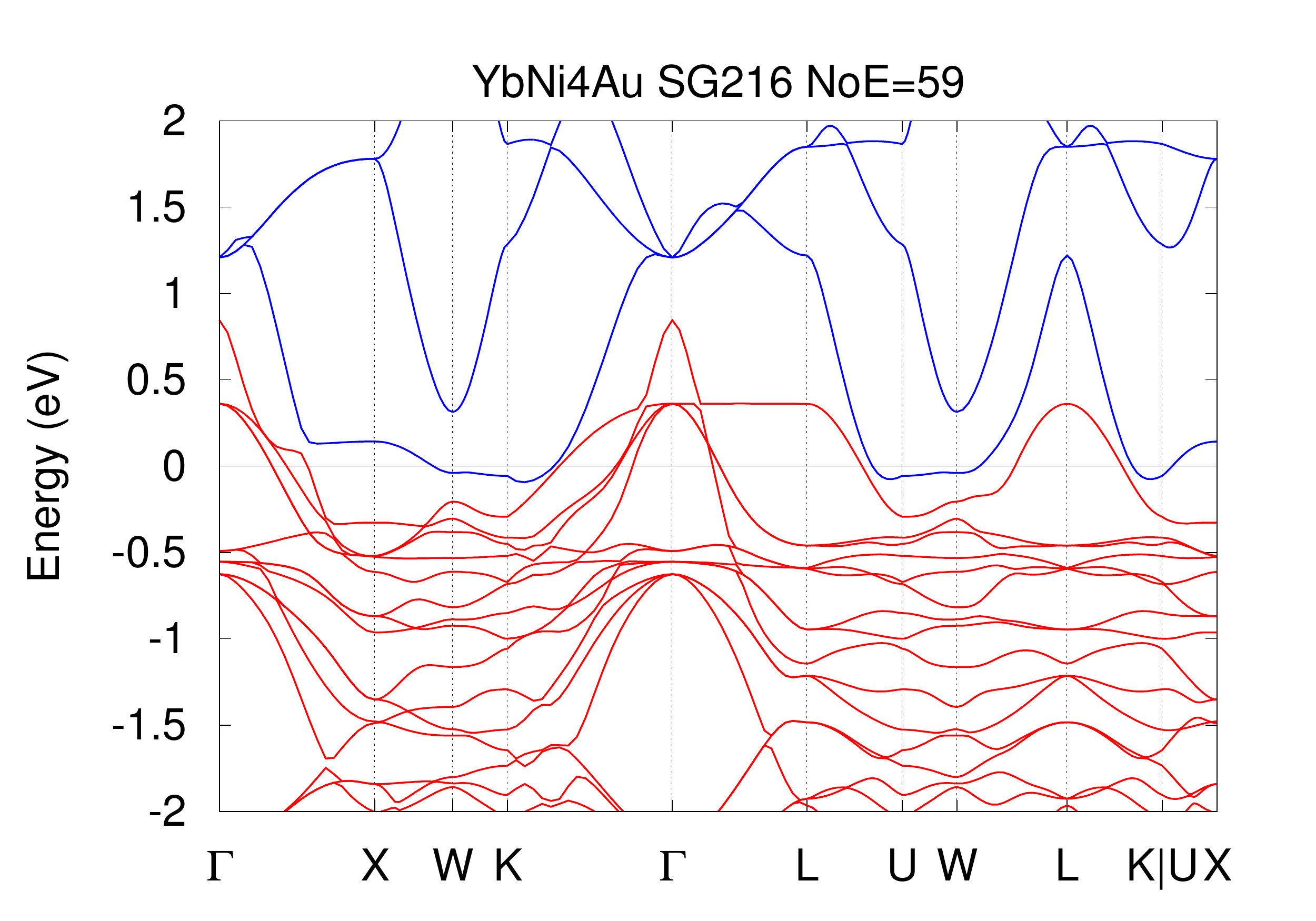}
}
\subfigure[Si$_{2}$Os SG12 NoA=6 NoE=32]{
\label{subfig:647776}
\includegraphics[scale=0.32,angle=270]{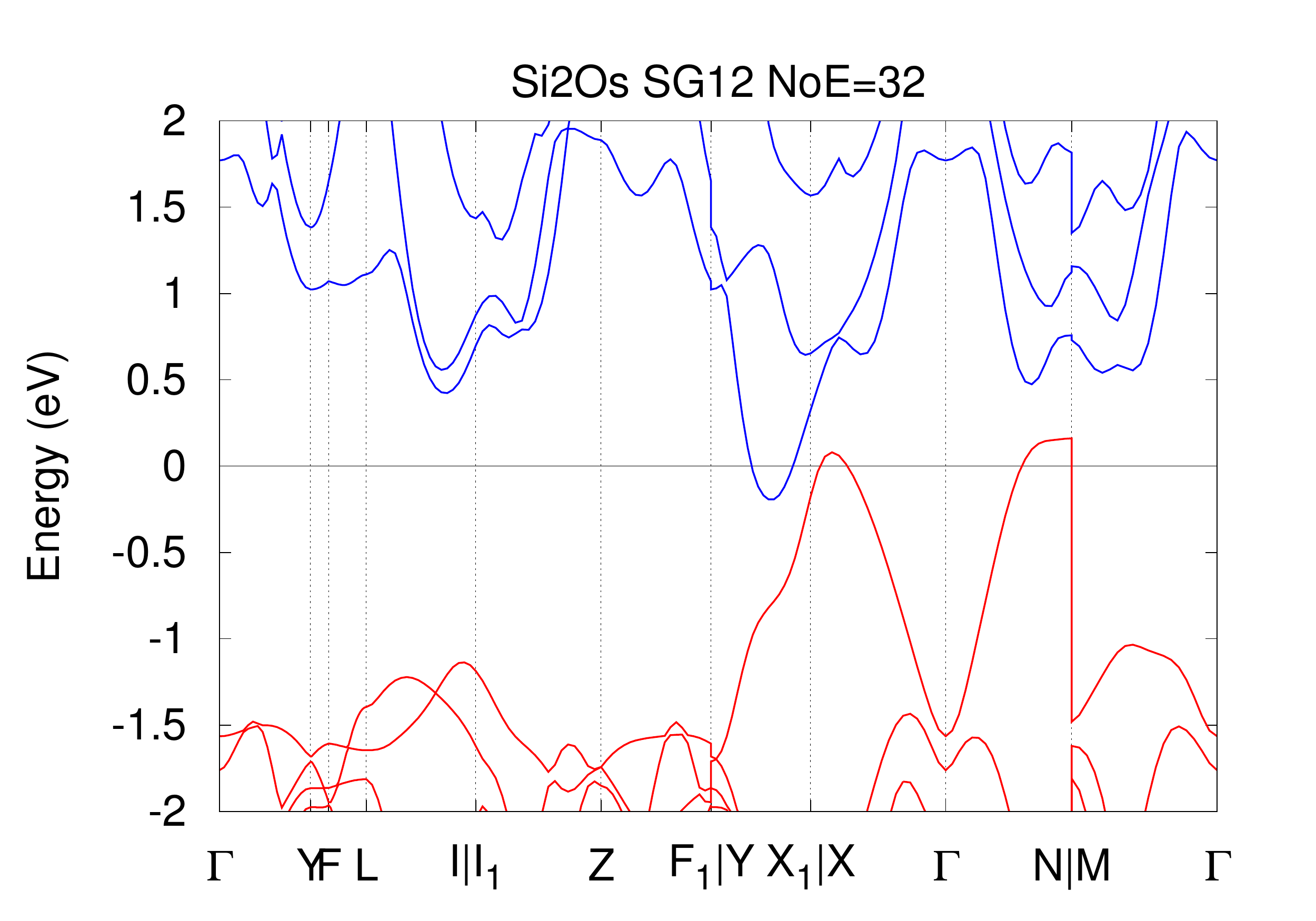}
}
\caption{\hyperref[tab:electride]{back to the table}}
\end{figure}

\begin{figure}[htp]
 \centering
\subfigure[Ge$_{2}$Os SG12 NoA=6 NoE=32]{
\label{subfig:43690}
\includegraphics[scale=0.32,angle=270]{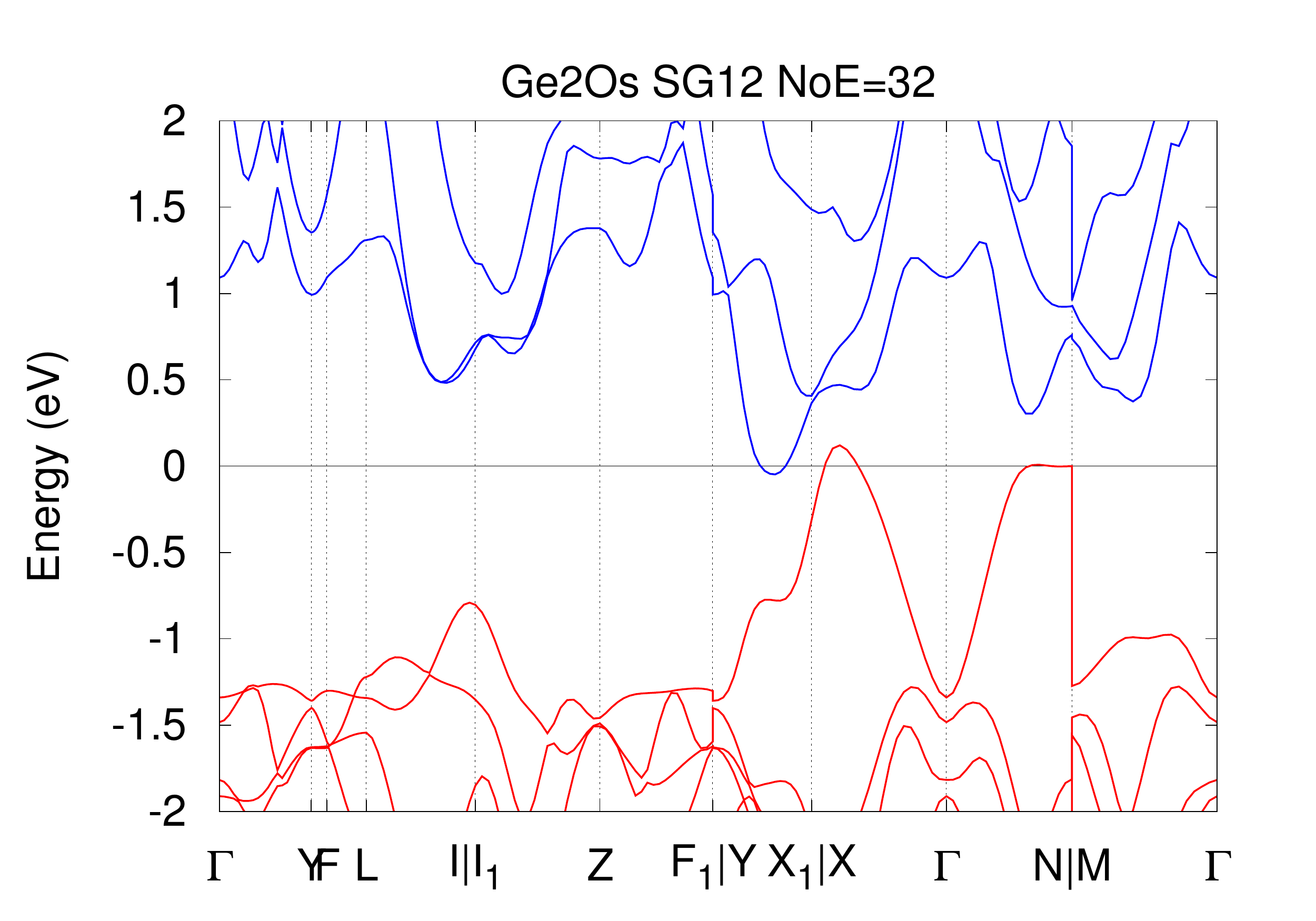}
}
\subfigure[NdMgCu$_{4}$ SG216 NoA=6 NoE=57]{
\label{subfig:194977}
\includegraphics[scale=0.32,angle=270]{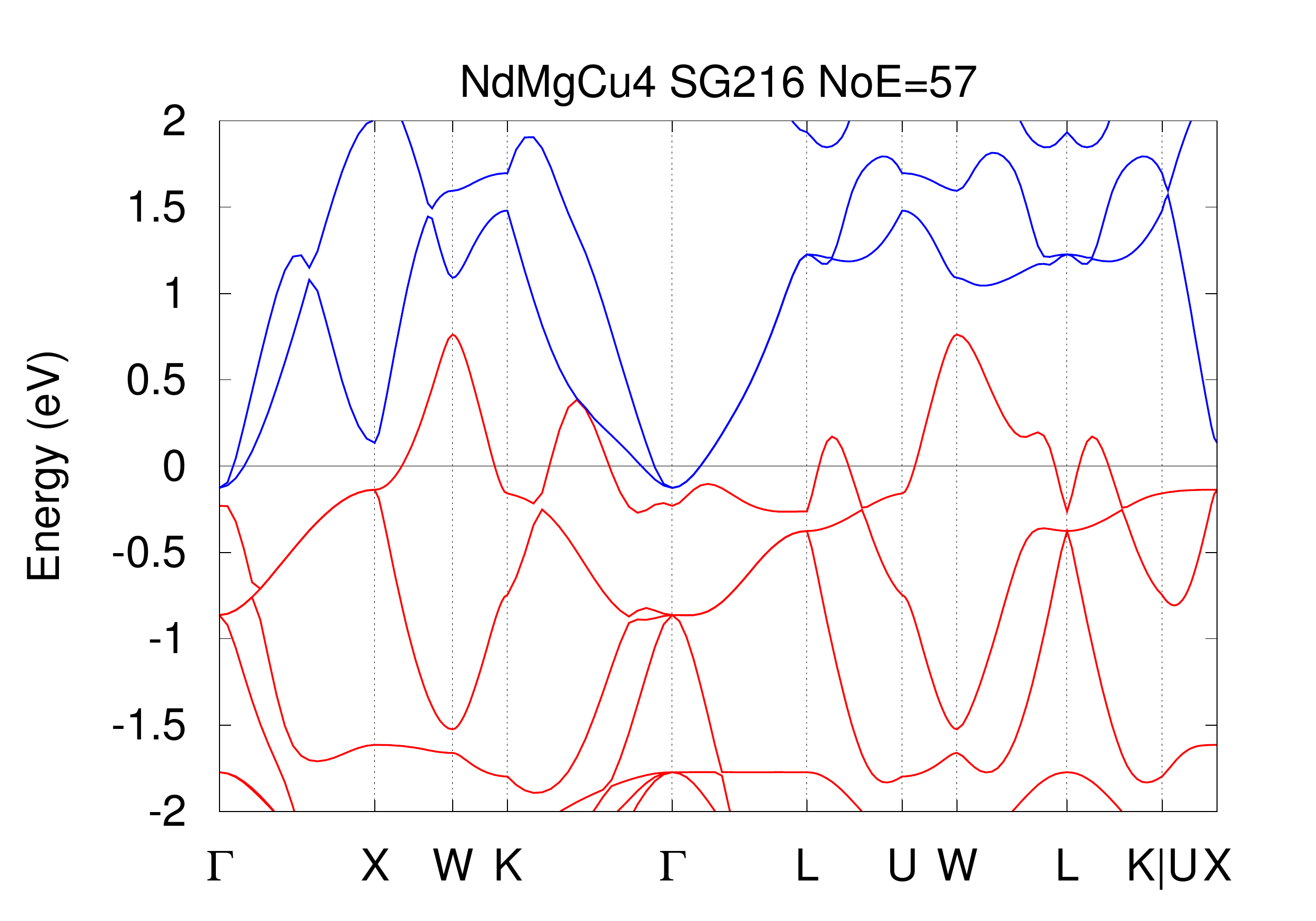}
}
\subfigure[Be$_{5}$Pd SG216 NoA=6 NoE=20]{
\label{subfig:616387}
\includegraphics[scale=0.32,angle=270]{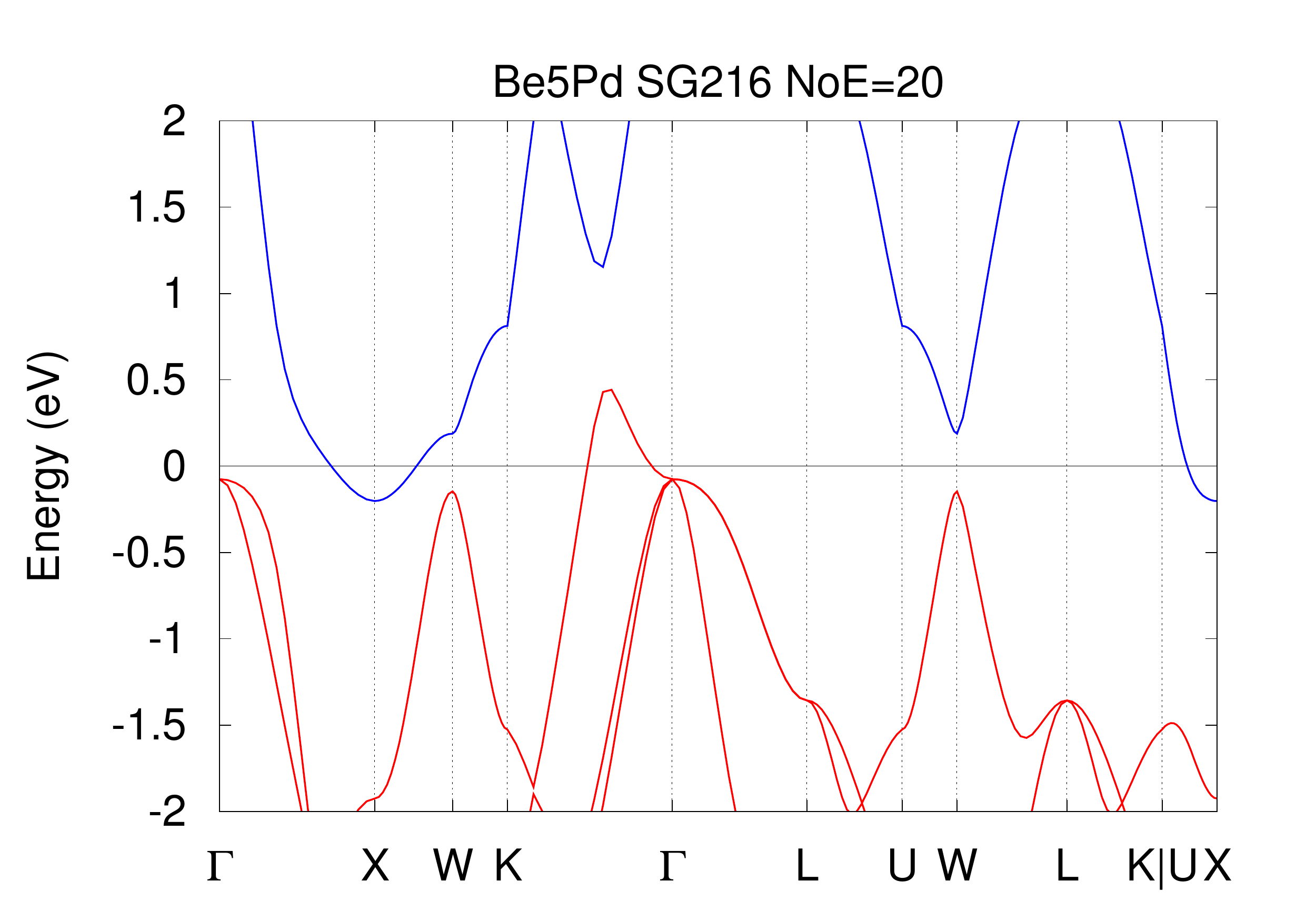}
}
\subfigure[Ga$_{2}$Os SG70 NoA=6 NoE=28]{
\label{subfig:103785}
\includegraphics[scale=0.32,angle=270]{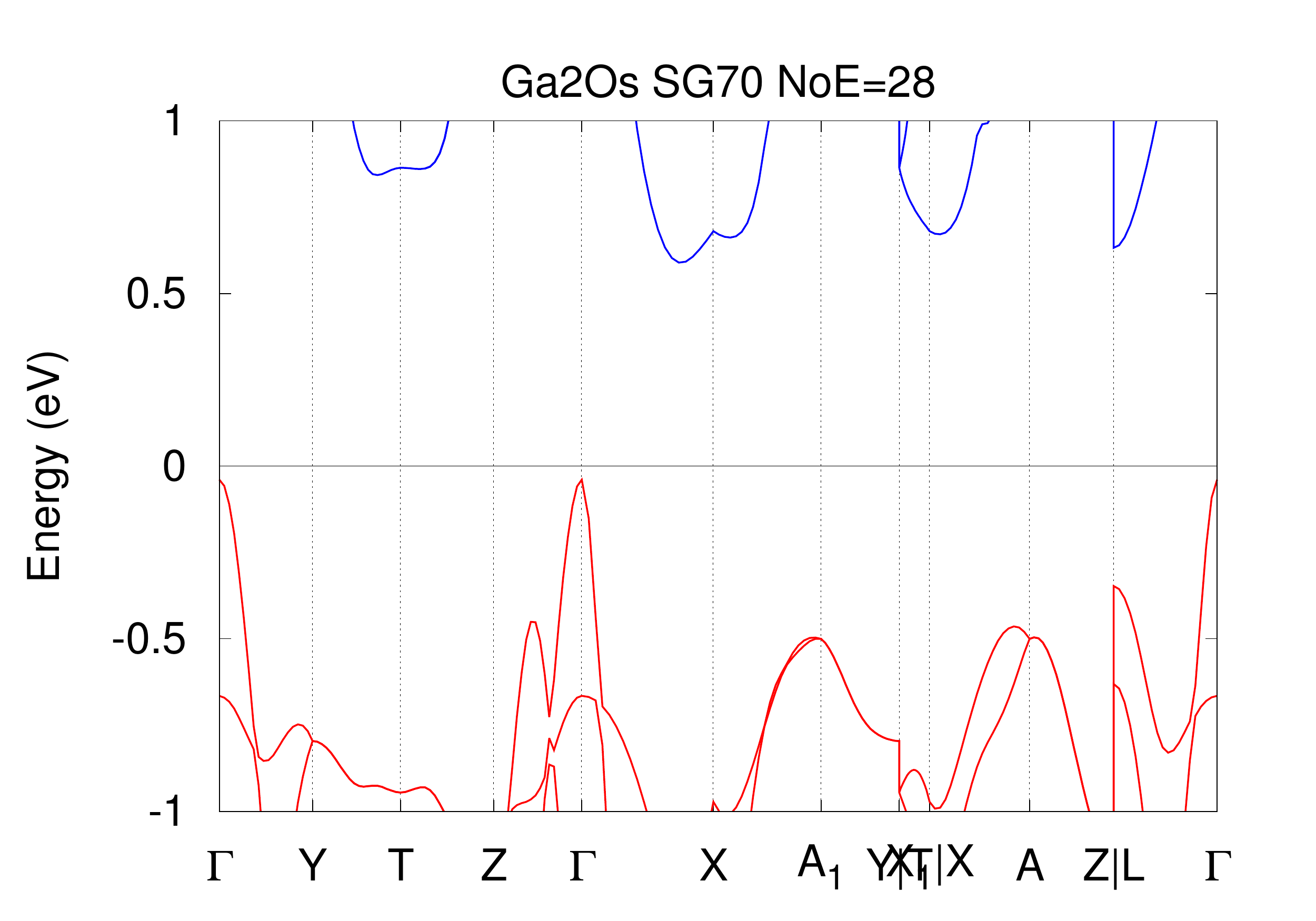}
}
\subfigure[Ga$_{2}$Ru SG70 NoA=6 NoE=28]{
\label{subfig:635228}
\includegraphics[scale=0.32,angle=270]{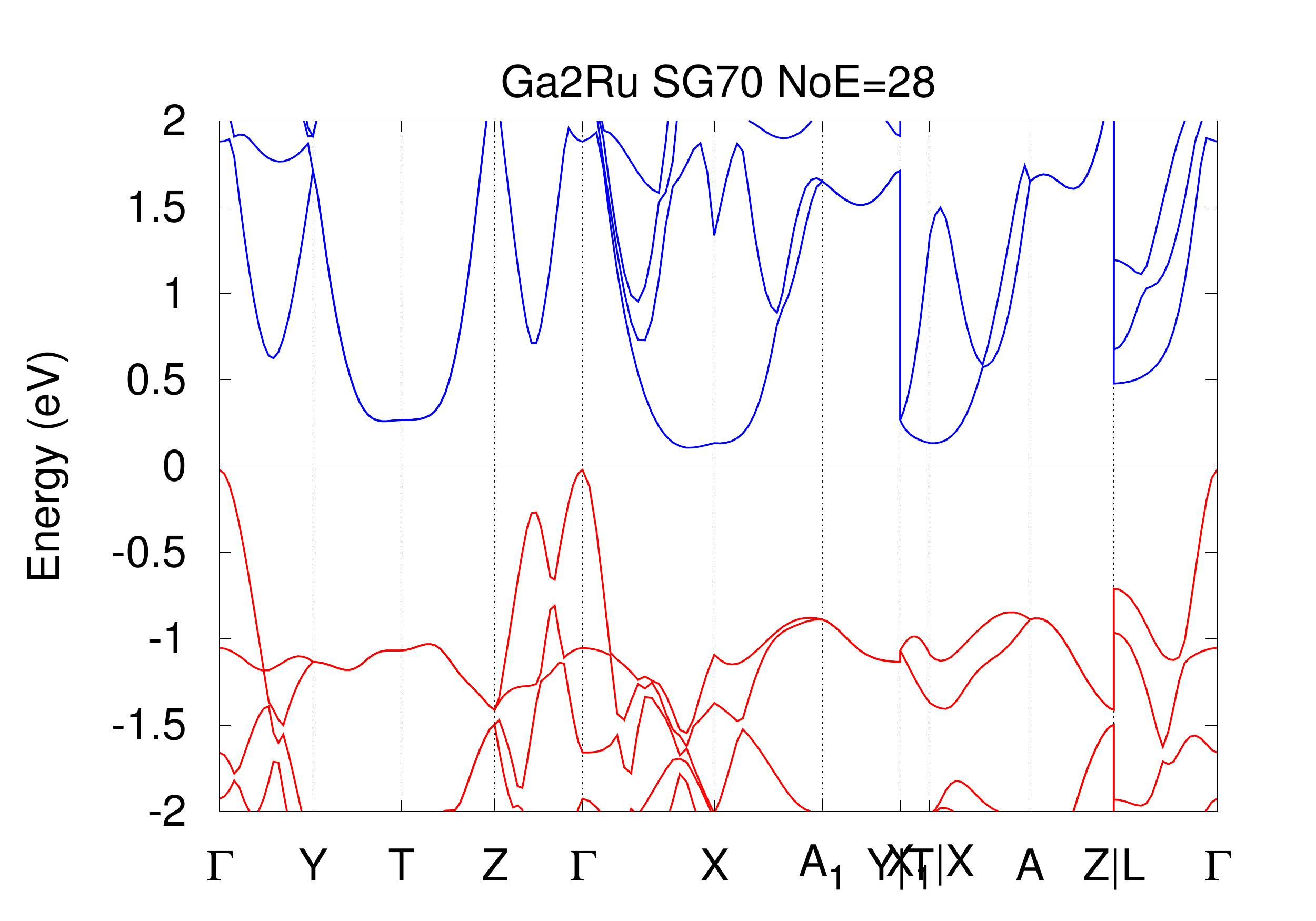}
}
\subfigure[NiP$_{2}$ SG15 NoA=6 NoE=40]{
\label{subfig:646107}
\includegraphics[scale=0.32,angle=270]{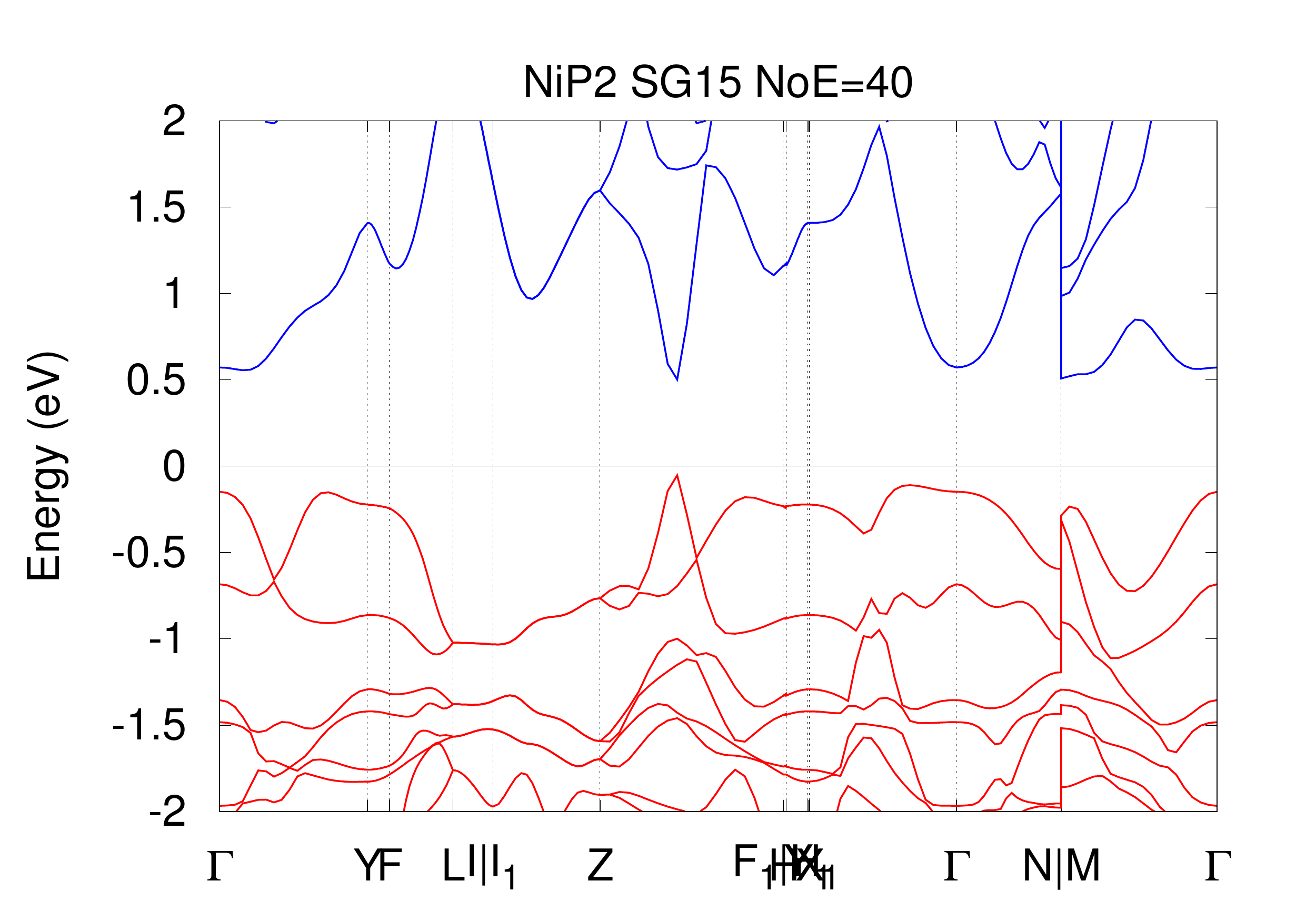}
}
\subfigure[PbSe$_{2}$ SG140 NoA=6 NoE=32]{
\label{subfig:174577}
\includegraphics[scale=0.32,angle=270]{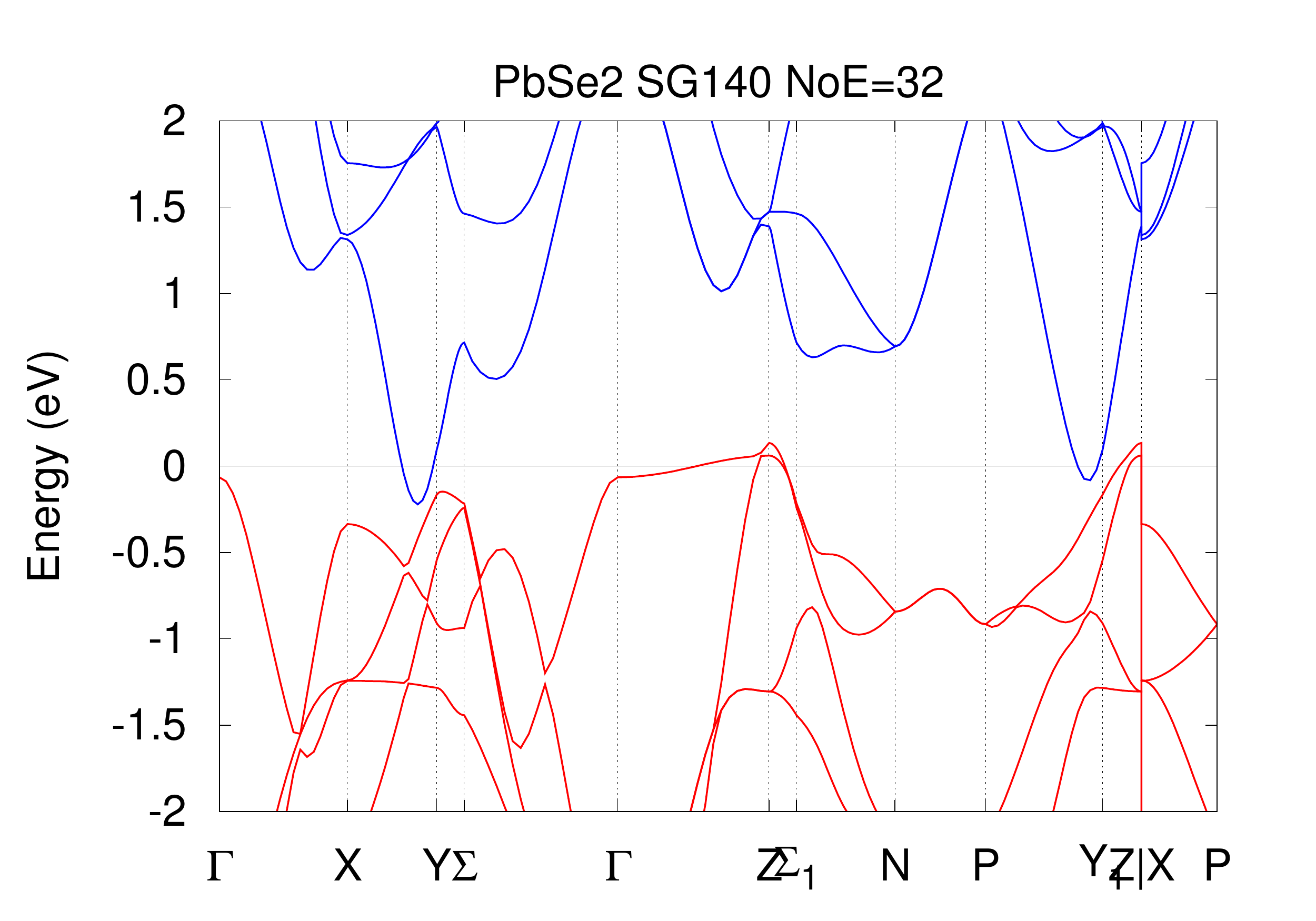}
}
\subfigure[CrAs$_{2}$ SG12 NoA=6 NoE=32]{
\label{subfig:43898}
\includegraphics[scale=0.32,angle=270]{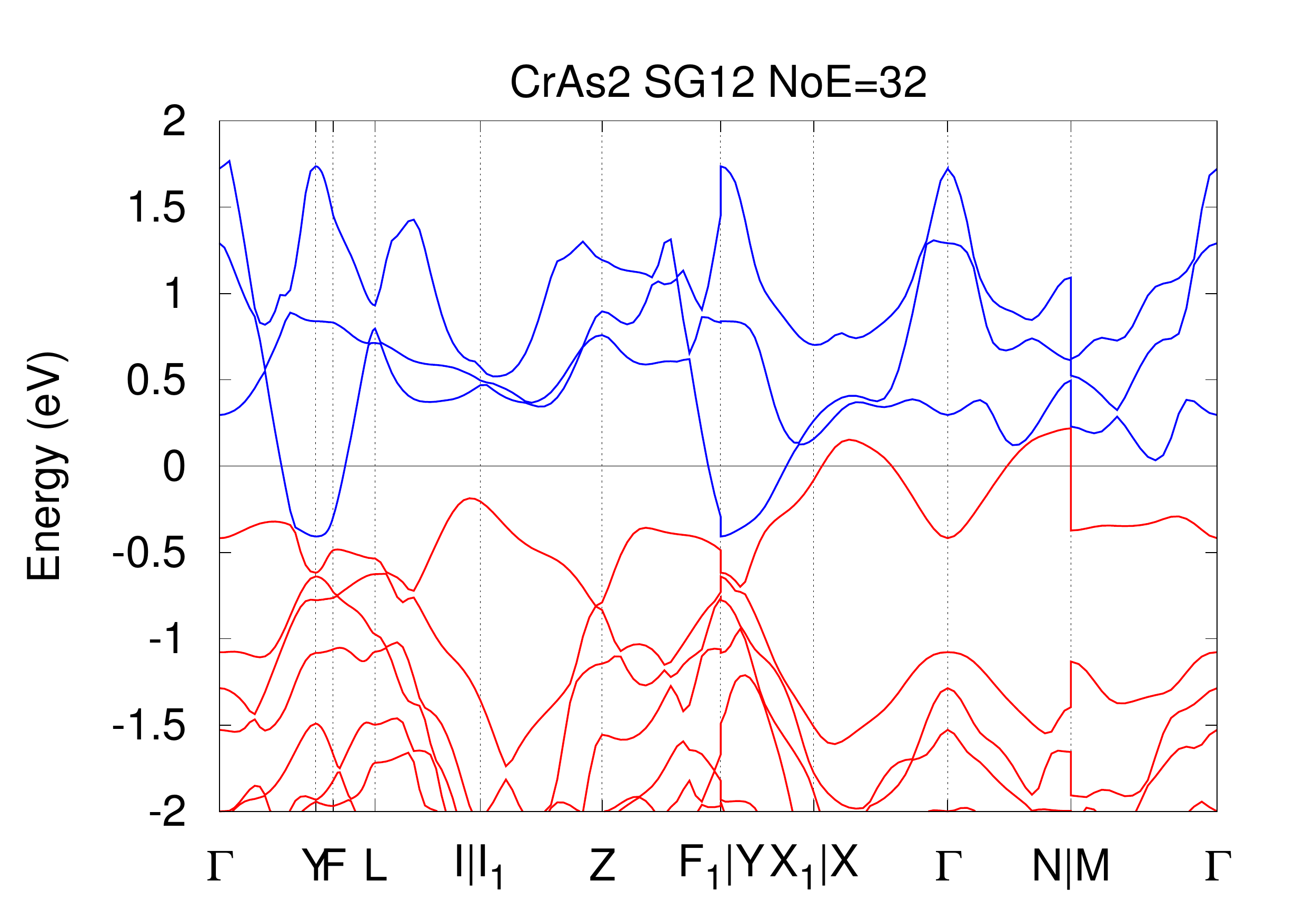}
}
\caption{\hyperref[tab:electride]{back to the table}}
\end{figure}

\begin{figure}[htp]
 \centering
\subfigure[FeP$_{2}$ SG58 NoA=6 NoE=36]{
\label{subfig:633072}
\includegraphics[scale=0.32,angle=270]{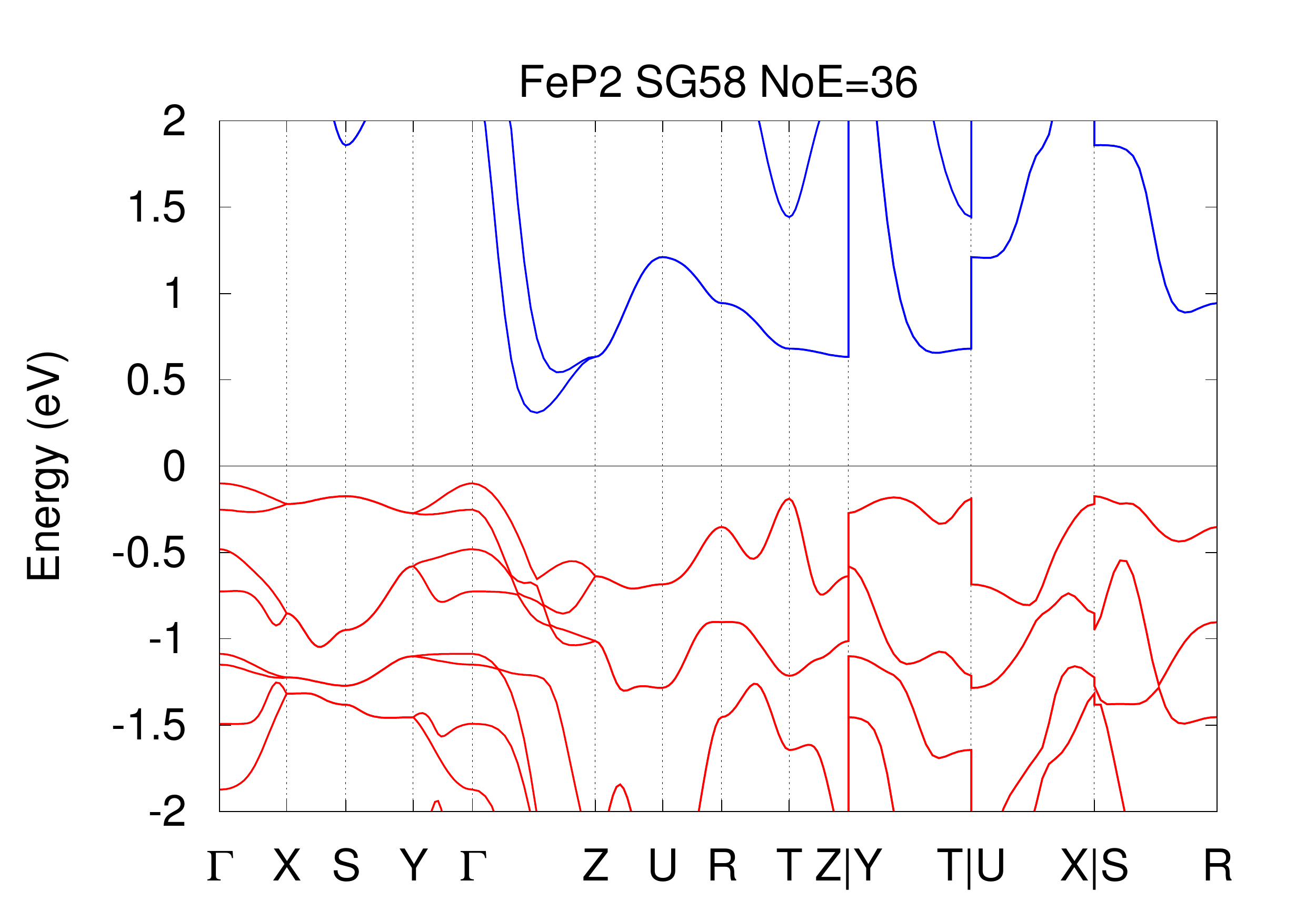}
}
\subfigure[Te$_{2}$Ru SG58 NoA=6 NoE=40]{
\label{subfig:106001}
\includegraphics[scale=0.32,angle=270]{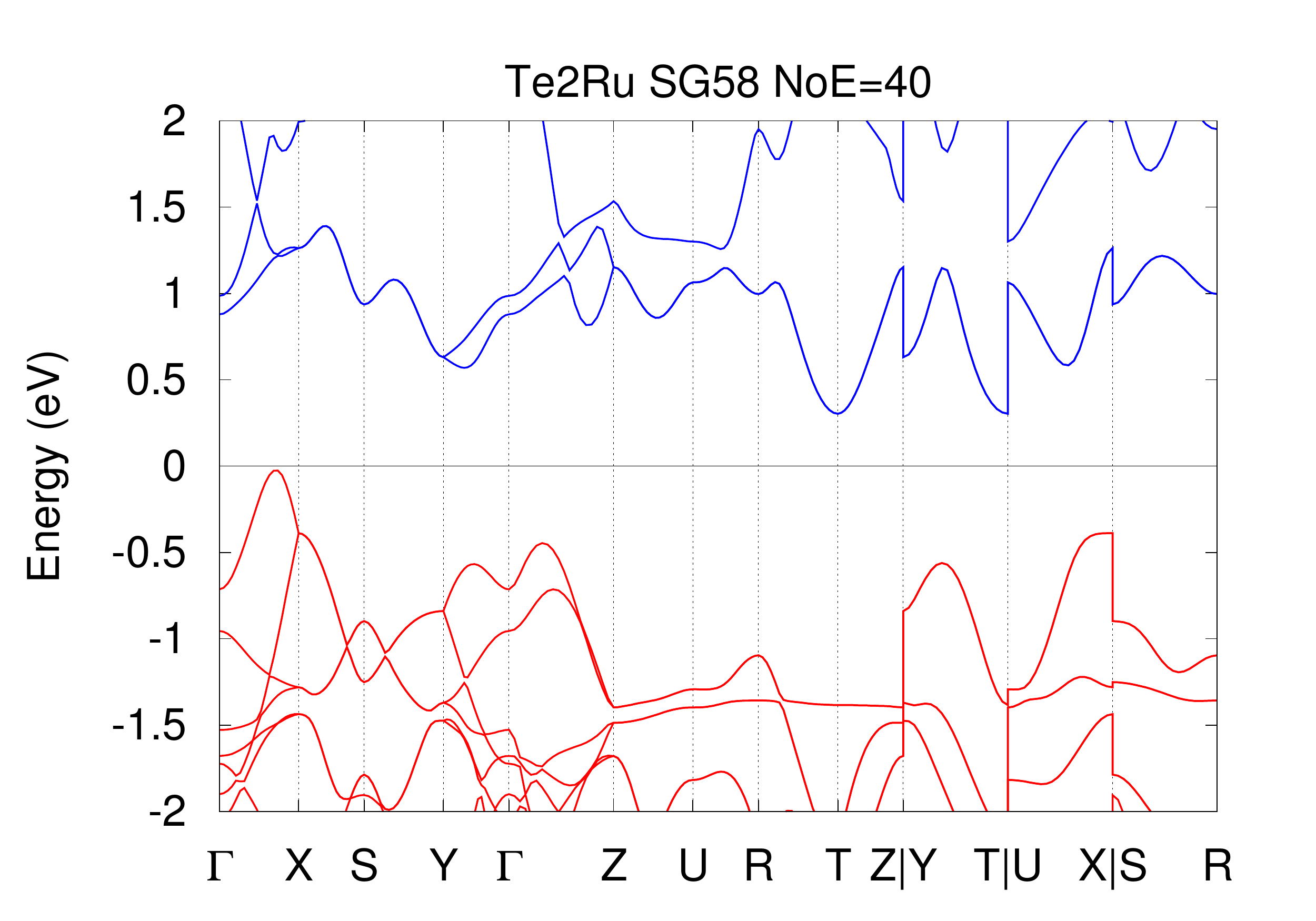}
}
\subfigure[CrP$_{2}$ SG12 NoA=6 NoE=32]{
\label{subfig:2526}
\includegraphics[scale=0.32,angle=270]{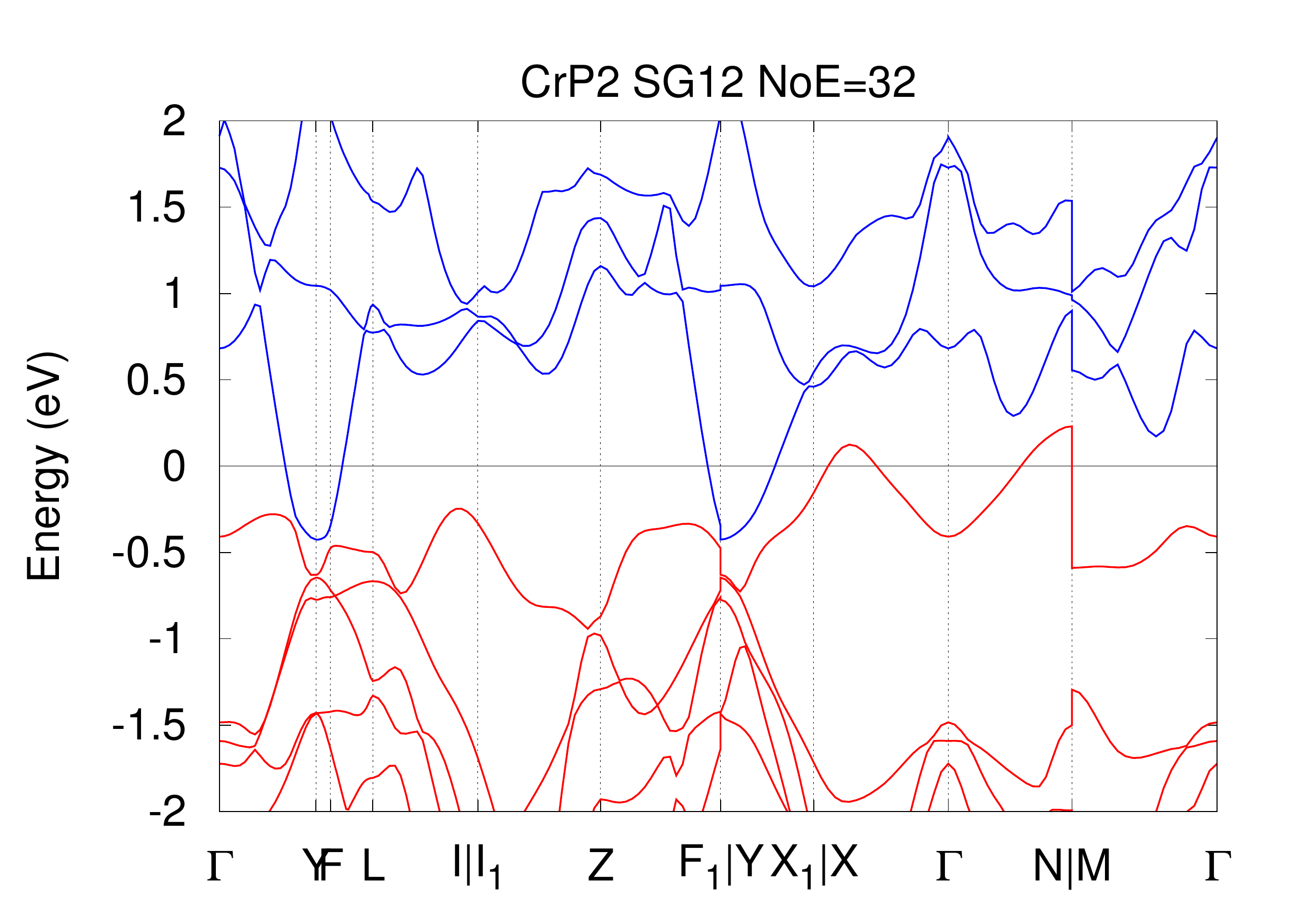}
}
\subfigure[BaTe$_{2}$ SG140 NoA=6 NoE=44]{
\label{subfig:75555}
\includegraphics[scale=0.32,angle=270]{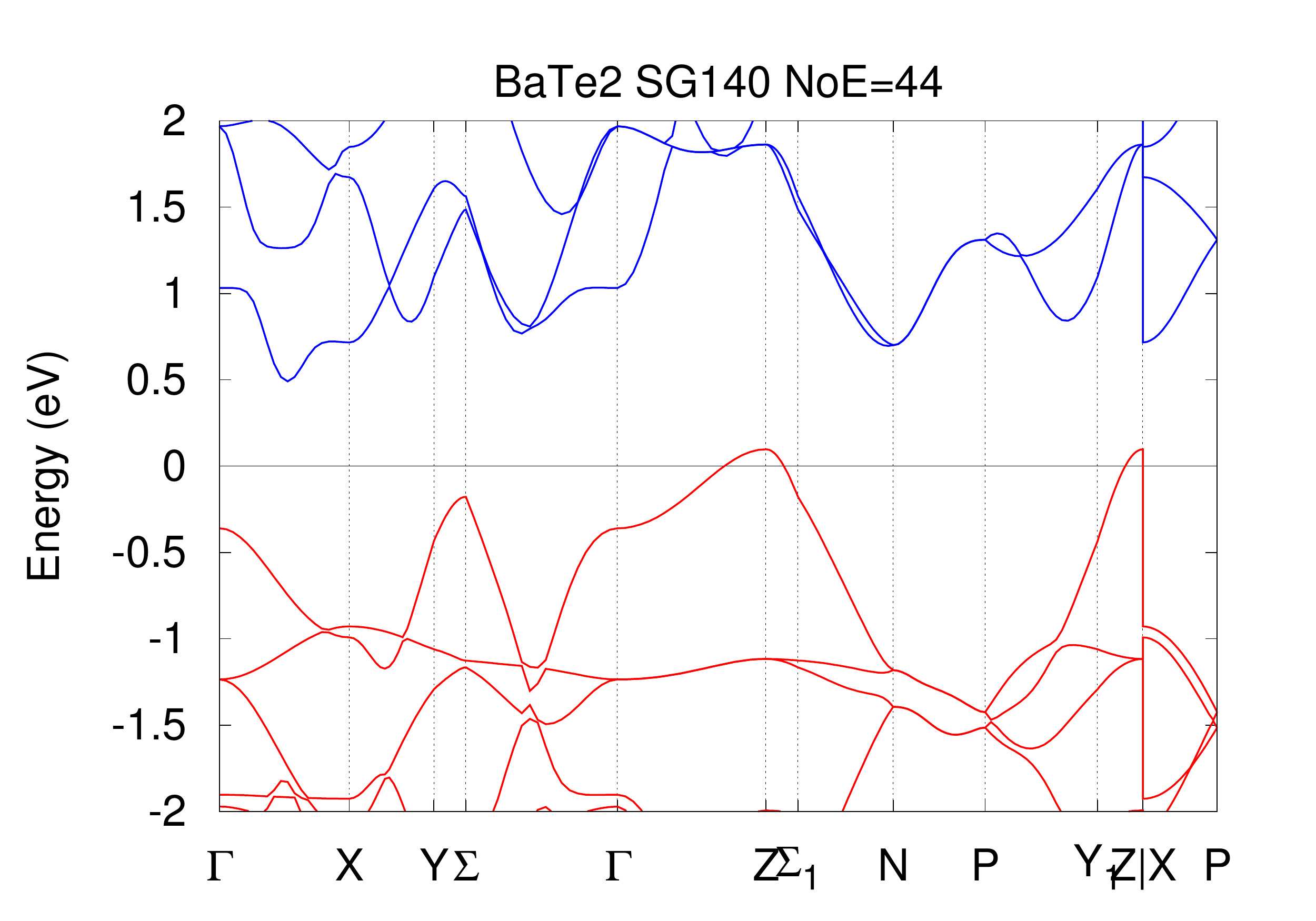}
}
\subfigure[Na$_{2}$Cl SG65 NoA=6 NoE=18]{
\label{subfig:190546}
\includegraphics[scale=0.32,angle=270]{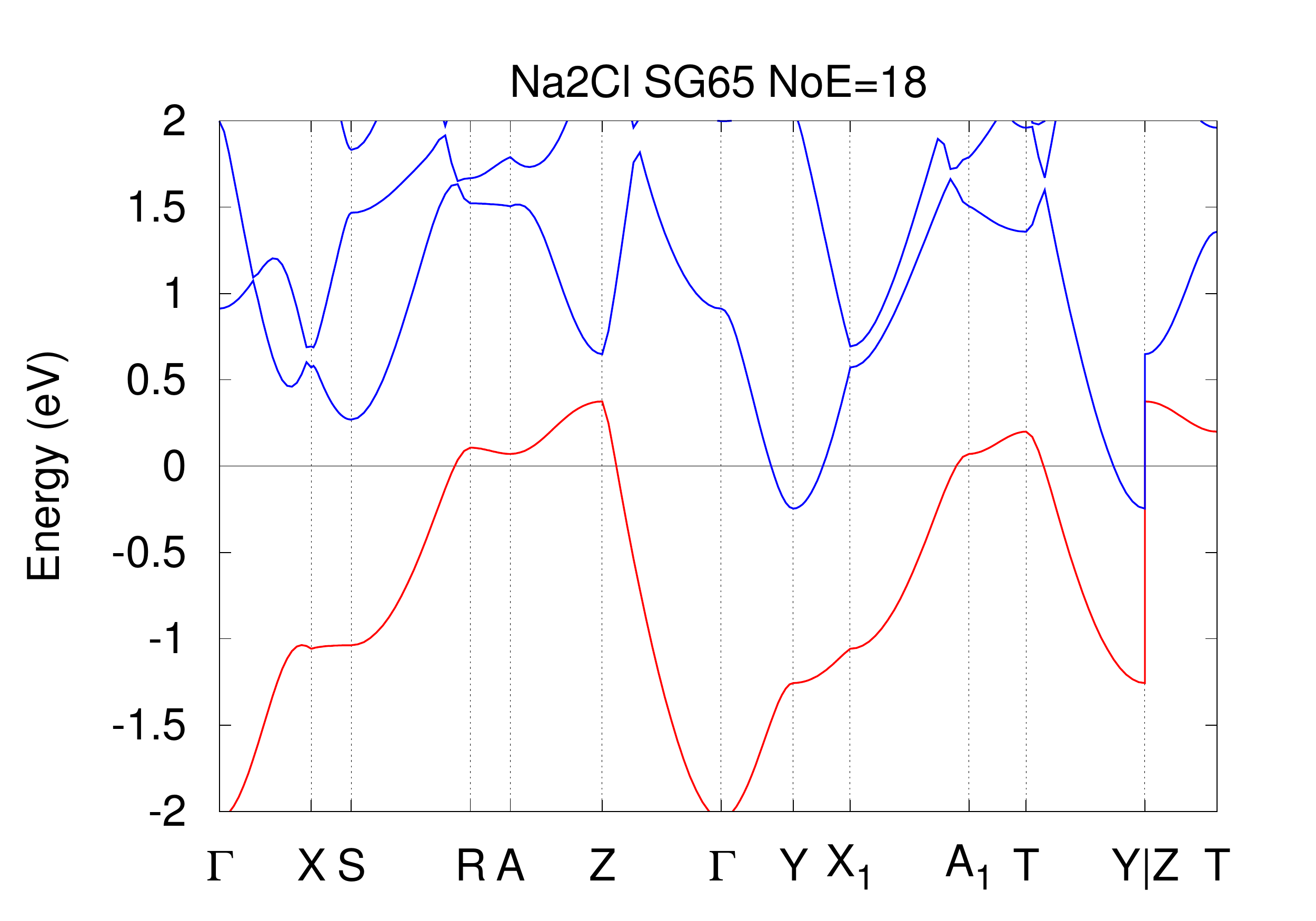}
}
\subfigure[YbInCu$_{4}$ SG216 NoA=6 NoE=55]{
\label{subfig:628189}
\includegraphics[scale=0.32,angle=270]{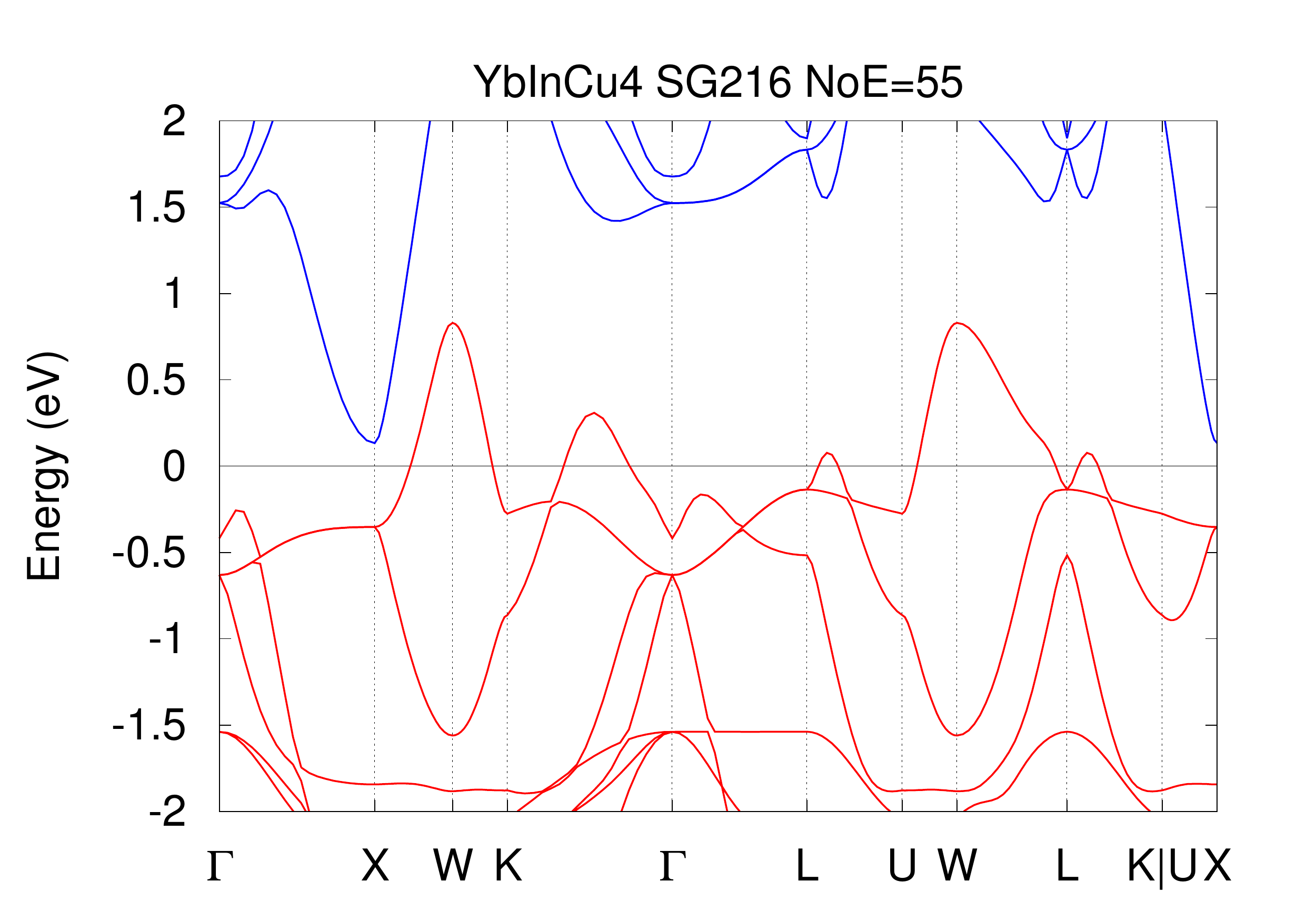}
}
\subfigure[MgInCu$_{4}$ SG216 NoA=6 NoE=49]{
\label{subfig:628018}
\includegraphics[scale=0.32,angle=270]{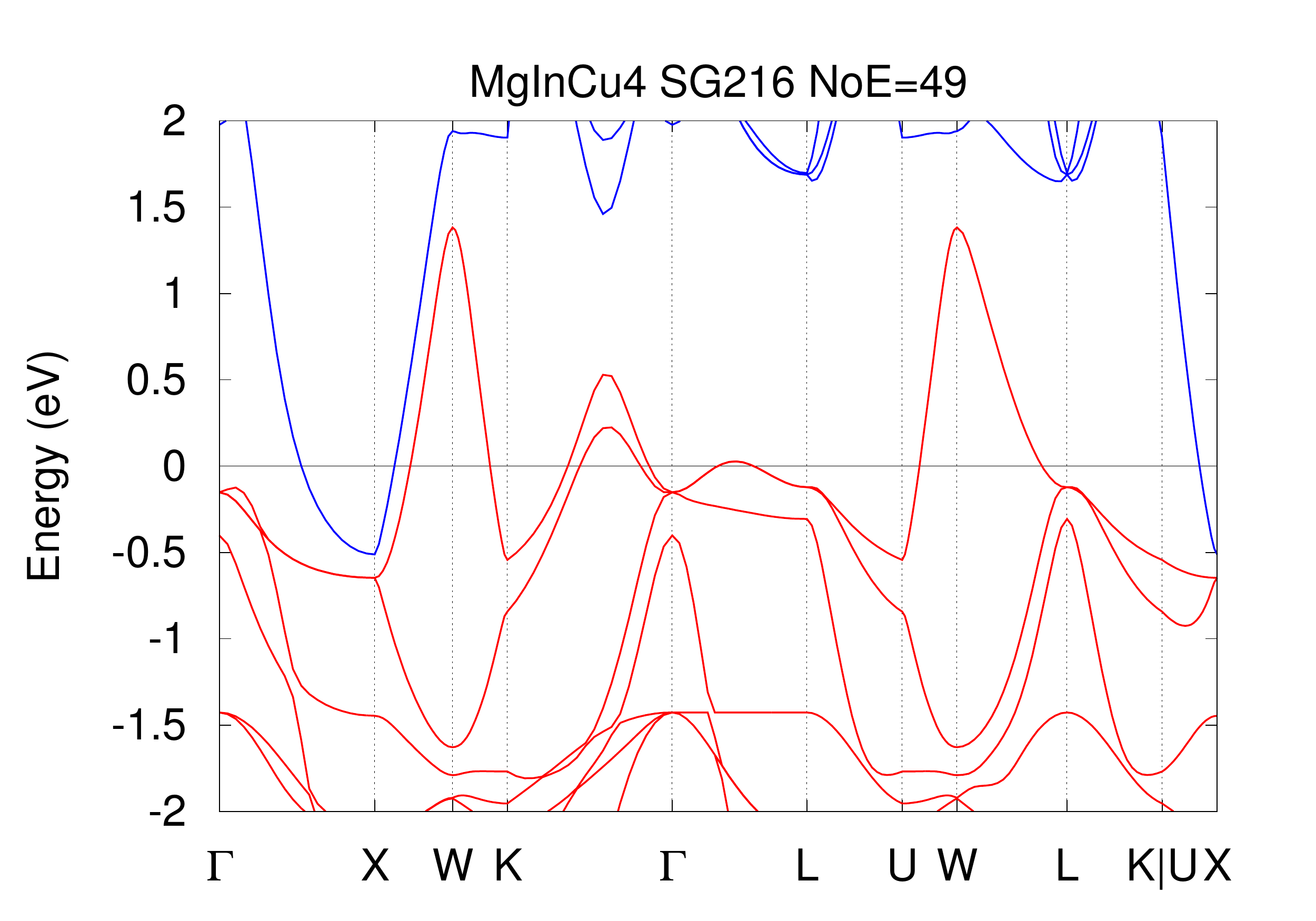}
}
\subfigure[ReN$_{2}$ SG12 NoA=6 NoE=34]{
\label{subfig:187441}
\includegraphics[scale=0.32,angle=270]{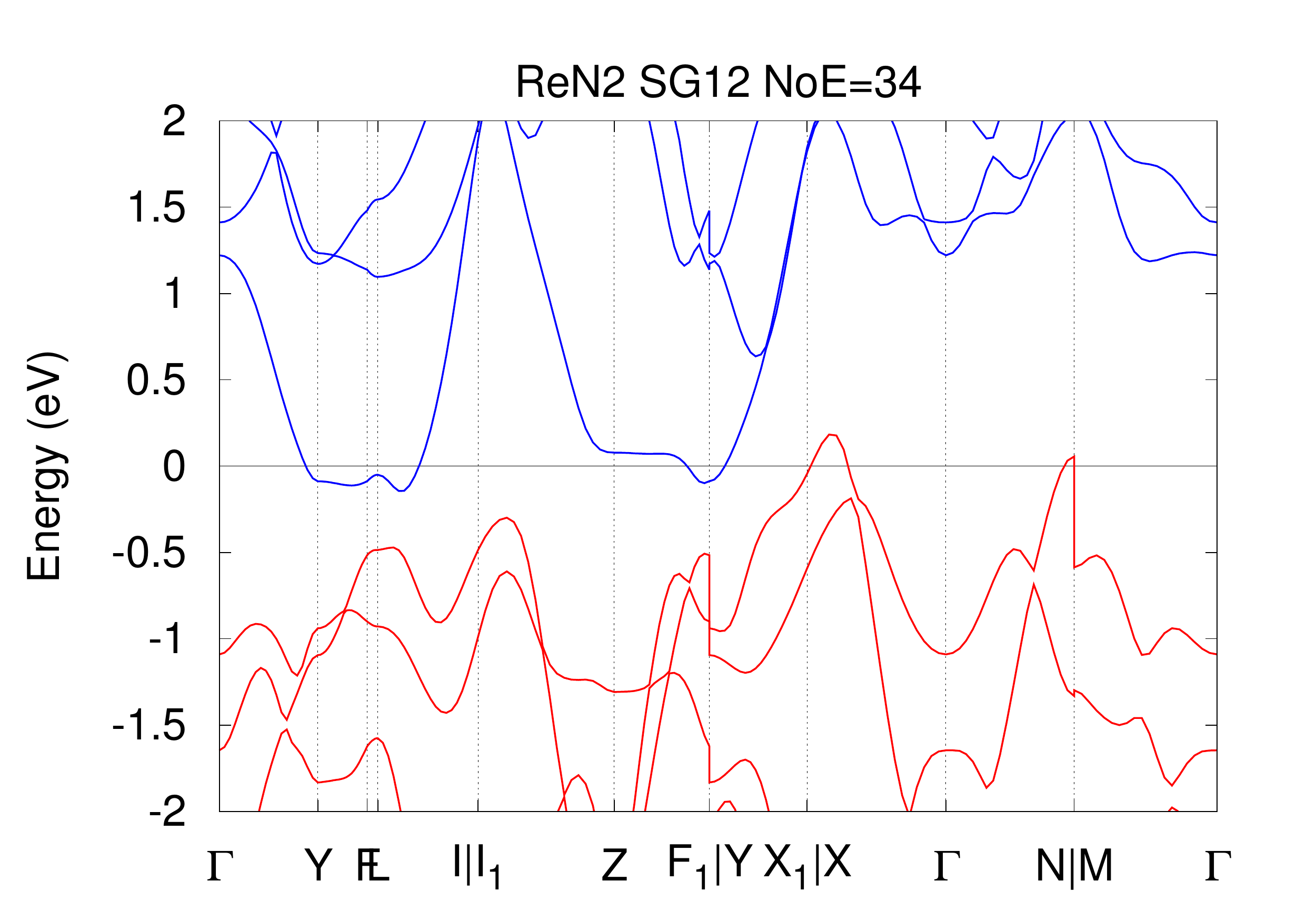}
}
\caption{\hyperref[tab:electride]{back to the table}}
\end{figure}

\begin{figure}[htp]
 \centering
\subfigure[LiYGa$_{4}$ SG187 NoA=6 NoE=24]{
\label{subfig:98666}
\includegraphics[scale=0.32,angle=270]{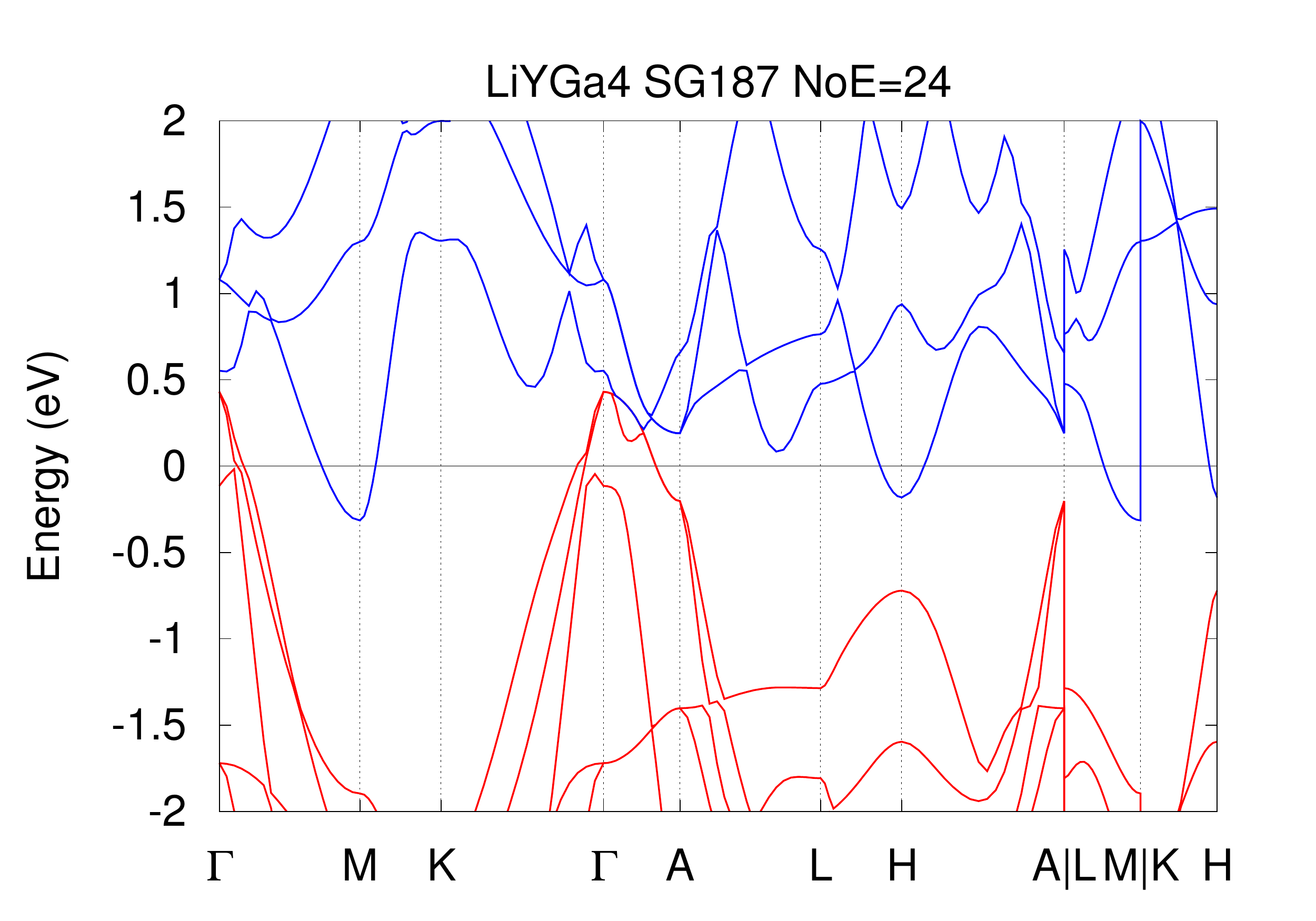}
}
\subfigure[FeAs$_{2}$ SG58 NoA=6 NoE=36]{
\label{subfig:65168}
\includegraphics[scale=0.32,angle=270]{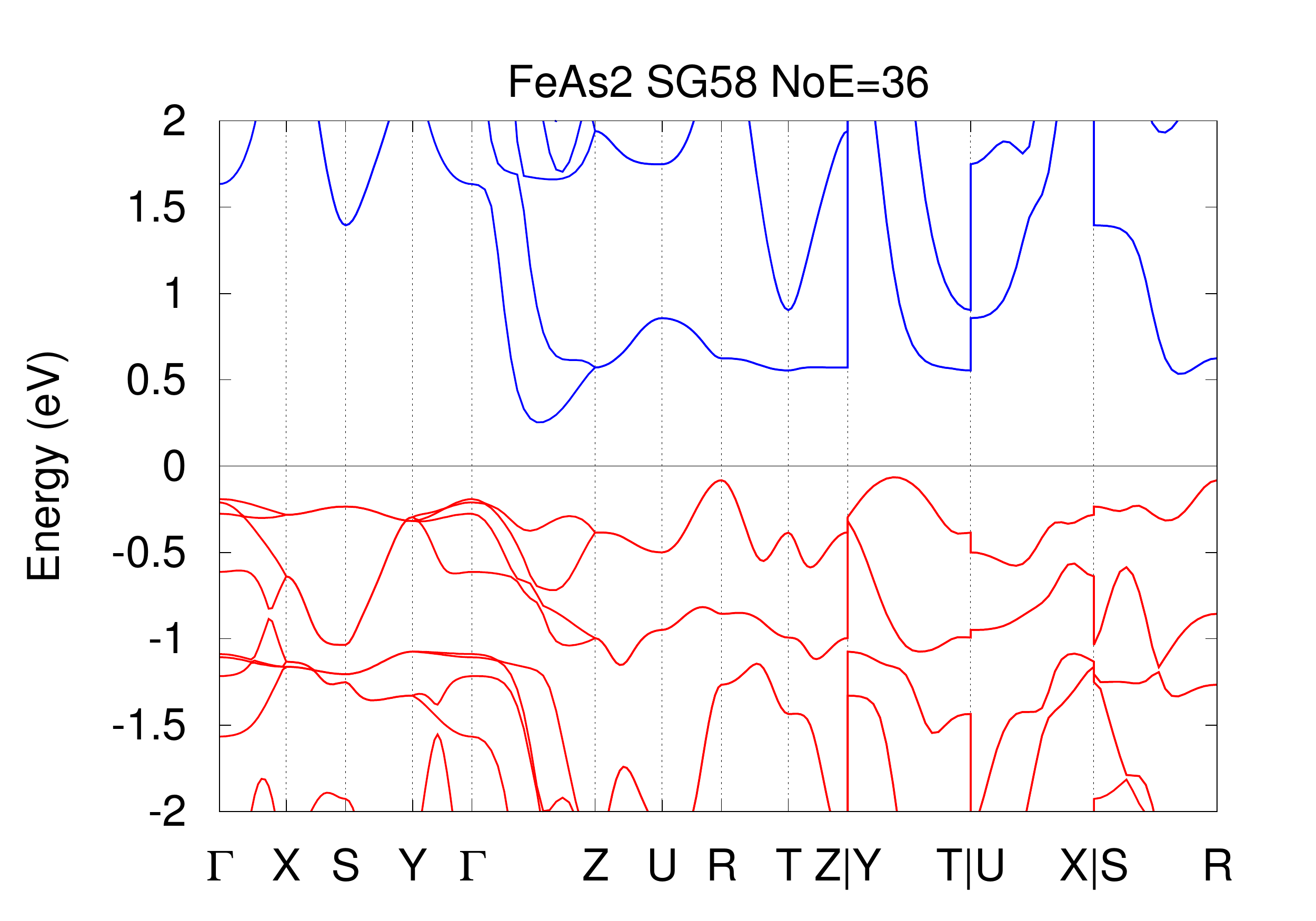}
}
\subfigure[MoAs$_{2}$ SG5 NoA=6 NoE=32]{
\label{subfig:16820}
\includegraphics[scale=0.32,angle=270]{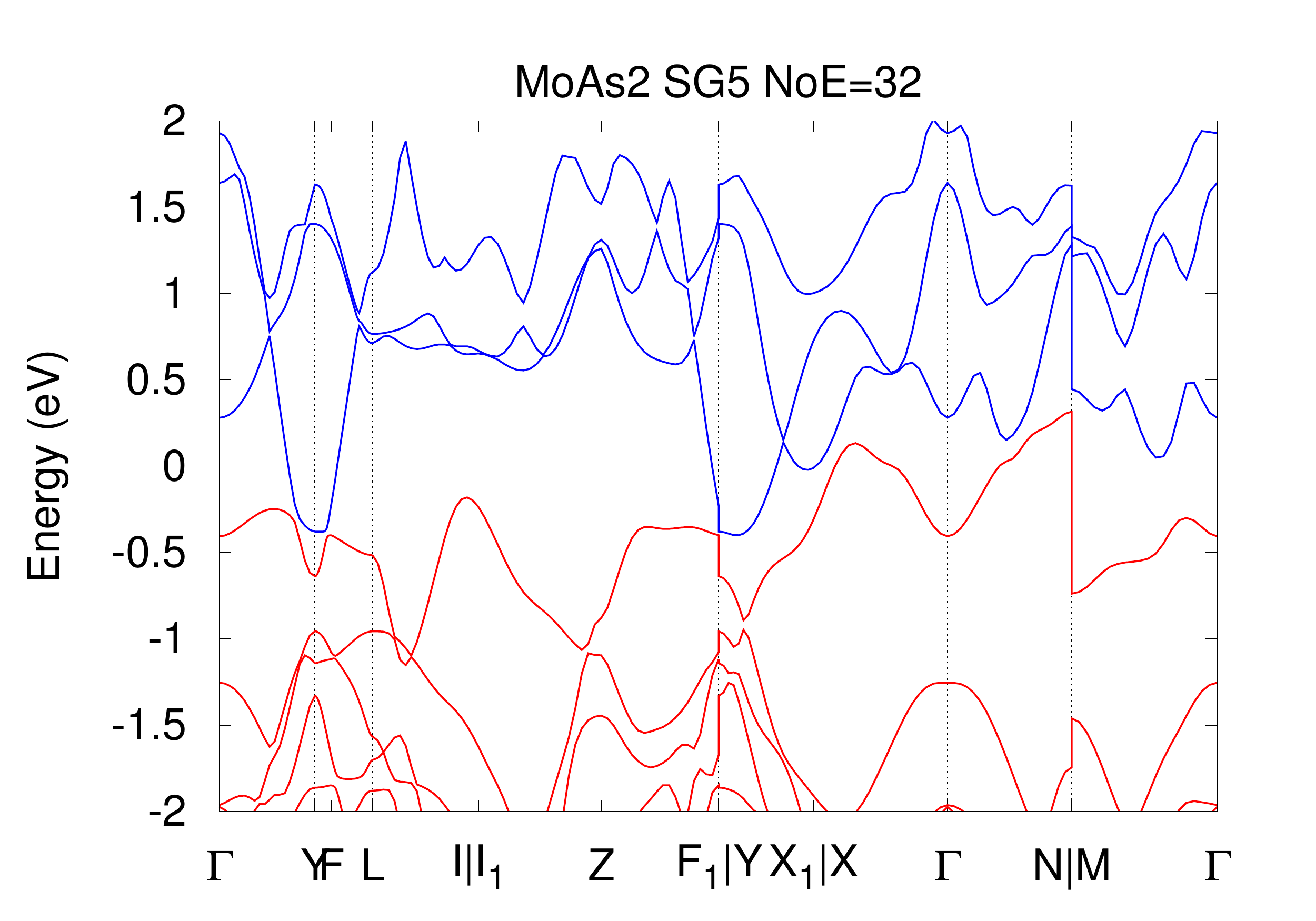}
}
\subfigure[YGaI SG164 NoA=6 NoE=42]{
\label{subfig:417149}
\includegraphics[scale=0.32,angle=270]{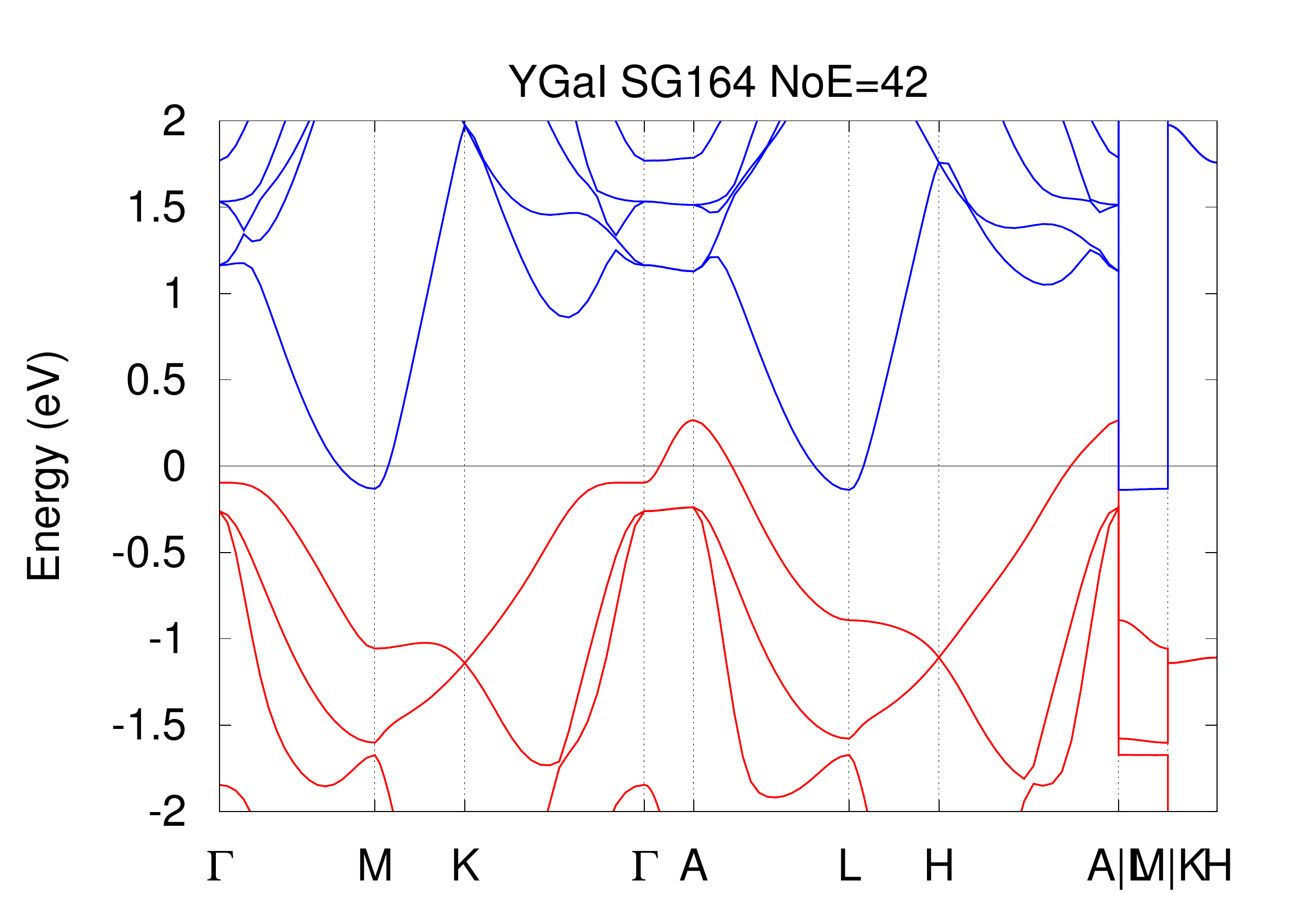}
}
\subfigure[ZrCu$_{5}$ SG216 NoA=6 NoE=67]{
\label{subfig:191404}
\includegraphics[scale=0.32,angle=270]{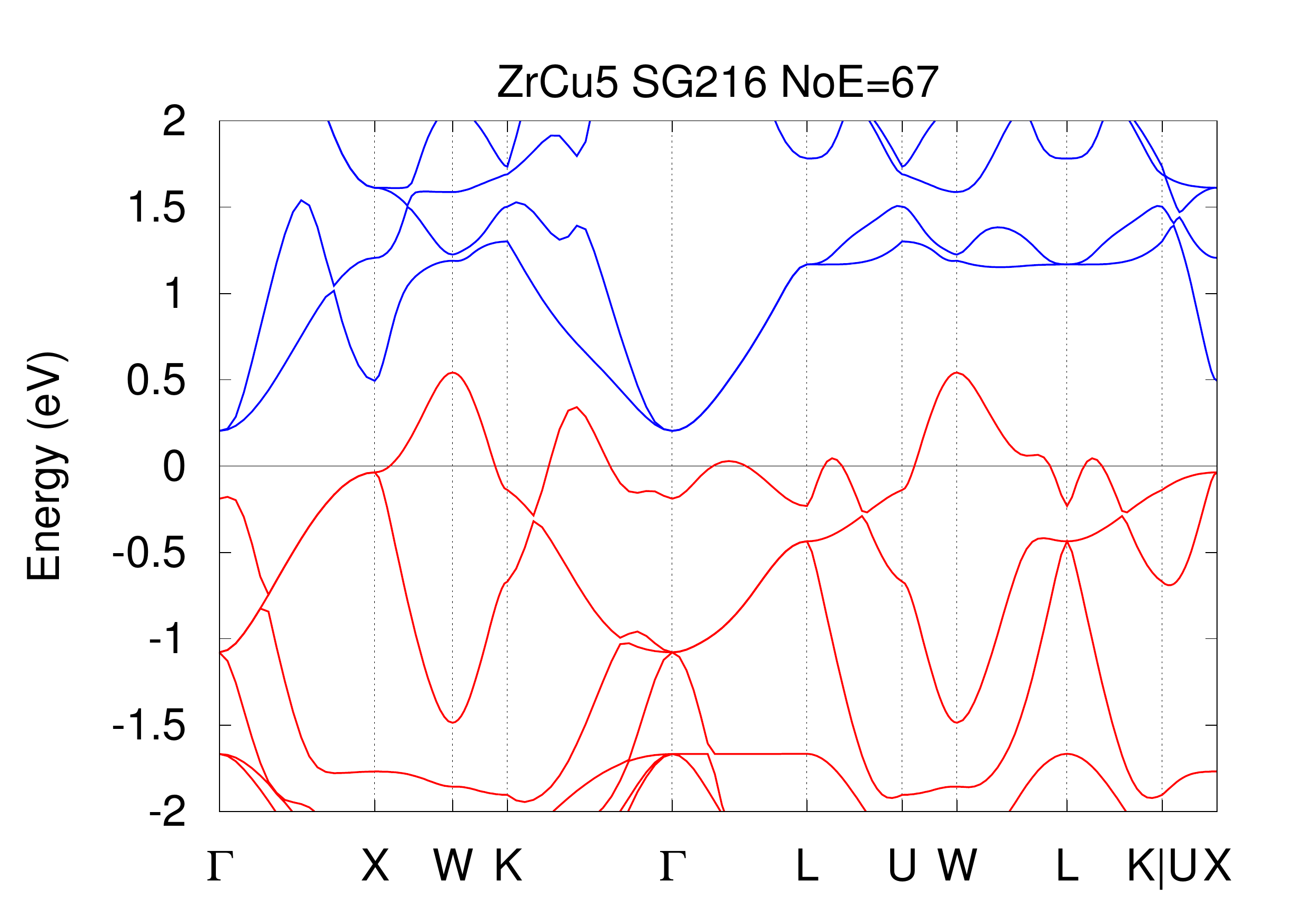}
}
\subfigure[FeTe$_{2}$ SG58 NoA=6 NoE=40]{
\label{subfig:633866}
\includegraphics[scale=0.32,angle=270]{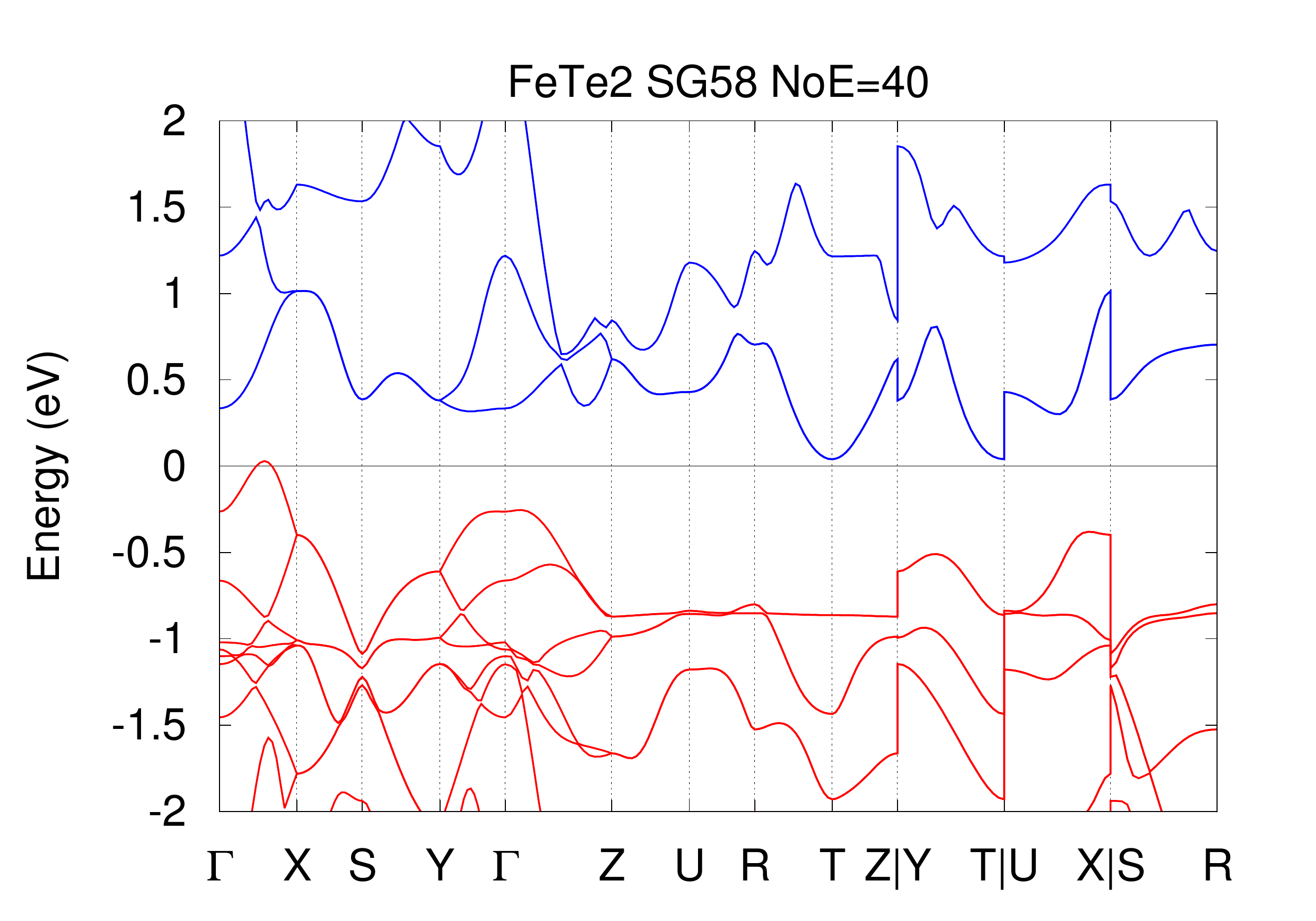}
}
\subfigure[CaInCu$_{4}$ SG216 NoA=6 NoE=57]{
\label{subfig:658914}
\includegraphics[scale=0.32,angle=270]{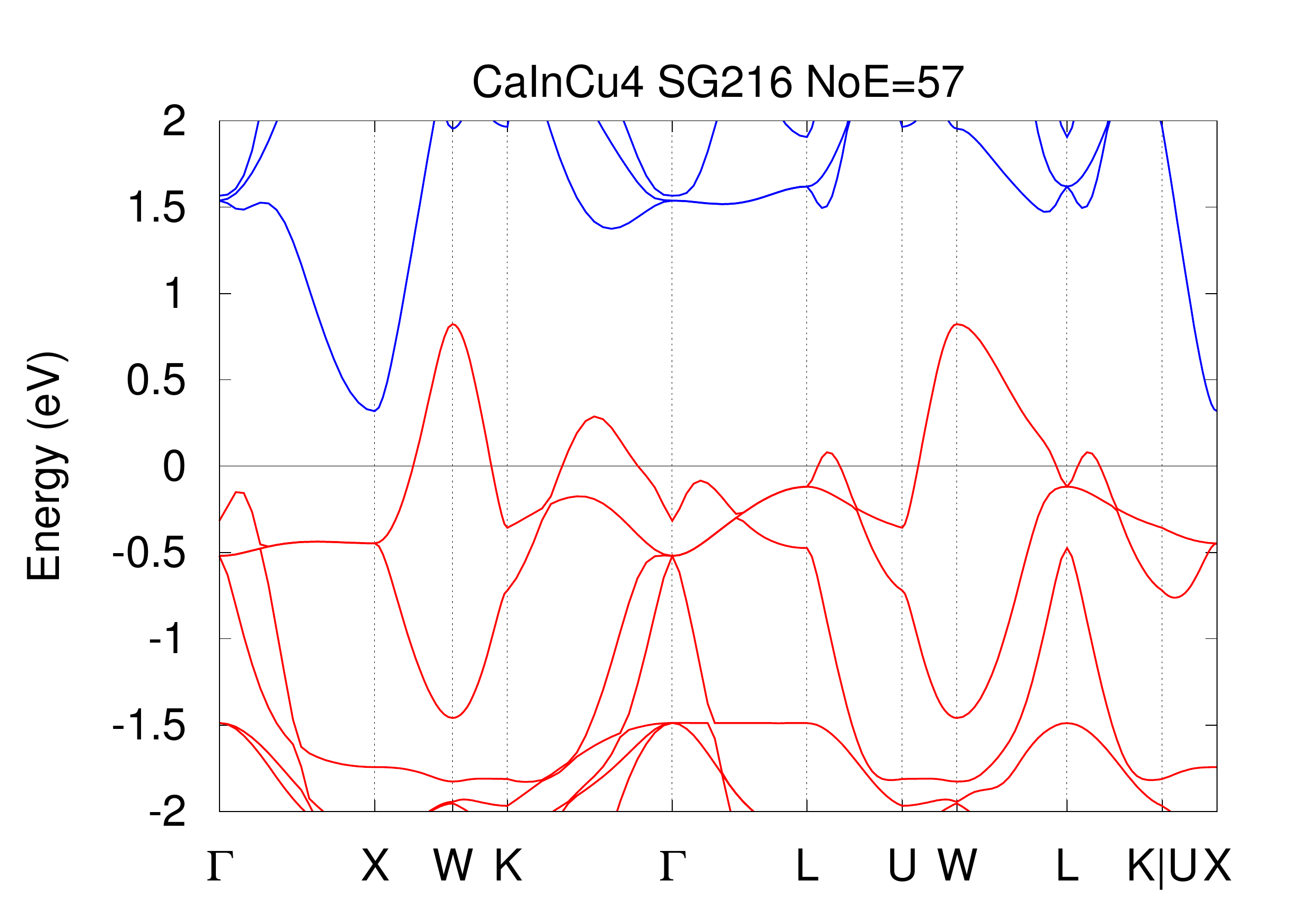}
}
\subfigure[As$_{2}$W SG12 NoA=6 NoE=32]{
\label{subfig:611576}
\includegraphics[scale=0.32,angle=270]{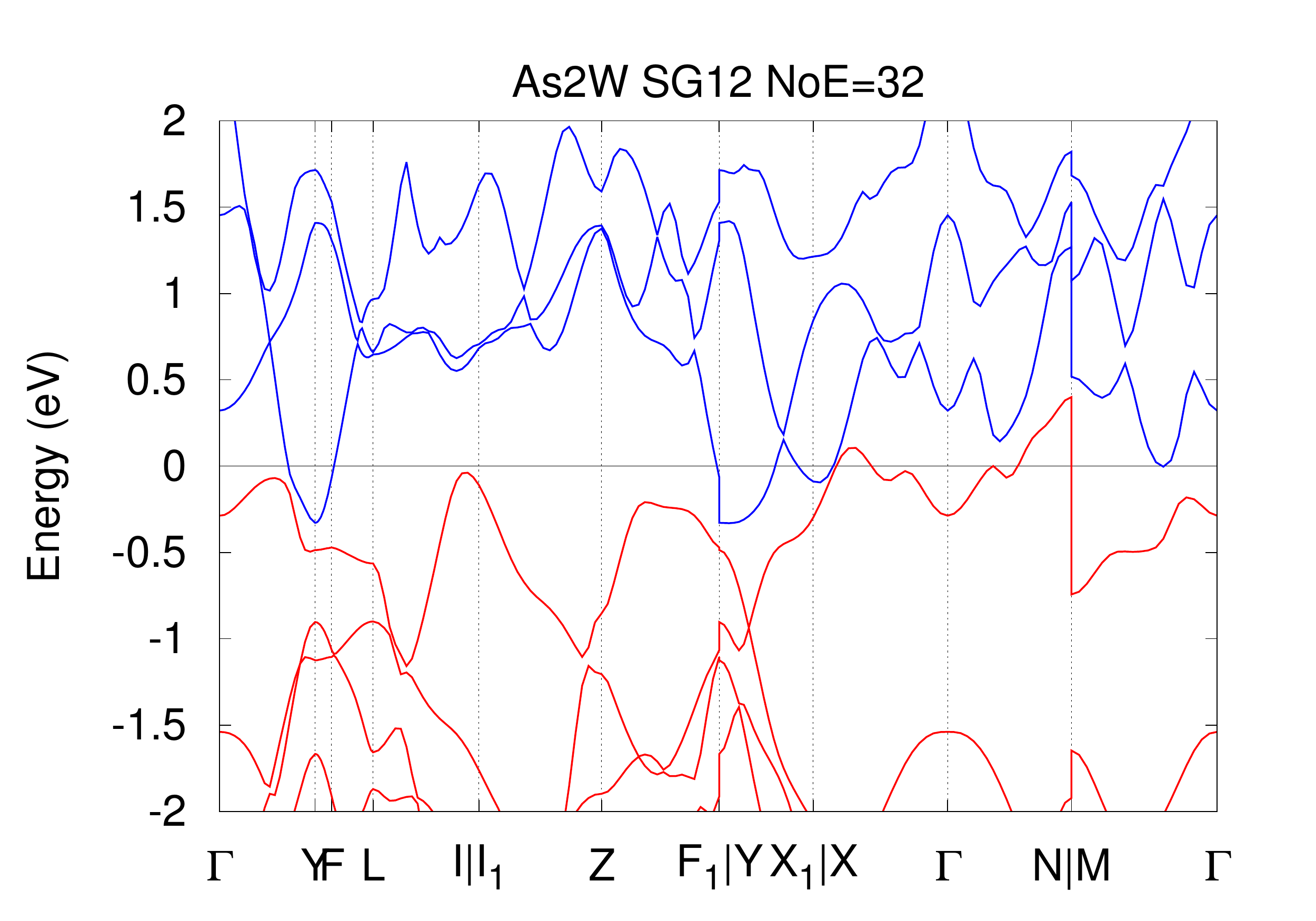}
}
\caption{\hyperref[tab:electride]{back to the table}}
\end{figure}

\begin{figure}[htp]
 \centering
\subfigure[As$_{2}$Ru SG58 NoA=6 NoE=36]{
\label{subfig:42578}
\includegraphics[scale=0.32,angle=270]{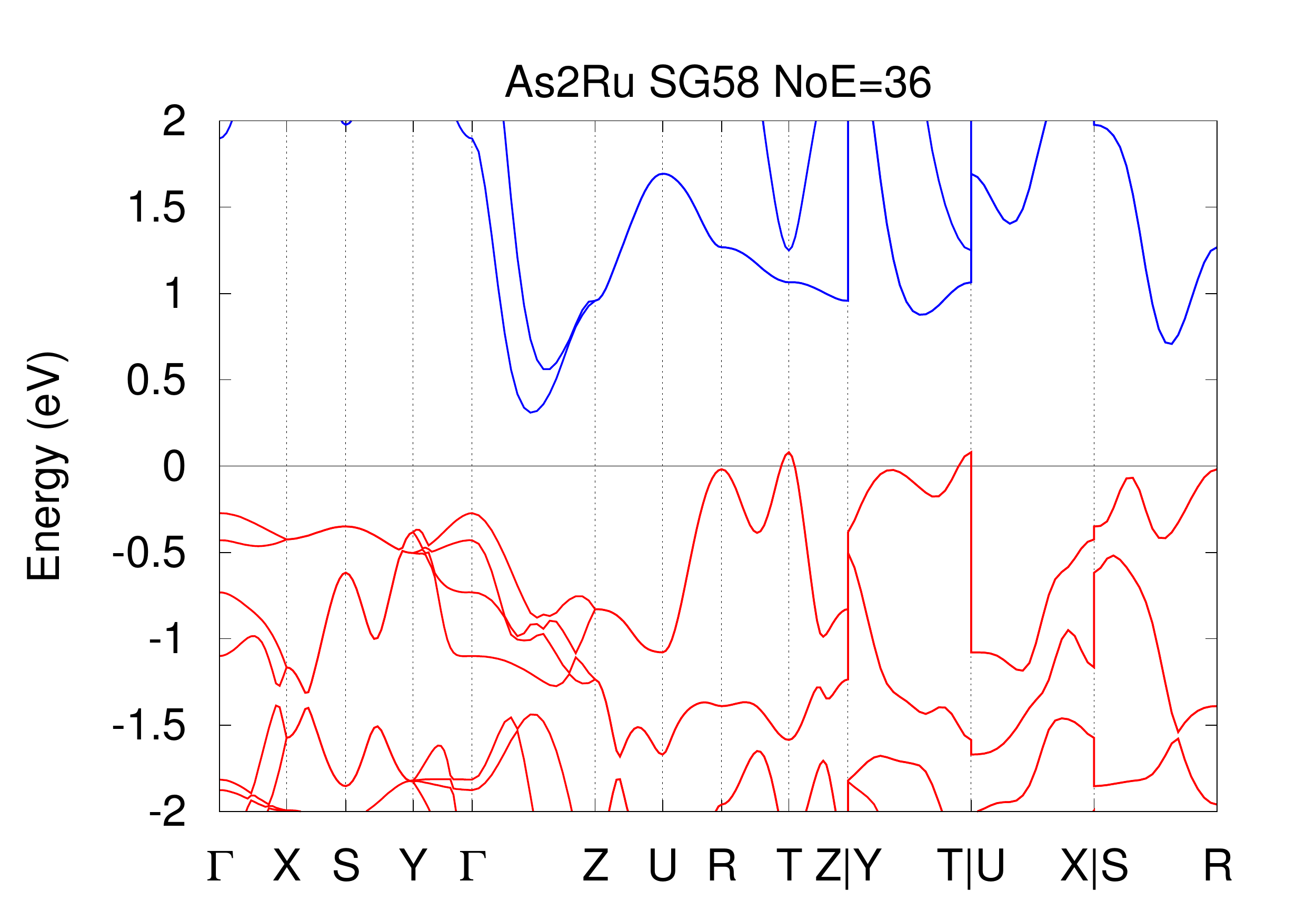}
}
\subfigure[TbInCu$_{4}$ SG216 NoA=6 NoE=56]{
\label{subfig:152560}
\includegraphics[scale=0.32,angle=270]{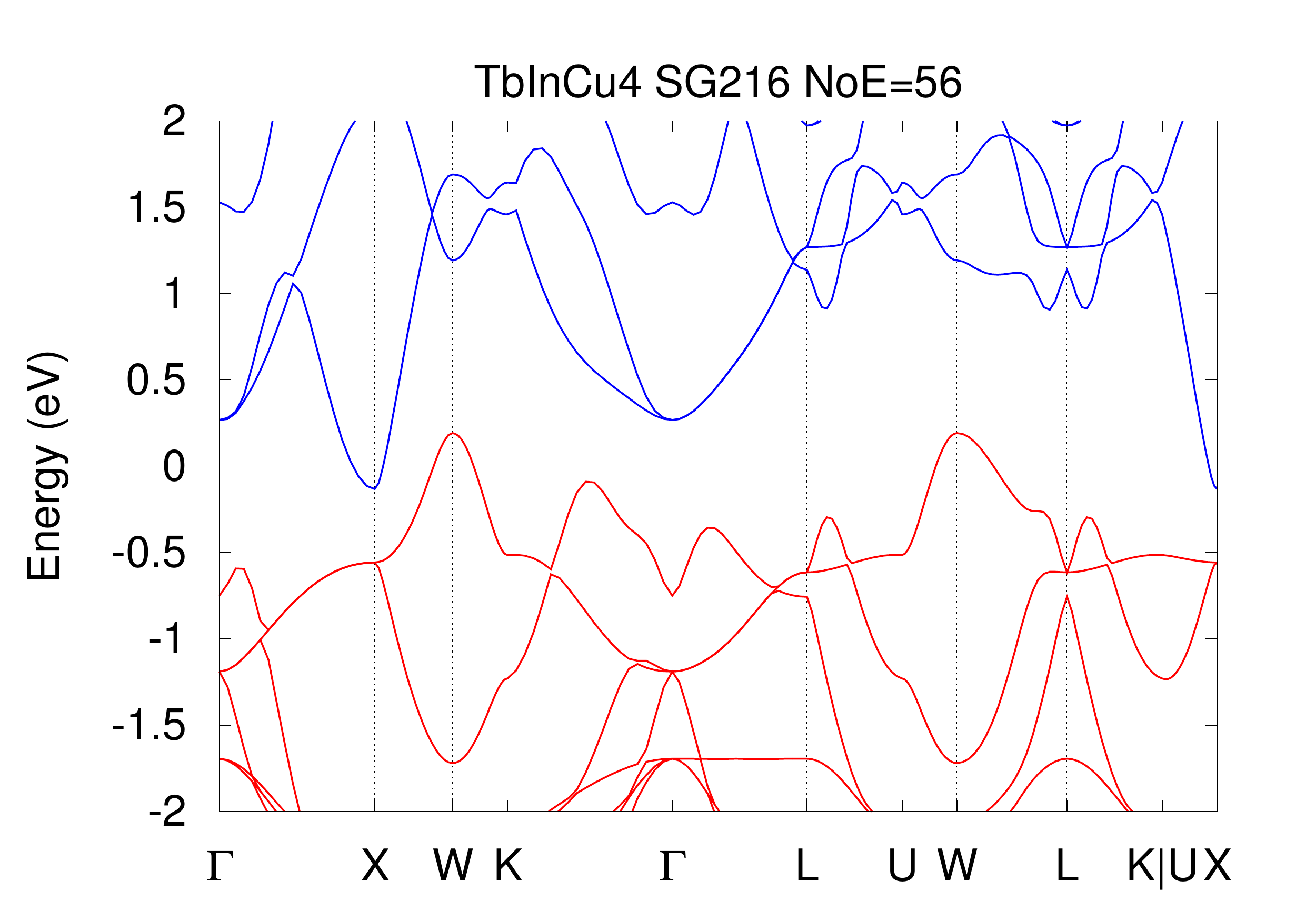}
}
\subfigure[KB$_{6}$ SG221 NoA=7 NoE=27]{
\label{subfig:98990}
\includegraphics[scale=0.32,angle=270]{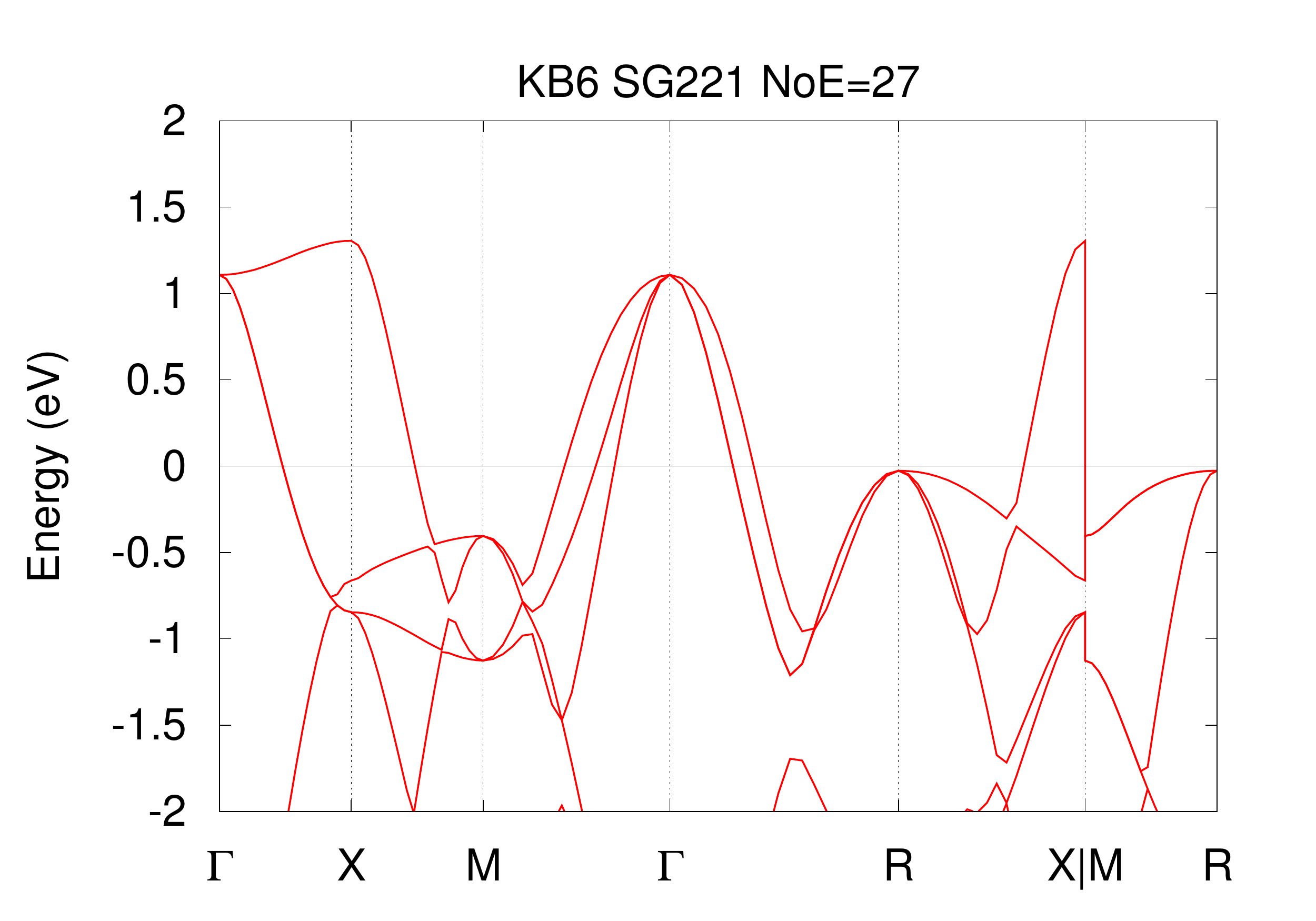}
}
\subfigure[Rb$_{2}$Te$_{5}$ SG12 NoA=7 NoE=48]{
\label{subfig:30734}
\includegraphics[scale=0.32,angle=270]{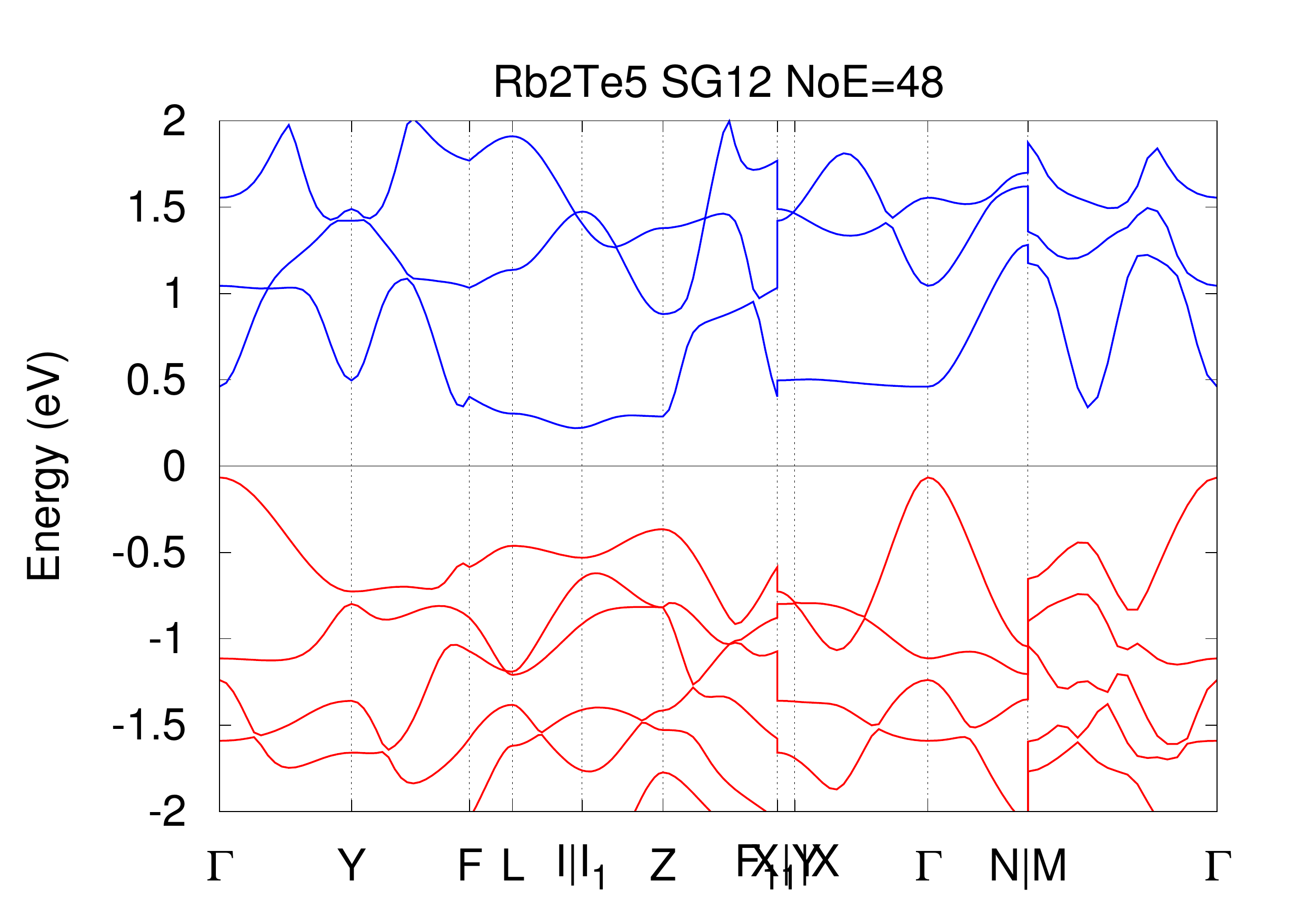}
}
\subfigure[SiB$_{6}$ SG221 NoA=7 NoE=22]{
\label{subfig:20240}
\includegraphics[scale=0.32,angle=270]{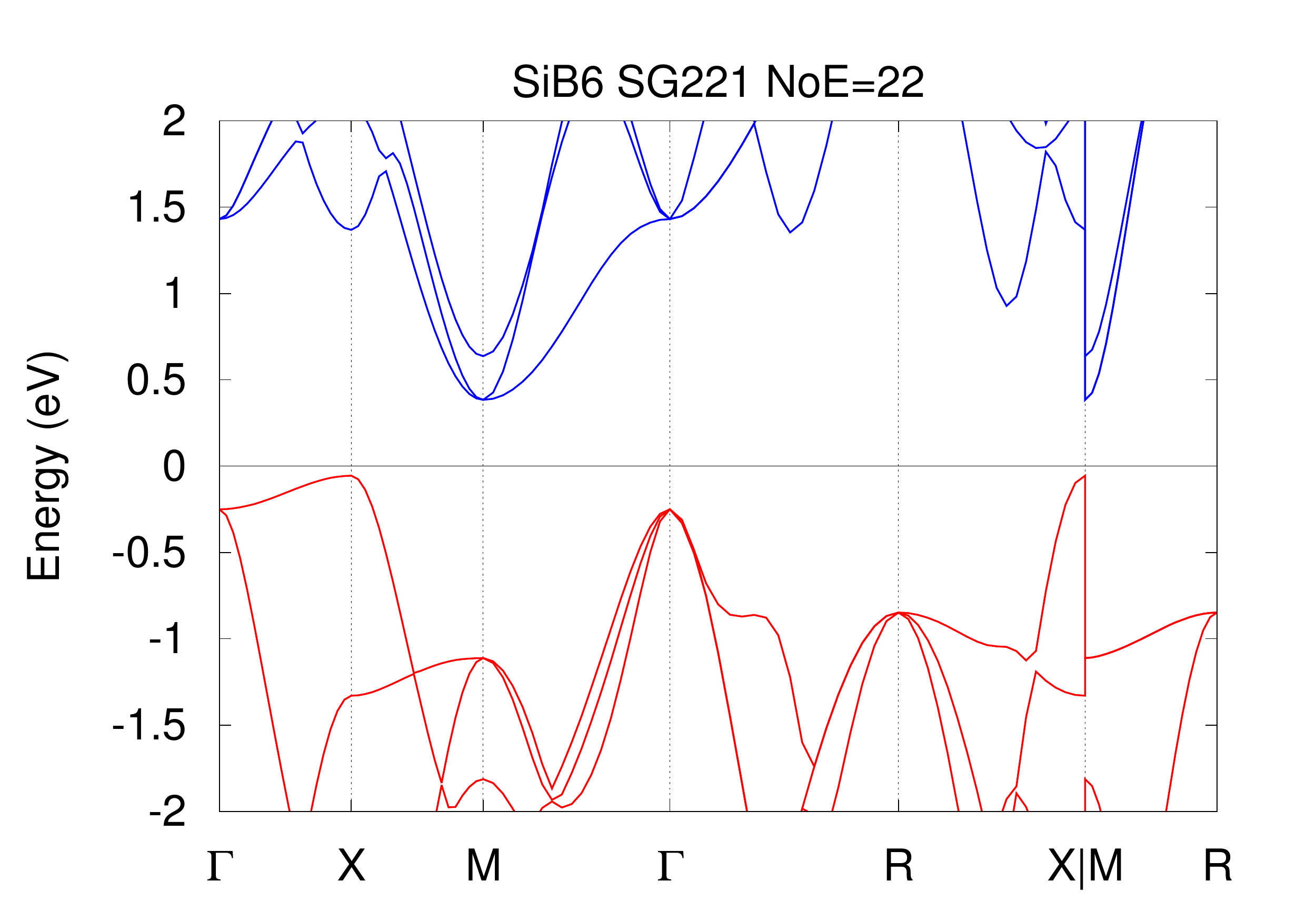}
}
\subfigure[V(MoS$_{2}$)$_{2}$ SG12 NoA=7 NoE=41]{
\label{subfig:201787}
\includegraphics[scale=0.32,angle=270]{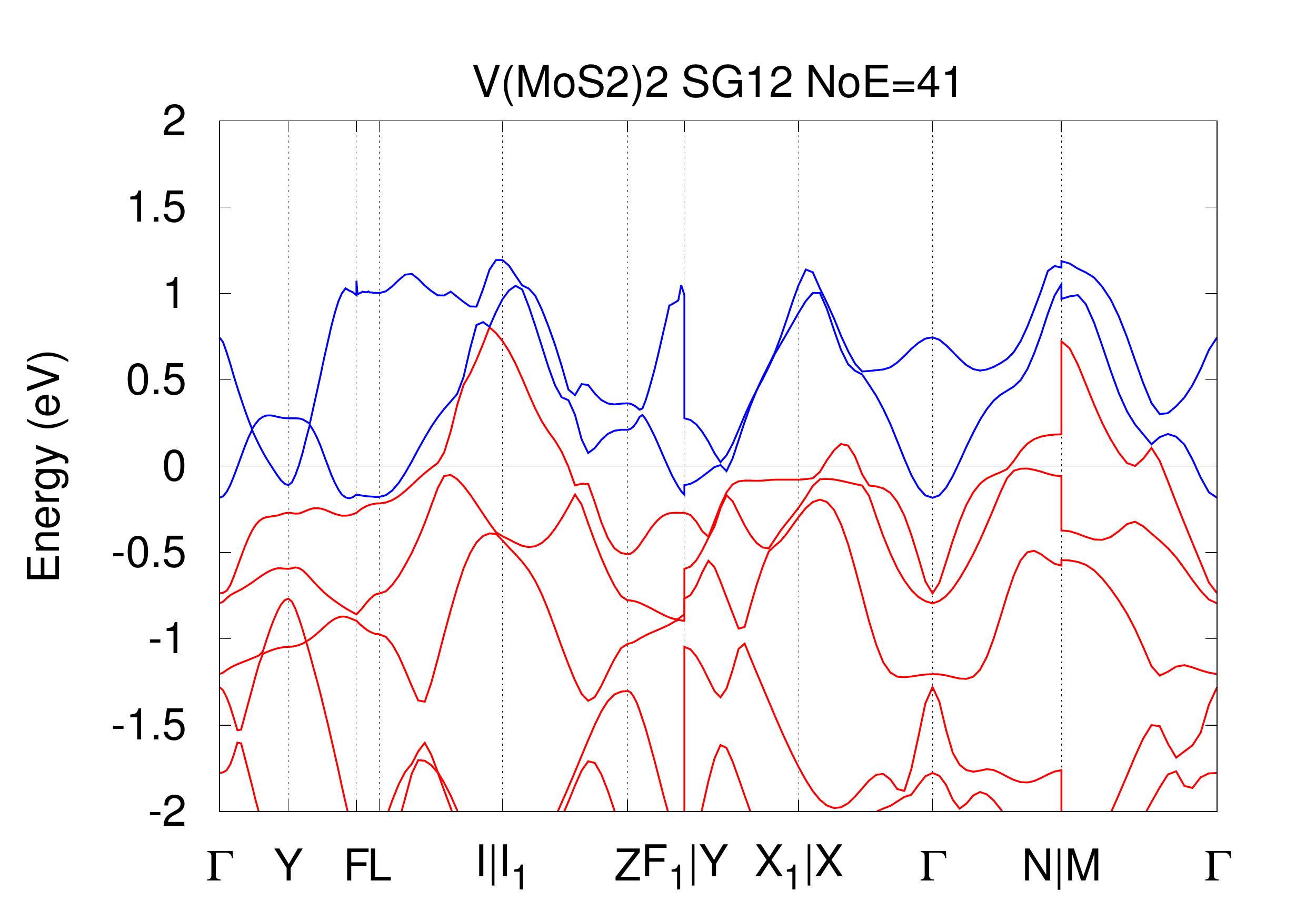}
}
\subfigure[Ge SG148 NoA=8 NoE=32]{
\label{subfig:245961}
\includegraphics[scale=0.32,angle=270]{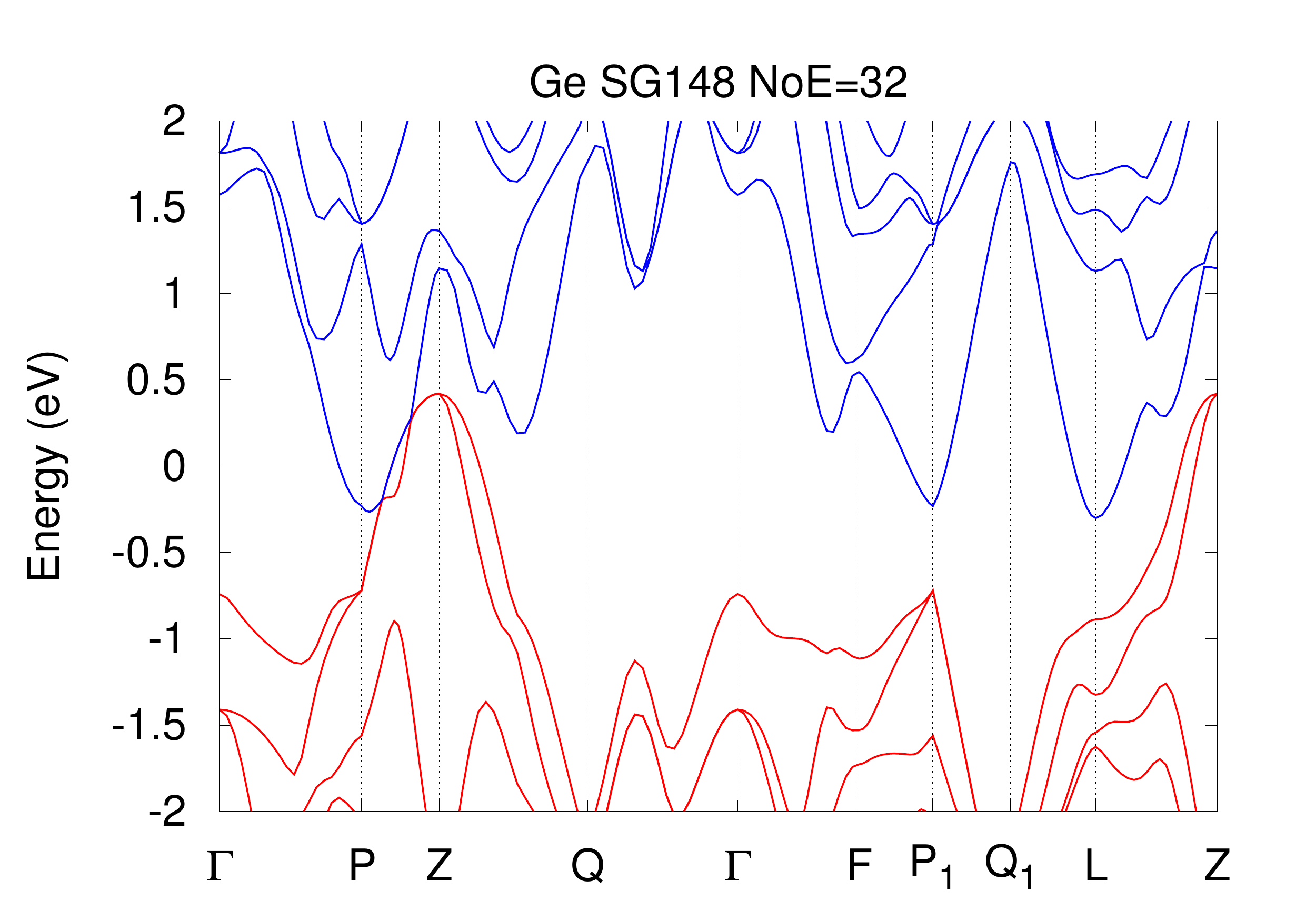}
}
\subfigure[CrB SG141 NoA=8 NoE=36]{
\label{subfig:613476}
\includegraphics[scale=0.32,angle=270]{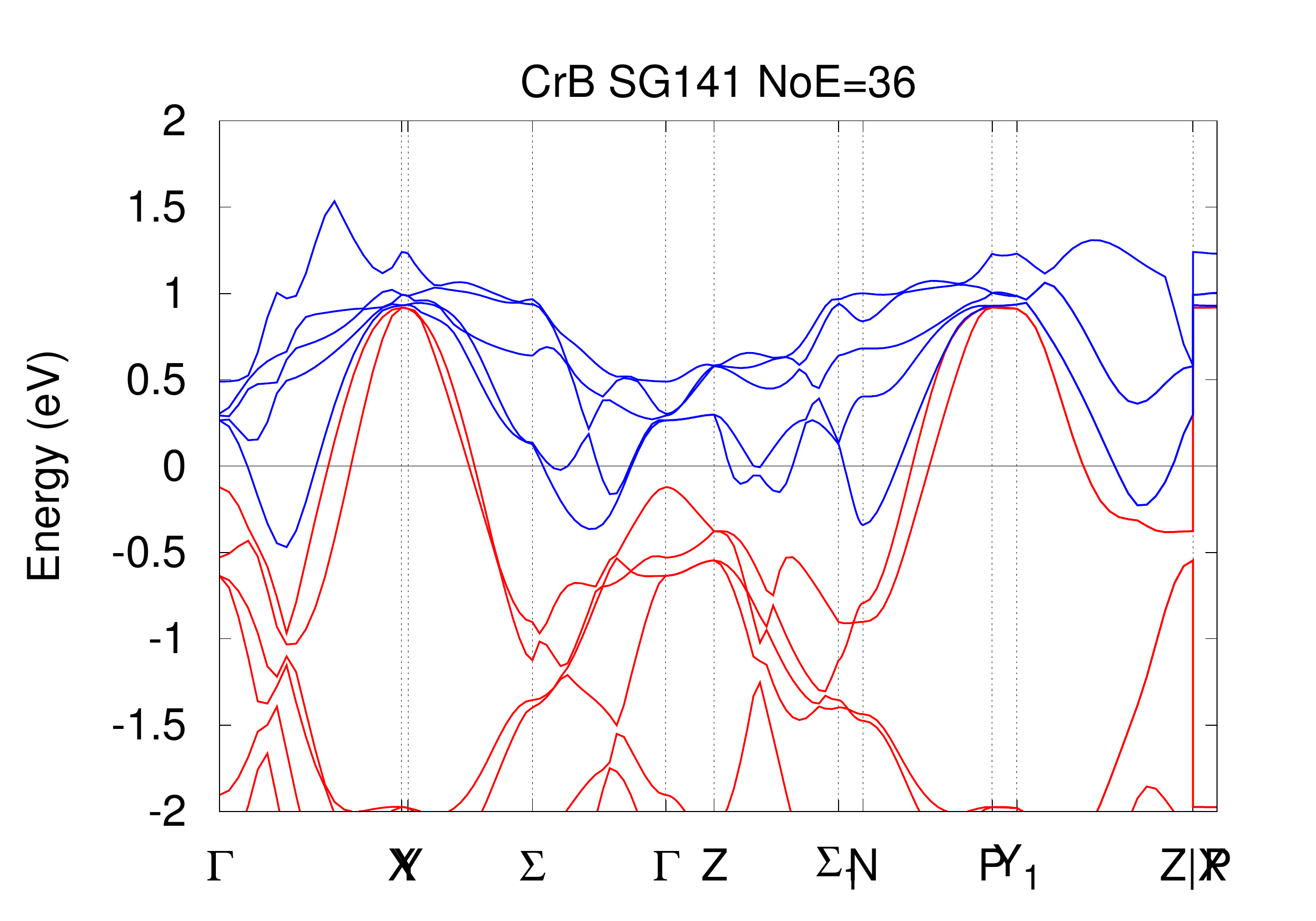}
}
\caption{\hyperref[tab:electride]{back to the table}}
\end{figure}

\begin{figure}[htp]
 \centering
\subfigure[KNbSe$_{2}$ SG194 NoA=8 NoE=64]{
\label{subfig:26288}
\includegraphics[scale=0.32,angle=270]{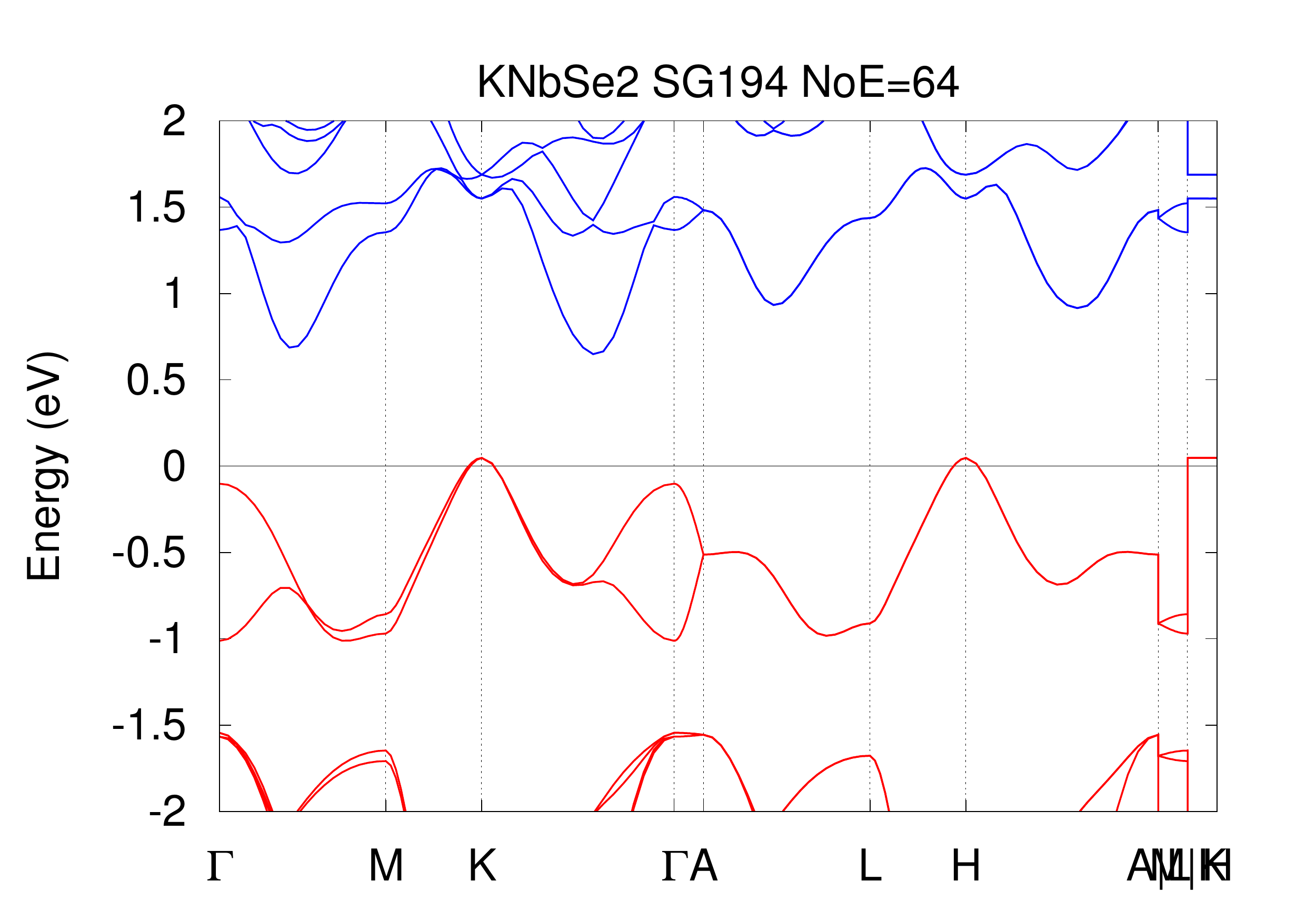}
}
\subfigure[Te$_{2}$AuI SG51 NoA=8 NoE=60]{
\label{subfig:16325}
\includegraphics[scale=0.32,angle=270]{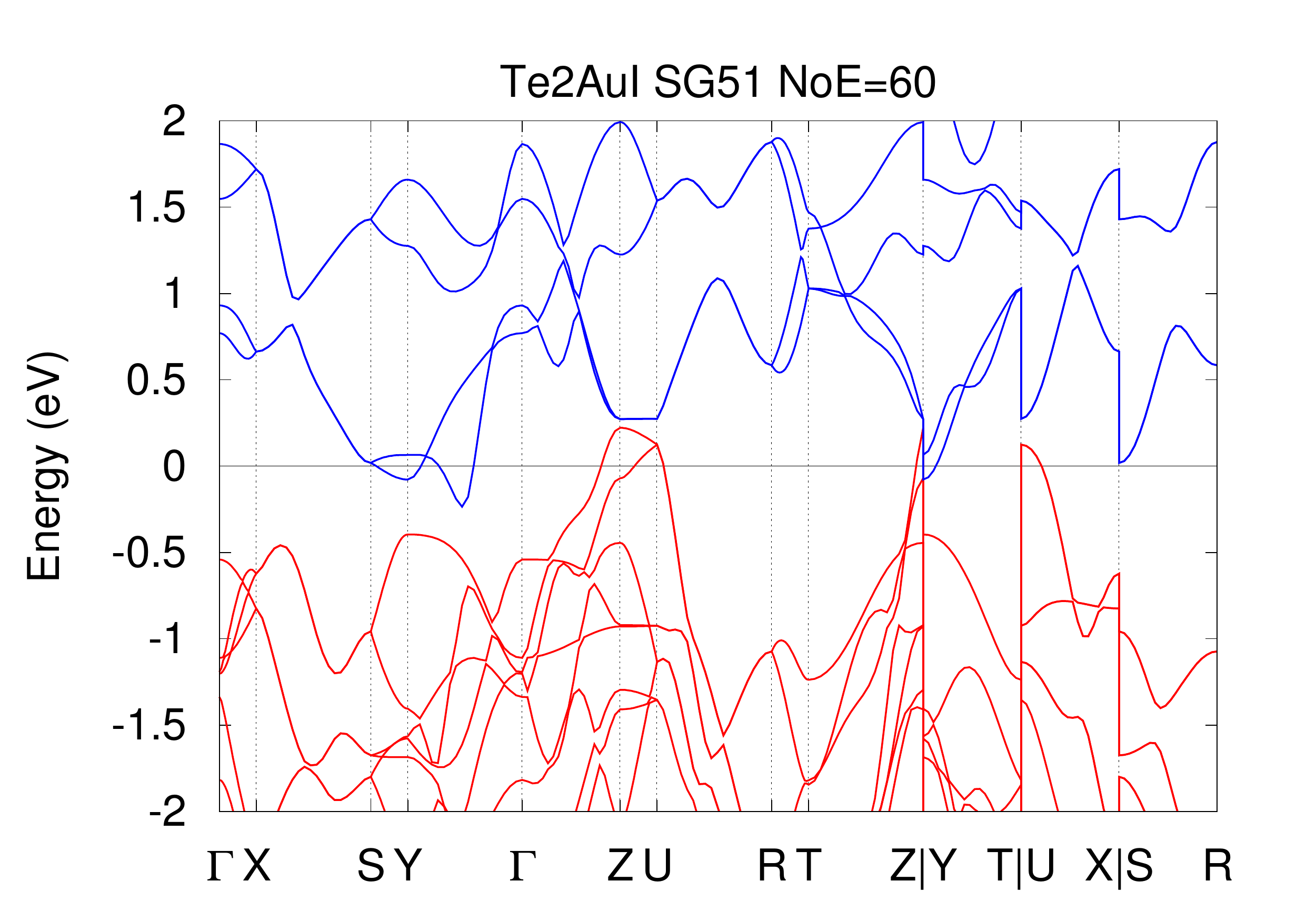}
}
\subfigure[BMo SG141 NoA=8 NoE=36]{
\label{subfig:614793}
\includegraphics[scale=0.32,angle=270]{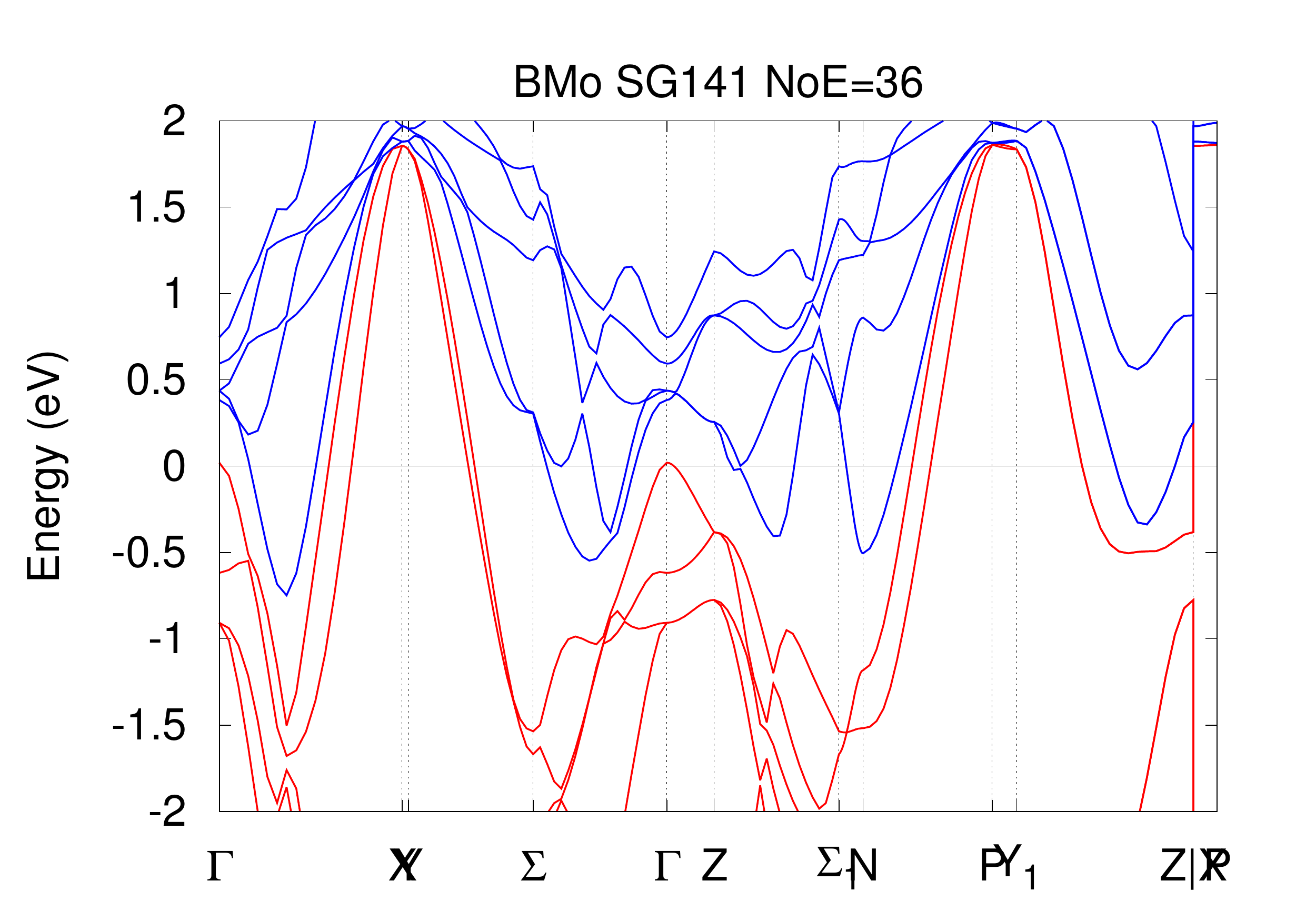}
}
\subfigure[NaNbS$_{2}$ SG194 NoA=8 NoE=48]{
\label{subfig:26285}
\includegraphics[scale=0.32,angle=270]{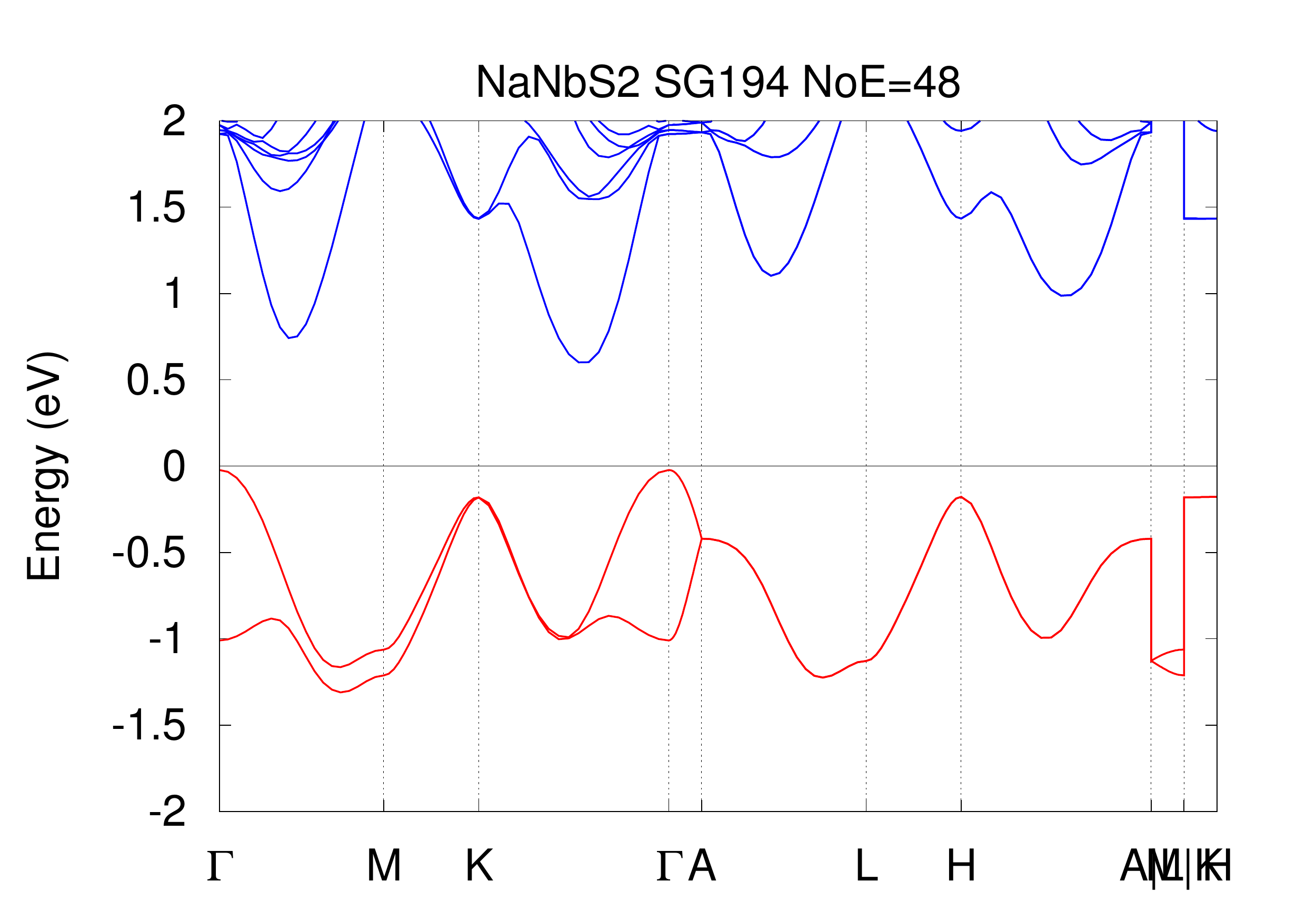}
}
\subfigure[NaCl$_{7}$ SG200 NoA=8 NoE=50]{
\label{subfig:190537}
\includegraphics[scale=0.32,angle=270]{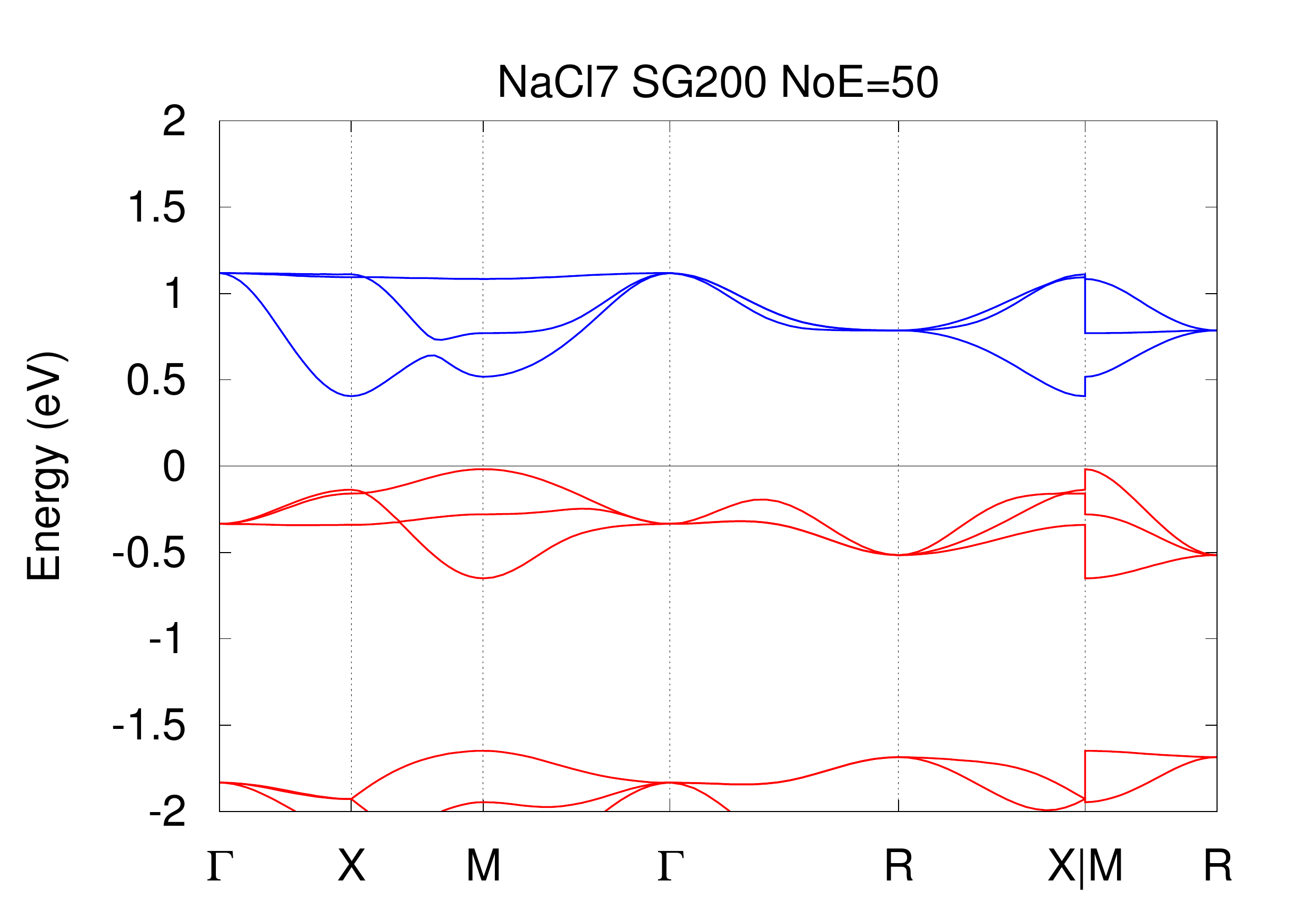}
}
\subfigure[LaSi SG63 NoA=8 NoE=60]{
\label{subfig:408030}
\includegraphics[scale=0.32,angle=270]{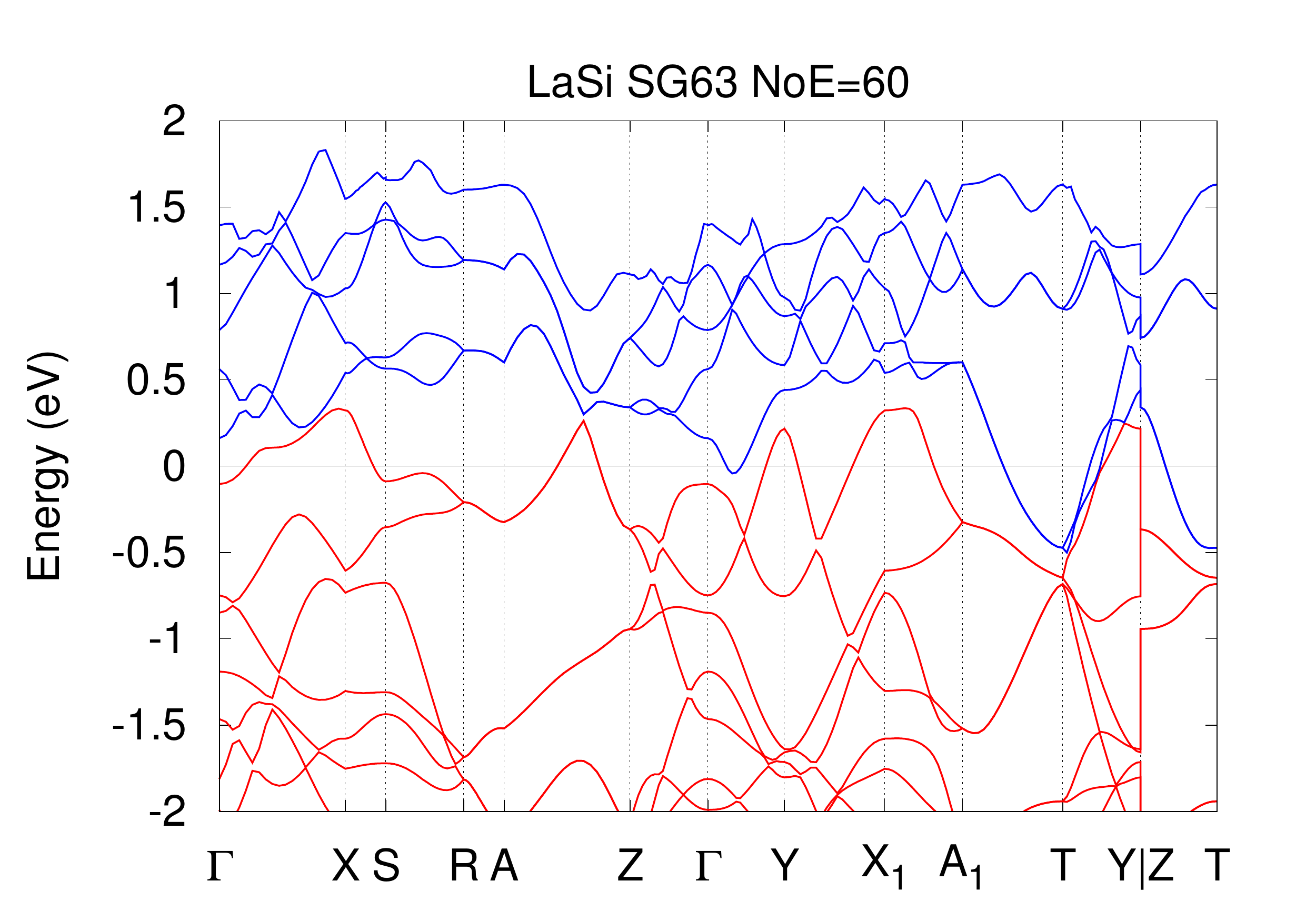}
}
\subfigure[TaInS$_{2}$ SG194 NoA=8 NoE=40]{
\label{subfig:640379}
\includegraphics[scale=0.32,angle=270]{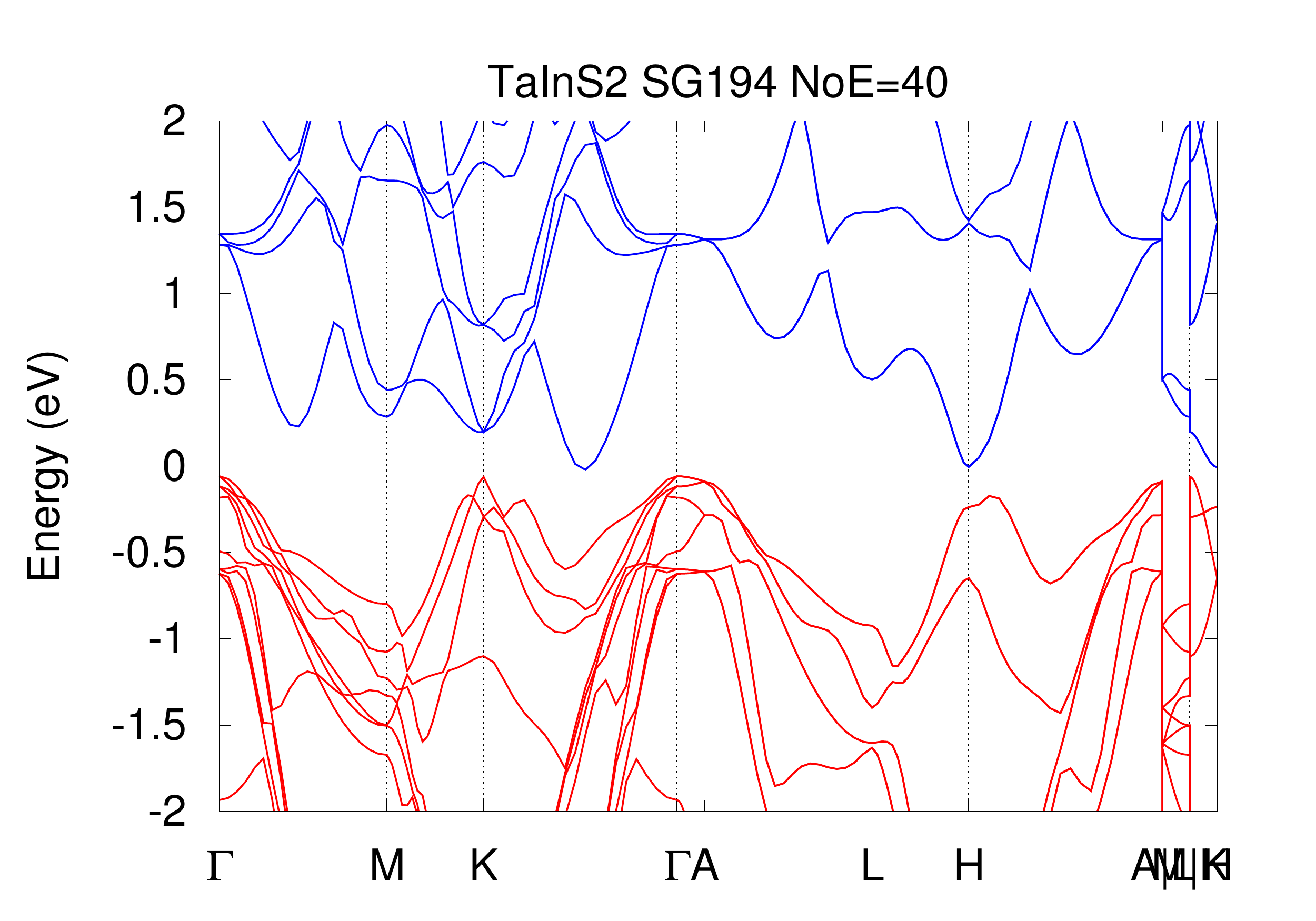}
}
\subfigure[SiRh SG14 NoA=8 NoE=52]{
\label{subfig:79235}
\includegraphics[scale=0.32,angle=270]{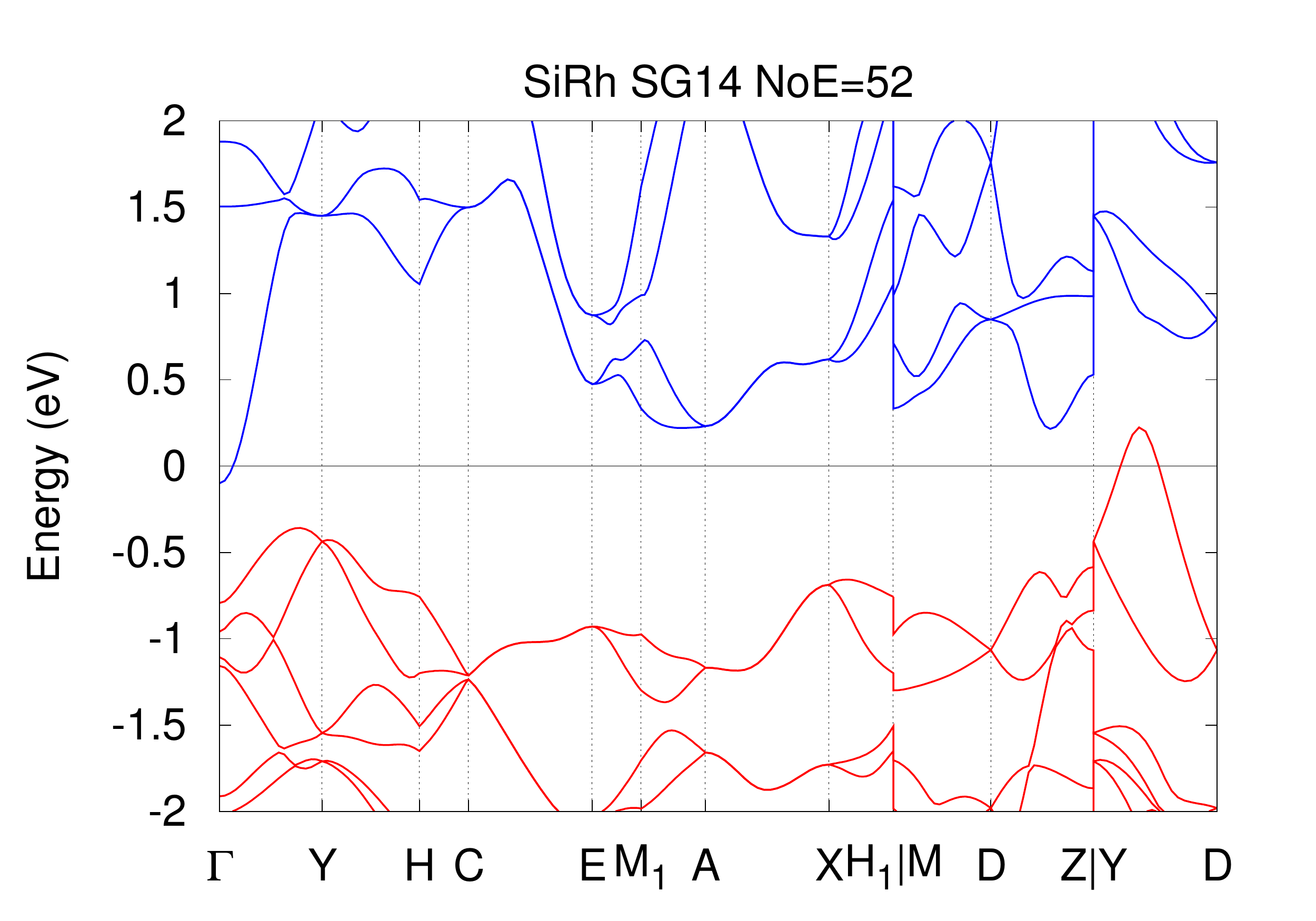}
}
\caption{\hyperref[tab:electride]{back to the table}}
\end{figure}

\begin{figure}[htp]
 \centering
\subfigure[VAuS$_{2}$ SG194 NoA=8 NoE=56]{
\label{subfig:96089}
\includegraphics[scale=0.32,angle=270]{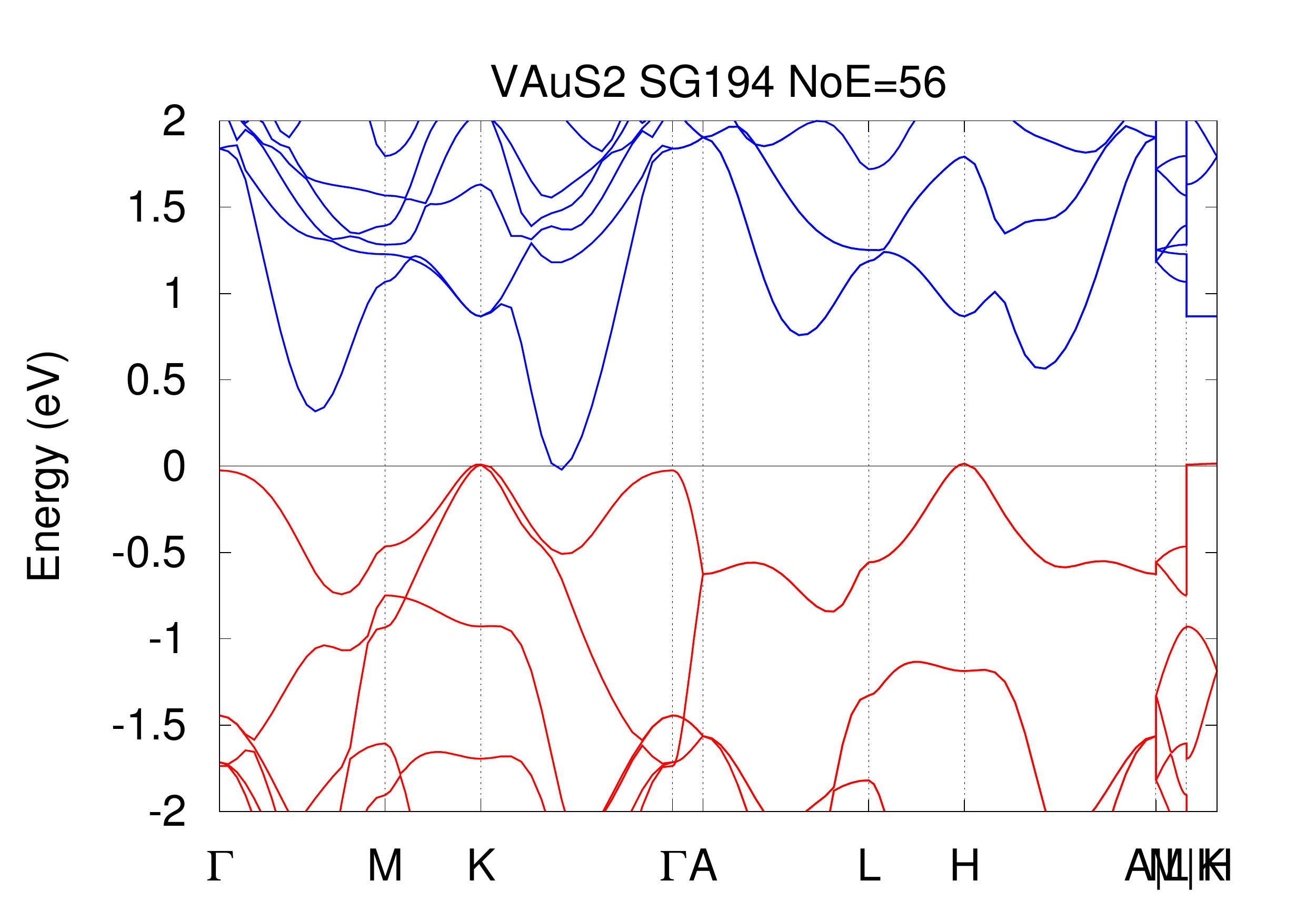}
}
\subfigure[InP$_{3}$ SG166 NoA=8 NoE=36]{
\label{subfig:37073}
\includegraphics[scale=0.32,angle=270]{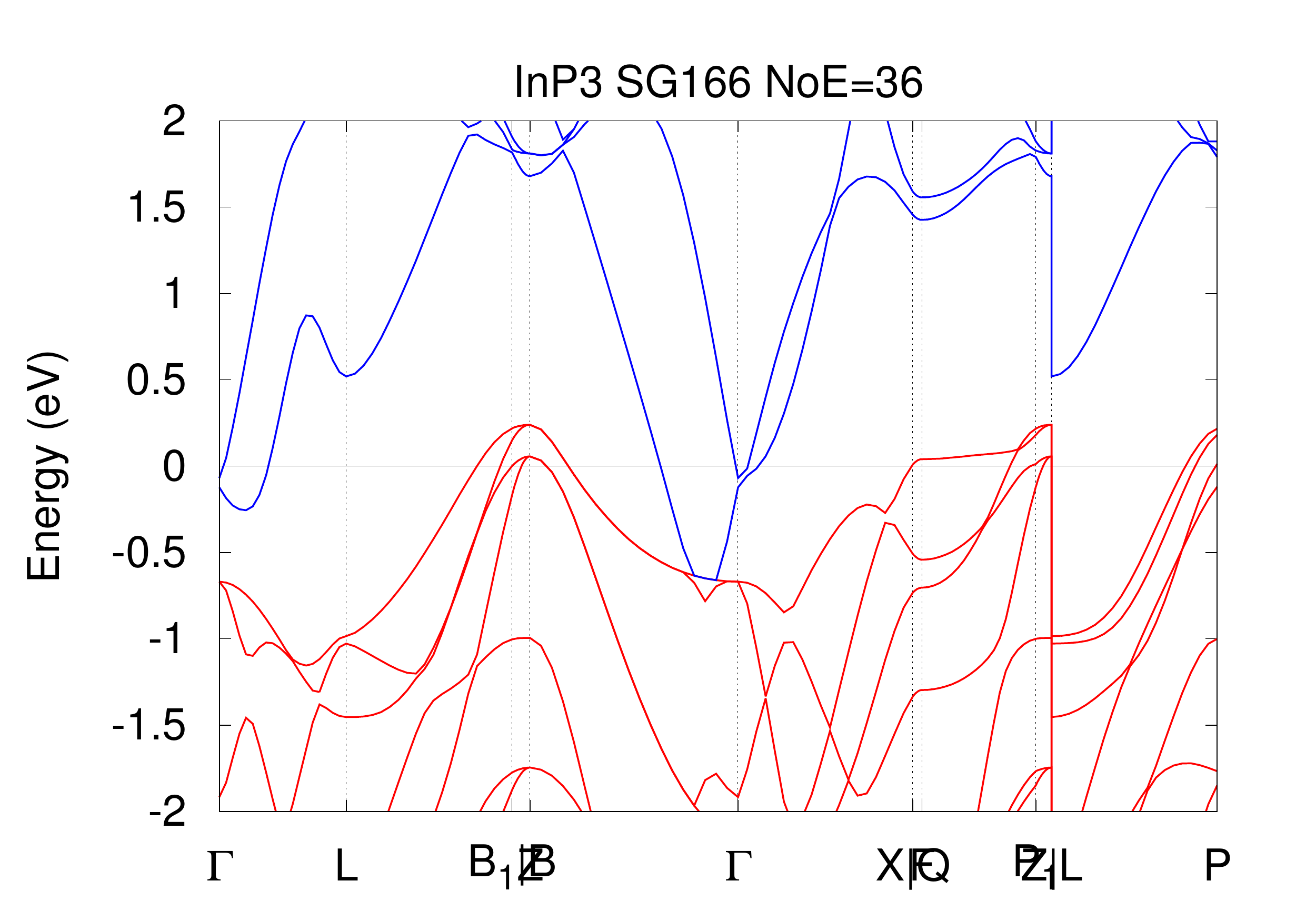}
}
\subfigure[LaGe SG63 NoA=8 NoE=60]{
\label{subfig:413736}
\includegraphics[scale=0.32,angle=270]{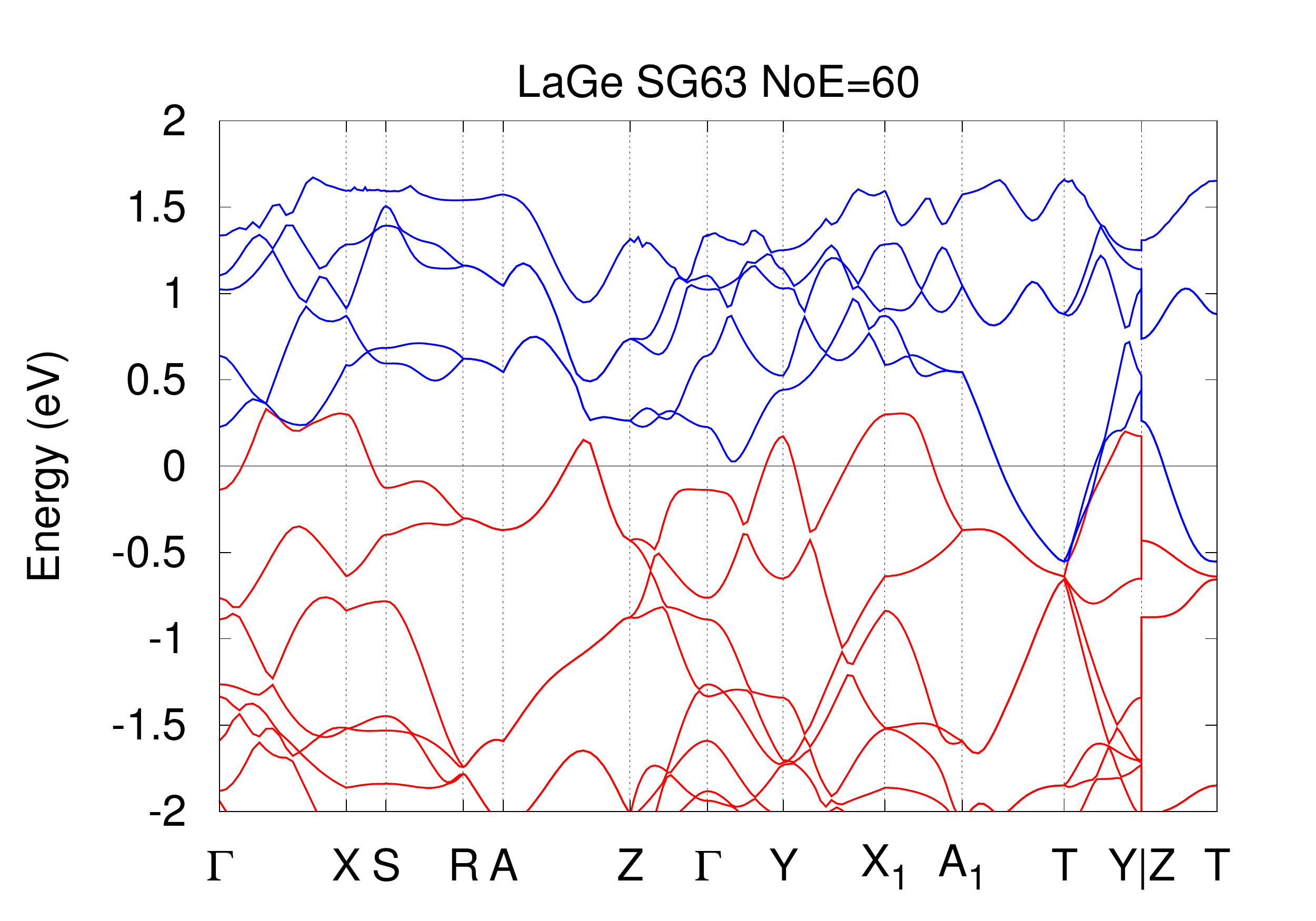}
}
\subfigure[InSe SG187 NoA=8 NoE=36]{
\label{subfig:640503}
\includegraphics[scale=0.32,angle=270]{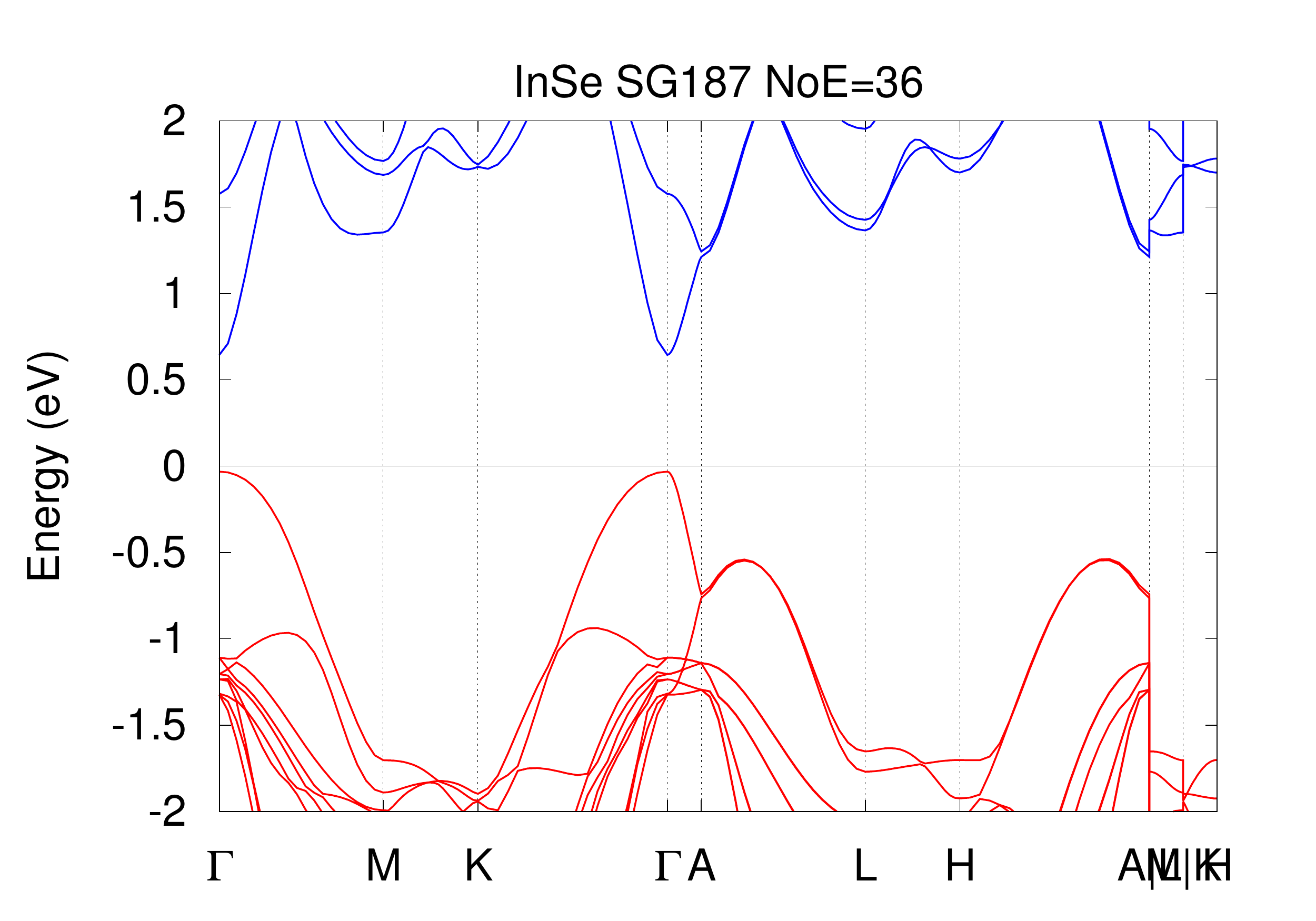}
}
\subfigure[SiPt$_{3}$ SG12 NoA=8 NoE=68]{
\label{subfig:246170}
\includegraphics[scale=0.32,angle=270]{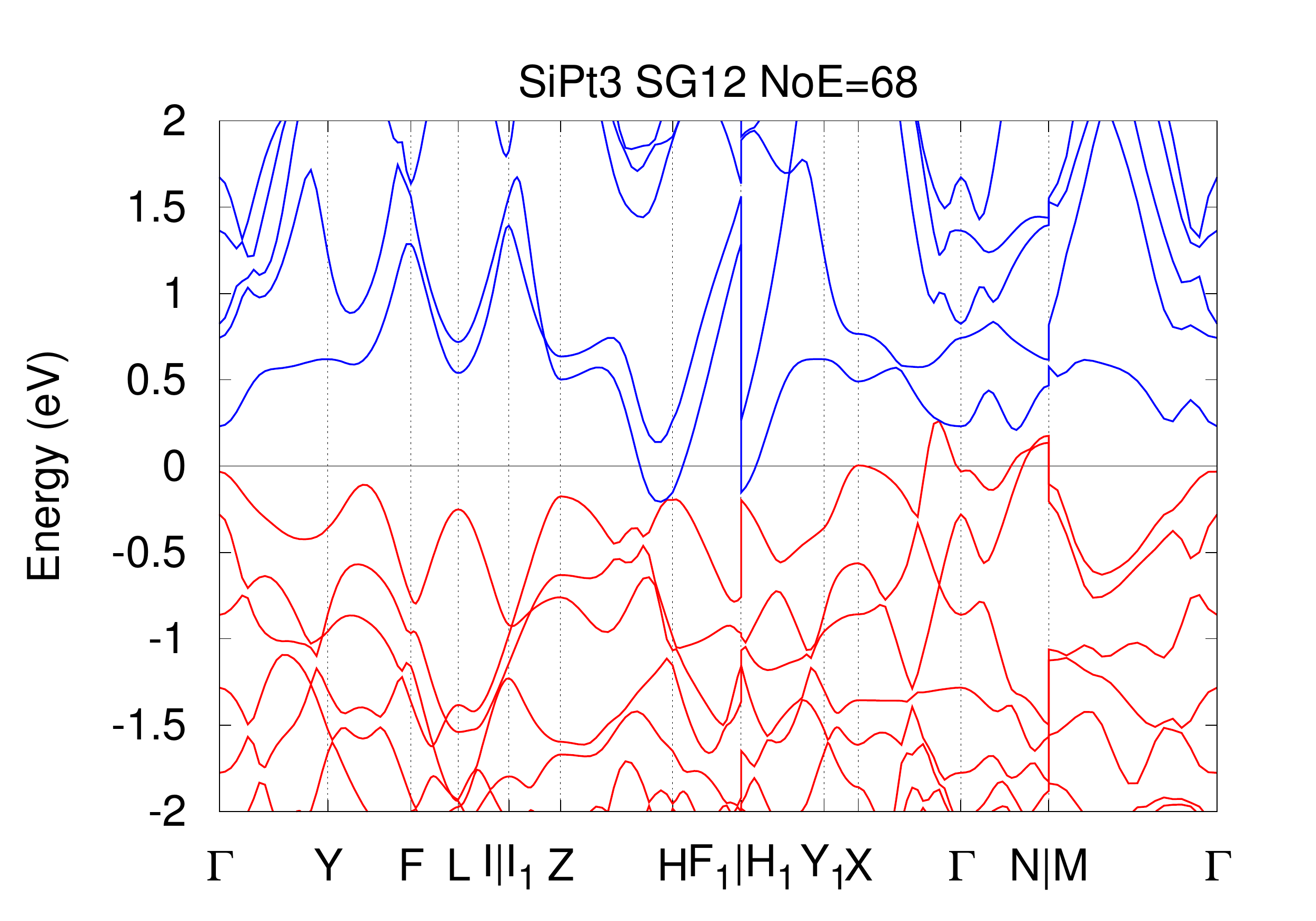}
}
\subfigure[NaSe SG194 NoA=8 NoE=28]{
\label{subfig:43408}
\includegraphics[scale=0.32,angle=270]{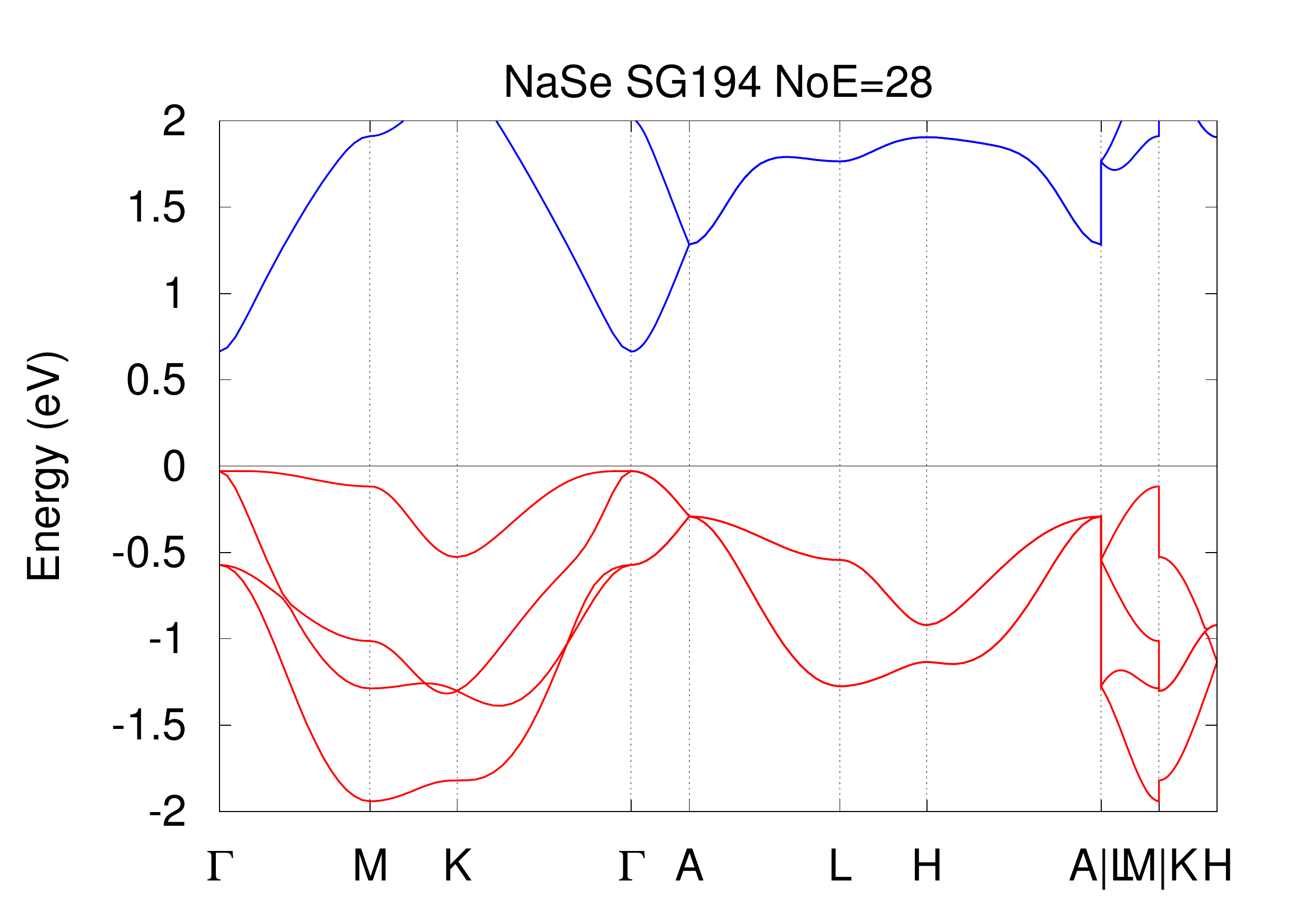}
}
\subfigure[InSe SG194 NoA=8 NoE=36]{
\label{subfig:185172}
\includegraphics[scale=0.32,angle=270]{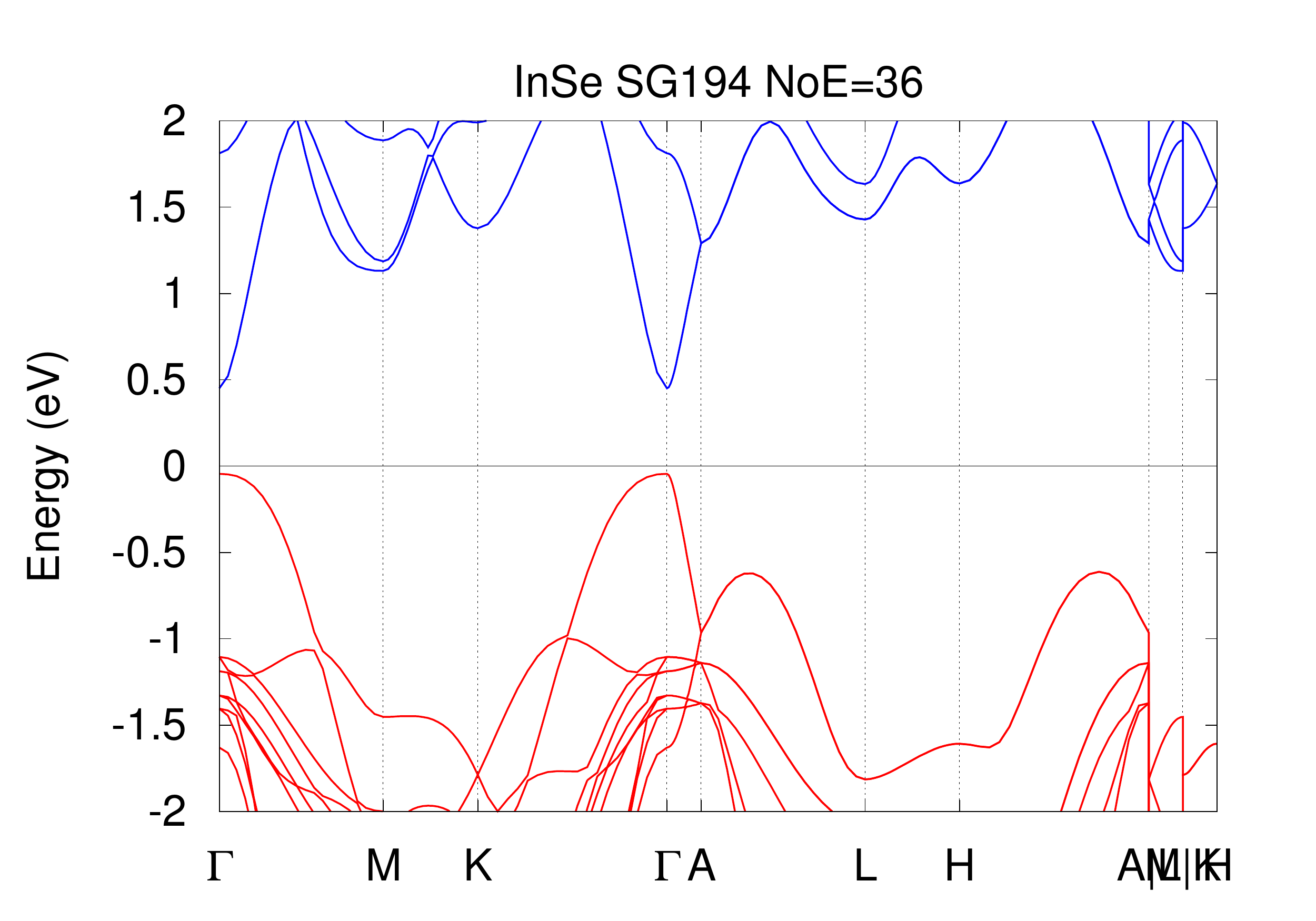}
}
\subfigure[Si SG148 NoA=8 NoE=32]{
\label{subfig:109036}
\includegraphics[scale=0.32,angle=270]{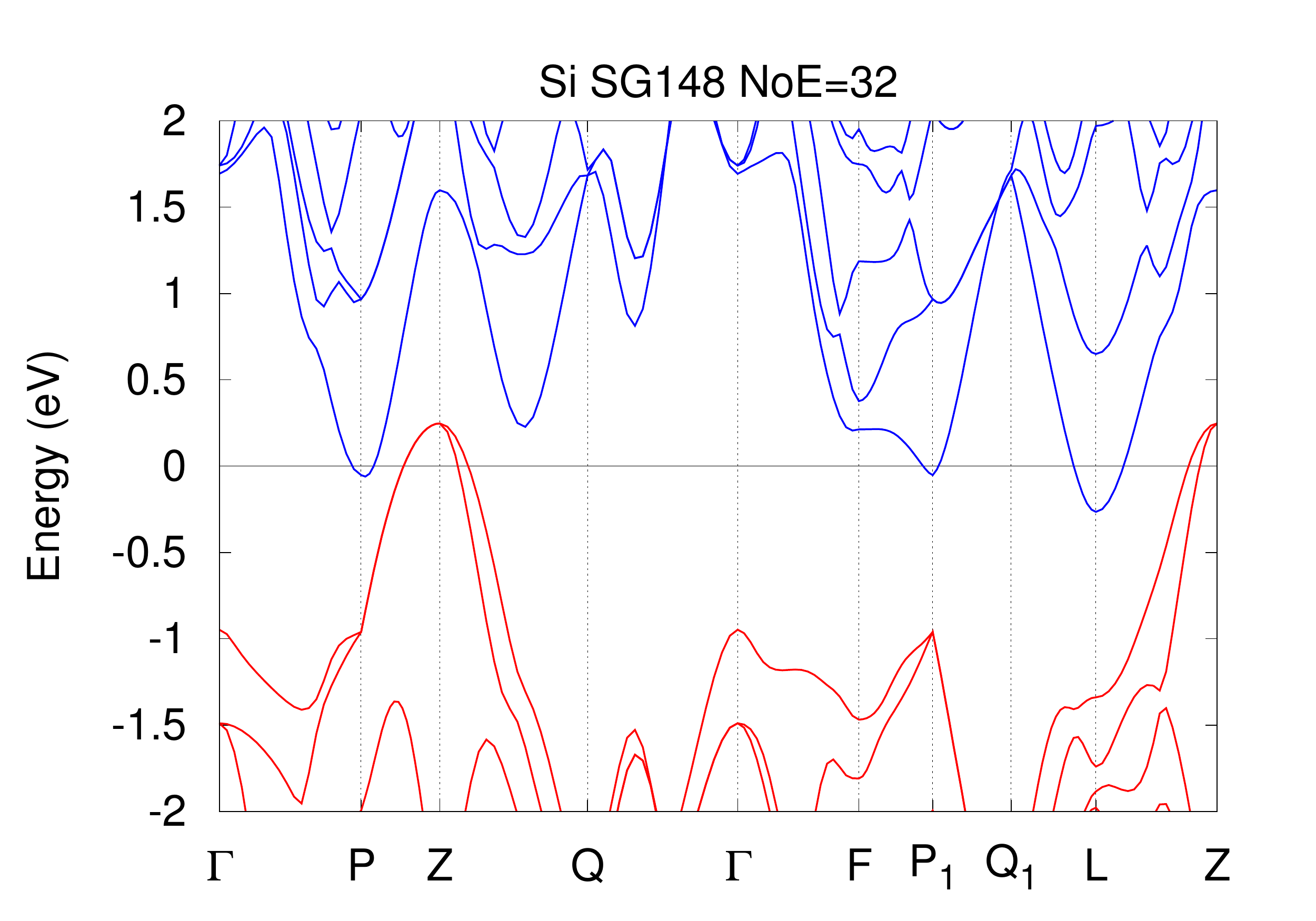}
}
\caption{\hyperref[tab:electride]{back to the table}}
\end{figure}

\begin{figure}[htp]
 \centering
\subfigure[LiNbS$_{2}$ SG194 NoA=8 NoE=48]{
\label{subfig:26284}
\includegraphics[scale=0.32,angle=270]{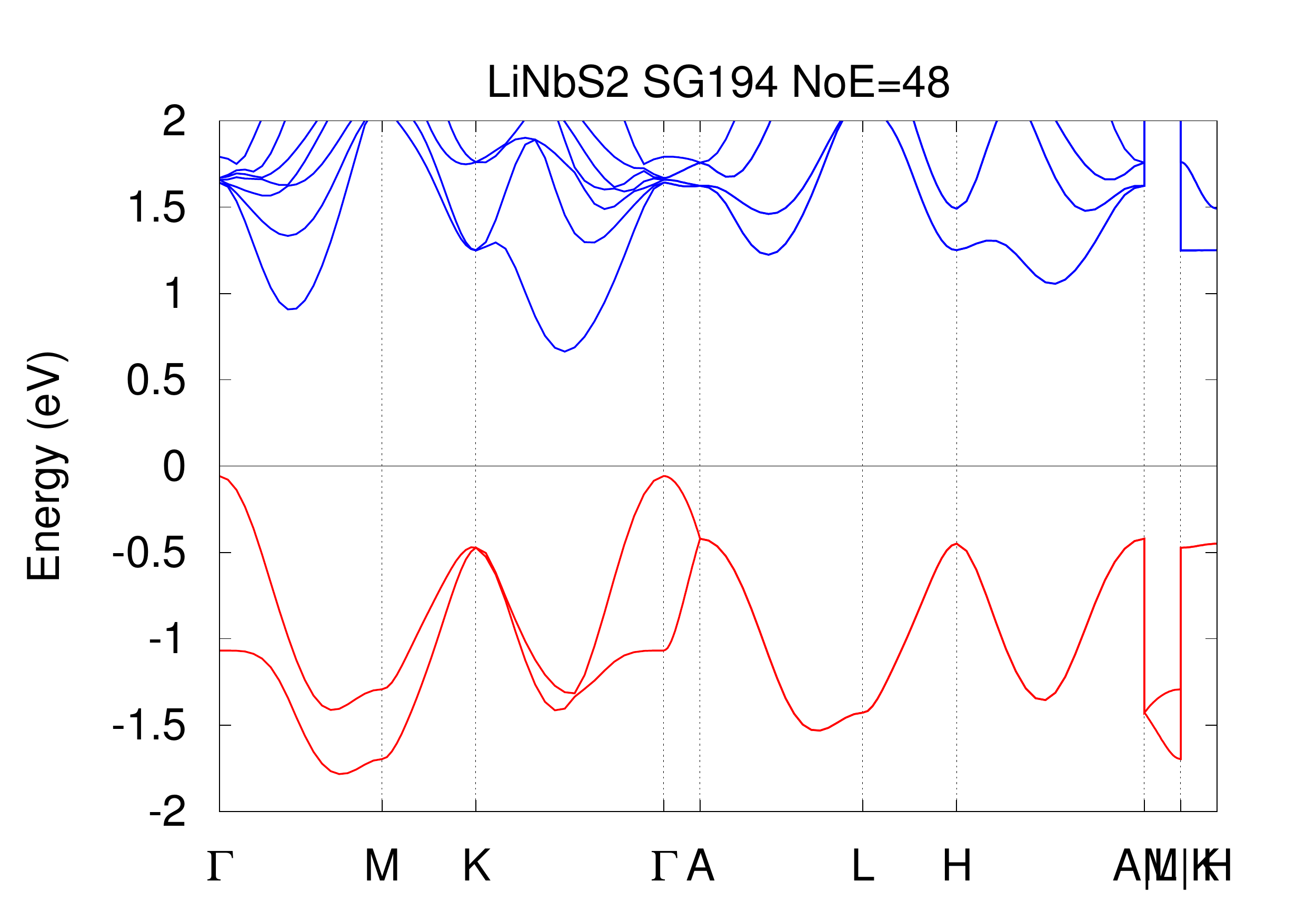}
}
\subfigure[NaNbSe$_{2}$ SG194 NoA=8 NoE=48]{
\label{subfig:26287}
\includegraphics[scale=0.32,angle=270]{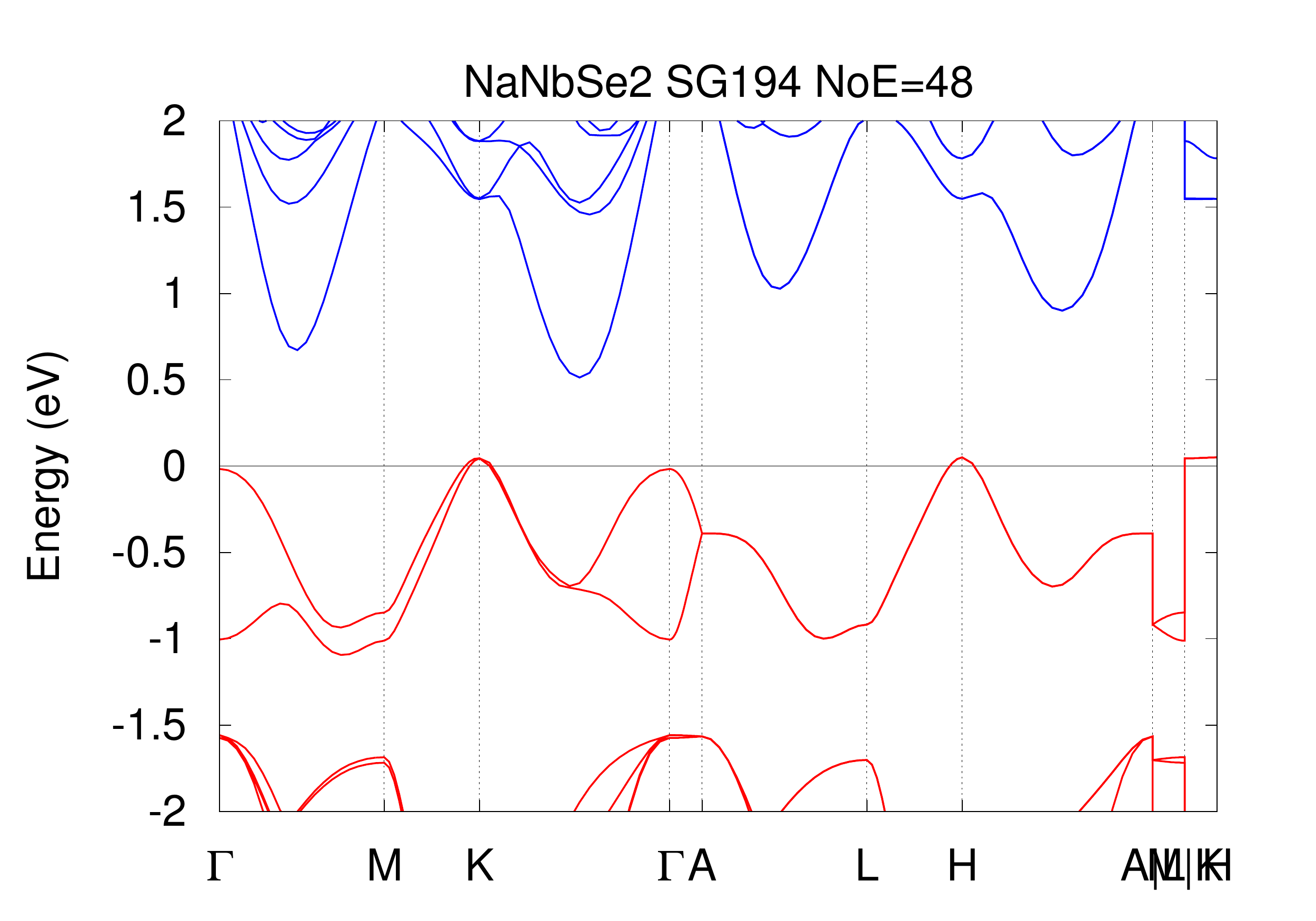}
}
\subfigure[BW SG141 NoA=8 NoE=36]{
\label{subfig:424240}
\includegraphics[scale=0.32,angle=270]{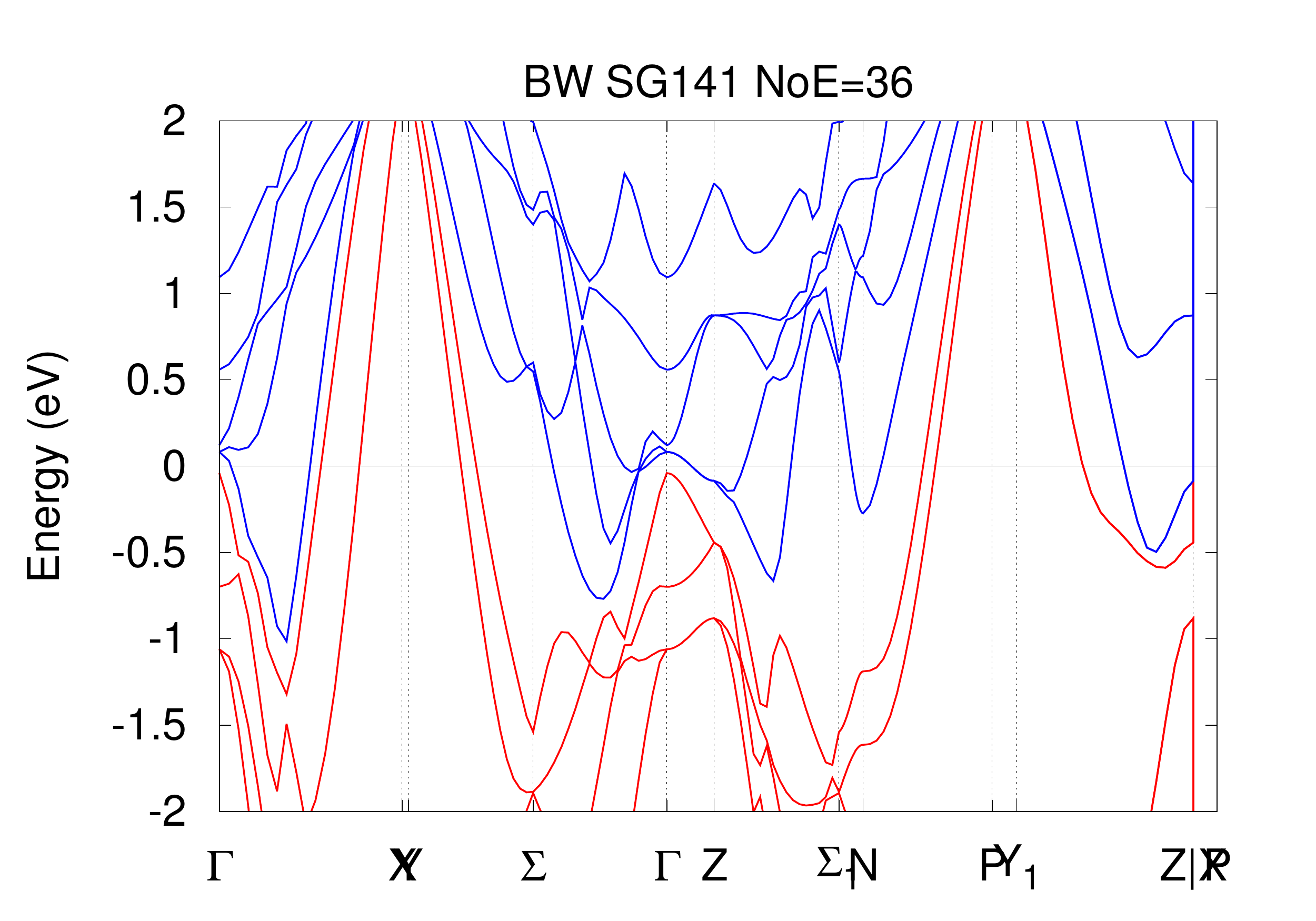}
}
\subfigure[Nb$_{5}$Sb$_{4}$ SG87 NoA=9 NoE=75]{
\label{subfig:154596}
\includegraphics[scale=0.32,angle=270]{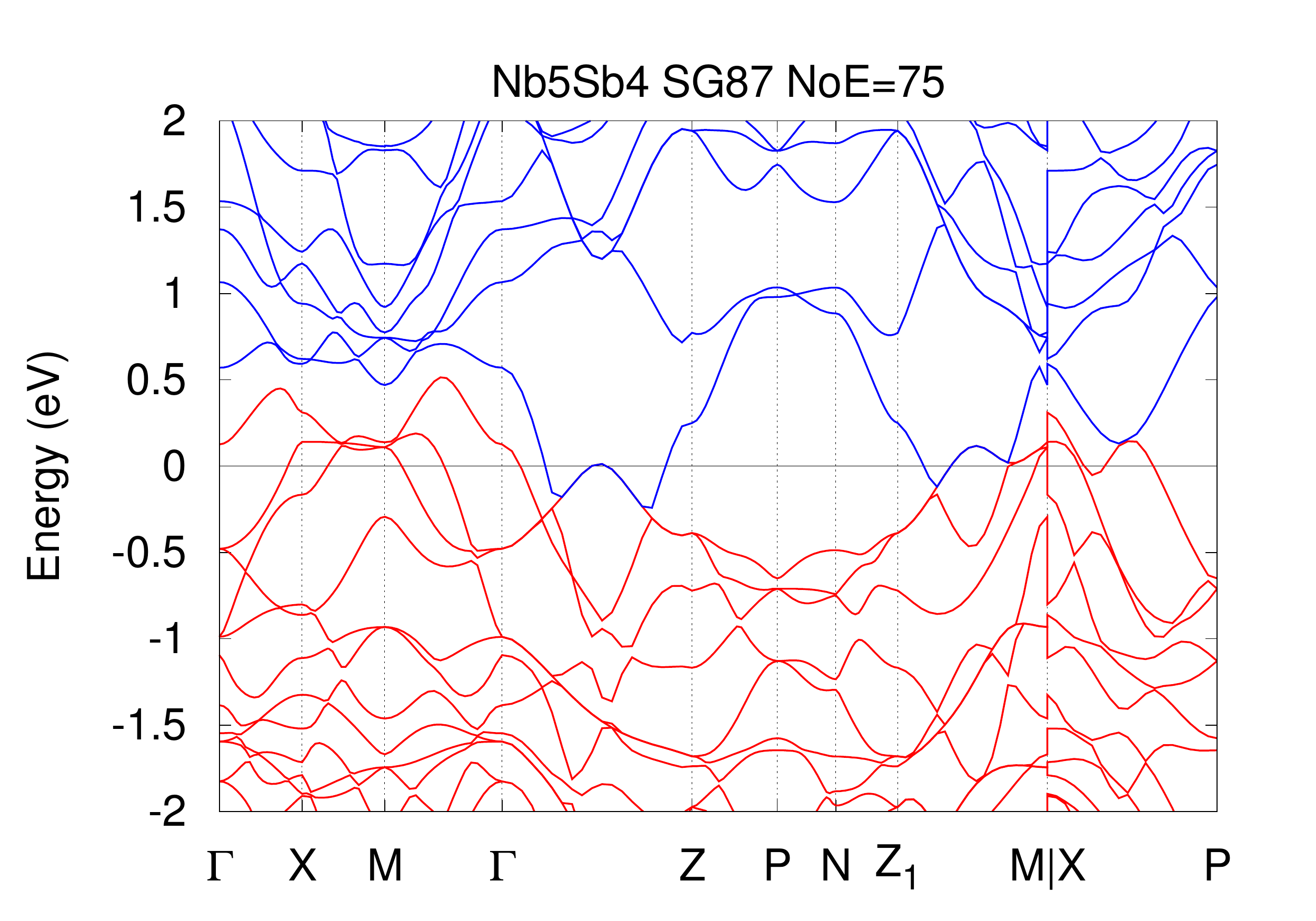}
}
\subfigure[Na(CuS)$_{4}$ SG164 NoA=9 NoE=69]{
\label{subfig:81306}
\includegraphics[scale=0.32,angle=270]{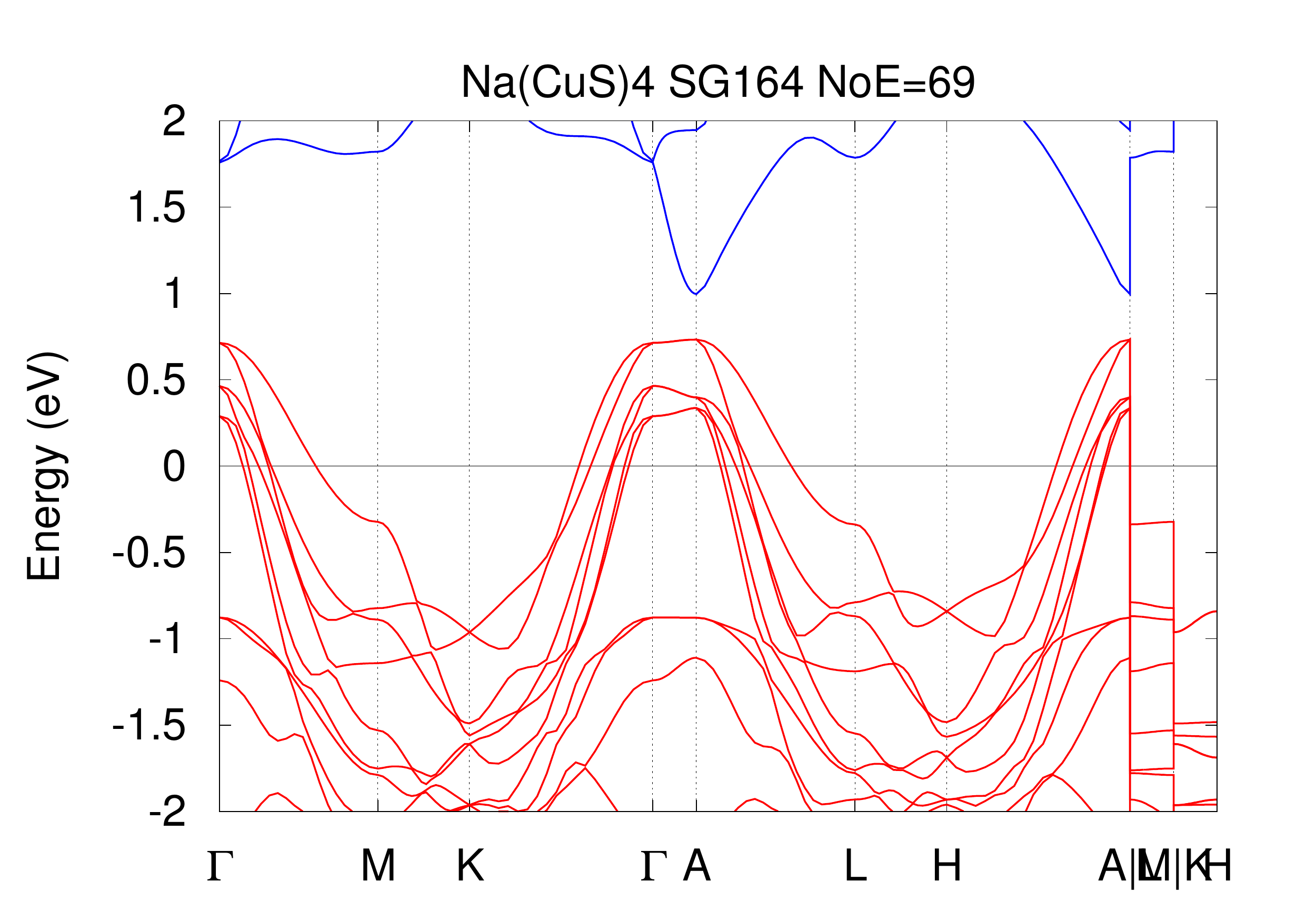}
}
\subfigure[Cs$_{2}$PdI$_{6}$ SG139 NoA=9 NoE=70]{
\label{subfig:280189}
\includegraphics[scale=0.32,angle=270]{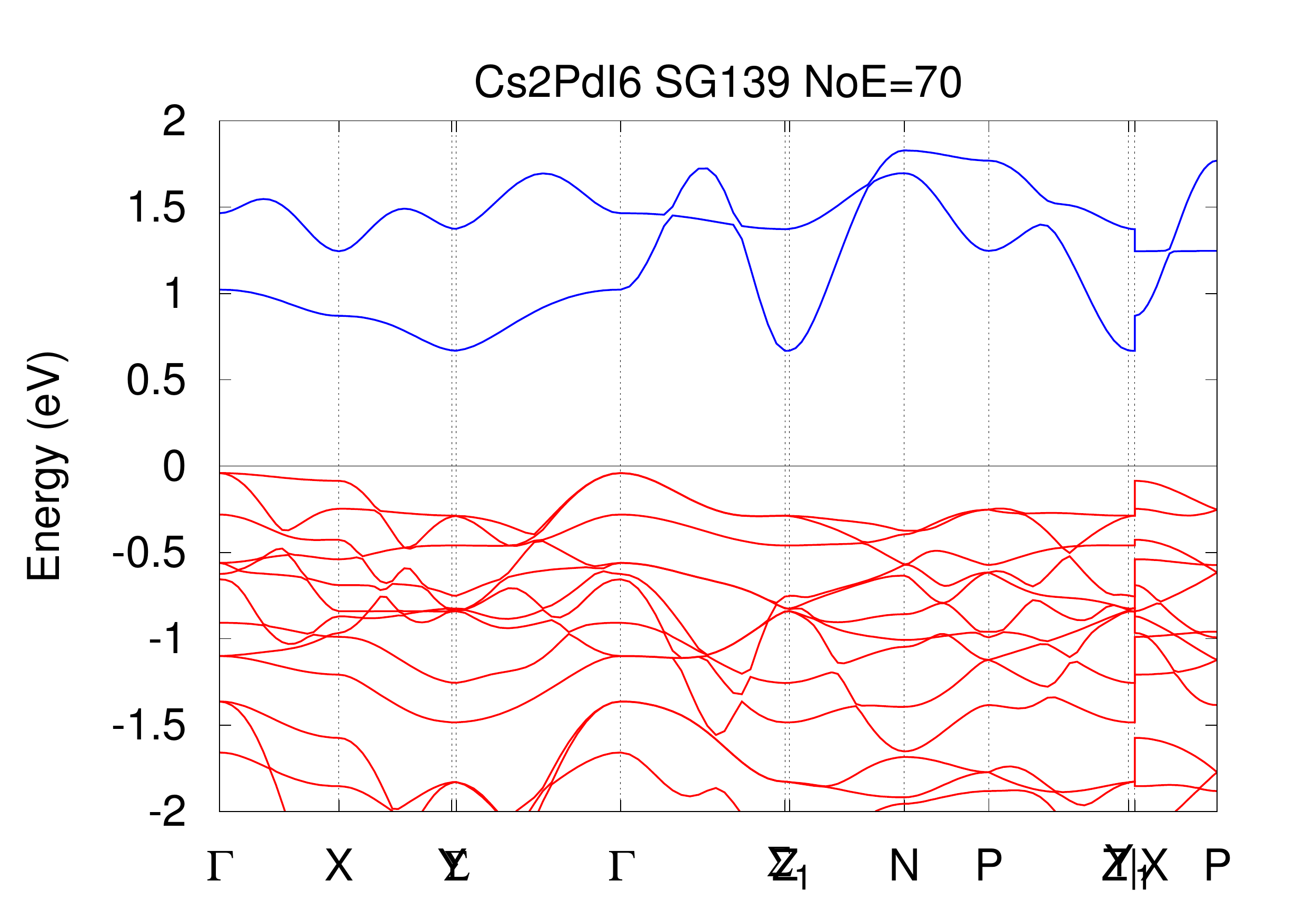}
}
\subfigure[CrSi$_{2}$ SG180 NoA=9 NoE=42]{
\label{subfig:626798}
\includegraphics[scale=0.32,angle=270]{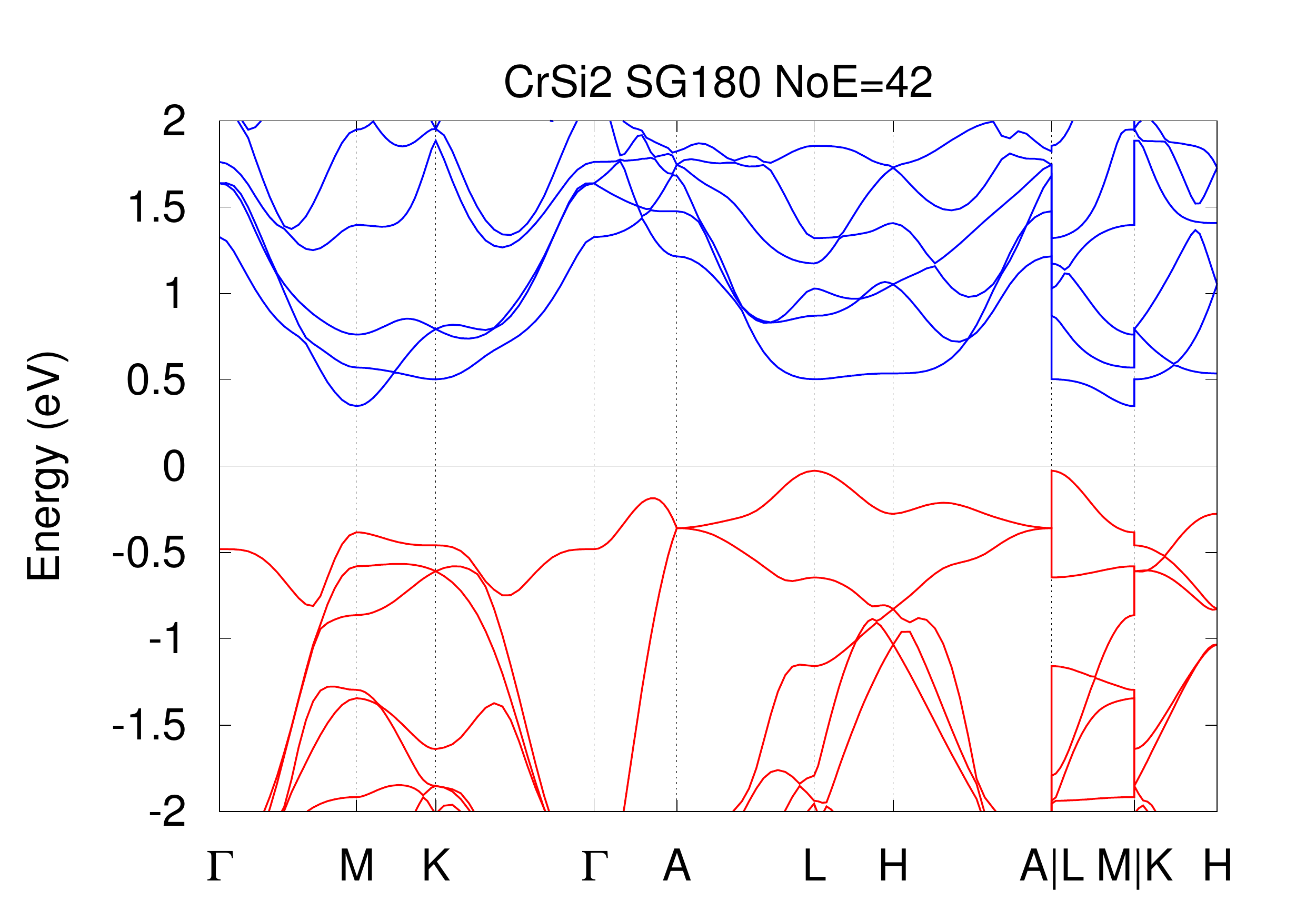}
}
\subfigure[Cs$_{2}$Pd(IBr$_{2}$)$_{2}$ SG139 NoA=9 NoE=70]{
\label{subfig:240481}
\includegraphics[scale=0.32,angle=270]{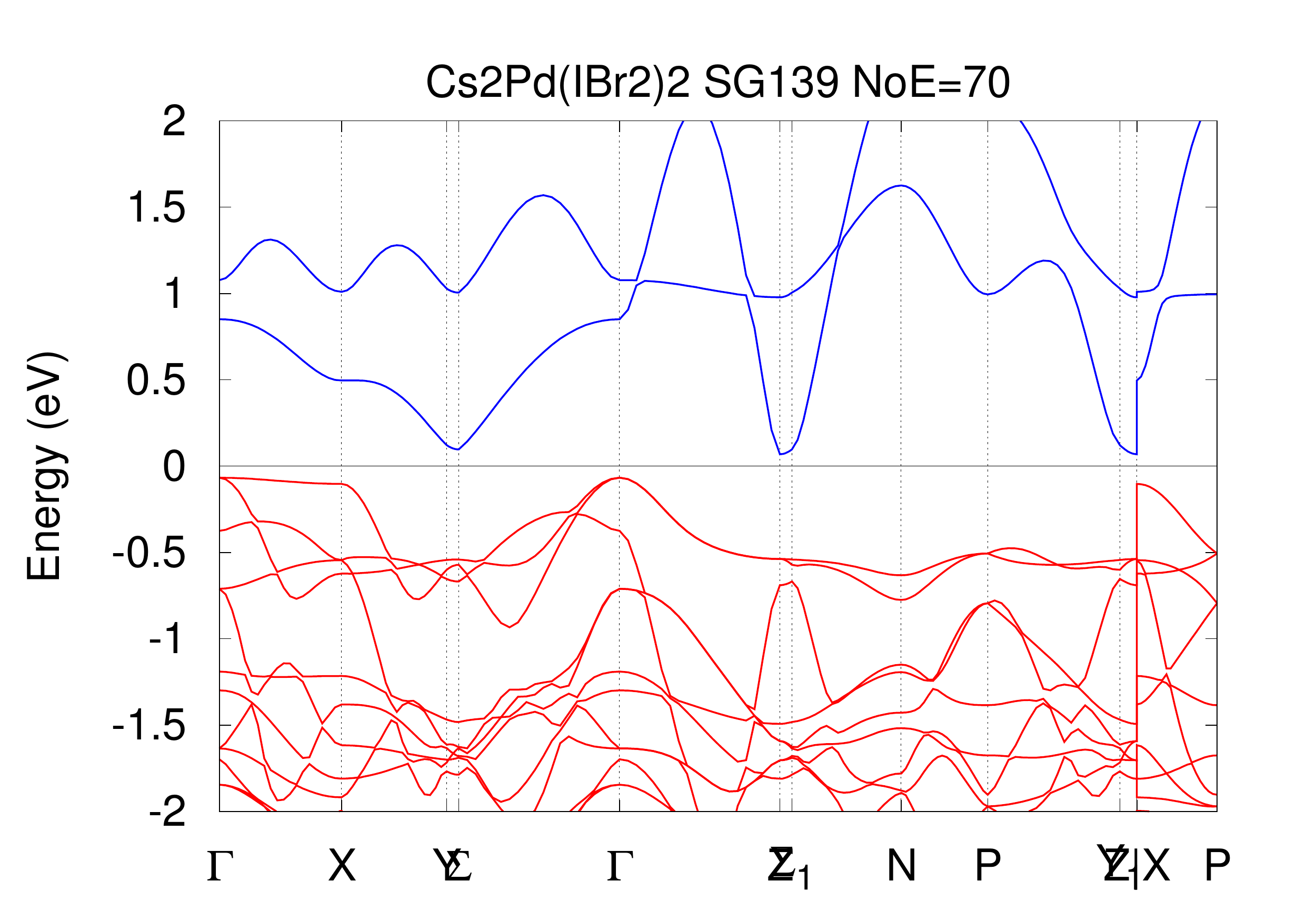}
}
\caption{\hyperref[tab:electride]{back to the table}}
\end{figure}

\begin{figure}[htp]
 \centering
\subfigure[Li$_{5}$In$_{4}$ SG164 NoA=9 NoE=17]{
\label{subfig:639879}
\includegraphics[scale=0.32,angle=270]{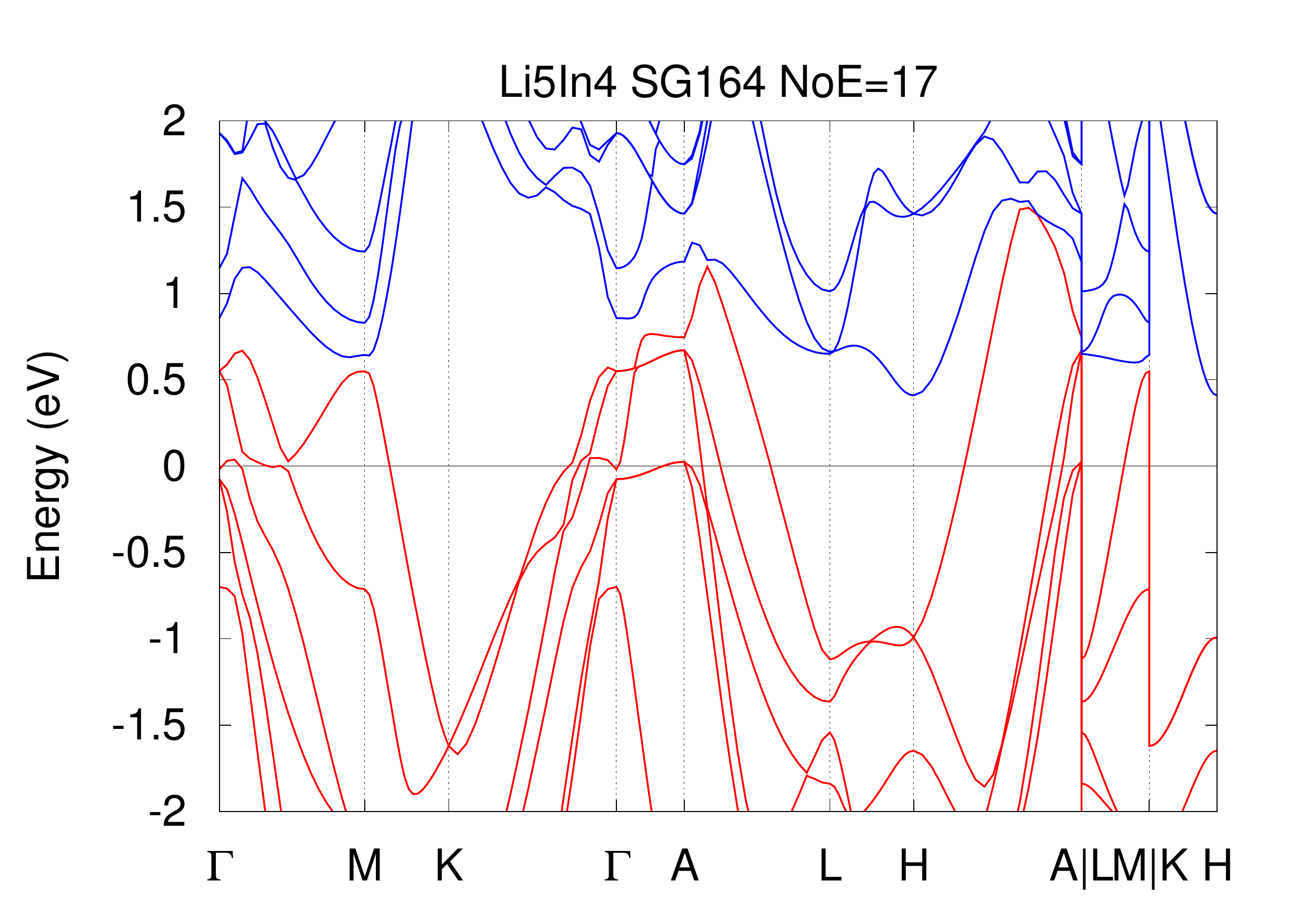}
}
\subfigure[Si$_{2}$Mo SG180 NoA=9 NoE=42]{
\label{subfig:182116}
\includegraphics[scale=0.32,angle=270]{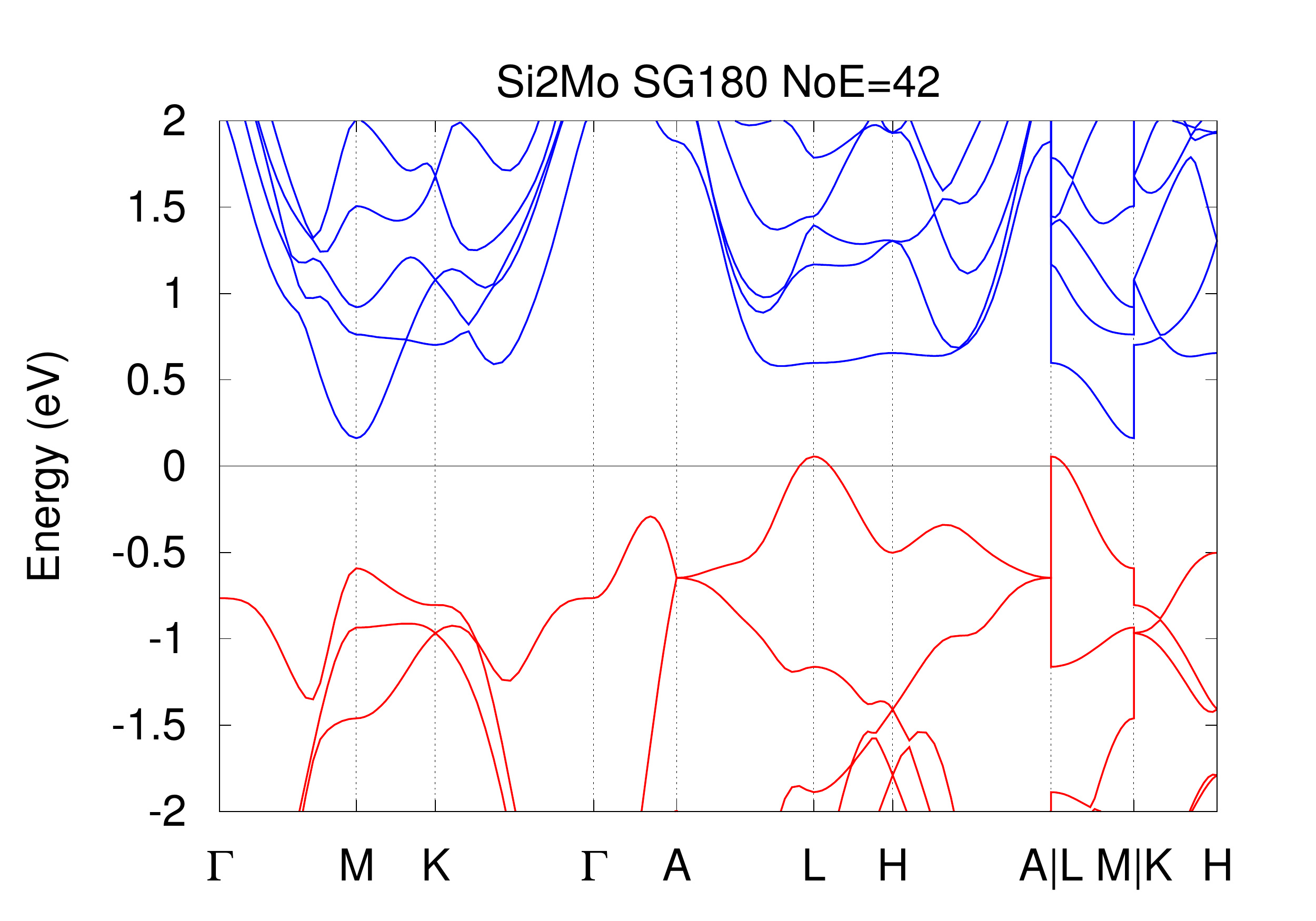}
}
\subfigure[CrSi$_{2}$ SG181 NoA=9 NoE=42]{
\label{subfig:96026}
\includegraphics[scale=0.32,angle=270]{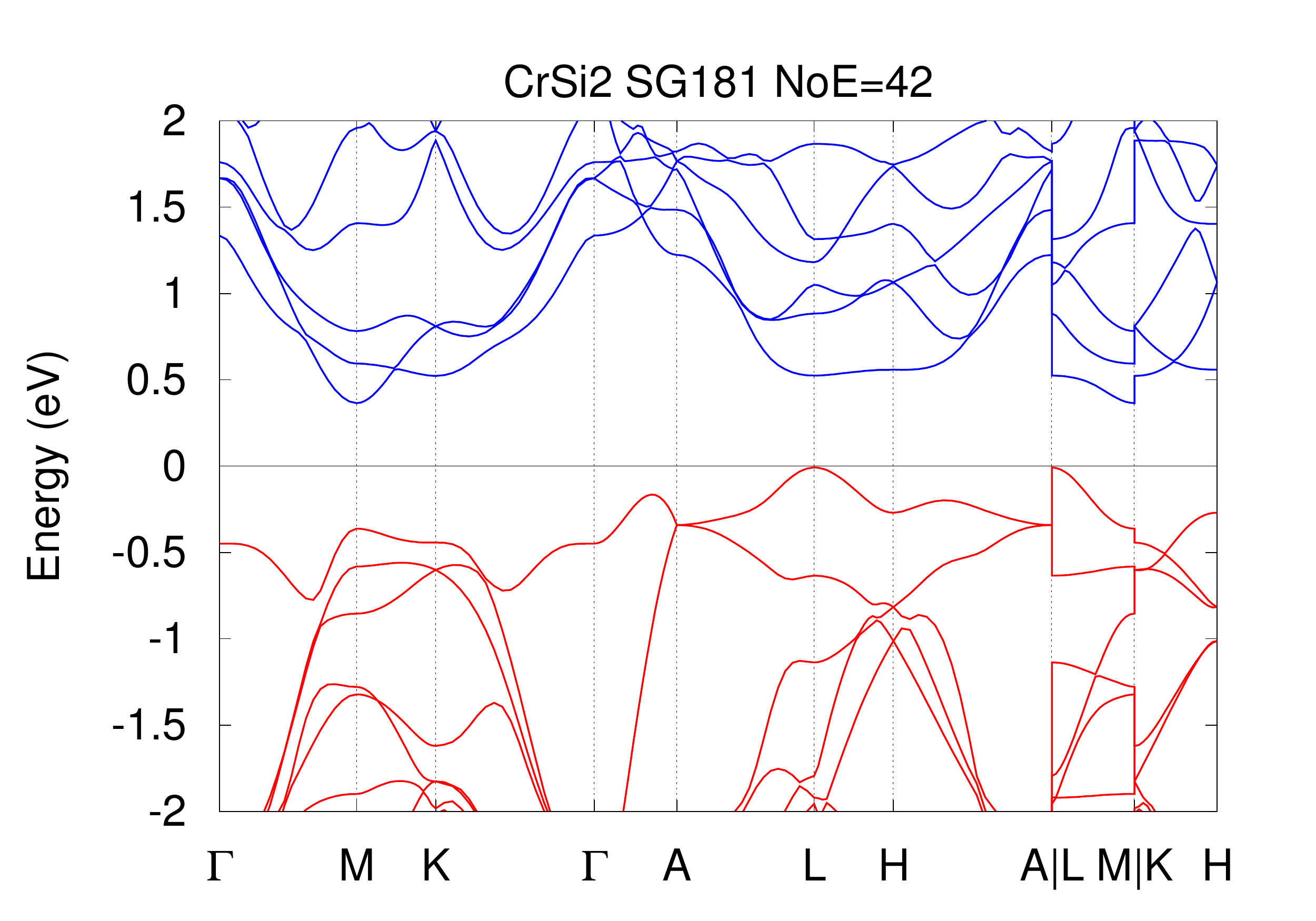}
}
\subfigure[Sr(RuO$_{3}$)$_{2}$ SG162 NoA=9 NoE=62]{
\label{subfig:248351}
\includegraphics[scale=0.32,angle=270]{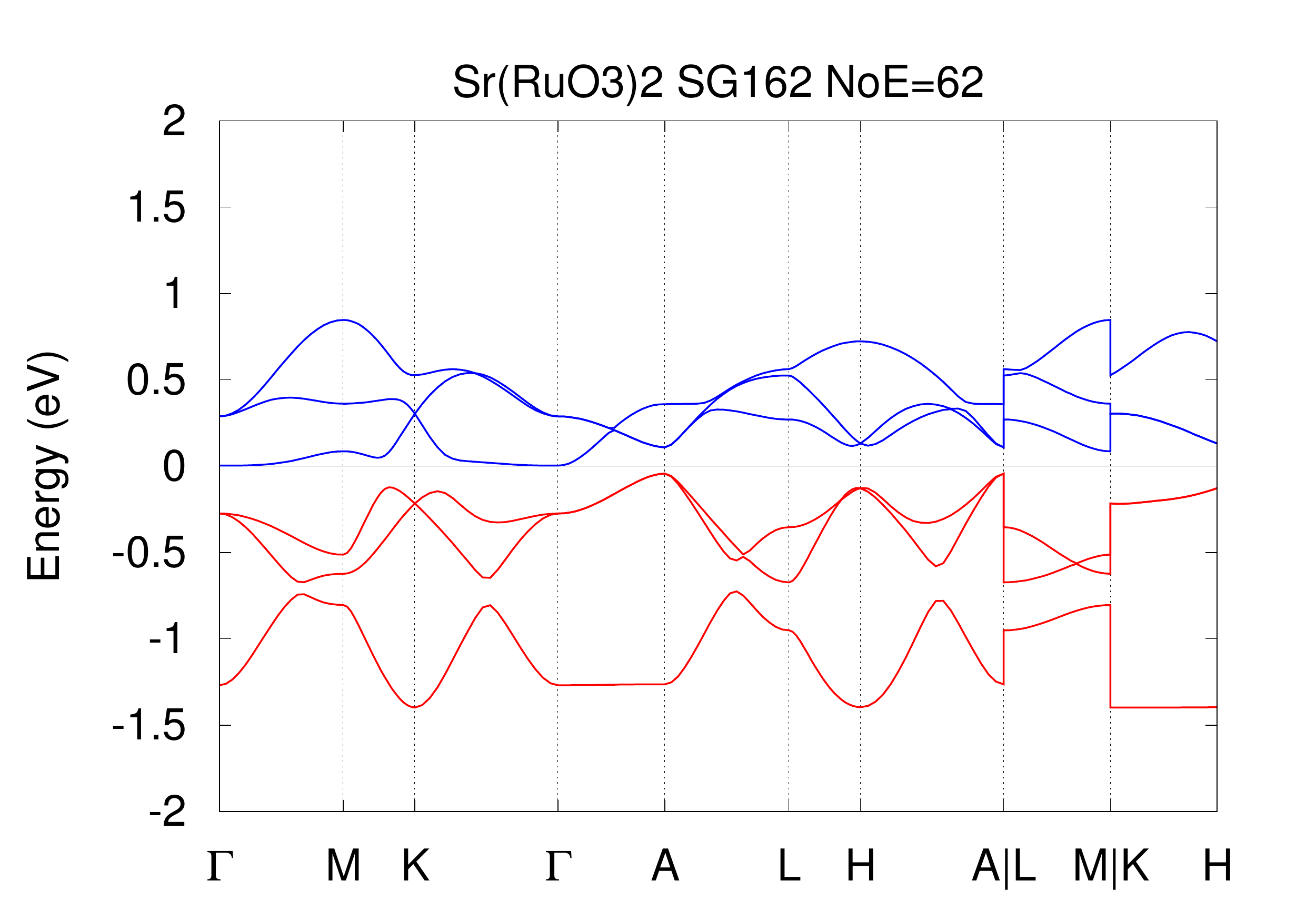}
}
\subfigure[Si$_{2}$W SG180 NoA=9 NoE=42]{
\label{subfig:652549}
\includegraphics[scale=0.32,angle=270]{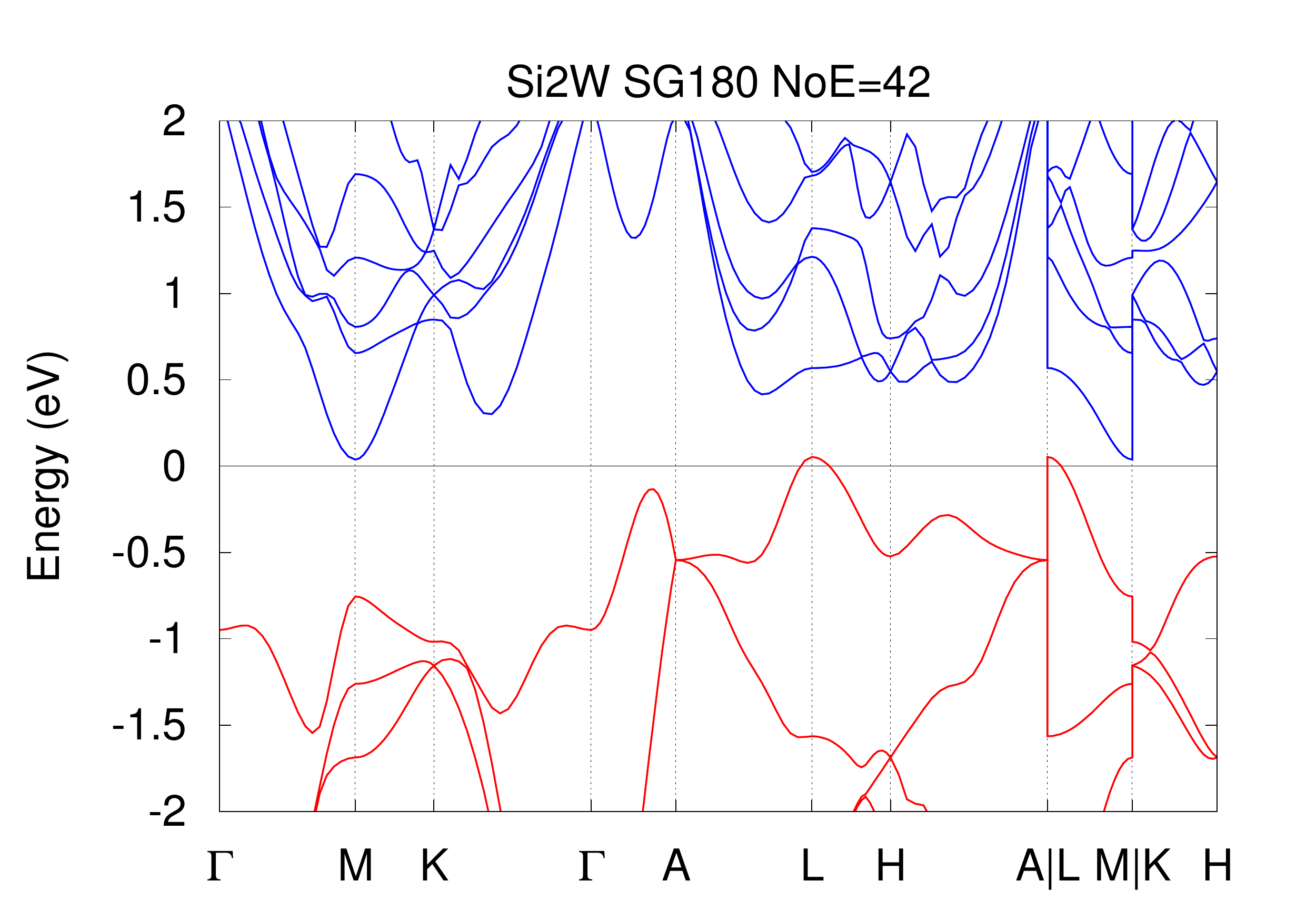}
}
\subfigure[Ca$_{6}$Ge$_{2}$O SG225 NoA=9 NoE=74]{
\label{subfig:181078}
\includegraphics[scale=0.32,angle=270]{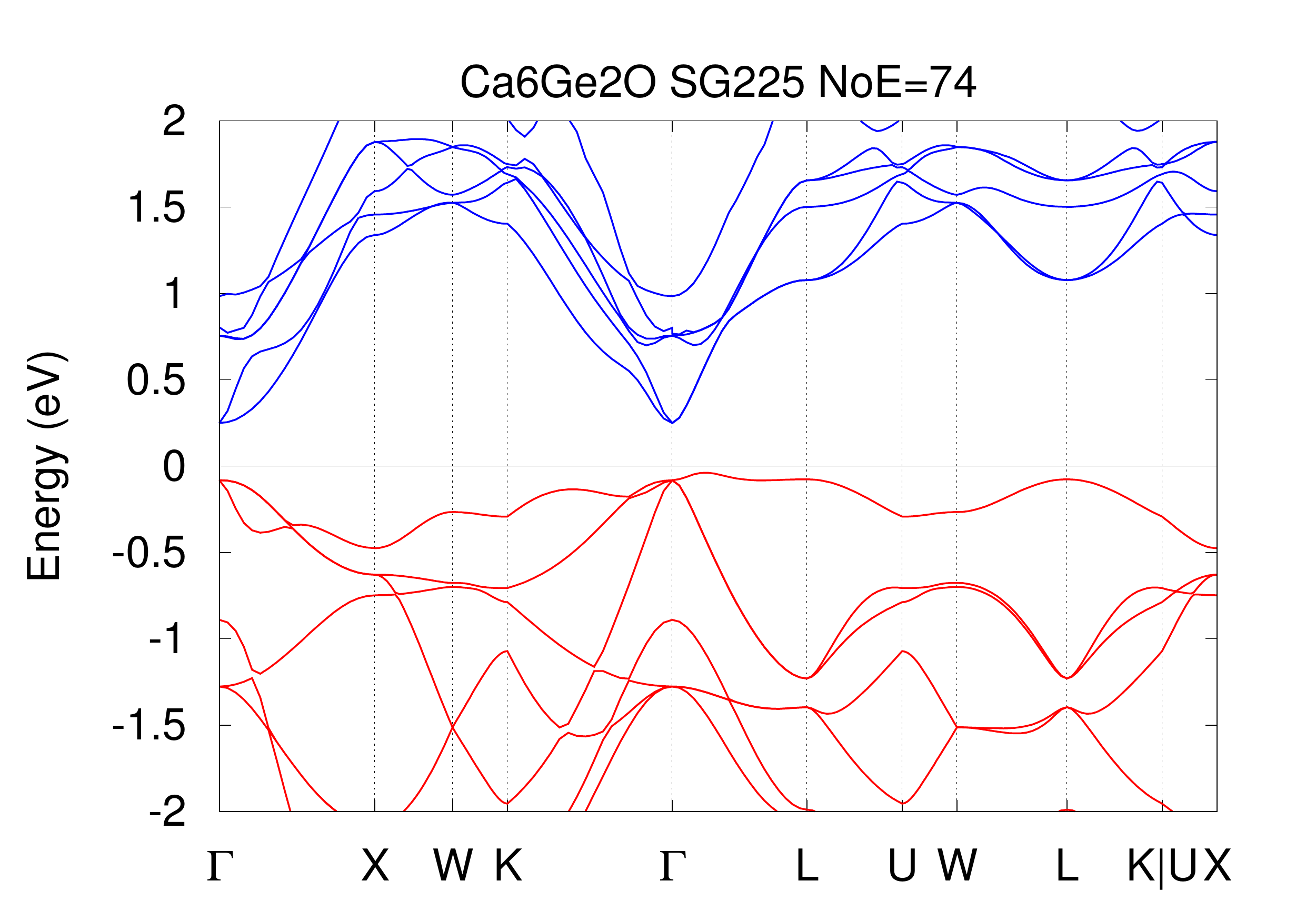}
}
\subfigure[Y$_{4}$CI$_{5}$ SG12 NoA=10 NoE=83]{
\label{subfig:68014}
\includegraphics[scale=0.32,angle=270]{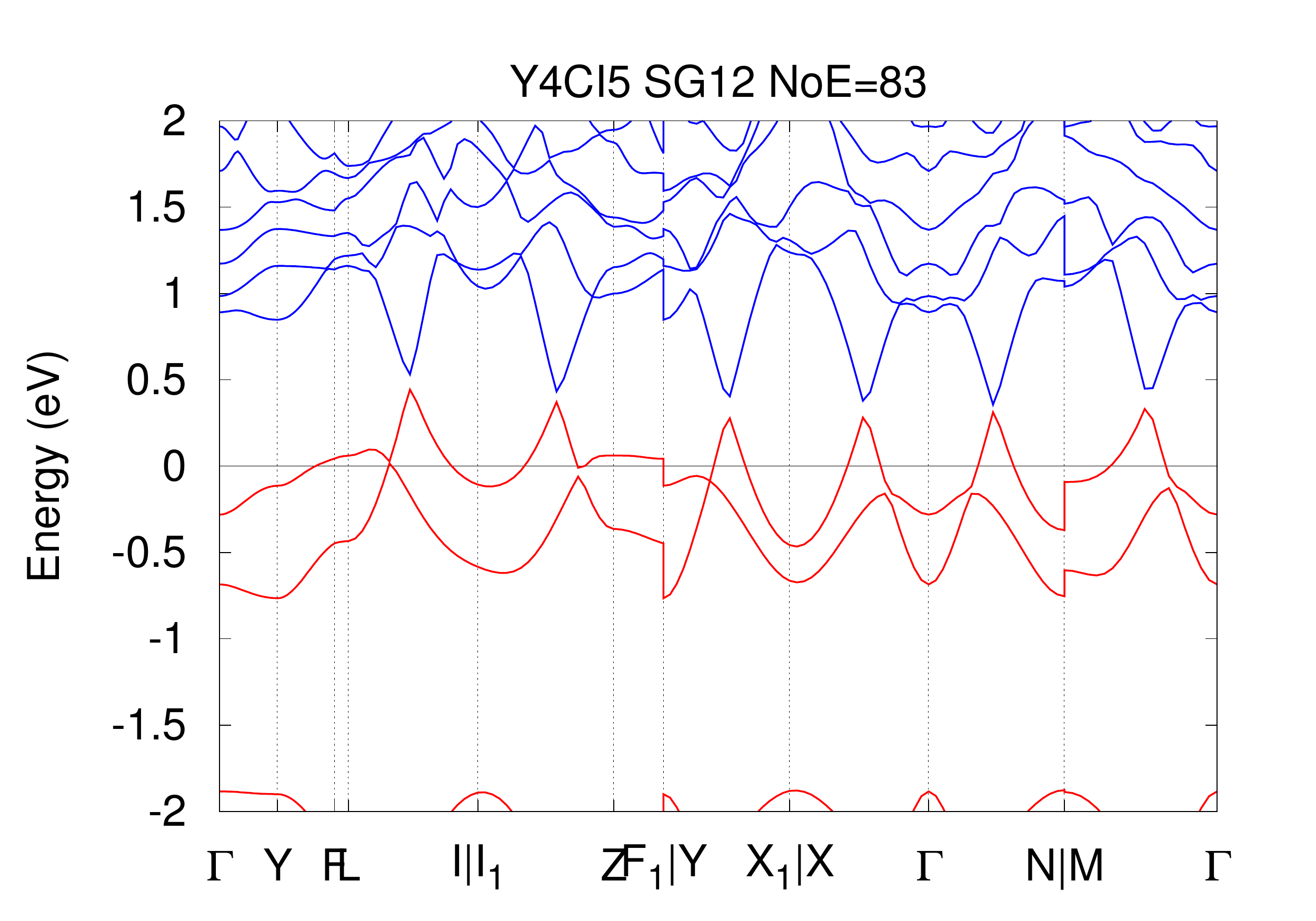}
}
\subfigure[Ho$_{3}$Ni$_{2}$ SG12 NoA=10 NoE=94]{
\label{subfig:639449}
\includegraphics[scale=0.32,angle=270]{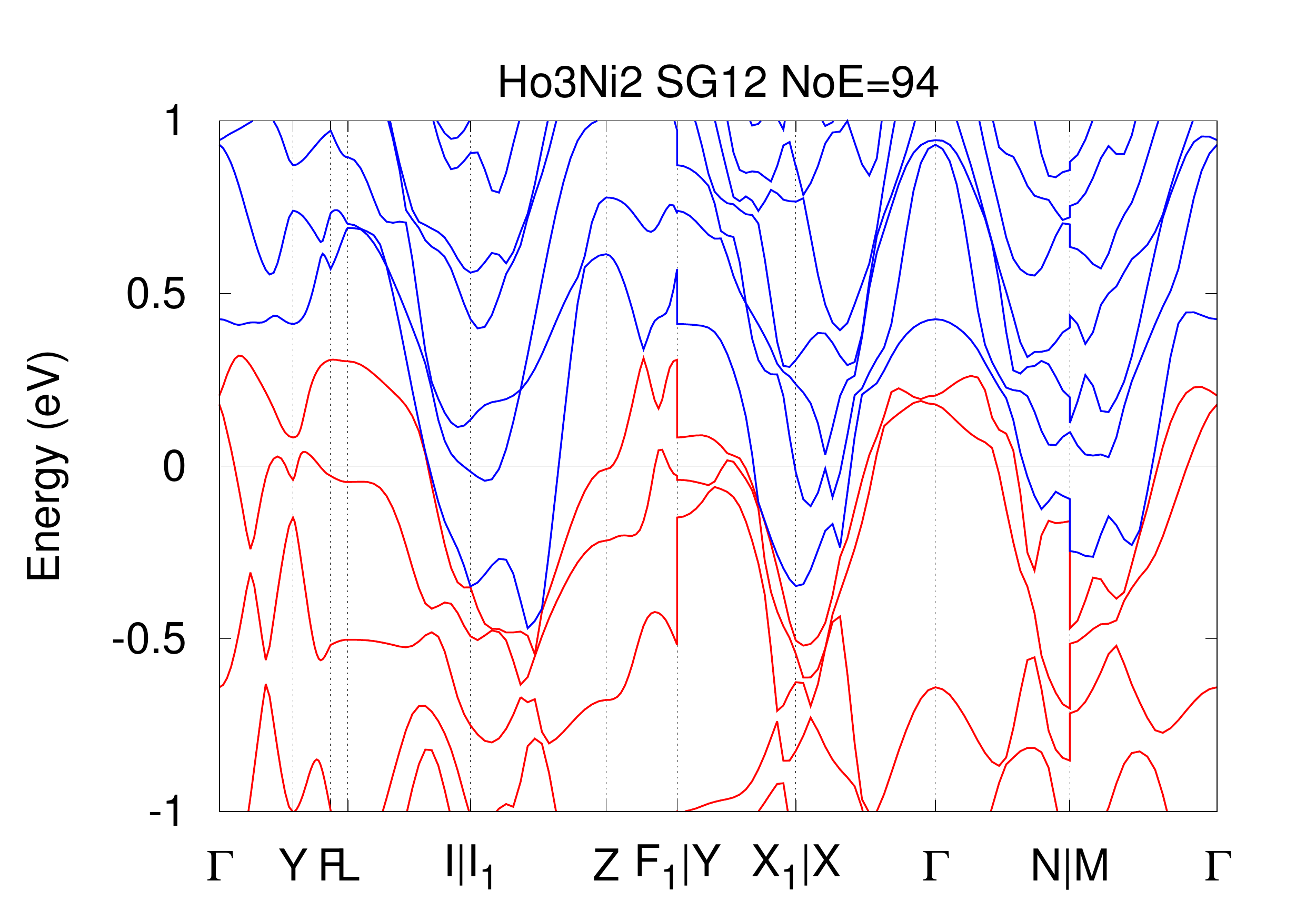}
}
\caption{\hyperref[tab:electride]{back to the table}}
\end{figure}

\begin{figure}[htp]
 \centering
\subfigure[NiPSe$_{3}$ SG12 NoA=10 NoE=66]{
\label{subfig:646145}
\includegraphics[scale=0.32,angle=270]{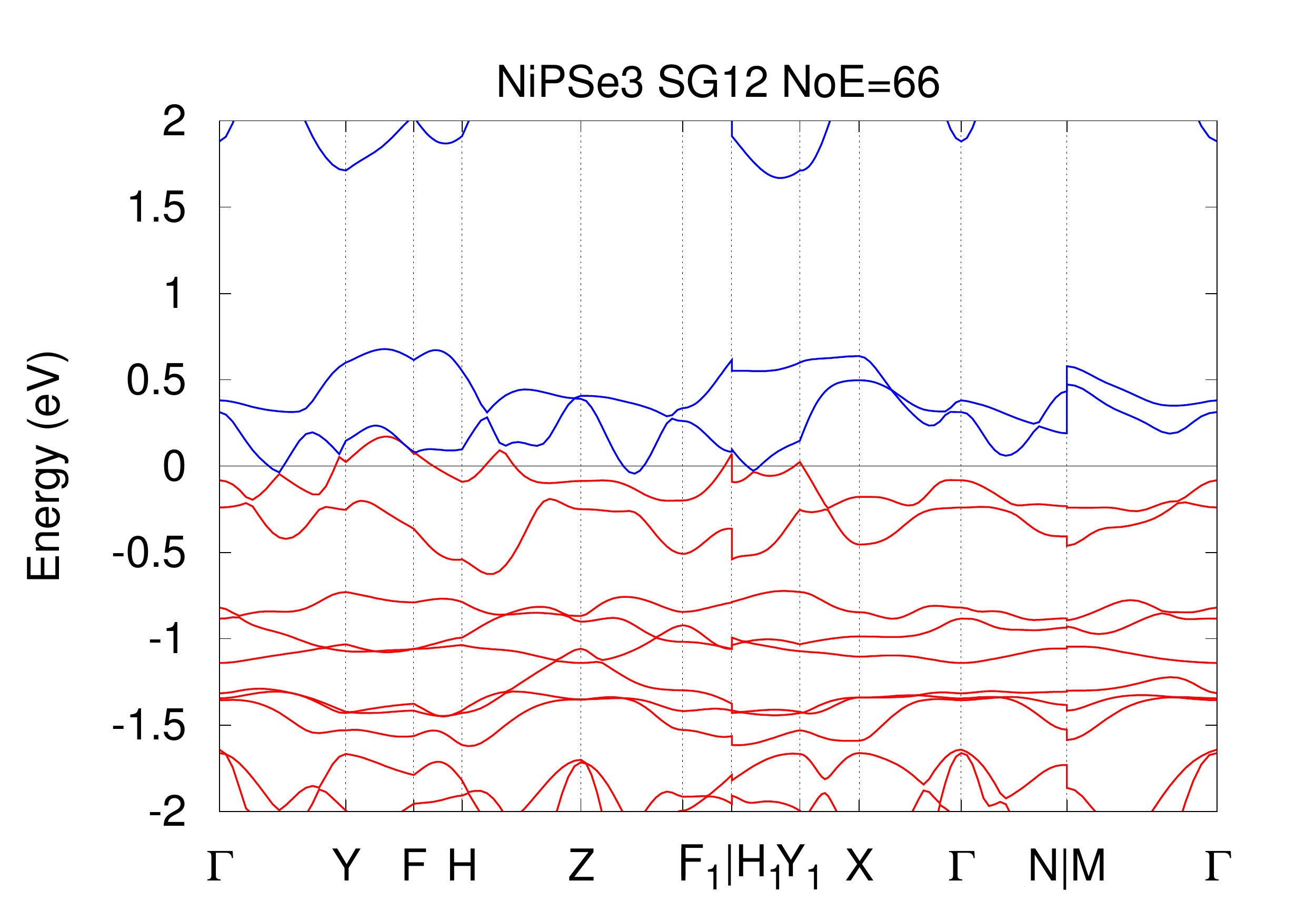}
}
\subfigure[Cs$_{2}$As$_{3}$ SG69 NoA=10 NoE=66]{
\label{subfig:409382}
\includegraphics[scale=0.32,angle=270]{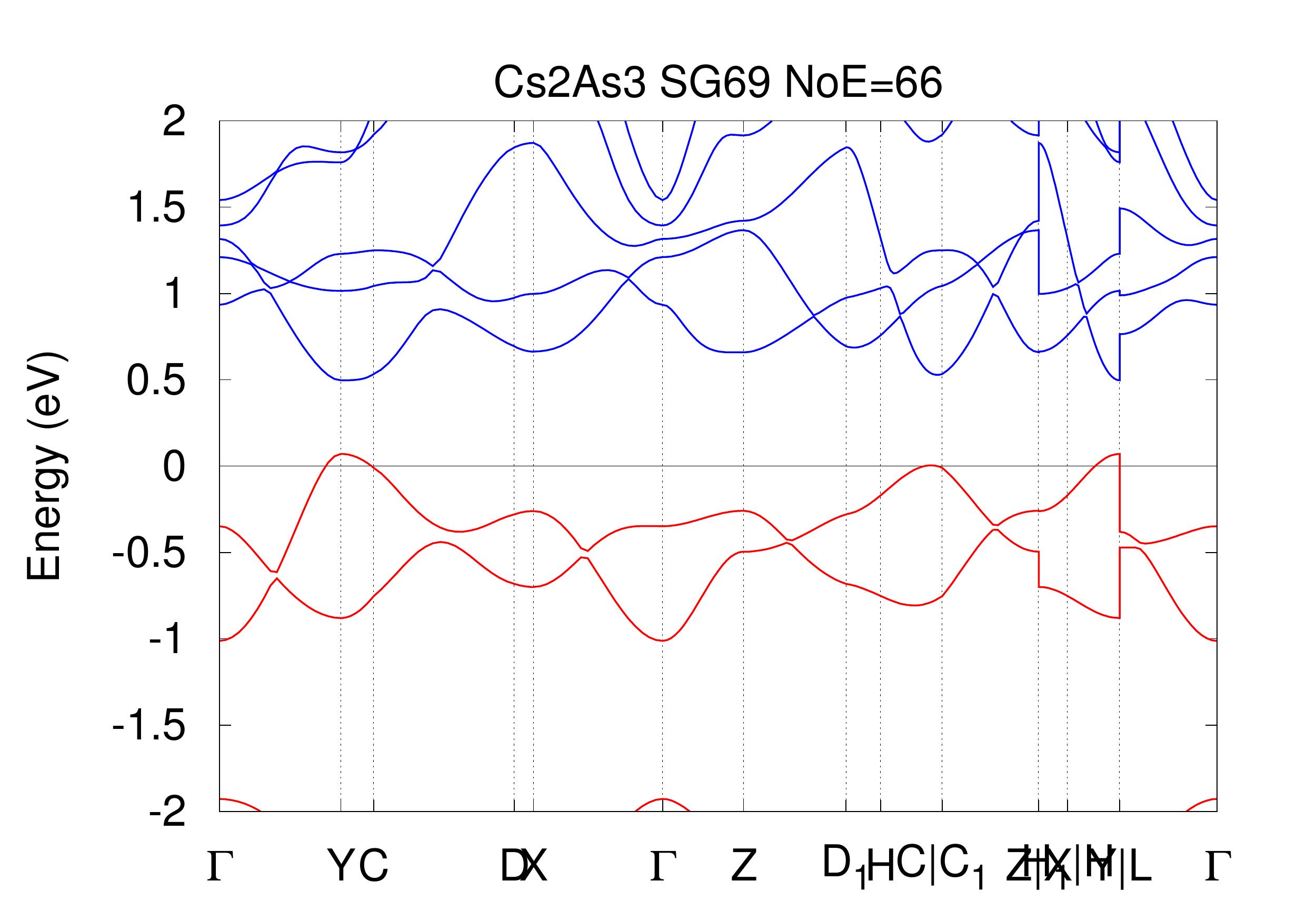}
}
\subfigure[NiPS$_{3}$ SG12 NoA=10 NoE=66]{
\label{subfig:602341}
\includegraphics[scale=0.32,angle=270]{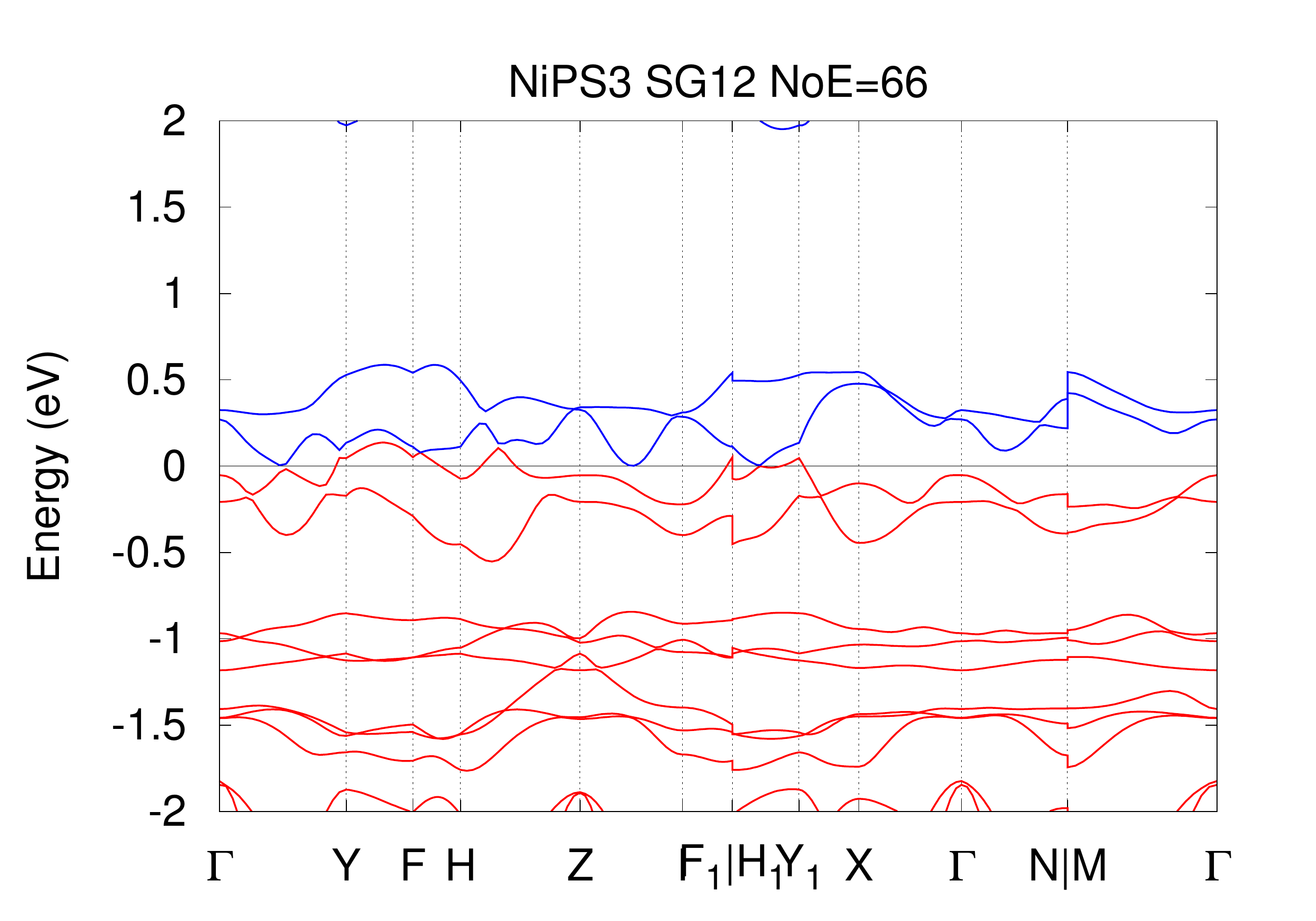}
}
\subfigure[Na$_{3}$Cl$_{2}$ SG83 NoA=10 NoE=34]{
\label{subfig:190543}
\includegraphics[scale=0.32,angle=270]{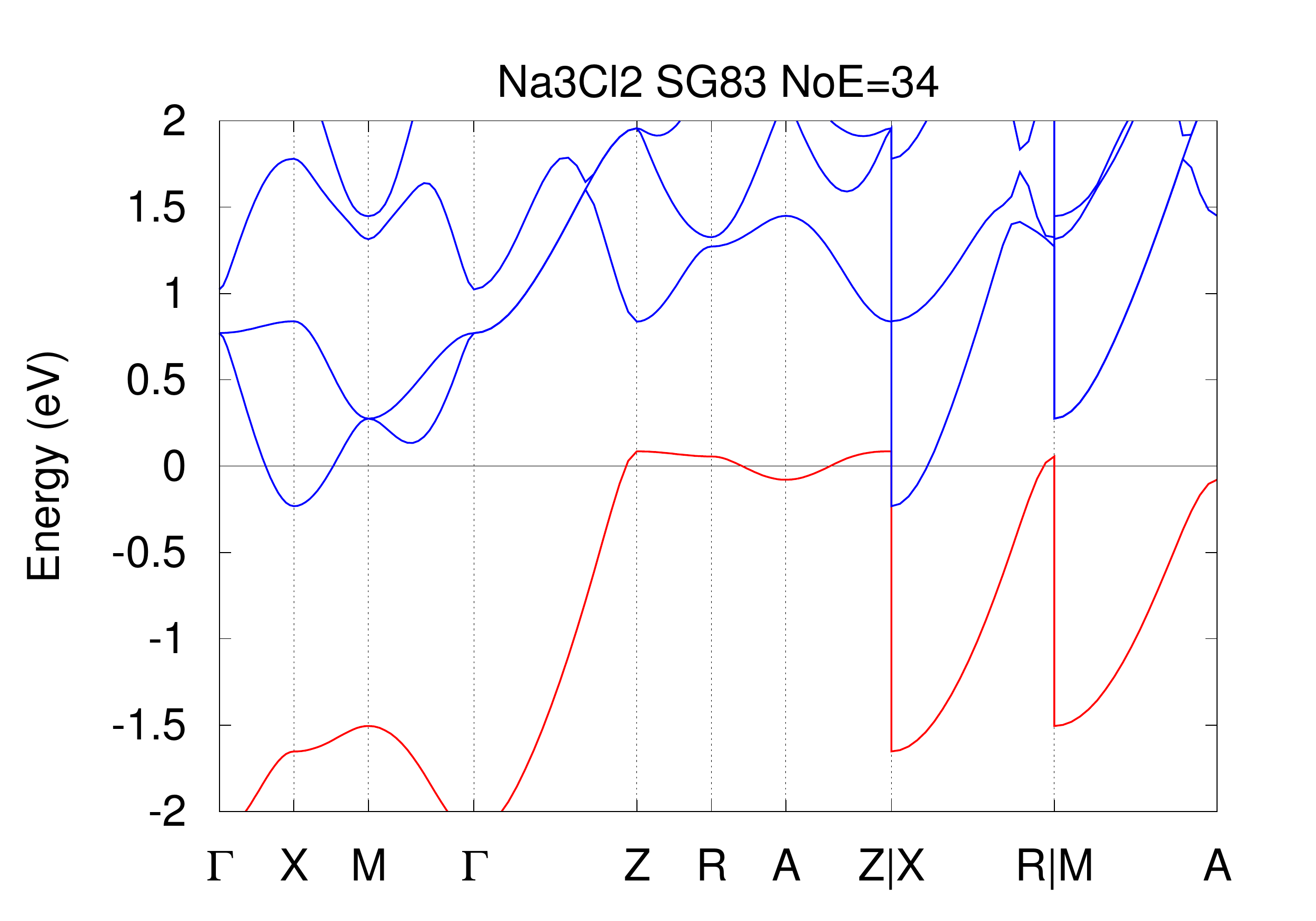}
}
\subfigure[Ca(GaP)$_{2}$ SG194 NoA=10 NoE=52]{
\label{subfig:422525}
\includegraphics[scale=0.32,angle=270]{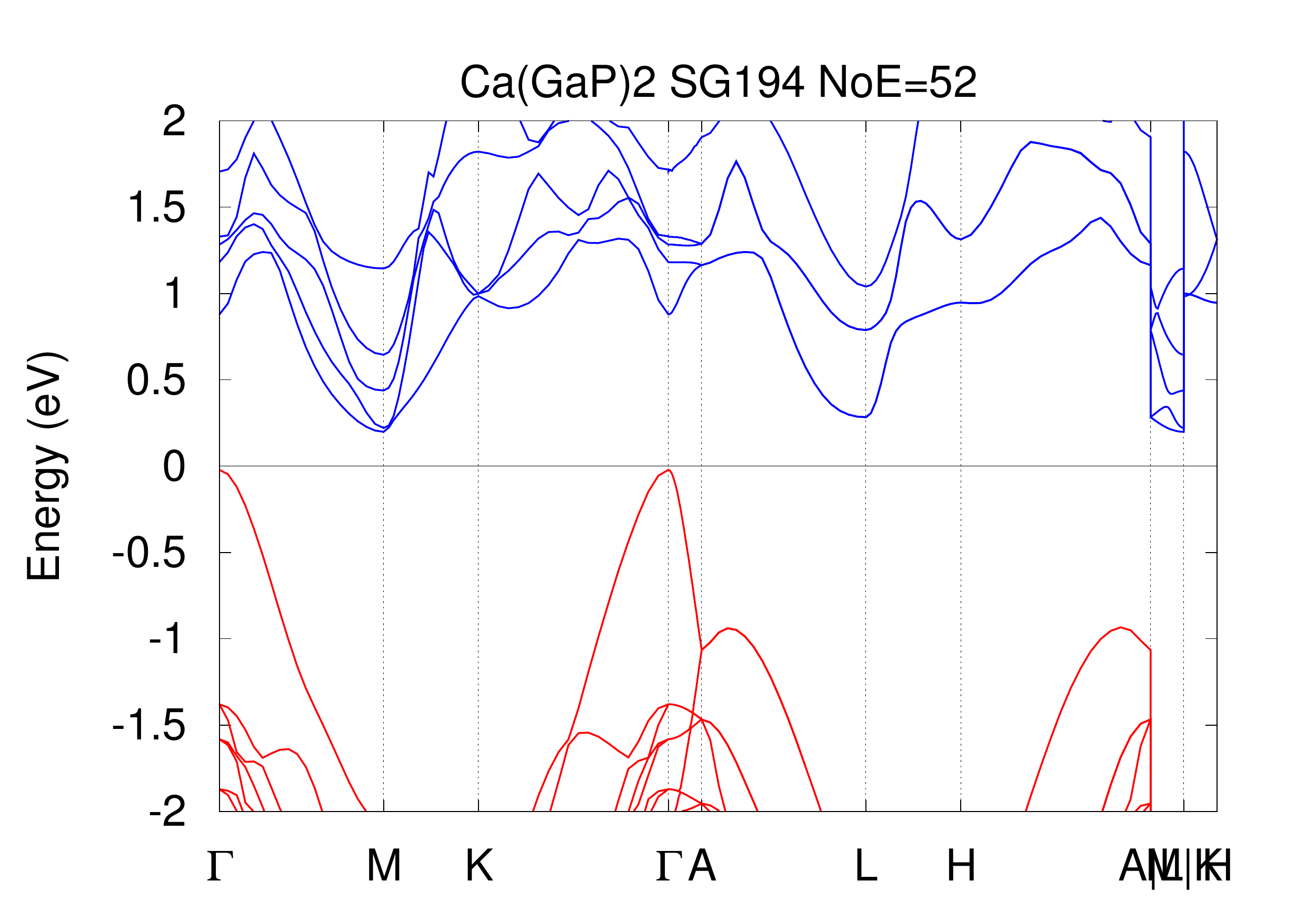}
}
\subfigure[Sr(InP)$_{2}$ SG194 NoA=10 NoE=52]{
\label{subfig:260563}
\includegraphics[scale=0.32,angle=270]{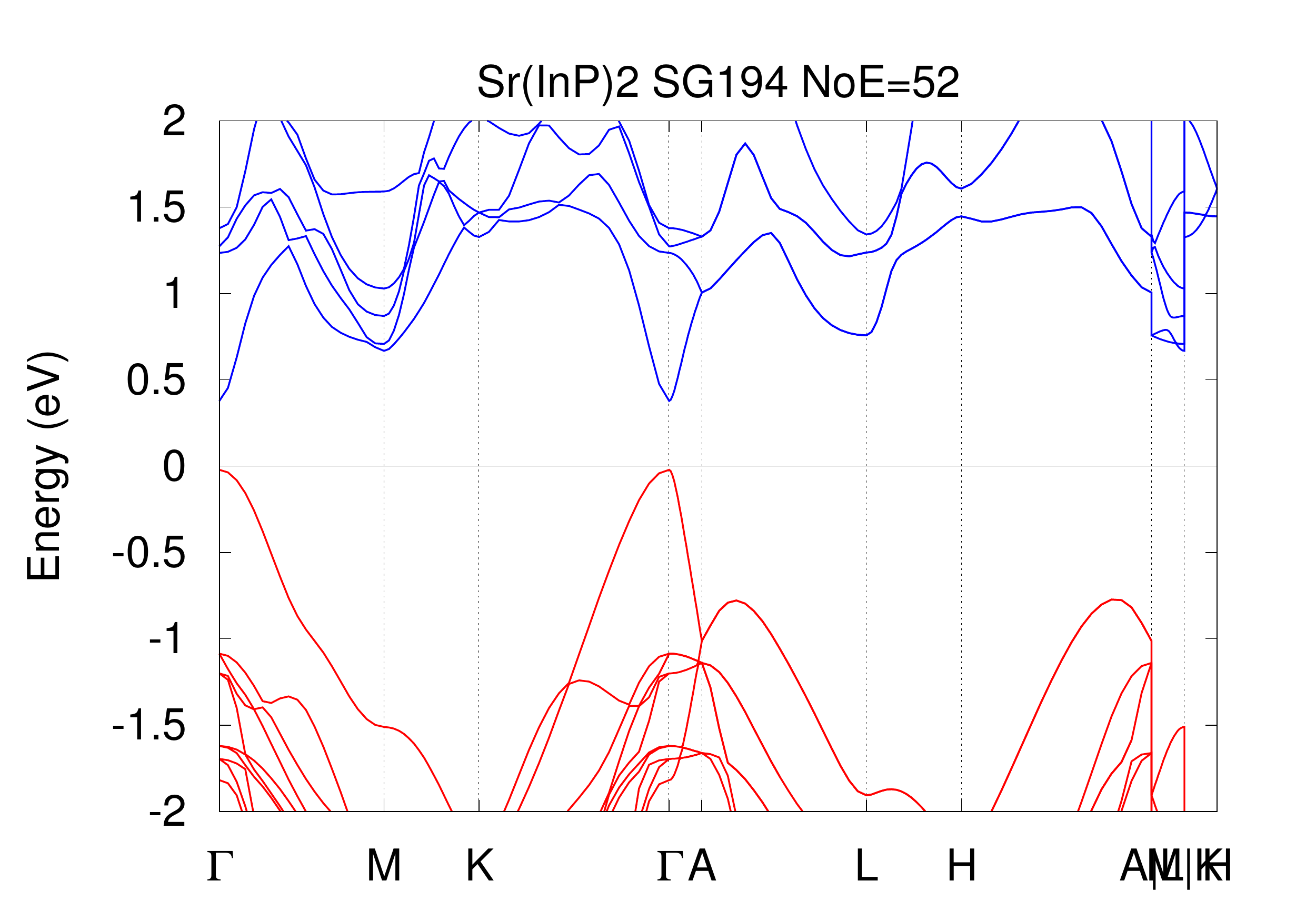}
}
\subfigure[Rb$_{2}$As$_{3}$ SG69 NoA=10 NoE=66]{
\label{subfig:409381}
\includegraphics[scale=0.32,angle=270]{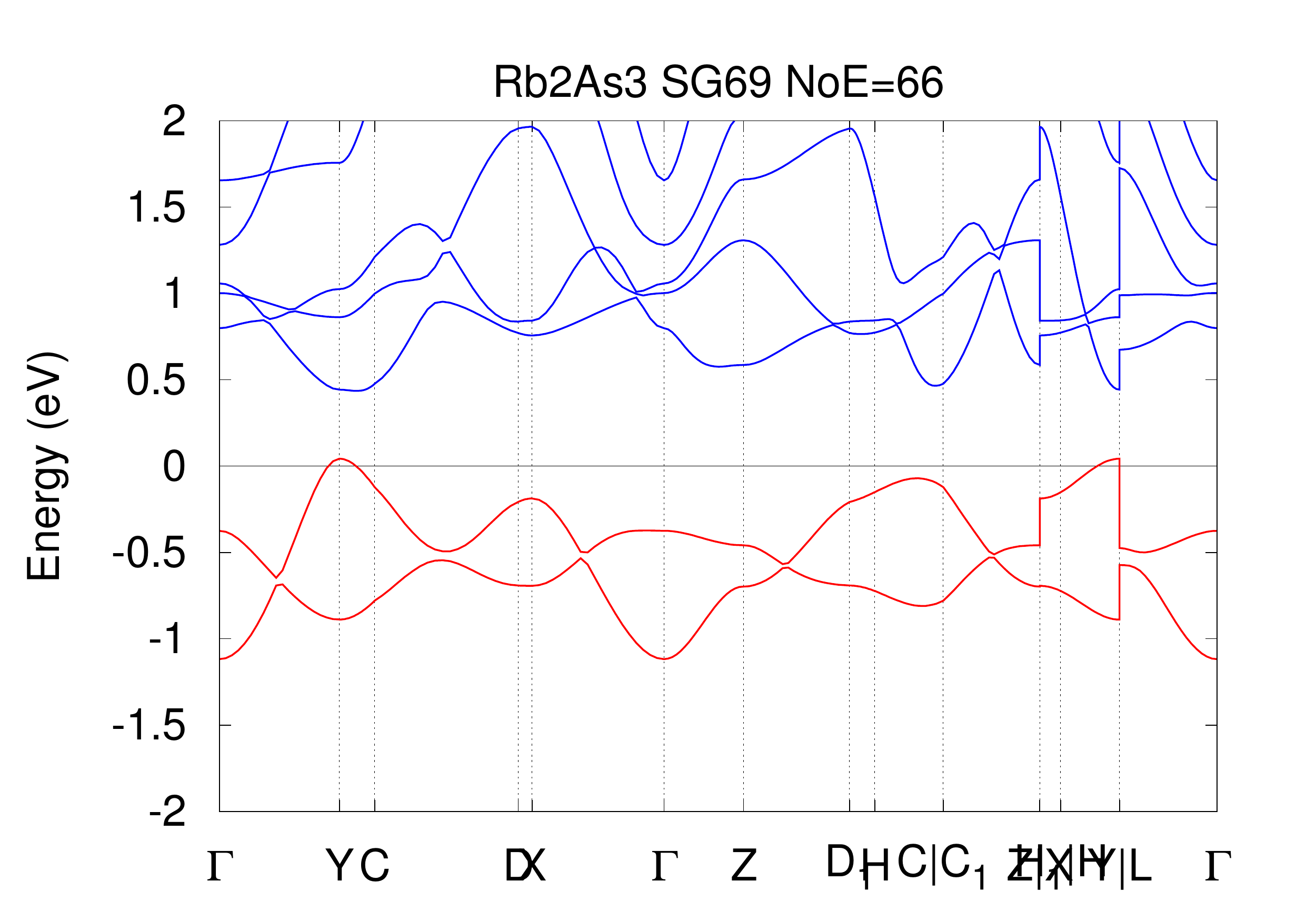}
}
\subfigure[Dy$_{3}$Ni$_{2}$ SG12 NoA=10 NoE=94]{
\label{subfig:2334}
\includegraphics[scale=0.32,angle=270]{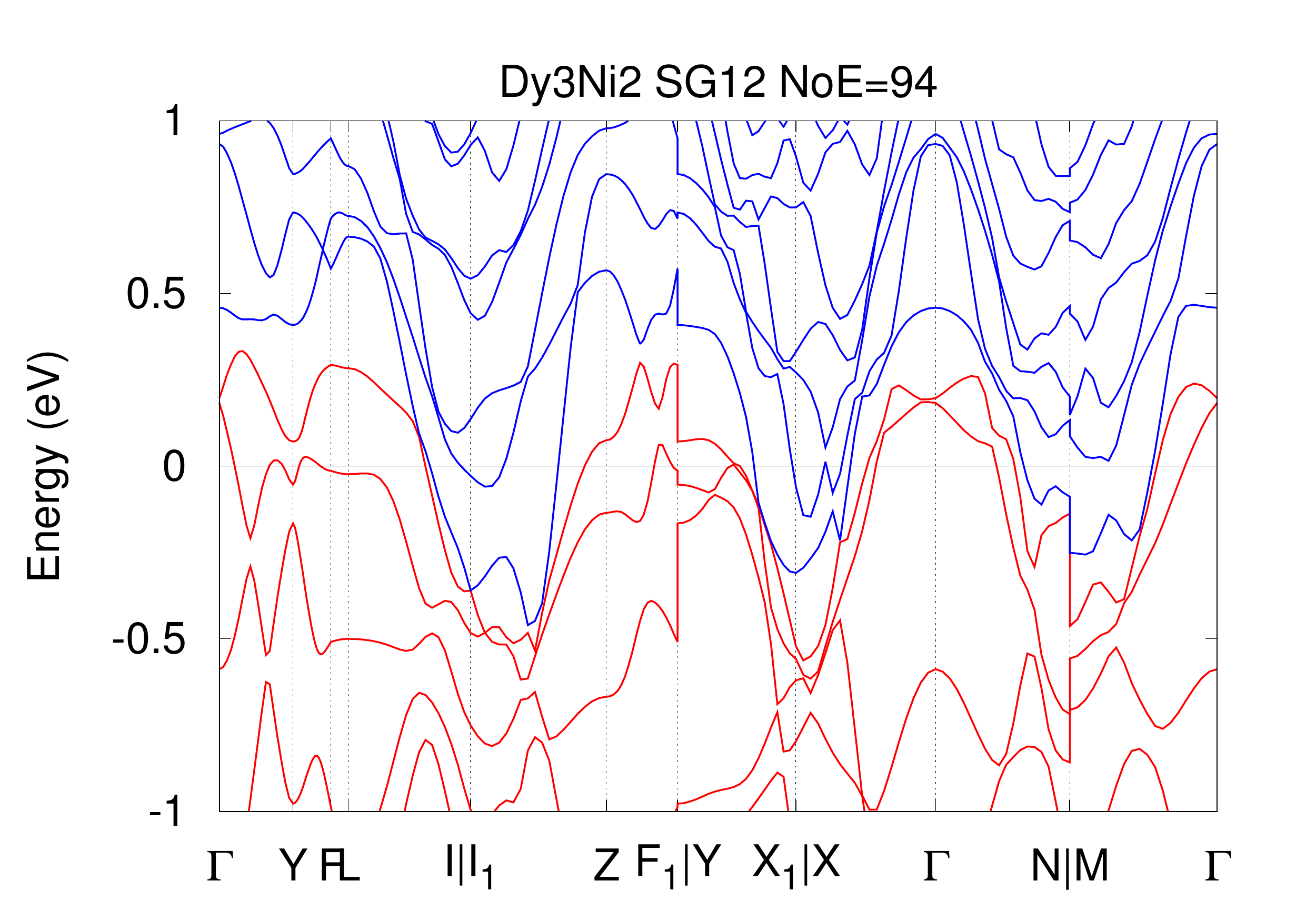}
}
\caption{\hyperref[tab:electride]{back to the table}}
\end{figure}

\begin{figure}[htp]
 \centering
\subfigure[FePSe$_{3}$ SG148 NoA=10 NoE=62]{
\label{subfig:633091}
\includegraphics[scale=0.32,angle=270]{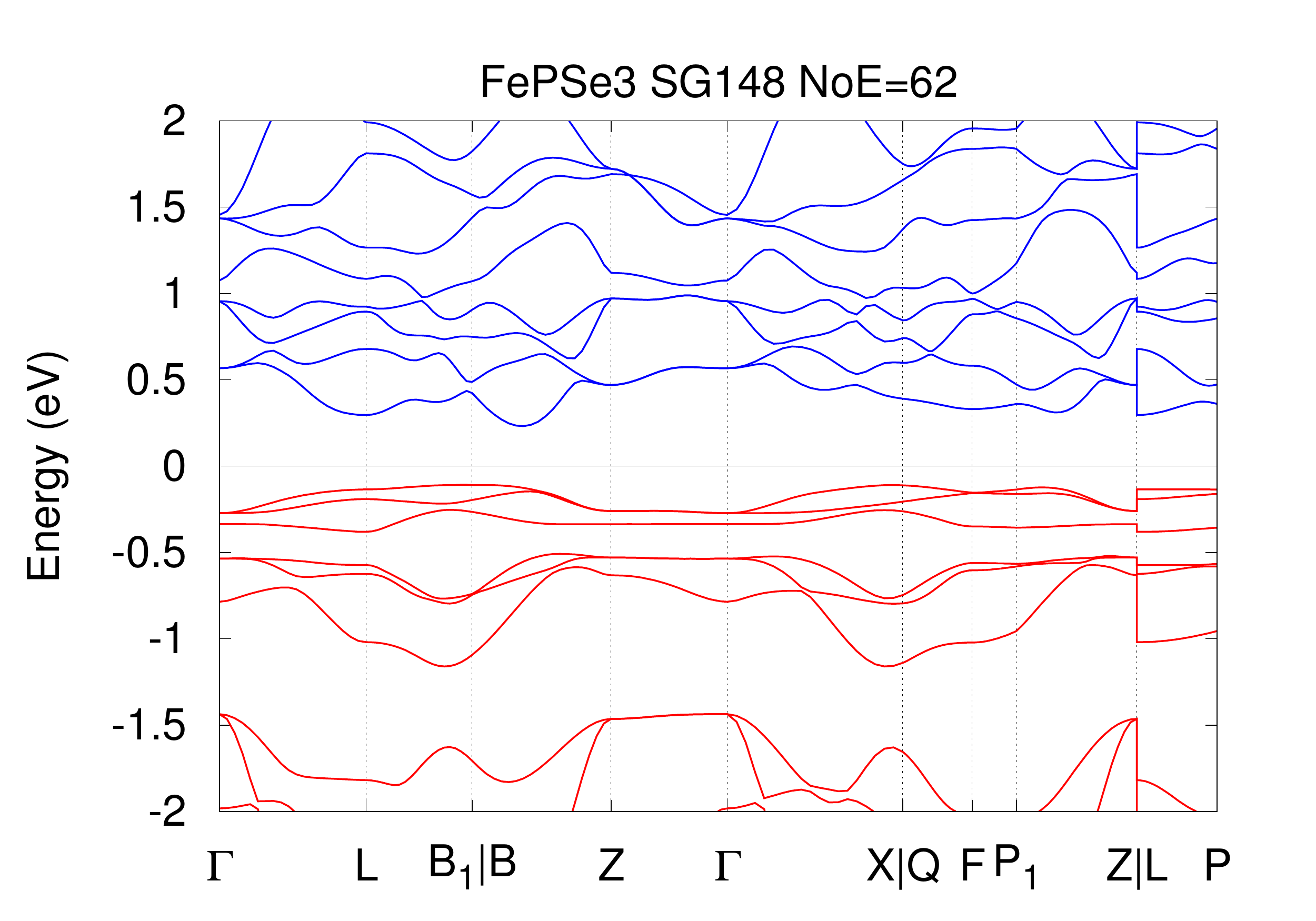}
}
\subfigure[K$_{2}$Ga$_{3}$ SG139 NoA=10 NoE=54]{
\label{subfig:160496}
\includegraphics[scale=0.32,angle=270]{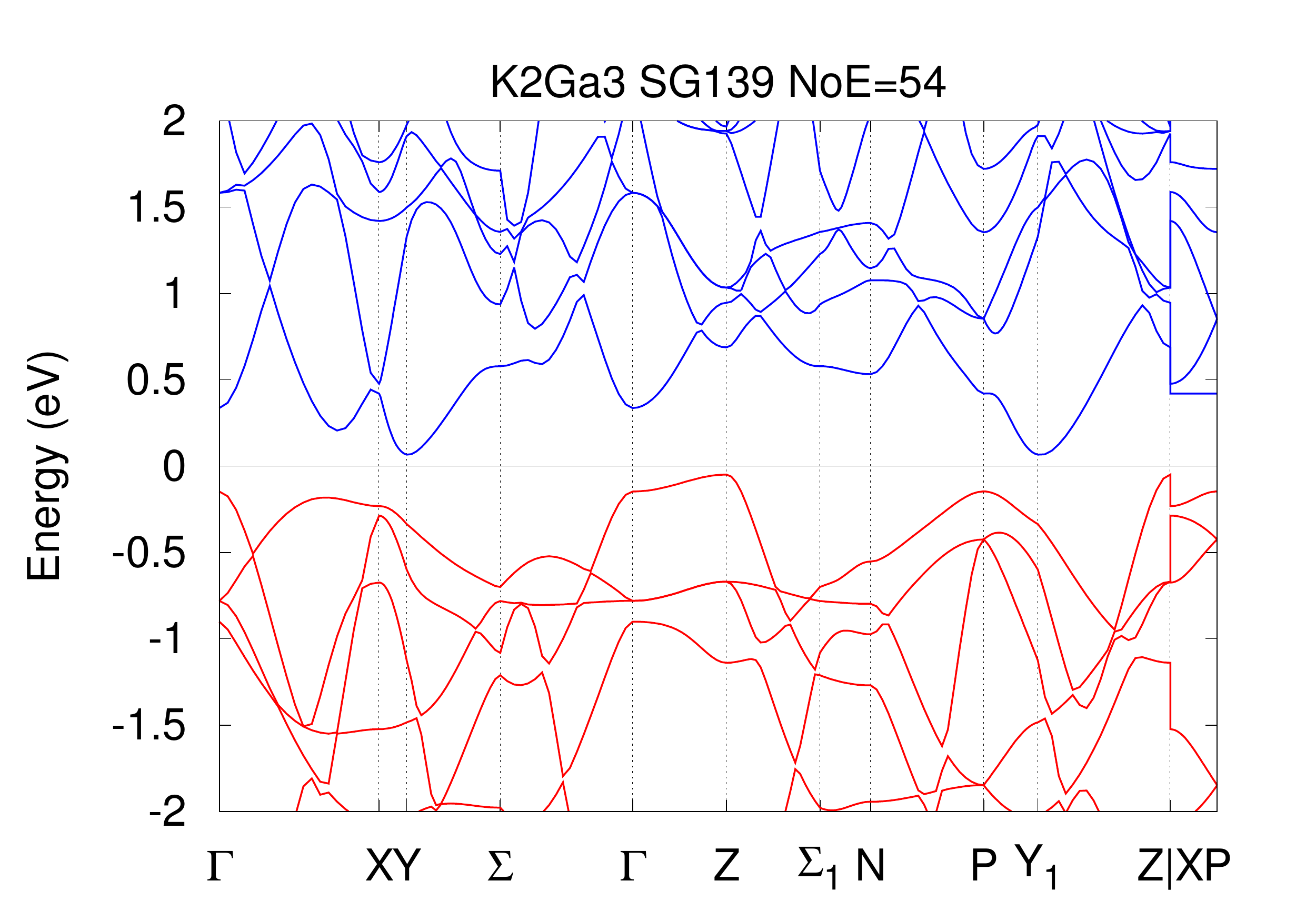}
}
\subfigure[Ca(InP)$_{2}$ SG194 NoA=10 NoE=52]{
\label{subfig:260562}
\includegraphics[scale=0.32,angle=270]{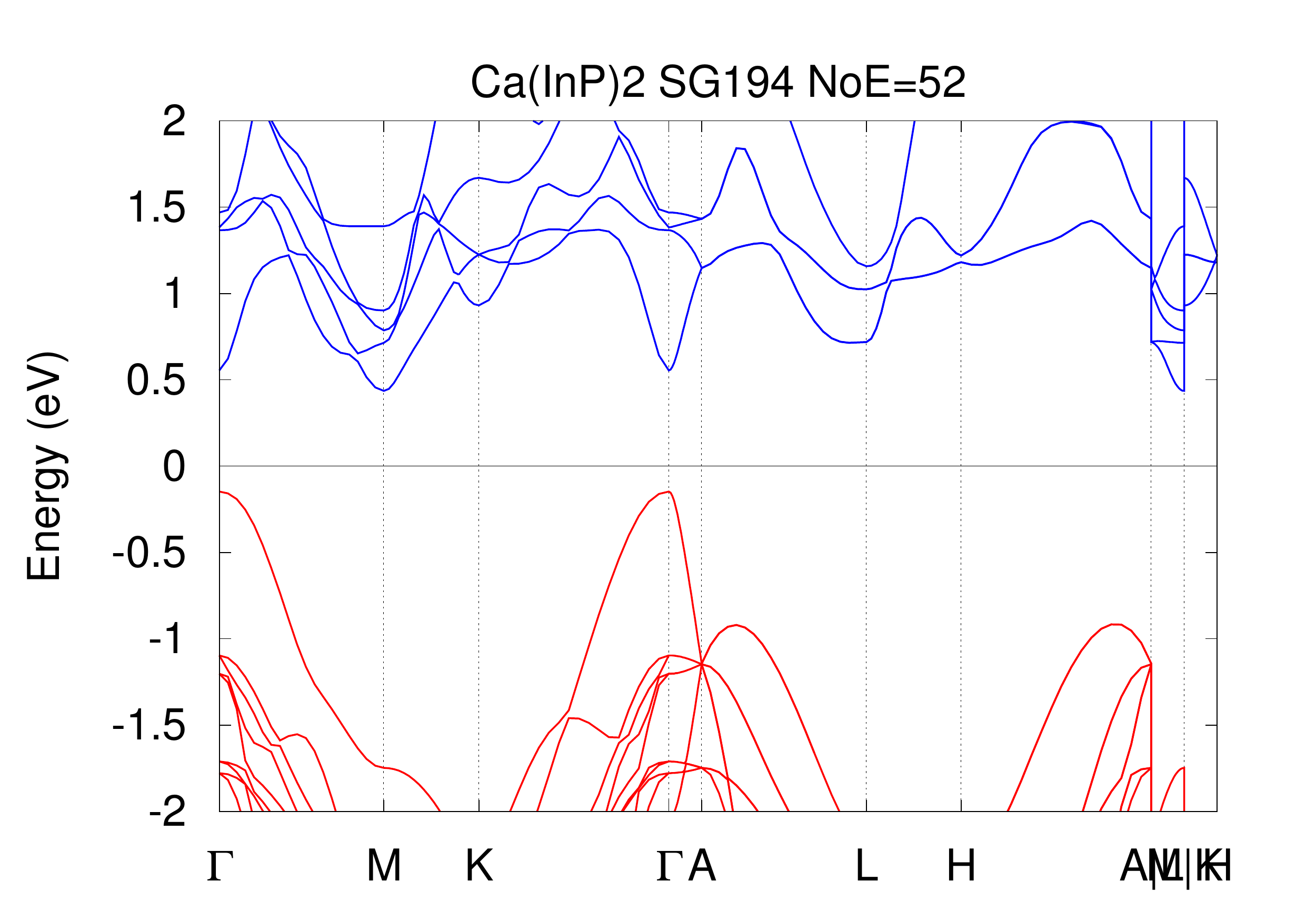}
}
\subfigure[Tb$_{3}$Ni$_{2}$ SG12 NoA=10 NoE=94]{
\label{subfig:646858}
\includegraphics[scale=0.32,angle=270]{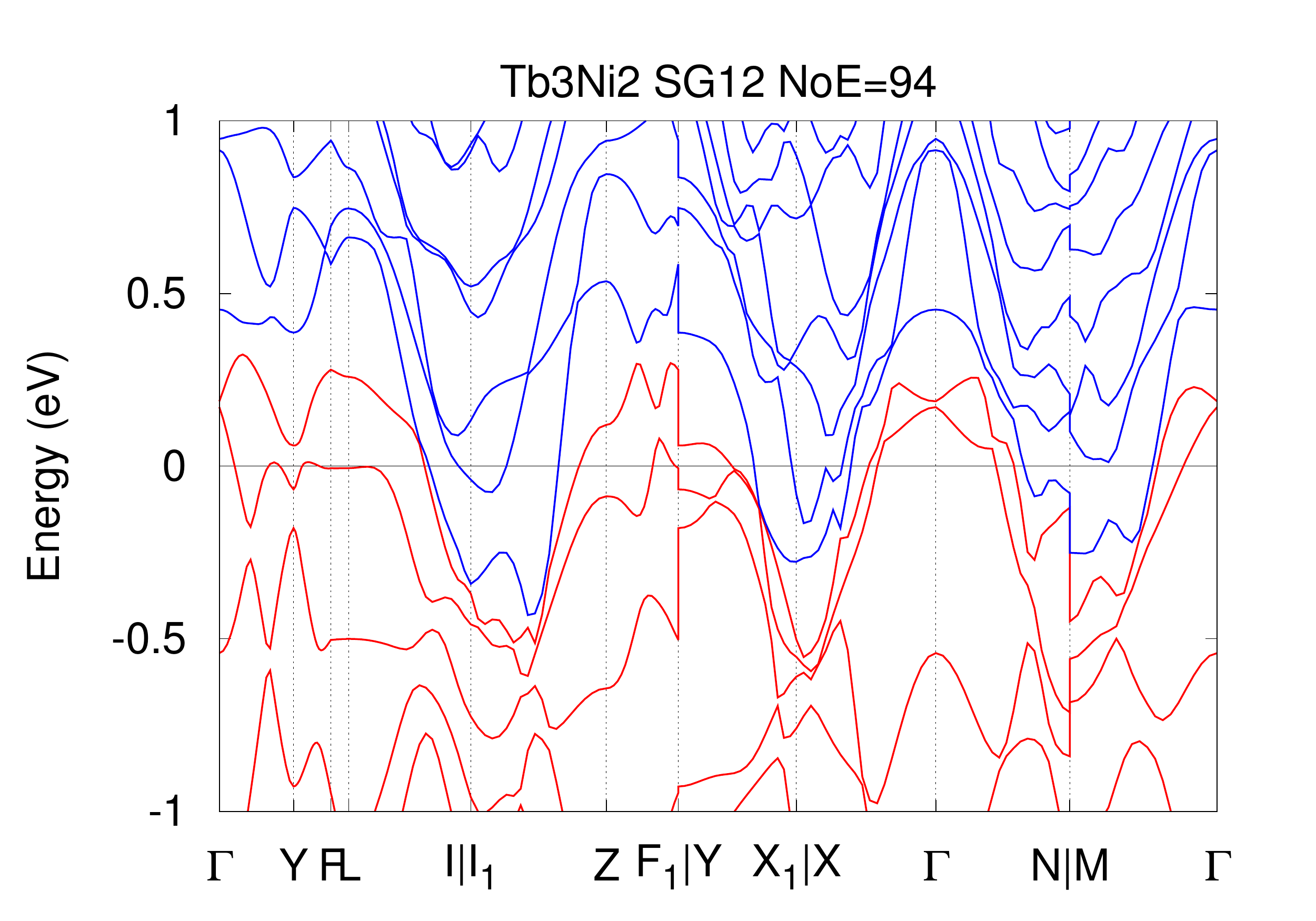}
}
\subfigure[Cs$_{2}$In$_{3}$ SG119 NoA=10 NoE=54]{
\label{subfig:102868}
\includegraphics[scale=0.32,angle=270]{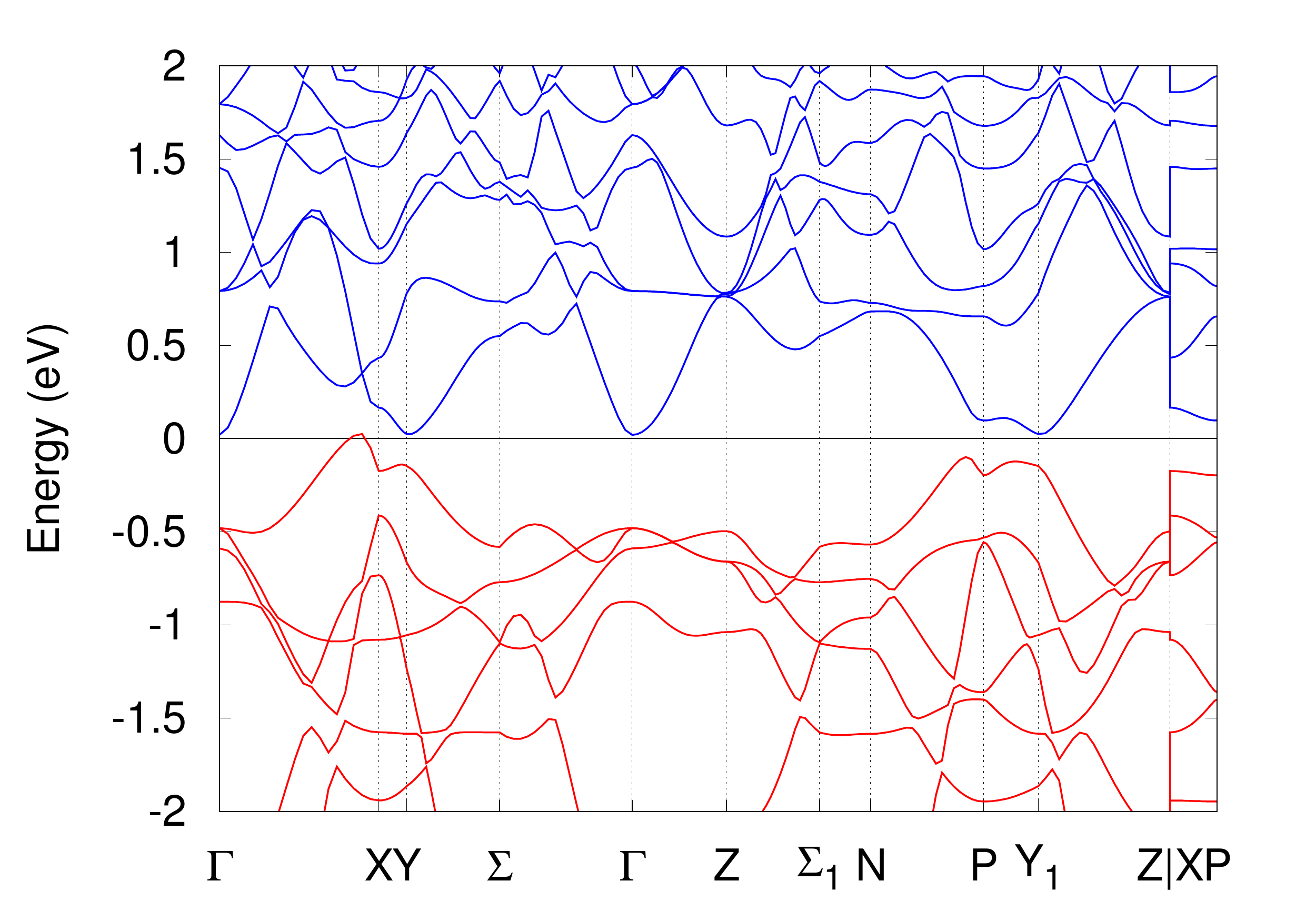}
}
\subfigure[FePS$_{3}$ SG12 NoA=10 NoE=62]{
\label{subfig:61392}
\includegraphics[scale=0.32,angle=270]{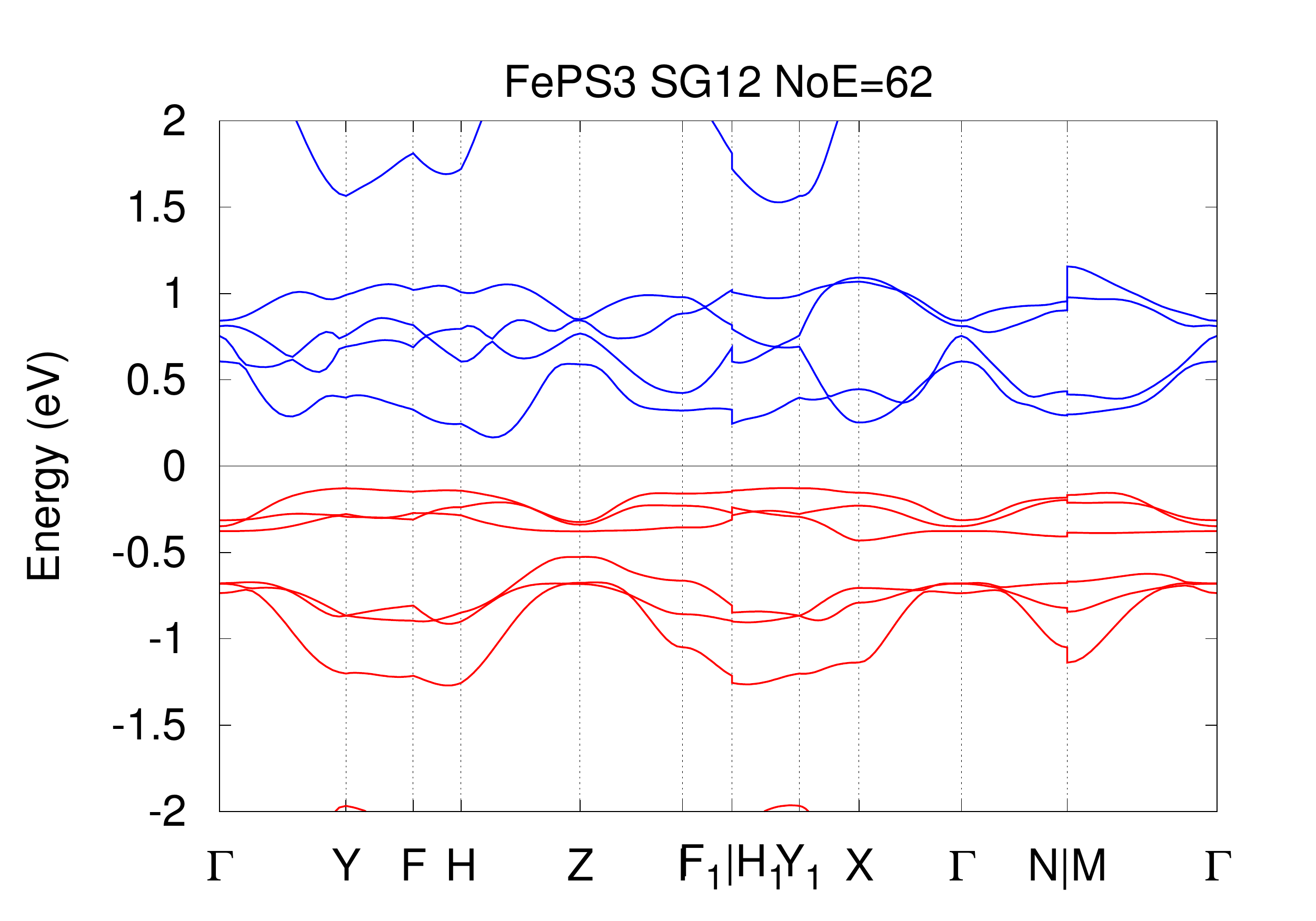}
}
\subfigure[Cu$_{2}$P$_{2}$O$_{7}$ SG12 NoA=11 NoE=74]{
\label{subfig:27436}
\includegraphics[scale=0.32,angle=270]{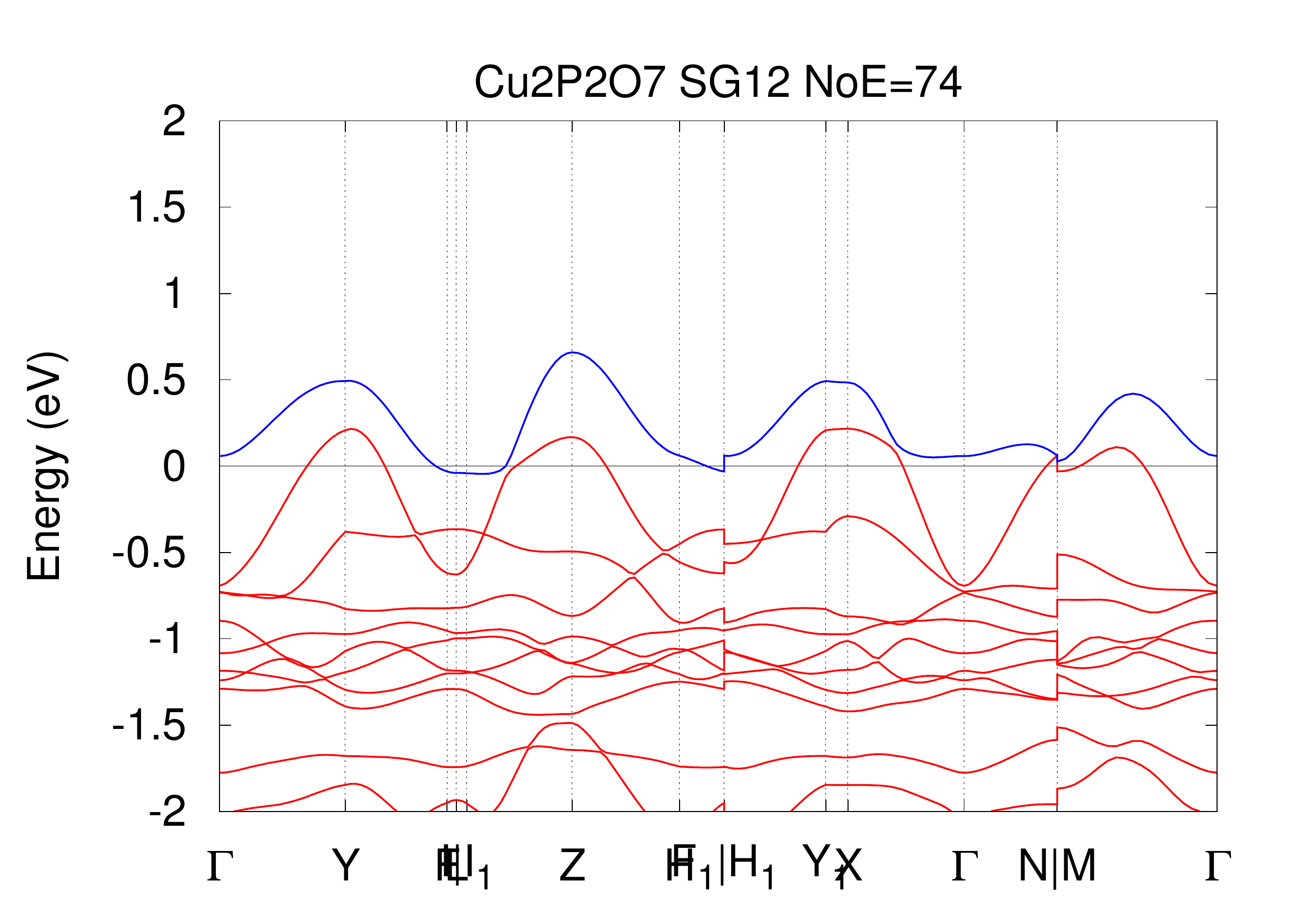}
}
\subfigure[Ir$_{3}$Se$_{8}$ SG148 NoA=11 NoE=75]{
\label{subfig:40823}
\includegraphics[scale=0.32,angle=270]{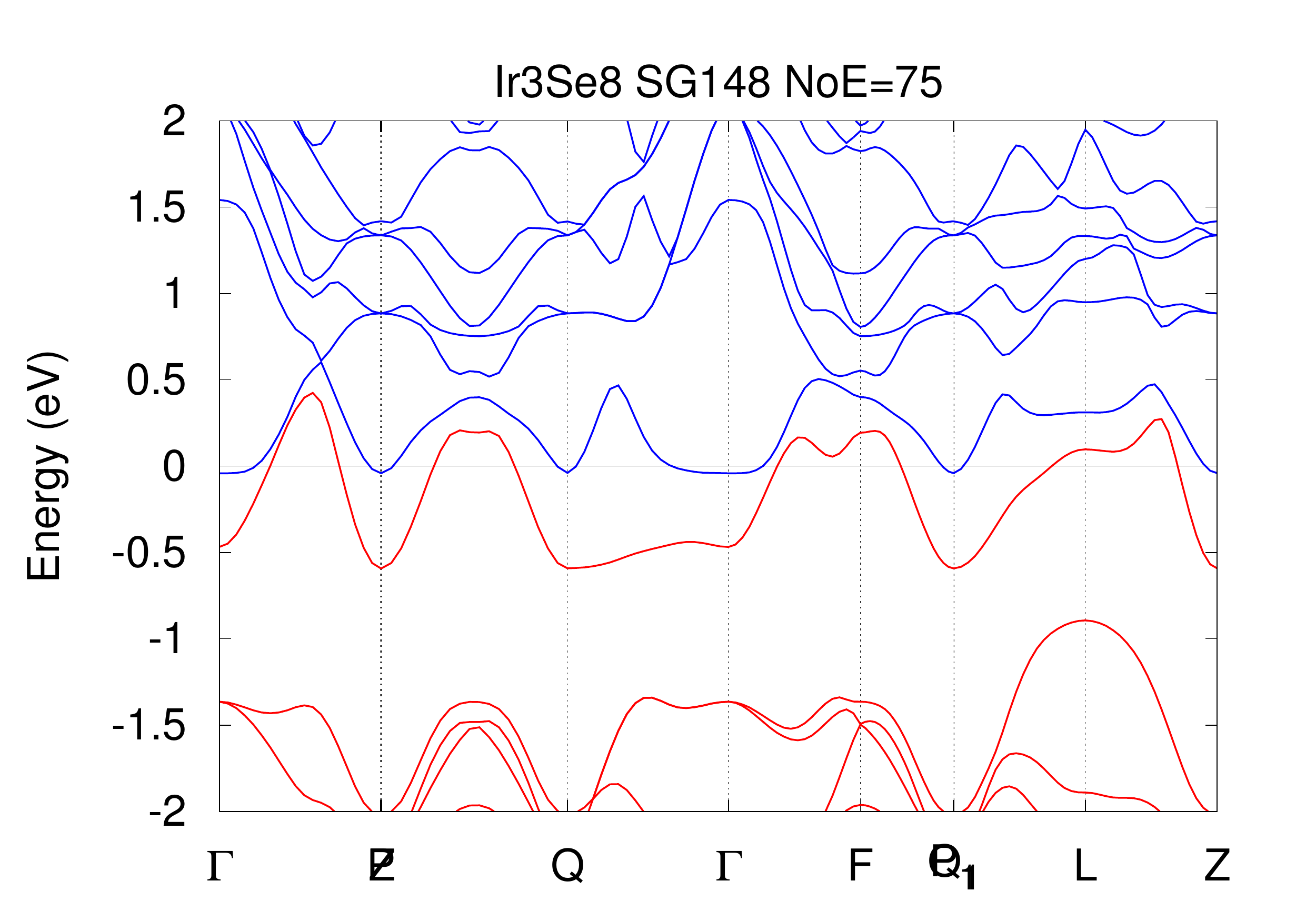}
}
\caption{\hyperref[tab:electride]{back to the table}}
\end{figure}

\begin{figure}[htp]
 \centering
\subfigure[Cu$_{2}$As$_{2}$O$_{7}$ SG12 NoA=11 NoE=74]{
\label{subfig:162061}
\includegraphics[scale=0.32,angle=270]{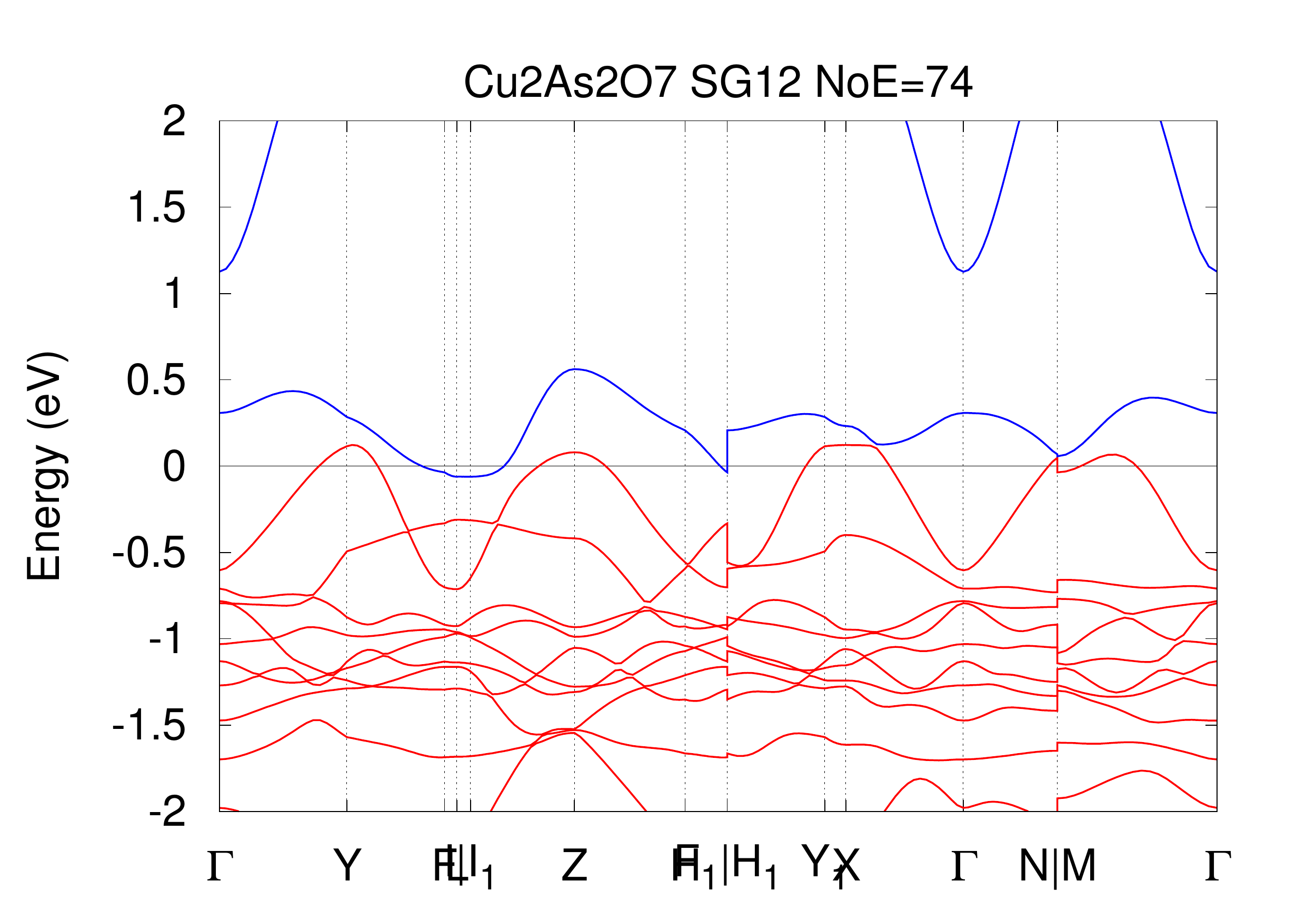}
}
\subfigure[Ba$_{3}$(LiAs)$_{4}$ SG71 NoA=11 NoE=54]{
\label{subfig:280027}
\includegraphics[scale=0.32,angle=270]{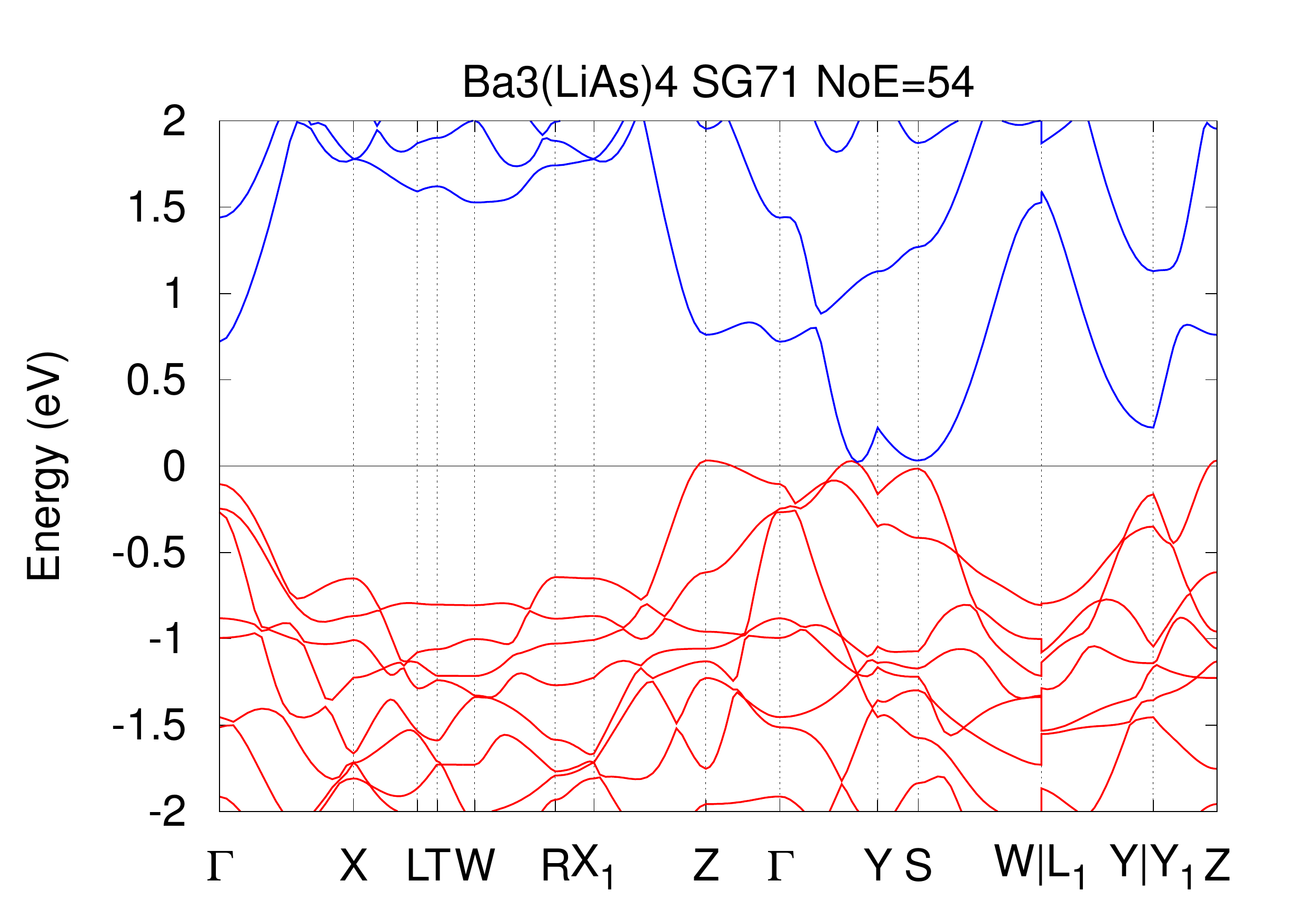}
}
\subfigure[Al$_{8}$Mo$_{3}$ SG12 NoA=11 NoE=42]{
\label{subfig:58001}
\includegraphics[scale=0.32,angle=270]{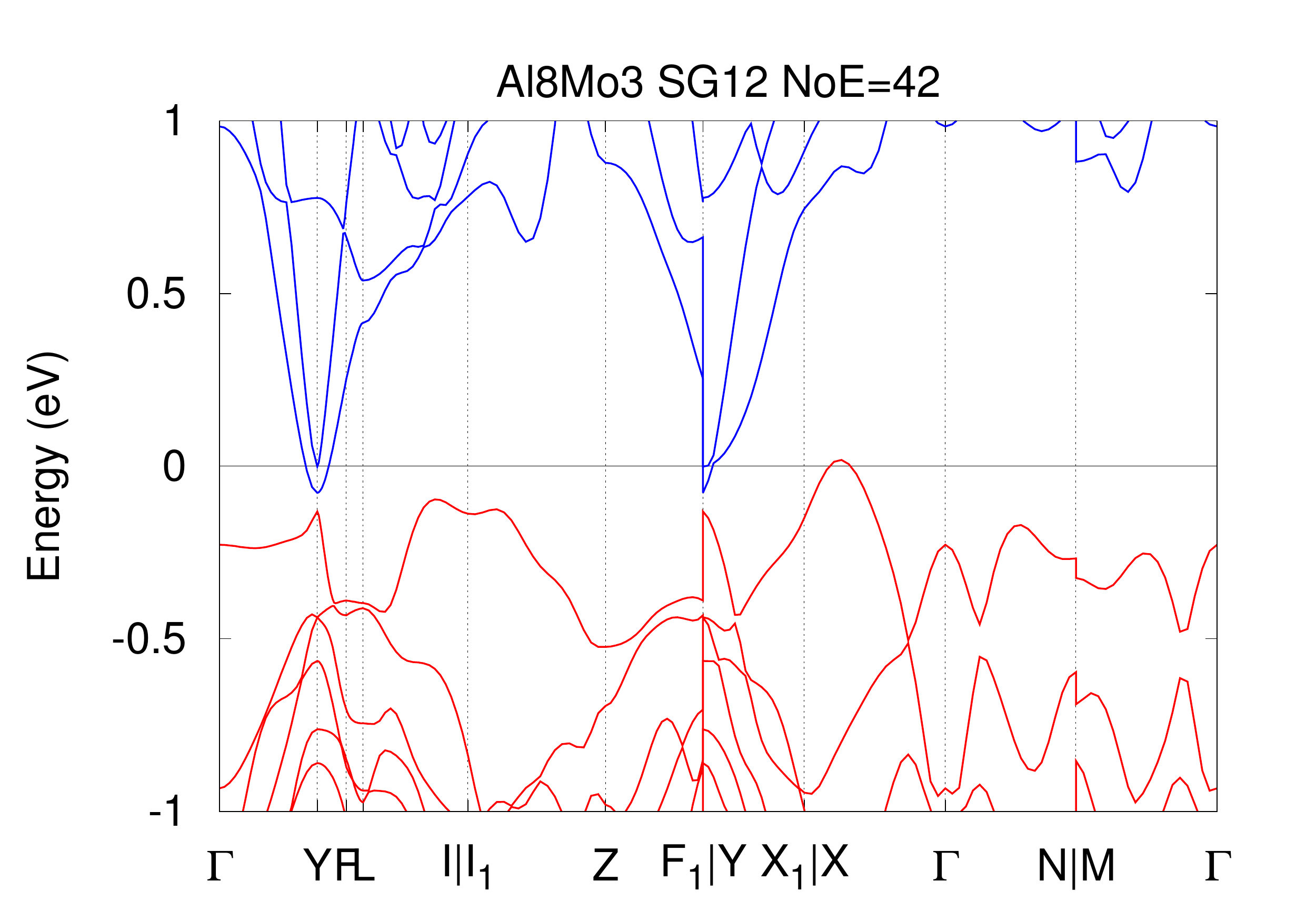}
}
\subfigure[ScCoSn SG62 NoA=12 NoE=64]{
\label{subfig:624977}
\includegraphics[scale=0.32,angle=270]{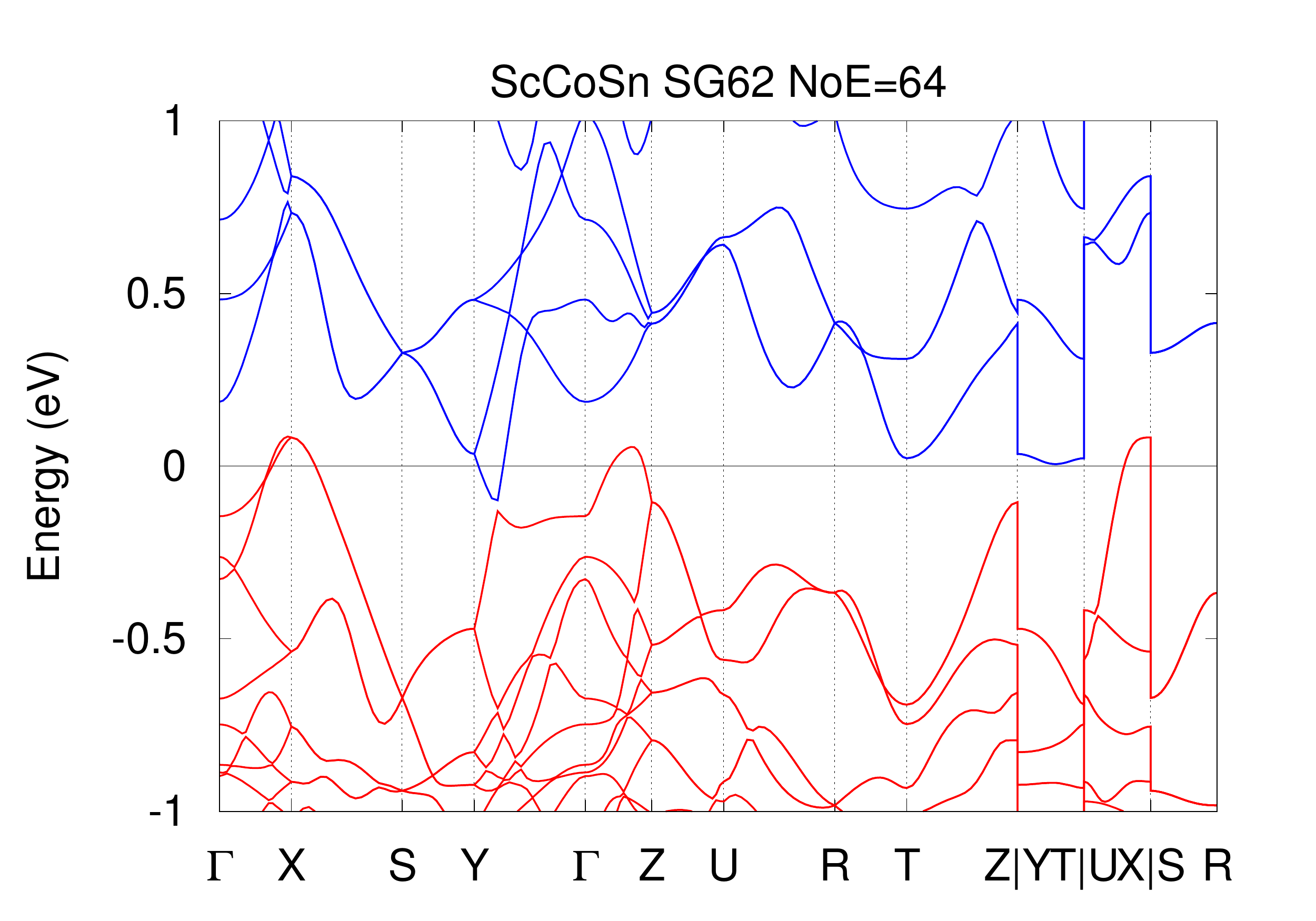}
}
\subfigure[CuBIr SG43 NoA=12 NoE=92]{
\label{subfig:75029}
\includegraphics[scale=0.32,angle=270]{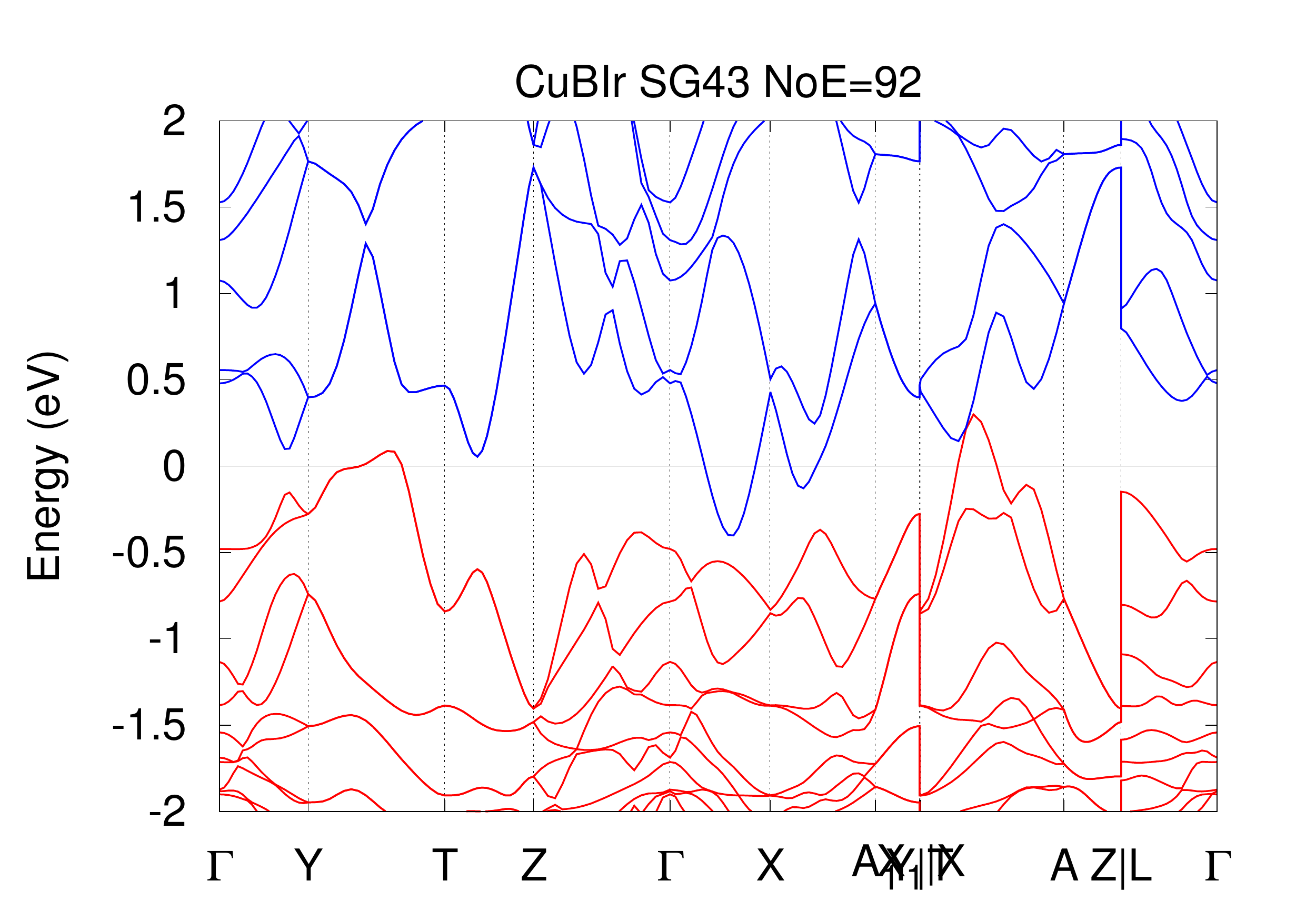}
}
\subfigure[RbPrTe$_{4}$ SG125 NoA=12 NoE=88]{
\label{subfig:412794}
\includegraphics[scale=0.32,angle=270]{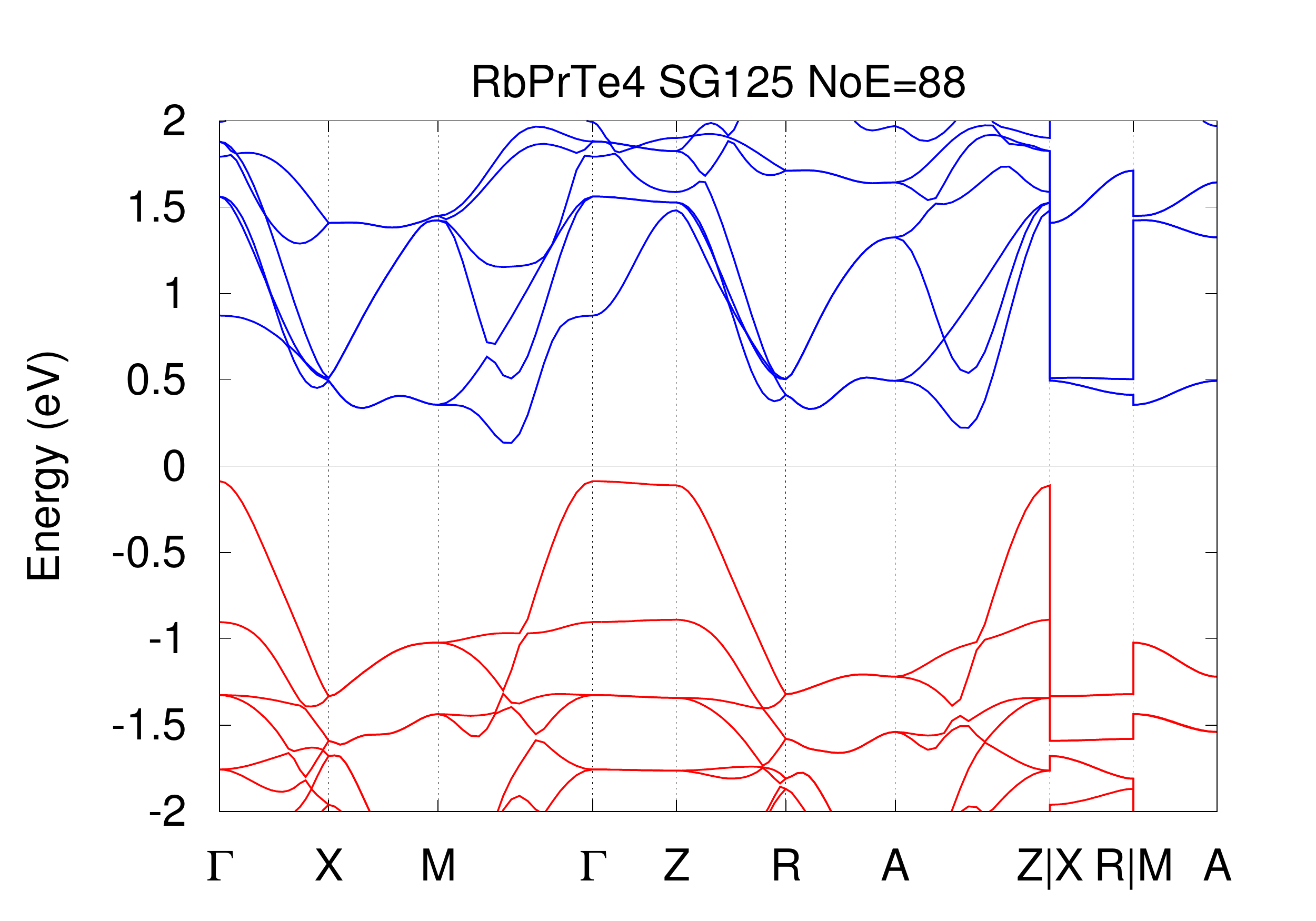}
}
\subfigure[NiAs$_{2}$ SG205 NoA=12 NoE=80]{
\label{subfig:42569}
\includegraphics[scale=0.32,angle=270]{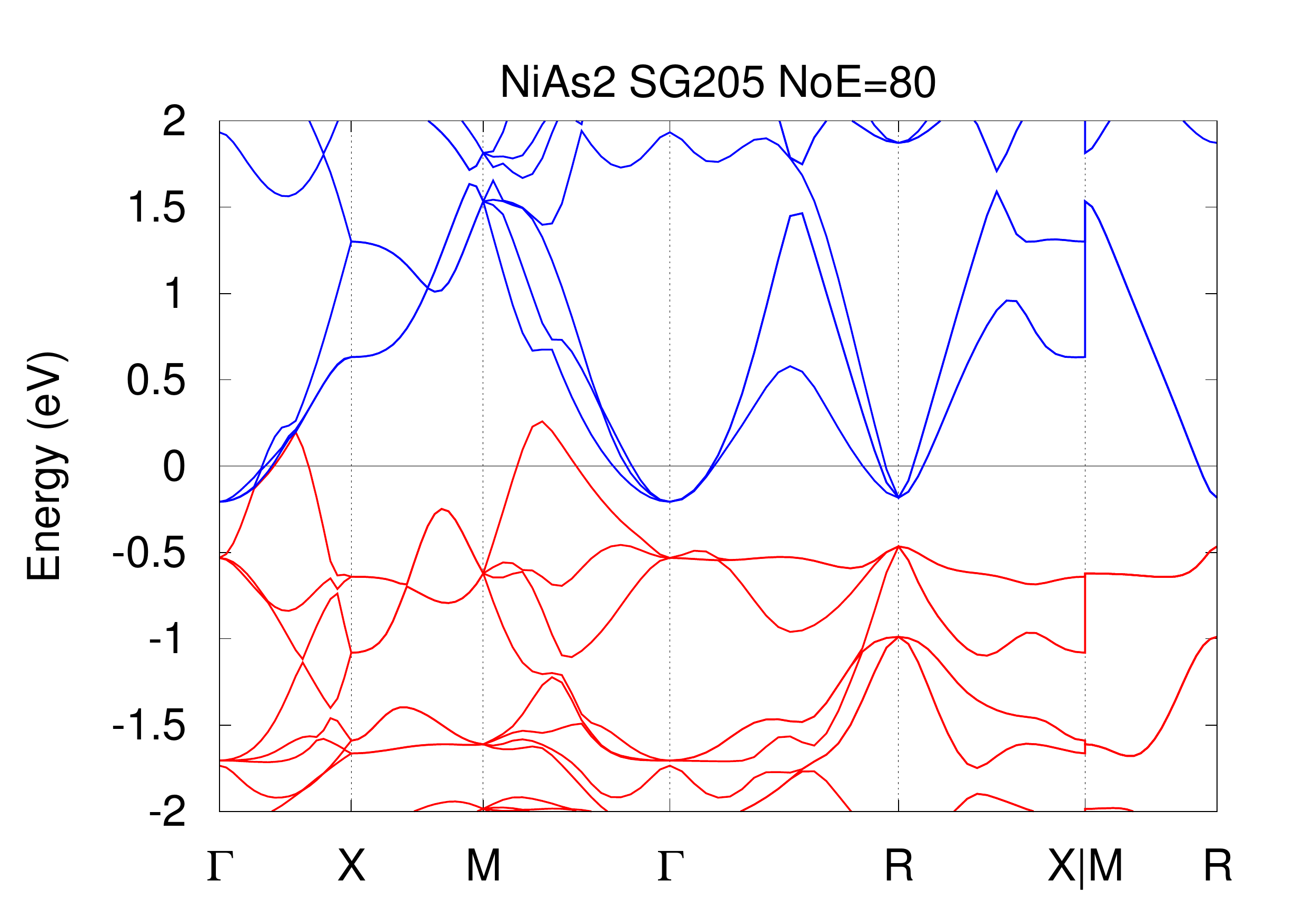}
}
\subfigure[OsSe$_{2}$ SG205 NoA=12 NoE=80]{
\label{subfig:24202}
\includegraphics[scale=0.32,angle=270]{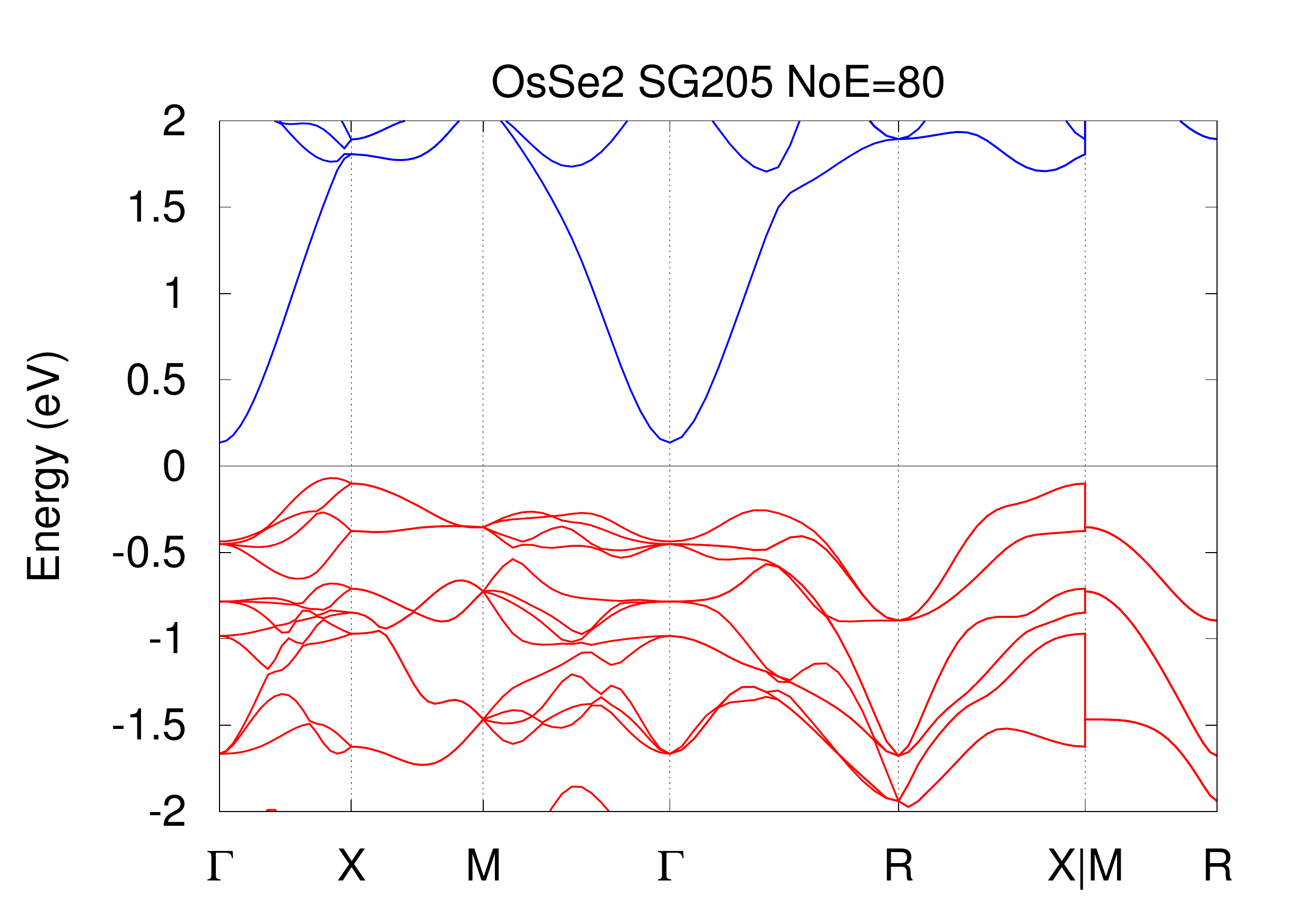}
}
\caption{\hyperref[tab:electride]{back to the table}}
\end{figure}

\begin{figure}[htp]
 \centering
\subfigure[RuSe$_{2}$ SG205 NoA=12 NoE=80]{
\label{subfig:650607}
\includegraphics[scale=0.32,angle=270]{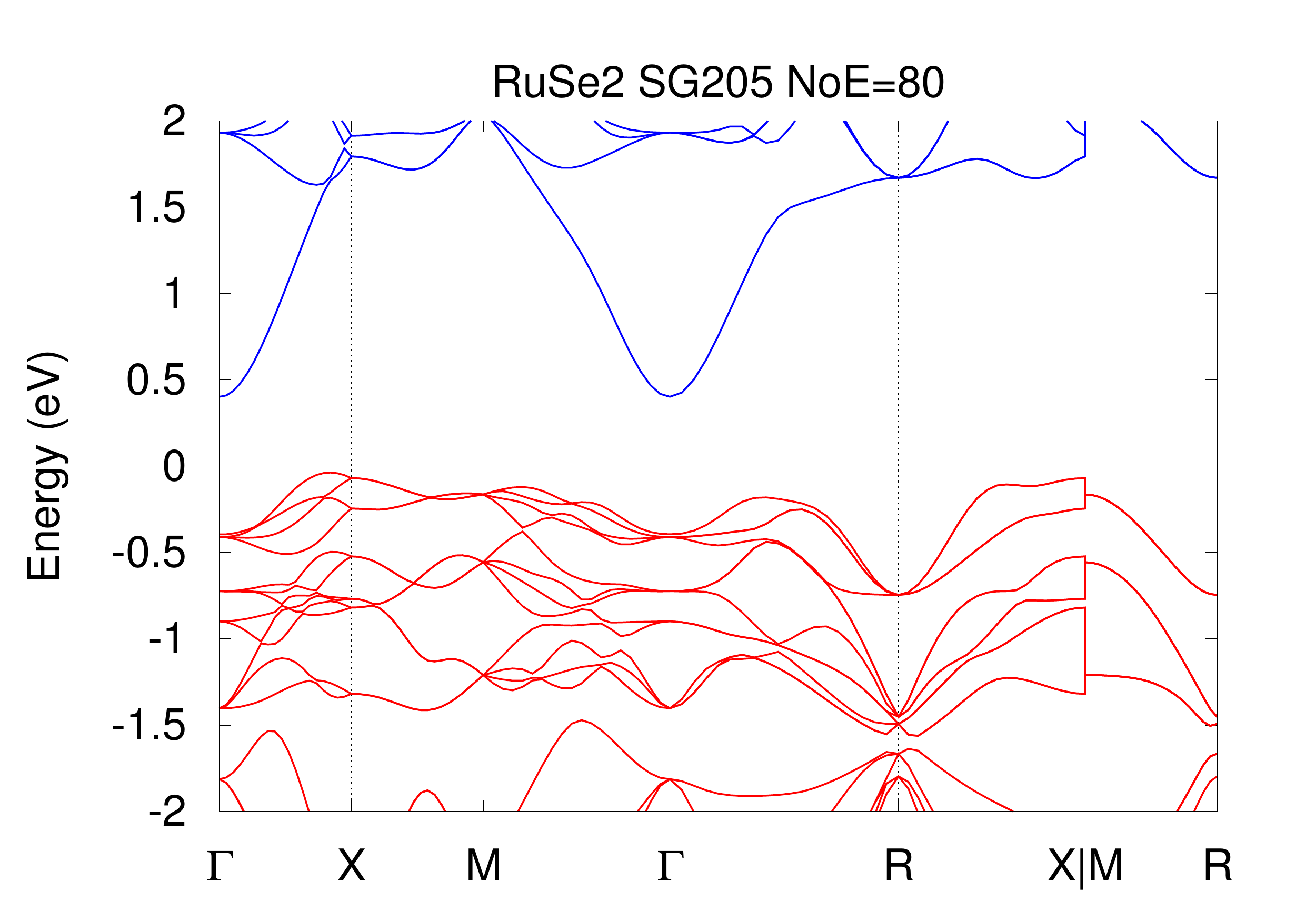}
}
\subfigure[OsS$_{2}$ SG205 NoA=12 NoE=80]{
\label{subfig:24187}
\includegraphics[scale=0.32,angle=270]{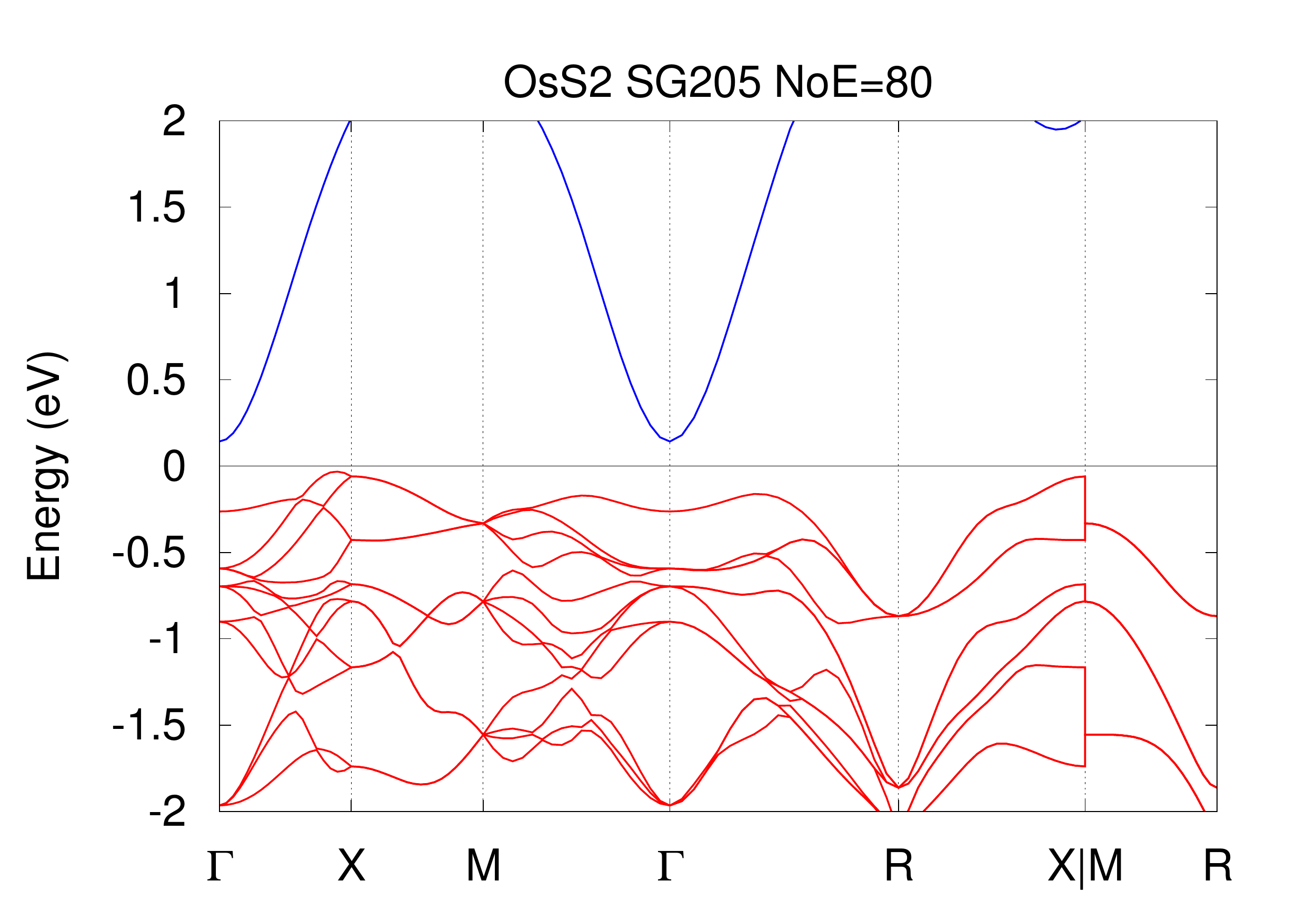}
}
\subfigure[NdS$_{2}$ SG14 NoA=12 NoE=92]{
\label{subfig:419345}
\includegraphics[scale=0.32,angle=270]{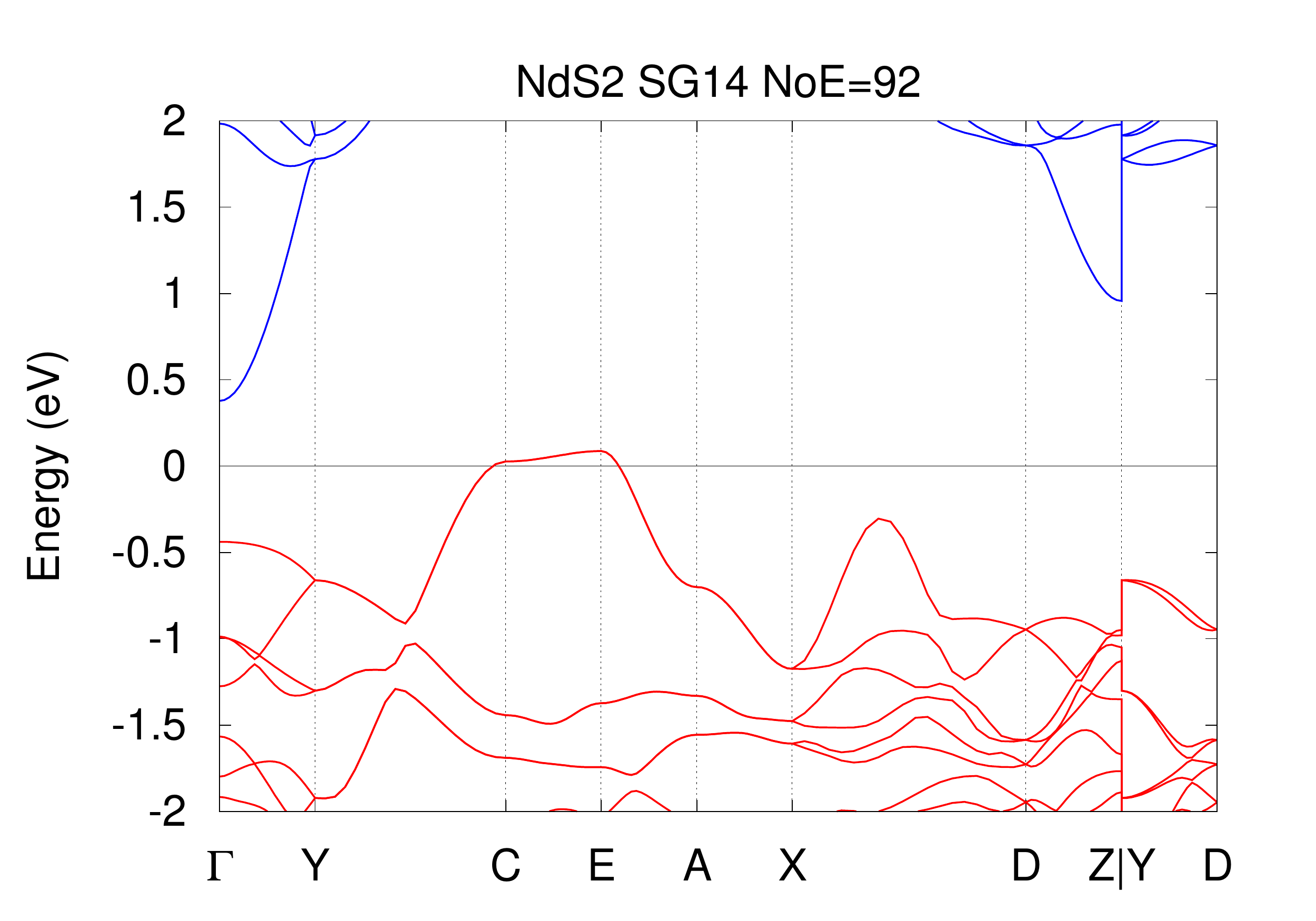}
}
\subfigure[As$_{2}$Pd SG205 NoA=12 NoE=80]{
\label{subfig:43101}
\includegraphics[scale=0.32,angle=270]{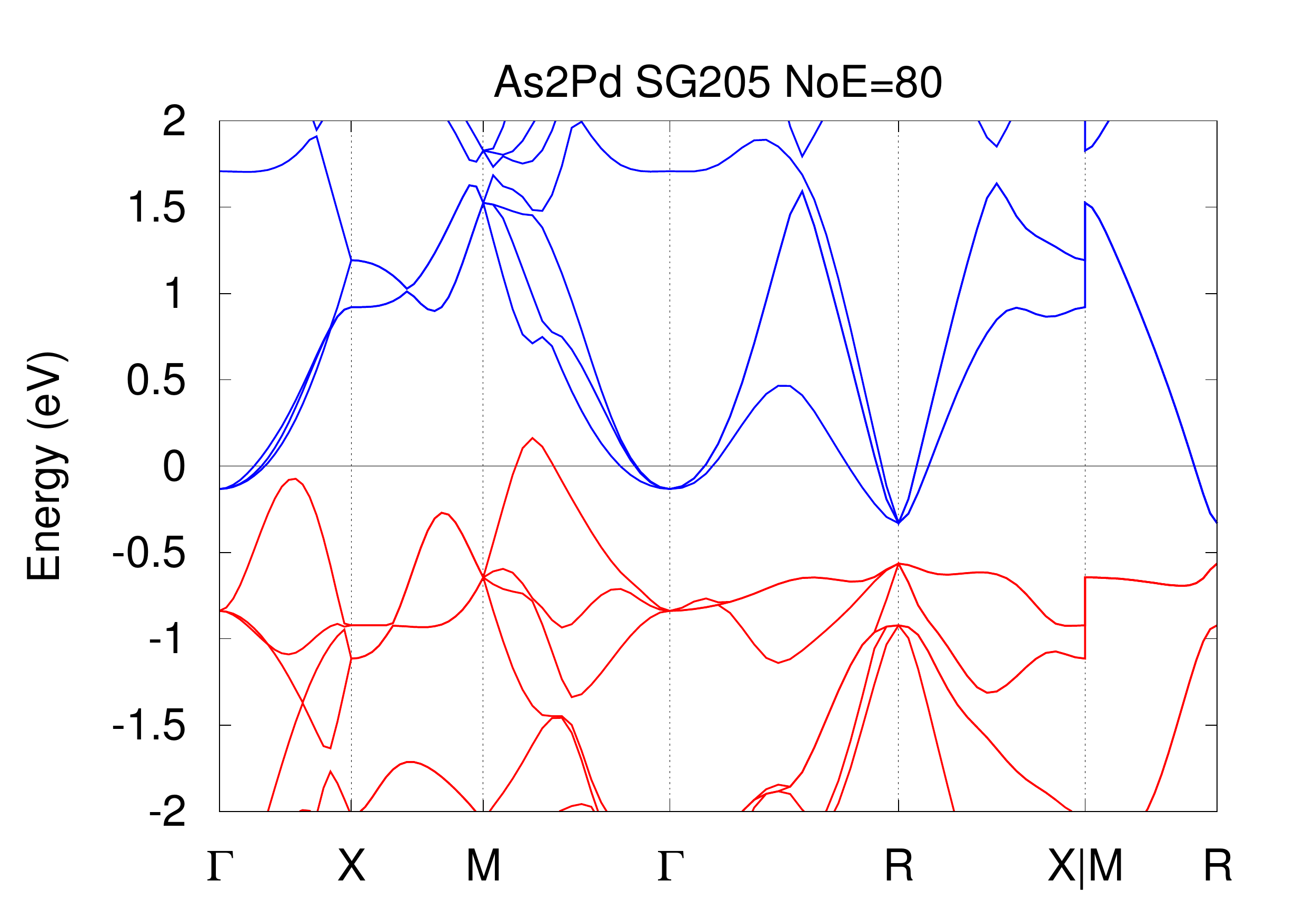}
}
\subfigure[Te$_{2}$Os SG205 NoA=12 NoE=80]{
\label{subfig:300225}
\includegraphics[scale=0.32,angle=270]{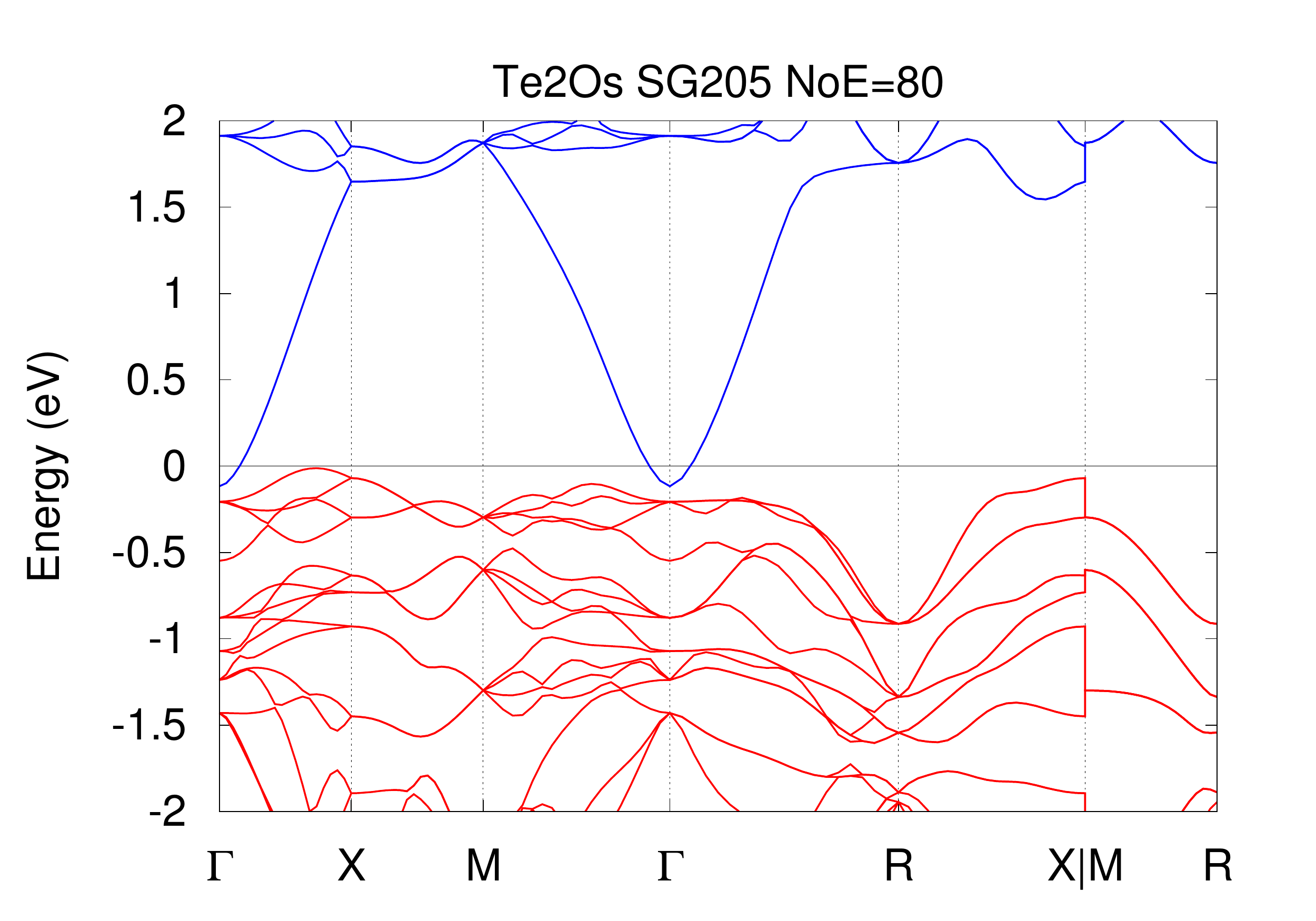}
}
\subfigure[KNdTe$_{4}$ SG125 NoA=12 NoE=88]{
\label{subfig:412792}
\includegraphics[scale=0.32,angle=270]{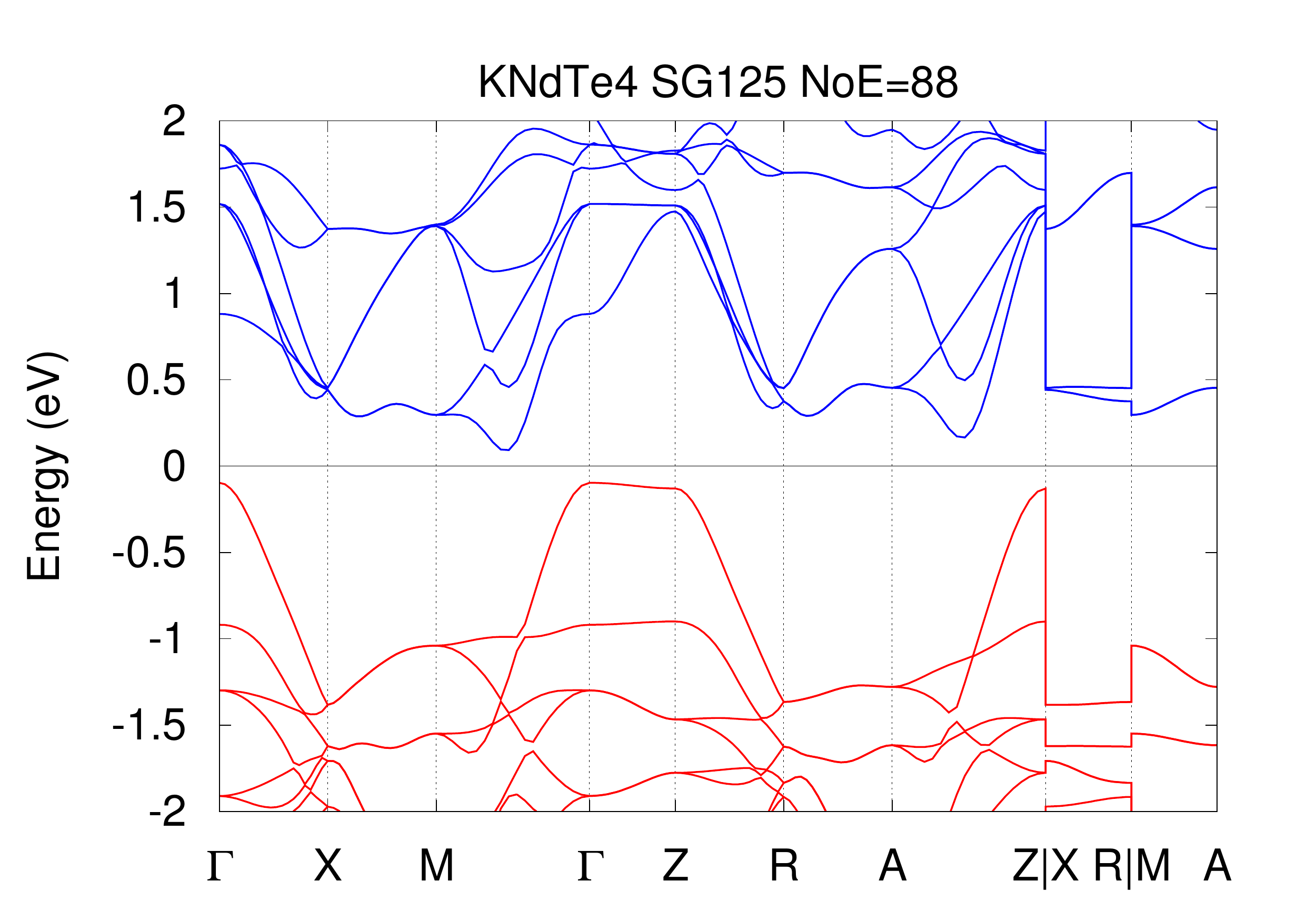}
}
\subfigure[DyCoSn SG62 NoA=12 NoE=88]{
\label{subfig:54415}
\includegraphics[scale=0.32,angle=270]{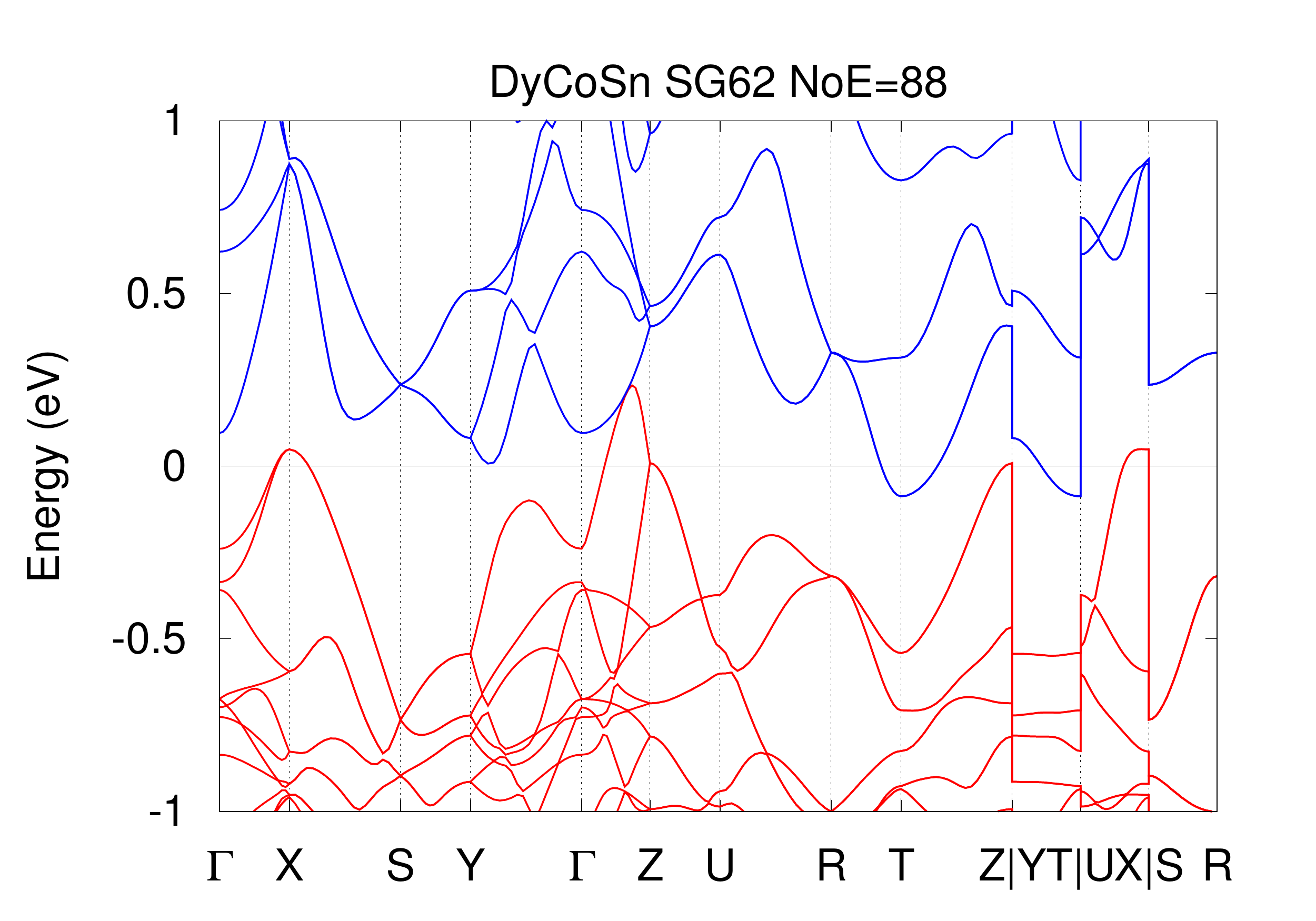}
}
\subfigure[PrAs$_{2}$ SG14 NoA=12 NoE=84]{
\label{subfig:611219}
\includegraphics[scale=0.32,angle=270]{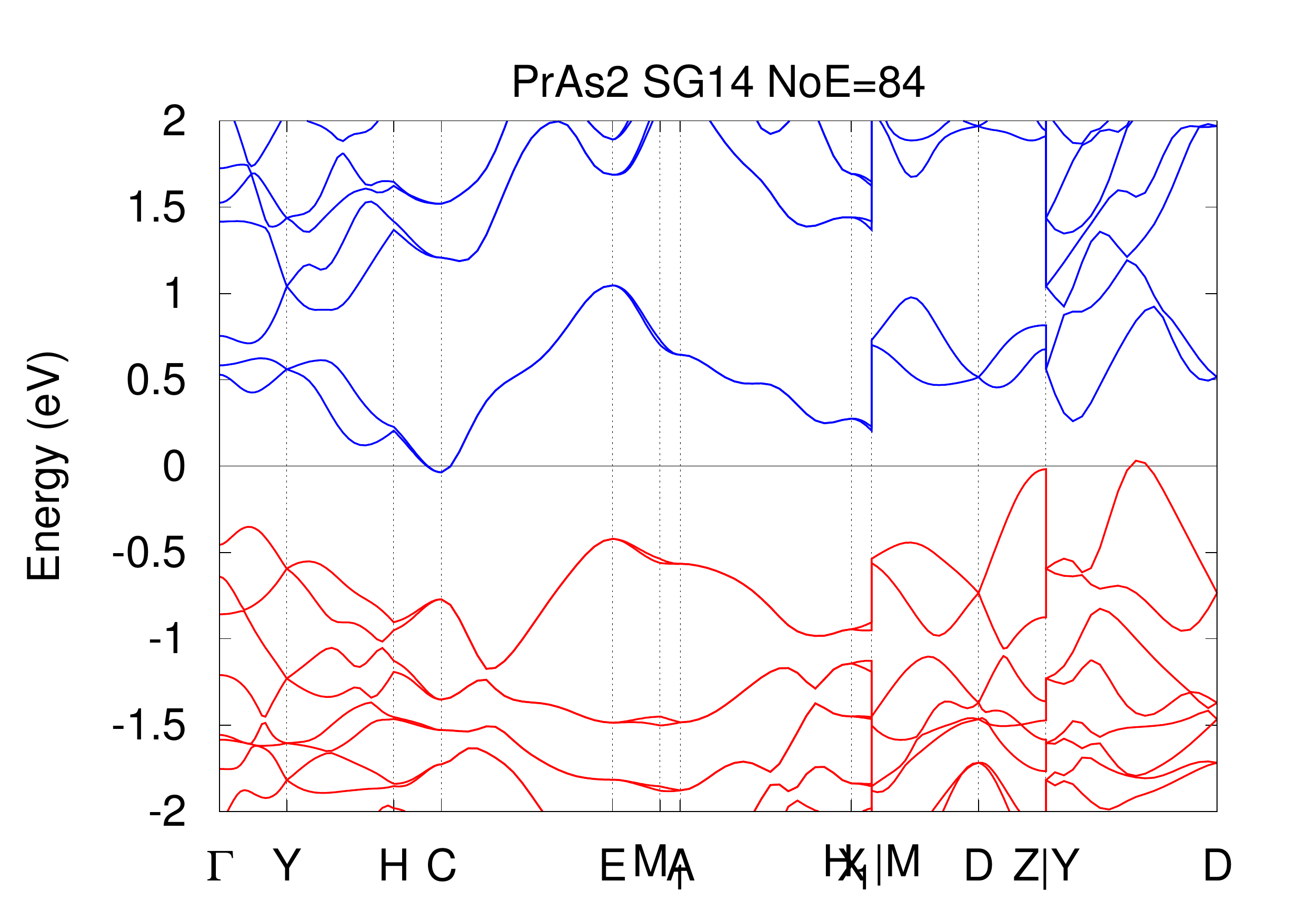}
}
\caption{\hyperref[tab:electride]{back to the table}}
\end{figure}

\begin{figure}[htp]
 \centering
\subfigure[CoAs$_{2}$ SG14 NoA=12 NoE=76]{
\label{subfig:610026}
\includegraphics[scale=0.32,angle=270]{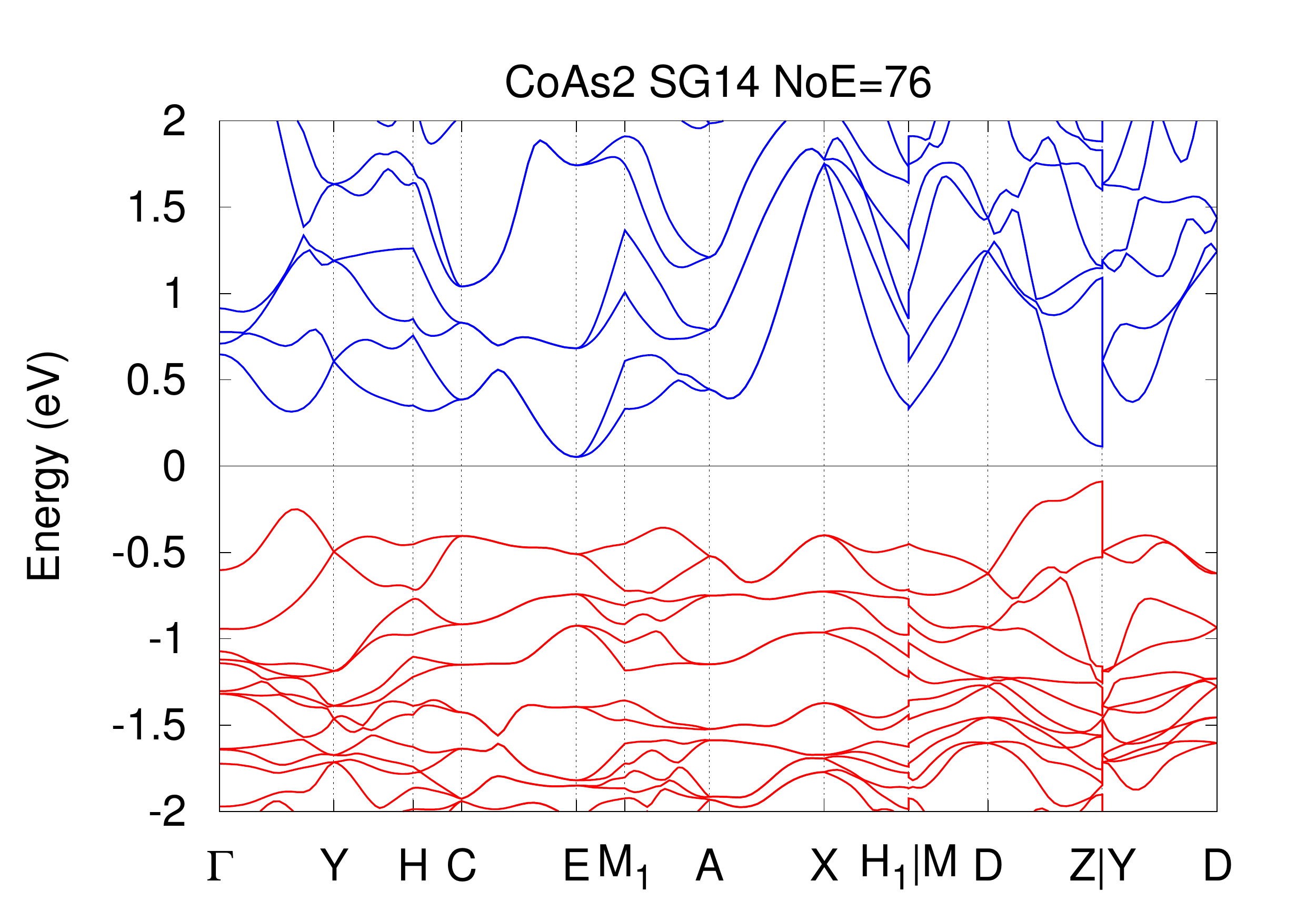}
}
\subfigure[Bi$_{2}$Ir SG14 NoA=12 NoE=76]{
\label{subfig:424397}
\includegraphics[scale=0.32,angle=270]{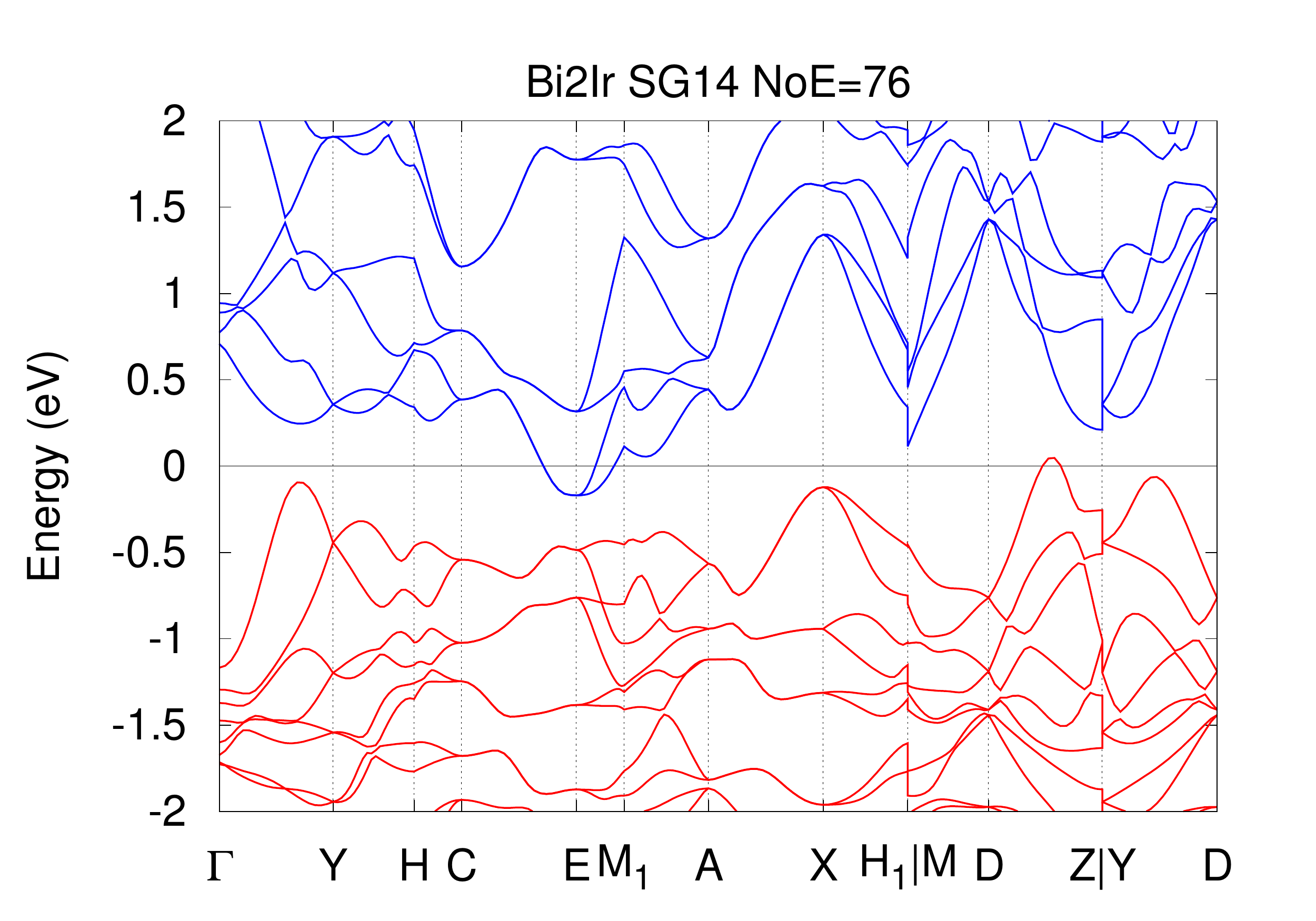}
}
\subfigure[Sb$_{2}$Ir SG14 NoA=12 NoE=76]{
\label{subfig:43502}
\includegraphics[scale=0.32,angle=270]{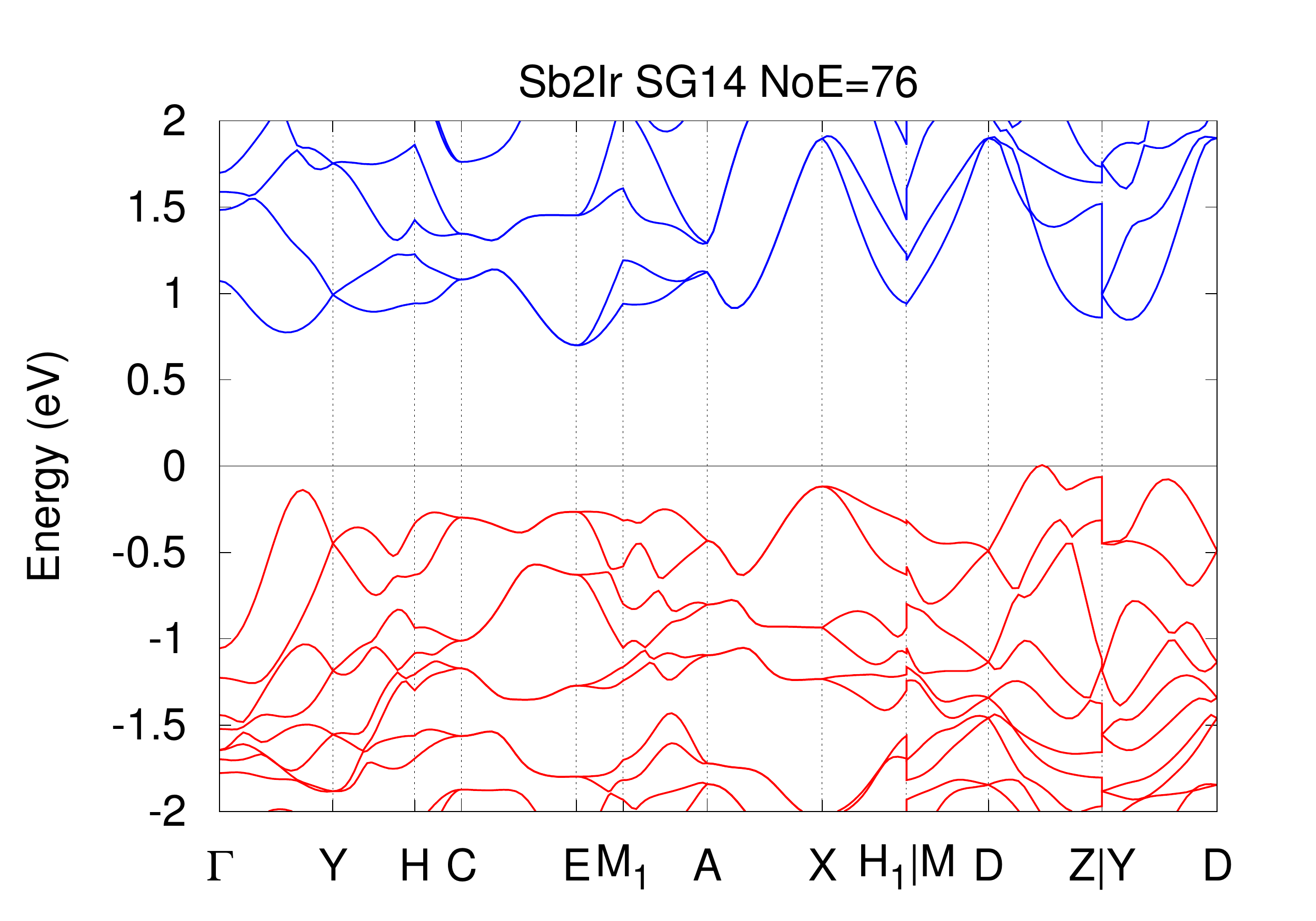}
}
\subfigure[NdGeRh SG62 NoA=12 NoE=96]{
\label{subfig:82549}
\includegraphics[scale=0.32,angle=270]{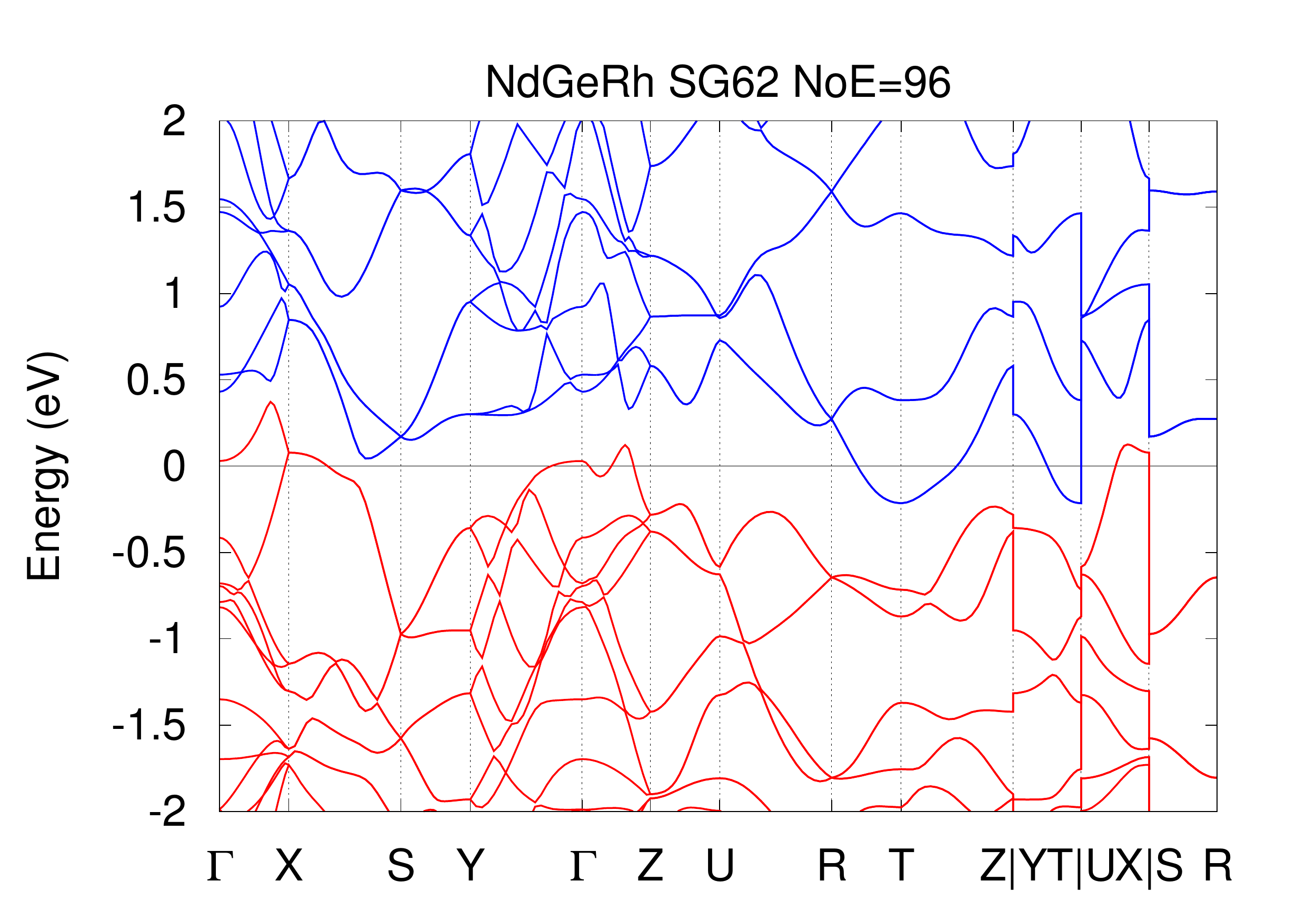}
}
\subfigure[TbGeIr SG62 NoA=12 NoE=88]{
\label{subfig:636739}
\includegraphics[scale=0.32,angle=270]{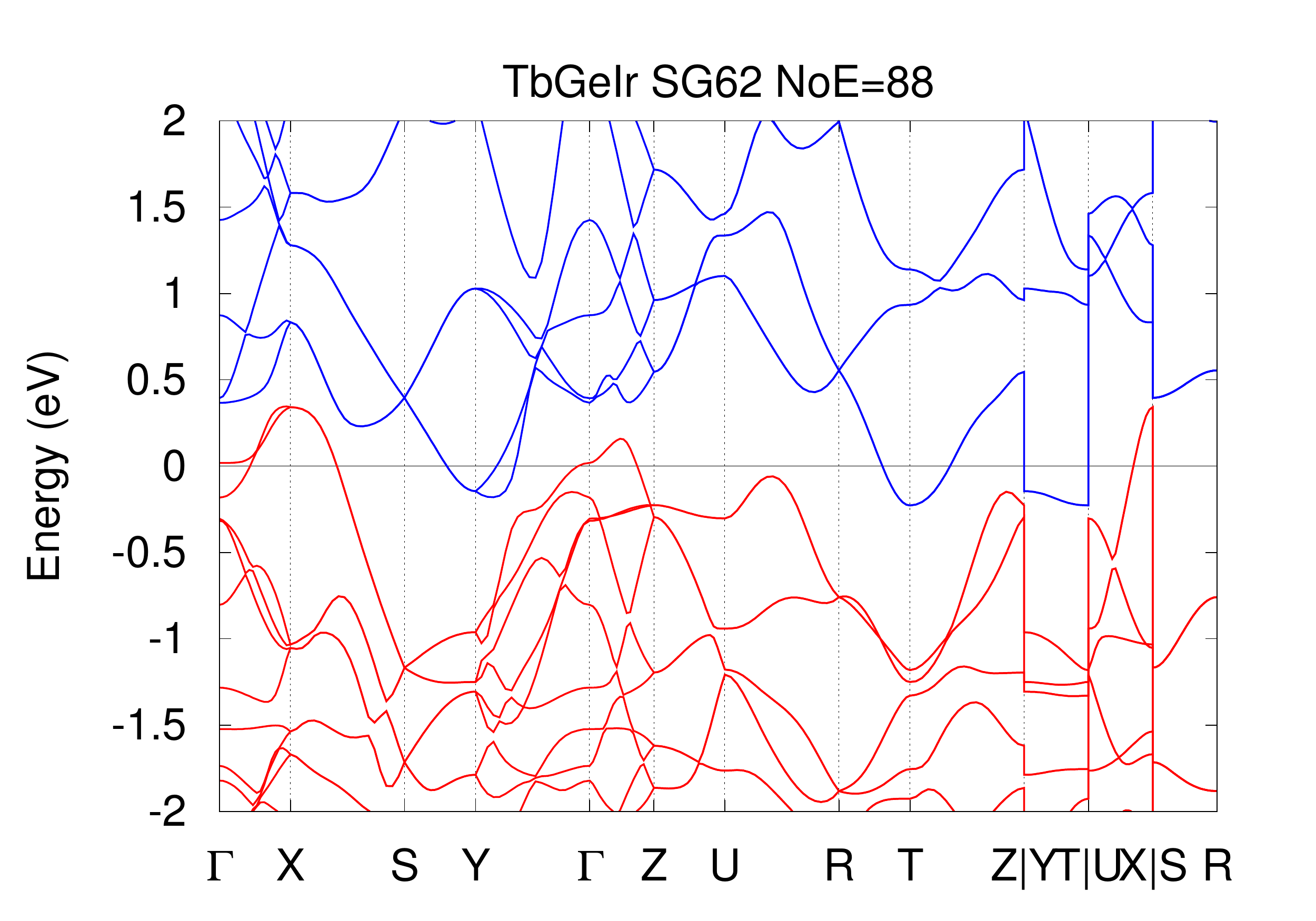}
}
\subfigure[DyCoSi SG62 NoA=12 NoE=88]{
\label{subfig:88272}
\includegraphics[scale=0.32,angle=270]{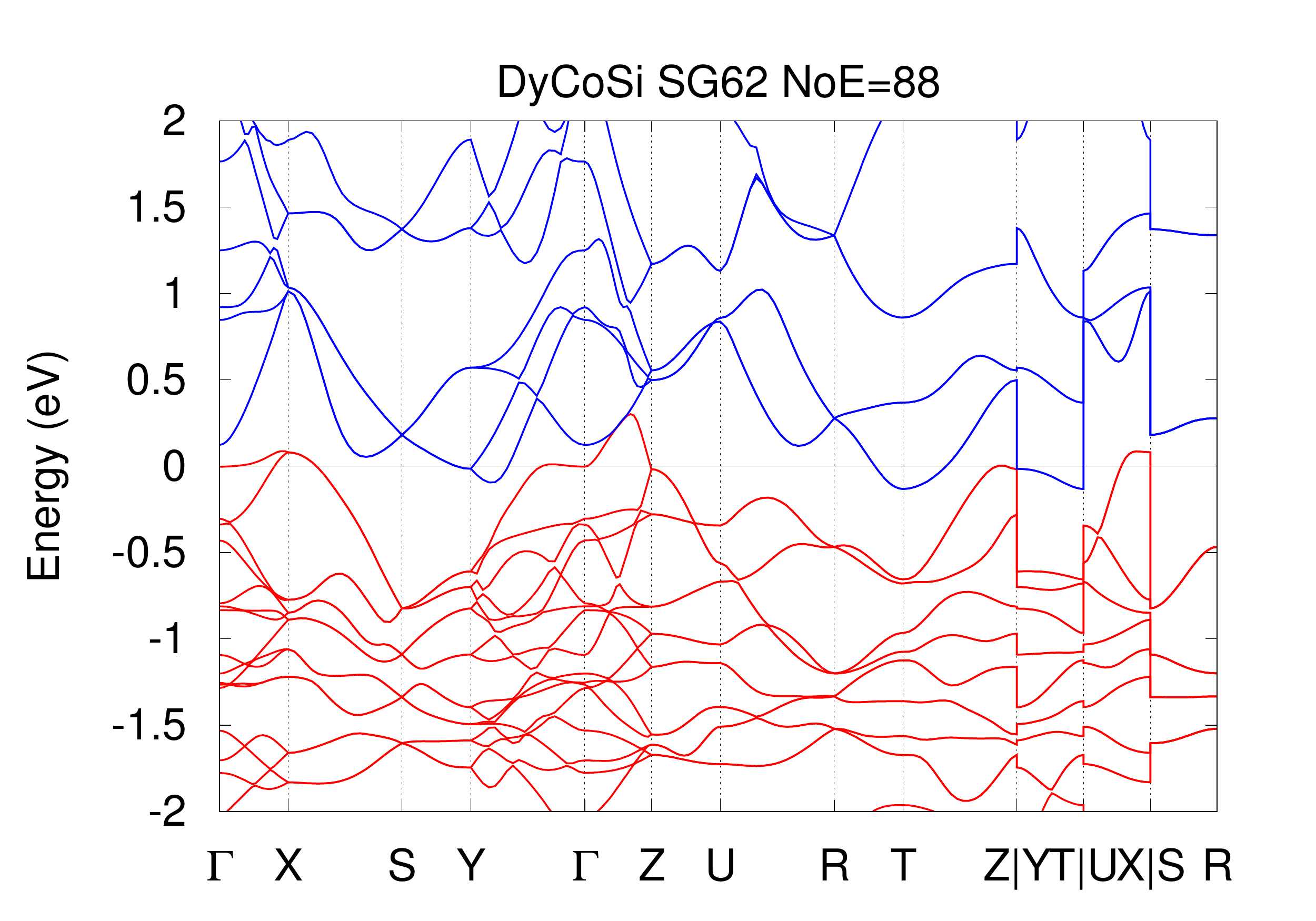}
}
\subfigure[Sb$_{2}$Pt SG205 NoA=12 NoE=80]{
\label{subfig:43105}
\includegraphics[scale=0.32,angle=270]{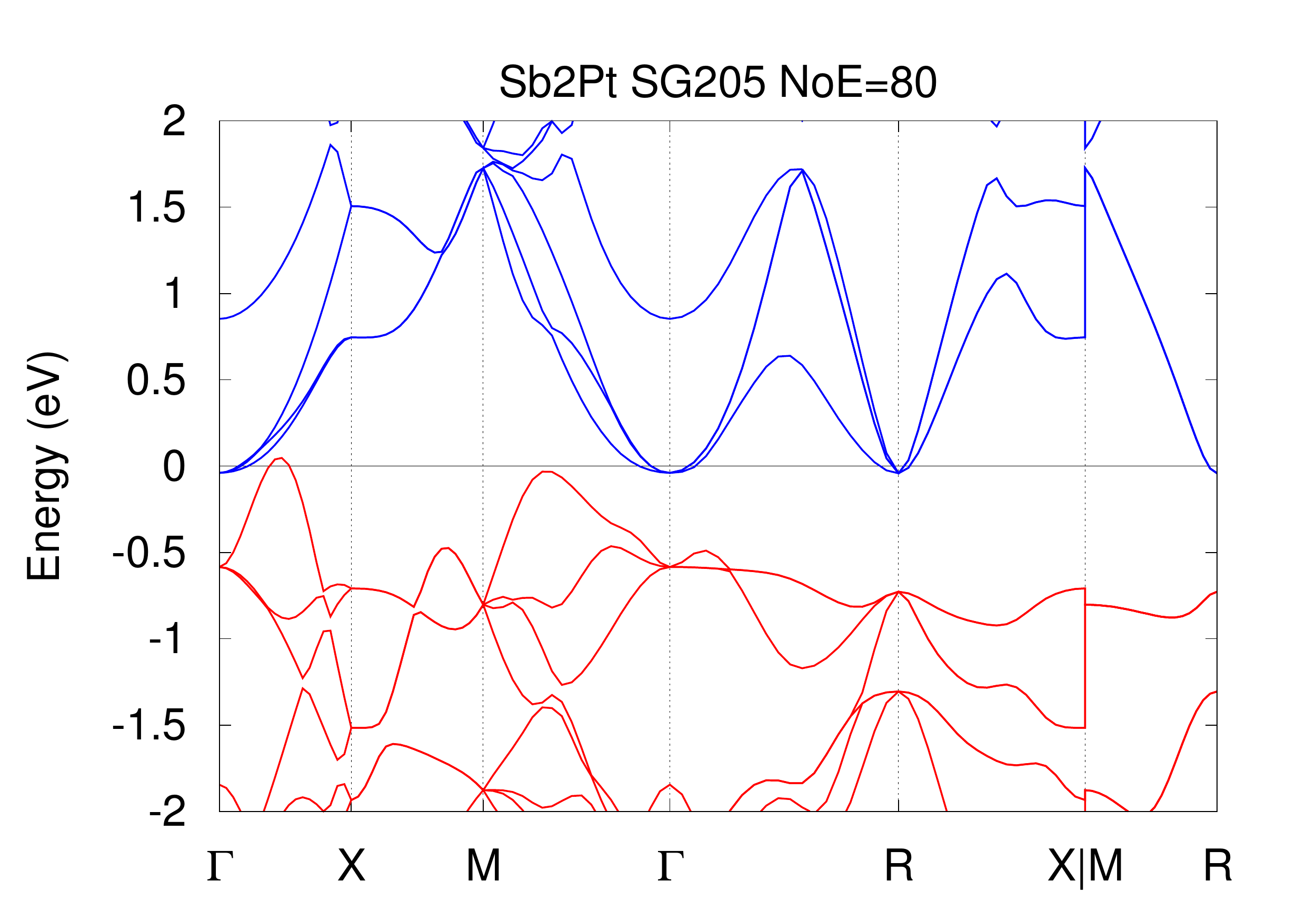}
}
\subfigure[Sb$_{2}$Rh SG14 NoA=12 NoE=76]{
\label{subfig:43501}
\includegraphics[scale=0.32,angle=270]{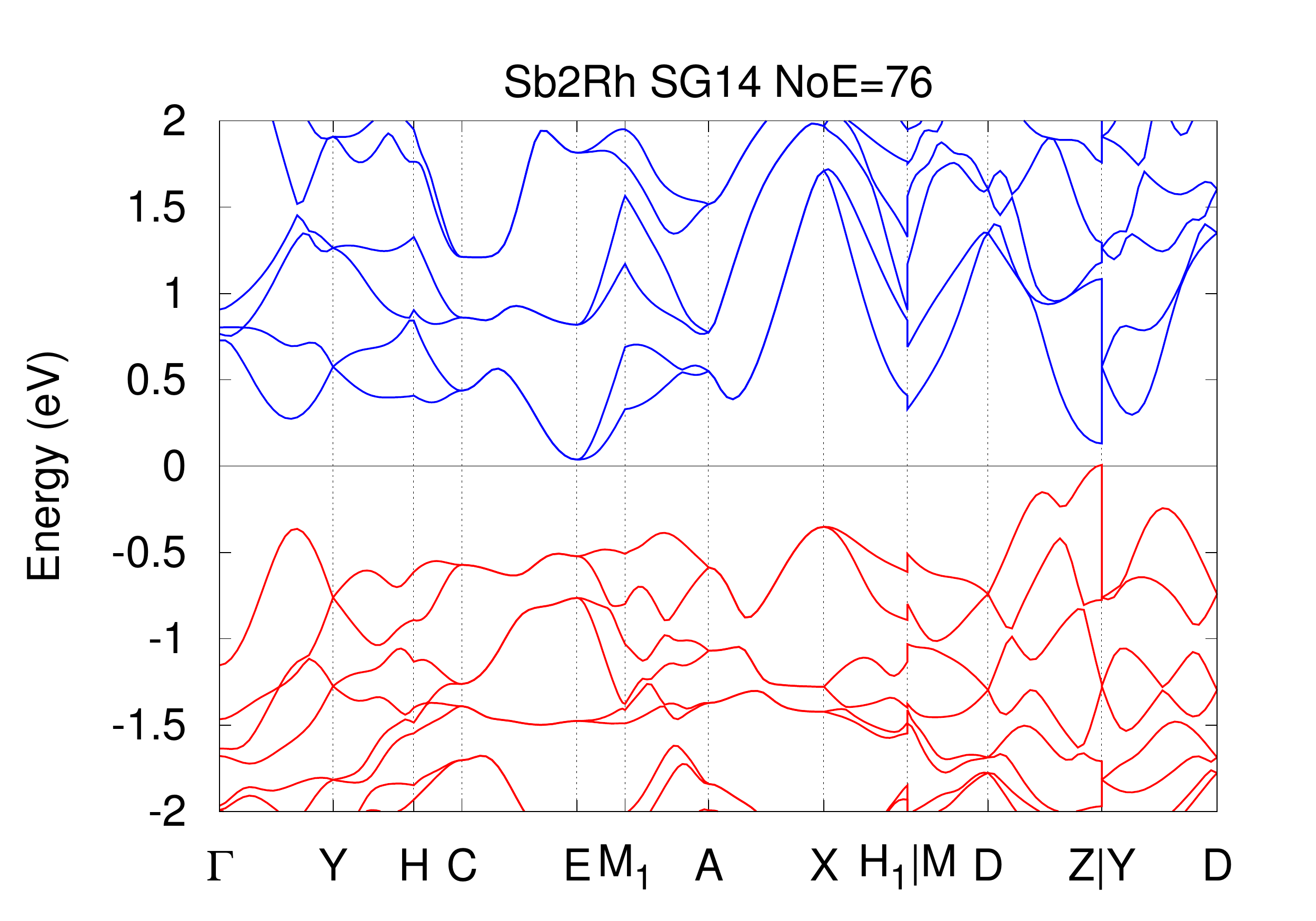}
}
\caption{\hyperref[tab:electride]{back to the table}}
\end{figure}

\begin{figure}[htp]
 \centering
\subfigure[Bi$_{3}$Te$_{2}$S SG164 NoA=12 NoE=66]{
\label{subfig:107587}
\includegraphics[scale=0.32,angle=270]{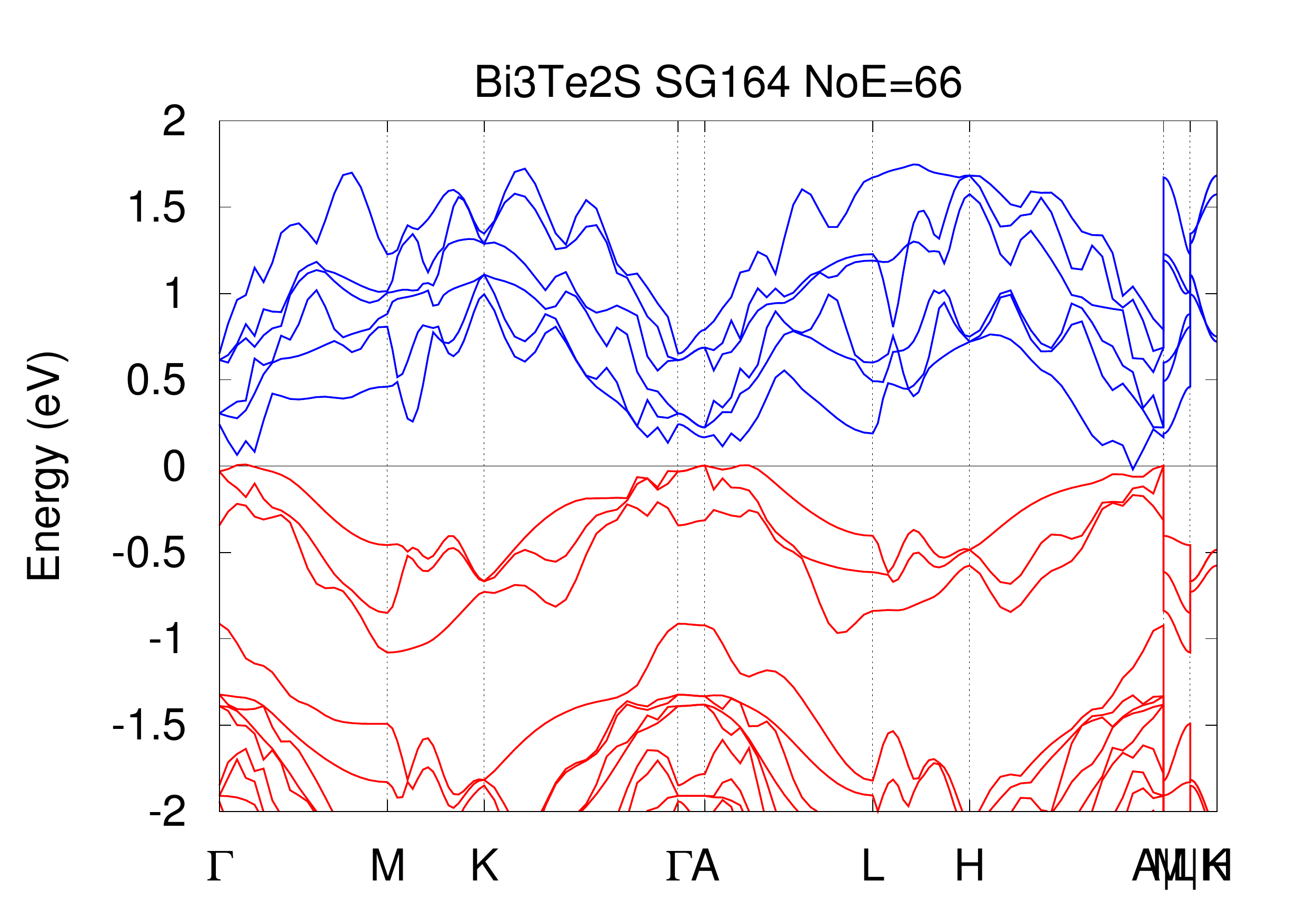}
}
\subfigure[Te$_{2}$Ru SG205 NoA=12 NoE=80]{
\label{subfig:65169}
\includegraphics[scale=0.32,angle=270]{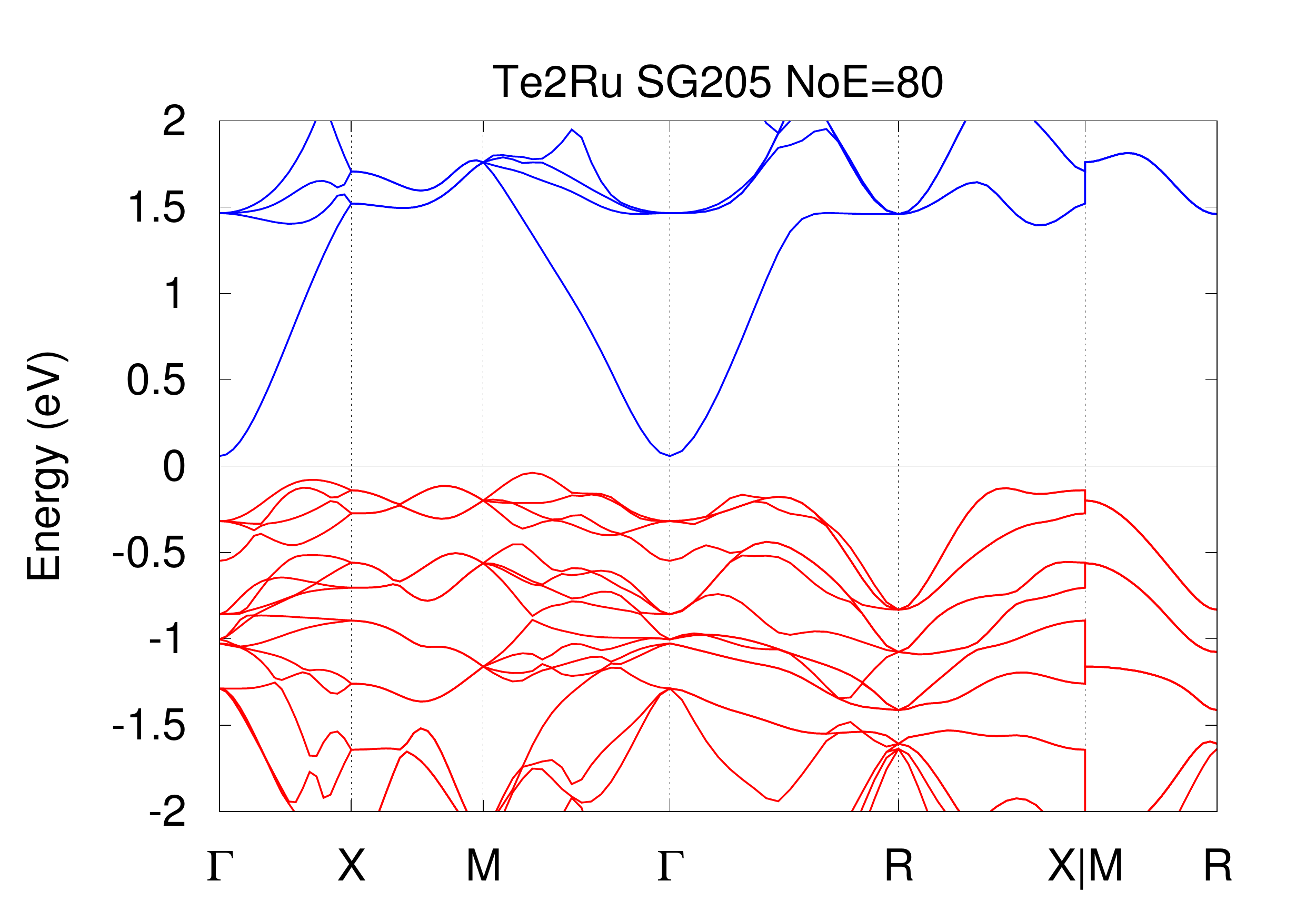}
}
\subfigure[CaGaGe SG194 NoA=12 NoE=68]{
\label{subfig:66002}
\includegraphics[scale=0.32,angle=270]{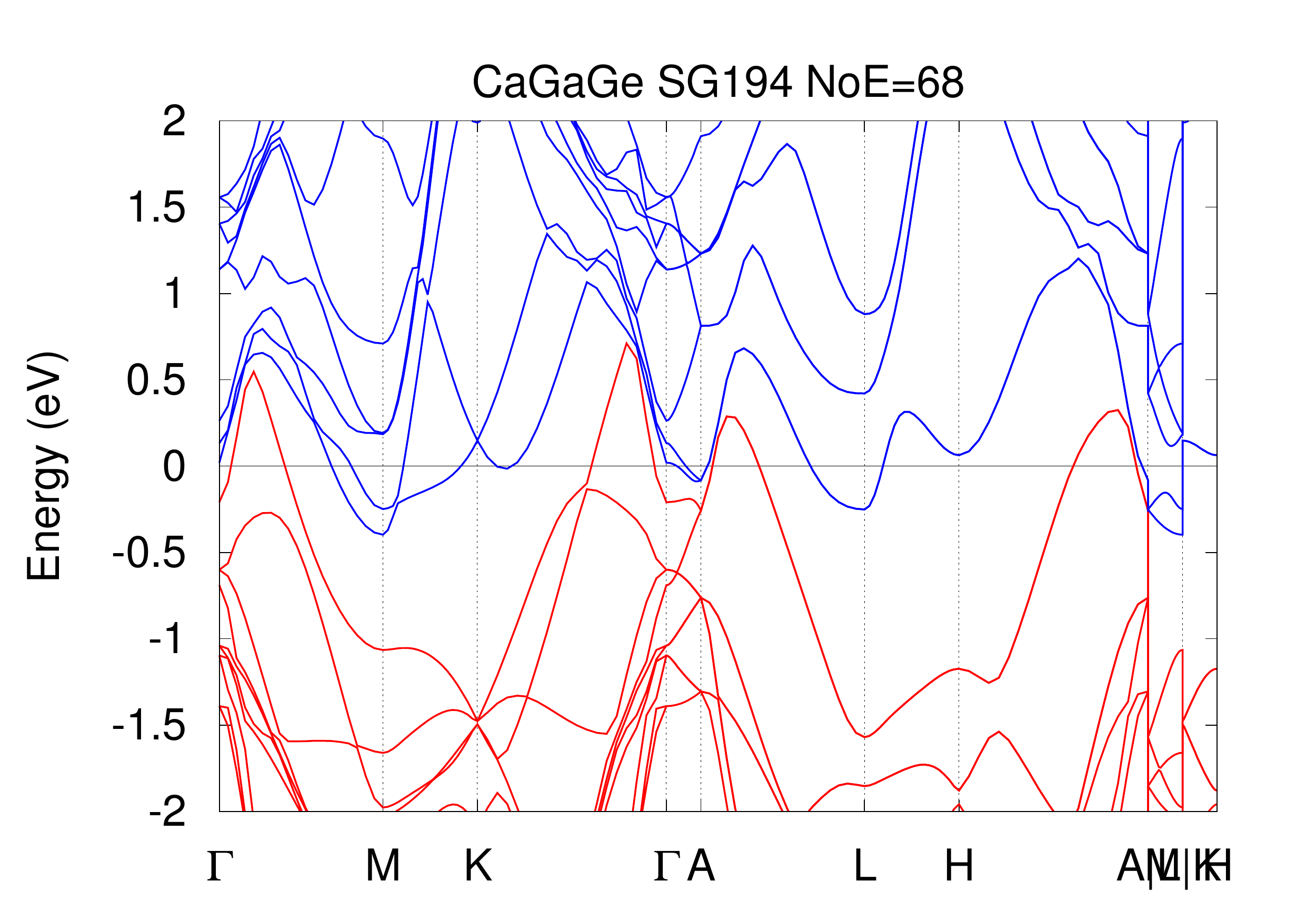}
}
\subfigure[ZrFeSi SG62 NoA=12 NoE=96]{
\label{subfig:633674}
\includegraphics[scale=0.32,angle=270]{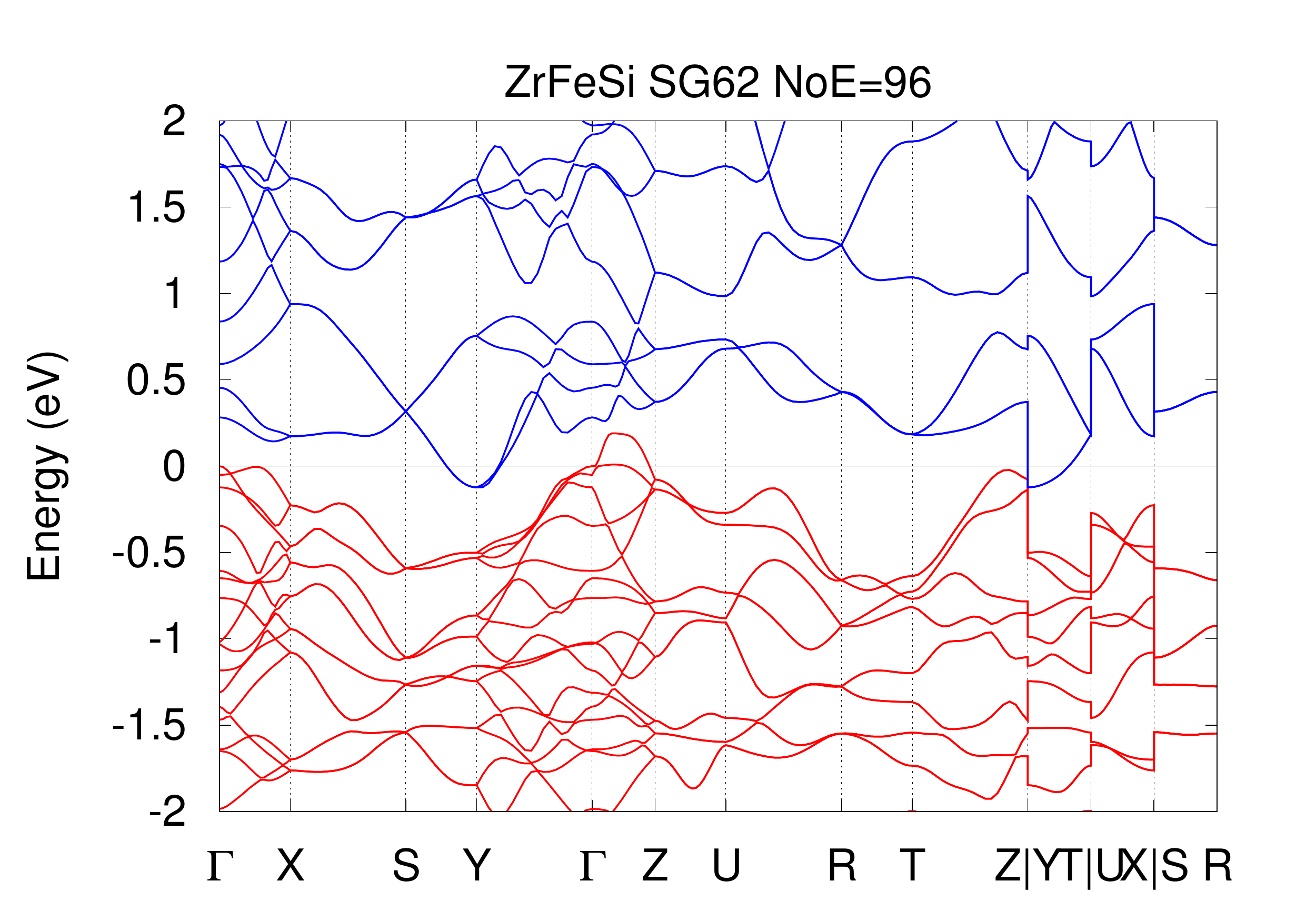}
}
\subfigure[TbCoSi SG62 NoA=12 NoE=88]{
\label{subfig:88213}
\includegraphics[scale=0.32,angle=270]{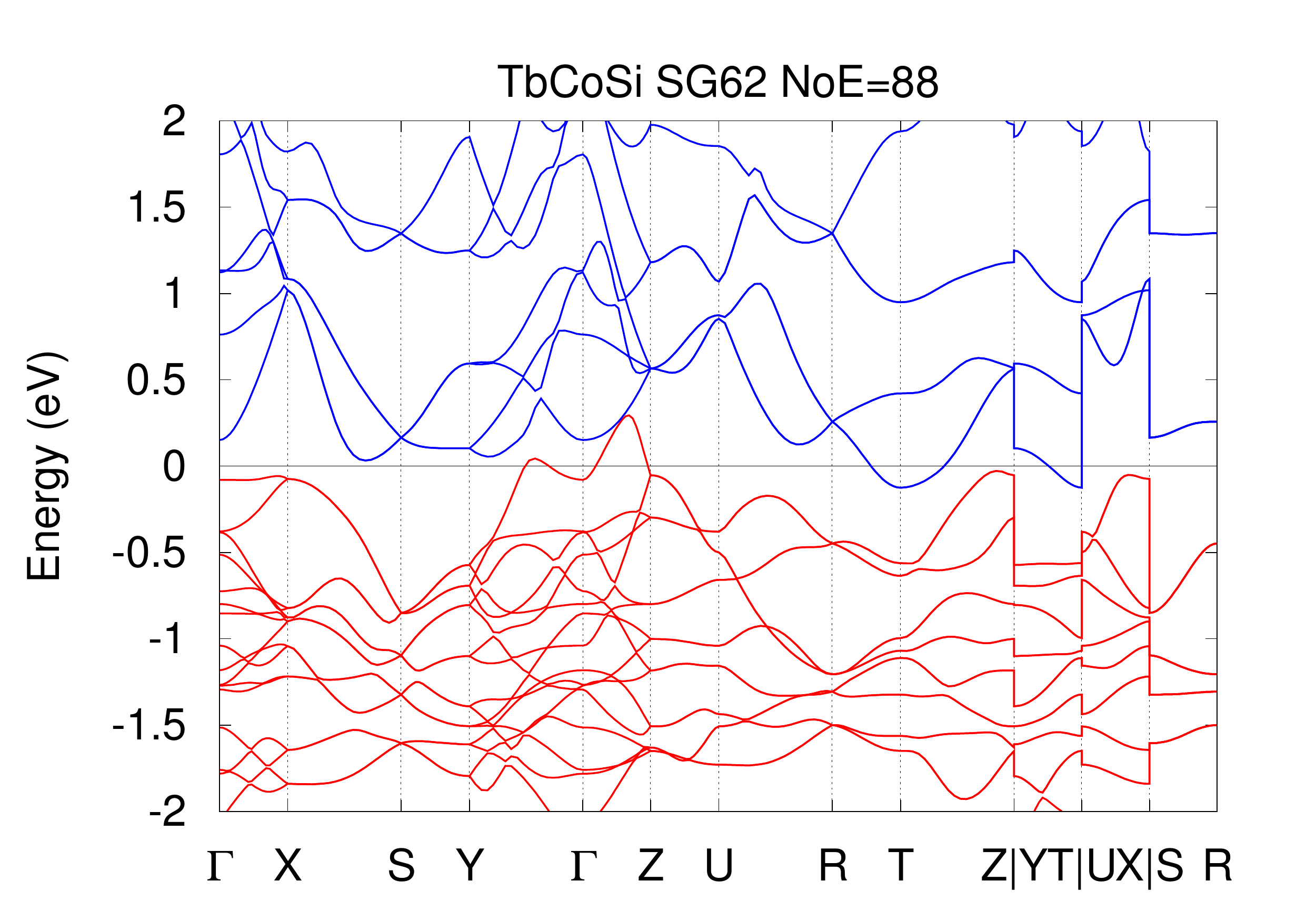}
}
\subfigure[KPrTe$_{4}$ SG125 NoA=12 NoE=88]{
\label{subfig:391204}
\includegraphics[scale=0.32,angle=270]{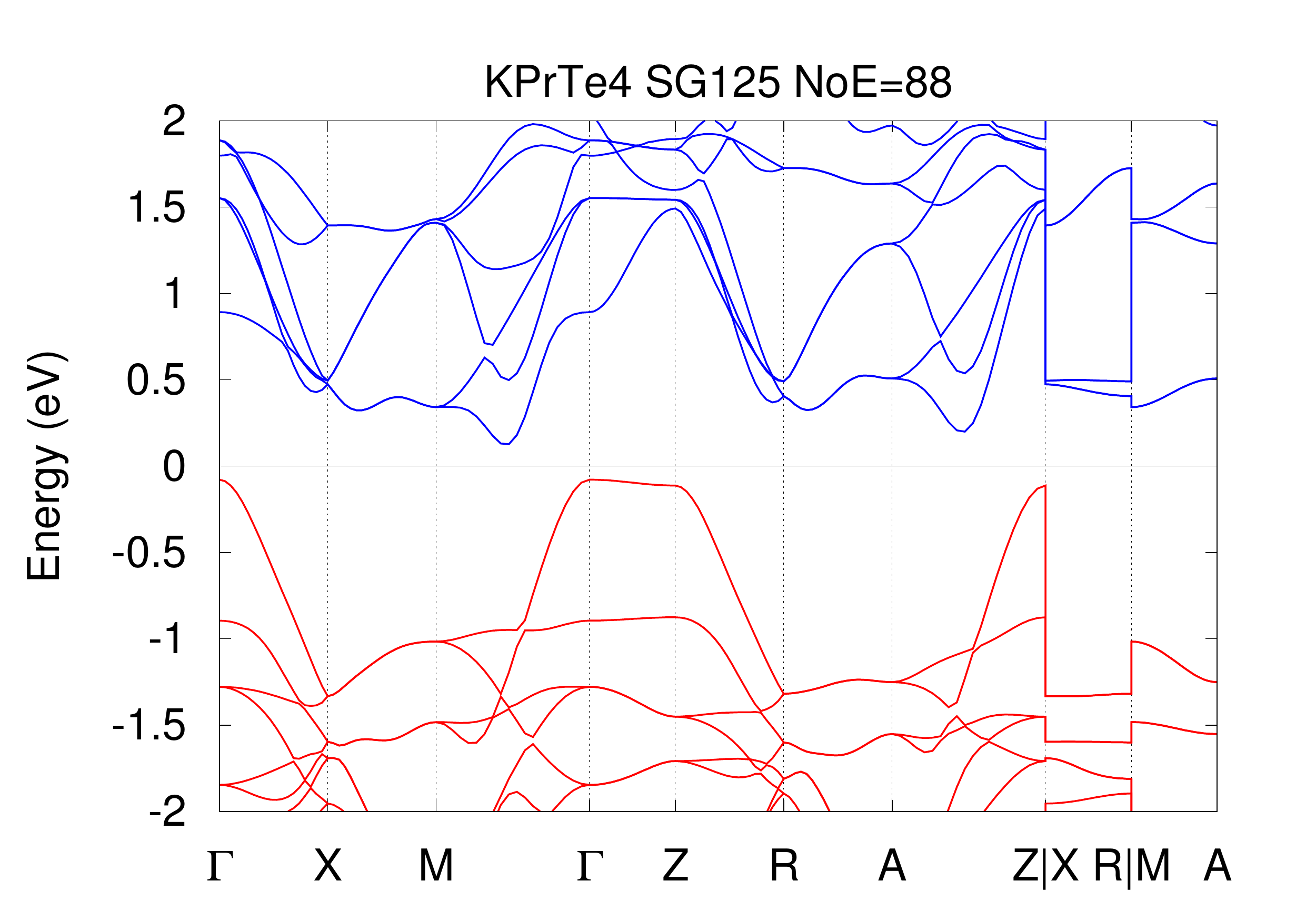}
}
\subfigure[SiP$_{2}$ SG205 NoA=12 NoE=56]{
\label{subfig:30692}
\includegraphics[scale=0.32,angle=270]{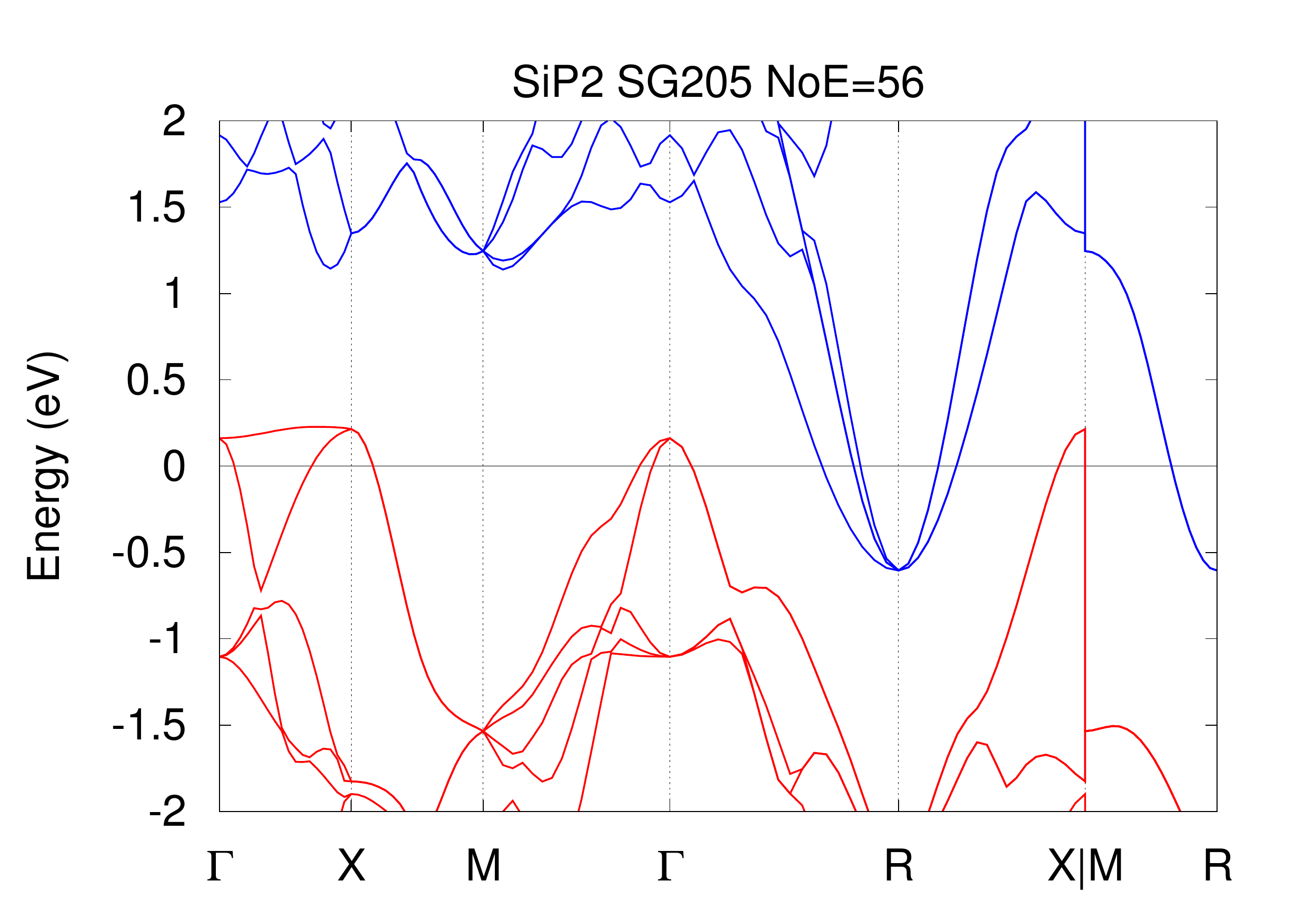}
}
\subfigure[RbNdTe$_{4}$ SG125 NoA=12 NoE=88]{
\label{subfig:412793}
\includegraphics[scale=0.32,angle=270]{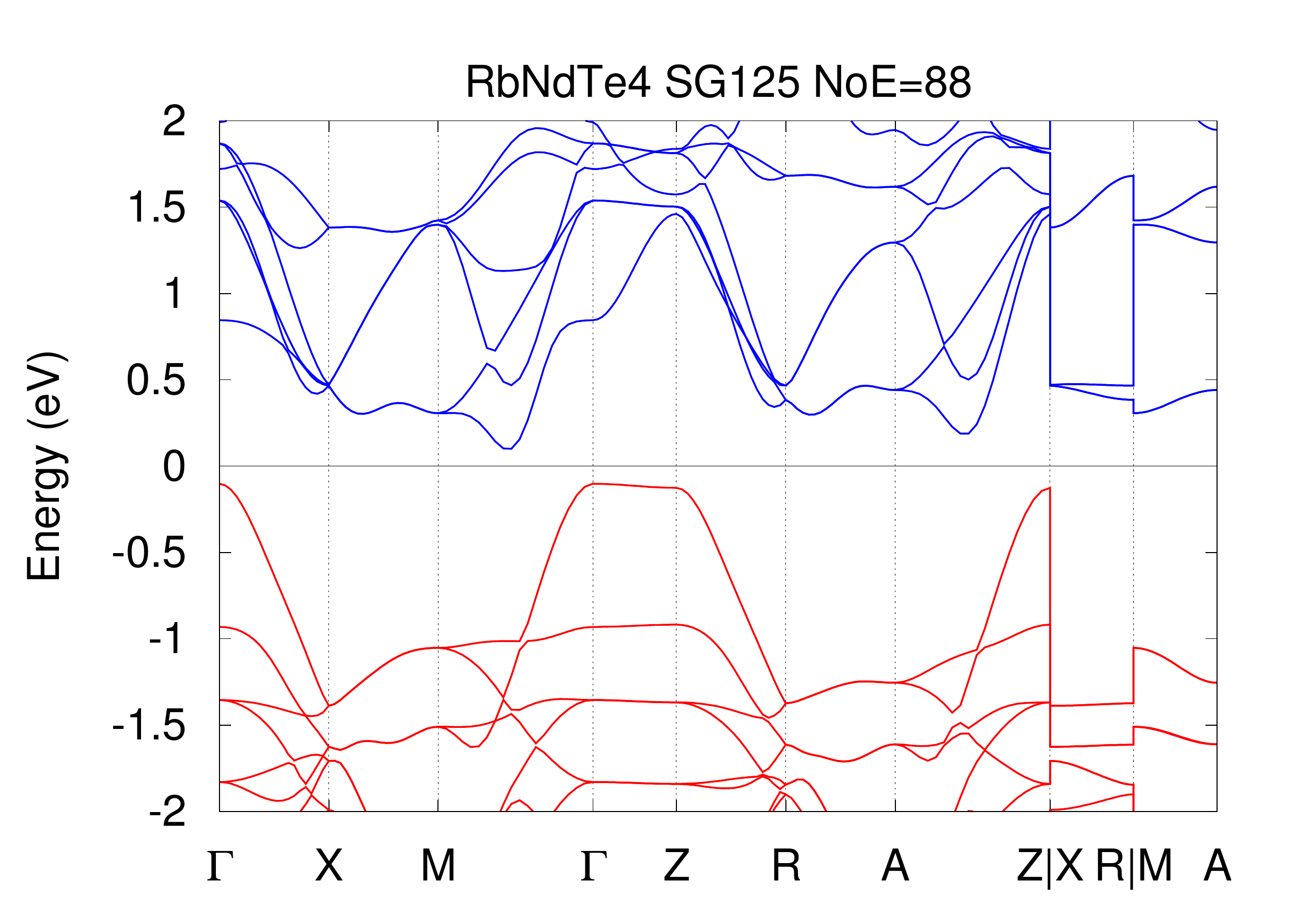}
}
\caption{\hyperref[tab:electride]{back to the table}}
\end{figure}

\begin{figure}[htp]
 \centering
\subfigure[As$_{2}$Rh SG14 NoA=12 NoE=76]{
\label{subfig:611271}
\includegraphics[scale=0.32,angle=270]{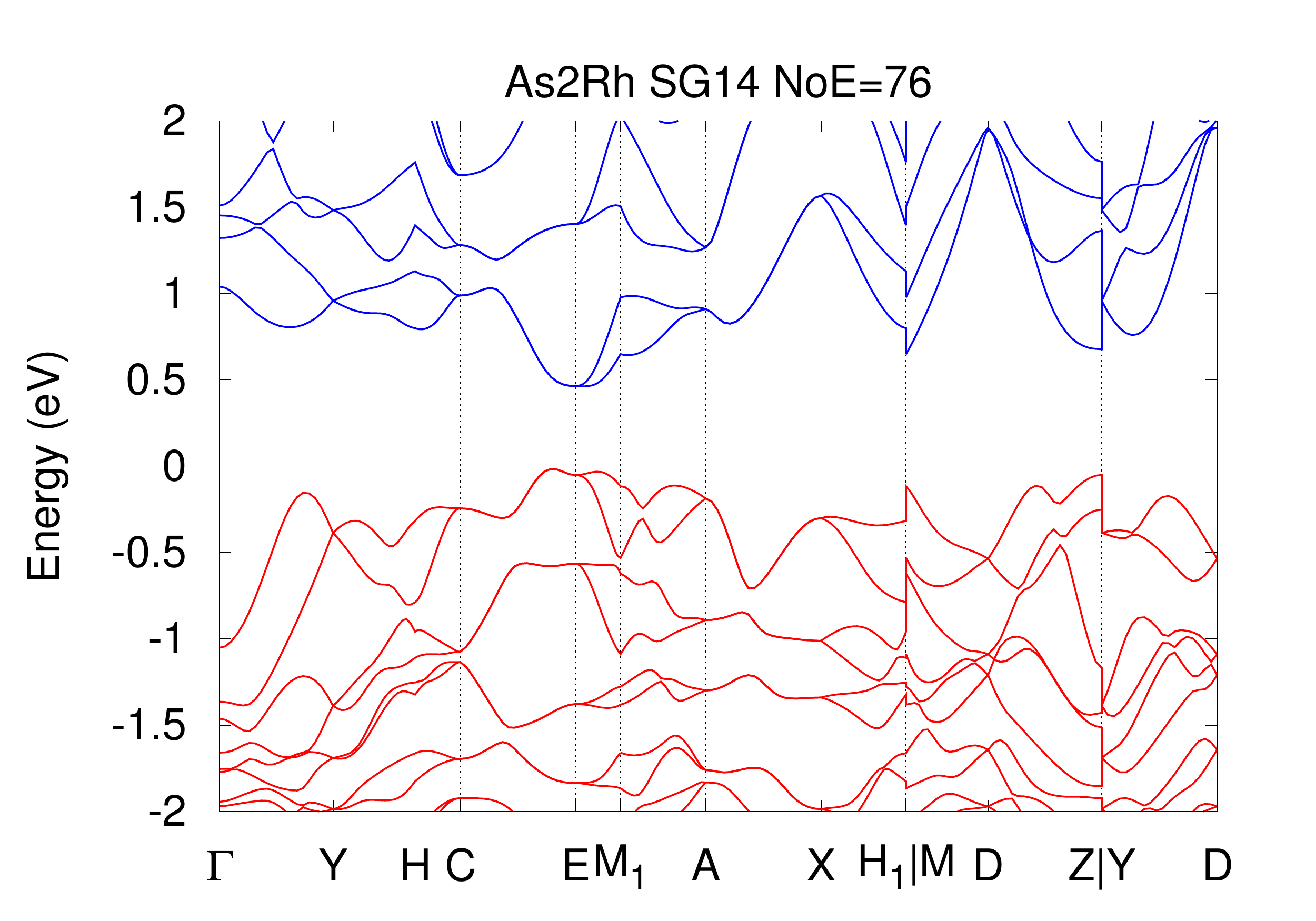}
}
\subfigure[ErGeIr SG62 NoA=12 NoE=88]{
\label{subfig:88167}
\includegraphics[scale=0.32,angle=270]{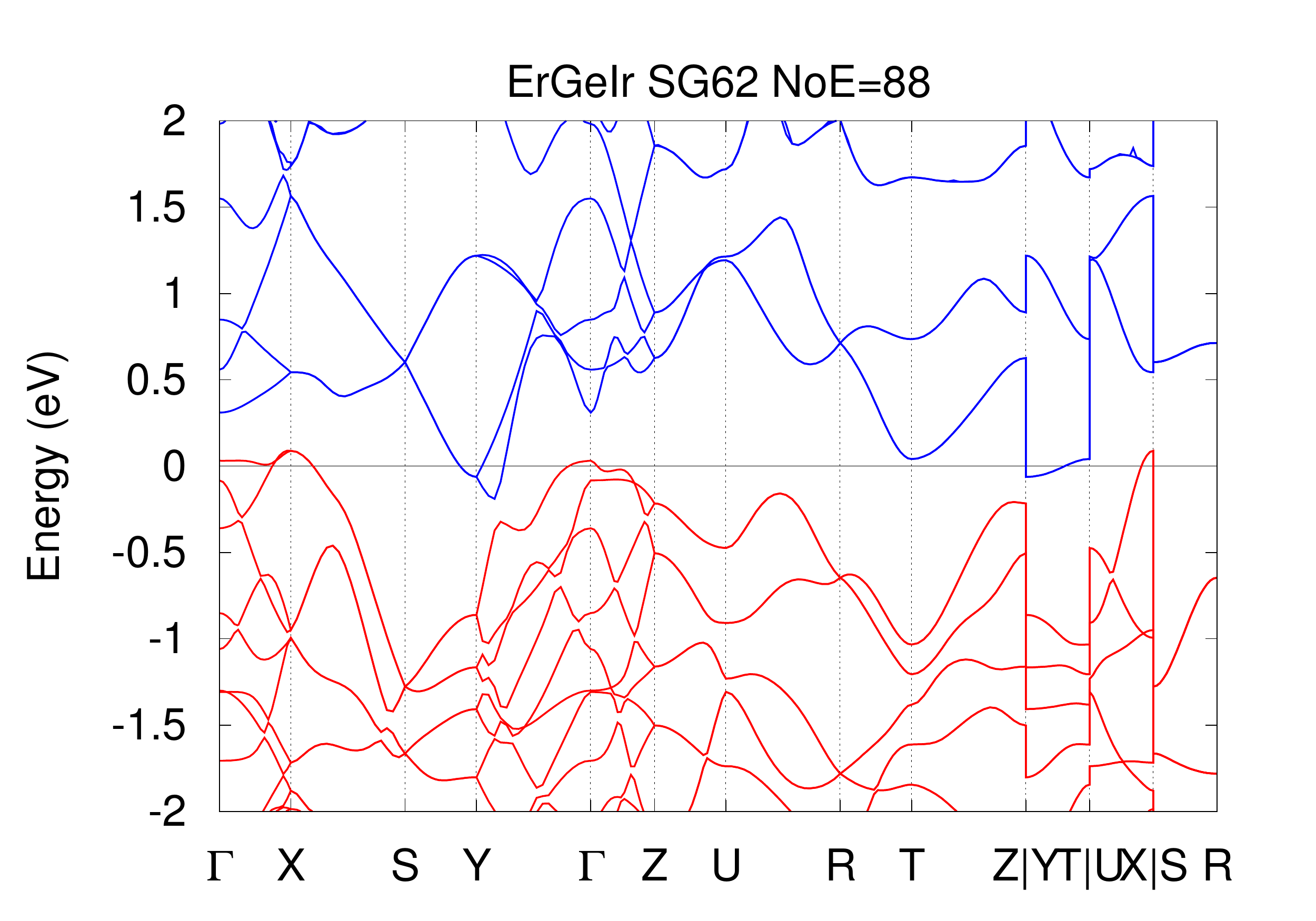}
}
\subfigure[LaAs$_{2}$ SG14 NoA=12 NoE=84]{
\label{subfig:610769}
\includegraphics[scale=0.32,angle=270]{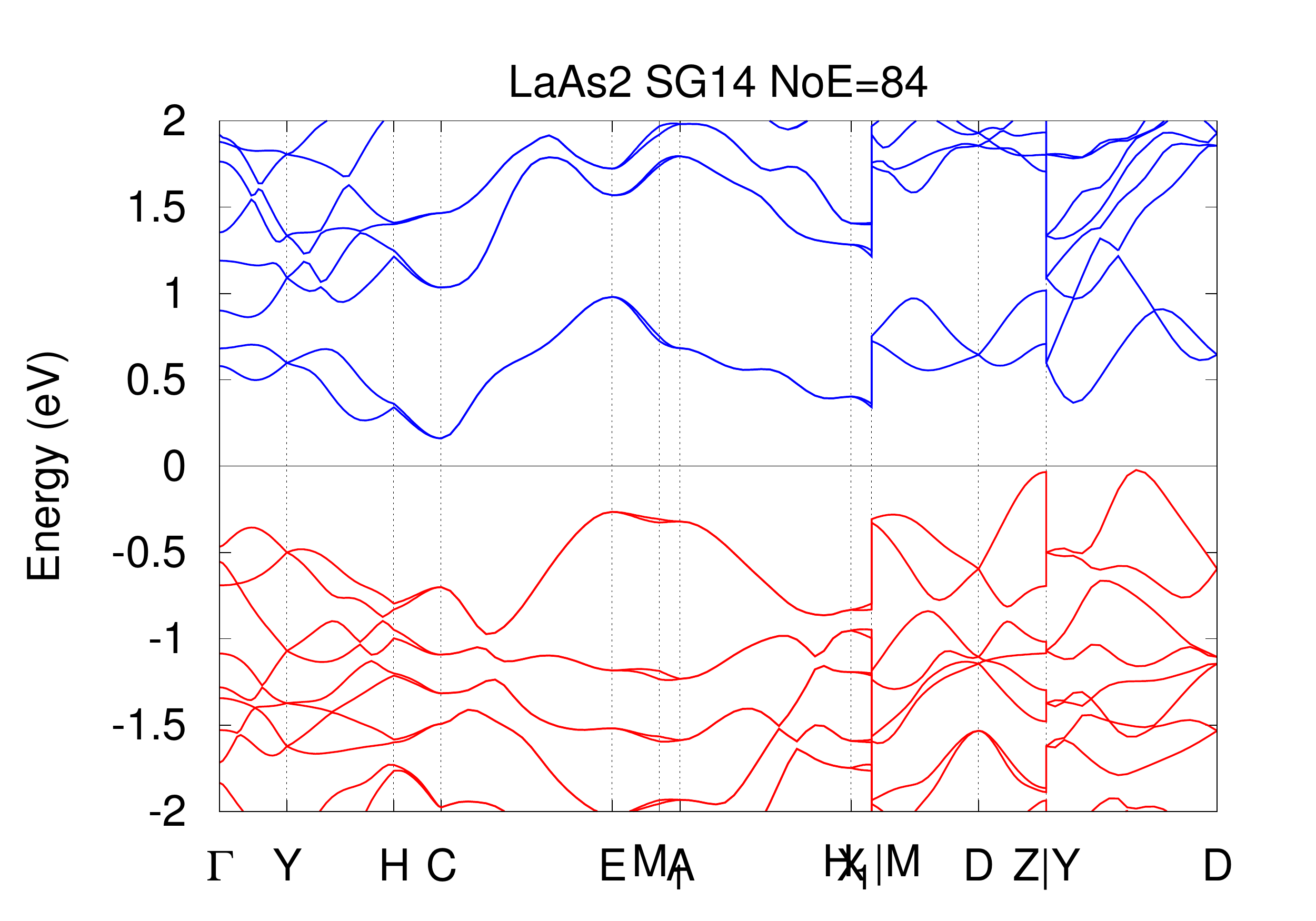}
}
\subfigure[LaTeAs SG62 NoA=12 NoE=88]{
\label{subfig:280231}
\includegraphics[scale=0.32,angle=270]{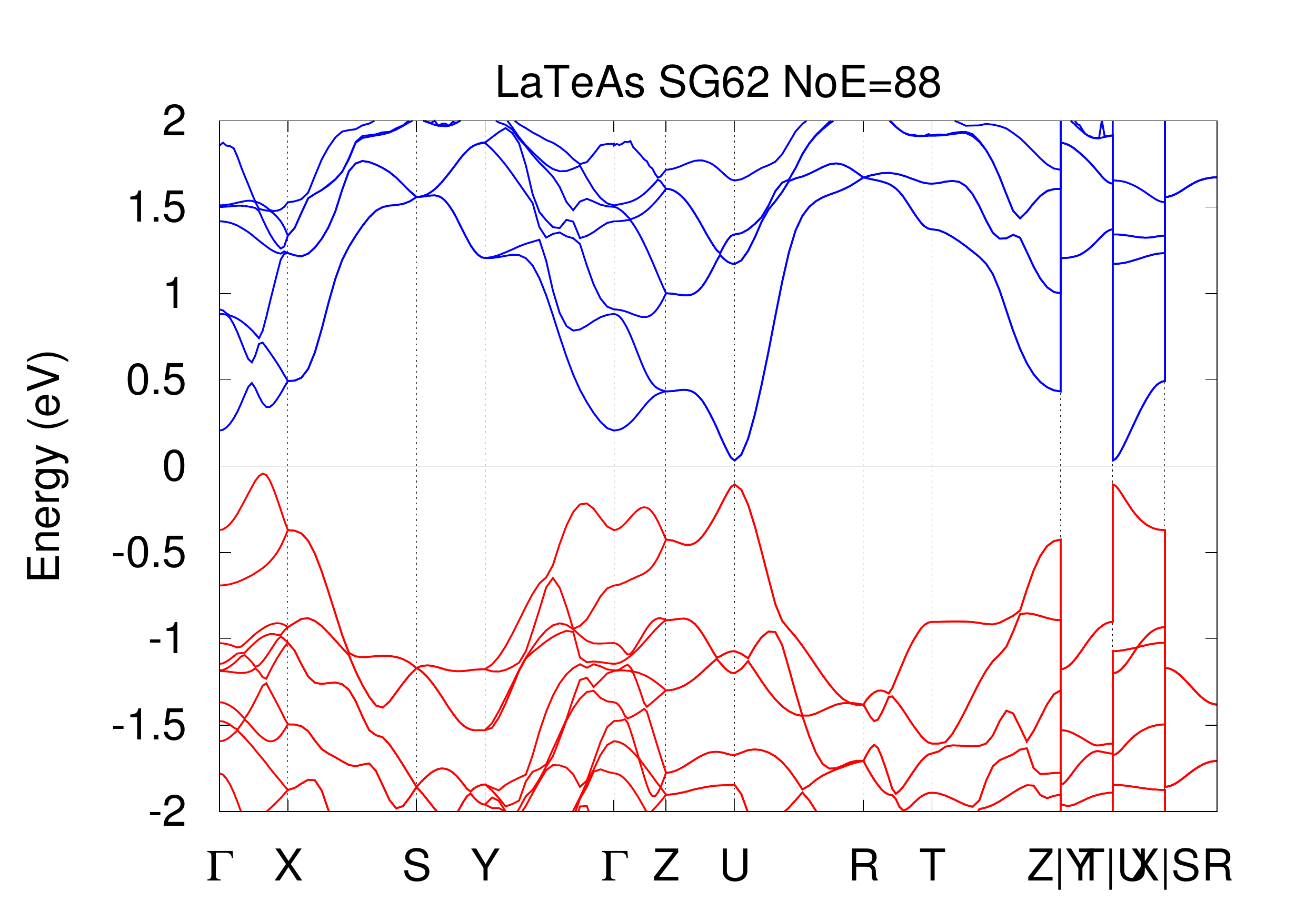}
}
\subfigure[HoGeIr SG62 NoA=12 NoE=88]{
\label{subfig:88166}
\includegraphics[scale=0.32,angle=270]{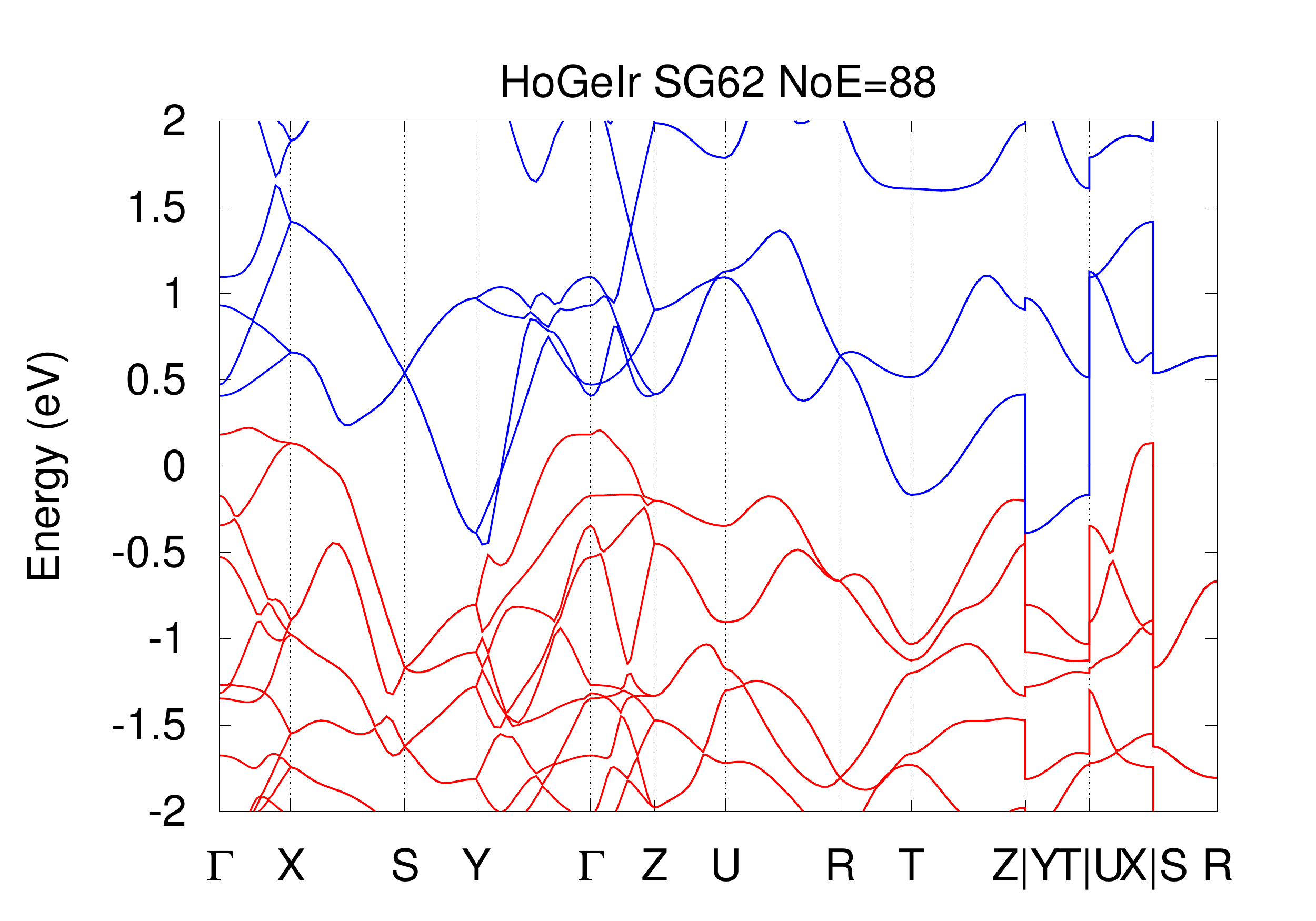}
}
\subfigure[Si SG63 NoA=12 NoE=48]{
\label{subfig:291479}
\includegraphics[scale=0.32,angle=270]{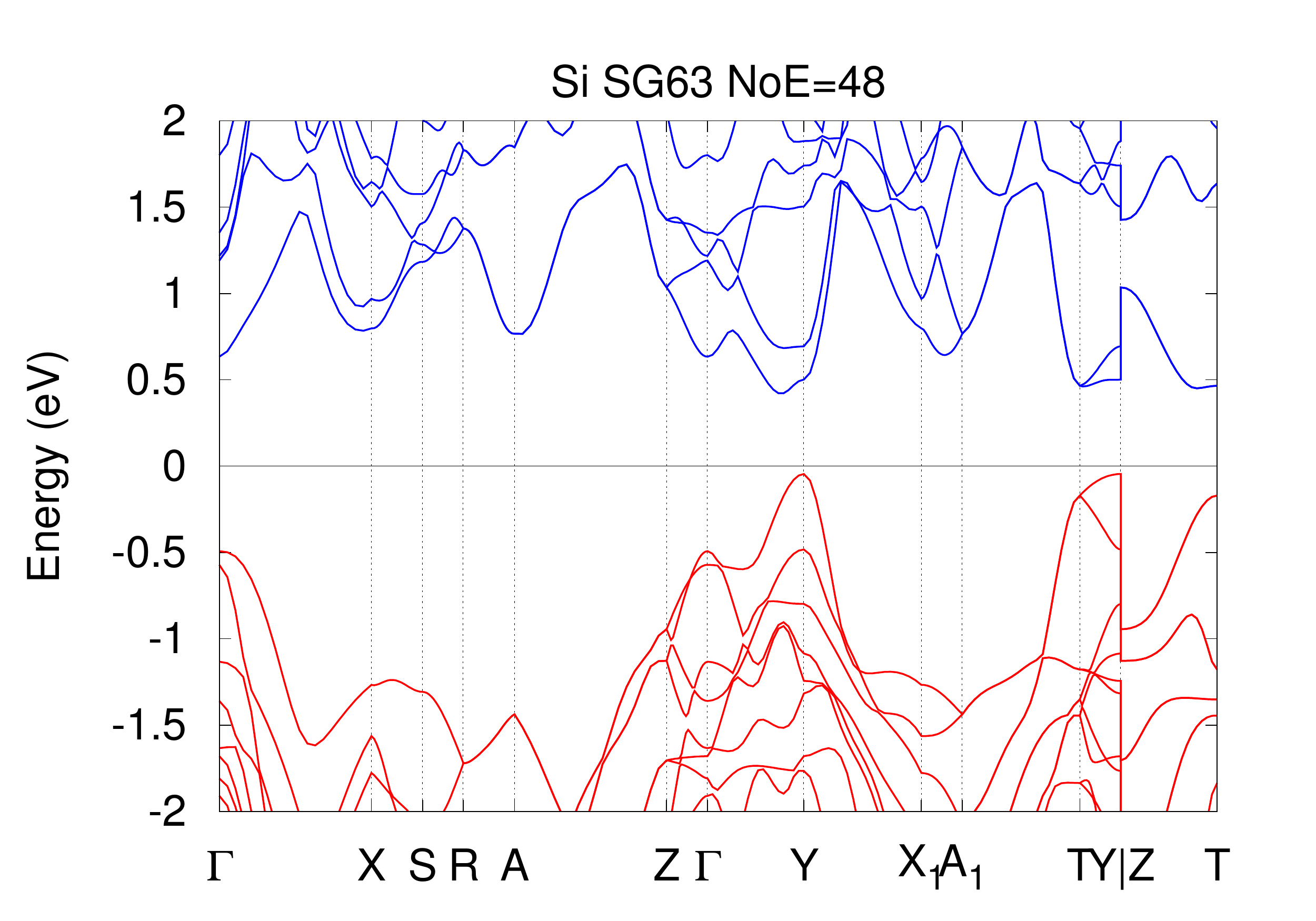}
}
\subfigure[LuGeIr SG62 NoA=12 NoE=88]{
\label{subfig:412708}
\includegraphics[scale=0.32,angle=270]{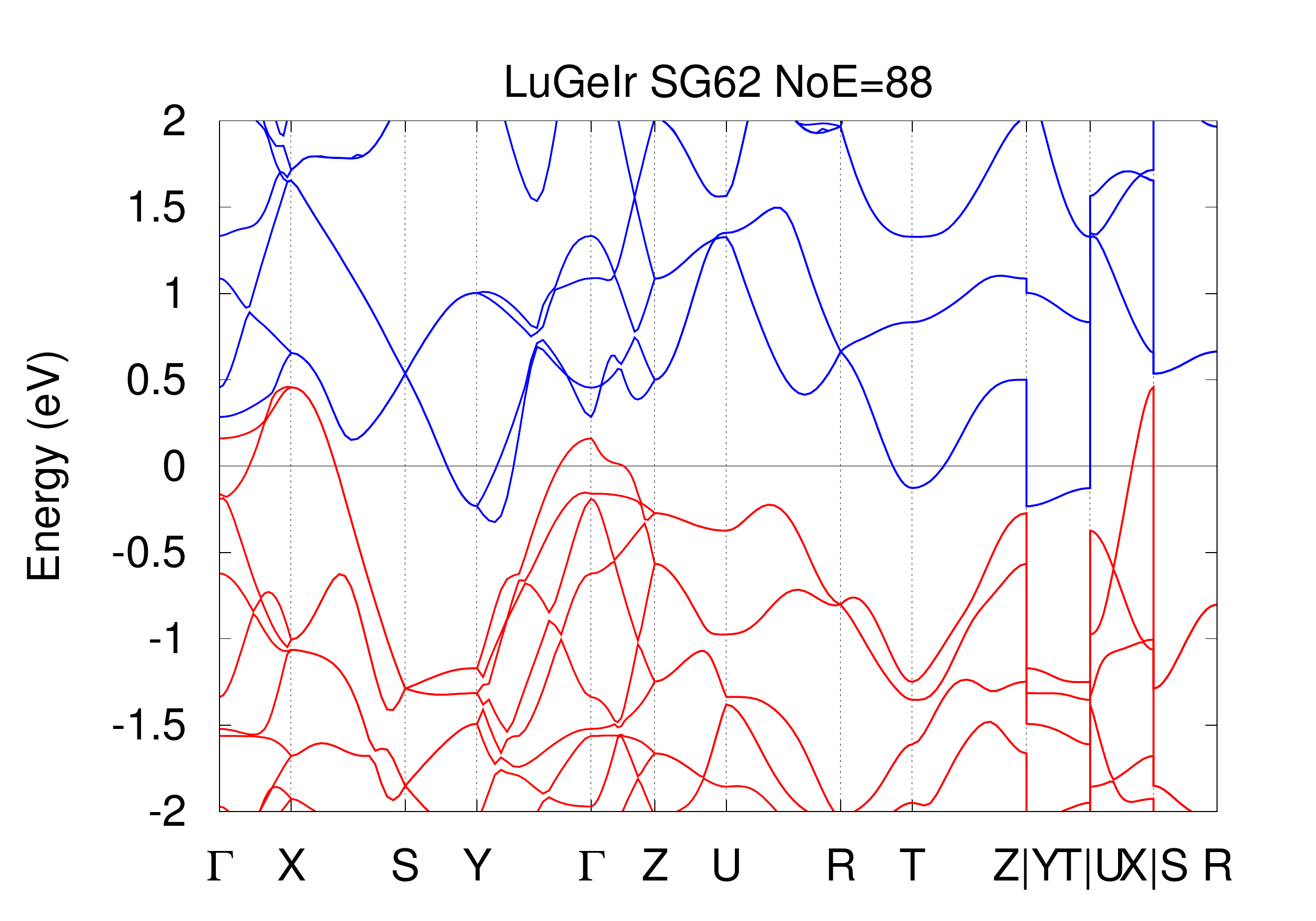}
}
\subfigure[ScSiRh SG62 NoA=12 NoE=64]{
\label{subfig:79596}
\includegraphics[scale=0.32,angle=270]{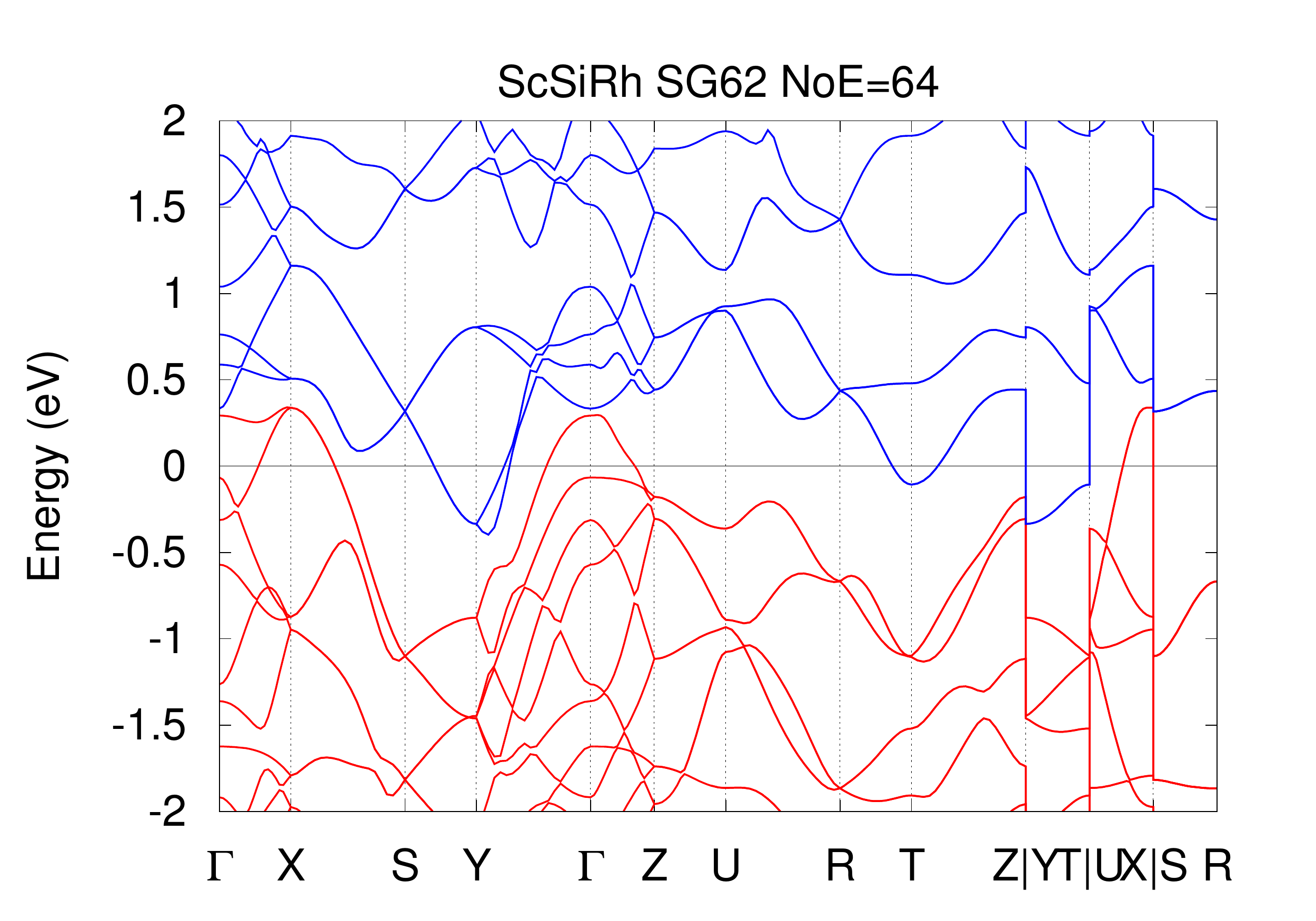}
}
\caption{\hyperref[tab:electride]{back to the table}}
\end{figure}

\begin{figure}[htp]
 \centering
\subfigure[SrNiGe SG62 NoA=12 NoE=96]{
\label{subfig:237303}
\includegraphics[scale=0.32,angle=270]{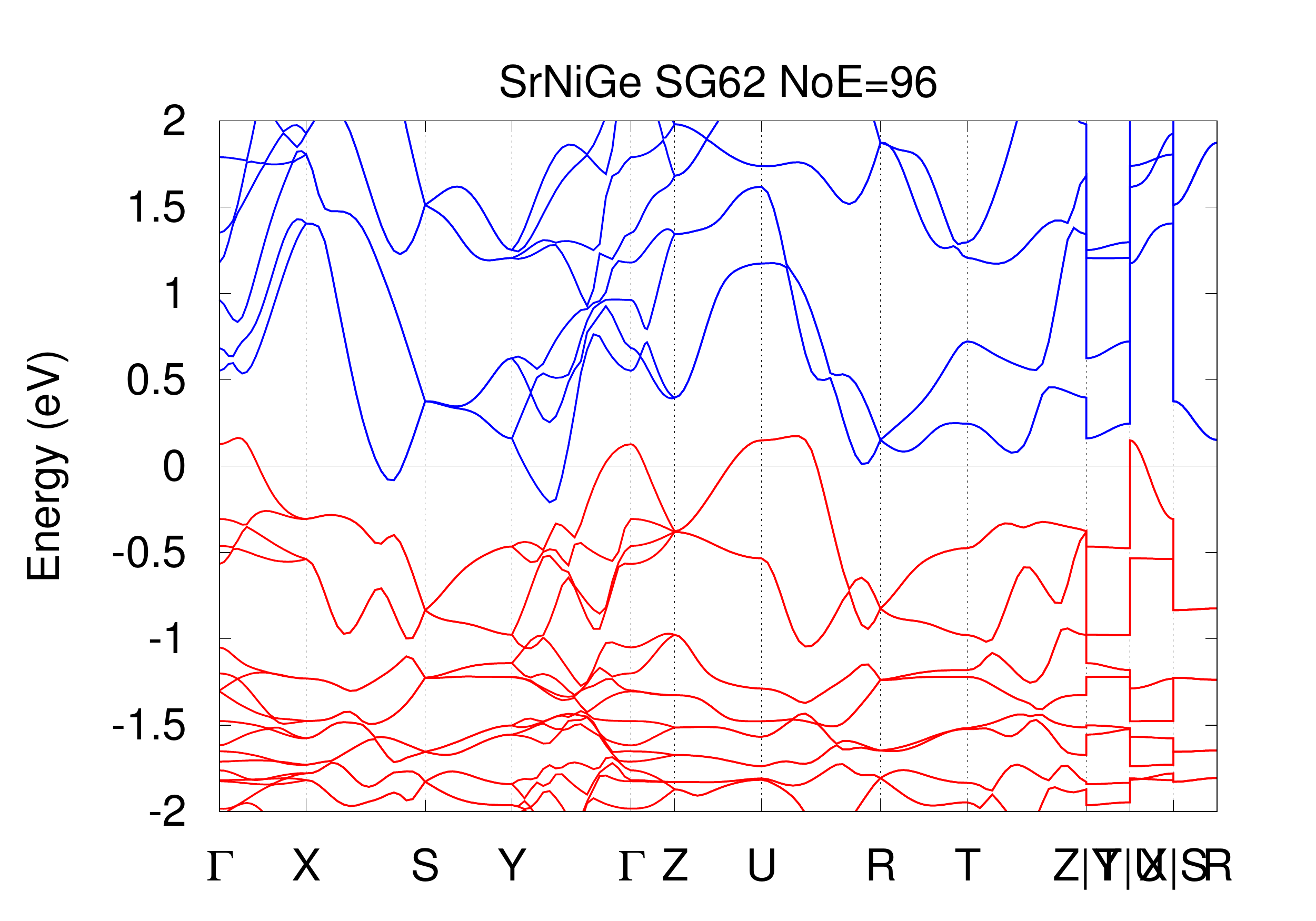}
}
\subfigure[FeSbTe SG14 NoA=12 NoE=76]{
\label{subfig:633405}
\includegraphics[scale=0.32,angle=270]{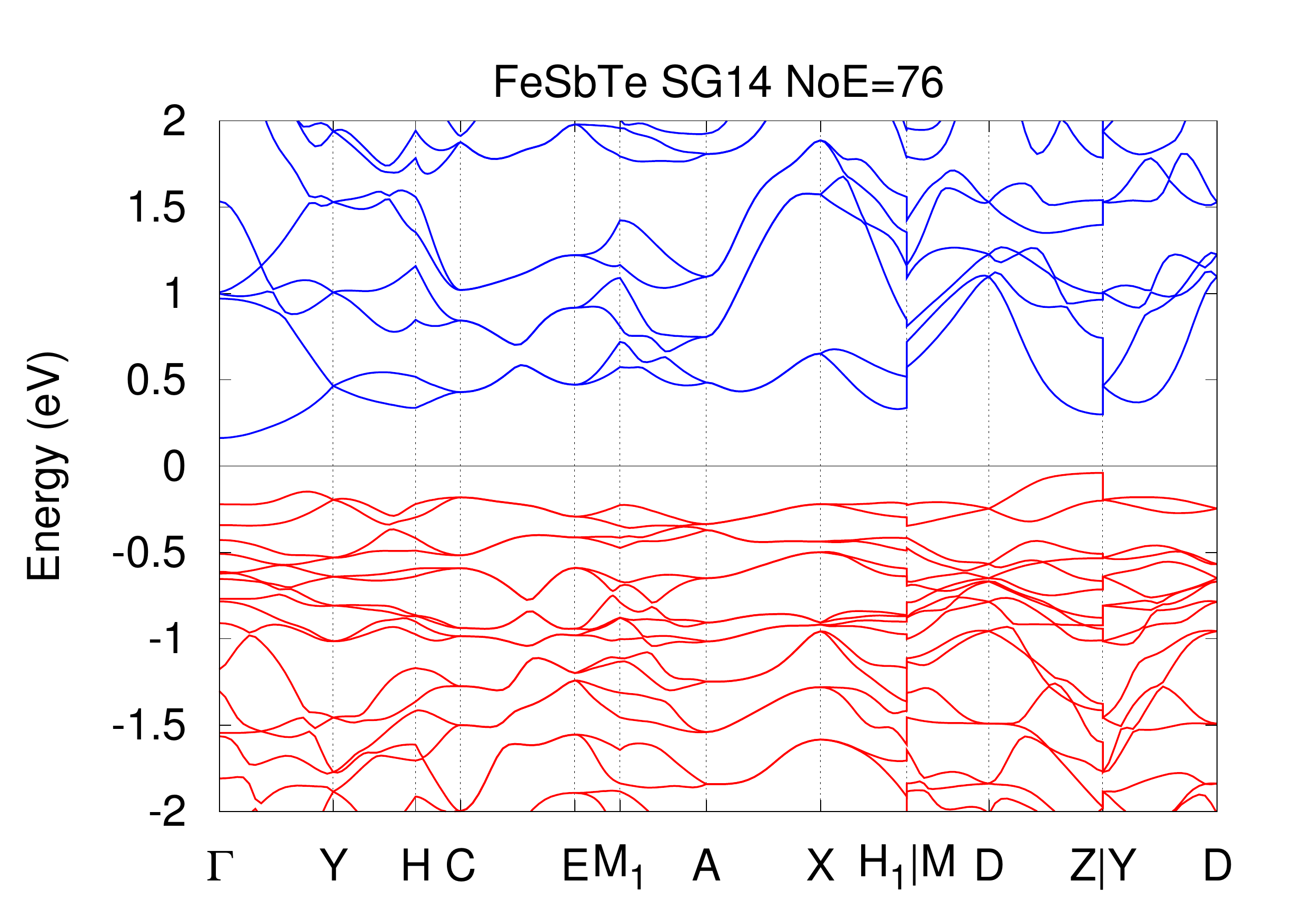}
}
\subfigure[TmGeIr SG62 NoA=12 NoE=88]{
\label{subfig:636748}
\includegraphics[scale=0.32,angle=270]{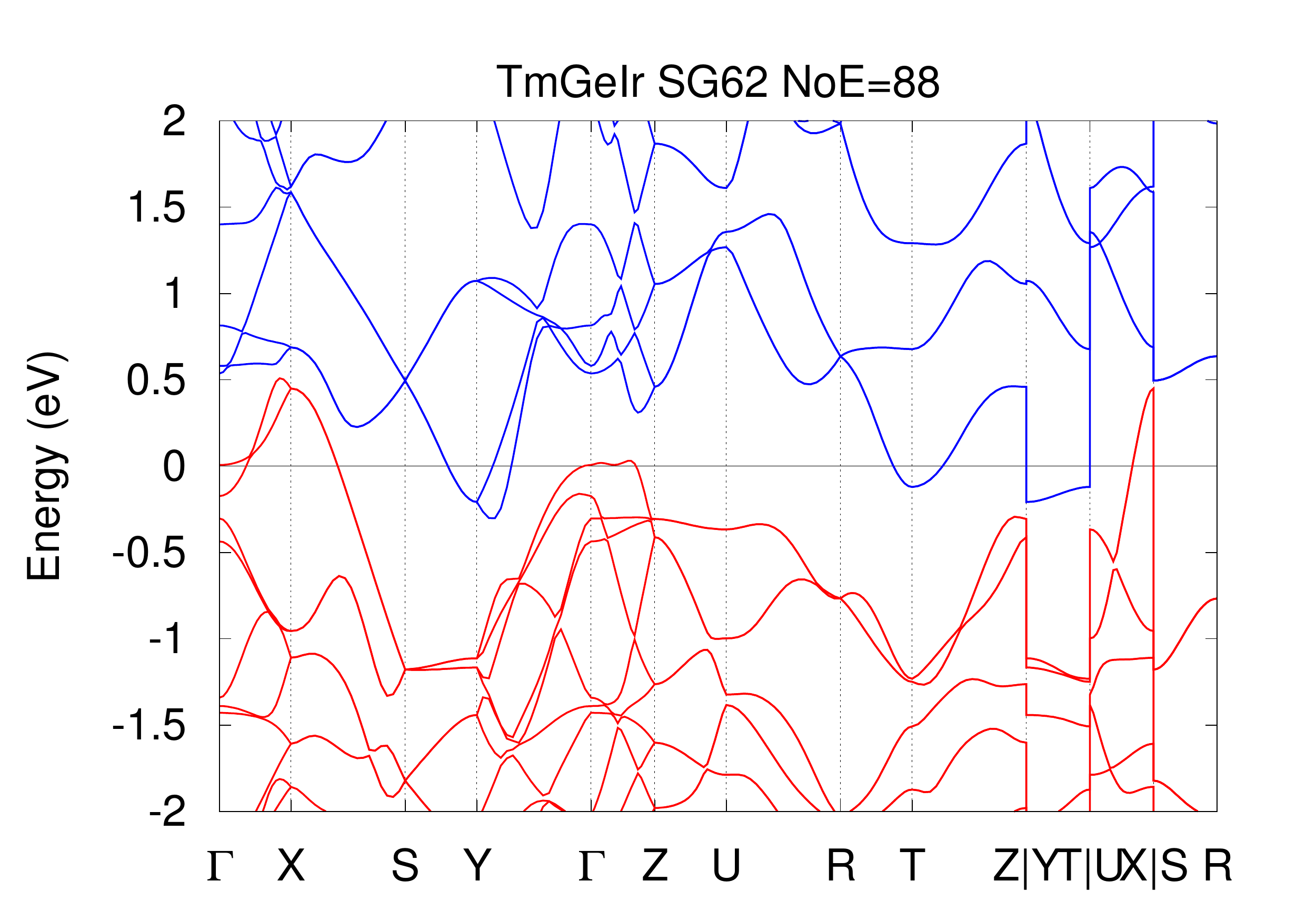}
}
\subfigure[ScGeRh SG62 NoA=12 NoE=64]{
\label{subfig:428475}
\includegraphics[scale=0.32,angle=270]{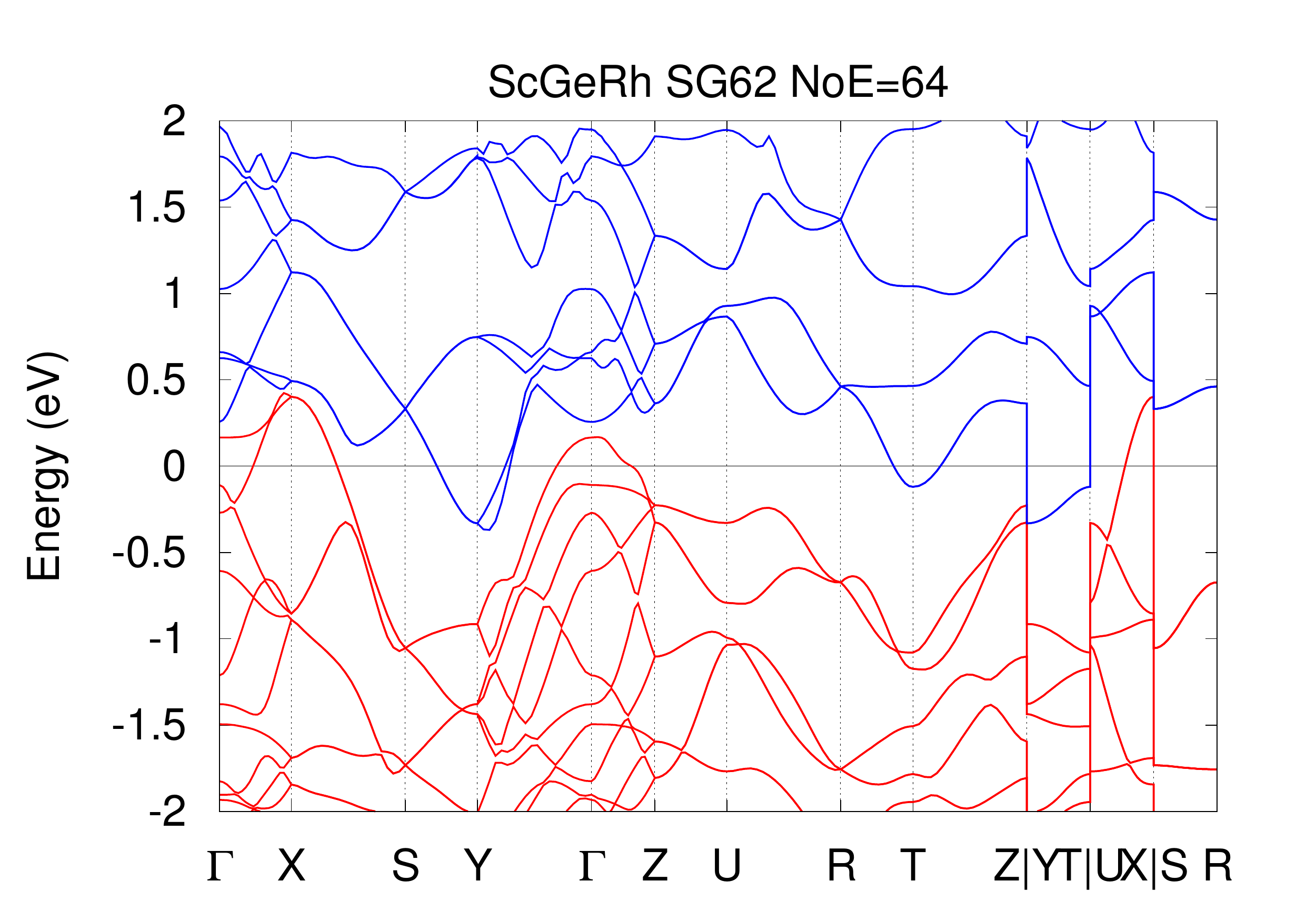}
}
\subfigure[YbGaGe SG194 NoA=12 NoE=60]{
\label{subfig:152569}
\includegraphics[scale=0.32,angle=270]{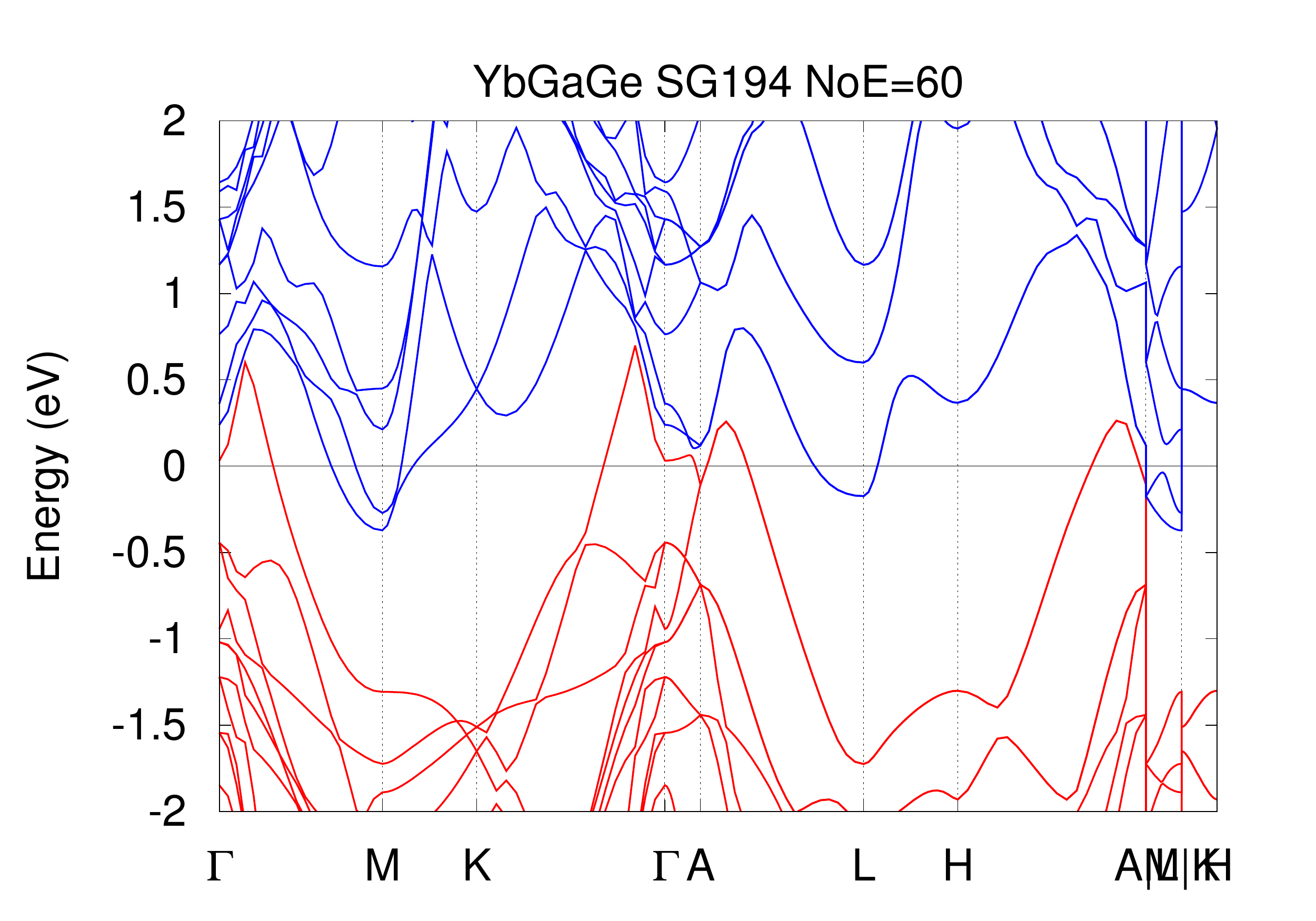}
}
\subfigure[CeZnGe SG194 NoA=12 NoE=108]{
\label{subfig:163343}
\includegraphics[scale=0.32,angle=270]{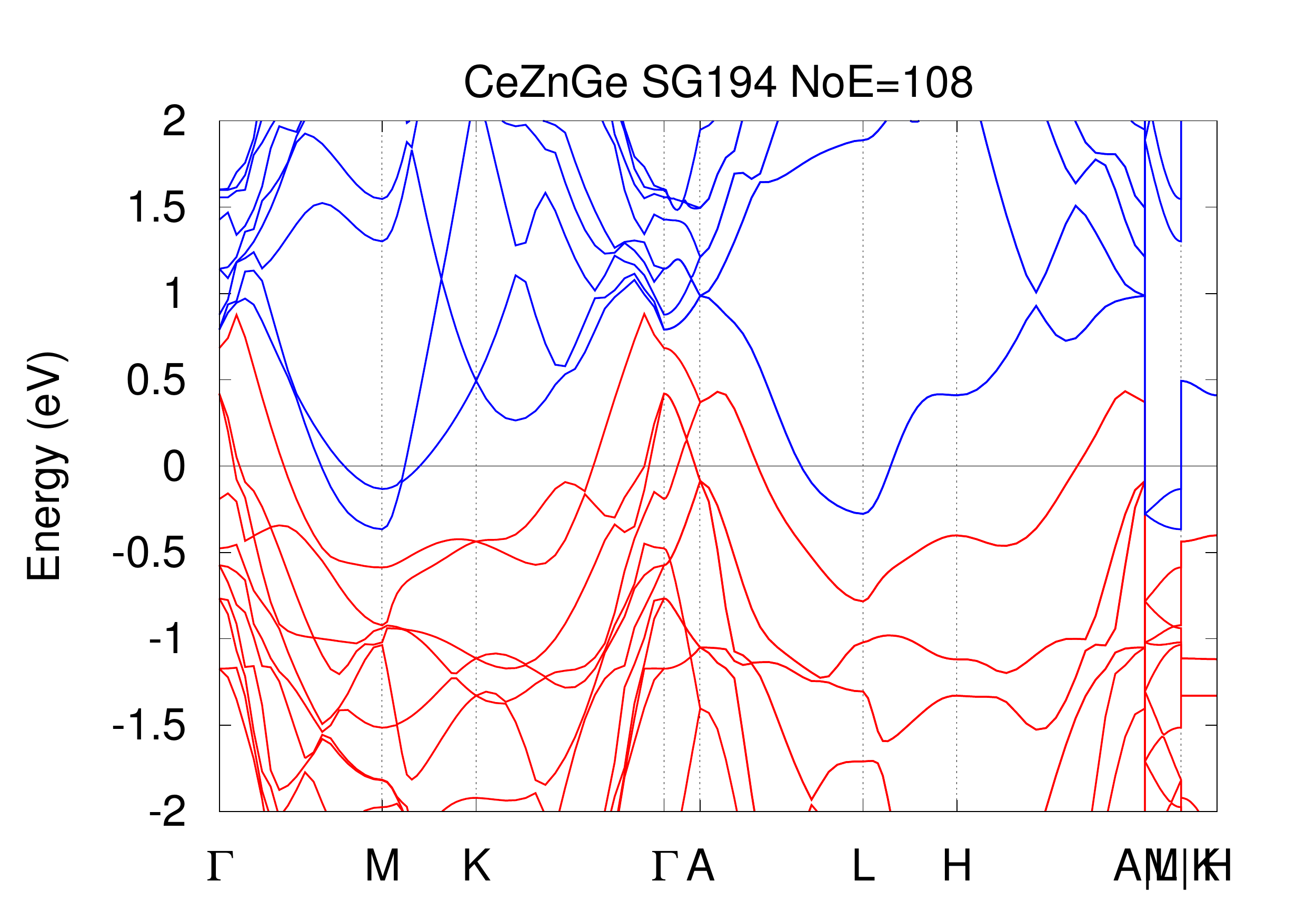}
}
\subfigure[As$_{2}$Pt SG205 NoA=12 NoE=80]{
\label{subfig:24203}
\includegraphics[scale=0.32,angle=270]{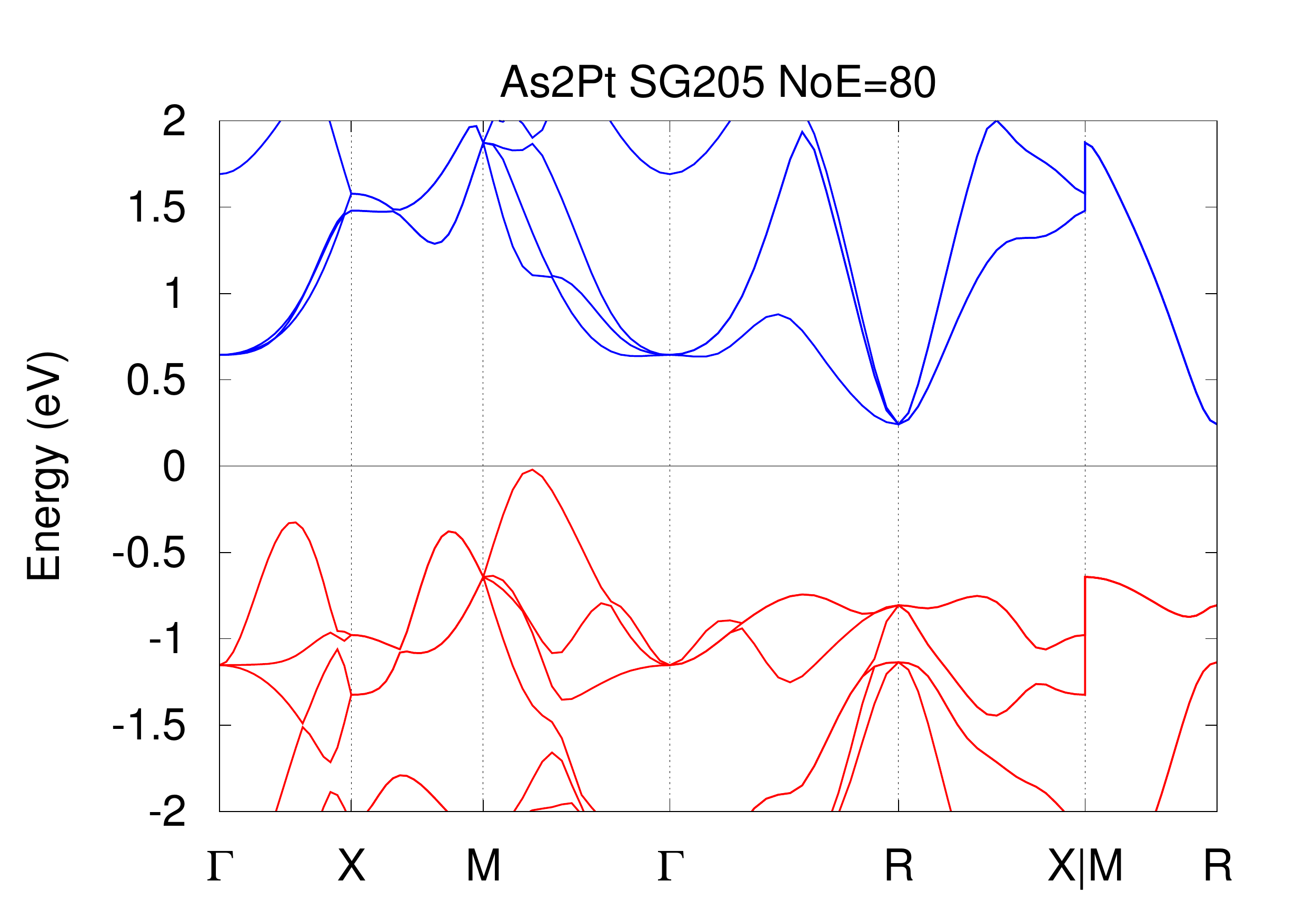}
}
\subfigure[SiAs$_{2}$ SG205 NoA=12 NoE=56]{
\label{subfig:24801}
\includegraphics[scale=0.32,angle=270]{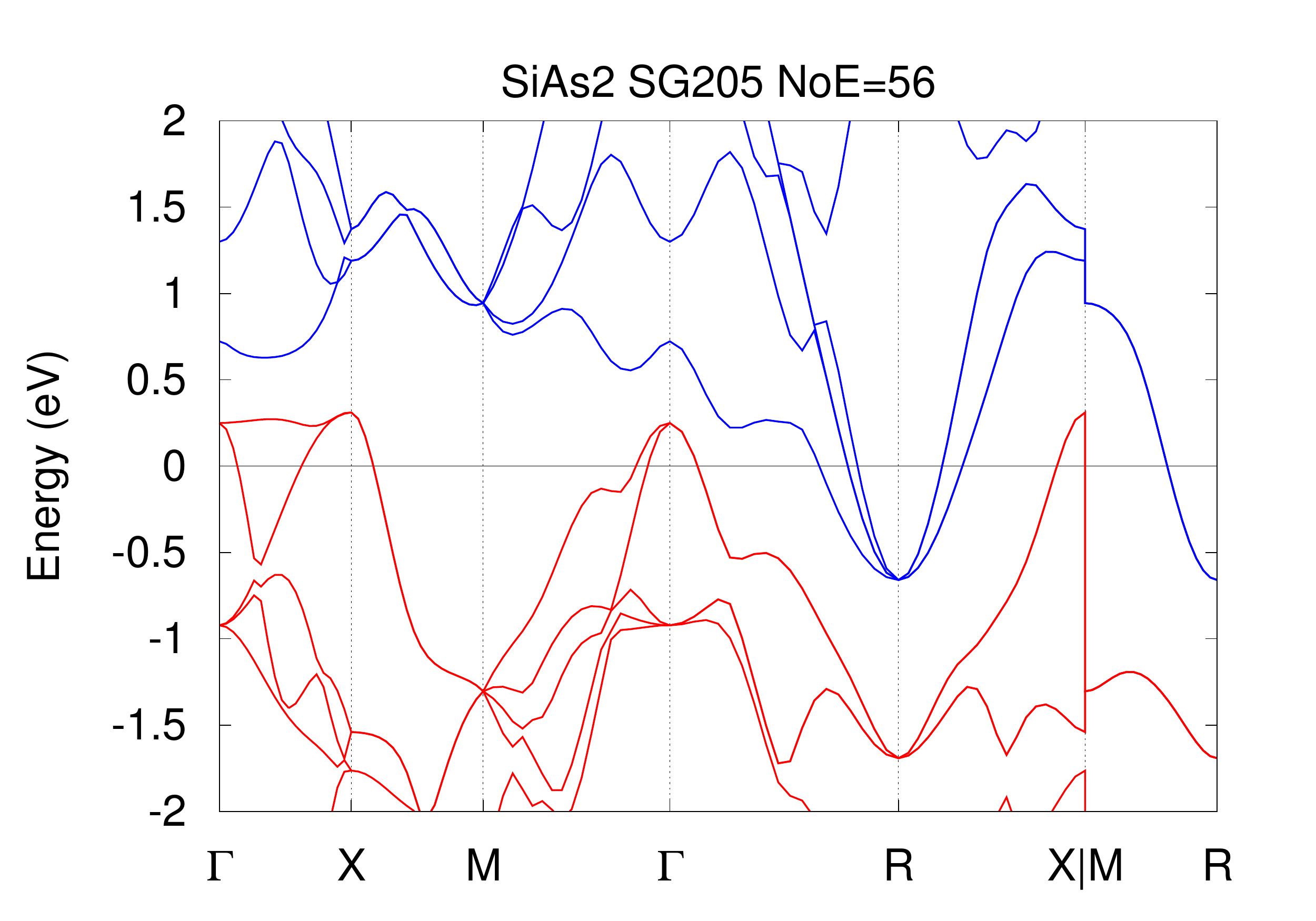}
}
\caption{\hyperref[tab:electride]{back to the table}}
\end{figure}

\begin{figure}[htp]
 \centering
\subfigure[ScCoGe SG62 NoA=12 NoE=64]{
\label{subfig:428472}
\includegraphics[scale=0.32,angle=270]{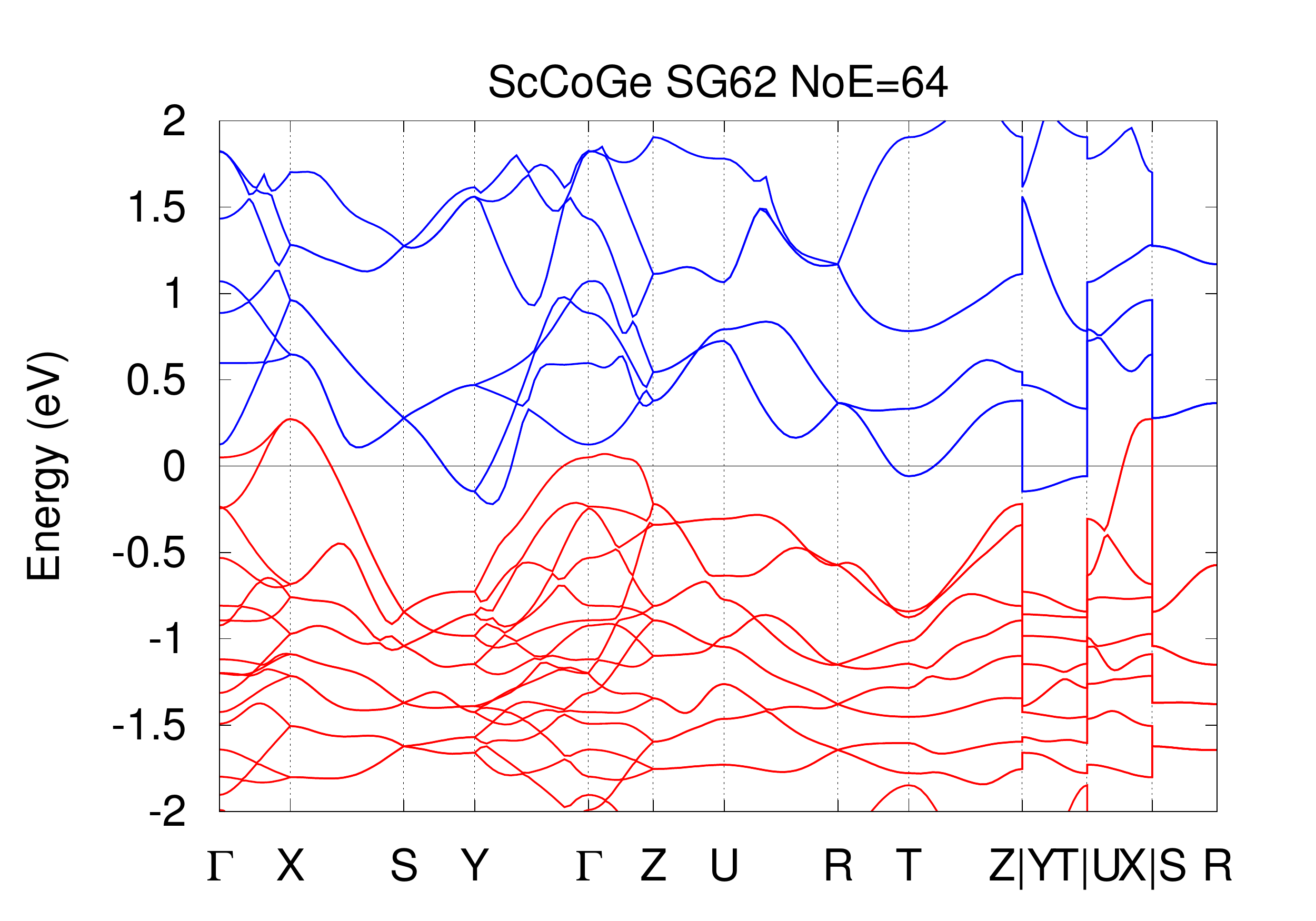}
}
\subfigure[TbSiIr SG62 NoA=12 NoE=88]{
\label{subfig:93221}
\includegraphics[scale=0.32,angle=270]{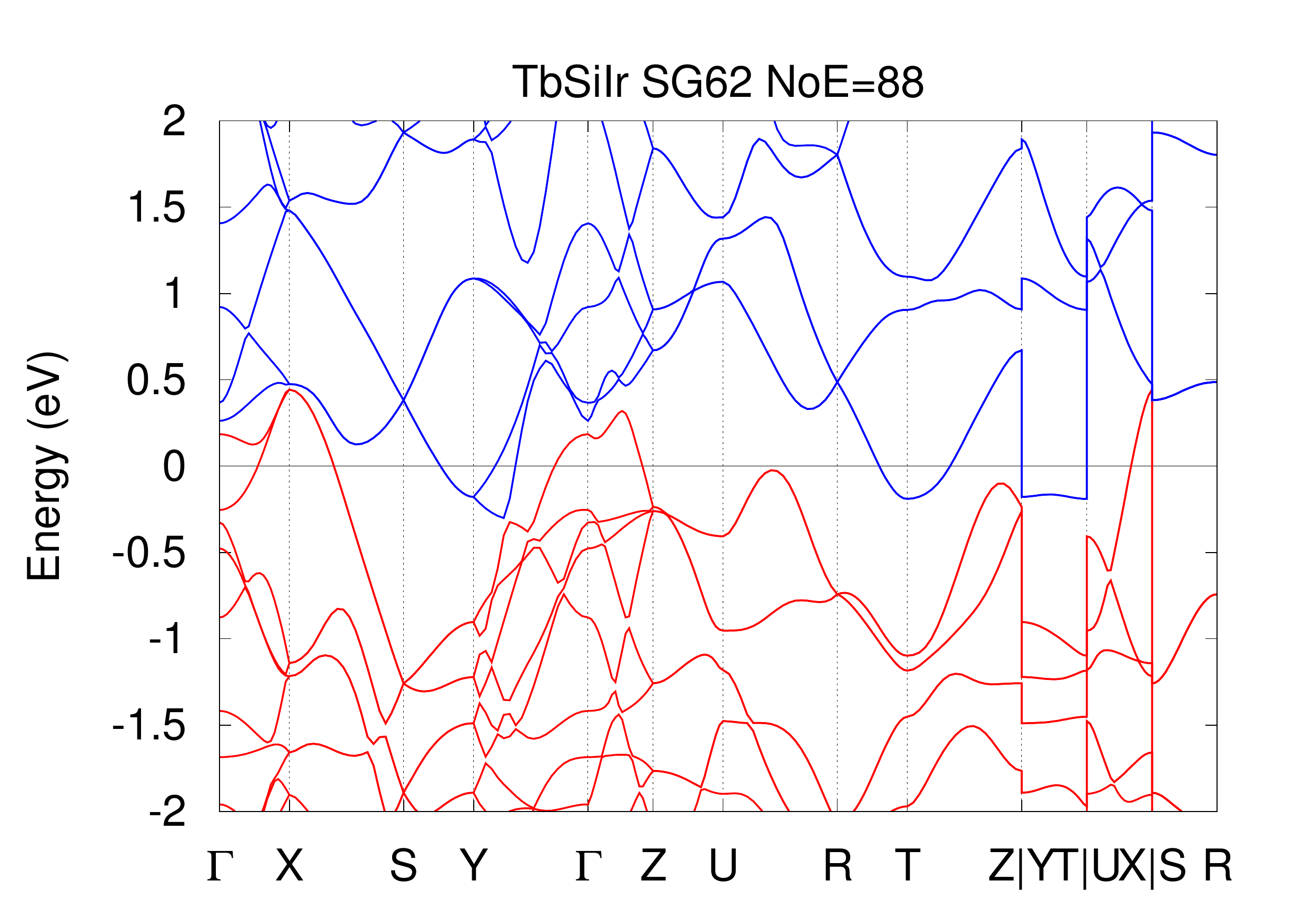}
}
\subfigure[PrGeIr SG62 NoA=12 NoE=96]{
\label{subfig:636724}
\includegraphics[scale=0.32,angle=270]{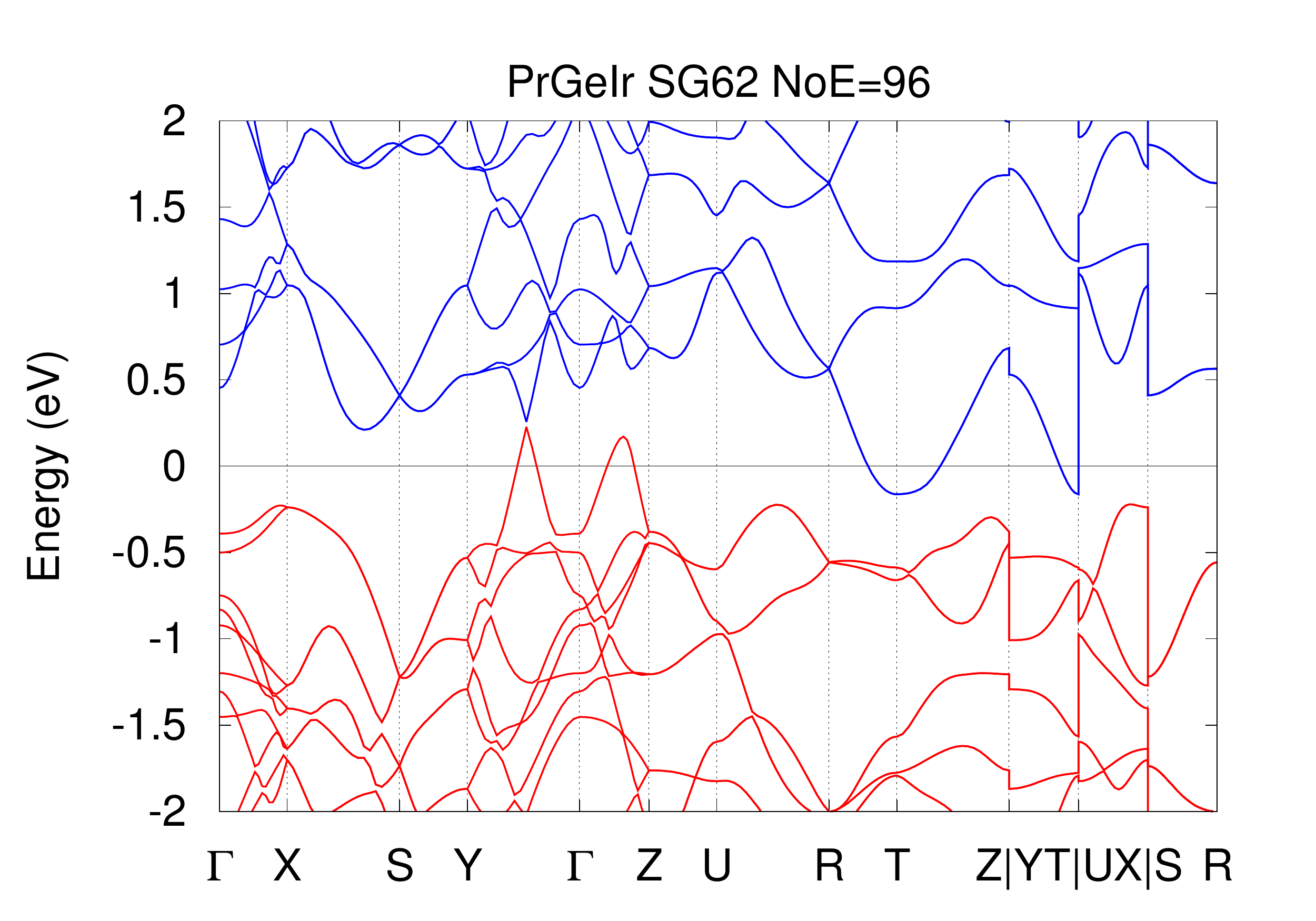}
}
\subfigure[ScCoSi SG62 NoA=12 NoE=64]{
\label{subfig:420415}
\includegraphics[scale=0.32,angle=270]{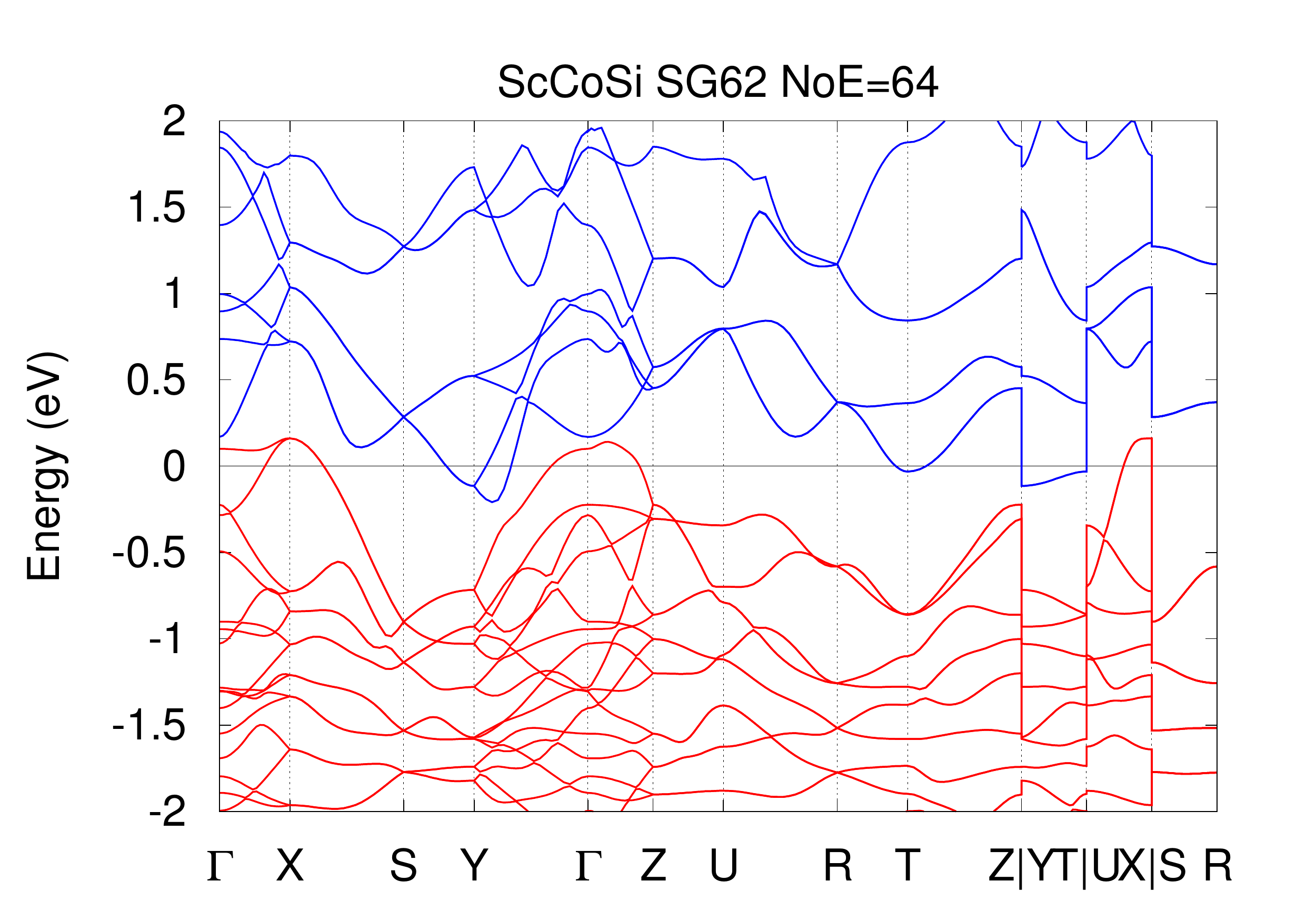}
}
\subfigure[PRuSe SG14 NoA=12 NoE=76]{
\label{subfig:648028}
\includegraphics[scale=0.32,angle=270]{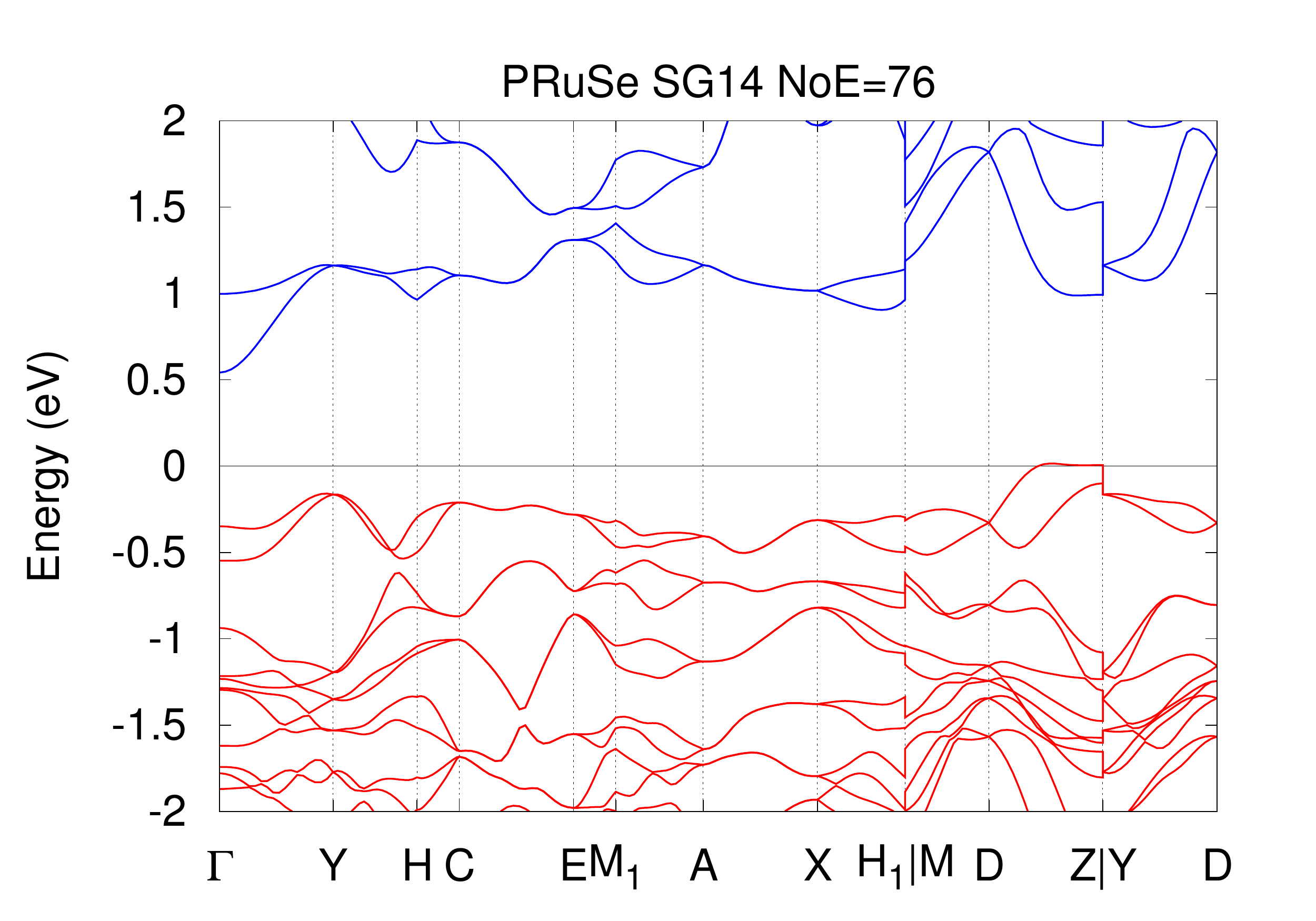}
}
\subfigure[NdAs$_{2}$ SG14 NoA=12 NoE=84]{
\label{subfig:610996}
\includegraphics[scale=0.32,angle=270]{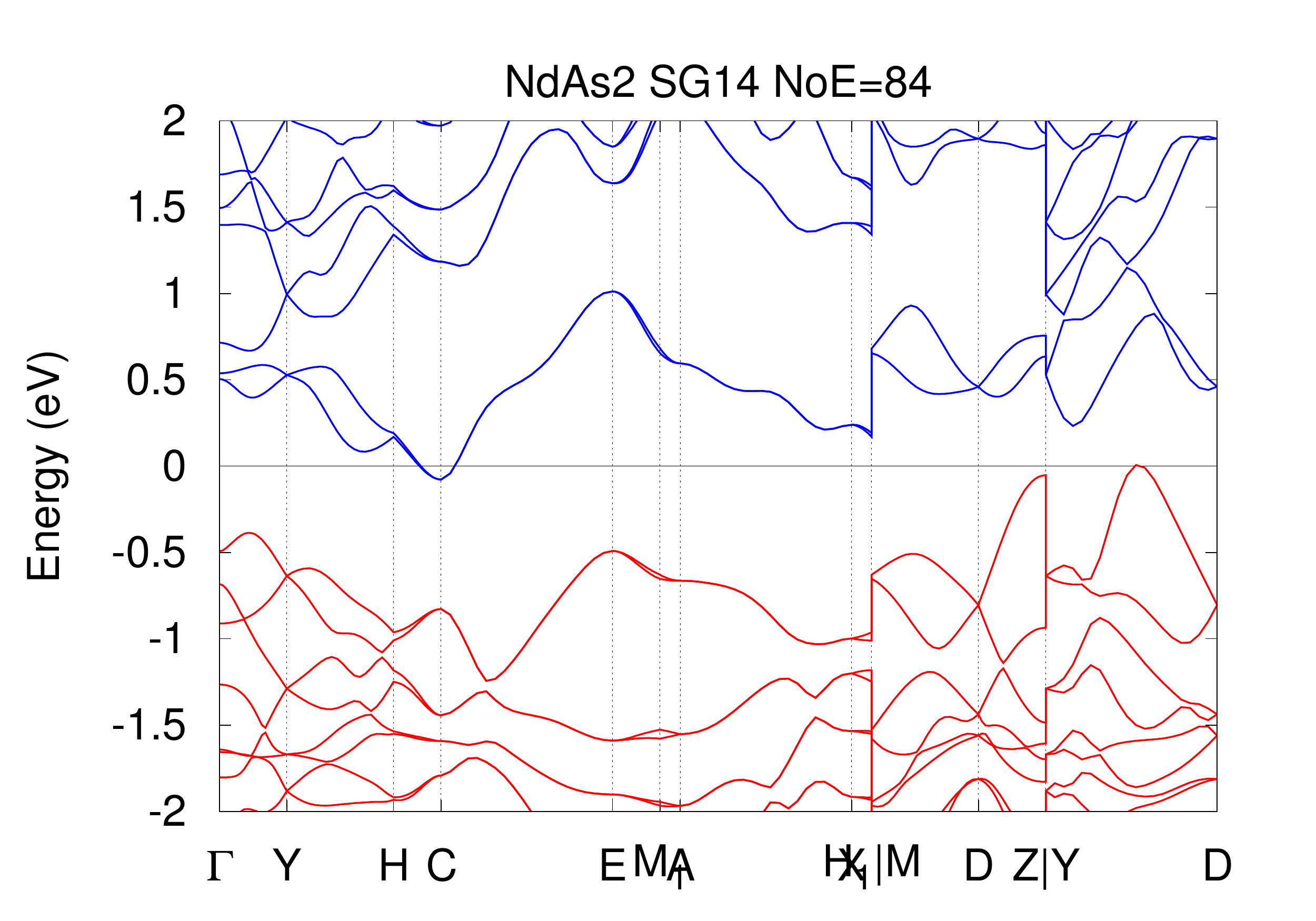}
}
\subfigure[PdSe$_{2}$ SG61 NoA=12 NoE=88]{
\label{subfig:648826}
\includegraphics[scale=0.32,angle=270]{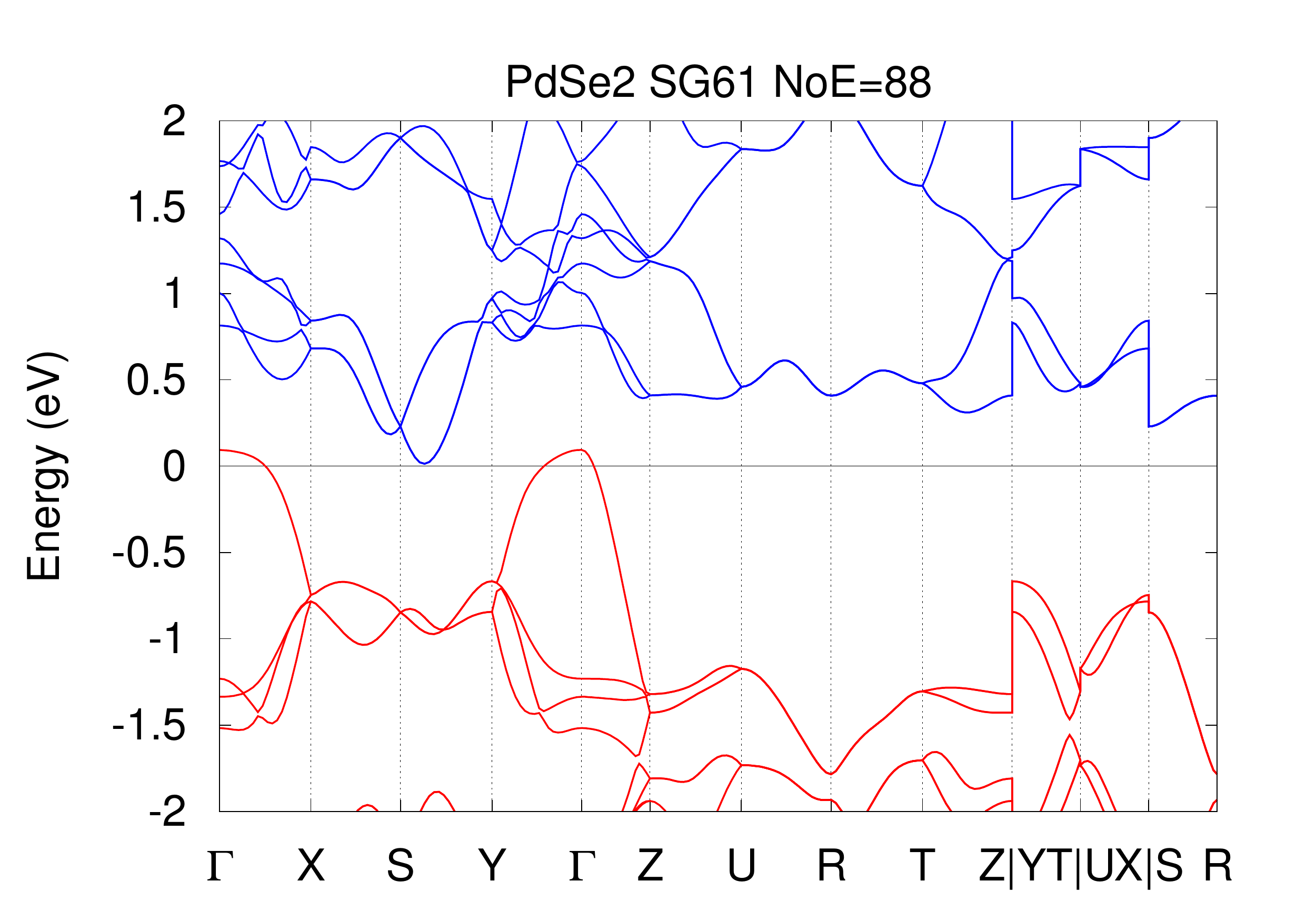}
}
\subfigure[FePSe SG14 NoA=12 NoE=76]{
\label{subfig:633093}
\includegraphics[scale=0.32,angle=270]{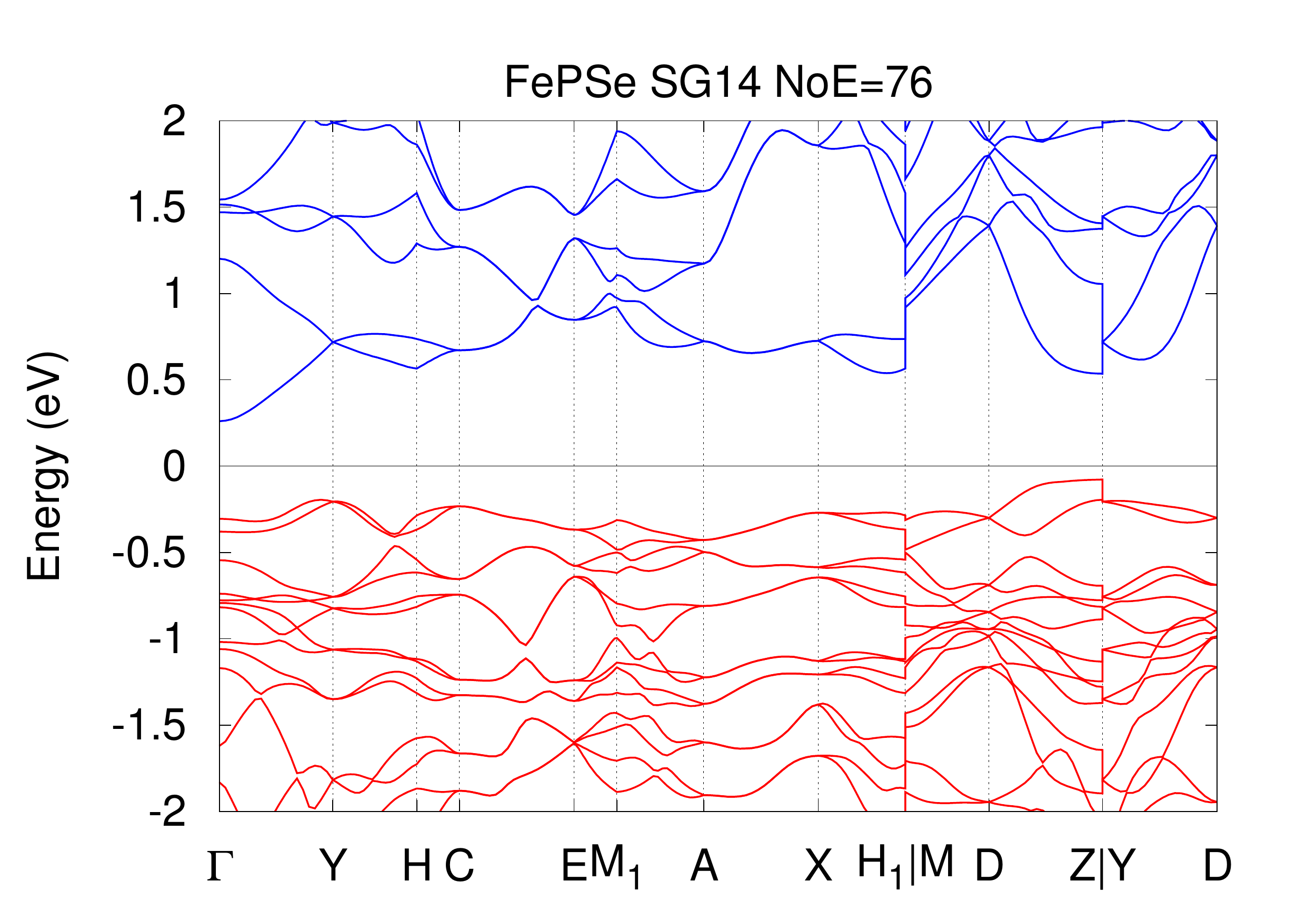}
}
\caption{\hyperref[tab:electride]{back to the table}}
\end{figure}

\begin{figure}[htp]
 \centering
\subfigure[CoP$_{2}$ SG14 NoA=12 NoE=76]{
\label{subfig:38316}
\includegraphics[scale=0.32,angle=270]{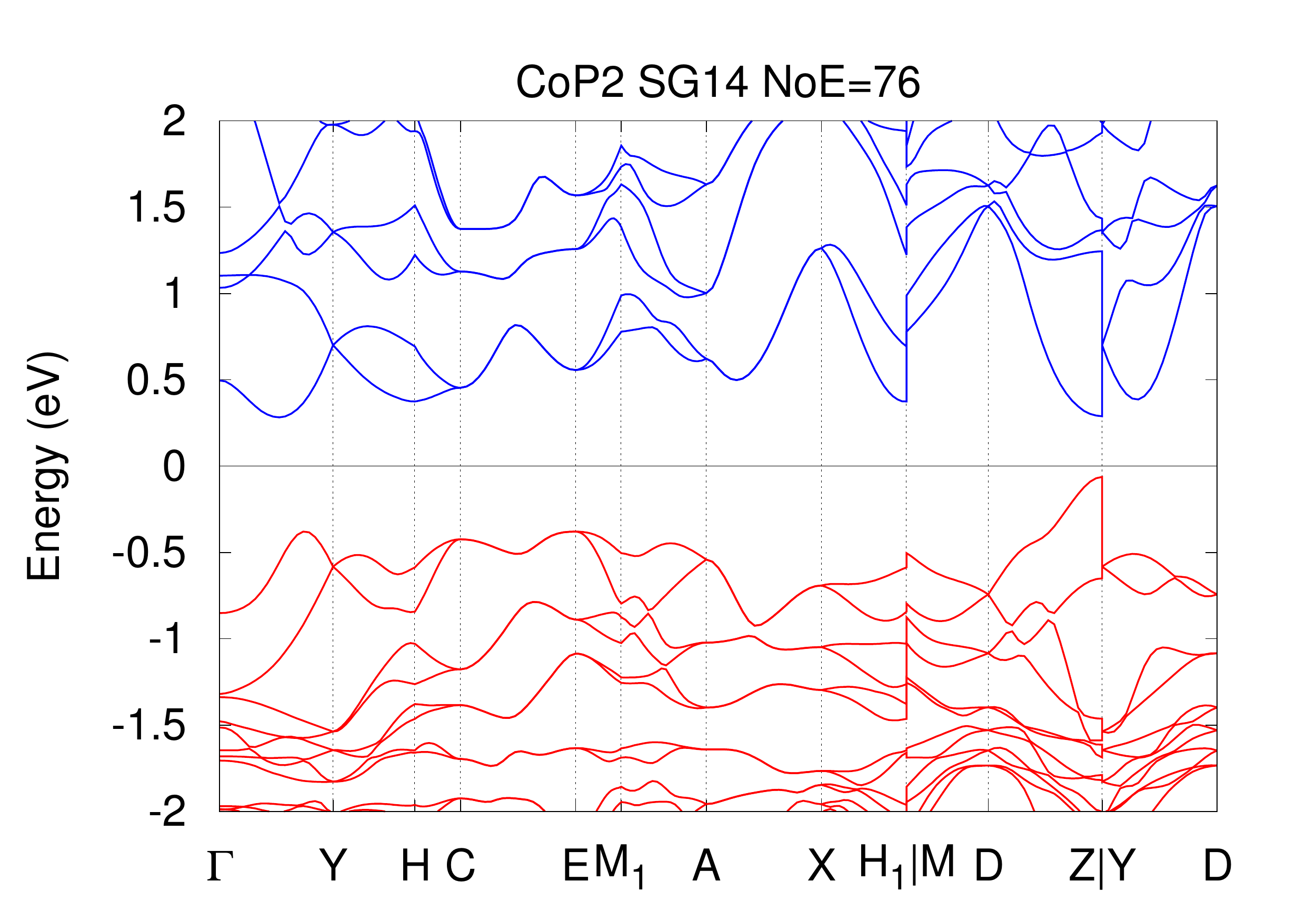}
}
\subfigure[GdS$_{2}$ SG14 NoA=12 NoE=84]{
\label{subfig:421335}
\includegraphics[scale=0.32,angle=270]{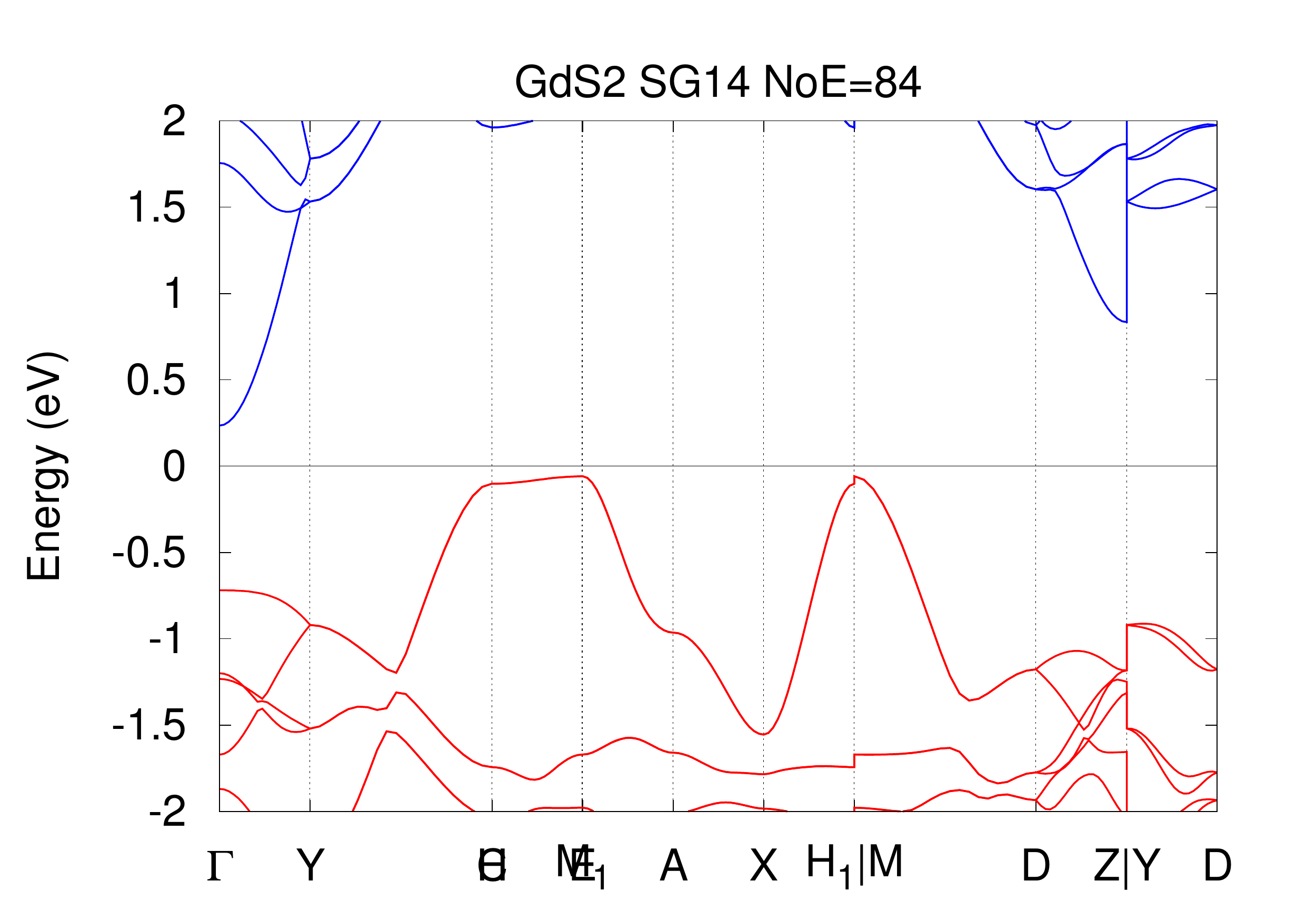}
}
\subfigure[SmP$_{5}$ SG11 NoA=12 NoE=72]{
\label{subfig:409183}
\includegraphics[scale=0.32,angle=270]{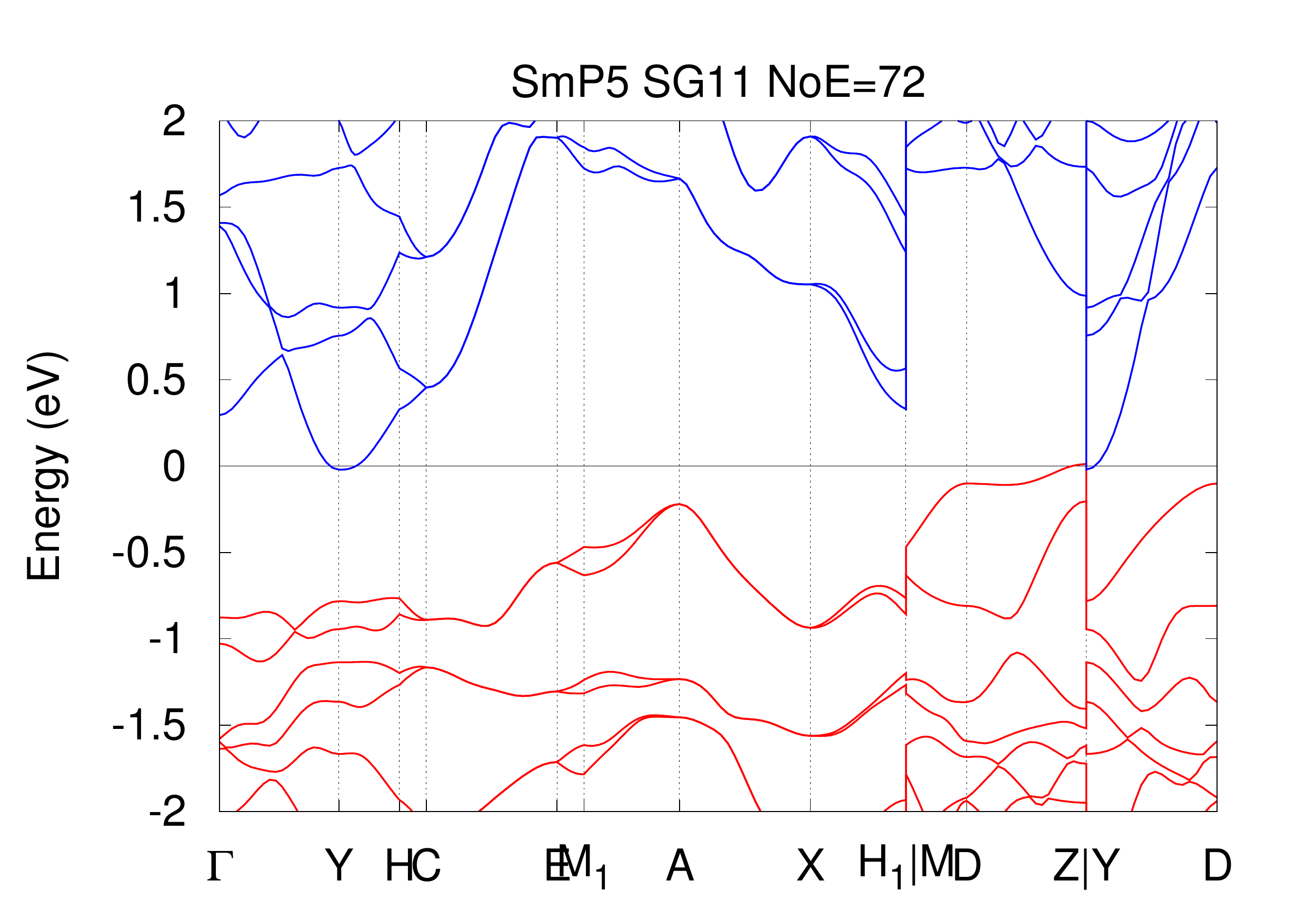}
}
\subfigure[ReTeS SG216 NoA=12 NoE=76]{
\label{subfig:82721}
\includegraphics[scale=0.32,angle=270]{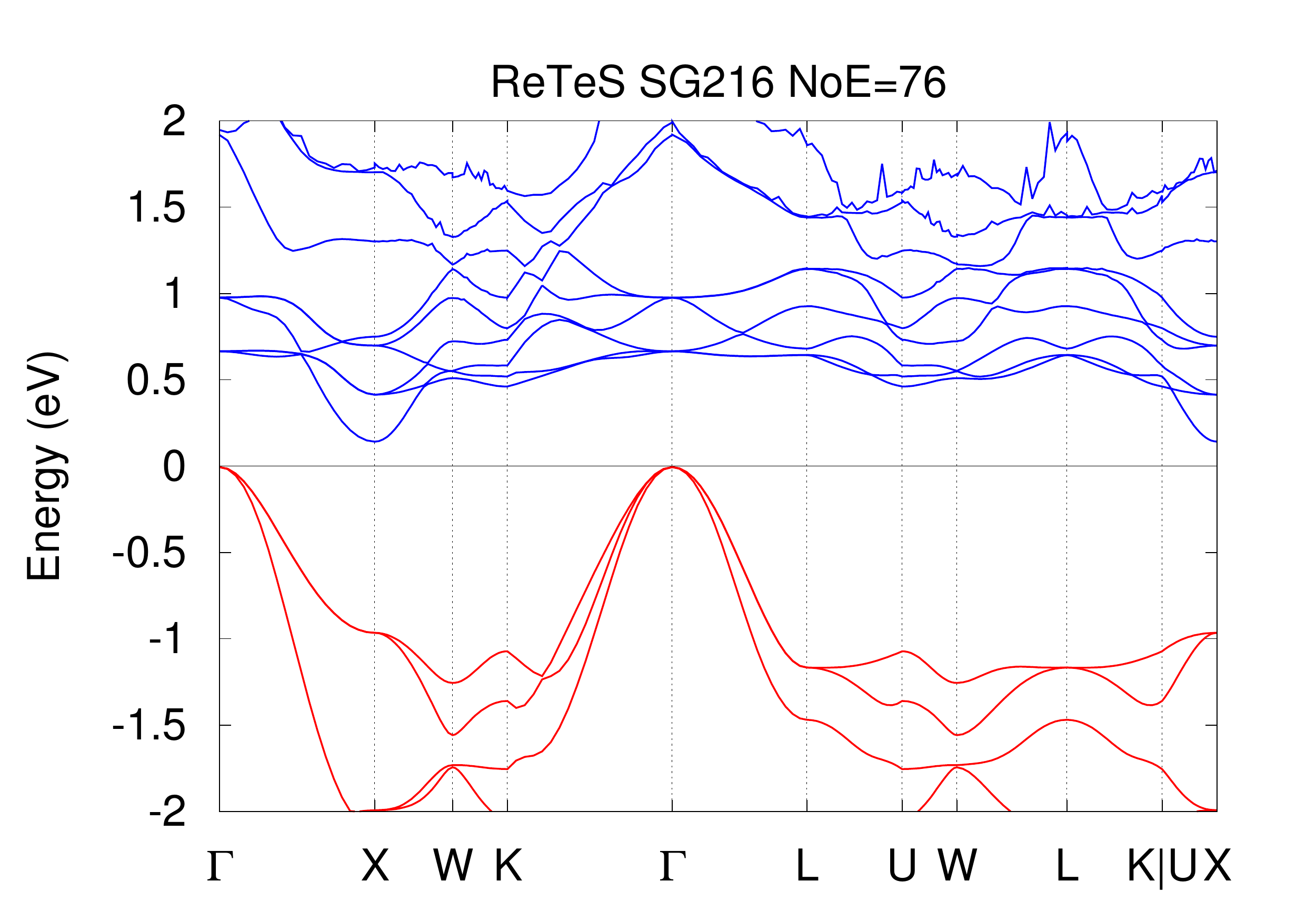}
}
\subfigure[SrGaSn SG194 NoA=12 NoE=68]{
\label{subfig:166387}
\includegraphics[scale=0.32,angle=270]{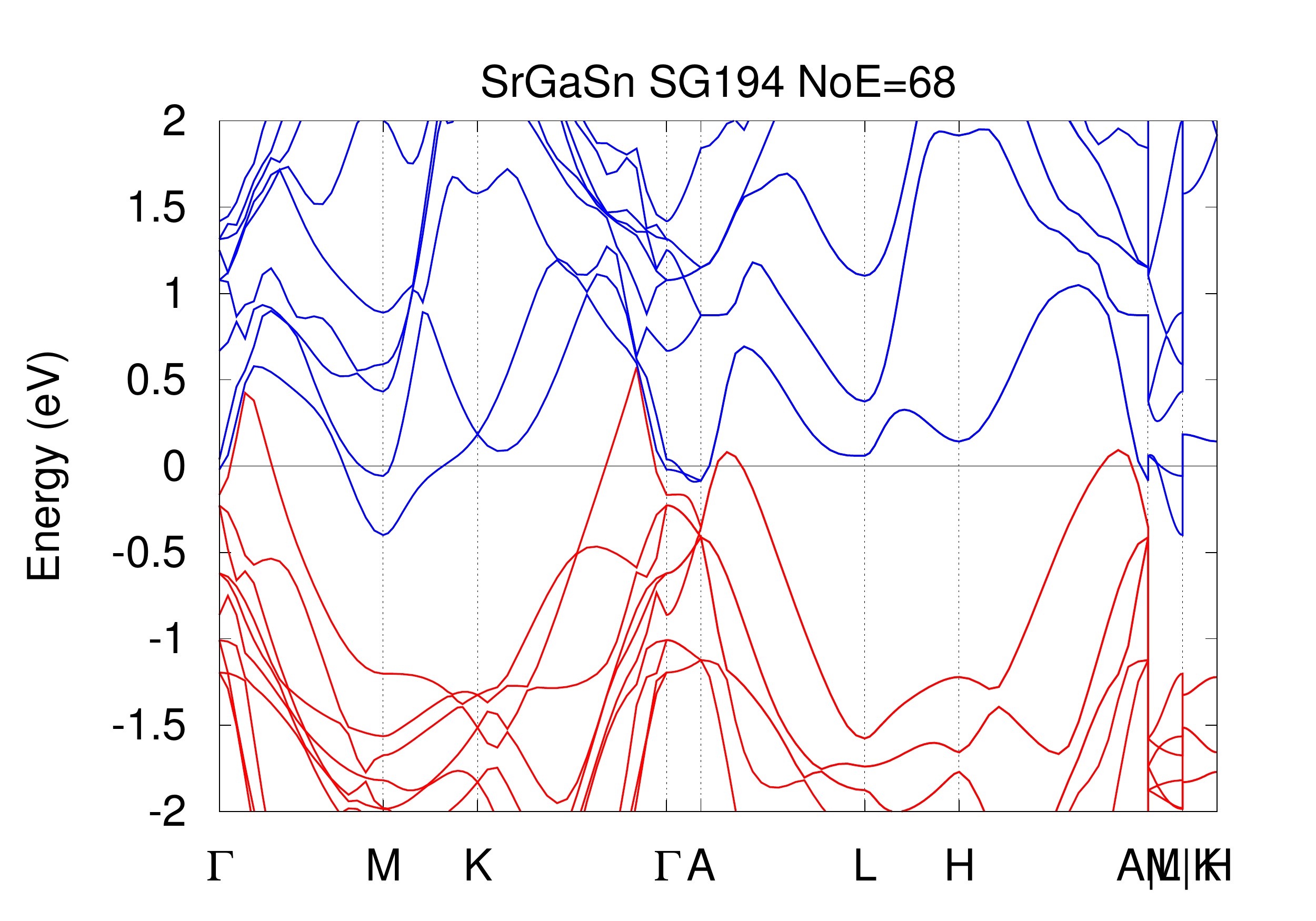}
}
\subfigure[NdP$_{5}$ SG11 NoA=12 NoE=72]{
\label{subfig:358}
\includegraphics[scale=0.32,angle=270]{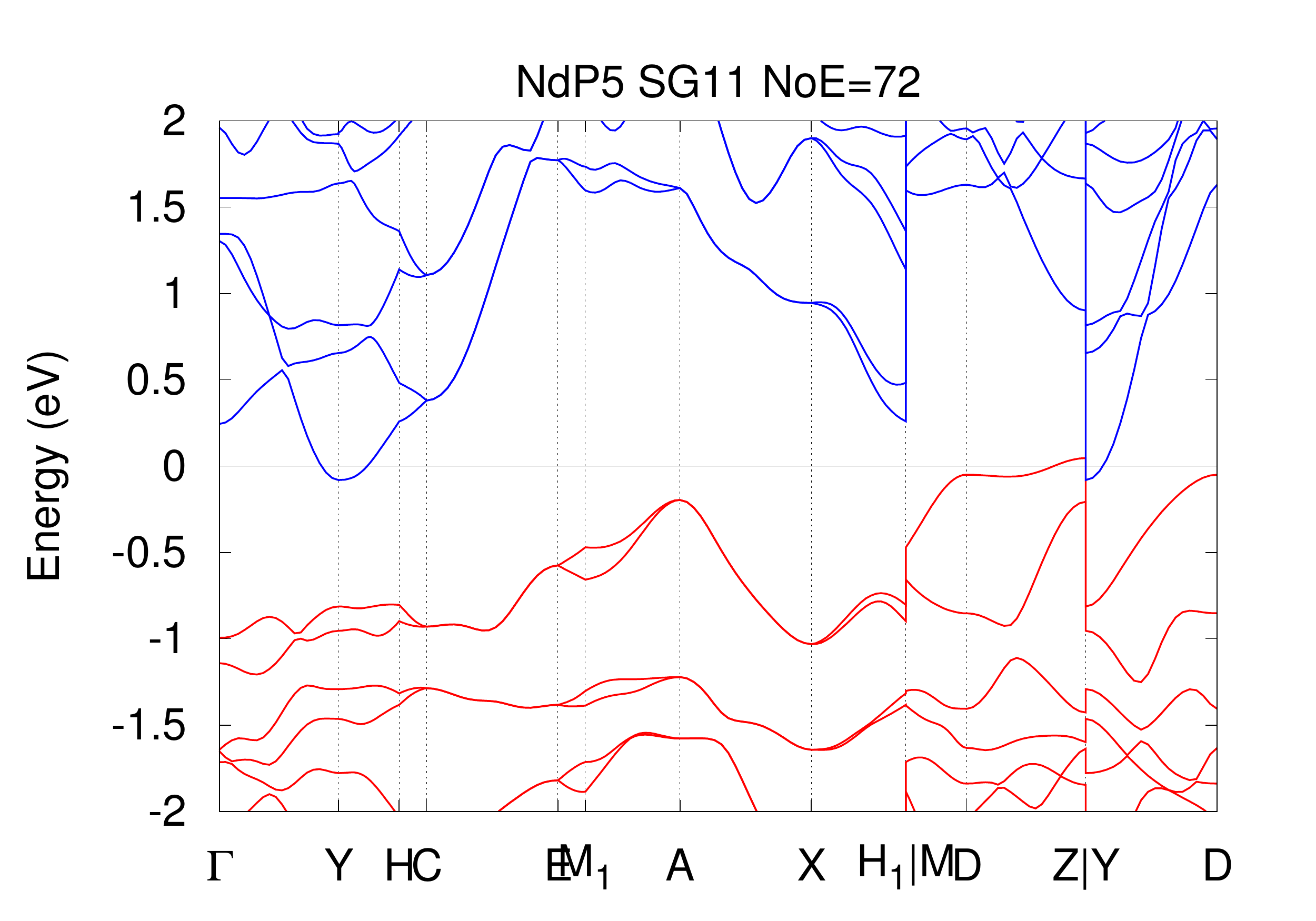}
}
\subfigure[CdSe$_{2}$ SG205 NoA=12 NoE=96]{
\label{subfig:620416}
\includegraphics[scale=0.32,angle=270]{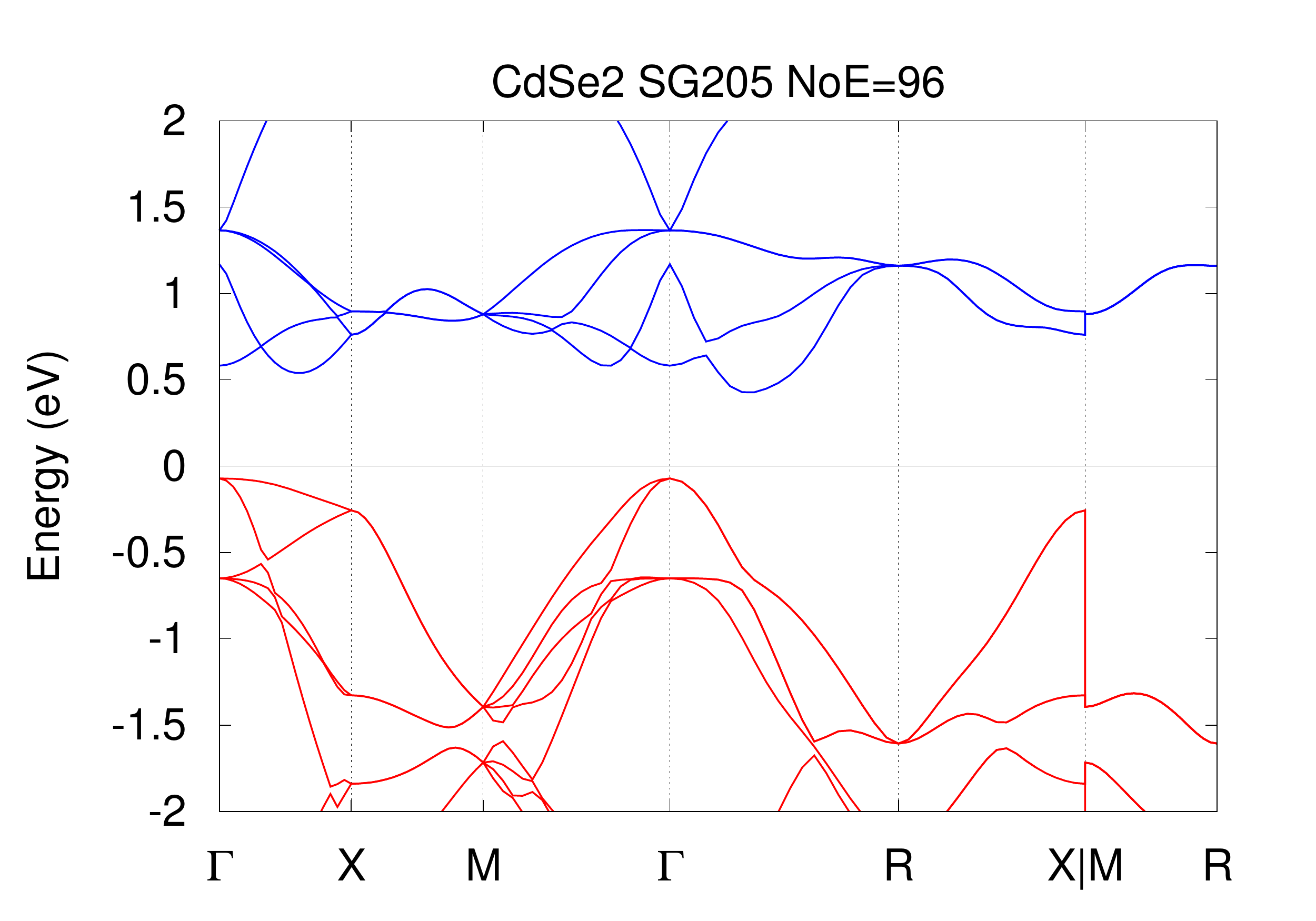}
}
\subfigure[IrN$_{2}$ SG14 NoA=12 NoE=76]{
\label{subfig:240755}
\includegraphics[scale=0.32,angle=270]{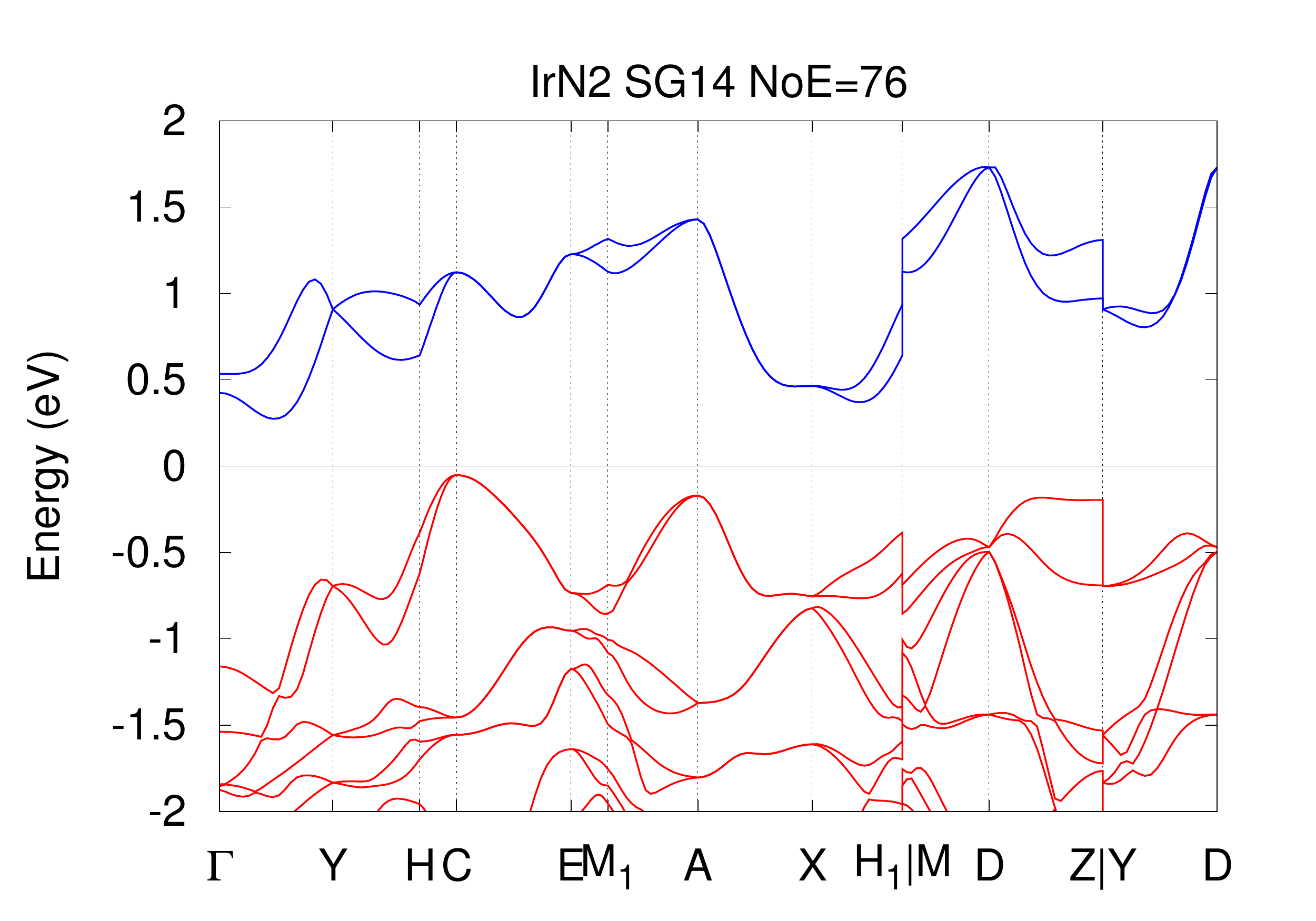}
}
\caption{\hyperref[tab:electride]{back to the table}}
\end{figure}

\begin{figure}[htp]
 \centering
\subfigure[KGa$_{3}$ SG119 NoA=12 NoE=54]{
\label{subfig:20664}
\includegraphics[scale=0.32,angle=270]{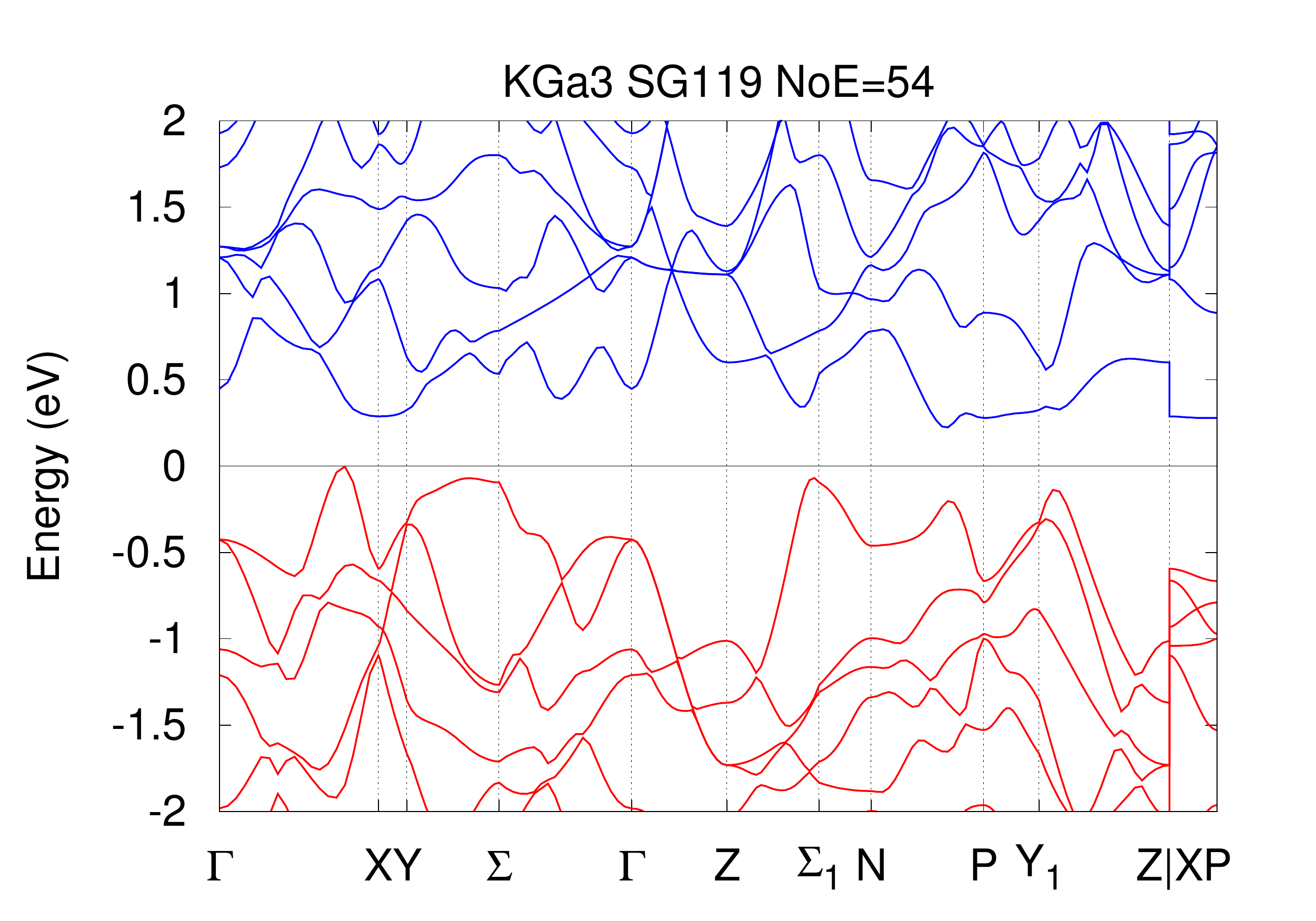}
}
\subfigure[BiOsSe SG14 NoA=12 NoE=76]{
\label{subfig:616892}
\includegraphics[scale=0.32,angle=270]{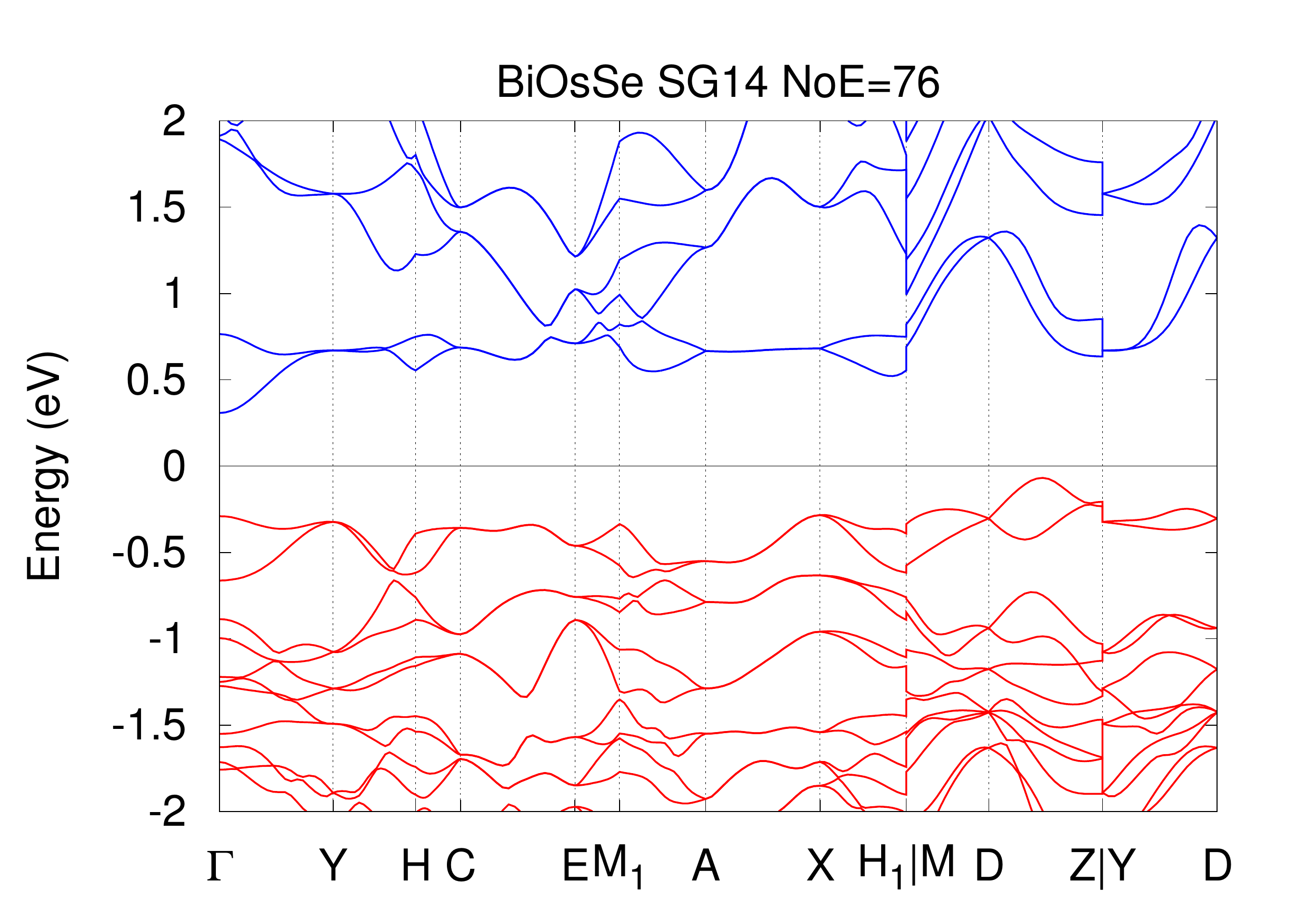}
}
\subfigure[DyP$_{5}$ SG11 NoA=12 NoE=68]{
\label{subfig:409184}
\includegraphics[scale=0.32,angle=270]{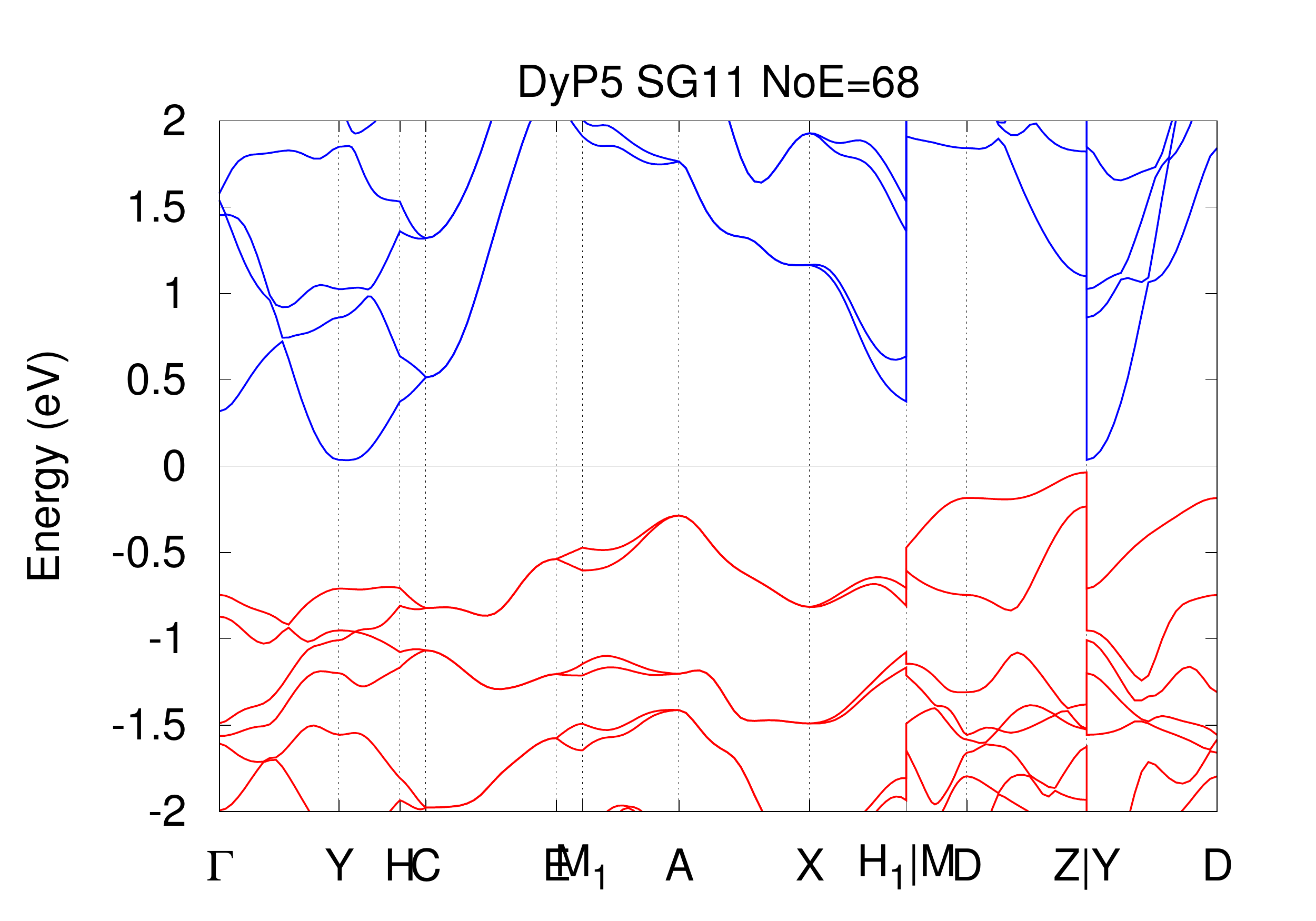}
}
\subfigure[FePS SG14 NoA=12 NoE=76]{
\label{subfig:633086}
\includegraphics[scale=0.32,angle=270]{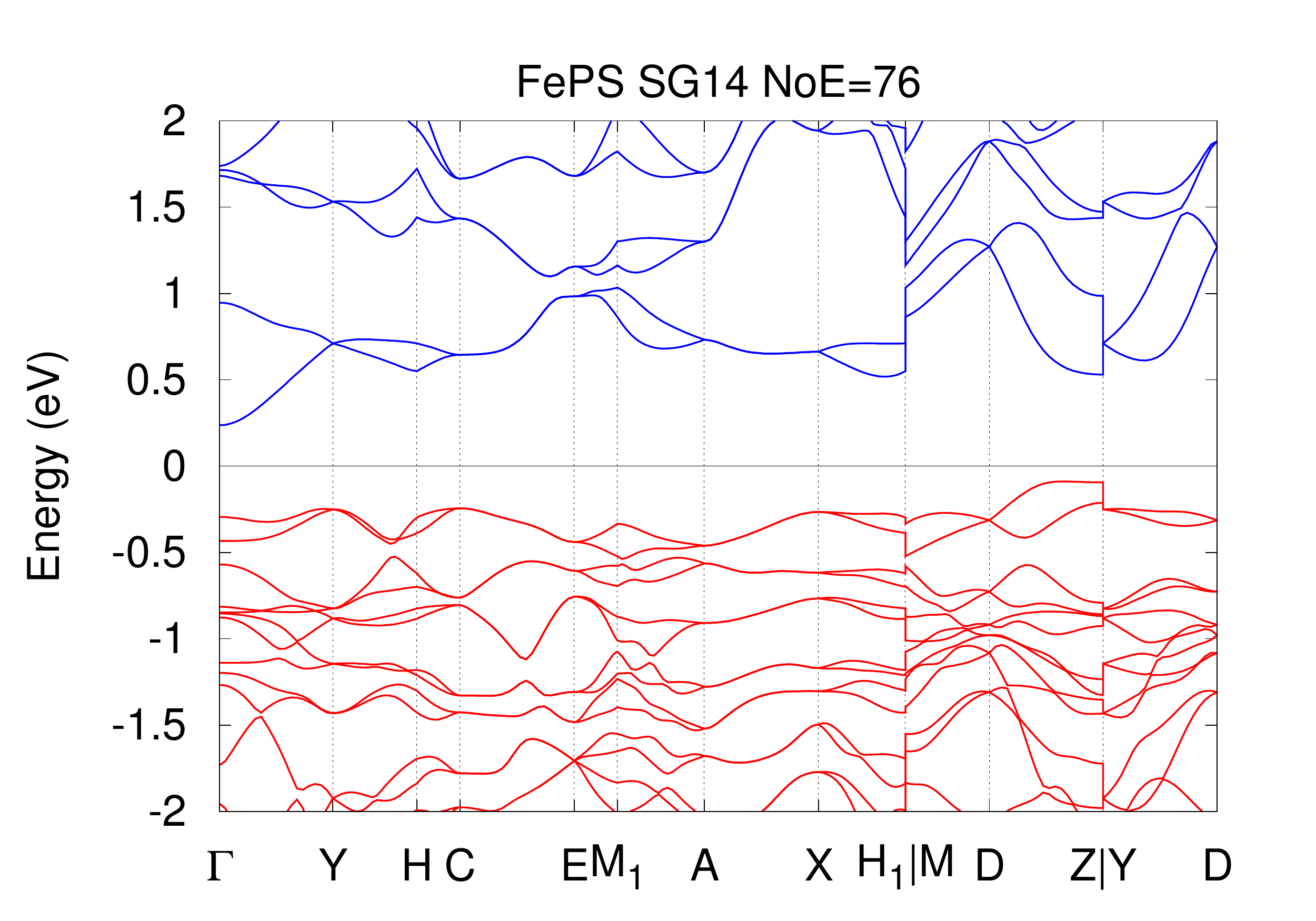}
}
\subfigure[ZnSe$_{2}$ SG205 NoA=12 NoE=96]{
\label{subfig:652213}
\includegraphics[scale=0.32,angle=270]{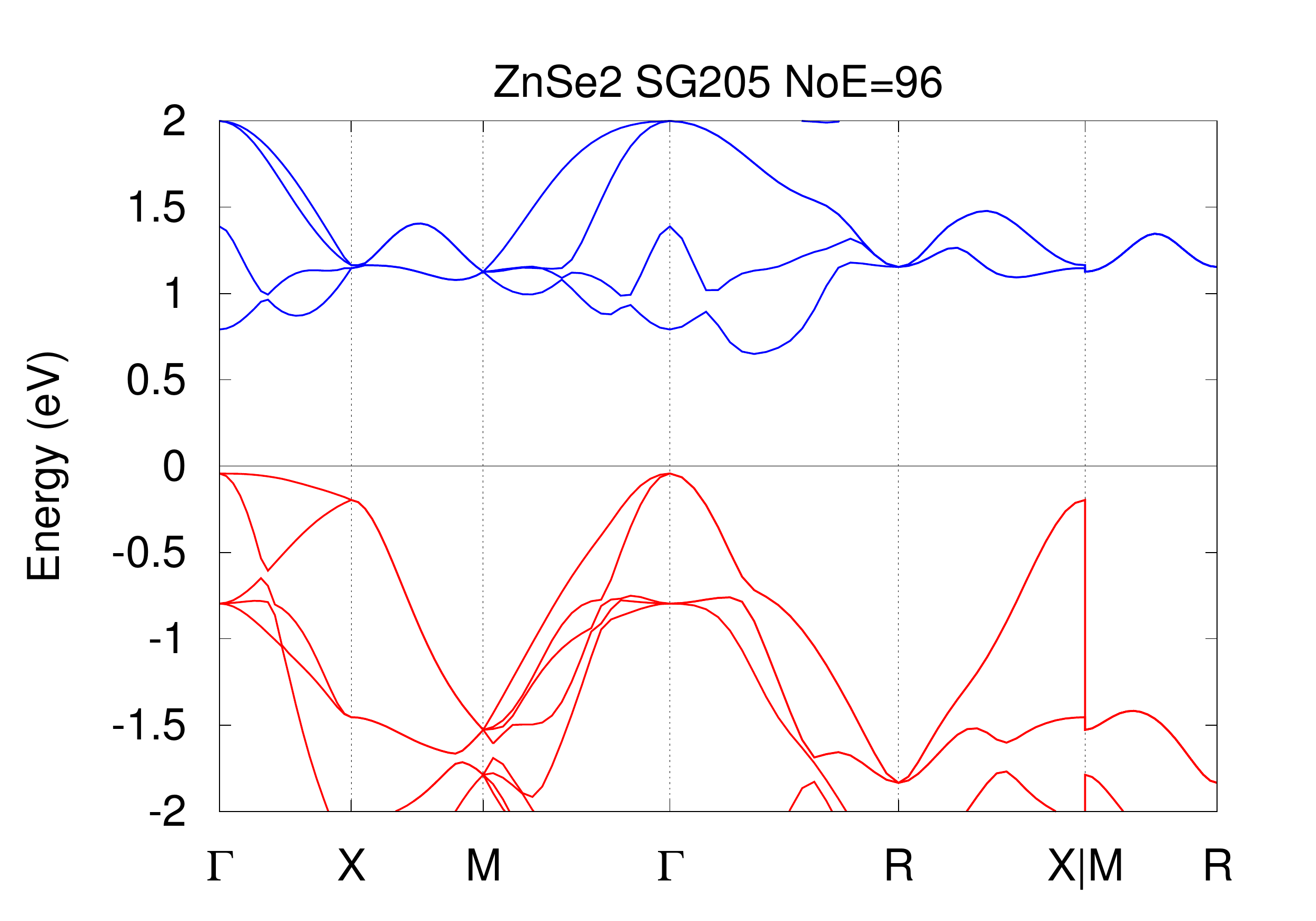}
}
\subfigure[TmP$_{5}$ SG11 NoA=12 NoE=68]{
\label{subfig:409186}
\includegraphics[scale=0.32,angle=270]{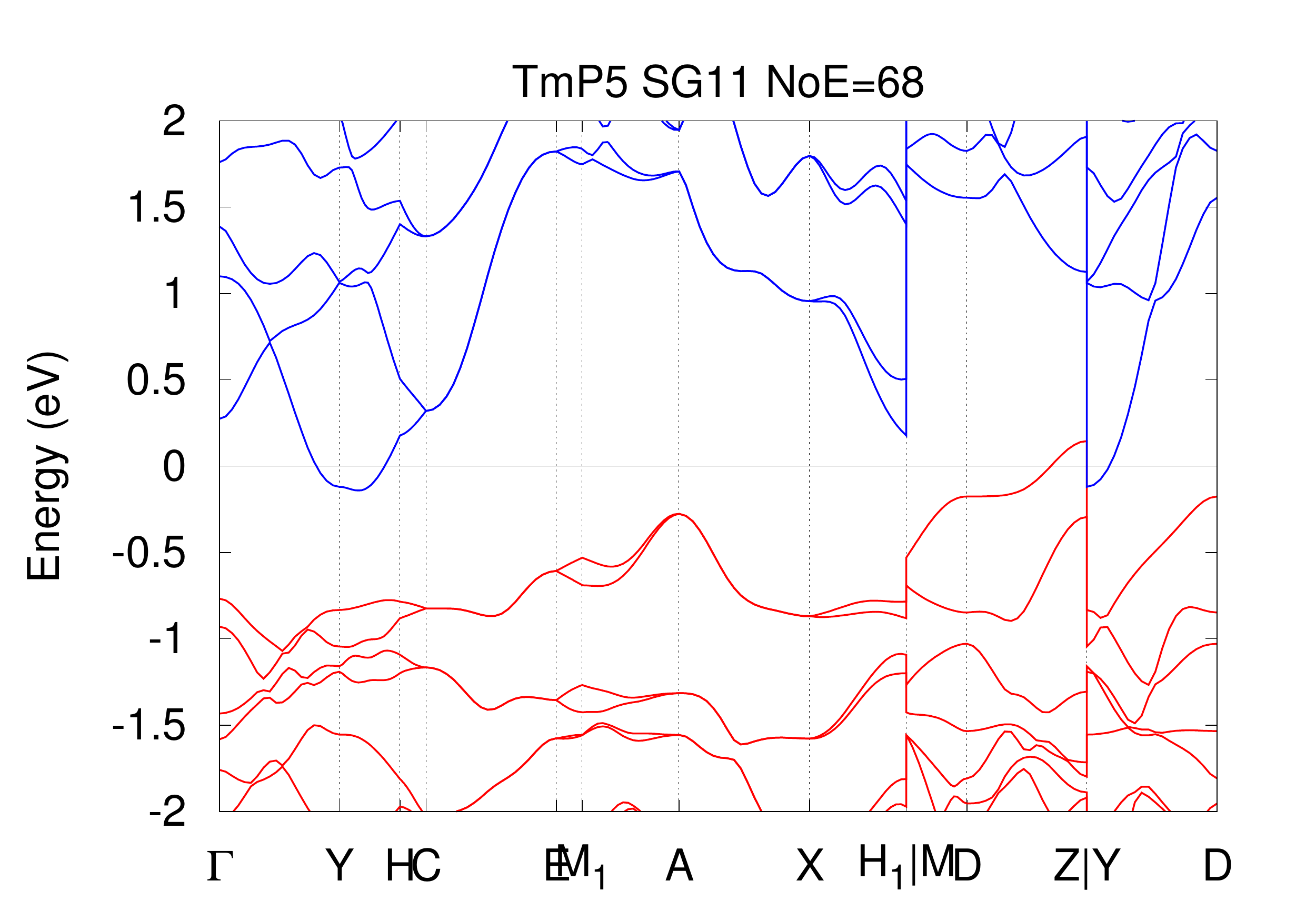}
}
\subfigure[CuAsS SG62 NoA=12 NoE=88]{
\label{subfig:34450}
\includegraphics[scale=0.32,angle=270]{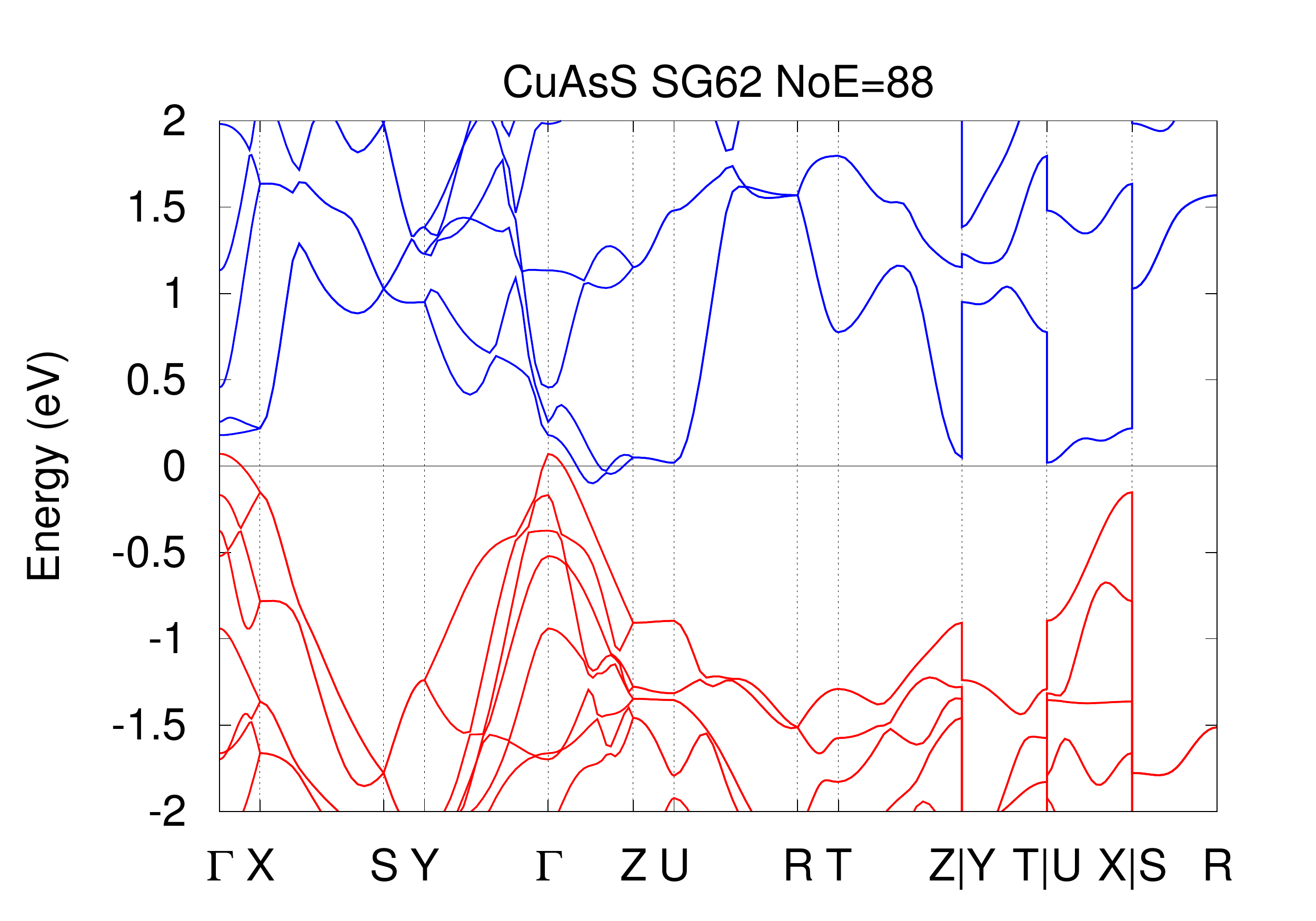}
}
\subfigure[CoAsS SG14 NoA=12 NoE=80]{
\label{subfig:41858}
\includegraphics[scale=0.32,angle=270]{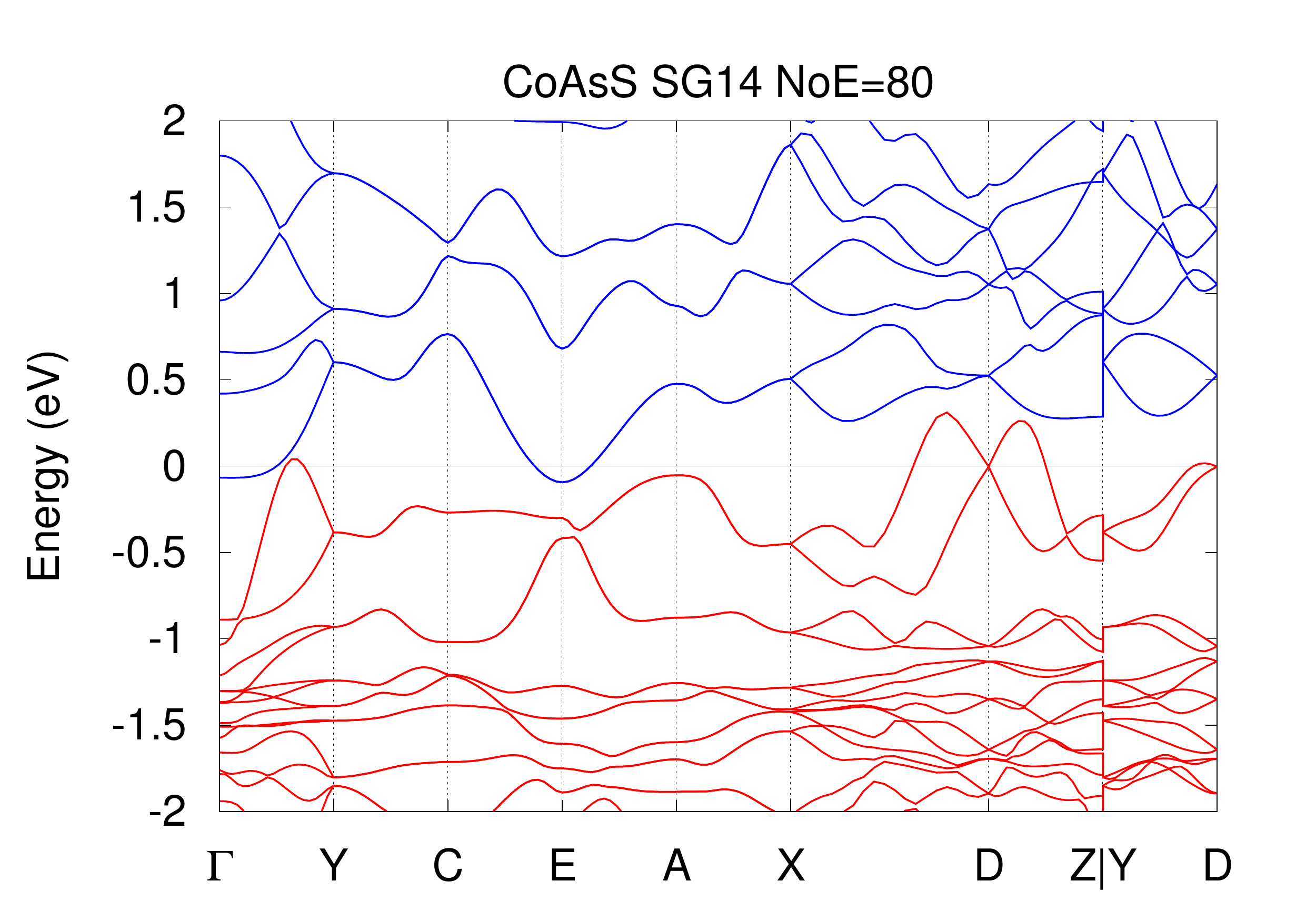}
}
\caption{\hyperref[tab:electride]{back to the table}}
\end{figure}

\begin{figure}[htp]
 \centering
\subfigure[ErS$_{2}$ SG14 NoA=12 NoE=84]{
\label{subfig:422673}
\includegraphics[scale=0.32,angle=270]{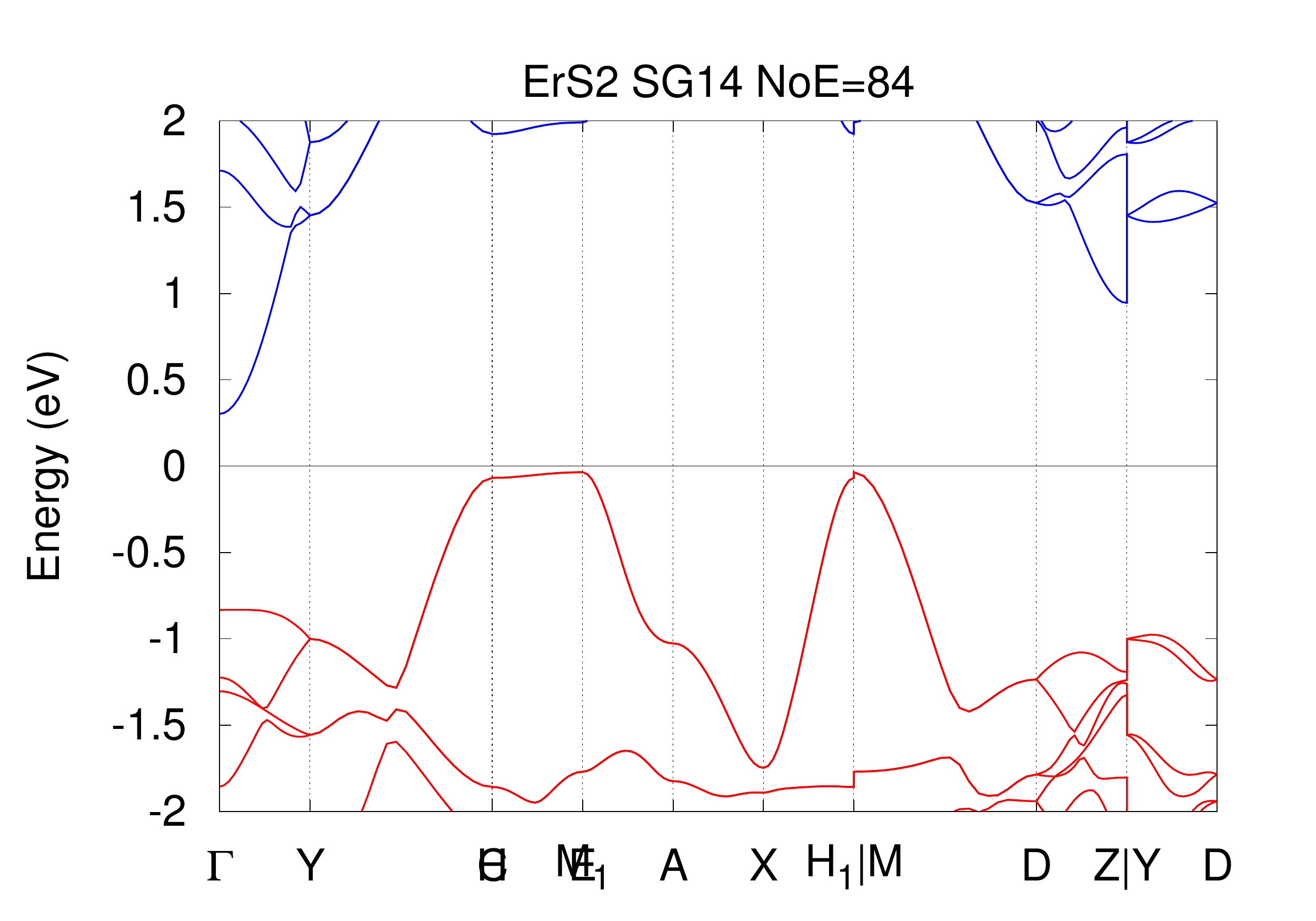}
}
\subfigure[NdSe$_{2}$ SG14 NoA=12 NoE=92]{
\label{subfig:414619}
\includegraphics[scale=0.32,angle=270]{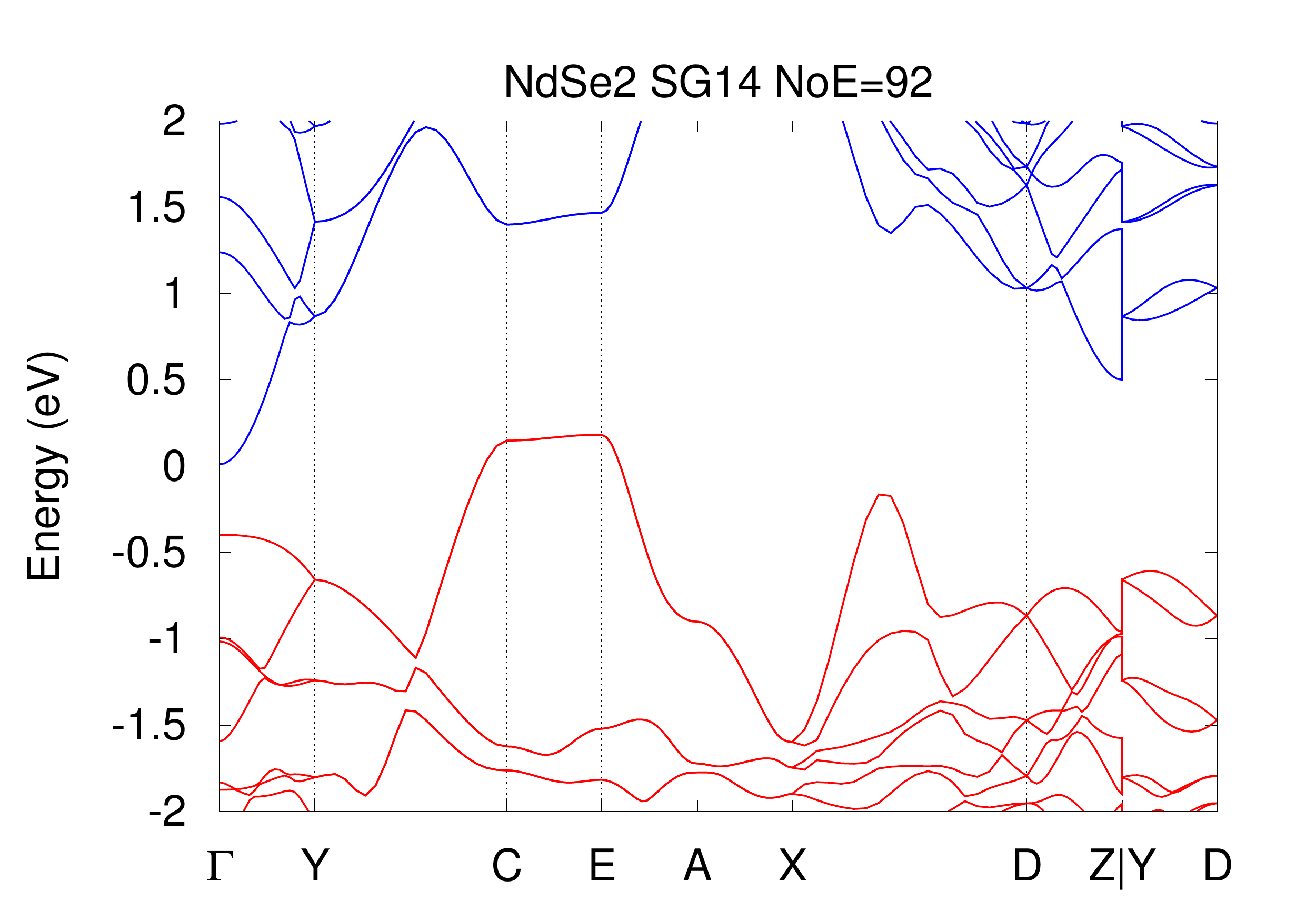}
}
\subfigure[FeAsSe SG14 NoA=12 NoE=76]{
\label{subfig:610526}
\includegraphics[scale=0.32,angle=270]{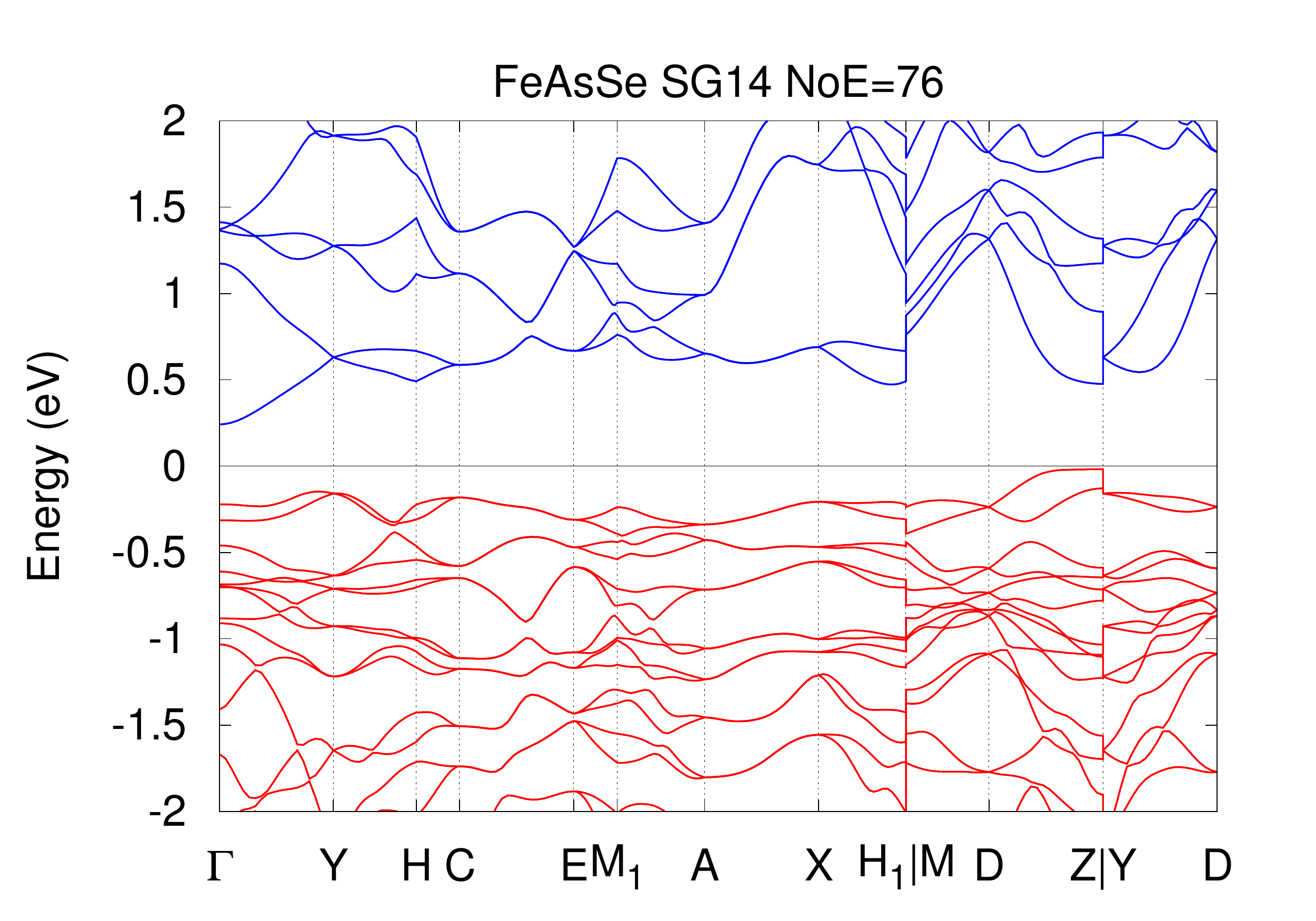}
}
\subfigure[TbS$_{2}$ SG14 NoA=12 NoE=84]{
\label{subfig:422672}
\includegraphics[scale=0.32,angle=270]{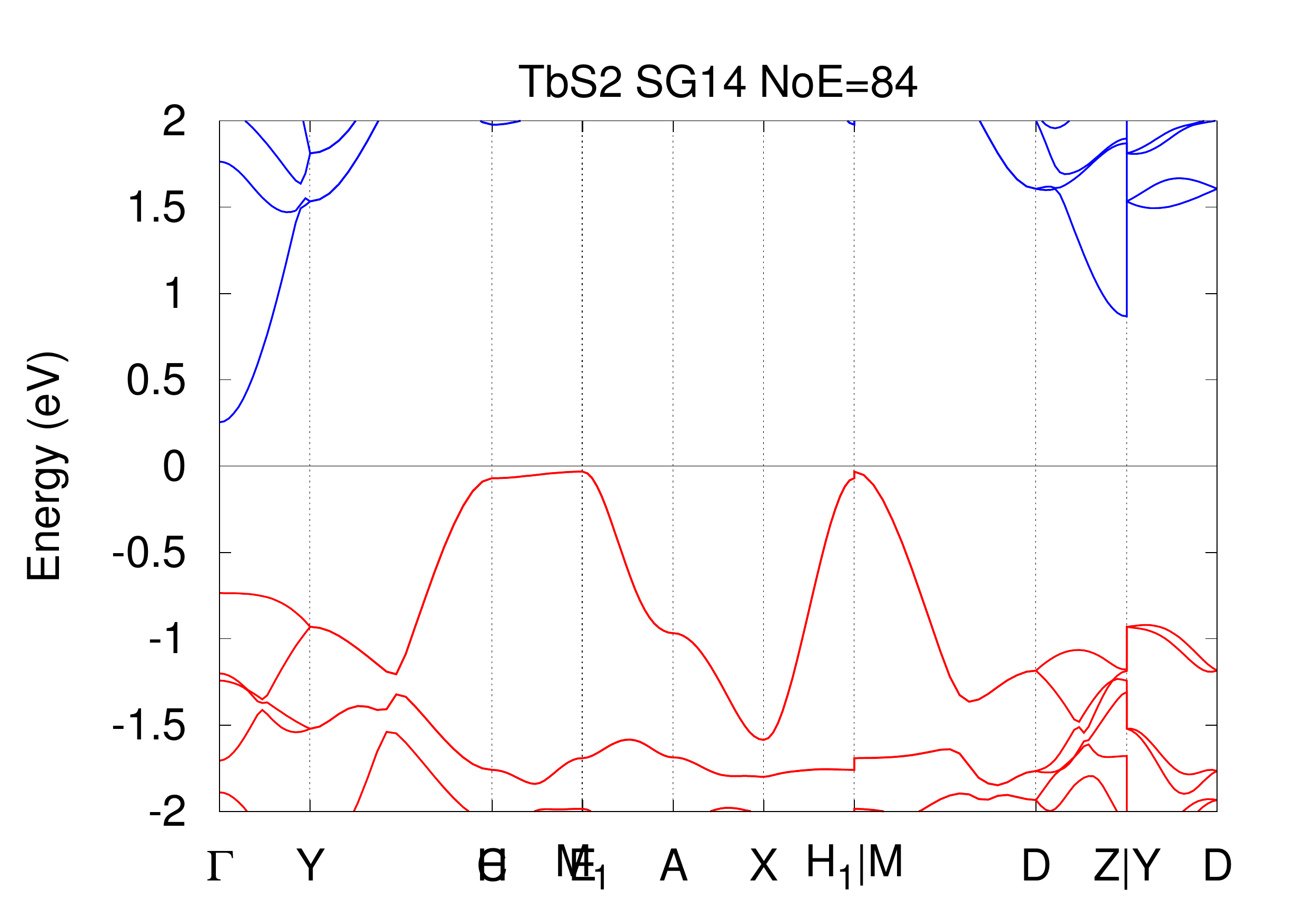}
}
\subfigure[FeS$_{2}$ SG205 NoA=12 NoE=80]{
\label{subfig:15012}
\includegraphics[scale=0.32,angle=270]{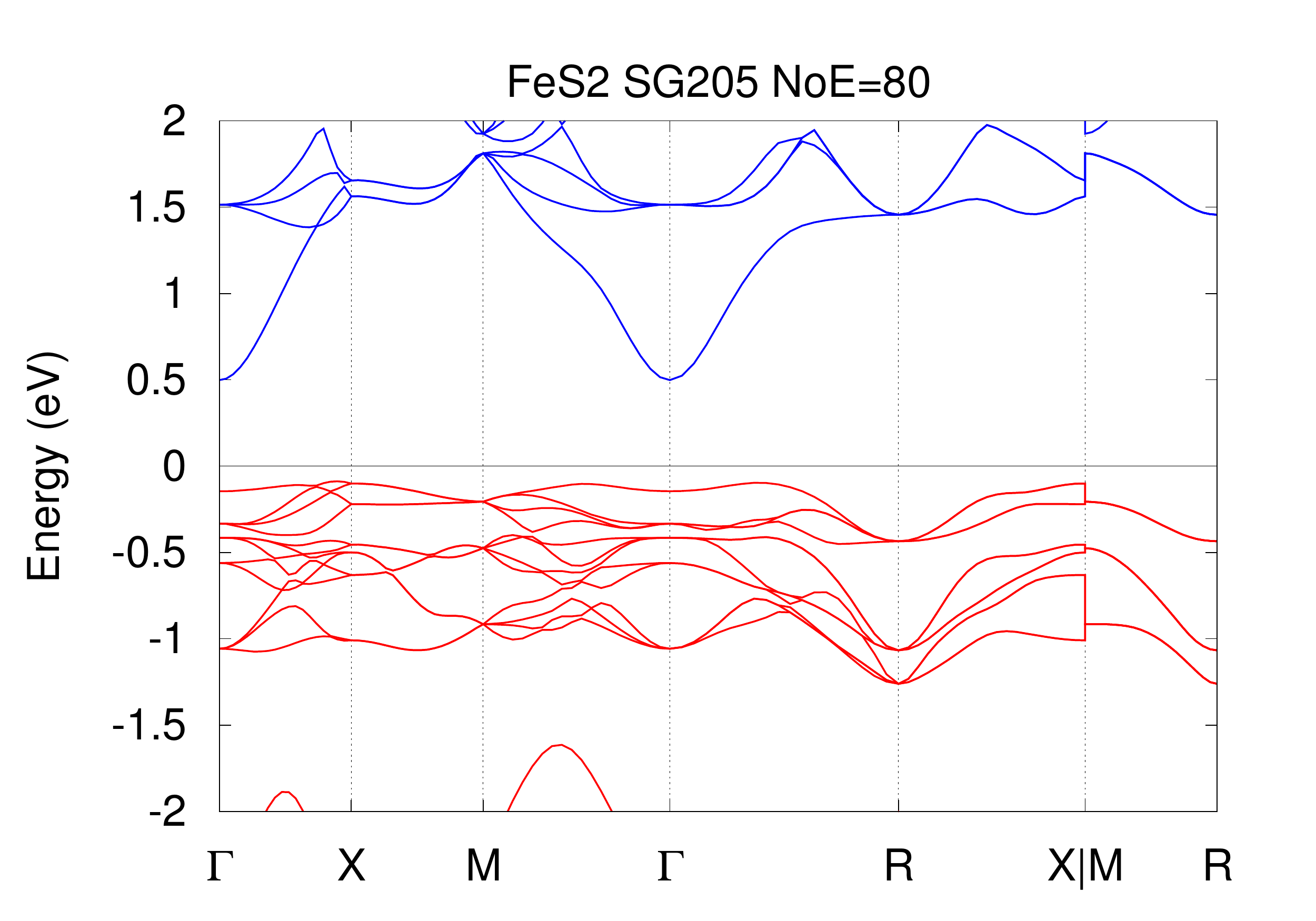}
}
\subfigure[PrP$_{2}$ SG14 NoA=12 NoE=84]{
\label{subfig:647944}
\includegraphics[scale=0.32,angle=270]{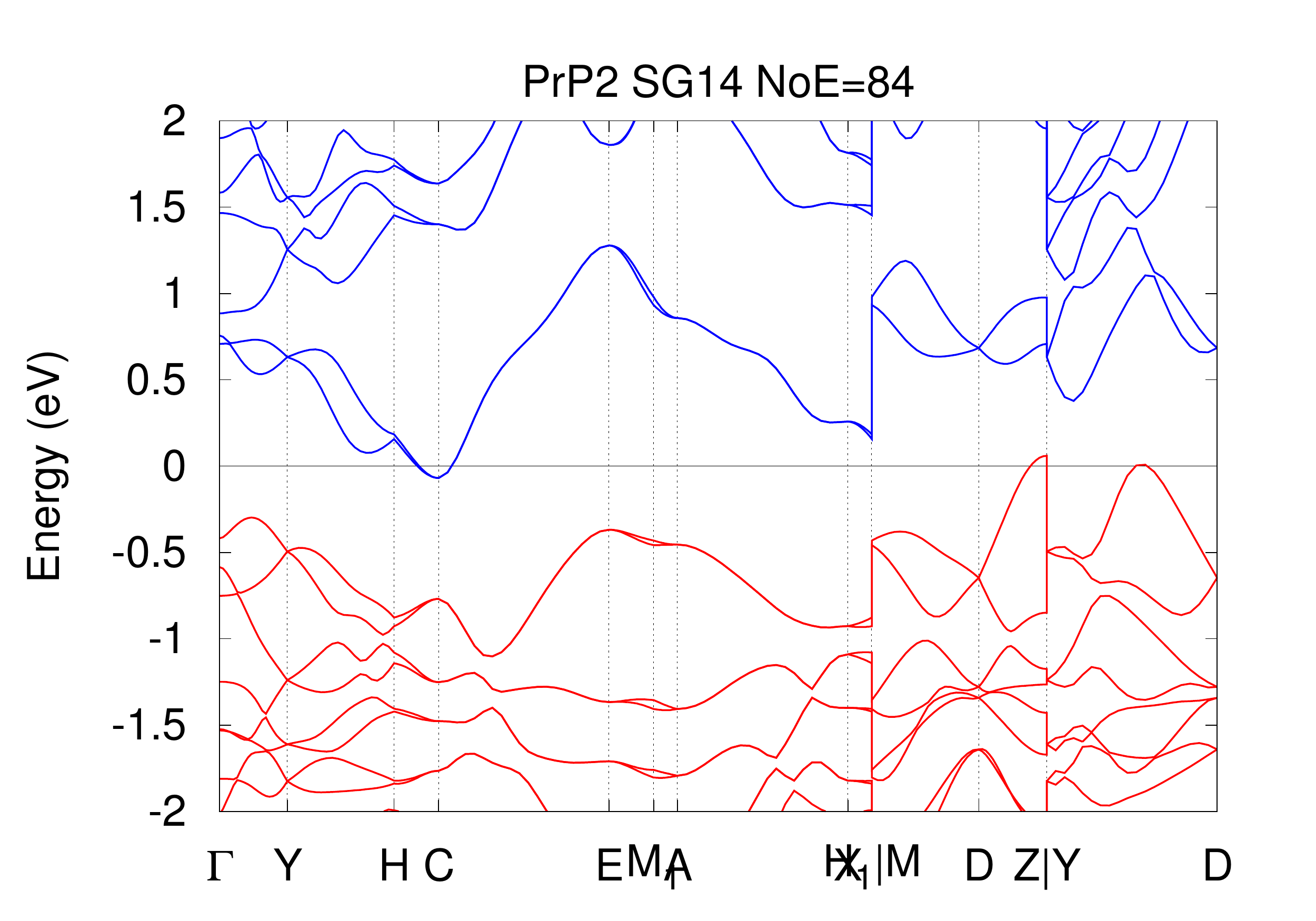}
}
\subfigure[RbGa$_{3}$ SG119 NoA=12 NoE=54]{
\label{subfig:103943}
\includegraphics[scale=0.32,angle=270]{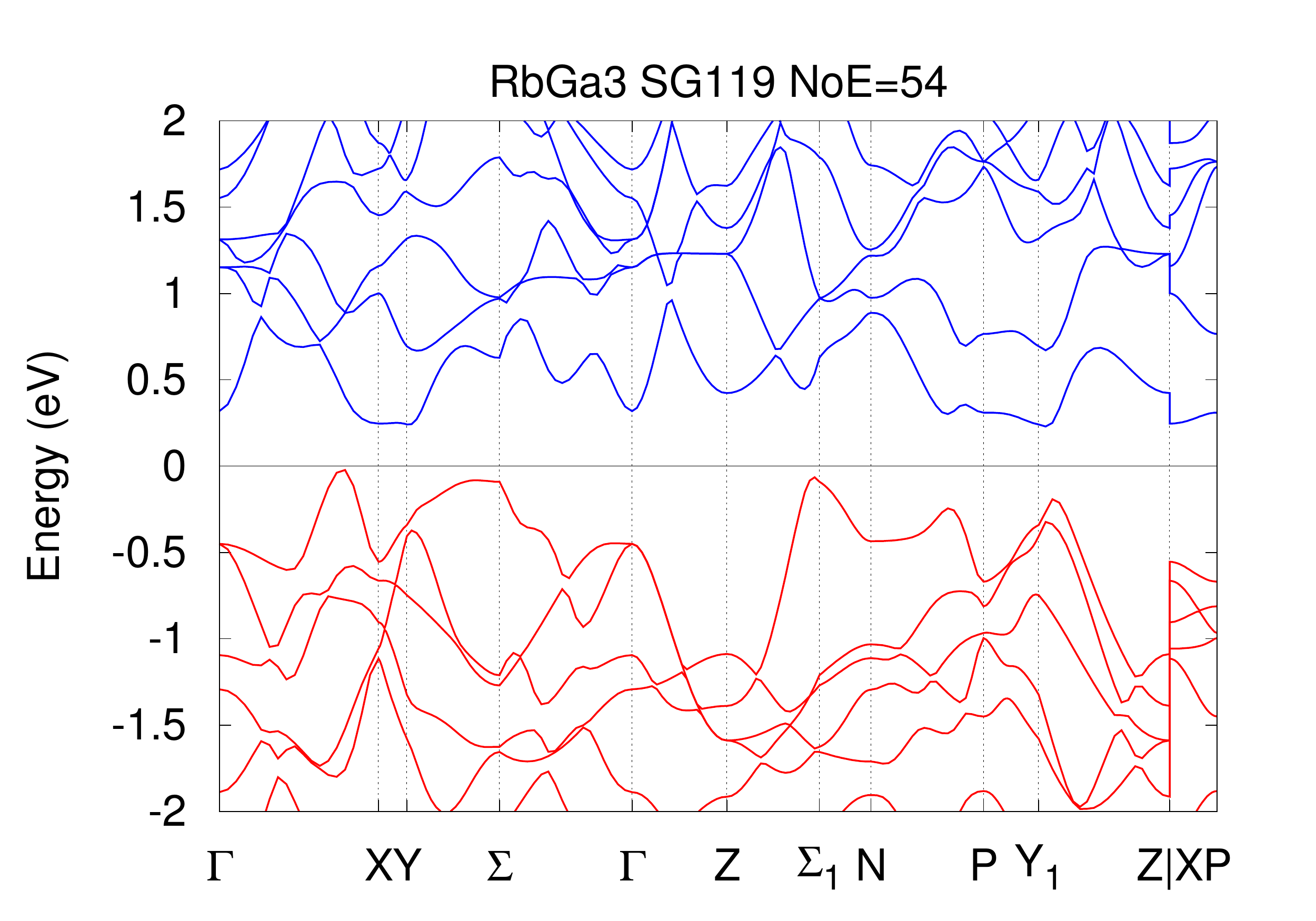}
}
\subfigure[HoP$_{5}$ SG11 NoA=12 NoE=68]{
\label{subfig:409185}
\includegraphics[scale=0.32,angle=270]{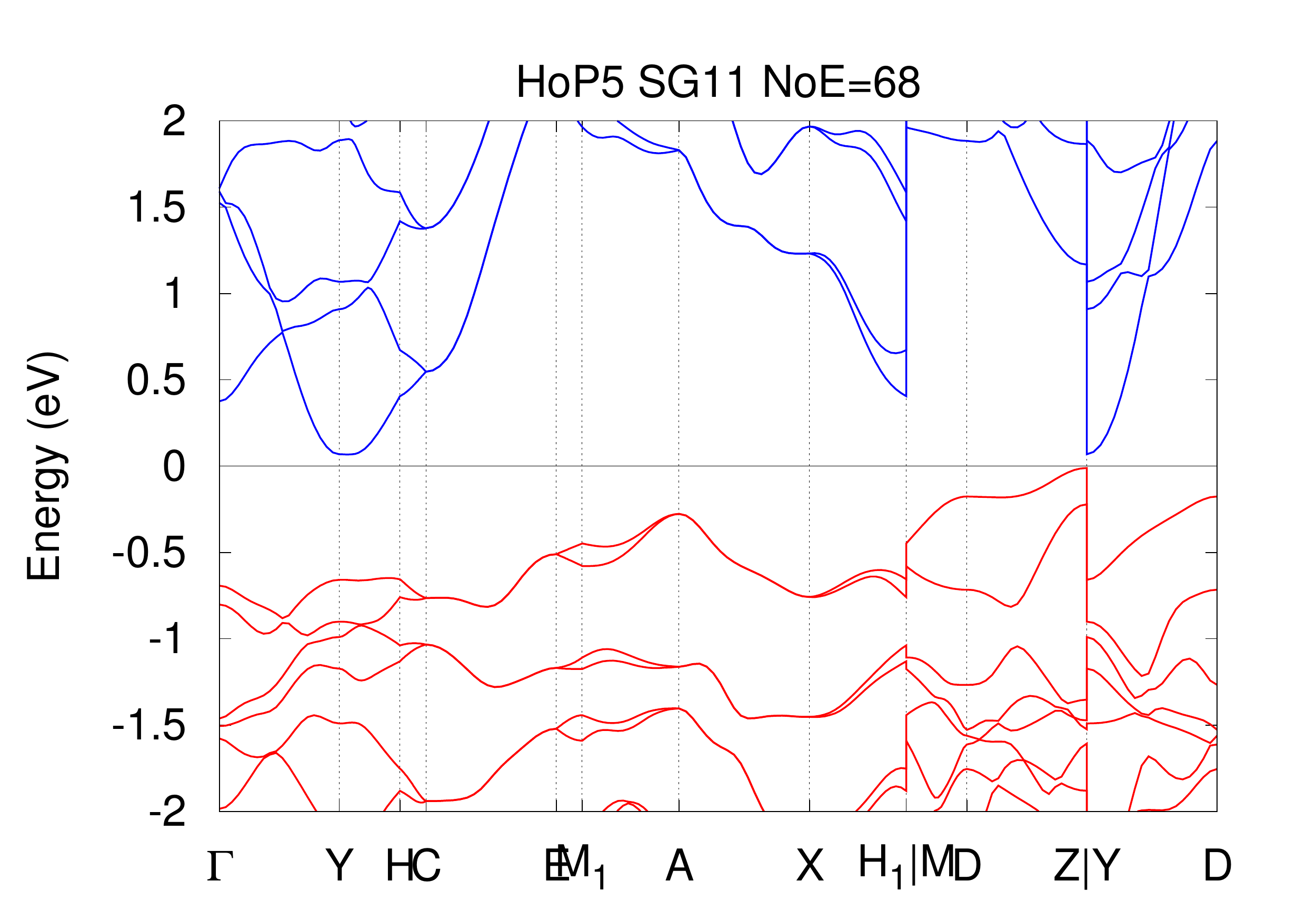}
}
\caption{\hyperref[tab:electride]{back to the table}}
\end{figure}

\begin{figure}[htp]
 \centering
\subfigure[LaZnSn SG194 NoA=12 NoE=108]{
\label{subfig:152623}
\includegraphics[scale=0.32,angle=270]{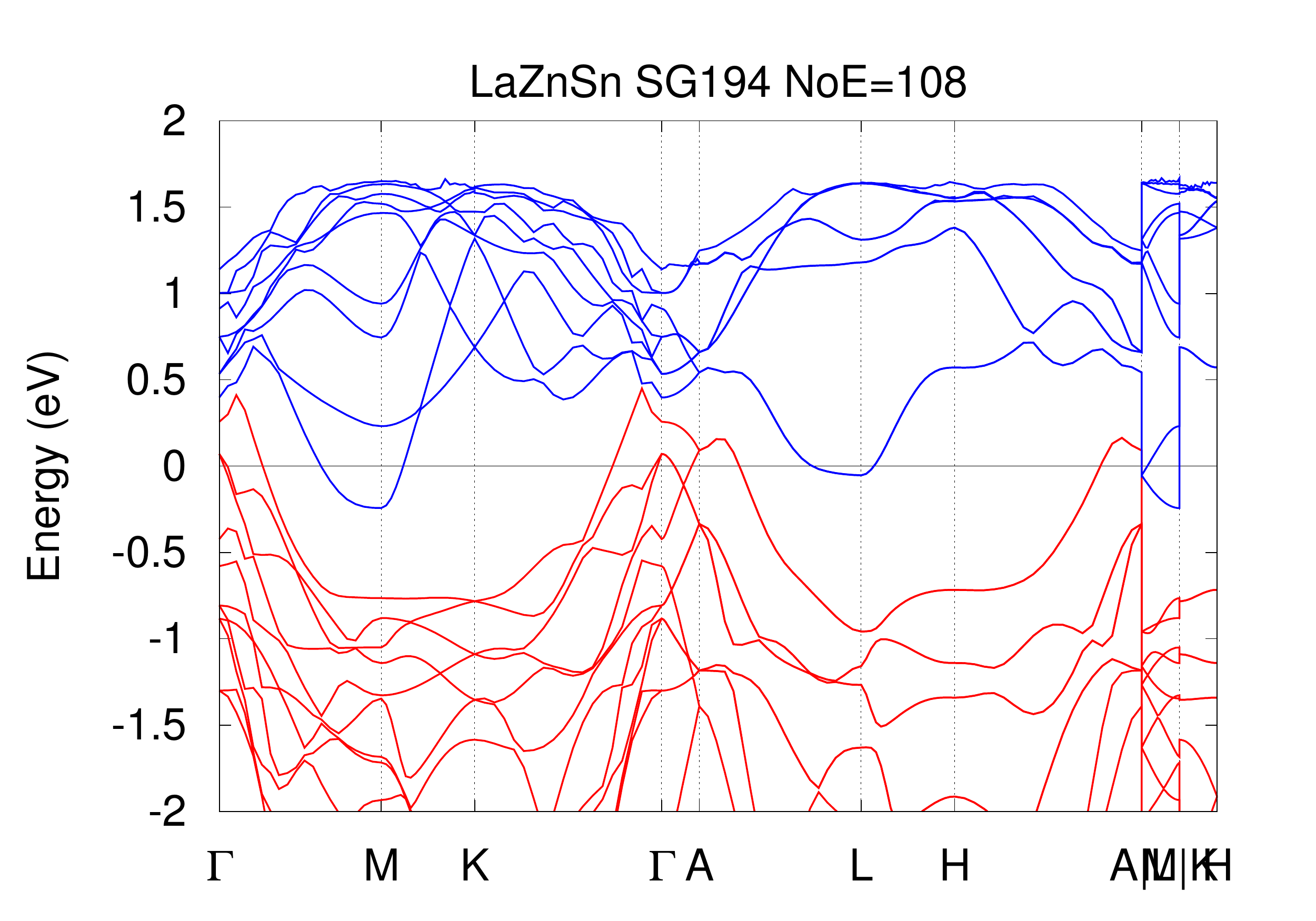}
}
\subfigure[HgO$_{2}$ SG61 NoA=12 NoE=96]{
\label{subfig:24774}
\includegraphics[scale=0.32,angle=270]{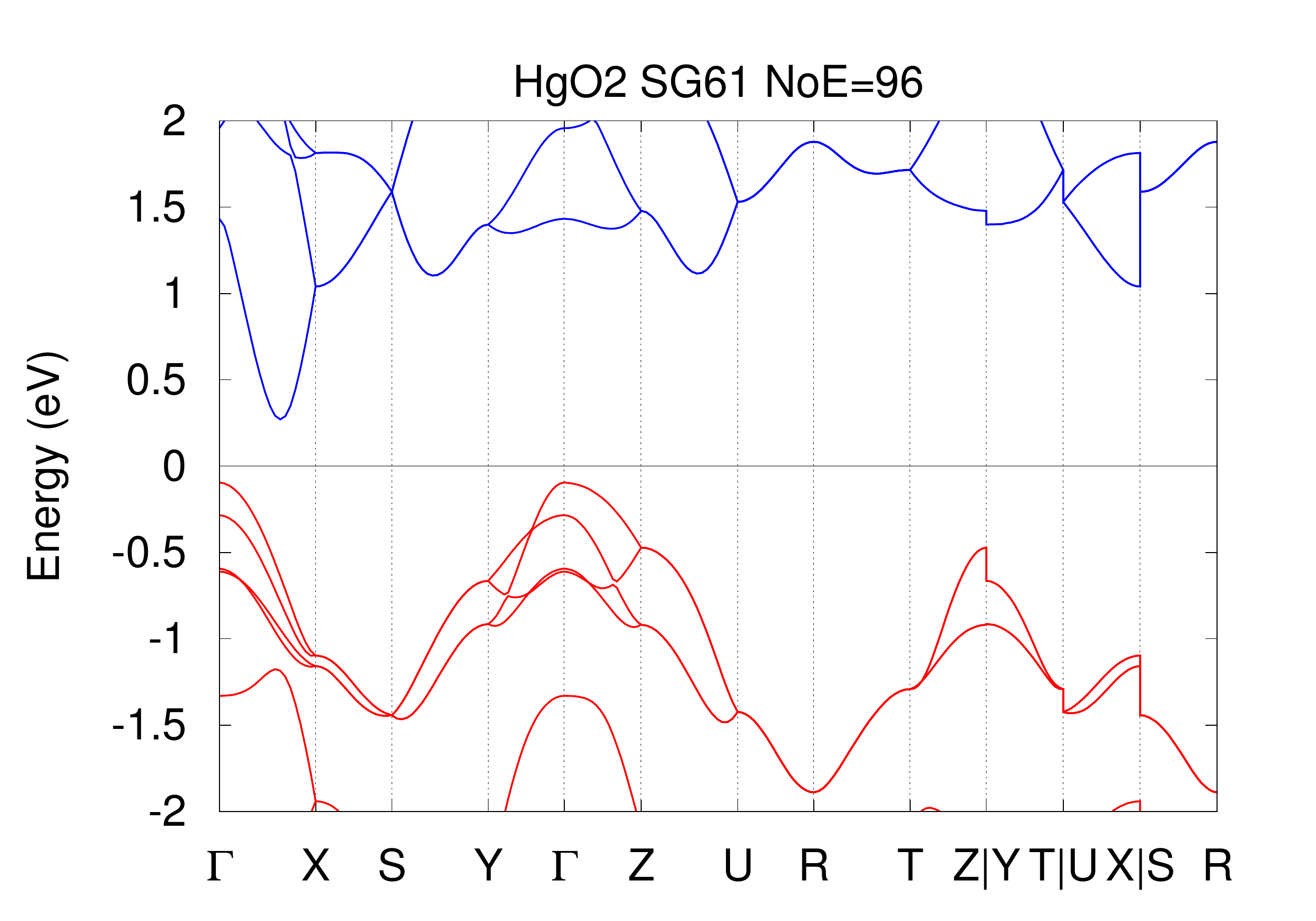}
}
\subfigure[PdN$_{2}$ SG205 NoA=12 NoE=80]{
\label{subfig:191245}
\includegraphics[scale=0.32,angle=270]{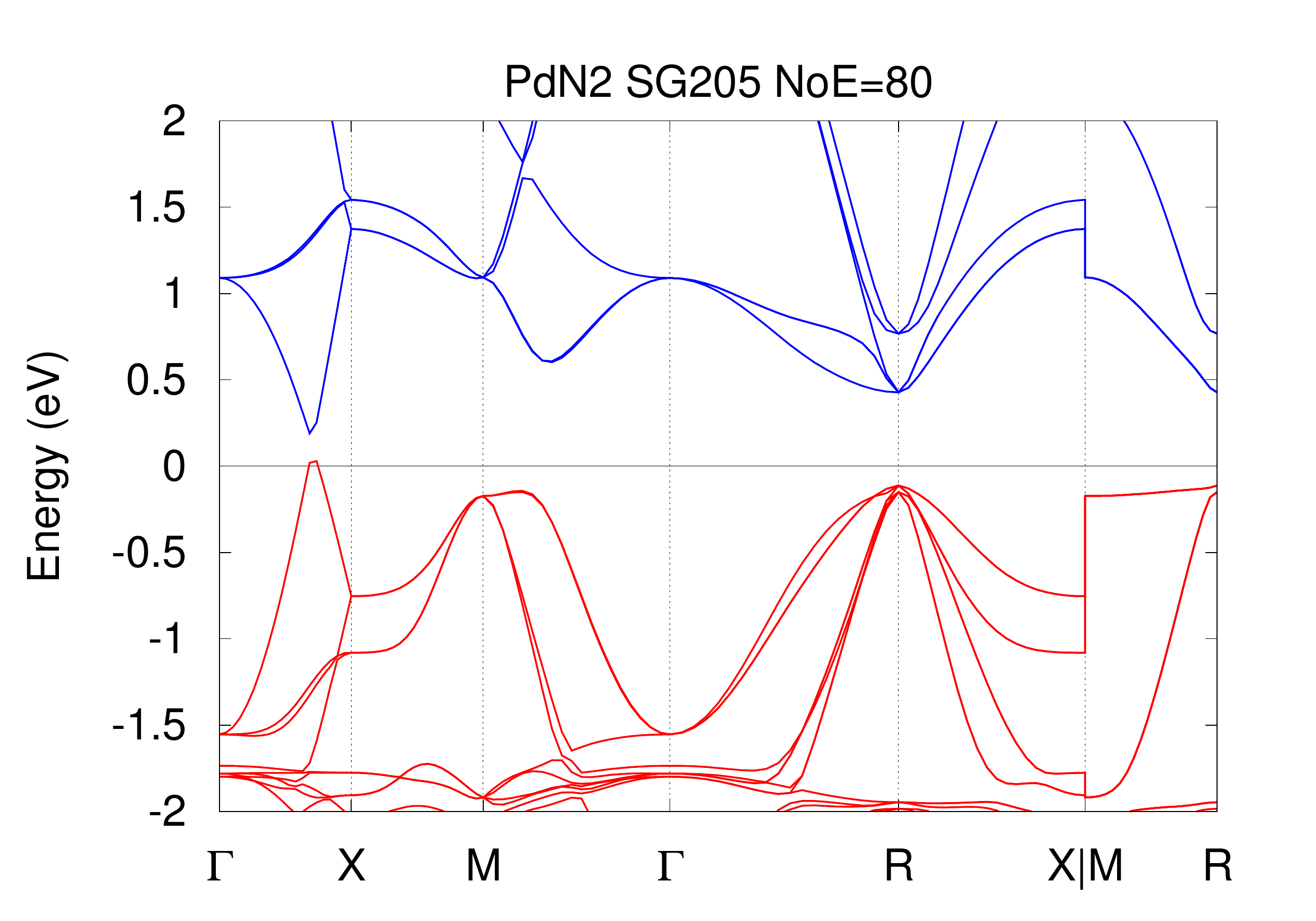}
}
\subfigure[RhN$_{2}$ SG14 NoA=12 NoE=76]{
\label{subfig:160624}
\includegraphics[scale=0.32,angle=270]{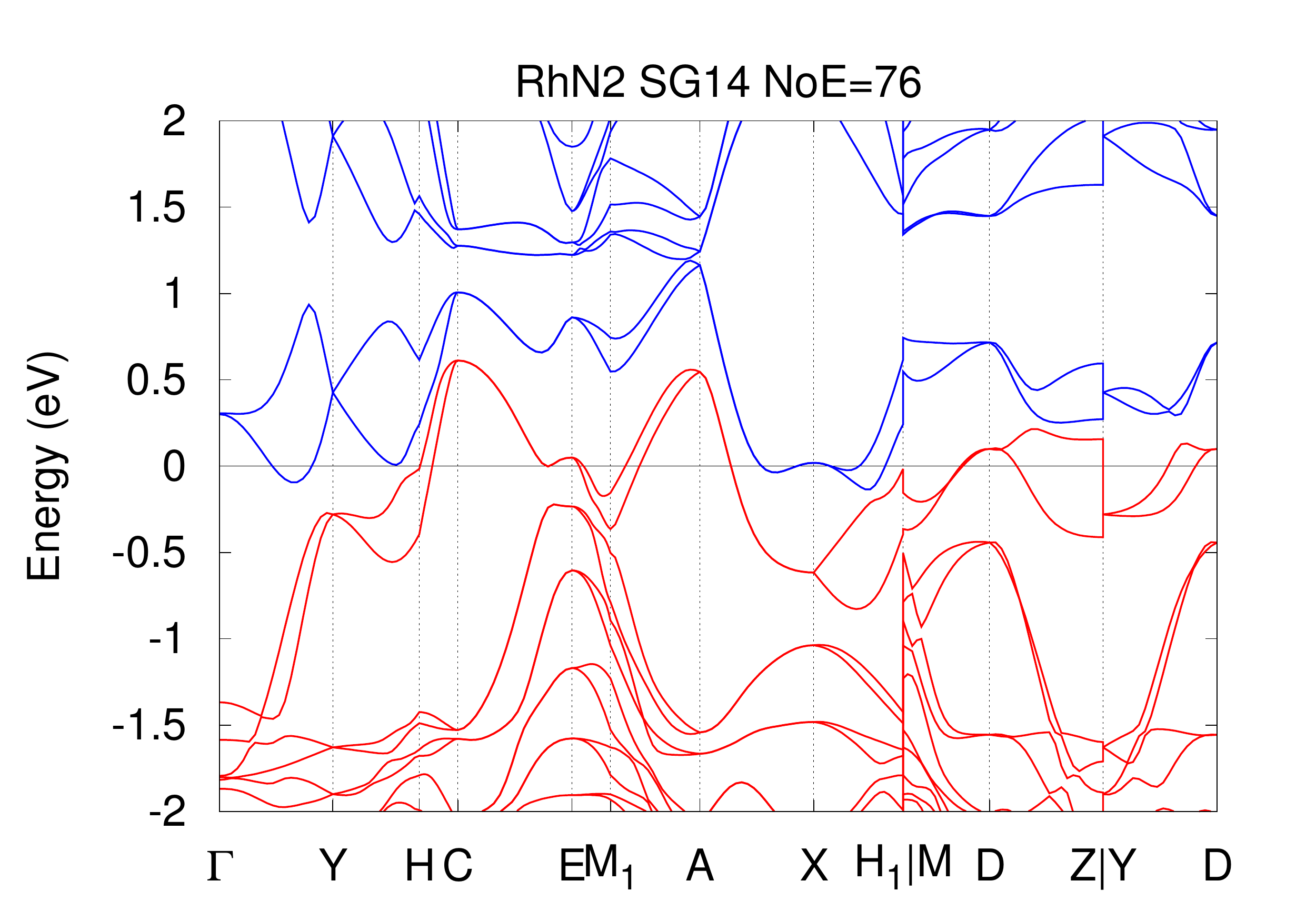}
}
\subfigure[RuS$_{2}$ SG205 NoA=12 NoE=80]{
\label{subfig:600680}
\includegraphics[scale=0.32,angle=270]{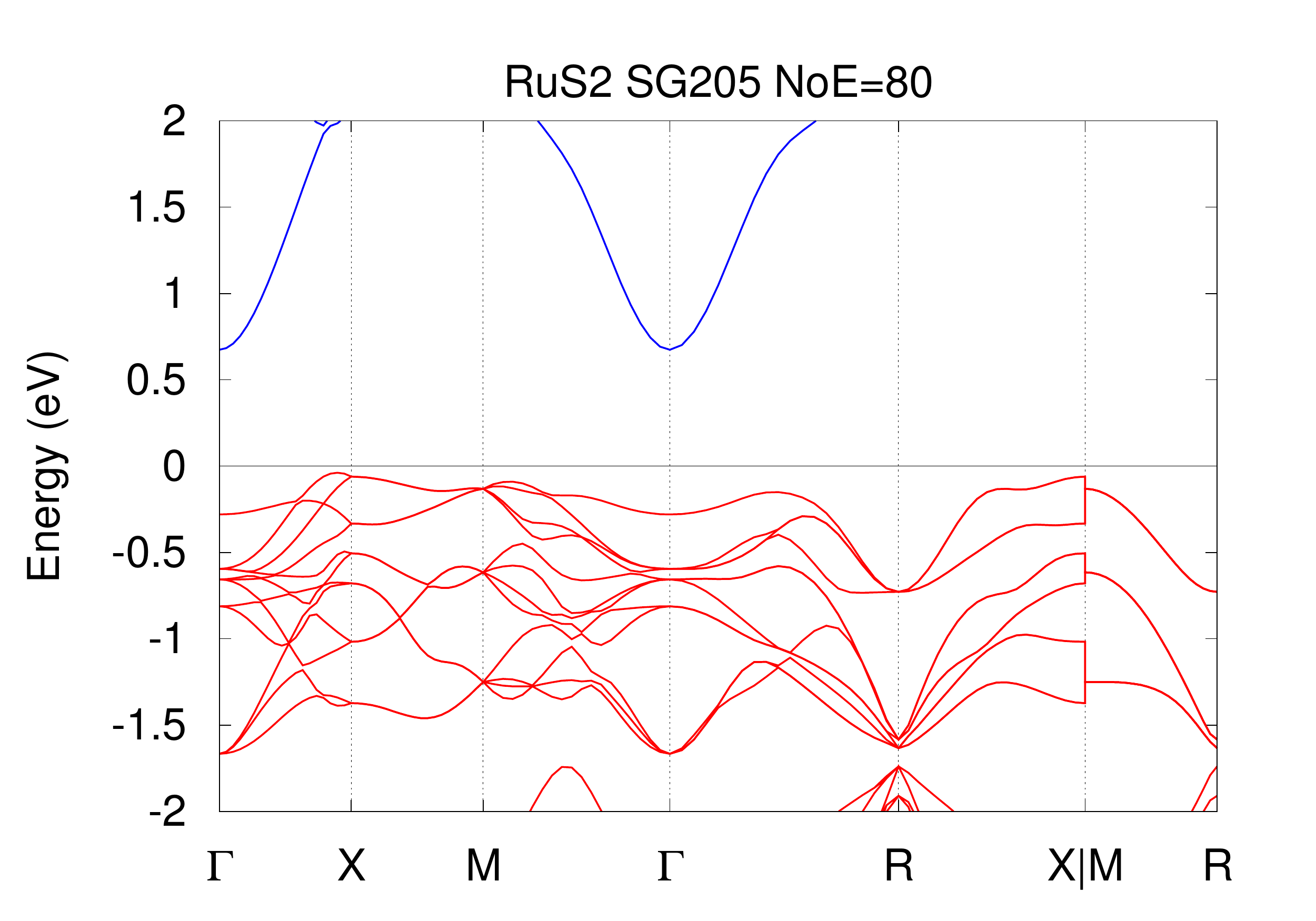}
}
\subfigure[PrP$_{5}$ SG11 NoA=12 NoE=72]{
\label{subfig:409182}
\includegraphics[scale=0.32,angle=270]{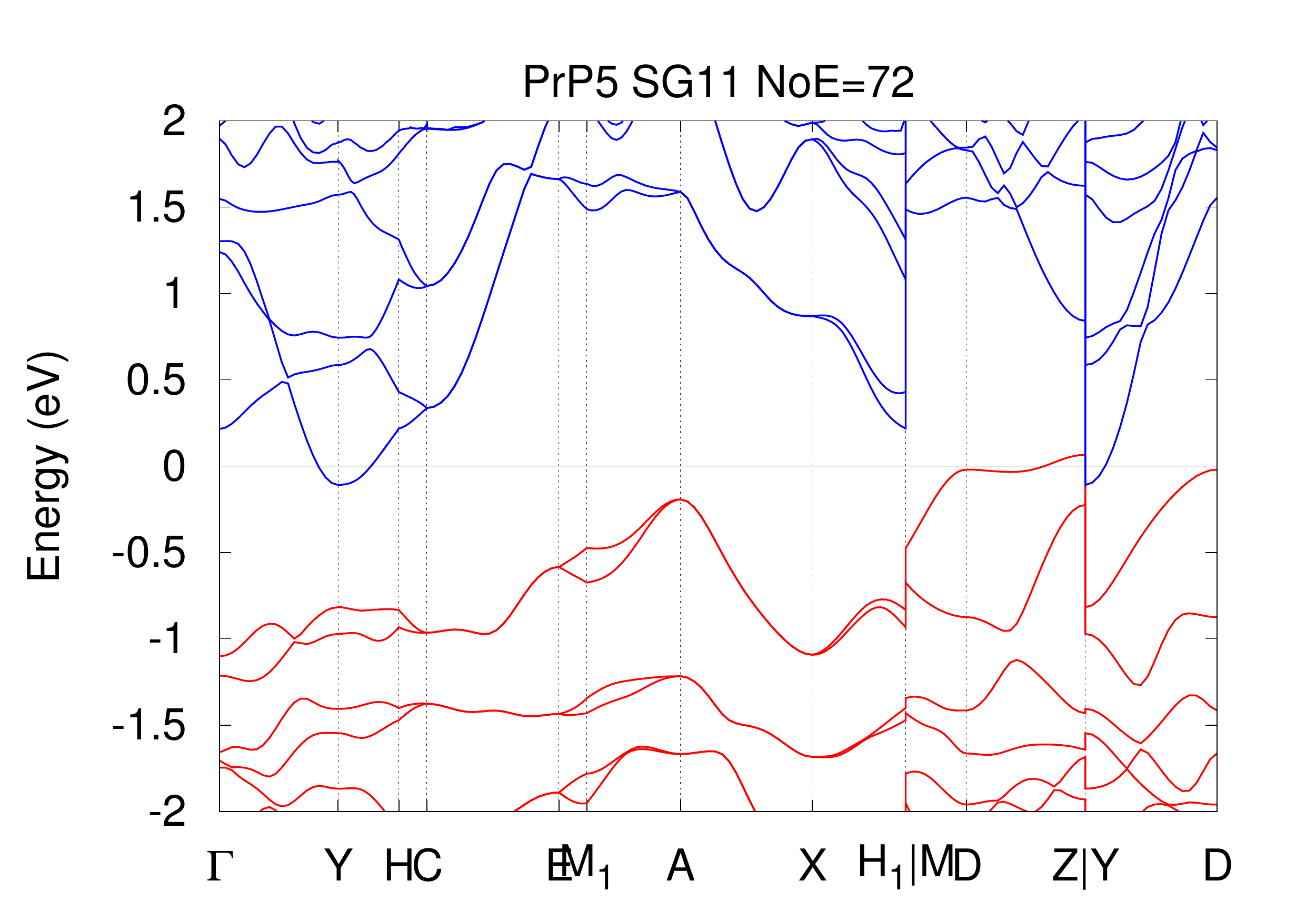}
}
\subfigure[FeSe$_{2}$ SG205 NoA=12 NoE=80]{
\label{subfig:633475}
\includegraphics[scale=0.32,angle=270]{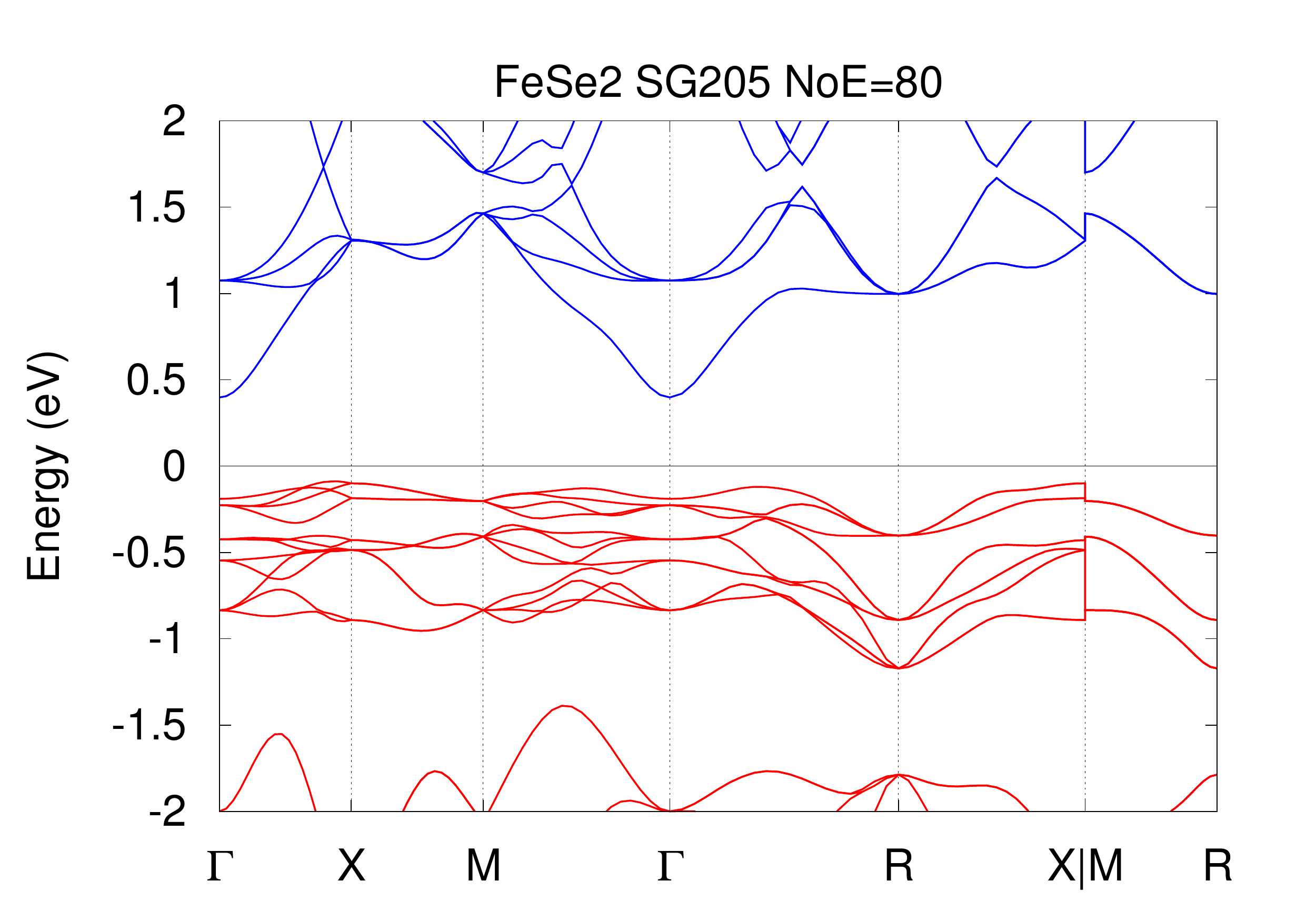}
}
\subfigure[CsGa$_{3}$ SG119 NoA=12 NoE=54]{
\label{subfig:102863}
\includegraphics[scale=0.32,angle=270]{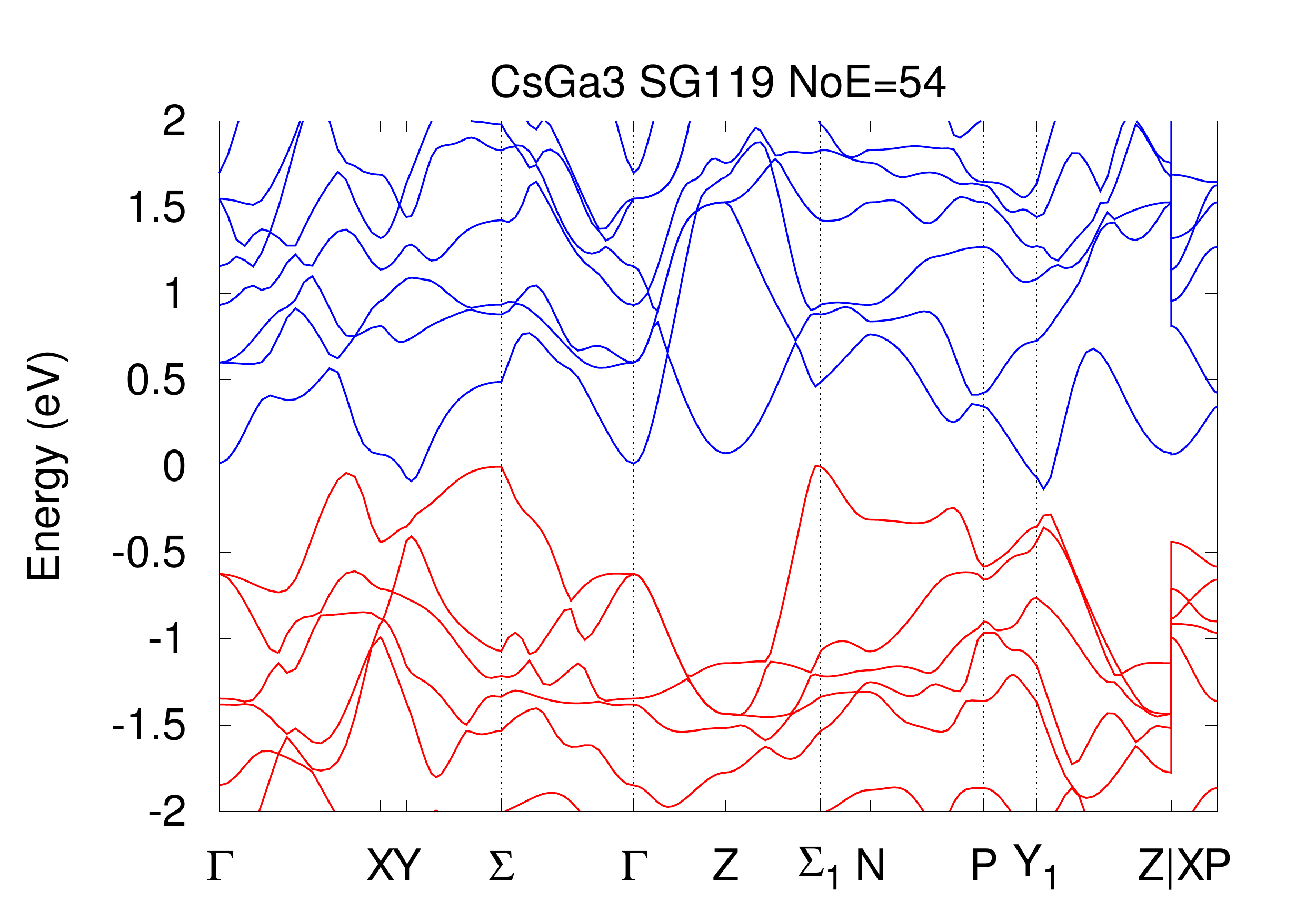}
}
\caption{\hyperref[tab:electride]{back to the table}}
\end{figure}

\begin{figure}[htp]
 \centering
\subfigure[FeSbSe SG14 NoA=12 NoE=76]{
\label{subfig:633399}
\includegraphics[scale=0.32,angle=270]{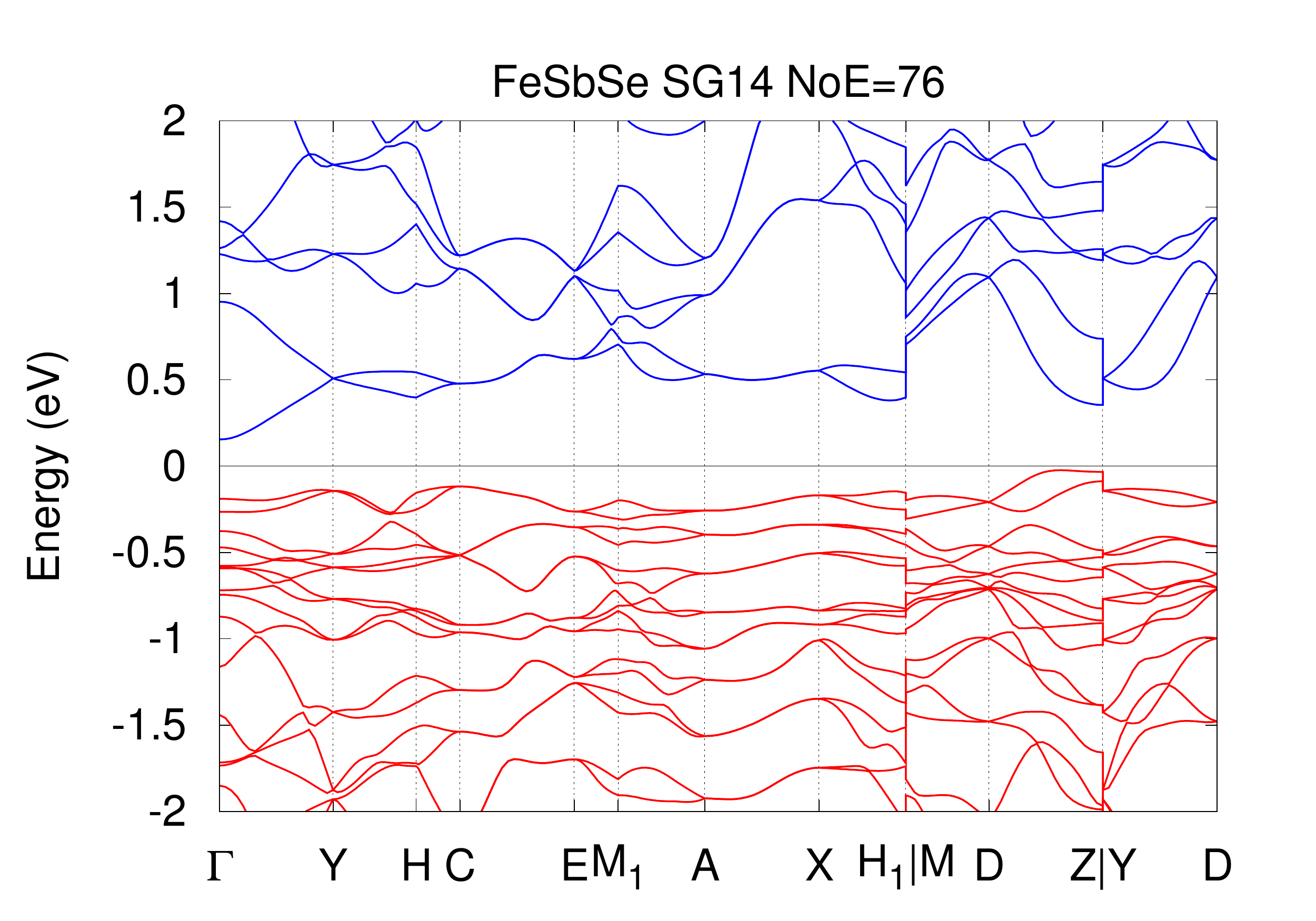}
}
\subfigure[PrSe$_{2}$ SG14 NoA=12 NoE=92]{
\label{subfig:413527}
\includegraphics[scale=0.32,angle=270]{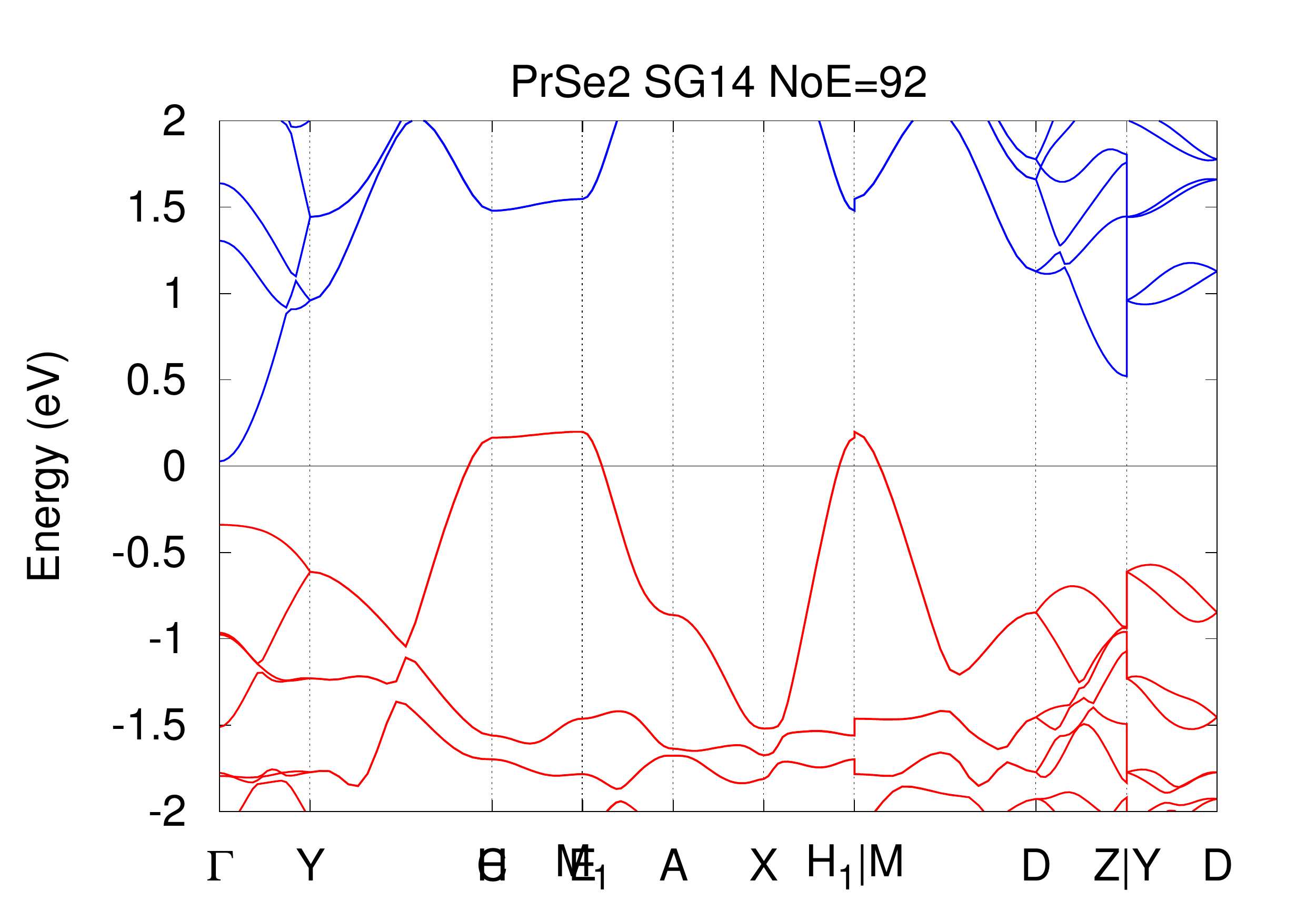}
}
\subfigure[NiP$_{2}$ SG205 NoA=12 NoE=80]{
\label{subfig:22221}
\includegraphics[scale=0.32,angle=270]{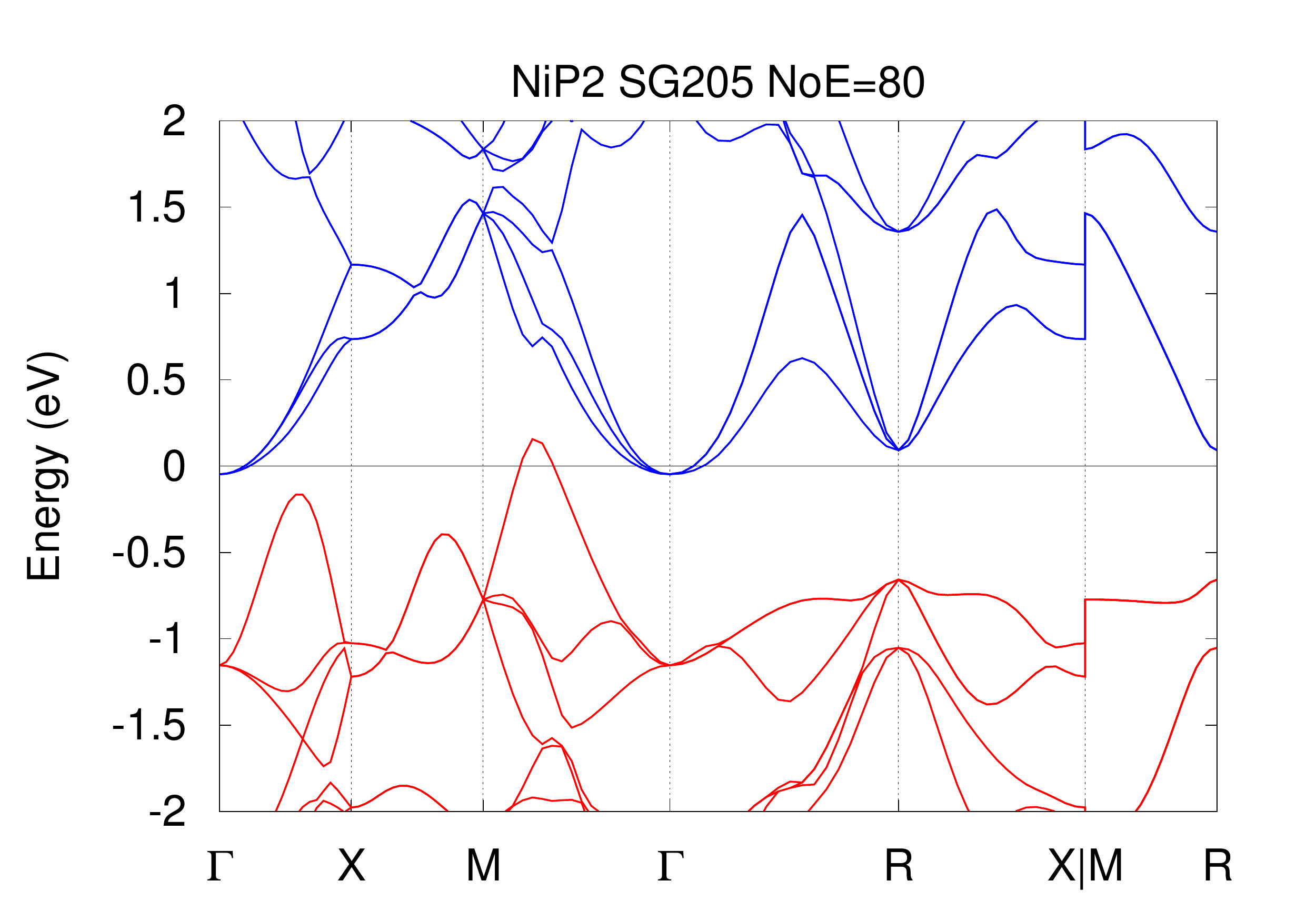}
}
\subfigure[LuP$_{5}$ SG11 NoA=12 NoE=68]{
\label{subfig:409187}
\includegraphics[scale=0.32,angle=270]{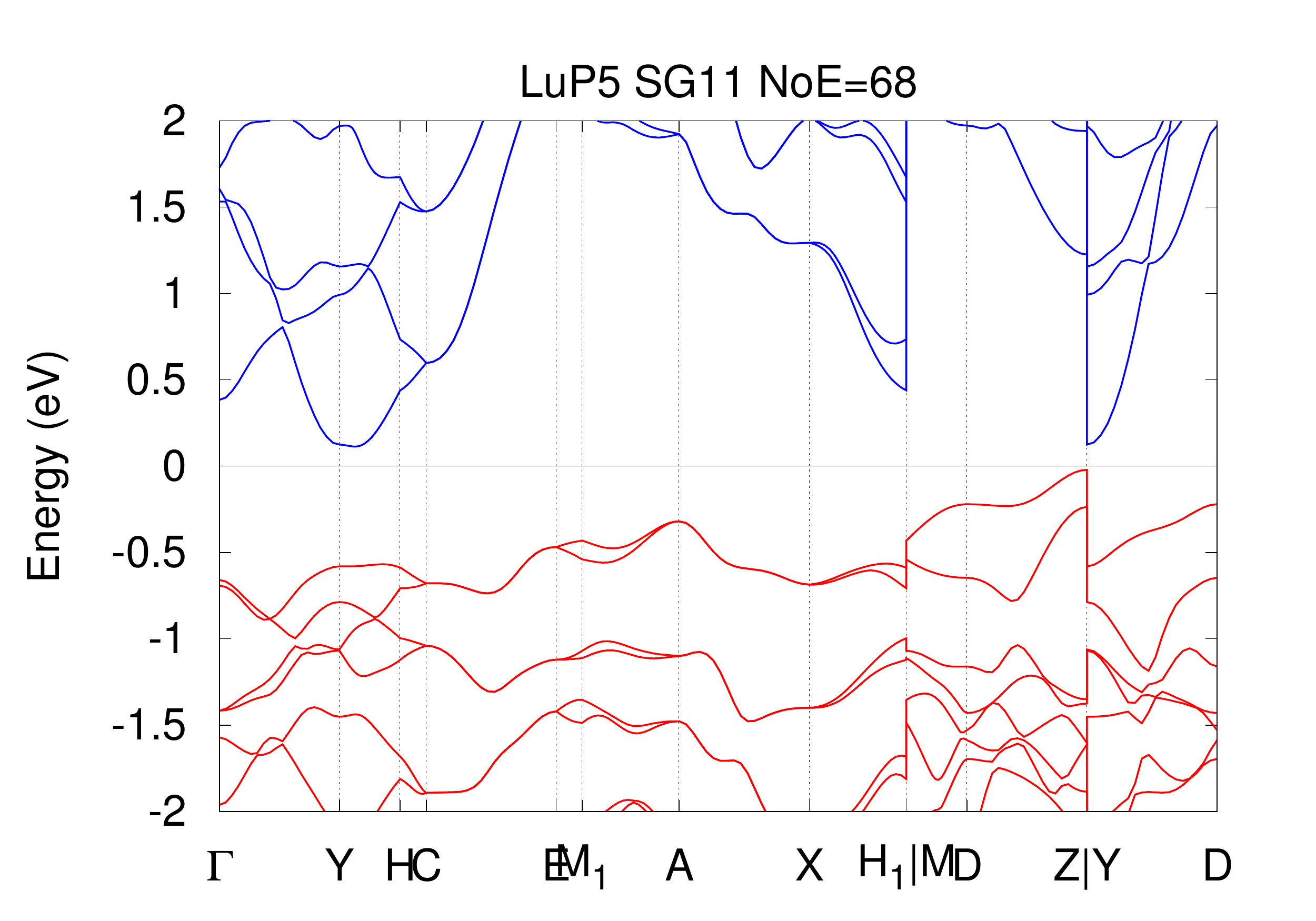}
}
\subfigure[LaSe$_{2}$ SG14 NoA=12 NoE=92]{
\label{subfig:32530}
\includegraphics[scale=0.32,angle=270]{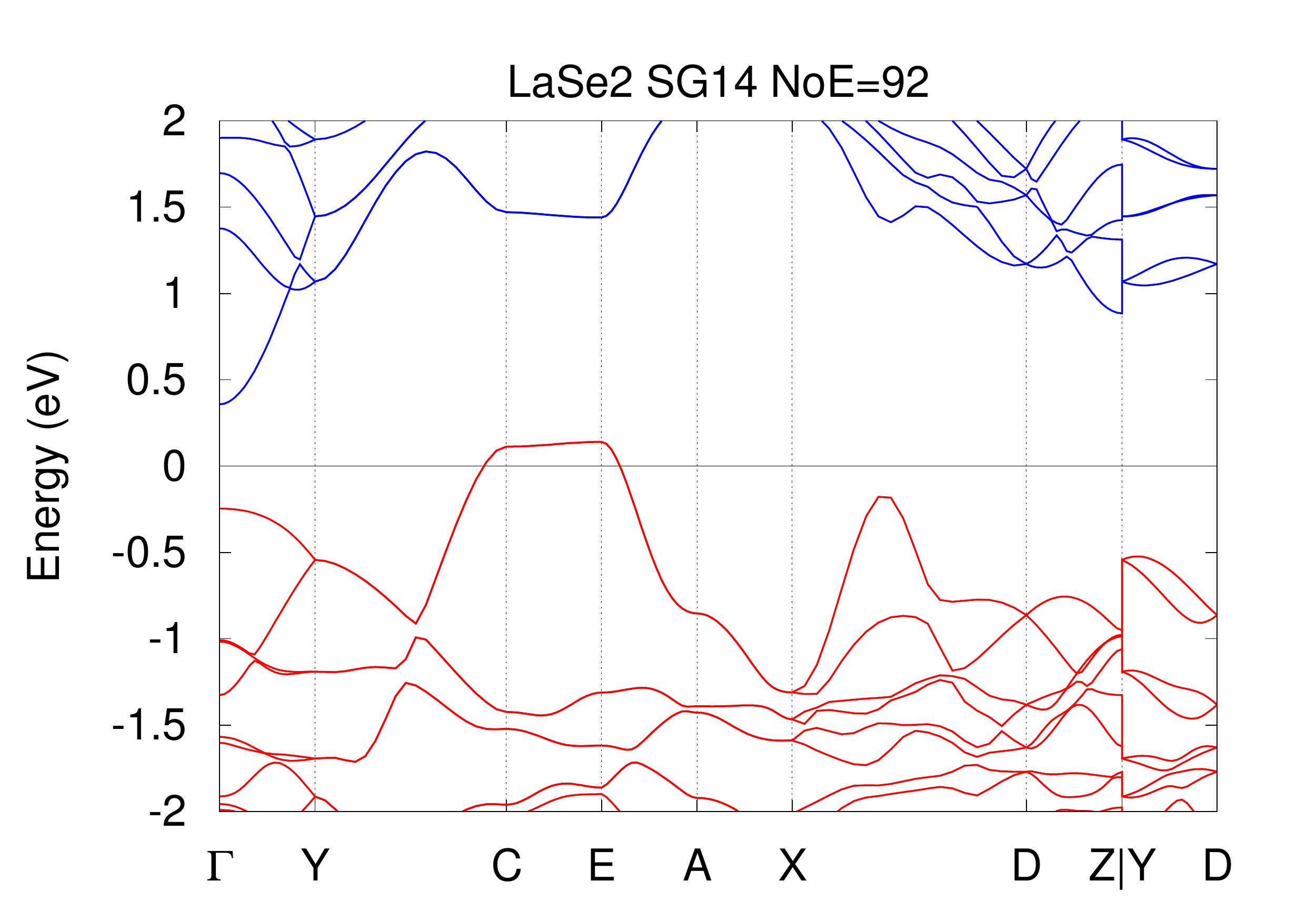}
}
\subfigure[YP$_{5}$ SG11 NoA=12 NoE=72]{
\label{subfig:409188}
\includegraphics[scale=0.32,angle=270]{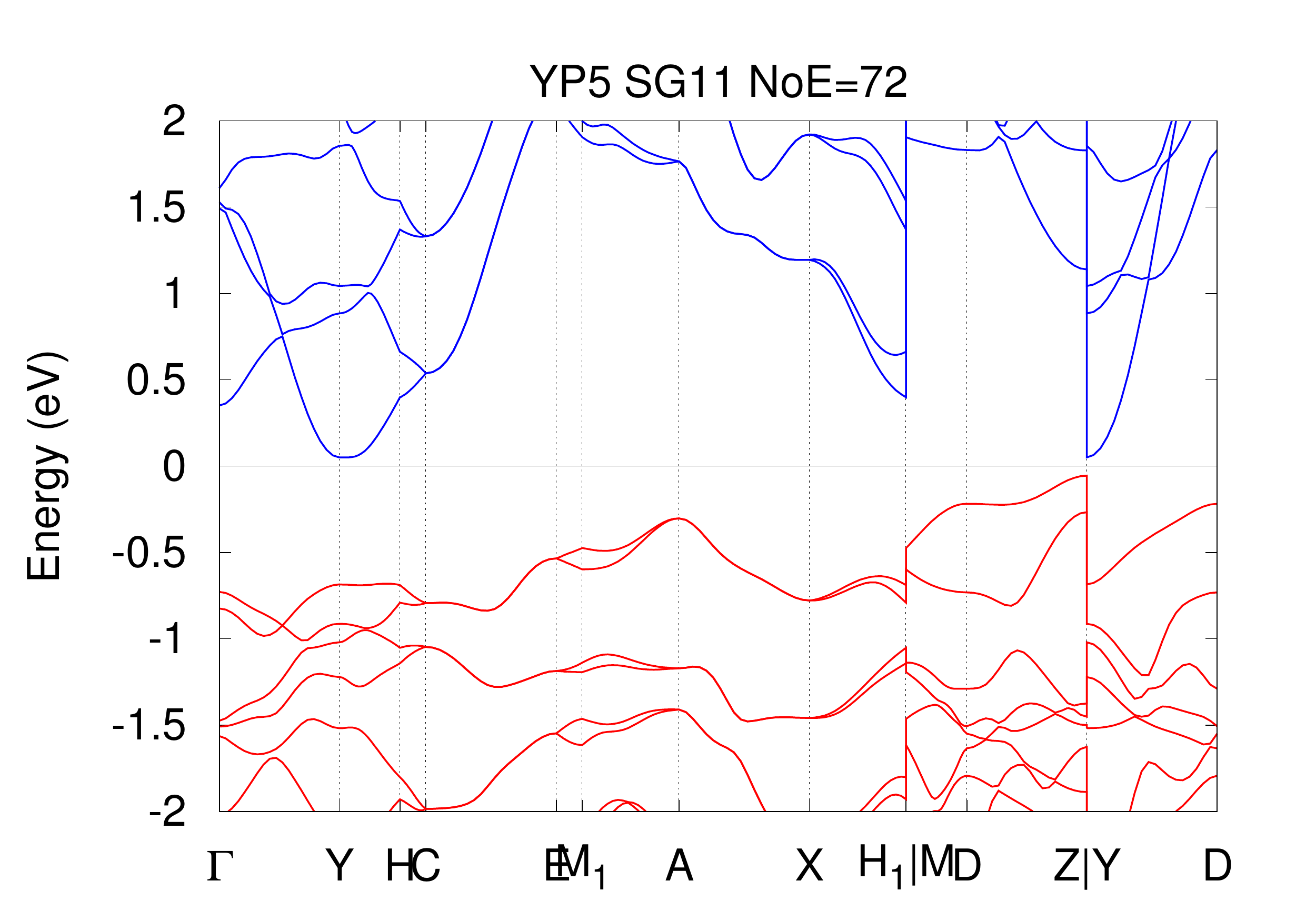}
}
\subfigure[CsIn$_{3}$ SG119 NoA=12 NoE=54]{
\label{subfig:102867}
\includegraphics[scale=0.32,angle=270]{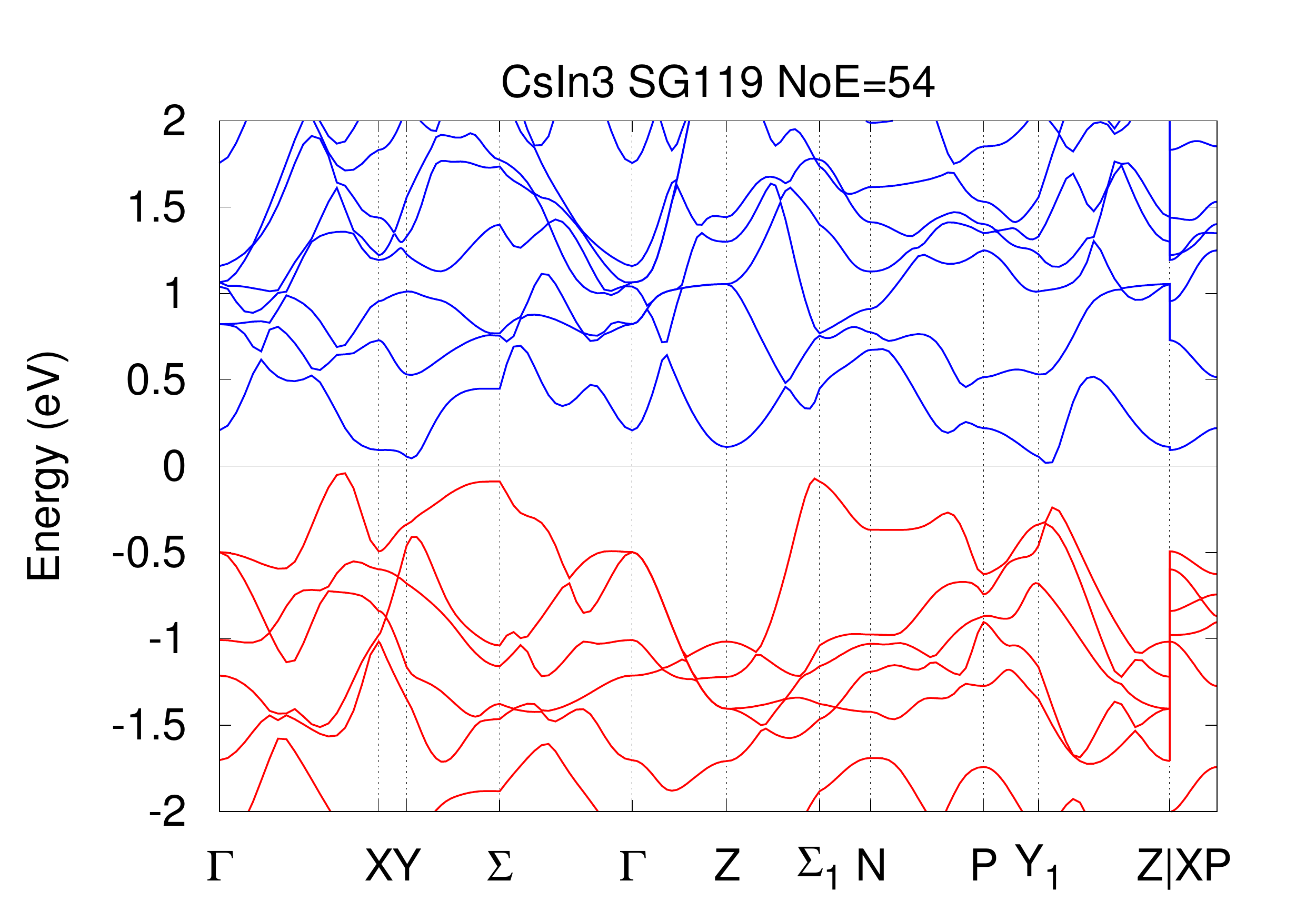}
}
\subfigure[PrS$_{2}$ SG14 NoA=12 NoE=92]{
\label{subfig:92525}
\includegraphics[scale=0.32,angle=270]{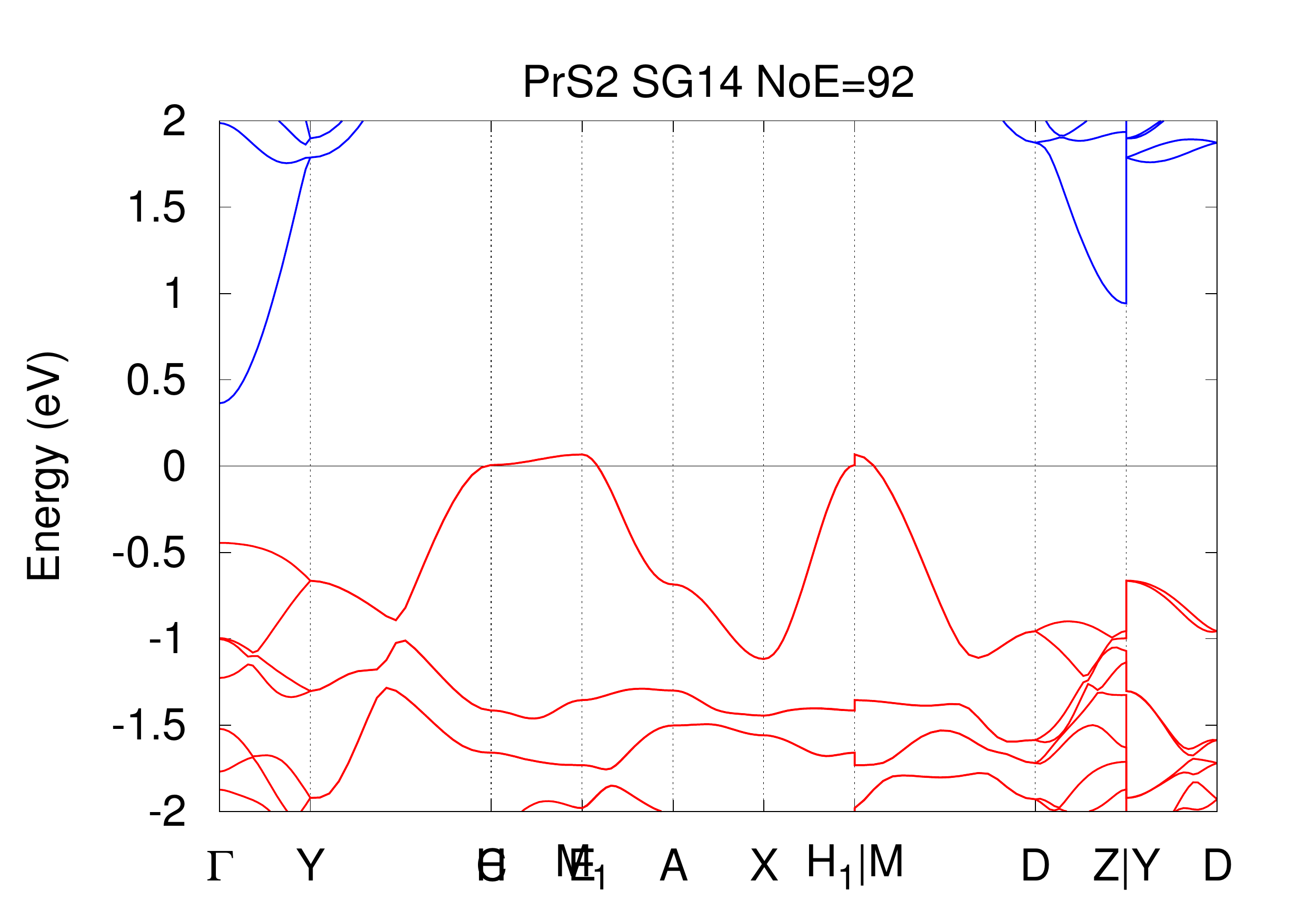}
}
\caption{\hyperref[tab:electride]{back to the table}}
\end{figure}

\begin{figure}[htp]
 \centering
\subfigure[YbB$_{12}$ SG225 NoA=13 NoE=44]{
\label{subfig:615734}
\includegraphics[scale=0.32,angle=270]{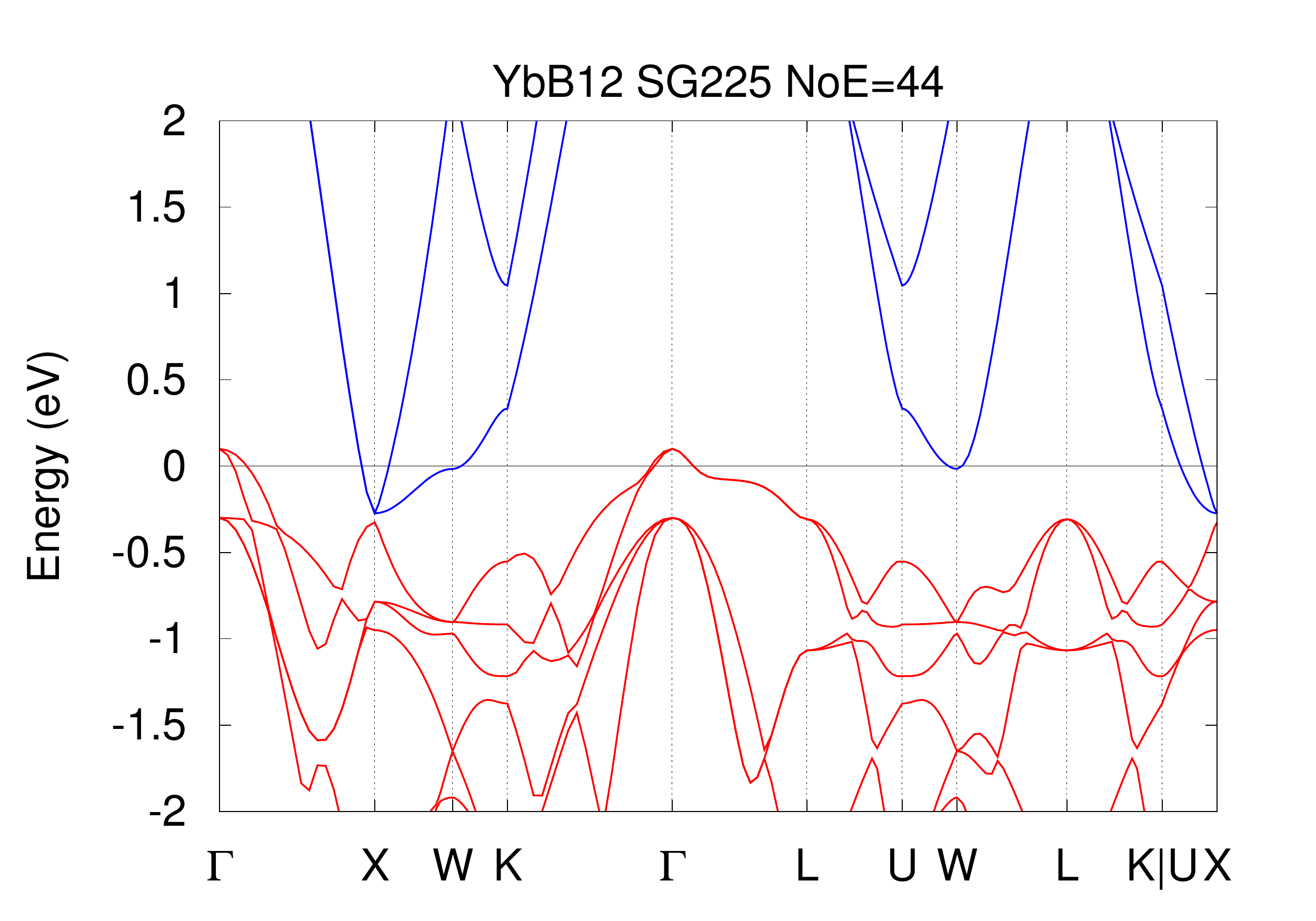}
}
\subfigure[Nb$_{7}$Co$_{6}$ SG166 NoA=13 NoE=131]{
\label{subfig:624284}
\includegraphics[scale=0.32,angle=270]{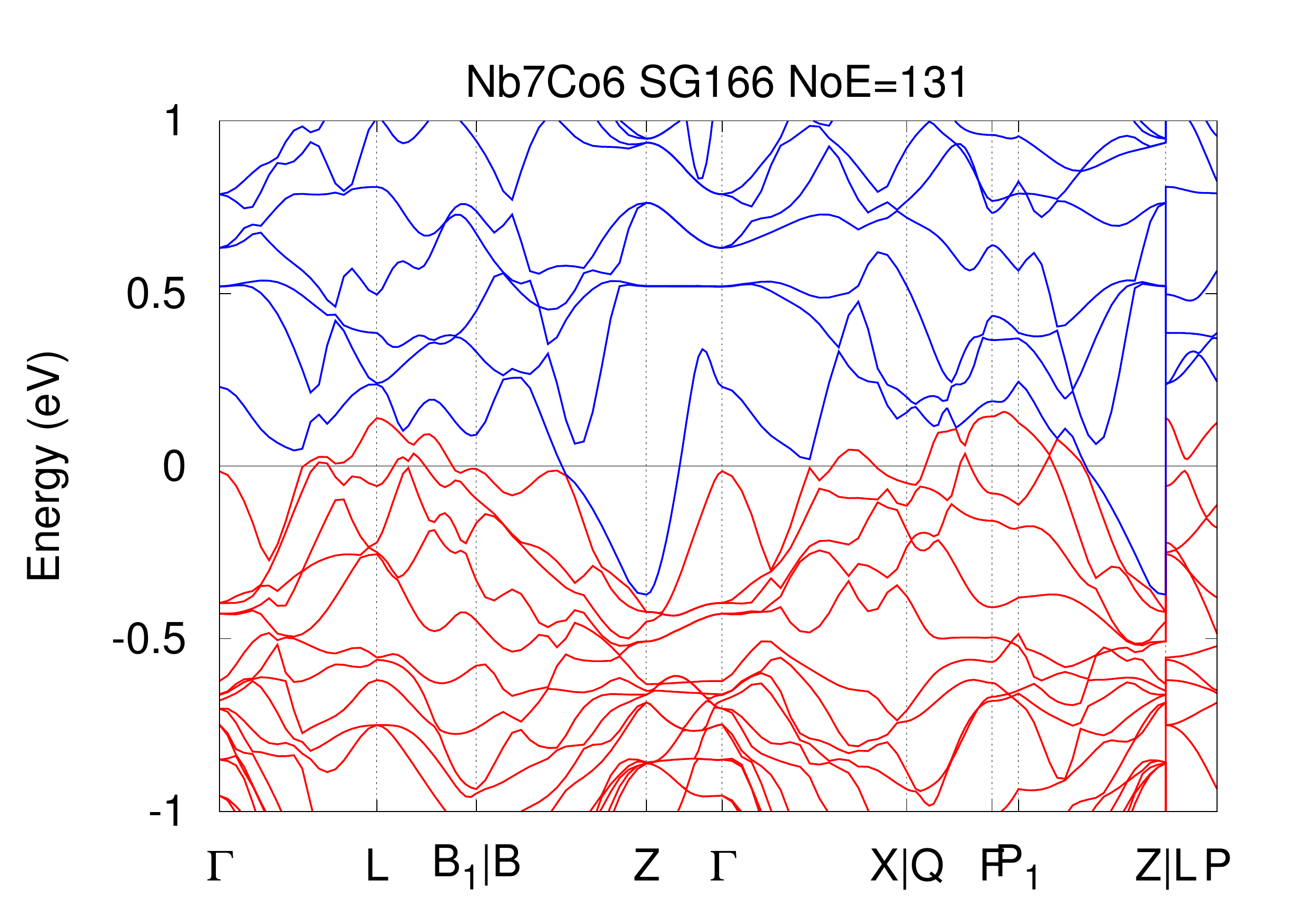}
}
\subfigure[Ba(P$_{2}$Au)$_{2}$ SG70 NoA=14 NoE=104]{
\label{subfig:425778}
\includegraphics[scale=0.32,angle=270]{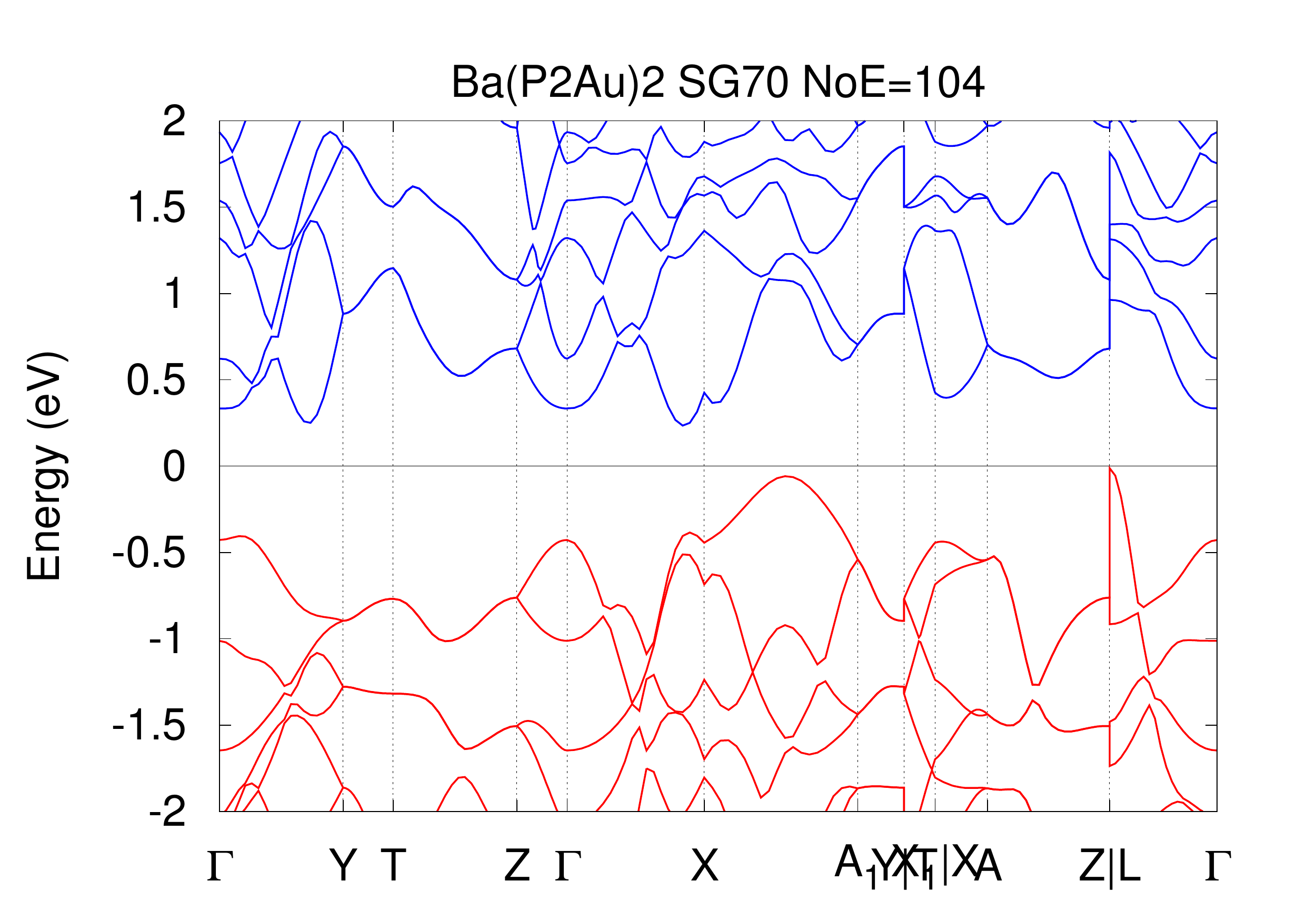}
}
\subfigure[Sr$_{2}$Zn$_{2}$As$_{3}$ SG12 NoA=14 NoE=118]{
\label{subfig:262413}
\includegraphics[scale=0.32,angle=270]{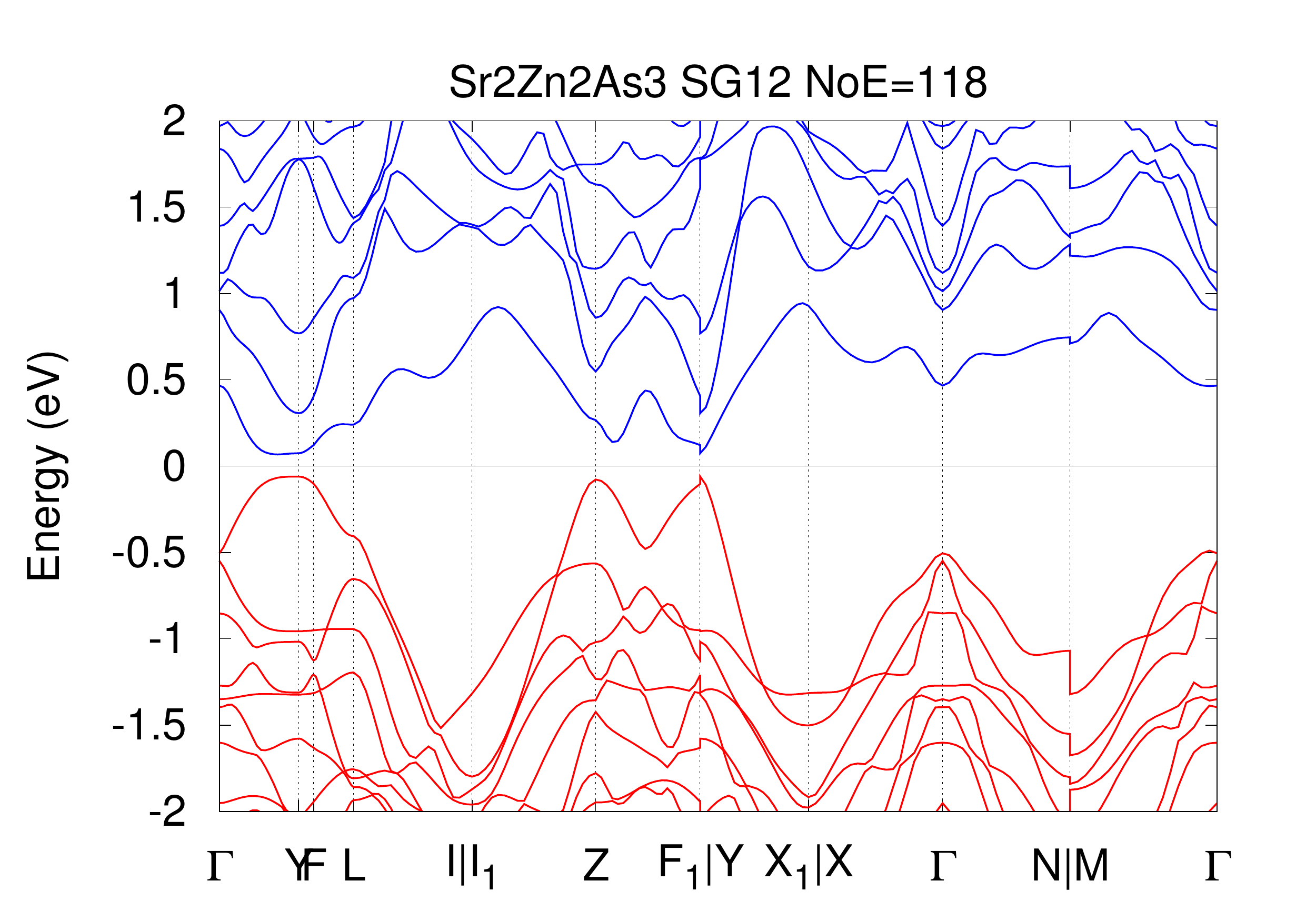}
}
\subfigure[Bi$_{4}$RuI$_{2}$ SG87 NoA=14 NoE=84]{
\label{subfig:406949}
\includegraphics[scale=0.32,angle=270]{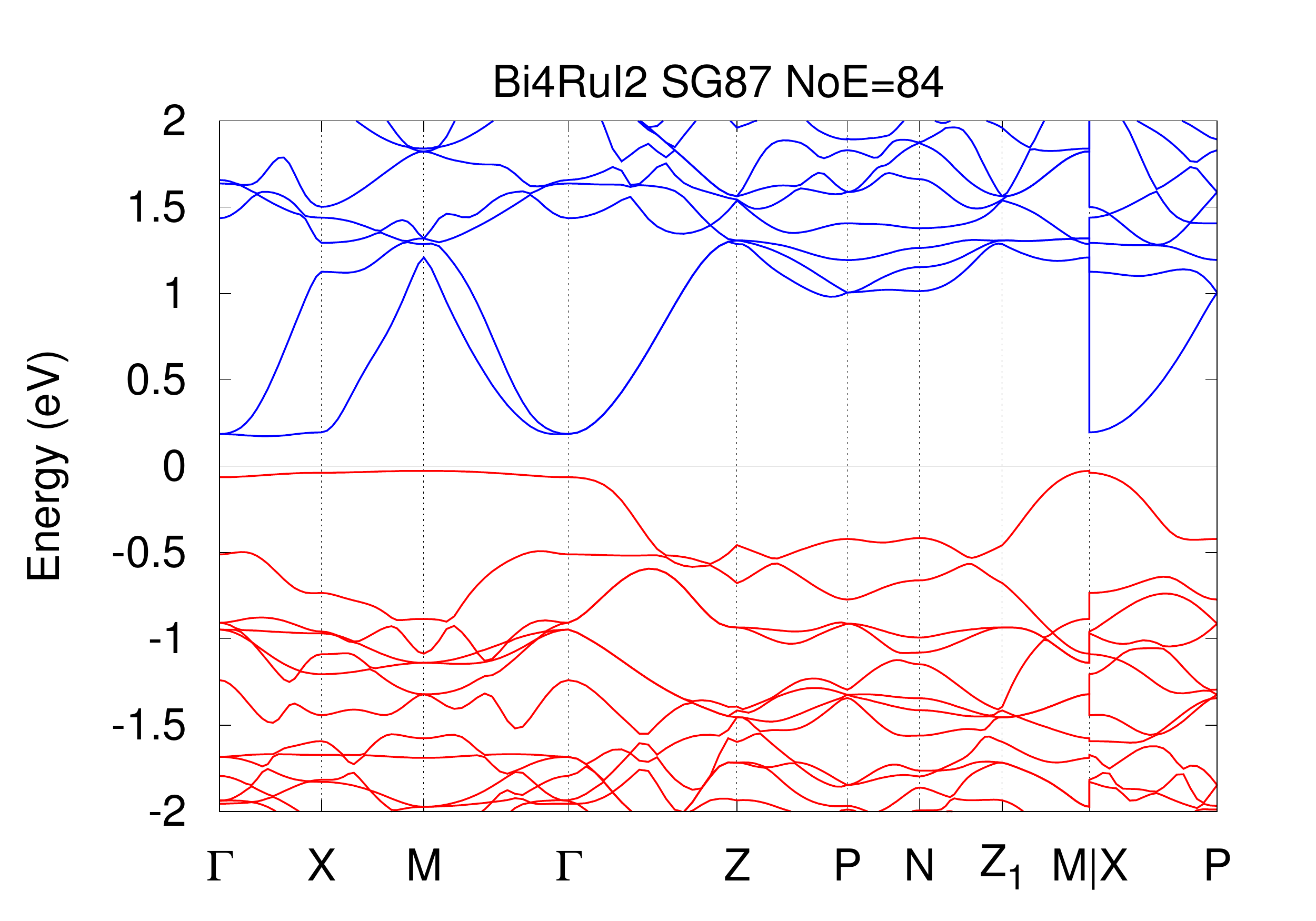}
}
\subfigure[Eu$_{2}$Zn$_{2}$P$_{3}$ SG12 NoA=14 NoE=114]{
\label{subfig:426082}
\includegraphics[scale=0.32,angle=270]{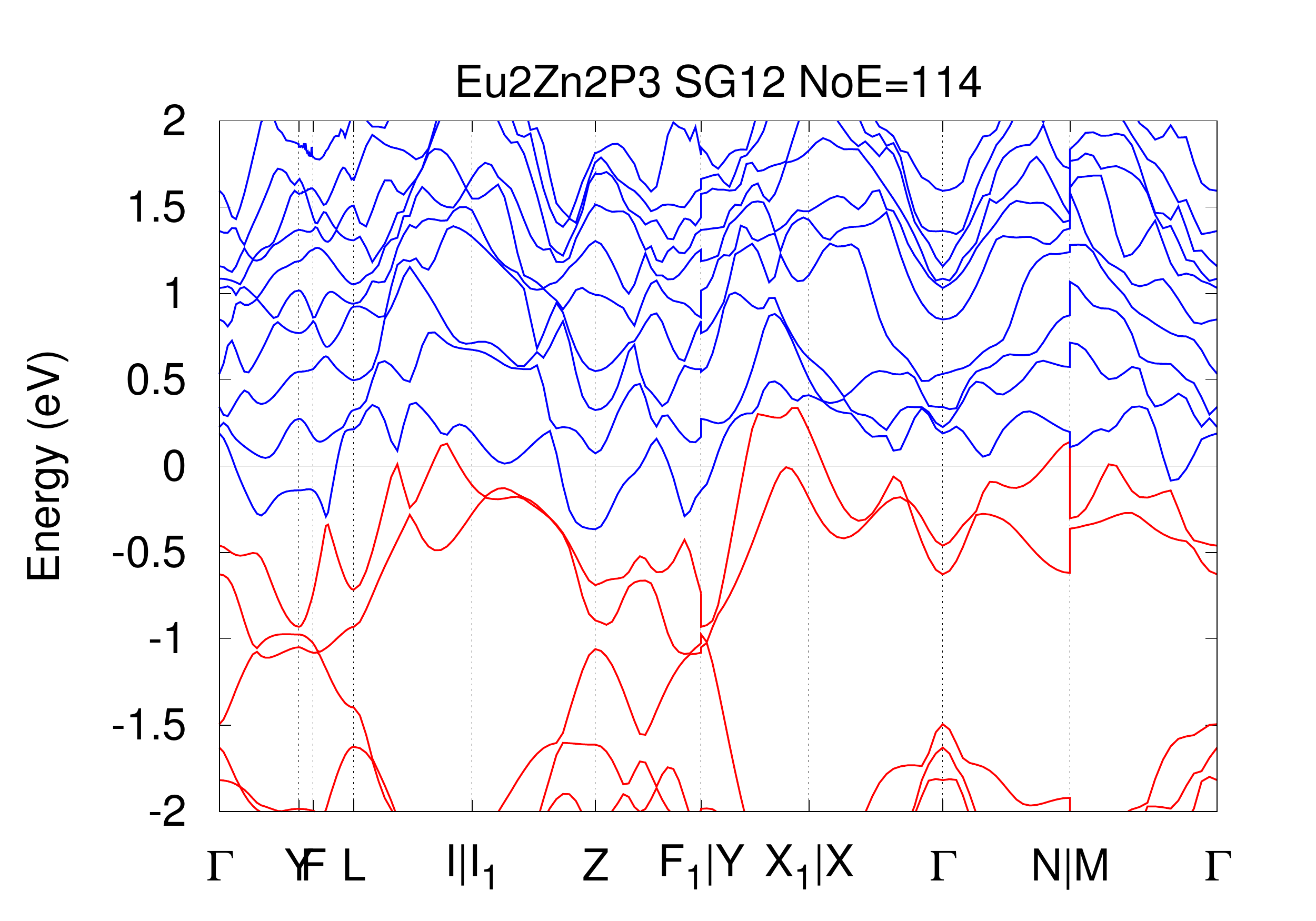}
}
\subfigure[Sr$_{3}$(GeN)$_{2}$ SG11 NoA=14 NoE=96]{
\label{subfig:82533}
\includegraphics[scale=0.32,angle=270]{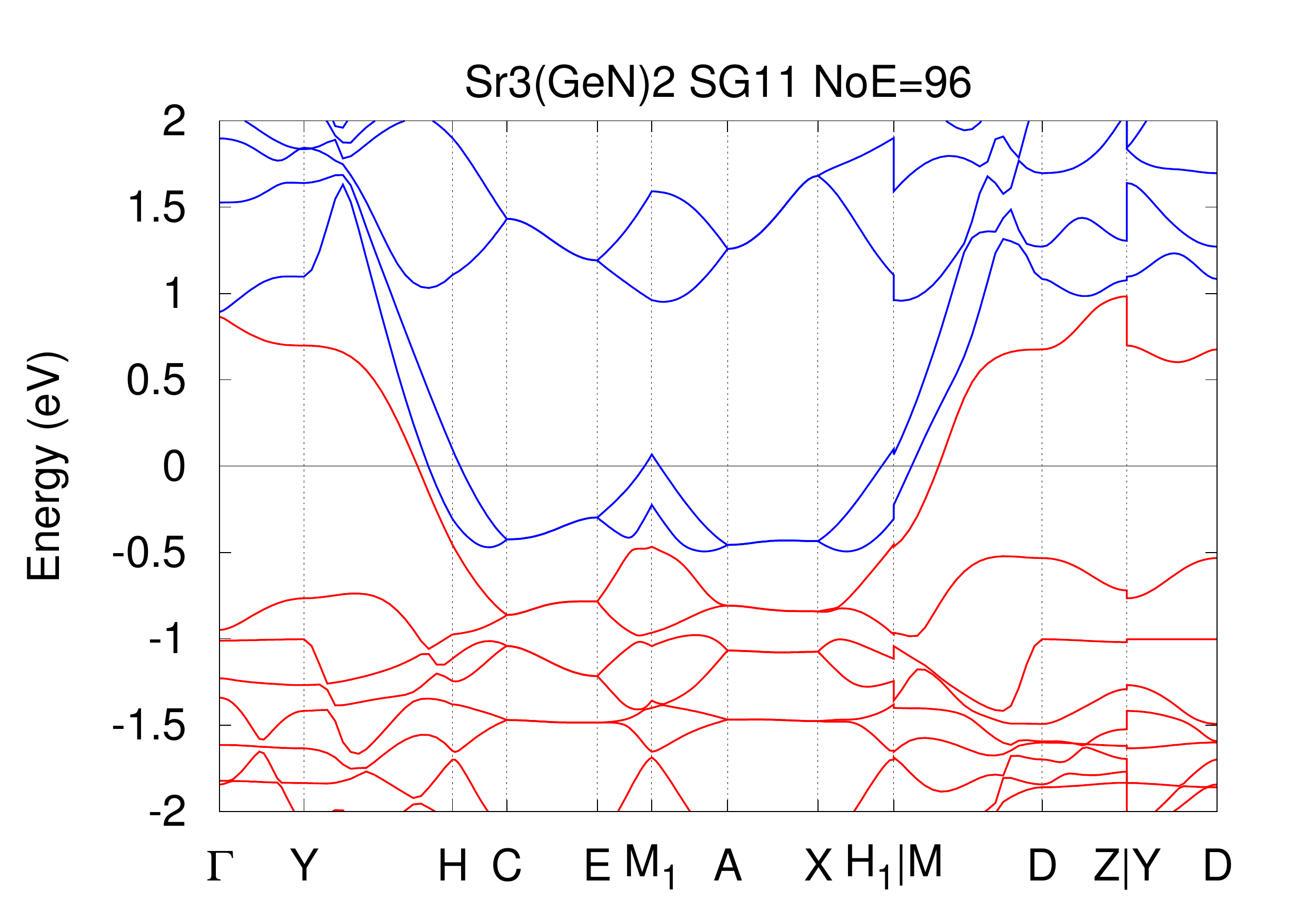}
}
\subfigure[Sc$_{2}$Si$_{2}$Pt$_{3}$ SG55 NoA=14 NoE=88]{
\label{subfig:247425}
\includegraphics[scale=0.32,angle=270]{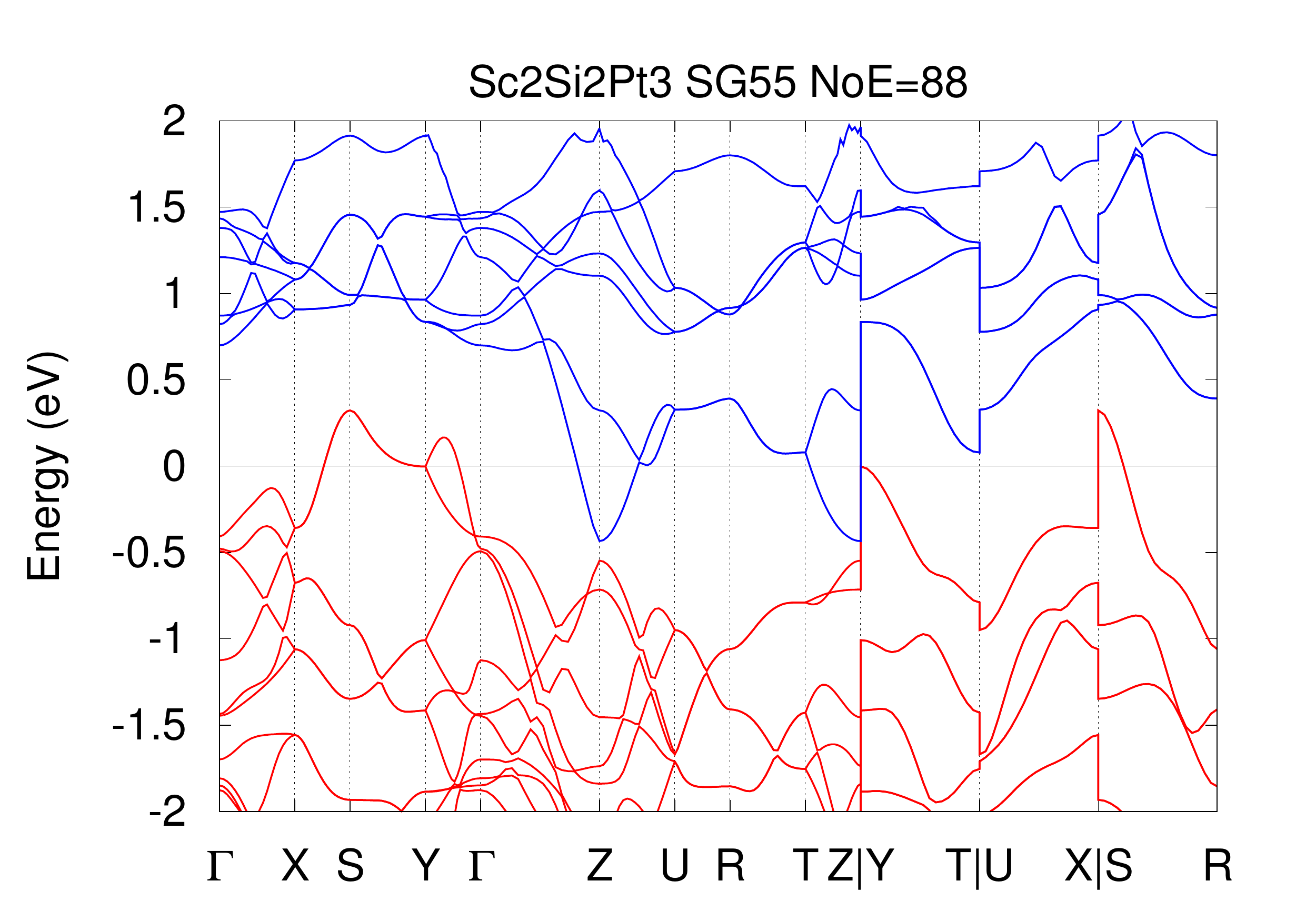}
}
\caption{\hyperref[tab:electride]{back to the table}}
\end{figure}

\begin{figure}[htp]
 \centering
\subfigure[Ho$_{3}$Ni$_{2}$ SG148 NoA=15 NoE=141]{
\label{subfig:639447}
\includegraphics[scale=0.32,angle=270]{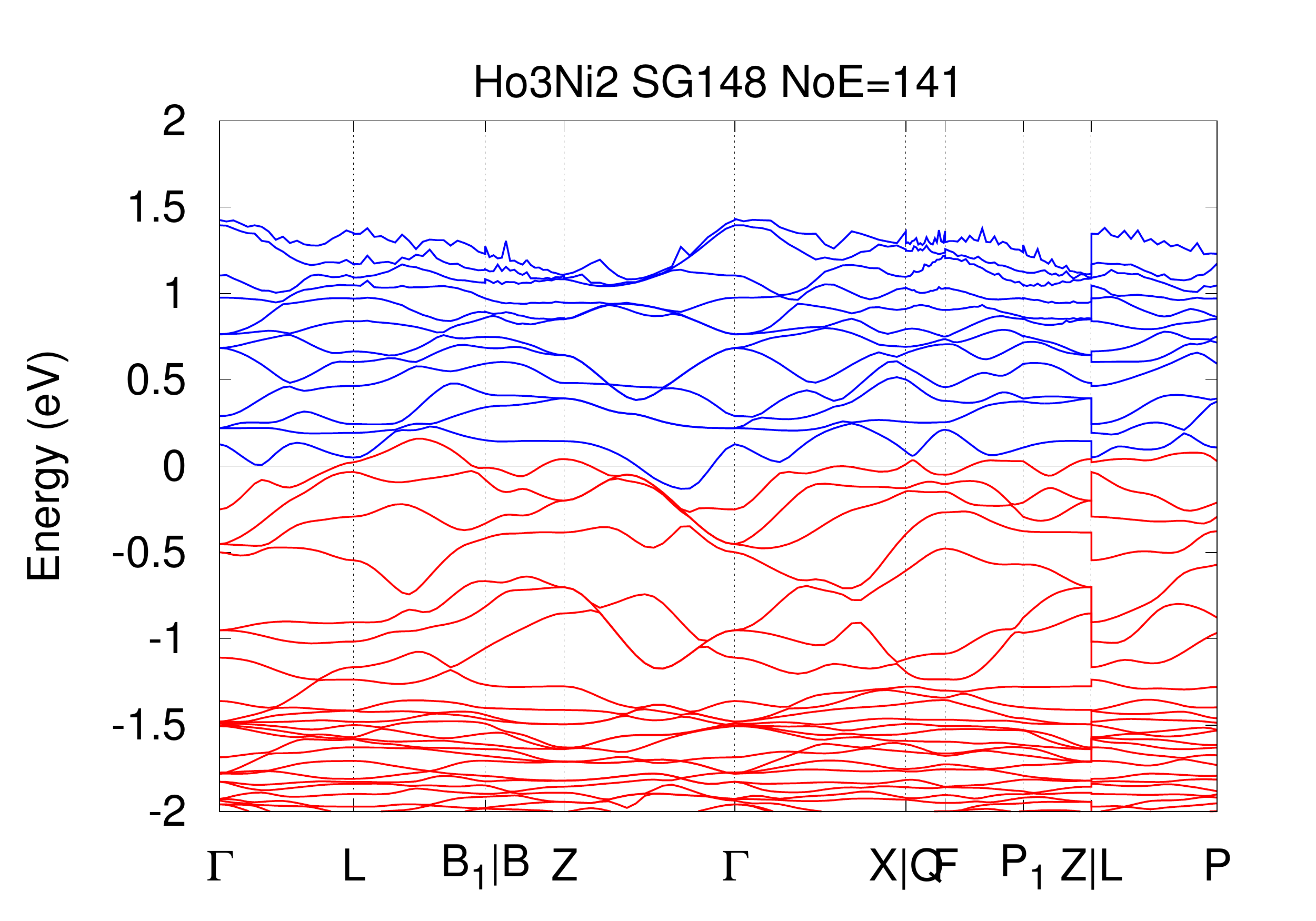}
}
\subfigure[Er$_{3}$Ni$_{2}$ SG148 NoA=15 NoE=141]{
\label{subfig:2150}
\includegraphics[scale=0.32,angle=270]{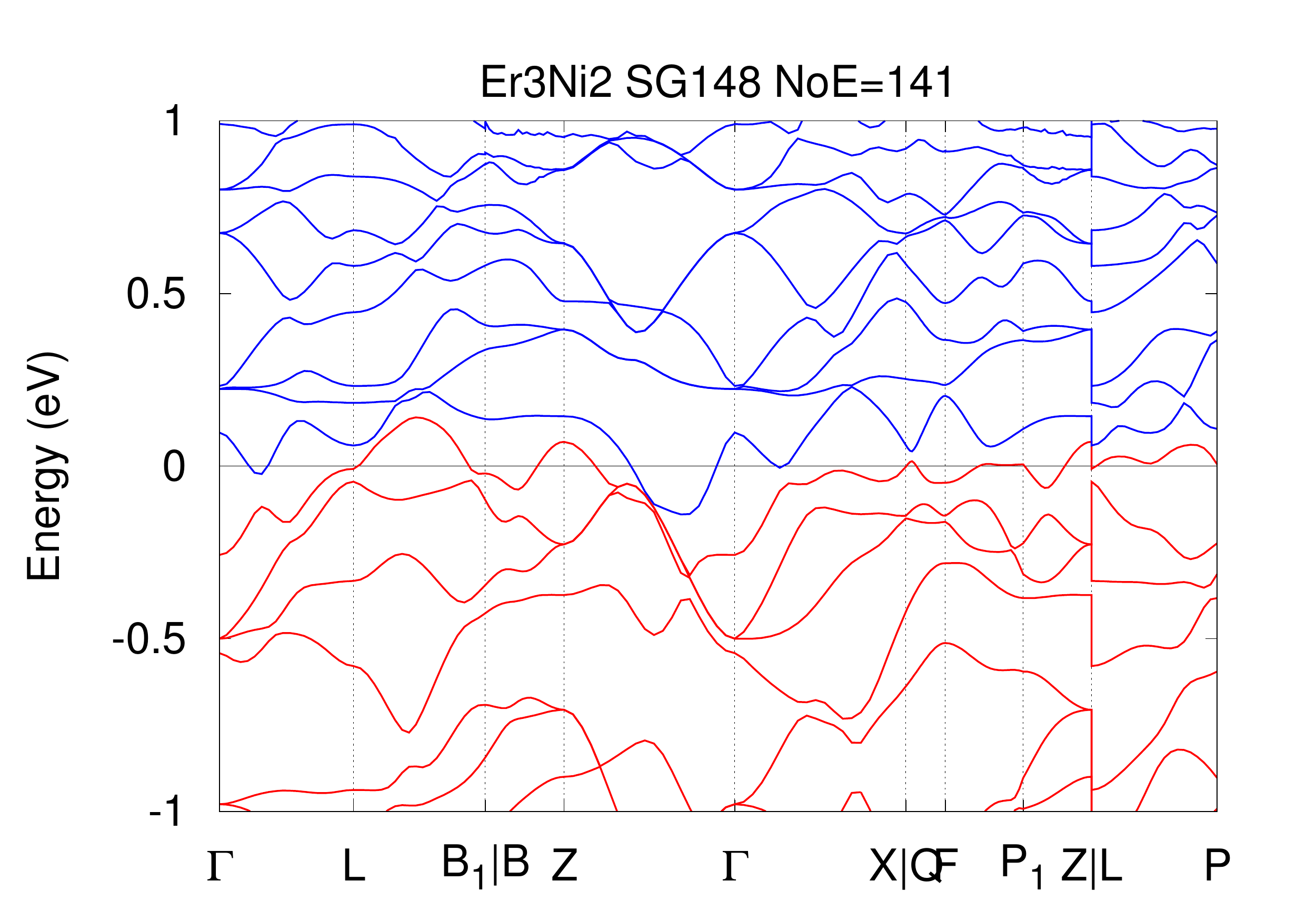}
}
\subfigure[Ca$_{6}$Cr$_{2}$HN$_{6}$ SG148 NoA=15 NoE=103]{
\label{subfig:281462}
\includegraphics[scale=0.32,angle=270]{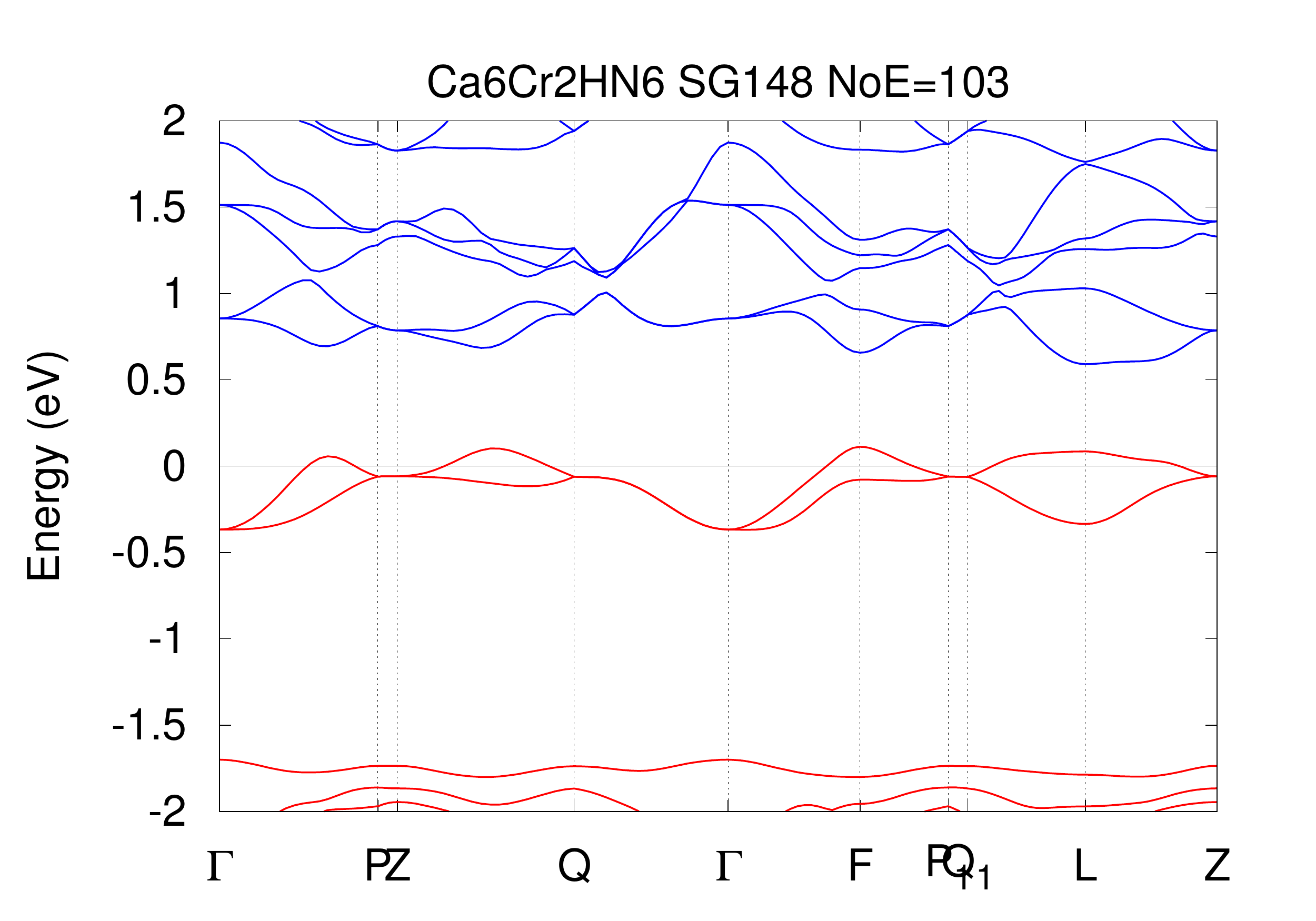}
}
\subfigure[Mn$_{4}$Al$_{11}$ SG2 NoA=15 NoE=61]{
\label{subfig:10509}
\includegraphics[scale=0.32,angle=270]{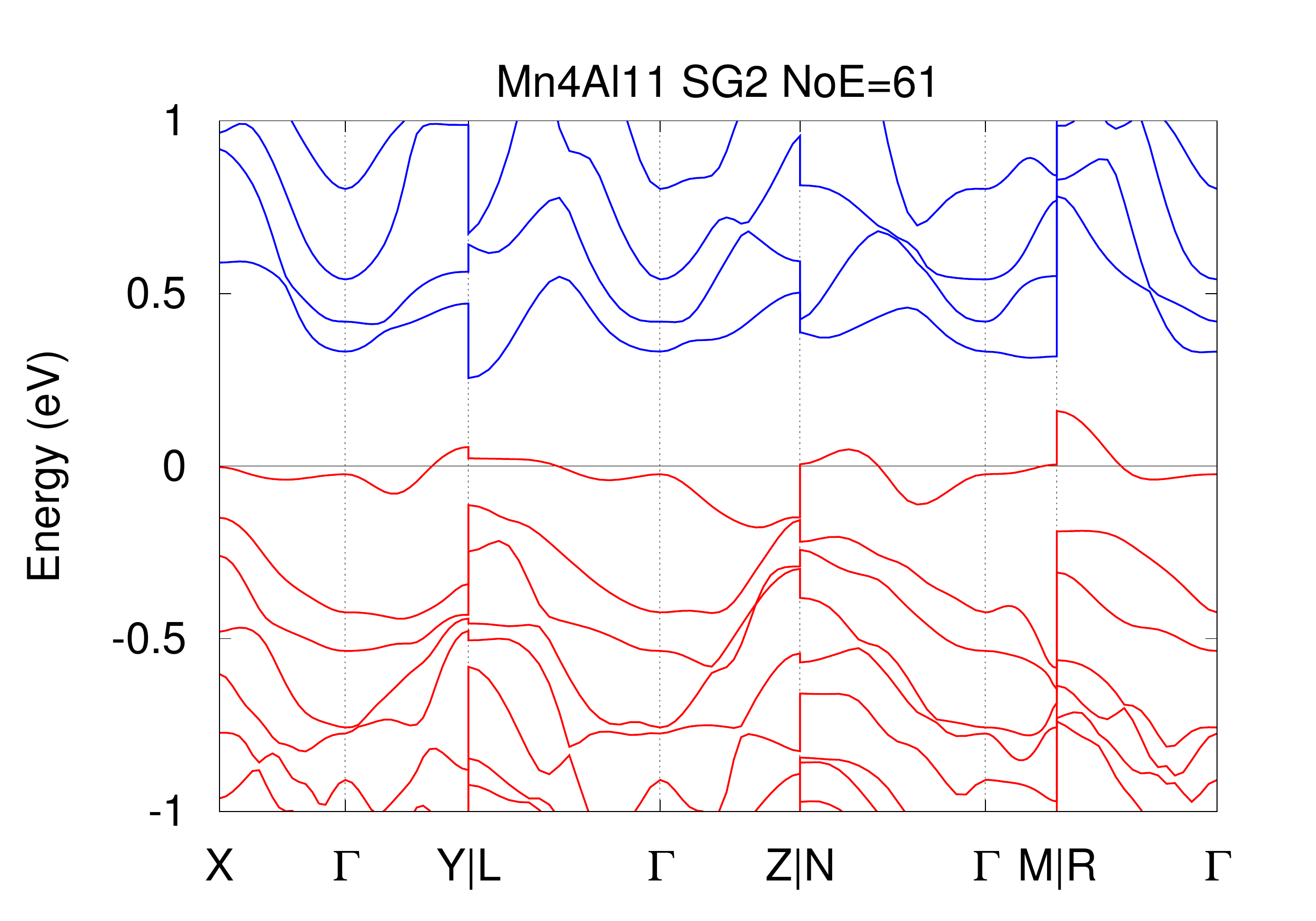}
}
\subfigure[Ba$_{4}$Fe$_{2}$S$_{4}$I$_{5}$ SG87 NoA=15 NoE=115]{
\label{subfig:173357}
\includegraphics[scale=0.32,angle=270]{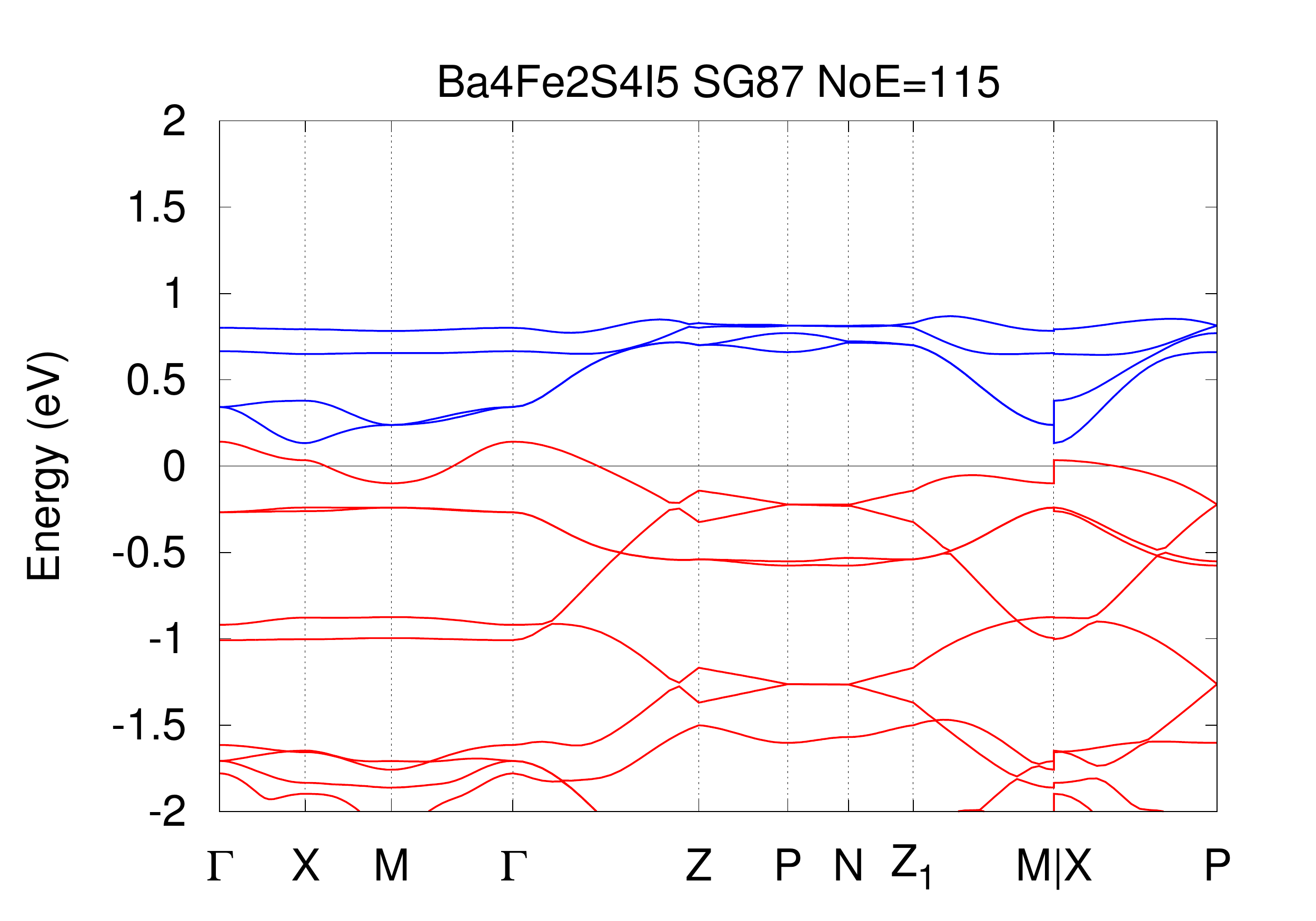}
}
\subfigure[Sb$_{2}$I$_{2}$F$_{11}$ SG5 NoA=15 NoE=101]{
\label{subfig:6031}
\includegraphics[scale=0.32,angle=270]{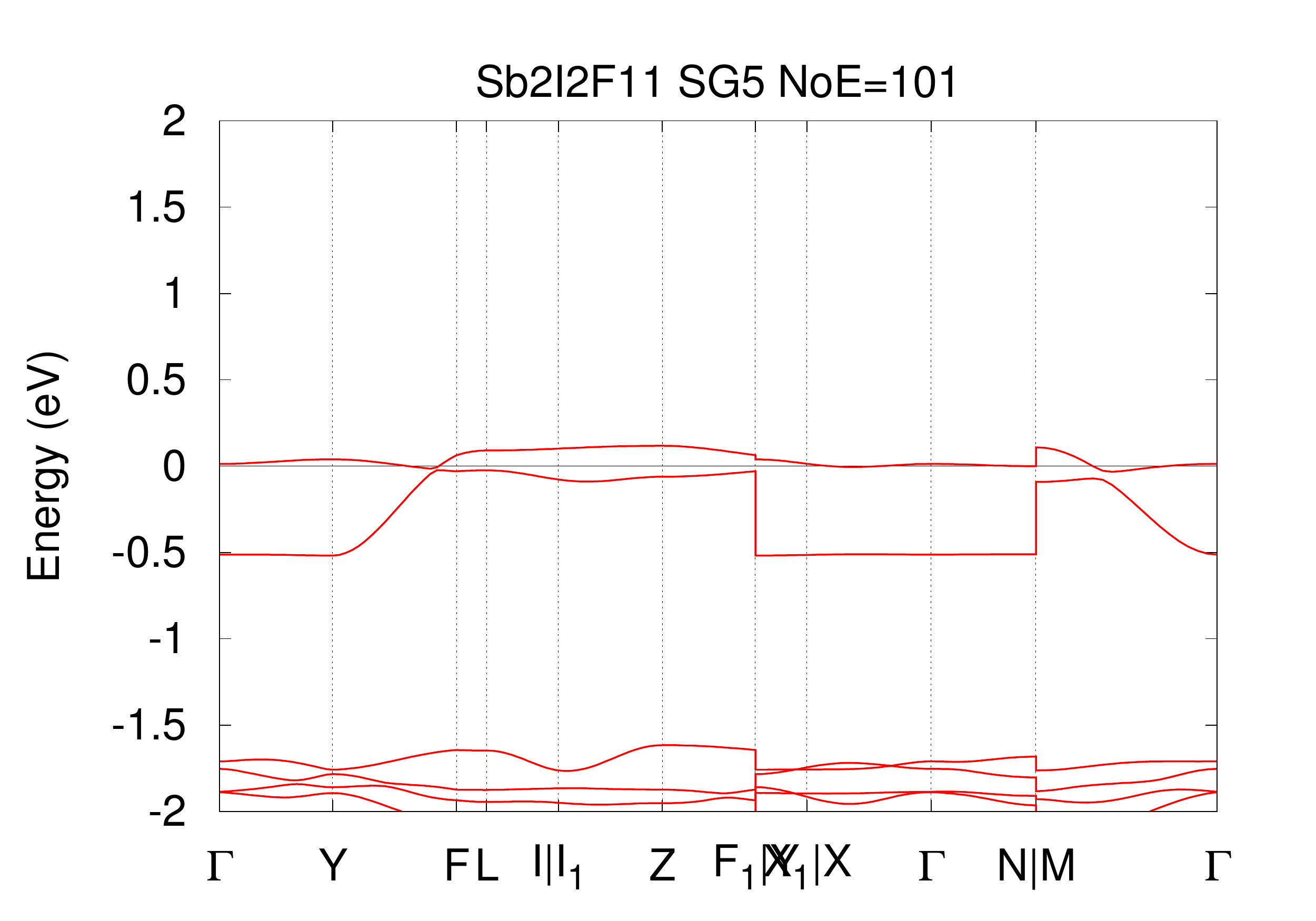}
}
\subfigure[K$_{5}$Te$_{3}$ SG87 NoA=16 NoE=126]{
\label{subfig:66024}
\includegraphics[scale=0.32,angle=270]{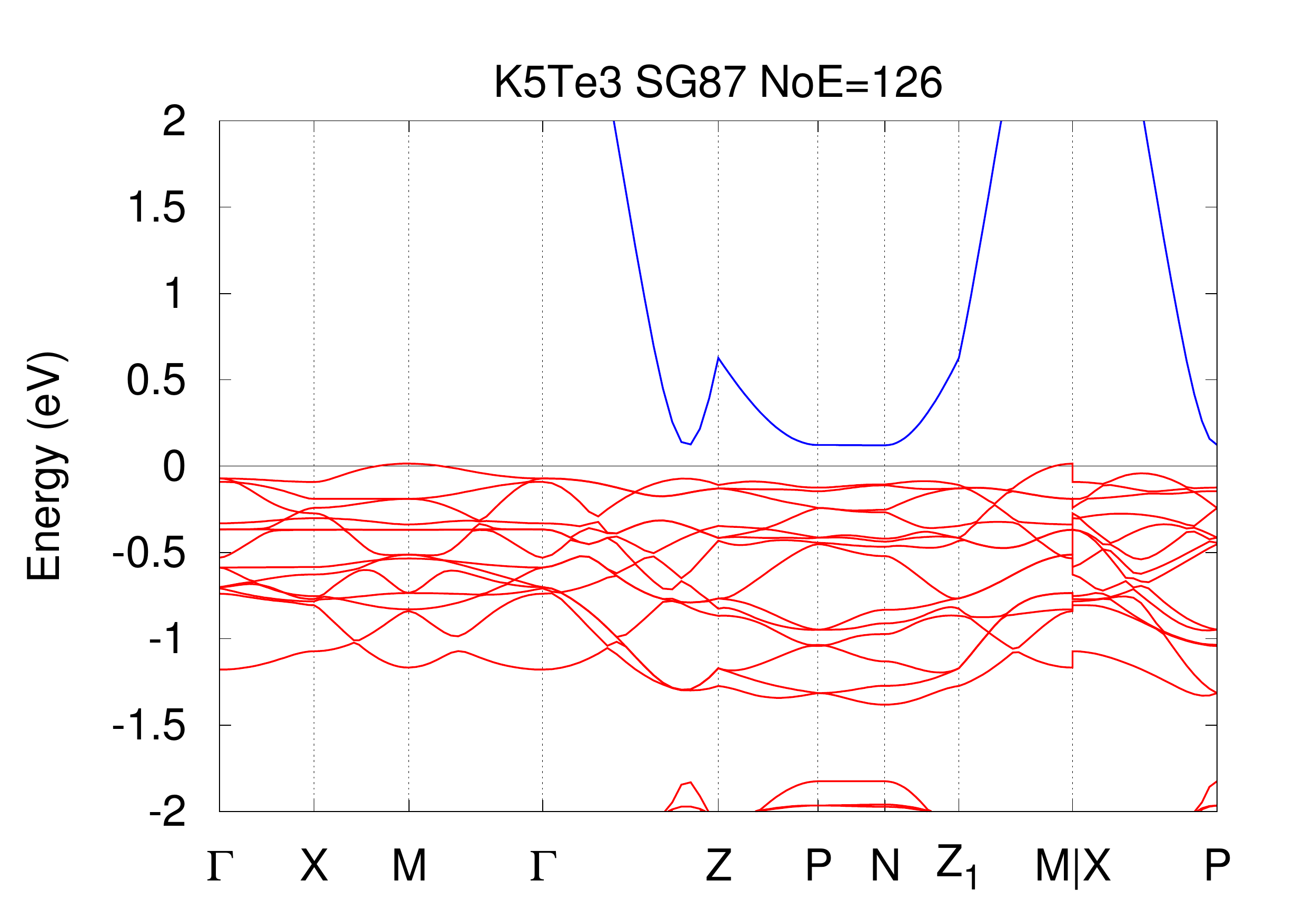}
}
\subfigure[NaSi SG15 NoA=16 NoE=40]{
\label{subfig:174081}
\includegraphics[scale=0.32,angle=270]{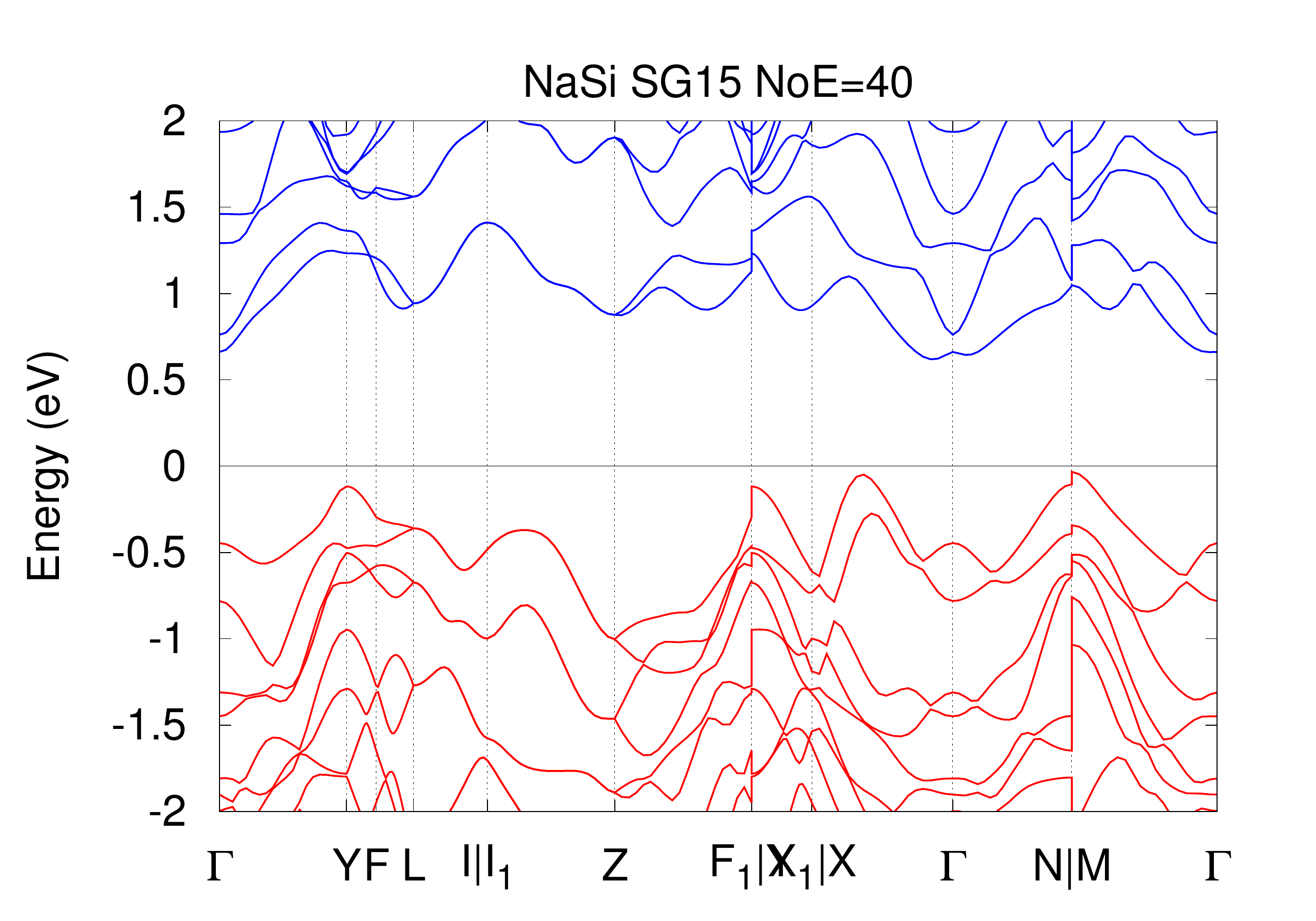}
}
\caption{\hyperref[tab:electride]{back to the table}}
\end{figure}

\begin{figure}[htp]
 \centering
\subfigure[Ga$_{3}$Ru SG136 NoA=16 NoE=68]{
\label{subfig:635229}
\includegraphics[scale=0.32,angle=270]{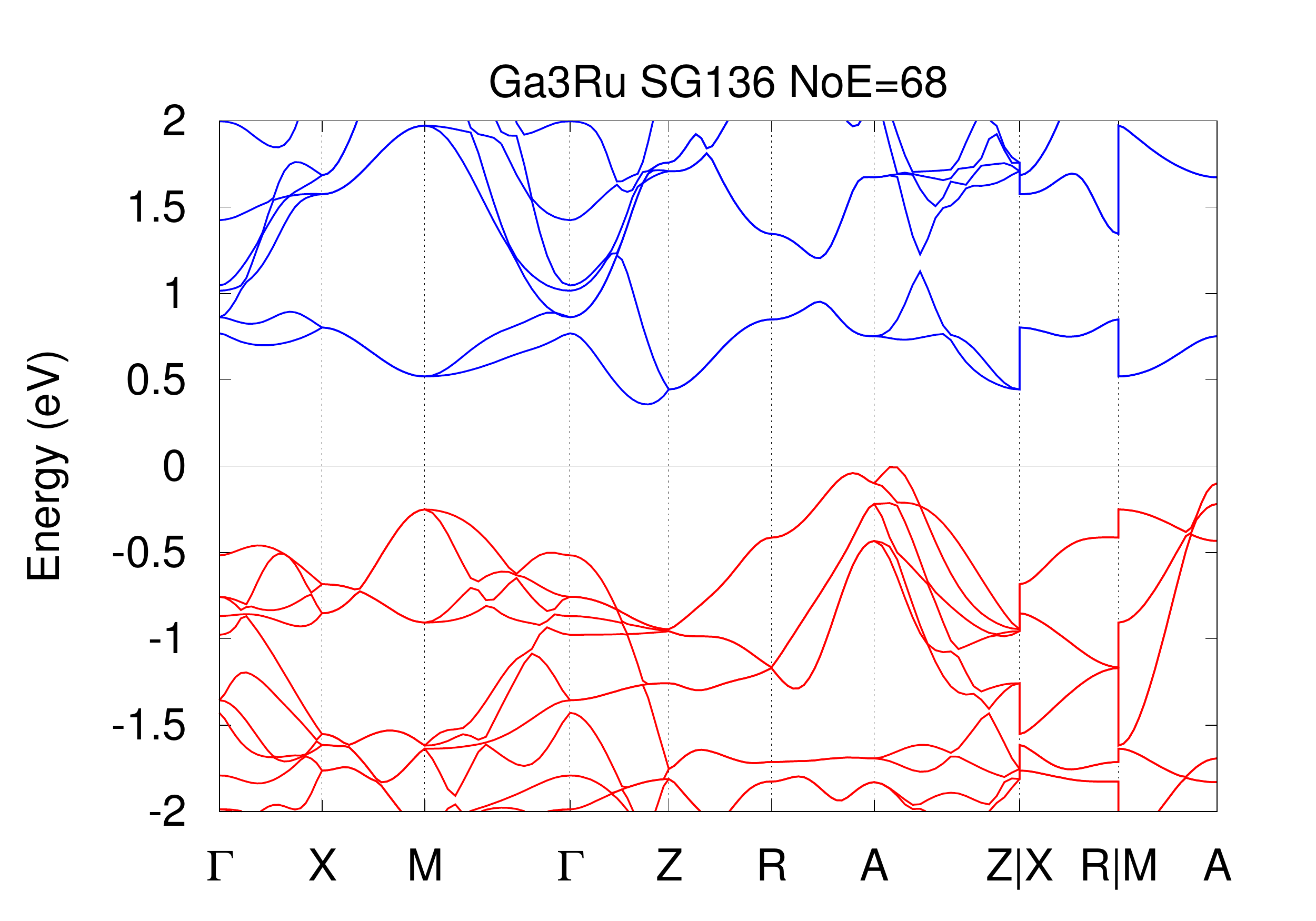}
}
\subfigure[CoSb$_{3}$ SG204 NoA=16 NoE=96]{
\label{subfig:34048}
\includegraphics[scale=0.32,angle=270]{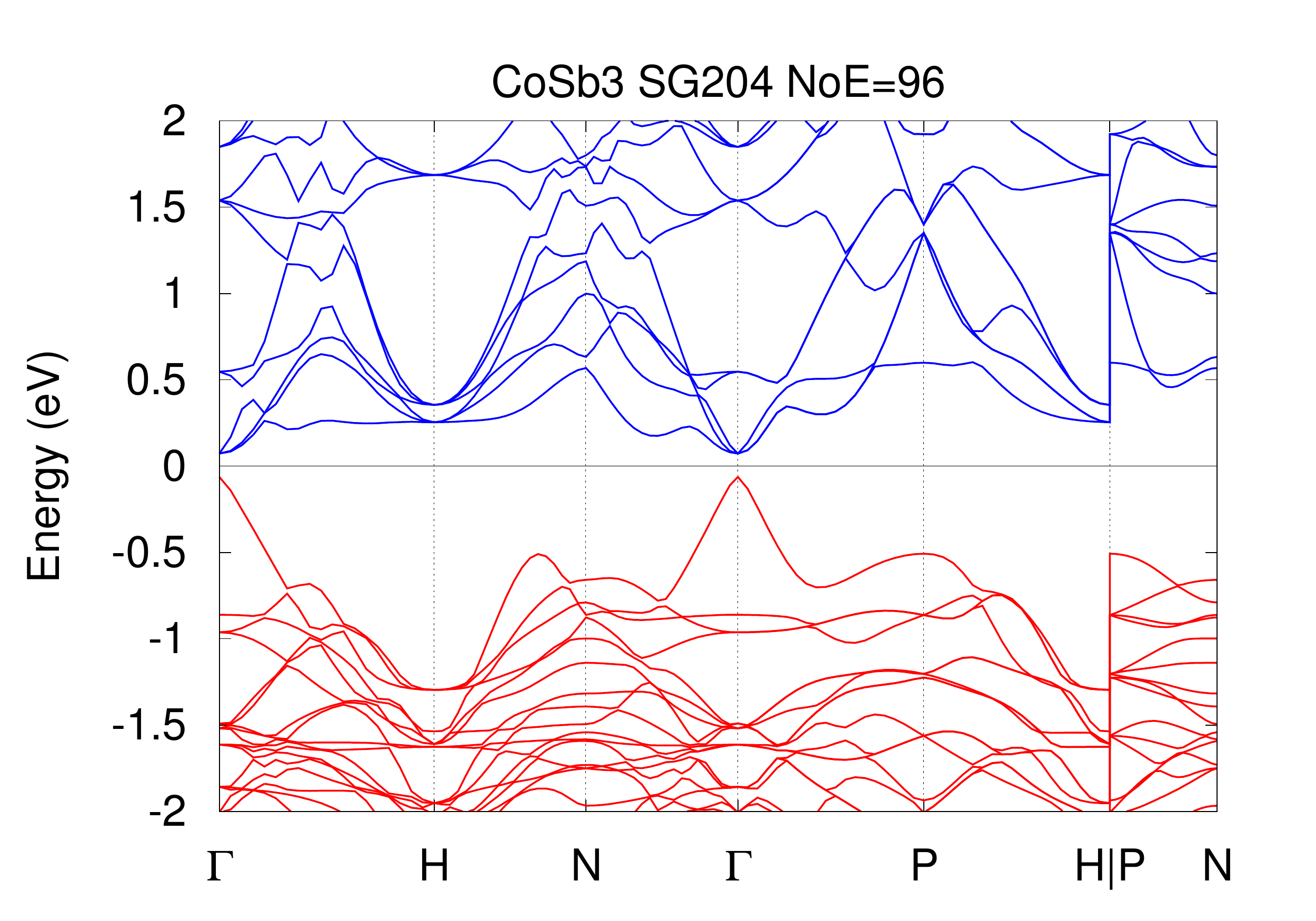}
}
\subfigure[In$_{3}$Ru SG136 NoA=16 NoE=68]{
\label{subfig:55514}
\includegraphics[scale=0.32,angle=270]{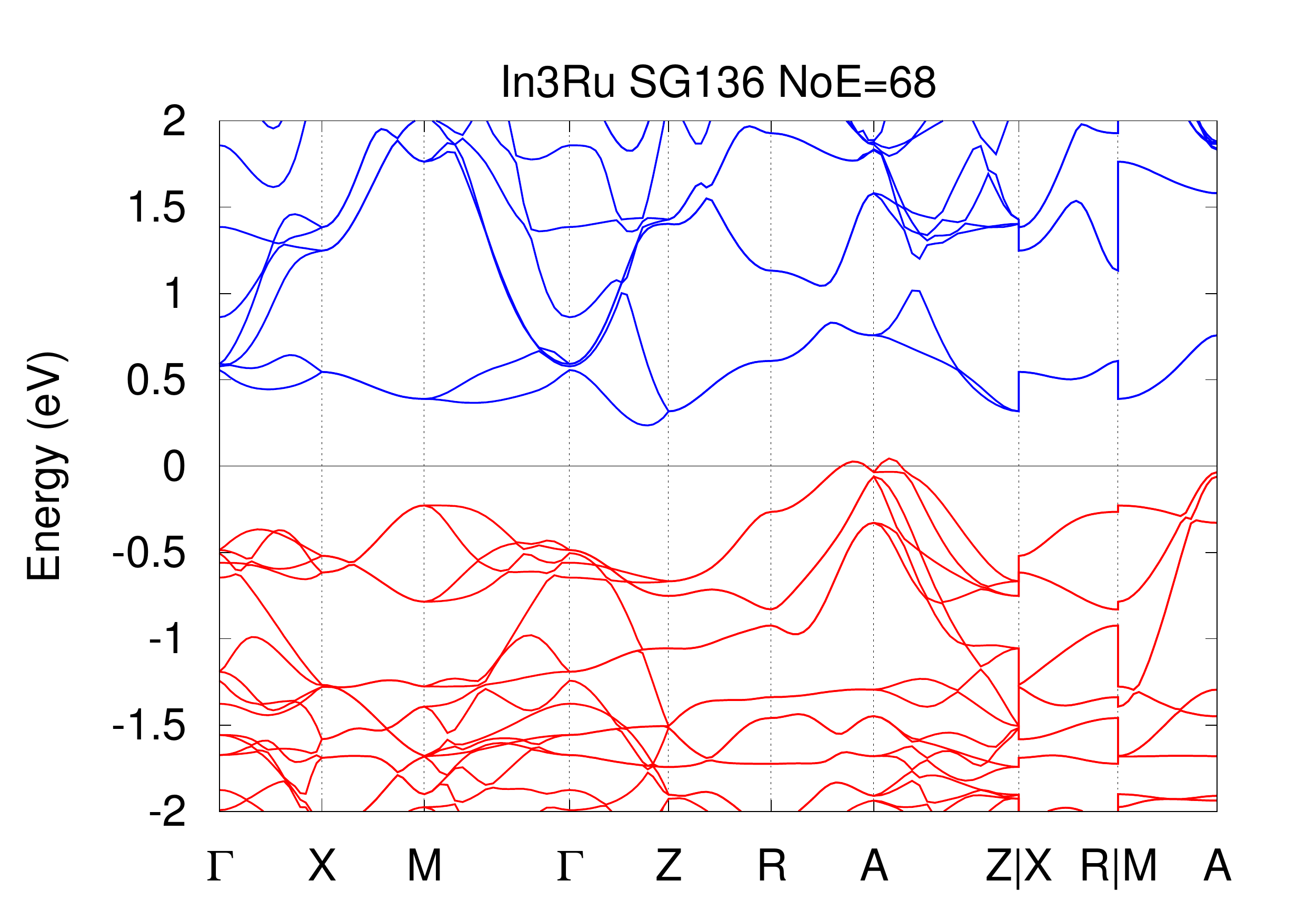}
}
\subfigure[TcP$_{3}$ SG62 NoA=16 NoE=88]{
\label{subfig:35200}
\includegraphics[scale=0.32,angle=270]{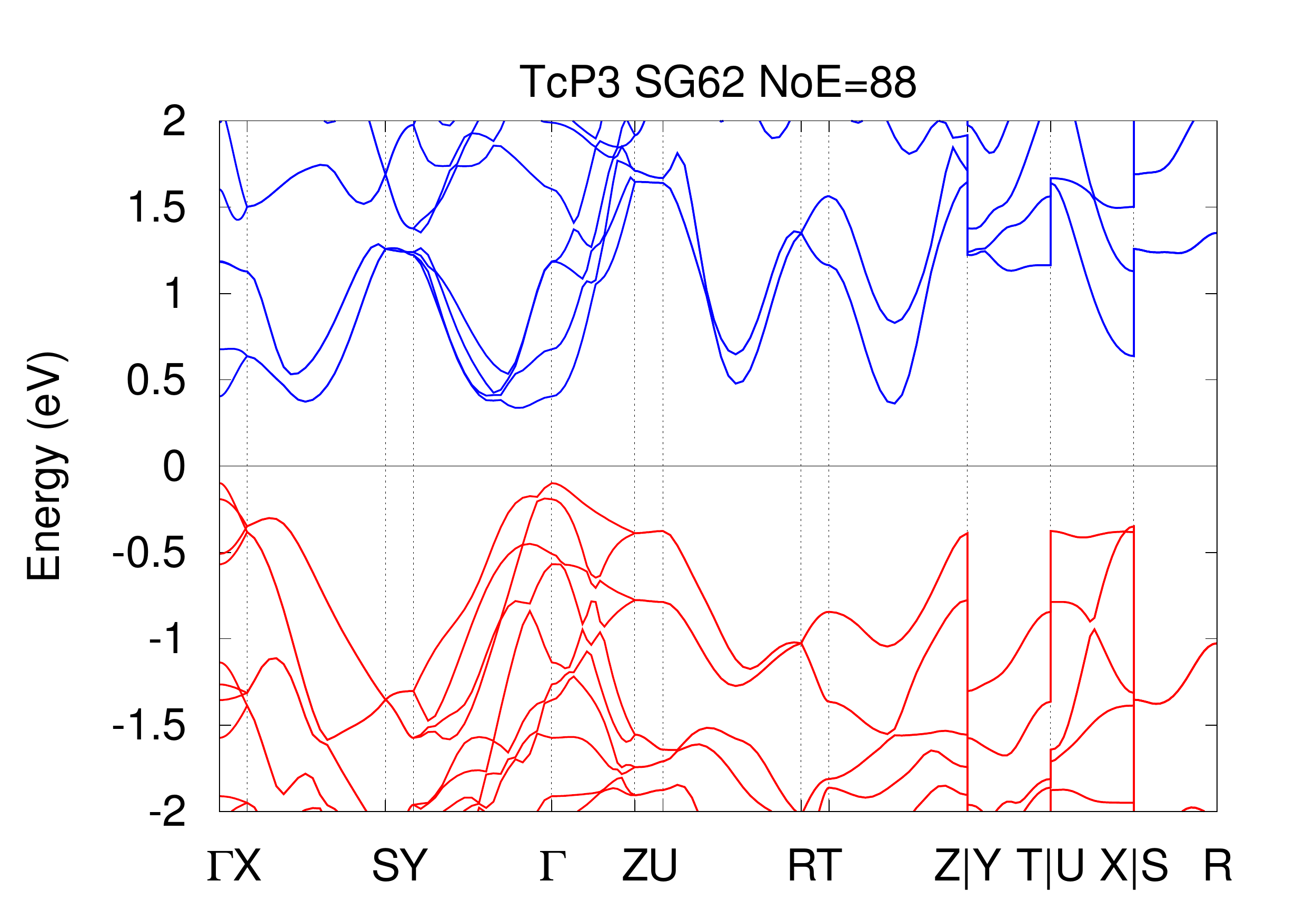}
}
\subfigure[LiSi SG88 NoA=16 NoE=40]{
\label{subfig:78364}
\includegraphics[scale=0.32,angle=270]{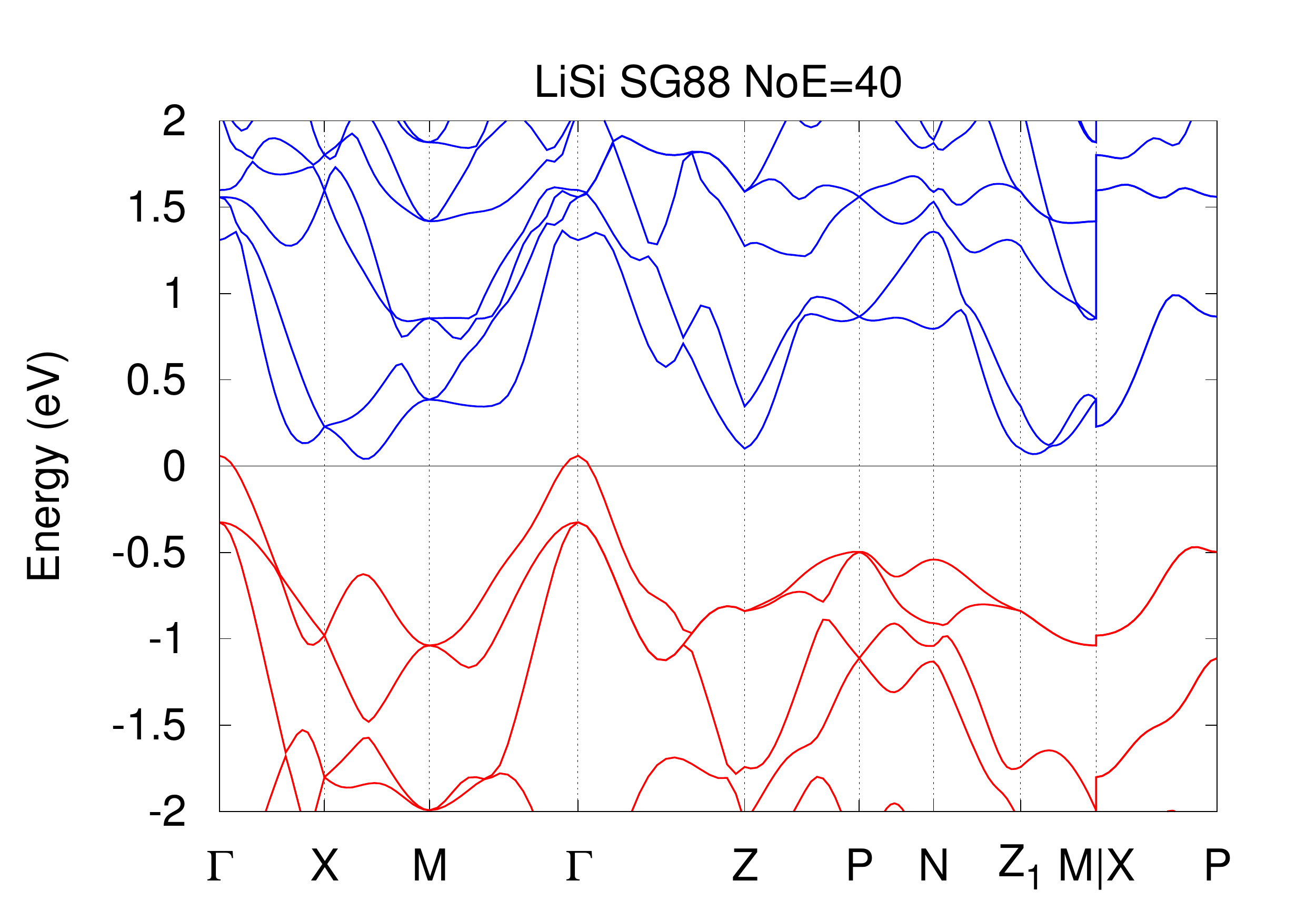}
}
\subfigure[ZnSb SG61 NoA=16 NoE=136]{
\label{subfig:601137}
\includegraphics[scale=0.32,angle=270]{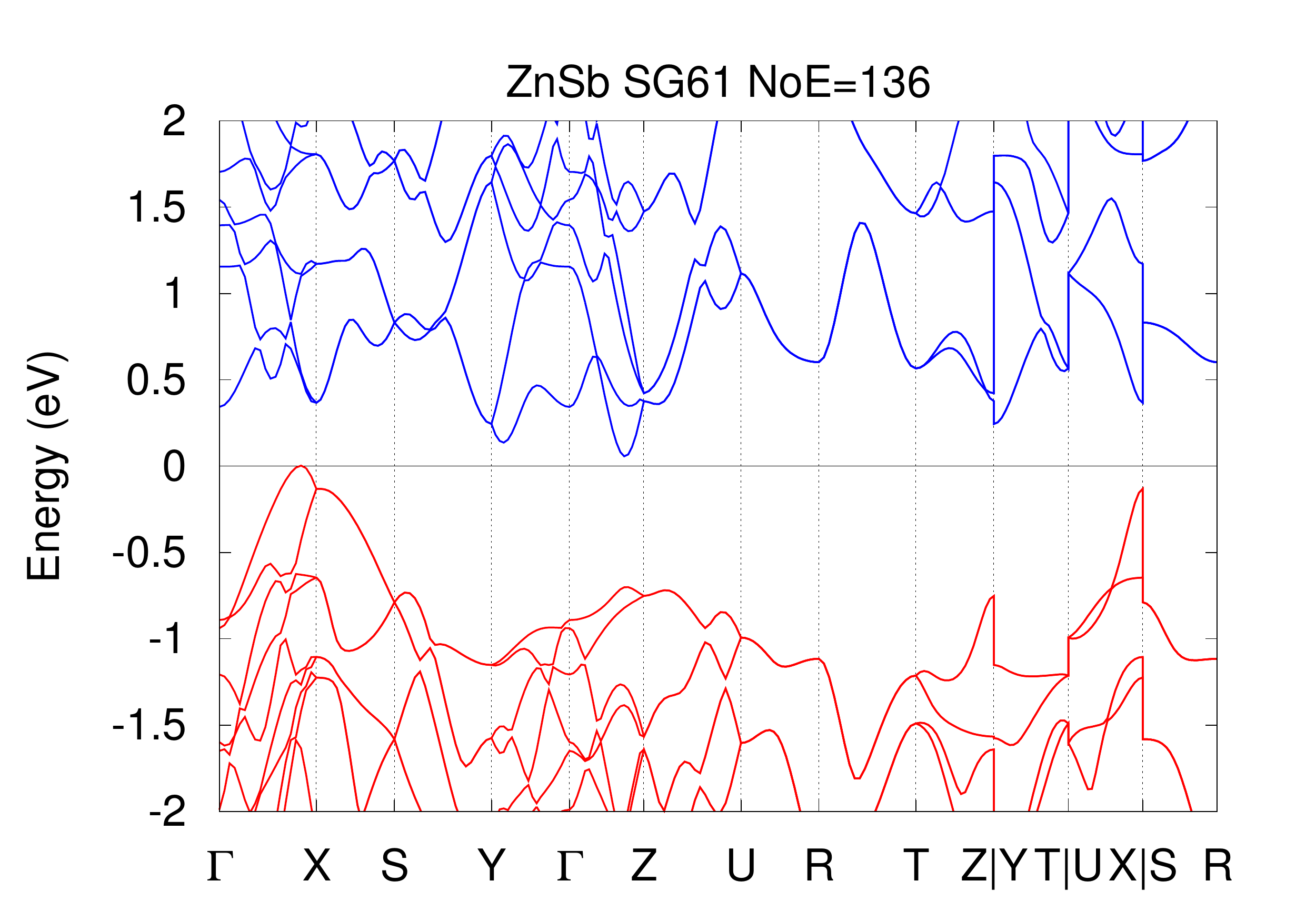}
}
\subfigure[CdSb SG61 NoA=16 NoE=136]{
\label{subfig:52831}
\includegraphics[scale=0.32,angle=270]{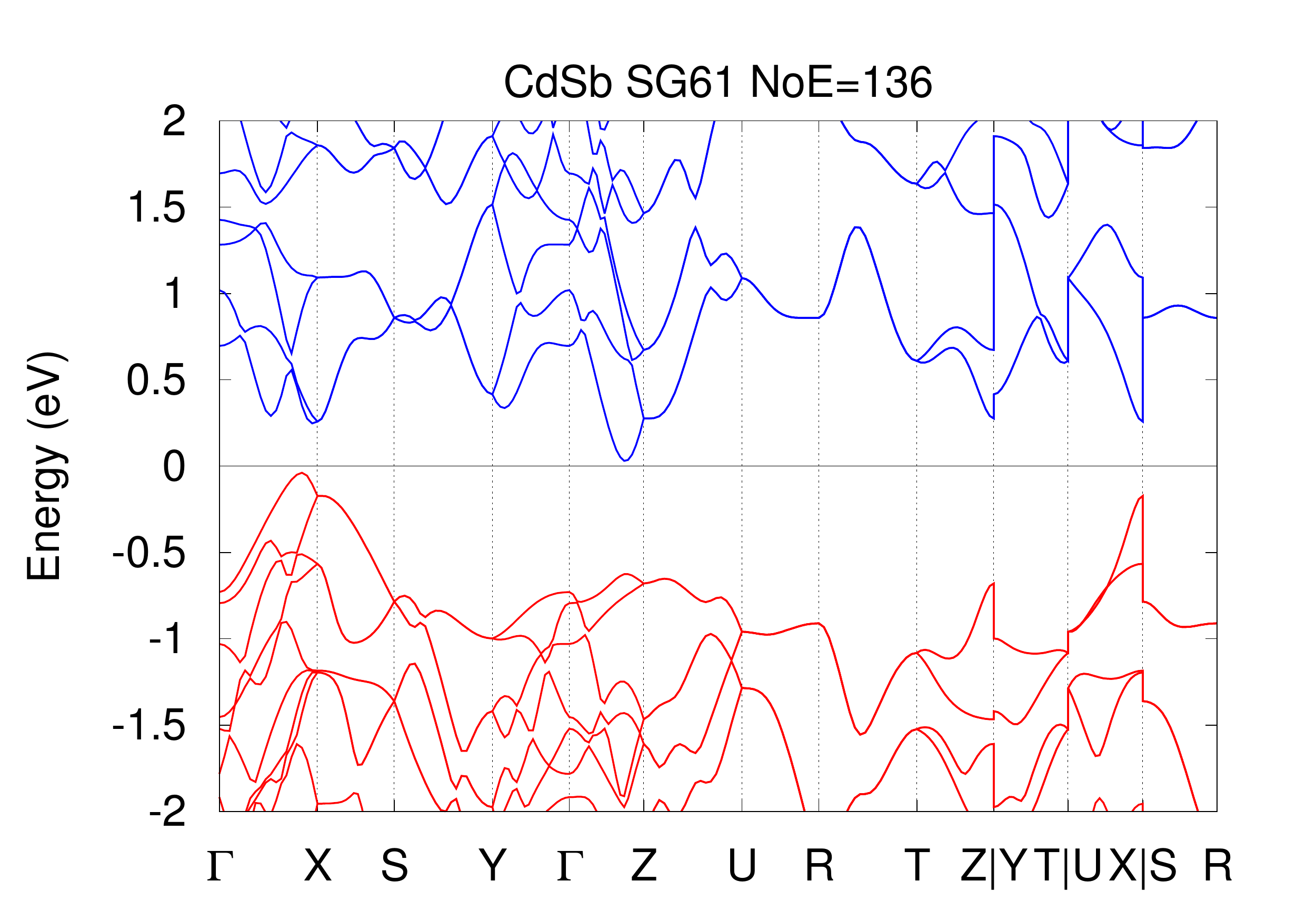}
}
\subfigure[Sb$_{3}$Ir SG204 NoA=16 NoE=96]{
\label{subfig:34050}
\includegraphics[scale=0.32,angle=270]{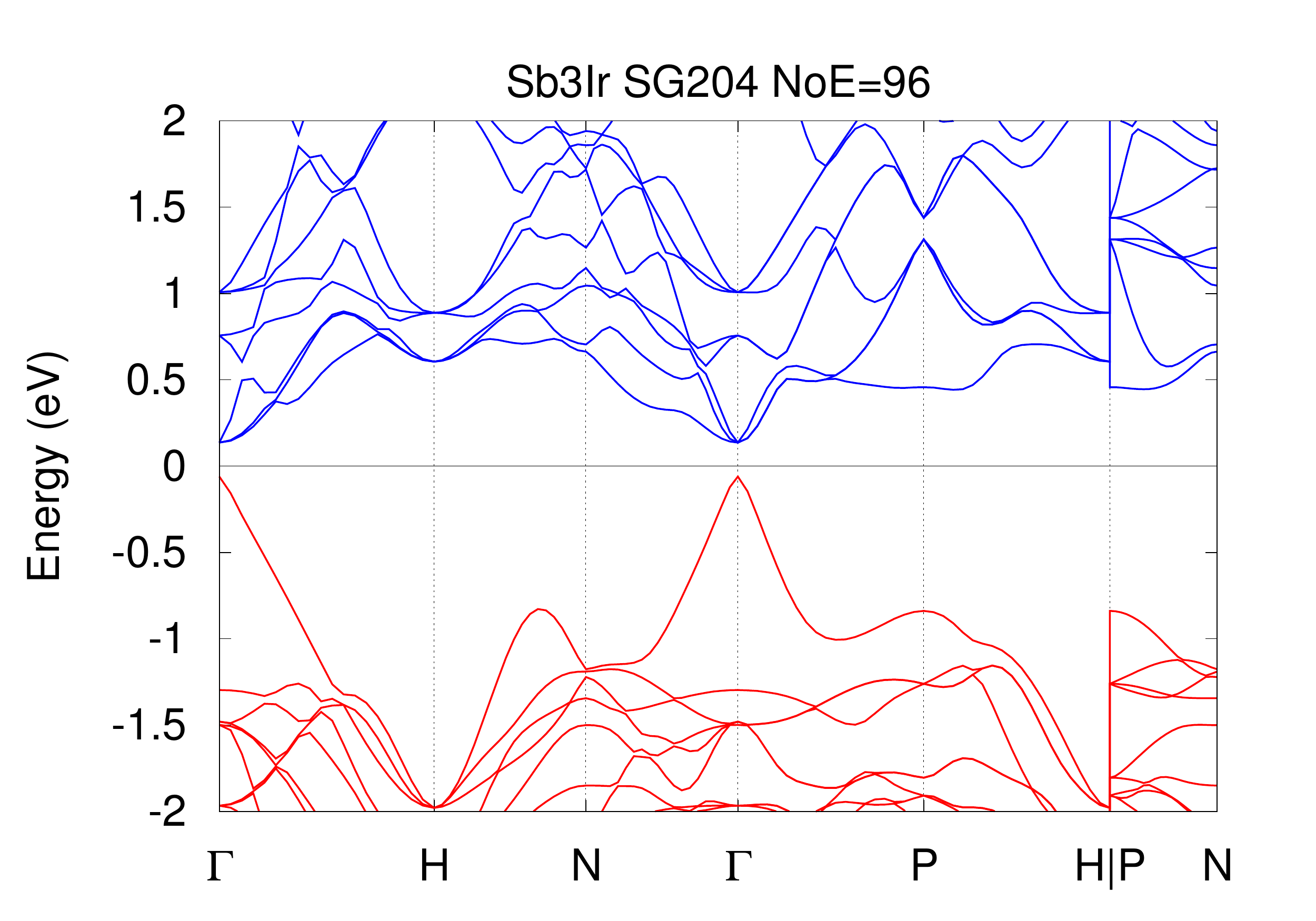}
}
\caption{\hyperref[tab:electride]{back to the table}}
\end{figure}

\begin{figure}[htp]
 \centering
\subfigure[SrP$_{3}$ SG12 NoA=16 NoE=100]{
\label{subfig:23628}
\includegraphics[scale=0.32,angle=270]{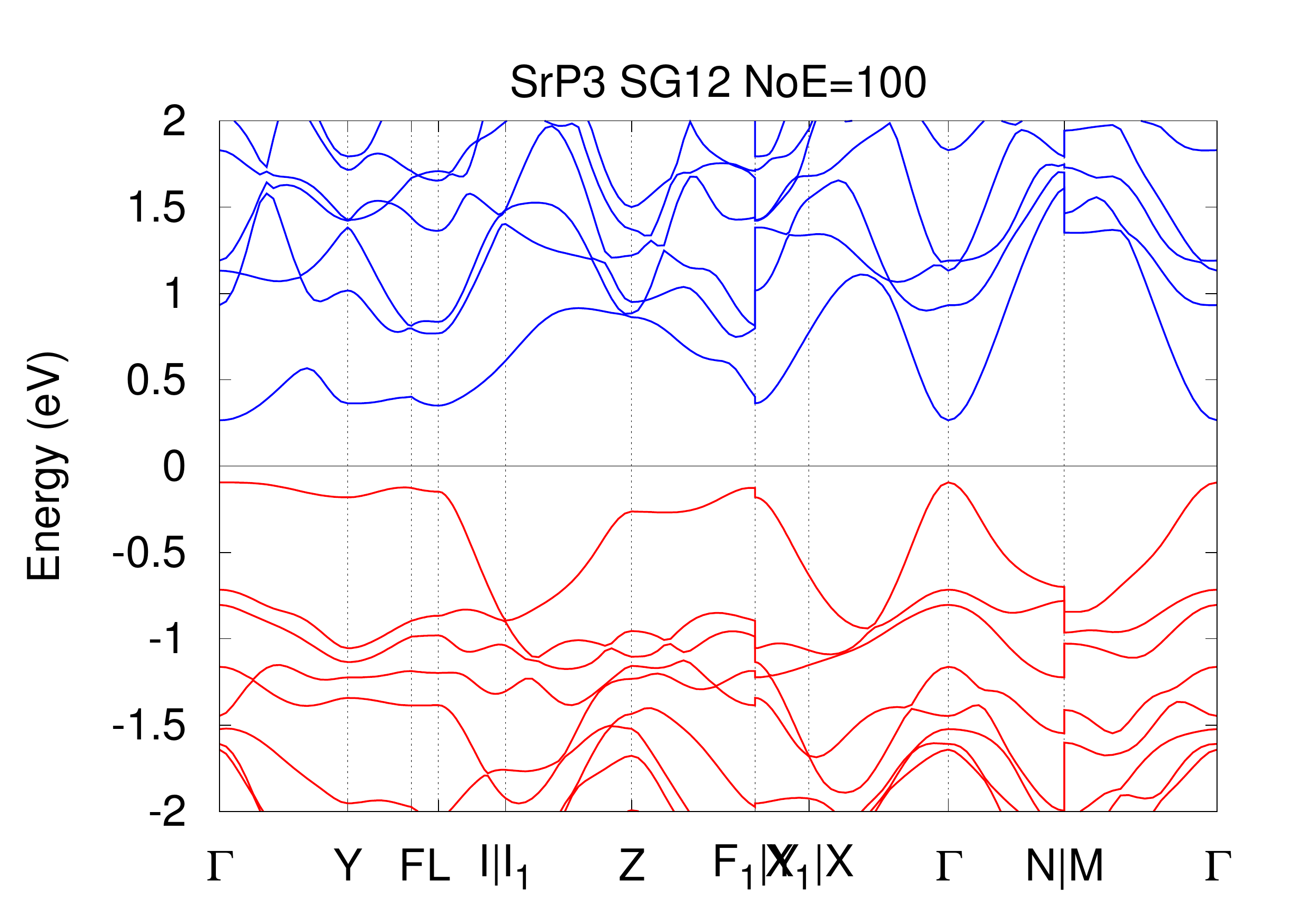}
}
\subfigure[RhN$_{3}$ SG204 NoA=16 NoE=96]{
\label{subfig:162107}
\includegraphics[scale=0.32,angle=270]{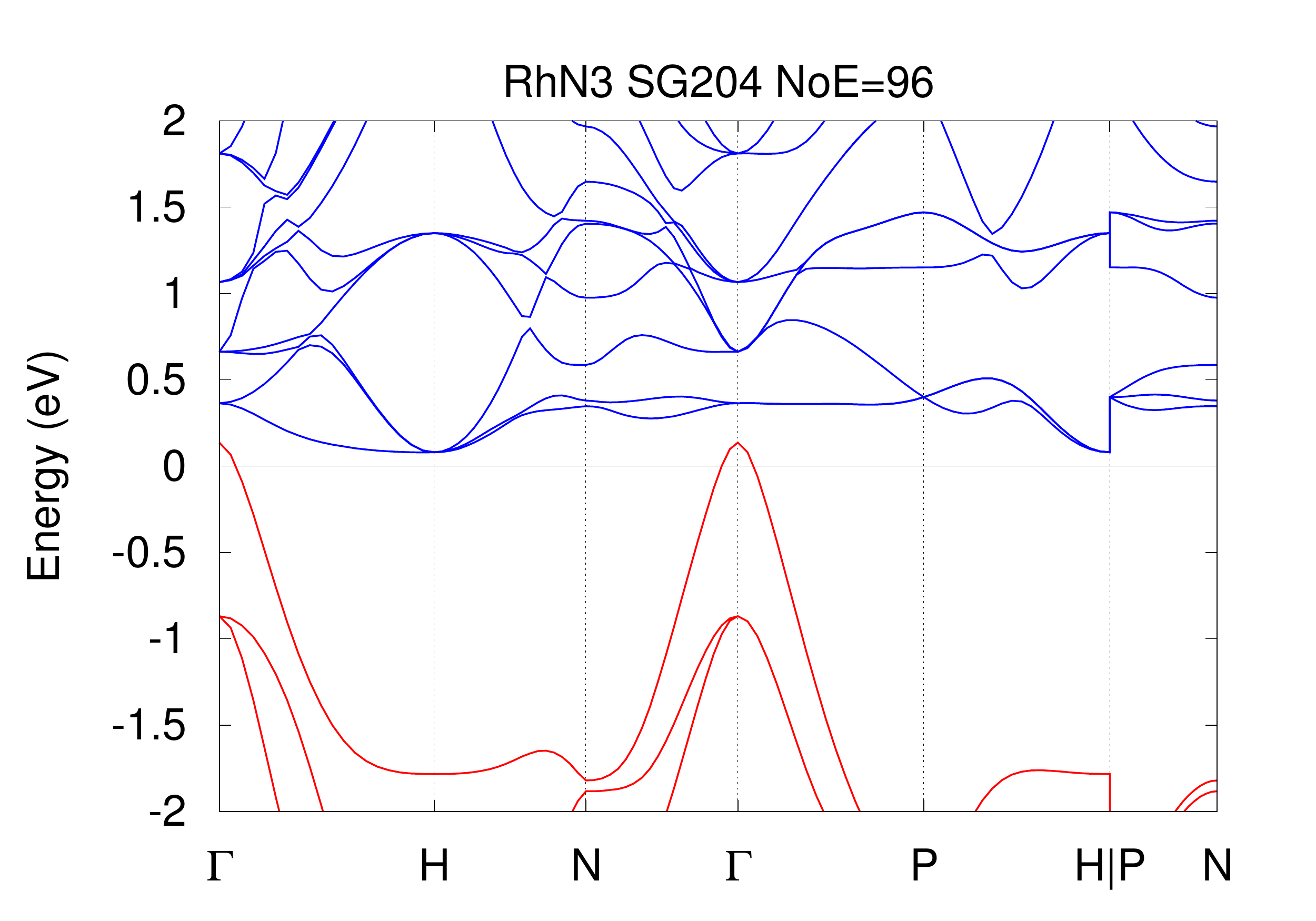}
}
\subfigure[Al$_{2}$(FeSi)$_{3}$ SG2 NoA=16 NoE=84]{
\label{subfig:83664}
\includegraphics[scale=0.32,angle=270]{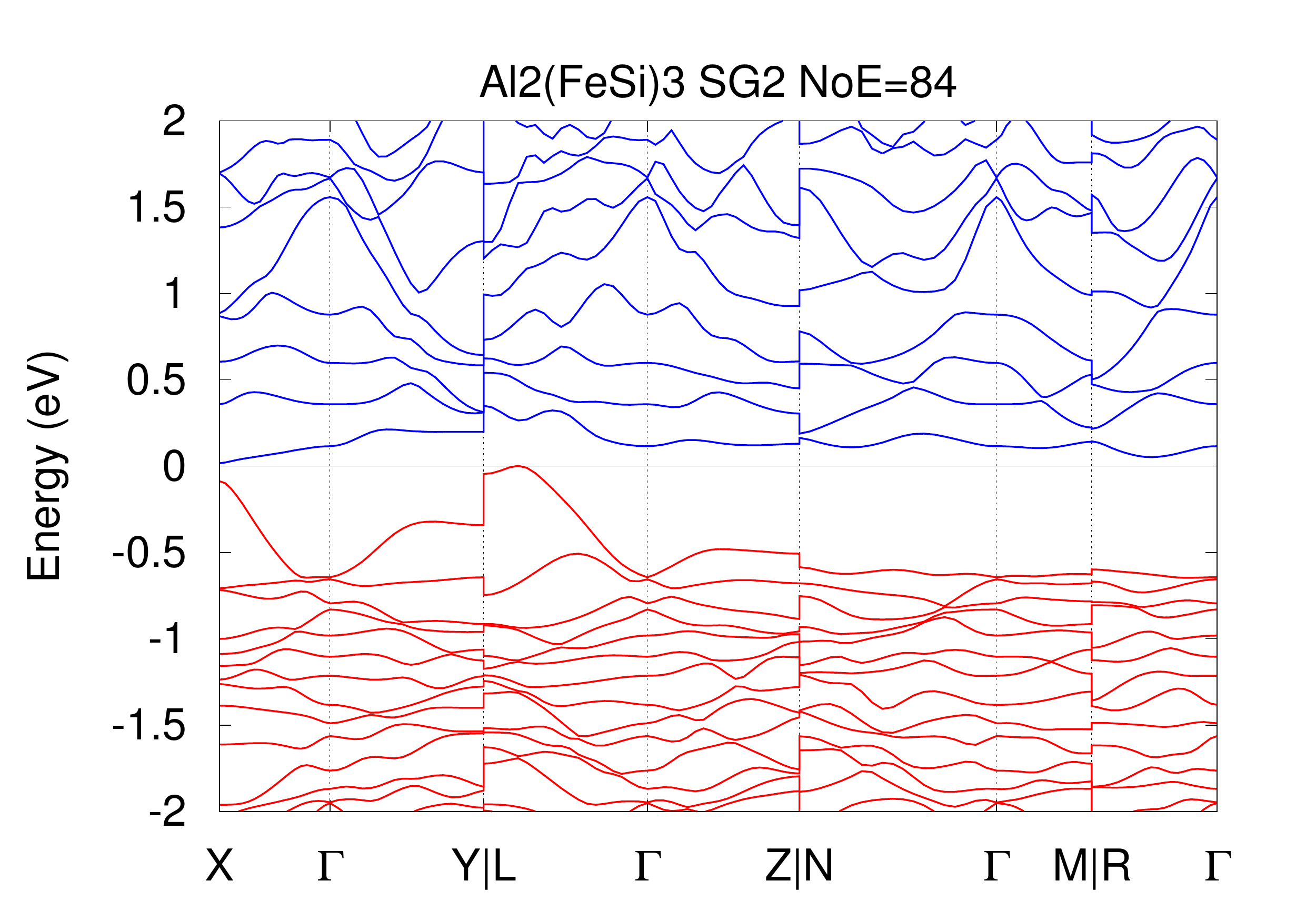}
}
\subfigure[ZnAs SG61 NoA=16 NoE=136]{
\label{subfig:427612}
\includegraphics[scale=0.32,angle=270]{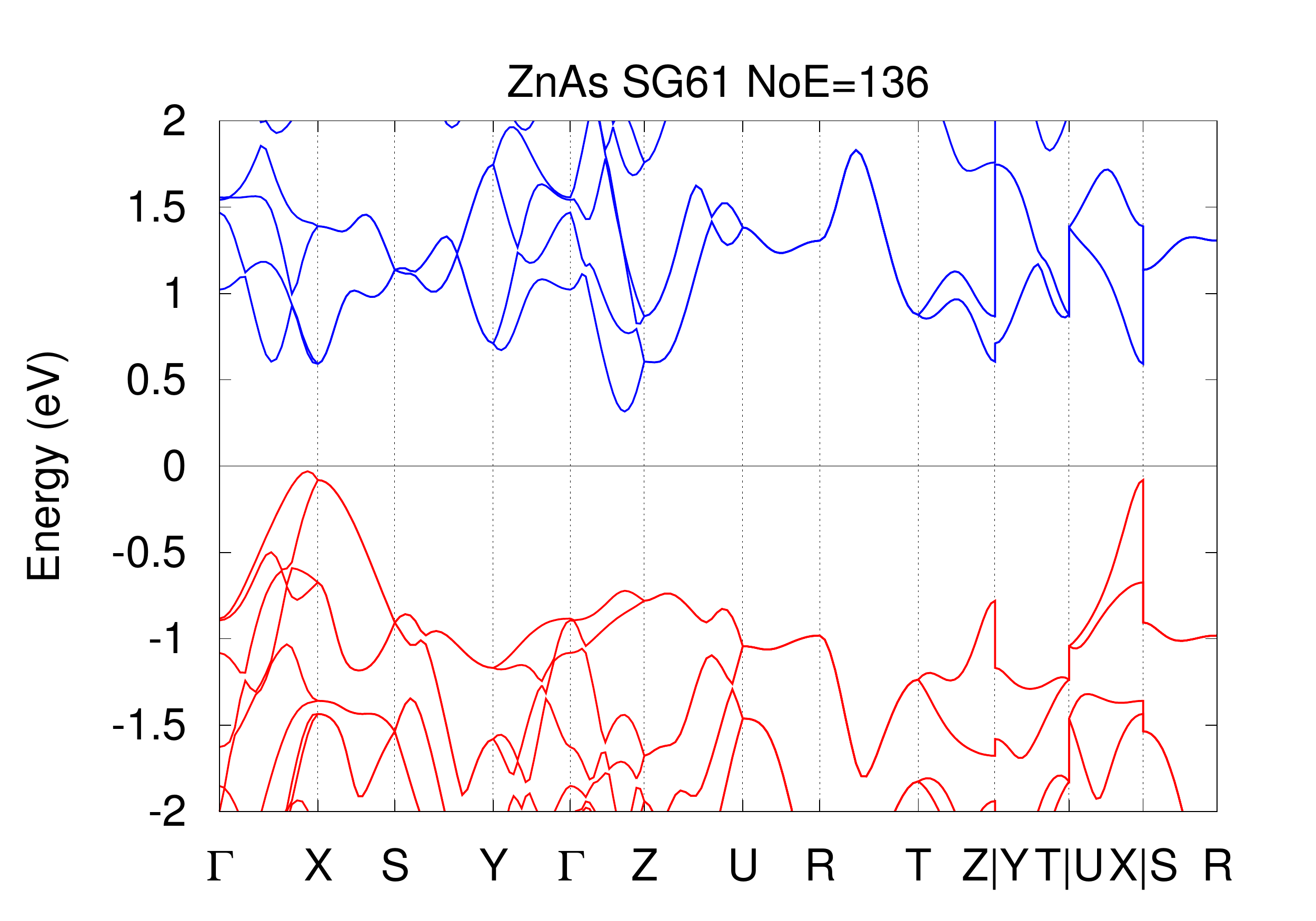}
}
\subfigure[Mg$_{3}$In SG166 NoA=16 NoE=36]{
\label{subfig:51976}
\includegraphics[scale=0.32,angle=270]{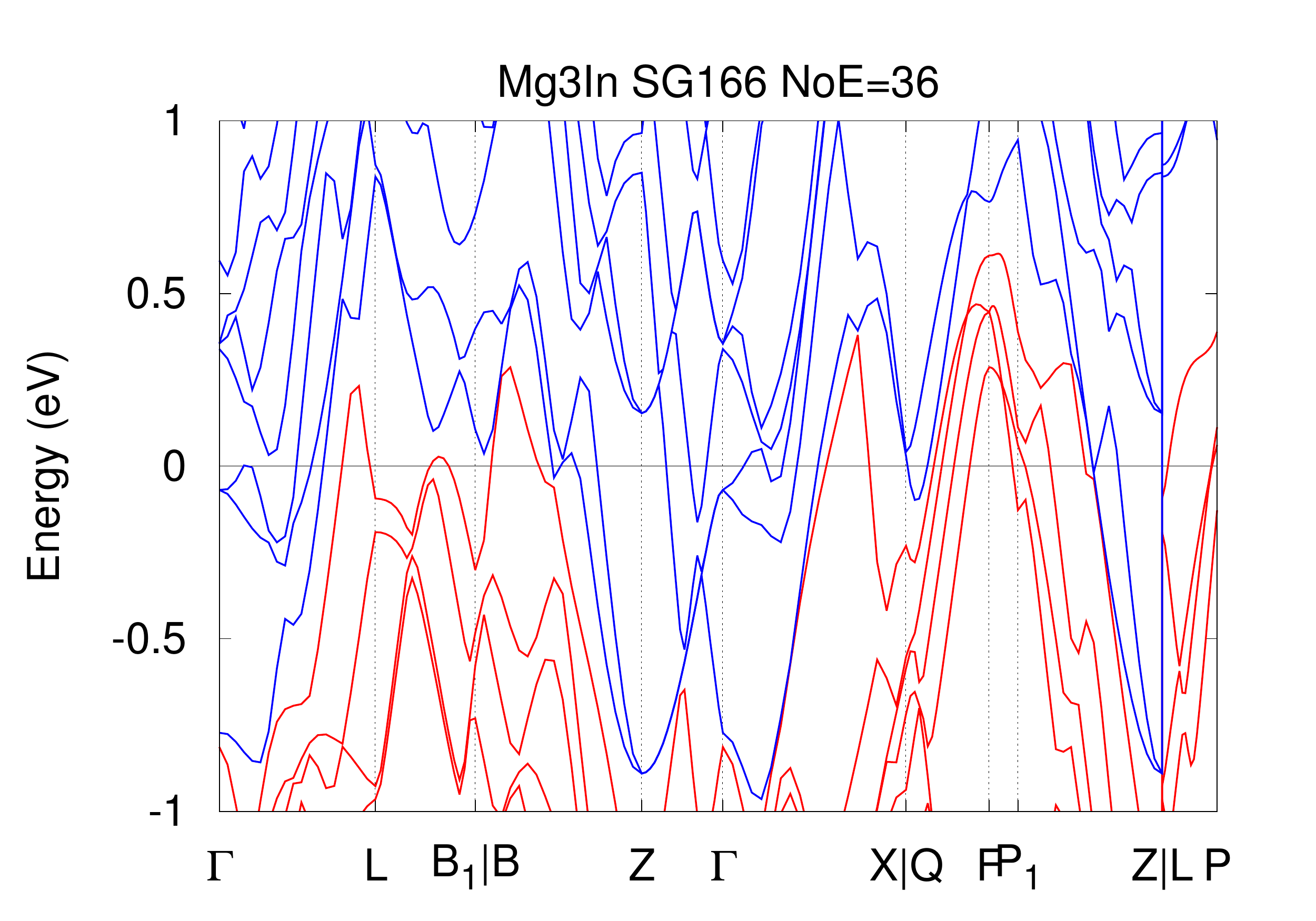}
}
\subfigure[ReP$_{3}$ SG62 NoA=16 NoE=88]{
\label{subfig:647985}
\includegraphics[scale=0.32,angle=270]{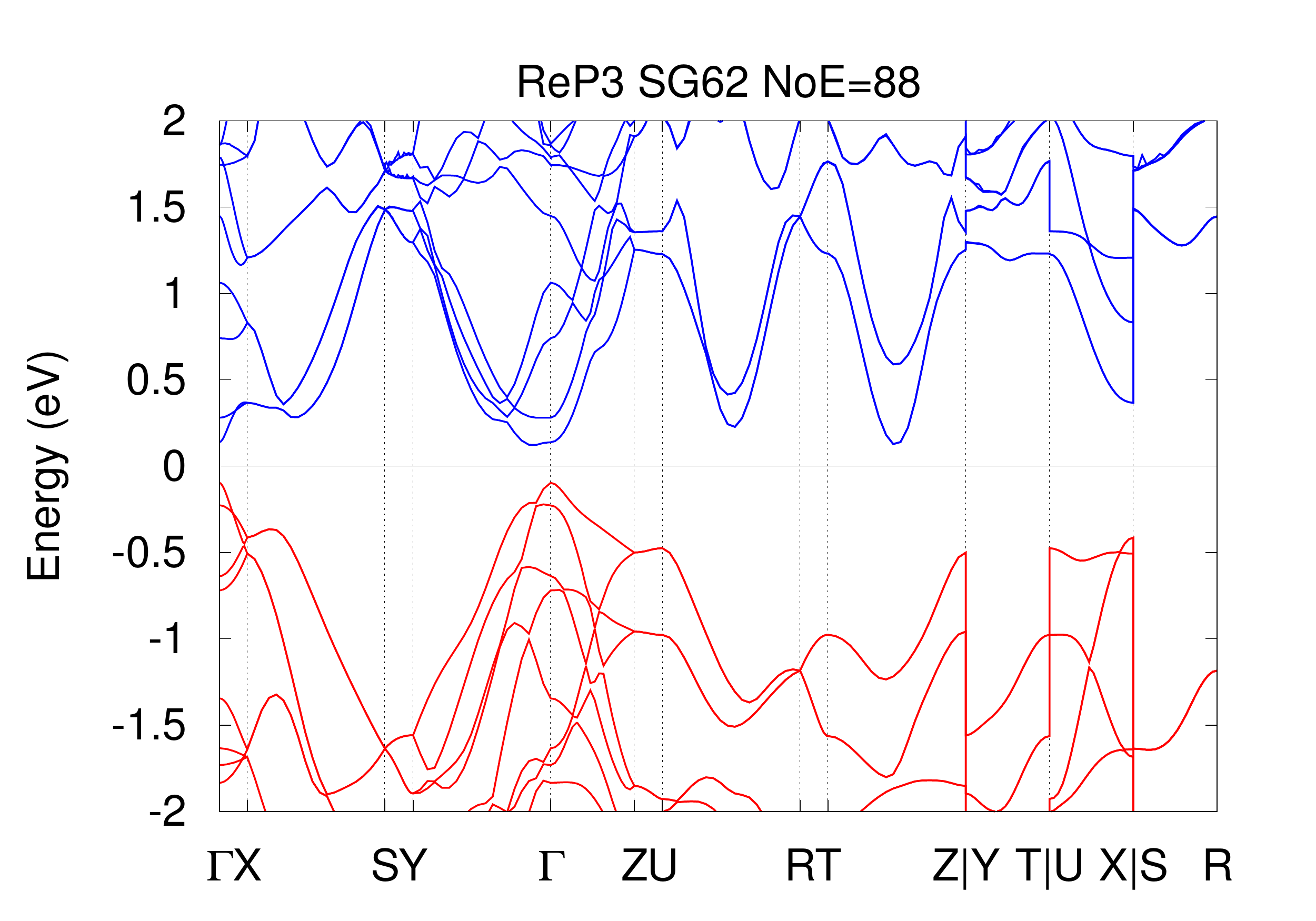}
}
\subfigure[Ga$_{3}$Fe SG136 NoA=16 NoE=68]{
\label{subfig:103448}
\includegraphics[scale=0.32,angle=270]{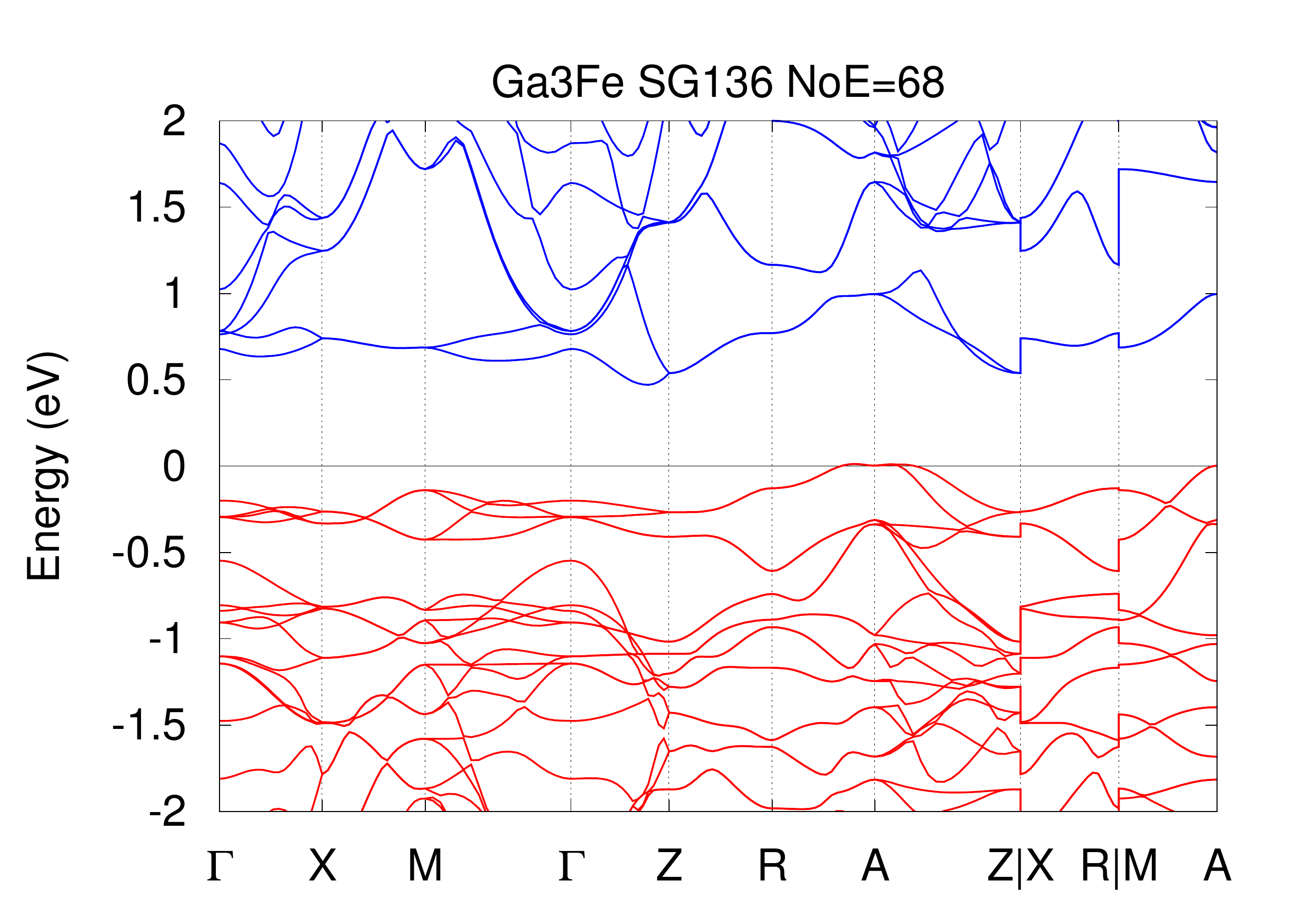}
}
\subfigure[Ga$_{3}$Os SG136 NoA=16 NoE=68]{
\label{subfig:635023}
\includegraphics[scale=0.32,angle=270]{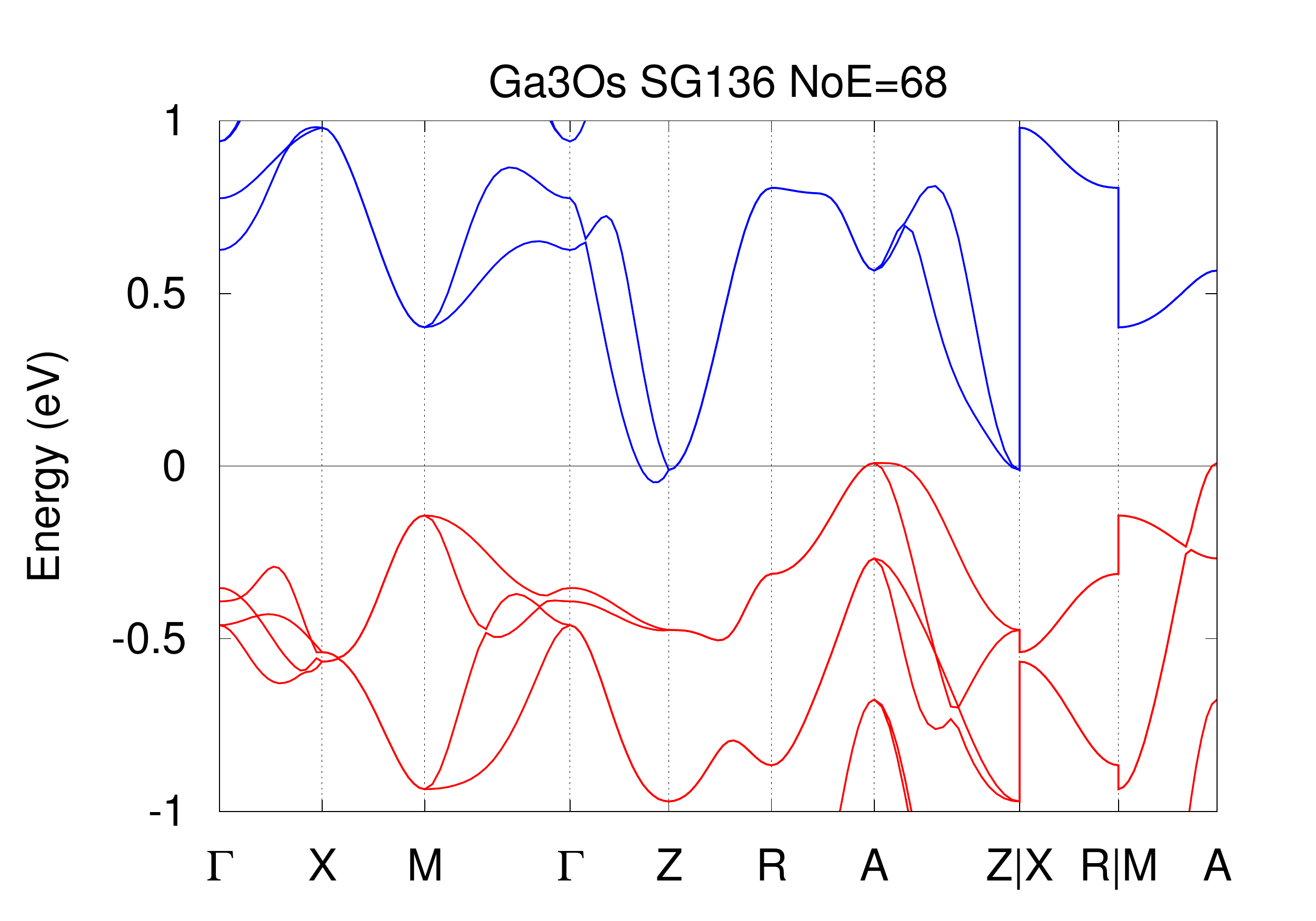}
}
\caption{\hyperref[tab:electride]{back to the table}}
\end{figure}

\begin{figure}[htp]
 \centering
\subfigure[Ga$_{3}$Ir SG136 NoA=16 NoE=72]{
\label{subfig:634441}
\includegraphics[scale=0.32,angle=270]{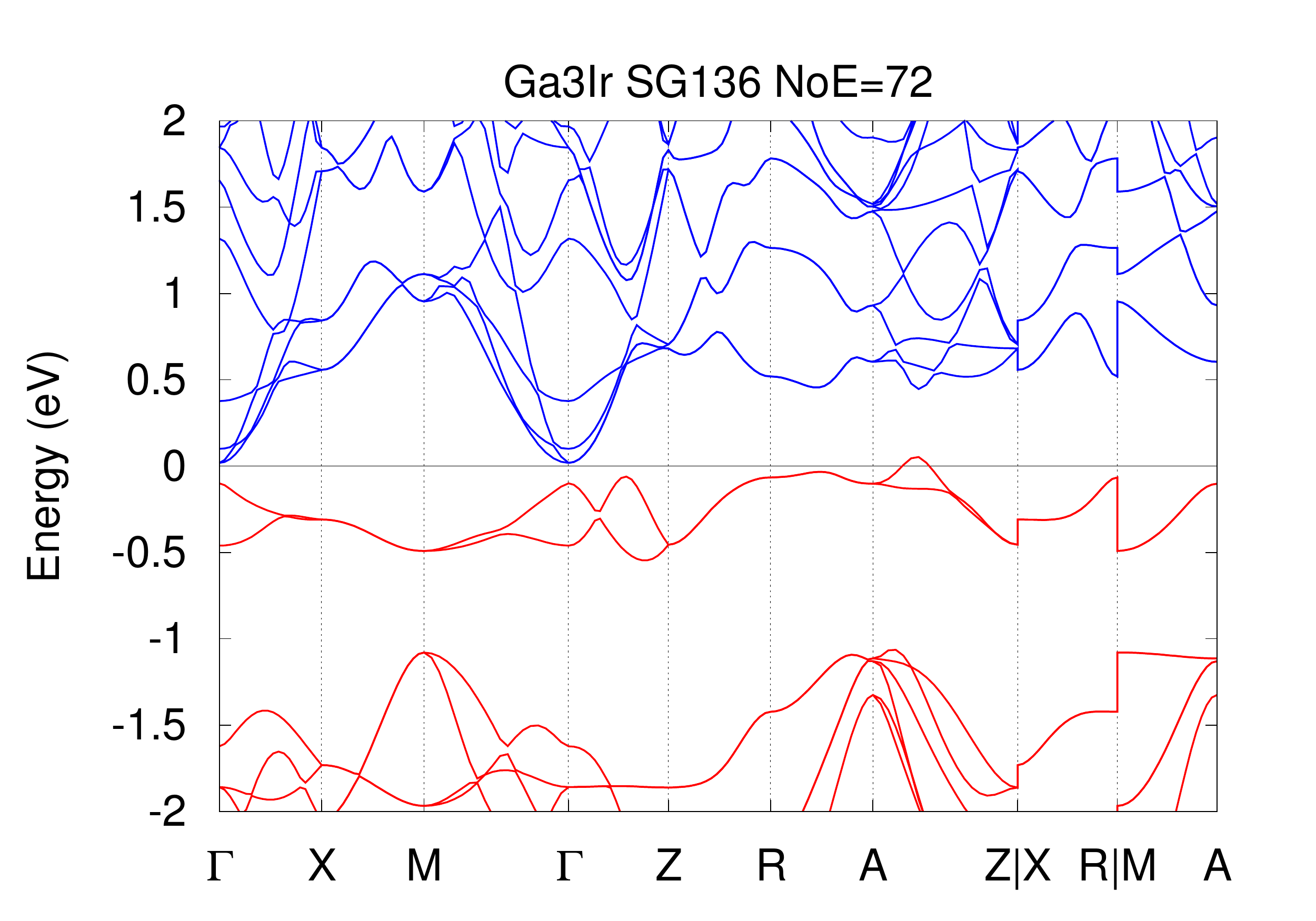}
}
\subfigure[SiPd$_{3}$ SG62 NoA=16 NoE=136]{
\label{subfig:648855}
\includegraphics[scale=0.32,angle=270]{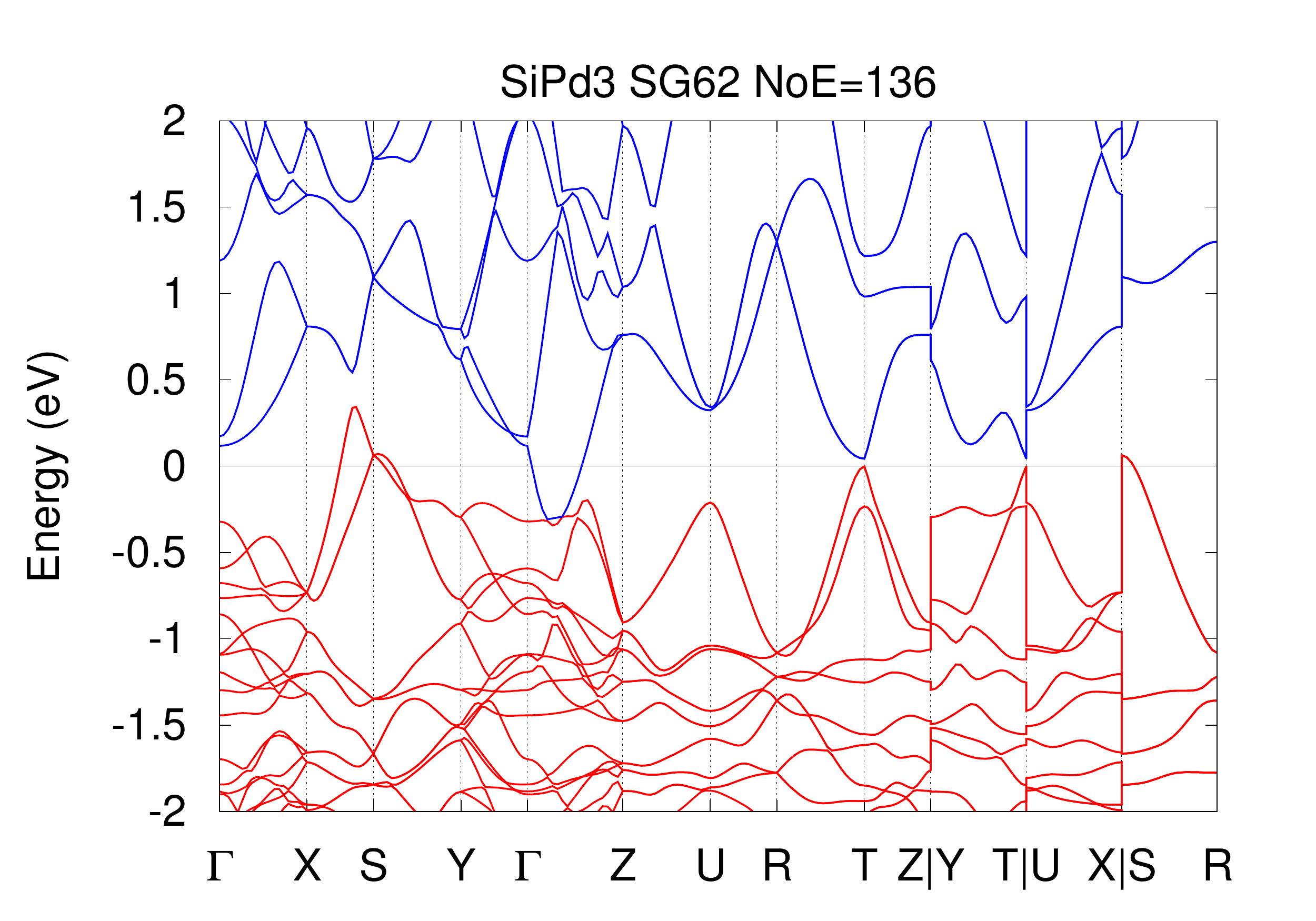}
}
\subfigure[Ga$_{3}$Fe SG136 NoA=16 NoE=68]{
\label{subfig:103448}
\includegraphics[scale=0.32,angle=270]{mater-band/103448}
}
\subfigure[Tl$_{2}$(CdSb)$_{3}$ SG12 NoA=16 NoE=114]{
\label{subfig:76500}
\includegraphics[scale=0.32,angle=270]{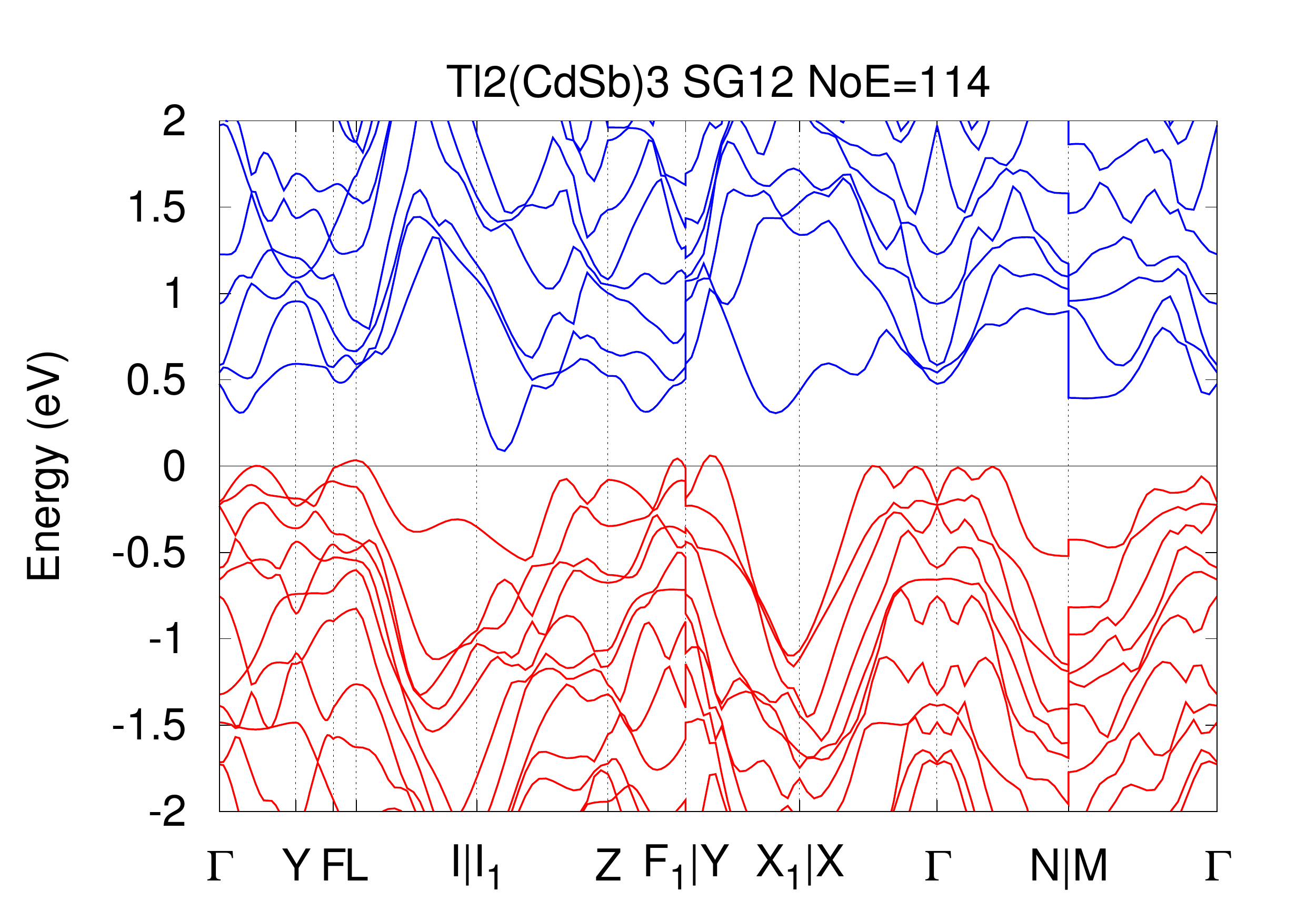}
}
\subfigure[P$_{3}$Ir SG204 NoA=16 NoE=96]{
\label{subfig:640899}
\includegraphics[scale=0.32,angle=270]{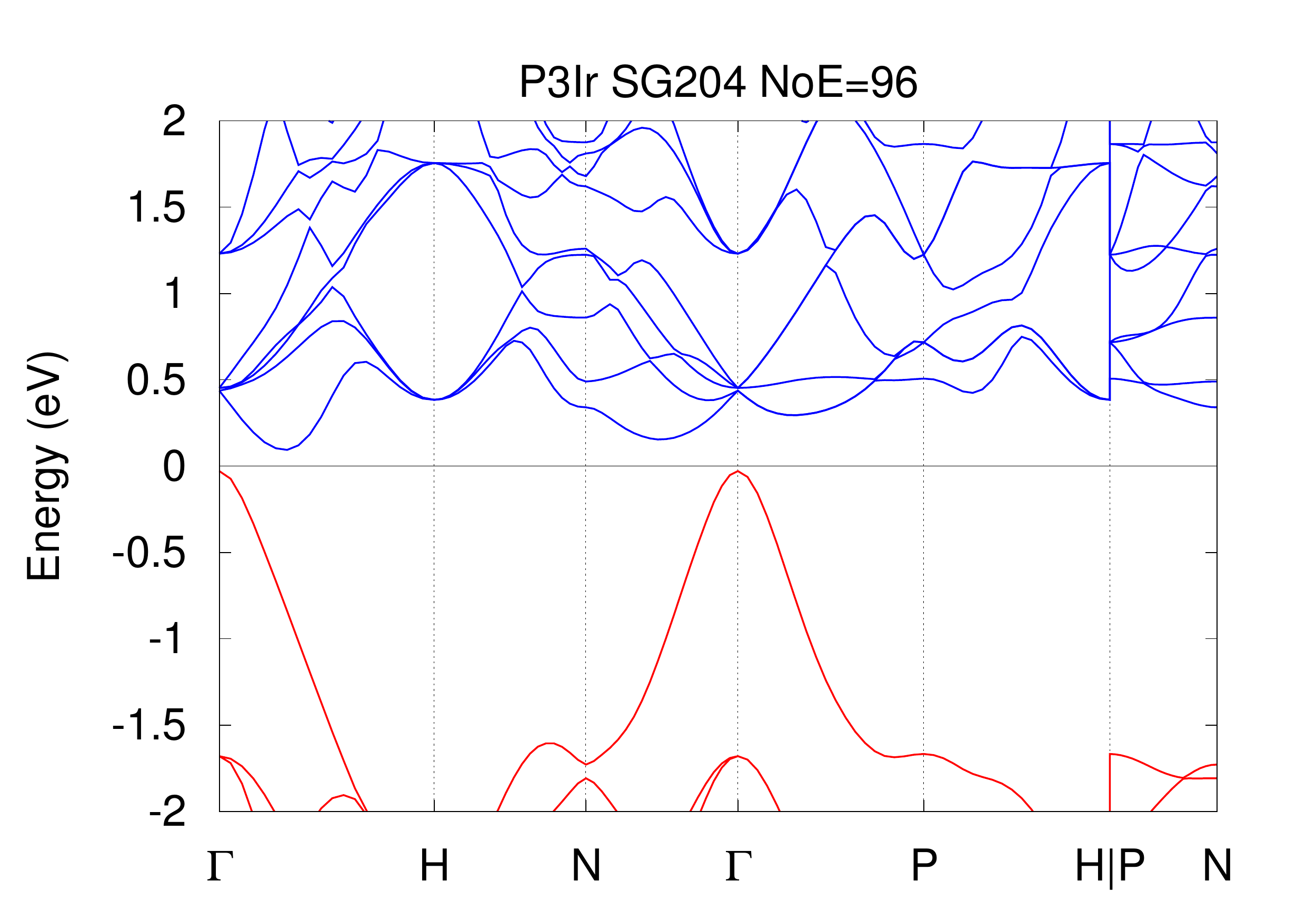}
}
\subfigure[LiSi SG88 NoA=16 NoE=40]{
\label{subfig:96509}
\includegraphics[scale=0.32,angle=270]{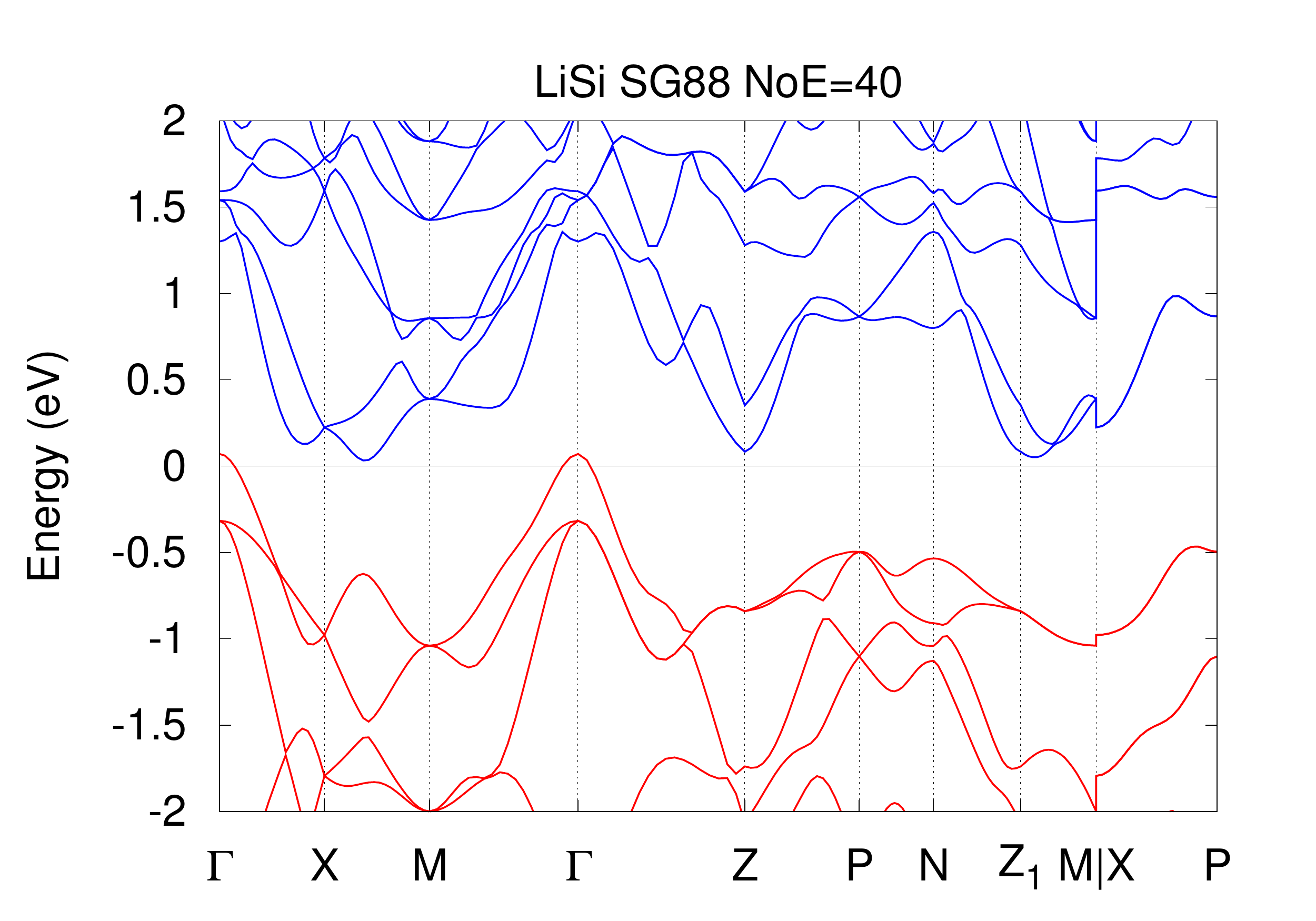}
}
\subfigure[NiP SG61 NoA=16 NoE=120]{
\label{subfig:27159}
\includegraphics[scale=0.32,angle=270]{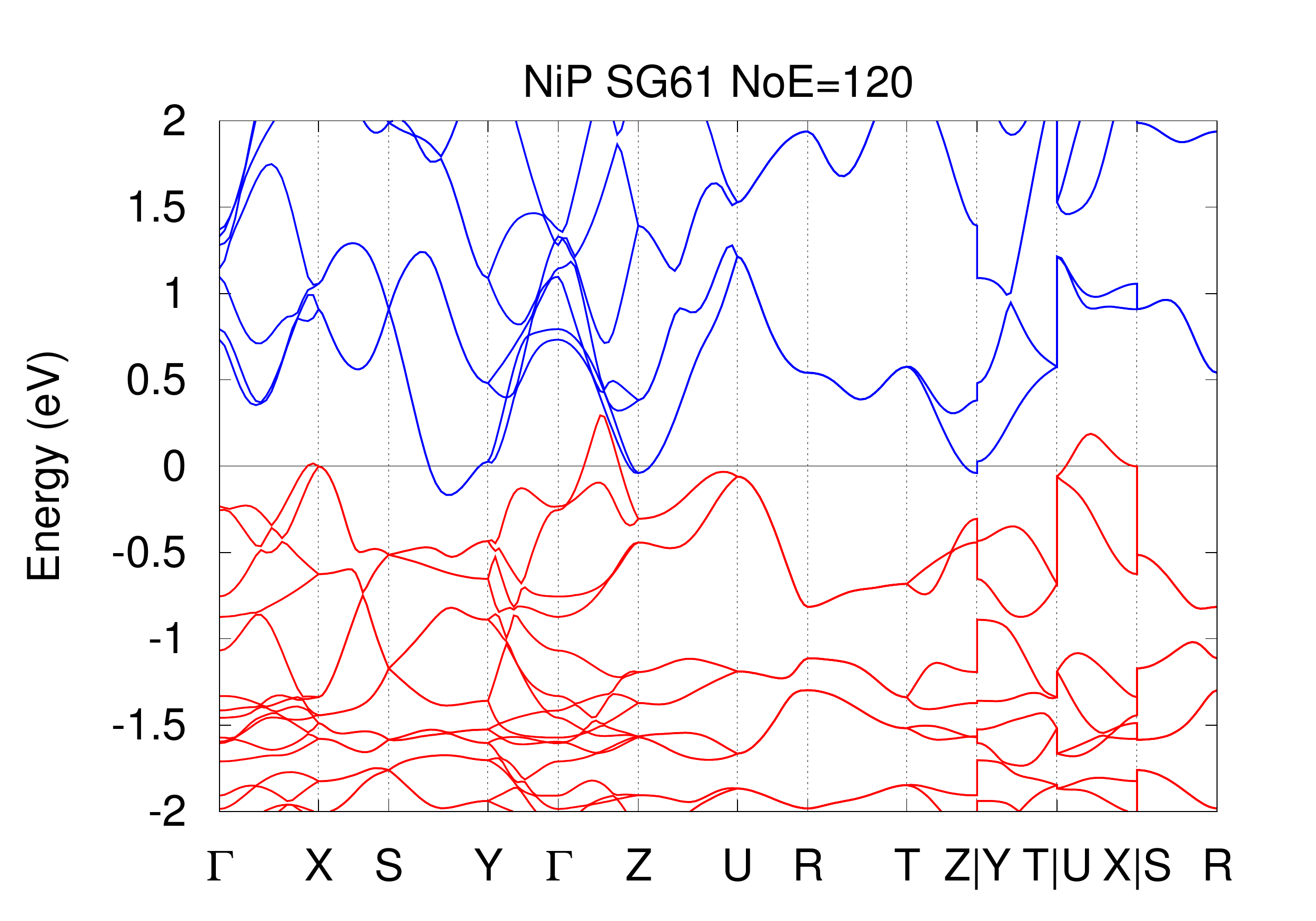}
}
\subfigure[Li$_{3}$CaMnN$_{3}$ SG148 NoA=16 NoE=70]{
\label{subfig:408324}
\includegraphics[scale=0.32,angle=270]{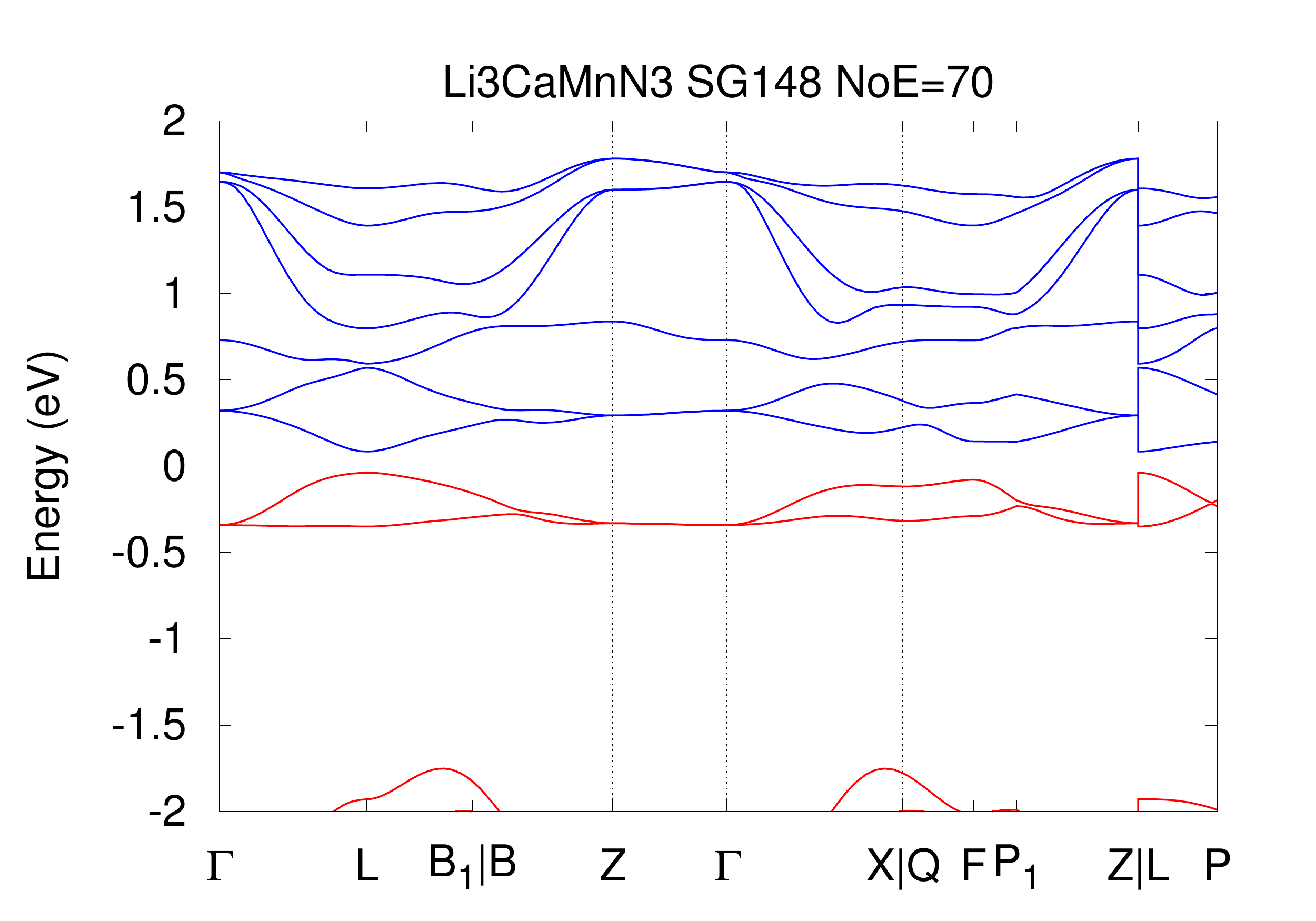}
}
\caption{\hyperref[tab:electride]{back to the table}}
\end{figure}

\begin{figure}[htp]
 \centering
\subfigure[Hf$_{3}$Sb SG82 NoA=16 NoE=68]{
\label{subfig:638878}
\includegraphics[scale=0.32,angle=270]{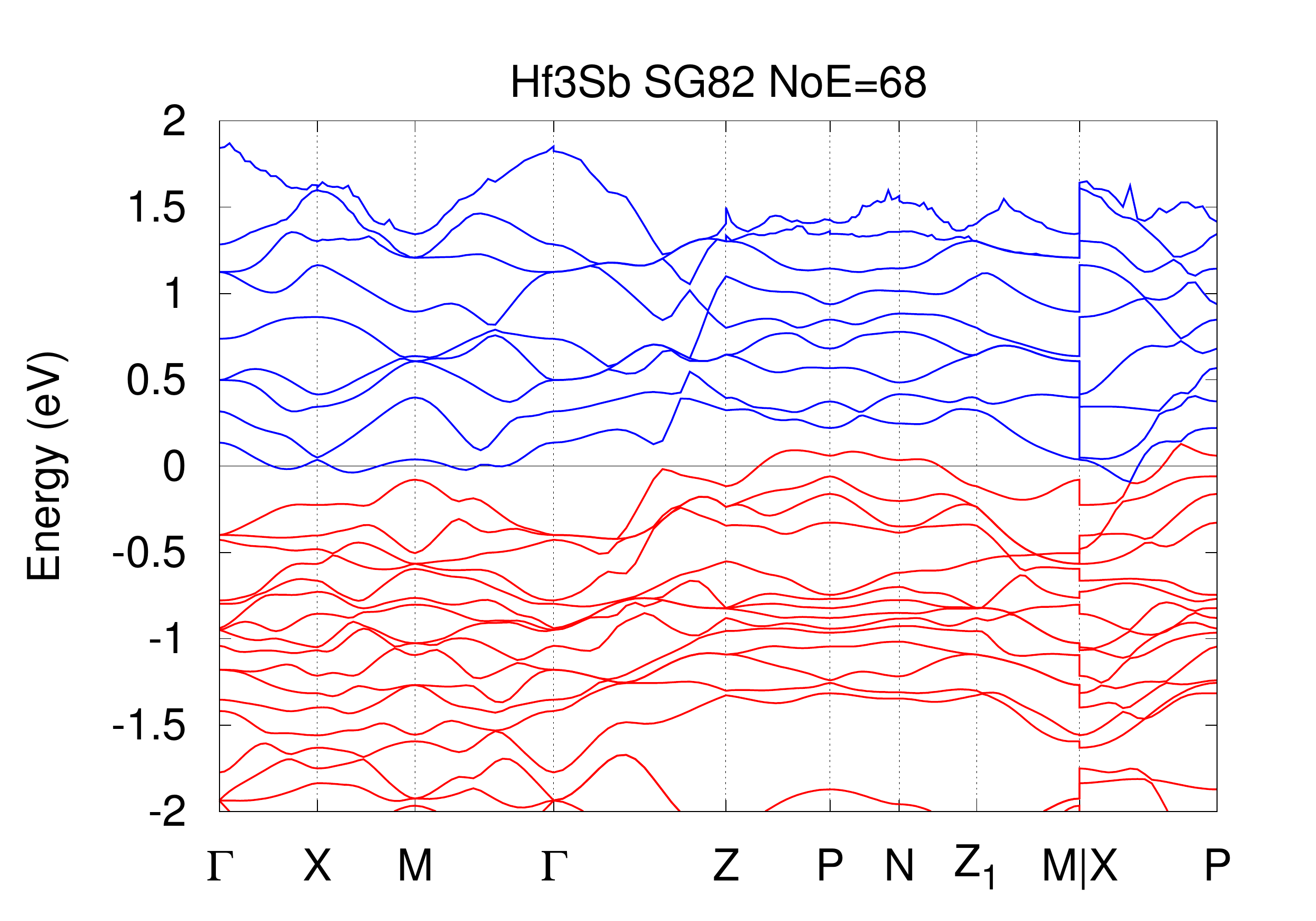}
}
\subfigure[Pr(FeAs$_{3}$)$_{4}$ SG204 NoA=17 NoE=103]{
\label{subfig:610521}
\includegraphics[scale=0.32,angle=270]{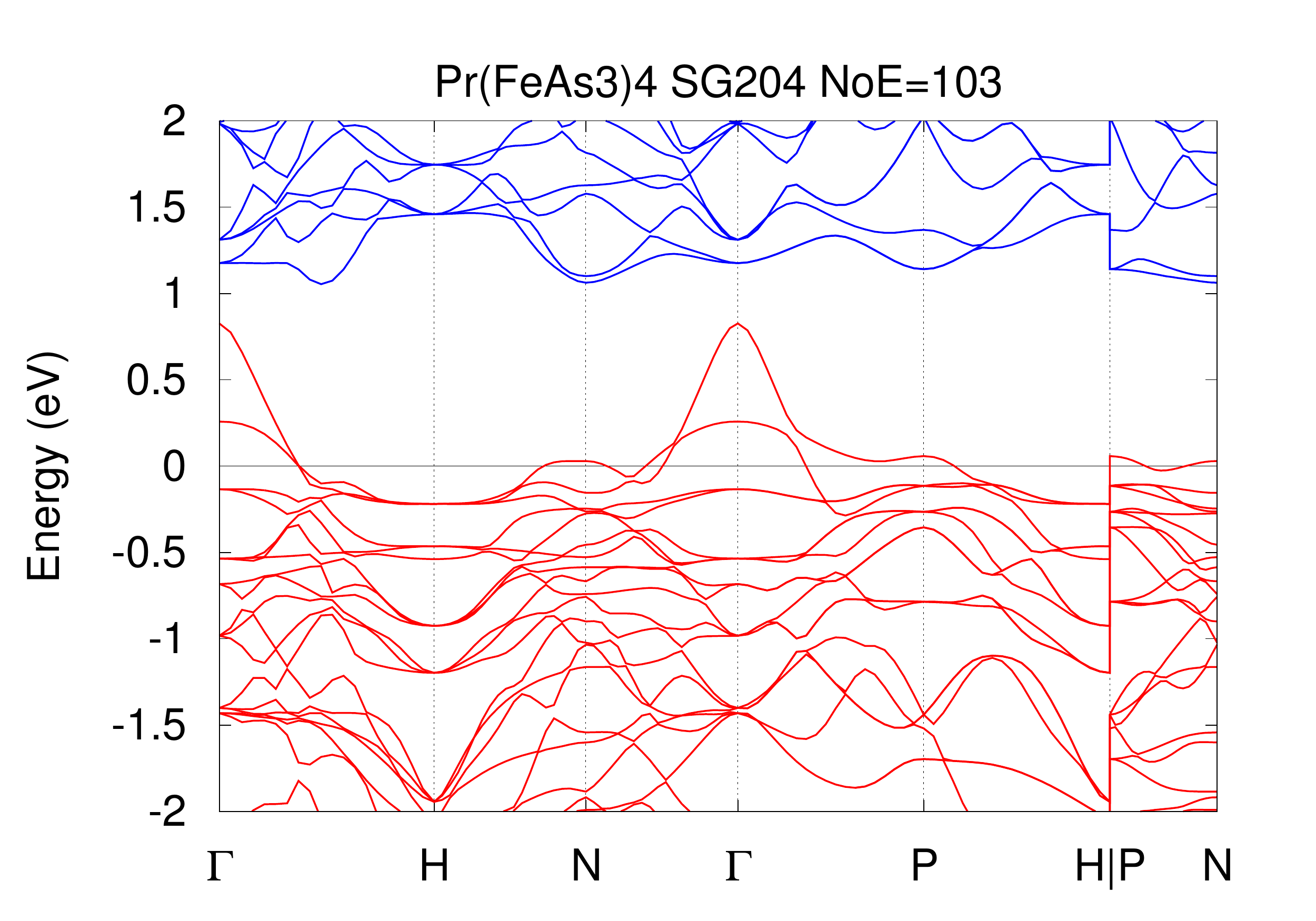}
}
\subfigure[Ce(FeSb$_{3}$)$_{4}$ SG204 NoA=17 NoE=103]{
\label{subfig:621065}
\includegraphics[scale=0.32,angle=270]{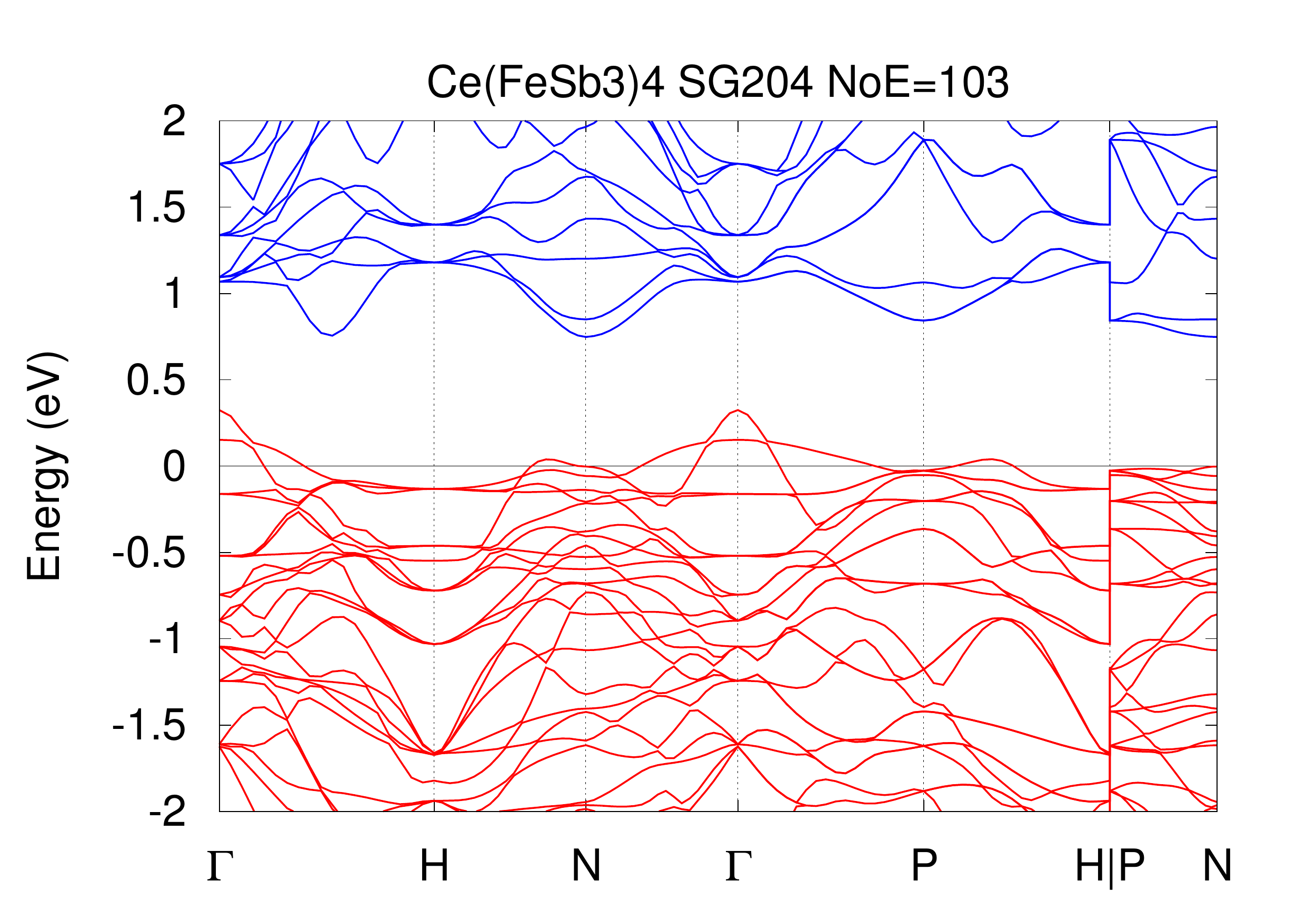}
}
\subfigure[Ce(Sb$_{3}$Ru)$_{4}$ SG204 NoA=17 NoE=103]{
\label{subfig:621988}
\includegraphics[scale=0.32,angle=270]{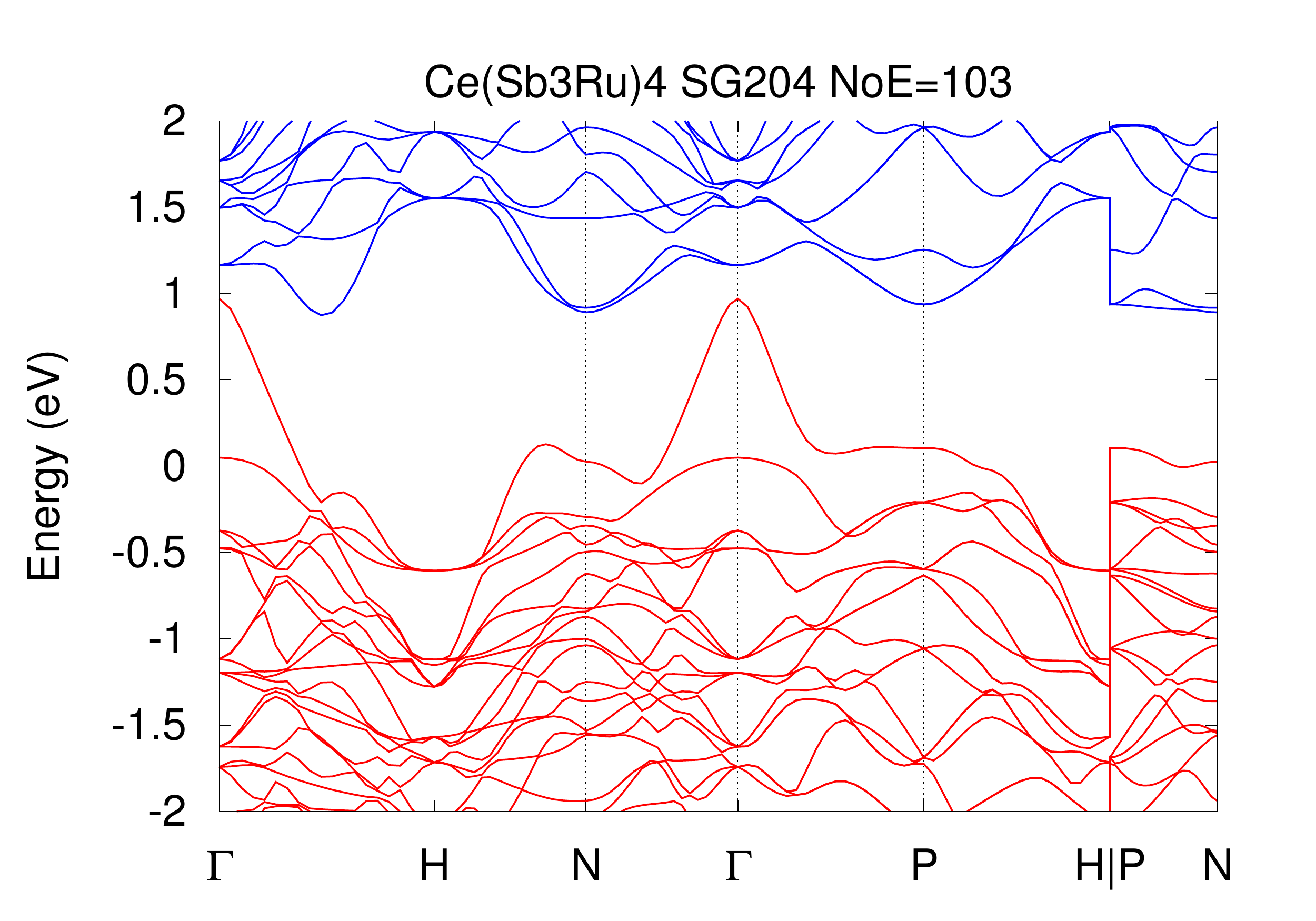}
}
\subfigure[Ce(As$_{3}$Os)$_{4}$ SG204 NoA=17 NoE=103]{
\label{subfig:610010}
\includegraphics[scale=0.32,angle=270]{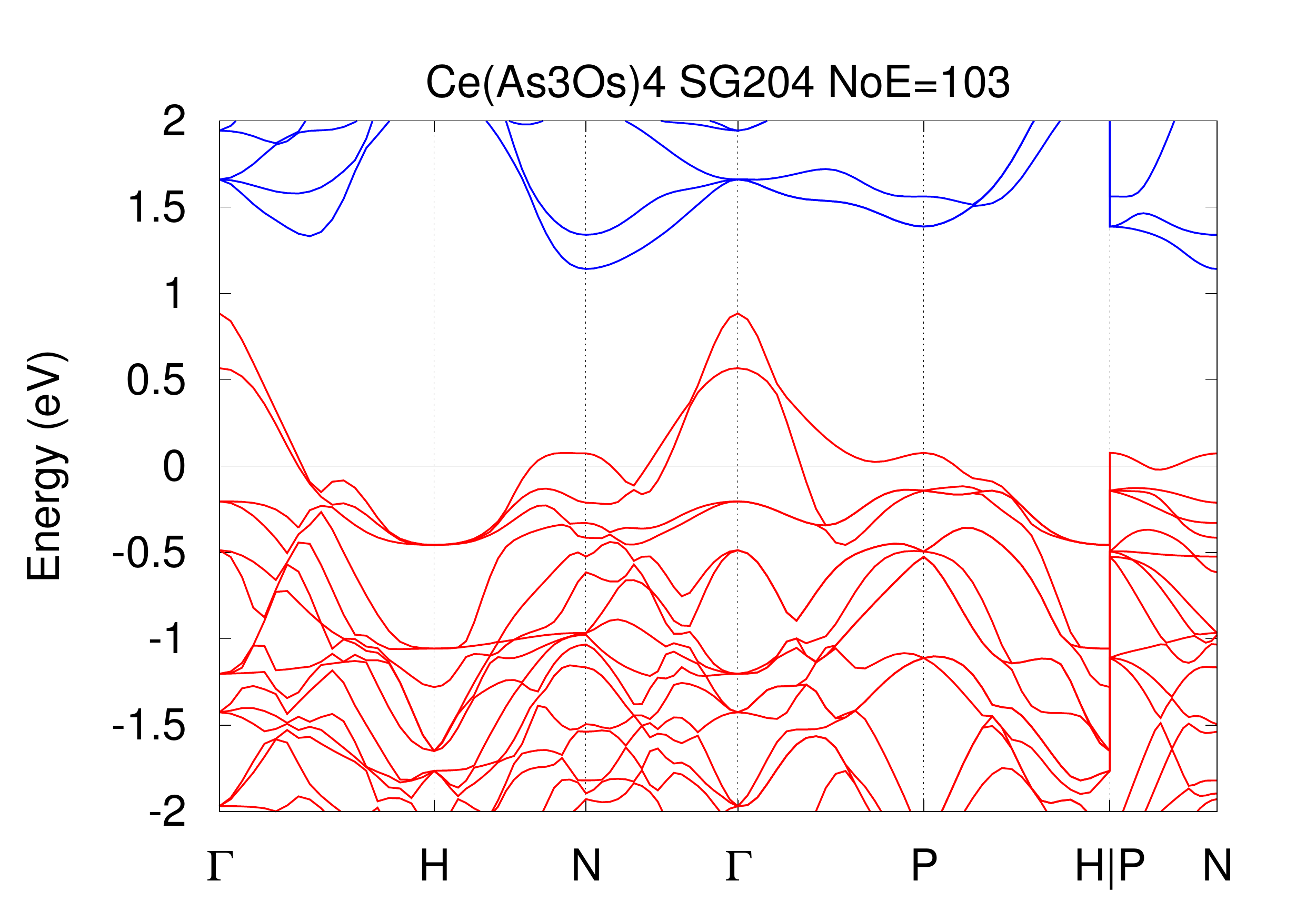}
}
\subfigure[La(As$_{3}$Os)$_{4}$ SG204 NoA=17 NoE=103]{
\label{subfig:610776}
\includegraphics[scale=0.32,angle=270]{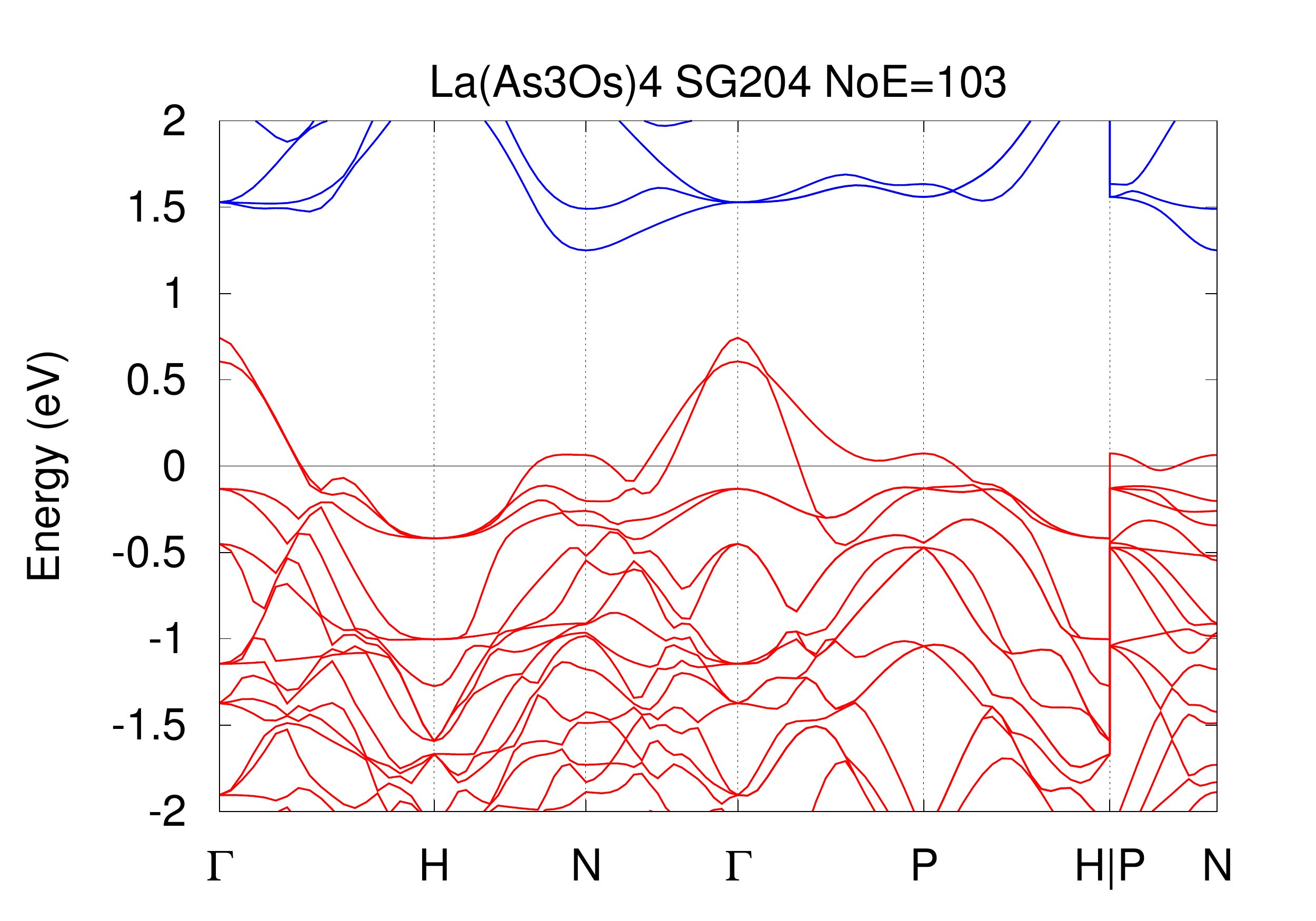}
}
\subfigure[Nd(As$_{3}$Os)$_{4}$ SG204 NoA=17 NoE=103]{
\label{subfig:611007}
\includegraphics[scale=0.32,angle=270]{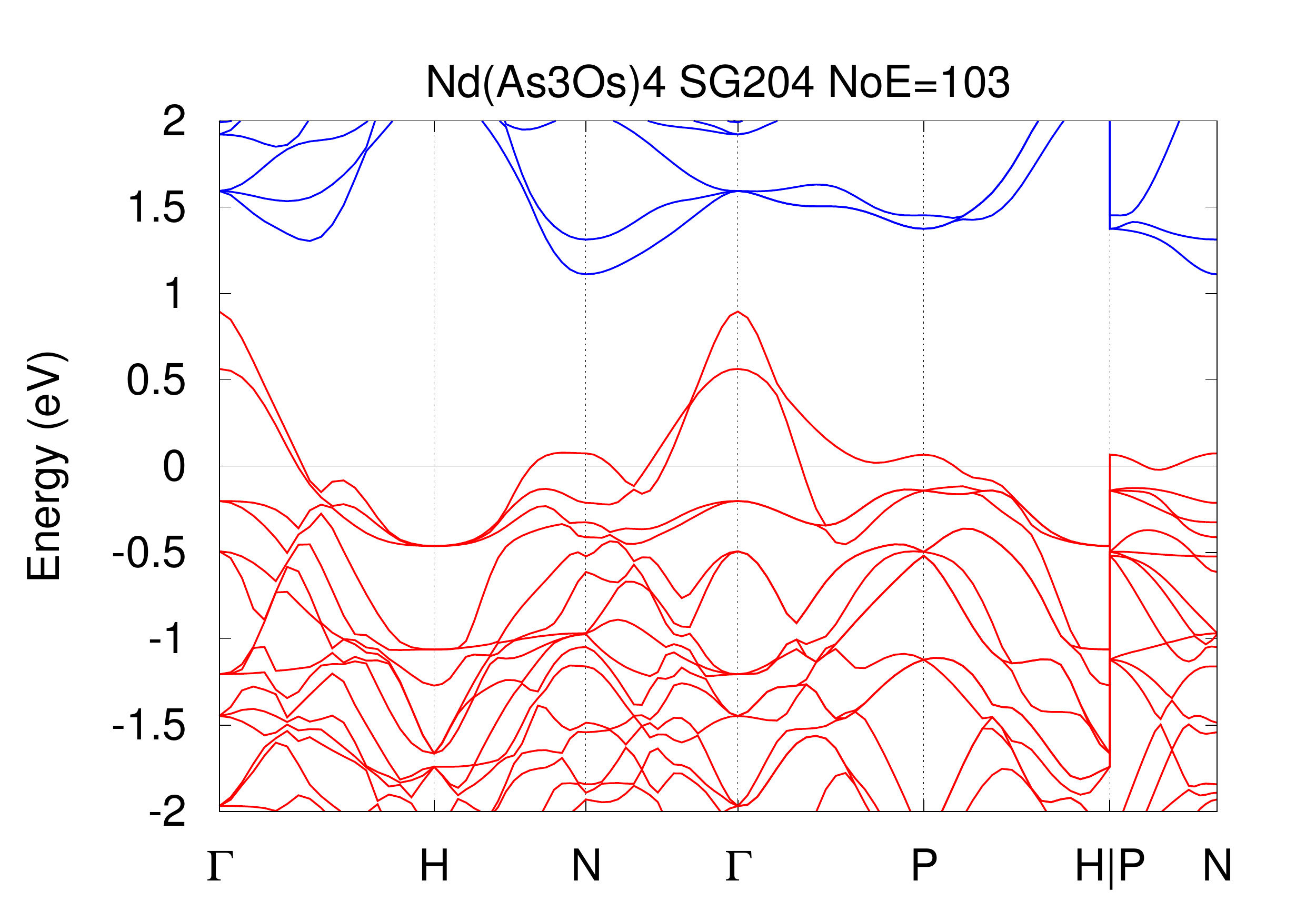}
}
\subfigure[La(FeAs$_{3}$)$_{4}$ SG204 NoA=17 NoE=103]{
\label{subfig:23080}
\includegraphics[scale=0.32,angle=270]{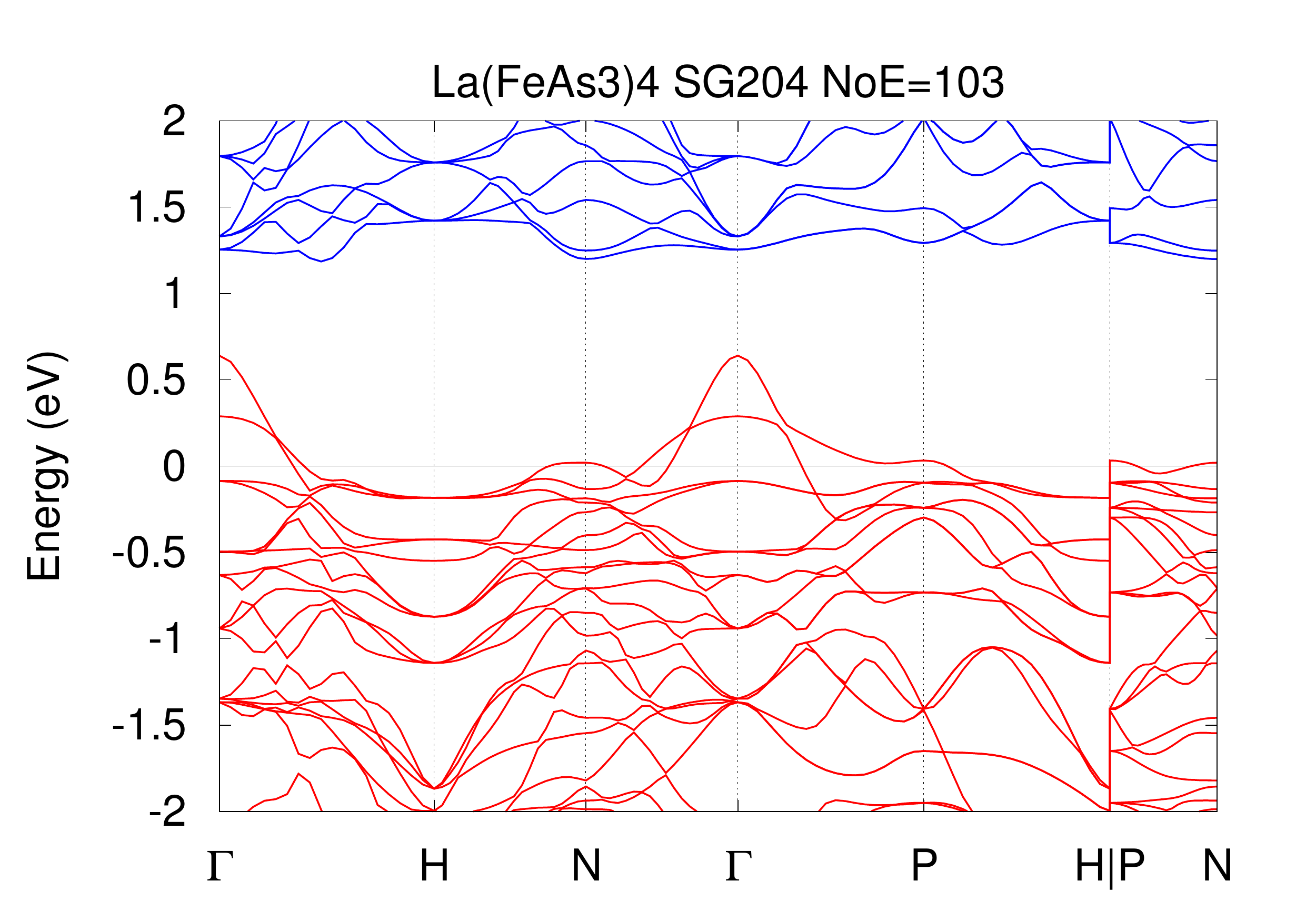}
}
\caption{\hyperref[tab:electride]{back to the table}}
\end{figure}

\begin{figure}[htp]
 \centering
\subfigure[Ce(FeAs$_{3}$)$_{4}$ SG204 NoA=17 NoE=103]{
\label{subfig:610003}
\includegraphics[scale=0.32,angle=270]{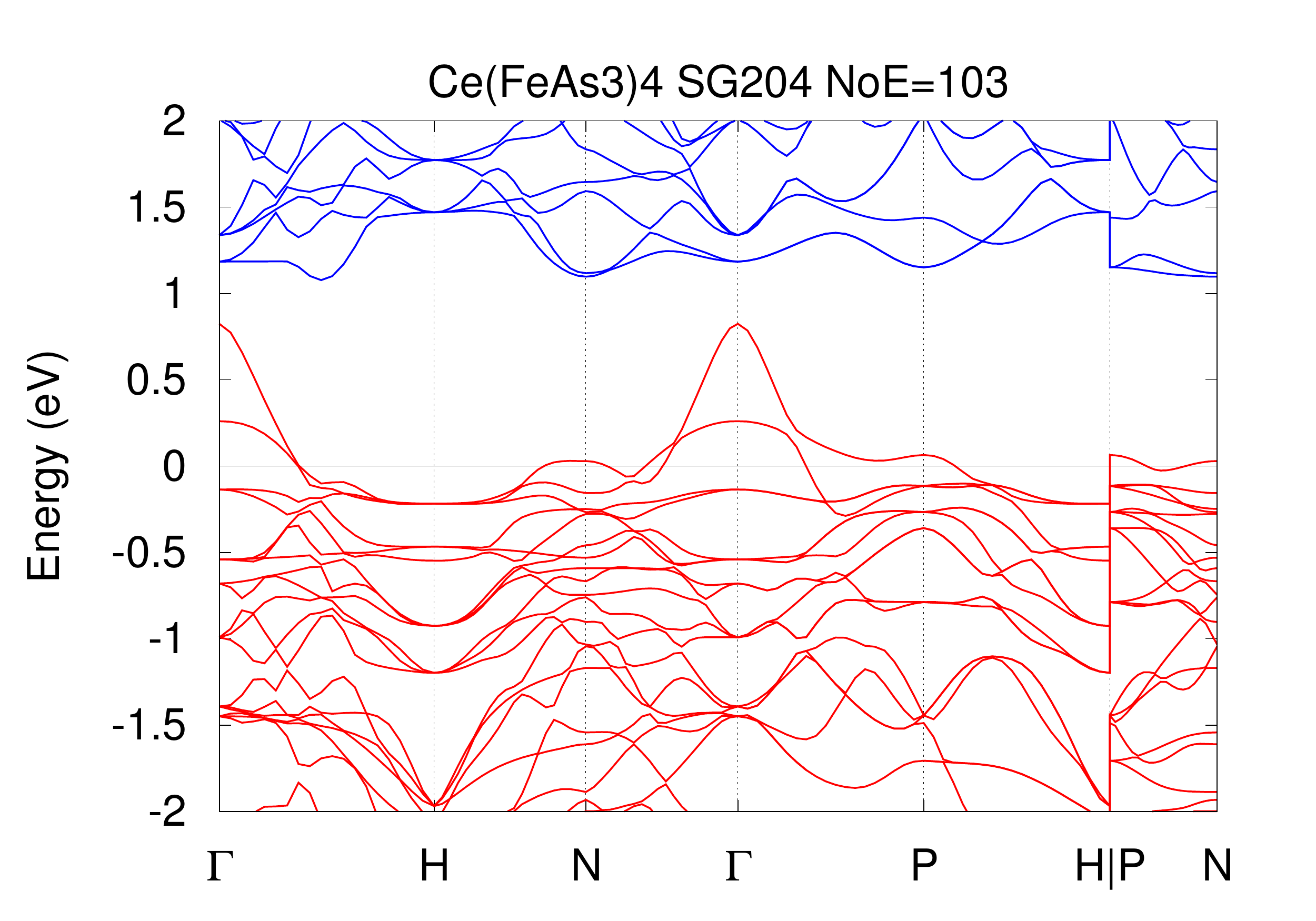}
}
\subfigure[Ce(Sb$_{3}$Os)$_{4}$ SG204 NoA=17 NoE=103]{
\label{subfig:621737}
\includegraphics[scale=0.32,angle=270]{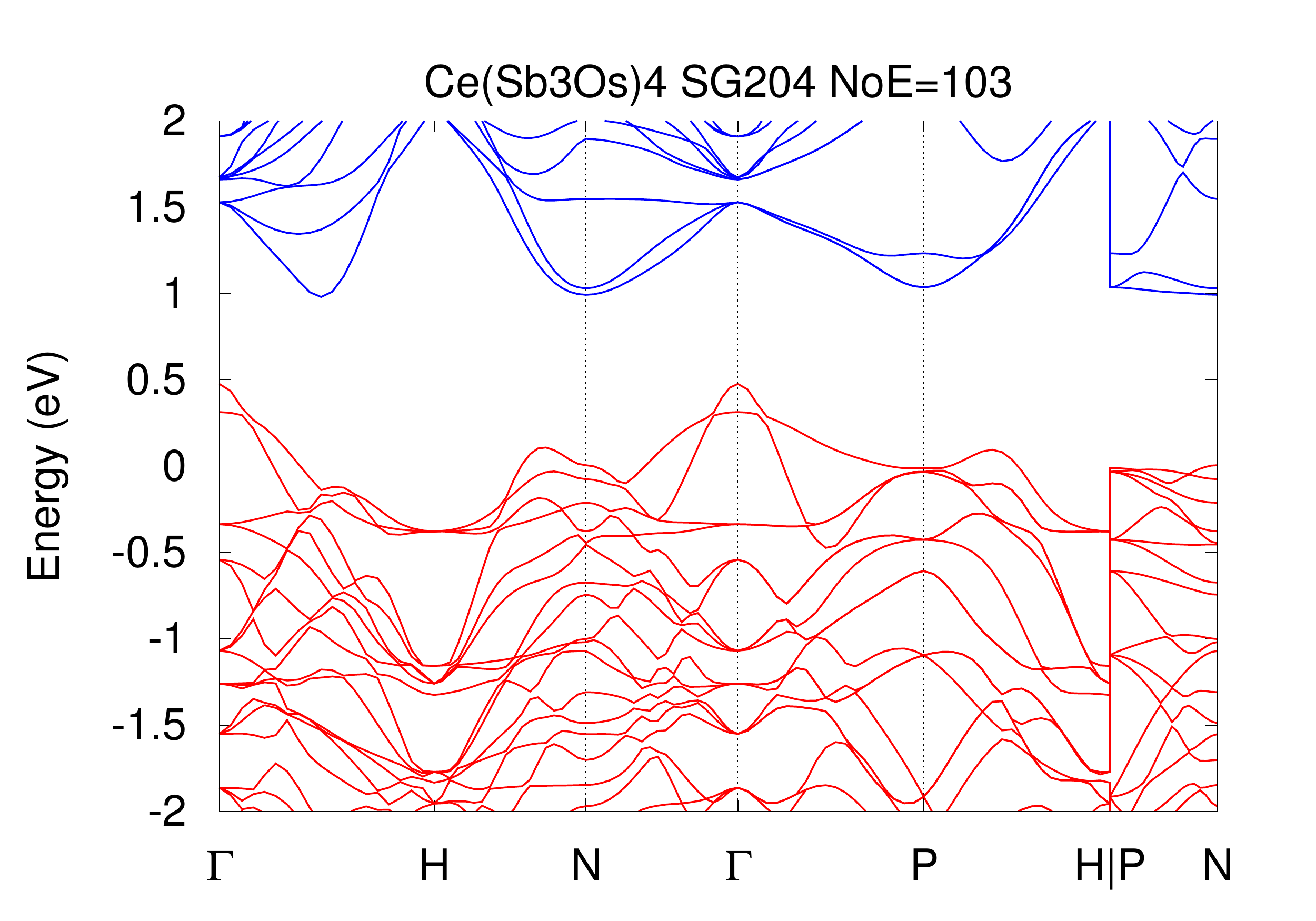}
}
\subfigure[Ce(As$_{3}$Ru)$_{4}$ SG204 NoA=17 NoE=103]{
\label{subfig:610013}
\includegraphics[scale=0.32,angle=270]{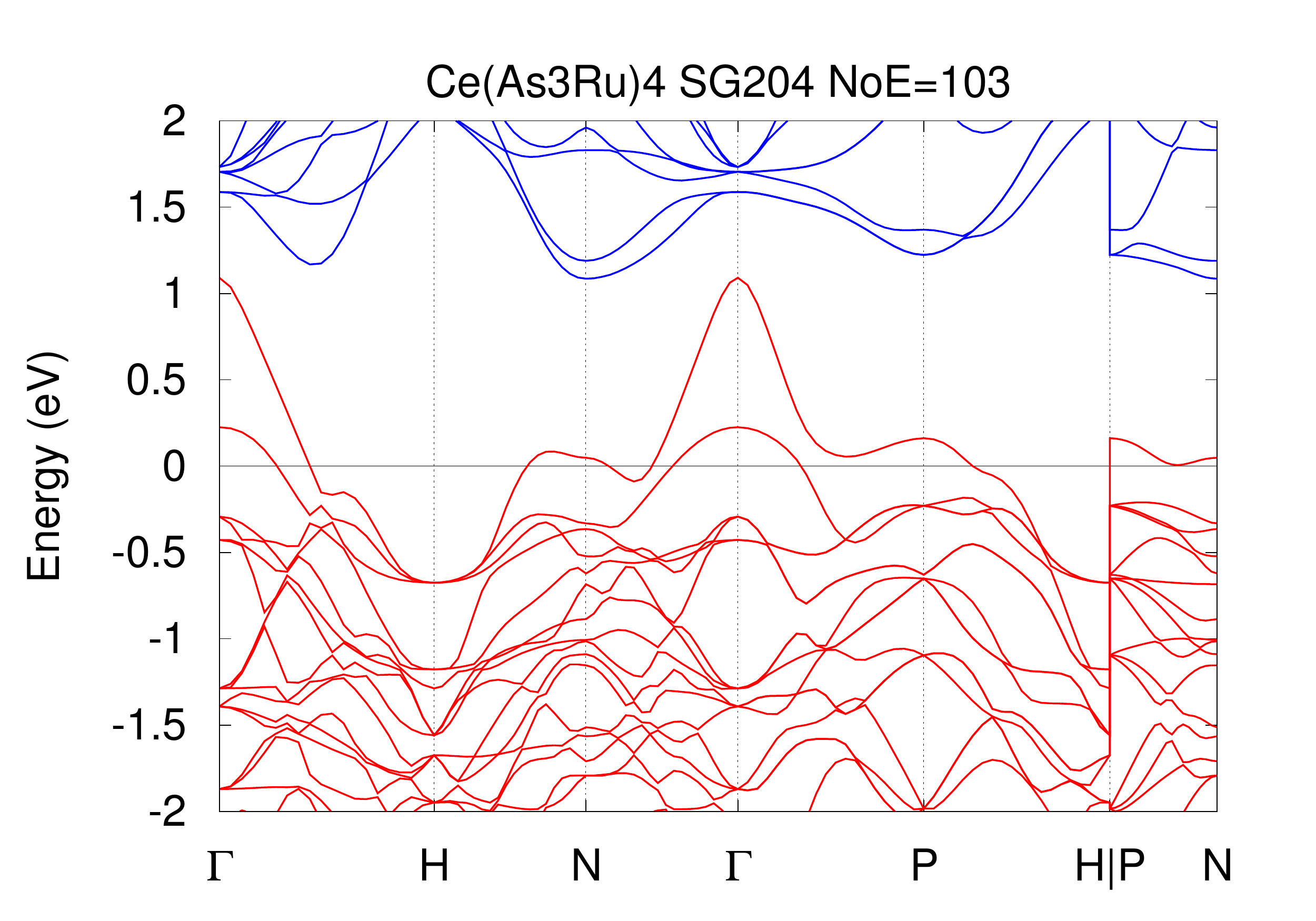}
}
\subfigure[Pr(P$_{3}$Os)$_{4}$ SG204 NoA=17 NoE=103]{
\label{subfig:647712}
\includegraphics[scale=0.32,angle=270]{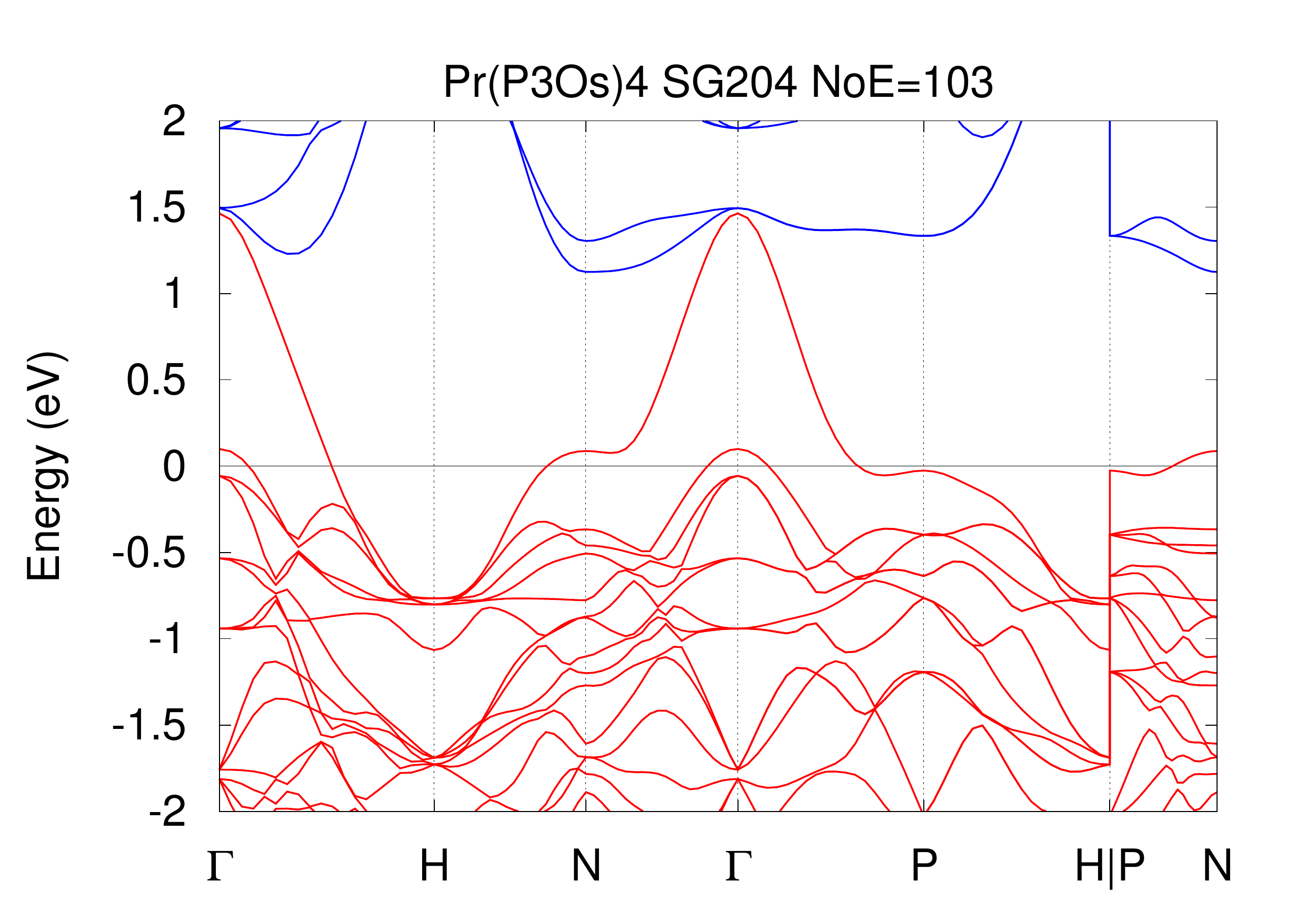}
}
\subfigure[Nd(Sb$_{3}$Os)$_{4}$ SG204 NoA=17 NoE=103]{
\label{subfig:183088}
\includegraphics[scale=0.32,angle=270]{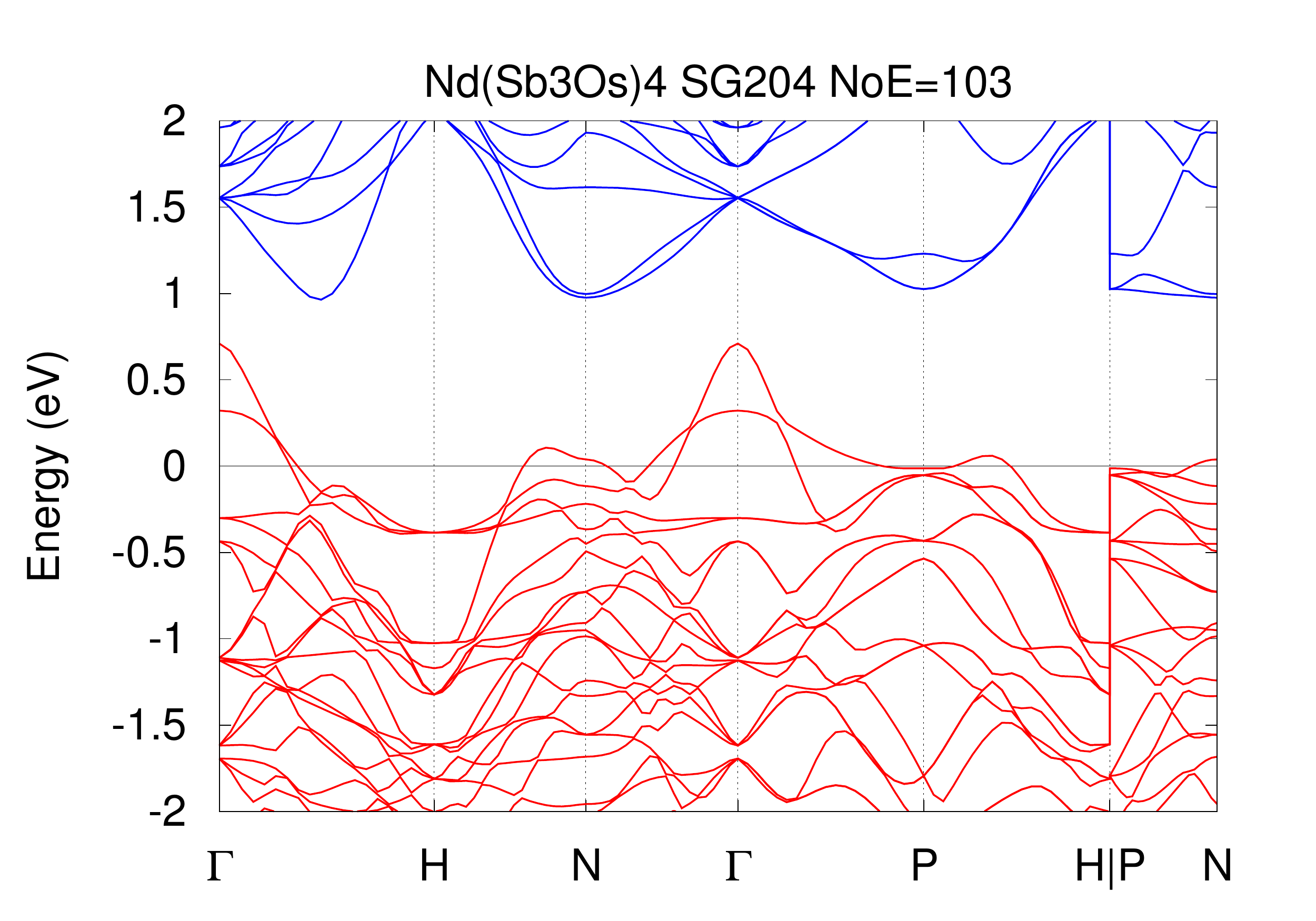}
}
\subfigure[Pr(As$_{3}$Ru)$_{4}$ SG204 NoA=17 NoE=103]{
\label{subfig:611222}
\includegraphics[scale=0.32,angle=270]{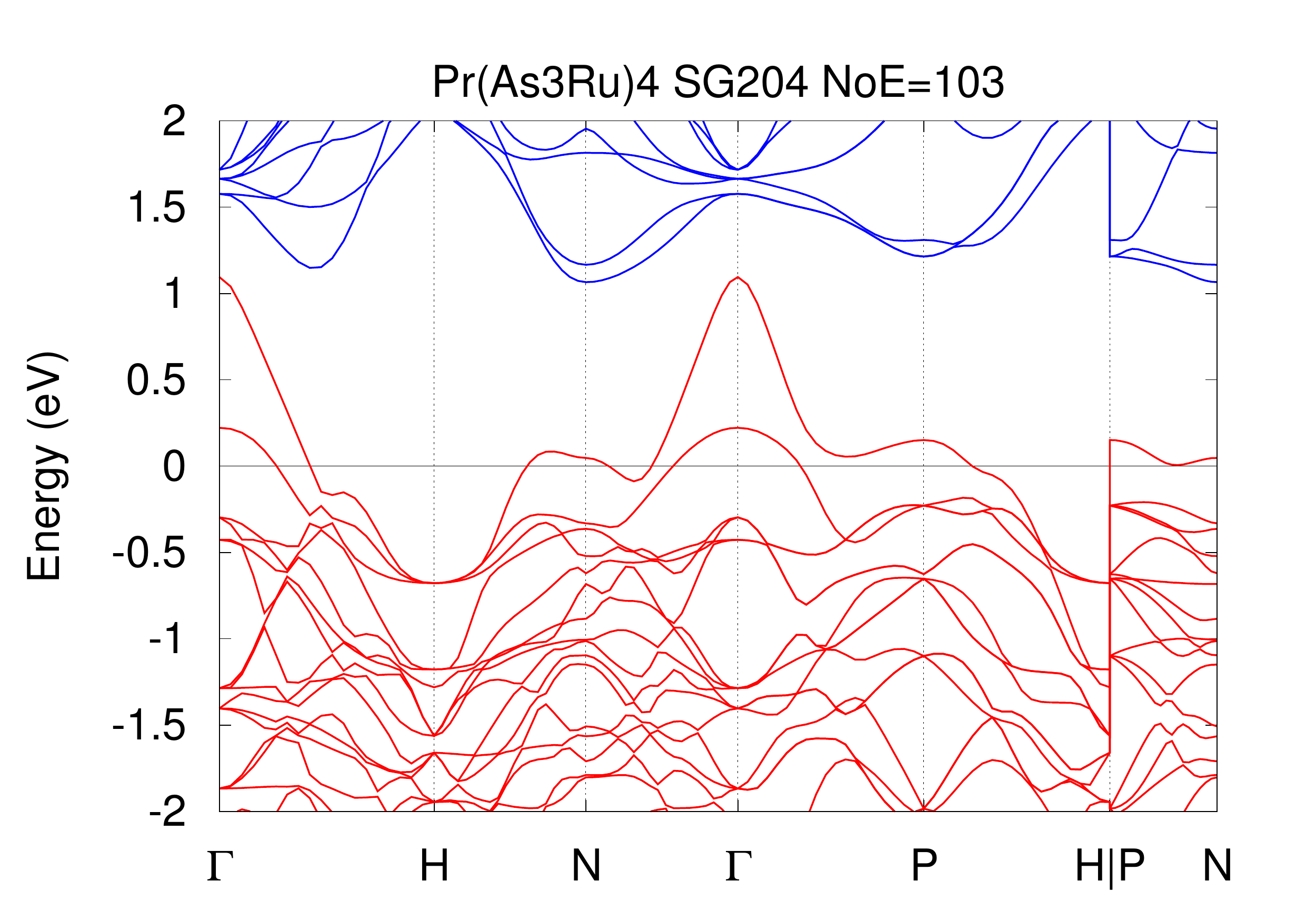}
}
\subfigure[La(FeP$_{3}$)$_{4}$ SG204 NoA=17 NoE=103]{
\label{subfig:1286}
\includegraphics[scale=0.32,angle=270]{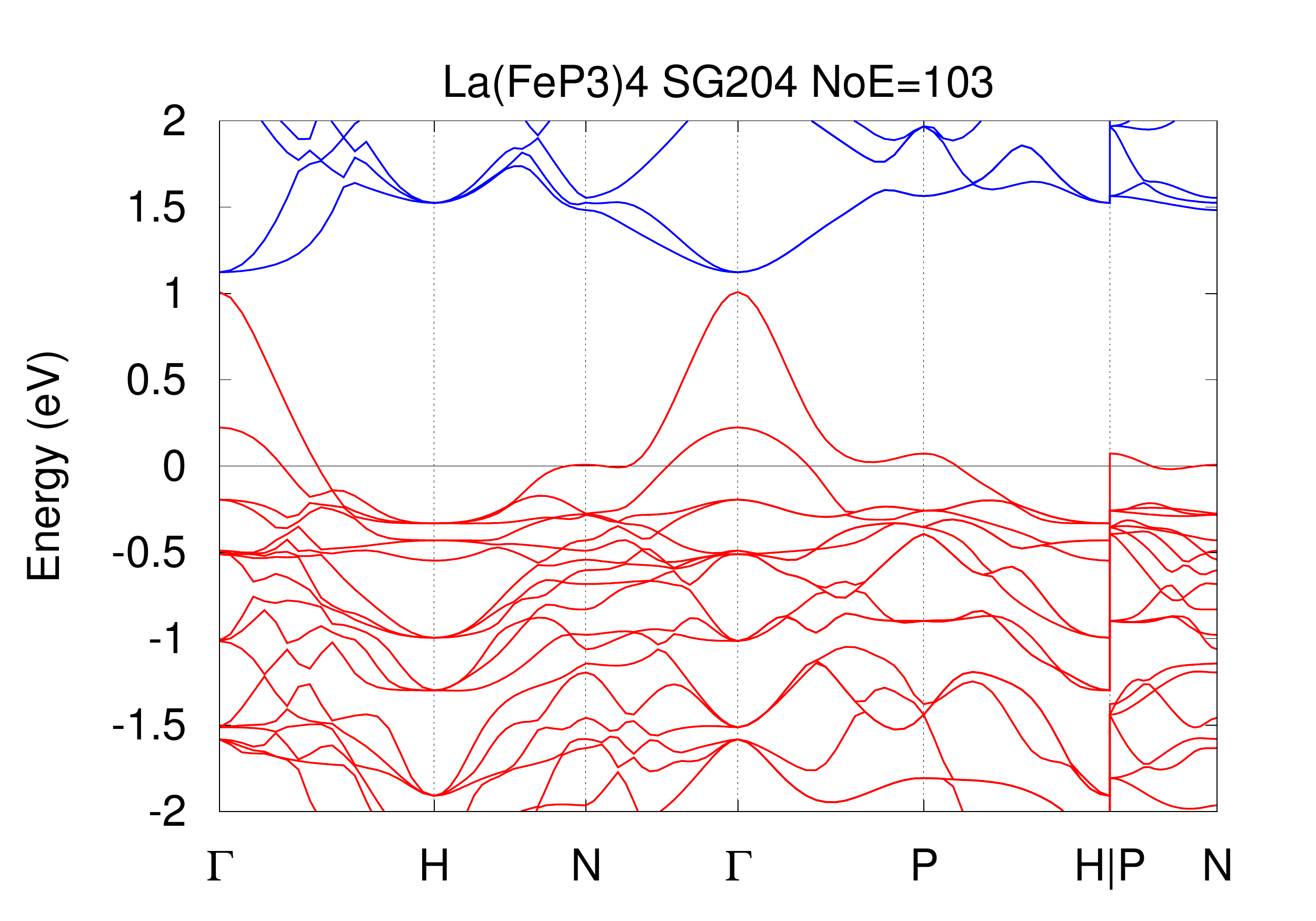}
}
\subfigure[Nd(FeSb$_{3}$)$_{4}$ SG204 NoA=17 NoE=103]{
\label{subfig:79927}
\includegraphics[scale=0.32,angle=270]{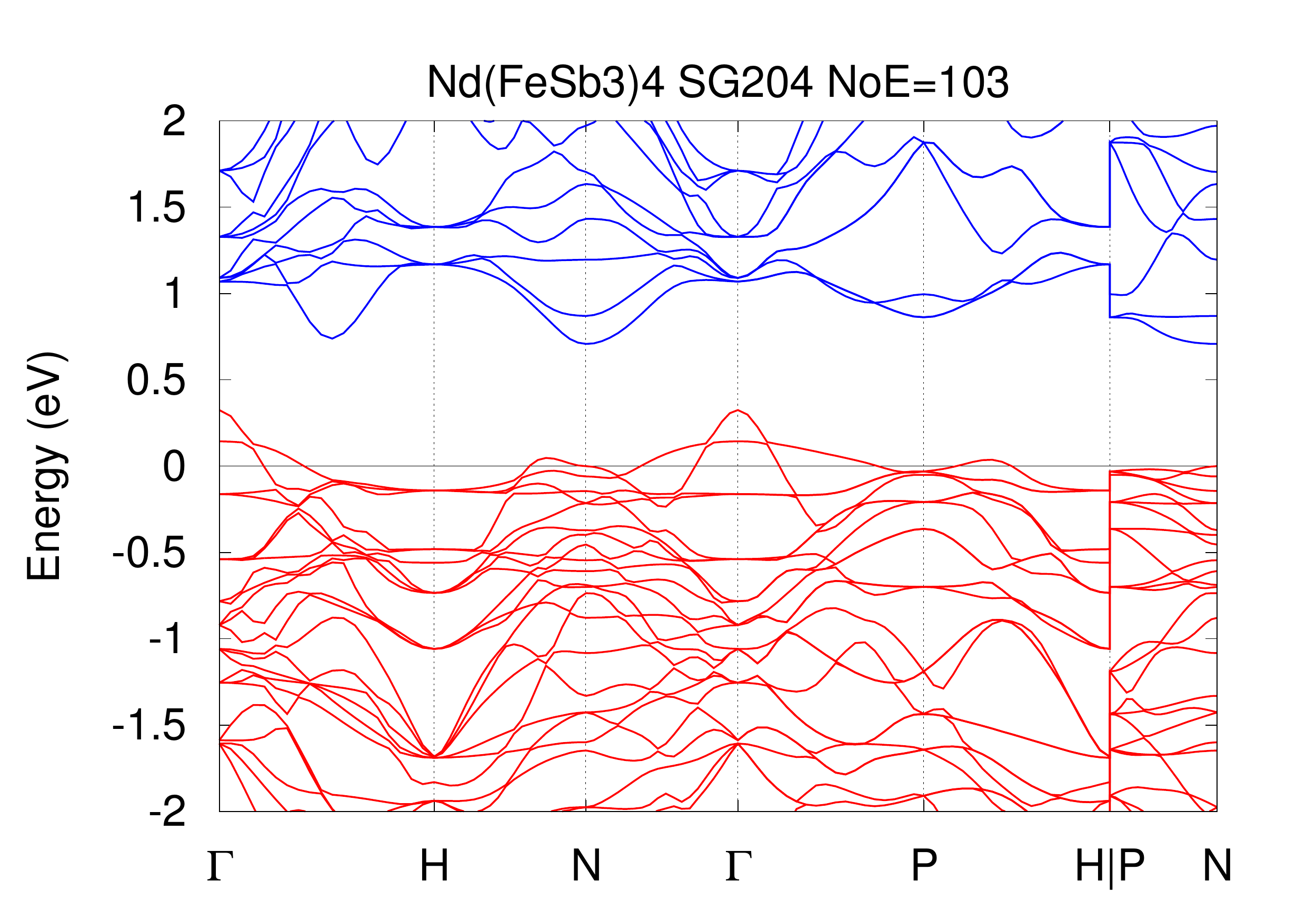}
}
\caption{\hyperref[tab:electride]{back to the table}}
\end{figure}

\begin{figure}[htp]
 \centering
\subfigure[Nd(Sb$_{3}$Ru)$_{4}$ SG204 NoA=17 NoE=103]{
\label{subfig:645809}
\includegraphics[scale=0.32,angle=270]{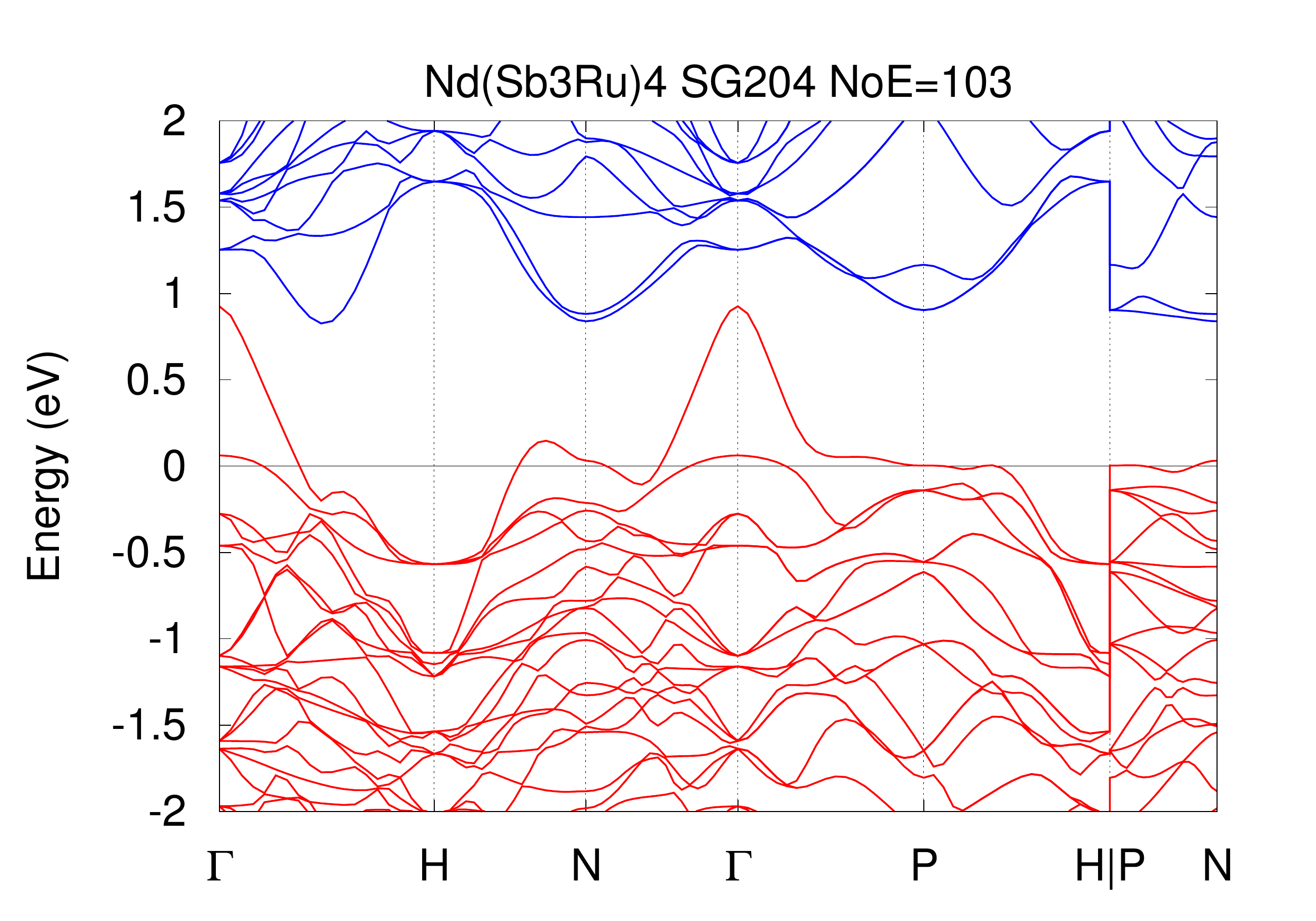}
}
\subfigure[Pr(Sb$_{3}$Os)$_{4}$ SG204 NoA=17 NoE=103]{
\label{subfig:155178}
\includegraphics[scale=0.32,angle=270]{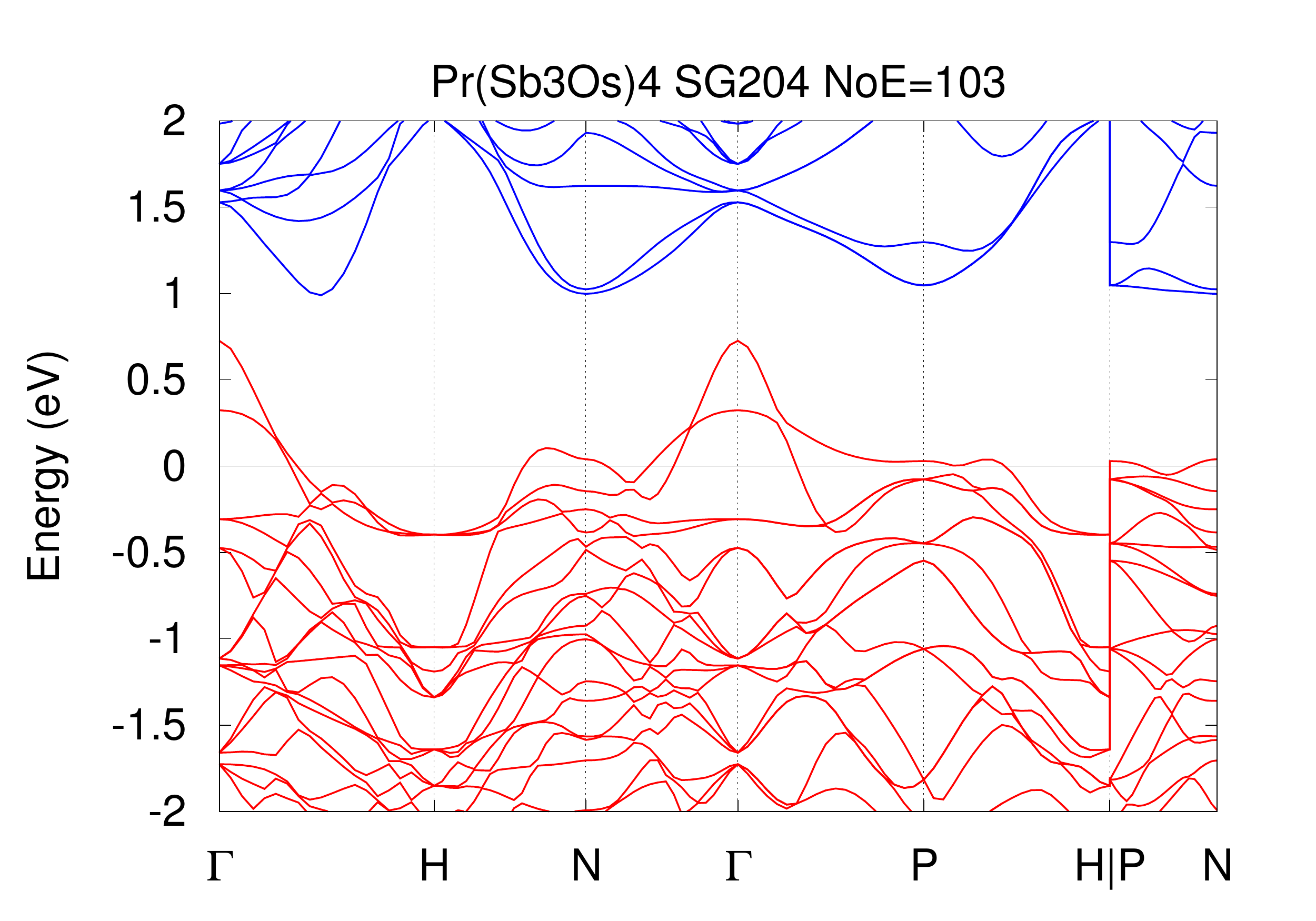}
}
\subfigure[La(FeSb$_{3}$)$_{4}$ SG204 NoA=17 NoE=103]{
\label{subfig:53490}
\includegraphics[scale=0.32,angle=270]{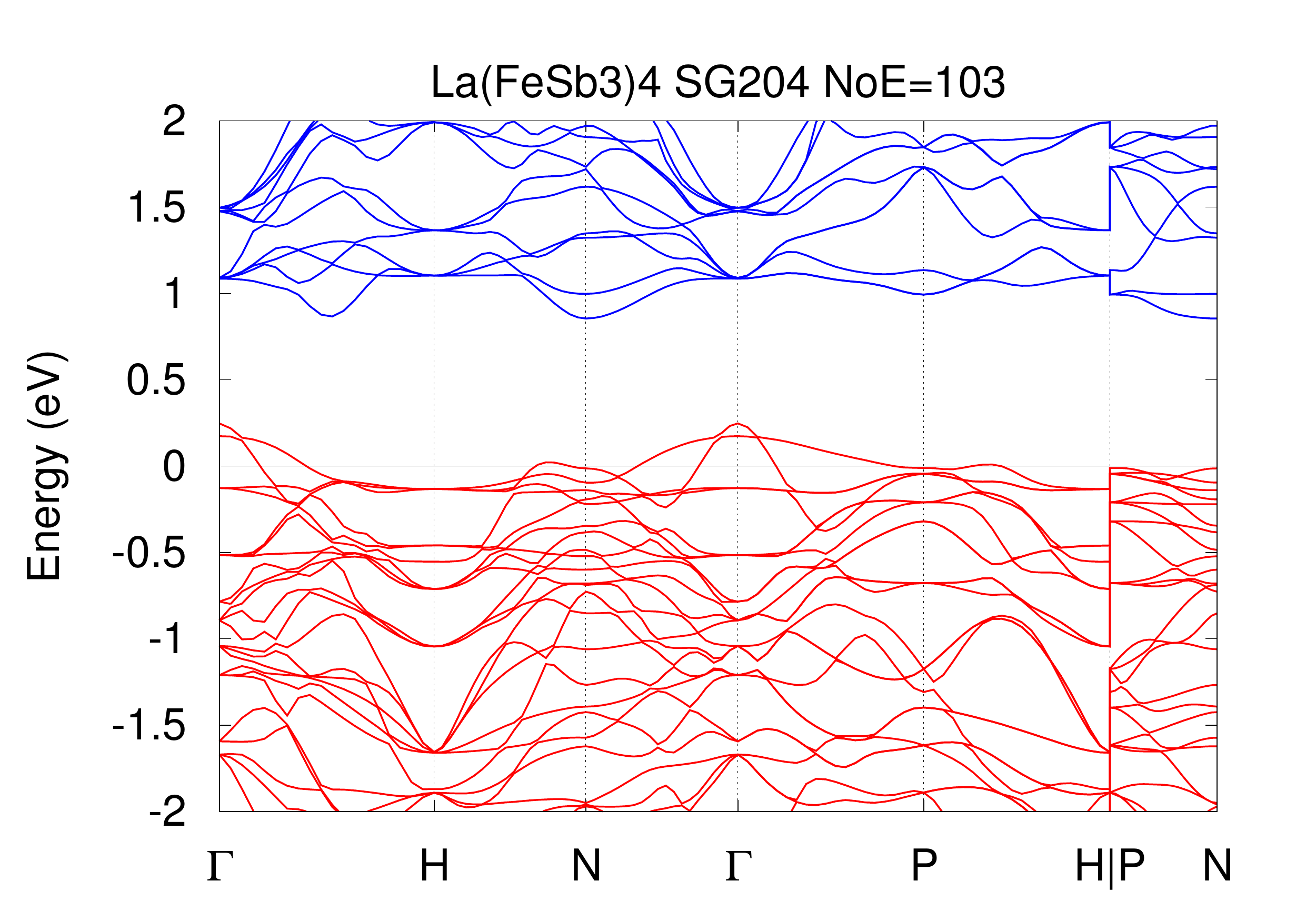}
}
\subfigure[Sm(Sb$_{3}$Os)$_{4}$ SG204 NoA=17 NoE=103]{
\label{subfig:647760}
\includegraphics[scale=0.32,angle=270]{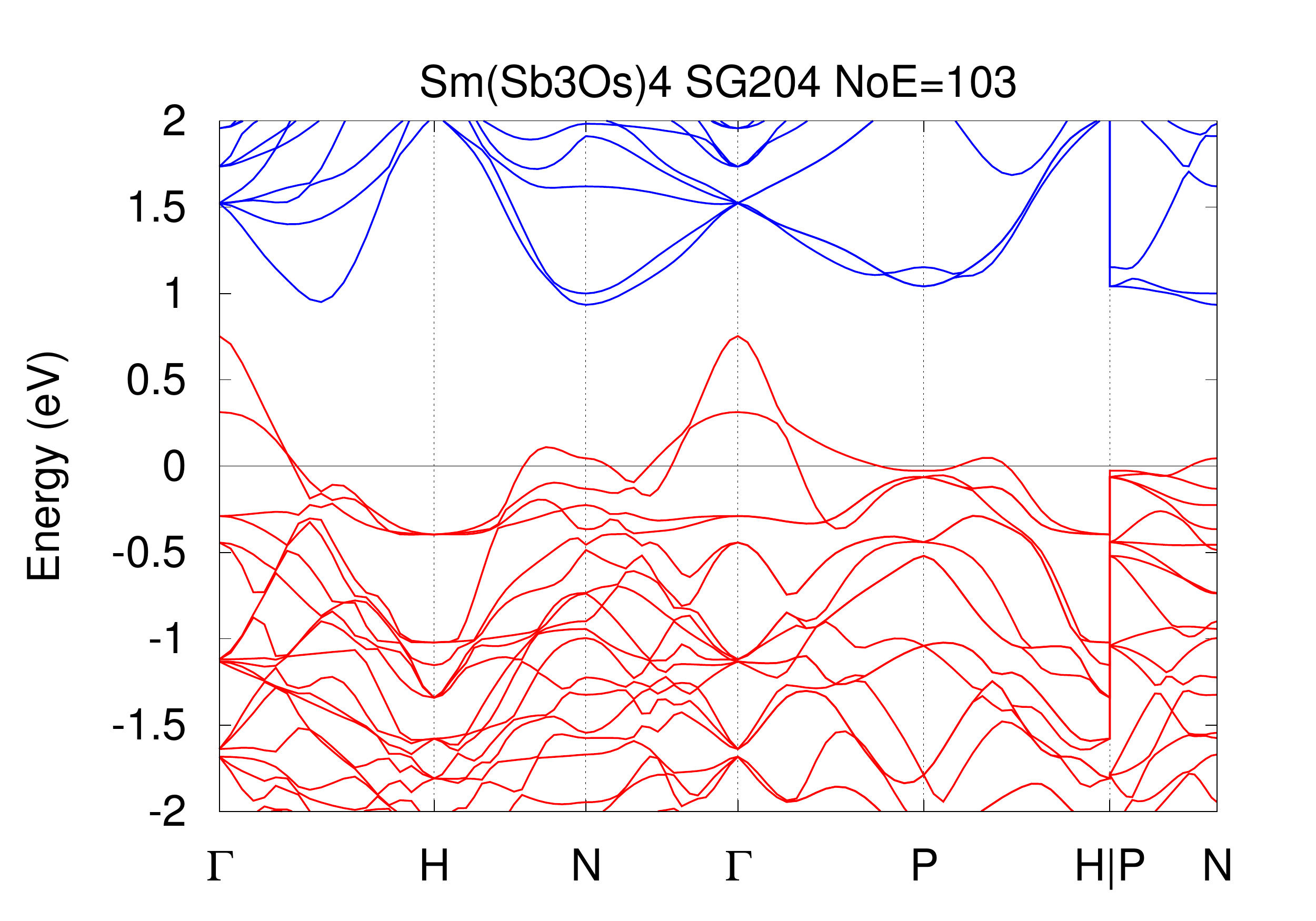}
}
\subfigure[La(Sb$_{3}$Os)$_{4}$ SG204 NoA=17 NoE=103]{
\label{subfig:183085}
\includegraphics[scale=0.32,angle=270]{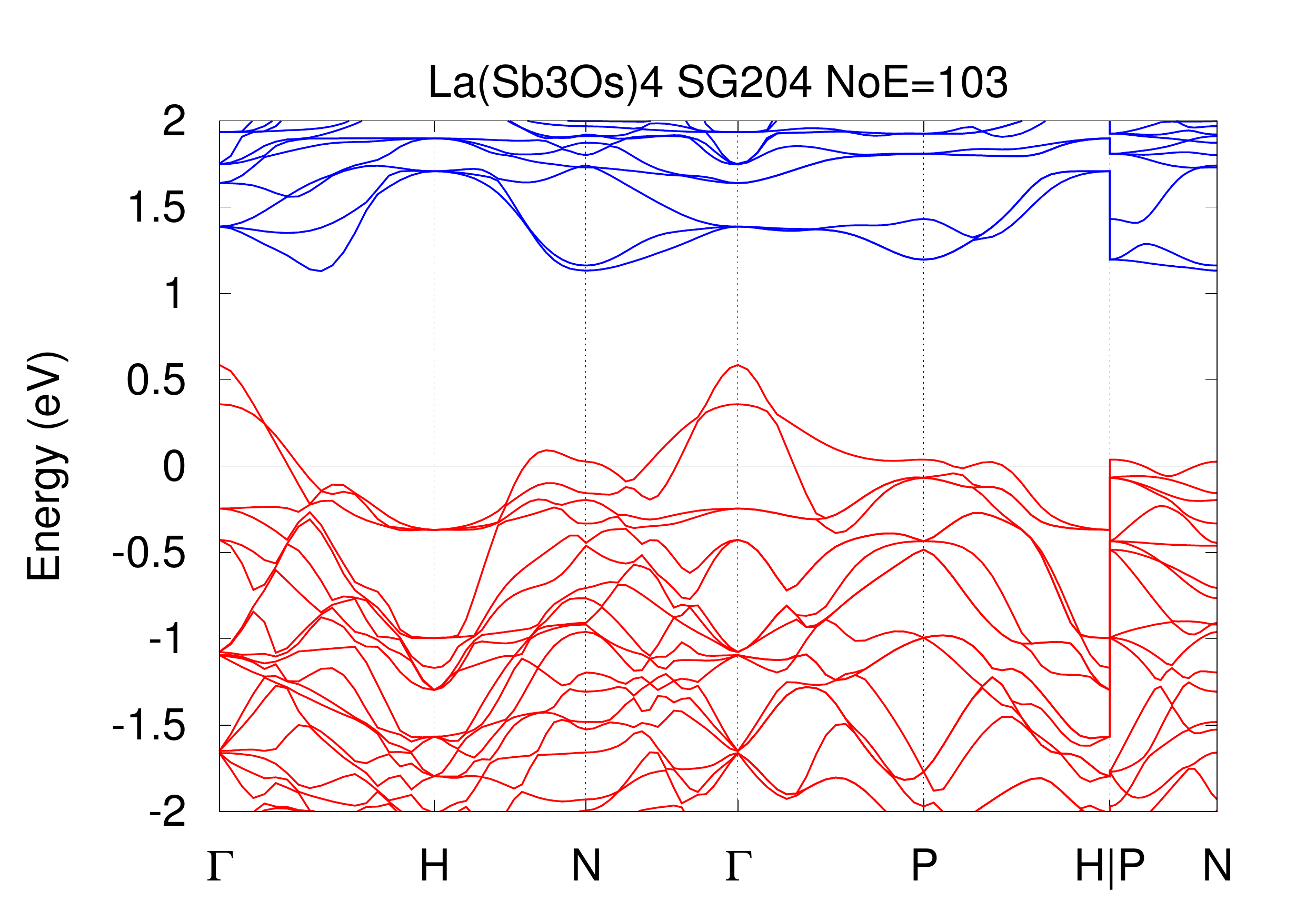}
}
\subfigure[Nd(P$_{3}$Os)$_{4}$ SG204 NoA=17 NoE=103]{
\label{subfig:645670}
\includegraphics[scale=0.32,angle=270]{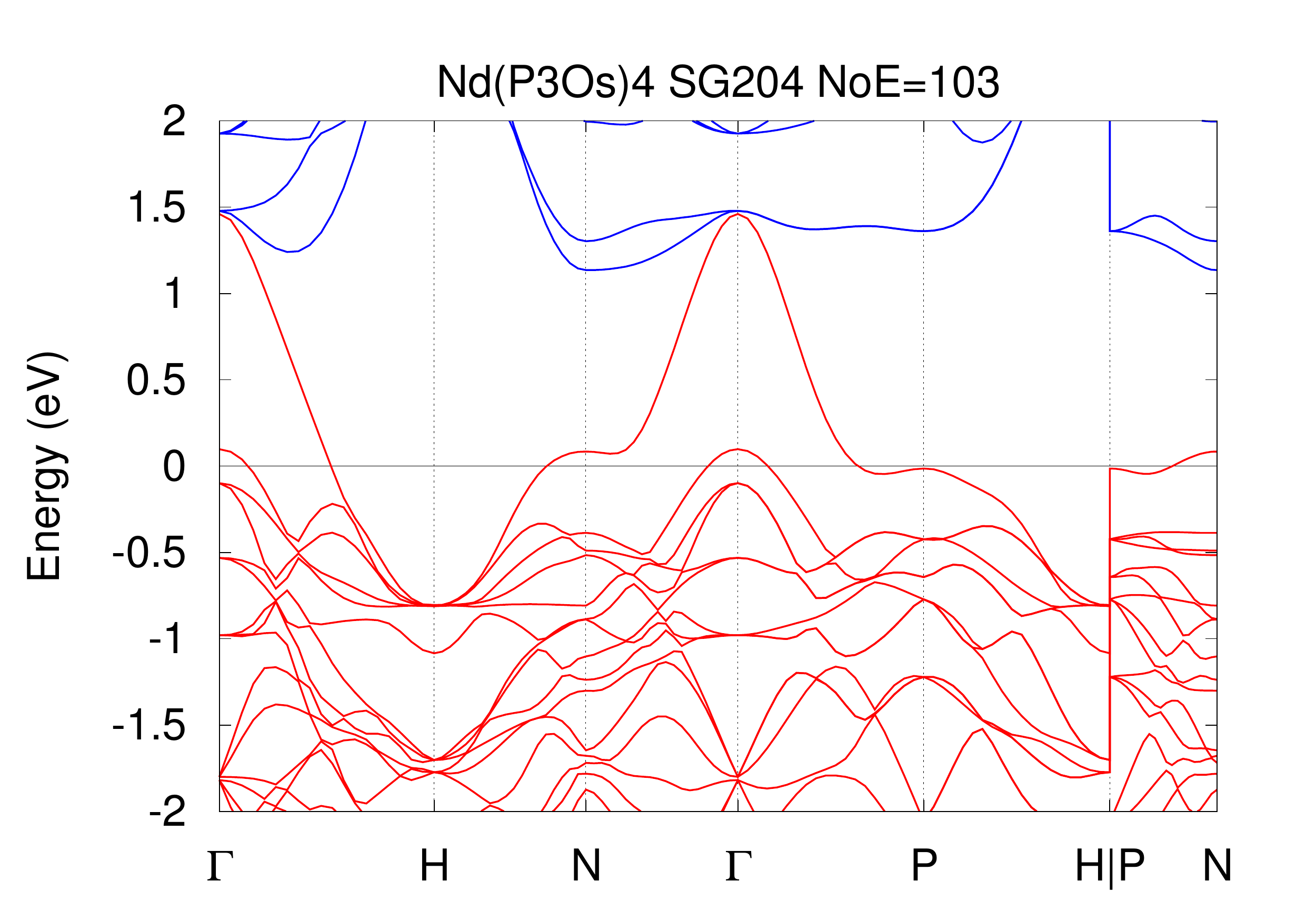}
}
\subfigure[La(P$_{3}$Os)$_{4}$ SG204 NoA=17 NoE=103]{
\label{subfig:641615}
\includegraphics[scale=0.32,angle=270]{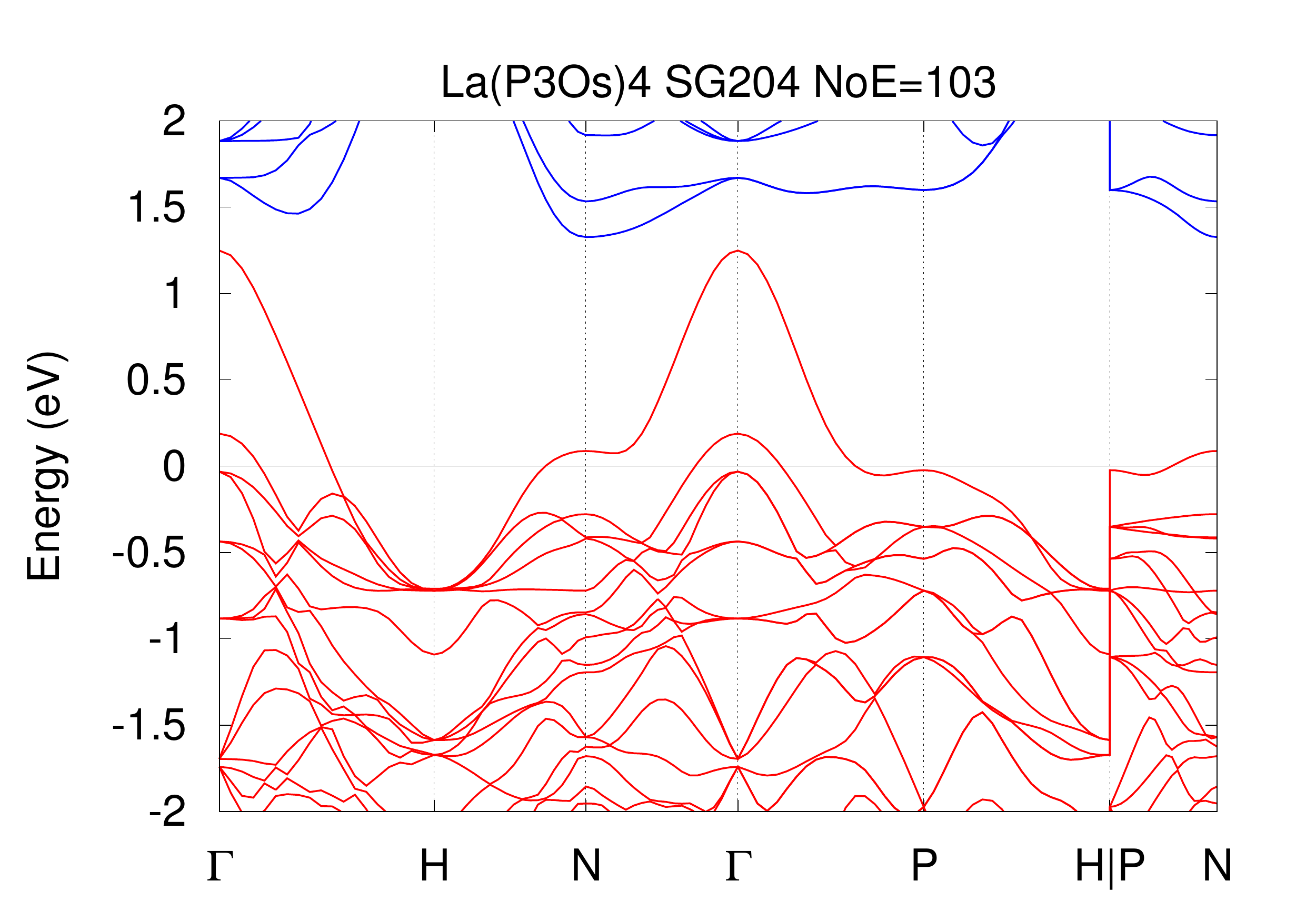}
}
\subfigure[Ba$_{5}$Sb$_{4}$ SG64 NoA=18 NoE=140]{
\label{subfig:280022}
\includegraphics[scale=0.32,angle=270]{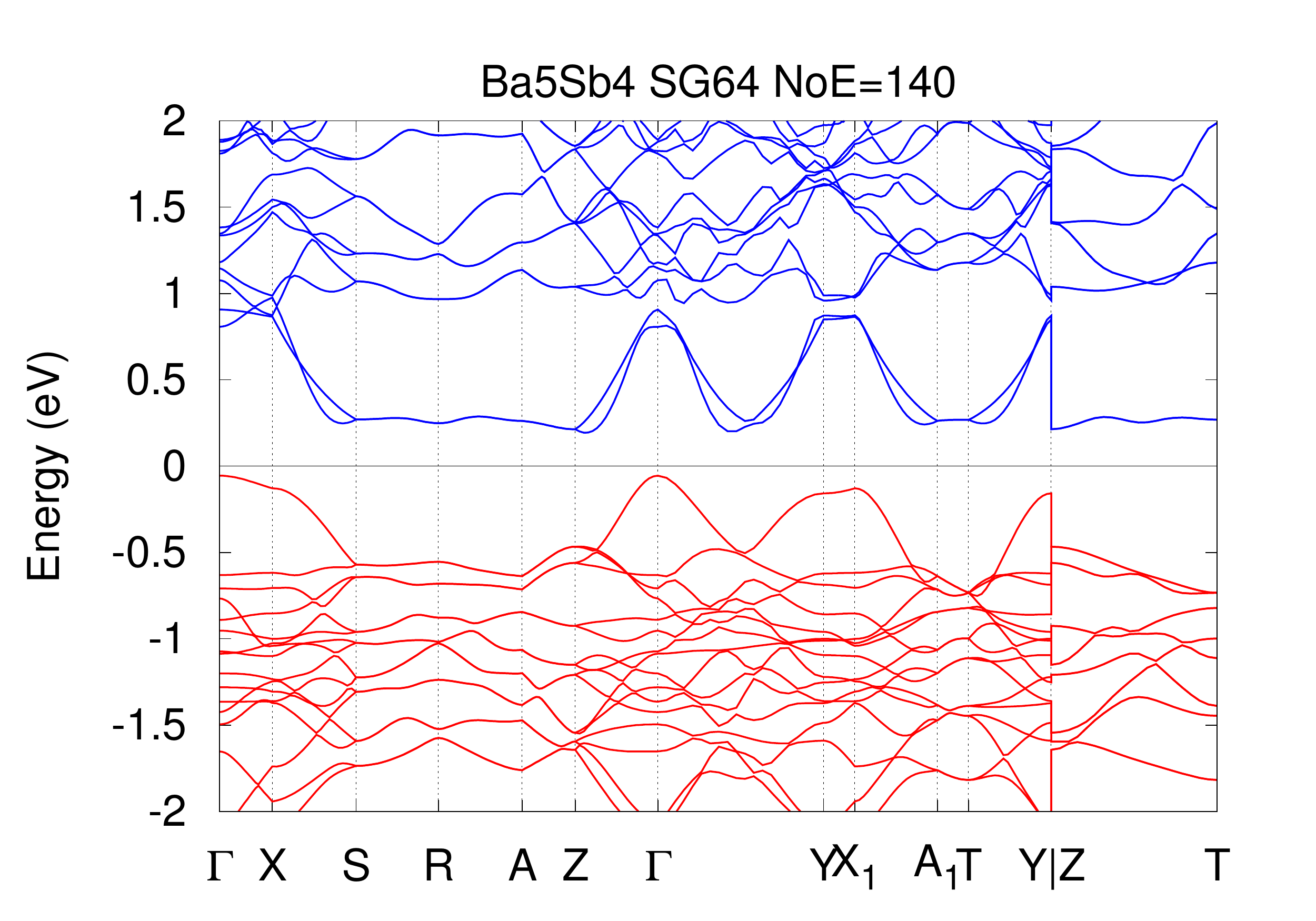}
}
\caption{\hyperref[tab:electride]{back to the table}}
\end{figure}

\begin{figure}[htp]
 \centering
\subfigure[Tm$_{5}$(ReO$_{6}$)$_{2}$ SG12 NoA=19 NoE=131]{
\label{subfig:280143}
\includegraphics[scale=0.32,angle=270]{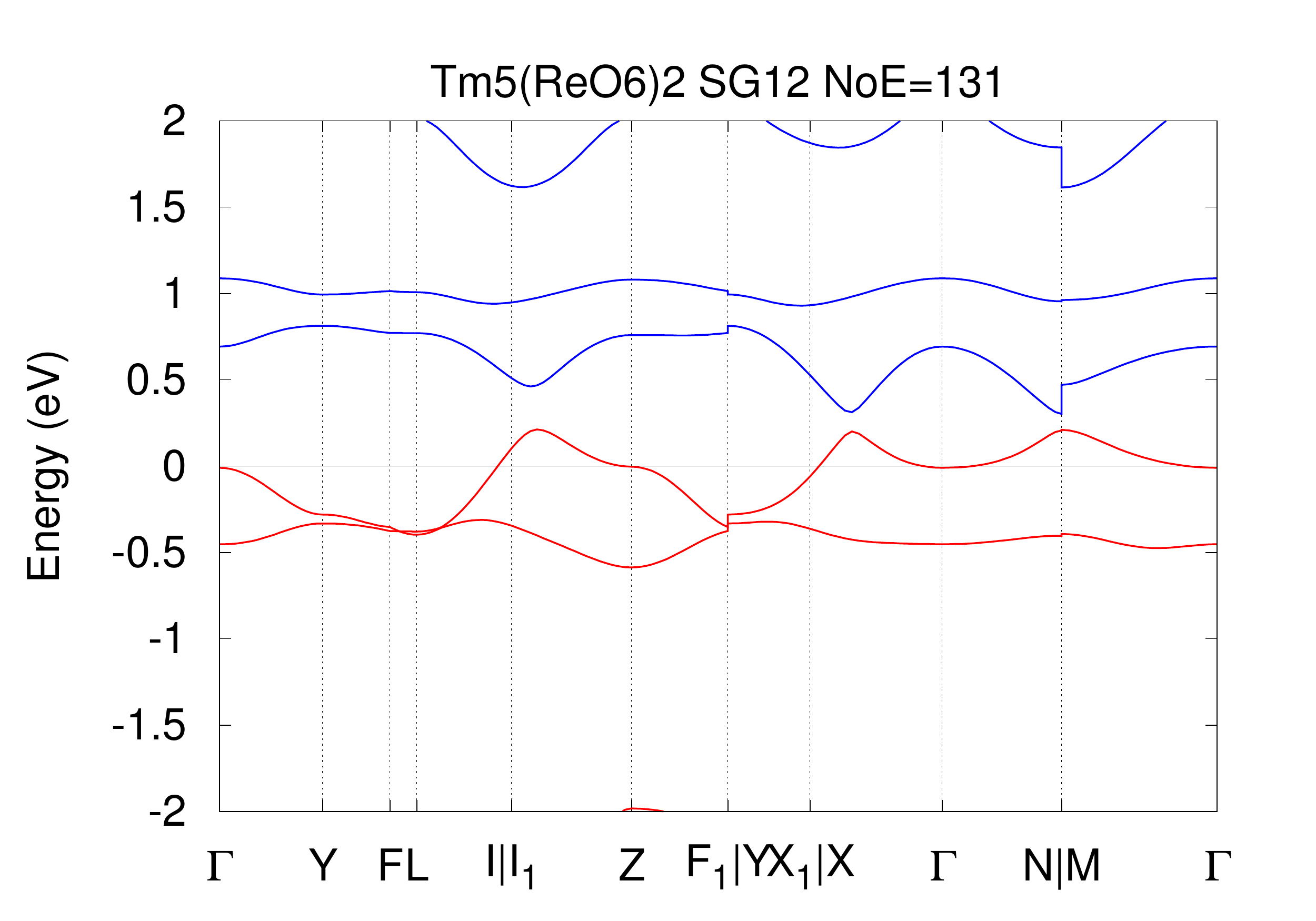}
}
\subfigure[BeSO$_{8}$ SG140 NoA=20 NoE=112]{
\label{subfig:36157}
\includegraphics[scale=0.32,angle=270]{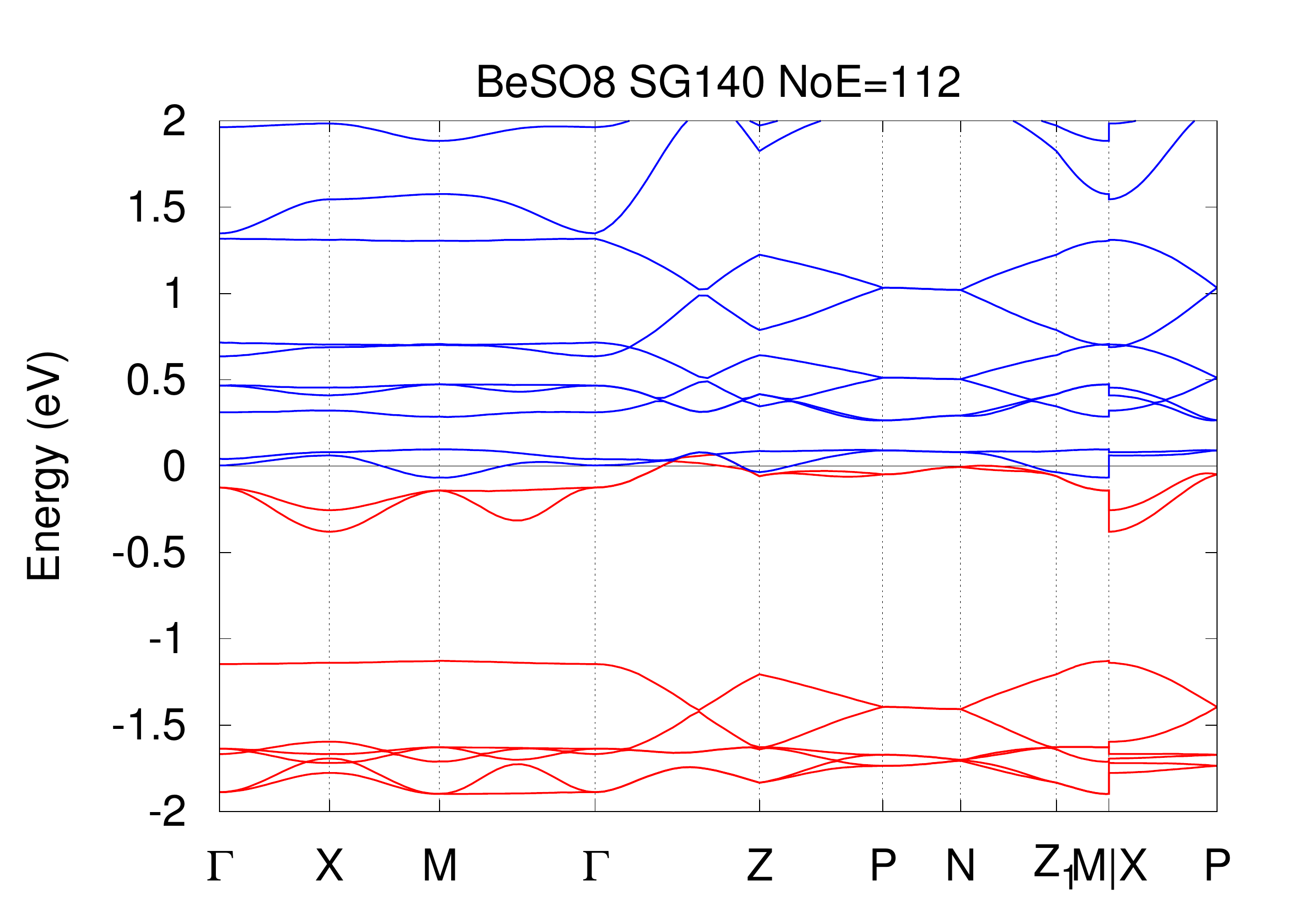}
}
\subfigure[Mg$_{3}$Nb$_{6}$O$_{11}$ SG164 NoA=20 NoE=138]{
\label{subfig:200210}
\includegraphics[scale=0.32,angle=270]{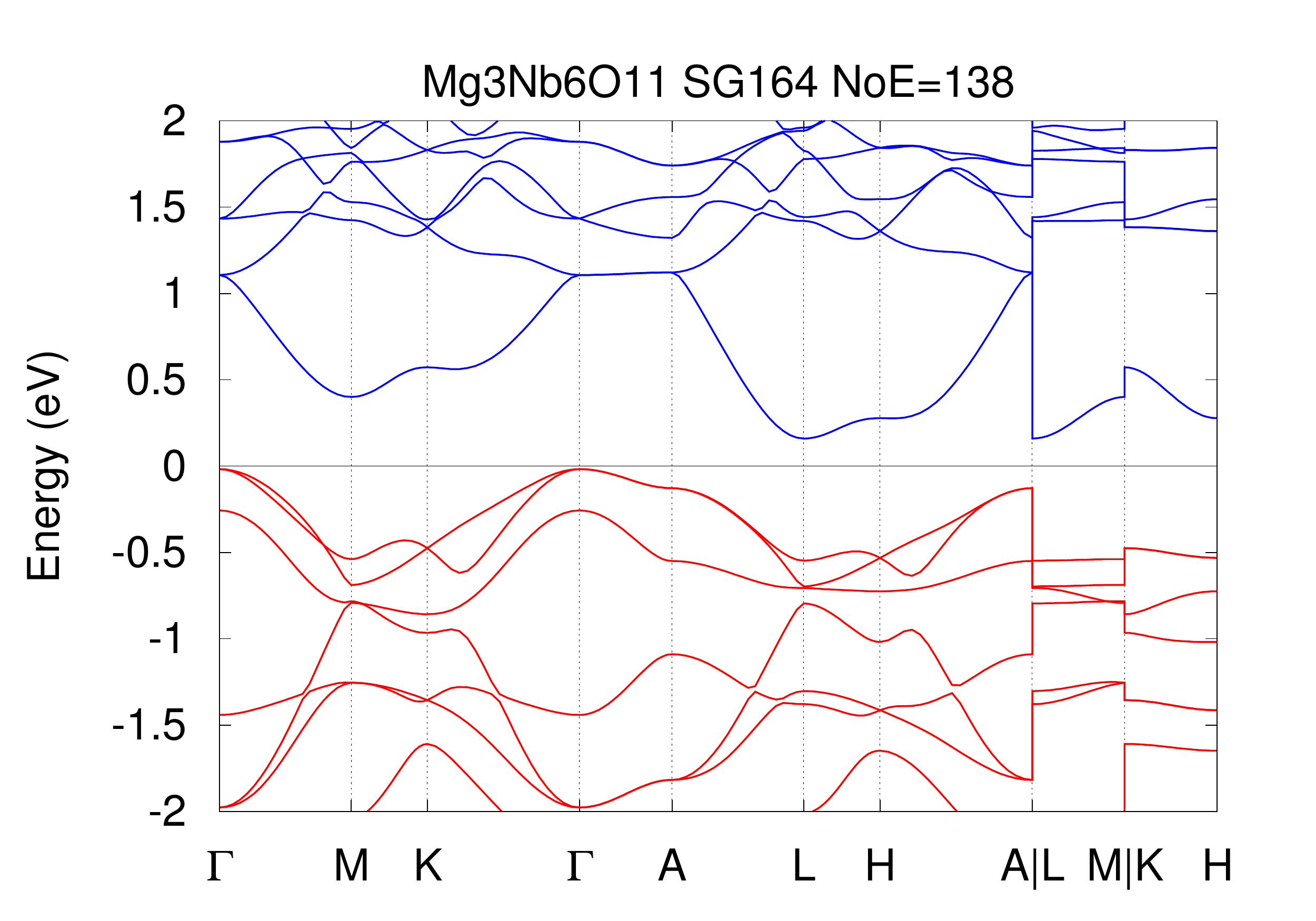}
}
\subfigure[Bi$_{4}$I SG12 NoA=20 NoE=108]{
\label{subfig:246145}
\includegraphics[scale=0.32,angle=270]{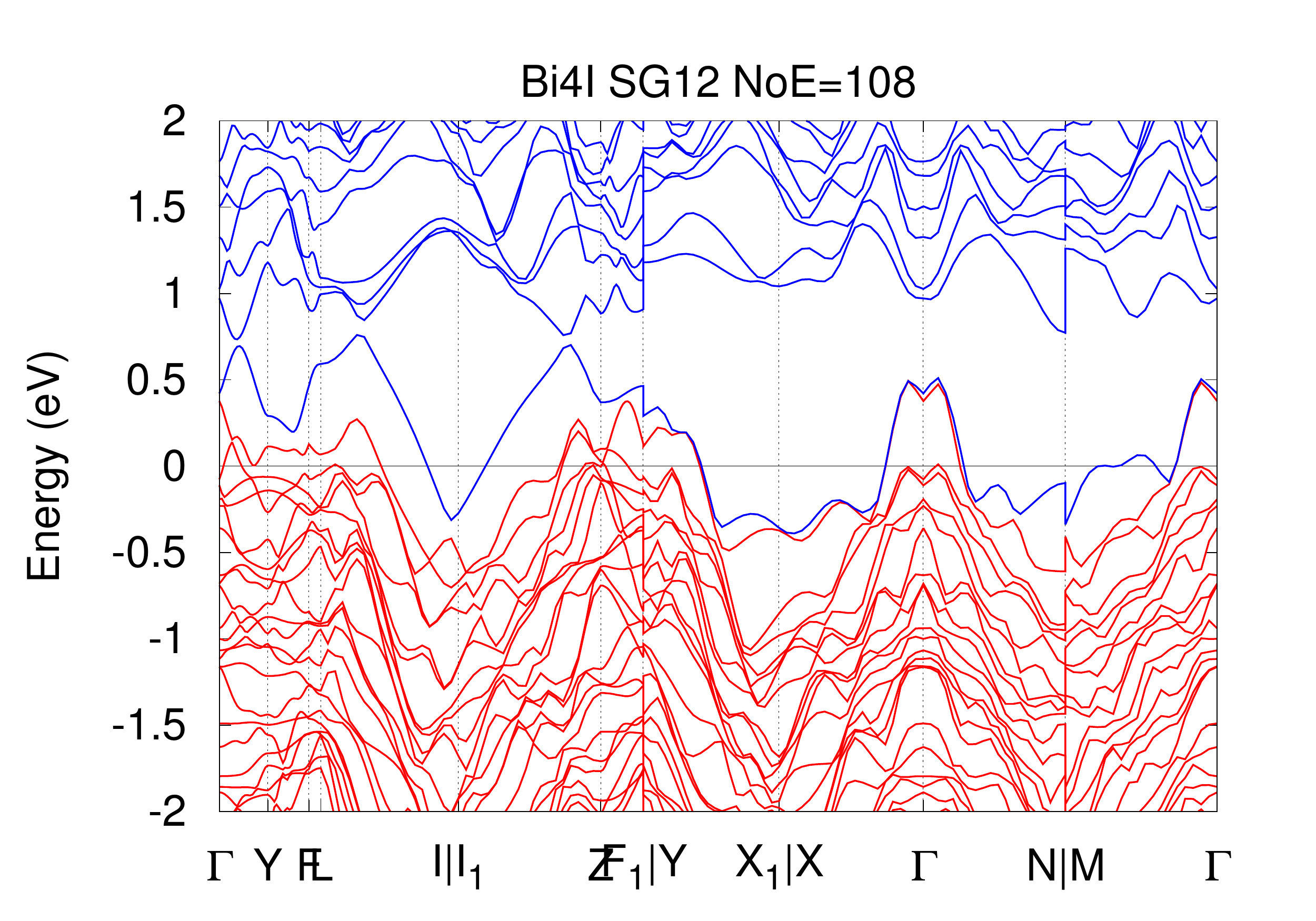}
}
\subfigure[Nd$_{2}$SbO$_{2}$ SG59 NoA=20 NoE=156]{
\label{subfig:262307}
\includegraphics[scale=0.32,angle=270]{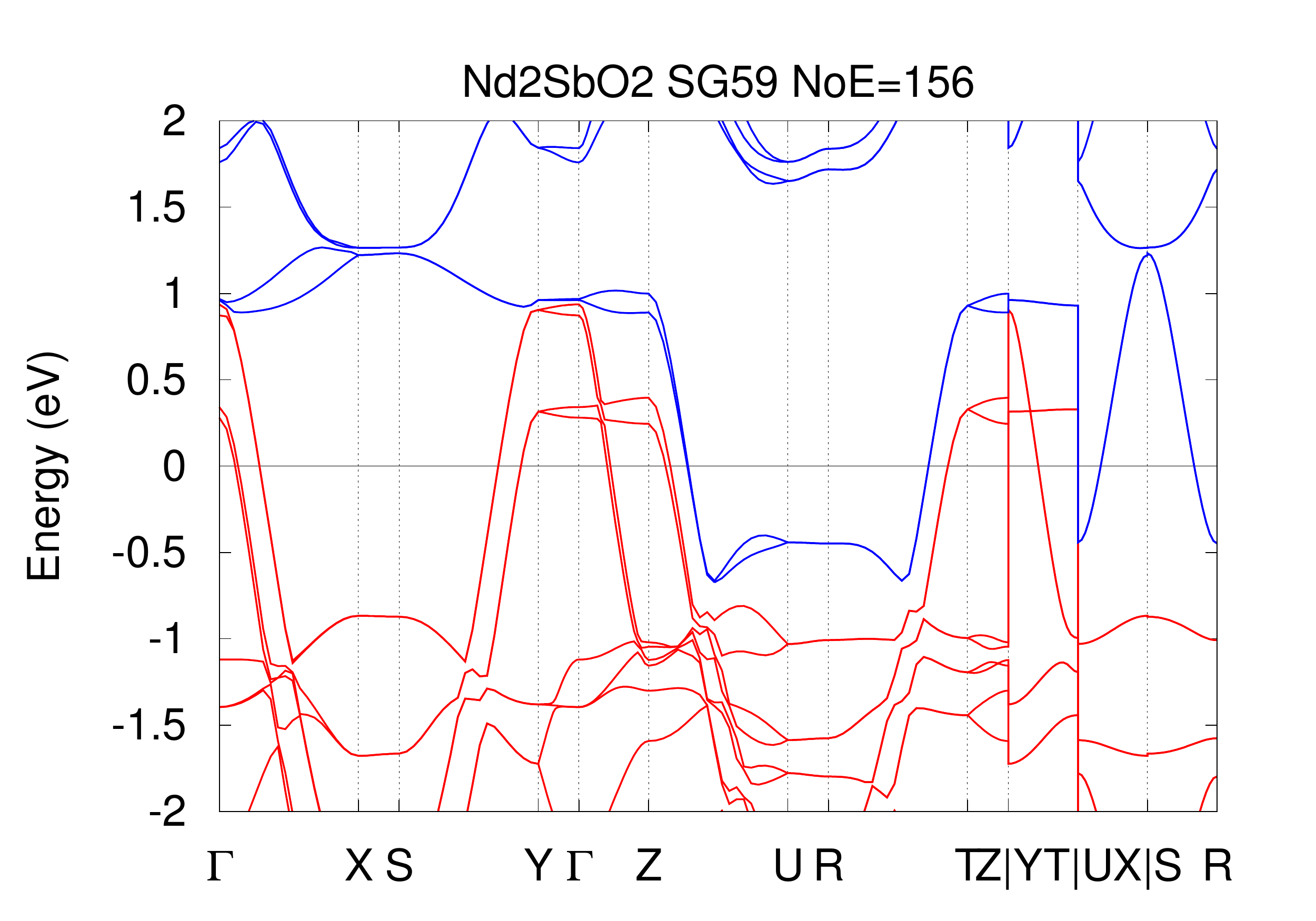}
}
\subfigure[Sr(GaAs)$_{2}$ SG10 NoA=20 NoE=104]{
\label{subfig:422527}
\includegraphics[scale=0.32,angle=270]{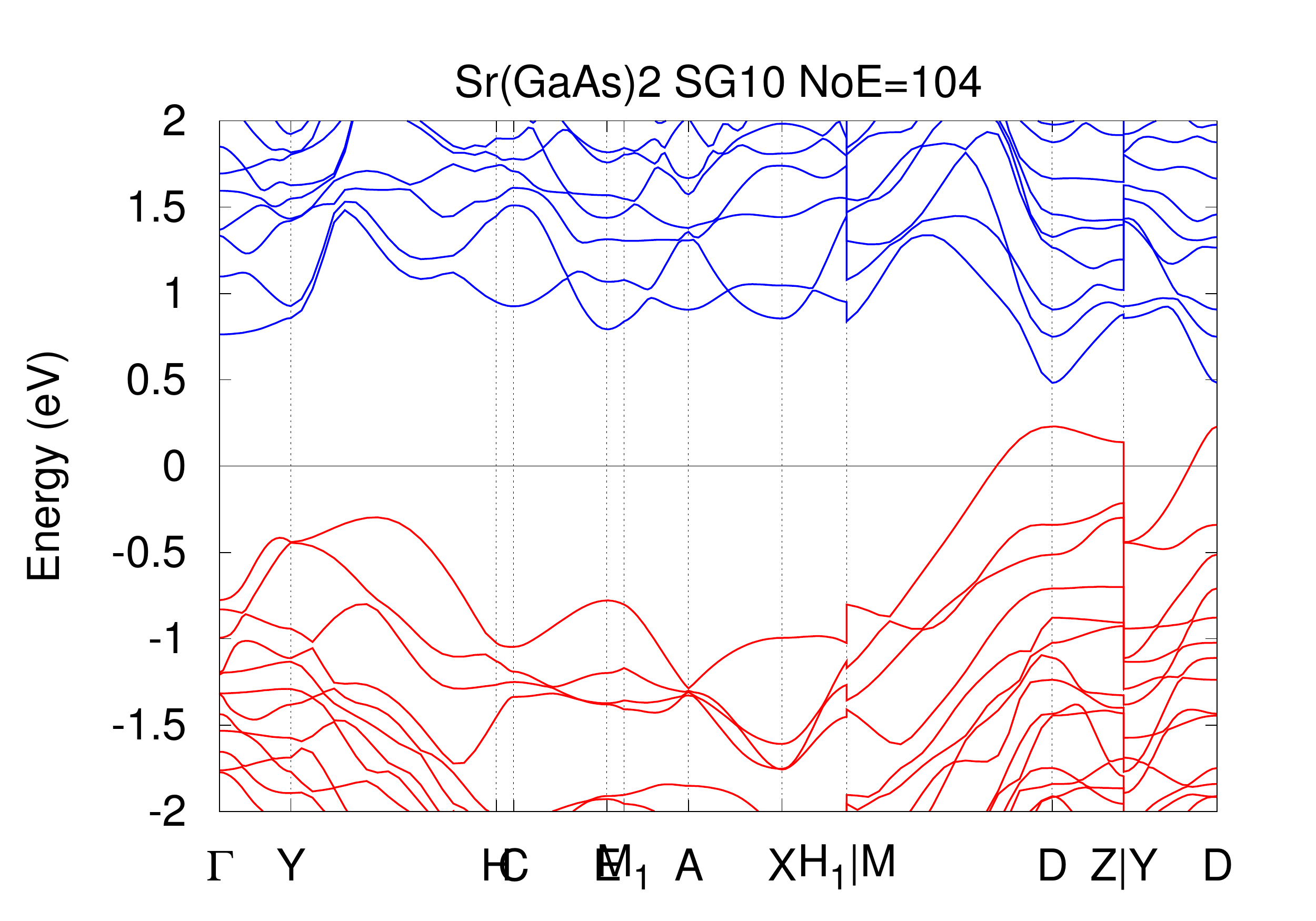}
}
\subfigure[Ba(GeP)$_{2}$ SG105 NoA=20 NoE=112]{
\label{subfig:26416}
\includegraphics[scale=0.32,angle=270]{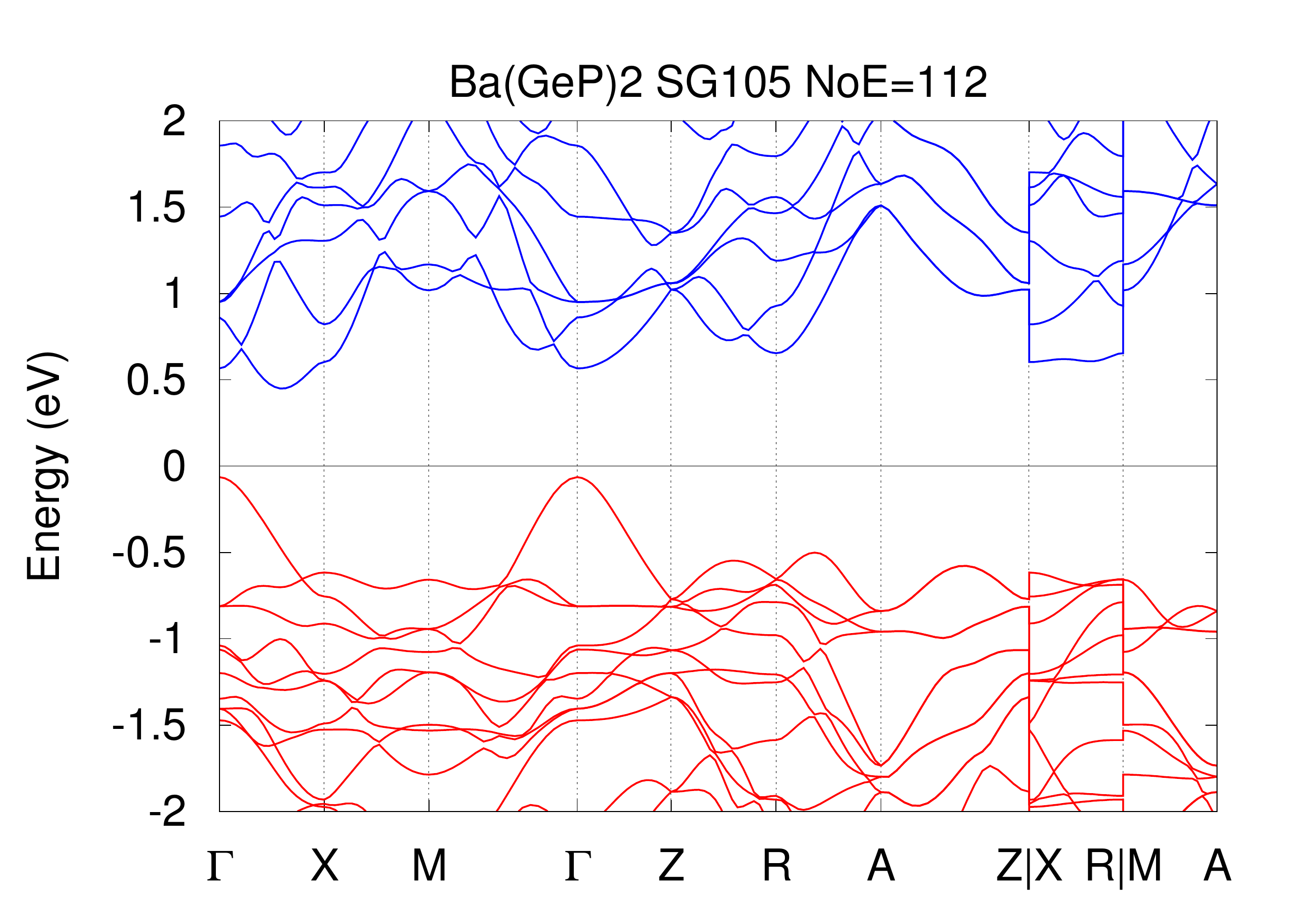}
}
\subfigure[AlSiP$_{3}$ SG62 NoA=20 NoE=88]{
\label{subfig:10032}
\includegraphics[scale=0.32,angle=270]{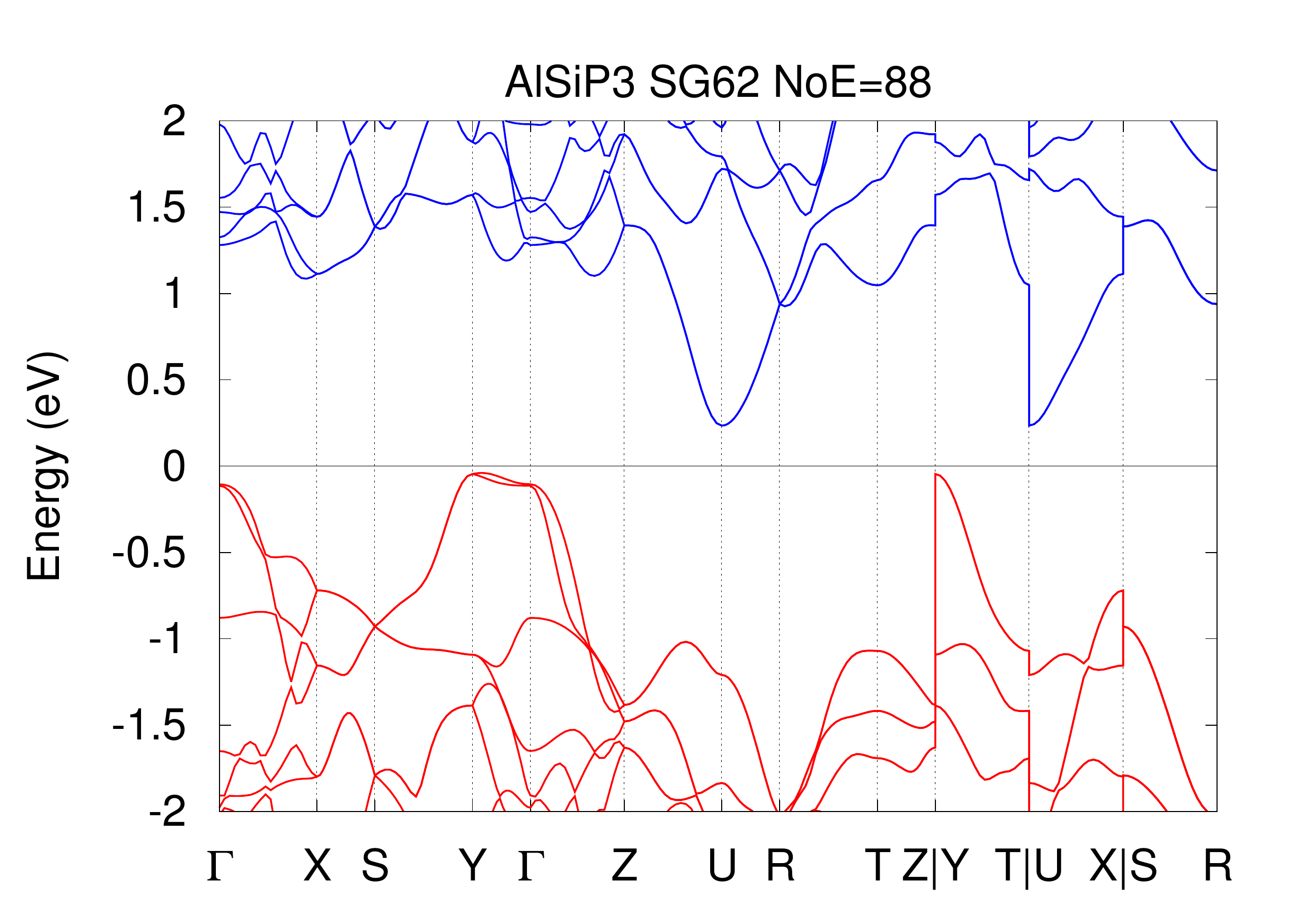}
}
\caption{\hyperref[tab:electride]{back to the table}}
\end{figure}

\begin{figure}[htp]
 \centering
\subfigure[TiNb$_{3}$O$_{6}$ SG148 NoA=20 NoE=146]{
\label{subfig:280002}
\includegraphics[scale=0.32,angle=270]{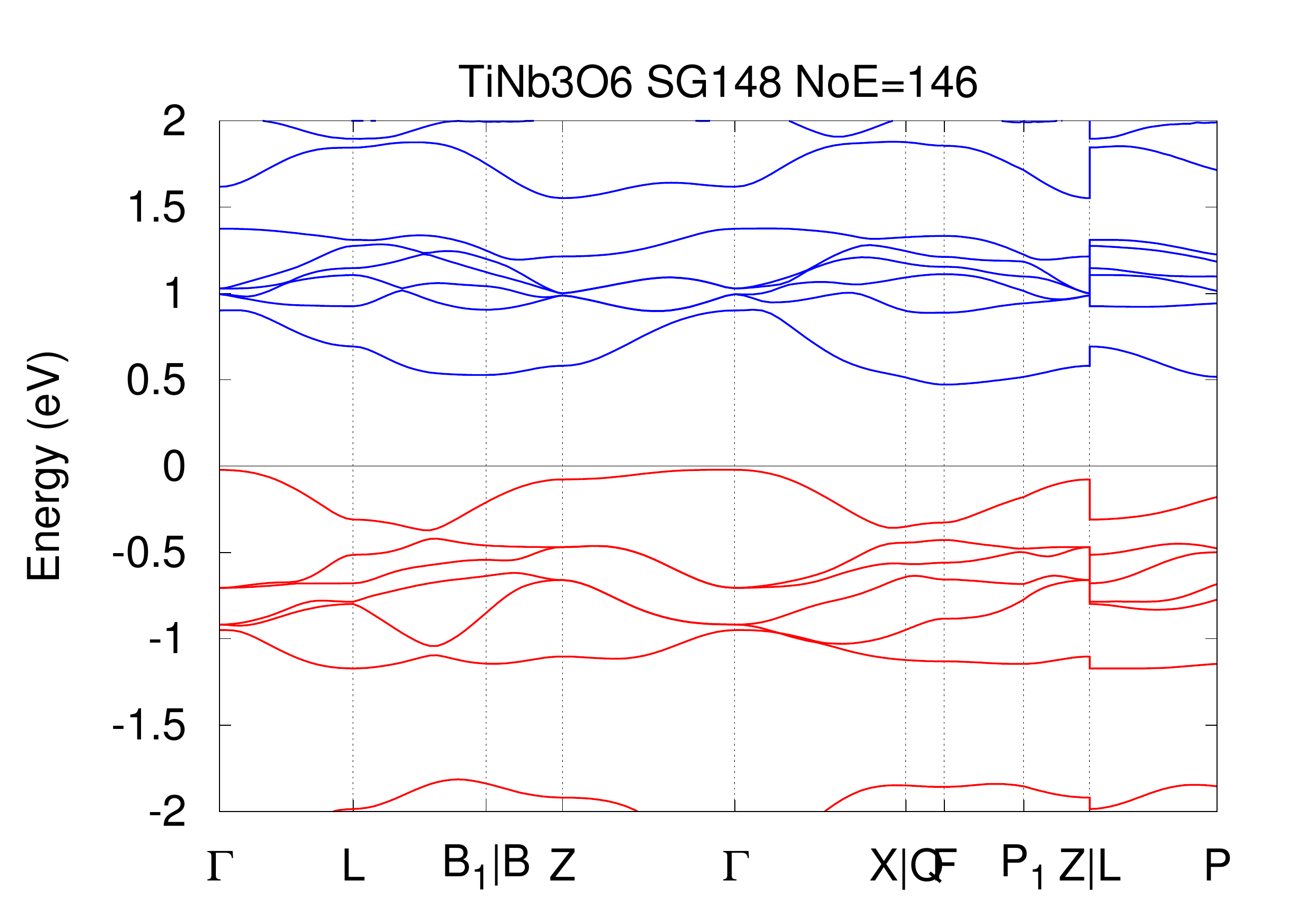}
}
\subfigure[Nb$_{2}$F$_{5}$ SG229 NoA=21 NoE=171]{
\label{subfig:415950}
\includegraphics[scale=0.32,angle=270]{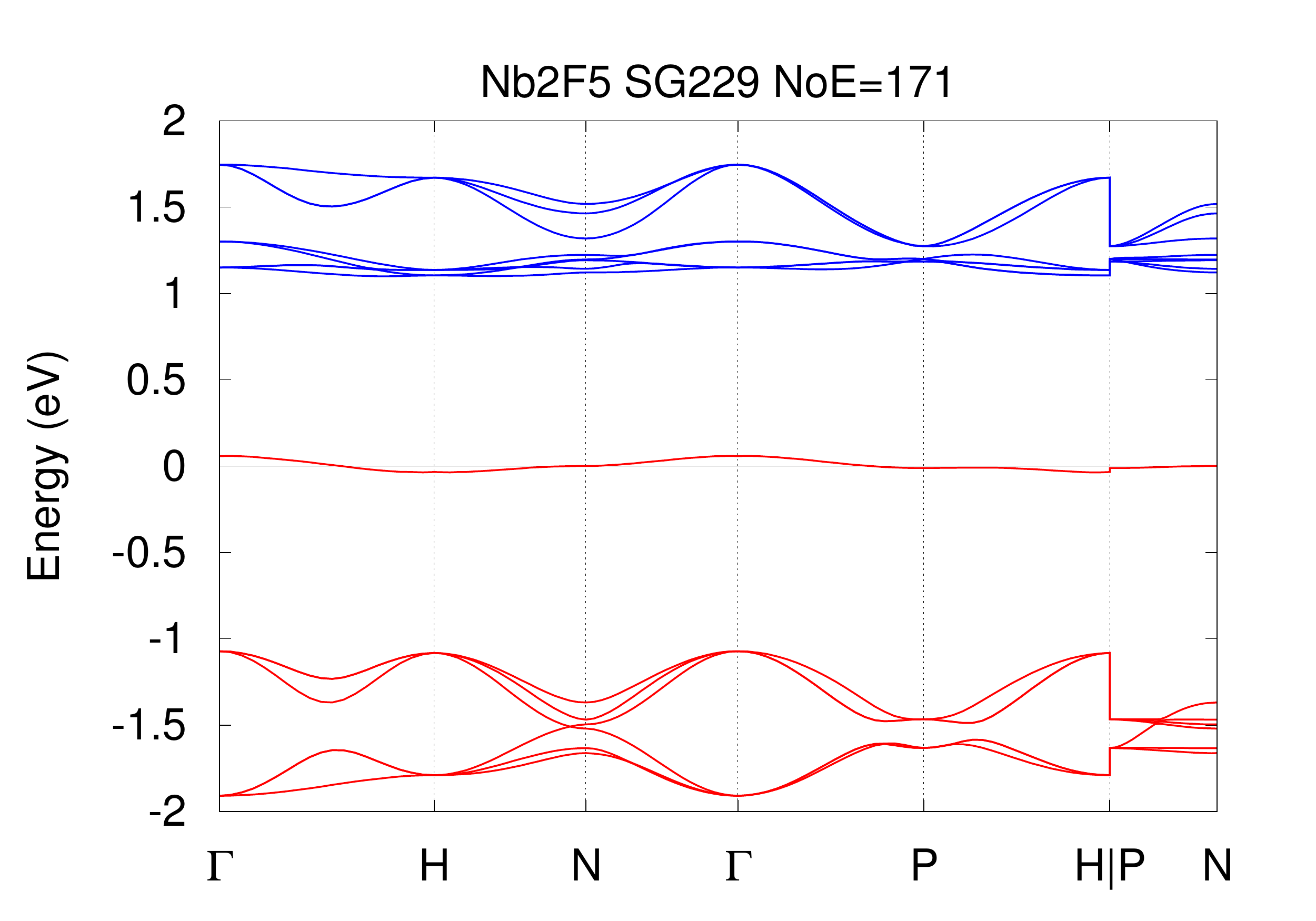}
}
\subfigure[Ba(As$_{3}$Pt$_{2}$)$_{2}$ SG15 NoA=22 NoE=160]{
\label{subfig:62519}
\includegraphics[scale=0.32,angle=270]{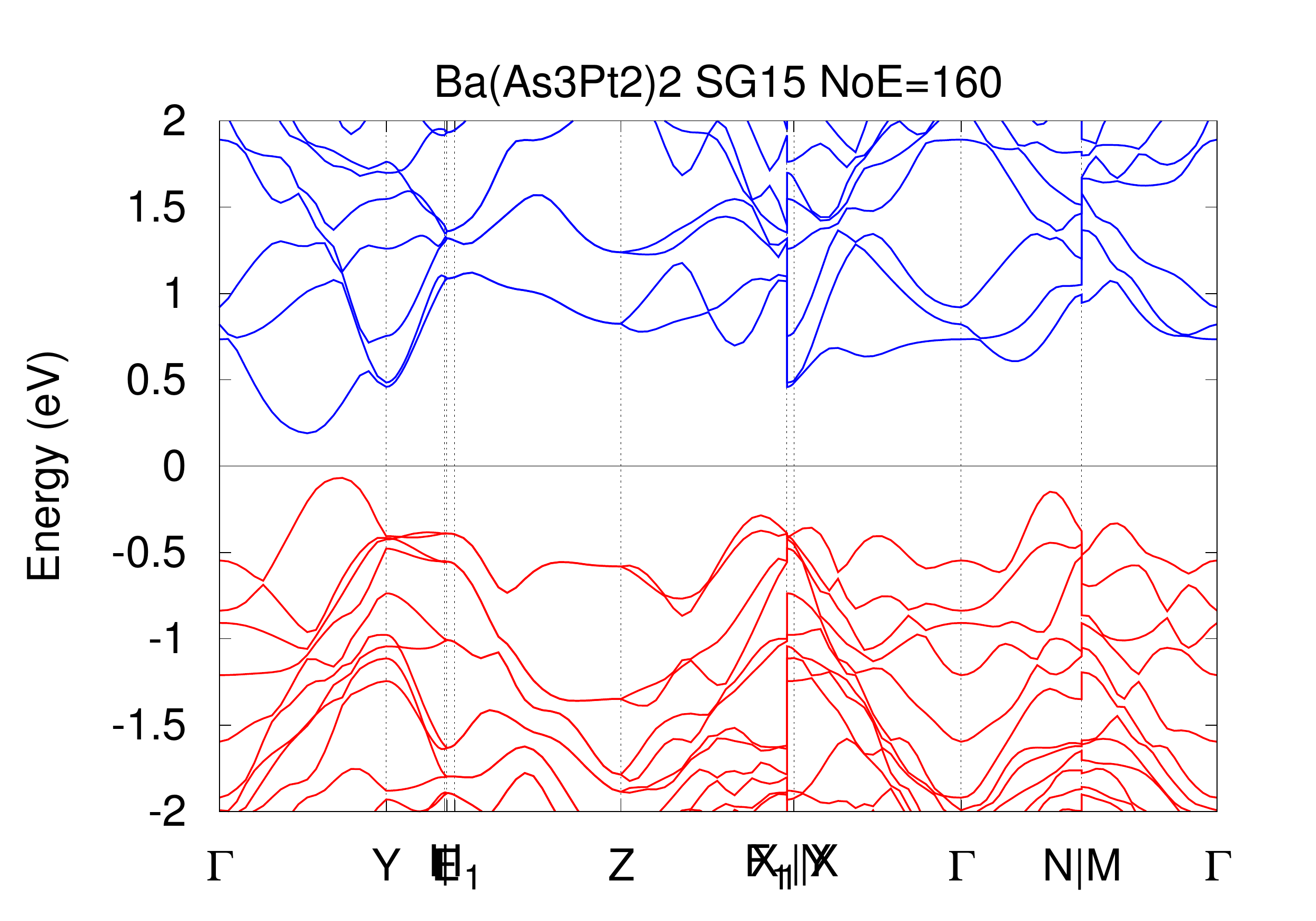}
}
\subfigure[K$_{2}$Fe(PS$_{3}$)$_{2}$ SG14 NoA=22 NoE=144]{
\label{subfig:657803}
\includegraphics[scale=0.32,angle=270]{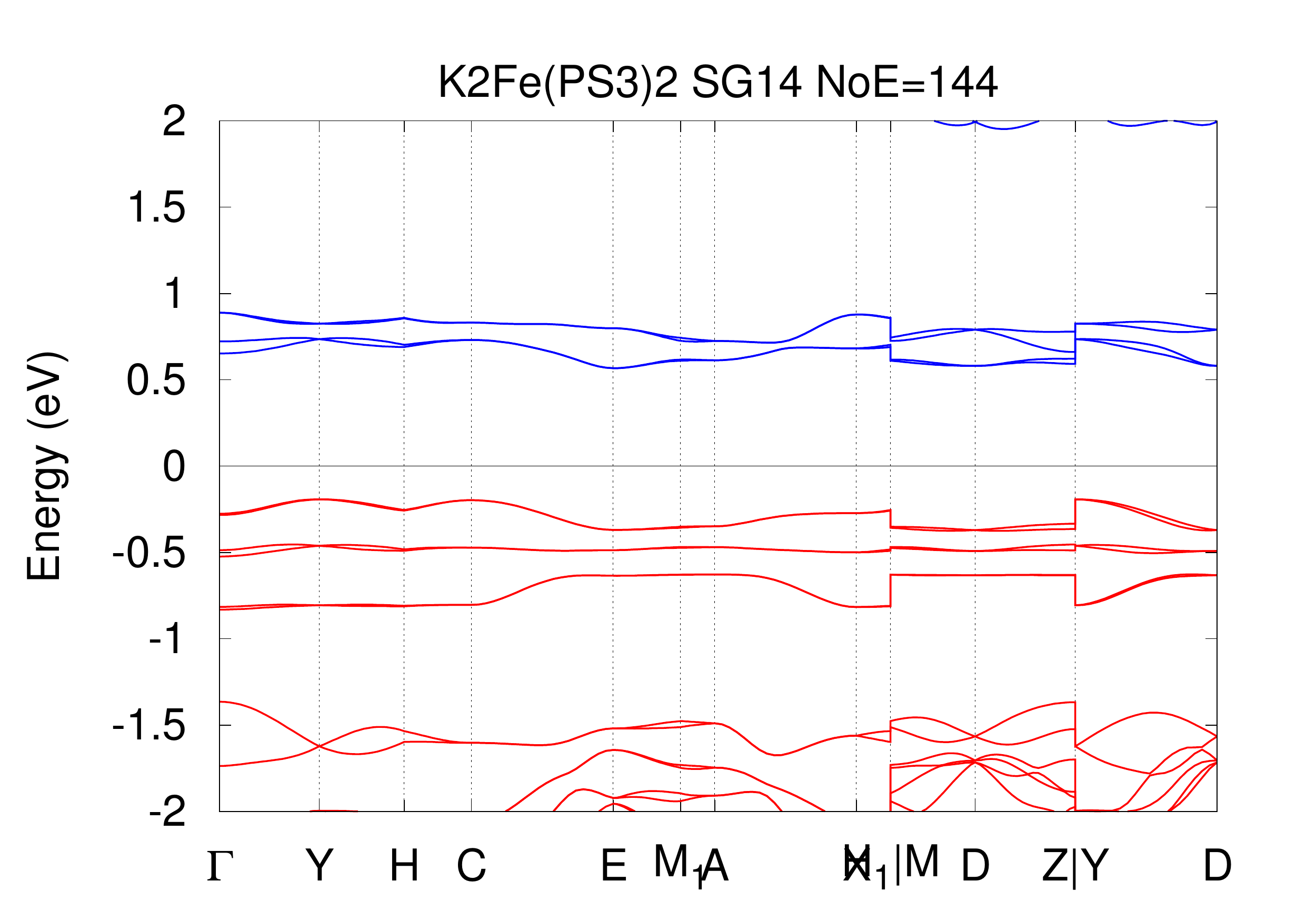}
}
\subfigure[Sr(P$_{3}$Pt$_{2}$)$_{2}$ SG15 NoA=22 NoE=160]{
\label{subfig:62517}
\includegraphics[scale=0.32,angle=270]{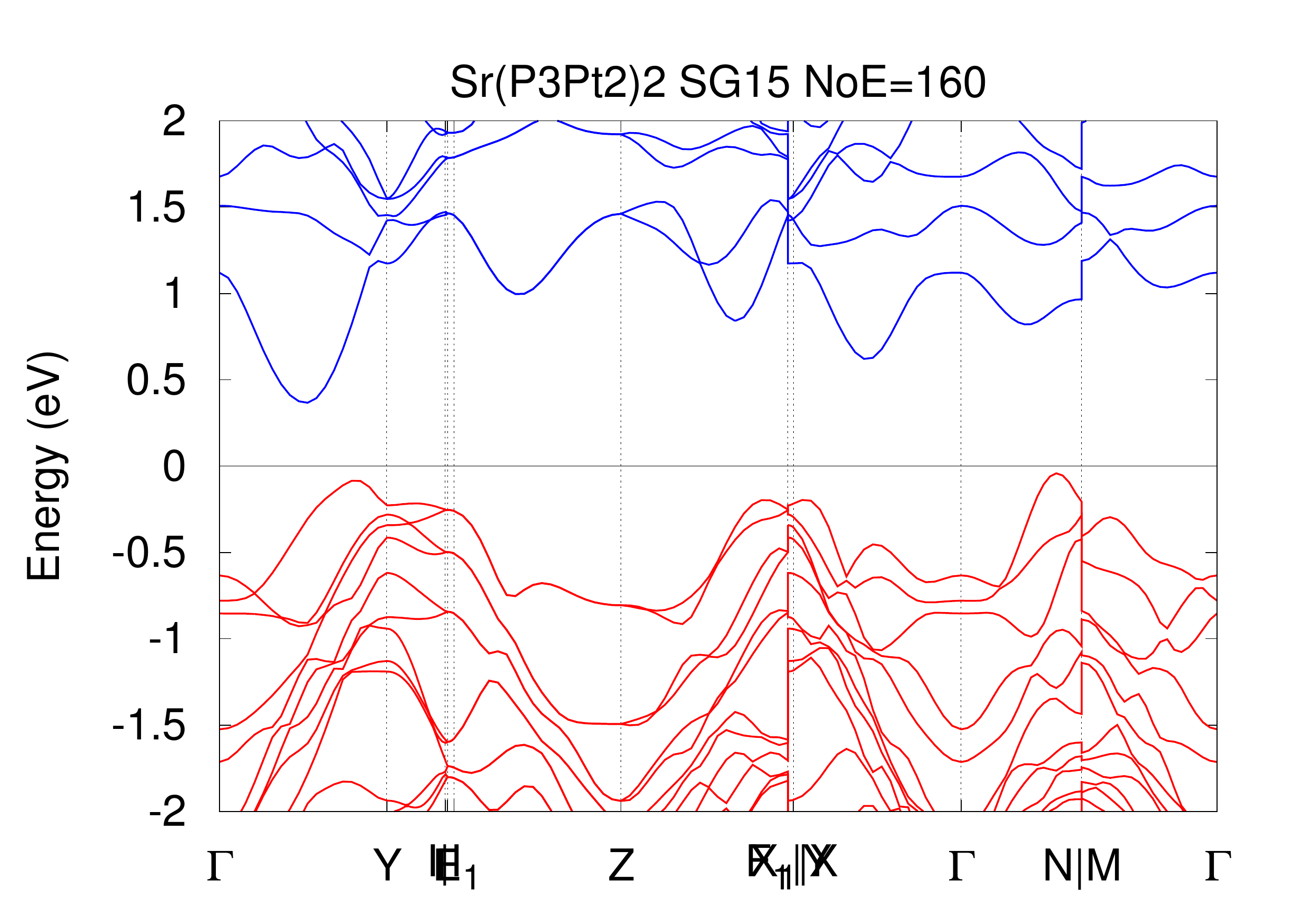}
}
\subfigure[Nb$_{3}$Br$_{8}$ SG166 NoA=22 NoE=178]{
\label{subfig:25766}
\includegraphics[scale=0.32,angle=270]{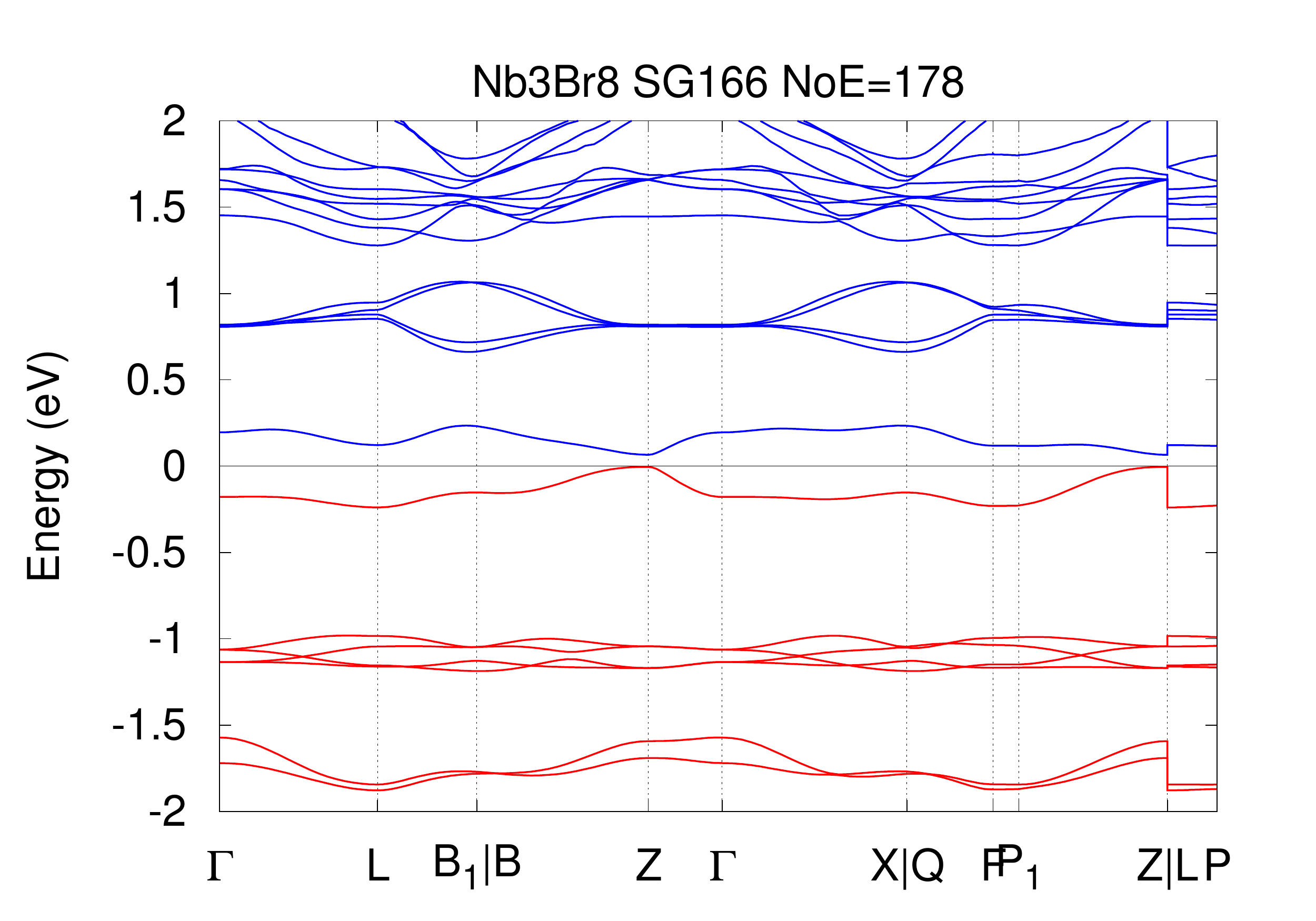}
}
\subfigure[La$_{2}$Mo$_{2}$O$_{7}$ SG58 NoA=22 NoE=152]{
\label{subfig:202189}
\includegraphics[scale=0.32,angle=270]{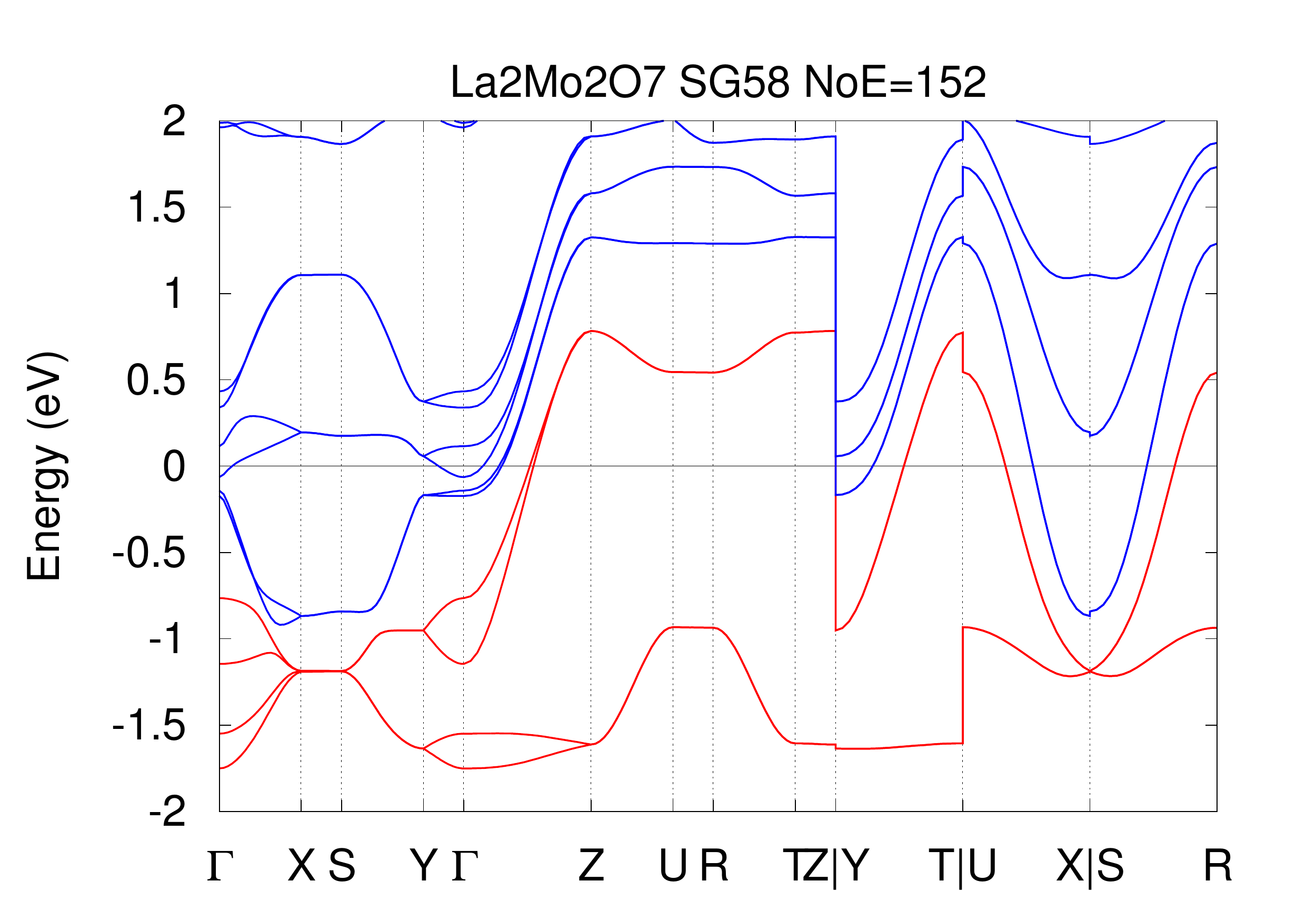}
}
\subfigure[Ba$_{5}$CrN$_{5}$ SG12 NoA=22 NoE=162]{
\label{subfig:82360}
\includegraphics[scale=0.32,angle=270]{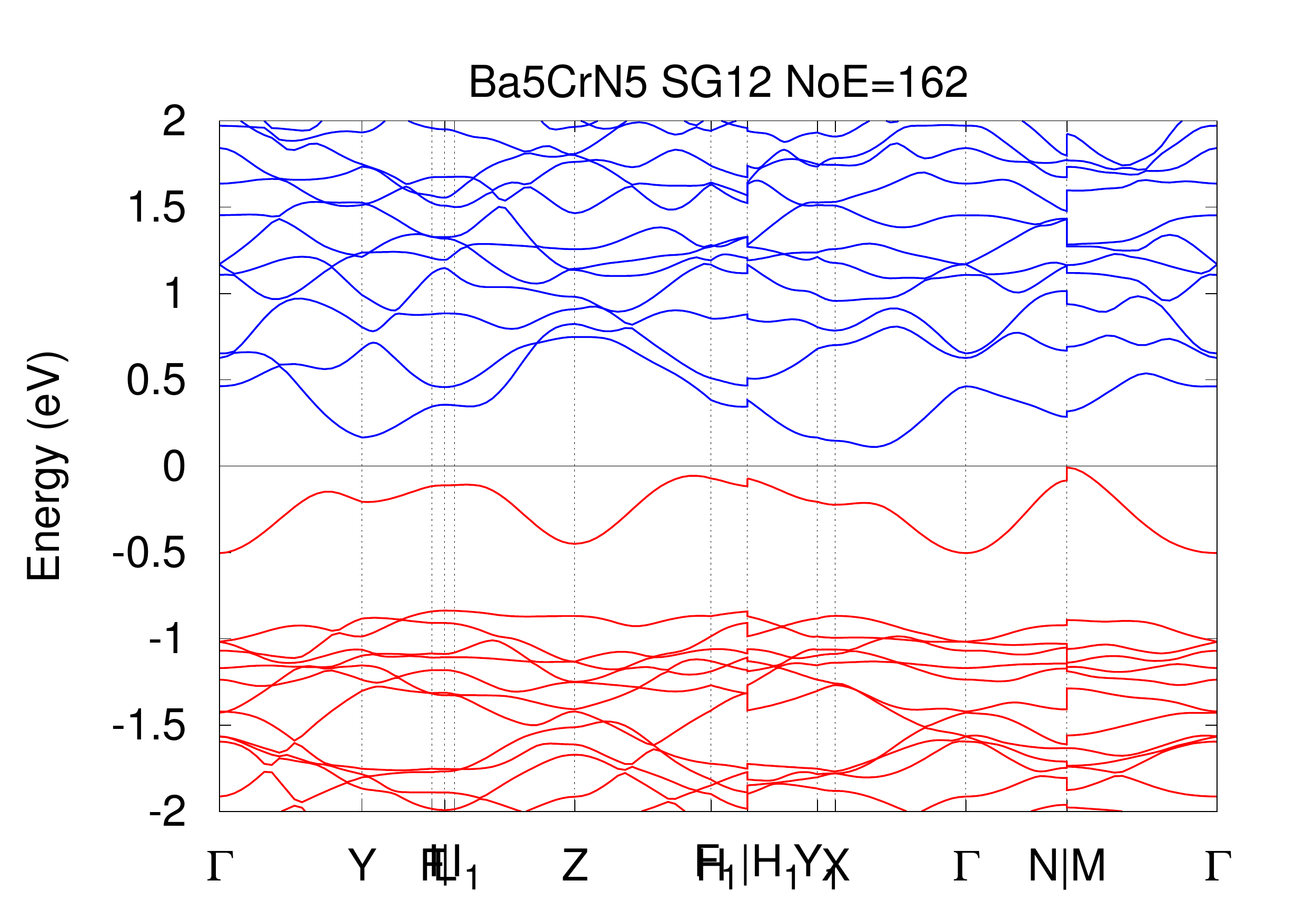}
}
\caption{\hyperref[tab:electride]{back to the table}}
\end{figure}

\begin{figure}[htp]
 \centering
\subfigure[Sr(As$_{3}$Pt$_{2}$)$_{2}$ SG15 NoA=22 NoE=160]{
\label{subfig:62518}
\includegraphics[scale=0.32,angle=270]{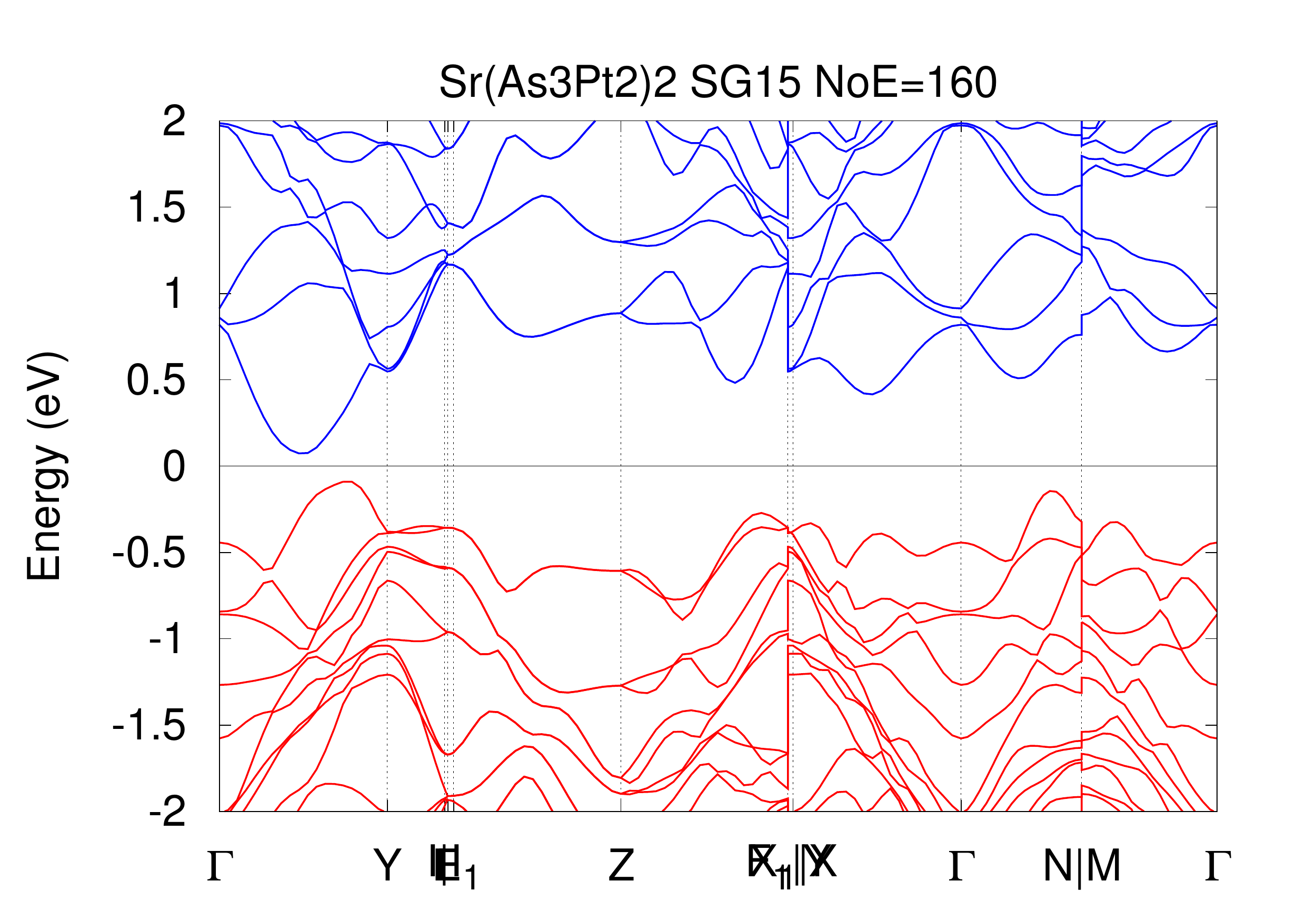}
}
\subfigure[Dy$_{4}$InRh SG216 NoA=24 NoE=192]{
\label{subfig:417514}
\includegraphics[scale=0.32,angle=270]{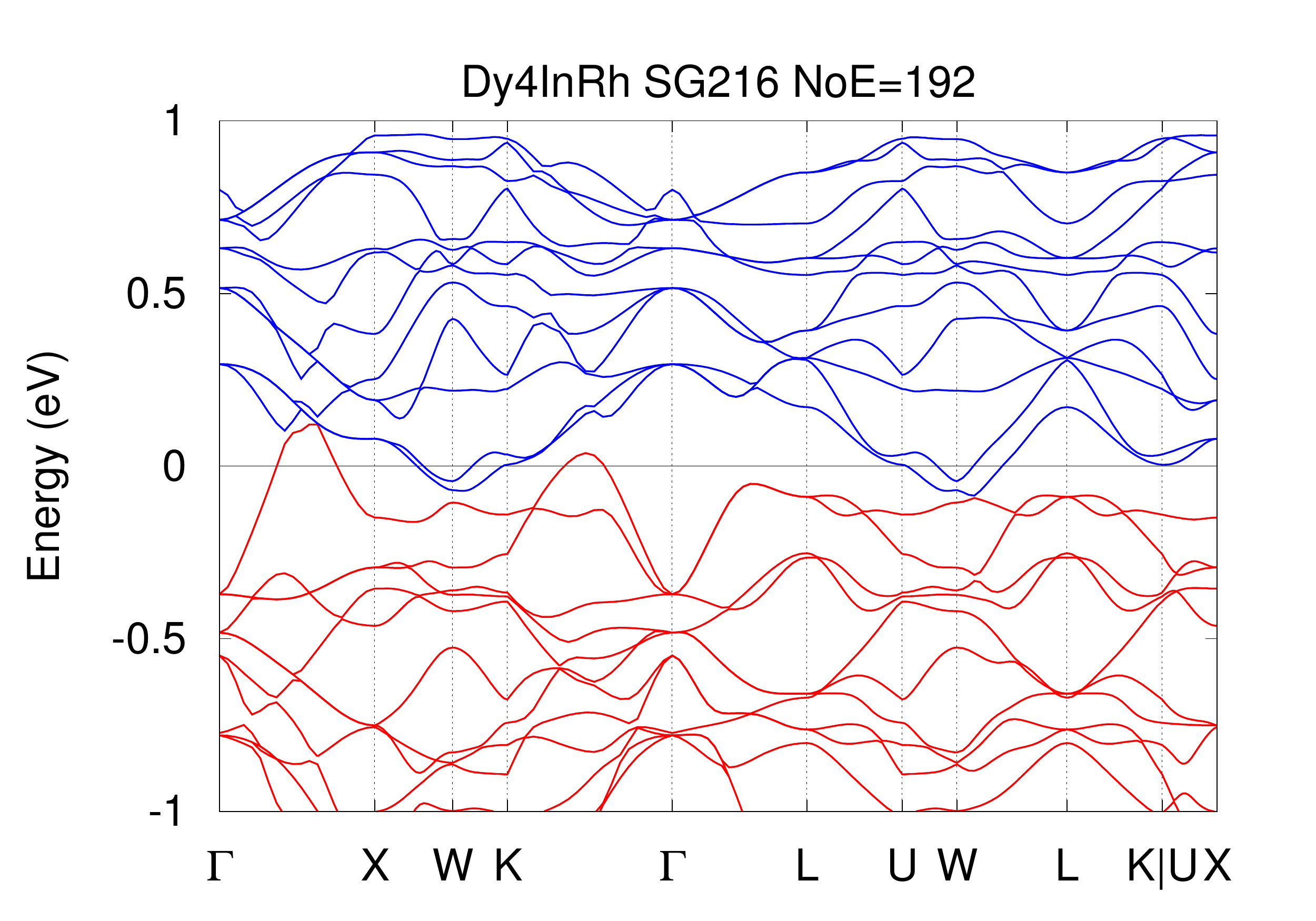}
}
\subfigure[Tb$_{4}$InRh SG216 NoA=24 NoE=192]{
\label{subfig:417518}
\includegraphics[scale=0.32,angle=270]{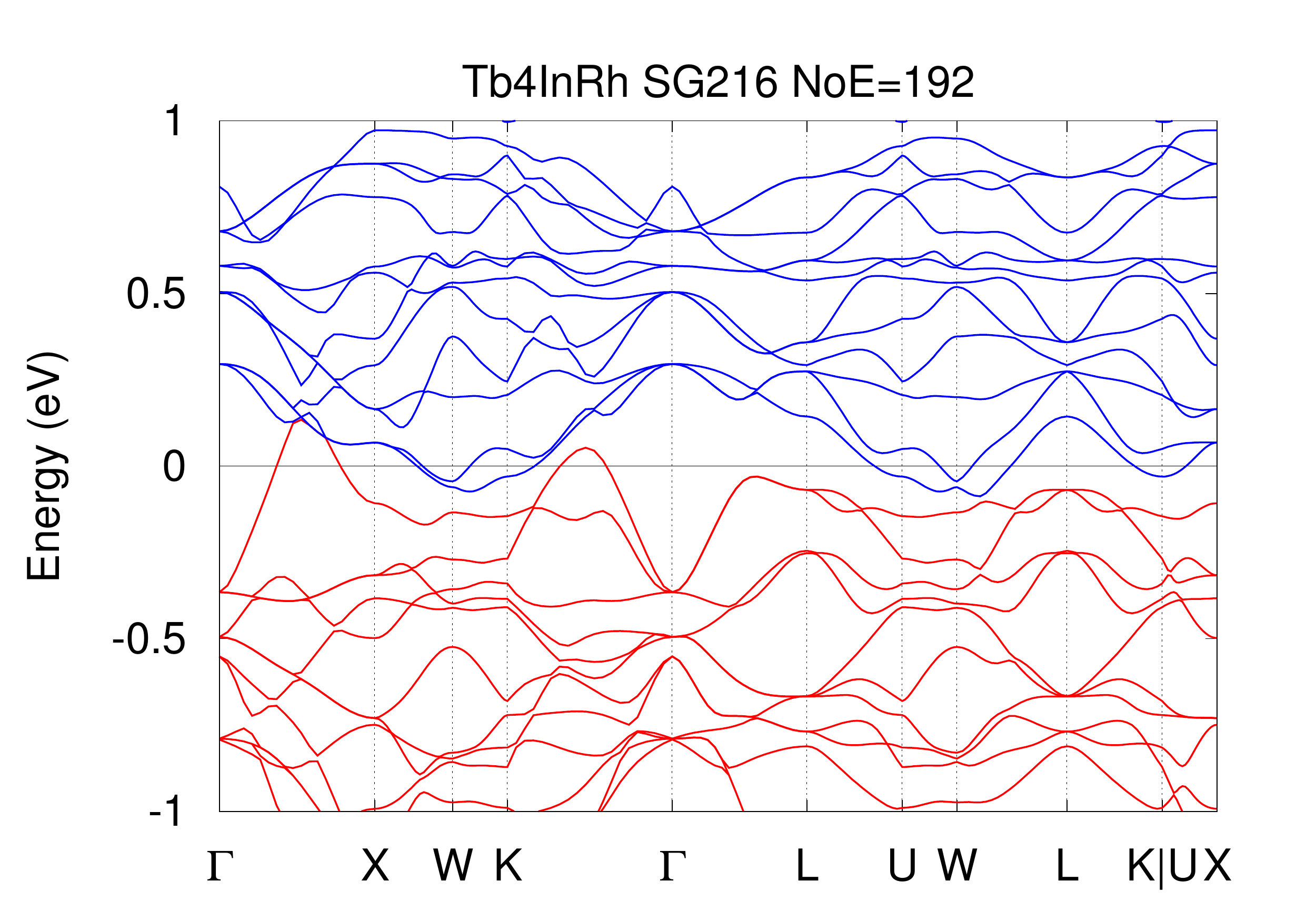}
}
\subfigure[Ho$_{4}$InRh SG216 NoA=24 NoE=192]{
\label{subfig:417517}
\includegraphics[scale=0.32,angle=270]{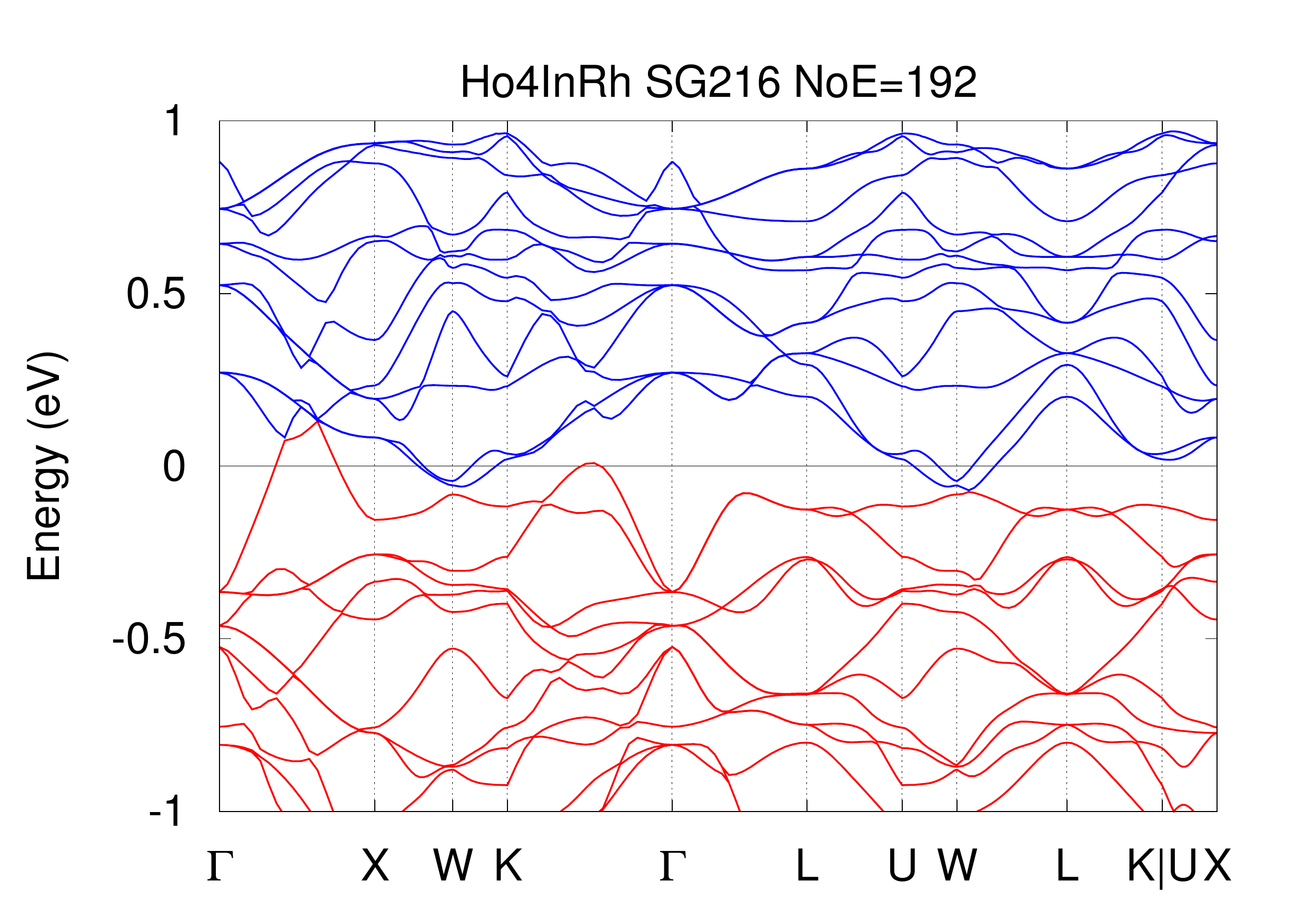}
}
\subfigure[Tm$_{4}$InRh SG216 NoA=24 NoE=192]{
\label{subfig:417519}
\includegraphics[scale=0.32,angle=270]{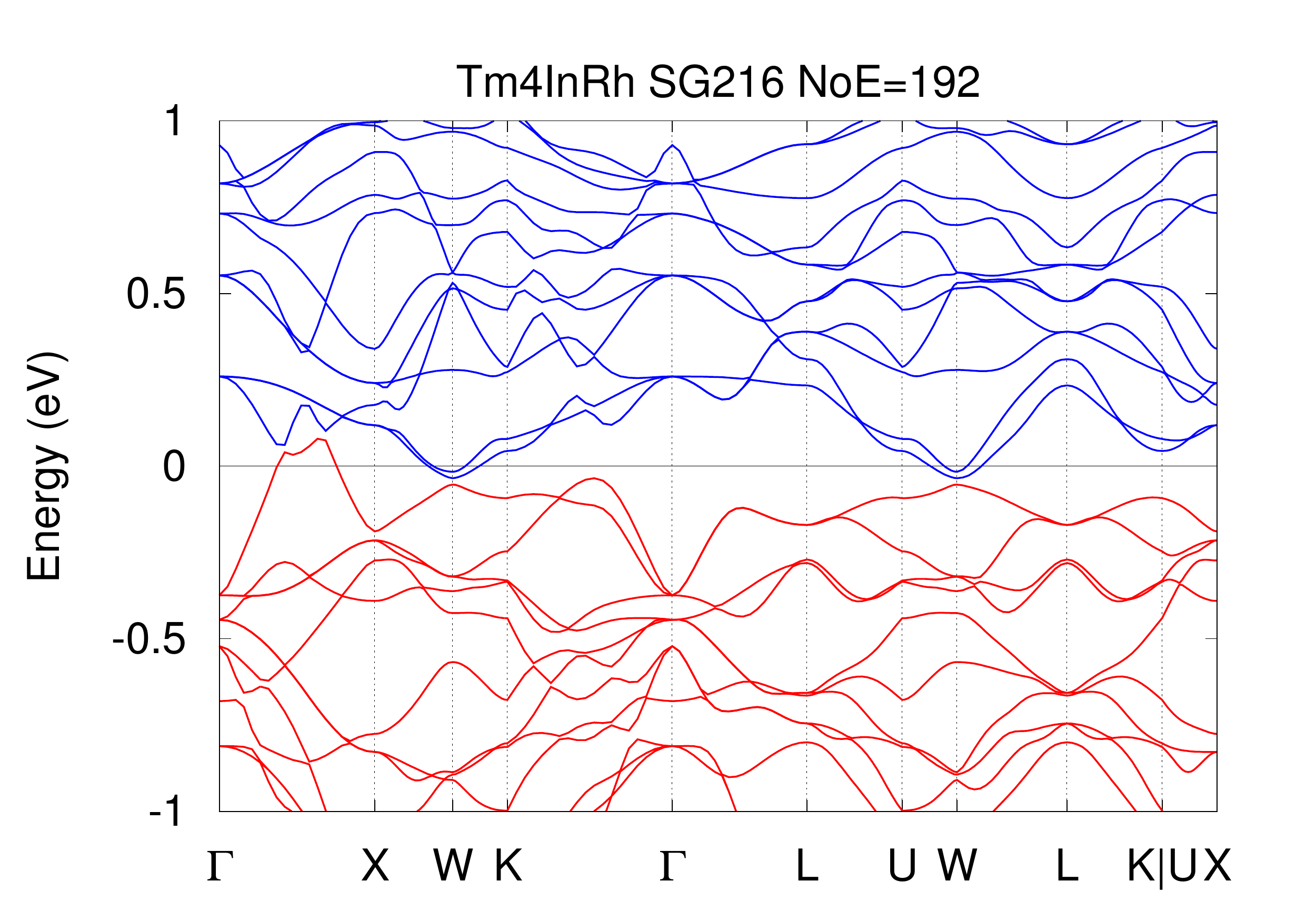}
}
\subfigure[Er$_{4}$InIr SG216 NoA=24 NoE=192]{
\label{subfig:418265}
\includegraphics[scale=0.32,angle=270]{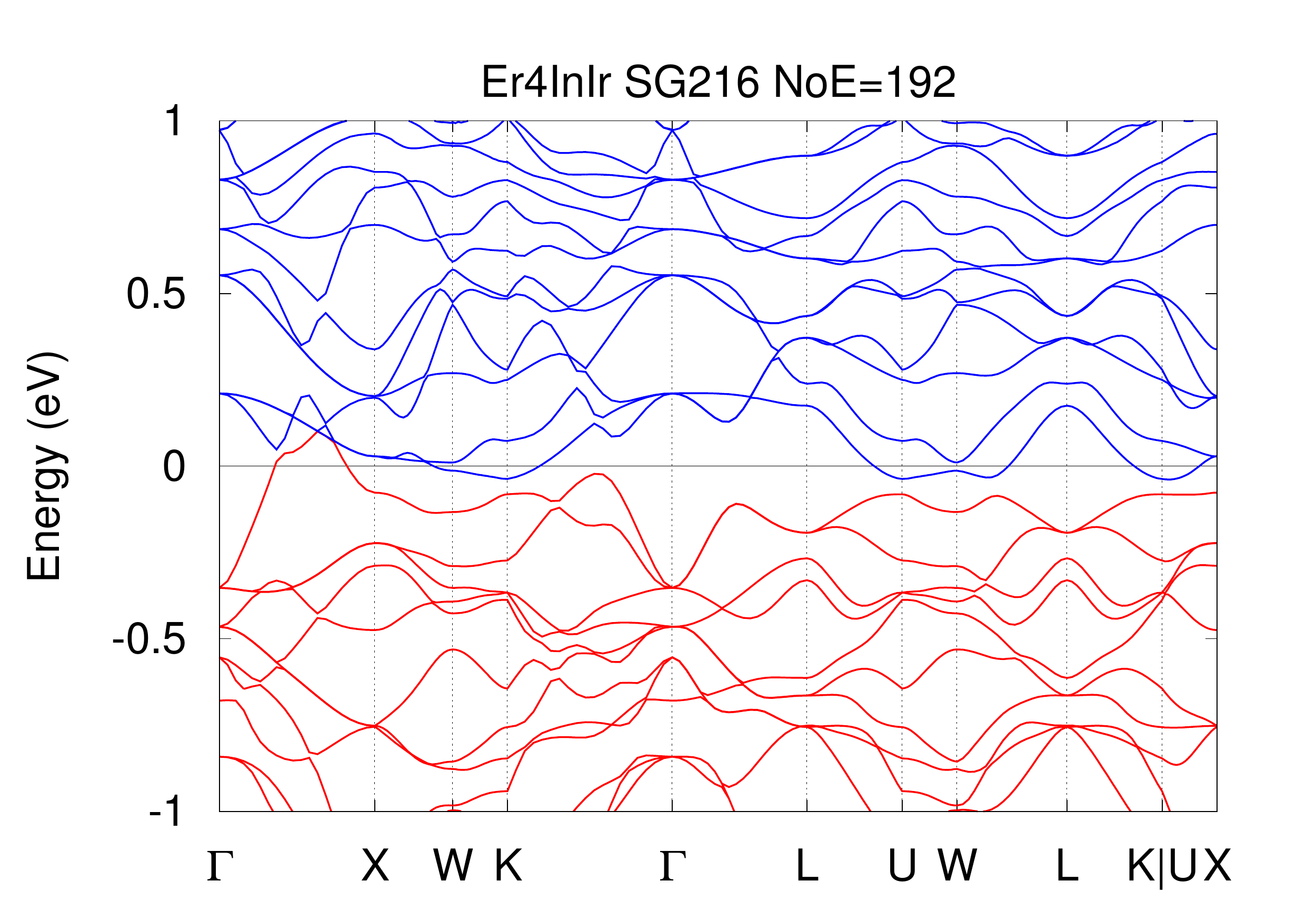}
}
\subfigure[CeCrB$_{4}$ SG55 NoA=24 NoE=116]{
\label{subfig:612757}
\includegraphics[scale=0.32,angle=270]{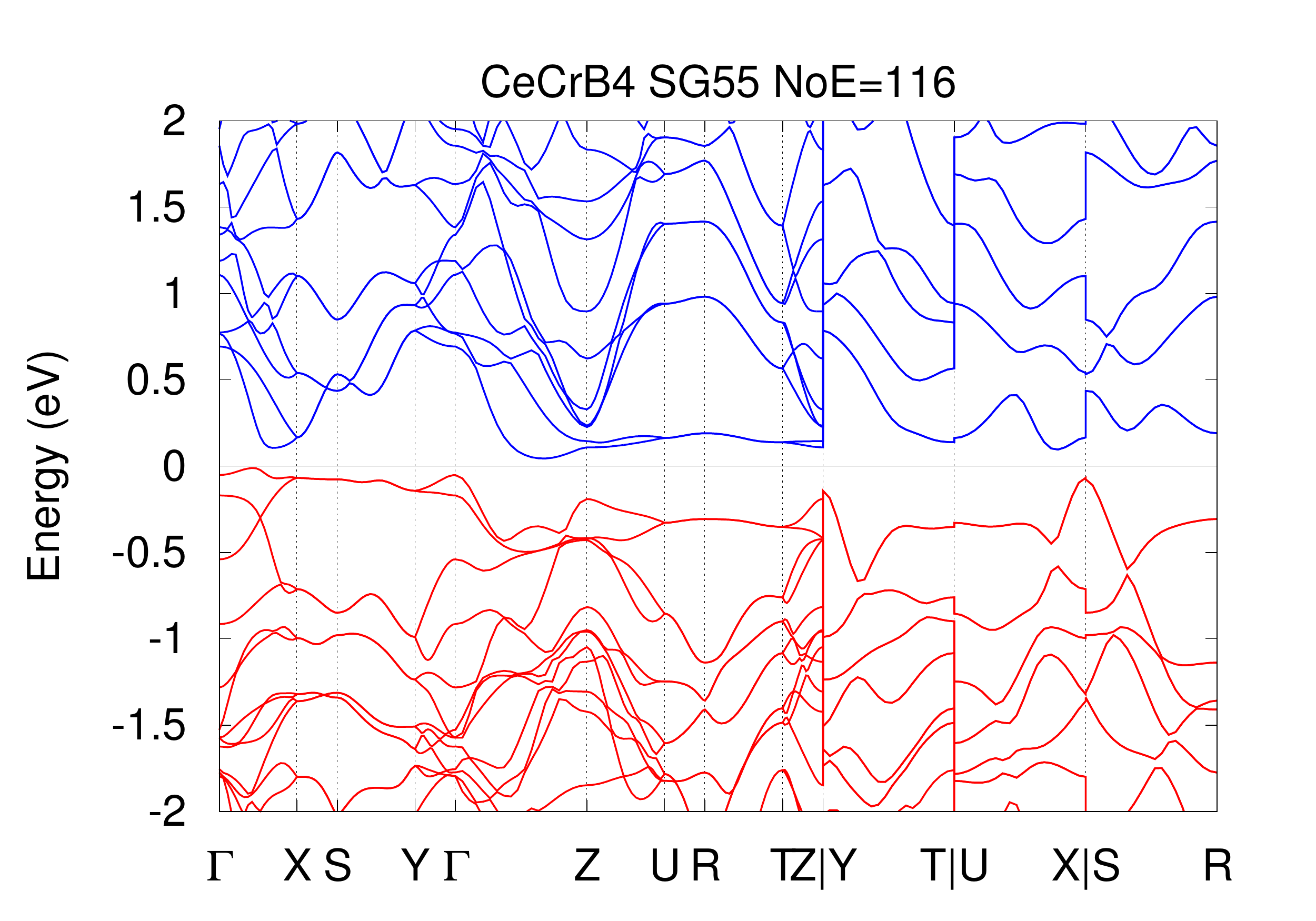}
}
\subfigure[Ho$_{4}$InIr SG216 NoA=24 NoE=192]{
\label{subfig:418270}
\includegraphics[scale=0.32,angle=270]{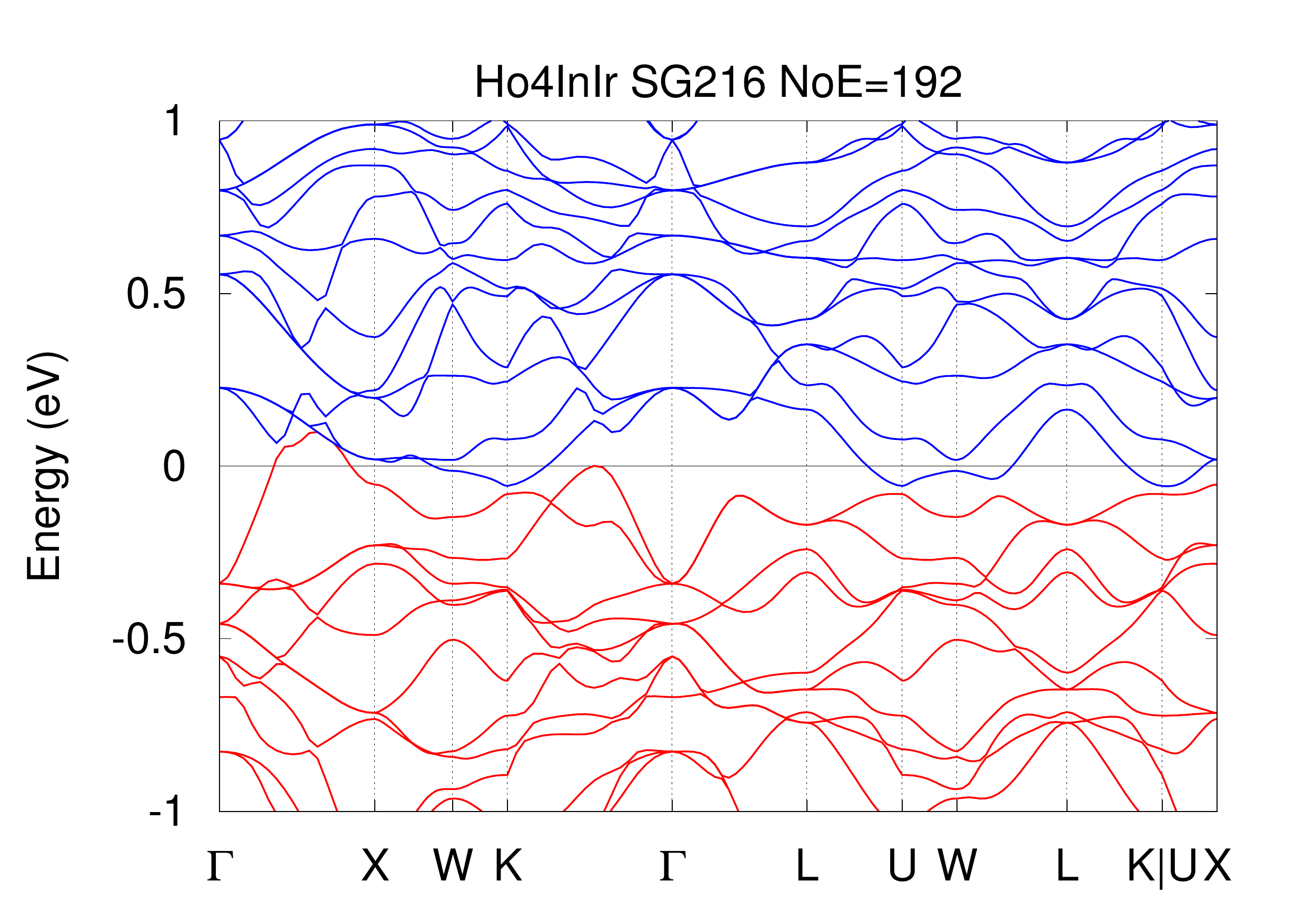}
}
\caption{\hyperref[tab:electride]{back to the table}}
\end{figure}

\begin{figure}[htp]
 \centering
\subfigure[Y$_{4}$InIr SG216 NoA=24 NoE=224]{
\label{subfig:418567}
\includegraphics[scale=0.32,angle=270]{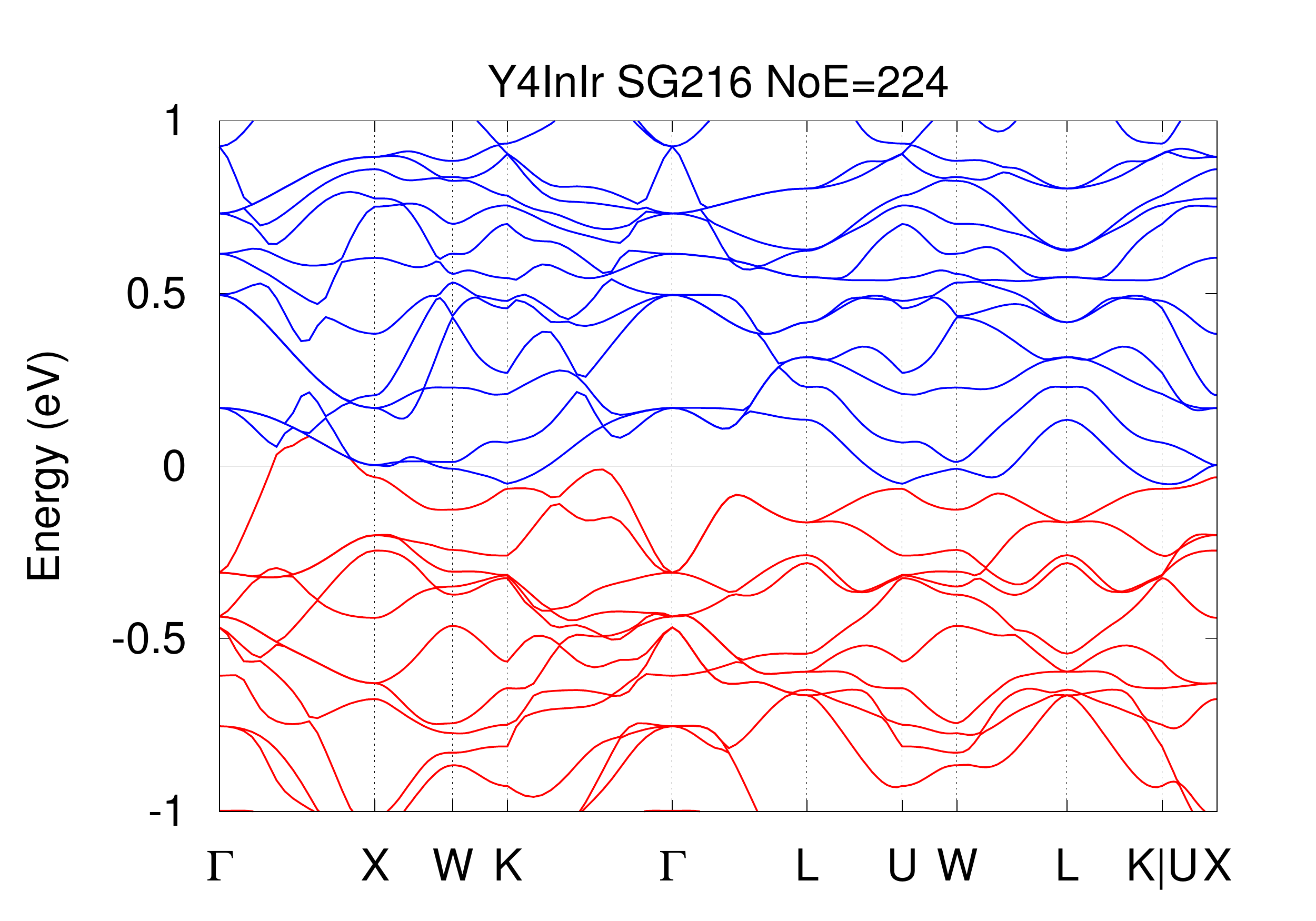}
}
\subfigure[Er$_{4}$InRh SG216 NoA=24 NoE=192]{
\label{subfig:417515}
\includegraphics[scale=0.32,angle=270]{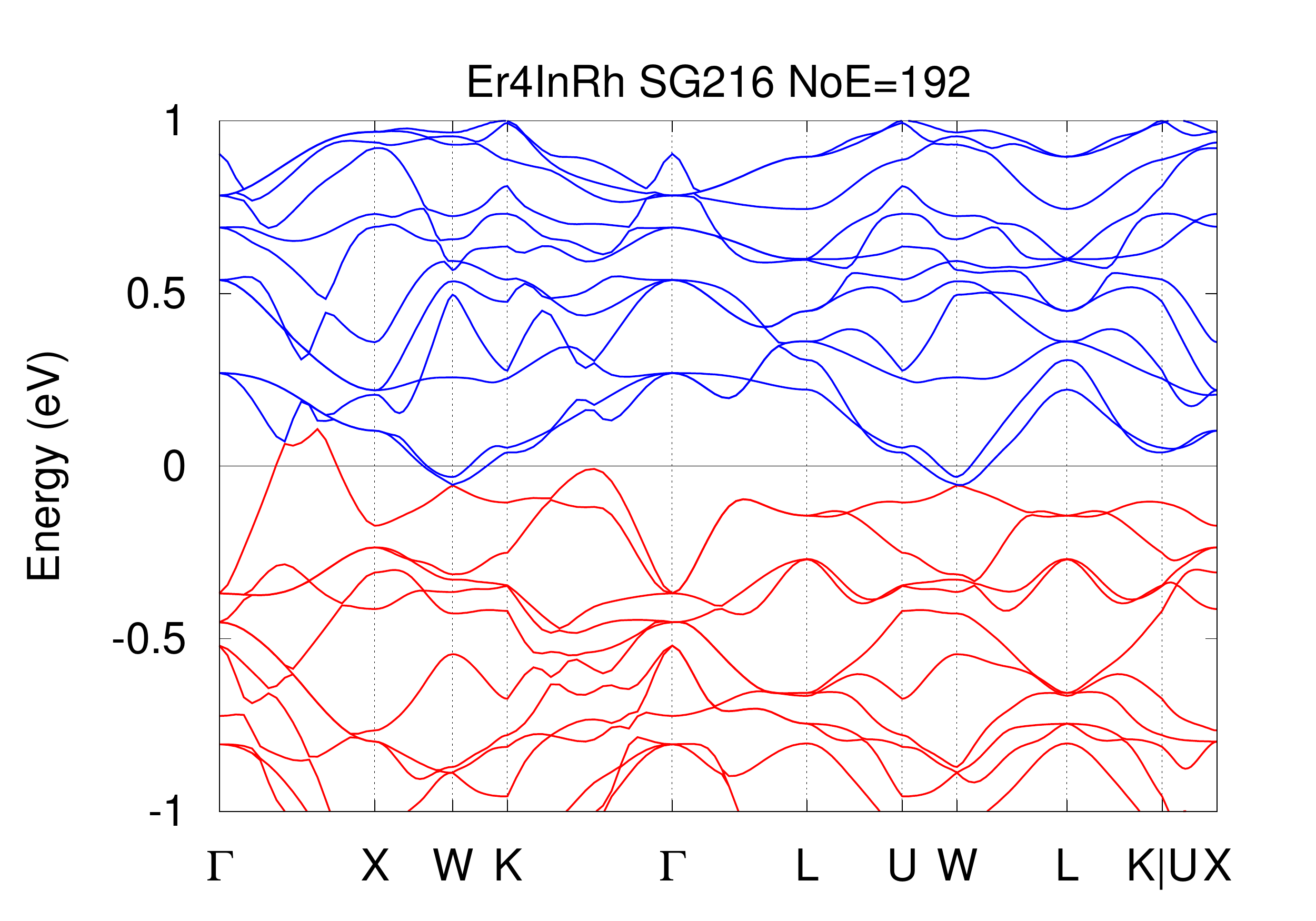}
}
\subfigure[YB$_{4}$Mo SG55 NoA=24 NoE=116]{
\label{subfig:20081}
\includegraphics[scale=0.32,angle=270]{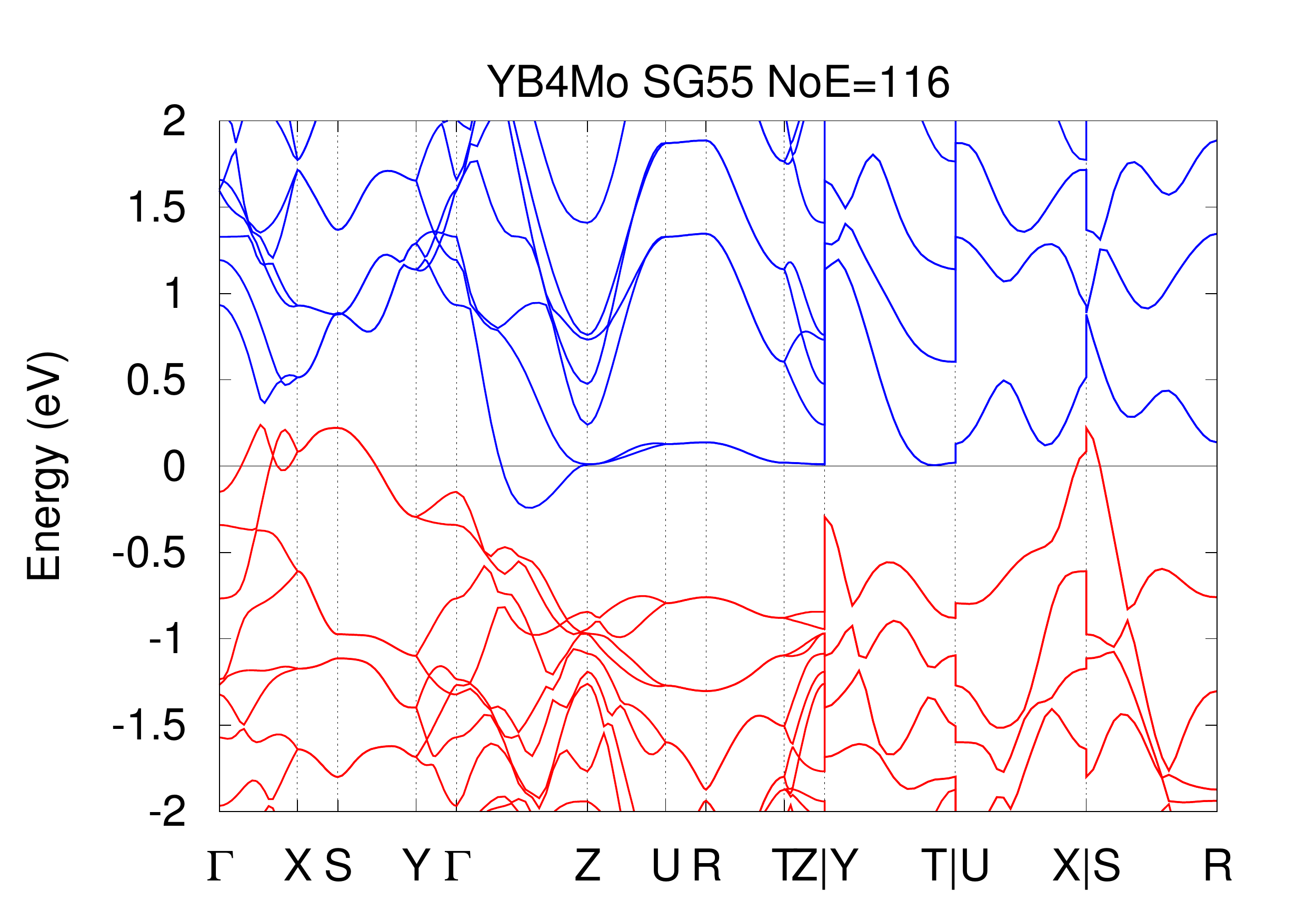}
}
\subfigure[HfBi$_{2}$ SG58 NoA=24 NoE=112]{
\label{subfig:616683}
\includegraphics[scale=0.32,angle=270]{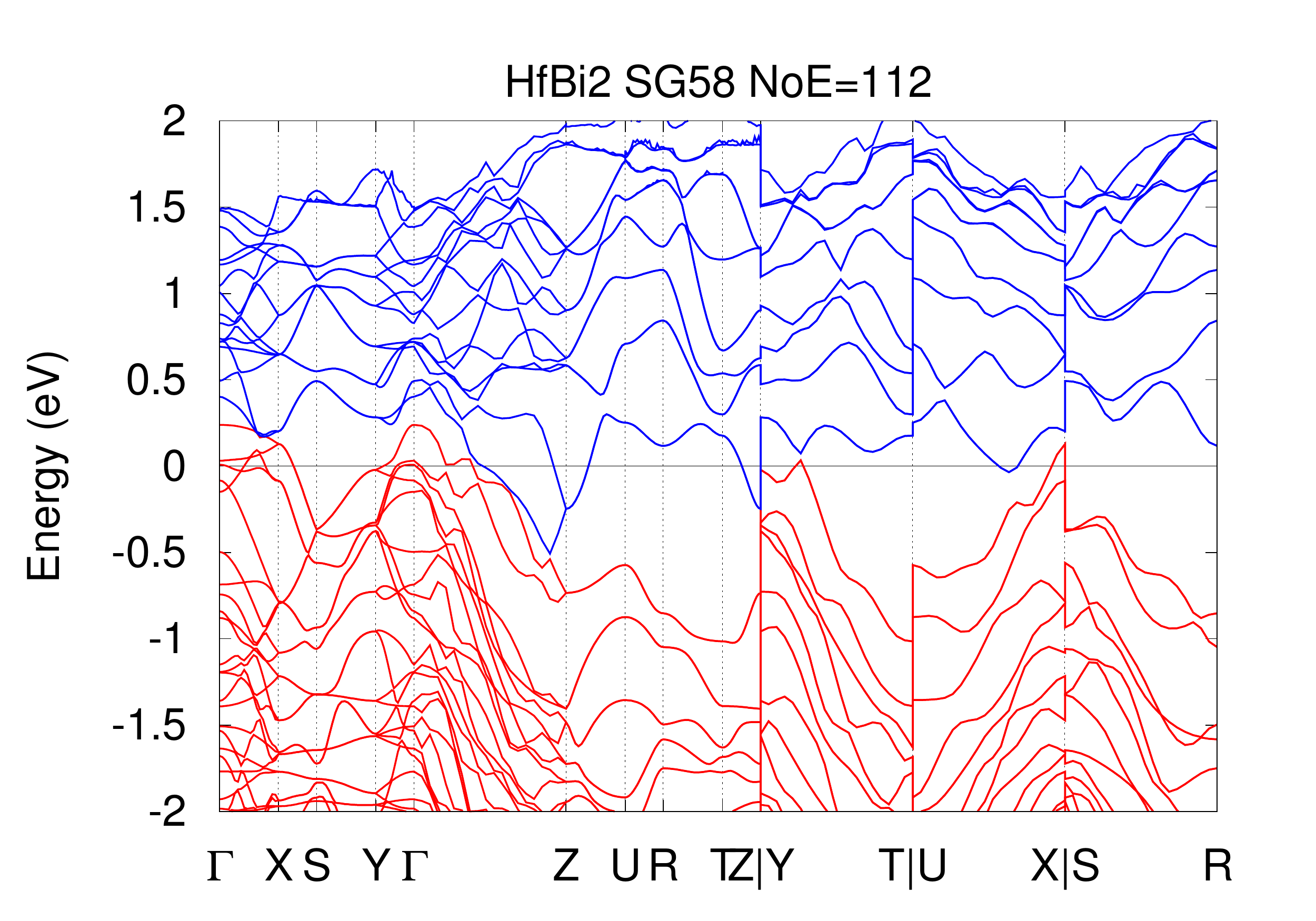}
}
\subfigure[TbCrB$_{4}$ SG55 NoA=24 NoE=108]{
\label{subfig:613559}
\includegraphics[scale=0.32,angle=270]{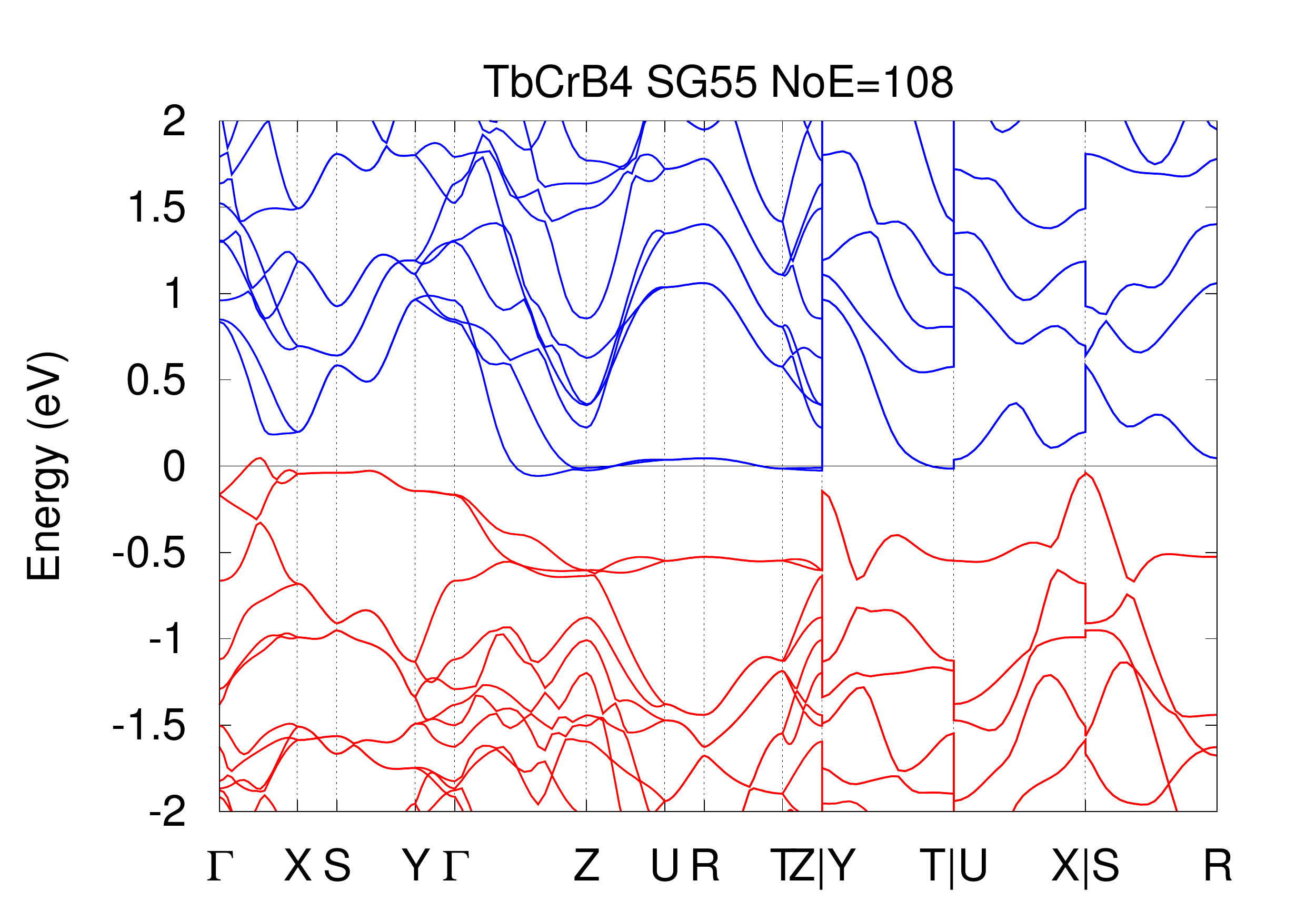}
}
\subfigure[TiCl$_{3}$ SG12 NoA=24 NoE=150]{
\label{subfig:39429}
\includegraphics[scale=0.32,angle=270]{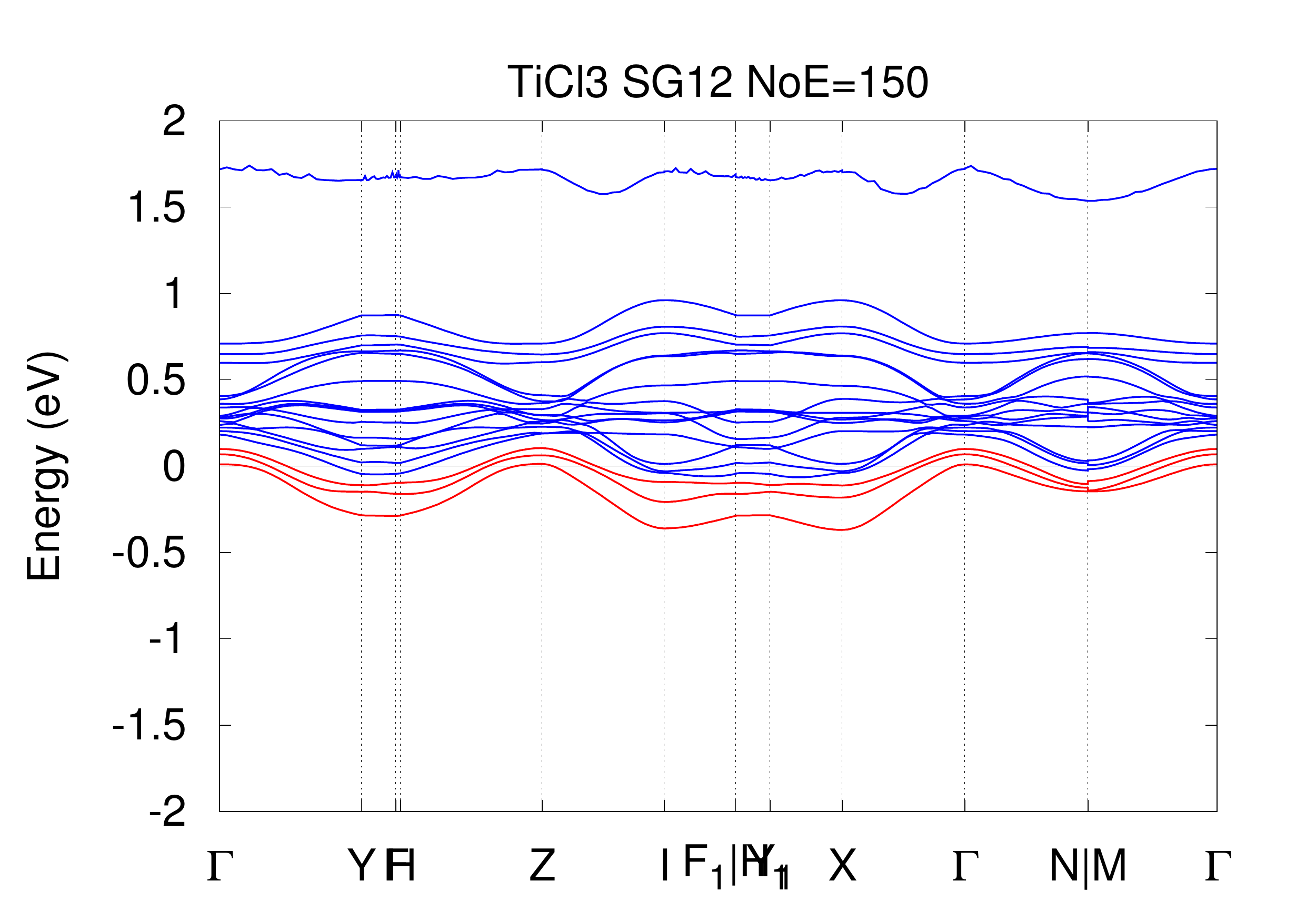}
}
\subfigure[HoCrB$_{4}$ SG55 NoA=24 NoE=108]{
\label{subfig:613514}
\includegraphics[scale=0.32,angle=270]{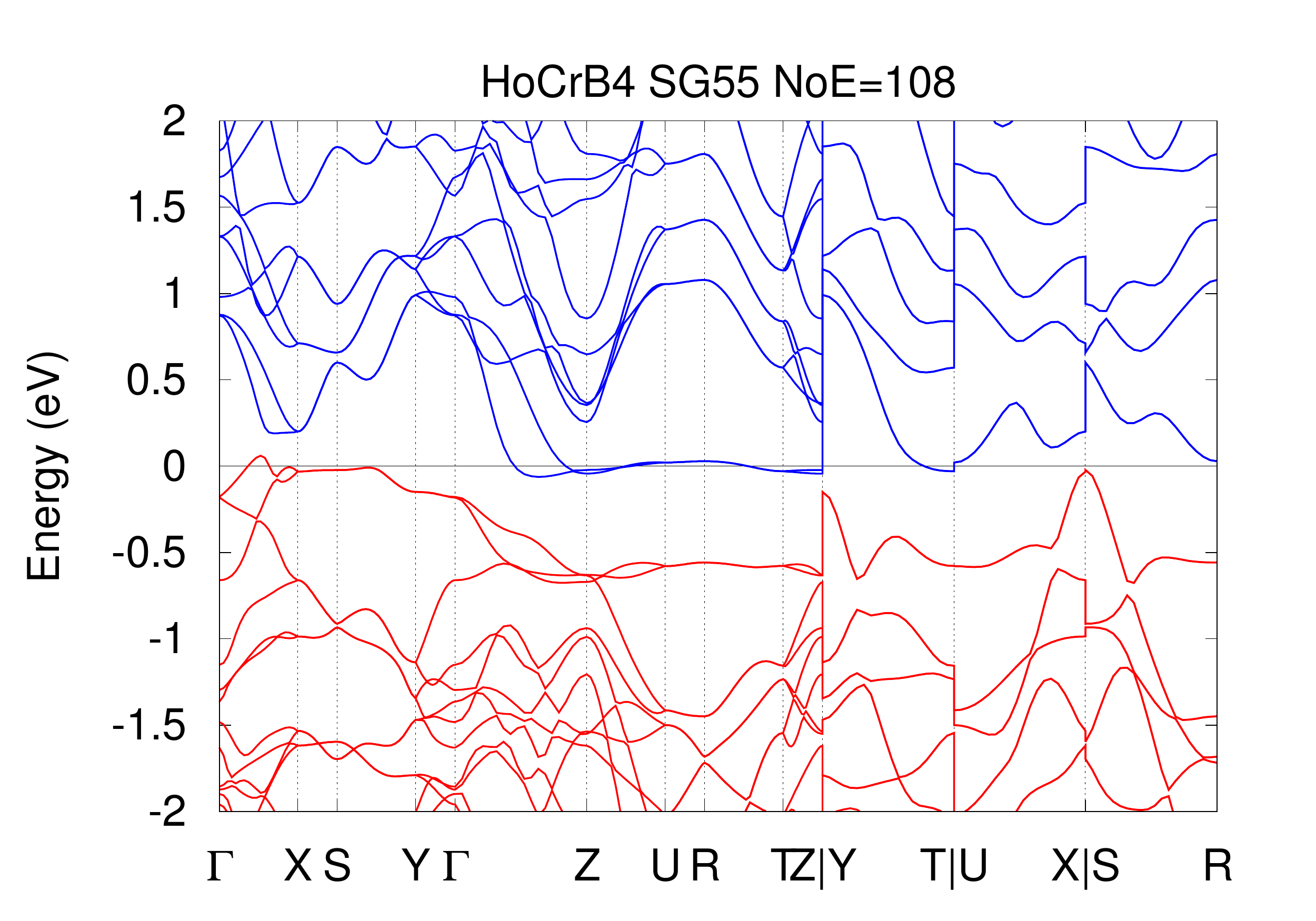}
}
\subfigure[TmPS SG62 NoA=24 NoE=160]{
\label{subfig:648073}
\includegraphics[scale=0.32,angle=270]{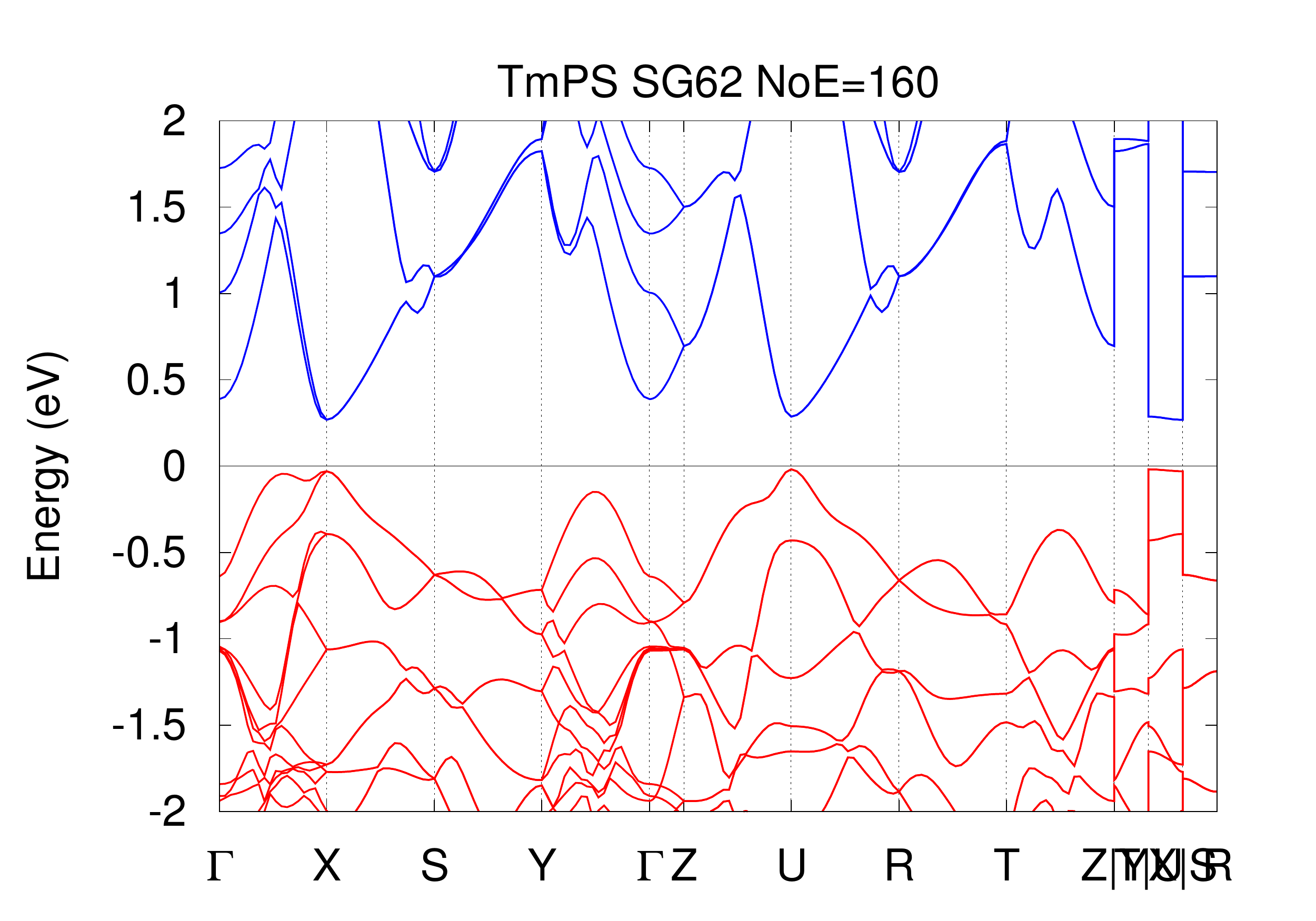}
}
\caption{\hyperref[tab:electride]{back to the table}}
\end{figure}

\begin{figure}[htp]
 \centering
\subfigure[KTl SG64 NoA=24 NoE=144]{
\label{subfig:262063}
\includegraphics[scale=0.32,angle=270]{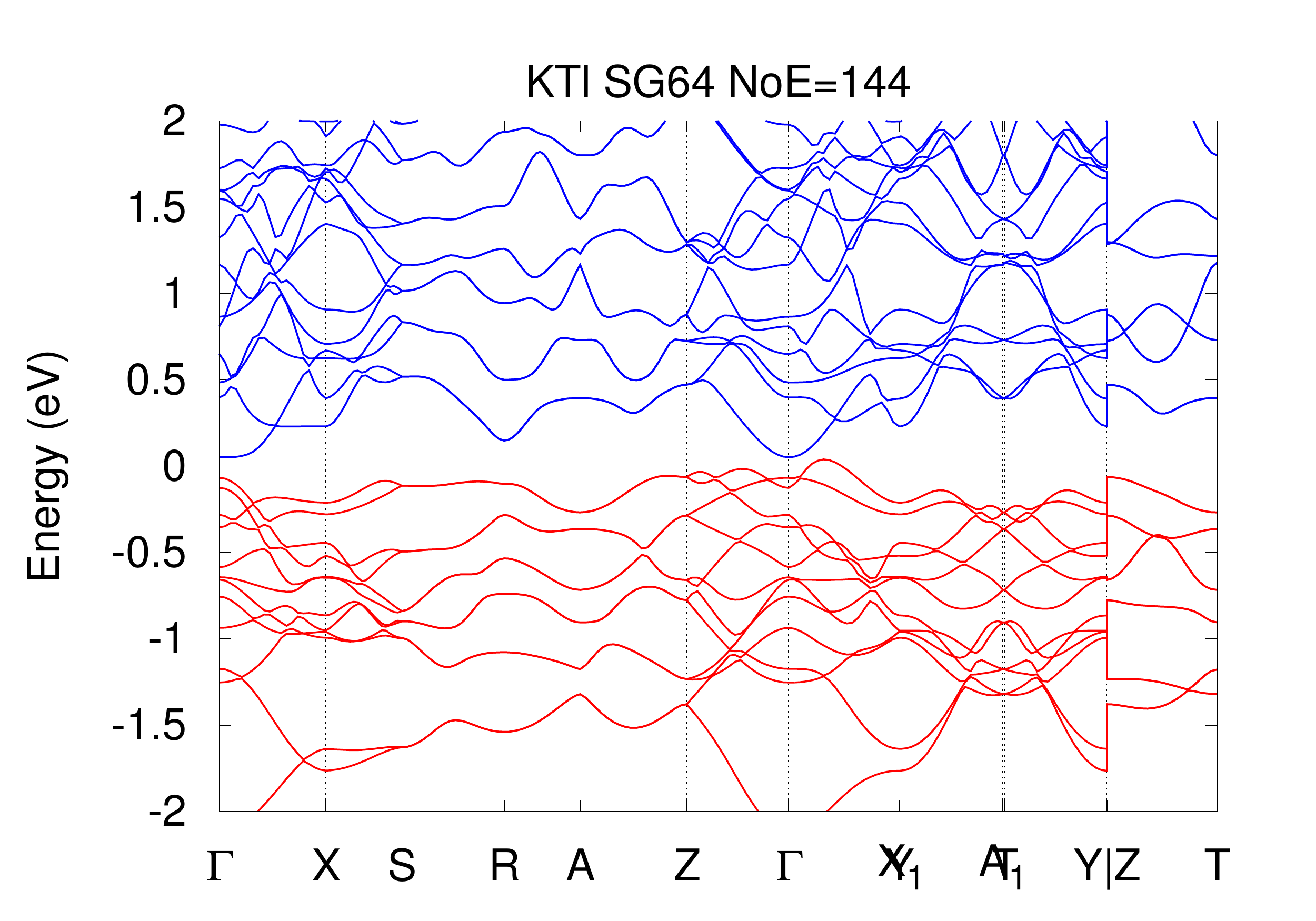}
}
\subfigure[FeS SG14 NoA=24 NoE=168]{
\label{subfig:89380}
\includegraphics[scale=0.32,angle=270]{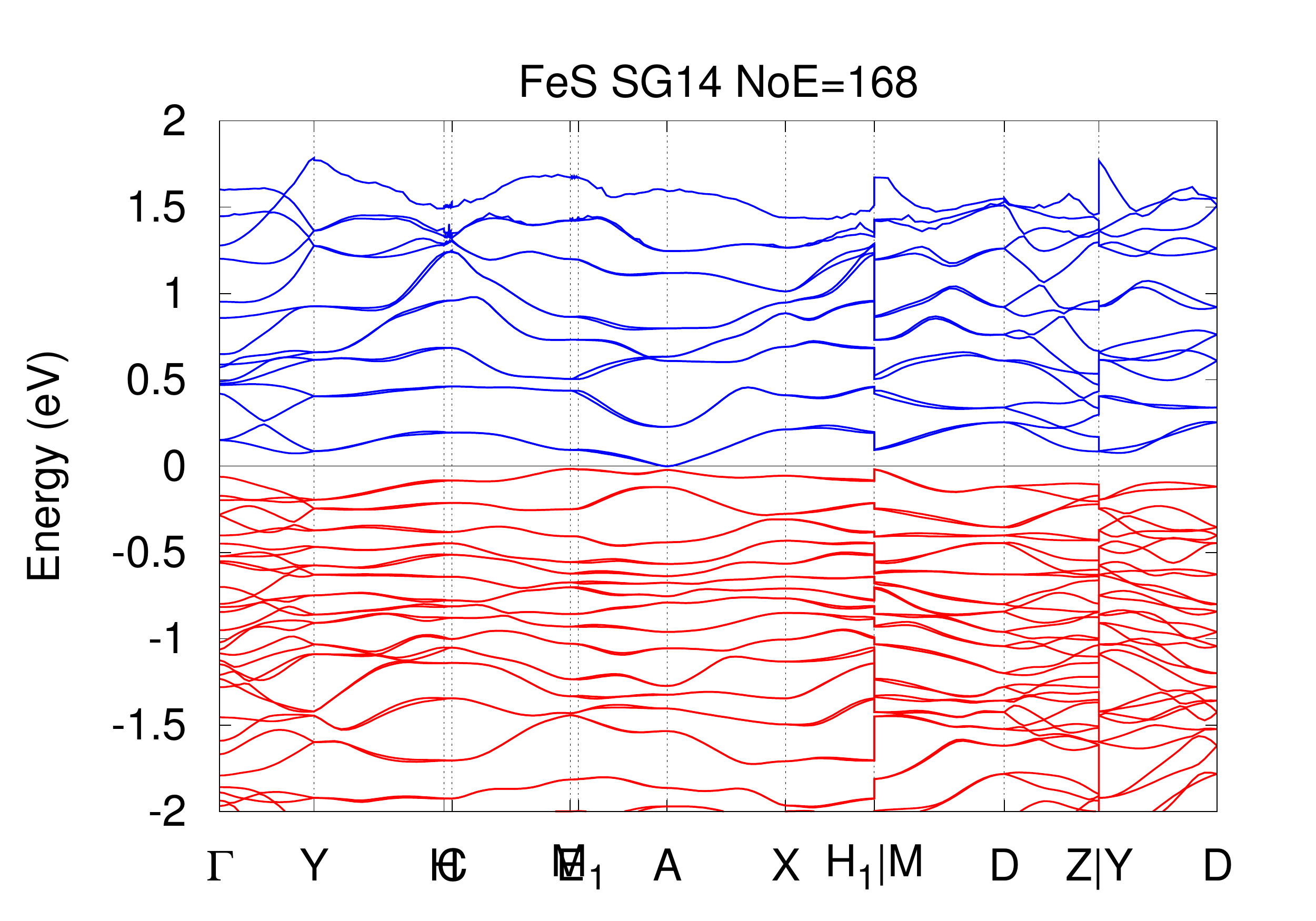}
}
\subfigure[CoGeTe SG61 NoA=24 NoE=152]{
\label{subfig:419780}
\includegraphics[scale=0.32,angle=270]{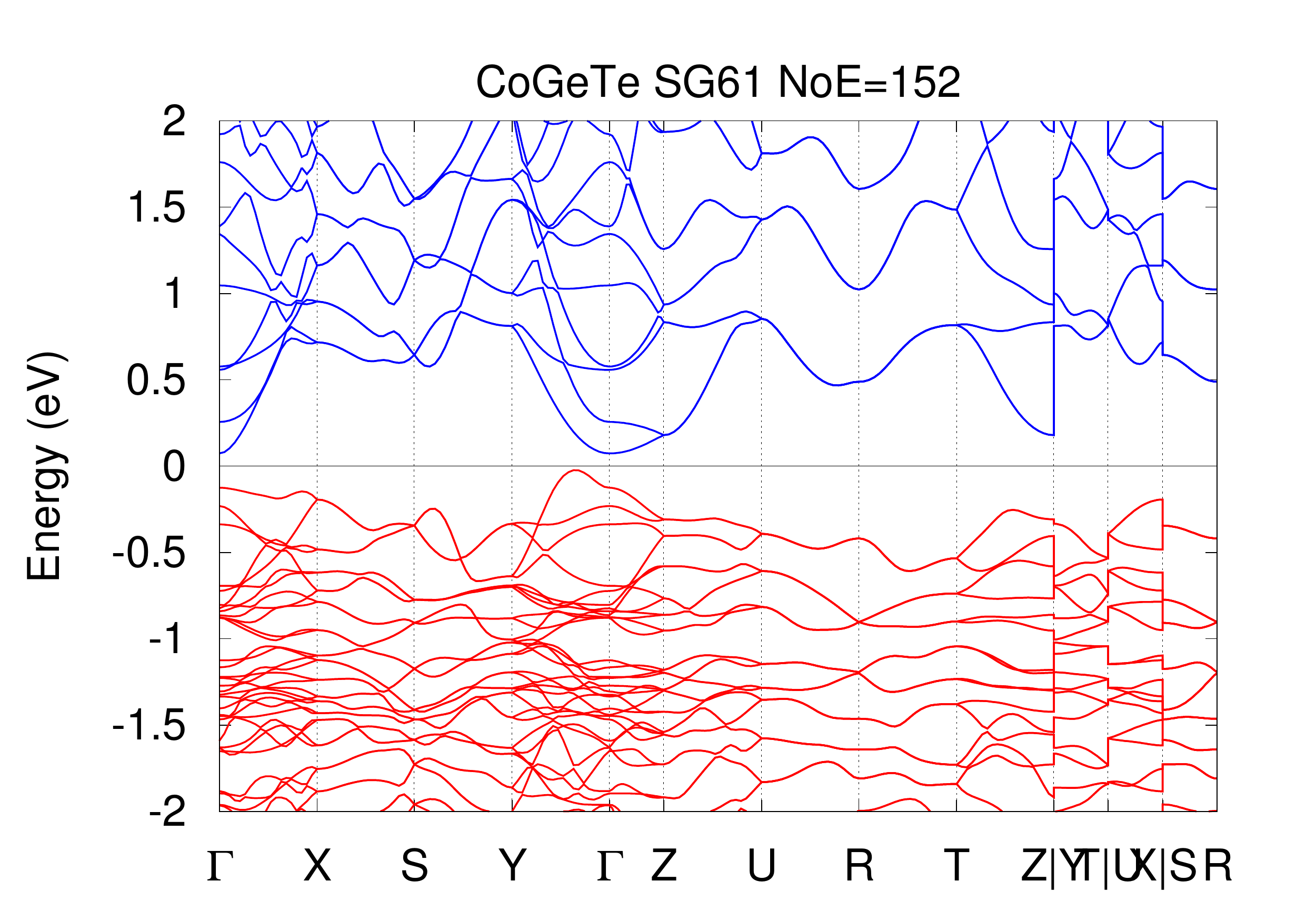}
}
\subfigure[TbPS SG62 NoA=24 NoE=160]{
\label{subfig:648063}
\includegraphics[scale=0.32,angle=270]{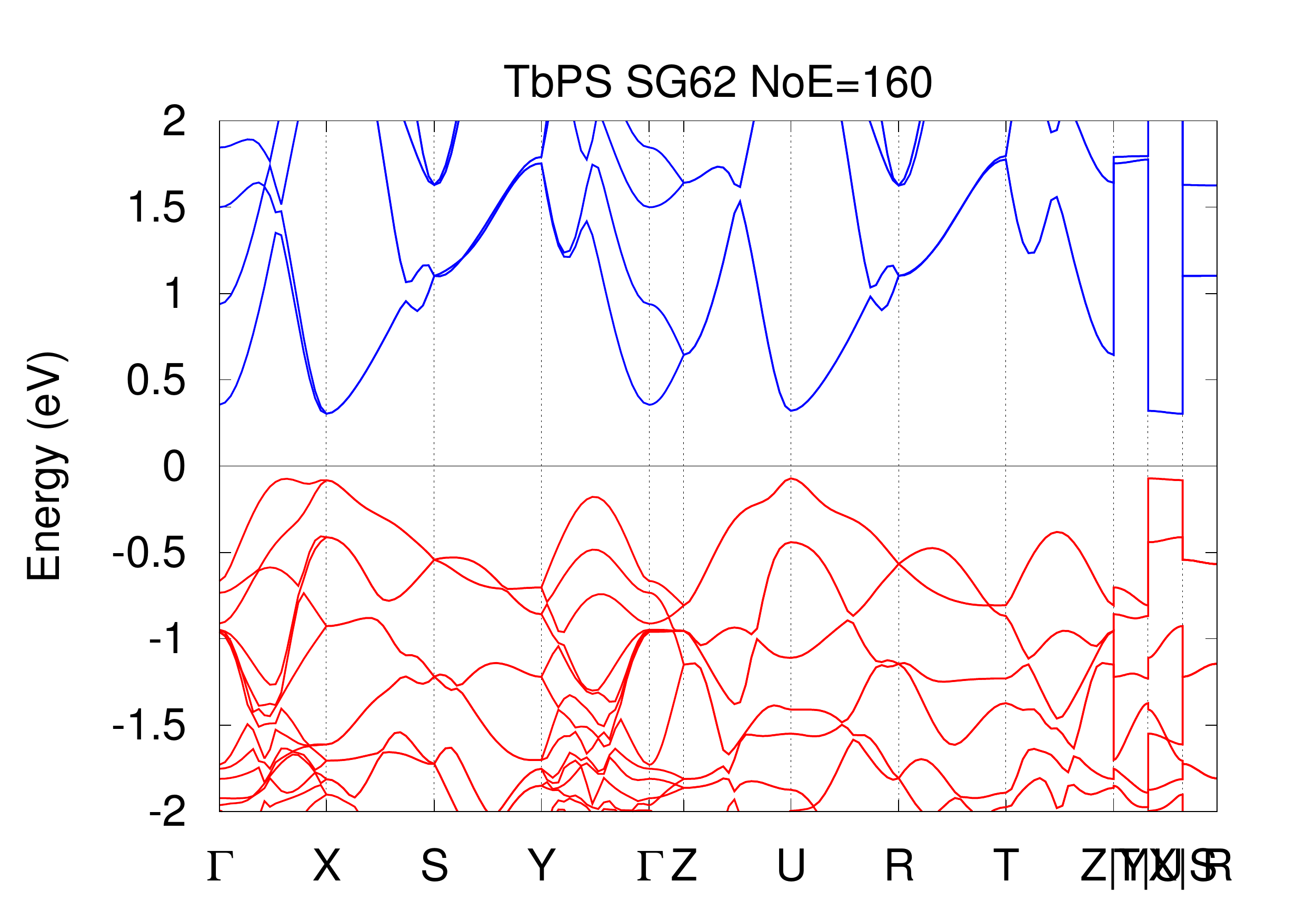}
}
\subfigure[PrPS SG62 NoA=24 NoE=176]{
\label{subfig:647960}
\includegraphics[scale=0.32,angle=270]{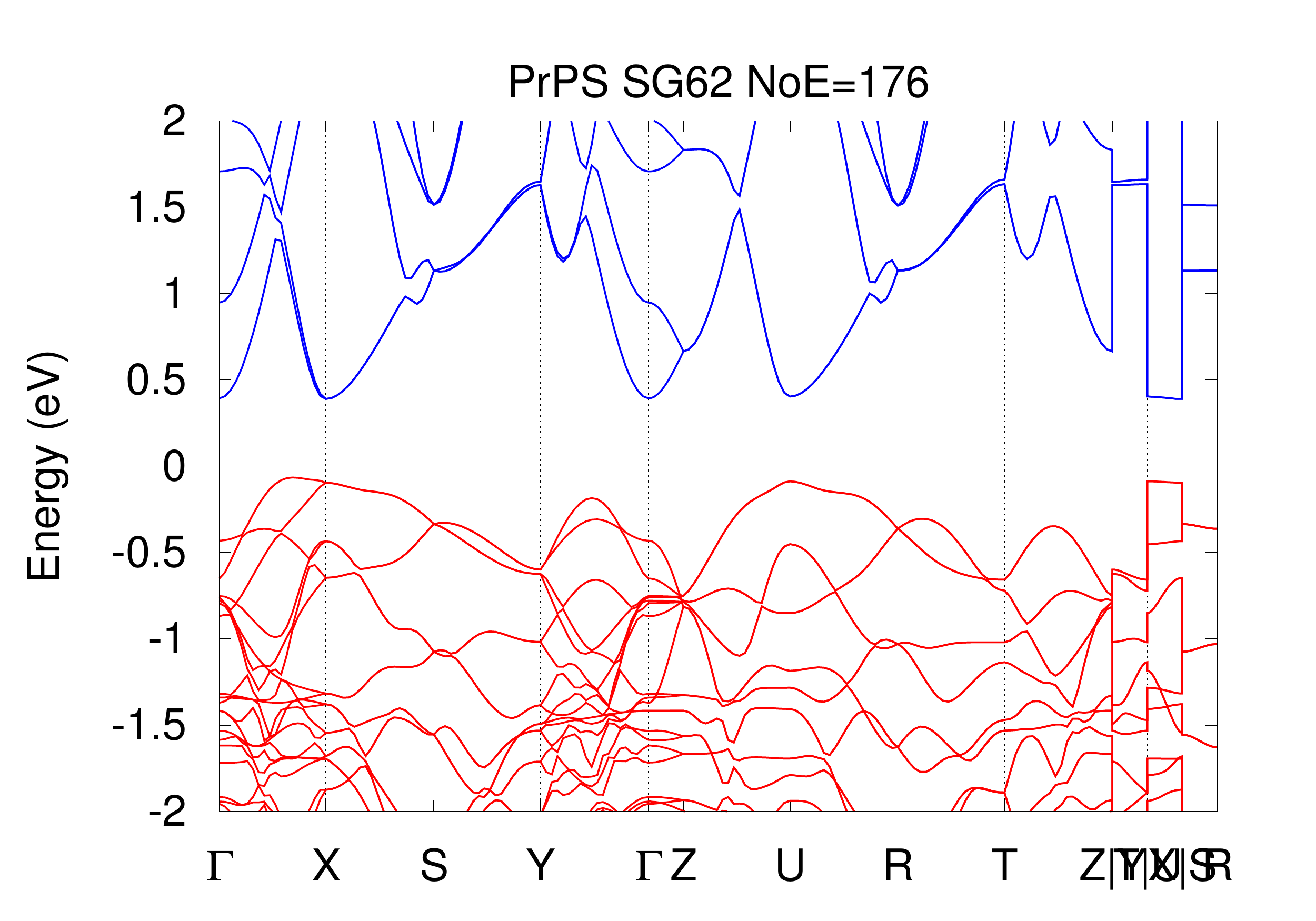}
}
\subfigure[ErPS SG62 NoA=24 NoE=160]{
\label{subfig:630913}
\includegraphics[scale=0.32,angle=270]{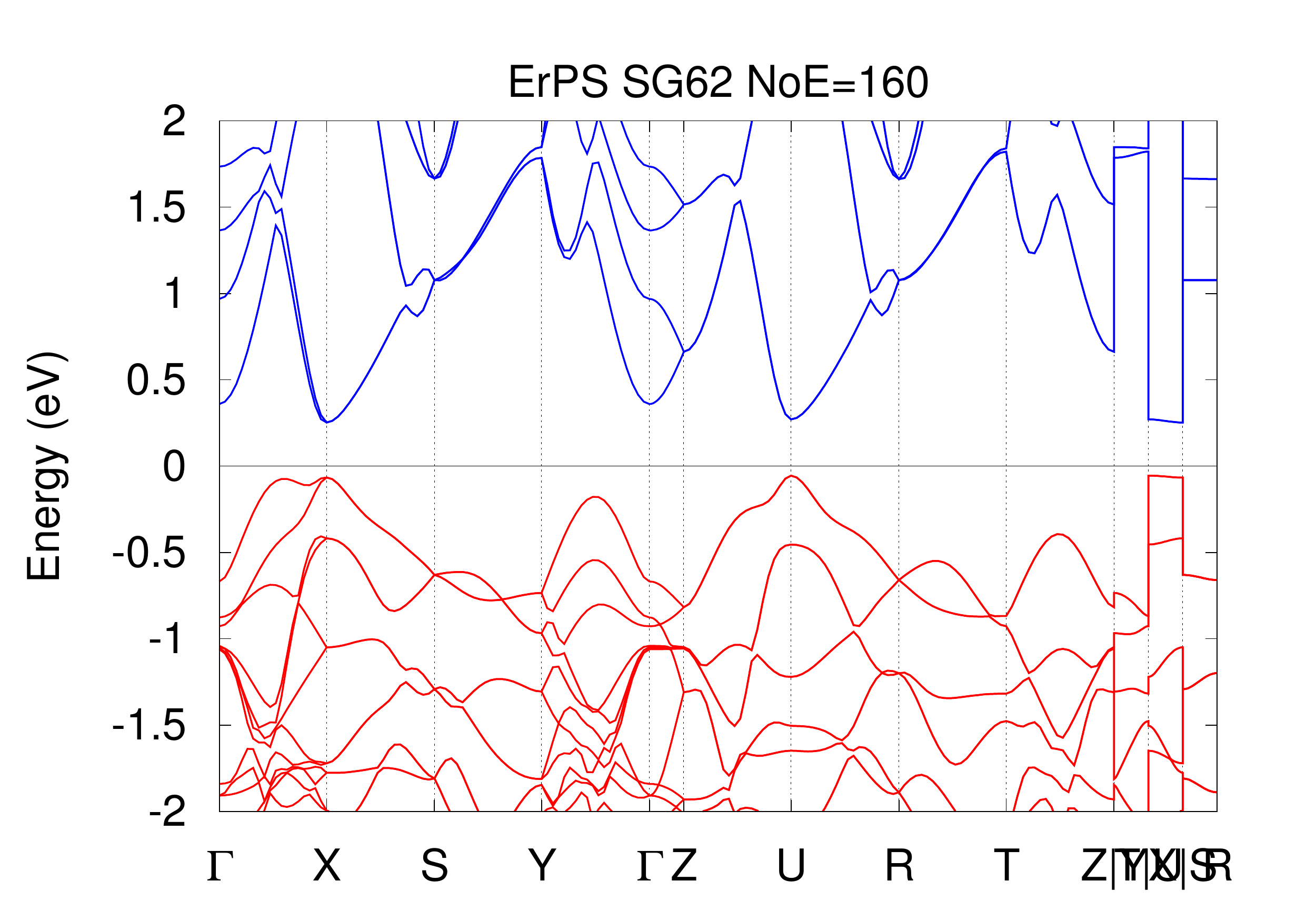}
}
\subfigure[DyCrB$_{4}$ SG55 NoA=24 NoE=108]{
\label{subfig:658658}
\includegraphics[scale=0.32,angle=270]{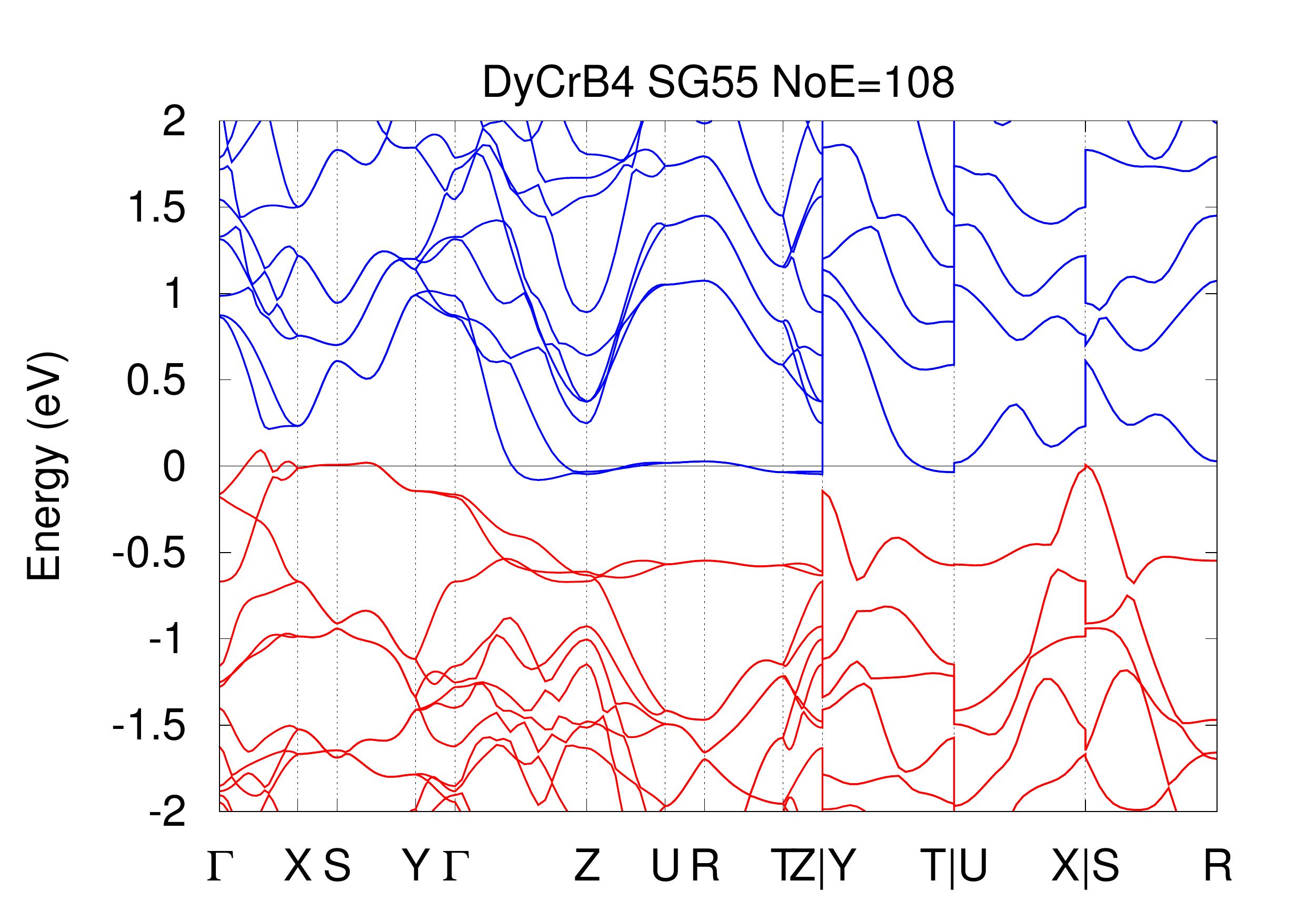}
}
\subfigure[TiAs$_{2}$ SG58 NoA=24 NoE=112]{
\label{subfig:611500}
\includegraphics[scale=0.32,angle=270]{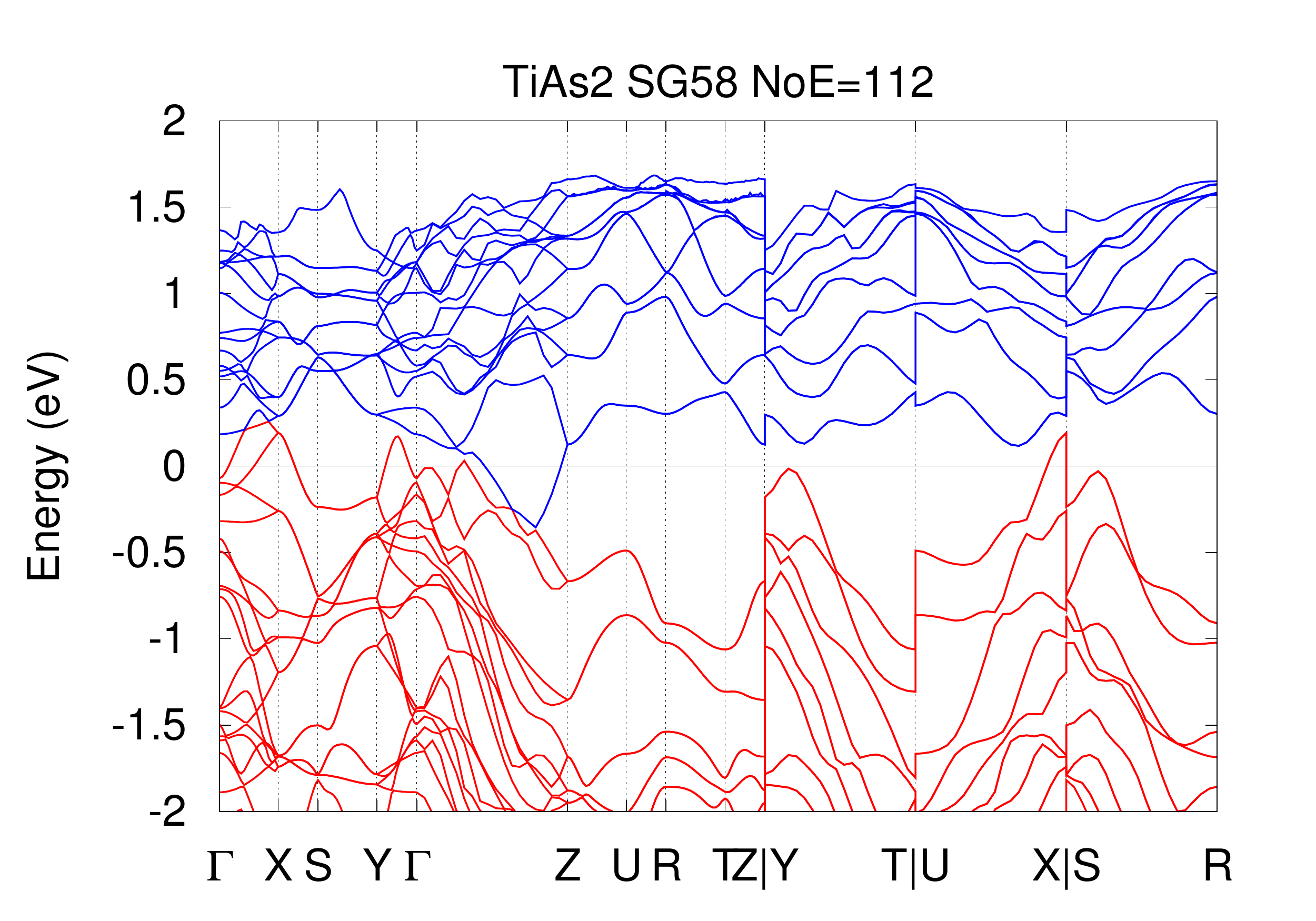}
}
\caption{\hyperref[tab:electride]{back to the table}}
\end{figure}

\begin{figure}[htp]
 \centering
\subfigure[YCrB$_{4}$ SG55 NoA=24 NoE=116]{
\label{subfig:16171}
\includegraphics[scale=0.32,angle=270]{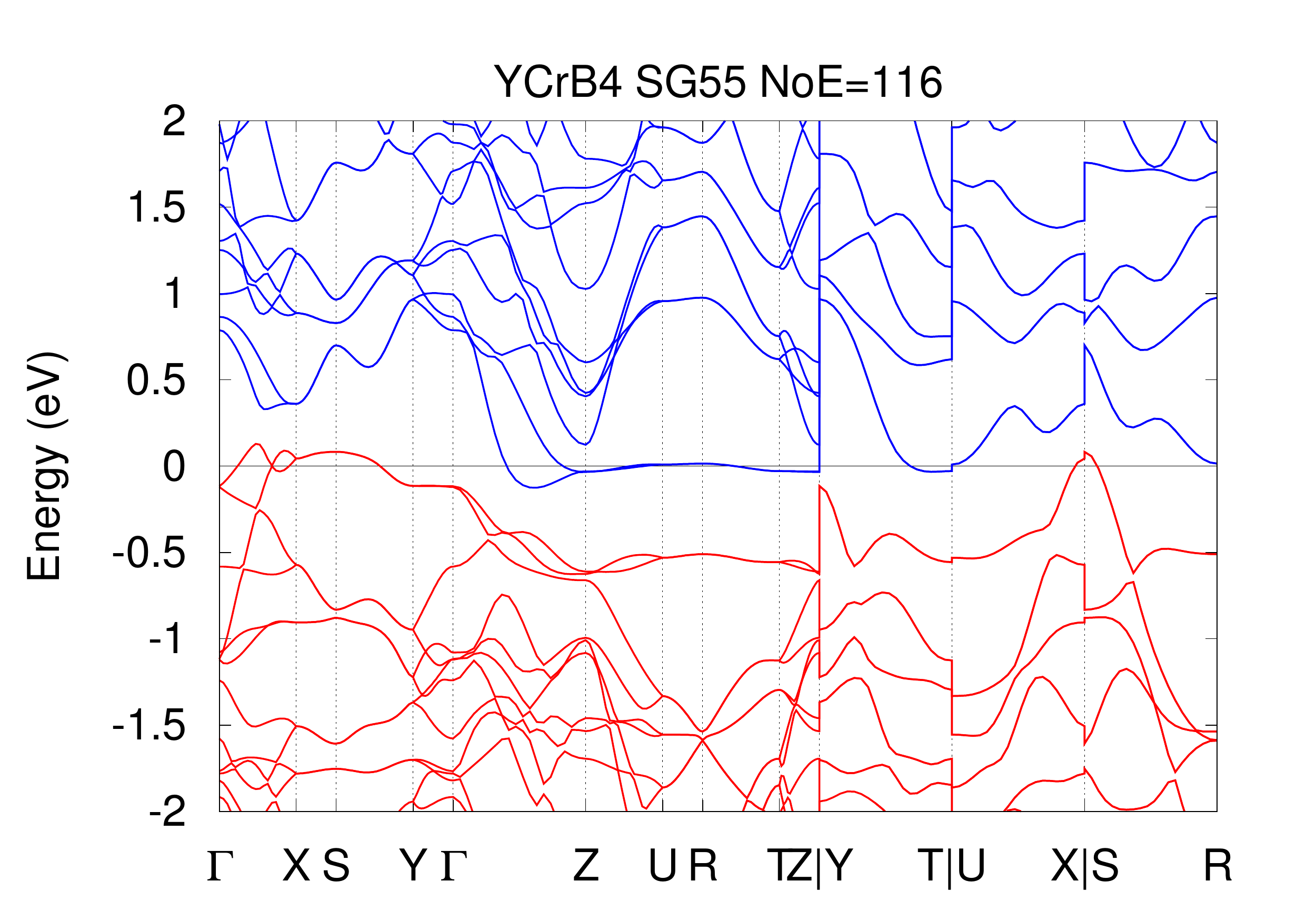}
}
\subfigure[YbNiB$_{4}$ SG55 NoA=24 NoE=120]{
\label{subfig:409820}
\includegraphics[scale=0.32,angle=270]{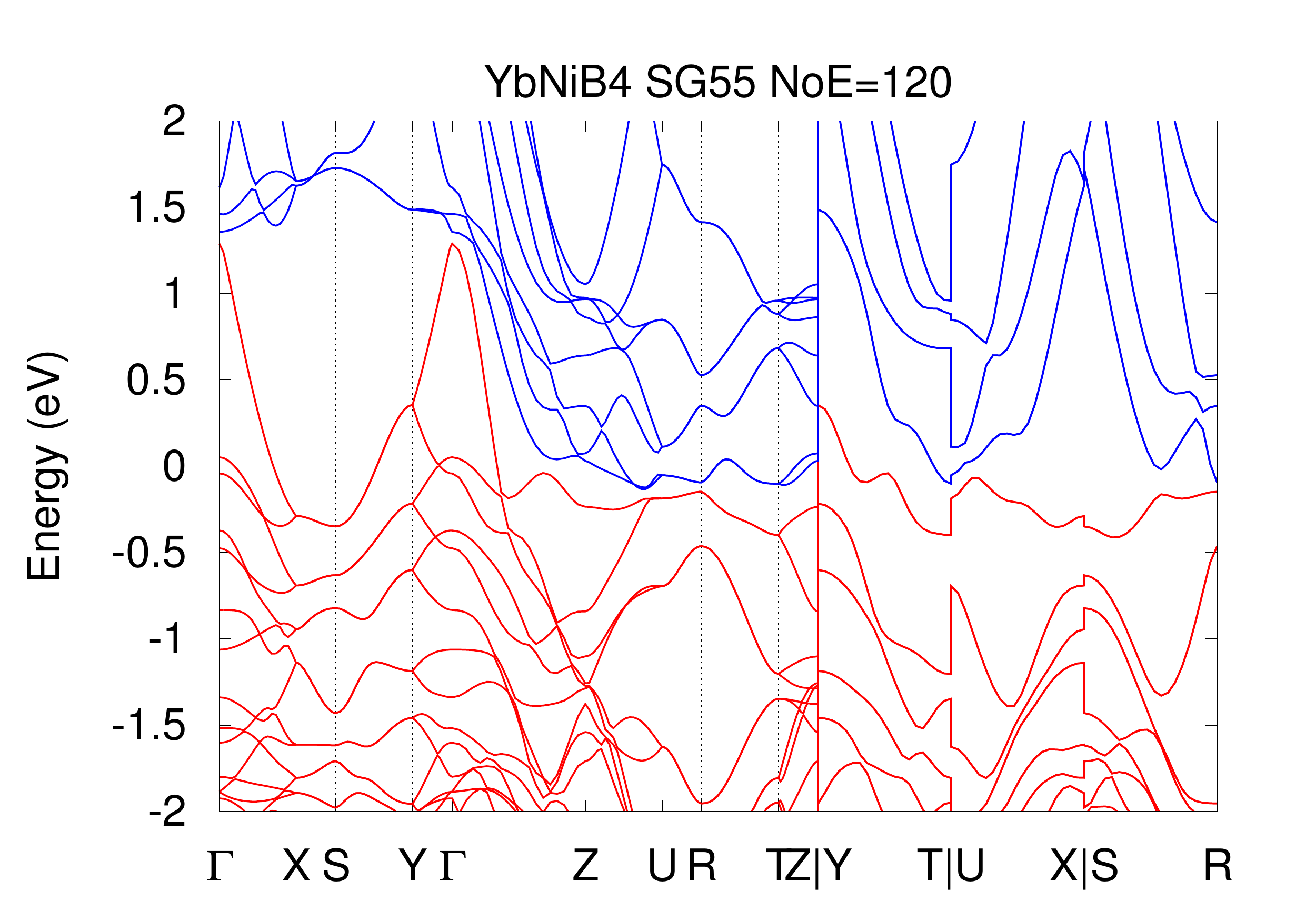}
}
\subfigure[LaPS SG62 NoA=24 NoE=176]{
\label{subfig:641637}
\includegraphics[scale=0.32,angle=270]{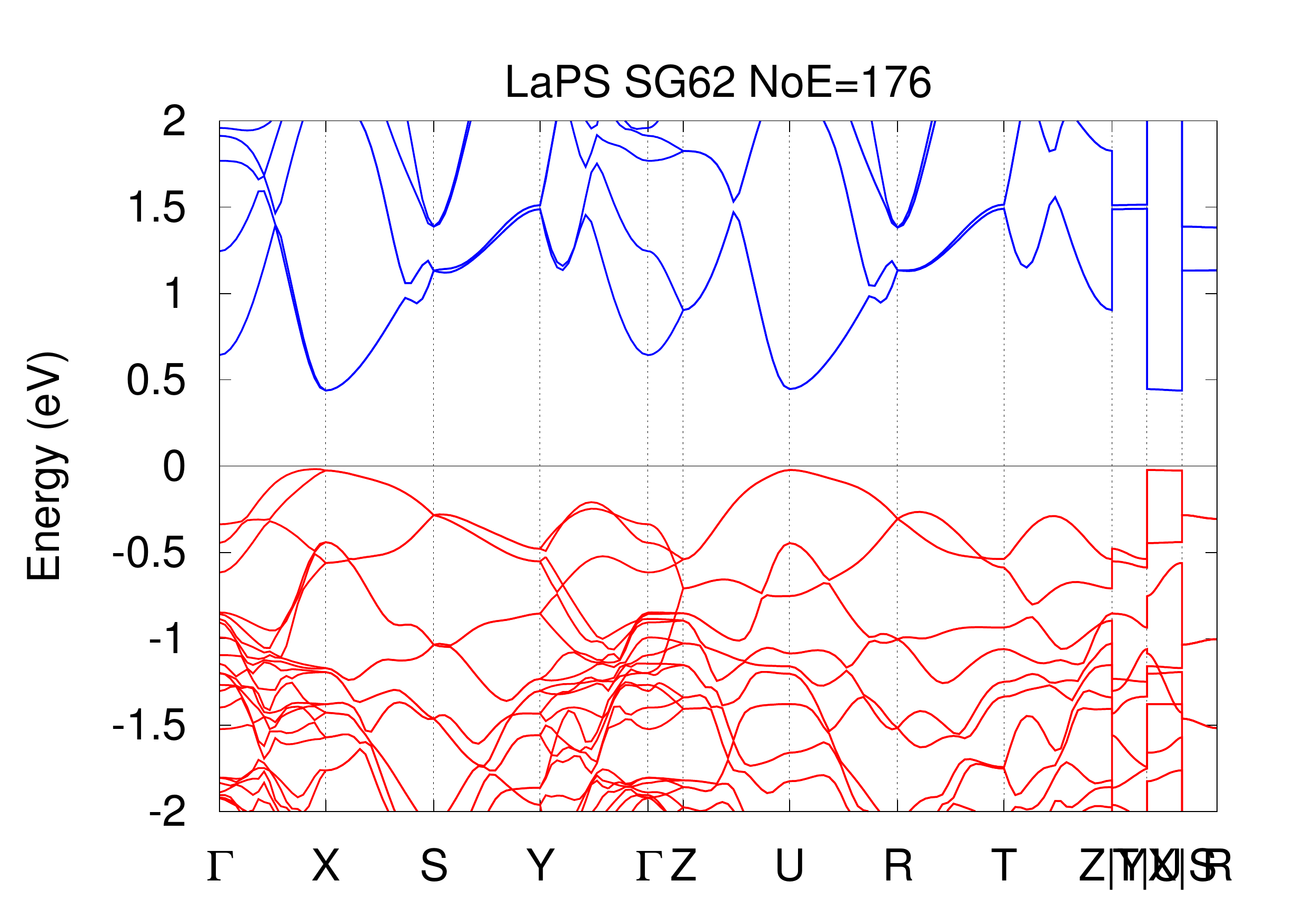}
}
\subfigure[LiGeTe$_{2}$ SG2 NoA=24 NoE=102]{
\label{subfig:35676}
\includegraphics[scale=0.32,angle=270]{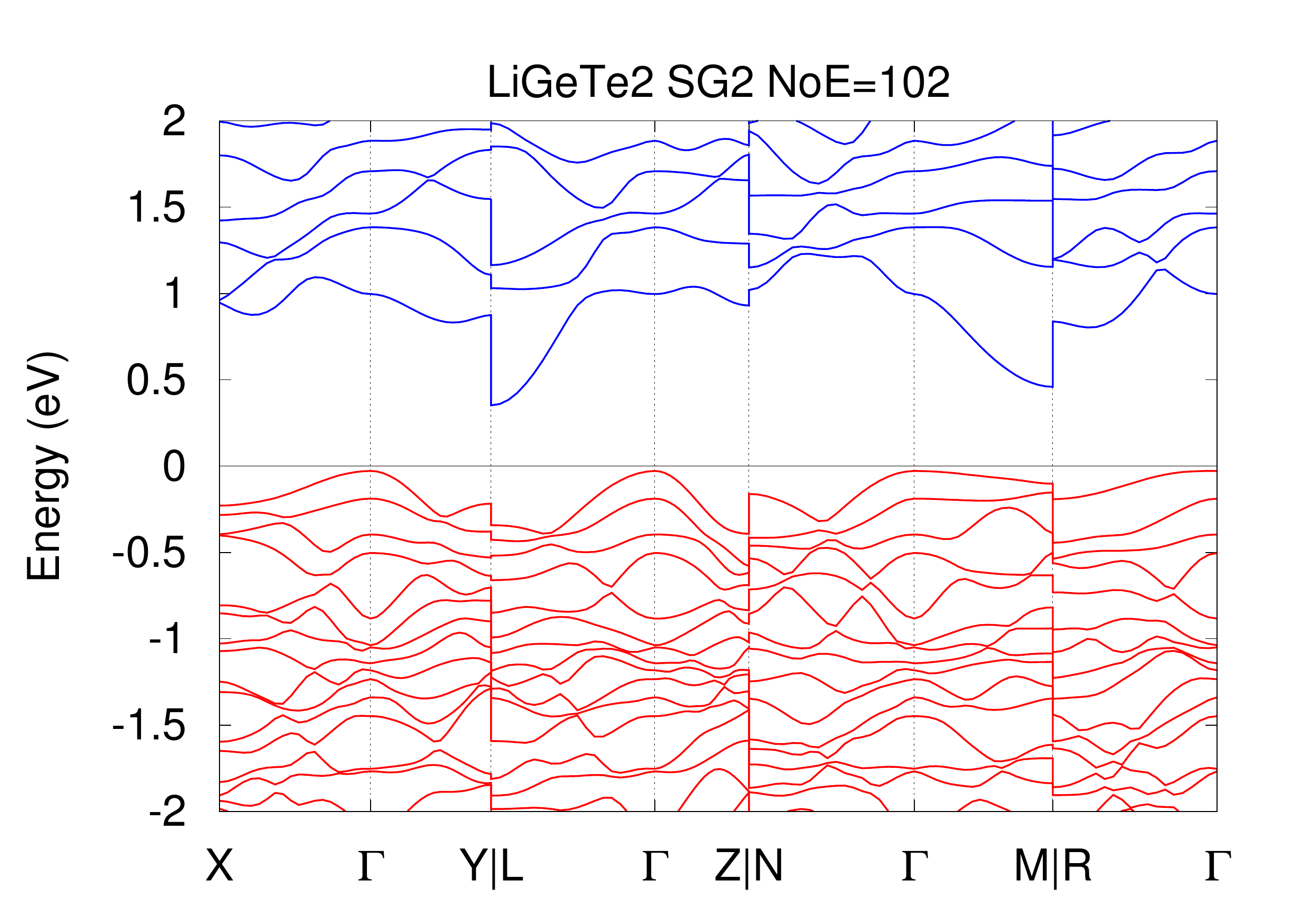}
}
\subfigure[GeTeRh SG61 NoA=24 NoE=152]{
\label{subfig:260373}
\includegraphics[scale=0.32,angle=270]{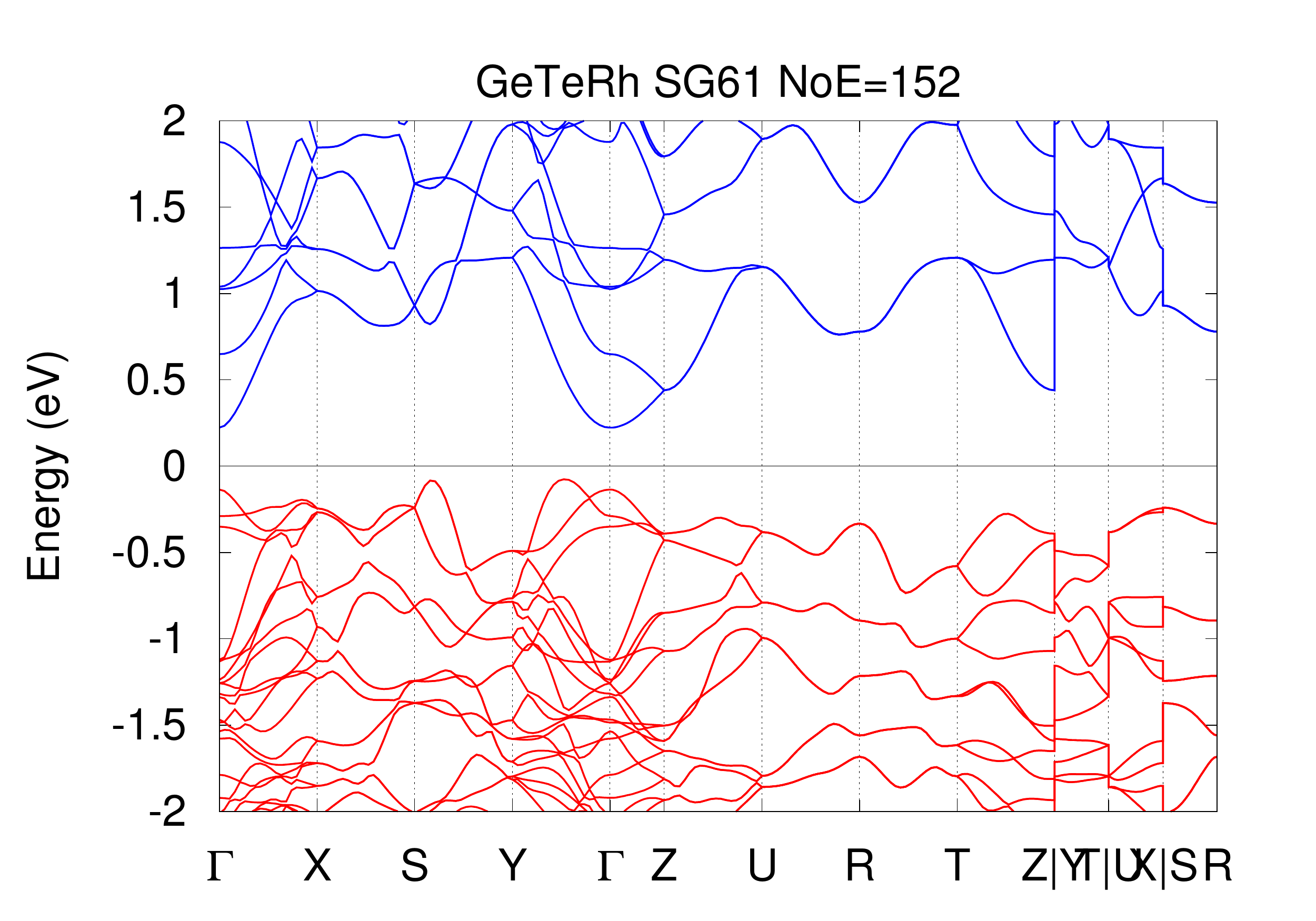}
}
\subfigure[Ba$_{3}$Sb$_{2}$O SG55 NoA=24 NoE=184]{
\label{subfig:280592}
\includegraphics[scale=0.32,angle=270]{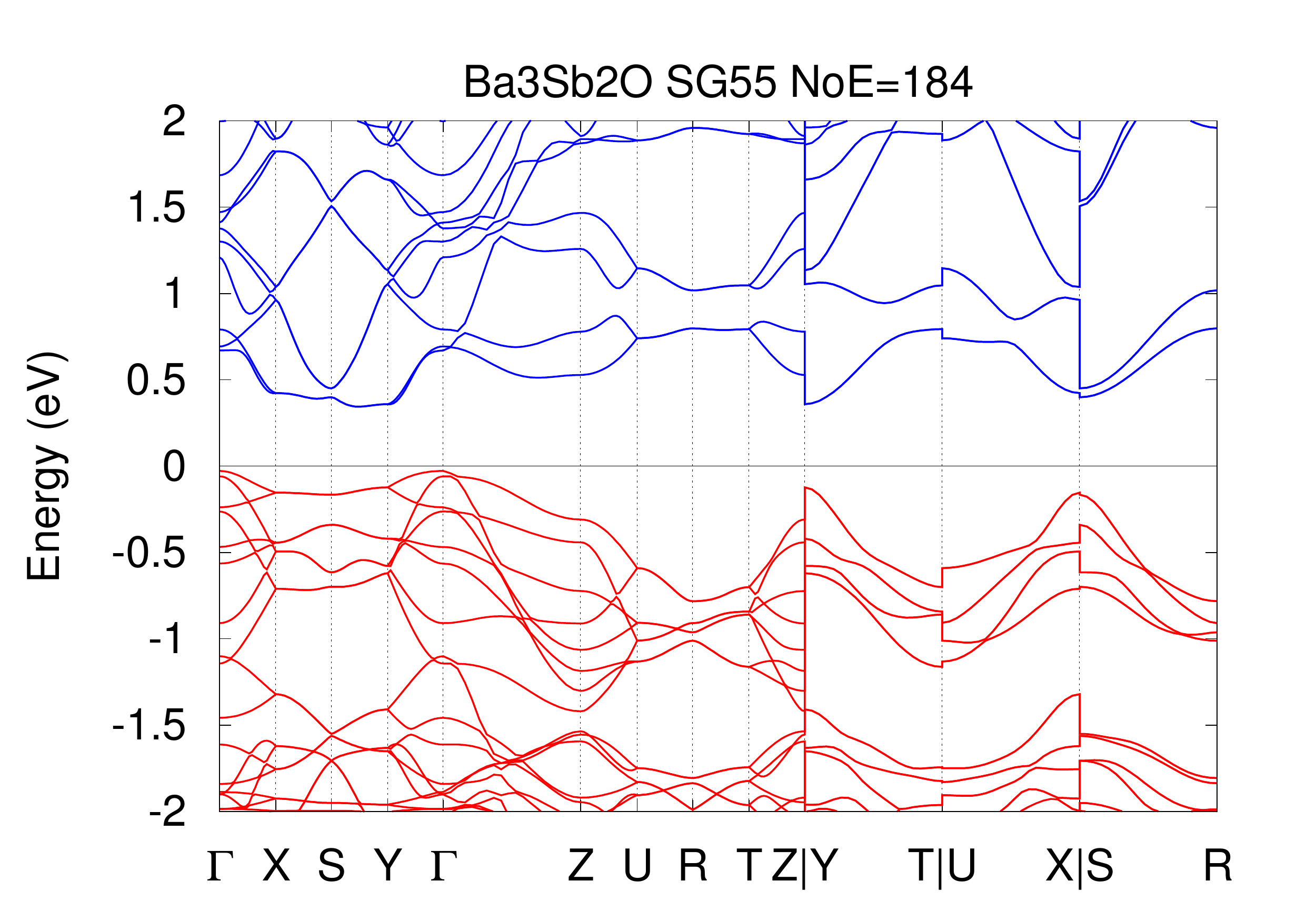}
}
\subfigure[NdPS SG62 NoA=24 NoE=176]{
\label{subfig:645690}
\includegraphics[scale=0.32,angle=270]{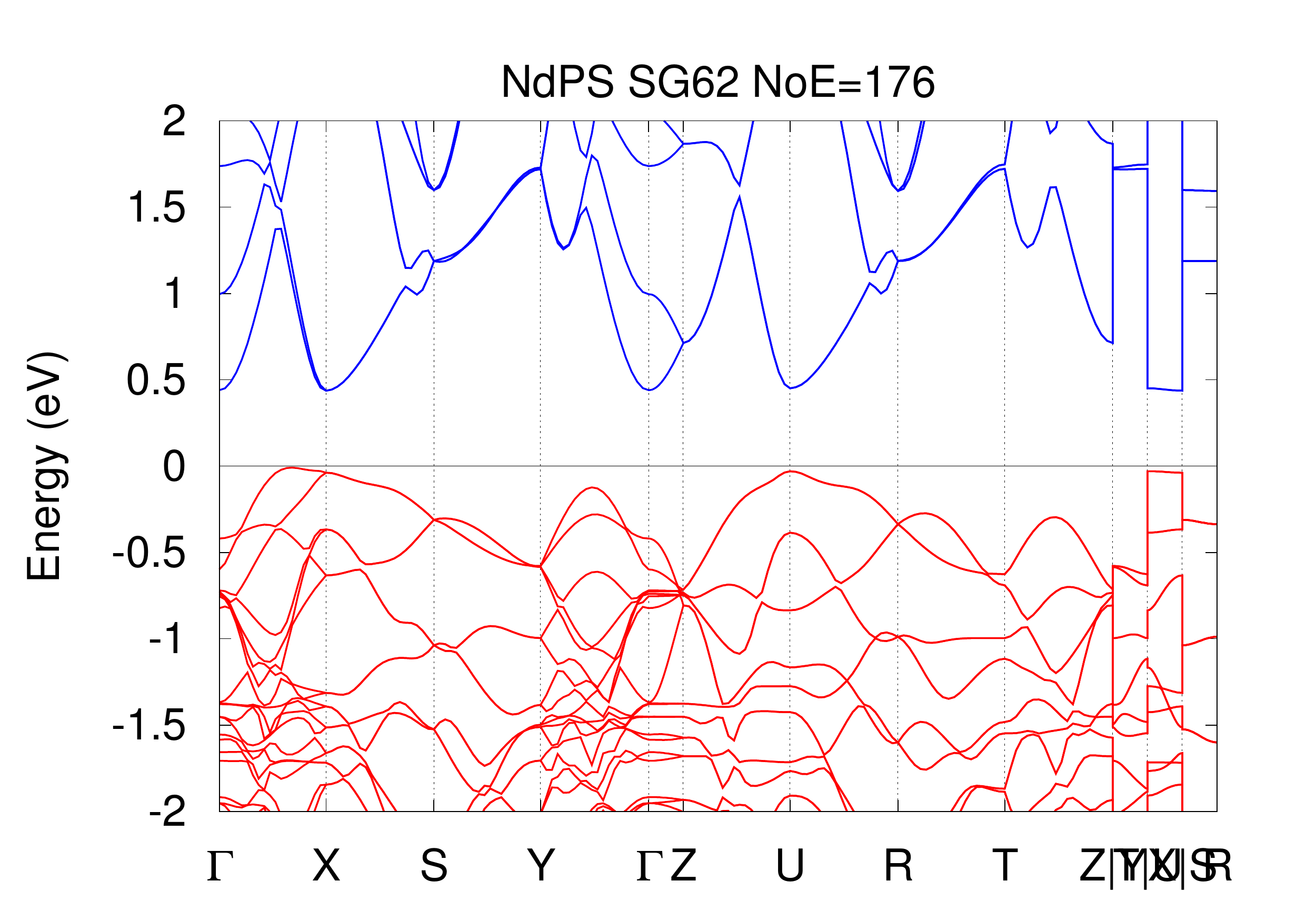}
}
\subfigure[HoPS SG62 NoA=24 NoE=160]{
\label{subfig:639545}
\includegraphics[scale=0.32,angle=270]{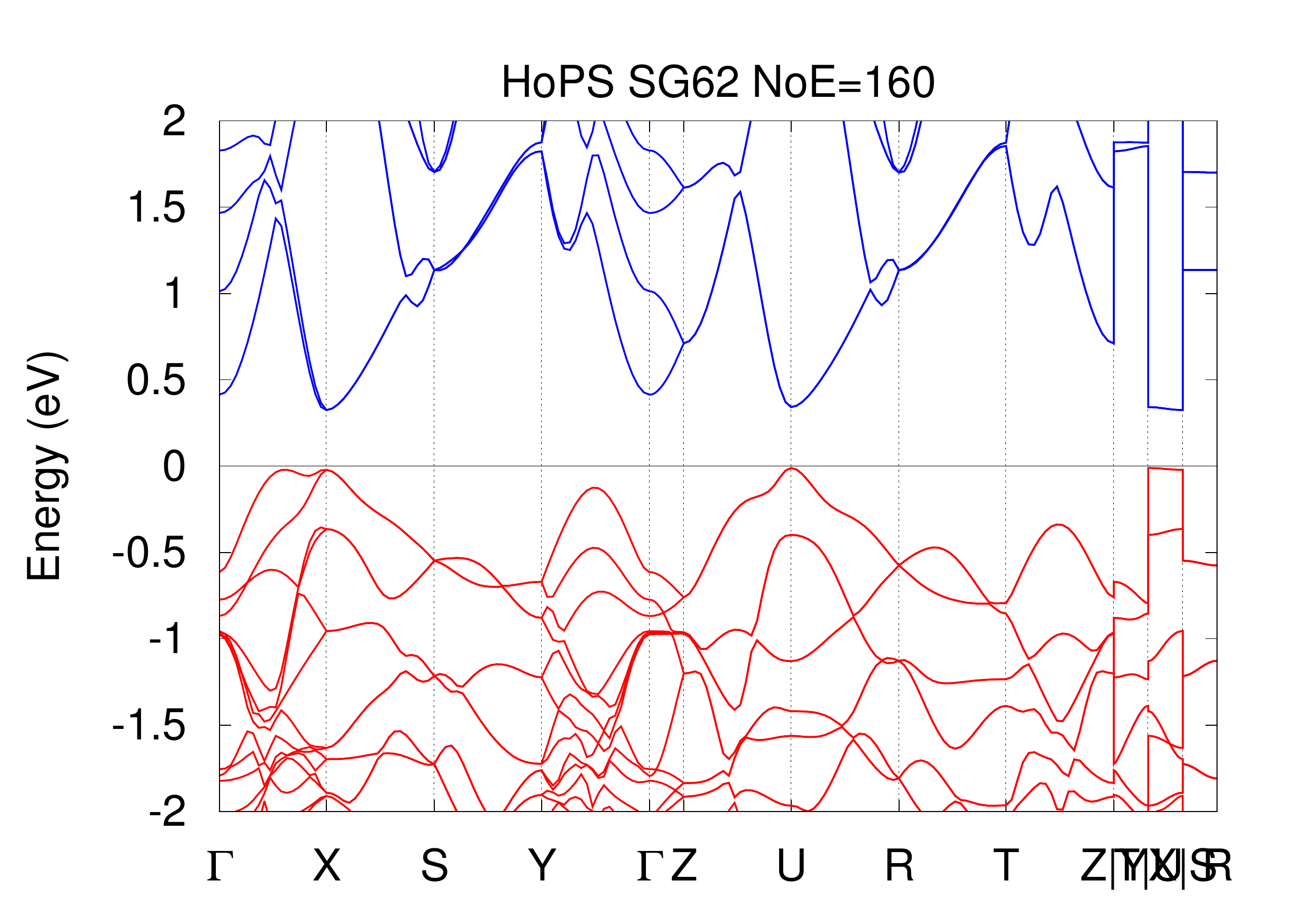}
}
\caption{\hyperref[tab:electride]{back to the table}}
\end{figure}

\begin{figure}[htp]
 \centering
\subfigure[ErCrB$_{4}$ SG55 NoA=24 NoE=108]{
\label{subfig:613495}
\includegraphics[scale=0.32,angle=270]{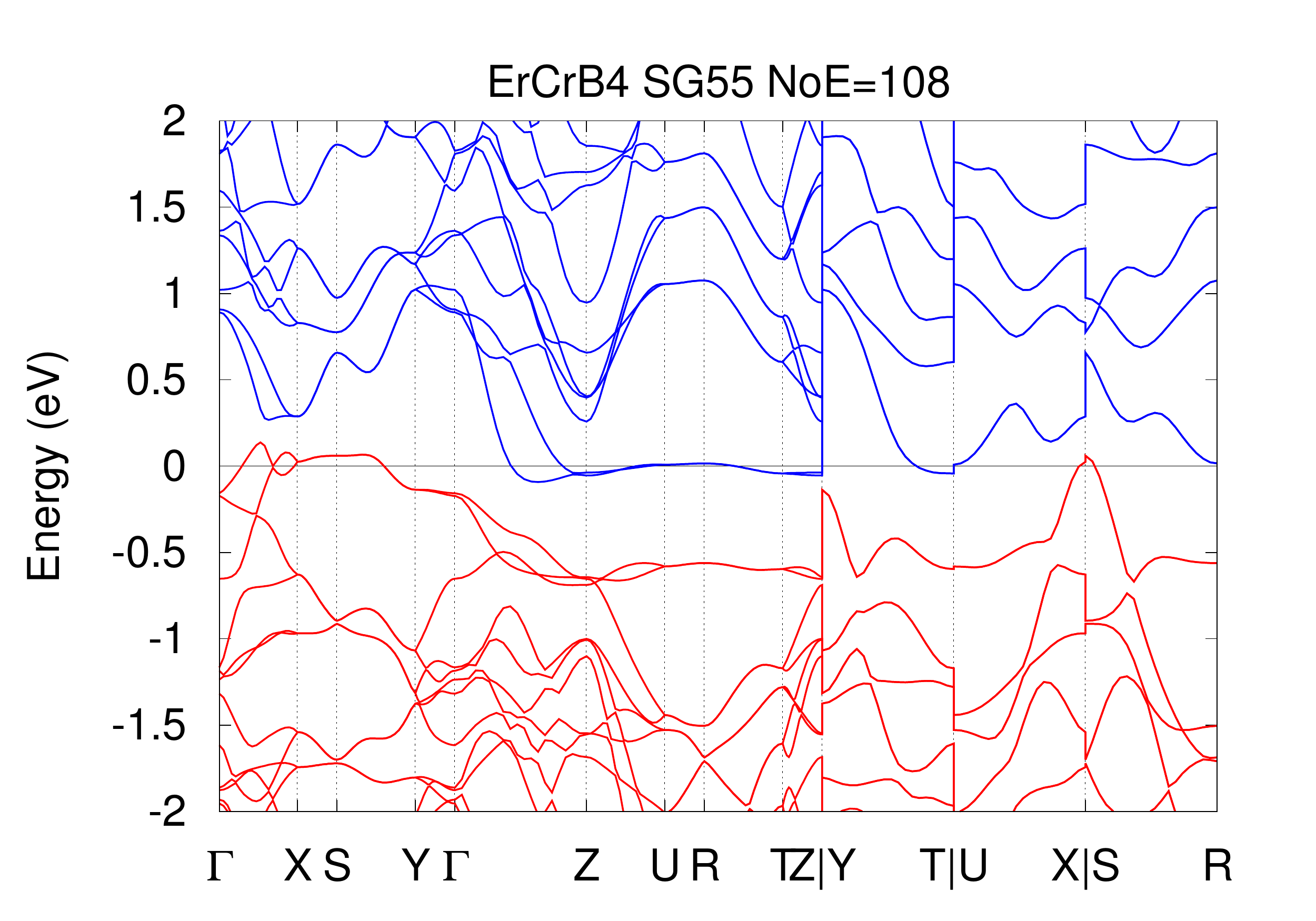}
}
\subfigure[DyPS SG62 NoA=24 NoE=160]{
\label{subfig:630062}
\includegraphics[scale=0.32,angle=270]{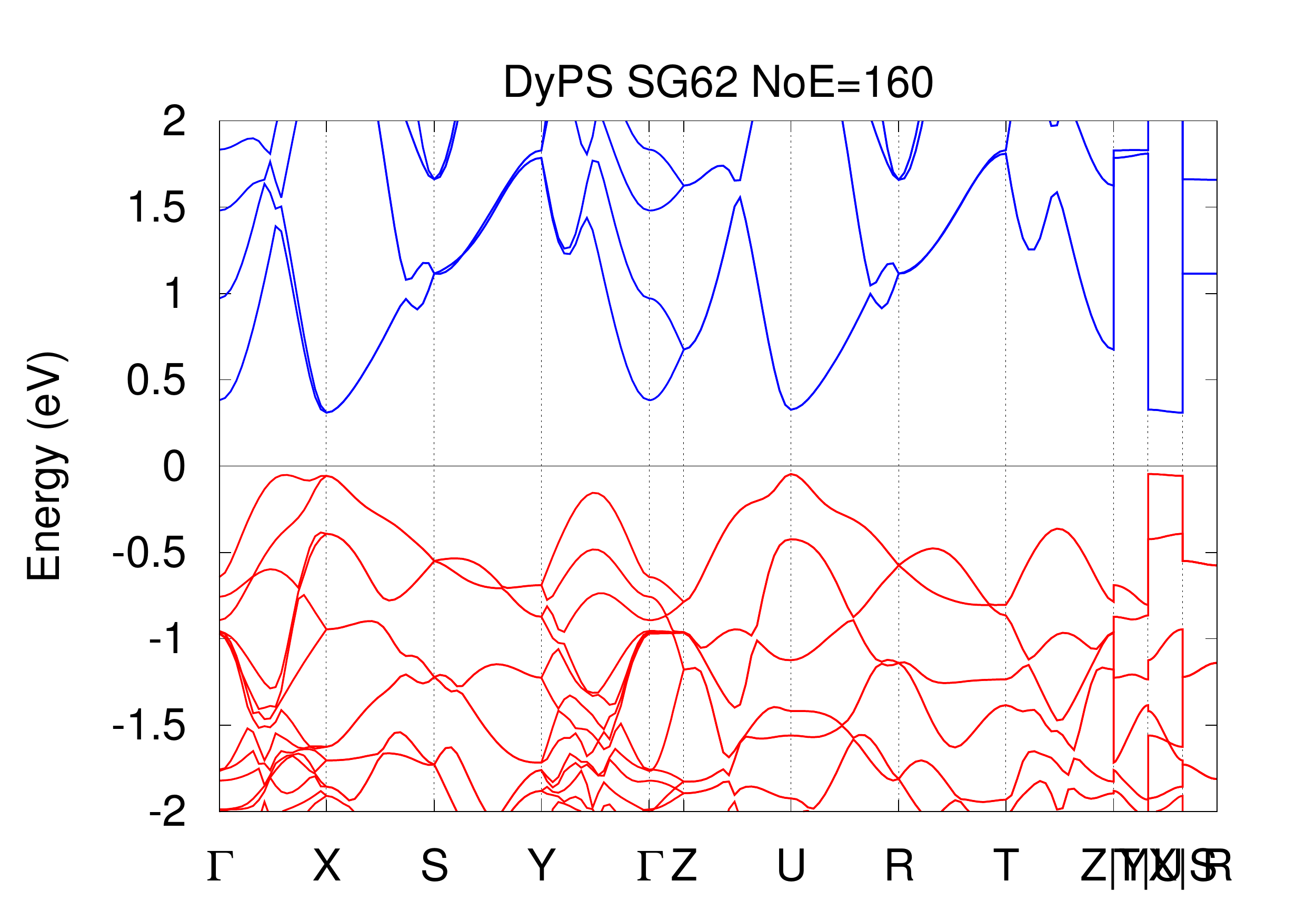}
}
\subfigure[SiSbPt SG61 NoA=24 NoE=152]{
\label{subfig:413194}
\includegraphics[scale=0.32,angle=270]{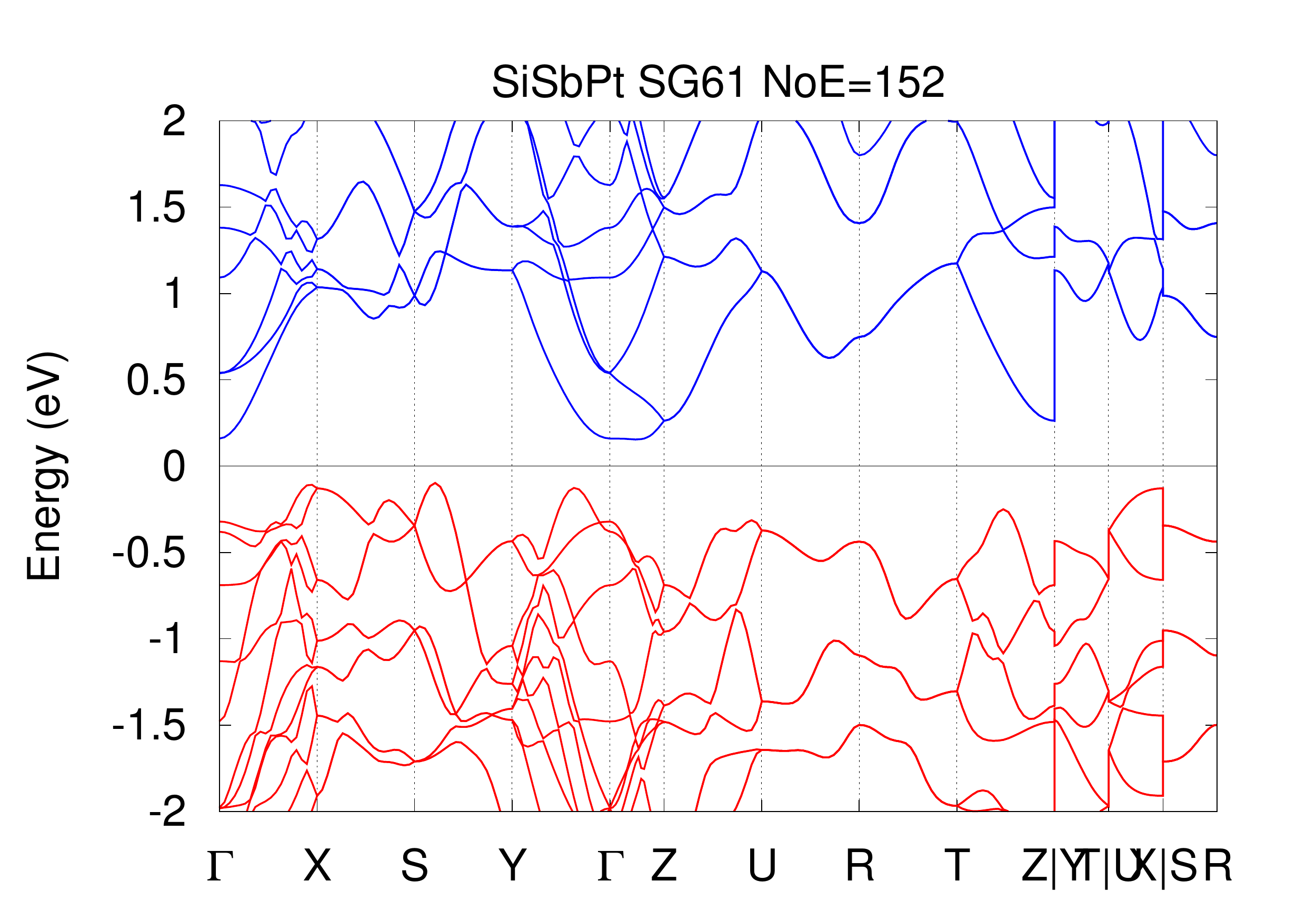}
}
\subfigure[Ba$_{2}$LiGe$_{3}$ SG70 NoA=24 NoE=132]{
\label{subfig:404705}
\includegraphics[scale=0.32,angle=270]{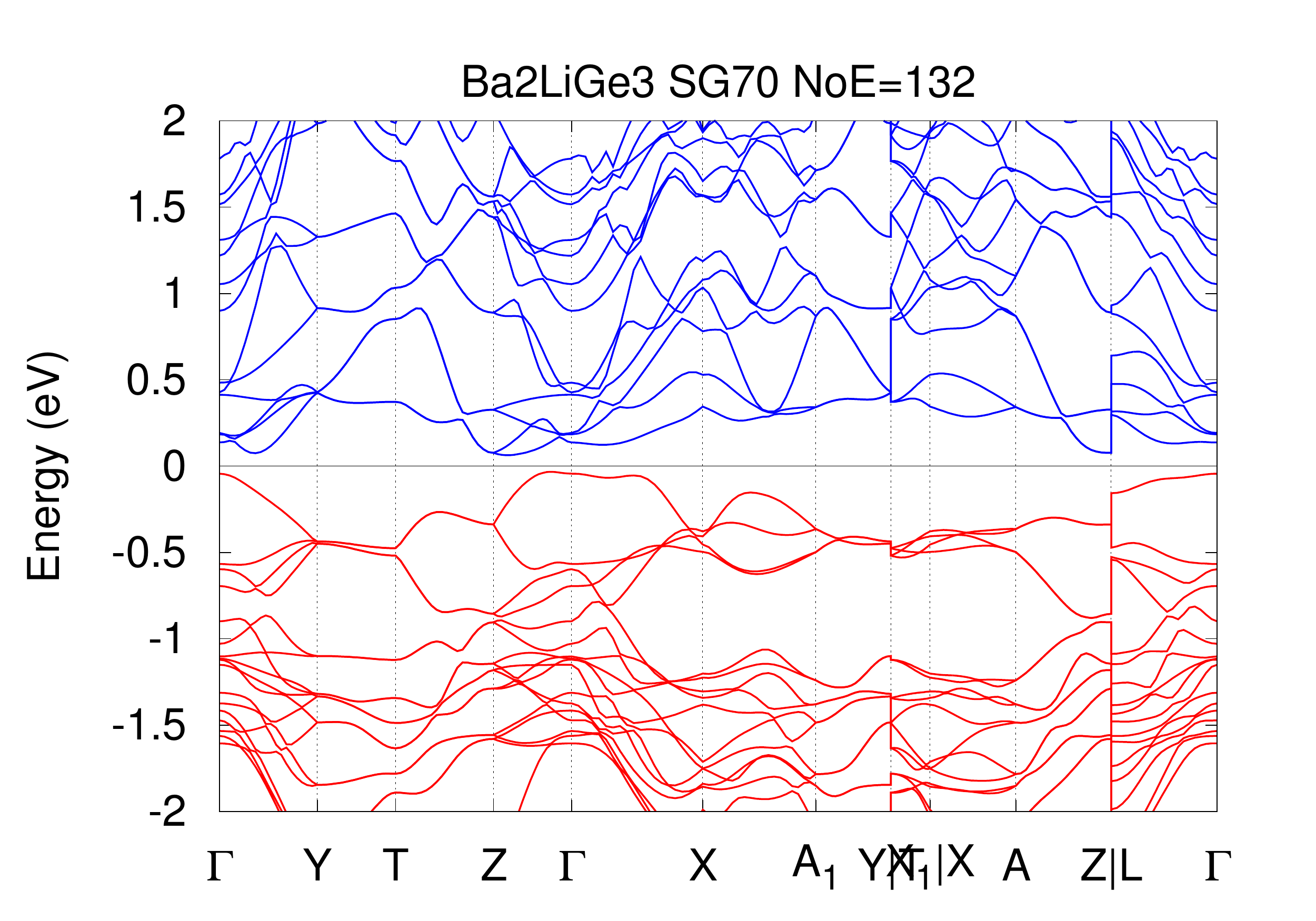}
}
\subfigure[YPS SG62 NoA=24 NoE=176]{
\label{subfig:648080}
\includegraphics[scale=0.32,angle=270]{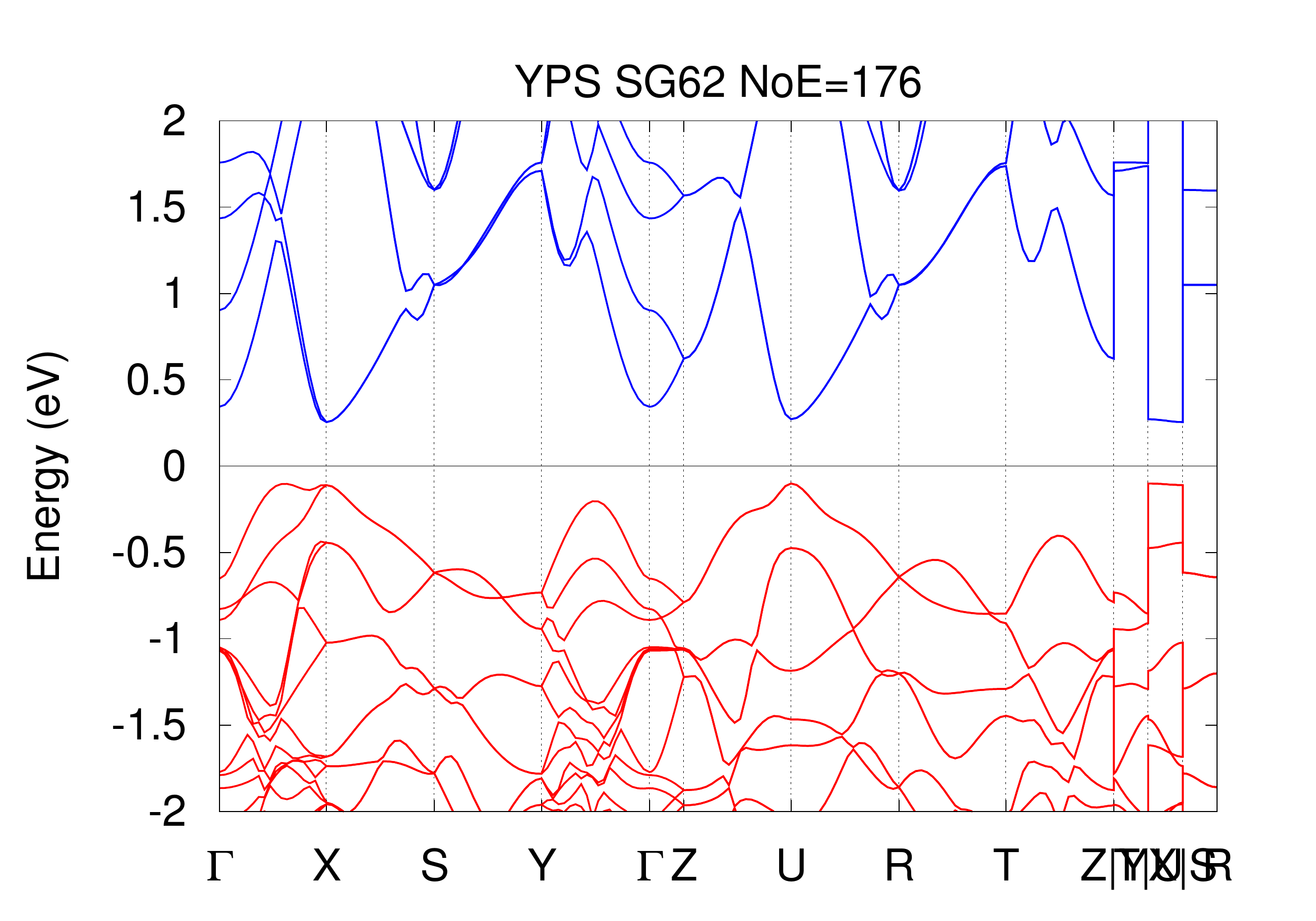}
}
\subfigure[KAlSb$_{4}$ SG62 NoA=24 NoE=128]{
\label{subfig:300157}
\includegraphics[scale=0.32,angle=270]{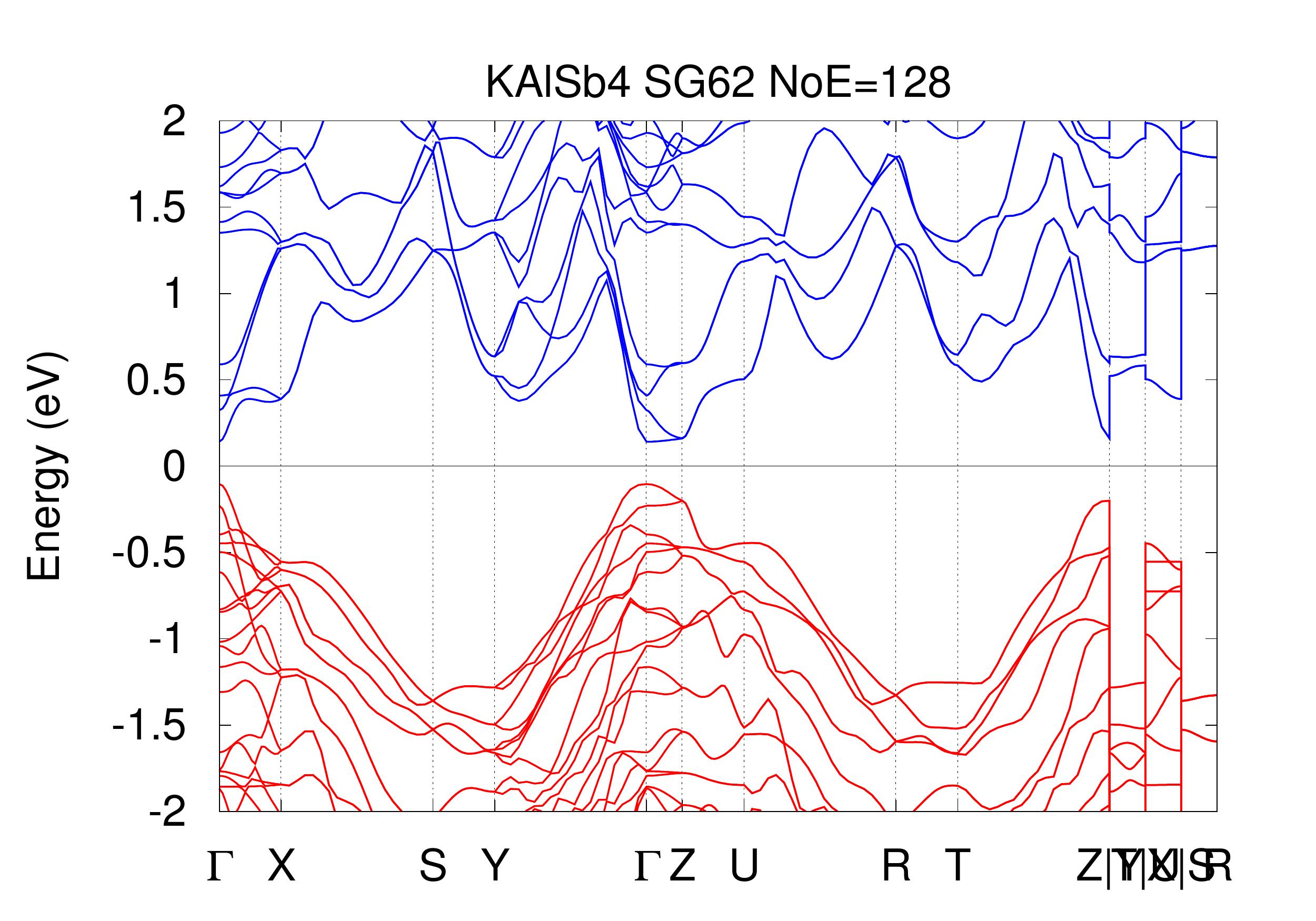}
}
\subfigure[Ba$_{2}$Si$_{3}$Ag SG70 NoA=24 NoE=172]{
\label{subfig:410520}
\includegraphics[scale=0.32,angle=270]{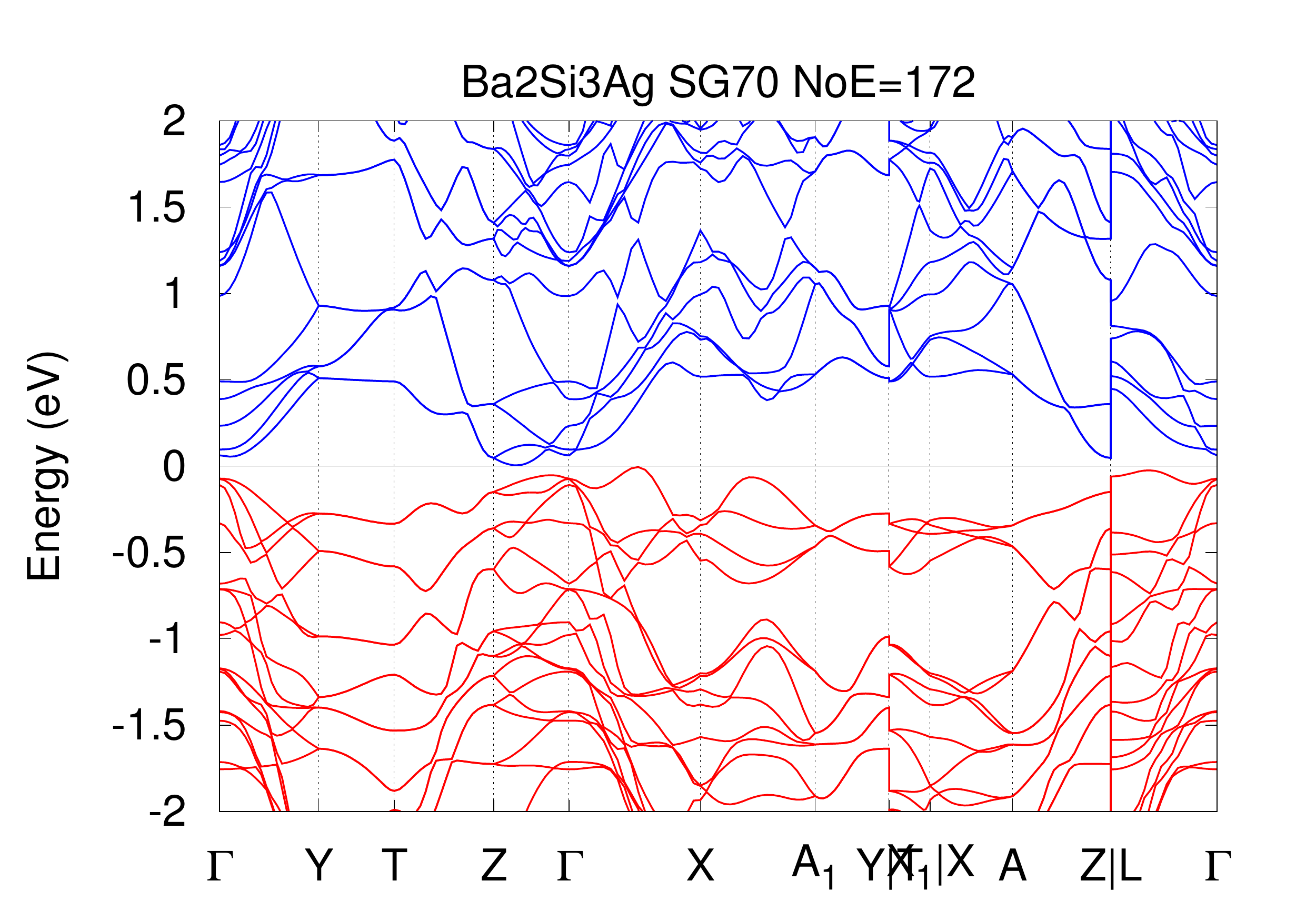}
}
\subfigure[HfSb$_{2}$ SG58 NoA=24 NoE=112]{
\label{subfig:638875}
\includegraphics[scale=0.32,angle=270]{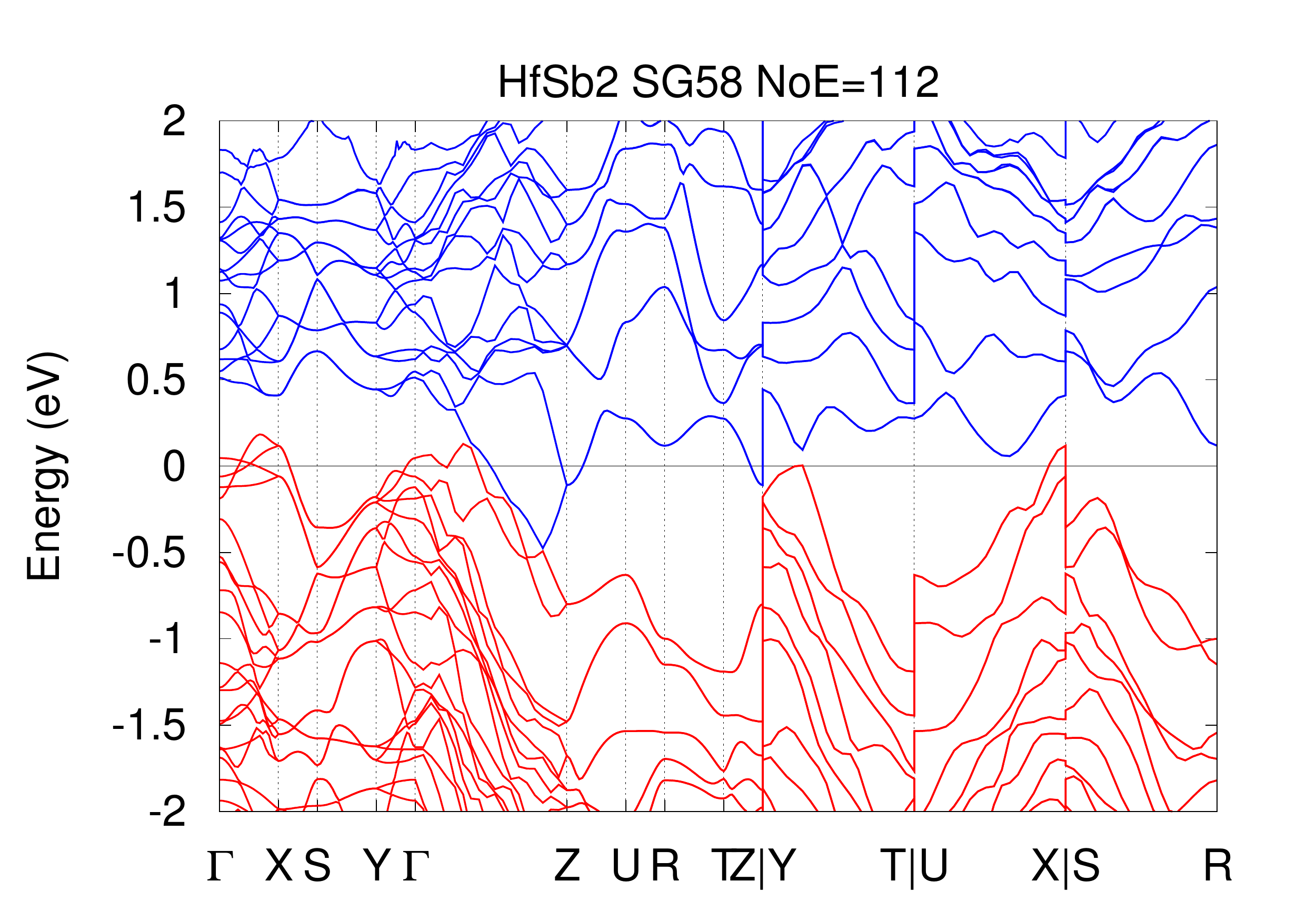}
}
\caption{\hyperref[tab:electride]{back to the table}}
\end{figure}

\begin{figure}[htp]
 \centering
\subfigure[Sr$_{2}$LiSi$_{3}$ SG70 NoA=24 NoE=132]{
\label{subfig:409295}
\includegraphics[scale=0.32,angle=270]{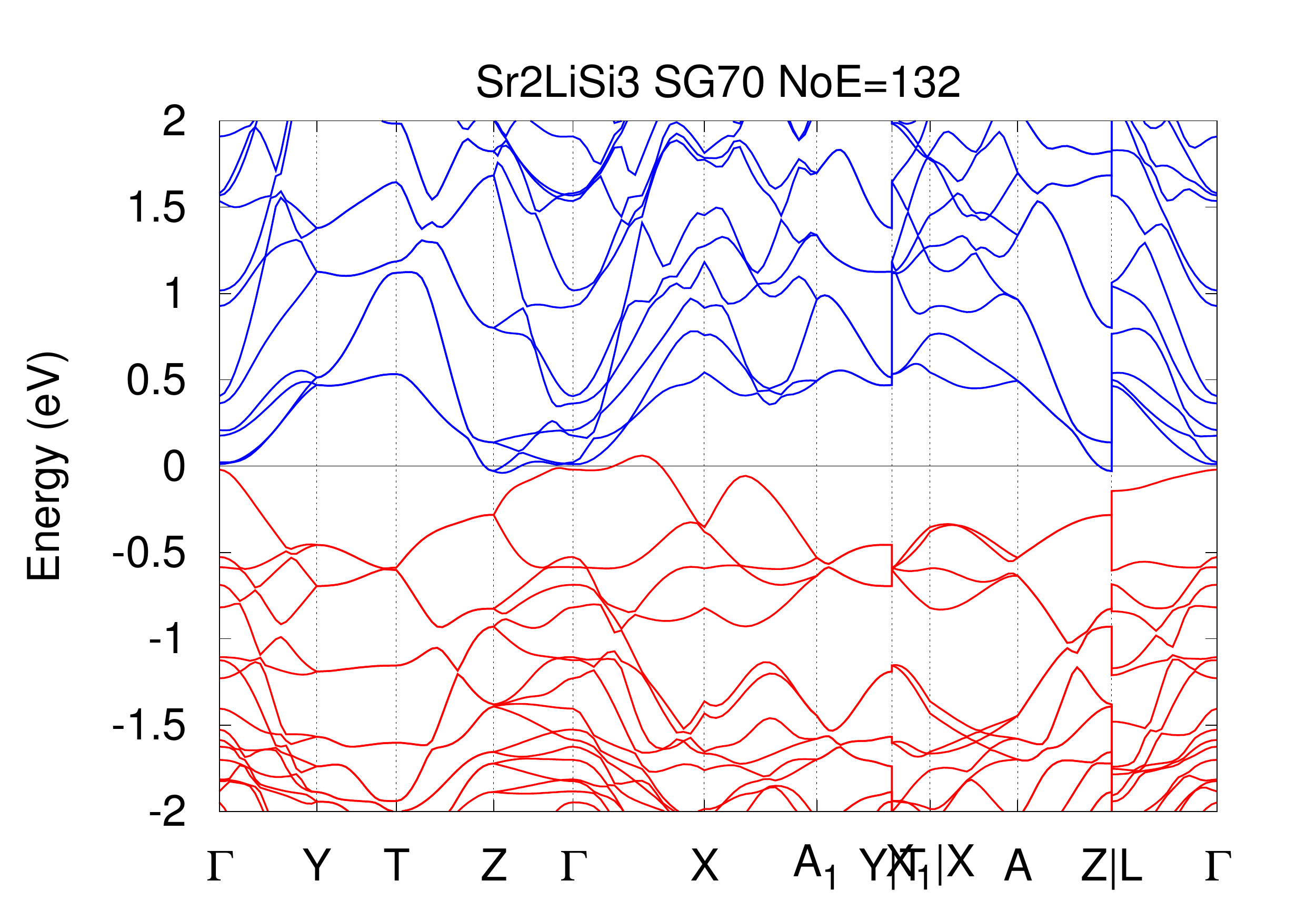}
}
\subfigure[SmPS SG62 NoA=24 NoE=176]{
\label{subfig:648052}
\includegraphics[scale=0.32,angle=270]{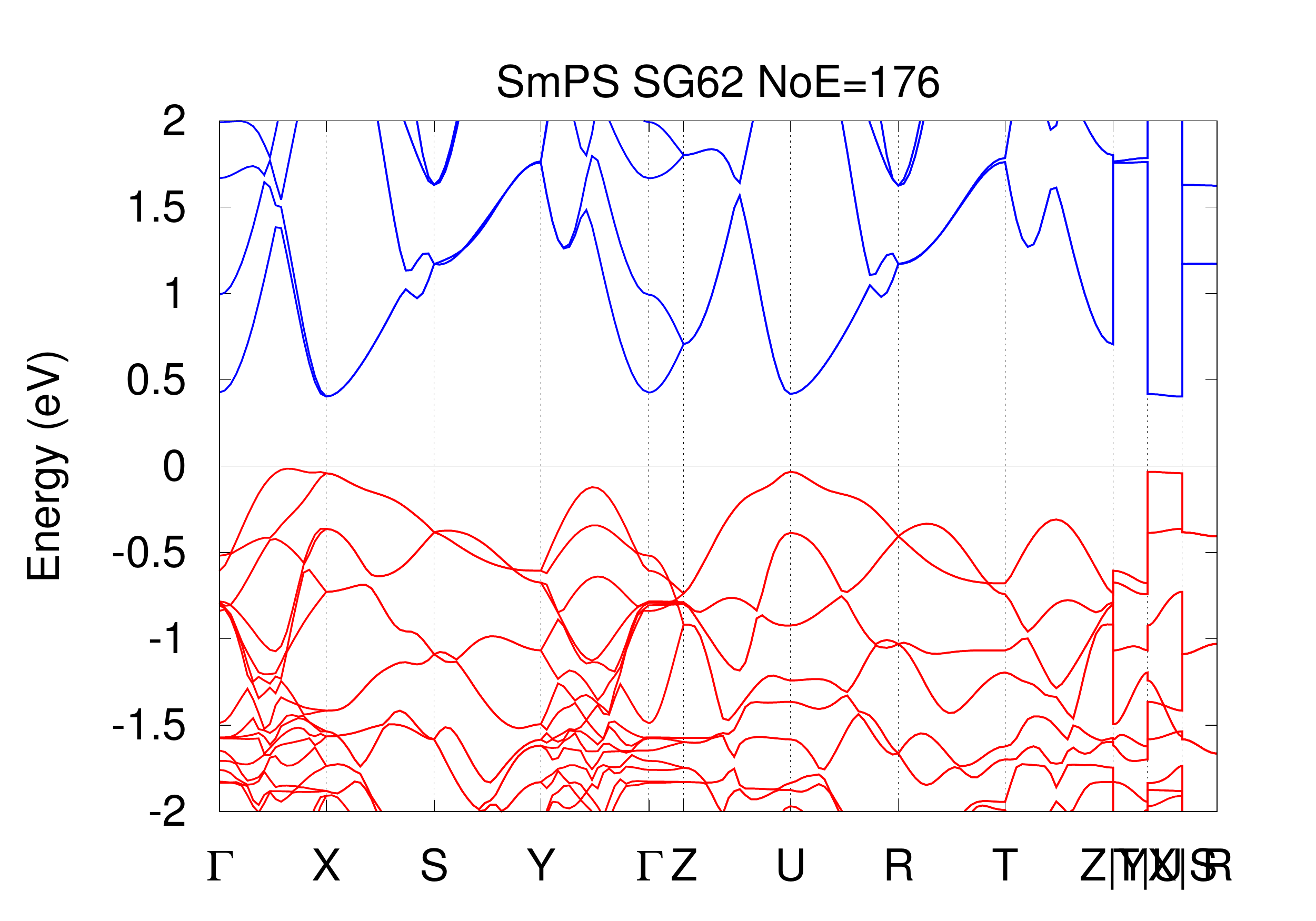}
}
\subfigure[Ba$_{2}$LiSi$_{3}$ SG70 NoA=24 NoE=132]{
\label{subfig:404707}
\includegraphics[scale=0.32,angle=270]{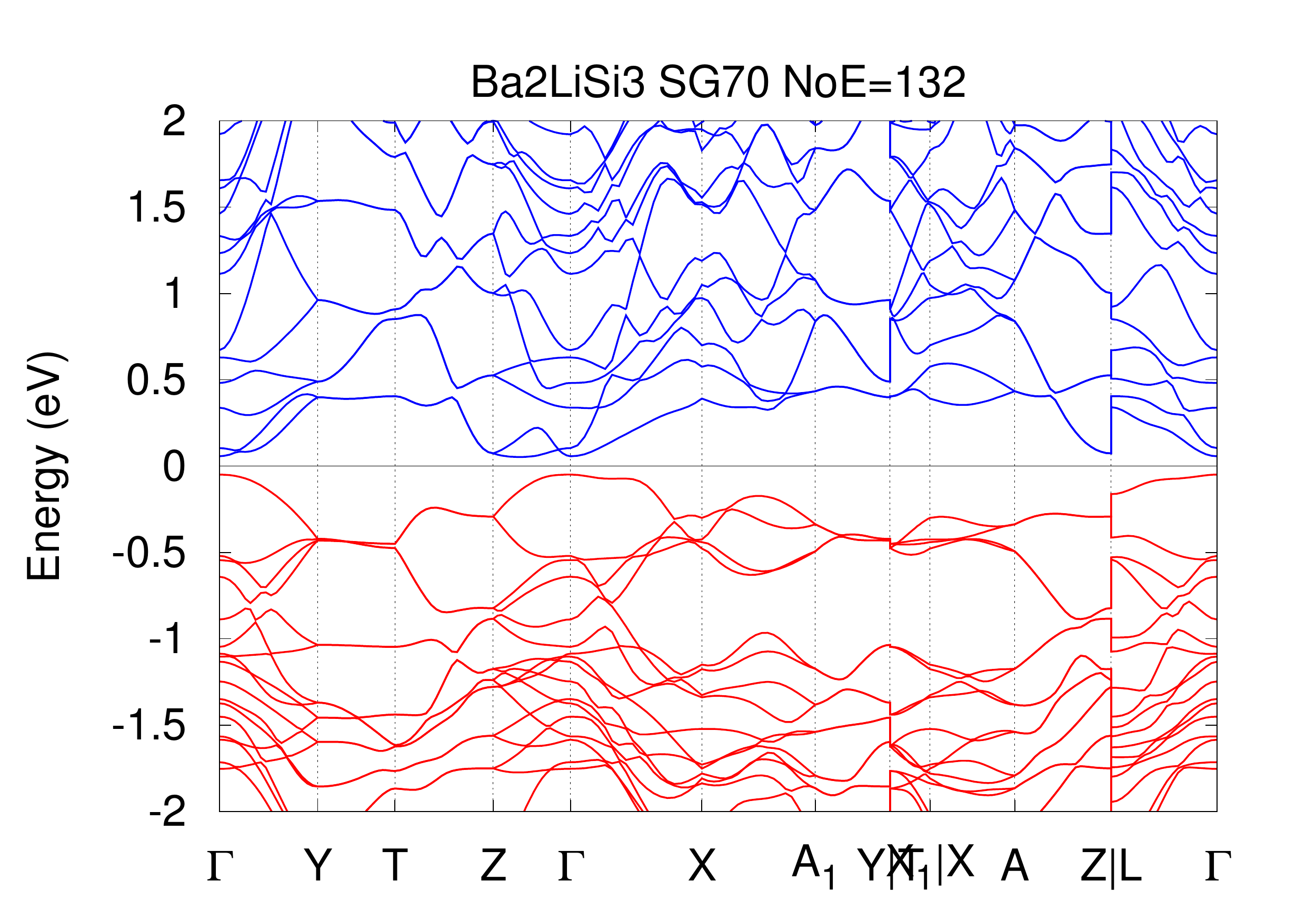}
}
\subfigure[KGaSb$_{4}$ SG62 NoA=24 NoE=128]{
\label{subfig:300158}
\includegraphics[scale=0.32,angle=270]{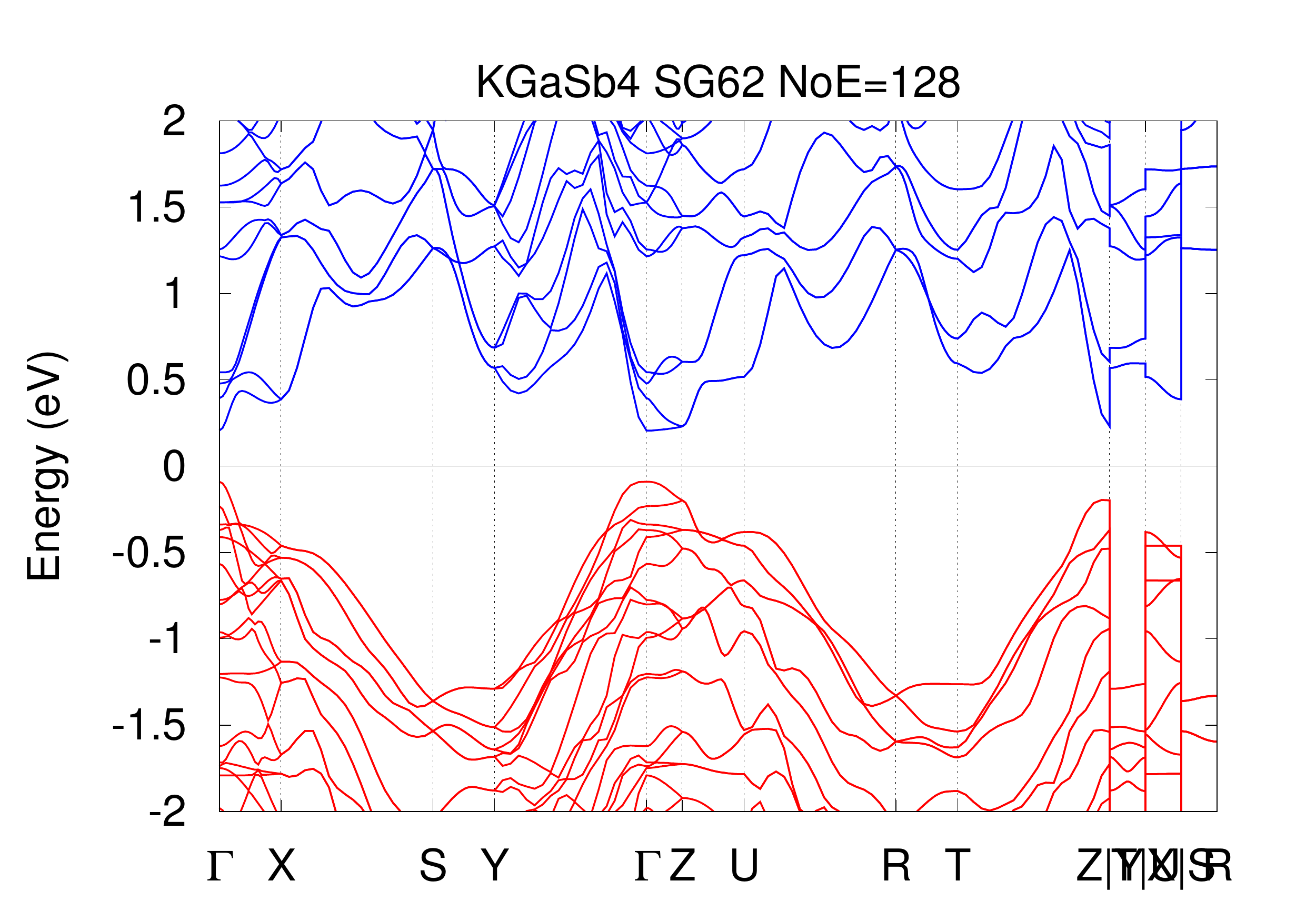}
}
\subfigure[BaP$_{3}$Pt$_{2}$ SG14 NoA=24 NoE=180]{
\label{subfig:62520}
\includegraphics[scale=0.32,angle=270]{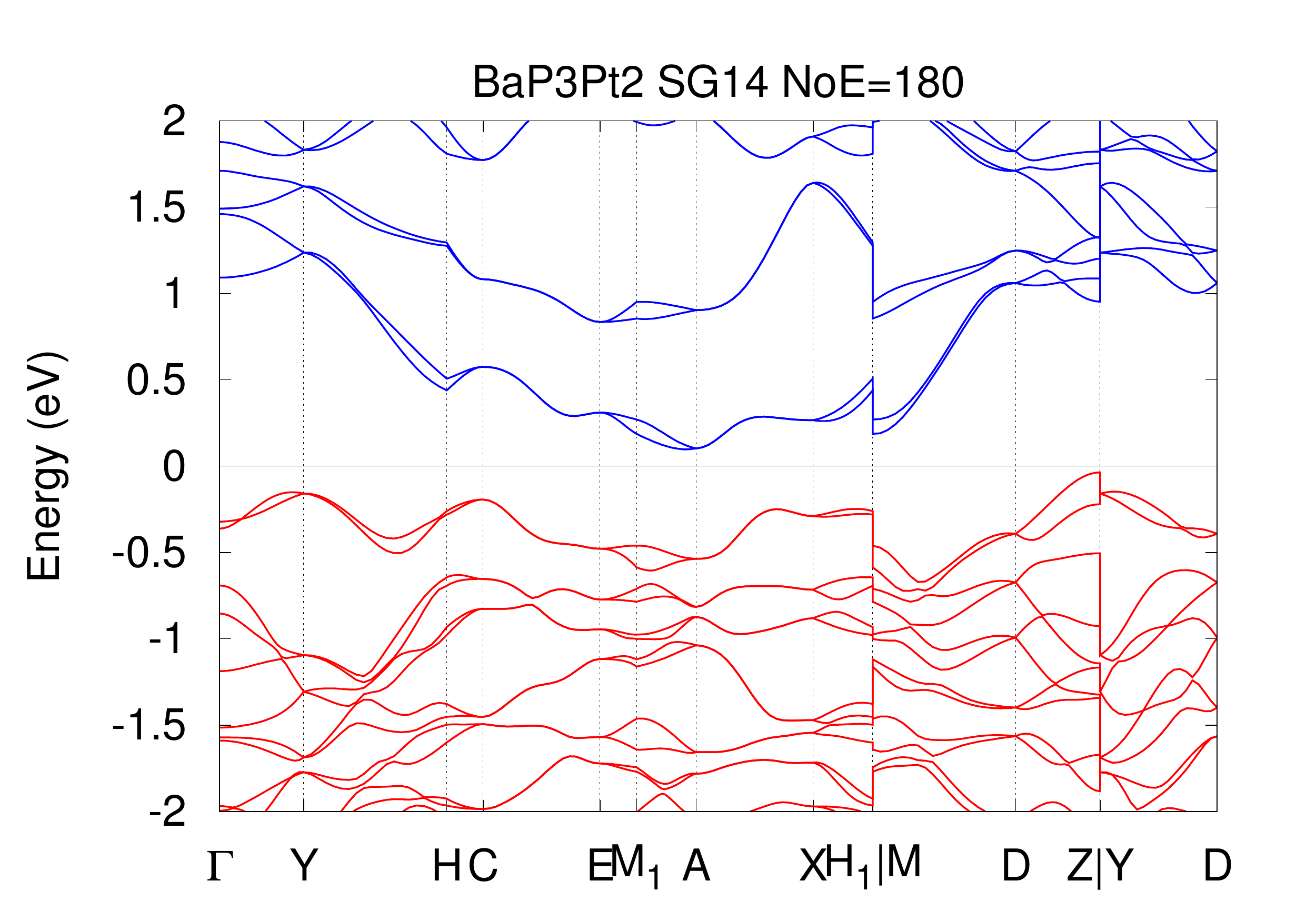}
}
\subfigure[ZrSb$_{2}$ SG58 NoA=24 NoE=176]{
\label{subfig:66779}
\includegraphics[scale=0.32,angle=270]{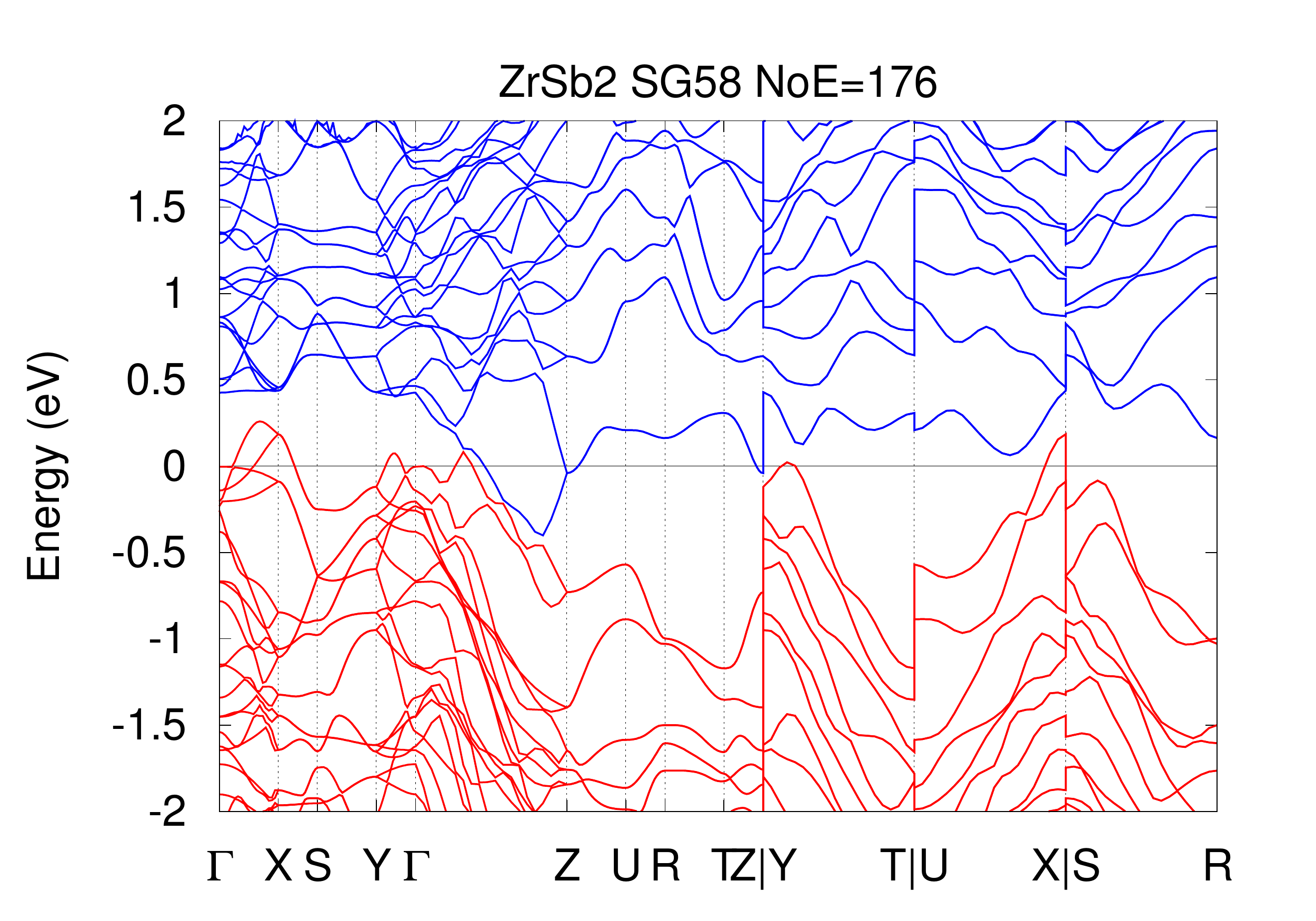}
}
\subfigure[LuCrB$_{4}$ SG55 NoA=24 NoE=108]{
\label{subfig:613517}
\includegraphics[scale=0.32,angle=270]{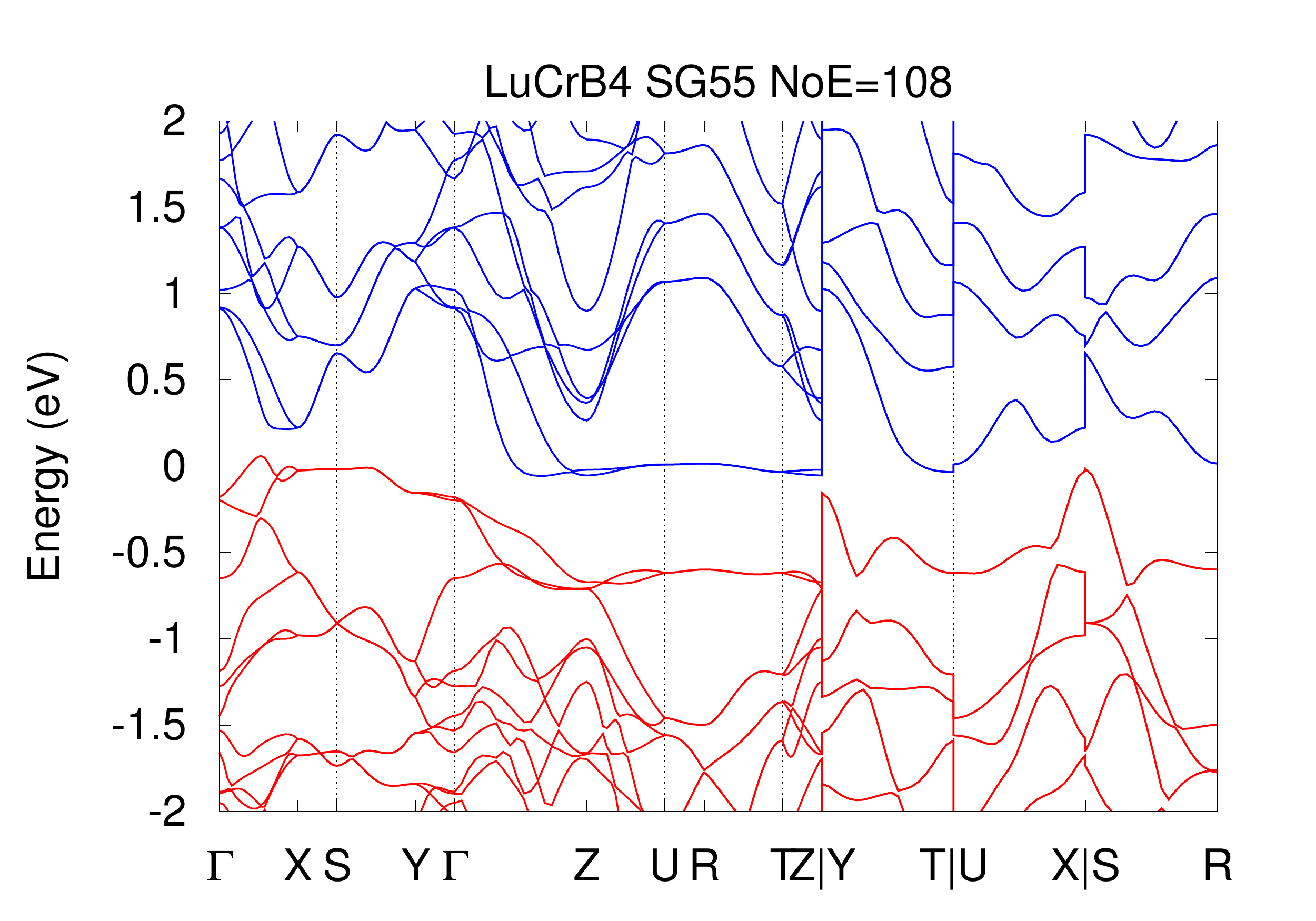}
}
\subfigure[ZrBi$_{2}$ SG58 NoA=24 NoE=176]{
\label{subfig:42880}
\includegraphics[scale=0.32,angle=270]{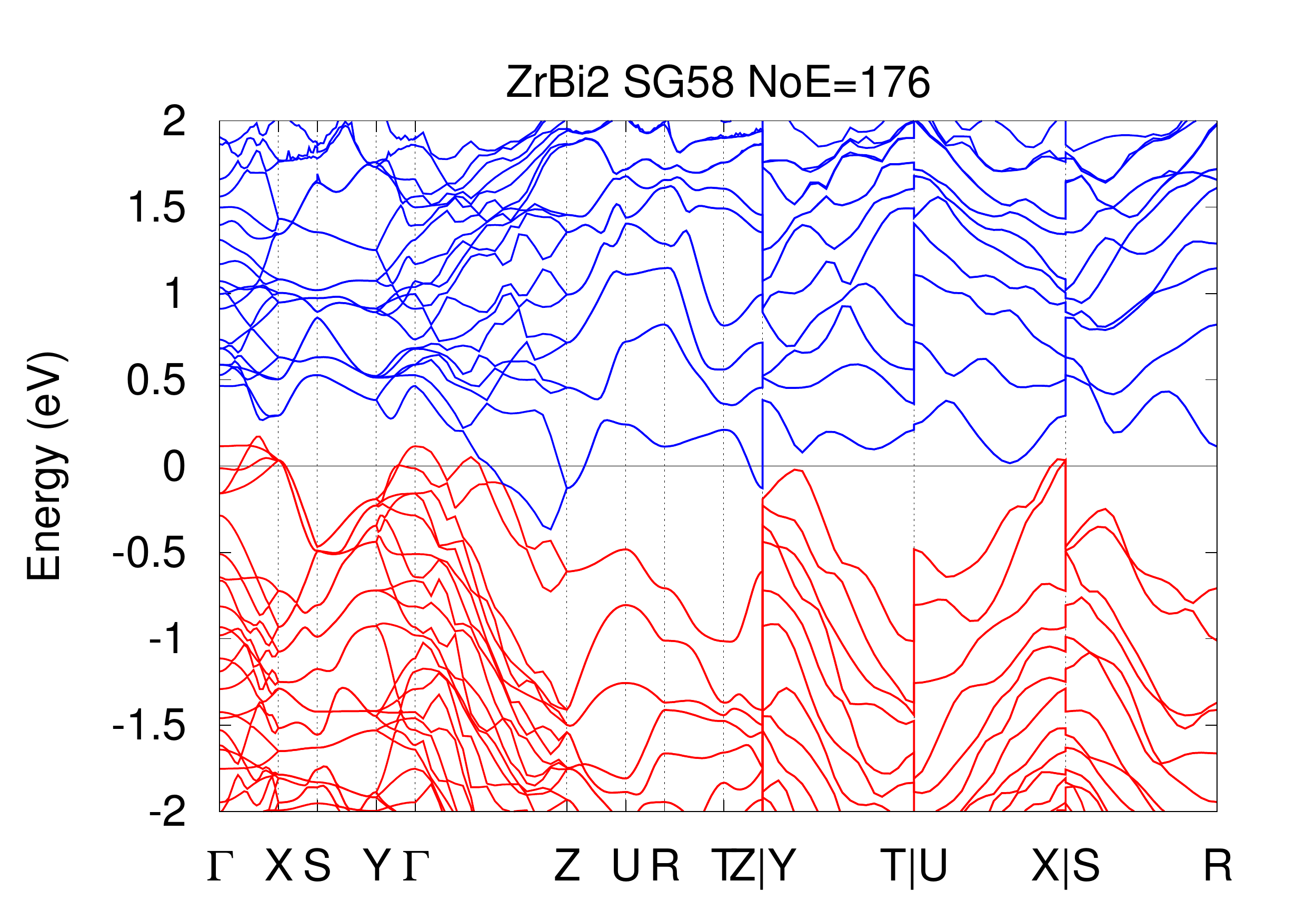}
}
\caption{\hyperref[tab:electride]{back to the table}}
\end{figure}

\begin{figure}[htp]
 \centering
\subfigure[K(MoO$_{3}$)$_{3}$ SG12 NoA=26 NoE=162]{
\label{subfig:1053}
\includegraphics[scale=0.32,angle=270]{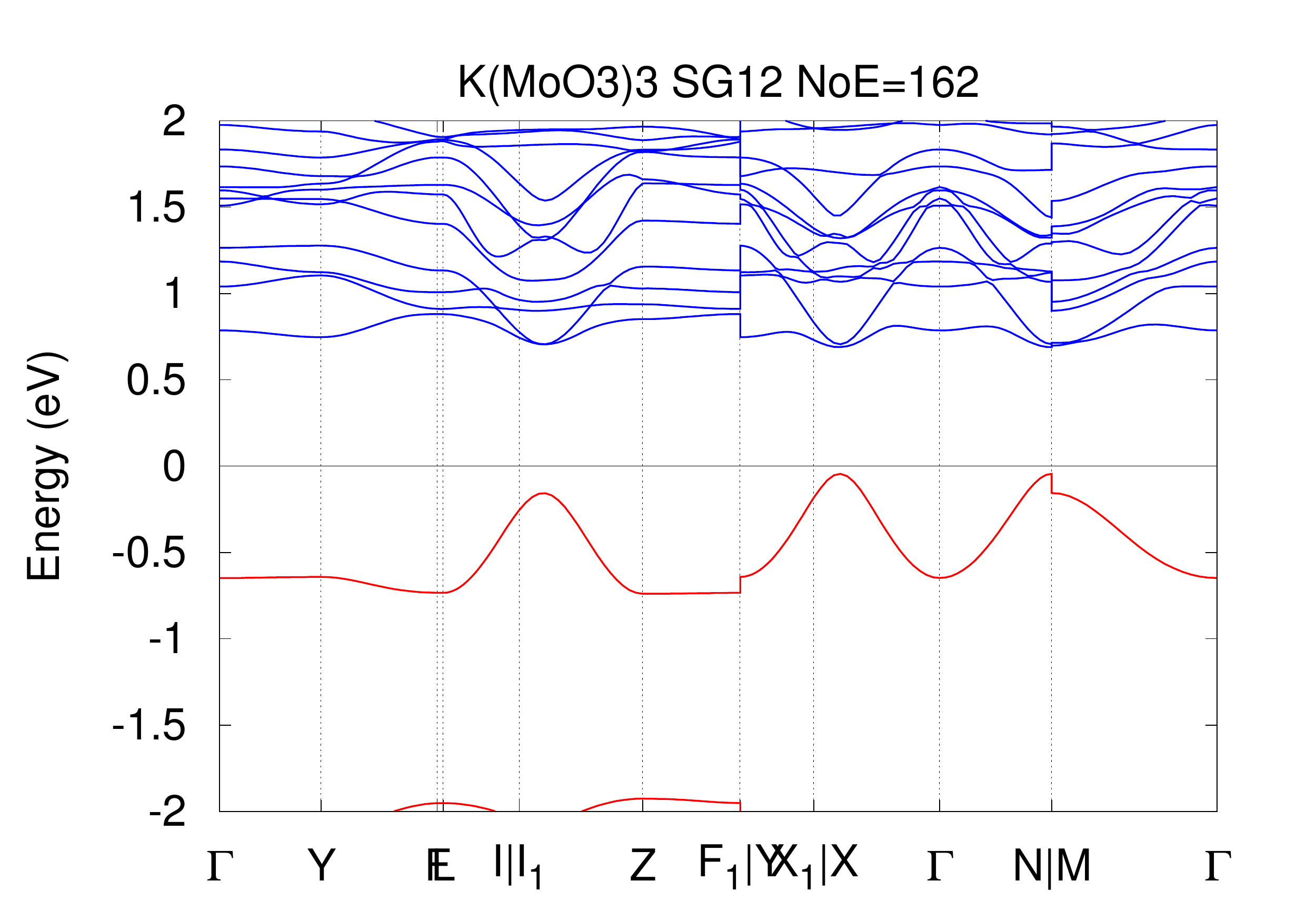}
}
\subfigure[Yb(Al$_{5}$Ru)$_{2}$ SG63 NoA=26 NoE=108]{
\label{subfig:186915}
\includegraphics[scale=0.32,angle=270]{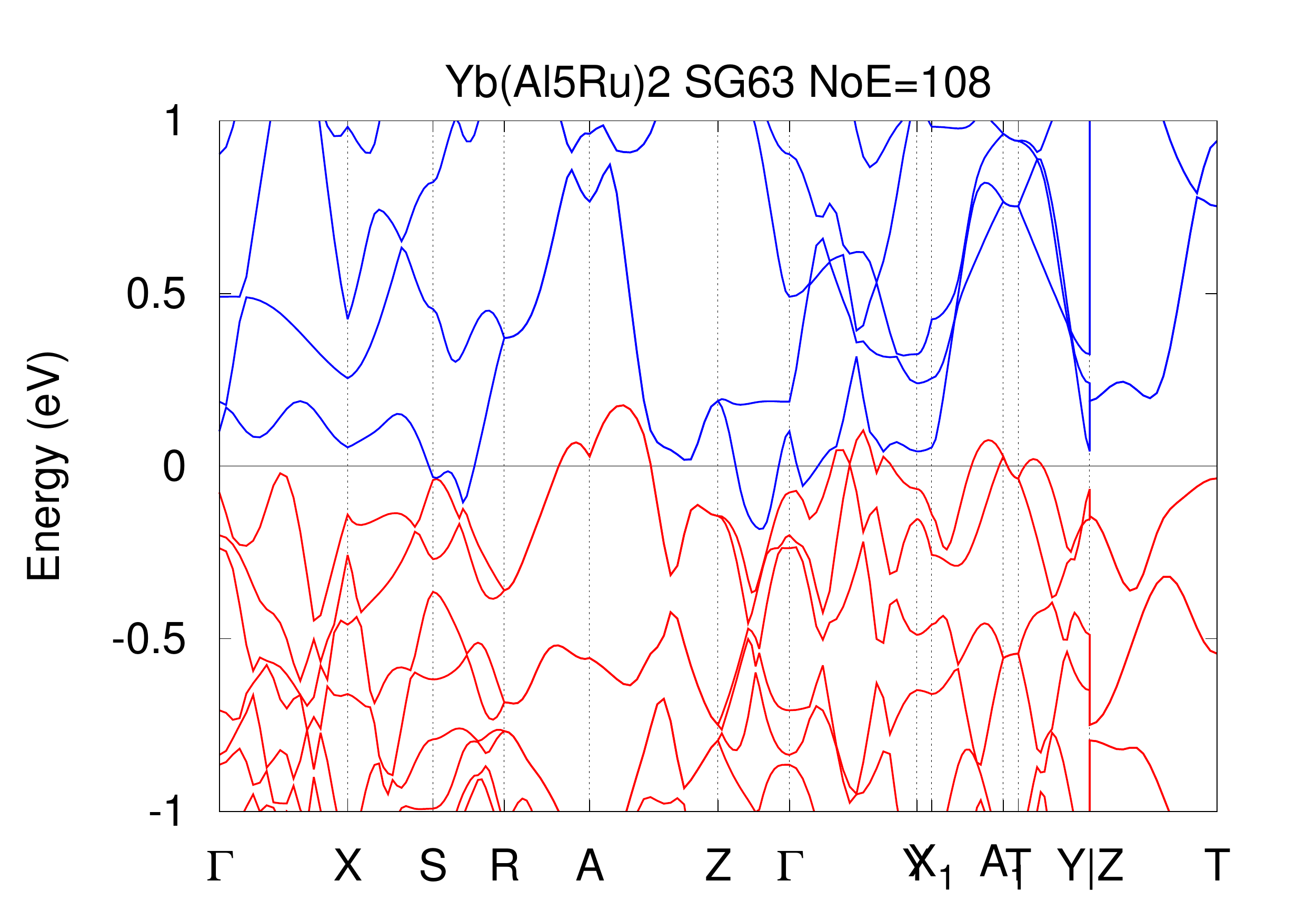}
}
\subfigure[Yb(Al$_{5}$Fe)$_{2}$ SG63 NoA=26 NoE=108]{
\label{subfig:151140}
\includegraphics[scale=0.32,angle=270]{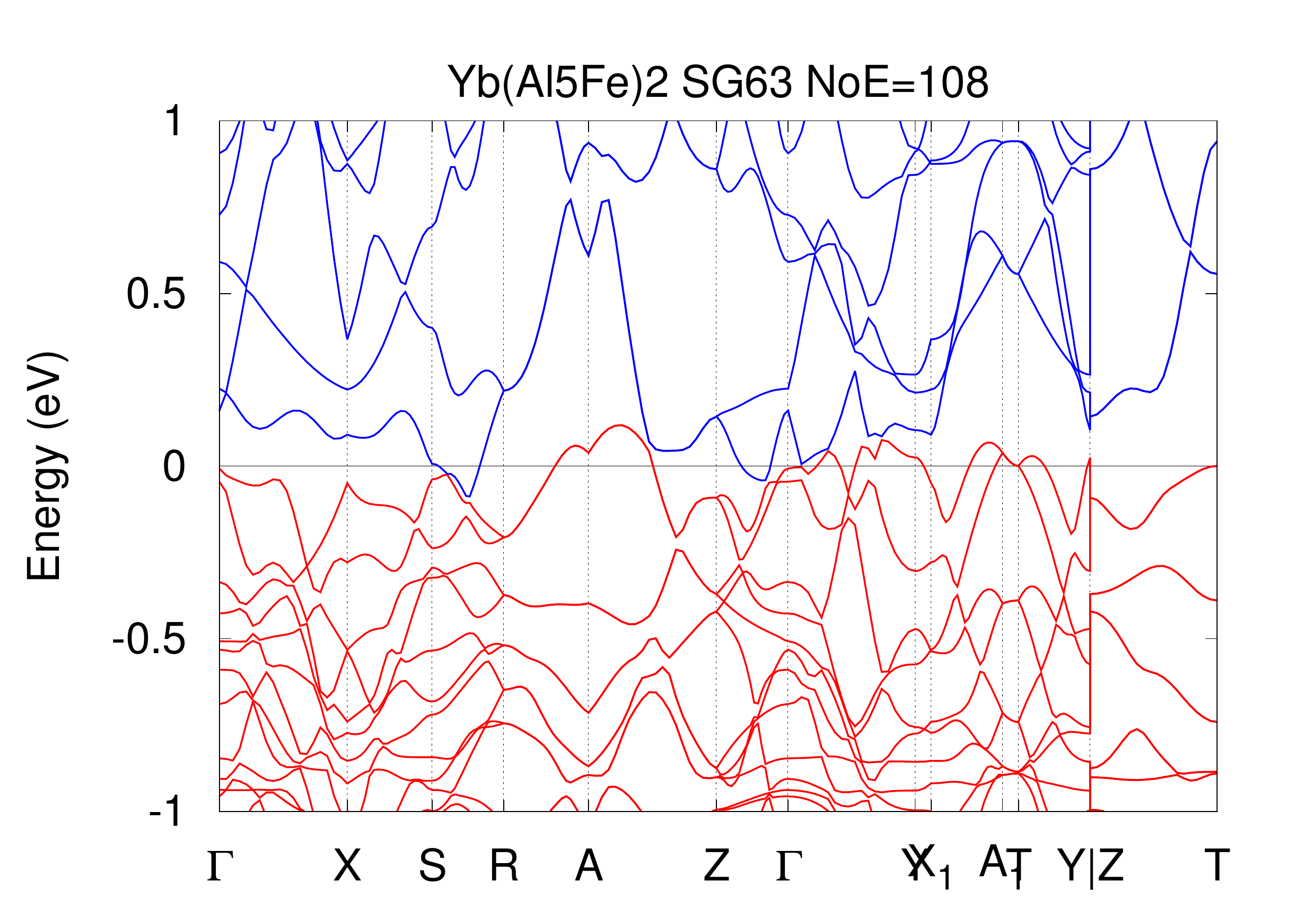}
}
\subfigure[Ba$_{3}$(Si$_{2}$P$_{3}$)$_{2}$ SG11 NoA=26 NoE=152]{
\label{subfig:29261}
\includegraphics[scale=0.32,angle=270]{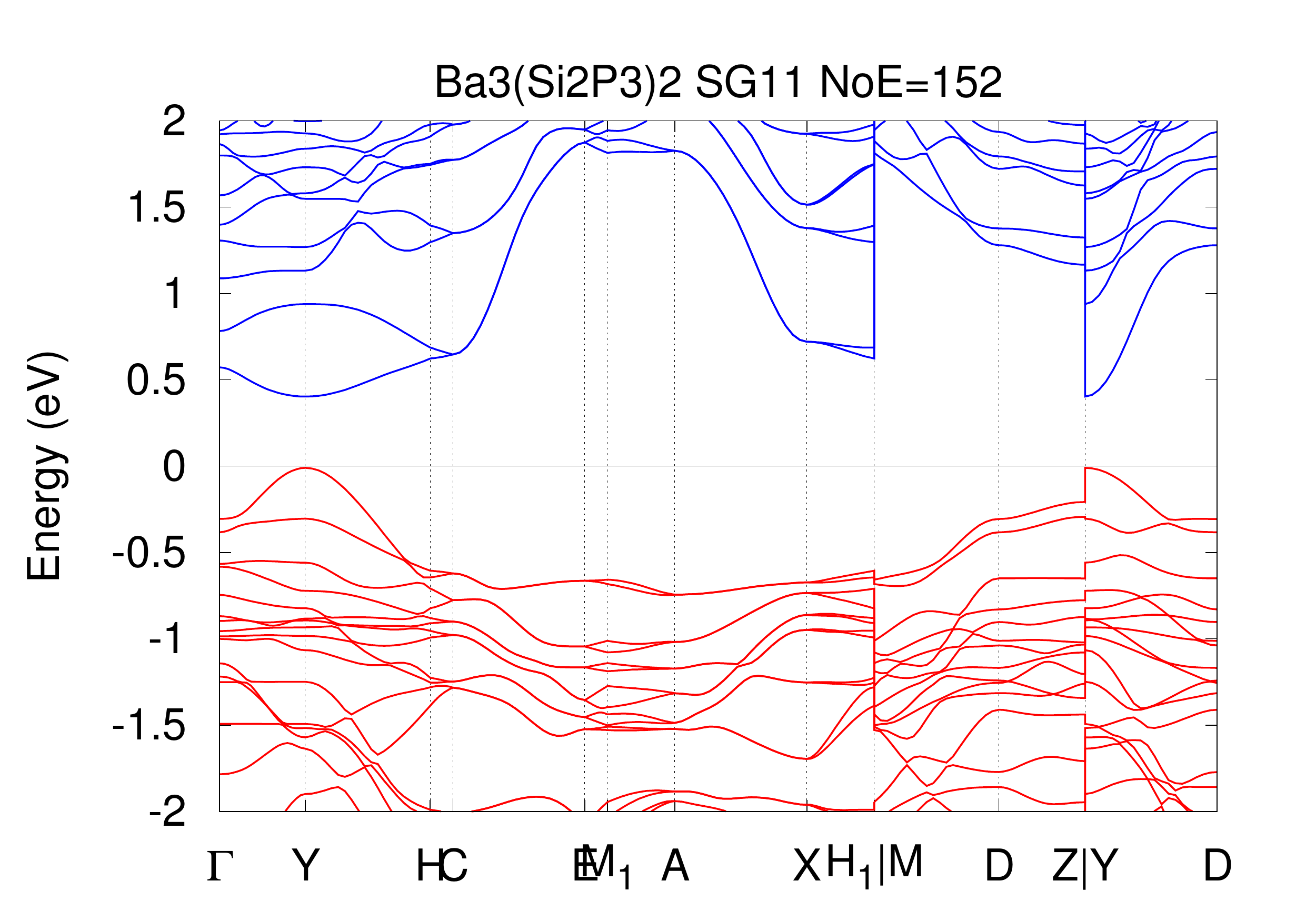}
}
\subfigure[Tl(MoO$_{3}$)$_{3}$ SG12 NoA=26 NoE=150]{
\label{subfig:62690}
\includegraphics[scale=0.32,angle=270]{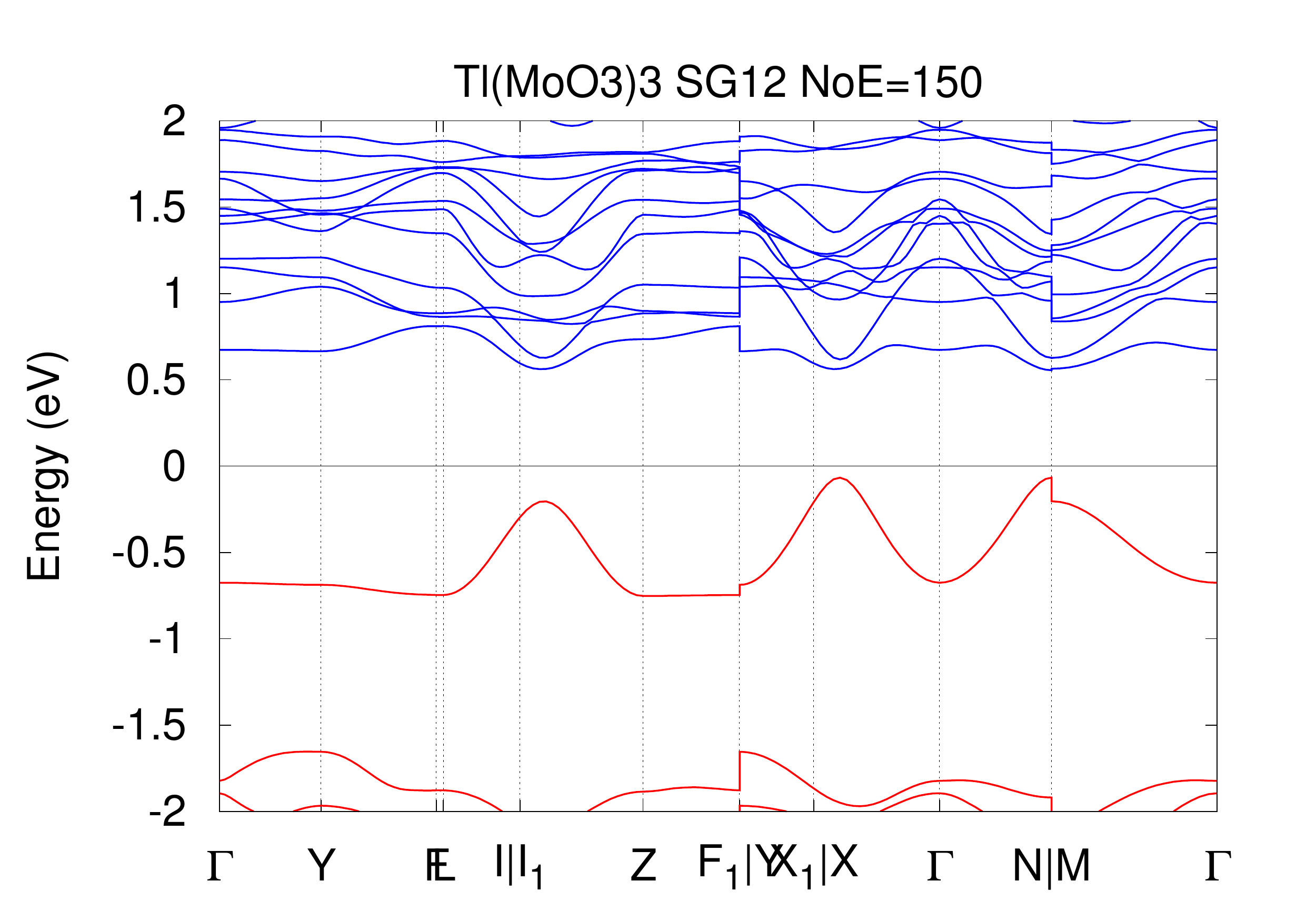}
}
\subfigure[Yb(Al$_{5}$Os)$_{2}$ SG63 NoA=26 NoE=108]{
\label{subfig:238040}
\includegraphics[scale=0.32,angle=270]{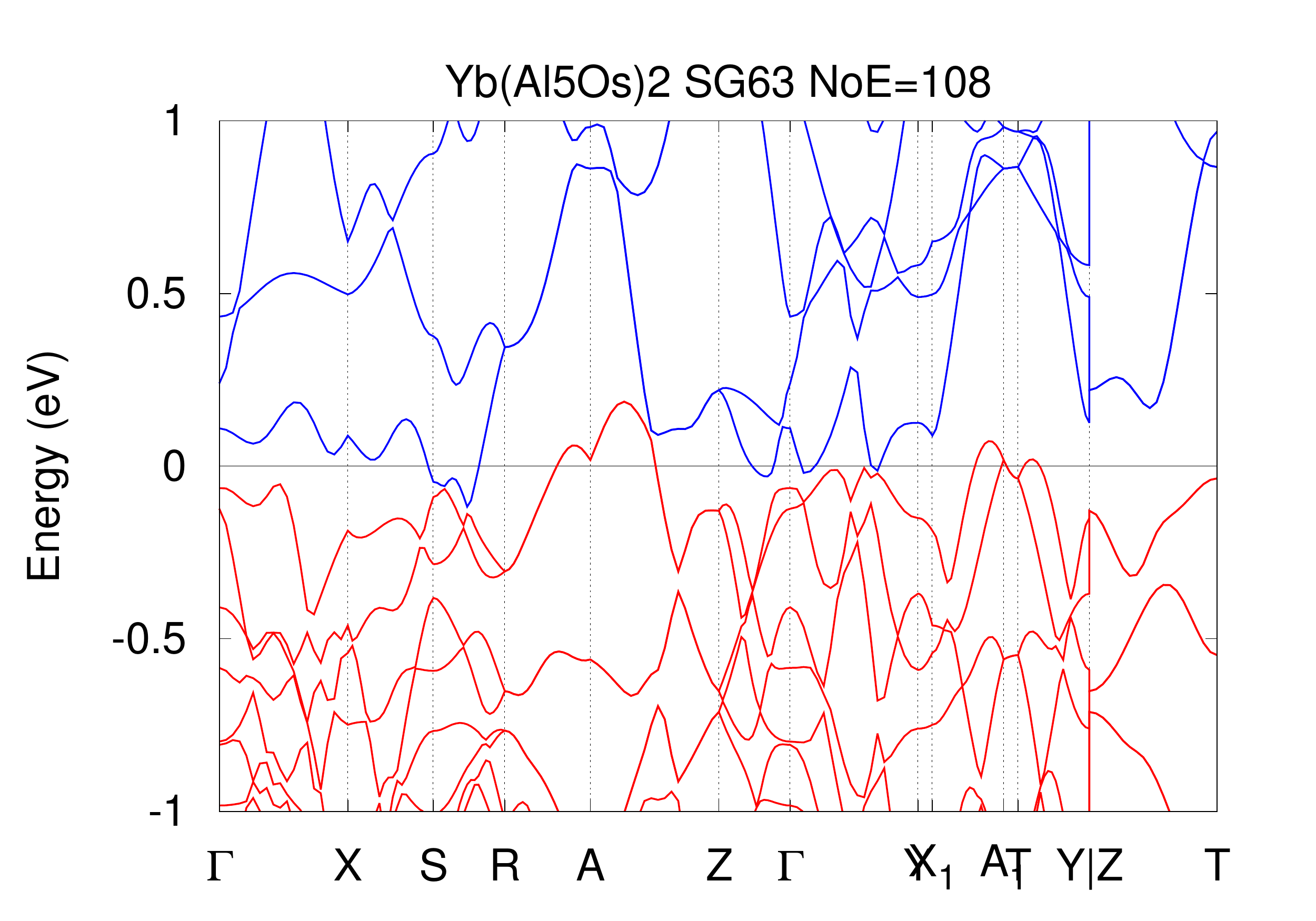}
}
\subfigure[Na$_{7}$Al$_{2}$Sb$_{5}$ SG11 NoA=28 NoE=76]{
\label{subfig:48168}
\includegraphics[scale=0.32,angle=270]{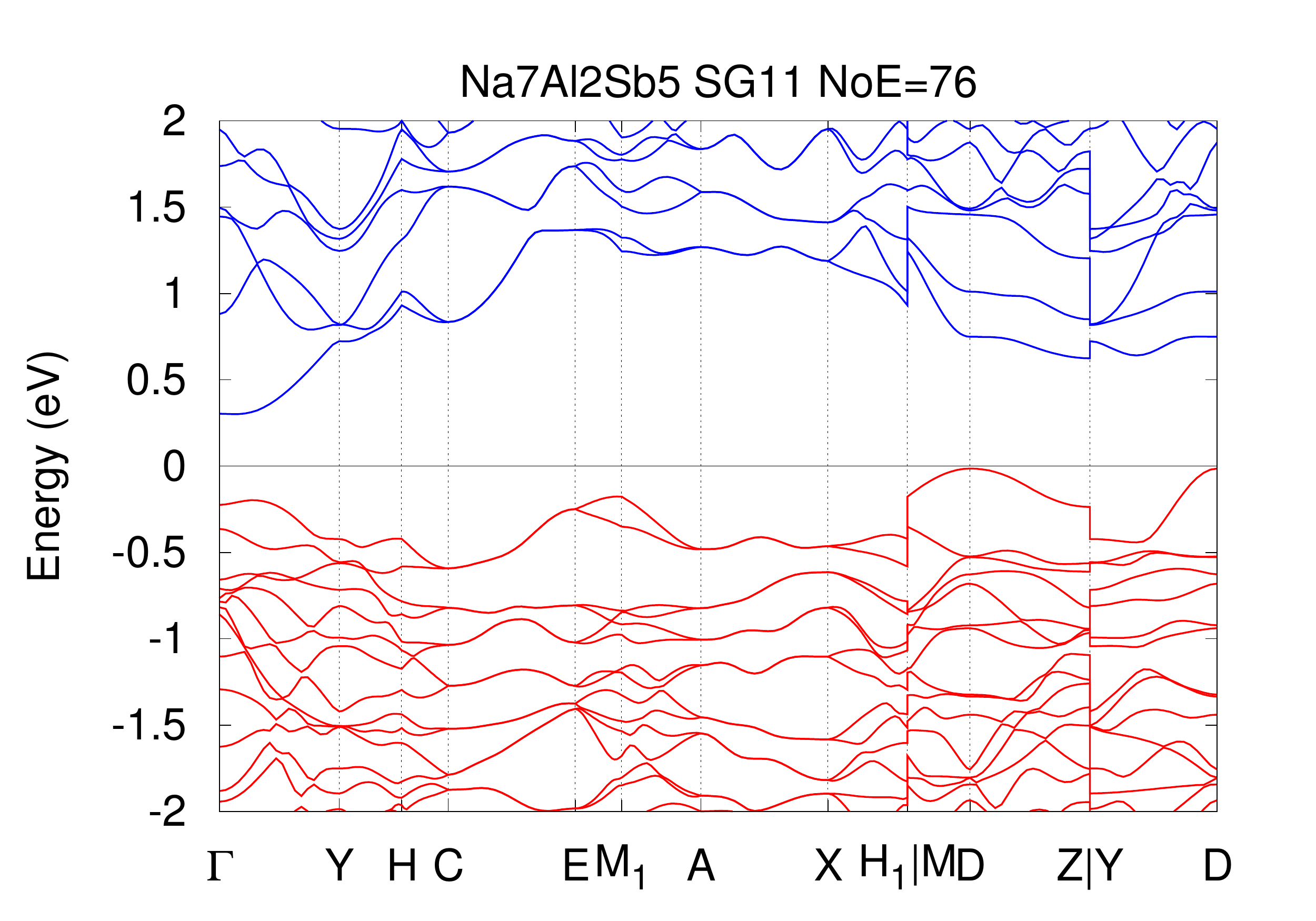}
}
\subfigure[Cs$_{3}$Zr$_{2}$I$_{9}$ SG194 NoA=28 NoE=228]{
\label{subfig:26565}
\includegraphics[scale=0.32,angle=270]{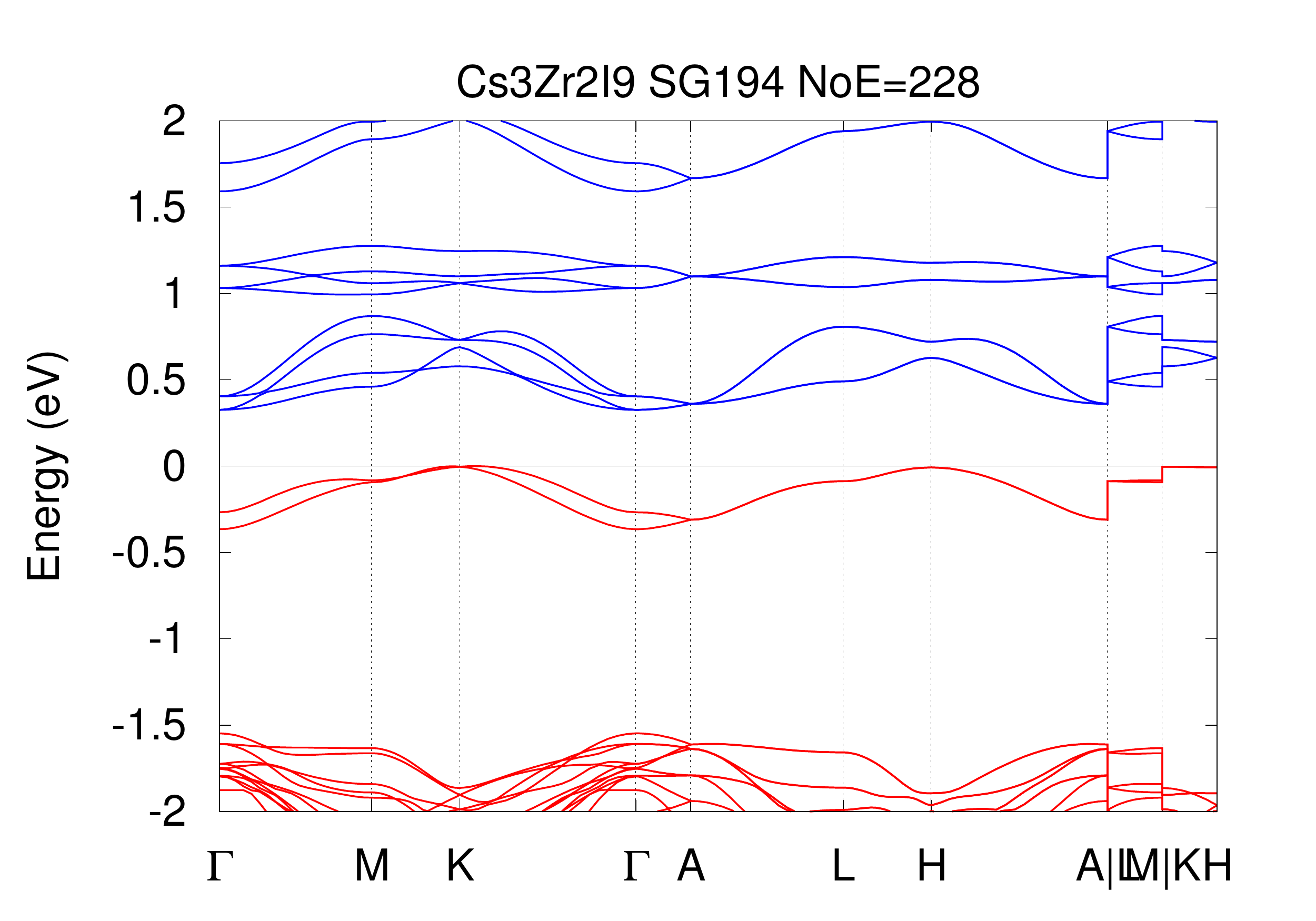}
}
\caption{\hyperref[tab:electride]{back to the table}}
\end{figure}

\begin{figure}[htp]
 \centering
\subfigure[CuAgPO$_{4}$ SG14 NoA=28 NoE=204]{
\label{subfig:165596}
\includegraphics[scale=0.32,angle=270]{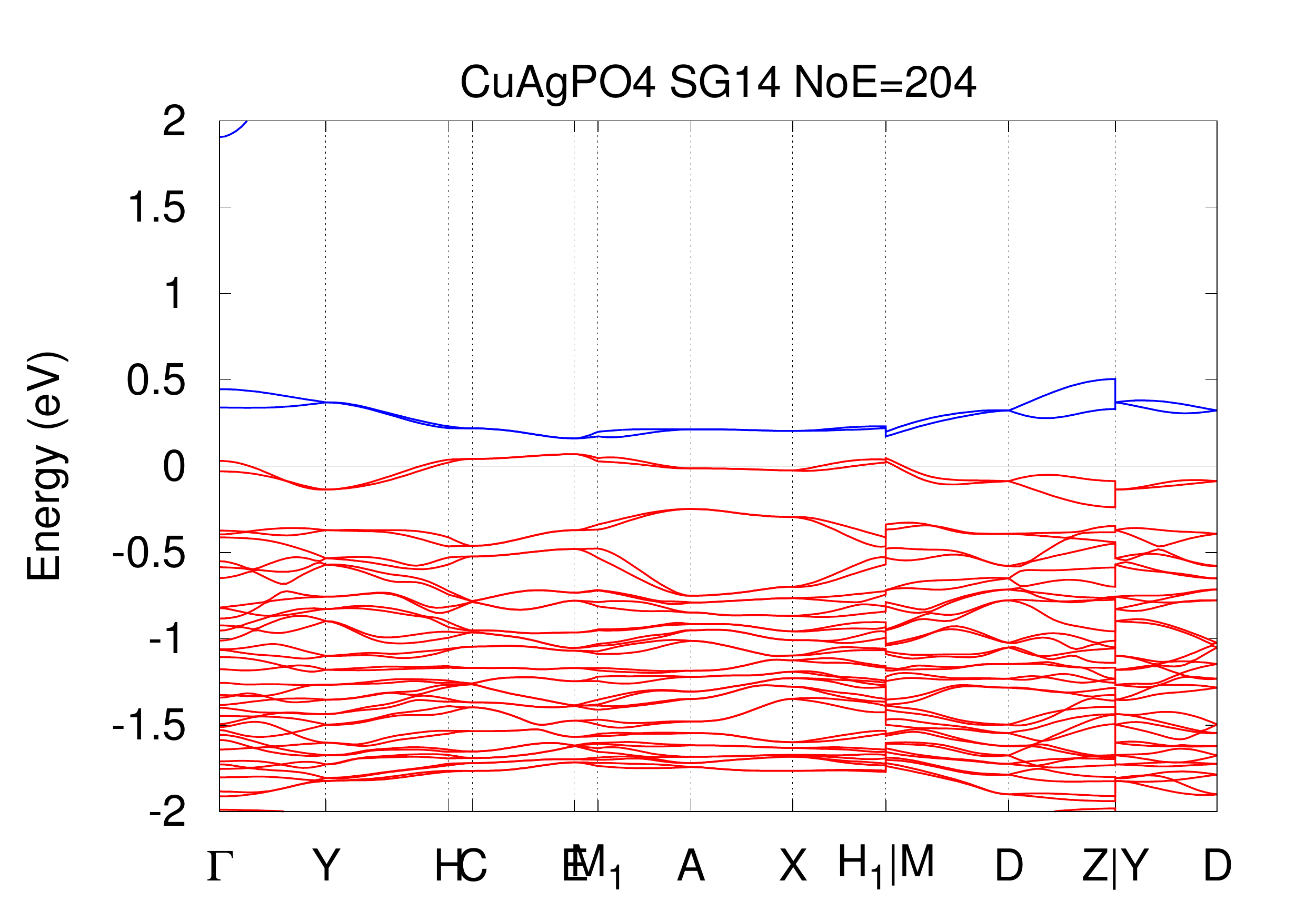}
}
\subfigure[La$_{3}$PI$_{3}$ SG214 NoA=28 NoE=236]{
\label{subfig:411801}
\includegraphics[scale=0.32,angle=270]{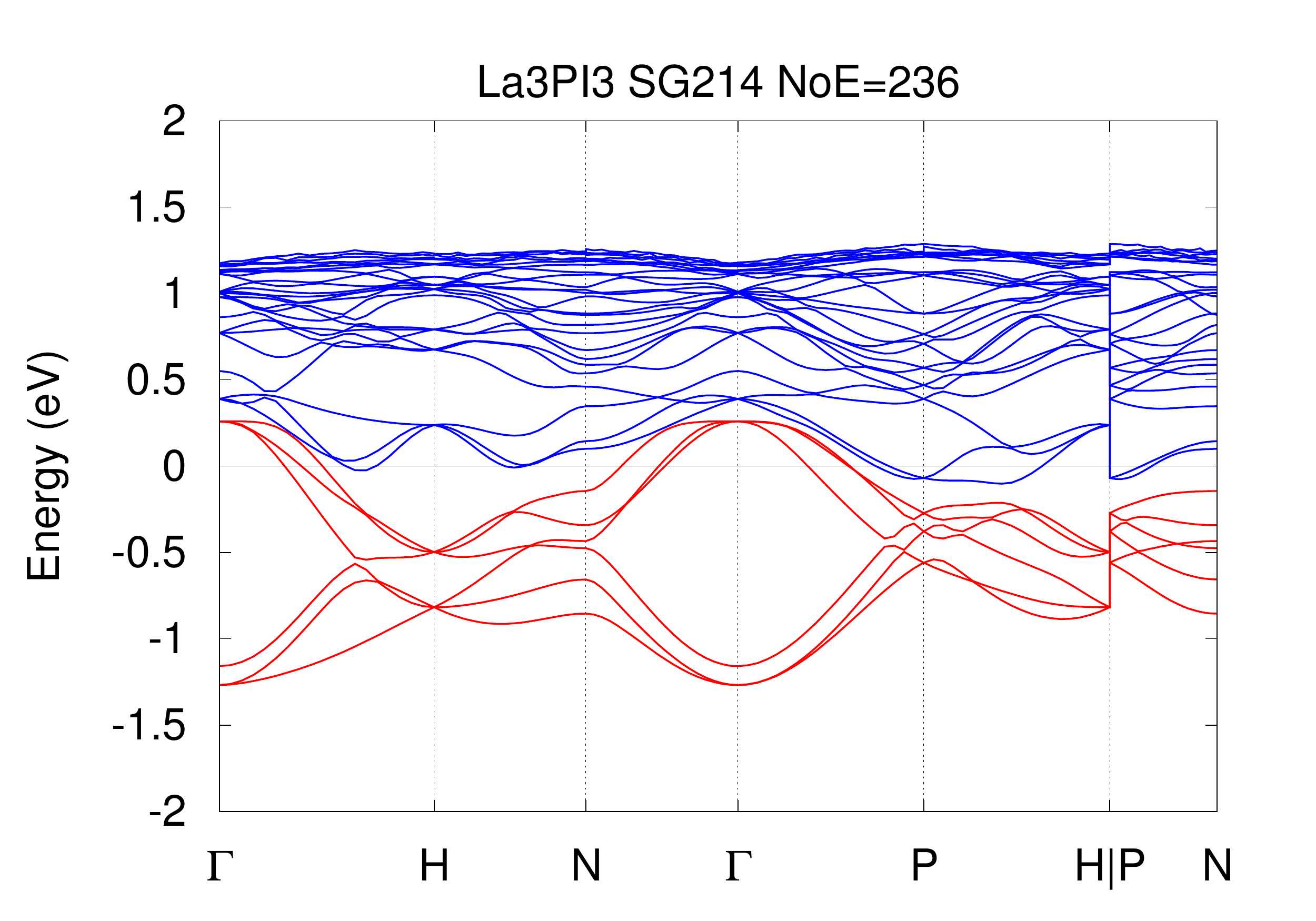}
}
\subfigure[La$_{3}$AsI$_{3}$ SG214 NoA=28 NoE=236]{
\label{subfig:411803}
\includegraphics[scale=0.32,angle=270]{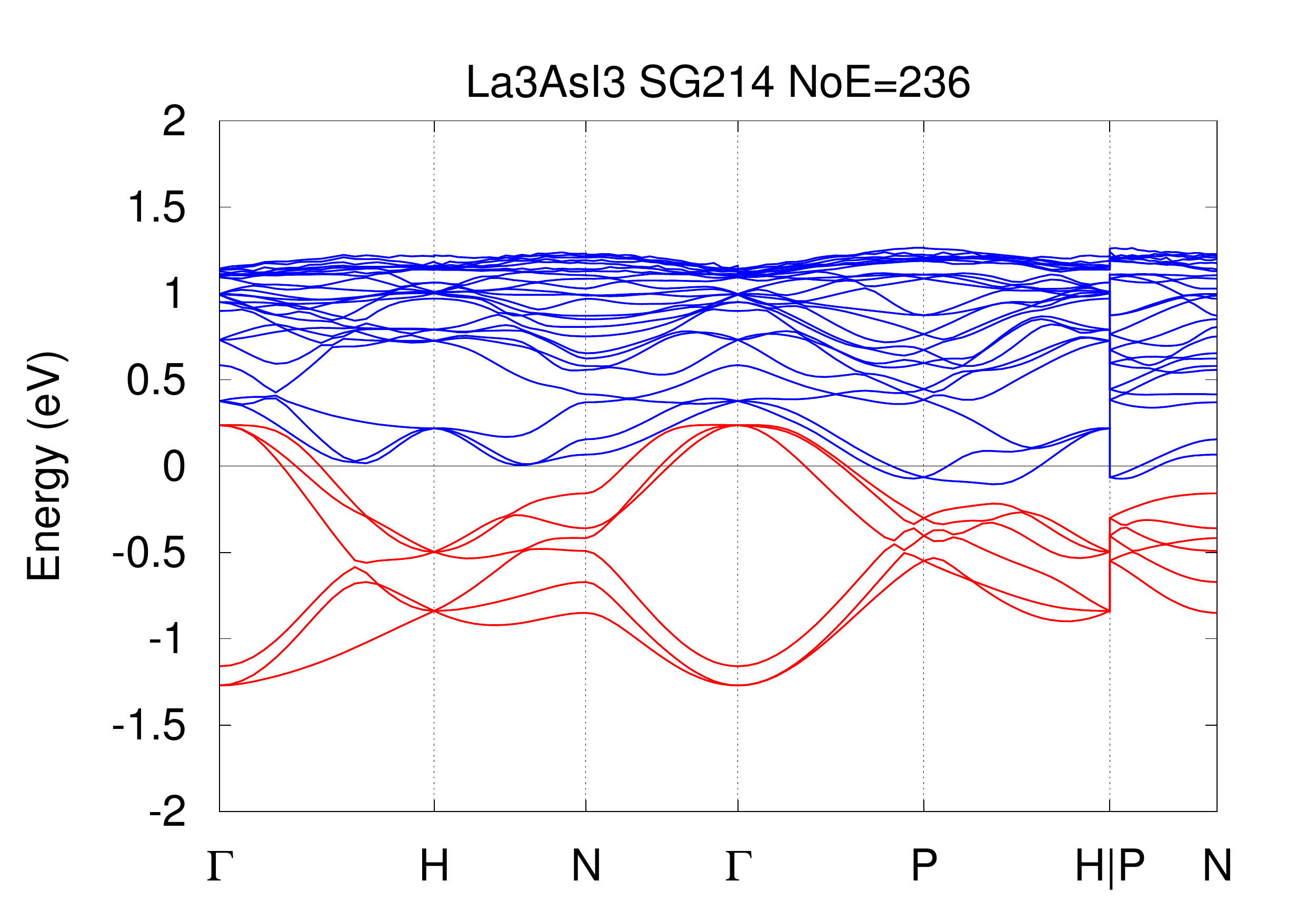}
}
\subfigure[Ba$_{11}$(CdSb$_{2}$)$_{6}$ SG12 NoA=29 NoE=242]{
\label{subfig:418886}
\includegraphics[scale=0.32,angle=270]{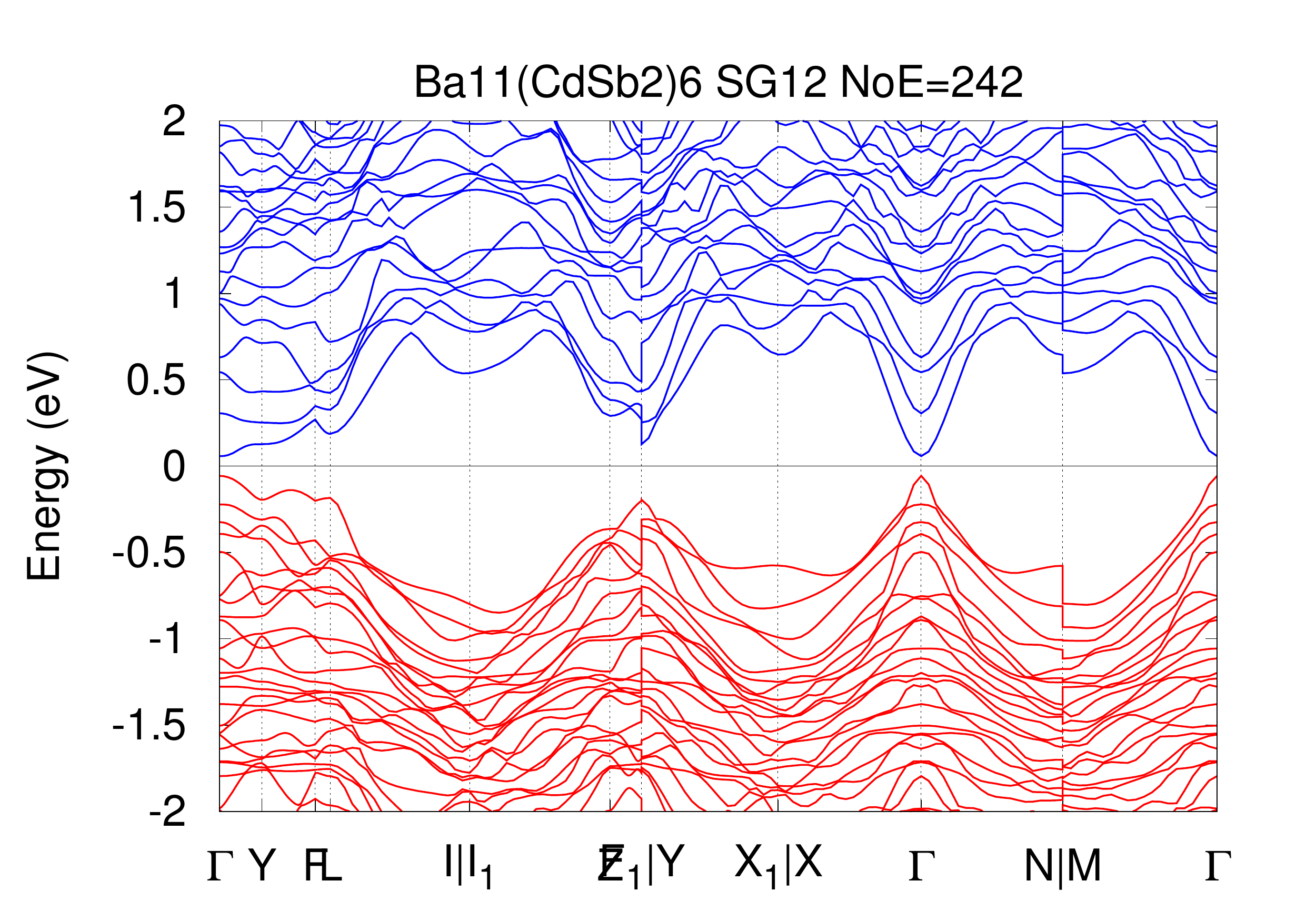}
}
\subfigure[Sr$_{11}$(CdSb$_{2}$)$_{6}$ SG12 NoA=29 NoE=242]{
\label{subfig:413701}
\includegraphics[scale=0.32,angle=270]{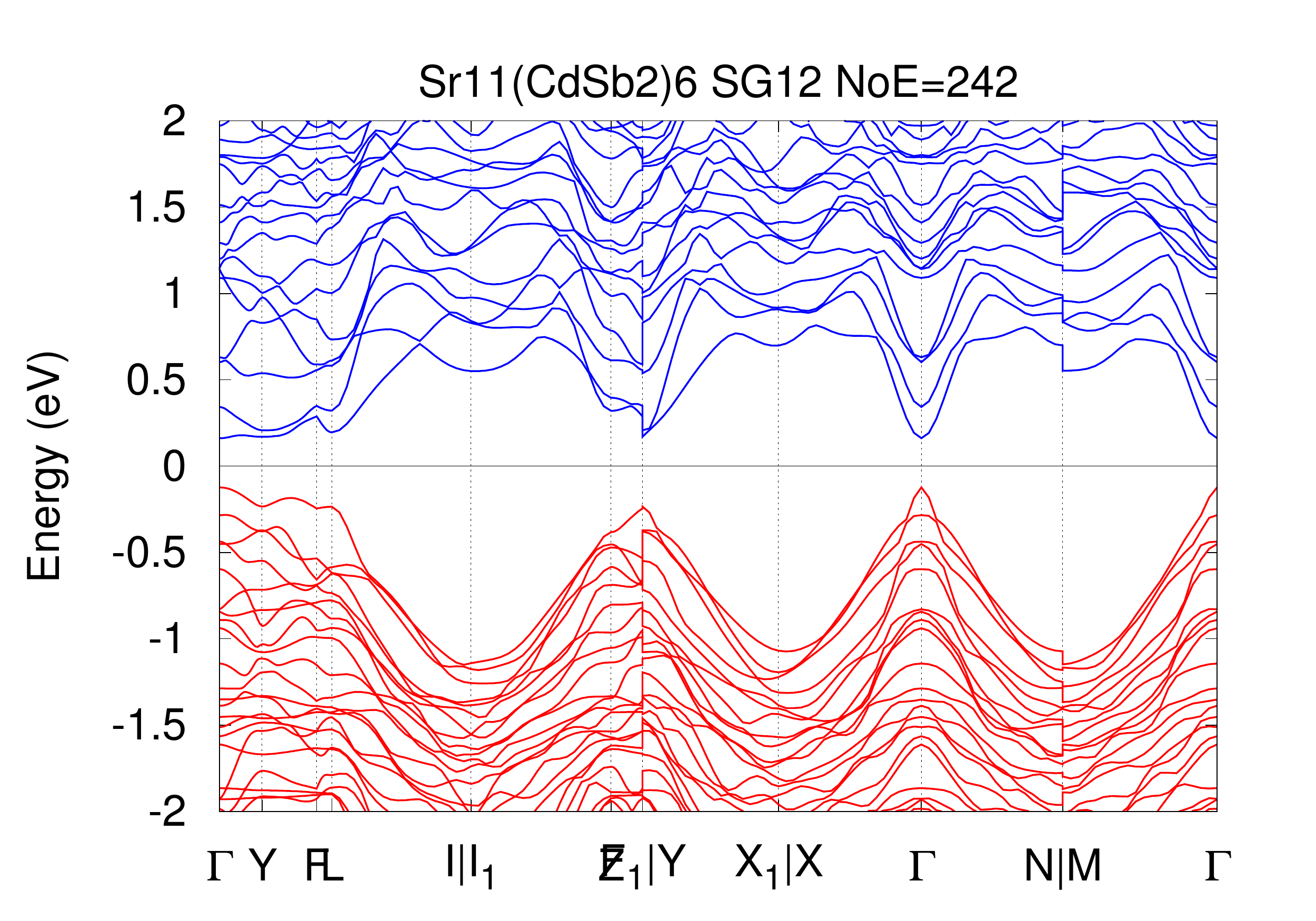}
}
\subfigure[Ta$_{6}$Fe$_{16}$Si$_{7}$ SG225 NoA=29 NoE=186]{
\label{subfig:107098}
\includegraphics[scale=0.32,angle=270]{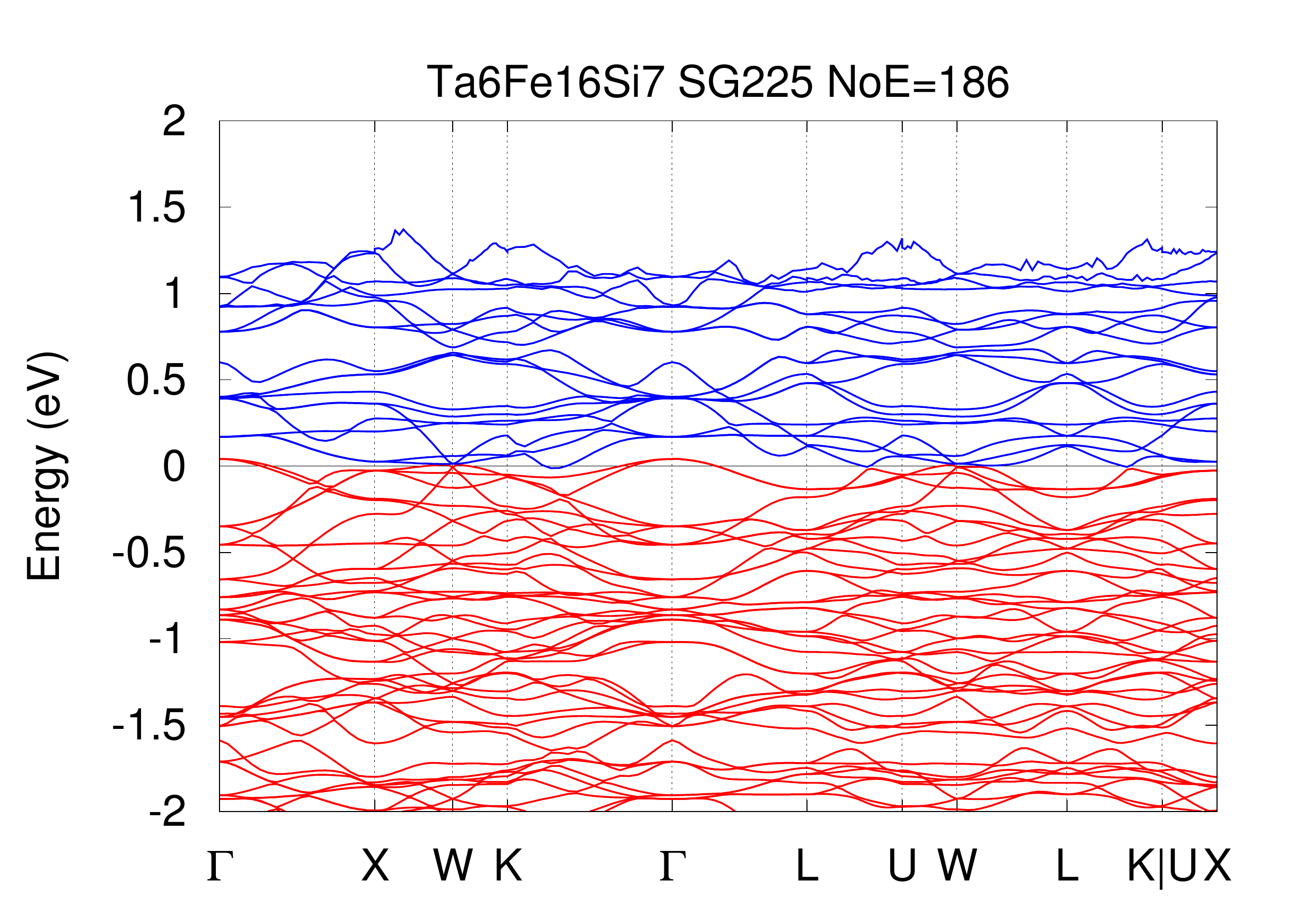}
}
\subfigure[Nb$_{6}$Fe$_{16}$Si$_{7}$ SG225 NoA=29 NoE=222]{
\label{subfig:107097}
\includegraphics[scale=0.32,angle=270]{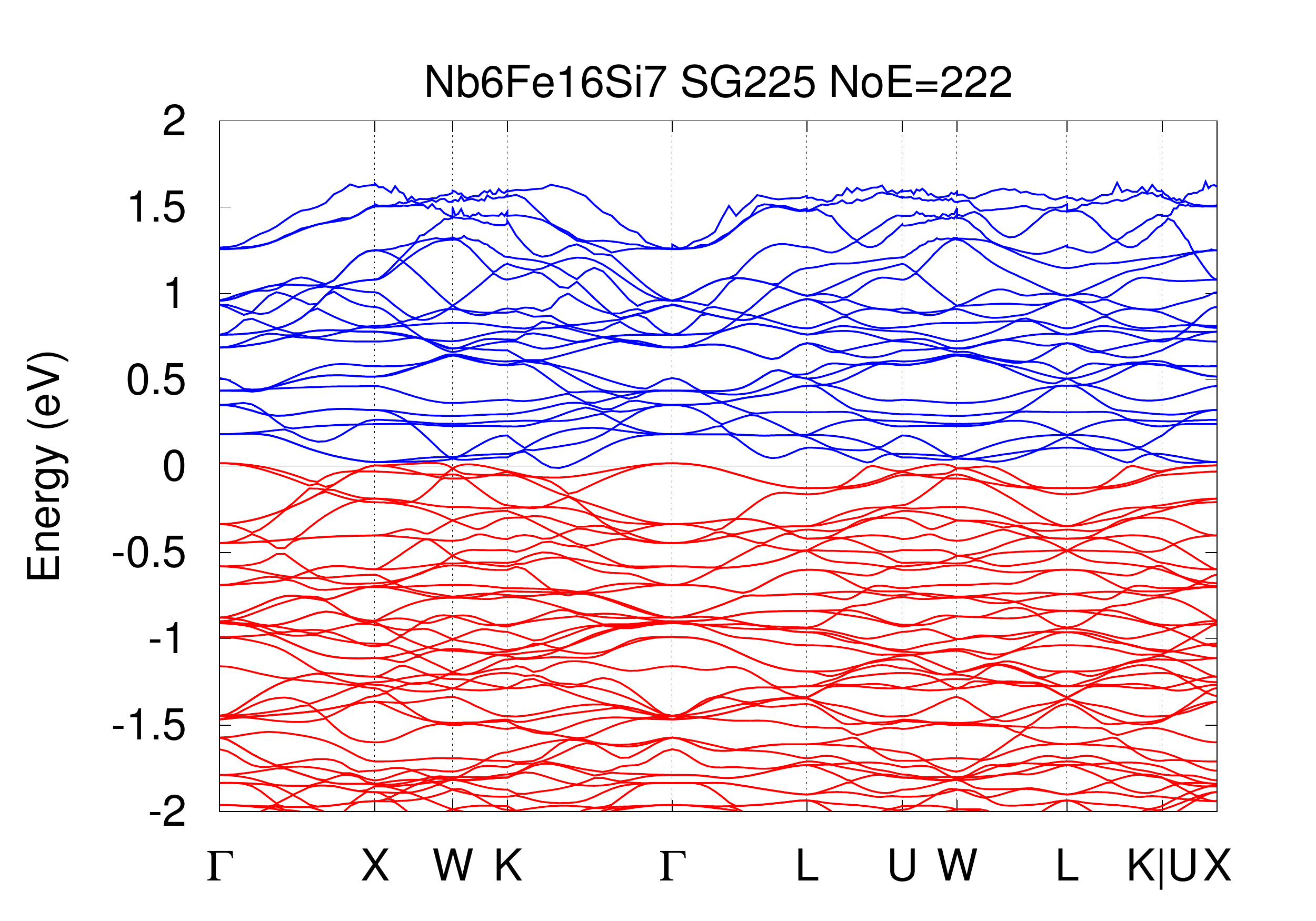}
}
\subfigure[Ba$_{4}$Li$_{2}$(CdAs$_{2}$)$_{3}$ SG63 NoA=30 NoE=216]{
\label{subfig:427778}
\includegraphics[scale=0.32,angle=270]{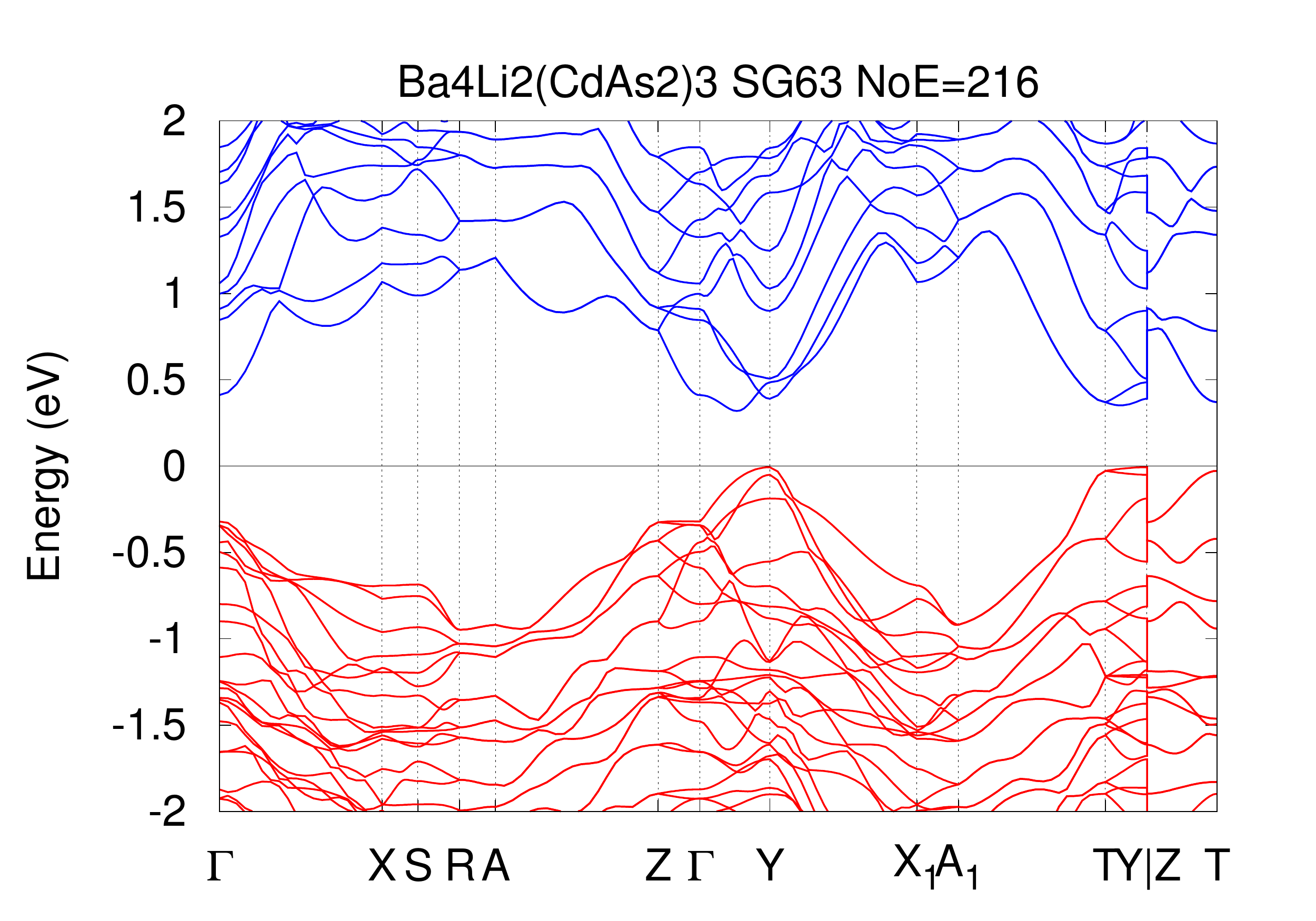}
}
\caption{\hyperref[tab:electride]{back to the table}}
\end{figure}

\begin{figure}[htp]
 \centering
\subfigure[Mn$_{2}$MoP$_{12}$ SG15 NoA=30 NoE=160]{
\label{subfig:68107}
\includegraphics[scale=0.32,angle=270]{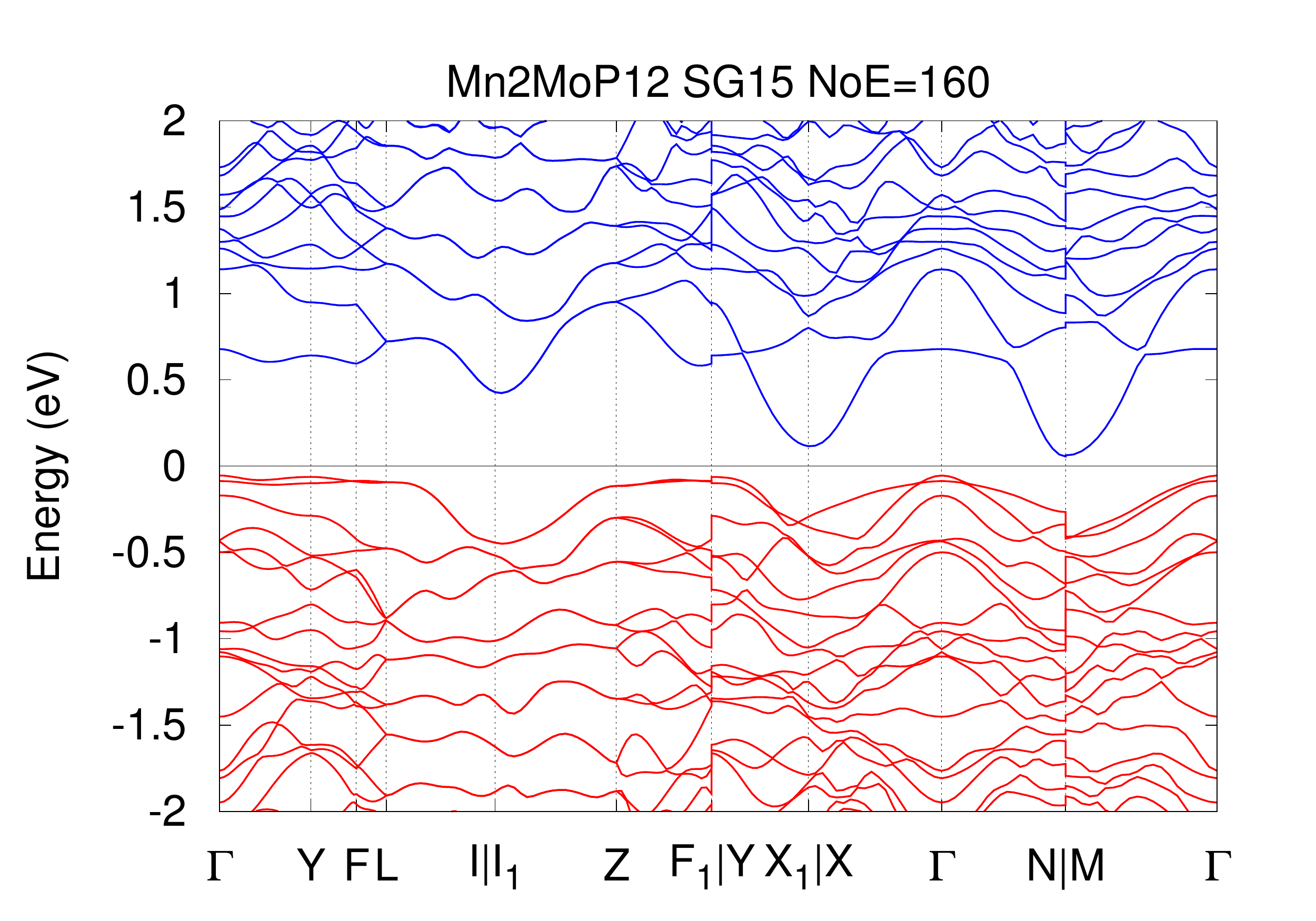}
}
\subfigure[Mn$_{2}$P$_{12}$W SG15 NoA=30 NoE=160]{
\label{subfig:643287}
\includegraphics[scale=0.32,angle=270]{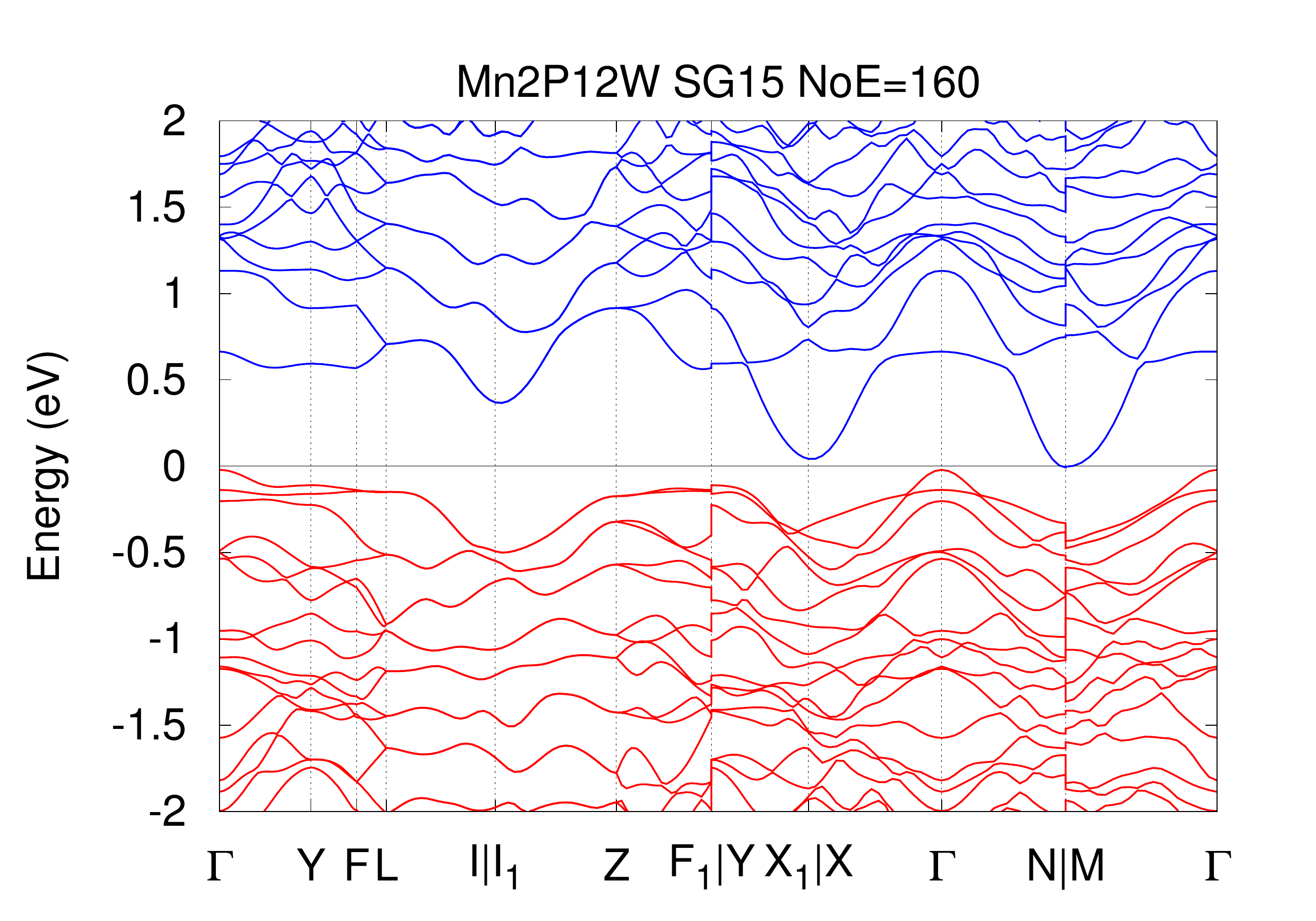}
}
\subfigure[Ta$_{3}$Ge SG86 NoA=32 NoE=152]{
\label{subfig:108870}
\includegraphics[scale=0.32,angle=270]{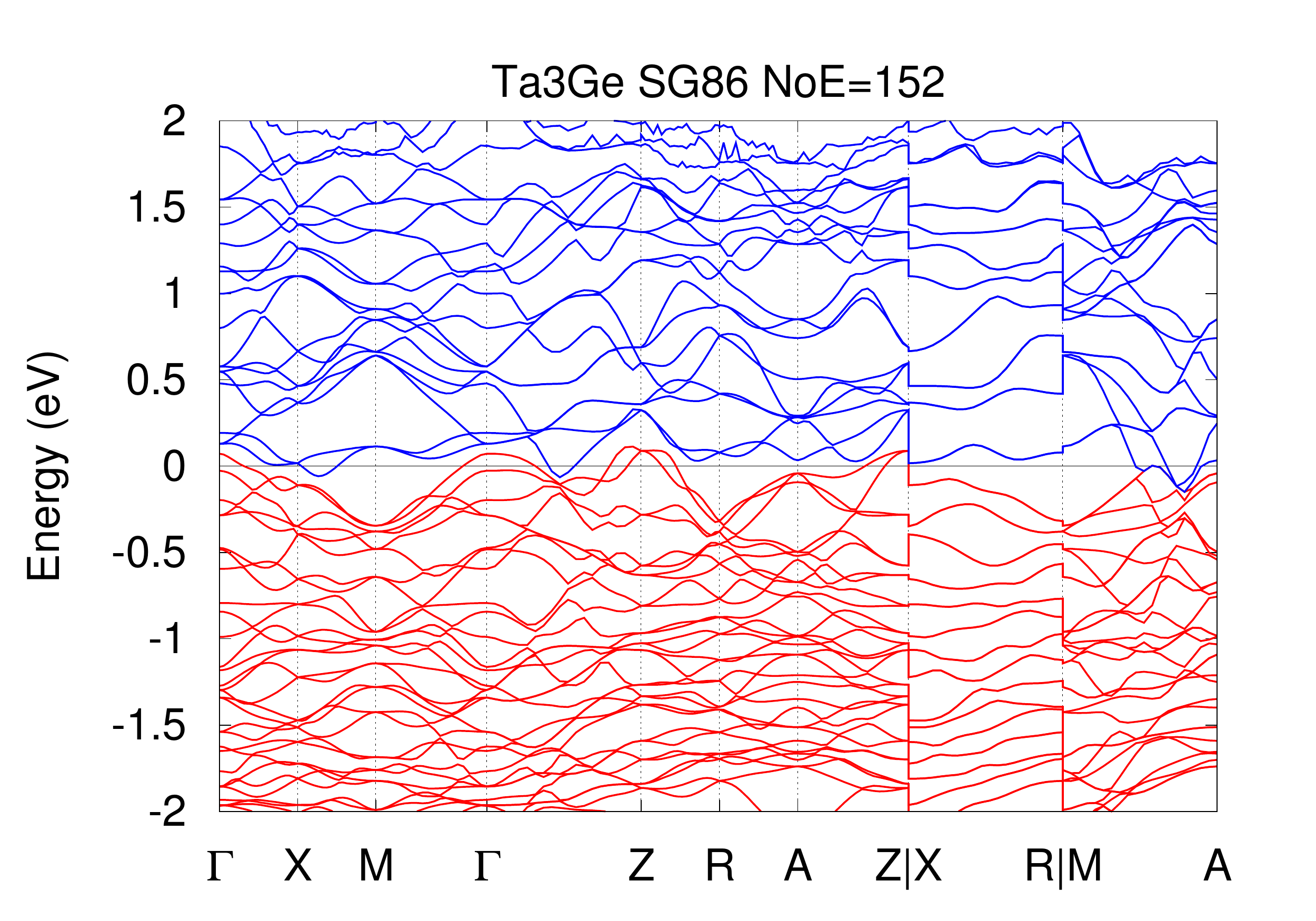}
}
\subfigure[SrSn$_{3}$Sb$_{4}$ SG62 NoA=32 NoE=168]{
\label{subfig:165617}
\includegraphics[scale=0.32,angle=270]{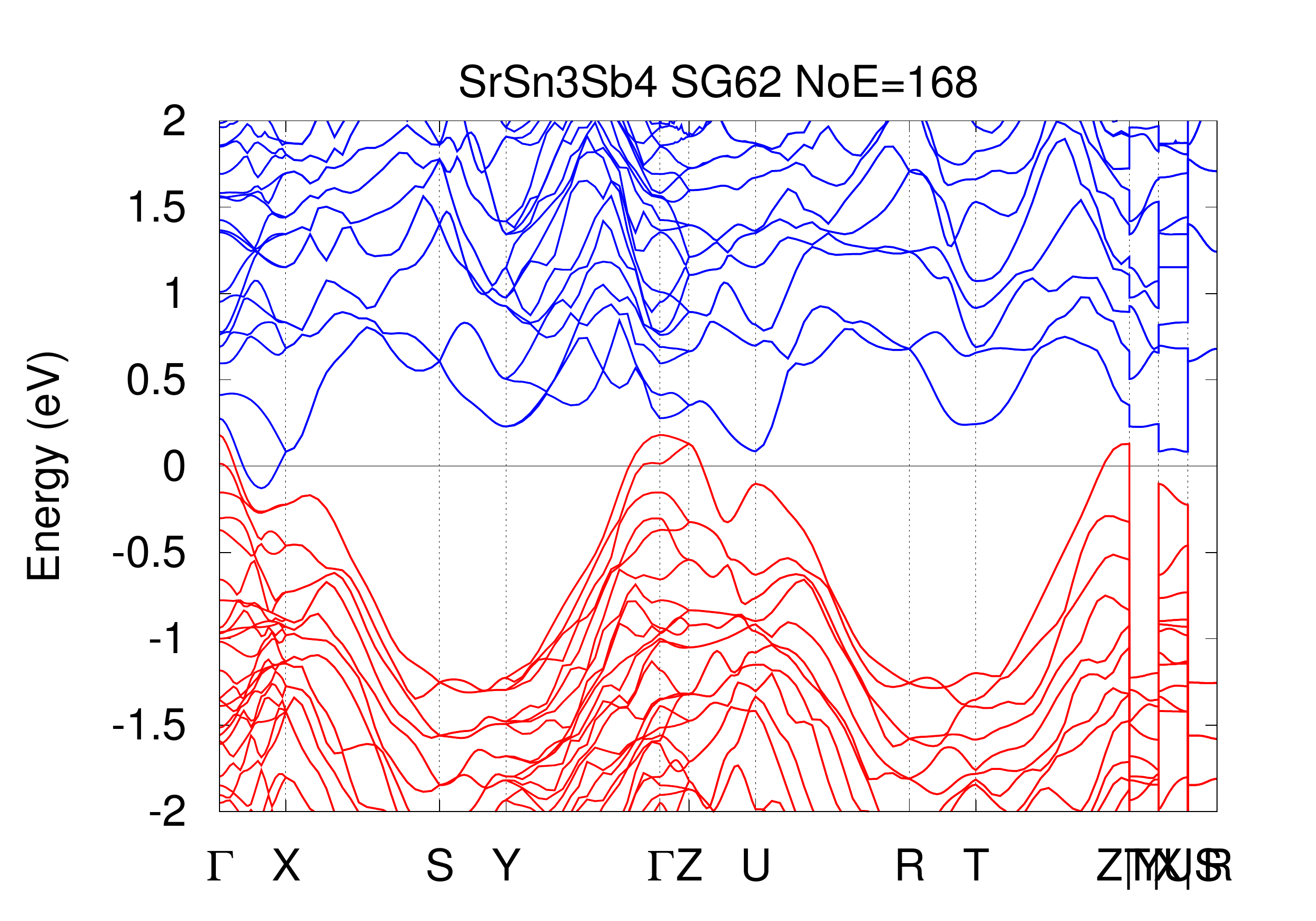}
}
\subfigure[KGe SG142 NoA=32 NoE=208]{
\label{subfig:636772}
\includegraphics[scale=0.32,angle=270]{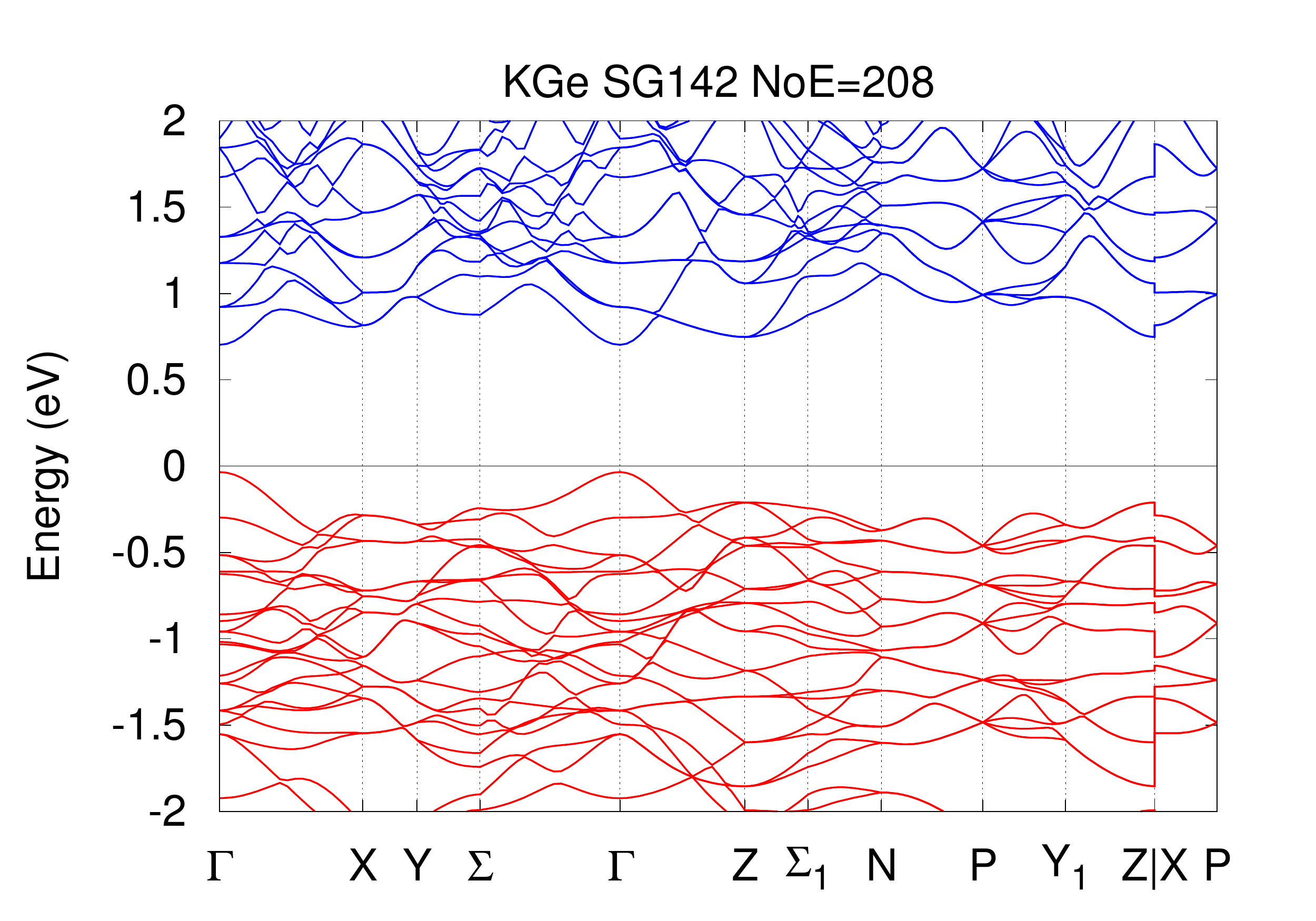}
}
\subfigure[NaSn SG142 NoA=32 NoE=80]{
\label{subfig:409434}
\includegraphics[scale=0.32,angle=270]{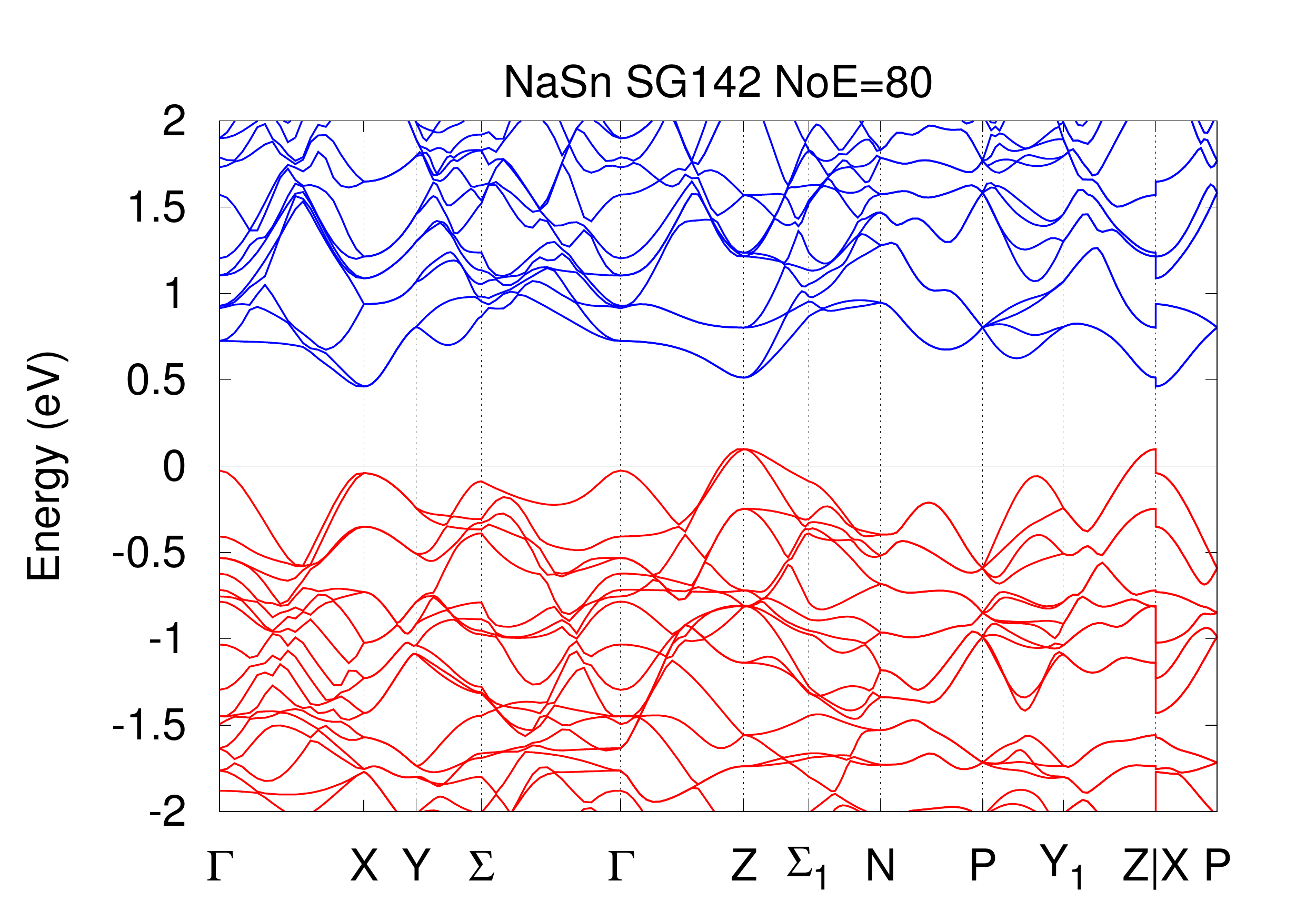}
}
\subfigure[CsSi SG142 NoA=32 NoE=208]{
\label{subfig:627104}
\includegraphics[scale=0.32,angle=270]{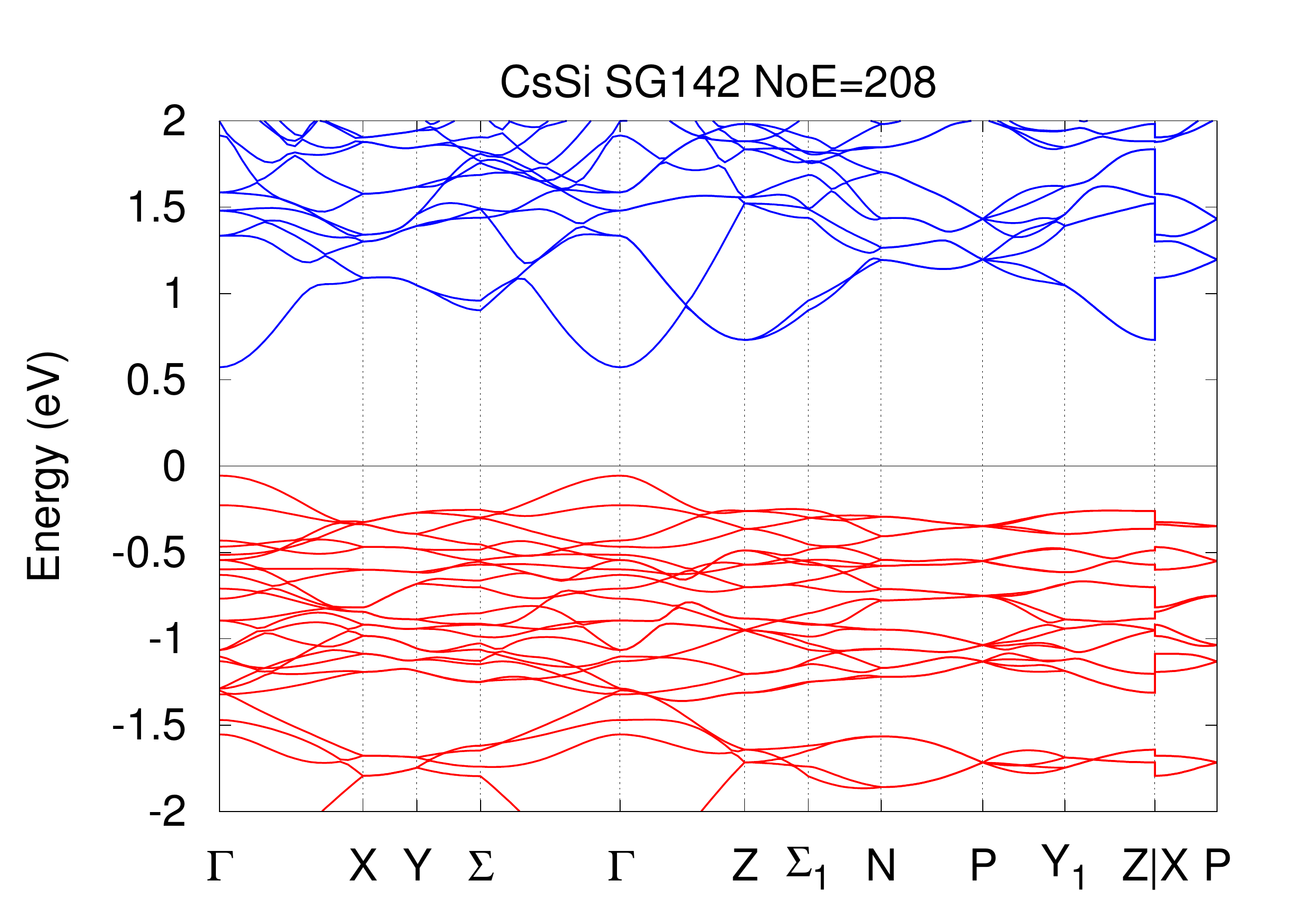}
}
\subfigure[CsSn SG142 NoA=32 NoE=208]{
\label{subfig:409439}
\includegraphics[scale=0.32,angle=270]{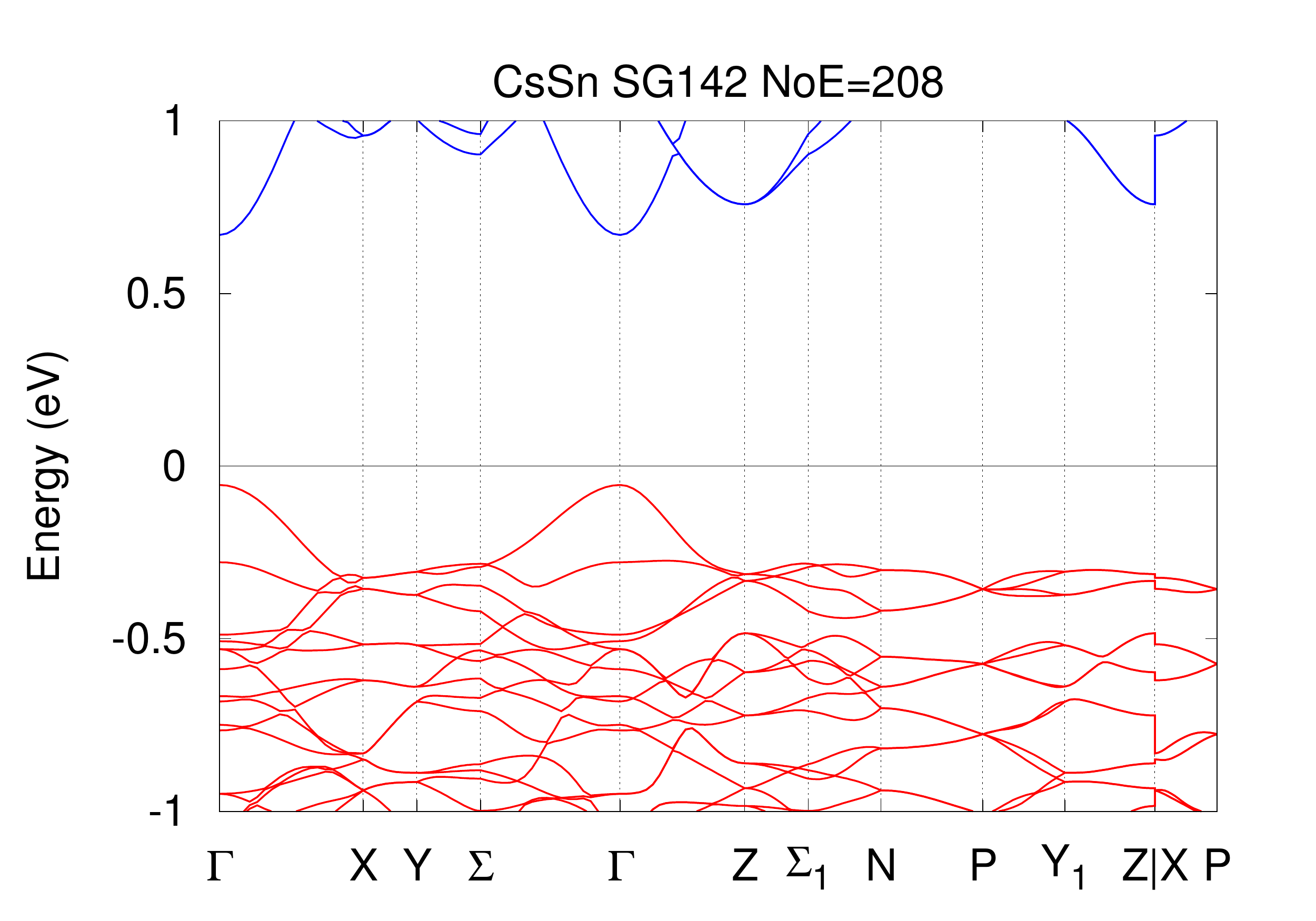}
}
\caption{\hyperref[tab:electride]{back to the table}}
\end{figure}

\begin{figure}[htp]
 \centering
\subfigure[NaPb SG142 NoA=32 NoE=80]{
\label{subfig:105156}
\includegraphics[scale=0.32,angle=270]{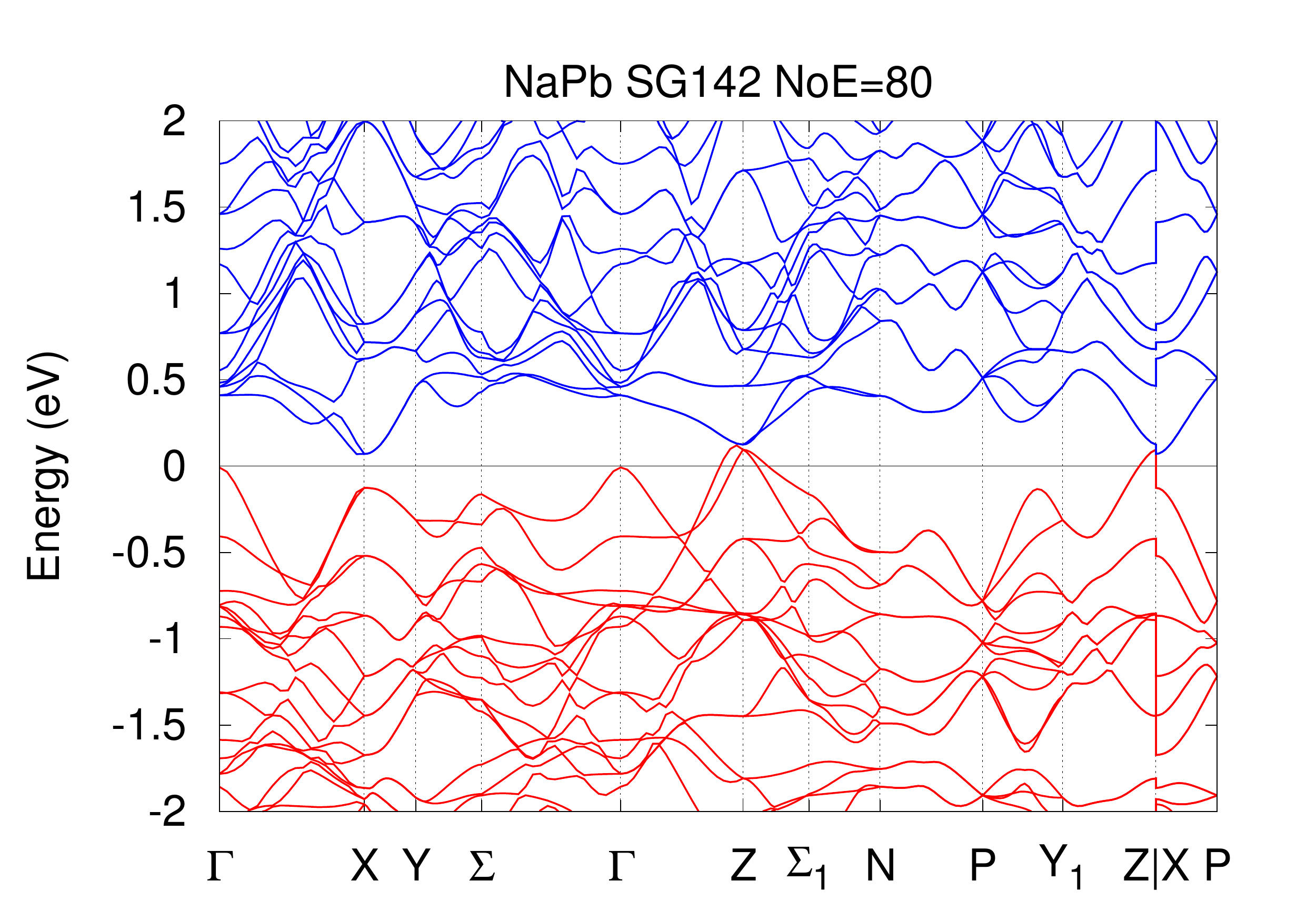}
}
\subfigure[KSn SG142 NoA=32 NoE=208]{
\label{subfig:409435}
\includegraphics[scale=0.32,angle=270]{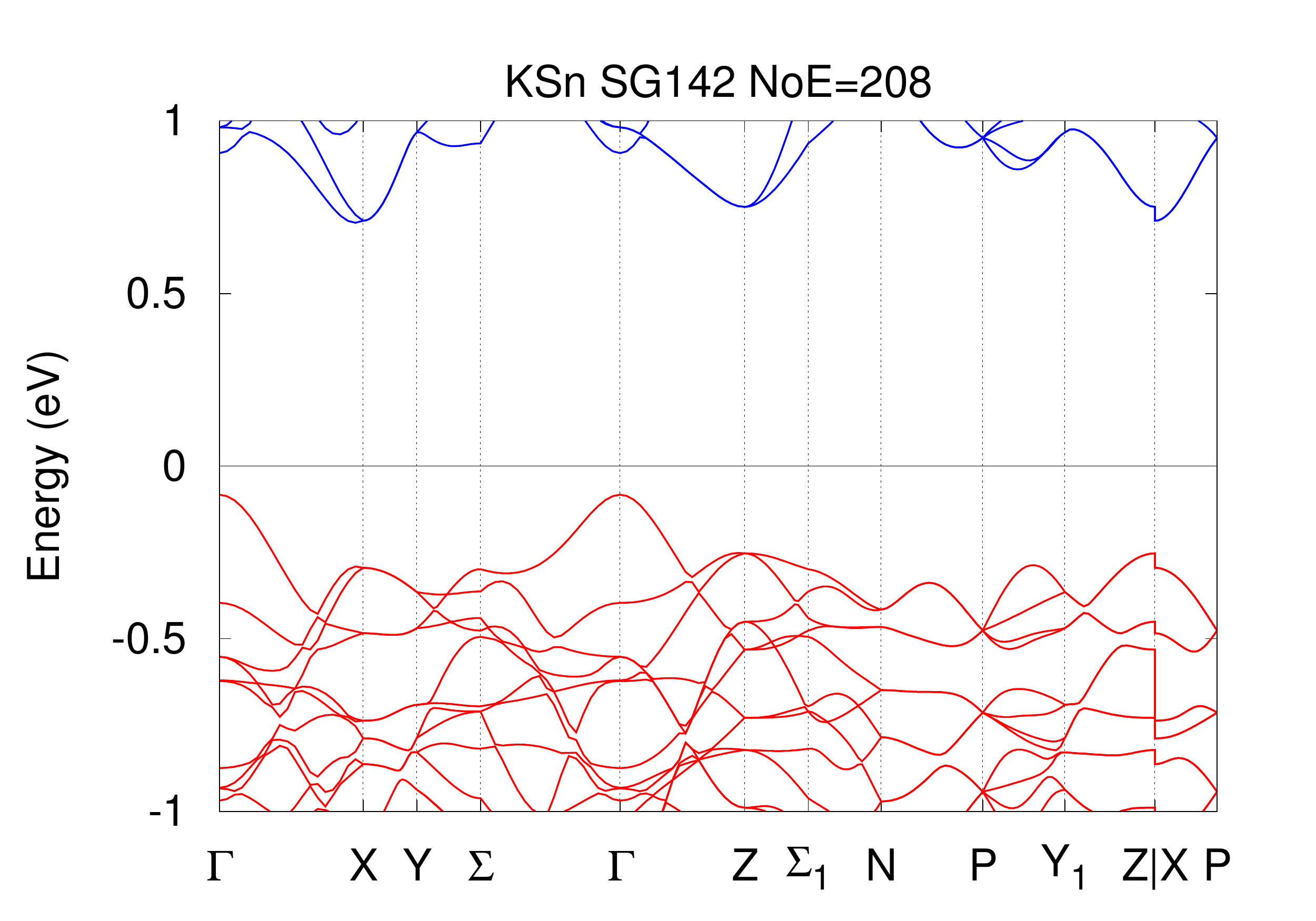}
}
\subfigure[Sb$_{2}$MoSe SG14 NoA=32 NoE=176]{
\label{subfig:280615}
\includegraphics[scale=0.32,angle=270]{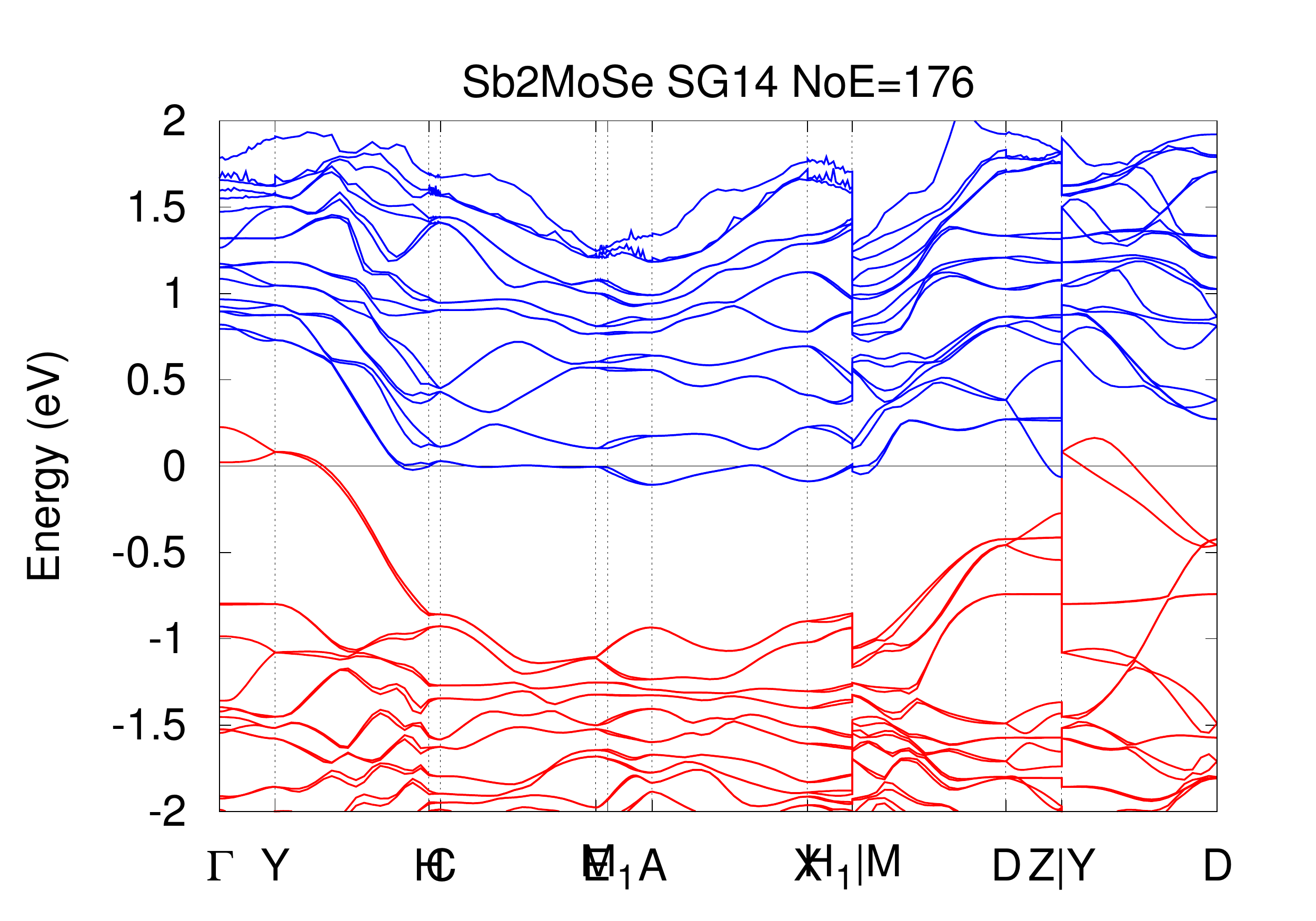}
}
\subfigure[RbNb$_{4}$Br$_{11}$ SG51 NoA=32 NoE=260]{
\label{subfig:380397}
\includegraphics[scale=0.32,angle=270]{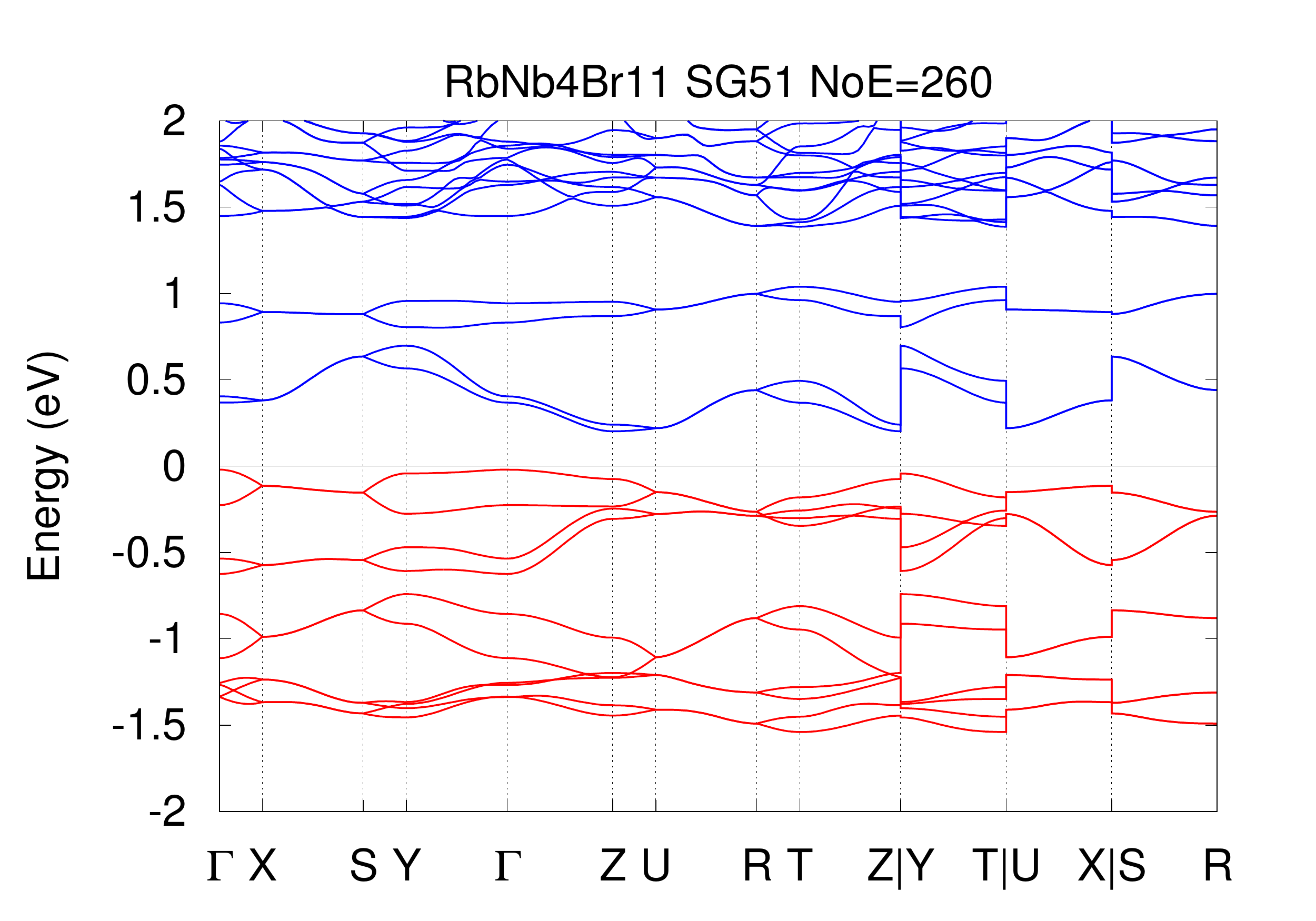}
}
\subfigure[Ta$_{4}$FeTe$_{4}$ SG55 NoA=36 NoE=208]{
\label{subfig:79796}
\includegraphics[scale=0.32,angle=270]{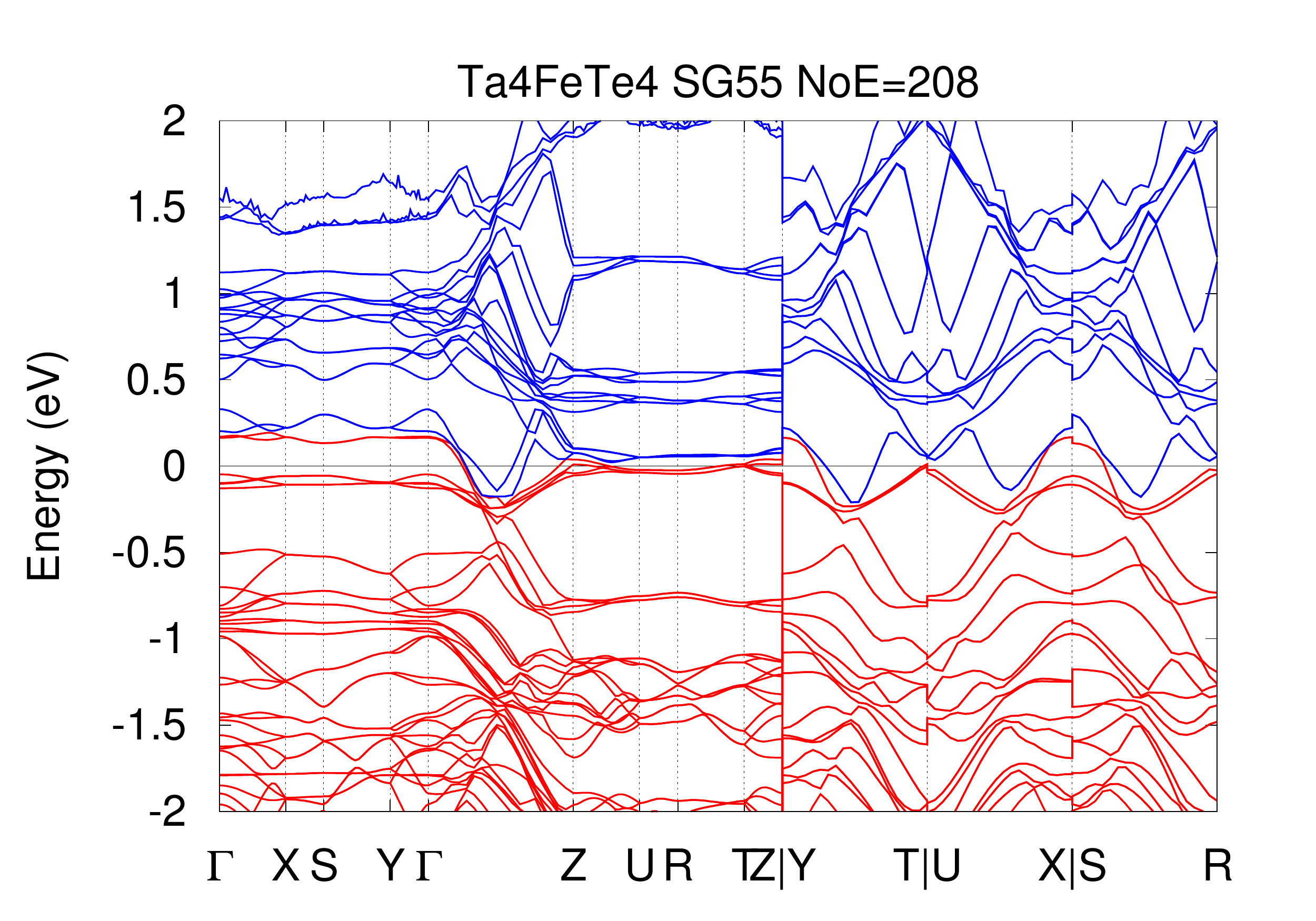}
}
\subfigure[TlCu$_{5}$Se$_{3}$ SG136 NoA=36 NoE=304]{
\label{subfig:88202}
\includegraphics[scale=0.32,angle=270]{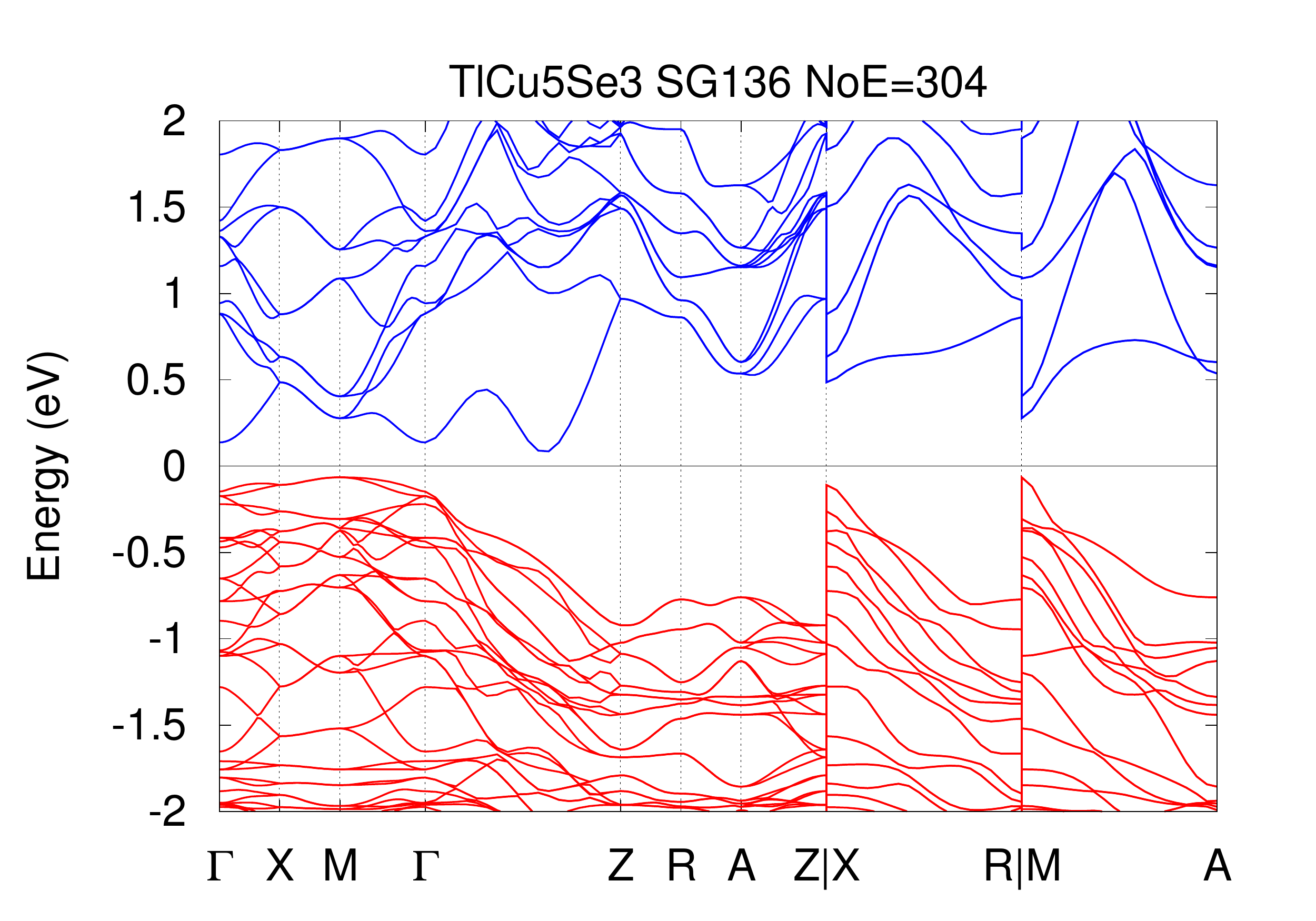}
}
\subfigure[K$_{6}$Na$_{14}$MgTl$_{18}$ SG200 NoA=39 NoE=124]{
\label{subfig:404695}
\includegraphics[scale=0.32,angle=270]{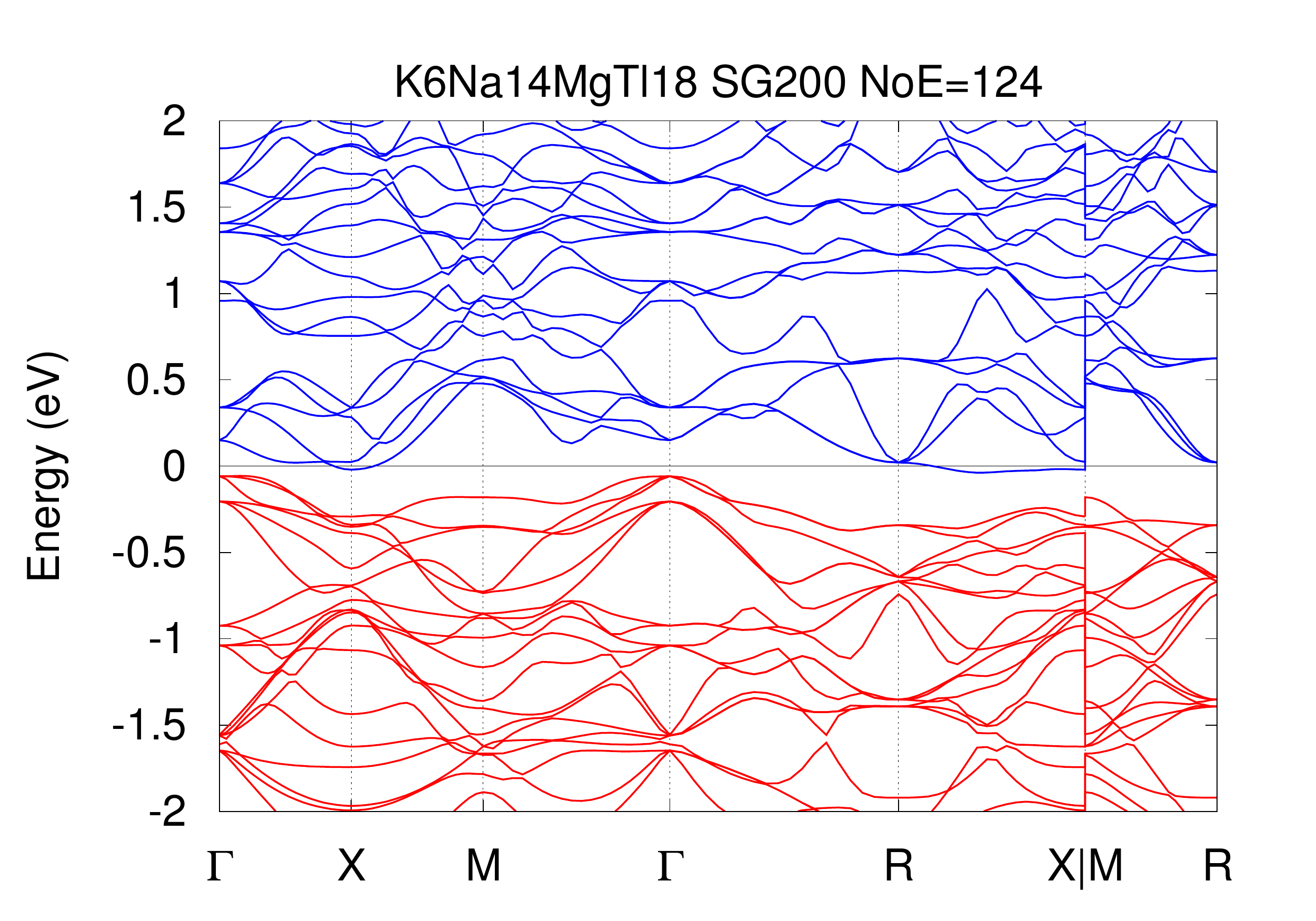}
}
\subfigure[K$_{6}$Na$_{14}$Tl$_{18}$Zn SG200 NoA=39 NoE=134]{
\label{subfig:236348}
\includegraphics[scale=0.32,angle=270]{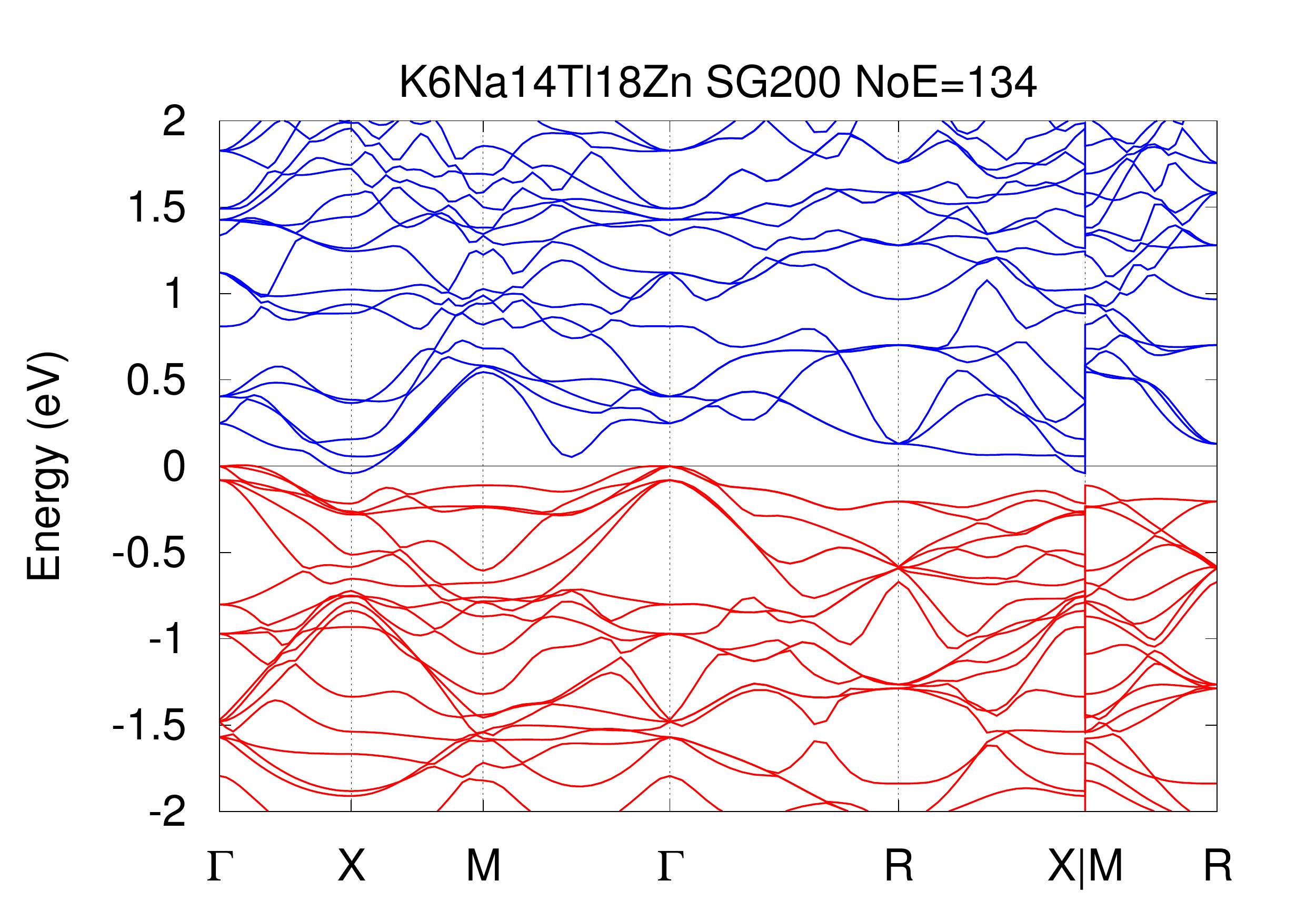}
}
\caption{\hyperref[tab:electride]{back to the table}}
\end{figure}

\begin{figure}[htp]
 \centering
\subfigure[Ge$_{3}$Os$_{2}$ SG60 NoA=40 NoE=224]{
\label{subfig:637466}
\includegraphics[scale=0.32,angle=270]{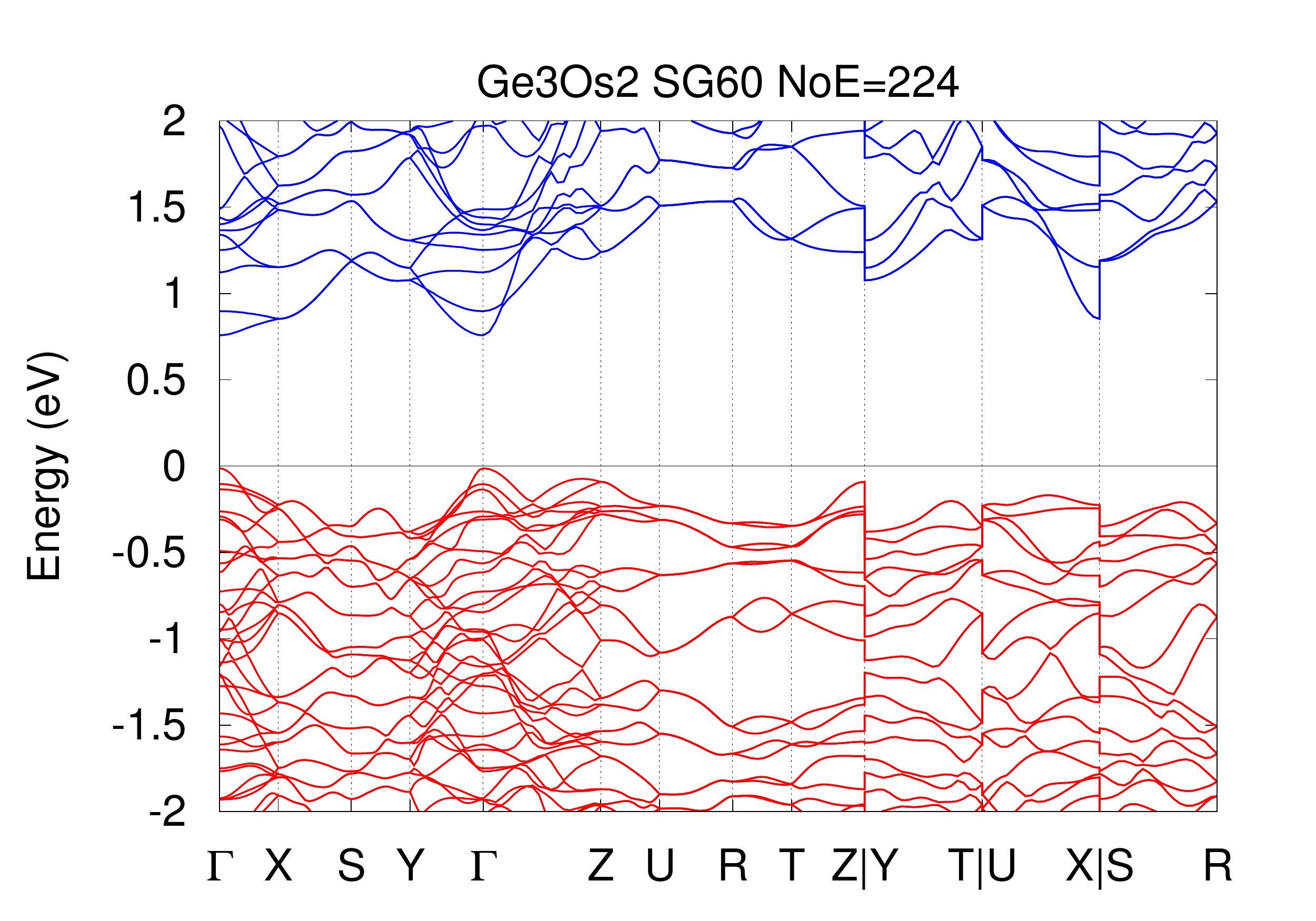}
}
\subfigure[Ge$_{3}$Ru$_{2}$ SG60 NoA=40 NoE=224]{
\label{subfig:637743}
\includegraphics[scale=0.32,angle=270]{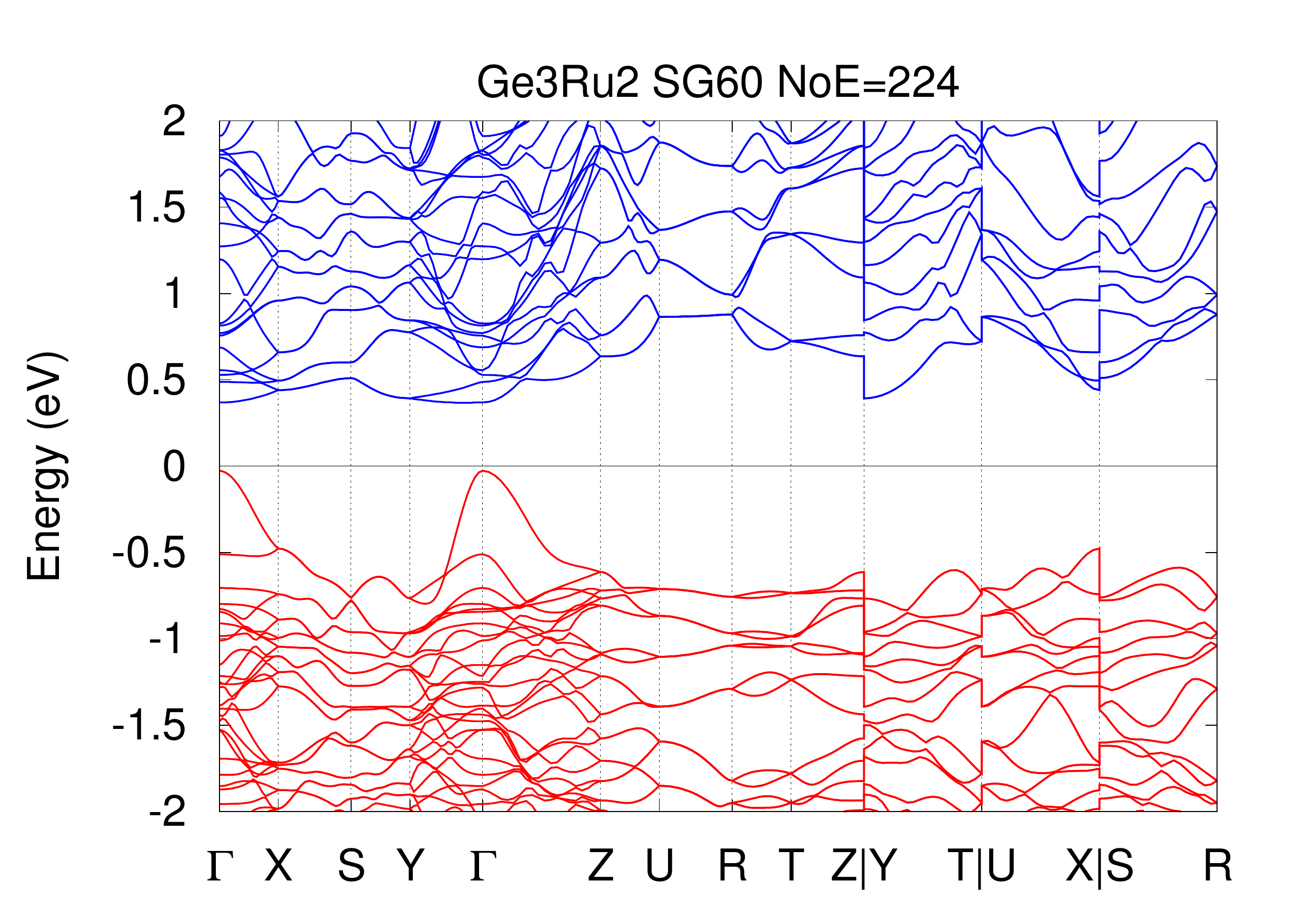}
}
\subfigure[Si$_{3}$Os$_{2}$ SG60 NoA=40 NoE=224]{
\label{subfig:647772}
\includegraphics[scale=0.32,angle=270]{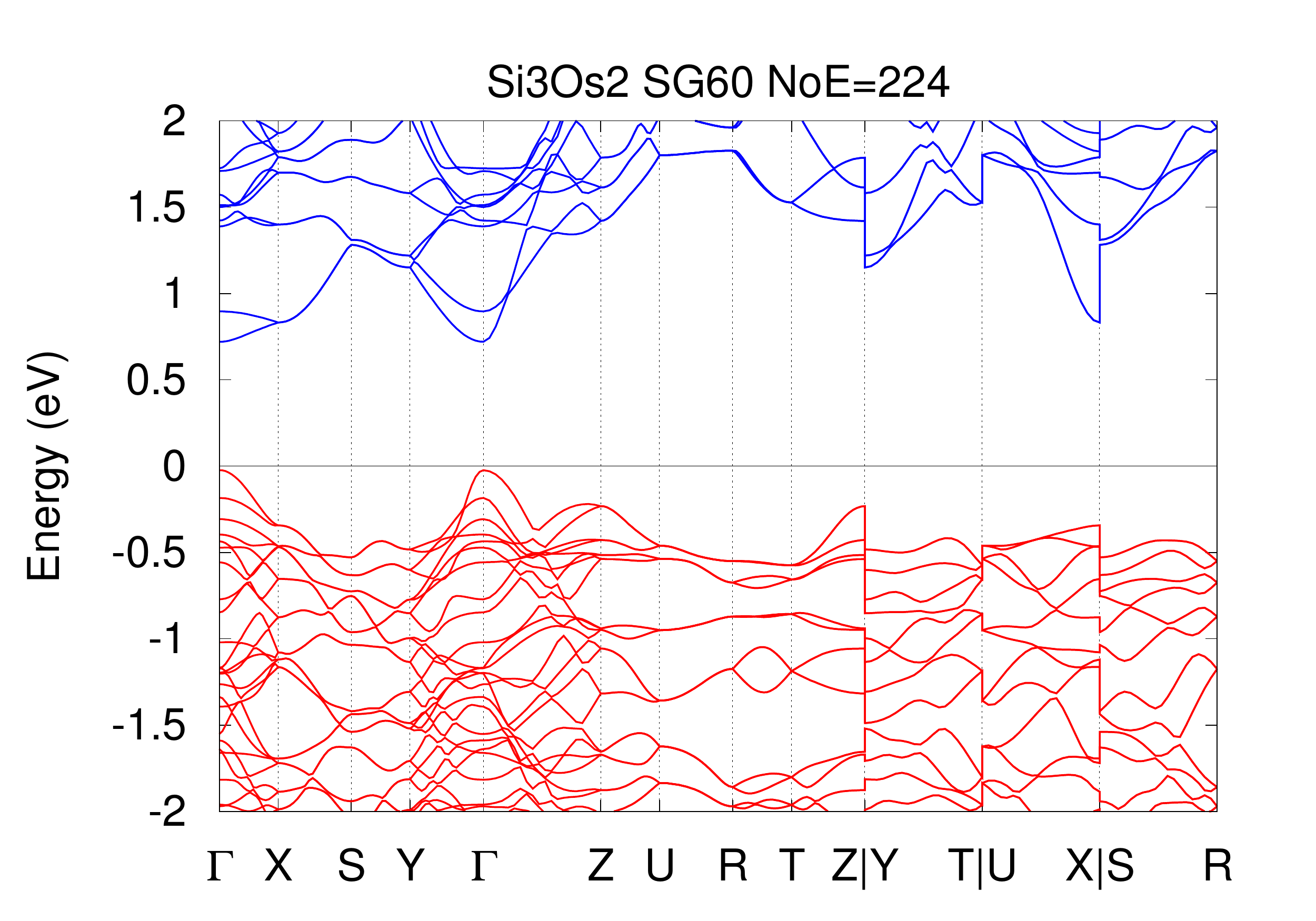}
}
\subfigure[Si$_{3}$Ru$_{2}$ SG60 NoA=40 NoE=224]{
\label{subfig:2344}
\includegraphics[scale=0.32,angle=270]{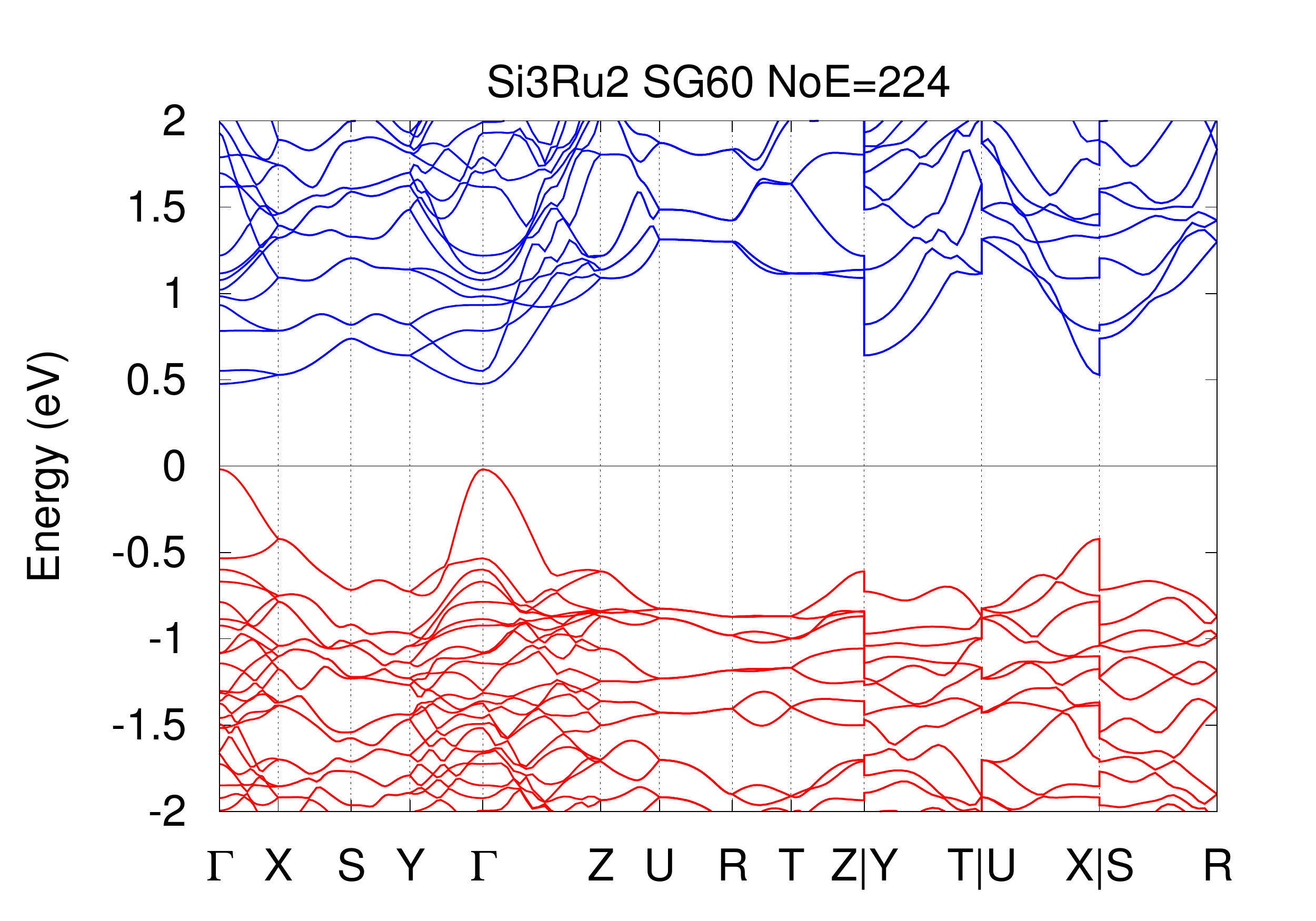}
}
\subfigure[Ba$_{11}$Sb$_{10}$ SG71 NoA=42 NoE=320]{
\label{subfig:413518}
\includegraphics[scale=0.32,angle=270]{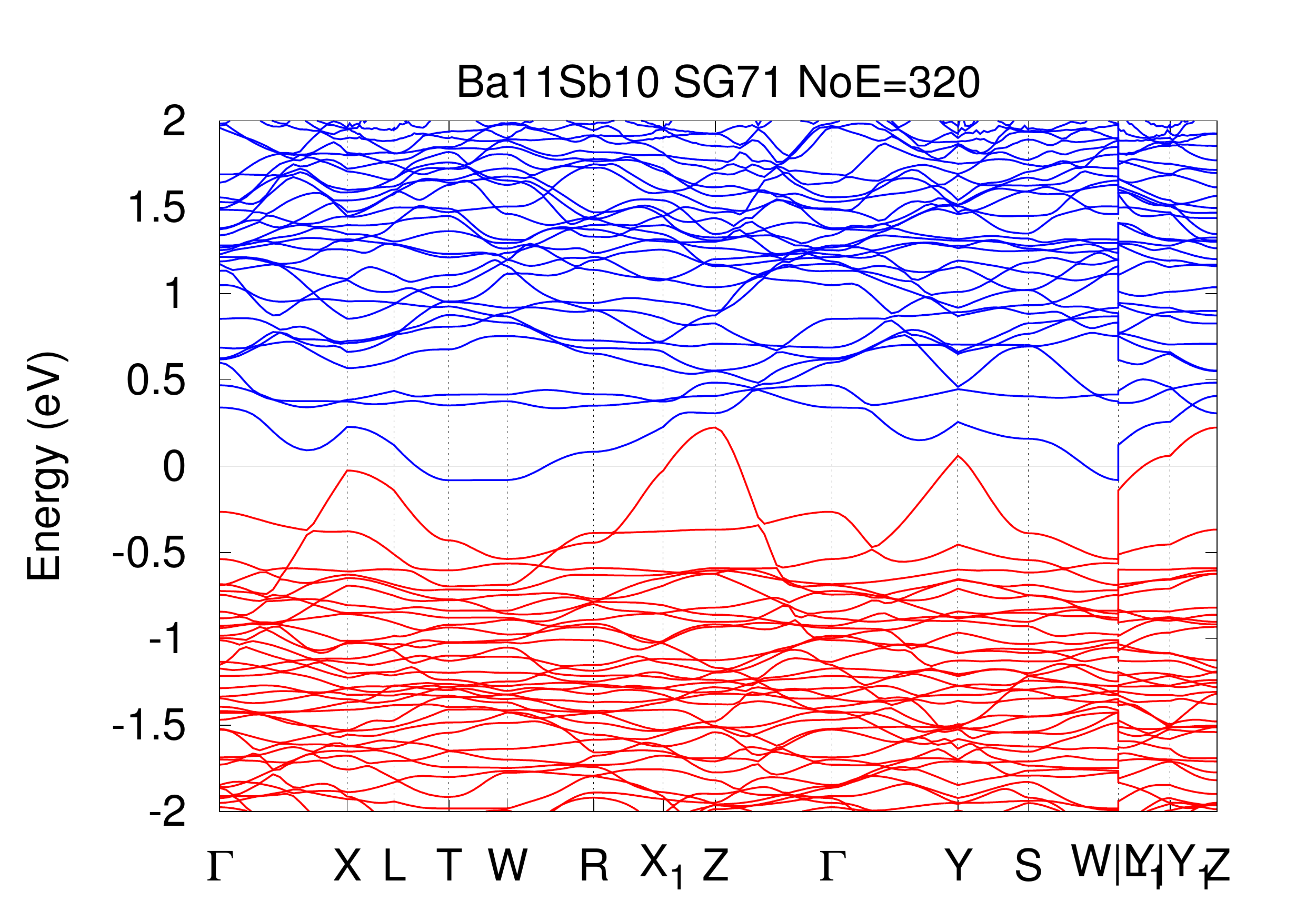}
}
\subfigure[Tb$_{4}$BBr$_{6}$ SG15 NoA=44 NoE=324]{
\label{subfig:410997}
\includegraphics[scale=0.32,angle=270]{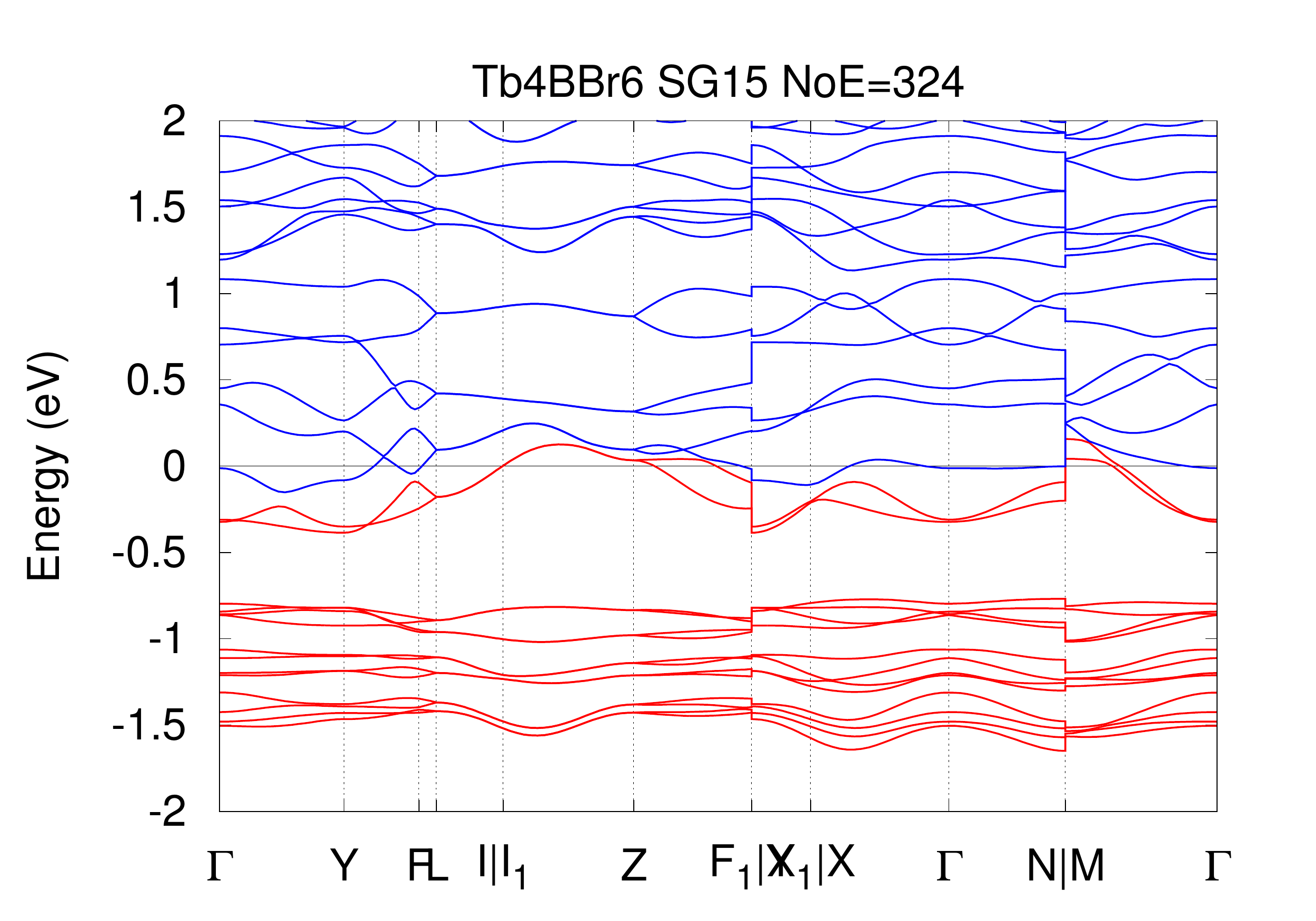}
}
\subfigure[Ba$_{2}$(SnSb$_{2}$)$_{3}$ SG62 NoA=44 NoE=248]{
\label{subfig:82529}
\includegraphics[scale=0.32,angle=270]{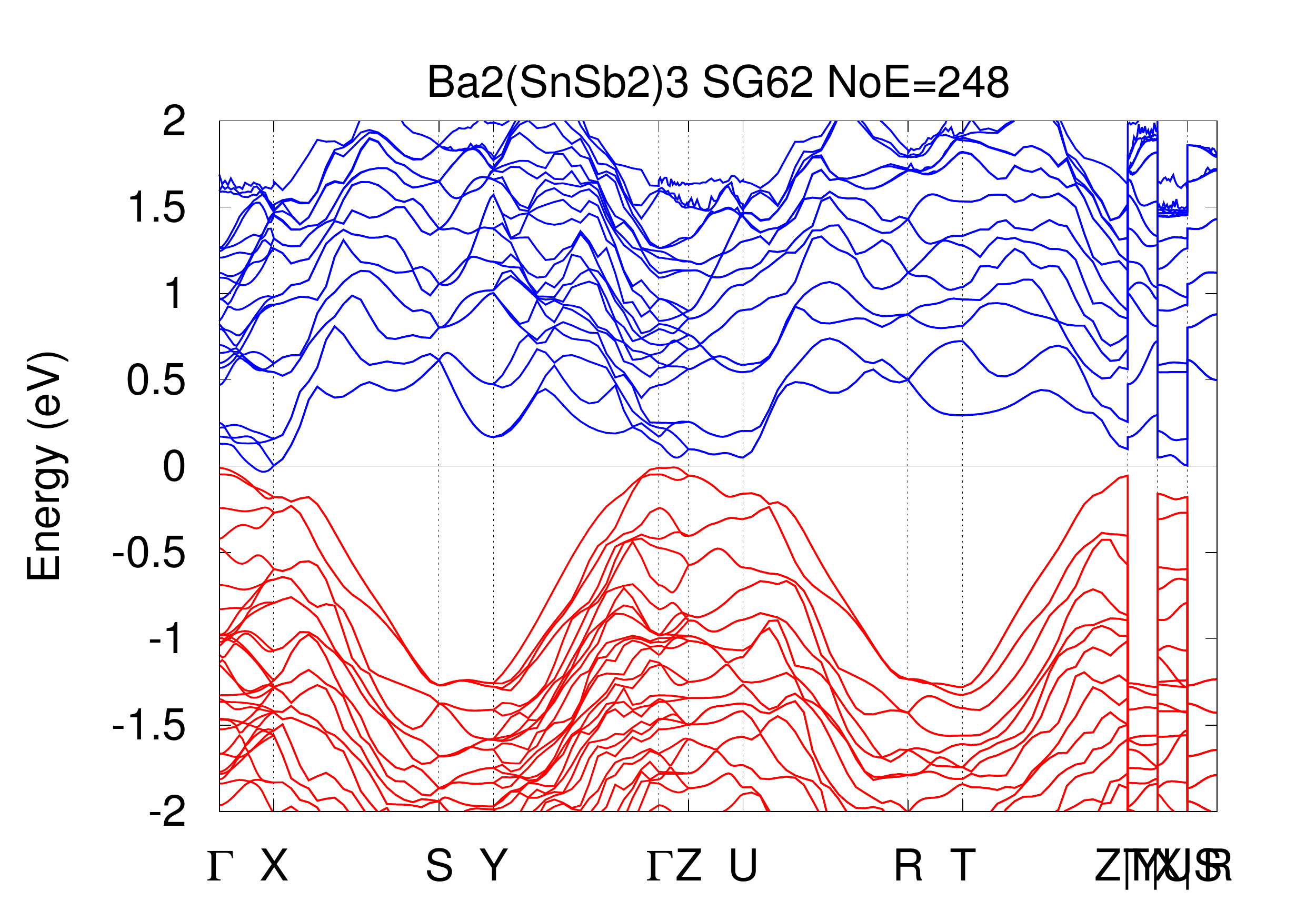}
}
\subfigure[YB$_{7}$Mo$_{3}$ SG62 NoA=44 NoE=200]{
\label{subfig:39873}
\includegraphics[scale=0.32,angle=270]{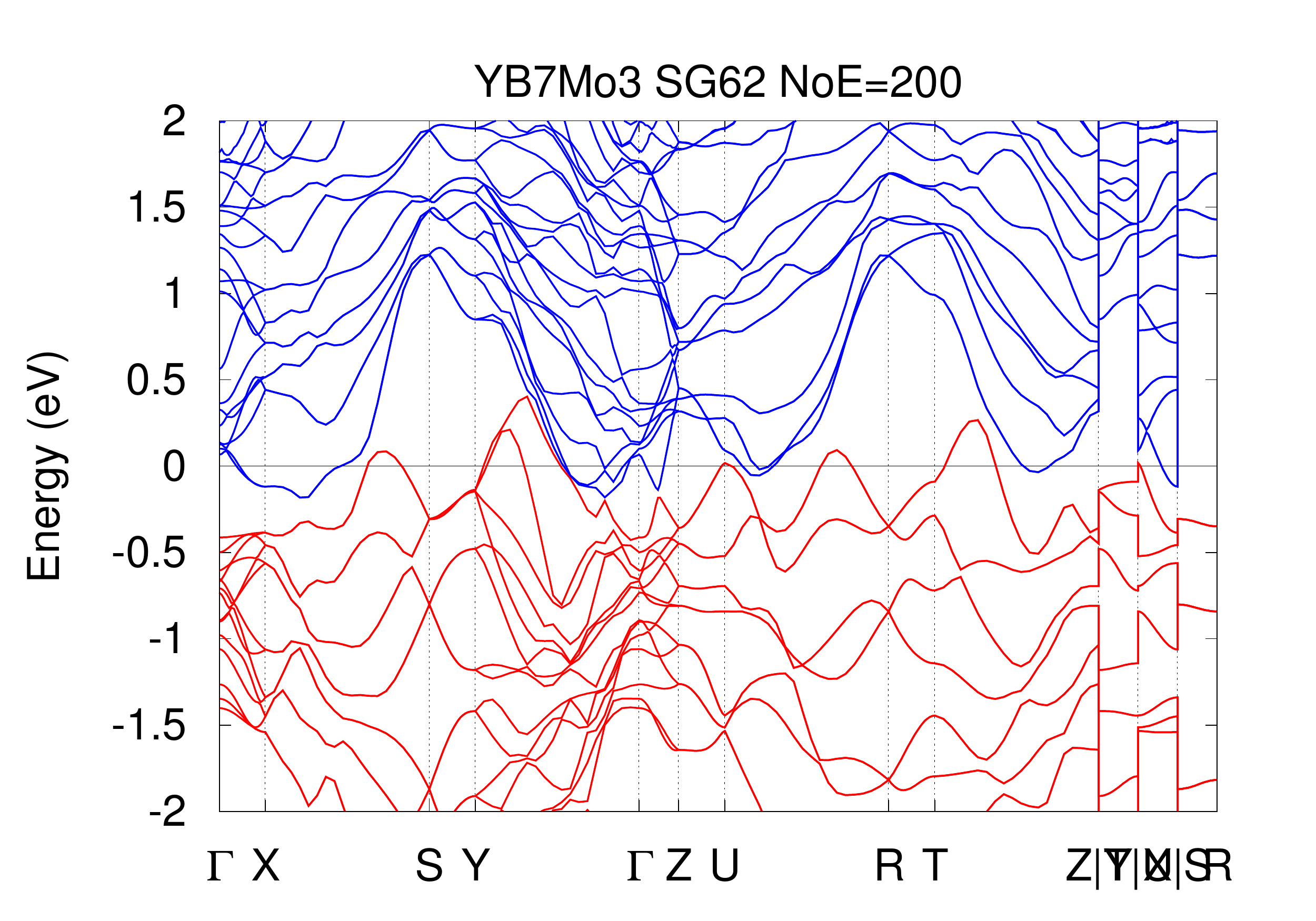}
}
\caption{\hyperref[tab:electride]{back to the table}}
\end{figure}

\begin{figure}[H]
 \centering
\subfigure[Y$_{4}$BBr$_{6}$ SG15 NoA=44 NoE=356]{
\label{subfig:402661}
\includegraphics[scale=0.32,angle=270]{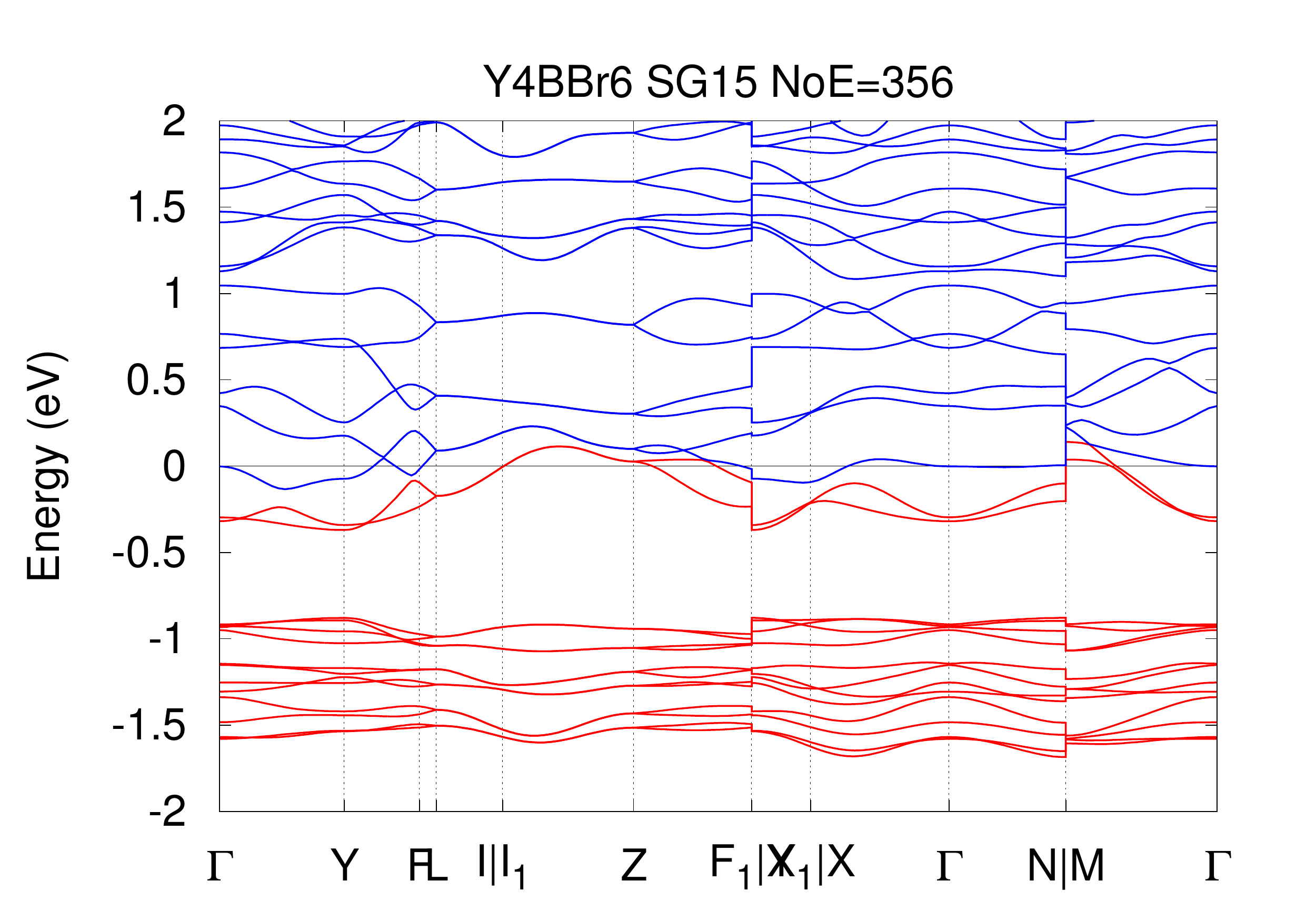}
}
\subfigure[SrNb$_{8}$O$_{14}$ SG55 NoA=46 NoE=364]{
\label{subfig:202674}
\includegraphics[scale=0.32,angle=270]{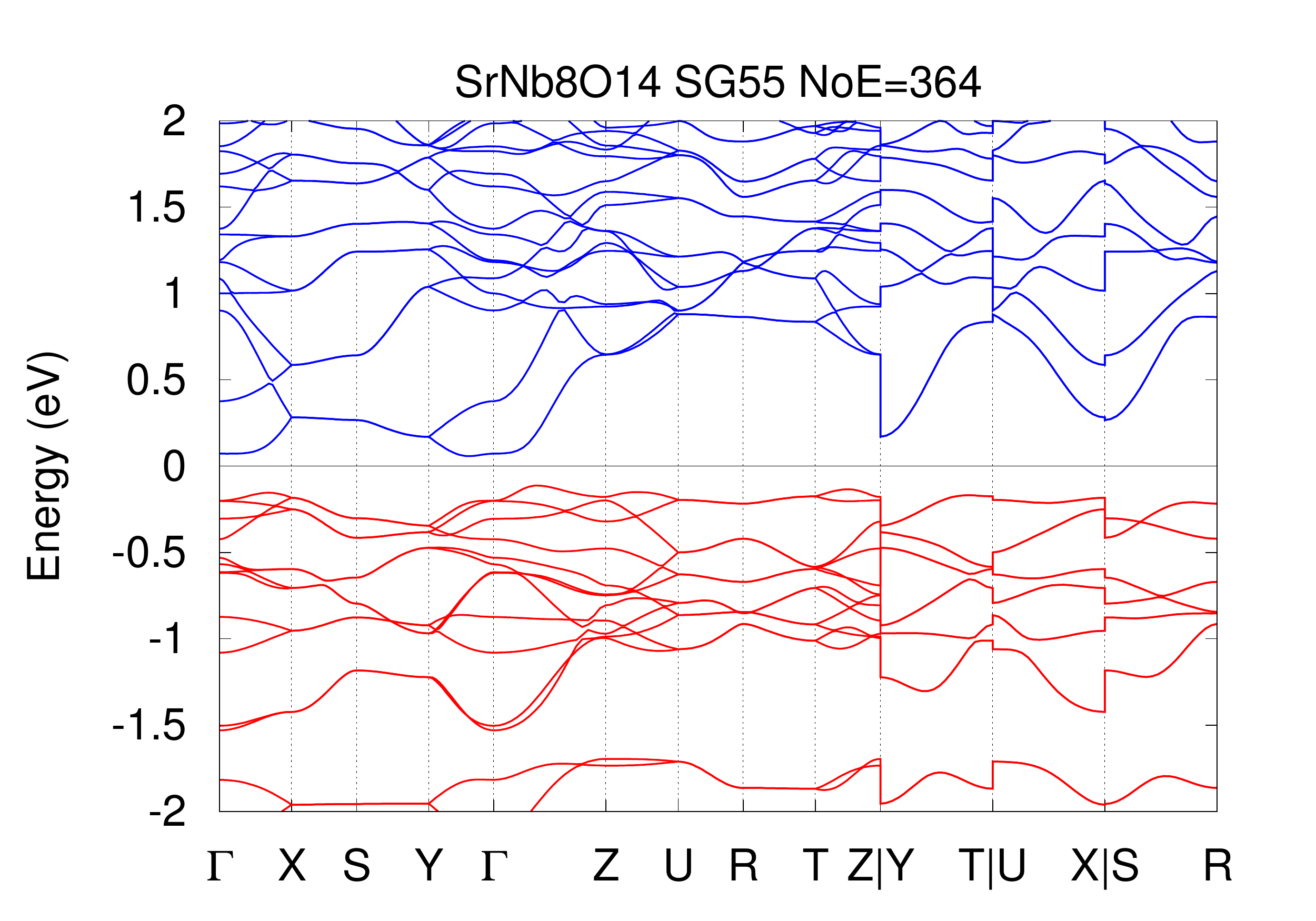}
}
\subfigure[BaNb$_{8}$O$_{14}$ SG55 NoA=46 NoE=364]{
\label{subfig:79976}
\includegraphics[scale=0.32,angle=270]{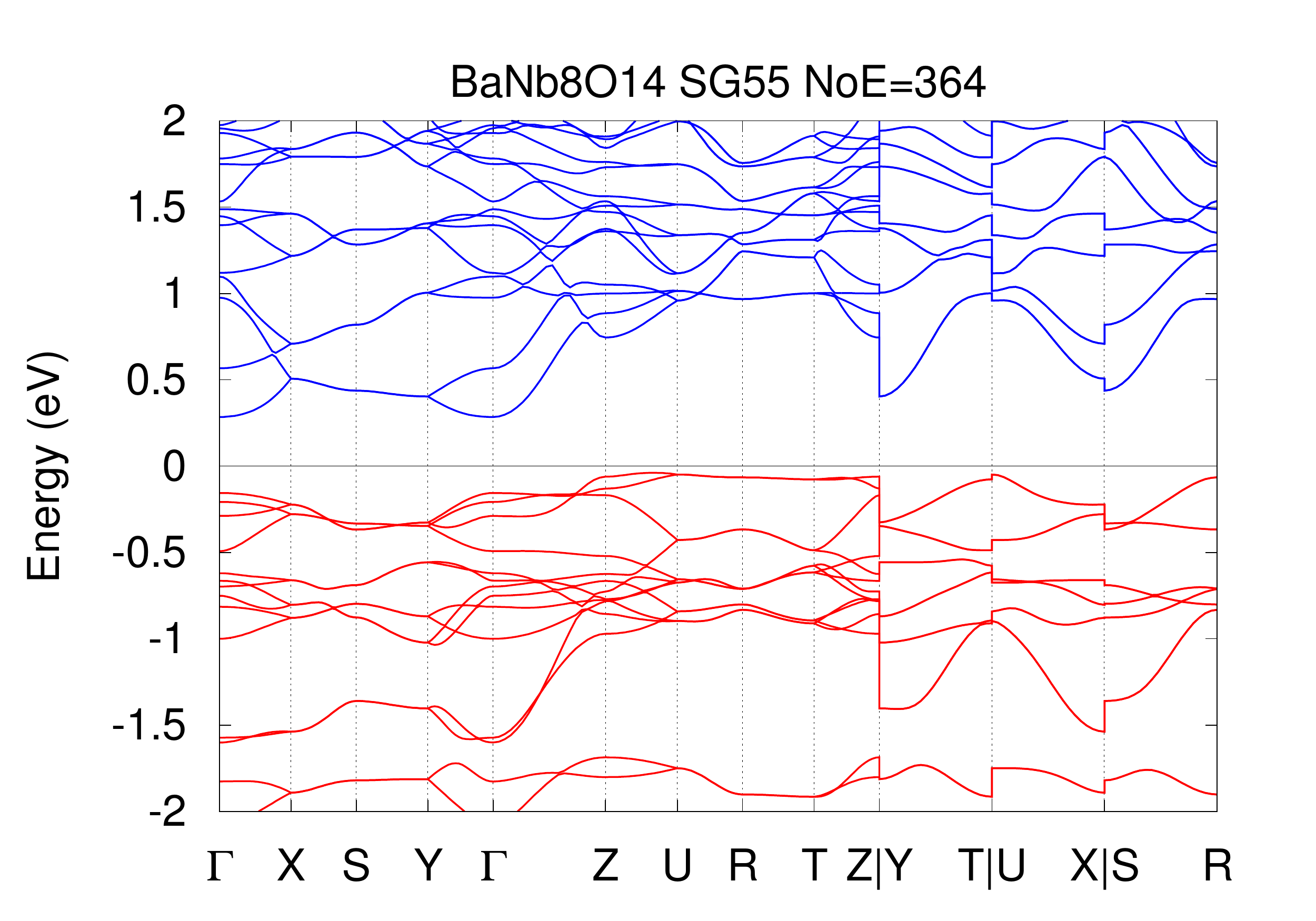}
}
\subfigure[K(NaIn$_{3}$)$_{3}$ SG64 NoA=52 NoE=156]{
\label{subfig:165180}
\includegraphics[scale=0.32,angle=270]{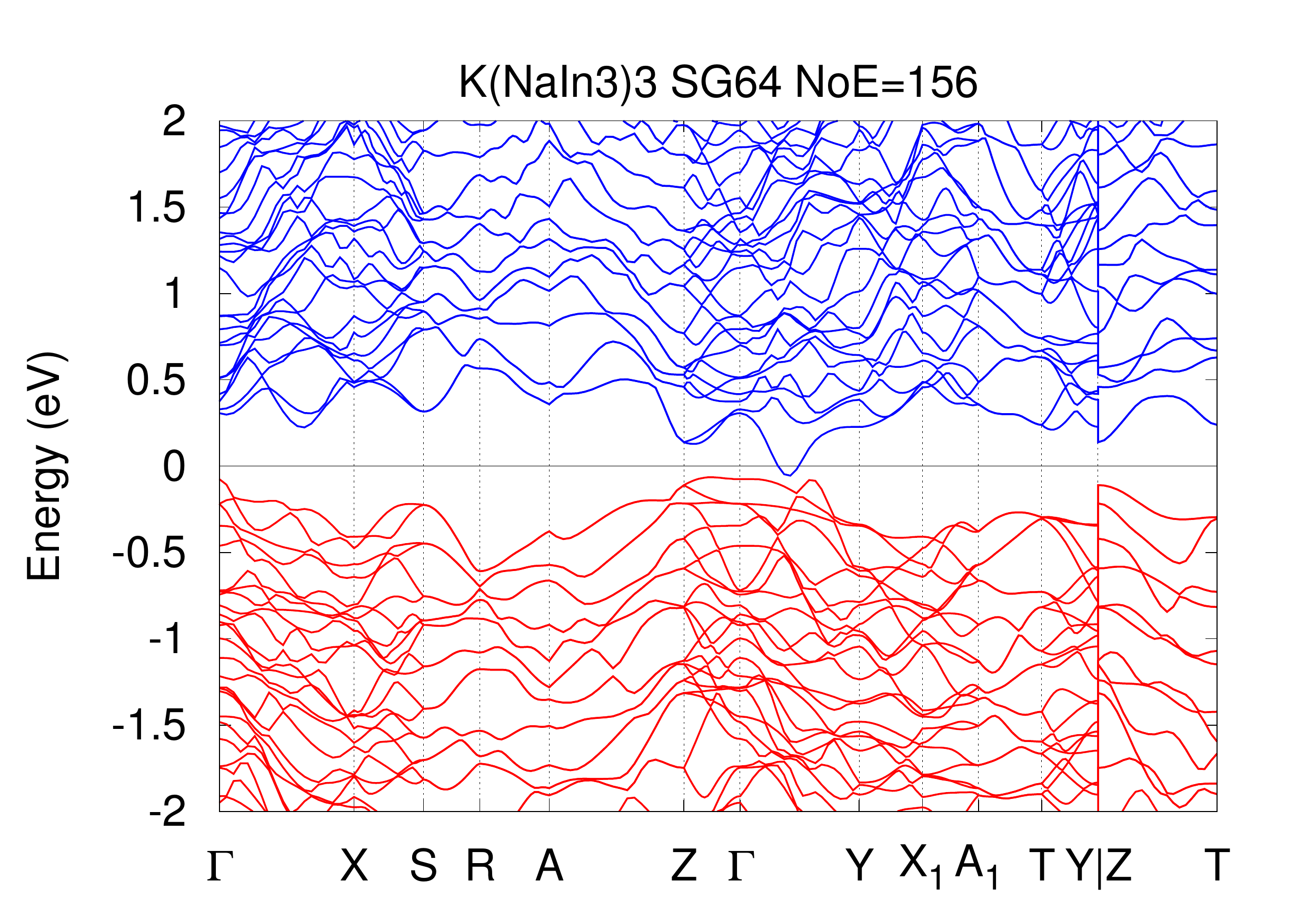}
}
\subfigure[Nb$_{3}$Se$_{12}$I SG128 NoA=64 NoE=448]{
\label{subfig:72021}
\includegraphics[scale=0.32,angle=270]{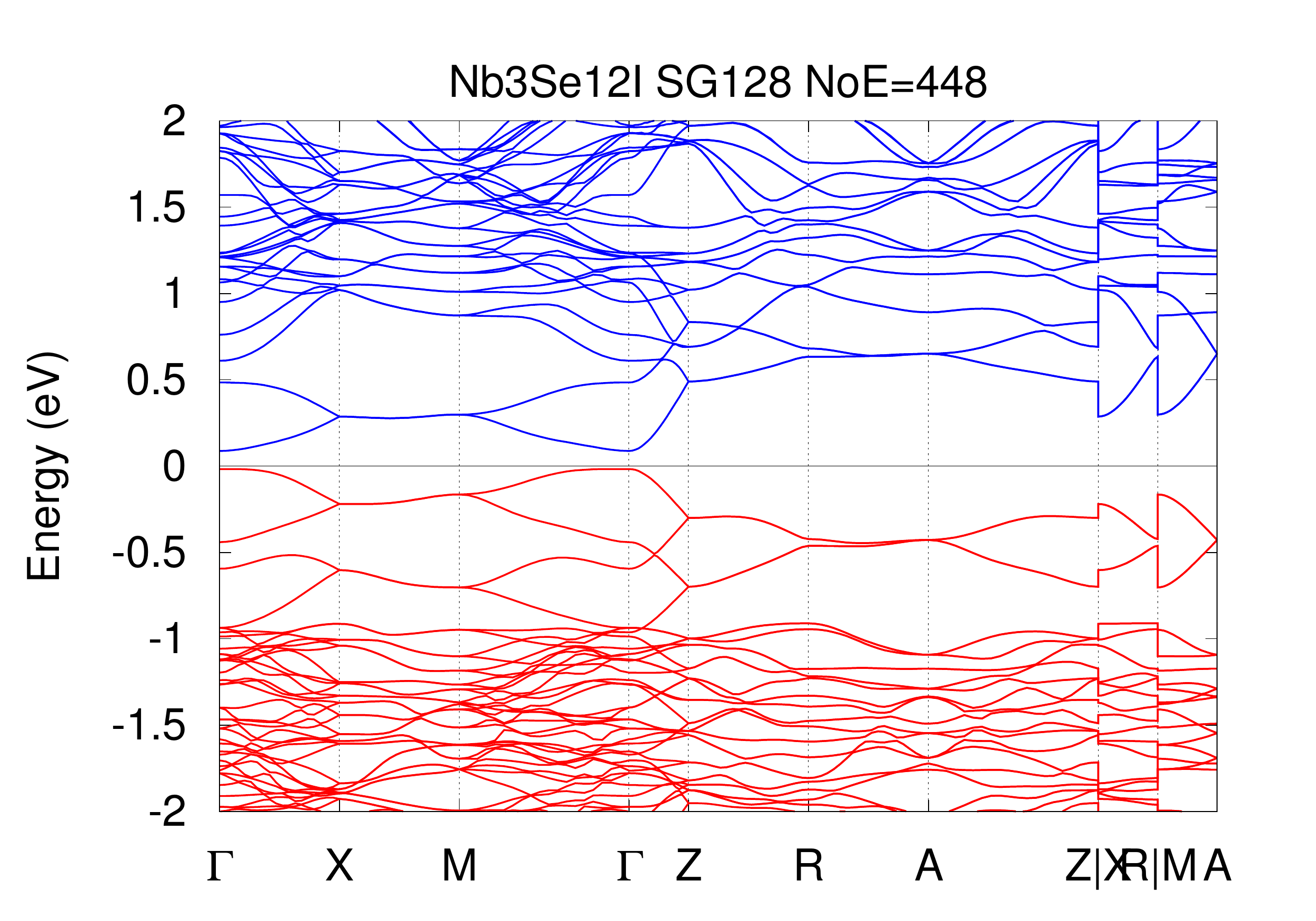}
}
\subfigure[K$_{3}$Ga$_{13}$ SG63 NoA=64 NoE=264]{
\label{subfig:103709}
\includegraphics[scale=0.32,angle=270]{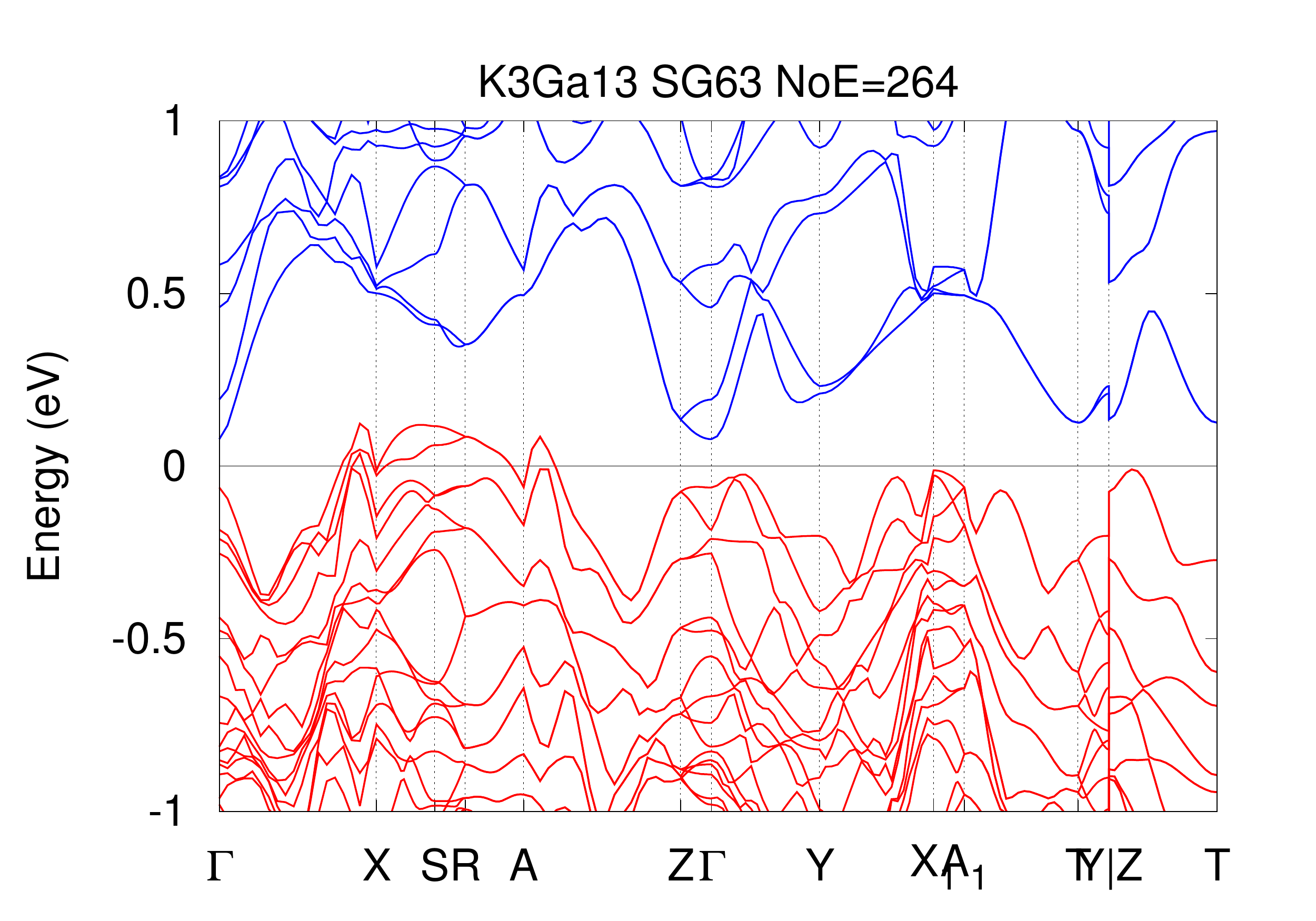}
}
\subfigure[Na$_{5}$Sn$_{13}$ SG63 NoA=144 NoE=456]{
\label{subfig:165613}
\includegraphics[scale=0.32,angle=270]{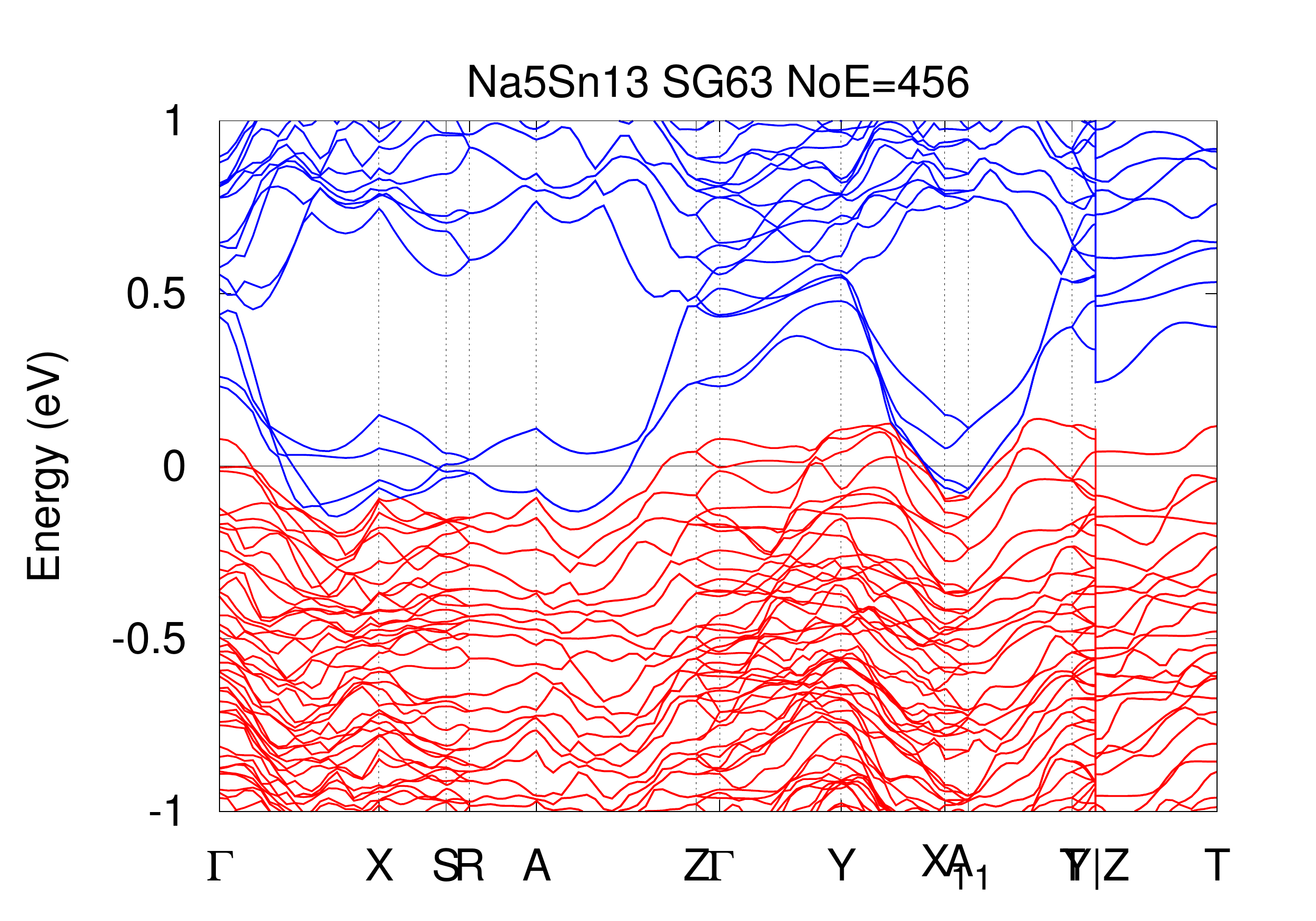}
}
\caption{\hyperref[tab:electride]{back to the table}}
\end{figure}

\clearpage
\subsection*{E. TQC files}
\label{tqcfile}
The \texttt{tqc.txt}/\texttt{tqc.data} files for the unconventional materials in the searching results are obtained. These files are generated by the program \webirvsp. For those with only one essential BR, the real space invariants at the empty site are calculated.
\subsubsection*{167876 TaN}
\label{sec:tqc167876}
\noindent Essential BR: $A1@4d$ \\
\noindent RSI:
\begin{flalign*}
&\delta_{1}@4d\equiv m(A1)-m(A2)-m(T2)+m(T1) = 1,&
\end{flalign*}
\lstset{language=bash, keywordstyle=\color{blue!70}, basicstyle=\ttfamily, frame=shadowbox}
\begin{lstlisting}
Computed bands:  1 -  5
GM: GM1 (1); GM4 (3); GM1 (1); [5]  ;
X : X3  (1); X1  (1); X5  (2); X3  (1); [5]  ;
L : L1  (1); L1  (1); L3  (2); L1  (1); [5]  ;
W : W4  (1); W2  (1); W1  (1); W3  (1); W3  (1); [5] ;
\end{lstlisting}
\hyperref[tab:electride]{Back to the table}

\subsubsection*{167870 TcN}
\label{sec:tqc167870}
\noindent Essential BR: $A1@4d$ \\
\noindent RSI:
\begin{flalign*}
&\delta_{1}@4d\equiv m(A1)-m(A2)-m(T2)+m(T1) = 1,&
\end{flalign*}
\lstset{language=bash, keywordstyle=\color{blue!70}, basicstyle=\ttfamily, frame=shadowbox}
\begin{lstlisting}
Computed bands:  1 -  6
GM: GM1 (1); GM4 (3); GM3 (2); [6]  ;
X : X3  (1); X1  (1); X5  (2); X3  (1); X2  (1); [6]  ;
L : L1  (1); L1  (1); L3  (2); L3  (2); [6]  ;
W : W4  (1); W2  (1); W1  (1); W3  (1); W3  (1); W2  (1); [6] ;
\end{lstlisting}
\hyperref[tab:electride]{Back to the table}

\subsubsection*{187182 MoN}
\label{sec:tqc187182}
\noindent Essential BR: $A1@4c$ \\
\noindent RSI:
\begin{flalign*}
&\delta_{1}@4c\equiv m(A1)-m(A2)-m(T2)+m(T1) = 1,&
\end{flalign*}
\lstset{language=bash, keywordstyle=\color{blue!70}, basicstyle=\ttfamily, frame=shadowbox}
\begin{lstlisting}
Computed bands:  1 -  6
GM: GM1 (1); GM4 (3); GM3 (2); [6]  ;
X : X3  (1); X1  (1); X5  (2); X3  (1); X2  (1); [6]  ;
L : L1  (1); L1  (1); L3  (2); L3  (2); [6]  ;
W : W3  (1); W2  (1); W1  (1); W4  (1); W4  (1); W2  (1); [6] ;
\end{lstlisting}
\hyperref[tab:electride]{Back to the table}

\subsubsection*{167875 HfN}
\label{sec:tqc167875}
\noindent Essential BR: $A1@4d$ \\
\noindent RSI:
\begin{flalign*}
&\delta_{1}@4d\equiv m(A1)-m(A2)-m(T2)+m(T1) = 1,&
\end{flalign*}
\lstset{language=bash, keywordstyle=\color{blue!70}, basicstyle=\ttfamily, frame=shadowbox}
\begin{lstlisting}
Computed bands:  1 -  5
GM: GM1 (1); GM4 (3); GM1 (1); [5]  ;
X : X3  (1); X1  (1); X5  (2); X3  (1); [5]  ;
L : L1  (1); L1  (1); L3  (2); L1  (1); [5]  ;
W : W4  (1); W2  (1); W1  (1); W3  (1); W3  (1); [5] ;
\end{lstlisting}
\hyperref[tab:electride]{Back to the table}

\subsubsection*{186876 MoP}
\label{sec:tqc186876}
\noindent Essential BR: $A1@4d$ \\
\noindent RSI:
\begin{flalign*}
&\delta_{1}@4d\equiv m(A1)-m(A2)-m(T2)+m(T1) = 1,&
\end{flalign*}
\lstset{language=bash, keywordstyle=\color{blue!70}, basicstyle=\ttfamily, frame=shadowbox}
\begin{lstlisting}
Computed bands:  1 -  6
GM: GM1 (1); GM4 (3); GM3 (2); [6]  ;
X : X3  (1); X1  (1); X5  (2); X3  (1); X2  (1); [6]  ;
L : L1  (1); L1  (1); L3  (2); L3  (2); [6]  ;
W : W4  (1); W2  (1); W1  (1); W3  (1); W3  (1); W2  (1); [6] ;
\end{lstlisting}
\hyperref[tab:electride]{Back to the table}

\subsubsection*{186243 IrN}
\label{sec:tqc186243}
\noindent Essential BR: $A1@4d$ \\
\noindent RSI:
\begin{flalign*}
&\delta_{1}@4d\equiv m(A1)-m(A2)-m(T2)+m(T1) = 1,&
\end{flalign*}
\lstset{language=bash, keywordstyle=\color{blue!70}, basicstyle=\ttfamily, frame=shadowbox}
\begin{lstlisting}
Computed bands:  1 -  7
GM: GM1 (1); GM4 (3); GM3 (2); GM1 (1); [7]  ;
X : X3  (1); X1  (1); X5  (2); X3  (1); X2  (1); X1  (1); [7]  ;
L : L1  (1); L1  (1); L3  (2); L3  (2); L1  (1); [7]  ;
W : W4  (1); W2  (1); W1  (1); W3  (1); W3  (1); W2  (1); W1  (1); [7] ;
\end{lstlisting}
\hyperref[tab:electride]{Back to the table}

\subsubsection*{183192 RhN}
\label{sec:tqc183192}
\noindent Essential BR: $A1@4d$ \\
\noindent RSI:
\begin{flalign*}
&\delta_{1}@4d\equiv m(A1)-m(A2)-m(T2)+m(T1) = 1,&
\end{flalign*}
\lstset{language=bash, keywordstyle=\color{blue!70}, basicstyle=\ttfamily, frame=shadowbox}
\begin{lstlisting}
Computed bands:  1 -  7
GM: GM1 (1); GM4 (3); GM3 (2); GM1 (1); [7]  ;
X : X3  (1); X1  (1); X5  (2); X3  (1); X2  (1); X1  (1); [7]  ;
L : L1  (1); L3  (2); L1  (1); L3  (2); L1  (1); [7]  ;
W : W4  (1); W2  (1); W1  (1); W3  (1); W3  (1); W2  (1); W1  (1); [7] ;
\end{lstlisting}
\hyperref[tab:electride]{Back to the table}

\subsubsection*{150682 Cu}
\label{sec:tqc150682}
\noindent Essential BR: $Ag@2b$ \\
\noindent RSI:
\begin{flalign*}
&\delta_{1}@2b\equiv -m(Ag)+m(Au)-m(Bg)+m(Bu) = -1,&
\end{flalign*}
\lstset{language=bash, keywordstyle=\color{blue!70}, basicstyle=\ttfamily, frame=shadowbox}
\begin{lstlisting}
Computed bands:  1 - 11
GM: GM1+(1); GM2-GM2-(2); GM1+GM1+(2); GM1+(1); GM1-GM1-(2); GM1-(1);
    GM2+GM2+(2); [11] ;
Y : Y1+ (1); Y2- Y2- (2); Y1+ Y1+ (2); Y1+ (1); Y1- Y1- (2); Y1- (1);
    Y2+ Y2+ (2); [11] ;
V : V1- V1- (2); V1+ (1); V1- (1); V1+ V1+ (2); V1+ V1+ (2); V1- V1- (2);
    V1- (1); [11] ;
L : L1- L1- (2); L1+ (1); L1- (1); L1+ L1+ (2); L1+ L1+ (2); L1- L1- (2);
    L1- (1); [11] ;
M : M1+ (1); M2- M2- (2); M1+ M1+ (2); M1+ (1); M1- M1- (2); M1- (1);
    M2+ M2+ (2); [11] ;
U : U2  U2  (2); U1  (1); U1  (1); U1  U1  (2); U2  U2  (2); U1  U1  (2);
    U1  (1); [11] ;
A : A1+ (1); A2- A2- (2); A1+ A1+ (2); A1+ (1); A1- A1- (2); A1- (1);
    A2+ A2+ (2); [11];
\end{lstlisting}
\hyperref[tab:electride]{Back to the table}

\subsubsection*{24981 GdO}
\label{sec:tqc24981}
\noindent Essential BR: $A1@4d$ \\
\noindent RSI:
\begin{flalign*}
&\delta_{1}@4d\equiv m(A1)-m(A2)-m(T2)+m(T1) = 1,&
\end{flalign*}
\lstset{language=bash, keywordstyle=\color{blue!70}, basicstyle=\ttfamily, frame=shadowbox}
\begin{lstlisting}
Computed bands:  1 -  8
GM: GM4 (3); GM1 (1); GM4 (3); GM1 (1); [8]  ;
X : X3  (1); X5  (2); X3  (1); X1  (1); X5  (2); X3  (1); [8]  ;
L : L1  (1); L3  (2); L1  (1); L1  (1); L3  (2); L1  (1); [8]  ;
W : W4  (1); W3  (1); W2  (1); W4  (1); W2  (1); W1  (1); W3  (1); W3  (1); [8] ;
\end{lstlisting}
\hyperref[tab:electride]{Back to the table}

\subsubsection*{183190 RuN}
\label{sec:tqc183190}
\noindent Essential BR: $A1@4d$ \\
\noindent RSI:
\begin{flalign*}
&\delta_{1}@4d\equiv m(A1)-m(A2)-m(T2)+m(T1) = 1,&
\end{flalign*}
\lstset{language=bash, keywordstyle=\color{blue!70}, basicstyle=\ttfamily, frame=shadowbox}
\begin{lstlisting}
Computed bands:  1 -  7
GM: GM1 (1); GM4 (3); GM3 (2); GM1 (1); [7]  ;
X : X3  (1); X1  (1); X5  (2); X3  (1); X2  (1); X1  (1); [7]  ;
L : L1  (1); L1  (1); L3  (2); L3  (2); L1  (1); [7]  ;
W : W4  (1); W2  (1); W1  (1); W3  (1); W3  (1); W2  (1); W1  (1); [7] ;
\end{lstlisting}
\hyperref[tab:electride]{Back to the table}

\subsubsection*{191171 MnSn}
\label{sec:tqc191171}
\noindent Essential BR: $A1@4d$ \\
\noindent RSI:
\begin{flalign*}
&\delta_{1}@4d\equiv m(A1)-m(A2)-m(T2)+m(T1) = 1,&
\end{flalign*}
\lstset{language=bash, keywordstyle=\color{blue!70}, basicstyle=\ttfamily, frame=shadowbox}
\begin{lstlisting}
Computed bands:  1 -  6
GM: GM1 (1); GM4 (3); GM3 (2); [6]  ;
X : X3  (1); X1  (1); X5  (2); X3  (1); X2  (1); [6]  ;
L : L1  (1); L1  (1); L3  (2); L3  (2); [6]  ;
W : W4  (1); W1  (1); W2  (1); W3  (1); W3  (1); W2  (1); [6] ;
\end{lstlisting}
\hyperref[tab:electride]{Back to the table}

\subsubsection*{183184 NbN}
\label{sec:tqc183184}
\noindent Essential BR: $A1@4d$ \\
\noindent RSI:
\begin{flalign*}
&\delta_{1}@4d\equiv m(A1)-m(A2)-m(T2)+m(T1) = 1,&
\end{flalign*}
\lstset{language=bash, keywordstyle=\color{blue!70}, basicstyle=\ttfamily, frame=shadowbox}
\begin{lstlisting}
Computed bands:  1 -  8
GM: GM4 (3); GM1 (1); GM4 (3); GM1 (1); [8]  ;
X : X3  (1); X5  (2); X3  (1); X1  (1); X5  (2); X3  (1); [8]  ;
L : L1  (1); L3  (2); L1  (1); L1  (1); L3  (2); L1  (1); [8]  ;
W : W4  (1); W3  (1); W2  (1); W4  (1); W2  (1); W1  (1); W3  (1); W3  (1); [8] ;
\end{lstlisting}
\hyperref[tab:electride]{Back to the table}

\subsubsection*{183182 ZrN}
\label{sec:tqc183182}
\noindent Essential BR: $A1@4d$ \\
\noindent RSI:
\begin{flalign*}
&\delta_{1}@4d\equiv m(A1)-m(A2)-m(T2)+m(T1) = 1,&
\end{flalign*}
\lstset{language=bash, keywordstyle=\color{blue!70}, basicstyle=\ttfamily, frame=shadowbox}
\begin{lstlisting}
Computed bands:  1 -  9
GM: GM1 (1); GM4 (3); GM1 (1); GM4 (3); GM1 (1); [9]  ;
X : X1  (1); X3  (1); X5  (2); X3  (1); X1  (1); X5  (2); X3  (1); [9]  ;
L : L1  (1); L1  (1); L3  (2); L1  (1); L1  (1); L3  (2); L1  (1); [9]  ;
W : W1  (1); W4  (1); W3  (1); W2  (1); W4  (1); W2  (1); W1  (1); W3  (1);
    W3  (1); [9] ;
\end{lstlisting}
\hyperref[tab:electride]{Back to the table}

\subsubsection*{41258 FeN}
\label{sec:tqc41258}
\noindent Essential BR: $A1@4d$ \\
\noindent RSI:
\begin{flalign*}
&\delta_{1}@4d\equiv m(A1)-m(A2)-m(T2)+m(T1) = 1,&
\end{flalign*}
\lstset{language=bash, keywordstyle=\color{blue!70}, basicstyle=\ttfamily, frame=shadowbox}
\begin{lstlisting}
Computed bands:  1 -  7
GM: GM1 (1); GM4 (3); GM3 (2); GM1 (1); [7]  ;
X : X3  (1); X1  (1); X5  (2); X3  (1); X2  (1); X1  (1); [7]  ;
L : L1  (1); L1  (1); L3  (2); L3  (2); L1  (1); [7]  ;
W : W4  (1); W2  (1); W1  (1); W3  (1); W3  (1); W2  (1); W1  (1); [7] ;
\end{lstlisting}
\hyperref[tab:electride]{Back to the table}

\subsubsection*{60389 Si}
\label{sec:tqc60389}
\noindent Essential BR: $A1g@16c$ \\
\noindent RSI:
\begin{flalign*}
&\delta_{1}@16c\equiv -m(Eg)+m(Eu) = 0,&
\\
&\delta_{2}@16c\equiv -m(A1g)+m(A1u)-m(A2g)+m(A2u) = -1,&
\end{flalign*}
\lstset{language=bash, keywordstyle=\color{blue!70}, basicstyle=\ttfamily, frame=shadowbox}
\begin{lstlisting}
Computed bands:  1 -  4
GM: GM1+(1); GM5+(3); [4]  ;
X : X1  (2); X3  (2); [4]  ;
L : L2- (1); L1+ (1); L3- (2); [4]  ;
W : W1  (2); W2  (2); [4] ;
\end{lstlisting}
\hyperref[tab:electride]{Back to the table}

\subsubsection*{30334 SiP}
\label{sec:tqc30334}
\noindent Essential BR: $A1@4d$ \\
\noindent RSI:
\begin{flalign*}
&\delta_{1}@4d\equiv m(A1)-m(A2)-m(T2)+m(T1) = 1,&
\end{flalign*}
\lstset{language=bash, keywordstyle=\color{blue!70}, basicstyle=\ttfamily, frame=shadowbox}
\begin{lstlisting}
Computed bands:  1 -  5
GM: GM1 (1); GM4 (3); GM1 (1); [5]  ;
X : X3  (1); X1  (1); X5  (2); X3  (1); [5]  ;
L : L1  (1); L1  (1); L3  (2); L1  (1); [5]  ;
W : W4  (1); W1  (1); W2  (1); W3  (1); W3  (1); [5] ;
\end{lstlisting}
\hyperref[tab:electride]{Back to the table}

\subsubsection*{181079 CrN}
\label{sec:tqc181079}
\noindent Essential BR: $A1@4b$ \\
\noindent RSI:
\begin{flalign*}
&\delta_{1}@4b\equiv m(A1)-m(A2)-m(T2)+m(T1) = 1,&
\end{flalign*}
\lstset{language=bash, keywordstyle=\color{blue!70}, basicstyle=\ttfamily, frame=shadowbox}
\begin{lstlisting}
Computed bands:  1 -  6
GM: GM1 (1); GM4 (3); GM3 (2); [6]  ;
X : X1  (1); X3  (1); X5  (2); X1  (1); X4  (1); [6]  ;
L : L1  (1); L1  (1); L3  (2); L3  (2); [6]  ;
W : W1  (1); W3  (1); W4  (1); W2  (1); W2  (1); W3  (1); [6] ;
\end{lstlisting}
\hyperref[tab:electride]{Back to the table}

\subsubsection*{236788 MnN}
\label{sec:tqc236788}
\noindent Essential BR: $A1@4d$ \\
\noindent RSI:
\begin{flalign*}
&\delta_{1}@4d\equiv m(A1)-m(A2)-m(T2)+m(T1) = 1,&
\end{flalign*}
\lstset{language=bash, keywordstyle=\color{blue!70}, basicstyle=\ttfamily, frame=shadowbox}
\begin{lstlisting}
Computed bands:  1 -  6
GM: GM1 (1); GM4 (3); GM3 (2); [6]  ;
X : X3  (1); X1  (1); X5  (2); X3  (1); X2  (1); [6]  ;
L : L1  (1); L1  (1); L3  (2); L3  (2); [6]  ;
W : W4  (1); W2  (1); W1  (1); W3  (1); W3  (1); W2  (1); [6] ;
\end{lstlisting}
\hyperref[tab:electride]{Back to the table}

\subsubsection*{167879 OsN}
\label{sec:tqc167879}
\noindent Essential BR: $A1@4d$ \\
\noindent RSI:
\begin{flalign*}
&\delta_{1}@4d\equiv m(A1)-m(A2)-m(T2)+m(T1) = 1,&
\end{flalign*}
\lstset{language=bash, keywordstyle=\color{blue!70}, basicstyle=\ttfamily, frame=shadowbox}
\begin{lstlisting}
Computed bands:  1 -  7
GM: GM1 (1); GM4 (3); GM3 (2); GM1 (1); [7]  ;
X : X3  (1); X1  (1); X5  (2); X3  (1); X2  (1); X1  (1); [7]  ;
L : L1  (1); L1  (1); L3  (2); L3  (2); L1  (1); [7]  ;
W : W4  (1); W2  (1); W1  (1); W3  (1); W3  (1); W2  (1); W1  (1); [7] ;
\end{lstlisting}
\hyperref[tab:electride]{Back to the table}

\subsubsection*{191788 MnP}
\label{sec:tqc191788}
\noindent Essential BR: $A1@4d$ \\
\noindent RSI:
\begin{flalign*}
&\delta_{1}@4d\equiv m(A1)-m(A2)-m(T2)+m(T1) = 1,&
\end{flalign*}
\lstset{language=bash, keywordstyle=\color{blue!70}, basicstyle=\ttfamily, frame=shadowbox}
\begin{lstlisting}
Computed bands:  1 -  6
GM: GM1 (1); GM4 (3); GM3 (2); [6]  ;
X : X3  (1); X1  (1); X5  (2); X3  (1); X2  (1); [6]  ;
L : L1  (1); L1  (1); L3  (2); L3  (2); [6]  ;
W : W4  (1); W2  (1); W1  (1); W3  (1); W3  (1); W2  (1); [6] ;
\end{lstlisting}
\hyperref[tab:electride]{Back to the table}

\subsubsection*{414330 Sr$_{2}$N}
\label{sec:tqc414330}
\noindent Essential BR: $A1g@3b$ \\
\noindent RSI:
\begin{flalign*}
&\delta_{1}@3b\equiv -m(Eg)+m(Eu) = 0,&
\\
&\delta_{2}@3b\equiv -m(A1g)+m(A1u)-m(A2g)+m(A2u) = -1,&
\end{flalign*}
\lstset{language=bash, keywordstyle=\color{blue!70}, basicstyle=\ttfamily, frame=shadowbox}
\begin{lstlisting}
Computed bands:  1 - 13
GM: GM1+(1); GM2-(1); GM1+(1); GM3+(2); GM2-(1); GM3-(2); GM1+(1); GM2-(1);
    GM1+(1); GM3-(2); [13] ;
T : T1+ (1); T2- (1); T1+ (1); T2- (1); T3+ (2); T3- (2); T1+ (1); T2- (1);
    T3- (2); T2- (1); [13] ;
F : F2- (1); F1+ (1); F1+ (1); F2- (1); F1- (1); F2- (1); F1+ (1); F2+ (1);
    F1+ (1); F2- (1); F2- (1); F1- (1); F1+ (1); [13] ;
L : L2- (1); L1+ (1); L1+ (1); L2- (1); L1- (1); L2- (1); L1+ (1); L2+ (1);
    L1+ (1); L2- (1); L2- (1); L1- (1); L2- (1); [13];
\end{lstlisting}
\hyperref[tab:electride]{Back to the table}

\subsubsection*{187135 LiFeP}
\label{sec:tqc187135}
\noindent Essential BR: $A@4b$ \\
\noindent RSI:
\begin{flalign*}
&\delta_{1}@4b\equiv -m(A)-m(A2)+m(B2)+m(B1) = -1,&
\end{flalign*}
\lstset{language=bash, keywordstyle=\color{blue!70}, basicstyle=\ttfamily, frame=shadowbox}
\begin{lstlisting}
Computed bands:  1 -  7
GM: GM1 (1); GM1 (1); GM2 (1); GM5 (2); GM1 (1); GM3 (1); [7]  ;
M : M1  (1); M1  (1); M2  (1); M5  (2); M1  (1); M3  (1); [7]  ;
P : P1  (1); P3  P4  (2); P1  (1); P1  (1); P3  P4  (2); [7]  ;
X : X1  (1); X3  (1); X4  (1); X1  (1); X4  (1); X1  (1); X3  (1); [7]  ;
N : N1  (1); N1  (1); N1  (1); N2  (1); N1  (1); N1  (1); N2  (1); [7] ;
\end{lstlisting}
\hyperref[tab:electride]{Back to the table}

\subsubsection*{54255 MnNiSb}
\label{sec:tqc54255}
\noindent Essential BR: $A1@4b$ \\
\noindent RSI:
\begin{flalign*}
&\delta_{1}@4b\equiv m(A1)-m(A2)-m(T2)+m(T1) = 1,&
\end{flalign*}
\lstset{language=bash, keywordstyle=\color{blue!70}, basicstyle=\ttfamily, frame=shadowbox}
\begin{lstlisting}
Computed bands:  1 - 11
GM: GM1 (1); GM4 (3); GM3 (2); GM4 (3); GM3 (2); [11] ;
X : X3  (1); X1  (1); X5  (2); X3  (1); X1  (1); X5  (2); X2  (1); X4  (1);
    X1  (1); [11] ;
L : L1  (1); L1  (1); L3  (2); L1  (1); L3  (2); L3  (2); L3  (2); [11] ;
W : W3  (1); W1  (1); W2  (1); W4  (1); W4  (1); W2  (1); W1  (1); W3  (1);
    W2  (1); W2  (1); W4  (1); [11];
\end{lstlisting}
\hyperref[tab:electride]{Back to the table}

\subsubsection*{22231 Ca$_{2}$N}
\label{sec:tqc22231}
\noindent Essential BR: $A1g@3b$ \\
\noindent RSI:
\begin{flalign*}
&\delta_{1}@3b\equiv -m(Eg)+m(Eu) = 0,&
\\
&\delta_{2}@3b\equiv -m(A1g)+m(A1u)-m(A2g)+m(A2u) = -1,&
\end{flalign*}
\lstset{language=bash, keywordstyle=\color{blue!70}, basicstyle=\ttfamily, frame=shadowbox}
\begin{lstlisting}
Computed bands:  1 - 13
GM: GM1+(1); GM2-(1); GM1+(1); GM3+(2); GM2-(1); GM3-(2); GM1+(1); GM2-(1);
    GM1+(1); GM3-(2); [13] ;
T : T1+ (1); T2- (1); T1+ (1); T3+ (2); T2- (1); T3- (2); T1+ (1); T2- (1);
    T3- (2); T2- (1); [13] ;
F : F2- (1); F1+ (1); F1+ (1); F2- (1); F1- (1); F2- (1); F1+ (1); F2+ (1);
    F1+ (1); F2- (1); F2- (1); F1- (1); F1+ (1); [13] ;
L : L2- (1); L1+ (1); L1+ (1); L2- (1); L1- (1); L2- (1); L1+ (1); L2+ (1);
    L1+ (1); L2- (1); L2- (1); L1- (1); L2- (1); [13];
\end{lstlisting}
\hyperref[tab:electride]{Back to the table}

\subsubsection*{647338 RbO$_{2}$}
\label{sec:tqc647338}
\noindent Essential BR: $A1g@2b$ \\
\noindent RSI:
\begin{flalign*}
&\delta_{1}@2b\equiv -m(A1g)+m(A1u)-m(A2g)+m(A2u)-m(Eg)+m(Eu) = -1,&
\\
&\delta_{2}@2b\equiv m(A1g)-m(A1u)-m(B1g)+m(B1u)+m(A2g)-m(A2u)-m(B2g)+m(B2u) = 1,&
\end{flalign*}
\lstset{language=bash, keywordstyle=\color{blue!70}, basicstyle=\ttfamily, frame=shadowbox}
\begin{lstlisting}
Computed bands:  1 - 11
GM: GM1+(1); GM1+(1); GM3-(1); GM5-(2); GM3-(1); GM1+(1); GM5-(2); GM5+(2);
    [11] ;
M : M1+ (1); M1+ (1); M3- (1); M5- (2); M3- (1); M1+ (1); M5- (2); M5+ (2);
    [11] ;
P : P1  (1); P3  (1); P1  (1); P5  (2); P3  (1); P3  (1); P5  (2); P5  (2);
    [11] ;
X : X1+ (1); X1+ (1); X2- (1); X4- (1); X3- (1); X2- (1); X1+ (1); X4- (1);
    X3- (1); X4+ (1); X3+ (1); [11] ;
N : N1+ (1); N2- (1); N1+ (1); N2- (1); N2- (1); N1- (1); N2- (1); N1+ (1);
    N2+ (1); N2- (1); N1- (1); [11];
\end{lstlisting}
\hyperref[tab:electride]{Back to the table}

\subsubsection*{290430 TaN$_{2}$}
\label{sec:tqc290430}
\lstset{language=bash, keywordstyle=\color{blue!70}, basicstyle=\ttfamily, frame=shadowbox}
\begin{lstlisting}
Computed bands:  1 -  8
A : A1  (1); A3  (1); A5  (2); A1  (1); A6  (2); A3  (1); [8]  ;
GM: GM1 (1); GM3 (1); GM1 (1); GM5 (2); GM6 (2); GM1 (1); [8]  ;
H : H1  (1); H2  (1); H3  (1); H1  (1); H5  (1); H6  (1); H4  (1); H4  (1);
    [8]  ;
K : K1  (1); K2  (1); K1  (1); K5  (1); K3  (1); K4  (1); K3  (1); K6  (1);
    [8]  ;
L : L1  (1); L3  (1); L1  (1); L2  (1); L1  (1); L3  (1); L4  (1); L3  (1);
    [8]  ;
M : M1  (1); M3  (1); M1  (1); M1  (1); M2  (1); M4  (1); M3  (1); M1  (1); [8] ;
\end{lstlisting}
\hyperref[tab:electride]{Back to the table}

\subsubsection*{54465 MnSnAu}
\label{sec:tqc54465}
\noindent Essential BR: $A1@4d$ \\
\noindent RSI:
\begin{flalign*}
&\delta_{1}@4d\equiv m(A1)-m(A2)-m(T2)+m(T1) = 1,&
\end{flalign*}
\lstset{language=bash, keywordstyle=\color{blue!70}, basicstyle=\ttfamily, frame=shadowbox}
\begin{lstlisting}
Computed bands:  1 - 11
GM: GM1 (1); GM4 (3); GM3 (2); GM4 (3); GM3 (2); [11] ;
X : X1  (1); X3  (1); X5  (2); X1  (1); X4  (1); X3  (1); X5  (2); X3  (1);
    X2  (1); [11] ;
L : L1  (1); L1  L3  (3); L3  (2); L1  (1); L3  (2); L3  (2); [11] ;
W : W1  (1); W3  (1); W4  (1); W2  (1); W3  (1); W4  (1); W1  (1); W2  (1);
    W3  (1); W3  (1); W2  (1); [11];
\end{lstlisting}
\hyperref[tab:electride]{Back to the table}

\subsubsection*{67443 NbS$_{2}$}
\label{sec:tqc67443}
\noindent Essential BR: $A1'@1c$ \\
\noindent RSI:
\begin{flalign*}
&\delta_{1}@1c\equiv m(A1')+m(A2')-m(A2'')-m(A1'')-m(E')+m(E'') = 1,&
\end{flalign*}
\lstset{language=bash, keywordstyle=\color{blue!70}, basicstyle=\ttfamily, frame=shadowbox}
\begin{lstlisting}
Computed bands:  1 - 12
A : A3  (1); A5  (2); A1  (1); A3  (1); A1  (1); A5  (2); A3  (1); A6  (2);
    A1  (1); [12] ;
GM: GM3 (1); GM5 (2); GM1 (1); GM3 (1); GM1 (1); GM5 (2); GM3 (1); GM6 (2);
    GM1 (1); [12] ;
H : H5  (1); H3  (1); H2  (1); H5  (1); H6  (1); H4  (1); H1  (1); H3  (1);
    H5  (1); H2  (1); H6  (1); H3  (1); [12] ;
K : K5  (1); K3  (1); K2  (1); K5  (1); K6  (1); K4  (1); K1  (1); K3  (1);
    K5  (1); K2  (1); K6  (1); K3  (1); [12] ;
L : L1  (1); L2  L3  (2); L1  (1); L3  (1); L1  (1); L3  (1); L1  (1); L4  (1);
    L3  (1); L2  (1); L1  (1); [12] ;
M : M1  (1); M2  M3  (2); M1  (1); M3  (1); M1  (1); M3  (1); M1  (1); M4  (1);
    M3  (1); M2  (1); M1  (1); [12];
\end{lstlisting}
\hyperref[tab:electride]{Back to the table}

\subsubsection*{58108 Al$_{2}$Os}
\label{sec:tqc58108}
\noindent Essential BR: $Ag@4c$ \\
\noindent RSI:
\begin{flalign*}
&\delta_{1}@4c\equiv -m(Ag)+m(Au)-m(B1g)+m(B1u)-m(B3g)+m(B3u)-m(B2g)+m(B2u) = -1,&
\end{flalign*}
\lstset{language=bash, keywordstyle=\color{blue!70}, basicstyle=\ttfamily, frame=shadowbox}
\begin{lstlisting}
Computed bands:  1 -  7
GM: GM1+(1); GM1+(1); GM2+(1); GM5+(2); GM3-(1); GM4+(1); [7]  ;
M : M1+ (1); M3- (1); M2+ (1); M1+ (1); M5+ (2); M4+ (1); [7]  ;
P : P3  (1); P5  (2); P1  (1); P5  (2); P3  (1); [7]  ;
X : X1+ (1); X4+ (1); X4- (1); X3+ (1); X1+ (1); X2- (1); X3- (1); [7]  ;
N : N1+ (1); N2- (1); N1+ (1); N1+ (1); N2- (1); N2+ (1); N2+ (1); [7] ;
\end{lstlisting}
\hyperref[tab:electride]{Back to the table}

\subsubsection*{38245 KO$_{2}$}
\label{sec:tqc38245}
\noindent Essential BR: $A1g@2a$ \\
\noindent RSI:
\begin{flalign*}
&\delta_{1}@2a\equiv -m(A1g)+m(A1u)-m(A2g)+m(A2u)-m(Eg)+m(Eu) = -1,&
\\
&\delta_{2}@2a\equiv m(A1g)-m(A1u)-m(B1g)+m(B1u)+m(A2g)-m(A2u)-m(B2g)+m(B2u) = 1,&
\end{flalign*}
\lstset{language=bash, keywordstyle=\color{blue!70}, basicstyle=\ttfamily, frame=shadowbox}
\begin{lstlisting}
Computed bands:  1 - 11
GM: GM1+(1); GM1+(1); GM3-(1); GM5-(2); GM3-(1); GM1+(1); GM5-(2); GM5+(2);
    [11] ;
M : M1+ (1); M1+ (1); M3- (1); M5- (2); M3- (1); M1+ (1); M5- (2); M5+ (2);
    [11] ;
P : P3  (1); P1  (1); P3  (1); P5  (2); P1  (1); P1  (1); P5  (2); P5  (2);
    [11] ;
X : X1+ (1); X1+ (1); X2- (1); X4- (1); X3- (1); X2- (1); X1+ (1); X4- (1);
    X3- (1); X4+ (1); X3+ (1); [11] ;
N : N2- (1); N1+ (1); N2- (1); N1+ (1); N1+ (1); N2+ (1); N1+ (1); N2- (1);
    N1- (1); N1+ (1); N2+ (1); [11];
\end{lstlisting}
\hyperref[tab:electride]{Back to the table}

\subsubsection*{290427 HfN$_{2}$}
\label{sec:tqc290427}
\noindent Essential BR: $A1'@1c$ \\
\noindent RSI:
\begin{flalign*}
&\delta_{1}@1c\equiv m(A1')+m(A2')-m(A2'')-m(A1'')-m(E')+m(E'') = 1,&
\end{flalign*}
\lstset{language=bash, keywordstyle=\color{blue!70}, basicstyle=\ttfamily, frame=shadowbox}
\begin{lstlisting}
Computed bands:  1 -  7
A : A1  (1); A3  (1); A5  (2); A1  (1); A6  (2); [7]  ;
GM: GM1 (1); GM3 (1); GM1 (1); GM5 (2); GM6 (2); [7]  ;
H : H3  (1); H4  (1); H5  (1); H1  (1); H3  (1); H2  (1); H6  (1); [7]  ;
K : K3  (1); K4  (1); K3  (1); K1  (1); K5  (1); K6  (1); K2  (1); [7]  ;
L : L1  (1); L3  (1); L1  (1); L2  (1); L1  (1); L3  (1); L4  (1); [7]  ;
M : M1  (1); M3  (1); M1  (1); M1  (1); M2  (1); M4  (1); M3  (1); [7] ;
\end{lstlisting}
\hyperref[tab:electride]{Back to the table}

\subsubsection*{48214 HgO$_{2}$}
\label{sec:tqc48214}
\noindent Essential BR: $Ag@2d$ \\
\noindent RSI:
\begin{flalign*}
&\delta_{1}@2d\equiv -m(Ag)+m(Au)-m(Bg)+m(Bu) = -1,&
\end{flalign*}
\lstset{language=bash, keywordstyle=\color{blue!70}, basicstyle=\ttfamily, frame=shadowbox}
\begin{lstlisting}
Computed bands:  1 - 12
GM: GM1+(1); GM2-(1); GM1+(1); GM1+(1); GM2+(1); GM1+(1); GM2+(1); GM1+(1);
    GM2-(1); GM1-(1); GM1+(1); GM2+(1); [12] ;
Y : Y1+ (1); Y2- (1); Y1+ (1); Y1+ (1); Y2+ (1); Y1+ (1); Y2- (1); Y1- (1);
    Y2+ (1); Y1+ (1); Y1+ (1); Y2+ (1); [12] ;
V : V1- (1); V1+ (1); V1+ (1); V1- (1); V1+ (1); V1+ (1); V1+ (1); V1+ (1);
    V1+ (1); V1+ (1); V1- (1); V1- (1); [12] ;
L : L1+ (1); L1- (1); L1+ (1); L1+ (1); L1+ (1); L1+ (1); L1+ (1); L1- (1);
    L1- (1); L1+ (1); L1+ (1); L1+ (1); [12] ;
M : M2- (1); M1+ (1); M1+ (1); M2+ (1); M1+ (1); M2+ (1); M1+ (1); M2- (1);
    M2+ (1); M1+ (1); M1- (1); M2- (1); [12] ;
U : U2  (1); U1  (1); U1  (1); U2  (1); U2  (1); U1  (1); U1  (1); U2  (1);
    U1  (1); U2  (1); U2  (1); U1  (1); [12] ;
A : A2- (1); A1+ (1); A1+ (1); A2+ (1); A1+ (1); A2+ (1); A1+ (1); A2- (1);
    A1+ (1); A2+ (1); A2- (1); A1- (1); [12];
\end{lstlisting}
\hyperref[tab:electride]{Back to the table}

\subsubsection*{251718 CrSe$_{2}$}
\label{sec:tqc251718}
\lstset{language=bash, keywordstyle=\color{blue!70}, basicstyle=\ttfamily, frame=shadowbox}
\begin{lstlisting}
Computed bands:  1 -  9
GM: GM1+(1); GM2-(1); GM1+(1); GM2+(1); GM1+(1); GM2-(1); GM2+(1); GM1+(1);
    GM1+(1); [9]  ;
Y : Y2- (1); Y1+ (1); Y1+ (1); Y2- (1); Y2+ (1); Y1+ (1); Y2- (1); Y1- (1);
    Y1+ (1); [9]  ;
V : V1- (1); V1+ (1); V1+ (1); V1- (1); V1+ (1); V1+ (1); V1- (1); V1- (1);
    V1+ (1); [9]  ;
L : L1- (1); L1+ (1); L1+ (1); L1- (1); L1+ (1); L1+ (1); L1- (1); L1- (1);
    L1+ (1); [9]  ;
M : M2- (1); M1+ (1); M1+ (1); M2- (1); M2+ (1); M1+ (1); M2- (1); M1- (1);
    M1+ (1); [9]  ;
U : U1  (1); U2  (1); U1  (1); U2  (1); U1  (1); U1  (1); U2  (1); U2  (1);
    U1  (1); [9]  ;
A : A1+ (1); A2- (1); A1+ (1); A2+ (1); A1+ (1); A2- (1); A2+ (1); A1+ (1);
    A1+ (1); [9] ;
\end{lstlisting}
\hyperref[tab:electride]{Back to the table}

\subsubsection*{631850 VGaFe$_{2}$}
\label{sec:tqc631850}
\noindent Essential BR: $Ag@24d$ \\
\noindent RSI:
\begin{flalign*}
&\delta_{1}@24d\equiv -m(Ag)+m(Au)-m(B1g)+m(B1u)-m(B3g)+m(B3u)-m(B2g)+m(B2u) = -1,&
\end{flalign*}
\lstset{language=bash, keywordstyle=\color{blue!70}, basicstyle=\ttfamily, frame=shadowbox}
\begin{lstlisting}
Computed bands:  1 - 12
GM: GM1+(1); GM4-(3); GM3+(2); GM5+(3); GM5+(3); [12] ;
X : X1+ (1); X3- (1); X4+ (1); X1+ (1); X4- (1); X5+ (2); X5- (2); X5- (2);
    X2- (1); [12] ;
L : L1+ (1); L2- (1); L3- (2); L3- (2); L2- (1); L3+ (2); L3+ (2); L1+ (1);
    [12] ;
W : W1  (1); W5  (2); W2  (1); W2  (1); W5  (2); W3  (1); W5  (2); W1  (1);
    W4  (1); [12];
\end{lstlisting}
\hyperref[tab:electride]{Back to the table}

\subsubsection*{30101 Si}
\label{sec:tqc30101}
\lstset{language=bash, keywordstyle=\color{blue!70}, basicstyle=\ttfamily, frame=shadowbox}
\begin{lstlisting}
Computed bands:  1 -  8
A : A1  (2); A1  (2); A3  (4); [8]  ;
GM: GM1+(1); GM3+(1); GM4-(1); GM5+(2); GM1+(1); GM6+(2); [8]  ;
H : H2  (2); H1  (2); H3  (2); H2  (2); [8]  ;
K : K5  (2); K6  (2); K1  (1); K5  (2); K2  (1); [8]  ;
L : L1  (2); L1  (2); L1  (2); L2  (2); [8]  ;
M : M4- (1); M1+ (1); M2- (1); M3+ (1); M1+ (1); M3- (1); M4- (1); M1- (1); [8] ;
\end{lstlisting}
\hyperref[tab:electride]{Back to the table}

\subsubsection*{633766 TiFe$_{2}$Sn}
\label{sec:tqc633766}
\noindent Essential BR: $Ag@24d$ \\
\noindent RSI:
\begin{flalign*}
&\delta_{1}@24d\equiv -m(Ag)+m(Au)-m(B1g)+m(B1u)-m(B3g)+m(B3u)-m(B2g)+m(B2u) = -1,&
\end{flalign*}
\lstset{language=bash, keywordstyle=\color{blue!70}, basicstyle=\ttfamily, frame=shadowbox}
\begin{lstlisting}
Computed bands:  1 - 12
GM: GM1+(1); GM4-(3); GM3+(2); GM5+(3); GM5+(3); [12] ;
X : X1+ (1); X3- (1); X4+ (1); X1+ (1); X5- (2); X4- (1); X5+ (2); X5- (2);
    X2- (1); [12] ;
L : L2- (1); L1+ (1); L3+ (2); L1+ (1); L3+ (2); L3- (2); L3- (2); L2- (1);
    [12] ;
W : W2  (1); W5  (2); W1  (1); W1  (1); W5  (2); W4  (1); W5  (2); W2  (1);
    W3  (1); [12];
\end{lstlisting}
\hyperref[tab:electride]{Back to the table}

\subsubsection*{57827 TiAlFe$_{2}$}
\label{sec:tqc57827}
\noindent Essential BR: $Ag@24d$ \\
\noindent RSI:
\begin{flalign*}
&\delta_{1}@24d\equiv -m(Ag)+m(Au)-m(B1g)+m(B1u)-m(B3g)+m(B3u)-m(B2g)+m(B2u) = -1,&
\end{flalign*}
\lstset{language=bash, keywordstyle=\color{blue!70}, basicstyle=\ttfamily, frame=shadowbox}
\begin{lstlisting}
Computed bands:  1 - 12
GM: GM1+(1); GM3+(2); GM4-(3); GM5+(3); GM5+(3); [12] ;
X : X1+ (1); X3- (1); X4+ (1); X1+ (1); X4- (1); X5+ (2); X5- (2); X2- X5- (3);
    [12] ;
L : L1+ (1); L2- (1); L3- (2); L3- (2); L2- (1); L3+ (2); L3+ (2); L1+ (1);
    [12] ;
W : W1  (1); W5  (2); W2  (1); W2  (1); W5  (2); W3  (1); W5  (2); W4  (1);
    W1  (1); [12];
\end{lstlisting}
\hyperref[tab:electride]{Back to the table}

\subsubsection*{35701 ZrCl}
\label{sec:tqc35701}
\noindent Essential BR: $A1g@1b$ \\
\noindent RSI:
\begin{flalign*}
&\delta_{1}@1b\equiv -m(Eg)+m(Eu) = 0,&
\\
&\delta_{2}@1b\equiv -m(A1g)+m(A1u)-m(A2g)+m(A2u) = -1,&
\end{flalign*}
\lstset{language=bash, keywordstyle=\color{blue!70}, basicstyle=\ttfamily, frame=shadowbox}
\begin{lstlisting}
Computed bands:  1 - 19
A : A2- (1); A1+ (1); A2- (1); A1+ (1); A3- (2); A3+ (2); A1+ (1); A2- (1);
    A1+ (1); A2- (1); A3+ (2); A3- (2); A2- (1); A1+ (1); A2- (1); [19] ;
GM: GM1+(1); GM2-(1); GM1+(1); GM2-(1); GM3+(2); GM3-(2); GM1+(1); GM2-(1);
    GM1+(1); GM2-(1); GM3+(2); GM3-(2); GM1+(1); GM2-(1); GM1+(1); [19] ;
H : H3  (2); H3  (2); H2  (1); H1  (1); H3  (2); H3  (2); H1  (1); H2  (1);
    H3  (2); H3  (2); H2  (1); H3  (2); [19] ;
K : K3  (2); K3  (2); K1  (1); K2  (1); K3  (2); K3  (2); K1  (1); K2  (1);
    K3  (2); K3  (2); K1  (1); K3  (2); [19] ;
L : L1+ (1); L2- (1); L2- (1); L1+ (1); L1+ (1); L2- (1); L2+ (1); L1- (1);
    L2- (1); L1+ (1); L1+ (1); L2- (1); L2- (1); L1+ (1); L1- (1); L2+ (1);
    L2- (1); L1+ (1); L2- (1); [19] ;
M : M2- (1); M1+ (1); M1+ (1); M2- (1); M2- (1); M1+ (1); M1- (1); M2+ (1);
    M2- (1); M1+ (1); M1+ (1); M2- (1); M2- (1); M1+ (1); M1- (1); M2+ (1);
    M1+ (1); M2- (1); M1+ (1); [19];
\end{lstlisting}
\hyperref[tab:electride]{Back to the table}

\subsubsection*{633246 FeSiRu$_{2}$}
\label{sec:tqc633246}
\noindent Essential BR: $Ag@24d$ \\
\noindent RSI:
\begin{flalign*}
&\delta_{1}@24d\equiv -m(Ag)+m(Au)-m(B1g)+m(B1u)-m(B3g)+m(B3u)-m(B2g)+m(B2u) = -1,&
\end{flalign*}
\lstset{language=bash, keywordstyle=\color{blue!70}, basicstyle=\ttfamily, frame=shadowbox}
\begin{lstlisting}
Computed bands:  1 - 14
GM: GM1+(1); GM4-(3); GM3+(2); GM5+(3); GM5+(3); GM3+(2); [14] ;
X : X1+ (1); X3- (1); X1+ (1); X4+ (1); X4- (1); X5- (2); X5+ (2); X2+ (1);
    X5- (2); X2- (1); X1+ (1); [14] ;
L : L2- (1); L1+ (1); L3+ (2); L1+ (1); L3- (2); L3+ (2); L3- (2); L2- (1);
    L3+ (2); [14] ;
W : W2  (1); W5  (2); W1  (1); W1  (1); W5  (2); W4  (1); W5  (2); W2  (1);
    W3  (1); W2  (1); W1  (1); [14];
\end{lstlisting}
\hyperref[tab:electride]{Back to the table}

\subsubsection*{57832 AlVFe$_{2}$}
\label{sec:tqc57832}
\noindent Essential BR: $Ag@24d$ \\
\noindent RSI:
\begin{flalign*}
&\delta_{1}@24d\equiv -m(Ag)+m(Au)-m(B1g)+m(B1u)-m(B3g)+m(B3u)-m(B2g)+m(B2u) = -1,&
\end{flalign*}
\lstset{language=bash, keywordstyle=\color{blue!70}, basicstyle=\ttfamily, frame=shadowbox}
\begin{lstlisting}
Computed bands:  1 - 12
GM: GM1+(1); GM3+(2); GM4-(3); GM5+(3); GM5+(3); [12] ;
X : X1+ (1); X3- (1); X4+ (1); X1+ (1); X5+ (2); X4- (1); X5- (2); X5- (2);
    X2- (1); [12] ;
L : L1+ (1); L2- (1); L3- (2); L3- (2); L2- (1); L3+ (2); L3+ (2); L1+ (1);
    [12] ;
W : W1  (1); W5  (2); W2  (1); W2  (1); W5  (2); W3  (1); W5  (2); W4  (1);
    W1  (1); [12];
\end{lstlisting}
\hyperref[tab:electride]{Back to the table}

\subsubsection*{240110 LiAl}
\label{sec:tqc240110}
\noindent Essential BR: $A1g@16d$ \\
\noindent RSI:
\begin{flalign*}
&\delta_{1}@16d\equiv -m(Eg)+m(Eu) = 0,&
\\
&\delta_{2}@16d\equiv -m(A1g)+m(A1u)-m(A2g)+m(A2u) = -1,&
\end{flalign*}
\lstset{language=bash, keywordstyle=\color{blue!70}, basicstyle=\ttfamily, frame=shadowbox}
\begin{lstlisting}
Computed bands:  1 -  4
GM: GM1+(1); GM5+(3); [4]  ;
X : X1  (2); X3  (2); [4]  ;
L : L1+ (1); L2- (1); L3+ (2); [4]  ;
W : W2  (2); W1  (2); [4] ;
\end{lstlisting}
\hyperref[tab:electride]{Back to the table}

\subsubsection*{51975 Mg$_{3}$In}
\label{sec:tqc51975}
\noindent Essential BR: $A1g@1b$ \\
\noindent RSI:
\begin{flalign*}
&\delta_{1}@1b\equiv -m(A1g)+m(A1u)-m(T1g)+m(T1u) = -1,&
\\
&\delta_{2}@1b\equiv -m(A2g)+m(A2u)-m(T2g)+m(T2u) = 0,&
\\
&\delta_{3}@1b\equiv m(A1g)-m(A1u)+m(A2g)-m(A2u)-m(Eg)+m(Eu) = 1,&
\end{flalign*}
\lstset{language=bash, keywordstyle=\color{blue!70}, basicstyle=\ttfamily, frame=shadowbox}
\begin{lstlisting}
Computed bands:  1 -  5
GM: GM1+(1); GM1+(1); GM4-(3); [5]  ;
R : R1+ (1); R4- (3); R2- (1); [5]  ;
M : M1+ (1); M5- (2); M4+ (1); M3- (1); [5]  ;
X : X1+ (1); X3- (1); X5- (2); X3- (1); [5] ;
\end{lstlisting}
\hyperref[tab:electride]{Back to the table}

\subsubsection*{620612 ZrCd}
\label{sec:tqc620612}
\lstset{language=bash, keywordstyle=\color{blue!70}, basicstyle=\ttfamily, frame=shadowbox}
\begin{lstlisting}
Computed bands:  1 - 24
A : A3  (2); A2  (2); A1  (2); A3  (2); A1  (2); A3  (2); A2  (2); A3  (2);
    A4  (2); A3  (2); A1  (2); A2  (2); [24] ;
GM: GM1+(1); GM3-(1); GM1+(1); GM3-(1); GM5+(2); GM5-(2); GM4-(1); GM3-(1);
    GM2+(1); GM5+(2); GM1+(1); GM4+(1); GM5-(2); GM2-(1); GM1+(1); GM3-(1);
    GM1+(1); GM3-(1); GM4-(1); GM2+(1); [24] ;
M : M3  (2); M2  (2); M1  (2); M3  (2); M2  (2); M3  (2); M1  (2); M3  (2);
    M4  (2); M3  (2); M2  (2); M1  (2); [24] ;
Z : Z1+ (1); Z3- (1); Z1+ (1); Z3- (1); Z5+ (2); Z5- (2); Z2+ (1); Z1+ (1);
    Z4- (1); Z5- (2); Z3- (1); Z2- (1); Z5+ (2); Z4+ (1); Z3- (1); Z1+ (1);
    Z3- (1); Z1+ (1); Z4- (1); Z2+ (1); [24] ;
R : R1  (2); R1  (2); R1  (2); R2  (2); R1  (2); R2  (2); R1  (2); R1  (2);
    R2  (2); R1  (2); R1  (2); R1  (2); [24] ;
X : X1  (2); X1  (2); X1  (2); X2  (2); X1  (2); X2  (2); X1  (2); X1  (2);
    X2  (2); X1  (2); X1  (2); X1  (2); [24];
\end{lstlisting}
\hyperref[tab:electride]{Back to the table}

\subsubsection*{616286 HfBe}
\label{sec:tqc616286}
\noindent Essential BR: $Ag@4b$ \\
\noindent RSI:
\begin{flalign*}
&\delta_{1}@4b\equiv -m(Ag)+m(Au)-m(Bg)+m(Bu) = -1,&
\end{flalign*}
\lstset{language=bash, keywordstyle=\color{blue!70}, basicstyle=\ttfamily, frame=shadowbox}
\begin{lstlisting}
Computed bands:  1 -  6
GM: GM1+(1); GM4-(1); GM4-(1); GM1+(1); GM3+(1); GM1+(1); [6]  ;
T : T1  (2); T1  (2); T1  (2); [6]  ;
Y : Y1+ (1); Y4- (1); Y1+ (1); Y1+ (1); Y3+ (1); Y4- (1); [6]  ;
Z : Z1  (2); Z1  (2); Z1  (2); [6]  ;
R : R1  (2); R1  (2); R1  (2); [6]  ;
S : S2- (1); S1+ (1); S2- (1); S1- (1); S1+ (1); S2- (1); [6] ;
\end{lstlisting}
\hyperref[tab:electride]{Back to the table}

\subsubsection*{249592 SrMgIn$_{3}$}
\label{sec:tqc249592}
\lstset{language=bash, keywordstyle=\color{blue!70}, basicstyle=\ttfamily, frame=shadowbox}
\begin{lstlisting}
Computed bands:  1 - 11
GM: GM1 (1); GM2 (1); GM5 (2); GM1 (1); GM2 (1); GM1 (1); GM5 (2); GM1 (1);
    GM2 (1); [11] ;
M : M1  (1); M2  (1); M5  (2); M2  (1); M1  (1); M5  (2); M2  (1); M1  (1);
    M1  (1); [11] ;
P : P1  (1); P2  (1); P3  P4  (2); P2  (1); P4  (1); P1  (1); P3  (1); P2  (1);
    P3  (1); P4  (1); [11] ;
X : X1  (1); X2  (1); X3  X4  (2); X1  (1); X3  (1); X2  (1); X3  (1); X1  (1);
    X4  (1); X4  (1); [11] ;
N : N1  (1); N1  (1); N1  (1); N2  (1); N1  (1); N1  (1); N1  (1); N1  (1);
    N2  (1); N1  (1); N1  (1); [11];
\end{lstlisting}
\hyperref[tab:electride]{Back to the table}

\subsubsection*{416528 ScInCu$_{4}$}
\label{sec:tqc416528}
\noindent Essential BR: $A1@4d$ \\
\noindent RSI:
\begin{flalign*}
&\delta_{1}@4d\equiv m(A1)-m(A2)-m(T2)+m(T1) = 1,&
\end{flalign*}
\lstset{language=bash, keywordstyle=\color{blue!70}, basicstyle=\ttfamily, frame=shadowbox}
\begin{lstlisting}
Computed bands:  1 - 25
GM: GM1 (1); GM1 (1); GM4 (3); GM4 (3); GM3 (2); GM5 (3); GM3 (2); GM4 (3);
    GM5 (3); GM4 (3); GM1 (1); [25] ;
X : X3  (1); X1  (1); X3  (1); X1  (1); X5  (2); X1  (1); X3  (1); X5  (2);
    X1  (1); X2  (1); X3  (1); X5  (2); X4  (1); X5  (2); X2  (1); X5  (2);
    X4  (1); X3  (1); X5  (2); [25] ;
L : L1  (1); L1  (1); L1  (1); L3  (2); L1  (1); L3  (2); L1  (1); L3  (2);
    L1  (1); L3  (2); L3  (2); L2  (1); L3  (2); L3  (2); L2  (1); L3  (2);
    L1  (1); [25] ;
W : W4  (1); W2  (1); W3  (1); W1  (1); W2  (1); W1  (1); W3  (1); W4  (1);
    W1  (1); W4  (1); W2  (1); W3  (1); W2  (1); W3  (1); W1  (1); W4  (1);
    W1  (1); W4  (1); W3  (1); W2  (1); W3  (1); W2  (1); W4  (1); W1  (1);
    W3  (1); [25];
\end{lstlisting}
\hyperref[tab:electride]{Back to the table}

\subsubsection*{103055 MgCu$_{4}$Sn}
\label{sec:tqc103055}
\noindent Essential BR: $A1@4b$ \\
\noindent RSI:
\begin{flalign*}
&\delta_{1}@4b\equiv m(A1)-m(A2)-m(T2)+m(T1) = 1,&
\end{flalign*}
\lstset{language=bash, keywordstyle=\color{blue!70}, basicstyle=\ttfamily, frame=shadowbox}
\begin{lstlisting}
Computed bands:  1 - 25
GM: GM1 (1); GM1 (1); GM4 (3); GM4 (3); GM3 (2); GM5 (3); GM3 (2); GM4 (3);
    GM5 (3); GM1 (1); GM4 (3); [25] ;
X : X1  (1); X3  (1); X1  (1); X5  (2); X3  (1); X3  (1); X5  (2); X1  (1);
    X3  (1); X1  (1); X5  (2); X4  (1); X2  (1); X5  (2); X5  (2); X4  (1);
    X2  (1); X1  (1); X5  (2); [25] ;
L : L1  (1); L1  (1); L1  (1); L3  (2); L1  (1); L3  (2); L3  (2); L1  (1);
    L1  (1); L3  (2); L3  (2); L2  (1); L3  (2); L3  (2); L2  (1); L1  (1);
    L3  (2); [25] ;
W : W1  (1); W3  (1); W2  (1); W4  (1); W2  (1); W3  (1); W4  (1); W1  (1);
    W4  (1); W1  (1); W2  (1); W3  (1); W3  (1); W2  (1); W4  (1); W1  (1);
    W4  (1); W1  (1); W3  (1); W2  (1); W3  (1); W2  (1); W1  (1); W4  (1);
    W2  (1); [25];
\end{lstlisting}
\hyperref[tab:electride]{Back to the table}

\subsubsection*{42607 P$_{2}$Ru}
\label{sec:tqc42607}
\lstset{language=bash, keywordstyle=\color{blue!70}, basicstyle=\ttfamily, frame=shadowbox}
\begin{lstlisting}
Computed bands:  1 - 18
GM: GM1+(1); GM2+(1); GM3-(1); GM4-(1); GM2+(1); GM2+(1); GM1+(1); GM1+(1);
    GM3-(1); GM2+(1); GM4-(1); GM1+(1); GM4+(1); GM3+(1); GM1+(1); GM2+(1);
    GM3+(1); GM4+(1); [18] ;
R : R2  (2); R1  (2); R2  (2); R1  (2); R2  (2); R1  (2); R2  (2); R1  (2);
    R2  (2); [18] ;
S : S3- S4- (2); S1+ S2+ (2); S1+ S2+ (2); S3- S4- (2); S1+ S2+ (2); S3+ S4+ (2);
    S1+ S2+ (2); S3- S4- (2); S3+ S4+ (2); [18] ;
T : T1- (2); T1+ (2); T1+ (2); T1+ (2); T1- (2); T1+ T1- (4); T1+ (2); T1- (2);
    [18] ;
U : U1+ (2); U1- (2); U1+ (2); U1- (2); U1+ (2); U1+ (2); U1- (2); U1+ (2);
    U1+ (2); [18] ;
X : X1  (2); X1  (2); X1  (2); X1  (2); X1  (2); X2  (2); X1  (2); X1  (2);
    X2  (2); [18] ;
Y : Y2  (2); Y2  (2); Y2  (2); Y2  (2); Y2  (2); Y1  (2); Y2  (2); Y1  (2);
    Y2  (2); [18] ;
Z : Z1  (2); Z2  (2); Z1  (2); Z2  (2); Z1  (2); Z2  (2); Z1  (2); Z2  (2);
    Z1  (2); [18];
\end{lstlisting}
\hyperref[tab:electride]{Back to the table}

\subsubsection*{43652 Sb$_{2}$Ru}
\label{sec:tqc43652}
\lstset{language=bash, keywordstyle=\color{blue!70}, basicstyle=\ttfamily, frame=shadowbox}
\begin{lstlisting}
Computed bands:  1 - 18
GM: GM1+(1); GM2+(1); GM3-(1); GM4-(1); GM2+(1); GM2+(1); GM1+(1); GM3-(1);
    GM1+(1); GM4-(1); GM2+(1); GM1+(1); GM3+(1); GM4+(1); GM1+(1); GM3+(1);
    GM2+(1); GM4+(1); [18] ;
R : R2  (2); R1  (2); R2  (2); R1  (2); R2  (2); R1  (2); R2  (2); R1  (2);
    R2  (2); [18] ;
S : S3- S4- (2); S1+ S2+ (2); S1+ S2+ (2); S3- S4- (2); S1+ S2+ (2); S1+ S2+ (2);
    S3+ S4+ (2); S3- S4- (2); S3+ S4+ (2); [18] ;
T : T1- (2); T1+ (2); T1+ (2); T1+ (2); T1- (2); T1+ (2); T1- (2); T1+ (2);
    T1- (2); [18] ;
U : U1+ (2); U1- (2); U1+ (2); U1- (2); U1+ (2); U1+ (2); U1+ (2); U1- (2);
    U1+ (2); [18] ;
X : X1  (2); X1  (2); X1  (2); X1  (2); X1  (2); X2  (2); X1  (2); X1  (2);
    X2  (2); [18] ;
Y : Y2  (2); Y2  (2); Y2  (2); Y2  (2); Y2  (2); Y2  (2); Y1  (2); Y1  (2);
    Y2  (2); [18] ;
Z : Z1  (2); Z2  (2); Z1  (2); Z2  (2); Z1  (2); Z2  (2); Z1  (2); Z2  (2);
    Z1  (2); [18];
\end{lstlisting}
\hyperref[tab:electride]{Back to the table}

\subsubsection*{163696 YMgCu$_{4}$}
\label{sec:tqc163696}
\noindent Essential BR: $A1@4b$ \\
\noindent RSI:
\begin{flalign*}
&\delta_{1}@4b\equiv m(A1)-m(A2)-m(T2)+m(T1) = 1,&
\end{flalign*}
\lstset{language=bash, keywordstyle=\color{blue!70}, basicstyle=\ttfamily, frame=shadowbox}
\begin{lstlisting}
Computed bands:  1 - 29
GM: GM1 (1); GM4 (3); GM1 (1); GM1 (1); GM4 (3); GM4 (3); GM3 (2); GM5 (3);
    GM3 (2); GM4 (3); GM5 (3); GM4 (3); GM1 (1); [29] ;
X : X3  (1); X1  (1); X5  (2); X3  (1); X1  (1); X1  (1); X3  (1); X3  (1);
    X5  (2); X1  (1); X5  (2); X3  (1); X4  (1); X1  (1); X5  (2); X2  (1);
    X5  (2); X4  (1); X2  (1); X5  (2); X1  (1); X5  (2); [29] ;
L : L1  (1); L1  (1); L3  (2); L1  (1); L1  (1); L1  (1); L3  (2); L1  (1);
    L3  (2); L1  (1); L1  (1); L3  (2); L3  (2); L3  (2); L2  (1); L3  (2);
    L3  (2); L2  (1); L3  (2); L1  (1); [29] ;
W : W4  (1); W1  W2  (2); W3  (1); W3  (1); W2  (1); W1  (1); W4  (1); W3  (1);
    W4  (1); W1  (1); W4  (1); W2  (1); W1  (1); W3  (1); W2  (1); W3  (1);
    W2  (1); W4  (1); W1  (1); W2  (1); W4  (1); W3  (1); W1  (1); W2  (1);
    W3  (1); W1  (1); W4  (1); W2  (1); [29];
\end{lstlisting}
\hyperref[tab:electride]{Back to the table}

\subsubsection*{186627 FeSb$_{2}$}
\label{sec:tqc186627}
\lstset{language=bash, keywordstyle=\color{blue!70}, basicstyle=\ttfamily, frame=shadowbox}
\begin{lstlisting}
Computed bands:  1 - 18
GM: GM1+(1); GM2+(1); GM3-(1); GM4-(1); GM2+(1); GM2+(1); GM3-(1); GM4-(1);
    GM1+(1); GM1+(1); GM2+(1); GM3+(1); GM4+(1); GM1+(1); GM1+(1); GM3+(1);
    GM4+(1); GM2+(1); [18] ;
R : R2  (2); R1  (2); R2  (2); R1  (2); R2  (2); R1  (2); R2  (2); R1  (2);
    R2  (2); [18] ;
S : S3- S4- (2); S1+ S2+ (2); S1+ S2+ (2); S3- S4- (2); S1+ S2+ (2); S1+ S2+ (2);
    S3+ S4+ (2); S3- S4- (2); S3+ S4+ (2); [18] ;
T : T1- (2); T1+ (2); T1+ (2); T1- (2); T1+ (2); T1- (2); T1+ (2); T1+ (2);
    T1- (2); [18] ;
U : U1+ (2); U1- (2); U1+ (2); U1- (2); U1+ (2); U1+ (2); U1- (2); U1+ (2);
    U1+ (2); [18] ;
X : X1  (2); X1  (2); X1  (2); X1  (2); X1  (2); X2  (2); X1  (2); X2  (2);
    X1  (2); [18] ;
Y : Y2  (2); Y2  (2); Y2  (2); Y2  (2); Y2  (2); Y2  (2); Y1  (2); Y1  (2);
    Y2  (2); [18] ;
Z : Z1  (2); Z2  (2); Z1  (2); Z2  (2); Z1  (2); Z1  (2); Z2  (2); Z2  (2);
    Z1  (2); [18];
\end{lstlisting}
\hyperref[tab:electride]{Back to the table}

\subsubsection*{616209 Be$_{5}$Co}
\label{sec:tqc616209}
\noindent Essential BR: $A1@4b$ \\
\noindent RSI:
\begin{flalign*}
&\delta_{1}@4b\equiv m(A1)-m(A2)-m(T2)+m(T1) = 1,&
\end{flalign*}
\lstset{language=bash, keywordstyle=\color{blue!70}, basicstyle=\ttfamily, frame=shadowbox}
\begin{lstlisting}
Computed bands:  1 - 10
GM: GM1 (1); GM1 (1); GM4 (3); GM3 (2); GM4 (3); [10] ;
X : X1  (1); X3  (1); X3  (1); X5  (2); X1  (1); X2  (1); X5  (2); X1  (1);
    [10] ;
L : L1  (1); L1  (1); L1  (1); L3  (2); L3  (2); L1  (1); L3  (2); [10] ;
W : W2  (1); W3  (1); W4  (1); W1  (1); W3  (1); W1  (1); W4  (1); W2  (1);
    W2  (1); W2  (1); [10];
\end{lstlisting}
\hyperref[tab:electride]{Back to the table}

\subsubsection*{616395 Be$_{5}$Pt}
\label{sec:tqc616395}
\noindent Essential BR: $A1@4b$ \\
\noindent RSI:
\begin{flalign*}
&\delta_{1}@4b\equiv m(A1)-m(A2)-m(T2)+m(T1) = 1,&
\end{flalign*}
\lstset{language=bash, keywordstyle=\color{blue!70}, basicstyle=\ttfamily, frame=shadowbox}
\begin{lstlisting}
Computed bands:  1 - 10
GM: GM1 (1); GM4 (3); GM3 (2); GM1 (1); GM4 (3); [10] ;
X : X1  (1); X3  (1); X3  (1); X1  (1); X2  (1); X5  (2); X5  (2); X1  (1);
    [10] ;
L : L1  (1); L1  (1); L3  (2); L1  (1); L3  (2); L1  (1); L3  (2); [10] ;
W : W2  (1); W3  (1); W1  (1); W4  (1); W1  (1); W2  (1); W3  (1); W4  (1);
    W2  (1); W2  (1); [10];
\end{lstlisting}
\hyperref[tab:electride]{Back to the table}

\subsubsection*{156265 HfGaAu}
\label{sec:tqc156265}
\lstset{language=bash, keywordstyle=\color{blue!70}, basicstyle=\ttfamily, frame=shadowbox}
\begin{lstlisting}
Computed bands:  1 - 18
A : A1  (1); A3  (1); A5  (2); A6  (2); A3  (1); A6  (2); A1  (1); A5  (2);
    A3  (1); A1  (1); A5  (2); A6  (2); [18] ;
GM: GM1 (1); GM3 (1); GM1 (1); GM6 (2); GM5 (2); GM5 (2); GM1 (1); GM6 (2);
    GM3 (1); GM5 (2); GM1 (1); GM6 (2); [18] ;
H : H5  (1); H6  (1); H2  (1); H4  (1); H6  (1); H3  (1); H1  H4  (2); H2  (1);
    H1  (1); H5  (1); H3  (1); H5  (1); H6  (1); H2  (1); H1  (1); H5  (1);
    H4  (1); [18] ;
K : K5  (1); K6  (1); K3  (1); K1  (1); K1  (1); K5  (1); K4  (1); K3  (1);
    K2  (1); K6  (1); K2  (1); K5  (1); K1  (1); K6  (1); K4  (1); K5  (1);
    K3  (1); K2  (1); [18] ;
L : L1  (1); L3  (1); L1  (1); L3  (1); L3  (1); L4  (1); L3  (1); L2  L4  (2);
    L3  (1); L1  (1); L2  (1); L1  (1); L1  (1); L2  (1); L4  (1); L1  (1);
    L3  (1); [18] ;
M : M1  (1); M3  (1); M1  (1); M1  (1); M1  (1); M2  (1); M2  (1); M4  (1);
    M3  (1); M1  (1); M3  (1); M4  (1); M3  (1); M3  (1); M1  (1); M4  (1);
    M1  (1); M2  (1); [18];
\end{lstlisting}
\hyperref[tab:electride]{Back to the table}

\subsubsection*{648286 P$_{2}$W}
\label{sec:tqc648286}
\lstset{language=bash, keywordstyle=\color{blue!70}, basicstyle=\ttfamily, frame=shadowbox}
\begin{lstlisting}
Computed bands:  1 - 16
GM: GM1+(1); GM1+(1); GM2-(1); GM2-(1); GM1+(1); GM2-(1); GM1+(1); GM1+(1);
    GM1+(1); GM2-(1); GM1-(1); GM2+(1); GM2+(1); GM1+(1); GM2-(1); GM1-(1);
    [16] ;
Y : Y2- (1); Y1+ (1); Y2- (1); Y1+ (1); Y2- (1); Y2- (1); Y1+ (1); Y1+ (1);
    Y1- (1); Y1+ (1); Y2- (1); Y1+ (1); Y1- (1); Y2+ (1); Y1+ (1); Y2+ (1);
    [16] ;
V : V1+ (1); V1- (1); V1+ (1); V1- (1); V1+ (1); V1- (1); V1+ (1); V1- (1);
    V1- (1); V1- (1); V1+ (1); V1+ (1); V1- (1); V1+ (1); V1+ (1); V1+ (1);
    [16] ;
L : L1+ (1); L1+ (1); L1- (1); L1- (1); L1+ (1); L1- (1); L1+ (1); L1- (1);
    L1+ (1); L1- (1); L1+ (1); L1- (1); L1+ (1); L1- (1); L1+ (1); L1- (1);
    [16] ;
M : M2- (1); M1+ (1); M1+ (1); M2- (1); M1+ (1); M2- (1); M2- (1); M1+ (1);
    M2- (1); M2+ (1); M1- (1); M1+ (1); M1+ (1); M2- (1); M1- (1); M2+ (1);
    [16] ;
U : U1  (1); U2  (1); U1  (1); U2  (1); U2  (1); U1  (1); U1  (1); U2  (1);
    U2  (1); U1  (1); U2  (1); U1  (1); U2  (1); U1  (1); U1  (1); U2  (1);
    [16] ;
A : A1+ (1); A2- (1); A1+ (1); A2- (1); A1+ (1); A2- (1); A2- (1); A1+ (1);
    A1+ (1); A2+ (1); A2- (1); A1- (1); A2- (1); A1+ (1); A2+ (1); A1- (1); [16];
\end{lstlisting}
\hyperref[tab:electride]{Back to the table}

\subsubsection*{58156 Al$_{2}$Ru}
\label{sec:tqc58156}
\lstset{language=bash, keywordstyle=\color{blue!70}, basicstyle=\ttfamily, frame=shadowbox}
\begin{lstlisting}
Computed bands:  1 - 14
GM: GM1+(1); GM1-(1); GM1-(1); GM2+(1); GM2-(1); GM4+(1); GM4-(1); GM3+(1);
    GM1+(1); GM2+(1); GM3-(1); GM1+(1); GM4+(1); GM3+(1); [14] ;
Z : Z1  (2); Z1  (2); Z1  (2); Z2  (2); Z2  (2); Z1  (2); Z2  (2); [14] ;
H : H1  (2); H1  (2); H1  (2); H1  (2); H1  (2); H1  (2); H1  (2); [14] ;
Y : Y1  (2); Y1  (2); Y2  (2); Y1  (2); Y2  (2); Y2  (2); Y1  (2); [14] ;
L : L1- (1); L1+ (1); L1- (1); L1- (1); L1+ (1); L1+ (1); L1+ (1); L1- (1);
    L1- (1); L1+ (1); L1+ (1); L1- (1); L1+ (1); L1+ (1); [14] ;
T : T1  (2); T2  (2); T1  (2); T2  (2); T1  (2); T2  (2); T1  (2); [14];
\end{lstlisting}
\hyperref[tab:electride]{Back to the table}

\subsubsection*{415195 HoCdCu$_{4}$}
\label{sec:tqc415195}
\noindent Essential BR: $A1@4b$ \\
\noindent RSI:
\begin{flalign*}
&\delta_{1}@4b\equiv m(A1)-m(A2)-m(T2)+m(T1) = 1,&
\end{flalign*}
\lstset{language=bash, keywordstyle=\color{blue!70}, basicstyle=\ttfamily, frame=shadowbox}
\begin{lstlisting}
Computed bands:  1 - 33
GM: GM4 (3); GM4 (3); GM3 (2); GM1 (1); GM1 (1); GM4 (3); GM4 (3); GM3 (2);
    GM5 (3); GM3 (2); GM4 (3); GM5 (3); GM4 (3); GM1 (1); [33] ;
X : X1  (1); X5  (2); X3  (1); X1  (1); X5  (2); X2  (1); X3  (1); X1  (1);
    X1  (1); X3  (1); X5  (2); X3  (1); X1  (1); X5  (2); X3  (1); X4  (1);
    X5  (2); X1  (1); X2  (1); X5  (2); X4  (1); X5  (2); X2  (1); X1  (1);
    X5  (2); [33] ;
L : L1  (1); L3  (2); L1  (1); L3  (2); L3  (2); L1  (1); L1  (1); L1  (1);
    L3  (2); L1  (1); L3  (2); L1  (1); L1  (1); L3  (2); L3  (2); L3  (2);
    L2  (1); L3  (2); L3  (2); L2  (1); L1  (1); L3  (2); [33] ;
W : W1  W2  (2); W4  (1); W4  (1); W2  (1); W3  (1); W1  (1); W2  (1); W1  (1);
    W4  (1); W2  (1); W3  (1); W4  (1); W1  (1); W3  (1); W3  (1); W1  (1);
    W2  (1); W4  (1); W2  (1); W4  (1); W3  (1); W2  (1); W1  (1); W3  (1);
    W1  (1); W2  (1); W4  (1); W2  (1); W4  (1); W1  (1); W3  (1); W2  (1); [33];
\end{lstlisting}
\hyperref[tab:electride]{Back to the table}

\subsubsection*{628179 YInCu$_{4}$}
\label{sec:tqc628179}
\noindent Essential BR: $A1@4d$ \\
\noindent RSI:
\begin{flalign*}
&\delta_{1}@4d\equiv m(A1)-m(A2)-m(T2)+m(T1) = 1,&
\end{flalign*}
\lstset{language=bash, keywordstyle=\color{blue!70}, basicstyle=\ttfamily, frame=shadowbox}
\begin{lstlisting}
Computed bands:  1 - 29
GM: GM1 (1); GM4 (3); GM1 (1); GM1 (1); GM4 (3); GM4 (3); GM3 (2); GM5 (3);
    GM3 (2); GM4 (3); GM5 (3); GM4 (3); GM1 (1); [29] ;
X : X1  (1); X3  (1); X5  (2); X3  (1); X1  (1); X3  (1); X5  (2); X1  (1);
    X1  (1); X3  (1); X5  (2); X1  (1); X2  (1); X5  (2); X3  (1); X4  (1);
    X5  (2); X2  (1); X5  (2); X4  (1); X3  (1); X5  (2); [29] ;
L : L1  (1); L1  (1); L3  (2); L1  (1); L1  (1); L1  (1); L3  (2); L1  (1);
    L3  (2); L1  (1); L1  (1); L3  (2); L3  (2); L3  (2); L2  (1); L3  (2);
    L3  (2); L2  (1); L1  (1); L3  (2); [29] ;
W : W1  (1); W4  (1); W3  (1); W2  (1); W4  (1); W2  (1); W3  (1); W1  (1);
    W2  (1); W1  (1); W4  (1); W1  (1); W3  (1); W4  (1); W2  (1); W3  (1);
    W2  (1); W1  (1); W3  (1); W4  (1); W1  (1); W3  (1); W4  (1); W2  (1);
    W3  (1); W2  (1); W4  (1); W1  (1); W3  (1); [29];
\end{lstlisting}
\hyperref[tab:electride]{Back to the table}

\subsubsection*{156264 ZrGaAu}
\label{sec:tqc156264}
\lstset{language=bash, keywordstyle=\color{blue!70}, basicstyle=\ttfamily, frame=shadowbox}
\begin{lstlisting}
Computed bands:  1 - 26
A : A3  (1); A1  (1); A6  (2); A5  (2); A1  (1); A3  (1); A1  (1); A3  (1);
    A5  (2); A6  (2); A3  (1); A6  (2); A1  (1); A5  (2); A3  (1); A1  (1);
    A5  (2); A6  (2); [26] ;
GM: GM1 (1); GM1 (1); GM3 (1); GM5 (2); GM5 (2); GM3 (1); GM1 (1); GM3 (1);
    GM1 (1); GM6 (2); GM5 (2); GM5 (2); GM1 (1); GM6 (2); GM3 (1); GM5 (2);
    GM1 (1); GM6 (2); [26] ;
H : H2  (1); H1  (1); H4  H6  (2); H5  (1); H3  (1); H1  (1); H2  (1); H5  (1);
    H6  (1); H4  (1); H3  (1); H2  (1); H6  (1); H4  (1); H2  (1); H1  (1);
    H1  (1); H5  (1); H3  (1); H5  (1); H6  (1); H2  (1); H1  (1); H5  (1);
    H4  (1); [26] ;
K : K1  (1); K1  (1); K2  (1); K3  (1); K5  (1); K5  (1); K3  (1); K2  (1);
    K5  (1); K6  (1); K3  (1); K1  (1); K4  (1); K5  (1); K1  (1); K3  (1);
    K2  (1); K6  (1); K2  (1); K5  (1); K6  (1); K1  (1); K4  (1); K5  (1);
    K3  (1); K2  (1); [26] ;
L : L3  (1); L1  (1); L3  (1); L1  (1); L4  (1); L2  (1); L1  (1); L3  (1);
    L1  (1); L3  (1); L1  (1); L3  (1); L3  (1); L4  (1); L2  (1); L3  (1);
    L4  (1); L1  (1); L2  (1); L3  (1); L1  (1); L1  (1); L2  (1); L4  (1);
    L1  (1); L3  (1); [26] ;
M : M1  (1); M1  (1); M3  (1); M1  (1); M2  (1); M1  (1); M2  (1); M3  (1);
    M1  (1); M3  (1); M1  (1); M1  (1); M1  (1); M2  (1); M2  (1); M4  (1);
    M3  (1); M1  (1); M3  (1); M4  (1); M3  (1); M3  (1); M1  (1); M4  (1);
    M1  (1); M2  (1); [26];
\end{lstlisting}
\hyperref[tab:electride]{Back to the table}

\subsubsection*{609407 SrAl$_{2}$}
\label{sec:tqc609407}
\lstset{language=bash, keywordstyle=\color{blue!70}, basicstyle=\ttfamily, frame=shadowbox}
\begin{lstlisting}
Computed bands:  1 - 16
GM: GM1+(1); GM2-(1); GM3+(1); GM1+(1); GM2-(1); GM4+(1); GM3-(1); GM4-(1);
    GM1+(1); GM4-(1); GM3+(1); GM1+(1); GM2-(1); GM1-(1); GM4+(1); GM1+(1);
    [16] ;
X : X1+ (1); X2- (1); X3+ (1); X2- (1); X3- (1); X1+ (1); X4+ (1); X4- (1);
    X2- (1); X3+ (1); X1+ (1); X2- (1); X3- (1); X1+ (1); X2+ (1); X3+ (1);
    [16] ;
R : R2- (1); R1+ (1); R1- (1); R1+ (1); R2- (1); R1+ (1); R2- (1); R2+ (1);
    R1+ (1); R2- (1); R2+ (1); R1- (1); R1+ (1); R2- (1); R2+ (1); R1+ (1);
    [16] ;
S : S1+ (1); S2- (1); S1+ (1); S2- (1); S1+ (1); S2- (1); S2+ (1); S1- (1);
    S2- (1); S1+ (1); S1+ (1); S2- (1); S2- (1); S1- (1); S2+ (1); S1+ (1);
    [16] ;
T : T1  (2); T1  (2); T1  (2); T1  (2); T1  (2); T1  (2); T1  (2); T1  (2);
    [16] ;
W : W1  (2); W1  (2); W1  (2); W1  (2); W1  (2); W1  (2); W1  (2); W1  (2); [16];
\end{lstlisting}
\hyperref[tab:electride]{Back to the table}

\subsubsection*{290428 HfN$_{2}$}
\label{sec:tqc290428}
\noindent Essential BR: $A1g@2a$ \\
\noindent RSI:
\begin{flalign*}
&\delta_{1}@2a\equiv -m(Eg)+m(Eu) = 0,&
\\
&\delta_{2}@2a\equiv -m(A1g)+m(A1u)-m(A2g)+m(A2u) = -1,&
\end{flalign*}
\lstset{language=bash, keywordstyle=\color{blue!70}, basicstyle=\ttfamily, frame=shadowbox}
\begin{lstlisting}
Computed bands:  1 - 14
A : A1  (2); A1  (2); A1  (2); A3  (4); A3  (4); [14] ;
GM: GM1+(1); GM3+(1); GM4-(1); GM2-(1); GM1+(1); GM6-(2); GM5-(2); GM3+(1);
    GM5+(2); GM6+(2); [14] ;
H : H3  (2); H3  (2); H2  (2); H3  (2); H1  (2); H2  (2); H1  (2); [14] ;
K : K1  (1); K2  (1); K4  (1); K3  (1); K1  (1); K6  (2); K5  (2); K2  (1);
    K5  (2); K6  (2); [14] ;
L : L1  (2); L1  (2); L1  (2); L1  (2); L2  (2); L1  (2); L2  (2); [14] ;
M : M1+ (1); M3+ (1); M4- (1); M2- (1); M1+ (1); M4- (1); M2- (1); M1- (1);
    M3+ (1); M3- (1); M1+ (1); M4+ (1); M3+ (1); M2+ (1); [14];
\end{lstlisting}
\hyperref[tab:electride]{Back to the table}

\subsubsection*{80945 KSb$_{2}$}
\label{sec:tqc80945}
\noindent Essential BR: $Ag@2a$ \\
\noindent RSI:
\begin{flalign*}
&\delta_{1}@2a\equiv -m(Ag)+m(Au)-m(Bg)+m(Bu) = -1,&
\end{flalign*}
\lstset{language=bash, keywordstyle=\color{blue!70}, basicstyle=\ttfamily, frame=shadowbox}
\begin{lstlisting}
Computed bands:  1 - 19
GM: GM1+(1); GM2-(1); GM1+(1); GM1+(1); GM2-(1); GM2+GM2-(2); GM1-(1); GM1+(1);
    GM2-(1); GM1+(1); GM2-(1); GM1+(1); GM2-(1); GM1+(1); GM1-(1); GM2+(1);
    GM2-(1); GM1+(1); [19] ;
Y : Y1+ Y2- (2); Y1+ (1); Y2- (1); Y1- (1); Y2- (1); Y1+ Y2+ (2); Y1+ (1);
    Y2- (1); Y2- (1); Y1+ (1); Y1+ (1); Y1- (1); Y2- (1); Y2- (1); Y1+ Y2+ (2);
    Y1+ (1); [19] ;
V : V1- (1); V1+ (1); V1+ (1); V1- (1); V1- (1); V1+ (1); V1+ V1- (2); V1+ (1);
    V1- (1); V1+ (1); V1- (1); V1- (1); V1+ (1); V1+ V1- (2); V1+ (1); V1- (1);
    V1+ (1); [19] ;
L : L1+ (1); L1- (1); L1- (1); L1+ (1); L1+ (1); L1- (1); L1- (1); L1+ (1);
    L1+ (1); L1- (1); L1+ (1); L1- (1); L1- (1); L1+ (1); L1- (1); L1+ (1);
    L1+ (1); L1- (1); L1+ (1); [19] ;
M : M1+ M2- (2); M2- (1); M1+ (1); M2+ (1); M1+ (1); M2- (1); M1- (1); M1+ (1);
    M2- (1); M2- (1); M1+ (1); M1+ (1); M1- (1); M2- (1); M2- (1); M2+ (1);
    M1+ (1); M1+ (1); [19] ;
U : U2  (1); U1  (1); U2  (1); U1  (1); U2  (1); U1  (1); U1  (1); U2  (1);
    U1  (1); U1  (1); U2  (1); U2  (1); U1  (1); U2  (1); U1  (1); U2  (1);
    U1  (1); U2  (1); U1  (1); [19] ;
A : A2- (1); A1+ (1); A2- (1); A1+ (1); A2- (1); A1+ (1); A1- (1); A2+ (1);
    A1+ (1); A2- (1); A1+ (1); A2- (1); A1+ (1); A2- (1); A1- (1); A2+ (1);
    A1+ (1); A2- (1); A1+ (1); [19];
\end{lstlisting}
\hyperref[tab:electride]{Back to the table}

\subsubsection*{415196 ErCdCu$_{4}$}
\label{sec:tqc415196}
\noindent Essential BR: $A1@4b$ \\
\noindent RSI:
\begin{flalign*}
&\delta_{1}@4b\equiv m(A1)-m(A2)-m(T2)+m(T1) = 1,&
\end{flalign*}
\lstset{language=bash, keywordstyle=\color{blue!70}, basicstyle=\ttfamily, frame=shadowbox}
\begin{lstlisting}
Computed bands:  1 - 33
GM: GM4 (3); GM4 (3); GM3 (2); GM1 (1); GM1 (1); GM4 (3); GM4 (3); GM3 (2);
    GM5 (3); GM3 (2); GM4 (3); GM5 (3); GM4 (3); GM1 (1); [33] ;
X : X1  (1); X5  (2); X3  (1); X1  (1); X5  (2); X2  (1); X3  (1); X1  (1);
    X1  (1); X3  (1); X5  (2); X3  (1); X1  (1); X5  (2); X3  (1); X4  (1);
    X5  (2); X1  (1); X2  (1); X5  (2); X4  (1); X5  (2); X2  (1); X1  (1);
    X5  (2); [33] ;
L : L1  (1); L3  (2); L1  (1); L3  (2); L3  (2); L1  (1); L1  (1); L1  (1);
    L3  (2); L1  (1); L3  (2); L1  (1); L1  (1); L3  (2); L3  (2); L3  (2);
    L2  (1); L3  (2); L3  (2); L2  (1); L1  (1); L3  (2); [33] ;
W : W1  W2  (2); W4  (1); W2  W4  (2); W3  (1); W1  (1); W2  (1); W1  (1);
    W4  (1); W2  (1); W3  (1); W4  (1); W1  (1); W3  (1); W3  (1); W1  W2  (2);
    W4  (1); W2  (1); W4  (1); W3  (1); W2  (1); W1  (1); W3  (1); W1  (1);
    W2  (1); W4  (1); W2  (1); W4  (1); W1  (1); W3  (1); W2  (1); [33];
\end{lstlisting}
\hyperref[tab:electride]{Back to the table}

\subsubsection*{71998 ScSiAu}
\label{sec:tqc71998}
\lstset{language=bash, keywordstyle=\color{blue!70}, basicstyle=\ttfamily, frame=shadowbox}
\begin{lstlisting}
Computed bands:  1 - 18
A : A3  (1); A1  (1); A6  (2); A5  (2); A1  (1); A3  (1); A5  (2); A6  (2);
    A1  (1); A3  (1); A6  (2); A5  (2); [18] ;
GM: GM1 (1); GM3 (1); GM5 (2); GM6 (2); GM1 (1); GM5 (2); GM1 (1); GM6 (2);
    GM3 (1); GM5 (2); GM1 (1); GM6 (2); [18] ;
H : H4  (1); H3  (1); H6  (1); H5  (1); H1  (1); H3  (1); H2  (1); H5  (1);
    H1  H2  (2); H4  (1); H6  (1); H4  (1); H3  (1); H1  (1); H2  (1); H4  (1);
    H5  (1); [18] ;
K : K3  (1); K4  (1); K5  (1); K1  (1); K3  (1); K6  (1); K1  (1); K5  (1);
    K2  (1); K4  (1); K2  (1); K3  (1); K1  (1); K4  (1); K6  (1); K3  (1);
    K2  (1); K5  (1); [18] ;
L : L3  (1); L1  (1); L3  (1); L1  (1); L1  (1); L2  (1); L4  (1); L1  (1);
    L2  (1); L3  (1); L1  (1); L4  (1); L3  (1); L3  (1); L3  (1); L4  (1);
    L2  (1); L1  (1); [18] ;
M : M1  (1); M3  (1); M1  (1); M1  (1); M2  (1); M1  (1); M3  (1); M4  (1);
    M2  (1); M1  (1); M3  (1); M4  (1); M3  (1); M3  (1); M1  (1); M4  (1);
    M2  (1); M1  (1); [18];
\end{lstlisting}
\hyperref[tab:electride]{Back to the table}

\subsubsection*{16358 BaSe$_{2}$}
\label{sec:tqc16358}
\noindent Essential BR: $Ag@4d$ \\
\noindent RSI:
\begin{flalign*}
&\delta_{1}@4d\equiv -m(Ag)+m(Au) = -1,&
\end{flalign*}
\lstset{language=bash, keywordstyle=\color{blue!70}, basicstyle=\ttfamily, frame=shadowbox}
\begin{lstlisting}
Computed bands:  1 - 22
GM: GM1+(1); GM1-(1); GM2+(1); GM1+(1); GM2+(1); GM1-(1); GM1+(1); GM2-(1);
    GM2-(1); GM2+(1); GM1-(1); GM2-(1); GM2+(1); GM1+(1); GM1-(1); GM2-(1);
    GM1-(1); GM2-(1); GM1+(1); GM1+(1); GM2+(1); GM2+(1); [22] ;
Y : Y1+ Y1- (2); Y2- (1); Y1- (1); Y2+ (1); Y1+ (1); Y2+ (1); Y2- (1); Y1- (1);
    Y2- (1); Y1+ (1); Y2+ (1); Y1+ (1); Y1- (1); Y2- (1); Y1+ (1); Y2+ (1);
    Y2+ (1); Y1- (1); Y2- (1); Y1- (1); Y2- (1); [22] ;
L : L1- (1); L1+ (1); L1+ (1); L1+ (1); L1- (1); L1- (1); L1- (1); L1+ (1);
    L1+ (1); L1+ (1); L1- (1); L1- (1); L1+ (1); L1- (1); L1+ (1); L1- (1);
    L1- (1); L1- (1); L1+ (1); L1+ (1); L1+ (1); L1+ (1); [22] ;
M : M1  (2); M1  (2); M1  (2); M1  (2); M1  (2); M1  (2); M1  (2); M1  (2);
    M1  (2); M1  (2); M1  (2); [22] ;
V : V1+ (1); V1- (1); V1- (1); V1+ (1); V1- (1); V1+ (1); V1+ (1); V1- (1);
    V1+ (1); V1- (1); V1- (1); V1+ (1); V1- (1); V1+ (1); V1+ (1); V1- (1);
    V1- (1); V1+ (1); V1+ (1); V1+ (1); V1- (1); V1- (1); [22] ;
U : U1  U2  (2); U1  U2  (2); U1  U2  (2); U1  U2  (2); U1  U2  (2); U1  U2  (2);
    U1  U2  (2); U1  U2  (2); U1  U2  (2); U1  U2  (2); U1  U2  (2); [22] ;
A : A1  (2); A1  (2); A1  (2); A1  (2); A1  (2); A1  (2); A1  (2); A1  (2);
    A1  (2); A1  (2); A1  (2); [22];
\end{lstlisting}
\hyperref[tab:electride]{Back to the table}

\subsubsection*{44751 FeSe$_{2}$}
\label{sec:tqc44751}
\noindent Essential BR: $Ag@2c$ \\
\noindent RSI:
\begin{flalign*}
&\delta_{1}@2c\equiv -m(Ag)+m(Au)-m(Bg)+m(Bu) = -1,&
\end{flalign*}
\lstset{language=bash, keywordstyle=\color{blue!70}, basicstyle=\ttfamily, frame=shadowbox}
\begin{lstlisting}
Computed bands:  1 - 20
GM: GM1+(1); GM2+(1); GM3-(1); GM4-(1); GM2+(1); GM2+(1); GM1+(1); GM3-(1);
    GM1-(1); GM1+(1); GM3+(1); GM4-(1); GM4+(1); GM2+(1); GM3+(1); GM1+(1);
    GM1+(1); GM2-(1); GM4+(1); GM2+(1); [20] ;
R : R2  (2); R1  (2); R2  (2); R1  (2); R1  (2); R2  (2); R2  (2); R1  (2);
    R2  (2); R1  (2); [20] ;
S : S3- S4- (2); S1+ S2+ (2); S1+ S2+ (2); S3- S4- (2); S3+ S4+ (2); S1+ S2+ (2);
    S1- S2- (2); S3- S4- (2); S1+ S2+ (2); S3+ S4+ (2); [20] ;
T : T1- (2); T1+ (2); T1+ (2); T1- (2); T1+ (2); T1- (2); T1- (2); T1+ (2);
    T1+ (2); T1+ (2); [20] ;
U : U1+ (2); U1- (2); U1+ (2); U1- (2); U1+ (2); U1- (2); U1+ (2); U1+ (2);
    U1+ (2); U1+ (2); [20] ;
X : X1  (2); X1  (2); X1  (2); X2  (2); X1  (2); X1  (2); X2  (2); X1  (2);
    X1  (2); X2  (2); [20] ;
Y : Y2  (2); Y2  (2); Y2  (2); Y2  (2); Y1  (2); Y2  (2); Y1  (2); Y2  (2);
    Y2  (2); Y1  (2); [20] ;
Z : Z1  (2); Z2  (2); Z1  (2); Z1  (2); Z2  (2); Z1  (2); Z2  (2); Z2  (2);
    Z1  (2); Z1  (2); [20];
\end{lstlisting}
\hyperref[tab:electride]{Back to the table}

\subsubsection*{419402 RbSb$_{2}$}
\label{sec:tqc419402}
\noindent Essential BR: $Ag@2a$ \\
\noindent RSI:
\begin{flalign*}
&\delta_{1}@2a\equiv -m(Ag)+m(Au)-m(Bg)+m(Bu) = -1,&
\end{flalign*}
\lstset{language=bash, keywordstyle=\color{blue!70}, basicstyle=\ttfamily, frame=shadowbox}
\begin{lstlisting}
Computed bands:  1 - 19
GM: GM1+(1); GM2-(1); GM1+(1); GM2-(1); GM1+(1); GM2-(1); GM2+(1); GM1-(1);
    GM1+(1); GM2-(1); GM1+(1); GM2-(1); GM1+(1); GM2-(1); GM1+(1); GM1-(1);
    GM2+(1); GM2-(1); GM1+(1); [19] ;
Y : Y1+ Y2- (2); Y1+ (1); Y2- (1); Y2- (1); Y1+ (1); Y1- (1); Y2+ (1); Y1+ (1);
    Y2- (1); Y1+ (1); Y2- (1); Y1+ (1); Y2- (1); Y1- (1); Y1+ (1); Y2+ (1);
    Y2- (1); Y1+ (1); [19] ;
V : V1+ V1- (2); V1- (1); V1+ (1); V1+ (1); V1- (1); V1+ (1); V1- (1); V1+ (1);
    V1- (1); V1+ (1); V1- (1); V1- (1); V1+ (1); V1- (1); V1+ (1); V1+ (1);
    V1- (1); V1+ (1); [19] ;
L : L1+ L1- (2); L1+ (1); L1- (1); L1- (1); L1+ (1); L1+ (1); L1- (1); L1+ (1);
    L1- (1); L1+ (1); L1- (1); L1- (1); L1+ (1); L1- (1); L1+ (1); L1+ (1);
    L1- (1); L1+ (1); [19] ;
M : M1+ M2- (2); M1+ (1); M2- (1); M1+ (1); M2- (1); M2+ (1); M1- (1); M1+ (1);
    M2- (1); M1+ (1); M2- (1); M1+ (1); M2- (1); M1- (1); M2- (1); M2+ (1);
    M1+ (1); M1+ (1); [19] ;
U : U1  U2  (2); U2  (1); U1  (1); U2  (1); U1  (1); U1  (1); U2  (1); U1  (1);
    U1  (1); U2  (1); U2  (1); U2  (1); U1  (1); U1  (1); U2  (1); U1  (1);
    U1  (1); U2  (1); [19] ;
A : A1+ A2- (2); A1+ (1); A2- (1); A2- (1); A1+ (1); A1- A2+ (2); A1+ (1);
    A2- (1); A1+ (1); A2- (1); A1+ (1); A2- (1); A1- (1); A2+ (1); A2- (1);
    A1+ (1); A1+ (1); [19];
\end{lstlisting}
\hyperref[tab:electride]{Back to the table}

\subsubsection*{196260 ZrCu$_{4}$Ag}
\label{sec:tqc196260}
\noindent Essential BR: $A1@4b$ \\
\noindent RSI:
\begin{flalign*}
&\delta_{1}@4b\equiv m(A1)-m(A2)-m(T2)+m(T1) = 1,&
\end{flalign*}
\lstset{language=bash, keywordstyle=\color{blue!70}, basicstyle=\ttfamily, frame=shadowbox}
\begin{lstlisting}
Computed bands:  1 - 34
GM: GM1 (1); GM4 (3); GM1 (1); GM4 (3); GM3 (2); GM1 (1); GM4 (3); GM4 (3);
    GM3 (2); GM5 (3); GM3 (2); GM4 (3); GM5 (3); GM4 (3); GM1 (1); [34] ;
X : X3  (1); X1  (1); X5  (2); X3  (1); X1  (1); X5  (2); X3  (1); X2  (1);
    X1  (1); X1  (1); X3  (1); X5  (2); X3  (1); X1  (1); X4  (1); X5  (2);
    X3  (1); X5  (2); X5  (2); X2  (1); X1  (1); X4  (1); X5  (2); X2  (1);
    X1  (1); X5  (2); [34] ;
L : L1  (1); L1  (1); L3  (2); L1  (1); L1  (1); L3  (2); L1  (1); L3  (2);
    L1  (1); L3  (2); L1  (1); L3  (2); L1  (1); L3  (2); L3  (2); L1  (1);
    L3  (2); L2  (1); L3  (2); L3  (2); L2  (1); L3  (2); L1  (1); [34] ;
W : W4  (1); W1  W2  (2); W3  (1); W3  (1); W2  (1); W4  (1); W1  (1); W2  (1);
    W1  (1); W3  (1); W2  (1); W4  (1); W3  (1); W4  (1); W1  (1); W1  (1);
    W3  (1); W4  (1); W2  (1); W3  (1); W2  (1); W4  (1); W2  (1); W1  (1);
    W4  (1); W1  (1); W3  (1); W2  (1); W2  (1); W3  (1); W1  (1); W4  (1);
    W2  (1); [34];
\end{lstlisting}
\hyperref[tab:electride]{Back to the table}

\subsubsection*{166463 PtN$_{2}$}
\label{sec:tqc166463}
\noindent Essential BR: $Ag@2c$ \\
\noindent RSI:
\begin{flalign*}
&\delta_{1}@2c\equiv -m(Ag)+m(Au)-m(Bg)+m(Bu) = -1,&
\end{flalign*}
\lstset{language=bash, keywordstyle=\color{blue!70}, basicstyle=\ttfamily, frame=shadowbox}
\begin{lstlisting}
Computed bands:  1 - 20
GM: GM1+(1); GM2+(1); GM3-(1); GM4-(1); GM2+(1); GM1+(1); GM1+(1); GM2+(1);
    GM3-(1); GM4+(1); GM3+(1); GM1-(1); GM4-(1); GM2-(1); GM1+(1); GM3+(1);
    GM2+(1); GM2+(1); GM4+(1); GM1+(1); [20] ;
R : R2  (2); R1  (2); R2  (2); R1  (2); R2  (2); R1  (2); R2  (2); R1  (2);
    R1  (2); R2  (2); [20] ;
S : S3- S4- (2); S1+ S2+ (2); S1+ S2+ (2); S3+ S4+ (2); S3- S4- (2); S1+ S2+ (2);
    S3+ S4+ (2); S1+ S2+ (2); S1- S2- (2); S3- S4- (2); [20] ;
T : T1- (2); T1+ (2); T1+ (2); T1+ (2); T1- (2); T1+ (2); T1+ (2); T1+ (2);
    T1- (2); T1- (2); [20] ;
U : U1+ (2); U1- (2); U1+ (2); U1- (2); U1+ (2); U1- (2); U1+ (2); U1+ (2);
    U1+ (2); U1+ (2); [20] ;
X : X1  (2); X1  (2); X1  (2); X2  (2); X1  (2); X1  (2); X2  (2); X1  (2);
    X1  (2); X2  (2); [20] ;
Y : Y2  (2); Y2  (2); Y2  (2); Y2  (2); Y1  (2); Y1  (2); Y2  (2); Y2  (2);
    Y1  (2); Y2  (2); [20] ;
Z : Z1  (2); Z2  (2); Z1  (2); Z2  (2); Z1  (2); Z1  (2); Z2  (2); Z1  (2);
    Z1  (2); Z2  (2); [20];
\end{lstlisting}
\hyperref[tab:electride]{Back to the table}

\subsubsection*{238254 Sb$_{2}$Os}
\label{sec:tqc238254}
\lstset{language=bash, keywordstyle=\color{blue!70}, basicstyle=\ttfamily, frame=shadowbox}
\begin{lstlisting}
Computed bands:  1 - 18
GM: GM1+(1); GM2+(1); GM3-(1); GM4-(1); GM2+(1); GM2+(1); GM1+(1); GM3-(1);
    GM1+(1); GM4-(1); GM2+(1); GM1+(1); GM3+(1); GM4+(1); GM1+(1); GM3+(1);
    GM4+(1); GM2+(1); [18] ;
R : R2  (2); R1  (2); R1  (2); R2  (2); R2  (2); R1  (2); R2  (2); R1  (2);
    R2  (2); [18] ;
S : S3- S4- (2); S1+ S2+ (2); S1+ S2+ (2); S3- S4- (2); S1+ S2+ (2); S1+ S2+ (2);
    S3+ S4+ (2); S3- S4- (2); S3+ S4+ (2); [18] ;
T : T1- (2); T1+ (2); T1+ (2); T1+ (2); T1- (2); T1+ (2); T1- (2); T1+ (2);
    T1- (2); [18] ;
U : U1+ (2); U1- (2); U1+ (2); U1- (2); U1+ (2); U1+ (2); U1- (2); U1+ (2);
    U1+ (2); [18] ;
X : X1  (2); X1  (2); X1  (2); X1  (2); X1  (2); X1  (2); X2  (2); X1  (2);
    X2  (2); [18] ;
Y : Y2  (2); Y2  (2); Y2  (2); Y2  (2); Y2  (2); Y2  (2); Y1  (2); Y1  (2);
    Y2  (2); [18] ;
Z : Z1  (2); Z2  (2); Z1  (2); Z2  (2); Z1  (2); Z1  (2); Z2  (2); Z2  (2);
    Z1  (2); [18];
\end{lstlisting}
\hyperref[tab:electride]{Back to the table}

\subsubsection*{612234 YbNi$_{4}$Au}
\label{sec:tqc612234}
\lstset{language=bash, keywordstyle=\color{blue!70}, basicstyle=\ttfamily, frame=shadowbox}
\begin{lstlisting}
Computed bands:  1 - 30
GM: GM4 (3); GM1 (1); GM4 (3); GM3 (2); GM1 (1); GM4 (3); GM4 (3); GM3 (2);
    GM4 (3); GM5 (3); GM3 (2); GM5 (3); GM1 (1); [30] ;
X : X3  (1); X5  (2); X3  (1); X1  (1); X3  (1); X4  (1); X5  (2); X1  (1);
    X3  (1); X1  (1); X5  (2); X3  (1); X1  (1); X5  (2); X2  (1); X1  (1);
    X5  (2); X4  (1); X5  (2); X3  (1); X2  (1); X5  (2); X4  (1); [30] ;
L : L1  (1); L3  (2); L1  (1); L1  (1); L3  (2); L1  (1); L3  (2); L1  (1);
    L1  (1); L3  (2); L1  (1); L3  (2); L3  (2); L3  (2); L1  (1); L3  (2);
    L3  (2); L2  (1); L3  (2); L2  (1); [30] ;
W : W3  W4  (2); W2  (1); W2  (1); W3  (1); W4  (1); W1  (1); W4  (1); W3  (1);
    W3  (1); W2  (1); W1  (1); W2  (1); W1  (1); W4  (1); W4  (1); W3  (1);
    W2  (1); W1  (1); W3  (1); W2  (1); W1  (1); W1  (1); W4  (1); W3  (1);
    W4  (1); W2  (1); W3  (1); W3  (1); W2  (1); [30];
\end{lstlisting}
\hyperref[tab:electride]{Back to the table}

\subsubsection*{647776 Si$_{2}$Os}
\label{sec:tqc647776}
\lstset{language=bash, keywordstyle=\color{blue!70}, basicstyle=\ttfamily, frame=shadowbox}
\begin{lstlisting}
Computed bands:  1 - 16
GM: GM1+(1); GM1+(1); GM2-(1); GM1+(1); GM2-(1); GM2-(1); GM1+(1); GM1+(1);
    GM2-(1); GM1-(1); GM1+(1); GM2+(1); GM2+(1); GM1+(1); GM2-(1); GM1-(1);
    [16] ;
Y : Y2- (1); Y1+ (1); Y2- (1); Y1+ (1); Y2- (1); Y2- (1); Y1+ (1); Y2- (1);
    Y1+ (1); Y1- (1); Y1+ (1); Y1+ (1); Y1+ (1); Y2+ (1); Y1- (1); Y2+ (1);
    [16] ;
V : V1+ (1); V1- (1); V1- (1); V1+ (1); V1- (1); V1+ (1); V1+ (1); V1- (1);
    V1- (1); V1- (1); V1+ (1); V1- (1); V1+ (1); V1+ (1); V1+ (1); V1+ (1);
    [16] ;
L : L1- (1); L1+ (1); L1- (1); L1+ (1); L1- (1); L1+ (1); L1- (1); L1+ (1);
    L1- (1); L1- (1); L1+ (1); L1+ (1); L1- (1); L1+ (1); L1- (1); L1+ (1);
    [16] ;
M : M1+ (1); M2- (1); M2- (1); M1+ (1); M2- (1); M1+ (1); M1+ (1); M2- (1);
    M1+ (1); M2- (1); M1- (1); M2+ (1); M1+ (1); M2+ (1); M1- (1); M2- (1);
    [16] ;
U : U2  (1); U1  (1); U2  (1); U1  (1); U1  (1); U2  (1); U2  (1); U1  (1);
    U2  (1); U1  (1); U2  (1); U1  (1); U1  (1); U2  (1); U2  (1); U1  (1);
    [16] ;
A : A2- (1); A1+ (1); A2- (1); A1+ (1); A2- (1); A1+ (1); A1+ (1); A2- (1);
    A1+ (1); A2- (1); A1- (1); A1+ (1); A2+ (1); A2- (1); A1- A2+ (2); [16];
\end{lstlisting}
\hyperref[tab:electride]{Back to the table}

\subsubsection*{43690 Ge$_{2}$Os}
\label{sec:tqc43690}
\lstset{language=bash, keywordstyle=\color{blue!70}, basicstyle=\ttfamily, frame=shadowbox}
\begin{lstlisting}
Computed bands:  1 - 16
GM: GM1+(1); GM1+(1); GM2-(1); GM2-(1); GM1+(1); GM2-(1); GM1+(1); GM1+(1);
    GM2-(1); GM1+(1); GM1-(1); GM2+(1); GM2+(1); GM1+(1); GM2-(1); GM1-(1);
    [16] ;
Y : Y2- (1); Y1+ (1); Y2- (1); Y1+ (1); Y2- (1); Y2- (1); Y1+ (1); Y1+ (1);
    Y2- (1); Y1- (1); Y1+ (1); Y1+ (1); Y1+ (1); Y2+ (1); Y2+ (1); Y1- (1);
    [16] ;
V : V1+ (1); V1- (1); V1- (1); V1+ (1); V1- (1); V1+ (1); V1+ (1); V1- (1);
    V1- (1); V1- (1); V1+ (1); V1- (1); V1+ (1); V1+ (1); V1+ (1); V1+ (1);
    [16] ;
L : L1- (1); L1- (1); L1+ (1); L1+ (1); L1- (1); L1+ (1); L1- (1); L1+ (1);
    L1- (1); L1- (1); L1+ (1); L1+ (1); L1- (1); L1+ (1); L1- (1); L1+ (1);
    [16] ;
M : M1+ (1); M2- (1); M2- (1); M1+ (1); M2- (1); M1+ (1); M1+ (1); M2- (1);
    M1+ (1); M2- (1); M1- (1); M2+ (1); M1+ (1); M2+ (1); M1- (1); M2- (1);
    [16] ;
U : U2  (1); U1  (1); U2  (1); U1  (1); U1  (1); U2  (1); U2  (1); U1  (1);
    U1  (1); U2  (1); U2  (1); U1  (1); U1  (1); U2  (1); U2  (1); U1  (1);
    [16] ;
A : A2- (1); A1+ (1); A2- (1); A1+ (1); A2- (1); A1+ (1); A1+ (1); A2- (1);
    A1+ (1); A2- (1); A1- (1); A1+ (1); A2+ (1); A2- (1); A1- (1); A2+ (1); [16];
\end{lstlisting}
\hyperref[tab:electride]{Back to the table}

\subsubsection*{194977 NdMgCu$_{4}$}
\label{sec:tqc194977}
\noindent Essential BR: $A1@4d$ \\
\noindent RSI:
\begin{flalign*}
&\delta_{1}@4d\equiv m(A1)-m(A2)-m(T2)+m(T1) = 1,&
\end{flalign*}
\lstset{language=bash, keywordstyle=\color{blue!70}, basicstyle=\ttfamily, frame=shadowbox}
\begin{lstlisting}
Computed bands:  1 - 29
GM: GM1 (1); GM4 (3); GM1 (1); GM1 (1); GM4 (3); GM4 (3); GM3 (2); GM5 (3);
    GM4 (3); GM3 (2); GM5 (3); GM4 (3); GM1 (1); [29] ;
X : X1  (1); X3  (1); X5  (2); X3  (1); X1  (1); X3  (1); X5  (2); X3  (1);
    X1  (1); X1  (1); X5  (2); X1  (1); X4  (1); X3  (1); X2  (1); X5  (2);
    X5  (2); X4  (1); X5  (2); X2  (1); X3  (1); X5  (2); [29] ;
L : L1  (1); L1  (1); L3  (2); L1  (1); L1  (1); L3  (2); L1  (1); L1  (1);
    L3  (2); L3  (2); L1  (1); L1  (1); L3  (2); L3  (2); L2  (1); L3  (2);
    L3  (2); L2  (1); L3  (2); L1  (1); [29] ;
W : W1  (1); W3  (1); W4  (1); W2  (1); W3  (1); W2  (1); W4  (1); W1  (1);
    W2  (1); W4  (1); W3  (1); W1  (1); W4  (1); W1  (1); W3  (1); W2  (1);
    W4  (1); W1  (1); W3  (1); W2  (1); W3  (1); W2  (1); W1  (1); W4  (1);
    W3  (1); W2  (1); W4  (1); W1  (1); W3  (1); [29];
\end{lstlisting}
\hyperref[tab:electride]{Back to the table}

\subsubsection*{616387 Be$_{5}$Pd}
\label{sec:tqc616387}
\noindent Essential BR: $A1@4b$ \\
\noindent RSI:
\begin{flalign*}
&\delta_{1}@4b\equiv m(A1)-m(A2)-m(T2)+m(T1) = 1,&
\end{flalign*}
\lstset{language=bash, keywordstyle=\color{blue!70}, basicstyle=\ttfamily, frame=shadowbox}
\begin{lstlisting}
Computed bands:  1 - 10
GM: GM1 (1); GM4 (3); GM3 (2); GM1 (1); GM4 (3); [10] ;
X : X1  (1); X3  (1); X3  (1); X1  (1); X2  (1); X5  (2); X5  (2); X1  (1);
    [10] ;
L : L1  (1); L1  (1); L3  (2); L1  (1); L3  (2); L1  (1); L3  (2); [10] ;
W : W2  (1); W3  (1); W4  (1); W1  (1); W1  (1); W2  (1); W3  (1); W4  (1);
    W2  (1); W2  (1); [10];
\end{lstlisting}
\hyperref[tab:electride]{Back to the table}

\subsubsection*{103785 Ga$_{2}$Os}
\label{sec:tqc103785}
\noindent Essential BR: $Ag@16d$ \\
\noindent RSI:
\begin{flalign*}
&\delta_{1}@16d\equiv -m(Ag)+m(Au) = -1,&
\end{flalign*}
\lstset{language=bash, keywordstyle=\color{blue!70}, basicstyle=\ttfamily, frame=shadowbox}
\begin{lstlisting}
Computed bands:  1 - 14
GM: GM1+(1); GM1-(1); GM1-(1); GM2+(1); GM4+(1); GM2-(1); GM4-(1); GM3+(1);
    GM1+(1); GM2+(1); GM3-(1); GM4+(1); GM1+(1); GM3+(1); [14] ;
Z : Z1  (2); Z1  (2); Z1  (2); Z2  (2); Z2  (2); Z2  (2); Z1  (2); [14] ;
H : H1  (2); H1  (2); H1  (2); H1  (2); H1  (2); H1  (2); H1  (2); [14] ;
Y : Y1  (2); Y1  (2); Y2  (2); Y1  (2); Y2  (2); Y2  (2); Y1  (2); [14] ;
L : L1- (1); L1+ (1); L1- (1); L1+ (1); L1- (1); L1+ (1); L1+ (1); L1- (1);
    L1- (1); L1+ (1); L1- (1); L1+ (1); L1+ (1); L1+ (1); [14] ;
T : T1  (2); T2  (2); T1  (2); T2  (2); T1  (2); T2  (2); T1  (2); [14];
\end{lstlisting}
\hyperref[tab:electride]{Back to the table}

\subsubsection*{635228 Ga$_{2}$Ru}
\label{sec:tqc635228}
\lstset{language=bash, keywordstyle=\color{blue!70}, basicstyle=\ttfamily, frame=shadowbox}
\begin{lstlisting}
Computed bands:  1 - 14
GM: GM1+(1); GM1-(1); GM2+(1); GM1-(1); GM4+(1); GM2-(1); GM4-(1); GM3+(1);
    GM1+(1); GM2+(1); GM3-(1); GM4+(1); GM1+(1); GM3+(1); [14] ;
Z : Z1  (2); Z1  (2); Z1  (2); Z2  (2); Z2  (2); Z2  (2); Z1  (2); [14] ;
H : H1  (2); H1  (2); H1  (2); H1  (2); H1  (2); H1  (2); H1  (2); [14] ;
Y : Y1  (2); Y1  (2); Y2  (2); Y1  (2); Y2  (2); Y2  (2); Y1  (2); [14] ;
L : L1- (1); L1+ (1); L1- (1); L1+ (1); L1- (1); L1+ (1); L1+ (1); L1- (1);
    L1- (1); L1+ (1); L1- (1); L1+ (1); L1+ (1); L1+ (1); [14] ;
T : T1  (2); T2  (2); T1  (2); T2  (2); T1  (2); T2  (2); T1  (2); [14];
\end{lstlisting}
\hyperref[tab:electride]{Back to the table}

\subsubsection*{646107 NiP$_{2}$}
\label{sec:tqc646107}
\lstset{language=bash, keywordstyle=\color{blue!70}, basicstyle=\ttfamily, frame=shadowbox}
\begin{lstlisting}
Computed bands:  1 - 20
GM: GM1+(1); GM1-(1); GM2+(1); GM2-(1); GM2+(1); GM1+(1); GM2+(1); GM1-(1);
    GM1+(1); GM1+(1); GM2-(1); GM1-(1); GM2+(1); GM1+(1); GM2+(1); GM1+(1);
    GM2+(1); GM1+(1); GM1+(1); GM2+(1); [20] ;
Y : Y1- (1); Y2- (1); Y1+ (1); Y2+ (1); Y1- (1); Y1+ (1); Y1- (1); Y2- (1);
    Y2+ (1); Y2- (1); Y1+ (1); Y1- (1); Y2- (1); Y1- (1); Y1- (1); Y2- (1);
    Y2- (1); Y1- (1); Y1- (1); Y2- (1); [20] ;
L : L1- (1); L1+ (1); L1- (1); L1+ (1); L1+ (1); L1+ (1); L1+ (1); L1- (1);
    L1- (1); L1+ (1); L1+ (1); L1- (1); L1+ (1); L1+ (1); L1+ (1); L1+ (1);
    L1+ (1); L1- (1); L1- (1); L1+ (1); [20] ;
M : M1  (2); M1  (2); M1  (2); M1  (2); M1  (2); M1  (2); M1  (2); M1  (2);
    M1  (2); M1  (2); [20] ;
V : V1- (1); V1+ (1); V1+ (1); V1- (1); V1- (1); V1+ (1); V1+ (1); V1- (1);
    V1+ (1); V1- (1); V1+ (1); V1- (1); V1- (1); V1+ (1); V1+ (1); V1- (1);
    V1- (1); V1+ (1); V1+ (1); V1- (1); [20] ;
U : U1  U2  (2); U1  U2  (2); U1  U2  (2); U1  U2  (2); U1  U2  (2); U1  U2  (2);
    U1  U2  (2); U1  U2  (2); U1  U2  (2); U1  U2  (2); [20] ;
A : A1  (2); A1  (2); A1  (2); A1  (2); A1  (2); A1  (2); A1  (2); A1  (2);
    A1  (2); A1  (2); [20];
\end{lstlisting}
\hyperref[tab:electride]{Back to the table}

\subsubsection*{174577 PbSe$_{2}$}
\label{sec:tqc174577}
\lstset{language=bash, keywordstyle=\color{blue!70}, basicstyle=\ttfamily, frame=shadowbox}
\begin{lstlisting}
Computed bands:  1 - 16
GM: GM1+(1); GM4+(1); GM5-(2); GM1+(1); GM1-(1); GM4+(1); GM3+(1); GM5-(2);
    GM2-(1); GM5+(2); GM3-(1); GM1+(1); GM2+(1); [16] ;
M : M3+ (1); M2+ (1); M5- (2); M1+ (1); M1- (1); M3+ (1); M2+ (1); M5- (2);
    M1- (1); M4- (1); M5+ (2); M4+ (1); M1+ (1); [16] ;
P : P2  P4  (2); P5  (2); P5  (2); P2  P4  (2); P2  P4  (2); P5  (2);
    P1  P3  (2); P5  (2); [16] ;
X : X4- (1); X4- (1); X1+ (1); X2+ (1); X1- (1); X1+ (1); X4- (1); X3+ (1);
    X2+ (1); X4- (1); X3- (1); X3+ (1); X1+ (1); X2- (1); X3- (1); X1- (1);
    [16] ;
N : N1  (2); N1  (2); N1  (2); N1  (2); N1  (2); N1  (2); N1  (2); N1  (2); [16];
\end{lstlisting}
\hyperref[tab:electride]{Back to the table}

\subsubsection*{43898 CrAs$_{2}$}
\label{sec:tqc43898}
\lstset{language=bash, keywordstyle=\color{blue!70}, basicstyle=\ttfamily, frame=shadowbox}
\begin{lstlisting}
Computed bands:  1 - 16
GM: GM1+(1); GM1+(1); GM2-(1); GM2-(1); GM1+(1); GM2-(1); GM1+(1); GM1+(1);
    GM2+(1); GM2-(1); GM1-(1); GM1+(1); GM2+(1); GM1+(1); GM2-(1); GM1-(1);
    [16] ;
Y : Y2- (1); Y1+ (1); Y2- (1); Y1+ (1); Y1+ (1); Y2- (1); Y2- (1); Y1+ (1);
    Y1- (1); Y2- (1); Y1+ (1); Y1- (1); Y1+ (1); Y2+ (1); Y1+ (1); Y2+ (1);
    [16] ;
V : V1+ (1); V1- (1); V1+ (1); V1- (1); V1+ (1); V1- (1); V1- (1); V1+ (1);
    V1- (1); V1+ (1); V1- (1); V1+ (1); V1- (1); V1+ (1); V1+ (1); V1+ (1);
    [16] ;
L : L1+ (1); L1+ (1); L1- (1); L1- (1); L1+ (1); L1- (1); L1- (1); L1+ (1);
    L1+ (1); L1- (1); L1+ (1); L1+ (1); L1- (1); L1- (1); L1+ (1); L1- (1);
    [16] ;
M : M1+ (1); M2- (1); M1+ (1); M2- (1); M1+ (1); M2- (1); M1- (1); M1+ (1);
    M2+ (1); M2- (1); M2- (1); M1+ (1); M2- (1); M1+ (1); M1- (1); M2+ (1);
    [16] ;
U : U1  (1); U2  (1); U1  (1); U2  (1); U2  (1); U1  (1); U1  (1); U2  (1);
    U2  (1); U1  (1); U2  (1); U2  (1); U1  (1); U1  (1); U1  (1); U2  (1);
    [16] ;
A : A1+ (1); A2- (1); A1+ (1); A2- (1); A1+ (1); A2- (1); A2- (1); A1+ (1);
    A2+ (1); A1+ (1); A2- (1); A1- (1); A2- (1); A1+ (1); A2+ (1); A1- (1); [16];
\end{lstlisting}
\hyperref[tab:electride]{Back to the table}

\subsubsection*{633072 FeP$_{2}$}
\label{sec:tqc633072}
\lstset{language=bash, keywordstyle=\color{blue!70}, basicstyle=\ttfamily, frame=shadowbox}
\begin{lstlisting}
Computed bands:  1 - 18
GM: GM1+(1); GM2+(1); GM3-(1); GM4-(1); GM2+(1); GM2+(1); GM1+(1); GM3-(1);
    GM4-(1); GM1+(1); GM2+(1); GM1+(1); GM4+(1); GM3+(1); GM1+(1); GM2+(1);
    GM3+(1); GM4+(1); [18] ;
R : R2  (2); R1  (2); R2  (2); R1  (2); R2  (2); R1  (2); R2  (2); R1  (2);
    R2  (2); [18] ;
S : S3- S4- (2); S1+ S2+ (2); S1+ S2+ (2); S3- S4- (2); S1+ S2+ (2); S1+ S2+ (2);
    S3+ S4+ (2); S3- S4- (2); S3+ S4+ (2); [18] ;
T : T1- (2); T1+ (2); T1+ (2); T1+ (2); T1- (2); T1- (2); T1+ (2); T1+ (2);
    T1- (2); [18] ;
U : U1+ (2); U1- (2); U1- (2); U1+ (2); U1+ (2); U1+ (2); U1- (2); U1+ (2);
    U1+ (2); [18] ;
X : X1  (2); X1  (2); X1  (2); X1  (2); X1  (2); X1  (2); X2  (2); X1  (2);
    X2  (2); [18] ;
Y : Y2  (2); Y2  (2); Y2  (2); Y2  (2); Y2  (2); Y2  (2); Y1  (2); Y2  (2);
    Y1  (2); [18] ;
Z : Z1  (2); Z2  (2); Z1  (2); Z2  (2); Z1  (2); Z1  (2); Z2  (2); Z2  (2);
    Z1  (2); [18];
\end{lstlisting}
\hyperref[tab:electride]{Back to the table}

\subsubsection*{106001 Te$_{2}$Ru}
\label{sec:tqc106001}
\noindent Essential BR: $Ag@2c$ \\
\noindent RSI:
\begin{flalign*}
&\delta_{1}@2c\equiv -m(Ag)+m(Au)-m(Bg)+m(Bu) = -1,&
\end{flalign*}
\lstset{language=bash, keywordstyle=\color{blue!70}, basicstyle=\ttfamily, frame=shadowbox}
\begin{lstlisting}
Computed bands:  1 - 20
GM: GM1+(1); GM2+(1); GM3-(1); GM4-(1); GM2+(1); GM2+(1); GM1+(1); GM3+(1);
    GM3-(1); GM4+(1); GM1+(1); GM1-(1); GM4-(1); GM2+(1); GM3+(1); GM4+(1);
    GM1+(1); GM1+(1); GM2+(1); GM2-(1); [20] ;
R : R2  (2); R1  (2); R2  (2); R1  (2); R1  (2); R2  (2); R2  (2); R1  (2);
    R1  (2); R2  (2); [20] ;
S : S3- S4- (2); S1+ S2+ (2); S1+ S2+ (2); S3+ S4+ (2); S3- S4- (2); S1+ S2+ (2);
    S1+ S2+ (2); S3+ S4+ (2); S1- S2- (2); S3- S4- (2); [20] ;
T : T1- (2); T1+ (2); T1+ (2); T1- (2); T1+ (2); T1- (2); T1+ (2); T1- (2);
    T1+ (2); T1+ (2); [20] ;
U : U1+ (2); U1- (2); U1+ (2); U1- (2); U1+ (2); U1+ (2); U1- (2); U1+ (2);
    U1+ (2); U1+ (2); [20] ;
X : X1  (2); X1  (2); X1  (2); X2  (2); X1  (2); X1  (2); X2  (2); X1  (2);
    X1  (2); X2  (2); [20] ;
Y : Y2  (2); Y2  (2); Y2  (2); Y2  (2); Y1  (2); Y2  (2); Y1  (2); Y2  (2);
    Y2  (2); Y1  (2); [20] ;
Z : Z1  (2); Z2  (2); Z1  (2); Z2  (2); Z1  (2); Z1  (2); Z2  (2); Z2  (2);
    Z1  (2); Z1  (2); [20];
\end{lstlisting}
\hyperref[tab:electride]{Back to the table}

\subsubsection*{2526 CrP$_{2}$}
\label{sec:tqc2526}
\lstset{language=bash, keywordstyle=\color{blue!70}, basicstyle=\ttfamily, frame=shadowbox}
\begin{lstlisting}
Computed bands:  1 - 16
GM: GM1+(1); GM1+(1); GM2-(1); GM2-(1); GM1+(1); GM1+(1); GM2-(1); GM1+(1);
    GM1+GM2-(2); GM1-(1); GM2+(1); GM2+(1); GM2-(1); GM1+(1); GM1-(1); [16] ;
Y : Y2- (1); Y1+ (1); Y2- (1); Y1+ (1); Y2- (1); Y1+ (1); Y2- (1); Y1+ (1);
    Y1- (1); Y1+ (1); Y2- (1); Y1- (1); Y1+ (1); Y2+ (1); Y1+ (1); Y2+ (1);
    [16] ;
V : V1+ (1); V1- (1); V1- (1); V1+ (1); V1+ (1); V1- (1); V1- (1); V1+ (1);
    V1- (1); V1+ (1); V1- (1); V1+ (1); V1- (1); V1+ (1); V1+ (1); V1+ (1);
    [16] ;
L : L1+ (1); L1+ (1); L1- (1); L1- (1); L1+ (1); L1- (1); L1+ (1); L1- (1);
    L1+ (1); L1- (1); L1+ (1); L1+ (1); L1- (1); L1- (1); L1+ (1); L1- (1);
    [16] ;
M : M2- (1); M1+ (1); M1+ (1); M2- (1); M2- (1); M1+ (1); M1- (1); M1+ (1);
    M2- (1); M2+ (1); M2- (1); M1+ (1); M1+ (1); M2- (1); M1- (1); M2+ (1);
    [16] ;
U : U1  (1); U2  (1); U1  (1); U2  (1); U2  (1); U1  (1); U1  (1); U2  (1);
    U1  (1); U2  (1); U2  (1); U2  (1); U1  (1); U1  (1); U1  (1); U2  (1);
    [16] ;
A : A1+ (1); A2- (1); A1+ (1); A2- (1); A1+ (1); A2- (1); A2- (1); A1+ (1);
    A1+ (1); A2+ (1); A2- (1); A1- (1); A2- (1); A1+ (1); A2+ (1); A1- (1); [16];
\end{lstlisting}
\hyperref[tab:electride]{Back to the table}

\subsubsection*{75555 BaTe$_{2}$}
\label{sec:tqc75555}
\lstset{language=bash, keywordstyle=\color{blue!70}, basicstyle=\ttfamily, frame=shadowbox}
\begin{lstlisting}
Computed bands:  1 - 22
GM: GM1+(1); GM1-(1); GM3+(1); GM5+(2); GM5-(2); GM3-(1); GM1+(1); GM4+(1);
    GM5-(2); GM4+(1); GM1+(1); GM2-(1); GM5-(2); GM3-(1); GM3+(1); GM5+(2);
    GM2+(1); [22] ;
M : M1+ (1); M1- (1); M3+ (1); M5- (2); M5+ (2); M3- (1); M3+ (1); M2+ (1);
    M5- (2); M1- (1); M2+ (1); M1+ (1); M5- (2); M3+ (1); M4- (1); M5+ (2);
    M4+ (1); [22] ;
P : P5  (2); P2  P4  (2); P5  (2); P1  P3  (2); P2  P4  (2); P5  (2);
    P2  P4  (2); P5  (2); P2  P4  (2); P1  P3  (2); P5  (2); [22] ;
X : X1+ (1); X1- (1); X4- (1); X2+ (1); X3+ (1); X3- (1); X4+ (1); X2- (1);
    X4- (1); X4- (1); X1+ (1); X2+ (1); X4- (1); X4- (1); X1+ (1); X3+ (1);
    X2+ (1); X3+ (1); X1- (1); X3- (1); X3- (1); X2- (1); [22] ;
N : N1  (2); N1  (2); N1  (2); N1  (2); N1  (2); N1  (2); N1  (2); N1  (2);
    N1  (2); N1  (2); N1  (2); [22];
\end{lstlisting}
\hyperref[tab:electride]{Back to the table}

\subsubsection*{190546 Na$_{2}$Cl}
\label{sec:tqc190546}
\noindent Essential BR: $Ag@2c$ \\
\noindent RSI:
\begin{flalign*}
&\delta_{1}@2c\equiv -m(Ag)+m(Au)-m(B1g)+m(B1u)-m(B3g)+m(B3u)-m(B2g)+m(B2u) = -1,&
\end{flalign*}
\lstset{language=bash, keywordstyle=\color{blue!70}, basicstyle=\ttfamily, frame=shadowbox}
\begin{lstlisting}
Computed bands:  1 -  9
GM: GM1+(1); GM4-(1); GM1+(1); GM2+(1); GM4-(1); GM2-GM3-(2); GM3+(1); GM1+(1);
    [9]  ;
T : T4- (1); T1+ (1); T3+ (1); T2- (1); T1+ (1); T3- (1); T4- (1); T2+ (1);
    T2- (1); [9]  ;
Y : Y4- (1); Y1+ (1); Y1+ (1); Y3- (1); Y4- (1); Y3+ (1); Y2+ Y2- (2); Y1+ (1);
    [9]  ;
Z : Z1+ (1); Z4- (1); Z2- (1); Z3+ (1); Z1+ (1); Z2+ (1); Z4- (1); Z3- (1);
    Z2- (1); [9]  ;
R : R1+ (1); R2- (1); R1- (1); R2+ (1); R1+ (1); R2- (1); R1+ (1); R2- (1);
    R2+ (1); [9]  ;
S : S1+ (1); S2- (1); S1+ (1); S2- (1); S1+ (1); S2- (1); S1- (1); S2+ (1);
    S2- (1); [9] ;
\end{lstlisting}
\hyperref[tab:electride]{Back to the table}

\subsubsection*{628189 YbInCu$_{4}$}
\label{sec:tqc628189}
\noindent Essential BR: $A1@4d$ \\
\noindent RSI:
\begin{flalign*}
&\delta_{1}@4d\equiv m(A1)-m(A2)-m(T2)+m(T1) = 1,&
\end{flalign*}
\lstset{language=bash, keywordstyle=\color{blue!70}, basicstyle=\ttfamily, frame=shadowbox}
\begin{lstlisting}
Computed bands:  1 - 28
GM: GM4 (3); GM1 (1); GM1 (1); GM4 (3); GM4 (3); GM3 (2); GM5 (3); GM3 (2);
    GM4 (3); GM5 (3); GM4 (3); GM1 (1); [28] ;
X : X3  (1); X5  (2); X3  (1); X1  (1); X3  (1); X1  (1); X5  (2); X1  (1);
    X3  (1); X5  (2); X1  (1); X2  (1); X5  (2); X3  (1); X4  (1); X5  (2);
    X2  (1); X5  (2); X4  (1); X3  (1); X5  (2); [28] ;
L : L1  (1); L3  (2); L1  (1); L1  (1); L1  (1); L3  (2); L1  (1); L3  (2);
    L1  (1); L1  (1); L3  (2); L3  (2); L3  (2); L2  (1); L3  (2); L3  (2);
    L2  (1); L1  (1); L3  (2); [28] ;
W : W4  (1); W3  (1); W2  (1); W4  (1); W2  (1); W3  (1); W1  (1); W2  (1);
    W1  (1); W4  (1); W3  (1); W1  (1); W4  (1); W2  (1); W3  (1); W2  (1);
    W3  (1); W1  (1); W4  (1); W1  (1); W4  (1); W3  (1); W2  (1); W3  (1);
    W2  (1); W4  (1); W1  (1); W3  (1); [28];
\end{lstlisting}
\hyperref[tab:electride]{Back to the table}

\subsubsection*{628018 MgInCu$_{4}$}
\label{sec:tqc628018}
\noindent Essential BR: $A1@4d$ \\
\noindent RSI:
\begin{flalign*}
&\delta_{1}@4d\equiv m(A1)-m(A2)-m(T2)+m(T1) = 1,&
\end{flalign*}
\lstset{language=bash, keywordstyle=\color{blue!70}, basicstyle=\ttfamily, frame=shadowbox}
\begin{lstlisting}
Computed bands:  1 - 25
GM: GM1 (1); GM1 (1); GM4 (3); GM4 (3); GM3 (2); GM5 (3); GM3 (2); GM4 (3);
    GM5 (3); GM1 (1); GM4 (3); [25] ;
X : X3  (1); X1  (1); X3  (1); X1  (1); X5  (2); X1  (1); X5  (2); X3  (1);
    X1  (1); X3  (1); X5  (2); X2  (1); X4  (1); X5  (2); X5  (2); X2  (1);
    X4  (1); X3  (1); X5  (2); [25] ;
L : L1  (1); L1  (1); L1  (1); L1  (1); L3  (2); L3  (2); L3  (2); L1  (1);
    L1  (1); L3  (2); L3  (2); L2  (1); L3  (2); L3  (2); L2  (1); L1  (1);
    L3  (2); [25] ;
W : W4  (1); W2  (1); W3  (1); W1  (1); W2  (1); W3  (1); W1  (1); W4  (1);
    W1  (1); W4  (1); W3  (1); W2  (1); W2  (1); W3  (1); W1  (1); W4  (1);
    W1  (1); W4  (1); W2  (1); W3  (1); W2  (1); W3  (1); W4  (1); W1  (1);
    W3  (1); [25];
\end{lstlisting}
\hyperref[tab:electride]{Back to the table}

\subsubsection*{187441 ReN$_{2}$}
\label{sec:tqc187441}
\noindent Essential BR: $Ag@2c$ \\
\noindent RSI:
\begin{flalign*}
&\delta_{1}@2c\equiv -m(Ag)+m(Au)-m(Bg)+m(Bu) = -1,&
\end{flalign*}
\lstset{language=bash, keywordstyle=\color{blue!70}, basicstyle=\ttfamily, frame=shadowbox}
\begin{lstlisting}
Computed bands:  1 - 17
GM: GM1+(1); GM1+(1); GM2-(1); GM2-(1); GM1+(1); GM1+(1); GM2-(1); GM1+(1);
    GM1-(1); GM2-(1); GM2+(1); GM1+(1); GM1-(1); GM2+(1); GM1+(1); GM2-(1);
    GM2-(1); [17] ;
Y : Y1+ (1); Y2- (1); Y1+ (1); Y2- (1); Y1+ (1); Y2- (1); Y1- (1); Y2- (1);
    Y1+ (1); Y2- (1); Y1+ (1); Y1- (1); Y2+ (1); Y1+ (1); Y2- (1); Y2+ (1);
    Y1+ (1); [17] ;
V : V1+ (1); V1- (1); V1+ (1); V1- (1); V1+ (1); V1- (1); V1- (1); V1+ (1);
    V1- (1); V1- (1); V1+ (1); V1- (1); V1+ (1); V1+ (1); V1+ (1); V1- (1);
    V1+ (1); [17] ;
L : L1- (1); L1- (1); L1+ (1); L1+ (1); L1+ (1); L1- (1); L1+ (1); L1- (1);
    L1+ (1); L1- (1); L1- (1); L1+ (1); L1- (1); L1+ (1); L1- (1); L1- (1);
    L1+ (1); [17] ;
M : M2- (1); M2- (1); M1+ (1); M1+ (1); M1+ (1); M2- (1); M1+ (1); M2+ (1);
    M2- (1); M2- (1); M1- (1); M1+ (1); M2+ (1); M2- (1); M1+ (1); M2- (1);
    M1- (1); [17] ;
U : U2  (1); U1  (1); U2  (1); U1  (1); U2  (1); U1  (1); U1  (1); U2  (1);
    U2  (1); U1  (1); U1  (1); U2  (1); U2  (1); U1  (1); U2  (1); U2  (1);
    U1  (1); [17] ;
A : A2- (1); A1+ (1); A2- (1); A1+ (1); A2- (1); A1+ (1); A1+ (1); A2- (1);
    A2+ (1); A1+ (1); A1- (1); A1- (1); A1+ (1); A2+ A2- (2); A2- (1); A2- (1);
    [17];
\end{lstlisting}
\hyperref[tab:electride]{Back to the table}

\subsubsection*{98666 LiYGa$_{4}$}
\label{sec:tqc98666}
\lstset{language=bash, keywordstyle=\color{blue!70}, basicstyle=\ttfamily, frame=shadowbox}
\begin{lstlisting}
Computed bands:  1 - 12
A : A1  (1); A3  (1); A5  (2); A3  (1); A1  (1); A1  (1); A3  (1); A5  (2);
    A6  (2); [12] ;
GM: GM1 (1); GM3 (1); GM5 (2); GM1 (1); GM3 (1); GM1 (1); GM5 (2); GM1 (1);
    GM6 (2); [12] ;
H : H1  (1); H5  (1); H3  (1); H2  (1); H4  (1); H5  (1); H6  (1); H3  (1);
    H1  (1); H2  (1); H4  (1); H5  (1); [12] ;
K : K1  (1); K5  (1); K3  (1); K2  (1); K3  (1); K5  (1); K6  (1); K4  (1);
    K1  (1); K5  (1); K2  (1); K3  (1); [12] ;
L : L1  (1); L1  (1); L3  (1); L2  (1); L3  (1); L1  (1); L1  (1); L3  (1);
    L4  (1); L3  (1); L1  (1); L2  (1); [12] ;
M : M1  (1); M1  (1); M3  (1); M2  (1); M1  (1); M1  (1); M3  (1); M3  (1);
    M1  (1); M1  (1); M2  (1); M4  (1); [12];
\end{lstlisting}
\hyperref[tab:electride]{Back to the table}

\subsubsection*{65168 FeAs$_{2}$}
\label{sec:tqc65168}
\lstset{language=bash, keywordstyle=\color{blue!70}, basicstyle=\ttfamily, frame=shadowbox}
\begin{lstlisting}
Computed bands:  1 - 18
GM: GM1+(1); GM2+(1); GM3-(1); GM4-(1); GM2+(1); GM2+(1); GM3-(1); GM1+(1);
    GM4-(1); GM1+(1); GM2+(1); GM1+(1); GM4+(1); GM3+(1); GM1+(1); GM3+(1);
    GM2+(1); GM4+(1); [18] ;
R : R2  (2); R1  (2); R2  (2); R1  (2); R2  (2); R1  (2); R2  (2); R1  (2);
    R2  (2); [18] ;
S : S3- S4- (2); S1+ S2+ (2); S1+ S2+ (2); S3- S4- (2); S1+ S2+ (2); S1+ S2+ (2);
    S3+ S4+ (2); S3- S4- (2); S3+ S4+ (2); [18] ;
T : T1- (2); T1+ (2); T1+ (2); T1- (2); T1+ (2); T1- (2); T1+ (2); T1+ (2);
    T1- (2); [18] ;
U : U1+ (2); U1- (2); U1- (2); U1+ (2); U1+ (2); U1+ (2); U1- (2); U1+ (2);
    U1+ (2); [18] ;
X : X1  (2); X1  (2); X1  (2); X1  (2); X1  (2); X2  (2); X1  (2); X1  (2);
    X2  (2); [18] ;
Y : Y2  (2); Y2  (2); Y2  (2); Y2  (2); Y2  (2); Y2  (2); Y1  (2); Y1  (2);
    Y2  (2); [18] ;
Z : Z1  (2); Z2  (2); Z1  (2); Z2  (2); Z1  (2); Z1  (2); Z2  (2); Z2  (2);
    Z1  (2); [18];
\end{lstlisting}
\hyperref[tab:electride]{Back to the table}

\subsubsection*{16820 MoAs$_{2}$}
\label{sec:tqc16820}
\lstset{language=bash, keywordstyle=\color{blue!70}, basicstyle=\ttfamily, frame=shadowbox}
\begin{lstlisting}
Computed bands:  1 - 16
GM: GM1 (1); GM1 (1); GM2 (1); GM2 (1); GM1 (1); GM2 (1); GM1 (1); GM1 (1);
    GM2 (1); GM1 (1); GM1 (1); GM2 (1); GM2 (1); GM1 (1); GM2 (1); GM1 (1);
    [16] ;
U : U1  (1); U2  (1); U1  (1); U2  (1); U2  (1); U1  (1); U1  (1); U2  (1);
    U2  (1); U1  (1); U2  (1); U2  (1); U1  (1); U1  (1); U1  (1); U2  (1);
    [16] ;
Y : Y2  (1); Y1  (1); Y2  (1); Y1  (1); Y2  (1); Y2  (1); Y1  (1); Y1  (1);
    Y1  (1); Y2  (1); Y1  (1); Y1  (1); Y1  (1); Y2  (1); Y1  (1); Y2  (1);
    [16] ;
V : V1  (1); V1  (1); V1  (1); V1  (1); V1  (1); V1  (1); V1  (1); V1  (1);
    V1  (1); V1  (1); V1  (1); V1  (1); V1  (1); V1  (1); V1  (1); V1  (1);
    [16] ;
L : L1  (1); L1  (1); L1  (1); L1  (1); L1  (1); L1  (1); L1  (1); L1  (1);
    L1  (1); L1  (1); L1  (1); L1  (1); L1  (1); L1  (1); L1  (1); L1  (1);
    [16] ;
M : M2  (1); M1  (1); M1  (1); M2  (1); M1  (1); M2  (1); M1  (1); M2  (1);
    M1  (1); M2  (1); M2  (1); M1  (1); M2  (1); M1  (1); M1  (1); M2  (1);
    [16] ;
U : U1  (1); U2  (1); U1  (1); U2  (1); U2  (1); U1  (1); U1  (1); U2  (1);
    U2  (1); U1  (1); U2  (1); U2  (1); U1  (1); U1  (1); U1  (1); U2  (1);
    [16] ;
A : A1  (1); A2  (1); A1  (1); A2  (1); A1  (1); A2  (1); A2  (1); A1  (1);
    A1  (1); A2  (1); A2  (1); A1  (1); A2  (1); A1  (1); A2  (1); A1  (1); [16];
\end{lstlisting}
\hyperref[tab:electride]{Back to the table}

\subsubsection*{417149 YGaI}
\label{sec:tqc417149}
\noindent Essential BR: $Ag@3e$ \\
\noindent RSI:
\begin{flalign*}
&\delta_{1}@3e\equiv -m(Ag)+m(Au)-m(Bg)+m(Bu) = -1,&
\end{flalign*}
\lstset{language=bash, keywordstyle=\color{blue!70}, basicstyle=\ttfamily, frame=shadowbox}
\begin{lstlisting}
Computed bands:  1 - 21
A : A1+ (1); A2- (1); A1+ (1); A2- (1); A3- (2); A3+ (2); A2- (1); A1+ (1);
    A1+ (1); A2- (1); A1+ (1); A3- (2); A3+ (2); A2- (1); A3+ (2); A1+ (1);
    [21] ;
GM: GM1+(1); GM2-(1); GM1+(1); GM2-(1); GM3-(2); GM3+(2); GM1+(1); GM2-(1);
    GM1+(1); GM2-(1); GM1+(1); GM3+(2); GM3-(2); GM2-(1); GM3+(2); GM1+(1);
    [21] ;
H : H1  (1); H2  (1); H1  (1); H3  (2); H3  (2); H2  (1); H3  (2); H3  (2);
    H3  (2); H2  (1); H1  (1); H3  (2); H1  (1); H3  (2); [21] ;
K : K1  (1); K2  (1); K1  (1); K3  (2); K3  (2); K2  (1); K3  (2); K3  (2);
    K1  (1); K3  (2); K2  (1); K3  (2); K1  (1); K3  (2); [21] ;
L : L1+ (1); L2- (1); L1+ (1); L2- (1); L1+ (1); L2- (1); L1- (1); L2+ (1);
    L1+ (1); L2- (1); L2- (1); L1+ (1); L2- (1); L1+ (1); L1+ (1); L2- (1);
    L2+ (1); L1- (1); L1+ (1); L1- (1); L2- (1); [21] ;
M : M1+ (1); M2- (1); M1+ (1); M2- (1); M1+ (1); M2- (1); M1- (1); M2+ (1);
    M2- (1); M1+ (1); M2- (1); M1+ (1); M2- (1); M1+ (1); M2- (1); M1+ (1);
    M1- (1); M2+ (1); M1+ (1); M1- (1); M2- (1); [21];
\end{lstlisting}
\hyperref[tab:electride]{Back to the table}

\subsubsection*{191404 ZrCu$_{5}$}
\label{sec:tqc191404}
\noindent Essential BR: $A1@4d$ \\
\noindent RSI:
\begin{flalign*}
&\delta_{1}@4d\equiv m(A1)-m(A2)-m(T2)+m(T1) = 1,&
\end{flalign*}
\lstset{language=bash, keywordstyle=\color{blue!70}, basicstyle=\ttfamily, frame=shadowbox}
\begin{lstlisting}
Computed bands:  1 - 34
GM: GM1 (1); GM4 (3); GM1 (1); GM1 (1); GM4 (3); GM4 (3); GM3 (2); GM3 (2);
    GM4 (3); GM5 (3); GM3 (2); GM4 (3); GM5 (3); GM4 (3); GM1 (1); [34] ;
X : X1  (1); X3  (1); X5  (2); X1  (1); X3  (1); X3  (1); X1  (1); X1  (1);
    X3  (1); X5  (2); X5  (2); X1  (1); X3  (1); X4  (1); X2  (1); X5  (2);
    X5  (2); X1  (1); X4  (1); X5  (2); X2  (1); X5  (2); X4  (1); X3  (1);
    X3  (1); X5  (2); [34] ;
L : L1  (1); L1  (1); L3  (2); L1  (1); L1  (1); L1  (1); L3  (2); L1  (1);
    L3  (2); L1  (1); L3  (2); L3  (2); L1  (1); L3  (2); L3  (2); L2  (1);
    L3  (2); L1  (1); L3  (2); L3  (2); L2  (1); L3  (2); L1  (1); [34] ;
W : W1  (1); W3  W4  (2); W2  (1); W2  (1); W3  (1); W4  (1); W1  (1); W1  (1);
    W2  (1); W4  (1); W3  (1); W1  (1); W4  (1); W3  (1); W2  (1); W4  (1);
    W2  (1); W3  (1); W1  (1); W2  (1); W3  (1); W1  (1); W4  (1); W1  (1);
    W3  (1); W2  (1); W4  (1); W3  (1); W2  (1); W3  (1); W4  (1); W1  (1);
    W3  (1); [34];
\end{lstlisting}
\hyperref[tab:electride]{Back to the table}

\subsubsection*{633866 FeTe$_{2}$}
\label{sec:tqc633866}
\noindent Essential BR: $Ag@2c$ \\
\noindent RSI:
\begin{flalign*}
&\delta_{1}@2c\equiv -m(Ag)+m(Au)-m(Bg)+m(Bu) = -1,&
\end{flalign*}
\lstset{language=bash, keywordstyle=\color{blue!70}, basicstyle=\ttfamily, frame=shadowbox}
\begin{lstlisting}
Computed bands:  1 - 20
GM: GM1+(1); GM2+(1); GM3-(1); GM4-(1); GM2+(1); GM2+(1); GM3-(1); GM1+(1);
    GM1-(1); GM4-(1); GM3+(1); GM1+(1); GM4+(1); GM2+(1); GM3+(1); GM1+(1);
    GM4+(1); GM1+(1); GM2+(1); GM2-(1); [20] ;
R : R2  (2); R1  (2); R1  (2); R2  (2); R1  (2); R2  (2); R2  (2); R1  (2);
    R1  (2); R2  (2); [20] ;
S : S3- S4- (2); S1+ S2+ (2); S1+ S2+ (2); S3- S4- (2); S3+ S4+ (2); S1+ S2+ (2);
    S1+ S2+ (2); S1- S2- (2); S3- S4- (2); S3+ S4+ (2); [20] ;
T : T1- (2); T1+ (2); T1+ (2); T1- (2); T1- (2); T1+ (2); T1- (2); T1+ (2);
    T1+ (2); T1+ (2); [20] ;
U : U1+ (2); U1- (2); U1+ (2); U1- (2); U1+ (2); U1- (2); U1+ (2); U1+ (2);
    U1+ (2); U1+ (2); [20] ;
X : X1  (2); X1  (2); X1  (2); X2  (2); X1  (2); X1  (2); X2  (2); X1  (2);
    X1  (2); X2  (2); [20] ;
Y : Y2  (2); Y2  (2); Y2  (2); Y2  (2); Y1  (2); Y2  (2); Y1  (2); Y2  (2);
    Y2  (2); Y1  (2); [20] ;
Z : Z1  (2); Z2  (2); Z1  (2); Z1  (2); Z2  (2); Z1  (2); Z2  (2); Z2  (2);
    Z1  (2); Z1  (2); [20];
\end{lstlisting}
\hyperref[tab:electride]{Back to the table}

\subsubsection*{658914 CaInCu$_{4}$}
\label{sec:tqc658914}
\noindent Essential BR: $A1@4d$ \\
\noindent RSI:
\begin{flalign*}
&\delta_{1}@4d\equiv m(A1)-m(A2)-m(T2)+m(T1) = 1,&
\end{flalign*}
\lstset{language=bash, keywordstyle=\color{blue!70}, basicstyle=\ttfamily, frame=shadowbox}
\begin{lstlisting}
Computed bands:  1 - 29
GM: GM1 (1); GM4 (3); GM1 (1); GM1 (1); GM4 (3); GM4 (3); GM3 (2); GM5 (3);
    GM3 (2); GM4 (3); GM5 (3); GM4 (3); GM1 (1); [29] ;
X : X1  (1); X3  (1); X5  (2); X3  (1); X1  (1); X3  (1); X5  (2); X1  (1);
    X1  (1); X3  (1); X5  (2); X1  (1); X2  (1); X3  X5  (3); X4  (1); X5  (2);
    X2  (1); X5  (2); X4  (1); X3  (1); X5  (2); [29] ;
L : L1  (1); L1  (1); L3  (2); L1  (1); L1  (1); L1  (1); L3  (2); L1  (1);
    L3  (2); L1  (1); L1  (1); L3  (2); L3  (2); L3  (2); L2  (1); L3  (2);
    L3  (2); L2  (1); L1  (1); L3  (2); [29] ;
W : W1  (1); W4  (1); W3  (1); W2  (1); W4  (1); W2  (1); W3  (1); W1  (1);
    W2  (1); W1  (1); W4  (1); W3  (1); W1  (1); W4  (1); W3  (1); W2  (1);
    W2  (1); W3  (1); W1  (1); W4  (1); W1  (1); W4  (1); W3  (1); W2  (1);
    W3  (1); W2  (1); W4  (1); W1  (1); W3  (1); [29];
\end{lstlisting}
\hyperref[tab:electride]{Back to the table}

\subsubsection*{611576 As$_{2}$W}
\label{sec:tqc611576}
\lstset{language=bash, keywordstyle=\color{blue!70}, basicstyle=\ttfamily, frame=shadowbox}
\begin{lstlisting}
Computed bands:  1 - 16
GM: GM1+(1); GM1+(1); GM2-(1); GM2-(1); GM1+(1); GM2-(1); GM1+(1); GM1+(1);
    GM2-(1); GM1+(1); GM1-(1); GM2+(1); GM1+(1); GM2+(1); GM2-(1); GM1-(1);
    [16] ;
Y : Y2- (1); Y1+ (1); Y2- (1); Y1+ (1); Y1+ (1); Y2- (1); Y2- (1); Y1+ (1);
    Y1- (1); Y2- (1); Y1+ (1); Y1+ (1); Y1- (1); Y2+ (1); Y1+ (1); Y2+ (1);
    [16] ;
V : V1+ (1); V1- (1); V1+ (1); V1- (1); V1+ (1); V1- (1); V1- (1); V1+ (1);
    V1- (1); V1- (1); V1+ (1); V1+ (1); V1- (1); V1+ (1); V1+ (1); V1+ (1);
    [16] ;
L : L1- (1); L1- (1); L1+ (1); L1+ (1); L1- (1); L1+ (1); L1+ (1); L1- (1);
    L1- (1); L1+ (1); L1- (1); L1- (1); L1+ (1); L1+ (1); L1- (1); L1+ (1);
    [16] ;
M : M1+ (1); M2- (1); M2- (1); M1+ (1); M2- (1); M1+ (1); M2- (1); M1+ (1);
    M1+ (1); M2+ (1); M1- (1); M2- (1); M2- (1); M1+ (1); M2+ (1); M1- (1);
    [16] ;
U : U2  (1); U1  (1); U2  (1); U1  (1); U1  (1); U2  (1); U2  (1); U1  (1);
    U2  (1); U1  (1); U1  (1); U1  (1); U2  (1); U2  (1); U2  (1); U1  (1);
    [16] ;
A : A2- (1); A1+ (1); A2- (1); A1+ (1); A2- (1); A1+ (1); A1+ (1); A2- (1);
    A2- (1); A1+ (1); A1- (1); A2+ (1); A1+ (1); A2- (1); A1- (1); A2+ (1); [16];
\end{lstlisting}
\hyperref[tab:electride]{Back to the table}

\subsubsection*{42578 As$_{2}$Ru}
\label{sec:tqc42578}
\lstset{language=bash, keywordstyle=\color{blue!70}, basicstyle=\ttfamily, frame=shadowbox}
\begin{lstlisting}
Computed bands:  1 - 18
GM: GM1+(1); GM2+(1); GM3-(1); GM4-(1); GM2+(1); GM2+(1); GM1+(1); GM3-(1);
    GM1+(1); GM4-(1); GM2+(1); GM1+(1); GM4+(1); GM3+(1); GM1+(1); GM2+(1);
    GM3+(1); GM4+(1); [18] ;
R : R2  (2); R1  (2); R2  (2); R1  (2); R2  (2); R1  (2); R2  (2); R1  (2);
    R2  (2); [18] ;
S : S3- S4- (2); S1+ S2+ (2); S1+ S2+ (2); S3- S4- (2); S1+ S2+ (2); S3+ S4+ (2);
    S1+ S2+ (2); S3- S4- (2); S3+ S4+ (2); [18] ;
T : T1- (2); T1+ (2); T1+ (2); T1+ (2); T1- (2); T1- (2); T1+ (2); T1+ (2);
    T1- (2); [18] ;
U : U1+ (2); U1- (2); U1+ (2); U1- (2); U1+ (2); U1+ (2); U1- (2); U1+ (2);
    U1+ (2); [18] ;
X : X1  (2); X1  (2); X1  (2); X1  (2); X1  (2); X2  (2); X1  (2); X1  (2);
    X2  (2); [18] ;
Y : Y2  (2); Y2  (2); Y2  (2); Y2  (2); Y2  (2); Y2  (2); Y1  (2); Y1  (2);
    Y2  (2); [18] ;
Z : Z1  (2); Z2  (2); Z1  (2); Z2  (2); Z1  (2); Z2  (2); Z1  (2); Z2  (2);
    Z1  (2); [18];
\end{lstlisting}
\hyperref[tab:electride]{Back to the table}

\subsubsection*{152560 TbInCu$_{4}$}
\label{sec:tqc152560}
\noindent Essential BR: $A1@4d$ \\
\noindent RSI:
\begin{flalign*}
&\delta_{1}@4d\equiv m(A1)-m(A2)-m(T2)+m(T1) = 1,&
\end{flalign*}
\lstset{language=bash, keywordstyle=\color{blue!70}, basicstyle=\ttfamily, frame=shadowbox}
\begin{lstlisting}
Computed bands:  1 - 28
GM: GM4 (3); GM1 (1); GM1 (1); GM4 (3); GM4 (3); GM3 (2); GM5 (3); GM4 (3);
    GM3 (2); GM5 (3); GM4 (3); GM1 (1); [28] ;
X : X3  (1); X5  (2); X3  (1); X1  (1); X3  (1); X5  (2); X1  (1); X3  (1);
    X5  (2); X1  (1); X1  (1); X3  (1); X4  (1); X2  (1); X5  (2); X5  (2);
    X4  (1); X5  (2); X2  (1); X3  (1); X5  (2); [28] ;
L : L1  (1); L3  (2); L1  (1); L1  (1); L1  (1); L3  (2); L1  (1); L3  (2);
    L3  (2); L1  (1); L1  (1); L3  (2); L3  (2); L2  (1); L3  (2); L3  (2);
    L2  (1); L1  (1); L3  (2); [28] ;
W : W4  (1); W3  (1); W2  (1); W4  (1); W3  (1); W2  (1); W1  (1); W3  (1);
    W4  (1); W2  (1); W1  (1); W4  (1); W1  (1); W2  (1); W3  (1); W4  (1);
    W2  (1); W3  (1); W1  (1); W3  (1); W2  (1); W1  (1); W4  (1); W3  (1);
    W2  (1); W4  (1); W1  (1); W3  (1); [28];
\end{lstlisting}
\hyperref[tab:electride]{Back to the table}

\subsubsection*{98990 KB$_{6}$}
\label{sec:tqc98990}
\lstset{language=bash, keywordstyle=\color{blue!70}, basicstyle=\ttfamily, frame=shadowbox}
\begin{lstlisting}
Computed bands:  1 - 14
GM: GM1+(1); GM1+(1); GM4-(3); GM4-(3); GM1+(1); GM3+(2); GM5+(3); [14] ;
R : R2- (1); R1+ (1); R5+ (3); R4- (3); R5+ (3); R4- (3); [14] ;
M : M4+ (1); M1+ (1); M5- (2); M2- (1); M5- (2); M1+ (1); M3- (1); M4+ (1);
    M5+ (2); M5- (2); [14] ;
X : X3- (1); X1+ (1); X5+ (2); X1+ (1); X3- (1); X5- (2); X1+ (1); X2+ (1);
    X5+ (2); X3- (1); X4+ (1); [14];
\end{lstlisting}
\hyperref[tab:electride]{Back to the table}

\subsubsection*{30734 Rb$_{2}$Te$_{5}$}
\label{sec:tqc30734}
\lstset{language=bash, keywordstyle=\color{blue!70}, basicstyle=\ttfamily, frame=shadowbox}
\begin{lstlisting}
Computed bands:  1 - 24
GM: GM1+(1); GM2-(1); GM1+(1); GM1-(1); GM2+(1); GM1+(1); GM2-(1); GM1+(1);
    GM2-(1); GM1-(1); GM1+(1); GM2+(1); GM2-(1); GM1+(1); GM1-(1); GM2-(1);
    GM1+(1); GM2-(1); GM2+(1); GM1-(1); GM2-(1); GM1-(1); GM1+(1); GM2+(1);
    [24] ;
Y : Y1+ (1); Y2- (1); Y1+ (1); Y1- (1); Y1+ (1); Y2- (1); Y2+ (1); Y2- (1);
    Y1+ (1); Y1- (1); Y1+ (1); Y2- (1); Y2+ (1); Y1- (1); Y2- (1); Y1- (1);
    Y1+ (1); Y2- (1); Y1+ (1); Y2+ (1); Y1- (1); Y2- (1); Y1+ (1); Y2+ (1);
    [24] ;
V : V1- (1); V1+ (1); V1+ (1); V1- (1); V1- (1); V1+ (1); V1- (1); V1+ (1);
    V1- (1); V1+ (1); V1+ (1); V1- (1); V1+ (1); V1- (1); V1- (1); V1+ (1);
    V1- (1); V1+ (1); V1- (1); V1- (1); V1+ (1); V1- (1); V1+ (1); V1+ (1);
    [24] ;
L : L1- (1); L1+ (1); L1+ (1); L1- (1); L1+ (1); L1- (1); L1+ (1); L1- (1);
    L1+ (1); L1- (1); L1+ (1); L1- (1); L1+ (1); L1- (1); L1- (1); L1+ (1);
    L1- (1); L1+ (1); L1+ (1); L1- (1); L1- (1); L1- (1); L1+ (1); L1+ (1);
    [24] ;
M : M1+ (1); M2- (1); M2- (1); M1+ (1); M2+ (1); M1+ (1); M1- (1); M1+ (1);
    M2- (1); M2+ (1); M2- (1); M1- (1); M1+ (1); M2- (1); M2- (1); M1- (1);
    M1+ (1); M2+ (1); M2+ (1); M2- (1); M1+ (1); M2- (1); M1- (1); M2+ (1);
    [24] ;
U : U1  (1); U2  (1); U2  (1); U1  (1); U2  (1); U1  (1); U1  (1); U1  (1);
    U2  (1); U2  (1); U2  (1); U1  (1); U1  (1); U2  (1); U2  (1); U1  (1);
    U2  (1); U1  (1); U2  (1); U1  (1); U2  (1); U2  (1); U1  (1); U2  (1);
    [24] ;
A : A1+ A2- (2); A1+ (1); A2+ (1); A2- (1); A1- (1); A1+ (1); A2- (1); A2+ (1);
    A1+ (1); A2- (1); A1- (1); A1+ (1); A2- (1); A2+ (1); A1- (1); A2- (1);
    A1+ (1); A2+ (1); A2- (1); A1- (1); A1+ A2- (2); A2+ (1); [24];
\end{lstlisting}
\hyperref[tab:electride]{Back to the table}

\subsubsection*{20240 SiB$_{6}$}
\label{sec:tqc20240}
\lstset{language=bash, keywordstyle=\color{blue!70}, basicstyle=\ttfamily, frame=shadowbox}
\begin{lstlisting}
Computed bands:  1 - 11
GM: GM1+(1); GM1+(1); GM4-(3); GM3+(2); GM1+(1); GM5+(3); [11] ;
R : R2- (1); R5+ (3); R1+ (1); R4- (3); R5+ (3); [11] ;
M : M4+ (1); M5- (2); M1+ (1); M2- (1); M4+ (1); M5+ (2); M1+ (1); M5- (2);
    [11] ;
X : X3- (1); X1+ (1); X3- (1); X5+ (2); X1+ (1); X4- (1); X5- (2); X1+ (1);
    X2- (1); [11];
\end{lstlisting}
\hyperref[tab:electride]{Back to the table}

\subsubsection*{201787 V(MoS$_{2}$)$_{2}$}
\label{sec:tqc201787}
\lstset{language=bash, keywordstyle=\color{blue!70}, basicstyle=\ttfamily, frame=shadowbox}
\begin{lstlisting}
Computed bands:  1 - 21
GM: GM1+(1); GM2-(1); GM1+GM2-(2); GM1+(1); GM2-(1); GM1+(1); GM2-(1); GM1+(1);
    GM1-(1); GM2+(1); GM1+(1); GM2+(1); GM2-(1); GM1+(1); GM2-(1); GM2-(1);
    GM1-(1); GM2+(1); GM1+(1); GM1+(1); [21] ;
Y : Y2- (1); Y1+ (1); Y1+ (1); Y2- (1); Y1+ (1); Y2- (1); Y1+ (1); Y2- (1);
    Y1+ (1); Y2- (1); Y2+ (1); Y1- (1); Y2- (1); Y1+ (1); Y2+ (1); Y1+ (1);
    Y1- (1); Y1- (1); Y2- (1); Y1+ (1); Y2- (1); [21] ;
V : V1+ (1); V1- (1); V1+ (1); V1- (1); V1+ (1); V1- (1); V1- (1); V1+ (1);
    V1+ (1); V1+ (1); V1- (1); V1+ (1); V1- (1); V1- (1); V1+ (1); V1- (1);
    V1+ (1); V1- (1); V1+ (1); V1+ (1); V1+ (1); [21] ;
L : L1- (1); L1+ (1); L1- (1); L1+ (1); L1+ (1); L1- (1); L1+ (1); L1- (1);
    L1+ (1); L1- (1); L1+ (1); L1- (1); L1- (1); L1+ (1); L1- (1); L1+ (1);
    L1+ (1); L1- (1); L1+ (1); L1+ (1); L1+ (1); [21] ;
M : M1+ (1); M2- (1); M2- (1); M1+ (1); M2- (1); M1+ (1); M2- (1); M2- (1);
    M1+ (1); M1- (1); M2+ (1); M1+ (1); M1+ (1); M2+ (1); M2- (1); M1- (1);
    M2- (1); M1+ (1); M2+ (1); M1+ (1); M1+ (1); [21] ;
U : U2  (1); U1  (1); U2  (1); U1  (1); U1  (1); U2  (1); U2  (1); U1  (1);
    U1  (1); U2  (1); U1  (1); U2  (1); U2  (1); U1  (1); U2  (1); U1  (1);
    U1  (1); U2  (1); U1  (1); U2  (1); U1  (1); [21] ;
A : A1+ (1); A2- (1); A1+ (1); A2- (1); A1+ (1); A2- (1); A1+ (1); A1+ (1);
    A2- (1); A2+ (1); A1+ (1); A2- (1); A1- (1); A2+ (1); A2- (1); A1+ (1);
    A1- (1); A2- (1); A1+ (1); A1- (1); A2- (1); [21];
\end{lstlisting}
\hyperref[tab:electride]{Back to the table}

\subsubsection*{245961 Ge}
\label{sec:tqc245961}
\lstset{language=bash, keywordstyle=\color{blue!70}, basicstyle=\ttfamily, frame=shadowbox}
\begin{lstlisting}
Computed bands:  1 - 16
GM: GM1+(1); GM1-(1); GM2+GM3+(2); GM1+(1); GM2-GM3-(2); GM1-(1); GM2-GM3-(2);
    GM1+(1); GM2+GM3+(2); GM2+GM3+(2); GM1+(1); [16] ;
T : T2+ T3+ (2); T1- (1); T1+ (1); T2- T3- (2); T1+ (1); T1- (1); T2- T3- (2);
    T2+ T3+ (2); T1- (1); T1+ (1); T2+ T3+ (2); [16] ;
F : F1+ (1); F1- (1); F1- (1); F1+ (1); F1- (1); F1+ (1); F1+ (1); F1- (1);
    F1+ (1); F1- (1); F1+ (1); F1- (1); F1- (1); F1+ (1); F1+ (1); F1- (1);
    [16] ;
L : L1- (1); L1+ (1); L1- (1); L1+ (1); L1- (1); L1+ (1); L1- (1); L1+ (1);
    L1+ (1); L1- (1); L1+ (1); L1- (1); L1+ (1); L1- (1); L1- (1); L1- (1); [16];
\end{lstlisting}
\hyperref[tab:electride]{Back to the table}

\subsubsection*{613476 CrB}
\label{sec:tqc613476}
\lstset{language=bash, keywordstyle=\color{blue!70}, basicstyle=\ttfamily, frame=shadowbox}
\begin{lstlisting}
Computed bands:  1 - 18
GM: GM1+(1); GM2+(1); GM4-(1); GM3-(1); GM5+(2); GM5-(2); GM1+(1); GM4-(1);
    GM2+(1); GM1+(1); GM3-(1); GM2+(1); GM5+(2); GM1-(1); GM4+(1); [18] ;
M : M1  (2); M1  (2); M3  (2); M3  (2); M1  (2); M1  (2); M1  (2); M4  (2);
    M2  (2); [18] ;
P : P1  (2); P2  (2); P2  (2); P1  (2); P2  (2); P1  (2); P2  (2); P1  (2);
    P2  (2); [18] ;
X : X2  (2); X1  (2); X2  (2); X1  (2); X2  (2); X1  (2); X1  (2); X1  (2);
    X2  (2); [18] ;
N : N2- (1); N1+ (1); N2- (1); N1+ (1); N1- (1); N1+ (1); N2- (1); N1+ (1);
    N2- (1); N2+ (1); N1- (1); N2- (1); N2- (1); N1+ (1); N2+ (1); N1+ (1);
    N1+ (1); N2+ (1); [18];
\end{lstlisting}
\hyperref[tab:electride]{Back to the table}

\subsubsection*{26288 KNbSe$_{2}$}
\label{sec:tqc26288}
\lstset{language=bash, keywordstyle=\color{blue!70}, basicstyle=\ttfamily, frame=shadowbox}
\begin{lstlisting}
Computed bands:  1 - 32
A : A1  (2); A1  (2); A3  (4); A1  (2); A3  (4); A1  (2); A1  (2); A1  (2);
    A1  (2); A3  (4); A3  (4); A1  (2); [32] ;
GM: GM1+GM3+(2); GM2-GM3+(2); GM5+GM6-(4); GM4-(1); GM2-(1); GM5-GM6-(4);
    GM1+(1); GM4-(1); GM3+(1); GM2-(1); GM1+(1); GM4-(1); GM3+(1); GM2-(1);
    GM5+(2); GM6-(2); GM6+(2); GM5-(2); GM1+(1); GM4-(1); [32] ;
H : H3  (2); H1  (2); H2  (2); H3  (2); H1  H2  (4); H3  (2); H1  (2); H2  (2);
    H1  (2); H3  (2); H2  (2); H1  (2); H3  (2); H2  (2); H2  (2); [32] ;
K : K1  K2  (2); K5  (2); K5  (2); K2  K3  (2); K5  K6  (4); K3  K4  (2);
    K5  (2); K6  (2); K6  (2); K1  (1); K4  (1); K5  (2); K5  (2); K2  (1);
    K3  (1); K6  (2); K5  (2); [32] ;
L : L1  (2); L1  (2); L1  (2); L2  (2); L1  (2); L2  (2); L1  (2); L1  (2);
    L1  (2); L1  (2); L1  (2); L1  (2); L2  (2); L1  (2); L2  (2); L1  (2);
    [32] ;
M : M1+ M3+ (2); M1+ M4- (2); M2- M3+ (2); M2+ M3- (2); M2- M4- (2); M1- M3- (2);
    M4- (1); M2- (1); M1+ (1); M4- (1); M3+ (1); M2- (1); M1+ (1); M4- (1);
    M3+ (1); M2- (1); M1+ (1); M4- (1); M4+ (1); M1- (1); M3+ (1); M2- (1);
    M2+ (1); M3- (1); M1+ (1); M4- (1); [32];
\end{lstlisting}
\hyperref[tab:electride]{Back to the table}

\subsubsection*{16325 Te$_{2}$AuI}
\label{sec:tqc16325}
\noindent Essential BR: $Ag@2b$ \\
\noindent RSI:
\begin{flalign*}
&\delta_{1}@2b\equiv -m(Ag)+m(Au)-m(Bg)+m(Bu) = -1,&
\end{flalign*}
\lstset{language=bash, keywordstyle=\color{blue!70}, basicstyle=\ttfamily, frame=shadowbox}
\begin{lstlisting}
Computed bands:  1 - 30
GM: GM1+(1); GM4+(1); GM2-(1); GM1+(1); GM3-(1); GM2-(1); GM1+(1); GM4+(1);
    GM3+(1); GM2+(1); GM4+(1); GM1+(1); GM4+(1); GM3+(1); GM2+(1); GM1+(1);
    GM4+(1); GM1+(1); GM3-(1); GM2-(1); GM1+(1); GM4-(1); GM1-(1); GM4+(1);
    GM2-(1); GM1+(1); GM4+(1); GM3-(1); GM3+(1); GM4-(1); [30] ;
R : R2  (2); R1  (2); R2  (2); R1  (2); R2  (2); R2  (2); R1  (2); R1  (2);
    R2  (2); R1  (2); R1  (2); R2  (2); R1  (2); R2  (2); R1  (2); [30] ;
S : S2  (2); S1  (2); S2  (2); S1  (2); S2  (2); S1  (2); S2  (2); S1  (2);
    S1  (2); S2  (2); S2  (2); S1  (2); S2  (2); S1  (2); S1  (2); [30] ;
T : T4- (1); T1- (1); T2- (1); T1+ (1); T3+ (1); T2+ (1); T2- (1); T3- (1);
    T4- (1); T1- (1); T1+ (1); T4- (1); T3+ (1); T2- T3- (2); T1- (1); T4+ (1);
    T3- (1); T2- (1); T3- (1); T2+ (1); T2- (1); T4- (1); T4- (1); T1- (1);
    T4+ (1); T3+ (1); T2- (1); T1+ (1); T3- (1); [30] ;
U : U1  (2); U1  (2); U1  (2); U1  (2); U2  (2); U1  (2); U1  (2); U2  (2);
    U1  (2); U1  (2); U1  (2); U2  (2); U1  (2); U2  (2); U1  (2); [30] ;
X : X1  (2); X1  (2); X1  (2); X1  (2); X2  (2); X1  (2); X1  (2); X2  (2);
    X1  (2); X1  (2); X2  (2); X1  (2); X1  (2); X2  (2); X1  (2); [30] ;
Y : Y4- (1); Y1- (1); Y1+ (1); Y2- (1); Y3+ (1); Y2+ (1); Y4+ (1); Y1+ (1);
    Y3+ (1); Y2+ (1); Y4+ (1); Y1+ (1); Y4+ (1); Y3+ (1); Y1+ (1); Y2+ (1);
    Y4+ (1); Y4- (1); Y1- (1); Y1+ (1); Y3- (1); Y4- (1); Y2- (1); Y3+ (1);
    Y1- (1); Y2- (1); Y4- (1); Y1+ (1); Y3- (1); Y4+ (1); [30] ;
Z : Z1+ (1); Z4+ (1); Z2- (1); Z1+ (1); Z2- (1); Z3- (1); Z2- (1); Z4- (1);
    Z1- (1); Z3- (1); Z1+ (1); Z3- (1); Z4+ (1); Z2- (1); Z4- (1); Z1- (1);
    Z3- (1); Z2- (1); Z3- (1); Z2- (1); Z4+ (1); Z1+ (1); Z2- (1); Z3+ (1);
    Z2+ (1); Z3+ (1); Z4- (1); Z3- (1); Z1+ (1); Z4+ (1); [30];
\end{lstlisting}
\hyperref[tab:electride]{Back to the table}

\subsubsection*{614793 BMo}
\label{sec:tqc614793}
\lstset{language=bash, keywordstyle=\color{blue!70}, basicstyle=\ttfamily, frame=shadowbox}
\begin{lstlisting}
Computed bands:  1 - 18
GM: GM1+(1); GM2+(1); GM4-(1); GM3-(1); GM5+(2); GM4-(1); GM1+(1); GM5-(2);
    GM2+(1); GM3-(1); GM2+(1); GM1+(1); GM5+(2); GM1-(1); GM4+(1); [18] ;
M : M1  (2); M1  (2); M3  (2); M1  (2); M3  (2); M1  (2); M1  (2); M4  (2);
    M2  (2); [18] ;
P : P1  (2); P2  (2); P2  (2); P1  (2); P2  (2); P1  (2); P1  (2); P2  (2);
    P1  (2); [18] ;
X : X1  (2); X2  (2); X1  (2); X2  (2); X1  (2); X2  (2); X1  (2); X1  (2);
    X2  (2); [18] ;
N : N1+ (1); N1+ N2- (2); N2- (1); N2+ (1); N1+ (1); N2- (1); N1+ (1); N2- (1);
    N1- (1); N2+ (1); N1+ (1); N1+ (1); N2- (1); N1- (1); N2- (1); N2- (1);
    N1- (1); [18];
\end{lstlisting}
\hyperref[tab:electride]{Back to the table}

\subsubsection*{26285 NaNbS$_{2}$}
\label{sec:tqc26285}
\lstset{language=bash, keywordstyle=\color{blue!70}, basicstyle=\ttfamily, frame=shadowbox}
\begin{lstlisting}
Computed bands:  1 - 24
A : A1  (2); A3  (4); A1  (2); A1  (2); A1  (2); A1  (2); A3  (4); A3  (4);
    A1  (2); [24] ;
GM: GM2-GM3+(2); GM5+GM6-(4); GM1+(1); GM4-(1); GM3+(1); GM2-(1); GM1+(1);
    GM4-(1); GM3+(1); GM5+(2); GM6-(2); GM2-(1); GM6+(2); GM5-(2); GM1+(1);
    GM4-(1); [24] ;
H : H1  (2); H2  (2); H3  (2); H1  (2); H2  (2); H1  (2); H3  (2); H1  (2);
    H2  (2); H3  (2); H2  (2); H2  (2); [24] ;
K : K5  (2); K5  (2); K2  K3  (2); K5  (2); K6  (2); K6  (2); K1  (1); K4  (1);
    K5  (2); K5  (2); K2  (1); K3  (1); K6  (2); K5  (2); [24] ;
L : L1  (2); L1  (2); L2  (2); L1  (2); L1  (2); L1  (2); L1  (2); L1  (2);
    L2  (2); L1  (2); L2  (2); L1  (2); [24] ;
M : M1+ M4- (2); M2- M3+ (2); M2+ M3- (2); M4- (1); M1+ (1); M2- (1); M3+ (1);
    M4- (1); M1+ (1); M3+ (1); M2- (1); M1+ (1); M4- (1); M1- (1); M4+ (1);
    M2- (1); M3+ (1); M3- (1); M2+ (1); M4- (1); M1+ (1); [24];
\end{lstlisting}
\hyperref[tab:electride]{Back to the table}

\subsubsection*{190537 NaCl$_{7}$}
\label{sec:tqc190537}
\noindent Essential BR: $Ag@3d$ \\
\noindent RSI:
\begin{flalign*}
&\delta_{1}@3d\equiv -m(Ag)+m(Au)-m(B1g)+m(B1u)-m(B3g)+m(B3u)-m(B2g)+m(B2u) = -1,&
\end{flalign*}
\lstset{language=bash, keywordstyle=\color{blue!70}, basicstyle=\ttfamily, frame=shadowbox}
\begin{lstlisting}
Computed bands:  1 - 25
GM: GM1+(1); GM2+GM3+(2); GM4-(3); GM1+(1); GM2+GM3+(2); GM1+(1); GM4-(3);
    GM4-(3); GM4+(3); GM4+(3); GM4-(3); [25] ;
R : R4- (3); R4+ (3); R1- (1); R4- (3); R1+ (1); R4+ (3); R2+ R3+ (2); R1- (1);
    R4- (3); R2- R3- (2); R4+ (3); [25] ;
M : M3- (1); M1+ M4- (2); M2+ (1); M4- (1); M4+ (1); M2+ (1); M3- (1); M1+ (1);
    M4- (1); M1+ (1); M2- (1); M3+ (1); M2+ (1); M1+ (1); M3- (1); M1- (1);
    M3- (1); M2- (1); M3+ (1); M2+ (1); M1- (1); M4- (1); M3- (1); M1- (1);
    [25] ;
X : X1+ (1); X4- (1); X1+ (1); X4- (1); X2+ X2- (2); X4- (1); X4- (1); X1+ (1);
    X1+ (1); X3- (1); X1+ (1); X2- (1); X3+ (1); X4- (1); X3- (1); X4+ (1);
    X3+ (1); X3- (1); X2+ (1); X1- X3+ (2); X1+ (1); X3+ (1); X2+ (1); [25];
\end{lstlisting}
\hyperref[tab:electride]{Back to the table}

\subsubsection*{408030 LaSi}
\label{sec:tqc408030}
\noindent Essential BR: $Ag@4a$ \\
\noindent RSI:
\begin{flalign*}
&\delta_{1}@4a\equiv -m(Ag)+m(Au)-m(Bg)+m(Bu) = -1,&
\end{flalign*}
\lstset{language=bash, keywordstyle=\color{blue!70}, basicstyle=\ttfamily, frame=shadowbox}
\begin{lstlisting}
Computed bands:  1 - 30
GM: GM1+(1); GM4-(1); GM1+(1); GM4-(1); GM3+(1); GM4-(1); GM1+(1); GM2+(1);
    GM3+(1); GM3-(1); GM3-(1); GM2+(1); GM2-(1); GM1+(1); GM4-(1); GM2-(1);
    GM1+(1); GM4-(1); GM3+(1); GM2-(1); GM1+(1); GM4-(1); GM1-(1); GM1+(1);
    GM3-(1); GM2+(1); GM3+(1); GM4-(1); GM4+(1); GM1+(1); [30] ;
T : T1  (2); T1  (2); T1  (2); T1  (2); T2  (2); T1  (2); T2  (2); T1  (2);
    T1  (2); T1  (2); T2  (2); T1  (2); T1  (2); T1  (2); T2  (2); [30] ;
Y : Y4- (1); Y1+ (1); Y1+ (1); Y4- (1); Y3+ (1); Y2- (1); Y3- (1); Y1+ (1);
    Y4- (1); Y1+ (1); Y2+ (1); Y4- (1); Y3- (1); Y2+ (1); Y3+ (1); Y2- (1);
    Y1+ (1); Y4- (1); Y3+ (1); Y2- (1); Y4- (1); Y1+ (1); Y1- (1); Y2+ (1);
    Y3+ (1); Y3- (1); Y4- (1); Y1+ (1); Y1+ (1); Y4+ (1); [30] ;
Z : Z1  (2); Z1  (2); Z1  (2); Z1  (2); Z2  (2); Z2  (2); Z1  (2); Z1  (2);
    Z1  (2); Z1  (2); Z1  (2); Z2  (2); Z1  (2); Z1  (2); Z2  (2); [30] ;
R : R1  (2); R1  (2); R1  (2); R1  (2); R1  (2); R1  (2); R1  (2); R1  (2);
    R1  (2); R1  (2); R1  (2); R1  (2); R1  (2); R1  (2); R1  (2); [30] ;
S : S2- (1); S1+ (1); S2- (1); S1+ (1); S1+ (1); S1- (1); S2- (1); S1+ (1);
    S2- (1); S1+ (1); S2+ (1); S1- (1); S2- (1); S2- (1); S1+ (1); S2+ (1);
    S1+ (1); S2- (1); S2+ (1); S1- (1); S2- (1); S1+ (1); S1+ (1); S2- (1);
    S2+ (1); S1+ (1); S2+ (1); S1- (1); S1+ (1); S2- (1); [30];
\end{lstlisting}
\hyperref[tab:electride]{Back to the table}

\subsubsection*{640379 TaInS$_{2}$}
\label{sec:tqc640379}
\lstset{language=bash, keywordstyle=\color{blue!70}, basicstyle=\ttfamily, frame=shadowbox}
\begin{lstlisting}
Computed bands:  1 - 20
A : A1  (2); A1  (2); A1  (2); A1  (2); A1  (2); A3  (4); A1  (2); A3  (4);
    [20] ;
GM: GM1+(1); GM3+(1); GM4-(1); GM2-(1); GM1+(1); GM3+(1); GM4-(1); GM2-(1);
    GM1+(1); GM4-(1); GM6-(2); GM5+(2); GM3+(1); GM1+(1); GM5-(2); GM6+(2);
    [20] ;
H : H2  (2); H1  (2); H3  (2); H1  (2); H3  (2); H2  (2); H1  (2); H2  (2);
    H3  (2); H2  (2); [20] ;
K : K6  (2); K5  (2); K1  (1); K2  (1); K4  (1); K6  (2); K5  (2); K3  (1);
    K6  (2); K5  (2); K1  (1); K5  (2); K2  (1); [20] ;
L : L1  (2); L1  (2); L1  (2); L1  (2); L1  (2); L1  (2); L1  (2); L2  (2);
    L1  (2); L2  (2); [20] ;
M : M2- (1); M4- (1); M3+ (1); M1+ (1); M1+ (1); M3+ (1); M4- (1); M2- (1);
    M2- (1); M4- (1); M1+ (1); M3+ (1); M4- (1); M1- (1); M4+ (1); M3+ (1);
    M1+ (1); M1+ (1); M3- (1); M2+ (1); [20];
\end{lstlisting}
\hyperref[tab:electride]{Back to the table}

\subsubsection*{79235 SiRh}
\label{sec:tqc79235}
\noindent Essential BR: $Ag@2a$ \\
\noindent RSI:
\begin{flalign*}
&\delta_{1}@2a\equiv -m(Ag)+m(Au) = -1,&
\end{flalign*}
\lstset{language=bash, keywordstyle=\color{blue!70}, basicstyle=\ttfamily, frame=shadowbox}
\begin{lstlisting}
Computed bands:  1 - 26
A : A1  (2); A1  (2); A1  (2); A1  (2); A1  (2); A1  (2); A1  (2); A1  (2);
    A1  (2); A1  (2); A1  (2); A1  (2); A1  (2); [26] ;
B : B1  (2); B1  (2); B1  (2); B1  (2); B1  (2); B1  (2); B1  (2); B1  (2);
    B1  (2); B1  (2); B1  (2); B1  (2); B1  (2); [26] ;
C : C1  (2); C1  (2); C1  (2); C1  (2); C1  (2); C1  (2); C1  (2); C1  (2);
    C1  (2); C1  (2); C1  (2); C1  (2); C1  (2); [26] ;
D : D1- D2- (2); D1+ D2+ (2); D1- D2- (2); D1- D2- (2); D1+ D2+ (2); D1+ D2+ (2);
    D1- D2- (2); D1+ D2+ (2); D1- D2- (2); D1+ D2+ (2); D1+ D2+ (2); D1- D2- (2);
    D1+ D2+ (2); [26] ;
E : E1+ E2+ (2); E1- E2- (2); E1- E2- (2); E1+ E2+ (2); E1+ E2+ (2); E1- E2- (2);
    E1+ E2+ (2); E1- E2- (2); E1- E2- (2); E1- E2- (2); E1+ E2+ (2); E1+ E2+ (2);
    E1+ E2+ (2); [26] ;
GM: GM1+(1); GM2+(1); GM2-(1); GM1-(1); GM2+(1); GM1-(1); GM1+(1); GM2-(1);
    GM2+(1); GM2-(1); GM1+(1); GM2+(1); GM1-(1); GM1+(1); GM1-(1); GM2-(1);
    GM2+(1); GM1+(1); GM2+(1); GM2-(1); GM1-(1); GM1+(1); GM1+(1); GM1-(1);
    GM2+(1); GM2-(1); [26] ;
Y : Y1+ (1); Y2- (1); Y1- (1); Y2+ (1); Y1- (1); Y2+ (1); Y2- (1); Y1+ (1);
    Y2- (1); Y1- (1); Y2+ (1); Y1+ (1); Y1- (1); Y1+ (1); Y2+ (1); Y2- (1);
    Y1- (1); Y1+ (1); Y2- (1); Y1- (1); Y2+ (1); Y1+ (1); Y2- (1); Y2+ (1);
    Y1+ (1); Y2+ (1); [26] ;
Z : Z1  (2); Z1  (2); Z1  (2); Z1  (2); Z1  (2); Z1  (2); Z1  (2); Z1  (2);
    Z1  (2); Z1  (2); Z1  (2); Z1  (2); Z1  (2); [26];
\end{lstlisting}
\hyperref[tab:electride]{Back to the table}

\subsubsection*{96089 VAuS$_{2}$}
\label{sec:tqc96089}
\noindent Essential BR: $A1'@2d$ \\
\noindent RSI:
\begin{flalign*}
&\delta_{1}@2d\equiv m(A1')+m(A2')-m(A2'')-m(A1'')-m(E')+m(E'') = 1,&
\end{flalign*}
\lstset{language=bash, keywordstyle=\color{blue!70}, basicstyle=\ttfamily, frame=shadowbox}
\begin{lstlisting}
Computed bands:  1 - 28
A : A1  (2); A1  (2); A1  (2); A1  (2); A1  (2); A3  (4); A3  (4); A3  (4);
    A3  (4); A1  (2); [28] ;
GM: GM1+(1); GM4-(1); GM3+(1); GM2-(1); GM1+(1); GM4-(1); GM3+(1); GM1+(1);
    GM3+(1); GM5+(2); GM6+(2); GM5+GM6+(4); GM2-(1); GM6-(2); GM5+(2); GM1+(1);
    GM5-(2); GM6+(2); GM4-(1); [28] ;
H : H3  (2); H3  (2); H3  (2); H1  (2); H2  (2); H1  (2); H2  (2); H3  (2);
    H1  (2); H2  (2); H1  (2); H2  (2); H3  (2); H2  (2); [28] ;
K : K1  (1); K4  (1); K2  (1); K3  (1); K1  (1); K2  (1); K5  (2); K6  (2);
    K5  (2); K4  (1); K6  (2); K6  (2); K5  (2); K5  (2); K3  (1); K6  (2);
    K1  (1); K2  (1); K5  (2); [28] ;
L : L1  (2); L1  (2); L1  (2); L1  (2); L1  (2); L1  (2); L1  (2); L2  (2);
    L1  (2); L2  (2); L2  (2); L2  (2); L1  (2); L1  (2); [28] ;
M : M1+ (1); M4- (1); M3+ (1); M2- (1); M1+ (1); M1+ (1); M4- (1); M3+ (1);
    M3+ (1); M2- (1); M4- (1); M1+ (1); M3+ (1); M4+ (1); M2+ (1); M1+ (1);
    M3+ (1); M1- (1); M2- (1); M4+ (1); M2+ M4+ (2); M3- (1); M1+ (1); M2+ (1);
    M4- (1); M3+ (1); M1+ (1); [28];
\end{lstlisting}
\hyperref[tab:electride]{Back to the table}

\subsubsection*{37073 InP$_{3}$}
\label{sec:tqc37073}
\lstset{language=bash, keywordstyle=\color{blue!70}, basicstyle=\ttfamily, frame=shadowbox}
\begin{lstlisting}
Computed bands:  1 - 18
GM: GM1+(1); GM3-(2); GM3+(2); GM2-(1); GM1+(1); GM3+(2); GM1-(1); GM3-(2);
    GM2-(1); GM3-(2); GM1+(1); GM3+(2); [18] ;
T : T1+ (1); T3- (2); T3+ (2); T2- (1); T1+ (1); T2- (1); T1- (1); T3+ (2);
    T3- (2); T1+ (1); T3+ (2); T3- (2); [18] ;
F : F1+ (1); F1- (1); F2- (1); F1+ (1); F2+ (1); F2- (1); F1+ (1); F2- (1);
    F1+ (1); F1- (1); F2- (1); F1+ (1); F2+ (1); F1- (1); F1+ (1); F2- (1);
    F1- (1); F2+ (1); [18] ;
L : L1+ (1); L2- (1); L1- (1); L2+ (1); L1+ (1); L2- (1); L1+ (1); L2+ (1);
    L1- (1); L2- (1); L1+ (1); L2- (1); L1+ (1); L1- (1); L2- (1); L2+ (1);
    L1- (1); L1+ (1); [18];
\end{lstlisting}
\hyperref[tab:electride]{Back to the table}

\subsubsection*{413736 LaGe}
\label{sec:tqc413736}
\noindent Essential BR: $Ag@4b$ \\
\noindent RSI:
\begin{flalign*}
&\delta_{1}@4b\equiv -m(Ag)+m(Au)-m(Bg)+m(Bu) = -1,&
\end{flalign*}
\lstset{language=bash, keywordstyle=\color{blue!70}, basicstyle=\ttfamily, frame=shadowbox}
\begin{lstlisting}
Computed bands:  1 - 30
GM: GM1+(1); GM4-(1); GM1+(1); GM4-(1); GM3+(1); GM4-(1); GM1+(1); GM3+(1);
    GM2+(1); GM3-(1); GM3-(1); GM2+(1); GM2-(1); GM1+(1); GM4-(1); GM2-(1);
    GM1+(1); GM4-(1); GM3+(1); GM2-(1); GM1+(1); GM4-(1); GM1+(1); GM1-(1);
    GM3-(1); GM2+(1); GM3+(1); GM4-(1); GM4+(1); GM1+(1); [30] ;
T : T1  (2); T1  (2); T1  (2); T1  (2); T2  (2); T1  (2); T2  (2); T1  (2);
    T1  (2); T1  (2); T1  (2); T2  (2); T1  (2); T1  (2); T2  (2); [30] ;
Y : Y1+ (1); Y4- (1); Y1+ (1); Y4- (1); Y3+ (1); Y2- (1); Y3- (1); Y1+ (1);
    Y4- (1); Y1+ (1); Y4- (1); Y2+ (1); Y3- (1); Y2+ (1); Y3+ (1); Y2- (1);
    Y1+ (1); Y4- (1); Y3+ (1); Y2- (1); Y4- (1); Y1+ (1); Y1- (1); Y2+ (1);
    Y3+ (1); Y4- (1); Y1+ (1); Y3- (1); Y1+ (1); Y4+ (1); [30] ;
Z : Z1  (2); Z1  (2); Z1  (2); Z1  (2); Z2  (2); Z2  (2); Z1  (2); Z1  (2);
    Z1  (2); Z1  (2); Z1  (2); Z2  (2); Z1  (2); Z1  (2); Z2  (2); [30] ;
R : R1  (2); R1  (2); R1  (2); R1  (2); R1  (2); R1  (2); R1  (2); R1  (2);
    R1  (2); R1  (2); R1  (2); R1  (2); R1  (2); R1  (2); R1  (2); [30] ;
S : S1+ (1); S2- (1); S1+ (1); S2- (1); S2+ (1); S2- (1); S1+ (1); S2- (1);
    S1+ (1); S2- (1); S1- (1); S1+ (1); S2+ (1); S1+ (1); S2- (1); S1- (1);
    S2- (1); S1+ (1); S1- (1); S2+ (1); S1+ (1); S2- (1); S2- (1); S1+ (1);
    S1- (1); S1- (1); S2- (1); S2+ (1); S2- (1); S1+ (1); [30];
\end{lstlisting}
\hyperref[tab:electride]{Back to the table}

\subsubsection*{640503 InSe}
\label{sec:tqc640503}
\lstset{language=bash, keywordstyle=\color{blue!70}, basicstyle=\ttfamily, frame=shadowbox}
\begin{lstlisting}
Computed bands:  1 - 18
A : A3  (1); A1  (1); A1  (1); A3  (1); A3  (1); A1  (1); A1  (1); A3  (1);
    A5  A6  (4); A5  A6  (4); A3  (1); A1  (1); [18] ;
GM: GM1 (1); GM3 (1); GM1 (1); GM3 (1); GM1 (1); GM1 (1); GM3 (1); GM3 (1);
    GM5 (2); GM1 (1); GM5 (2); GM6 (2); GM6 (2); GM1 (1); [18] ;
H : H6  (1); H3  (1); H5  (1); H4  (1); H2  H5  (2); H6  (1); H1  (1); H4  (1);
    H1  (1); H3  (1); H2  (1); H6  (1); H3  (1); H5  (1); H2  (1); H4  (1);
    H5  (1); [18] ;
K : K5  (1); K3  (1); K6  (1); K4  (1); K1  K5  (2); K6  (1); K2  (1); K3  (1);
    K1  (1); K4  (1); K2  (1); K5  (1); K3  (1); K1  (1); K5  (1); K6  (1);
    K4  (1); [18] ;
L : L3  (1); L1  (1); L1  (1); L3  (1); L3  (1); L1  (1); L3  (1); L1  (1);
    L1  (1); L3  (1); L4  (1); L2  (1); L3  (1); L1  (1); L2  (1); L4  (1);
    L1  (1); L3  (1); [18] ;
M : M1  (1); M1  M3  (2); M3  (1); M1  (1); M1  (1); M3  (1); M3  (1); M1  (1);
    M1  (1); M2  (1); M3  (1); M2  (1); M4  (1); M3  (1); M4  (1); M1  (1);
    M1  (1); [18];
\end{lstlisting}
\hyperref[tab:electride]{Back to the table}

\subsubsection*{246170 SiPt$_{3}$}
\label{sec:tqc246170}
\noindent Essential BR: $Ag@4f$ \\
\noindent RSI:
\begin{flalign*}
&\delta_{1}@4f\equiv -m(Ag)+m(Au) = -1,&
\end{flalign*}
\lstset{language=bash, keywordstyle=\color{blue!70}, basicstyle=\ttfamily, frame=shadowbox}

\hyperref[tab:electride]{Back to the table}

\subsubsection*{43408 NaSe}
\label{sec:tqc43408}
\noindent Essential BR: $A1'@2d$ \\
\noindent RSI:
\begin{flalign*}
&\delta_{1}@2d\equiv m(A1')+m(A2')-m(A2'')-m(A1'')-m(E')+m(E'') = 1,&
\end{flalign*}
\lstset{language=bash, keywordstyle=\color{blue!70}, basicstyle=\ttfamily, frame=shadowbox}
\begin{lstlisting}
Computed bands:  1 - 14
A : A1  (2); A1  (2); A1  (2); A3  (4); A3  (4); [14] ;
GM: GM1+(1); GM4-(1); GM3+(1); GM2-(1); GM1+(1); GM4-(1); GM5+(2); GM6-(2);
    GM6+(2); GM5-(2); [14] ;
H : H2  (2); H1  (2); H2  (2); H3  (2); H1  (2); H3  (2); H2  (2); [14] ;
K : K5  (2); K6  (2); K5  (2); K1  (1); K4  (1); K5  (2); K2  (1); K6  (2);
    K3  (1); [14] ;
L : L1  (2); L1  (2); L1  (2); L1  (2); L2  (2); L1  (2); L2  (2); [14] ;
M : M4- (1); M1+ (1); M2- (1); M3+ (1); M4- (1); M1+ (1); M4- (1); M1+ (1);
    M3- (1); M2+ (1); M3+ (1); M2- (1); M1- (1); M4+ (1); [14];
\end{lstlisting}
\hyperref[tab:electride]{Back to the table}

\subsubsection*{185172 InSe}
\label{sec:tqc185172}
\noindent Essential BR: $A1'@2c$ \\
\noindent RSI:
\begin{flalign*}
&\delta_{1}@2c\equiv m(A1')+m(A2')-m(A2'')-m(A1'')-m(E')+m(E'') = 1,&
\end{flalign*}
\lstset{language=bash, keywordstyle=\color{blue!70}, basicstyle=\ttfamily, frame=shadowbox}
\begin{lstlisting}
Computed bands:  1 - 18
A : A1  (2); A1  (2); A1  (2); A1  (2); A3  (4); A3  (4); A1  (2); [18] ;
GM: GM1+(1); GM3+(1); GM4-(1); GM2-(1); GM1+(1); GM4-(1); GM3+(1); GM2-(1);
    GM1+(1); GM5+(2); GM6+(2); GM6-(2); GM5-(2); GM4-(1); [18] ;
H : H2  (2); H1  (2); H1  (2); H2  (2); H3  (2); H3  (2); H2  (2); H1  (2);
    H1  (2); [18] ;
K : K5  (2); K6  (2); K5  (2); K6  (2); K1  (1); K2  (1); K4  (1); K3  (1);
    K5  (2); K6  (2); K5  (2); [18] ;
L : L1  (2); L1  (2); L1  (2); L1  (2); L1  (2); L2  (2); L1  (2); L2  (2);
    L1  (2); [18] ;
M : M4- (1); M2- (1); M1+ (1); M3+ (1); M1+ (1); M4- (1); M3+ (1); M2- (1);
    M4- (1); M1+ (1); M2- (1); M3- (1); M1- (1); M2+ (1); M1+ (1); M3+ (1);
    M4+ (1); M4- (1); [18];
\end{lstlisting}
\hyperref[tab:electride]{Back to the table}

\subsubsection*{109036 Si}
\label{sec:tqc109036}
\lstset{language=bash, keywordstyle=\color{blue!70}, basicstyle=\ttfamily, frame=shadowbox}
\begin{lstlisting}
Computed bands:  1 - 16
GM: GM1+(1); GM1-(1); GM1+(1); GM2+GM3+(2); GM2-GM3-(2); GM1-(1); GM2-GM3-(2);
    GM1+(1); GM2+GM3+(2); GM2+GM3+(2); GM1+(1); [16] ;
T : T1- (1); T2+ T3+ (2); T1+ (1); T2- T3- (2); T1+ (1); T2- T3- (2); T1- (1);
    T2+ T3+ (2); T1+ (1); T1- (1); T2+ T3+ (2); [16] ;
F : F1+ (1); F1- (1); F1+ (1); F1- (1); F1- (1); F1+ (1); F1+ (1); F1- (1);
    F1+ (1); F1- (1); F1- (1); F1+ (1); F1- (1); F1+ (1); F1+ (1); F1- (1);
    [16] ;
L : L1- (1); L1+ (1); L1- (1); L1+ (1); L1- (1); L1+ (1); L1+ (1); L1- (1);
    L1+ (1); L1- (1); L1- (1); L1+ (1); L1- (1); L1+ (1); L1- (1); L1- (1); [16];
\end{lstlisting}
\hyperref[tab:electride]{Back to the table}

\subsubsection*{26284 LiNbS$_{2}$}
\label{sec:tqc26284}
\lstset{language=bash, keywordstyle=\color{blue!70}, basicstyle=\ttfamily, frame=shadowbox}
\begin{lstlisting}
Computed bands:  1 - 24
A : A1  (2); A3  (4); A1  (2); A1  (2); A1  (2); A1  (2); A3  (4); A3  (4);
    A1  (2); [24] ;
GM: GM2-GM3+(2); GM5+GM6-(4); GM1+(1); GM4-(1); GM3+(1); GM2-(1); GM1+(1);
    GM4-(1); GM3+(1); GM5+(2); GM6-(2); GM6+(2); GM5-(2); GM2-(1); GM1+(1);
    GM4-(1); [24] ;
H : H1  (2); H2  (2); H3  (2); H1  (2); H2  (2); H1  (2); H3  (2); H2  (2);
    H1  (2); H3  (2); H2  (2); H2  (2); [24] ;
K : K5  (2); K5  (2); K2  K3  (2); K5  (2); K6  (2); K6  (2); K1  (1); K4  (1);
    K5  (2); K5  (2); K2  (1); K3  (1); K6  (2); K5  (2); [24] ;
L : L1  (2); L1  (2); L2  (2); L1  (2); L1  (2); L1  (2); L1  (2); L1  (2);
    L2  (2); L1  (2); L2  (2); L1  (2); [24] ;
M : M1+ M4- (2); M2- M3+ (2); M2+ M3- (2); M4- (1); M1+ (1); M2- (1); M3+ (1);
    M4- (1); M1+ (1); M3+ (1); M2- (1); M4- (1); M1+ (1); M2- (1); M1- (1);
    M4+ (1); M3+ (1); M3- (1); M2+ (1); M4- (1); M1+ (1); [24];
\end{lstlisting}
\hyperref[tab:electride]{Back to the table}

\subsubsection*{26287 NaNbSe$_{2}$}
\label{sec:tqc26287}
\lstset{language=bash, keywordstyle=\color{blue!70}, basicstyle=\ttfamily, frame=shadowbox}
\begin{lstlisting}
Computed bands:  1 - 24
A : A1  (2); A3  (4); A1  (2); A1  (2); A1  (2); A1  (2); A3  (4); A3  (4);
    A1  (2); [24] ;
GM: GM2-GM3+(2); GM5+GM6-(4); GM1+(1); GM4-(1); GM3+(1); GM2-(1); GM1+(1);
    GM4-(1); GM3+(1); GM2-(1); GM5+(2); GM6-(2); GM6+(2); GM5-(2); GM1+(1);
    GM4-(1); [24] ;
H : H1  (2); H2  (2); H3  (2); H1  (2); H2  (2); H1  (2); H3  (2); H2  (2);
    H1  (2); H3  (2); H2  (2); H2  (2); [24] ;
K : K5  (2); K5  (2); K2  K3  (2); K5  (2); K6  (2); K6  (2); K1  (1); K4  (1);
    K5  (2); K5  (2); K2  (1); K3  (1); K6  (2); K5  (2); [24] ;
L : L1  (2); L1  (2); L2  (2); L1  (2); L1  (2); L1  (2); L1  (2); L1  (2);
    L2  (2); L1  (2); L2  (2); L1  (2); [24] ;
M : M1+ M4- (2); M2- M3+ (2); M2+ M3- (2); M4- (1); M1+ (1); M2- (1); M3+ (1);
    M4- (1); M1+ (1); M3+ (1); M2- (1); M1+ M4- (2); M2- (1); M1- (1); M4+ (1);
    M3+ (1); M3- (1); M2+ (1); M4- (1); M1+ (1); [24];
\end{lstlisting}
\hyperref[tab:electride]{Back to the table}

\subsubsection*{424240 BW}
\label{sec:tqc424240}
\lstset{language=bash, keywordstyle=\color{blue!70}, basicstyle=\ttfamily, frame=shadowbox}
\begin{lstlisting}
Computed bands:  1 - 18
GM: GM1+(1); GM2+(1); GM4-(1); GM3-(1); GM1+(1); GM2+(1); GM4-(1); GM5+(2);
    GM5-(2); GM1+(1); GM3-(1); GM2+(1); GM5+(2); GM1-(1); GM4+(1); [18] ;
M : M1  (2); M1  (2); M1  (2); M1  (2); M3  (2); M3  (2); M1  (2); M4  (2);
    M2  (2); [18] ;
P : P1  (2); P2  (2); P1  (2); P2  (2); P1  (2); P2  (2); P2  (2); P1  (2);
    P2  (2); [18] ;
X : X2  (2); X1  (2); X2  (2); X1  (2); X1  (2); X2  (2); X1  (2); X1  (2);
    X2  (2); [18] ;
N : N2- (1); N1+ (1); N2- (1); N1+ (1); N2- (1); N2- (1); N1+ (1); N1- (1);
    N1+ (1); N2+ (1); N1- (1); N2- (1); N2- (1); N1+ (1); N2+ (1); N1+ (1);
    N1+ (1); N2+ (1); [18];
\end{lstlisting}
\hyperref[tab:electride]{Back to the table}

\subsubsection*{154596 Nb$_{5}$Sb$_{4}$}
\label{sec:tqc154596}
\noindent Essential BR: $Ag@2b$ \\
\noindent RSI:
\begin{flalign*}
&\delta_{1}@2b\equiv -m(Ag)+m(Au)-m(2Eg)+m(2Eu) = -1,&
\\
&\delta_{2}@2b\equiv m(Ag)-m(Au)-m(Bg)+m(Bu) = 1,&
\end{flalign*}
\lstset{language=bash, keywordstyle=\color{blue!70}, basicstyle=\ttfamily, frame=shadowbox}
\begin{lstlisting}
Computed bands:  1 - 38
GM: GM3-GM4-(2); GM2+(1); GM1+(1); GM1-(1); GM3-GM4-(2); GM1+(1); GM3+GM4+(2);
    GM2+GM2-(2); GM3-GM4-(2); GM1-(1); GM1+(1); GM3-GM4-(2); GM2+(1); GM2+(1);
    GM1+(1); GM3-GM4-(2); GM1+(1); GM2+(1); GM2-(1); GM3-GM4-(2); GM3+GM4+(2);
    GM1+(1); GM2+(1); GM1+(1); GM3-GM4-(2); GM3+GM4+(2); GM1-(1); [38] ;
M : M3- M4- (2); M2+ (1); M1- (1); M3- M4- (2); M1+ (1); M2- (1); M3+ M4+ (2);
    M1+ (1); M2+ (1); M3- M4- (2); M1- (1); M2+ (1); M3- M4- (2); M1+ (1);
    M1+ (1); M2+ (1); M2+ (1); M3- M4- (2); M1+ (1); M3- M4- (2); M1+ (1);
    M3+ M4+ (2); M2- (1); M2+ (1); M3- M4- (2); M1- (1); M3+ M4+ (2); M1+ (1);
    [38] ;
P : P3  P4  (2); P2  (1); P1  (1); P1  (1); P3  P4  (2); P2  (1); P2  (1);
    P3  P4  (2); P1  (1); P3  P4  (2); P2  (1); P2  (1); P3  P4  (2); P1  (1);
    P2  (1); P3  P4  (2); P1  (1); P1  (1); P3  P4  (2); P2  (1); P1  (1);
    P3  P4  (2); P2  (1); P2  (1); P1  (1); P3  P4  (2); P3  P4  (2); P2  (1);
    [38] ;
X : X2- (1); X2- (1); X1+ (1); X2- (1); X1- (1); X1+ (1); X2- (1); X1+ (1);
    X2+ (1); X1- (1); X2+ (1); X1+ (1); X2- (1); X2- (1); X1- (1); X2- (1);
    X1+ (1); X1+ (1); X2- (1); X2- (1); X1+ (1); X1+ (1); X1+ (1); X2- (1);
    X1+ (1); X2- (1); X1- (1); X2+ (1); X1+ (1); X2- (1); X1+ (1); X2+ (1);
    X2- (1); X2+ (1); X2- (1); X1+ (1); X1- (1); X2+ (1); [38] ;
N : N1- (1); N1- (1); N1- (1); N1- (1); N1+ (1); N1+ (1); N1+ (1); N1+ (1);
    N1- (1); N1+ (1); N1+ (1); N1- (1); N1- (1); N1- (1); N1- (1); N1+ (1);
    N1- (1); N1- (1); N1+ (1); N1- (1); N1+ (1); N1+ (1); N1- (1); N1+ (1);
    N1- (1); N1+ (1); N1+ (1); N1- (1); N1+ (1); N1- (1); N1+ (1); N1- (1);
    N1+ (1); N1+ (1); N1- (1); N1- (1); N1+ (1); N1- (1); [38];
\end{lstlisting}
\hyperref[tab:electride]{Back to the table}

\subsubsection*{81306 Na(CuS)$_{4}$}
\label{sec:tqc81306}
\noindent Essential BR: $A1g@1a$ \\
\noindent RSI:
\begin{flalign*}
&\delta_{1}@1a\equiv -m(Eg)+m(Eu) = 0,&
\\
&\delta_{2}@1a\equiv -m(A1g)+m(A1u)-m(A2g)+m(A2u) = -1,&
\end{flalign*}
\lstset{language=bash, keywordstyle=\color{blue!70}, basicstyle=\ttfamily, frame=shadowbox}
\begin{lstlisting}
Computed bands:  1 - 35
A : A1+ (1); A2- (1); A1+ (1); A2- (1); A1+ (1); A2- (1); A3- (2); A3- (2);
    A3+ (2); A1+ (1); A3+ (2); A3+ (2); A3- (2); A2- (1); A2- (1); A3- (2);
    A1+ (1); A3+ (2); A1+ (1); A3- (2); A3+ (2); A3- (2); A3+ (2); [35] ;
GM: GM1+(1); GM1+(1); GM2-(1); GM2-(1); GM1+(1); GM1+(1); GM3-(2); GM3+(2);
    GM3-(2); GM2-(1); GM3+(2); GM3+(2); GM3-(2); GM2-(1); GM1+(1); GM3-(2);
    GM3+(2); GM2-(1); GM1+(1); GM3-(2); GM3+(2); GM3-(2); GM3+(2); [35] ;
H : H1  (1); H3  (2); H2  (1); H1  (1); H3  (2); H3  (2); H2  (1); H3  (2);
    H1  (1); H3  (2); H2  (1); H3  (2); H1  (1); H2  (1); H1  (1); H3  (2);
    H3  (2); H2  (1); H1  (1); H3  (2); H3  (2); H2  (1); H1  (1); H3  (2);
    [35] ;
K : K1  (1); K3  (2); K2  (1); K1  (1); K3  (2); K3  (2); K2  (1); K1  (1);
    K3  (2); K3  (2); K2  (1); K1  (1); K3  (2); K2  (1); K1  (1); K3  (2);
    K3  (2); K2  (1); K1  K3  (3); K3  (2); K2  (1); K1  (1); K3  (2); [35] ;
L : L1+ (1); L1+ (1); L2- (1); L2- (1); L2- (1); L1+ (1); L2- (1); L1+ (1);
    L1+ (1); L1- (1); L2- (1); L1+ (1); L2+ (1); L1- (1); L2+ (1); L2- (1);
    L1- (1); L1+ (1); L2+ (1); L2- (1); L1+ L2- (2); L1- (1); L1+ (1); L2+ (1);
    L2- (1); L1+ (1); L2- (1); L1- (1); L1+ (1); L2+ (1); L2- (1); L1- (1);
    L1+ (1); L2+ (1); [35] ;
M : M1+ (1); M2- (1); M1+ (1); M2- (1); M2- (1); M1+ (1); M2- (1); M1+ (1);
    M1- (1); M1+ (1); M2- (1); M1- (1); M2+ (1); M1+ (1); M2+ (1); M2- (1);
    M1- (1); M1+ (1); M2+ (1); M2- (1); M1+ (1); M2- (1); M1- (1); M1+ (1);
    M2+ (1); M2- (1); M2- (1); M1+ (1); M1- (1); M1+ (1); M2+ (1); M2- (1);
    M1- (1); M1+ (1); M2+ (1); [35];
\end{lstlisting}
\hyperref[tab:electride]{Back to the table}

\subsubsection*{280189 Cs$_{2}$PdI$_{6}$}
\label{sec:tqc280189}
\lstset{language=bash, keywordstyle=\color{blue!70}, basicstyle=\ttfamily, frame=shadowbox}
\begin{lstlisting}
Computed bands:  1 - 35
GM: GM1+(1); GM4-(1); GM1+(1); GM1+(1); GM5-(2); GM4+(1); GM3-(1); GM2+(1);
    GM5-(2); GM5+(2); GM3-(1); GM1+(1); GM4+(1); GM1+(1); GM2+(1); GM5-(2);
    GM5+(2); GM3-(1); GM5-(2); GM5+(2); GM2+(1); GM1+(1); GM2-(1); GM5-(2);
    GM3+(1); GM5+(2); [35] ;
M : M2+ (1); M3- (1); M1+ (1); M1+ (1); M5- (2); M4+ (1); M3- (1); M1+ (1);
    M5+ (2); M5- (2); M4- (1); M1+ (1); M1+ M4+ (2); M2+ (1); M5- (2); M5+ (2);
    M3- (1); M5- (2); M2- (1); M2+ (1); M5+ (2); M5- (2); M1+ (1); M3+ (1);
    M5+ (2); [35] ;
P : P5  (2); P3  (1); P1  (1); P5  (2); P3  (1); P1  (1); P5  (2); P2  (1);
    P3  P4  (2); P1  (1); P3  (1); P1  (1); P3  (1); P2  (1); P5  (2); P5  (2);
    P1  (1); P5  (2); P3  (1); P5  (2); P1  (1); P2  (1); P5  (2); P4  (1);
    P5  (2); [35] ;
X : X4- (1); X4+ (1); X1+ (1); X1+ (1); X4- (1); X3- (1); X1+ (1); X2- (1);
    X3- (1); X2+ (1); X2- (1); X1- (1); X1+ (1); X3+ (1); X1+ (1); X1+ (1);
    X1+ (1); X2+ (1); X3- (1); X4+ (1); X4- (1); X3+ (1); X4- (1); X2- (1);
    X3- (1); X4+ (1); X2+ (1); X2- (1); X4- (1); X3+ (1); X1+ (1); X2+ (1);
    X3- (1); X3+ (1); X4+ (1); [35] ;
N : N1+ N2- (2); N2- (1); N1+ (1); N1- (1); N2- (1); N2+ (1); N1+ (1); N1+ (1);
    N2- (1); N2+ (1); N2- (1); N1- (1); N1+ (1); N2- (1); N1+ (1); N2+ (1);
    N1+ (1); N2+ (1); N1+ (1); N2- (1); N1- (1); N1+ (1); N2+ (1); N1+ (1);
    N2- (1); N1- (1); N2- (1); N1+ (1); N1- (1); N2+ (1); N2- (1); N1+ (1);
    N1- (1); N2+ (1); [35];
\end{lstlisting}
\hyperref[tab:electride]{Back to the table}

\subsubsection*{626798 CrSi$_{2}$}
\label{sec:tqc626798}
\lstset{language=bash, keywordstyle=\color{blue!70}, basicstyle=\ttfamily, frame=shadowbox}
\begin{lstlisting}
Computed bands:  1 - 21
A : A5  (2); A2  (1); A1  (1); A5  (2); A3  (1); A1  (1); A5  (2); A6  (2);
    A2  (1); A6  (2); A4  (1); A6  (2); A4  (1); A5  (2); [21] ;
GM: GM1 (1); GM5 (2); GM1 (1); GM5 (2); GM6 (2); GM5 (2); GM4 (1); GM2 (1);
    GM5 (2); GM6 (2); GM3 (1); GM6 (2); GM4 (1); GM1 (1); [21] ;
H : H3  (2); H1  (1); H3  (2); H2  (1); H1  (1); H3  (2); H2  (1); H3  (2);
    H3  (2); H1  (1); H2  (1); H3  (2); H3  (2); H2  (1); [21] ;
K : K1  (1); K3  (2); K3  (2); K2  (1); K1  (1); K3  (2); K1  (1); K3  (2);
    K3  (2); K2  (1); K3  (2); K1  (1); K3  (2); K2  (1); [21] ;
L : L2  (1); L4  (1); L1  (1); L3  (1); L4  (1); L1  (1); L2  (1); L3  (1);
    L2  (1); L4  (1); L1  (1); L3  (1); L2  (1); L3  (1); L4  (1); L3  (1);
    L2  (1); L1  (1); L4  (1); L1  (1); L3  (1); [21] ;
M : M1  (1); M4  (1); M3  (1); M1  (1); M2  (1); M4  (1); M1  (1); M3  (1);
    M3  (1); M2  (1); M1  (1); M4  (1); M3  (1); M1  (1); M4  (1); M2  (1);
    M3  (1); M4  (1); M1  (1); M2  (1); M4  (1); [21];
\end{lstlisting}
\hyperref[tab:electride]{Back to the table}

\subsubsection*{240481 Cs$_{2}$Pd(IBr$_{2}$)$_{2}$}
\label{sec:tqc240481}
\lstset{language=bash, keywordstyle=\color{blue!70}, basicstyle=\ttfamily, frame=shadowbox}
\begin{lstlisting}
Computed bands:  1 - 35
GM: GM1+(1); GM4-(1); GM1+(1); GM5-(2); GM4+(1); GM1+(1); GM3-(1); GM2+(1);
    GM5-(2); GM5+(2); GM3-(1); GM1+(1); GM4+(1); GM1+(1); GM5-(2); GM2+(1);
    GM5+(2); GM3-(1); GM5-(2); GM2-(1); GM5+(2); GM3+(1); GM2+(1); GM5-(2);
    GM1+(1); GM5+(2); [35] ;
M : M2+ (1); M3- (1); M1+ (1); M5- (2); M4+ (1); M1+ (1); M3- (1); M1+ (1);
    M5+ (2); M5- (2); M4- (1); M1+ (1); M4+ (1); M1+ (1); M2+ (1); M5- (2);
    M5+ (2); M3- (1); M2- (1); M5- (2); M2+ (1); M5+ (2); M5- (2); M3+ (1);
    M1+ (1); M5+ (2); [35] ;
P : P5  (2); P1  (1); P3  (1); P5  (2); P3  (1); P1  (1); P2  (1); P5  (2);
    P3  (1); P4  (1); P1  (1); P3  (1); P1  (1); P5  (2); P3  (1); P2  (1);
    P5  (2); P1  (1); P5  (2); P3  (1); P5  (2); P1  (1); P4  (1); P2  (1);
    P5  (2); P5  (2); [35] ;
X : X4- (1); X4+ (1); X1+ (1); X1+ (1); X4- (1); X3- (1); X1+ (1); X2- (1);
    X3- (1); X2+ (1); X2- (1); X1- (1); X1+ (1); X3+ (1); X1+ (1); X1+ (1);
    X1+ (1); X3- (1); X4+ (1); X2+ (1); X4- (1); X3+ (1); X4- (1); X2- (1);
    X3- (1); X2- (1); X4+ (1); X2+ (1); X3+ (1); X4- (1); X2+ (1); X3+ (1);
    X1+ X3- (2); X4+ (1); [35] ;
N : N1+ N2- (2); N1+ (1); N2- (1); N1- (1); N2+ (1); N2- (1); N1+ (1); N1+ (1);
    N2- (1); N2+ (1); N1- (1); N2- (1); N1+ (1); N1+ (1); N2+ (1); N2- (1);
    N1+ (1); N1+ (1); N2+ (1); N1- (1); N2- (1); N1+ (1); N2- (1); N1- (1);
    N2+ (1); N2- (1); N1+ (1); N1- (1); N2+ (1); N1+ (1); N2+ (1); N1+ (1);
    N2- (1); N1- (1); [35];
\end{lstlisting}
\hyperref[tab:electride]{Back to the table}

\subsubsection*{639879 Li$_{5}$In$_{4}$}
\label{sec:tqc639879}
\noindent Essential BR: $A1g@1a$ \\
\noindent RSI:
\begin{flalign*}
&\delta_{1}@1a\equiv -m(Eg)+m(Eu) = 0,&
\\
&\delta_{2}@1a\equiv -m(A1g)+m(A1u)-m(A2g)+m(A2u) = -1,&
\end{flalign*}
\lstset{language=bash, keywordstyle=\color{blue!70}, basicstyle=\ttfamily, frame=shadowbox}
\begin{lstlisting}
Computed bands:  1 -  9
A : A1+ (1); A2- (1); A1+ (1); A2- (1); A3- (2); A3+ (2); A1+ (1); [9]  ;
GM: GM1+(1); GM2-(1); GM1+(1); GM2-(1); GM3-(2); GM1+(1); GM3+(2); [9]  ;
H : H1  (1); H3  (2); H2  (1); H3  (2); H1  (1); H3  (2); [9]  ;
K : K1  (1); K3  (2); K2  (1); K3  (2); K1  (1); K3  (2); [9]  ;
L : L1+ (1); L2- (1); L1+ (1); L2- (1); L1+ (1); L1- (1); L2- (1); L2+ (1);
    L1+ (1); [9]  ;
M : M1+ (1); M2- (1); M1+ (1); M2- (1); M1+ (1); M2- (1); M1- (1); M2+ (1);
    M1+ (1); [9] ;
\end{lstlisting}
\hyperref[tab:electride]{Back to the table}

\subsubsection*{182116 Si$_{2}$Mo}
\label{sec:tqc182116}
\lstset{language=bash, keywordstyle=\color{blue!70}, basicstyle=\ttfamily, frame=shadowbox}
\begin{lstlisting}
Computed bands:  1 - 21
A : A5  (2); A1  (1); A2  (1); A5  (2); A4  (1); A5  (2); A6  (2); A2  (1);
    A3  (1); A6  (2); A1  (1); A6  (2); A3  (1); A5  (2); [21] ;
GM: GM1 (1); GM5 (2); GM1 (1); GM5 (2); GM6 (2); GM4 (1); GM5 (2); GM2 (1);
    GM5 (2); GM6 (2); GM6 (2); GM3 (1); GM4 (1); GM1 (1); [21] ;
H : H3  (2); H2  (1); H3  (2); H1  (1); H2  (1); H3  (2); H1  (1); H3  (2);
    H3  (2); H2  (1); H1  (1); H3  (2); H3  (2); H1  (1); [21] ;
K : K1  (1); K3  (2); K3  (2); K1  (1); K2  (1); K3  (2); K1  (1); K3  (2);
    K2  (1); K3  (2); K3  (2); K1  (1); K3  (2); K2  (1); [21] ;
L : L1  (1); L2  (1); L3  (1); L4  (1); L3  (1); L2  (1); L1  (1); L4  (1);
    L1  (1); L3  (1); L2  (1); L4  (1); L4  (1); L1  (1); L1  (1); L4  (1);
    L3  (1); L2  (1); L3  (1); L2  (1); L4  (1); [21] ;
M : M1  (1); M4  (1); M3  (1); M1  (1); M2  (1); M4  (1); M1  (1); M3  (1);
    M1  (1); M4  (1); M3  (1); M2  (1); M3  (1); M1  (1); M4  (1); M2  (1);
    M3  (1); M4  (1); M1  (1); M2  (1); M4  (1); [21];
\end{lstlisting}
\hyperref[tab:electride]{Back to the table}

\subsubsection*{96026 CrSi$_{2}$}
\label{sec:tqc96026}
\lstset{language=bash, keywordstyle=\color{blue!70}, basicstyle=\ttfamily, frame=shadowbox}
\begin{lstlisting}
Computed bands:  1 - 21
A : A5  (2); A1  (1); A2  (1); A5  (2); A4  (1); A2  (1); A5  (2); A6  (2);
    A1  (1); A6  (2); A3  (1); A6  (2); A3  (1); A5  (2); [21] ;
GM: GM1 (1); GM5 (2); GM1 (1); GM5 (2); GM6 (2); GM5 (2); GM4 (1); GM2 (1);
    GM5 (2); GM6 (2); GM3 (1); GM6 (2); GM4 (1); GM1 (1); [21] ;
H : H3  (2); H2  (1); H3  (2); H1  (1); H2  (1); H3  (2); H1  (1); H3  (2);
    H3  (2); H2  (1); H1  (1); H3  (2); H3  (2); H1  (1); [21] ;
K : K1  (1); K3  (2); K3  (2); K2  (1); K1  (1); K3  (2); K1  (1); K3  (2);
    K3  (2); K2  (1); K3  (2); K1  (1); K3  (2); K2  (1); [21] ;
L : L2  (1); L4  (1); L1  (1); L3  (1); L4  (1); L1  (1); L2  (1); L3  (1);
    L2  (1); L4  (1); L1  (1); L3  (1); L2  (1); L3  (1); L4  (1); L3  (1);
    L2  (1); L1  (1); L4  (1); L1  (1); L3  (1); [21] ;
M : M1  (1); M4  (1); M3  (1); M1  (1); M2  (1); M4  (1); M1  (1); M3  (1);
    M3  (1); M2  (1); M1  (1); M4  (1); M3  (1); M1  (1); M4  (1); M2  (1);
    M3  (1); M4  (1); M1  (1); M2  (1); M4  (1); [21];
\end{lstlisting}
\hyperref[tab:electride]{Back to the table}

\subsubsection*{248351 Sr(RuO$_{3}$)$_{2}$}
\label{sec:tqc248351}
\lstset{language=bash, keywordstyle=\color{blue!70}, basicstyle=\ttfamily, frame=shadowbox}
\begin{lstlisting}
Computed bands:  1 - 31
A : A1+ (1); A2- (1); A3+ (2); A3- (2); A1+ (1); A3- (2); A2- (1); A3- (2);
    A2- (1); A3+ (2); A2+ (1); A3- (2); A3+ (2); A1- (1); A2- (1); A1+ (1);
    A3+ (2); A1+ (1); A3- (2); A2+ (1); A3- (2); [31] ;
GM: GM1+(1); GM1+(1); GM3-(2); GM3+(2); GM2-(1); GM3-(2); GM2-(1); GM1+(1);
    GM3+(2); GM1-(1); GM3+(2); GM3-(2); GM3-(2); GM2+(1); GM1+(1); GM3+(2);
    GM2-(1); GM3-(2); GM2-(1); GM1-(1); GM3+(2); [31] ;
H : H1  (1); H3  (2); H1  (1); H3  (2); H1  (1); H3  (2); H1  (1); H1  (1);
    H2  (1); H3  (2); H3  (2); H3  (2); H1  (1); H3  (2); H1  (1); H3  (2);
    H2  (1); H3  (2); H1  (1); H2  (1); H3  (2); [31] ;
K : K1  (1); K3  (2); K1  (1); K3  (2); K1  (1); K1  (1); K3  (2); K1  (1);
    K2  (1); K3  (2); K3  (2); K3  (2); K1  (1); K3  (2); K1  (1); K3  (2);
    K2  (1); K1  (1); K3  (2); K2  (1); K3  (2); [31] ;
L : L1+ (1); L2+ (1); L2- (1); L2- (1); L1+ (1); L1- (1); L1+ (1); L2- (1);
    L2- (1); L1- (1); L2+ (1); L1+ (1); L2- (1); L1- (1); L2+ (1); L2- (1);
    L1+ (1); L2- (1); L2+ (1); L1- (1); L1+ (1); L2- (1); L2+ (1); L1+ (1);
    L1- (1); L2- (1); L1- (1); L1+ (1); L2+ (1); L1- (1); L2- (1); [31] ;
M : M1+ (1); M1- (1); M1+ (1); M2- (1); M1+ (1); M2- (1); M2+ (1); M1- (1);
    M2- (1); M2- (1); M1- (1); M1+ (1); M2+ (1); M2- (1); M1+ (1); M1- (1);
    M2- (1); M1+ (1); M1- (1); M2+ (1); M1+ (1); M2- (1); M1- (1); M2- (1);
    M2+ (1); M1+ (1); M2+ (1); M2- (1); M1- (1); M1+ (1); M2+ (1); [31];
\end{lstlisting}
\hyperref[tab:electride]{Back to the table}

\subsubsection*{652549 Si$_{2}$W}
\label{sec:tqc652549}
\lstset{language=bash, keywordstyle=\color{blue!70}, basicstyle=\ttfamily, frame=shadowbox}
\begin{lstlisting}
Computed bands:  1 - 21
A : A5  (2); A2  (1); A1  (1); A3  (1); A5  (2); A5  (2); A6  (2); A1  (1);
    A6  (2); A4  (1); A2  (1); A6  (2); A4  (1); A5  (2); [21] ;
GM: GM1 (1); GM5 (2); GM1 (1); GM5 (2); GM6 (2); GM4 (1); GM5 (2); GM2 (1);
    GM5 (2); GM6 (2); GM6 (2); GM3 (1); GM4 (1); GM1 (1); [21] ;
H : H3  (2); H1  (1); H3  (2); H2  (1); H1  (1); H3  (2); H2  (1); H3  (2);
    H3  (2); H1  (1); H2  (1); H3  (2); H3  (2); H2  (1); [21] ;
K : K1  (1); K3  (2); K3  (2); K1  (1); K2  (1); K3  (2); K1  (1); K3  (2);
    K2  (1); K3  (2); K3  (2); K1  (1); K3  (2); K2  (1); [21] ;
L : L2  (1); L1  (1); L4  (1); L3  (1); L4  (1); L1  (1); L2  (1); L3  (1);
    L2  (1); L4  (1); L1  (1); L3  (1); L3  (1); L2  (1); L2  (1); L3  (1);
    L4  (1); L1  (1); L4  (1); L1  (1); L3  (1); [21] ;
M : M1  (1); M4  (1); M1  (1); M3  (1); M2  (1); M4  (1); M1  (1); M3  (1);
    M4  (1); M1  (1); M3  (1); M2  (1); M3  (1); M4  (1); M1  (1); M2  (1);
    M3  (1); M4  (1); M1  (1); M2  (1); M4  (1); [21];
\end{lstlisting}
\hyperref[tab:electride]{Back to the table}

\subsubsection*{181078 Ca$_{6}$Ge$_{2}$O}
\label{sec:tqc181078}
\noindent Essential BR: $A1g@4b$ \\
\noindent RSI:
\begin{flalign*}
&\delta_{1}@4b\equiv -m(A1g)+m(A1u)-m(T1g)+m(T1u) = -1,&
\\
&\delta_{2}@4b\equiv -m(A2g)+m(A2u)-m(T2g)+m(T2u) = 0,&
\\
&\delta_{3}@4b\equiv m(A1g)-m(A1u)+m(A2g)-m(A2u)-m(Eg)+m(Eu) = 1,&
\end{flalign*}
\lstset{language=bash, keywordstyle=\color{blue!70}, basicstyle=\ttfamily, frame=shadowbox}
\begin{lstlisting}
Computed bands:  1 - 37
GM: GM1+(1); GM4-(3); GM3+(2); GM1+(1); GM4-(3); GM5+(3); GM5-(3); GM4+(3);
    GM3+GM4-(5); GM1+(1); GM1+(1); GM2-(1); GM4-(3); GM5+(3); GM1+(1); GM4-(3);
    [37] ;
X : X1+ (1); X3- X5- (3); X1+ (1); X2+ (1); X1+ (1); X4+ (1); X5- (2); X3- (1);
    X1+ X5+ (3); X5- (2); X3- (1); X5+ (2); X4- (1); X5- (2); X3+ (1); X2+ (1);
    X1+ (1); X3- (1); X4+ (1); X5- (2); X3- (1); X1+ (1); X5- (2); X2- (1);
    X5+ (2); X1+ (1); [37] ;
L : L1+ (1); L2- (1); L3- (2); L3+ (2); L1+ (1); L2- (1); L3- (2); L1+ L3+ (3);
    L2- (1); L3- (2); L3+ (2); L1- (1); L2+ (1); L3- (2); L3+ (2); L1+ (1);
    L2- (1); L1+ (1); L2- (1); L3- (2); L2- (1); L1+ (1); L3+ (2); L3- (2);
    L2- (1); [37] ;
W : W1  (1); W2  (1); W5  (2); W2  (1); W1  (1); W1  (1); W2  (1); W5  (2);
    W5  (2); W1  W4  (2); W5  (2); W2  (1); W3  (1); W5  (2); W5  (2); W2  (1);
    W1  (1); W1  (1); W5  (2); W2  (1); W5  (2); W2  (1); W1  (1); W5  (2);
    W3  (1); W4  (1); W2  (1); [37];
\end{lstlisting}
\hyperref[tab:electride]{Back to the table}

\subsubsection*{68014 Y$_{4}$CI$_{5}$}
\label{sec:tqc68014}
\lstset{language=bash, keywordstyle=\color{blue!70}, basicstyle=\ttfamily, frame=shadowbox}

\hyperref[tab:electride]{Back to the table}

\subsubsection*{639449 Ho$_{3}$Ni$_{2}$}
\label{sec:tqc639449}
\noindent Essential BR: $Ag@2c$ \\
\noindent RSI:
\begin{flalign*}
&\delta_{1}@2c\equiv -m(Ag)+m(Au)-m(Bg)+m(Bu) = -1,&
\end{flalign*}
\lstset{language=bash, keywordstyle=\color{blue!70}, basicstyle=\ttfamily, frame=shadowbox}

\hyperref[tab:electride]{Back to the table}

\subsubsection*{646145 NiPSe$_{3}$}
\label{sec:tqc646145}
\lstset{language=bash, keywordstyle=\color{blue!70}, basicstyle=\ttfamily, frame=shadowbox}

\hyperref[tab:electride]{Back to the table}

\subsubsection*{409382 Cs$_{2}$As$_{3}$}
\label{sec:tqc409382}
\noindent Essential BR: $B3u@4a$ \\
\noindent RSI:
\begin{flalign*}
&\delta_{1}@4a\equiv -m(Ag)+m(Au)-m(B1g)+m(B1u)-m(B3g)+m(B3u)-m(B2g)+m(B2u) = 1,&
\end{flalign*}
\lstset{language=bash, keywordstyle=\color{blue!70}, basicstyle=\ttfamily, frame=shadowbox}
\begin{lstlisting}
Computed bands:  1 - 33
GM: GM1+(1); GM1-(1); GM1+(1); GM4-(1); GM1+(1); GM2-(1); GM3-(1); GM4+(1);
    GM1+(1); GM3+(1); GM2+(1); GM2-(1); GM4-(1); GM3-(1); GM1+(1); GM4-(1);
    GM3+(1); GM2+(1); GM2-(1); GM3-(1); GM4+(1); GM2-(1); GM3-(1); GM1+(1);
    GM4-(1); GM3+(1); GM1+(1); GM4+(1); GM2-(1); GM3-(1); GM2+(1); GM1-(1);
    GM4-(1); [33] ;
Z : Z2- (1); Z2+ (1); Z1+ (1); Z4- (1); Z1+ (1); Z3- (1); Z2- (1); Z1+ (1);
    Z4+ (1); Z4- (1); Z3- (1); Z1+ Z1- (2); Z3+ (1); Z4+ (1); Z2+ (1); Z3+ (1);
    Z2- (1); Z4- (1); Z1+ (1); Z3- (1); Z2- (1); Z3- (1); Z1+ (1); Z4- (1);
    Z2- (1); Z1+ (1); Z4+ (1); Z3+ (1); Z2+ (1); Z3- (1); Z1- (1); Z4- (1);
    [33] ;
H : H3  (1); H4  (1); H1  (1); H3  (1); H1  (1); H4  (1); H1  (1); H1  (1);
    H4  (1); H3  (1); H1  (1); H2  (1); H2  (1); H1  (1); H3  (1); H1  (1);
    H4  (1); H3  (1); H4  (1); H2  (1); H1  (1); H1  (1); H4  (1); H1  (1);
    H3  (1); H4  (1); H2  (1); H1  (1); H3  (1); H4  (1); H1  (1); H3  (1);
    H2  (1); [33] ;
Y : Y4- (1); Y4+ (1); Y1+ (1); Y4- (1); Y1+ (1); Y3- (1); Y2- (1); Y1+ (1);
    Y4+ (1); Y3+ (1); Y2- (1); Y3- (1); Y1- Y2+ (2); Y1+ (1); Y2- (1); Y4- (1);
    Y3+ (1); Y3- (1); Y2+ (1); Y1+ (1); Y2- (1); Y3- (1); Y1+ (1); Y4- (1);
    Y2+ (1); Y3- (1); Y4+ (1); Y1+ Y3+ (2); Y2- (1); Y4- (1); Y1- (1); [33] ;
L : L1+ L1- (2); L1- (1); L1+ (1); L1+ (1); L1- (1); L1- (1); L1+ (1); L1+ (1);
    L1+ (1); L1- (1); L1- (1); L1- (1); L1+ (1); L1- (1); L1+ (1); L1- (1);
    L1+ (1); L1- (1); L1+ (1); L1+ (1); L1- (1); L1- (1); L1+ (1); L1- (1);
    L1+ (1); L1- (1); L1+ (1); L1+ (1); L1- (1); L1+ (1); L1- (1); L1- (1);
    [33] ;
T : T3+ (1); T3- (1); T1+ T4- (2); T1+ (1); T3- (1); T2- (1); T1+ (1); T4+ (1);
    T4- (1); T2- (1); T2+ (1); T1- (1); T1+ (1); T3+ (1); T2- (1); T4- (1);
    T3- (1); T2+ (1); T1+ (1); T4+ (1); T2- (1); T3- (1); T1+ (1); T4- (1);
    T3- (1); T3+ (1); T1+ (1); T4+ (1); T2- (1); T2+ (1); T4- (1); T1- (1); [33];
\end{lstlisting}
\hyperref[tab:electride]{Back to the table}

\subsubsection*{602341 NiPS$_{3}$}
\label{sec:tqc602341}
\lstset{language=bash, keywordstyle=\color{blue!70}, basicstyle=\ttfamily, frame=shadowbox}

\hyperref[tab:electride]{Back to the table}

\subsubsection*{190543 Na$_{3}$Cl$_{2}$}
\label{sec:tqc190543}
\noindent Essential BR: $Ag@1b$ \\
\noindent RSI:
\begin{flalign*}
&\delta_{1}@1b\equiv -m(Ag)+m(Au)-m(2Eg)+m(2Eu) = -1,&
\\
&\delta_{2}@1b\equiv m(Ag)-m(Au)-m(Bg)+m(Bu) = 1,&
\end{flalign*}
\lstset{language=bash, keywordstyle=\color{blue!70}, basicstyle=\ttfamily, frame=shadowbox}
\begin{lstlisting}
Computed bands:  1 - 17
A : A2+ (1); A3- A4- (2); A1+ (1); A2- (1); A3+ A4+ (2); A1- (1); A3- A4- (2);
    A1+ (1); A2+ (1); A3- A4- (2); A2+ (1); A1+ (1); A1- (1); [17] ;
GM: GM1+(1); GM3-GM4-(2); GM2+(1); GM1+(1); GM2+(1); GM3-GM4-(2); GM1+(1);
    GM2+(1); GM3-GM4-(2); GM1-(1); GM3+GM4+(2); GM2-(1); GM1+(1); [17] ;
M : M2+ (1); M3- M4- (2); M1+ (1); M1+ (1); M3- M4- (2); M2+ (1); M3- M4- (2);
    M2+ (1); M1+ (1); M2- (1); M3+ M4+ (2); M1- (1); M1+ (1); [17] ;
Z : Z1+ (1); Z3- Z4- (2); Z2+ (1); Z1- (1); Z3+ Z4+ (2); Z2- (1); Z1+ (1);
    Z2+ (1); Z3- Z4- (2); Z1+ (1); Z2+ (1); Z3- Z4- (2); Z1- (1); [17] ;
R : R2- (1); R1+ (1); R1+ (1); R2- (1); R2+ (1); R1- (1); R1- (1); R2- (1);
    R2+ (1); R1+ (1); R2- (1); R1+ (1); R1+ (1); R2- (1); R1+ (1); R2- (1);
    R1- (1); [17] ;
X : X2- (1); X1+ (1); X1+ (1); X2- (1); X2- (1); X1+ (1); X2- (1); X1+ (1);
    X1+ (1); X2- (1); X1+ (1); X2- (1); X2+ (1); X1- (1); X1- (1); X2+ (1);
    X1+ (1); [17];
\end{lstlisting}
\hyperref[tab:electride]{Back to the table}

\subsubsection*{422525 Ca(GaP)$_{2}$}
\label{sec:tqc422525}
\noindent Essential BR: $A1'@2c$ \\
\noindent RSI:
\begin{flalign*}
&\delta_{1}@2c\equiv m(A1')+m(A2')-m(A2'')-m(A1'')-m(E')+m(E'') = 1,&
\end{flalign*}
\lstset{language=bash, keywordstyle=\color{blue!70}, basicstyle=\ttfamily, frame=shadowbox}
\begin{lstlisting}
Computed bands:  1 - 26
A : A1  (2); A1  (2); A3  (4); A1  (2); A1  (2); A1  (2); A1  (2); A3  (4);
    A3  (4); A1  (2); [26] ;
GM: GM1+GM3+(2); GM2-GM4-(2); GM5-GM6-(4); GM1+(1); GM4-(1); GM3+(1); GM2-(1);
    GM1+(1); GM4-(1); GM3+(1); GM2-(1); GM1+(1); GM5+(2); GM6+(2); GM6-(2);
    GM5-(2); GM4-(1); [26] ;
H : H3  (2); H1  H2  (4); H3  (2); H2  (2); H1  (2); H1  (2); H2  (2); H3  (2);
    H3  (2); H2  (2); H1  (2); H1  (2); [26] ;
K : K1  K2  (2); K5  K6  (4); K3  K4  (2); K5  (2); K6  (2); K5  (2); K6  (2);
    K1  (1); K4  (1); K2  (1); K3  (1); K5  (2); K5  (2); K6  (2); [26] ;
L : L1  (2); L1  (2); L1  (2); L2  (2); L1  (2); L1  (2); L1  (2); L1  (2);
    L1  (2); L2  (2); L1  (2); L2  (2); L1  (2); [26] ;
M : M1+ M3+ (2); M2- M4- (2); M2- M4- (2); M1- M3- (2); M1+ (1); M4- (1);
    M3+ (1); M2- (1); M1+ (1); M4- (1); M3+ (1); M2- (1); M4- (1); M1+ (1);
    M2+ (1); M3- (1); M3+ (1); M1+ (1); M4+ (1); M2- (1); M1- (1); M4- (1); [26];
\end{lstlisting}
\hyperref[tab:electride]{Back to the table}

\subsubsection*{260563 Sr(InP)$_{2}$}
\label{sec:tqc260563}
\noindent Essential BR: $A1'@2d$ \\
\noindent RSI:
\begin{flalign*}
&\delta_{1}@2d\equiv m(A1')+m(A2')-m(A2'')-m(A1'')-m(E')+m(E'') = 1,&
\end{flalign*}
\lstset{language=bash, keywordstyle=\color{blue!70}, basicstyle=\ttfamily, frame=shadowbox}
\begin{lstlisting}
Computed bands:  1 - 26
A : A1  (2); A1  (2); A3  (4); A1  (2); A1  (2); A1  (2); A1  (2); A3  (4);
    A3  (4); A1  (2); [26] ;
GM: GM1+GM3+(2); GM2-GM4-(2); GM5-GM6-(4); GM1+(1); GM3+(1); GM4-(1); GM2-(1);
    GM1+(1); GM4-(1); GM3+(1); GM2-(1); GM1+(1); GM5+(2); GM6+(2); GM6-(2);
    GM5-(2); GM4-(1); [26] ;
H : H3  (2); H1  H2  (4); H3  (2); H1  (2); H2  (2); H2  (2); H1  (2); H3  (2);
    H3  (2); H1  (2); H2  (2); H2  (2); [26] ;
K : K1  K2  (2); K5  K6  (4); K3  K4  (2); K5  (2); K6  (2); K5  (2); K6  (2);
    K1  (1); K2  (1); K4  (1); K3  (1); K5  (2); K6  (2); K5  (2); [26] ;
L : L1  (2); L1  (2); L1  (2); L2  (2); L1  (2); L1  (2); L1  (2); L1  (2);
    L1  (2); L2  (2); L1  (2); L2  (2); L1  (2); [26] ;
M : M1+ M3+ (2); M2- M4- (2); M2- M4- (2); M1- M3- (2); M1+ (1); M3+ (1);
    M4- (1); M2- (1); M1+ (1); M4- (1); M3+ (1); M2- (1); M1+ (1); M4- (1);
    M3+ (1); M2+ (1); M4+ (1); M1+ (1); M3- (1); M2- (1); M1- (1); M4- (1); [26];
\end{lstlisting}
\hyperref[tab:electride]{Back to the table}

\subsubsection*{409381 Rb$_{2}$As$_{3}$}
\label{sec:tqc409381}
\noindent Essential BR: $B3u@4a$ \\
\noindent RSI:
\begin{flalign*}
&\delta_{1}@4a\equiv -m(Ag)+m(Au)-m(B1g)+m(B1u)-m(B3g)+m(B3u)-m(B2g)+m(B2u) = 1,&
\end{flalign*}
\lstset{language=bash, keywordstyle=\color{blue!70}, basicstyle=\ttfamily, frame=shadowbox}
\begin{lstlisting}
Computed bands:  1 - 33
GM: GM1+(1); GM1-(1); GM1+(1); GM4-(1); GM1+(1); GM2-(1); GM3-(1); GM3+(1);
    GM4+(1); GM2+(1); GM4-(1); GM3-(1); GM2-(1); GM1+(1); GM2+(1); GM4-(1);
    GM3+(1); GM3-(1); GM2-(1); GM1+(1); GM4+(1); GM2-(1); GM3-(1); GM1+(1);
    GM4-(1); GM3+(1); GM2+(1); GM1+(1); GM4+(1); GM2-(1); GM3-(1); GM1-(1);
    GM4-(1); [33] ;
Z : Z2- (1); Z2+ (1); Z1+ (1); Z4- (1); Z1+ (1); Z2- (1); Z3- (1); Z4- (1);
    Z3+ (1); Z1- Z4+ (2); Z1+ (1); Z3- (1); Z2+ (1); Z3+ (1); Z1+ (1); Z4- (1);
    Z3- (1); Z2- (1); Z1+ (1); Z4+ (1); Z2- (1); Z3- (1); Z1+ (1); Z2- (1);
    Z4- (1); Z1+ (1); Z2+ (1); Z3+ (1); Z4+ (1); Z3- (1); Z1- (1); Z4- (1);
    [33] ;
H : H3  (1); H4  (1); H3  (1); H1  (1); H1  (1); H4  (1); H1  (1); H1  (1);
    H3  (1); H2  (1); H2  (1); H4  (1); H1  (1); H3  (1); H3  (1); H1  H2  (2);
    H1  (1); H4  (1); H4  (1); H1  (1); H1  (1); H4  (1); H1  (1); H3  (1);
    H3  (1); H4  (1); H1  (1); H1  (1); H2  (1); H4  (1); H3  (1); H2  (1);
    [33] ;
Y : Y4- (1); Y4+ (1); Y4- (1); Y1+ (1); Y1+ (1); Y3- (1); Y2- (1); Y3+ (1);
    Y1+ (1); Y1- (1); Y2+ (1); Y2- (1); Y3- (1); Y3+ (1); Y4- (1); Y2- (1);
    Y2+ (1); Y1+ (1); Y3- (1); Y4+ (1); Y1+ (1); Y2- (1); Y3- (1); Y1+ (1);
    Y4- (1); Y3+ (1); Y2+ (1); Y3- (1); Y4+ (1); Y1+ (1); Y2- (1); Y4- (1);
    Y1- (1); [33] ;
L : L1+ L1- (2); L1+ L1- (2); L1+ (1); L1- (1); L1- (1); L1+ (1); L1+ (1);
    L1- (1); L1+ (1); L1- (1); L1- (1); L1+ (1); L1- (1); L1+ (1); L1- (1);
    L1+ (1); L1- (1); L1+ (1); L1+ (1); L1- (1); L1- (1); L1+ (1); L1- (1);
    L1- (1); L1+ (1); L1+ (1); L1+ (1); L1+ (1); L1- (1); L1- (1); L1- (1);
    [33] ;
T : T3+ (1); T3- (1); T4- (1); T1+ (1); T1+ (1); T2- (1); T3- (1); T1+ (1);
    T4+ (1); T4- (1); T2+ (1); T1- (1); T2- (1); T3+ (1); T4- (1); T2+ (1);
    T2- (1); T1+ (1); T3- (1); T1+ (1); T4+ (1); T2- (1); T3- (1); T1+ (1);
    T4- (1); T3+ (1); T2- (1); T1+ (1); T3- (1); T4+ (1); T2+ (1); T4- (1);
    T1- (1); [33];
\end{lstlisting}
\hyperref[tab:electride]{Back to the table}

\subsubsection*{2334 Dy$_{3}$Ni$_{2}$}
\label{sec:tqc2334}
\noindent Essential BR: $Ag@2c$ \\
\noindent RSI:
\begin{flalign*}
&\delta_{1}@2c\equiv -m(Ag)+m(Au)-m(Bg)+m(Bu) = -1,&
\end{flalign*}
\lstset{language=bash, keywordstyle=\color{blue!70}, basicstyle=\ttfamily, frame=shadowbox}

\hyperref[tab:electride]{Back to the table}

\subsubsection*{633091 FePSe$_{3}$}
\label{sec:tqc633091}
\noindent Essential BR: $A1g@3b$ \\
\noindent RSI:
\begin{flalign*}
&\delta_{1}@3b\equiv -m(2Eg)+m(2Eu) = 0,&
\\
&\delta_{2}@3b\equiv -m(A1g)+m(A1u) = -1,&
\end{flalign*}
\lstset{language=bash, keywordstyle=\color{blue!70}, basicstyle=\ttfamily, frame=shadowbox}
\begin{lstlisting}
Computed bands:  1 - 31
GM: GM1+(1); GM1-(1); GM2-GM3-(2); GM2+GM3+(2); GM1+(1); GM1-(1); GM1+(1);
    GM2-GM3-(2); GM2+GM3+(2); GM1-(1); GM1+(1); GM1+(1); GM2-GM3-(2);
    GM2+GM3+(2); GM2+GM3+(2); GM1-(1); GM2-GM3-(2); GM1+(1); GM2+GM3+(2);
    GM1-(1); GM2-GM3-(2); [31] ;
T : T1- (1); T1+ (1); T2+ T3+ (2); T2- T3- (2); T1- (1); T1+ (1); T1- (1);
    T2+ T3+ (2); T2- T3- (2); T1+ (1); T1+ (1); T1- (1); T1- (1); T2+ T3+ (2);
    T2- T3- (2); T2- T3- (2); T2+ T3+ (2); T1- (1); T2- T3- (2); T1+ (1);
    T2+ T3+ (2); [31] ;
F : F1+ (1); F1- (1); F1- (1); F1+ (1); F1- (1); F1+ (1); F1+ (1); F1- (1);
    F1+ (1); F1- (1); F1- (1); F1+ (1); F1+ (1); F1- (1); F1- (1); F1+ (1);
    F1- (1); F1+ (1); F1+ (1); F1- (1); F1- (1); F1+ (1); F1+ (1); F1- (1);
    F1+ (1); F1+ (1); F1- (1); F1- (1); F1+ (1); F1+ (1); F1- (1); [31] ;
L : L1- (1); L1+ (1); L1+ (1); L1- (1); L1+ (1); L1- (1); L1- (1); L1+ (1);
    L1- (1); L1+ (1); L1+ (1); L1- (1); L1- (1); L1+ (1); L1- (1); L1- (1);
    L1+ (1); L1+ (1); L1- (1); L1- (1); L1+ (1); L1+ (1); L1- (1); L1+ (1);
    L1- (1); L1- (1); L1+ (1); L1+ (1); L1- (1); L1+ (1); L1- (1); [31];
\end{lstlisting}
\hyperref[tab:electride]{Back to the table}

\subsubsection*{160496 K$_{2}$Ga$_{3}$}
\label{sec:tqc160496}
\lstset{language=bash, keywordstyle=\color{blue!70}, basicstyle=\ttfamily, frame=shadowbox}
\begin{lstlisting}
Computed bands:  1 - 27
GM: GM1+(1); GM3-(1); GM4-(1); GM1+(1); GM5+(2); GM5-(2); GM3-(1); GM1+(1);
    GM2+(1); GM3-(1); GM5-(2); GM5+(2); GM1+(1); GM3-(1); GM2+(1); GM5-(2);
    GM1+(1); GM1+(1); GM3-(1); GM5+(2); GM4+(1); [27] ;
M : M3- (1); M1+ (1); M2+ (1); M3- (1); M5- (2); M5+ (2); M1+ (1); M4- (1);
    M1+ (1); M3- (1); M5+ (2); M5- (2); M1+ (1); M3- (1); M2+ (1); M5- (2);
    M1+ (1); M1+ (1); M3- (1); M5+ (2); M4+ (1); [27] ;
P : P3  (1); P1  (1); P5  (2); P3  (1); P1  (1); P5  (2); P5  (2); P2  P4  (2);
    P3  (1); P5  (2); P1  (1); P1  (1); P5  (2); P3  (1); P1  (1); P3  (1);
    P5  (2); P5  (2); P3  (1); [27] ;
X : X1+ (1); X2- (1); X4+ X4- (2); X1+ (1); X2- (1); X3- (1); X3+ (1); X4- (1);
    X4+ (1); X1- X2+ (2); X1+ (1); X3- (1); X3+ (1); X2- (1); X1+ (1); X4- (1);
    X3- (1); X2- (1); X1+ (1); X1+ (1); X4+ (1); X3+ (1); X4- (1); X3- (1);
    X2- (1); [27] ;
N : N1+ N2- (2); N1+ N2- (2); N2+ (1); N1- (1); N2- (1); N1+ (1); N2- (1);
    N1+ (1); N2- (1); N1+ (1); N2- (1); N1+ (1); N2+ (1); N1- (1); N1+ (1);
    N2- (1); N2- (1); N1- (1); N1+ (1); N1+ (1); N1+ (1); N2+ (1); N2- (1);
    N2+ (1); N2- (1); [27];
\end{lstlisting}
\hyperref[tab:electride]{Back to the table}

\subsubsection*{260562 Ca(InP)$_{2}$}
\label{sec:tqc260562}
\noindent Essential BR: $A1'@2d$ \\
\noindent RSI:
\begin{flalign*}
&\delta_{1}@2d\equiv m(A1')+m(A2')-m(A2'')-m(A1'')-m(E')+m(E'') = 1,&
\end{flalign*}
\lstset{language=bash, keywordstyle=\color{blue!70}, basicstyle=\ttfamily, frame=shadowbox}
\begin{lstlisting}
Computed bands:  1 - 26
A : A1  (2); A1  (2); A3  (4); A1  (2); A1  (2); A1  (2); A1  (2); A3  (4);
    A3  (4); A1  (2); [26] ;
GM: GM1+GM3+(2); GM2-GM4-(2); GM5-GM6-(4); GM1+(1); GM3+(1); GM4-(1); GM2-(1);
    GM1+(1); GM4-(1); GM3+(1); GM2-(1); GM1+(1); GM5+(2); GM6+(2); GM6-(2);
    GM5-(2); GM4-(1); [26] ;
H : H3  (2); H3  (2); H1  H2  (4); H1  (2); H2  (2); H2  (2); H1  (2); H3  (2);
    H3  (2); H1  (2); H2  (2); H2  (2); [26] ;
K : K1  K2  (2); K3  K4  (2); K5  K6  (4); K5  (2); K6  (2); K5  (2); K6  (2);
    K1  (1); K2  (1); K4  (1); K3  (1); K5  (2); K6  (2); K5  (2); [26] ;
L : L1  (2); L1  (2); L1  (2); L2  (2); L1  (2); L1  (2); L1  (2); L1  (2);
    L1  (2); L1  (2); L2  (2); L2  (2); L1  (2); [26] ;
M : M1+ M3+ (2); M2- M4- (2); M2- M4- (2); M1- M3- (2); M1+ (1); M4- (1);
    M3+ (1); M2- (1); M1+ (1); M4- (1); M3+ (1); M2- (1); M4- (1); M1+ (1);
    M3+ (1); M2- (1); M2+ (1); M1+ (1); M3- (1); M4+ (1); M1- (1); M4- (1); [26];
\end{lstlisting}
\hyperref[tab:electride]{Back to the table}

\subsubsection*{646858 Tb$_{3}$Ni$_{2}$}
\label{sec:tqc646858}
\noindent Essential BR: $Ag@2c$ \\
\noindent RSI:
\begin{flalign*}
&\delta_{1}@2c\equiv -m(Ag)+m(Au)-m(Bg)+m(Bu) = -1,&
\end{flalign*}
\lstset{language=bash, keywordstyle=\color{blue!70}, basicstyle=\ttfamily, frame=shadowbox}

\hyperref[tab:electride]{Back to the table}

\subsubsection*{102868 Cs$_{2}$In$_{3}$}
\label{sec:tqc102868}
\lstset{language=bash, keywordstyle=\color{blue!70}, basicstyle=\ttfamily, frame=shadowbox}
\begin{lstlisting}
Computed bands:  1 - 27
GM: GM1 (1); GM2 (1); GM1 (1); GM1 (1); GM1 (1); GM5 (2); GM5 (2); GM2 (1);
    GM2 (1); GM5 (2); GM5 (2); GM2 (1); GM1 (1); GM2 (1); GM2 (1); GM5 (2);
    GM1 (1); GM1 (1); GM2 (1); GM4 (1); GM5 (2); [27] ;
M : M1  (1); M2  (1); M2  (1); M2  (1); M1  (1); M5  (2); M5  (2); M1  (1);
    M1  (1); M5  (2); M5  (2); M2  (1); M1  (1); M2  (1); M2  (1); M5  (2);
    M1  (1); M1  (1); M2  (1); M5  (2); M4  (1); [27] ;
P : P2  (1); P3  (1); P1  (1); P4  (1); P2  (1); P4  (1); P1  P3  (2); P4  (1);
    P3  (1); P2  (1); P1  (1); P2  (1); P4  (1); P3  (1); P1  (1); P1  (1);
    P4  (1); P3  (1); P2  (1); P1  (1); P2  (1); P4  (1); P3  (1); P4  (1);
    P3  (1); P2  (1); [27] ;
X : X1  (1); X3  (1); X2  (1); X3  (1); X1  (1); X4  (1); X2  (1); X4  (1);
    X3  (1); X1  (1); X3  (1); X2  (1); X1  (1); X4  (1); X4  (1); X2  (1);
    X1  (1); X3  (1); X4  (1); X2  (1); X1  (1); X1  (1); X3  (1); X4  (1);
    X3  (1); X4  (1); X2  (1); [27] ;
N : N1  (1); N1  (1); N1  (1); N1  (1); N1  (1); N1  (1); N2  (1); N2  (1);
    N1  (1); N1  (1); N1  (1); N1  (1); N2  (1); N1  (1); N2  (1); N1  (1);
    N1  (1); N1  (1); N1  (1); N2  (1); N1  (1); N1  (1); N1  (1); N2  (1);
    N1  (1); N2  (1); N1  (1); [27];
\end{lstlisting}
\hyperref[tab:electride]{Back to the table}

\subsubsection*{61392 FePS$_{3}$}
\label{sec:tqc61392}
\lstset{language=bash, keywordstyle=\color{blue!70}, basicstyle=\ttfamily, frame=shadowbox}

\hyperref[tab:electride]{Back to the table}

\subsubsection*{27436 Cu$_{2}$P$_{2}$O$_{7}$}
\label{sec:tqc27436}
\lstset{language=bash, keywordstyle=\color{blue!70}, basicstyle=\ttfamily, frame=shadowbox}

\hyperref[tab:electride]{Back to the table}

\subsubsection*{40823 Ir$_{3}$Se$_{8}$}
\label{sec:tqc40823}
\noindent Essential BR: $Ag@9d$ \\
\noindent RSI:
\begin{flalign*}
&\delta_{1}@9d\equiv -m(Ag)+m(Au) = -1,&
\end{flalign*}
\lstset{language=bash, keywordstyle=\color{blue!70}, basicstyle=\ttfamily, frame=shadowbox}
\begin{lstlisting}
Computed bands:  1 - 38
GM: GM1+(1); GM2+GM3+(2); GM1+(1); GM1-(1); GM2-GM3-(2); GM1-(1); GM1+(1);
    GM2+GM3+(2); GM2+GM3+(2); GM2-GM3-(2); GM1-(1); GM1+(1); GM2+GM3+(2);
    GM1+(1); GM2-GM3-(2); GM2+GM3+(2); GM1+(1); GM2+GM3+(2); GM1-(1);
    GM2-GM3-(2); GM1+(1); GM2+GM3+(2); GM1+(1); GM1+(1); GM2+GM3+(2); GM1-(1);
    [38] ;
T : T1- (1); T2- T3- (2); T1- (1); T1+ (1); T2+ T3+ (2); T1+ (1); T2+ T3+ (2);
    T1+ (1); T1+ (1); T2+ T3+ (2); T1- (1); T2- T3- (2); T1- (1); T2+ T3+ (2);
    T2+ T3+ (2); T1+ (1); T2- T3- (2); T1+ (1); T2+ T3+ (2); T1+ (1);
    T2- T3- (2); T2+ T3+ (2); T1- (1); T2- T3- (2); T1- (1); T1+ (1); [38] ;
F : F1- (1); F1- (1); F1+ (1); F1+ (1); F1- (1); F1- (1); F1+ (1); F1+ (1);
    F1- (1); F1+ (1); F1- (1); F1+ (1); F1- (1); F1- (1); F1+ (1); F1+ (1);
    F1+ (1); F1- (1); F1- (1); F1- (1); F1+ (1); F1- (1); F1+ (1); F1- (1);
    F1- (1); F1+ (1); F1- (1); F1- (1); F1+ (1); F1+ (1); F1- (1); F1- (1);
    F1+ (1); F1+ (1); F1- (1); F1+ (1); F1- (1); F1- (1); [38] ;
L : L1+ (1); L1- (1); L1+ (1); L1- (1); L1+ (1); L1- (1); L1- (1); L1+ (1);
    L1- (1); L1+ L1- (2); L1- (1); L1+ (1); L1+ (1); L1- (1); L1- (1); L1+ (1);
    L1- (1); L1+ (1); L1+ L1- (2); L1- (1); L1- (1); L1+ (1); L1+ (1); L1+ (1);
    L1- (1); L1- (1); L1- (1); L1- (1); L1+ (1); L1- (1); L1+ (1); L1- (1);
    L1+ (1); L1- (1); L1+ (1); L1+ (1); [38];
\end{lstlisting}
\hyperref[tab:electride]{Back to the table}

\subsubsection*{162061 Cu$_{2}$As$_{2}$O$_{7}$}
\label{sec:tqc162061}
\lstset{language=bash, keywordstyle=\color{blue!70}, basicstyle=\ttfamily, frame=shadowbox}

\hyperref[tab:electride]{Back to the table}

\subsubsection*{280027 Ba$_{3}$(LiAs)$_{4}$}
\label{sec:tqc280027}
\lstset{language=bash, keywordstyle=\color{blue!70}, basicstyle=\ttfamily, frame=shadowbox}
\begin{lstlisting}
Computed bands:  1 - 27
GM: GM1+(1); GM1+(1); GM2-(1); GM1+(1); GM4-(1); GM2-(1); GM3-(1); GM3+(1);
    GM3-(1); GM2-GM4+(2); GM4-(1); GM1+(1); GM1+(1); GM2-(1); GM4-(1); GM1+(1);
    GM2-(1); GM3+(1); GM3-(1); GM1+(1); GM4+(1); GM2+(1); GM3+(1); GM3-(1);
    GM2-(1); GM4-(1); [27] ;
X : X1+ (1); X1+ (1); X2- (1); X4- (1); X1+ (1); X3- (1); X2- (1); X3+ (1);
    X2- (1); X3- (1); X4+ (1); X4- (1); X1+ (1); X2- (1); X1+ (1); X4- (1);
    X1+ (1); X2- (1); X4- (1); X3- (1); X1+ (1); X3- (1); X2- (1); X3+ (1);
    X2+ (1); X4+ (1); X3+ (1); [27] ;
R : R1+ (1); R1+ (1); R2- (1); R2- (1); R1- (1); R2- (1); R1+ (1); R1+ (1);
    R2- (1); R2+ (1); R2- (1); R1- (1); R2- (1); R1+ (1); R2- (1); R2+ (1);
    R1+ (1); R2- (1); R1+ (1); R2- (1); R1+ (1); R1- (1); R2+ (1); R1+ (1);
    R1- (1); R2- (1); R1- (1); [27] ;
S : S1+ (1); S2- (1); S1+ (1); S2- (1); S2- (1); S1+ (1); S1- (1); S2- (1);
    S1+ (1); S2+ (1); S1- (1); S2- (1); S2- (1); S1+ (1); S2- (1); S1+ (1);
    S1+ (1); S2- (1); S2+ (1); S2- (1); S1+ (1); S1+ (1); S2- (1); S1- (1);
    S2+ (1); S2- (1); S1- (1); [27] ;
T : T1+ (1); T2- (1); T2+ (1); T1+ (1); T2- (1); T2- (1); T1- (1); T1- (1);
    T1+ (1); T2- (1); T2+ (1); T1- (1); T1+ (1); T1+ (1); T1- (1); T2- (1);
    T1+ (1); T2- (1); T2+ (1); T1- (1); T1+ (1); T2- (1); T2+ (1); T2- (1);
    T1+ (1); T2+ (1); T1- (1); [27] ;
W : W1  (1); W4  (1); W3  (1); W2  (1); W3  (1); W4  (1); W2  (1); W1  (1);
    W1  (1); W4  (1); W3  (1); W2  (1); W2  (1); W1  (1); W2  (1); W3  (1);
    W2  (1); W4  (1); W3  (1); W1  W4  (2); W1  (1); W4  (1); W3  (1); W1  (1);
    W4  (1); W2  (1); [27];
\end{lstlisting}
\hyperref[tab:electride]{Back to the table}

\subsubsection*{58001 Al$_{8}$Mo$_{3}$}
\label{sec:tqc58001}
\lstset{language=bash, keywordstyle=\color{blue!70}, basicstyle=\ttfamily, frame=shadowbox}
\begin{lstlisting}
Computed bands:  1 - 21
GM: GM1+(1); GM2-(1); GM1+(1); GM1+(1); GM2-(1); GM1+(1); GM1+(1); GM2-(1);
    GM1+(1); GM2-(1); GM2-(1); GM1+(1); GM2+(1); GM2+(1); GM1+(1); GM1-(1);
    GM1-(1); GM1+(1); GM2-(1); GM2+(1); GM2+(1); [21] ;
Y : Y2- (1); Y1+ (1); Y2- (1); Y1+ (1); Y2- (1); Y1+ (1); Y2- (1); Y1+ (1);
    Y1+ (1); Y1+ (1); Y2+ (1); Y2- (1); Y1- (1); Y1+ (1); Y2- (1); Y1+ (1);
    Y2+ (1); Y1+ (1); Y1- Y2+ (2); Y2+ (1); [21] ;
V : V1- (1); V1+ (1); V1+ (1); V1- (1); V1- (1); V1+ (1); V1+ (1); V1- (1);
    V1+ (1); V1- (1); V1- (1); V1+ (1); V1- (1); V1+ (1); V1+ (1); V1- (1);
    V1+ (1); V1- (1); V1+ (1); V1+ (1); V1+ (1); [21] ;
L : L1- (1); L1+ (1); L1+ (1); L1- (1); L1+ (1); L1+ (1); L1- (1); L1+ (1);
    L1- (1); L1- (1); L1+ (1); L1- (1); L1+ (1); L1+ (1); L1+ (1); L1- (1);
    L1+ (1); L1- (1); L1+ (1); L1- (1); L1- (1); [21] ;
M : M2- (1); M1+ (1); M2- (1); M1+ (1); M1+ (1); M2- (1); M1+ (1); M2- (1);
    M1+ (1); M2+ (1); M2- (1); M1+ (1); M2- (1); M1- (1); M1+ (1); M2- (1);
    M2+ (1); M1+ (1); M2+ (1); M1- (1); M2+ (1); [21] ;
U : U1  (1); U2  (1); U2  (1); U1  (1); U2  (1); U2  (1); U1  (1); U1  (1);
    U1  (1); U2  (1); U2  (1); U2  (1); U1  (1); U1  (1); U2  (1); U1  (1);
    U2  (1); U1  (1); U2  (1); U1  (1); U2  (1); [21] ;
A : A1+ (1); A2- (1); A1+ (1); A2- (1); A1+ (1); A2- (1); A1+ (1); A1+ (1);
    A2- (1); A2- (1); A1+ (1); A2+ (1); A2- (1); A2+ (1); A1- (1); A1+ (1);
    A1+ (1); A2- (1); A2+ (1); A1- (1); A2+ (1); [21];
\end{lstlisting}
\hyperref[tab:electride]{Back to the table}

\subsubsection*{624977 ScCoSn}
\label{sec:tqc624977}
\noindent Essential BR: $Ag@4b$ \\
\noindent RSI:
\begin{flalign*}
&\delta_{1}@4b\equiv -m(Ag)+m(Au) = -1,&
\end{flalign*}
\lstset{language=bash, keywordstyle=\color{blue!70}, basicstyle=\ttfamily, frame=shadowbox}
\begin{lstlisting}
Computed bands:  1 - 32
GM: GM1+(1); GM3-(1); GM4+(1); GM2-(1); GM4+(1); GM3-(1); GM4+(1); GM2-(1);
    GM1+(1); GM3+(1); GM1-(1); GM4-(1); GM2-(1); GM3-(1); GM2+(1); GM3-(1);
    GM1+(1); GM2+(1); GM3+(1); GM2-(1); GM1-(1); GM1+(1); GM3-(1); GM4+(1);
    GM1+(1); GM2+(1); GM4+(1); GM4+(1); GM4-(1); GM2-(1); GM3+(1); GM1+(1);
    [32] ;
R : R1  R2  (4); R1  R2  (4); R1  R2  (4); R1  R2  (4); R1  R2  (4); R1  R2  (4);
    R1  R2  (4); R1  R2  (4); [32] ;
S : S1  S2  (4); S1  S2  (4); S1  S2  (4); S1  S2  (4); S1  S2  (4); S1  S2  (4);
    S1  S2  (4); S1  S2  (4); [32] ;
T : T2  (2); T1  (2); T1  (2); T2  (2); T1  (2); T2  (2); T1  (2); T2  (2);
    T1  (2); T2  (2); T1  (2); T2  (2); T1  (2); T2  (2); T1  (2); T2  (2);
    [32] ;
U : U2- U3- (2); U1+ U4+ (2); U1+ U4+ (2); U2- U3- (2); U1- U4- (2); U2+ U3+ (2);
    U1+ U4+ (2); U2- U3- (2); U1- U4- (2); U1+ U4+ (2); U2- U3- (2); U2+ U3+ (2);
    U2- U3- (2); U1+ U4+ (2); U2- U3- (2); U1- U4- (2); [32] ;
X : X1  (2); X1  (2); X1  (2); X1  (2); X2  (2); X1  (2); X2  (2); X1  (2);
    X2  (2); X1  (2); X1  (2); X2  (2); X1  (2); X1  (2); X2  (2); X1  (2);
    [32] ;
Y : Y2  (2); Y1  (2); Y1  (2); Y2  (2); Y1  (2); Y2  (2); Y2  (2); Y1  (2);
    Y1  (2); Y2  (2); Y2  (2); Y1  (2); Y1  (2); Y1  (2); Y2  (2); Y2  (2);
    [32] ;
Z : Z1  (2); Z1  (2); Z1  (2); Z1  (2); Z1  (2); Z2  (2); Z2  (2); Z1  (2);
    Z1  (2); Z2  (2); Z1  (2); Z1  (2); Z2  (2); Z1  (2); Z2  (2); Z1  (2); [32];
\end{lstlisting}
\hyperref[tab:electride]{Back to the table}

\subsubsection*{75029 CuBIr}
\label{sec:tqc75029}
\noindent Essential BR: $A@8a$ \\
\noindent RSI:
\begin{flalign*}
&\delta_{1}@8a\equiv -m(A)+m(B) = -1,&
\end{flalign*}
\lstset{language=bash, keywordstyle=\color{blue!70}, basicstyle=\ttfamily, frame=shadowbox}

\hyperref[tab:electride]{Back to the table}

\subsubsection*{412794 RbPrTe$_{4}$}
\label{sec:tqc412794}
\lstset{language=bash, keywordstyle=\color{blue!70}, basicstyle=\ttfamily, frame=shadowbox}
\begin{lstlisting}
Computed bands:  1 - 44
A : A2  (2); A1  (2); A1  (2); A3  (2); A4  (2); A3  (2); A3  (2); A2  (2);
    A4  (2); A3  (2); A2  (2); A1  (2); A3  (2); A3  (2); A1  (2); A2  (2);
    A3  (2); A2  (2); A3  A4  (4); A4  (2); A1  (2); [44] ;
GM: GM1+(1); GM1-(1); GM1+(1); GM1-(1); GM3-(1); GM3+(1); GM5+(2); GM5-(2);
    GM1+(1); GM3-(1); GM2-(1); GM5+(2); GM4+(1); GM3+(1); GM5-(2); GM3-(1);
    GM5-(2); GM5+(2); GM4+(1); GM1+(1); GM2-(1); GM1-(1); GM1+(1); GM5+(2);
    GM5-(2); GM2-(1); GM3-(1); GM3+(1); GM3-(1); GM4+(1); GM5+(2); GM5-(2);
    GM2+(1); GM4-(1); [44] ;
M : M1  (2); M1  (2); M2  (2); M3  (2); M4  (2); M3  (2); M3  (2); M2  (2);
    M4  (2); M3  (2); M1  (2); M2  (2); M3  (2); M3  (2); M2  (2); M1  (2);
    M3  (2); M1  (2); M4  (2); M3  (2); M4  (2); M2  (2); [44] ;
Z : Z3- (1); Z3+ (1); Z1+ (1); Z1- (1); Z1+ (1); Z1- (1); Z5- (2); Z5+ (2);
    Z3- (1); Z1+ (1); Z4+ (1); Z5+ (2); Z2- (1); Z3+ (1); Z5- (2); Z3- (1);
    Z5- (2); Z5+ (2); Z2- (1); Z4+ (1); Z3- (1); Z3+ (1); Z5- (2); Z3- (1);
    Z5+ (2); Z1+ (1); Z4+ (1); Z1- (1); Z1+ (1); Z2- (1); Z5- (2); Z5+ (2);
    Z4- (1); Z2+ (1); [44] ;
R : R2  (2); R1  (2); R1  (2); R2  (2); R1  (2); R2  (2); R1  (2); R2  (2);
    R2  (2); R1  (2); R2  (2); R1  (2); R2  (2); R1  (2); R2  (2); R1  (2);
    R2  (2); R1  (2); R2  (2); R1  (2); R2  (2); R1  (2); [44] ;
X : X1  (2); X1  (2); X2  (2); X1  (2); X2  (2); X1  (2); X2  (2); X2  (2);
    X1  (2); X2  (2); X1  (2); X2  (2); X1  (2); X2  (2); X1  (2); X2  (2);
    X1  (2); X2  (2); X1  (2); X2  (2); X1  (2); X2  (2); [44];
\end{lstlisting}
\hyperref[tab:electride]{Back to the table}

\subsubsection*{42569 NiAs$_{2}$}
\label{sec:tqc42569}
\noindent Essential BR: $A1g@4b$ \\
\noindent RSI:
\begin{flalign*}
&\delta_{1}@4b\equiv -m(2Eg)+m(2Eu) = 0,&
\\
&\delta_{2}@4b\equiv -m(A1g)+m(A1u) = -1,&
\end{flalign*}
\lstset{language=bash, keywordstyle=\color{blue!70}, basicstyle=\ttfamily, frame=shadowbox}
\begin{lstlisting}
Computed bands:  1 - 40
GM: GM1+(1); GM4+(3); GM4-(3); GM1-(1); GM4+(3); GM4-(3); GM4+(3); GM2-GM3-(2);
    GM4+(3); GM2+GM3+(2); GM4+(3); GM1+(1); GM2+GM3+(2); GM4+(3); GM1+(1);
    GM4+(3); GM4-(3); [40] ;
R : R1- R3- (4); R1+ R3+ (4); R1+ R3+ (4); R2+ R2+ (4); R1- R3- (4); R2+ R2+ (4);
    R1+ R3+ (4); R1+ R3+ (4); R2- R2- (4); R1- R3- (4); [40] ;
M : M1  M2  (4); M1  M2  (4); M1  M2  (4); M1  M2  (4); M1  M2  (4); M1  M2  (4);
    M1  M2  (4); M1  M2  (4); M1  M2  (4); M1  M2  (4); [40] ;
X : X2  (2); X1  (2); X2  (2); X1  (2); X1  (2); X2  (2); X1  (2); X1  (2);
    X2  (2); X2  (2); X1  (2); X2  (2); X2  (2); X1  (2); X2  (2); X1  (2);
    X1  (2); X2  (2); X2  (2); X1  (2); [40];
\end{lstlisting}
\hyperref[tab:electride]{Back to the table}

\subsubsection*{24202 OsSe$_{2}$}
\label{sec:tqc24202}
\noindent Essential BR: $A1g@4b$ \\
\noindent RSI:
\begin{flalign*}
&\delta_{1}@4b\equiv -m(2Eg)+m(2Eu) = 0,&
\\
&\delta_{2}@4b\equiv -m(A1g)+m(A1u) = -1,&
\end{flalign*}
\lstset{language=bash, keywordstyle=\color{blue!70}, basicstyle=\ttfamily, frame=shadowbox}
\begin{lstlisting}
Computed bands:  1 - 40
GM: GM1+(1); GM4+(3); GM4-(3); GM1-(1); GM4+(3); GM4+(3); GM4-(3); GM4+(3);
    GM2+GM3+(2); GM2-GM3-(2); GM1+(1); GM4-(3); GM4+(3); GM2+GM3+(2); GM4+(3);
    GM4+(3); GM1+(1); [40] ;
R : R1- R3- (4); R1+ R3+ (4); R1+ R3+ (4); R2+ R2+ (4); R1- R3- (4); R2- R2- (4);
    R1+ R3+ (4); R1- R3- (4); R2+ R2+ (4); R1+ R3+ (4); [40] ;
M : M1  M2  (4); M1  M2  (4); M1  M2  (4); M1  M2  (4); M1  M2  (4); M1  M2  (4);
    M1  M2  (4); M1  M2  (4); M1  M2  (4); M1  M2  (4); [40] ;
X : X2  (2); X1  (2); X2  (2); X1  (2); X1  (2); X2  (2); X1  (2); X1  (2);
    X2  (2); X2  (2); X1  (2); X2  (2); X2  (2); X1  (2); X1  (2); X1  (2);
    X2  (2); X2  (2); X2  (2); X1  (2); [40];
\end{lstlisting}
\hyperref[tab:electride]{Back to the table}

\subsubsection*{650607 RuSe$_{2}$}
\label{sec:tqc650607}
\noindent Essential BR: $A1g@4b$ \\
\noindent RSI:
\begin{flalign*}
&\delta_{1}@4b\equiv -m(2Eg)+m(2Eu) = 0,&
\\
&\delta_{2}@4b\equiv -m(A1g)+m(A1u) = -1,&
\end{flalign*}
\lstset{language=bash, keywordstyle=\color{blue!70}, basicstyle=\ttfamily, frame=shadowbox}
\begin{lstlisting}
Computed bands:  1 - 40
GM: GM1+(1); GM4+(3); GM4-(3); GM1-(1); GM4+(3); GM4-(3); GM4+(3); GM2-GM3-(2);
    GM2+GM3+(2); GM4+(3); GM1+(1); GM4-(3); GM4+(3); GM2+GM3+(2); GM4+(3);
    GM4+(3); GM1+(1); [40] ;
R : R1- R3- (4); R1+ R3+ (4); R1+ R3+ (4); R2+ R2+ (4); R1- R3- (4); R2- R2- (4);
    R1+ R3+ (4); R1- R3- (4); R2+ R2+ (4); R1+ R3+ (4); [40] ;
M : M1  M2  (4); M1  M2  (4); M1  M2  (4); M1  M2  (4); M1  M2  (4); M1  M2  (4);
    M1  M2  (4); M1  M2  (4); M1  M2  (4); M1  M2  (4); [40] ;
X : X2  (2); X1  (2); X2  (2); X1  (2); X1  (2); X2  (2); X1  (2); X1  (2);
    X2  (2); X2  (2); X1  (2); X2  (2); X2  (2); X1  (2); X1  (2); X1  (2);
    X2  (2); X2  (2); X2  (2); X1  (2); [40];
\end{lstlisting}
\hyperref[tab:electride]{Back to the table}

\subsubsection*{24187 OsS$_{2}$}
\label{sec:tqc24187}
\noindent Essential BR: $A1g@4b$ \\
\noindent RSI:
\begin{flalign*}
&\delta_{1}@4b\equiv -m(2Eg)+m(2Eu) = 0,&
\\
&\delta_{2}@4b\equiv -m(A1g)+m(A1u) = -1,&
\end{flalign*}
\lstset{language=bash, keywordstyle=\color{blue!70}, basicstyle=\ttfamily, frame=shadowbox}
\begin{lstlisting}
Computed bands:  1 - 40
GM: GM1+(1); GM4+(3); GM4-(3); GM1-(1); GM4+(3); GM4+(3); GM4-(3); GM1+(1);
    GM2+GM3+GM4+(5); GM2-GM3-(2); GM4-(3); GM4+(3); GM2+GM3+(2); GM4+(3);
    GM4+(3); GM1+(1); [40] ;
R : R1- R3- (4); R1+ R3+ (4); R1+ R3+ (4); R2+ R2+ (4); R1- R3- (4); R2- R2- (4);
    R1- R3- (4); R1+ R3+ (4); R2+ R2+ (4); R1+ R3+ (4); [40] ;
M : M1  M2  (4); M1  M2  (4); M1  M2  (4); M1  M2  (4); M1  M2  (4); M1  M2  (4);
    M1  M2  (4); M1  M2  (4); M1  M2  (4); M1  M2  (4); [40] ;
X : X2  (2); X1  (2); X2  (2); X1  (2); X1  (2); X2  (2); X1  (2); X1  (2);
    X2  (2); X2  (2); X1  (2); X2  (2); X2  (2); X1  (2); X1  (2); X1  (2);
    X2  (2); X2  (2); X2  (2); X1  (2); [40];
\end{lstlisting}
\hyperref[tab:electride]{Back to the table}

\subsubsection*{419345 NdS$_{2}$}
\label{sec:tqc419345}
\noindent Essential BR: $Ag@2c$ \\
\noindent RSI:
\begin{flalign*}
&\delta_{1}@2c\equiv -m(Ag)+m(Au) = -1,&
\end{flalign*}
\lstset{language=bash, keywordstyle=\color{blue!70}, basicstyle=\ttfamily, frame=shadowbox}

\hyperref[tab:electride]{Back to the table}

\subsubsection*{43101 As$_{2}$Pd}
\label{sec:tqc43101}
\noindent Essential BR: $A1g@4b$ \\
\noindent RSI:
\begin{flalign*}
&\delta_{1}@4b\equiv -m(2Eg)+m(2Eu) = 0,&
\\
&\delta_{2}@4b\equiv -m(A1g)+m(A1u) = -1,&
\end{flalign*}
\lstset{language=bash, keywordstyle=\color{blue!70}, basicstyle=\ttfamily, frame=shadowbox}
\begin{lstlisting}
Computed bands:  1 - 40
GM: GM1+(1); GM4+(3); GM4-(3); GM1-(1); GM4+(3); GM4+(3); GM2+GM3+(2); GM4-(3);
    GM4+(3); GM4+(3); GM2+GM3+(2); GM1+(1); GM4+(3); GM2-GM3-(2); GM4+(3);
    GM1+(1); GM4-(3); [40] ;
R : R1- R3- (4); R1+ R3+ (4); R1+ R3+ (4); R2+ R2+ (4); R2+ R2+ (4); R1+ R3+ (4);
    R1+ R3+ (4); R1- R3- (4); R2- R2- (4); R1- R3- (4); [40] ;
M : M1  M2  (4); M1  M2  (4); M1  M2  (4); M1  M2  (4); M1  M2  (4); M1  M2  (4);
    M1  M2  (4); M1  M2  (4); M1  M2  (4); M1  M2  (4); [40] ;
X : X2  (2); X1  (2); X2  (2); X1  (2); X1  (2); X2  (2); X1  (2); X2  (2);
    X1  (2); X1  (2); X2  (2); X2  (2); X1  (2); X2  (2); X1  (2); X2  (2);
    X1  (2); X2  (2); X2  (2); X1  (2); [40];
\end{lstlisting}
\hyperref[tab:electride]{Back to the table}

\subsubsection*{300225 Te$_{2}$Os}
\label{sec:tqc300225}
\noindent Essential BR: $A1g@4b$ \\
\noindent RSI:
\begin{flalign*}
&\delta_{1}@4b\equiv -m(2Eg)+m(2Eu) = 0,&
\\
&\delta_{2}@4b\equiv -m(A1g)+m(A1u) = -1,&
\end{flalign*}
\lstset{language=bash, keywordstyle=\color{blue!70}, basicstyle=\ttfamily, frame=shadowbox}
\begin{lstlisting}
Computed bands:  1 - 40
GM: GM1+(1); GM4+(3); GM4-(3); GM1-(1); GM4+(3); GM4+(3); GM4-(3); GM2-GM3-(2);
    GM4+(3); GM2+GM3+(2); GM1+(1); GM4+(3); GM4-(3); GM2+GM3+(2); GM4+(3);
    GM1+(1); GM4+(3); [40] ;
R : R1- R3- (4); R1+ R3+ (4); R1+ R3+ (4); R2+ R2+ (4); R1- R3- (4); R1+ R3+ (4);
    R2- R2- (4); R2+ R2+ (4); R1- R3- (4); R1+ R3+ (4); [40] ;
M : M1  M2  (4); M1  M2  (4); M1  M2  (4); M1  M2  (4); M1  M2  (4); M1  M2  (4);
    M1  M2  (4); M1  M2  (4); M1  M2  (4); M1  M2  (4); [40] ;
X : X2  (2); X1  (2); X2  (2); X1  (2); X1  (2); X2  (2); X1  (2); X1  (2);
    X2  (2); X2  (2); X1  (2); X2  (2); X2  (2); X1  (2); X1  (2); X2  (2);
    X1  (2); X2  (2); X2  (2); X1  (2); [40];
\end{lstlisting}
\hyperref[tab:electride]{Back to the table}

\subsubsection*{412792 KNdTe$_{4}$}
\label{sec:tqc412792}
\lstset{language=bash, keywordstyle=\color{blue!70}, basicstyle=\ttfamily, frame=shadowbox}
\begin{lstlisting}
Computed bands:  1 - 44
A : A2  (2); A1  (2); A1  (2); A3  (2); A4  (2); A3  (2); A2  (2); A4  (2);
    A3  (2); A3  (2); A2  (2); A1  (2); A3  (2); A3  (2); A1  (2); A2  (2);
    A3  (2); A3  (2); A2  (2); A4  (2); A4  (2); A1  (2); [44] ;
GM: GM1+(1); GM1-(1); GM1+GM1-(2); GM3-(1); GM3+(1); GM5+(2); GM5-(2); GM3-(1);
    GM5+(2); GM3+(1); GM5-(2); GM1+(1); GM3-(1); GM2-(1); GM4+(1); GM5-(2);
    GM5+(2); GM4+(1); GM1+(1); GM2-(1); GM1-(1); GM1+(1); GM5+(2); GM5-(2);
    GM2-(1); GM3-(1); GM3+(1); GM3-(1); GM4+(1); GM5+(2); GM5-(2); GM2+(1);
    GM4-(1); [44] ;
M : M1  (2); M1  (2); M2  (2); M3  (2); M4  (2); M3  (2); M2  (2); M4  (2);
    M3  (2); M3  (2); M1  (2); M2  (2); M3  (2); M3  (2); M3  (2); M2  (2);
    M1  (2); M1  (2); M4  (2); M3  (2); M4  (2); M2  (2); [44] ;
Z : Z3- (1); Z3+ (1); Z1+ Z1- (2); Z1+ (1); Z1- (1); Z5- (2); Z5+ (2); Z3- (1);
    Z5+ (2); Z3+ (1); Z5- (2); Z3- (1); Z1+ (1); Z4+ (1); Z2- (1); Z5+ (2);
    Z5- (2); Z2- (1); Z4+ (1); Z3- (1); Z3+ (1); Z5- (2); Z3- (1); Z5+ (2);
    Z1+ (1); Z1- (1); Z4+ (1); Z1+ (1); Z2- (1); Z5- (2); Z5+ (2); Z4- (1);
    Z2+ (1); [44] ;
R : R2  (2); R1  (2); R1  (2); R2  (2); R1  (2); R2  (2); R1  (2); R2  (2);
    R2  (2); R1  (2); R2  (2); R1  (2); R2  (2); R1  (2); R2  (2); R1  (2);
    R2  (2); R1  (2); R2  (2); R1  (2); R2  (2); R1  (2); [44] ;
X : X1  (2); X1  (2); X2  (2); X1  (2); X2  (2); X2  (2); X1  (2); X2  (2);
    X1  (2); X2  (2); X1  (2); X2  (2); X1  (2); X2  (2); X1  (2); X2  (2);
    X1  (2); X2  (2); X1  (2); X2  (2); X1  (2); X2  (2); [44];
\end{lstlisting}
\hyperref[tab:electride]{Back to the table}

\subsubsection*{54415 DyCoSn}
\label{sec:tqc54415}
\noindent Essential BR: $Ag@4b$ \\
\noindent RSI:
\begin{flalign*}
&\delta_{1}@4b\equiv -m(Ag)+m(Au) = -1,&
\end{flalign*}
\lstset{language=bash, keywordstyle=\color{blue!70}, basicstyle=\ttfamily, frame=shadowbox}
\begin{lstlisting}
Computed bands:  1 - 44
GM: GM2-(1); GM4+(1); GM4+(1); GM1+(1); GM2+(1); GM3+(1); GM4-(1); GM1-(1);
    GM2-GM3-(2); GM3-(1); GM1+(1); GM1+(1); GM3-(1); GM4+(1); GM2-(1); GM4+(1);
    GM3-(1); GM4+(1); GM2-(1); GM1+(1); GM3+(1); GM1-(1); GM2-(1); GM4-(1);
    GM3-(1); GM2+(1); GM3-(1); GM1+(1); GM2-(1); GM2+(1); GM3+(1); GM1-(1);
    GM3-(1); GM1+(1); GM4+(1); GM4+(1); GM4+(1); GM1+(1); GM2+GM4-(2); GM3+(1);
    GM2-(1); GM1+(1); [44] ;
R : R1  R2  (4); R1  R2  (4); R1  R2  (4); R1  R2  (4); R1  R2  (4); R1  R2  (4);
    R1  R2  (4); R1  R2  (4); R1  R2  (4); R1  R2  (4); R1  R2  (4); [44] ;
S : S1  S2  (4); S1  S2  (4); S1  S2  (4); S1  S2  (4); S1  S2  (4); S1  S2  (4);
    S1  S2  (4); S1  S2  (4); S1  S2  (4); S1  S2  (4); S1  S2  (4); [44] ;
T : T2  (2); T2  (2); T1  (2); T2  (2); T1  (2); T1  (2); T2  (2); T1  (2);
    T1  (2); T2  (2); T1  (2); T2  (2); T1  (2); T2  (2); T2  (2); T1  (2);
    T1  (2); T1  (2); T2  (2); T2  (2); T1  (2); T2  (2); [44] ;
U : U1+ U4+ (2); U2- U3- (2); U2+ U3+ (2); U1+ U4+ (2); U1- U4- (2); U2- U3- (2);
    U2- U3- (2); U1+ U4+ (2); U2- U3- (2); U1+ U4+ (2); U1- U4- (2); U2+ U3+ (2);
    U1+ U4+ (2); U2- U3- (2); U1- U4- (2); U1+ U4+ (2); U2- U3- (2); U2+ U3+ (2);
    U2- U3- (2); U2- U3- (2); U1+ U4+ (2); U1- U4- (2); [44] ;
X : X1  (2); X1  (2); X2  (2); X1  (2); X2  (2); X1  (2); X1  (2); X1  (2);
    X1  (2); X1  (2); X2  (2); X1  (2); X1  (2); X2  (2); X2  (2); X1  (2);
    X1  (2); X2  (2); X1  (2); X1  (2); X2  (2); X1  (2); [44] ;
Y : Y1  (2); Y2  (2); Y1  (2); Y2  (2); Y1  (2); Y2  (2); Y2  (2); Y1  (2);
    Y1  (2); Y2  (2); Y2  (2); Y1  (2); Y2  (2); Y1  (2); Y1  (2); Y2  (2);
    Y2  (2); Y1  (2); Y1  (2); Y2  (2); Y1  (2); Y2  (2); [44] ;
Z : Z1  (2); Z1  (2); Z2  (2); Z2  (2); Z1  (2); Z1  (2); Z1  (2); Z1  (2);
    Z1  (2); Z1  (2); Z2  (2); Z1  (2); Z2  (2); Z1  (2); Z1  (2); Z2  (2);
    Z1  (2); Z1  (2); Z2  (2); Z1  (2); Z2  (2); Z1  (2); [44];
\end{lstlisting}
\hyperref[tab:electride]{Back to the table}

\subsubsection*{611219 PrAs$_{2}$}
\label{sec:tqc611219}
\noindent Essential BR: $Ag@2b$ \\
\noindent RSI:
\begin{flalign*}
&\delta_{1}@2b\equiv -m(Ag)+m(Au) = -1,&
\end{flalign*}
\lstset{language=bash, keywordstyle=\color{blue!70}, basicstyle=\ttfamily, frame=shadowbox}

\hyperref[tab:electride]{Back to the table}

\subsubsection*{610026 CoAs$_{2}$}
\label{sec:tqc610026}
\noindent Essential BR: $Ag@2d$ \\
\noindent RSI:
\begin{flalign*}
&\delta_{1}@2d\equiv -m(Ag)+m(Au) = -1,&
\end{flalign*}
\lstset{language=bash, keywordstyle=\color{blue!70}, basicstyle=\ttfamily, frame=shadowbox}

\hyperref[tab:electride]{Back to the table}

\subsubsection*{424397 Bi$_{2}$Ir}
\label{sec:tqc424397}
\noindent Essential BR: $Ag@2d$ \\
\noindent RSI:
\begin{flalign*}
&\delta_{1}@2d\equiv -m(Ag)+m(Au) = -1,&
\end{flalign*}
\lstset{language=bash, keywordstyle=\color{blue!70}, basicstyle=\ttfamily, frame=shadowbox}

\hyperref[tab:electride]{Back to the table}

\subsubsection*{43502 Sb$_{2}$Ir}
\label{sec:tqc43502}
\noindent Essential BR: $Ag@2d$ \\
\noindent RSI:
\begin{flalign*}
&\delta_{1}@2d\equiv -m(Ag)+m(Au) = -1,&
\end{flalign*}
\lstset{language=bash, keywordstyle=\color{blue!70}, basicstyle=\ttfamily, frame=shadowbox}

\hyperref[tab:electride]{Back to the table}

\subsubsection*{82549 NdGeRh}
\label{sec:tqc82549}
\noindent Essential BR: $Ag@4b$ \\
\noindent RSI:
\begin{flalign*}
&\delta_{1}@4b\equiv -m(Ag)+m(Au) = -1,&
\end{flalign*}
\lstset{language=bash, keywordstyle=\color{blue!70}, basicstyle=\ttfamily, frame=shadowbox}

\hyperref[tab:electride]{Back to the table}

\subsubsection*{636739 TbGeIr}
\label{sec:tqc636739}
\noindent Essential BR: $Ag@4b$ \\
\noindent RSI:
\begin{flalign*}
&\delta_{1}@4b\equiv -m(Ag)+m(Au) = -1,&
\end{flalign*}
\lstset{language=bash, keywordstyle=\color{blue!70}, basicstyle=\ttfamily, frame=shadowbox}
\begin{lstlisting}
Computed bands:  1 - 44
GM: GM2-(1); GM4+(1); GM4+(1); GM1+(1); GM2+(1); GM3+(1); GM4-(1); GM2-(1);
    GM1-GM3-(2); GM3-(1); GM1+(1); GM1+(1); GM4+(1); GM3-(1); GM2-(1); GM3-(1);
    GM4+(1); GM1+(1); GM3+(1); GM2-(1); GM4+(1); GM4-(1); GM2-(1); GM3-(1);
    GM1-(1); GM2+(1); GM3-(1); GM1+(1); GM3+(1); GM2+(1); GM2-(1); GM1-(1);
    GM1+(1); GM4+(1); GM4+(1); GM3-(1); GM4-(1); GM4+(1); GM1+(1); GM2+(1);
    GM3+(1); GM2-(1); GM1+(1); [44] ;
R : R1  R2  (4); R1  R2  (4); R1  R2  (4); R1  R2  (4); R1  R2  (4); R1  R2  (4);
    R1  R2  (4); R1  R2  (4); R1  R2  (4); R1  R2  (4); R1  R2  (4); [44] ;
S : S1  S2  (4); S1  S2  (4); S1  S2  (4); S1  S2  (4); S1  S2  (4); S1  S2  (4);
    S1  S2  (4); S1  S2  (4); S1  S2  (4); S1  S2  (4); S1  S2  (4); [44] ;
T : T2  (2); T2  (2); T1  (2); T2  (2); T1  (2); T1  (2); T1  (2); T2  (2);
    T1  (2); T2  (2); T1  (2); T2  (2); T1  (2); T2  (2); T1  (2); T2  (2);
    T1  (2); T2  (2); T1  (2); T2  (2); T1  (2); T2  (2); [44] ;
U : U1+ U4+ (2); U2- U3- (2); U2+ U3+ (2); U1+ U4+ (2); U1- U4- (2); U2- U3- (2);
    U2- U3- (2); U1+ U4+ (2); U2- U3- (2); U1+ U4+ (2); U2+ U3+ (2); U1- U4- (2);
    U1+ U4+ (2); U1- U4- (2); U2- U3- (2); U1+ U4+ (2); U2- U3- (2); U2+ U3+ (2);
    U1+ U4+ (2); U2- U3- (2); U2- U3- (2); U1- U4- (2); [44] ;
X : X1  (2); X1  (2); X1  X2  (4); X2  (2); X1  (2); X1  (2); X1  (2); X1  (2);
    X1  (2); X2  (2); X1  (2); X2  (2); X1  (2); X2  (2); X1  (2); X1  (2);
    X1  (2); X2  (2); X2  (2); X1  (2); X1  (2); [44] ;
Y : Y1  (2); Y2  (2); Y1  (2); Y2  (2); Y1  (2); Y2  (2); Y2  (2); Y1  (2);
    Y1  (2); Y2  (2); Y2  (2); Y1  (2); Y2  (2); Y1  (2); Y2  (2); Y1  (2);
    Y2  (2); Y1  (2); Y1  (2); Y2  (2); Y2  (2); Y1  (2); [44] ;
Z : Z1  (2); Z1  (2); Z2  (2); Z1  Z2  (4); Z1  (2); Z1  (2); Z1  (2); Z1  (2);
    Z1  (2); Z1  (2); Z2  (2); Z2  (2); Z1  (2); Z2  (2); Z1  (2); Z1  (2);
    Z1  (2); Z2  (2); Z1  (2); Z1  (2); Z2  (2); [44];
\end{lstlisting}
\hyperref[tab:electride]{Back to the table}

\subsubsection*{88272 DyCoSi}
\label{sec:tqc88272}
\noindent Essential BR: $Ag@4a$ \\
\noindent RSI:
\begin{flalign*}
&\delta_{1}@4a\equiv -m(Ag)+m(Au) = -1,&
\end{flalign*}
\lstset{language=bash, keywordstyle=\color{blue!70}, basicstyle=\ttfamily, frame=shadowbox}
\begin{lstlisting}
Computed bands:  1 - 44
GM: GM2-(1); GM4+(1); GM1+(1); GM4+(1); GM2+(1); GM3+(1); GM4-(1); GM2-(1);
    GM3-(1); GM1-(1); GM3-(1); GM1+(1); GM1+(1); GM4+(1); GM3-(1); GM2-(1);
    GM3-(1); GM4+(1); GM2-(1); GM4+(1); GM1+(1); GM3+(1); GM1-(1); GM2-(1);
    GM3-GM4-(2); GM2+(1); GM1+(1); GM3-(1); GM2+(1); GM2-(1); GM3+(1); GM1-(1);
    GM4+(1); GM3-(1); GM1+(1); GM4+(1); GM4-(1); GM4+(1); GM1+(1); GM3+(1);
    GM2-(1); GM2+(1); GM1+(1); [44] ;
R : R1  R2  (4); R1  R2  (4); R1  R2  (4); R1  R2  (4); R1  R2  (4); R1  R2  (4);
    R1  R2  (4); R1  R2  (4); R1  R2  (4); R1  R2  (4); R1  R2  (4); [44] ;
S : S1  S2  (4); S1  S2  (4); S1  S2  (4); S1  S2  (4); S1  S2  (4); S1  S2  (4);
    S1  S2  (4); S1  S2  (4); S1  S2  (4); S1  S2  (4); S1  S2  (4); [44] ;
T : T1  (2); T1  (2); T2  (2); T1  (2); T2  (2); T2  (2); T2  (2); T1  (2);
    T2  (2); T1  (2); T2  (2); T1  (2); T2  (2); T1  (2); T1  (2); T2  (2);
    T2  (2); T1  (2); T2  (2); T1  (2); T1  (2); T2  (2); [44] ;
U : U2- U3- (2); U1+ U4+ (2); U1- U4- (2); U2- U3- (2); U1+ U4+ (2); U2+ U3+ (2);
    U1+ U4+ (2); U2- U3- (2); U1+ U4+ (2); U2- U3- (2); U2+ U3+ (2); U1- U4- (2);
    U2- U3- (2); U1+ U4+ (2); U2+ U3+ (2); U2- U3- (2); U1+ U4+ (2); U1+ U4+ (2);
    U1- U4- (2); U1+ U4+ (2); U2- U3- (2); U2+ U3+ (2); [44] ;
X : X1  (2); X1  (2); X2  (2); X1  (2); X2  (2); X1  (2); X1  (2); X1  (2);
    X1  (2); X1  (2); X2  (2); X1  (2); X1  (2); X2  (2); X2  (2); X1  (2);
    X1  (2); X1  (2); X2  (2); X1  (2); X2  (2); X1  (2); [44] ;
Y : Y2  (2); Y1  (2); Y1  (2); Y2  (2); Y1  (2); Y2  (2); Y2  (2); Y1  (2);
    Y1  (2); Y2  (2); Y2  (2); Y1  (2); Y2  (2); Y1  (2); Y1  (2); Y2  (2);
    Y2  (2); Y1  (2); Y1  (2); Y2  (2); Y2  (2); Y1  (2); [44] ;
Z : Z1  (2); Z1  (2); Z2  (2); Z1  (2); Z2  (2); Z1  (2); Z1  (2); Z1  (2);
    Z1  (2); Z1  (2); Z2  (2); Z1  (2); Z2  (2); Z1  (2); Z2  (2); Z1  (2);
    Z1  (2); Z1  (2); Z2  (2); Z1  (2); Z2  (2); Z1  (2); [44];
\end{lstlisting}
\hyperref[tab:electride]{Back to the table}

\subsubsection*{43105 Sb$_{2}$Pt}
\label{sec:tqc43105}
\noindent Essential BR: $A1g@4b$ \\
\noindent RSI:
\begin{flalign*}
&\delta_{1}@4b\equiv -m(2Eg)+m(2Eu) = 0,&
\\
&\delta_{2}@4b\equiv -m(A1g)+m(A1u) = -1,&
\end{flalign*}
\lstset{language=bash, keywordstyle=\color{blue!70}, basicstyle=\ttfamily, frame=shadowbox}
\begin{lstlisting}
Computed bands:  1 - 40
GM: GM1+(1); GM4+(3); GM4-(3); GM1-(1); GM4+(3); GM4+(3); GM2+GM3+(2); GM4+(3);
    GM4-(3); GM4+(3); GM2+GM3+(2); GM4+(3); GM1+(1); GM2-GM3-(2); GM4+(3);
    GM1+(1); GM4-(3); [40] ;
R : R1- R3- (4); R1+ R3+ (4); R1+ R3+ (4); R2+ R2+ (4); R2+ R2+ (4); R1+ R3+ (4);
    R1+ R3+ (4); R1- R3- (4); R2- R2- (4); R1- R3- (4); [40] ;
M : M1  M2  (4); M1  M2  (4); M1  M2  (4); M1  M2  (4); M1  M2  (4); M1  M2  (4);
    M1  M2  (4); M1  M2  (4); M1  M2  (4); M1  M2  (4); [40] ;
X : X2  (2); X1  (2); X2  (2); X1  (2); X1  (2); X2  (2); X1  (2); X1  (2);
    X2  (2); X2  (2); X1  (2); X2  (2); X1  (2); X2  (2); X1  (2); X2  (2);
    X1  (2); X2  (2); X2  (2); X1  (2); [40];
\end{lstlisting}
\hyperref[tab:electride]{Back to the table}

\subsubsection*{43501 Sb$_{2}$Rh}
\label{sec:tqc43501}
\noindent Essential BR: $Ag@2d$ \\
\noindent RSI:
\begin{flalign*}
&\delta_{1}@2d\equiv -m(Ag)+m(Au) = -1,&
\end{flalign*}
\lstset{language=bash, keywordstyle=\color{blue!70}, basicstyle=\ttfamily, frame=shadowbox}

\hyperref[tab:electride]{Back to the table}

\subsubsection*{107587 Bi$_{3}$Te$_{2}$S}
\label{sec:tqc107587}
\noindent Essential BR: $Ag@3f$ \\
\noindent RSI:
\begin{flalign*}
&\delta_{1}@3f\equiv -m(Ag)+m(Au)-m(Bg)+m(Bu) = -1,&
\end{flalign*}
\lstset{language=bash, keywordstyle=\color{blue!70}, basicstyle=\ttfamily, frame=shadowbox}
\begin{lstlisting}
Computed bands:  1 - 33
A : A1+ (1); A2- (1); A1+ (1); A2- (1); A2- (1); A1+ (1); A2- (1); A1+ (1);
    A2- (1); A1+ (1); A2- (1); A1+ (1); A2- (1); A3+ A3- (4); A1+ (1); A2- (1);
    A1+ (1); A3+ (2); A3- (2); A1+ (1); A3- (2); A3+ (2); A2- (1); A2- (1);
    A3- (2); [33] ;
GM: GM1+(1); GM2-(1); GM1+(1); GM2-(1); GM1+(1); GM2-(1); GM1+(1); GM2-(1);
    GM1+(1); GM2-(1); GM1+(1); GM2-(1); GM1+(1); GM3+GM3-(4); GM2-(1); GM1+(1);
    GM3+(2); GM1+(1); GM3-(2); GM2-(1); GM3+(2); GM3-(2); GM2-(1); GM1+(1);
    GM3+(2); [33] ;
H : H3  (2); H3  (2); H3  (2); H1  (1); H2  (1); H2  (1); H1  (1); H3  (2);
    H2  (1); H1  (1); H3  (2); H3  (2); H2  (1); H3  (2); H3  (2); H1  (1);
    H3  (2); H2  (1); H1  (1); H3  (2); H2  (1); H3  (2); [33] ;
K : K3  (2); K3  (2); K3  (2); K1  (1); K2  (1); K1  (1); K2  (1); K3  (2);
    K1  (1); K2  (1); K3  (2); K3  (2); K3  (2); K1  (1); K2  (1); K3  (2);
    K3  (2); K1  (1); K2  (1); K3  (2); K1  (1); K3  (2); [33] ;
L : L1+ L2- (2); L2- (1); L1+ (1); L2- (1); L1+ (1); L1+ (1); L2- (1); L1+ (1);
    L2- (1); L1+ (1); L2- (1); L2- (1); L1+ (1); L2- (1); L1+ L1- (2); L2+ (1);
    L2- (1); L1+ (1); L1+ (1); L2- (1); L2- (1); L1- (1); L2+ (1); L1+ (1);
    L2+ (1); L2- (1); L1- (1); L1+ (1); L2- (1); L1+ (1); L2+ (1); [33] ;
M : M1+ M2- (2); M2- (1); M1+ (1); M2- (1); M1+ (1); M2- (1); M1+ (1); M2- (1);
    M1+ (1); M2- (1); M1+ (1); M1+ (1); M2- (1); M1+ (1); M2- (1); M1- (1);
    M2+ (1); M1+ (1); M2- (1); M1+ (1); M2- (1); M1- (1); M1+ (1); M2+ (1);
    M2- (1); M1- (1); M2- (1); M2+ (1); M1+ (1); M1+ (1); M2- (1); M1- (1); [33];
\end{lstlisting}
\hyperref[tab:electride]{Back to the table}

\subsubsection*{65169 Te$_{2}$Ru}
\label{sec:tqc65169}
\noindent Essential BR: $A1g@4b$ \\
\noindent RSI:
\begin{flalign*}
&\delta_{1}@4b\equiv -m(2Eg)+m(2Eu) = 0,&
\\
&\delta_{2}@4b\equiv -m(A1g)+m(A1u) = -1,&
\end{flalign*}
\lstset{language=bash, keywordstyle=\color{blue!70}, basicstyle=\ttfamily, frame=shadowbox}
\begin{lstlisting}
Computed bands:  1 - 40
GM: GM1+(1); GM4+(3); GM4-(3); GM1-(1); GM4+(3); GM4-(3); GM4+(3); GM2-GM3-(2);
    GM2+GM3+(2); GM4+(3); GM1+(1); GM4+(3); GM2+GM3+(2); GM4-(3); GM4+(3);
    GM1+(1); GM4+(3); [40] ;
R : R1- R3- (4); R1+ R3+ (4); R1+ R3+ (4); R2+ R2+ (4); R1- R3- (4); R1+ R3+ (4);
    R2- R2- (4); R2+ R2+ (4); R1- R3- (4); R1+ R3+ (4); [40] ;
M : M1  M2  (4); M1  M2  (4); M1  M2  (4); M1  M2  (4); M1  M2  (4); M1  M2  (4);
    M1  M2  (4); M1  M2  (4); M1  M2  (4); M1  M2  (4); [40] ;
X : X2  (2); X1  (2); X2  (2); X1  (2); X1  (2); X2  (2); X1  (2); X1  (2);
    X2  (2); X2  (2); X1  (2); X2  (2); X2  (2); X1  (2); X1  (2); X2  (2);
    X1  (2); X2  (2); X2  (2); X1  (2); [40];
\end{lstlisting}
\hyperref[tab:electride]{Back to the table}

\subsubsection*{66002 CaGaGe}
\label{sec:tqc66002}
\noindent Essential BR: $A1'@2c$ \\
\noindent RSI:
\begin{flalign*}
&\delta_{1}@2c\equiv m(A1')+m(A2')-m(A2'')-m(A1'')-m(E')+m(E'') = 1,&
\end{flalign*}
\lstset{language=bash, keywordstyle=\color{blue!70}, basicstyle=\ttfamily, frame=shadowbox}
\begin{lstlisting}
Computed bands:  1 - 34
A : A1  (2); A1  (2); A1  (2); A3  (4); A1  (2); A3  (4); A1  (2); A1  (2);
    A1  (2); A1  (2); A3  (4); A3  (4); A1  (2); [34] ;
GM: GM1+GM3+(2); GM1+GM4-(2); GM2-GM4-(2); GM5-GM6-(4); GM3+(1); GM2-(1);
    GM5+GM6-(4); GM1+(1); GM4-(1); GM3+(1); GM2-(1); GM1+(1); GM4-(1); GM3+(1);
    GM2-(1); GM5+(2); GM6-(2); GM6+(2); GM1+(1); GM5-(2); GM4-(1); [34] ;
H : H3  (2); H3  (2); H3  (2); H1  H2  (4); H3  (2); H1  (2); H2  (2); H2  (2);
    H1  (2); H1  (2); H2  (2); H3  (2); H3  (2); H2  (2); H1  (2); H1  (2);
    [34] ;
K : K1  K2  (2); K1  K4  (2); K3  K4  (2); K5  K6  (4); K2  K3  (2); K5  (2);
    K5  (2); K5  (2); K6  (2); K5  (2); K6  (2); K1  (1); K4  (1); K2  (1);
    K3  (1); K5  (2); K6  (2); K5  (2); [34] ;
L : L1  (2); L1  (2); L1  (2); L1  (2); L2  (2); L1  (2); L1  (2); L2  (2);
    L1  (2); L1  (2); L1  (2); L1  (2); L1  (2); L2  (2); L2  (2); L1  (2);
    L1  (2); [34] ;
M : M1+ M3+ (2); M1+ M4- (2); M2- M4- (2); M2- M4- (2); M1- M3- (2); M2- M3+ (2);
    M1+ M4- (2); M2+ M3- (2); M4- (1); M1+ (1); M2- (1); M3+ (1); M1+ (1);
    M4- (1); M3+ (1); M2- (1); M4- (1); M4+ (1); M1- M2+ (2); M3- (1); M1+ (1);
    M2- (1); M1+ (1); M3+ (1); M4- (1); [34];
\end{lstlisting}
\hyperref[tab:electride]{Back to the table}

\subsubsection*{633674 ZrFeSi}
\label{sec:tqc633674}
\noindent Essential BR: $Ag@4a$ \\
\noindent RSI:
\begin{flalign*}
&\delta_{1}@4a\equiv -m(Ag)+m(Au) = -1,&
\end{flalign*}
\lstset{language=bash, keywordstyle=\color{blue!70}, basicstyle=\ttfamily, frame=shadowbox}

\hyperref[tab:electride]{Back to the table}

\subsubsection*{88213 TbCoSi}
\label{sec:tqc88213}
\noindent Essential BR: $Ag@4b$ \\
\noindent RSI:
\begin{flalign*}
&\delta_{1}@4b\equiv -m(Ag)+m(Au) = -1,&
\end{flalign*}
\lstset{language=bash, keywordstyle=\color{blue!70}, basicstyle=\ttfamily, frame=shadowbox}
\begin{lstlisting}
Computed bands:  1 - 44
GM: GM4+(1); GM1+(1); GM2-(1); GM2+(1); GM4+(1); GM3+(1); GM3-GM4-(2); GM2-(1);
    GM3-(1); GM1-(1); GM1+(1); GM1+(1); GM4+(1); GM3-(1); GM2-(1); GM3-(1);
    GM4+(1); GM2-(1); GM4+(1); GM1+(1); GM3+(1); GM1-(1); GM2-(1); GM4-(1);
    GM3-(1); GM2+(1); GM1+(1); GM3-(1); GM2-(1); GM2+(1); GM3+(1); GM3-(1);
    GM4+(1); GM1-(1); GM1+(1); GM4+(1); GM4-(1); GM4+(1); GM1+(1); GM3+(1);
    GM2+(1); GM2-(1); GM1+(1); [44] ;
R : R1  R2  (4); R1  R2  (4); R1  R2  (4); R1  R2  (4); R1  R2  (4); R1  R2  (4);
    R1  R2  (4); R1  R2  (4); R1  R2  (4); R1  R2  (4); R1  R2  (4); [44] ;
S : S1  S2  (4); S1  S2  (4); S1  S2  (4); S1  S2  (4); S1  S2  (4); S1  S2  (4);
    S1  S2  (4); S1  S2  (4); S1  S2  (4); S1  S2  (4); S1  S2  (4); [44] ;
T : T2  (2); T1  (2); T2  (2); T2  (2); T1  (2); T1  (2); T1  (2); T2  (2);
    T1  (2); T2  (2); T1  (2); T2  (2); T1  (2); T2  (2); T2  (2); T1  (2);
    T1  (2); T2  (2); T1  (2); T2  (2); T2  (2); T1  (2); [44] ;
U : U1+ U4+ (2); U2+ U3+ (2); U2- U3- (2); U1+ U4+ (2); U2- U3- (2); U1- U4- (2);
    U2- U3- (2); U1+ U4+ (2); U2- U3- (2); U1+ U4+ (2); U1- U4- (2); U2+ U3+ (2);
    U1+ U4+ (2); U2- U3- (2); U1- U4- (2); U1+ U4+ (2); U2- U3- (2); U2- U3- (2);
    U2+ U3+ (2); U2- U3- (2); U1+ U4+ (2); U1- U4- (2); [44] ;
X : X1  (2); X2  (2); X1  (2); X1  (2); X2  (2); X1  (2); X1  (2); X1  (2);
    X1  (2); X1  (2); X2  (2); X1  (2); X1  (2); X2  (2); X2  (2); X1  (2);
    X1  (2); X1  (2); X2  (2); X1  (2); X2  (2); X1  (2); [44] ;
Y : Y2  (2); Y1  (2); Y1  (2); Y2  (2); Y1  (2); Y2  (2); Y2  (2); Y1  (2);
    Y1  (2); Y2  (2); Y2  (2); Y1  (2); Y2  (2); Y1  (2); Y1  (2); Y2  (2);
    Y2  (2); Y1  (2); Y1  (2); Y2  (2); Y2  (2); Y1  (2); [44] ;
Z : Z1  (2); Z1  (2); Z2  (2); Z1  (2); Z2  (2); Z1  (2); Z1  (2); Z1  (2);
    Z1  (2); Z1  (2); Z2  (2); Z1  (2); Z2  (2); Z1  (2); Z2  (2); Z1  (2);
    Z1  (2); Z1  (2); Z2  (2); Z1  (2); Z2  (2); Z1  (2); [44];
\end{lstlisting}
\hyperref[tab:electride]{Back to the table}

\subsubsection*{391204 KPrTe$_{4}$}
\label{sec:tqc391204}
\lstset{language=bash, keywordstyle=\color{blue!70}, basicstyle=\ttfamily, frame=shadowbox}
\begin{lstlisting}
Computed bands:  1 - 44
A : A2  (2); A1  (2); A1  (2); A3  (2); A4  (2); A3  (2); A2  (2); A4  (2);
    A3  (2); A3  (2); A2  (2); A1  (2); A3  (2); A3  (2); A1  (2); A2  (2);
    A3  (2); A3  (2); A4  (2); A2  (2); A4  (2); A1  (2); [44] ;
GM: GM1+(1); GM1-(1); GM1+GM1-(2); GM3-(1); GM3+(1); GM5+(2); GM5-(2); GM3-(1);
    GM5+(2); GM3+(1); GM5-(2); GM1+(1); GM3-(1); GM2-(1); GM4+(1); GM5-(2);
    GM5+(2); GM4+(1); GM1+(1); GM2-(1); GM1-(1); GM1+(1); GM5+(2); GM5-(2);
    GM2-(1); GM3-(1); GM3+(1); GM3-(1); GM4+(1); GM5+(2); GM5-(2); GM2+(1);
    GM4-(1); [44] ;
M : M1  (2); M1  (2); M2  (2); M3  (2); M4  (2); M3  (2); M2  (2); M4  (2);
    M3  (2); M3  (2); M1  (2); M2  (2); M3  (2); M3  (2); M3  (2); M2  (2);
    M1  (2); M1  (2); M4  (2); M3  (2); M4  (2); M2  (2); [44] ;
Z : Z3- (1); Z3+ (1); Z1+ Z1- (2); Z1+ (1); Z1- (1); Z5- (2); Z5+ (2); Z3- (1);
    Z5+ (2); Z3+ (1); Z5- (2); Z3- (1); Z1+ (1); Z4+ (1); Z2- (1); Z5+ (2);
    Z5- (2); Z2- (1); Z4+ (1); Z3- (1); Z3+ (1); Z5- (2); Z3- (1); Z5+ (2);
    Z1+ (1); Z4+ (1); Z1- (1); Z1+ (1); Z2- (1); Z5- (2); Z5+ (2); Z4- (1);
    Z2+ (1); [44] ;
R : R2  (2); R1  (2); R1  (2); R2  (2); R1  (2); R2  (2); R1  (2); R2  (2);
    R2  (2); R1  (2); R2  (2); R1  (2); R2  (2); R1  (2); R2  (2); R1  (2);
    R2  (2); R1  (2); R2  (2); R1  (2); R2  (2); R1  (2); [44] ;
X : X1  (2); X1  (2); X2  (2); X1  (2); X2  (2); X2  (2); X1  (2); X2  (2);
    X1  (2); X2  (2); X1  (2); X2  (2); X1  (2); X2  (2); X1  (2); X2  (2);
    X1  (2); X2  (2); X1  (2); X2  (2); X1  (2); X2  (2); [44];
\end{lstlisting}
\hyperref[tab:electride]{Back to the table}

\subsubsection*{30692 SiP$_{2}$}
\label{sec:tqc30692}
\noindent Essential BR: $A1g@4b$ \\
\noindent RSI:
\begin{flalign*}
&\delta_{1}@4b\equiv -m(2Eg)+m(2Eu) = 0,&
\\
&\delta_{2}@4b\equiv -m(A1g)+m(A1u) = -1,&
\end{flalign*}
\lstset{language=bash, keywordstyle=\color{blue!70}, basicstyle=\ttfamily, frame=shadowbox}
\begin{lstlisting}
Computed bands:  1 - 28
GM: GM1+(1); GM4+(3); GM4-(3); GM1-(1); GM4+(3); GM4-(3); GM2-GM3-(2); GM1+(1);
    GM4+(3); GM4-(3); GM4+(3); GM2+GM3+(2); [28] ;
R : R1- R3- (4); R1+ R3+ (4); R1+ R3+ (4); R1- R3- (4); R2- R2- (4); R1- R3- (4);
    R2+ R2+ (4); [28] ;
M : M1  M2  (4); M1  M2  (4); M1  M2  (4); M1  M2  (4); M1  M2  (4); M1  M2  (4);
    M1  M2  (4); [28] ;
X : X2  (2); X1  (2); X1  (2); X2  (2); X1  (2); X2  (2); X1  (2); X2  (2);
    X2  (2); X1  (2); X1  (2); X1  (2); X2  (2); X2  (2); [28];
\end{lstlisting}
\hyperref[tab:electride]{Back to the table}

\subsubsection*{412793 RbNdTe$_{4}$}
\label{sec:tqc412793}
\lstset{language=bash, keywordstyle=\color{blue!70}, basicstyle=\ttfamily, frame=shadowbox}
\begin{lstlisting}
Computed bands:  1 - 44
A : A2  (2); A1  (2); A1  (2); A3  (2); A4  (2); A3  (2); A3  (2); A2  (2);
    A4  (2); A3  (2); A2  (2); A1  (2); A3  (2); A3  (2); A1  (2); A2  (2);
    A3  (2); A2  (2); A4  (2); A3  (2); A4  (2); A1  (2); [44] ;
GM: GM1+(1); GM1-(1); GM1+(1); GM1-(1); GM3-(1); GM3+(1); GM5+(2); GM5-(2);
    GM1+(1); GM3-(1); GM2-(1); GM5+(2); GM4+(1); GM3+(1); GM5-(2); GM3-(1);
    GM5-(2); GM5+(2); GM4+(1); GM1+(1); GM2-(1); GM1-(1); GM1+(1); GM5+(2);
    GM5-(2); GM2-(1); GM3-(1); GM3+(1); GM3-(1); GM4+(1); GM5+(2); GM5-(2);
    GM2+(1); GM4-(1); [44] ;
M : M1  (2); M1  (2); M2  (2); M3  (2); M4  (2); M3  (2); M3  (2); M2  (2);
    M4  (2); M3  (2); M1  (2); M2  (2); M3  (2); M3  (2); M2  (2); M1  M3  (4);
    M1  (2); M4  (2); M3  (2); M4  (2); M2  (2); [44] ;
Z : Z3- (1); Z3+ (1); Z1+ (1); Z1- (1); Z1+ (1); Z1- (1); Z5- (2); Z5+ (2);
    Z3- (1); Z1+ (1); Z4+ (1); Z5+ (2); Z2- (1); Z3+ (1); Z5- (2); Z3- (1);
    Z5- (2); Z5+ (2); Z2- (1); Z4+ (1); Z3- (1); Z3+ (1); Z5- (2); Z3- (1);
    Z5+ (2); Z1+ (1); Z4+ (1); Z1- (1); Z1+ (1); Z2- (1); Z5- (2); Z5+ (2);
    Z4- (1); Z2+ (1); [44] ;
R : R2  (2); R1  (2); R1  (2); R2  (2); R1  (2); R2  (2); R1  (2); R2  (2);
    R2  (2); R1  (2); R2  (2); R1  (2); R2  (2); R1  (2); R2  (2); R1  (2);
    R2  (2); R1  (2); R2  (2); R1  (2); R2  (2); R1  (2); [44] ;
X : X1  (2); X1  (2); X2  (2); X1  (2); X2  (2); X1  (2); X2  (2); X2  (2);
    X1  (2); X2  (2); X1  (2); X2  (2); X1  (2); X2  (2); X1  (2); X2  (2);
    X1  (2); X2  (2); X1  (2); X2  (2); X1  (2); X2  (2); [44];
\end{lstlisting}
\hyperref[tab:electride]{Back to the table}

\subsubsection*{611271 As$_{2}$Rh}
\label{sec:tqc611271}
\noindent Essential BR: $Ag@2d$ \\
\noindent RSI:
\begin{flalign*}
&\delta_{1}@2d\equiv -m(Ag)+m(Au) = -1,&
\end{flalign*}
\lstset{language=bash, keywordstyle=\color{blue!70}, basicstyle=\ttfamily, frame=shadowbox}

\hyperref[tab:electride]{Back to the table}

\subsubsection*{88167 ErGeIr}
\label{sec:tqc88167}
\noindent Essential BR: $Ag@4a$ \\
\noindent RSI:
\begin{flalign*}
&\delta_{1}@4a\equiv -m(Ag)+m(Au) = -1,&
\end{flalign*}
\lstset{language=bash, keywordstyle=\color{blue!70}, basicstyle=\ttfamily, frame=shadowbox}
\begin{lstlisting}
Computed bands:  1 - 44
GM: GM2-(1); GM4+(1); GM2+(1); GM4+(1); GM1+(1); GM3+(1); GM4-(1); GM1-(1);
    GM3-(1); GM2-(1); GM3-(1); GM1+(1); GM1+(1); GM3-(1); GM2-(1); GM4+(1);
    GM4+(1); GM3-(1); GM2-(1); GM4+(1); GM3+(1); GM1+(1); GM2-(1); GM4-(1);
    GM1-(1); GM3-(1); GM3-(1); GM2+(1); GM1+(1); GM3+(1); GM2-(1); GM1+(1);
    GM2+(1); GM1-(1); GM4+(1); GM3-(1); GM4+(1); GM4-(1); GM4+(1); GM1+(1);
    GM2+(1); GM2-(1); GM3+(1); GM1+(1); [44] ;
R : R1  R2  (4); R1  R2  (4); R1  R2  (4); R1  R2  (4); R1  R2  (4); R1  R2  (4);
    R1  R2  (4); R1  R2  (4); R1  R2  (4); R1  R2  (4); R1  R2  (4); [44] ;
S : S1  S2  (4); S1  S2  (4); S1  S2  (4); S1  S2  (4); S1  S2  (4); S1  S2  (4);
    S1  S2  (4); S1  S2  (4); S1  S2  (4); S1  S2  (4); S1  S2  (4); [44] ;
T : T1  (2); T1  (2); T2  (2); T1  (2); T2  (2); T2  (2); T1  (2); T2  (2);
    T2  (2); T1  (2); T2  (2); T1  (2); T2  (2); T1  (2); T2  (2); T1  (2);
    T2  (2); T2  (2); T1  (2); T1  (2); T1  (2); T2  (2); [44] ;
U : U1+ U2- U3- U4+ (4); U1- U4- (2); U2- U3- (2); U2+ U3+ (2); U1+ U4+ (2);
    U1+ U4+ (2); U2- U3- (2); U1+ U4+ (2); U2- U3- (2); U2- U3- (2); U1- U4- (2);
    U2+ U3+ (2); U1+ U4+ (2); U2+ U3+ (2); U2- U3- (2); U1+ U4+ (2); U1- U4- (2);
    U2- U3- (2); U1+ U4+ (2); U1+ U4+ (2); U2+ U3+ (2); [44] ;
X : X1  (2); X2  (2); X1  (2); X1  (2); X2  (2); X1  (2); X1  (2); X1  (2);
    X1  (2); X1  (2); X2  (2); X1  (2); X1  (2); X2  (2); X2  (2); X1  (2);
    X1  (2); X1  (2); X2  (2); X2  (2); X1  (2); X1  (2); [44] ;
Y : Y1  (2); Y2  (2); Y1  (2); Y1  (2); Y2  (2); Y2  (2); Y2  (2); Y1  (2);
    Y1  (2); Y2  (2); Y2  (2); Y1  (2); Y2  (2); Y1  (2); Y2  (2); Y1  (2);
    Y1  (2); Y2  (2); Y1  (2); Y2  (2); Y1  (2); Y2  (2); [44] ;
Z : Z1  (2); Z2  (2); Z1  (2); Z2  (2); Z1  (2); Z1  (2); Z1  (2); Z1  (2);
    Z1  (2); Z1  (2); Z1  (2); Z2  (2); Z2  (2); Z1  (2); Z1  (2); Z2  (2);
    Z1  (2); Z1  (2); Z2  (2); Z1  (2); Z1  (2); Z2  (2); [44];
\end{lstlisting}
\hyperref[tab:electride]{Back to the table}

\subsubsection*{610769 LaAs$_{2}$}
\label{sec:tqc610769}
\noindent Essential BR: $Ag@2b$ \\
\noindent RSI:
\begin{flalign*}
&\delta_{1}@2b\equiv -m(Ag)+m(Au) = -1,&
\end{flalign*}
\lstset{language=bash, keywordstyle=\color{blue!70}, basicstyle=\ttfamily, frame=shadowbox}

\hyperref[tab:electride]{Back to the table}

\subsubsection*{280231 LaTeAs}
\label{sec:tqc280231}
\noindent Essential BR: $Ag@4a$ \\
\noindent RSI:
\begin{flalign*}
&\delta_{1}@4a\equiv -m(Ag)+m(Au) = -1,&
\end{flalign*}
\lstset{language=bash, keywordstyle=\color{blue!70}, basicstyle=\ttfamily, frame=shadowbox}
\begin{lstlisting}
Computed bands:  1 - 44
GM: GM1+(1); GM3-(1); GM2-GM4+(2); GM4+(1); GM4+(1); GM2-(1); GM3-(1); GM1+(1);
    GM1+(1); GM2-(1); GM3-(1); GM2+(1); GM4-(1); GM3+(1); GM1-(1); GM1+(1);
    GM4+(1); GM1+(1); GM3-(1); GM2-(1); GM4+(1); GM2-(1); GM3-(1); GM4+(1);
    GM3-(1); GM1+(1); GM2-(1); GM3+(1); GM2+(1); GM4+(1); GM2-(1); GM1+(1);
    GM3+(1); GM3-(1); GM1-(1); GM4-(1); GM2-(1); GM4+(1); GM2+(1); GM1+(1);
    GM3-(1); GM4+(1); GM1+(1); [44] ;
R : R1  R2  (4); R1  R2  (4); R1  R2  (4); R1  R2  (4); R1  R2  (4); R1  R2  (4);
    R1  R2  (4); R1  R2  (4); R1  R2  (4); R1  R2  (4); R1  R2  (4); [44] ;
S : S1  S2  (4); S1  S2  (4); S1  S2  (4); S1  S2  (4); S1  S2  (4); S1  S2  (4);
    S1  S2  (4); S1  S2  (4); S1  S2  (4); S1  S2  (4); S1  S2  (4); [44] ;
T : T1  (2); T2  (2); T2  (2); T1  (2); T2  (2); T1  (2); T2  (2); T1  (2);
    T1  (2); T1  (2); T2  (2); T2  (2); T2  (2); T1  (2); T1  (2); T2  (2);
    T1  (2); T2  (2); T1  (2); T2  (2); T2  (2); T1  (2); [44] ;
U : U1+ U2- U3- U4+ (4); U1+ U4+ (2); U2- U3- (2); U1+ U2- U3- U4+ (4);
    U1- U4- (2); U2+ U3+ (2); U1+ U4+ (2); U2- U3- (2); U1+ U4+ (2); U2- U3- (2);
    U2- U3- (2); U1+ U4+ (2); U2+ U3+ (2); U2- U3- (2); U2- U3- (2); U1- U4- (2);
    U1+ U4+ (2); U1+ U4+ (2); U2+ U3+ (2); U1+ U4+ (2); [44] ;
X : X1  (2); X1  (2); X1  (2); X1  (2); X1  (2); X1  (2); X2  (2); X2  (2);
    X1  (2); X1  (2); X1  (2); X1  (2); X1  (2); X2  (2); X1  (2); X1  (2);
    X1  (2); X2  (2); X1  (2); X2  (2); X1  (2); X1  (2); [44] ;
Y : Y2  (2); Y1  (2); Y2  (2); Y1  (2); Y1  (2); Y1  Y2  (4); Y2  (2); Y1  (2);
    Y2  (2); Y1  (2); Y2  (2); Y1  (2); Y2  (2); Y1  (2); Y2  (2); Y2  (2);
    Y1  (2); Y1  (2); Y2  (2); Y2  (2); Y1  (2); [44] ;
Z : Z1  (2); Z1  (2); Z1  (2); Z1  (2); Z1  (2); Z1  (2); Z2  (2); Z2  (2);
    Z1  (2); Z1  (2); Z1  (2); Z1  (2); Z1  (2); Z1  (2); Z2  (2); Z1  (2);
    Z1  (2); Z2  (2); Z1  (2); Z2  (2); Z1  (2); Z1  (2); [44];
\end{lstlisting}
\hyperref[tab:electride]{Back to the table}

\subsubsection*{88166 HoGeIr}
\label{sec:tqc88166}
\noindent Essential BR: $Ag@4a$ \\
\noindent RSI:
\begin{flalign*}
&\delta_{1}@4a\equiv -m(Ag)+m(Au) = -1,&
\end{flalign*}
\lstset{language=bash, keywordstyle=\color{blue!70}, basicstyle=\ttfamily, frame=shadowbox}
\begin{lstlisting}
Computed bands:  1 - 44
GM: GM2-(1); GM4+(1); GM4+(1); GM2+(1); GM1+(1); GM3+(1); GM4-(1); GM1-(1);
    GM3-(1); GM2-(1); GM3-(1); GM1+(1); GM1+(1); GM3-(1); GM2-(1); GM4+(1);
    GM4+(1); GM3-(1); GM2-(1); GM1+(1); GM4+(1); GM3+(1); GM2-(1); GM4-(1);
    GM1-(1); GM3-(1); GM3-(1); GM2+(1); GM2+(1); GM3+(1); GM1+(1); GM1+GM2-(2);
    GM1-GM4+(2); GM3-(1); GM4+(1); GM4+(1); GM1+(1); GM4-(1); GM2+(1); GM2-(1);
    GM3+(1); GM1+(1); [44] ;
R : R1  R2  (4); R1  R2  (4); R1  R2  (4); R1  R2  (4); R1  R2  (4); R1  R2  (4);
    R1  R2  (4); R1  R2  (4); R1  R2  (4); R1  R2  (4); R1  R2  (4); [44] ;
S : S1  S2  (4); S1  S2  (4); S1  S2  (4); S1  S2  (4); S1  S2  (4); S1  S2  (4);
    S1  S2  (4); S1  S2  (4); S1  S2  (4); S1  S2  (4); S1  S2  (4); [44] ;
T : T1  (2); T1  (2); T2  (2); T1  (2); T2  (2); T2  (2); T2  (2); T1  (2);
    T2  (2); T1  (2); T2  (2); T1  (2); T2  (2); T1  (2); T2  (2); T1  (2);
    T2  (2); T1  (2); T2  (2); T1  (2); T1  (2); T2  (2); [44] ;
U : U1+ U4+ (2); U2- U3- (2); U1- U4- (2); U2- U3- (2); U2+ U3+ (2); U1+ U4+ (2);
    U1+ U4+ (2); U2- U3- (2); U1+ U4+ (2); U2- U3- (2); U2- U3- (2); U1- U4- (2);
    U2+ U3+ (2); U1+ U4+ (2); U2+ U3+ (2); U2- U3- (2); U1+ U4+ (2); U1- U4- (2);
    U1+ U4+ (2); U2- U3- (2); U1+ U4+ (2); U2+ U3+ (2); [44] ;
X : X1  (2); X1  (2); X2  (2); X1  (2); X2  (2); X1  (2); X1  (2); X1  (2);
    X1  (2); X1  (2); X2  (2); X1  (2); X1  (2); X2  (2); X2  (2); X1  (2);
    X1  (2); X1  (2); X2  (2); X2  (2); X1  (2); X1  (2); [44] ;
Y : Y1  (2); Y2  (2); Y1  (2); Y1  (2); Y2  (2); Y2  (2); Y2  (2); Y1  (2);
    Y2  (2); Y1  (2); Y2  (2); Y1  (2); Y2  (2); Y1  (2); Y2  (2); Y1  (2);
    Y1  (2); Y2  (2); Y1  (2); Y2  (2); Y1  (2); Y2  (2); [44] ;
Z : Z1  (2); Z2  (2); Z1  (2); Z2  (2); Z1  (2); Z1  (2); Z1  (2); Z1  (2);
    Z1  (2); Z1  (2); Z1  (2); Z2  (2); Z2  (2); Z1  (2); Z2  (2); Z1  (2);
    Z1  (2); Z1  (2); Z2  (2); Z1  (2); Z1  (2); Z2  (2); [44];
\end{lstlisting}
\hyperref[tab:electride]{Back to the table}

\subsubsection*{291479 Si}
\label{sec:tqc291479}
\lstset{language=bash, keywordstyle=\color{blue!70}, basicstyle=\ttfamily, frame=shadowbox}
\begin{lstlisting}
Computed bands:  1 - 24
GM: GM1+(1); GM3+(1); GM4-(1); GM2-(1); GM1+(1); GM3+(1); GM1+(1); GM4-(1);
    GM2-(1); GM4-(1); GM3+(1); GM2-(1); GM4-(1); GM1+(1); GM3+(1); GM3-(1);
    GM2+(1); GM1-(1); GM4+(1); GM1+(1); GM2+(1); GM1+(1); GM4+(1); GM3+(1);
    [24] ;
T : T1  (2); T1  (2); T1  (2); T1  (2); T1  (2); T1  (2); T2  (2); T1  (2);
    T1  (2); T2  (2); T1  (2); T2  (2); [24] ;
Y : Y1+ (1); Y4- (1); Y3+ (1); Y2- (1); Y4- (1); Y2- (1); Y1+ (1); Y3+ (1);
    Y1+ (1); Y4- (1); Y3+ (1); Y3- (1); Y1- (1); Y4- (1); Y2- (1); Y1+ (1);
    Y2+ (1); Y3+ (1); Y1+ (1); Y4- (1); Y2- (1); Y4+ (1); Y3- (1); Y1- (1);
    [24] ;
Z : Z1  (2); Z1  (2); Z1  (2); Z1  (2); Z1  (2); Z1  (2); Z1  (2); Z2  (2);
    Z1  (2); Z2  (2); Z1  (2); Z2  (2); [24] ;
R : R1  (2); R1  (2); R1  (2); R1  (2); R1  (2); R1  (2); R1  (2); R1  (2);
    R1  (2); R1  (2); R1  (2); R1  (2); [24] ;
S : S1+ (1); S2+ (1); S2- (1); S1+ (1); S1- (1); S2- (1); S2+ (1); S2- (1);
    S1- (1); S1+ (1); S1- (1); S2+ (1); S2- (1); S1+ (1); S2- (1); S1+ (1);
    S1- (1); S2+ (1); S1+ (1); S1- (1); S2+ (1); S2- (1); S1+ (1); S2+ (1); [24];
\end{lstlisting}
\hyperref[tab:electride]{Back to the table}

\subsubsection*{412708 LuGeIr}
\label{sec:tqc412708}
\noindent Essential BR: $Ag@4a$ \\
\noindent RSI:
\begin{flalign*}
&\delta_{1}@4a\equiv -m(Ag)+m(Au) = -1,&
\end{flalign*}
\lstset{language=bash, keywordstyle=\color{blue!70}, basicstyle=\ttfamily, frame=shadowbox}
\begin{lstlisting}
Computed bands:  1 - 44
GM: GM2-(1); GM4+(1); GM4+(1); GM1+GM2+(2); GM3+(1); GM4-(1); GM1-(1); GM3-(1);
    GM2-(1); GM3-(1); GM1+(1); GM1+(1); GM4+(1); GM3-(1); GM2-(1); GM3-(1);
    GM4+(1); GM2-(1); GM4+(1); GM1+(1); GM3+(1); GM4-(1); GM2-(1); GM1-(1);
    GM3-(1); GM2+(1); GM3-(1); GM1+GM3+(2); GM2+(1); GM2-(1); GM1+(1); GM4+(1);
    GM1-(1); GM4+(1); GM3-(1); GM4-(1); GM4+(1); GM1+(1); GM2+(1); GM2-(1);
    GM3+(1); GM1+(1); [44] ;
R : R1  R2  (4); R1  R2  (4); R1  R2  (4); R1  R2  (4); R1  R2  (4); R1  R2  (4);
    R1  R2  (4); R1  R2  (4); R1  R2  (4); R1  R2  (4); R1  R2  (4); [44] ;
S : S1  S2  (4); S1  S2  (4); S1  S2  (4); S1  S2  (4); S1  S2  (4); S1  S2  (4);
    S1  S2  (4); S1  S2  (4); S1  S2  (4); S1  S2  (4); S1  S2  (4); [44] ;
T : T1  (2); T1  (2); T2  (2); T1  (2); T2  (2); T2  (2); T1  (2); T2  (2);
    T2  (2); T1  (2); T2  (2); T1  (2); T2  (2); T1  (2); T2  (2); T1  (2);
    T2  (2); T1  (2); T2  (2); T1  (2); T1  (2); T2  (2); [44] ;
U : U2- U3- (2); U1+ U4+ (2); U1- U4- (2); U2- U3- (2); U2+ U3+ (2); U1+ U4+ (2);
    U1+ U4+ (2); U2- U3- (2); U2- U3- (2); U1+ U4+ (2); U1- U4- (2); U2- U3- (2);
    U2+ U3+ (2); U1+ U4+ (2); U2+ U3+ (2); U2- U3- (2); U1+ U4+ (2); U1- U4- (2);
    U2- U3- (2); U1+ U4+ (2); U1+ U4+ (2); U2+ U3+ (2); [44] ;
X : X1  (2); X1  (2); X2  (2); X1  (2); X2  (2); X1  (2); X1  (2); X1  (2);
    X1  (2); X1  (2); X2  (2); X1  (2); X2  (2); X1  (2); X2  (2); X1  (2);
    X1  (2); X1  (2); X2  (2); X2  (2); X1  (2); X1  (2); [44] ;
Y : Y1  (2); Y2  (2); Y1  (2); Y2  (2); Y1  (2); Y2  (2); Y2  (2); Y1  (2);
    Y1  (2); Y2  (2); Y2  (2); Y1  (2); Y2  (2); Y1  (2); Y1  (2); Y2  (2);
    Y2  (2); Y1  (2); Y1  (2); Y2  (2); Y1  (2); Y2  (2); [44] ;
Z : Z1  (2); Z1  (2); Z2  (2); Z2  (2); Z1  (2); Z1  (2); Z1  (2); Z1  (2);
    Z1  (2); Z1  (2); Z1  (2); Z2  (2); Z2  (2); Z1  (2); Z2  (2); Z1  (2);
    Z1  (2); Z1  (2); Z2  (2); Z1  (2); Z1  (2); Z2  (2); [44];
\end{lstlisting}
\hyperref[tab:electride]{Back to the table}

\subsubsection*{79596 ScSiRh}
\label{sec:tqc79596}
\noindent Essential BR: $Ag@4a$ \\
\noindent RSI:
\begin{flalign*}
&\delta_{1}@4a\equiv -m(Ag)+m(Au) = -1,&
\end{flalign*}
\lstset{language=bash, keywordstyle=\color{blue!70}, basicstyle=\ttfamily, frame=shadowbox}
\begin{lstlisting}
Computed bands:  1 - 32
GM: GM1+(1); GM4+(1); GM3-(1); GM2-(1); GM3-(1); GM4+(1); GM2-(1); GM4+(1);
    GM1+(1); GM4-(1); GM3+(1); GM2-(1); GM1-(1); GM3-(1); GM2+(1); GM3-(1);
    GM3+(1); GM2+(1); GM1+(1); GM1+(1); GM2-(1); GM1-(1); GM4+(1); GM4+(1);
    GM3-(1); GM4-(1); GM4+(1); GM1+(1); GM2+(1); GM2-(1); GM3+(1); GM1+(1);
    [32] ;
R : R1  R2  (4); R1  R2  (4); R1  R2  (4); R1  R2  (4); R1  R2  (4); R1  R2  (4);
    R1  R2  (4); R1  R2  (4); [32] ;
S : S1  S2  (4); S1  S2  (4); S1  S2  (4); S1  S2  (4); S1  S2  (4); S1  S2  (4);
    S1  S2  (4); S1  S2  (4); [32] ;
T : T1  (2); T2  (2); T2  (2); T1  (2); T2  (2); T1  (2); T2  (2); T1  (2);
    T2  (2); T1  (2); T2  (2); T1  (2); T2  (2); T1  (2); T2  (2); T1  (2);
    [32] ;
U : U1+ U4+ (2); U2- U3- (2); U2- U3- (2); U1+ U4+ (2); U2- U3- (2); U1- U4- (2);
    U2+ U3+ (2); U2+ U3+ (2); U1+ U4+ (2); U2- U3- (2); U1+ U4+ (2); U1- U4- (2);
    U1+ U4+ (2); U2- U3- (2); U1+ U4+ (2); U2+ U3+ (2); [32] ;
X : X1  (2); X1  (2); X1  (2); X1  (2); X2  (2); X1  (2); X2  (2); X1  (2);
    X2  (2); X1  (2); X1  (2); X1  (2); X2  (2); X1  (2); X2  (2); X1  (2);
    [32] ;
Y : Y2  (2); Y1  (2); Y1  (2); Y2  (2); Y2  (2); Y1  (2); Y2  (2); Y1  (2);
    Y2  (2); Y1  (2); Y2  (2); Y1  (2); Y1  (2); Y1  (2); Y2  (2); Y2  (2);
    [32] ;
Z : Z1  (2); Z1  (2); Z1  (2); Z1  (2); Z1  (2); Z2  (2); Z2  (2); Z2  (2);
    Z1  (2); Z1  (2); Z1  (2); Z1  (2); Z2  (2); Z1  (2); Z1  (2); Z2  (2); [32];
\end{lstlisting}
\hyperref[tab:electride]{Back to the table}

\subsubsection*{237303 SrNiGe}
\label{sec:tqc237303}
\noindent Essential BR: $Ag@4a$ \\
\noindent RSI:
\begin{flalign*}
&\delta_{1}@4a\equiv -m(Ag)+m(Au) = -1,&
\end{flalign*}
\lstset{language=bash, keywordstyle=\color{blue!70}, basicstyle=\ttfamily, frame=shadowbox}

\hyperref[tab:electride]{Back to the table}

\subsubsection*{633405 FeSbTe}
\label{sec:tqc633405}
\noindent Essential BR: $Ag@2d$ \\
\noindent RSI:
\begin{flalign*}
&\delta_{1}@2d\equiv -m(Ag)+m(Au) = -1,&
\end{flalign*}
\lstset{language=bash, keywordstyle=\color{blue!70}, basicstyle=\ttfamily, frame=shadowbox}

\hyperref[tab:electride]{Back to the table}

\subsubsection*{636748 TmGeIr}
\label{sec:tqc636748}
\noindent Essential BR: $Ag@4b$ \\
\noindent RSI:
\begin{flalign*}
&\delta_{1}@4b\equiv -m(Ag)+m(Au) = -1,&
\end{flalign*}
\lstset{language=bash, keywordstyle=\color{blue!70}, basicstyle=\ttfamily, frame=shadowbox}
\begin{lstlisting}
Computed bands:  1 - 44
GM: GM2-(1); GM4+(1); GM4+(1); GM1+(1); GM2+(1); GM3+(1); GM4-(1); GM1-(1);
    GM2-(1); GM3-(1); GM3-(1); GM1+(1); GM1+(1); GM4+(1); GM3-(1); GM2-(1);
    GM3-(1); GM4+(1); GM1+(1); GM2-(1); GM4+(1); GM3+(1); GM4-(1); GM2-(1);
    GM1-(1); GM3-(1); GM2+(1); GM3-(1); GM3+(1); GM2+(1); GM1+(1); GM2-(1);
    GM1+(1); GM1-(1); GM4+(1); GM3-(1); GM4+(1); GM4-(1); GM4+(1); GM1+(1);
    GM2+(1); GM3+(1); GM2-(1); GM1+(1); [44] ;
R : R1  R2  (4); R1  R2  (4); R1  R2  (4); R1  R2  (4); R1  R2  (4); R1  R2  (4);
    R1  R2  (4); R1  R2  (4); R1  R2  (4); R1  R2  (4); R1  R2  (4); [44] ;
S : S1  S2  (4); S1  S2  (4); S1  S2  (4); S1  S2  (4); S1  S2  (4); S1  S2  (4);
    S1  S2  (4); S1  S2  (4); S1  S2  (4); S1  S2  (4); S1  S2  (4); [44] ;
T : T2  (2); T2  (2); T1  (2); T2  (2); T1  (2); T1  (2); T2  (2); T1  (2);
    T1  (2); T2  (2); T1  (2); T2  (2); T1  (2); T2  (2); T1  (2); T2  (2);
    T1  (2); T2  (2); T1  (2); T2  (2); T2  (2); T1  (2); [44] ;
U : U2- U3- (2); U1+ U4+ (2); U1+ U2+ U3+ U4+ (4); U1- U4- (2); U2- U3- (2);
    U2- U3- (2); U1+ U4+ (2); U1+ U4+ (2); U2- U3- (2); U2+ U3+ (2); U1+ U4+ (2);
    U1- U4- (2); U1- U4- (2); U2- U3- (2); U1+ U4+ (2); U2- U3- (2); U2+ U3+ (2);
    U1+ U4+ (2); U2- U3- (2); U2- U3- (2); U1- U4- (2); [44] ;
X : X1  (2); X1  (2); X1  (2); X2  (2); X2  (2); X1  (2); X1  (2); X1  (2);
    X1  (2); X1  (2); X2  (2); X1  (2); X2  (2); X1  (2); X2  (2); X1  (2);
    X1  (2); X1  (2); X2  (2); X1  (2); X2  (2); X1  (2); [44] ;
Y : Y1  (2); Y2  (2); Y1  (2); Y1  (2); Y2  (2); Y2  (2); Y2  (2); Y1  (2);
    Y1  (2); Y2  (2); Y2  (2); Y1  (2); Y2  (2); Y1  (2); Y2  (2); Y1  (2);
    Y2  (2); Y1  (2); Y1  (2); Y2  (2); Y1  (2); Y2  (2); [44] ;
Z : Z1  (2); Z1  (2); Z2  (2); Z2  (2); Z1  (2); Z1  (2); Z1  (2); Z1  (2);
    Z1  (2); Z1  (2); Z1  (2); Z2  (2); Z2  (2); Z1  (2); Z2  (2); Z1  (2);
    Z1  (2); Z1  (2); Z2  (2); Z1  (2); Z1  (2); Z2  (2); [44];
\end{lstlisting}
\hyperref[tab:electride]{Back to the table}

\subsubsection*{428475 ScGeRh}
\label{sec:tqc428475}
\noindent Essential BR: $Ag@4a$ \\
\noindent RSI:
\begin{flalign*}
&\delta_{1}@4a\equiv -m(Ag)+m(Au) = -1,&
\end{flalign*}
\lstset{language=bash, keywordstyle=\color{blue!70}, basicstyle=\ttfamily, frame=shadowbox}
\begin{lstlisting}
Computed bands:  1 - 32
GM: GM1+(1); GM3-(1); GM4+(1); GM2-(1); GM3-(1); GM4+(1); GM4+(1); GM2-(1);
    GM1+(1); GM4-(1); GM3+(1); GM2-(1); GM1-(1); GM3-(1); GM2+(1); GM3-(1);
    GM3+(1); GM2+(1); GM1+(1); GM1+(1); GM2-(1); GM1-(1); GM4+(1); GM4+(1);
    GM3-(1); GM4-(1); GM4+(1); GM1+(1); GM2+(1); GM2-(1); GM3+(1); GM1+(1);
    [32] ;
R : R1  R2  (4); R1  R2  (4); R1  R2  (4); R1  R2  (4); R1  R2  (4); R1  R2  (4);
    R1  R2  (4); R1  R2  (4); [32] ;
S : S1  S2  (4); S1  S2  (4); S1  S2  (4); S1  S2  (4); S1  S2  (4); S1  S2  (4);
    S1  S2  (4); S1  S2  (4); [32] ;
T : T1  (2); T2  (2); T2  (2); T1  (2); T2  (2); T1  (2); T2  (2); T1  (2);
    T2  (2); T1  (2); T2  (2); T1  (2); T2  (2); T1  (2); T2  (2); T1  (2);
    [32] ;
U : U1+ U4+ (2); U2- U3- (2); U2- U3- (2); U1+ U4+ (2); U2- U3- (2); U1- U4- (2);
    U2+ U3+ (2); U2+ U3+ (2); U1+ U4+ (2); U2- U3- (2); U1+ U4+ (2); U1- U4- (2);
    U1+ U4+ (2); U2- U3- (2); U1+ U4+ (2); U2+ U3+ (2); [32] ;
X : X1  (2); X1  (2); X1  (2); X1  (2); X2  (2); X1  (2); X2  (2); X1  (2);
    X2  (2); X1  (2); X1  (2); X1  (2); X2  (2); X1  (2); X2  (2); X1  (2);
    [32] ;
Y : Y2  (2); Y1  (2); Y2  (2); Y1  (2); Y2  (2); Y1  (2); Y2  (2); Y1  (2);
    Y2  (2); Y1  (2); Y2  (2); Y1  (2); Y1  (2); Y1  (2); Y2  (2); Y2  (2);
    [32] ;
Z : Z1  (2); Z1  (2); Z1  (2); Z1  (2); Z1  (2); Z2  (2); Z2  (2); Z2  (2);
    Z1  (2); Z1  (2); Z1  (2); Z1  (2); Z2  (2); Z1  (2); Z1  (2); Z2  (2); [32];
\end{lstlisting}
\hyperref[tab:electride]{Back to the table}

\subsubsection*{152569 YbGaGe}
\label{sec:tqc152569}
\noindent Essential BR: $A1'@2c$ \\
\noindent RSI:
\begin{flalign*}
&\delta_{1}@2c\equiv m(A1')+m(A2')-m(A2'')-m(A1'')-m(E')+m(E'') = 1,&
\end{flalign*}
\lstset{language=bash, keywordstyle=\color{blue!70}, basicstyle=\ttfamily, frame=shadowbox}
\begin{lstlisting}
Computed bands:  1 - 30
A : A1  (2); A3  (4); A1  (2); A3  (4); A1  (2); A1  (2); A1  (2); A1  (2);
    A3  (4); A3  (4); A1  (2); [30] ;
GM: GM2-(1); GM4-(1); GM5-GM6-(4); GM3+(1); GM2-(1); GM5+GM6-(4); GM1+(1);
    GM4-(1); GM3+(1); GM2-(1); GM1+(1); GM4-(1); GM3+(1); GM2-(1); GM5+(2);
    GM6-(2); GM6+(2); GM1+(1); GM5-(2); GM4-(1); [30] ;
H : H3  (2); H1  H2  (4); H3  (2); H1  (2); H2  (2); H2  (2); H1  (2); H1  (2);
    H2  (2); H3  (2); H3  (2); H1  (2); H2  (2); H1  (2); [30] ;
K : K3  (1); K4  (1); K5  K6  (4); K2  (1); K3  (1); K5  (2); K5  (2); K5  (2);
    K6  (2); K5  (2); K6  (2); K1  (1); K4  (1); K2  (1); K3  (1); K5  (2);
    K5  (2); K6  (2); [30] ;
L : L1  (2); L1  (2); L2  (2); L1  (2); L1  (2); L2  (2); L1  (2); L1  (2);
    L1  (2); L1  (2); L1  (2); L2  (2); L2  (2); L1  (2); L1  (2); [30] ;
M : M2- (1); M4- (1); M2- M4- (2); M1- M3- (2); M3+ (1); M2- (1); M1+ M4- (2);
    M2+ M3- (2); M1+ (1); M3+ (1); M4- (1); M2- (1); M1+ (1); M4- (1); M3+ (1);
    M2- (1); M4- (1); M4+ (1); M1+ (1); M1- (1); M2+ (1); M1+ (1); M3- (1);
    M2- (1); M3+ (1); M4- (1); [30];
\end{lstlisting}
\hyperref[tab:electride]{Back to the table}

\subsubsection*{163343 CeZnGe}
\label{sec:tqc163343}
\lstset{language=bash, keywordstyle=\color{blue!70}, basicstyle=\ttfamily, frame=shadowbox}
\begin{lstlisting}
Computed bands:  1 - 54
A : A1  (2); A1  (2); A1  (2); A1  (2); A3  (4); A3  (4); A1  (2); A1  (2);
    A3  (4); A3  (4); A3  (4); A3  (4); A1  (2); A1  (2); A1  (2); A1  (2);
    A3  (4); A3  (4); A1  (2); [54] ;
GM: GM1+GM3+(2); GM1+GM4-(2); GM2-(1); GM4-(1); GM3+(1); GM5-GM6-(4); GM2-(1);
    GM5+GM6-(4); GM1+(1); GM3+(1); GM4-(1); GM2-(1); GM6+(2); GM5-(2); GM5+(2);
    GM6-(2); GM5+(2); GM6-(2); GM6+(2); GM1+(1); GM5-(2); GM4-(1); GM3+(1);
    GM2-(1); GM1+(1); GM4-(1); GM3+(1); GM5+(2); GM6+(2); GM6-(2); GM1+(1);
    GM2-(1); GM5-(2); GM4-(1); [54] ;
H : H3  (2); H3  (2); H1  H2  H3  (6); H2  (2); H1  (2); H3  (2); H1  (2);
    H2  (2); H3  (2); H3  (2); H1  (2); H1  (2); H3  (2); H2  (2); H2  (2);
    H3  (2); H1  (2); H2  (2); H2  (2); H1  (2); H3  (2); H3  (2); H2  (2);
    H1  (2); H1  (2); [54] ;
K : K1  K2  (2); K1  K4  (2); K3  (1); K5  (2); K6  (2); K4  (1); K2  (1);
    K5  (2); K5  (2); K3  (1); K6  (2); K5  (2); K1  (1); K4  (1); K2  (1);
    K3  (1); K5  (2); K6  (2); K1  (1); K5  (2); K6  (2); K2  K4  (2); K3  (1);
    K6  (2); K5  (2); K6  (2); K5  (2); K1  (1); K4  (1); K2  (1); K3  (1);
    K5  (2); K6  (2); K5  (2); [54] ;
L : L1  (2); L1  (2); L1  (2); L1  (2); L2  (2); L1  (2); L1  (2); L2  (2);
    L1  (2); L1  (2); L2  (2); L1  L2  (4); L1  (2); L1  (2); L2  (2); L2  (2);
    L1  (2); L1  (2); L1  (2); L1  (2); L1  (2); L2  (2); L1  (2); L2  (2);
    L1  (2); L1  (2); [54] ;
M : M1+ M3+ (2); M1+ M4- (2); M4- (1); M2- (1); M2- (1); M4- (1); M1- M3- (2);
    M1+ (1); M4- (1); M3+ (1); M2- (1); M2+ M3- (2); M3+ (1); M1+ (1); M2- (1);
    M4- (1); M1- M4+ (2); M1+ (1); M2+ M3- (2); M3+ (1); M4- (1); M2- (1);
    M1+ (1); M4- (1); M2+ (1); M4+ (1); M1- M3+ M3- (3); M2- (1); M1+ (1);
    M3+ (1); M4- (1); M2- (1); M1+ (1); M3+ (1); M4- (1); M2- (1); M4+ (1);
    M1+ (1); M2+ (1); M1- (1); M3+ (1); M3- (1); M1+ (1); M2- (1); M4- (1);
    M4- (1); [54];
\end{lstlisting}
\hyperref[tab:electride]{Back to the table}

\subsubsection*{24203 As$_{2}$Pt}
\label{sec:tqc24203}
\noindent Essential BR: $A1g@4b$ \\
\noindent RSI:
\begin{flalign*}
&\delta_{1}@4b\equiv -m(2Eg)+m(2Eu) = 0,&
\\
&\delta_{2}@4b\equiv -m(A1g)+m(A1u) = -1,&
\end{flalign*}
\lstset{language=bash, keywordstyle=\color{blue!70}, basicstyle=\ttfamily, frame=shadowbox}
\begin{lstlisting}
Computed bands:  1 - 40
GM: GM1+(1); GM4+(3); GM4-(3); GM1-(1); GM4+(3); GM4+(3); GM2+GM3+(2); GM4+(3);
    GM4-(3); GM4+(3); GM2-GM3-(2); GM2+GM3+(2); GM1+(1); GM4+(3); GM1+(1);
    GM4+(3); GM4-(3); [40] ;
R : R1- R3- (4); R1+ R3+ (4); R1+ R3+ (4); R2+ R2+ (4); R2+ R2+ (4); R1+ R3+ (4);
    R1+ R3+ (4); R1- R3- (4); R2- R2- (4); R1- R3- (4); [40] ;
M : M1  M2  (4); M1  M2  (4); M1  M2  (4); M1  M2  (4); M1  M2  (4); M1  M2  (4);
    M1  M2  (4); M1  M2  (4); M1  M2  (4); M1  M2  (4); [40] ;
X : X2  (2); X1  (2); X2  (2); X1  (2); X1  (2); X2  (2); X1  (2); X2  (2);
    X1  (2); X1  (2); X2  (2); X2  (2); X1  (2); X2  (2); X2  (2); X1  (2);
    X1  (2); X2  (2); X2  (2); X1  (2); [40];
\end{lstlisting}
\hyperref[tab:electride]{Back to the table}

\subsubsection*{24801 SiAs$_{2}$}
\label{sec:tqc24801}
\noindent Essential BR: $A1g@4b$ \\
\noindent RSI:
\begin{flalign*}
&\delta_{1}@4b\equiv -m(2Eg)+m(2Eu) = 0,&
\\
&\delta_{2}@4b\equiv -m(A1g)+m(A1u) = -1,&
\end{flalign*}
\lstset{language=bash, keywordstyle=\color{blue!70}, basicstyle=\ttfamily, frame=shadowbox}
\begin{lstlisting}
Computed bands:  1 - 28
GM: GM1+(1); GM4+(3); GM4-(3); GM1-(1); GM4+(3); GM4-(3); GM2-GM3-(2); GM1+(1);
    GM4+(3); GM4-(3); GM4+(3); GM2+GM3+(2); [28] ;
R : R1- R3- (4); R1+ R3+ (4); R1+ R3+ (4); R1- R3- (4); R2- R2- (4); R1- R3- (4);
    R2+ R2+ (4); [28] ;
M : M1  M2  (4); M1  M2  (4); M1  M2  (4); M1  M2  (4); M1  M2  (4); M1  M2  (4);
    M1  M2  (4); [28] ;
X : X2  (2); X1  (2); X1  (2); X2  (2); X1  (2); X2  (2); X1  (2); X2  (2);
    X2  (2); X1  (2); X1  (2); X1  (2); X2  (2); X2  (2); [28];
\end{lstlisting}
\hyperref[tab:electride]{Back to the table}

\subsubsection*{428472 ScCoGe}
\label{sec:tqc428472}
\noindent Essential BR: $Ag@4a$ \\
\noindent RSI:
\begin{flalign*}
&\delta_{1}@4a\equiv -m(Ag)+m(Au) = -1,&
\end{flalign*}
\lstset{language=bash, keywordstyle=\color{blue!70}, basicstyle=\ttfamily, frame=shadowbox}
\begin{lstlisting}
Computed bands:  1 - 32
GM: GM1+(1); GM4+(1); GM3-(1); GM2-(1); GM4+(1); GM3-(1); GM4+(1); GM2-(1);
    GM1+(1); GM3+(1); GM4-(1); GM1-(1); GM2-(1); GM3-(1); GM2+(1); GM3-(1);
    GM1+(1); GM2+(1); GM3+(1); GM2-(1); GM1+(1); GM1-(1); GM3-(1); GM4+(1);
    GM4+(1); GM1+(1); GM4+(1); GM4-(1); GM2+(1); GM2-(1); GM3+(1); GM1+(1);
    [32] ;
R : R1  R2  (4); R1  R2  (4); R1  R2  (4); R1  R2  (4); R1  R2  (4); R1  R2  (4);
    R1  R2  (4); R1  R2  (4); [32] ;
S : S1  S2  (4); S1  S2  (4); S1  S2  (4); S1  S2  (4); S1  S2  (4); S1  S2  (4);
    S1  S2  (4); S1  S2  (4); [32] ;
T : T1  (2); T2  (2); T2  (2); T1  (2); T2  (2); T1  (2); T2  (2); T1  (2);
    T1  (2); T2  (2); T2  (2); T1  (2); T2  (2); T1  (2); T2  (2); T1  (2);
    [32] ;
U : U1+ U4+ (2); U2- U3- (2); U1+ U4+ (2); U2- U3- (2); U1- U4- (2); U2- U3- (2);
    U2+ U3+ (2); U1+ U4+ (2); U2+ U3+ (2); U2- U3- (2); U1+ U4+ (2); U1- U4- (2);
    U1+ U4+ (2); U2- U3- (2); U1+ U4+ (2); U2+ U3+ (2); [32] ;
X : X1  (2); X1  (2); X1  (2); X1  (2); X2  (2); X1  (2); X1  (2); X2  (2);
    X2  (2); X1  (2); X1  (2); X1  (2); X2  (2); X1  (2); X2  (2); X1  (2);
    [32] ;
Y : Y2  (2); Y1  (2); Y1  (2); Y2  (2); Y2  (2); Y1  (2); Y2  (2); Y1  (2);
    Y1  (2); Y2  (2); Y1  (2); Y2  (2); Y1  (2); Y1  (2); Y2  (2); Y2  (2);
    [32] ;
Z : Z1  (2); Z1  (2); Z1  (2); Z1  (2); Z1  (2); Z2  (2); Z2  (2); Z1  (2);
    Z2  (2); Z1  (2); Z1  (2); Z1  (2); Z2  (2); Z1  (2); Z2  (2); Z1  (2); [32];
\end{lstlisting}
\hyperref[tab:electride]{Back to the table}

\subsubsection*{93221 TbSiIr}
\label{sec:tqc93221}
\noindent Essential BR: $Ag@4a$ \\
\noindent RSI:
\begin{flalign*}
&\delta_{1}@4a\equiv -m(Ag)+m(Au) = -1,&
\end{flalign*}
\lstset{language=bash, keywordstyle=\color{blue!70}, basicstyle=\ttfamily, frame=shadowbox}
\begin{lstlisting}
Computed bands:  1 - 44
GM: GM2-(1); GM4+(1); GM1+(1); GM4+(1); GM2+(1); GM3+(1); GM4-(1); GM2-(1);
    GM3-(1); GM1-(1); GM3-(1); GM1+(1); GM1+(1); GM4+(1); GM3-(1); GM2-(1);
    GM3-(1); GM4+(1); GM2-(1); GM3+(1); GM1+(1); GM4+(1); GM4-(1); GM2-(1);
    GM3-(1); GM1-(1); GM2+(1); GM3-(1); GM1+(1); GM3+(1); GM2+(1); GM2-(1);
    GM1-(1); GM1+(1); GM4+(1); GM4+(1); GM3-(1); GM4-(1); GM4+(1); GM1+(1);
    GM2+(1); GM3+(1); GM2-(1); GM1+(1); [44] ;
R : R1  R2  (4); R1  R2  (4); R1  R2  (4); R1  R2  (4); R1  R2  (4); R1  R2  (4);
    R1  R2  (4); R1  R2  (4); R1  R2  (4); R1  R2  (4); R1  R2  (4); [44] ;
S : S1  S2  (4); S1  S2  (4); S1  S2  (4); S1  S2  (4); S1  S2  (4); S1  S2  (4);
    S1  S2  (4); S1  S2  (4); S1  S2  (4); S1  S2  (4); S1  S2  (4); [44] ;
T : T1  (2); T1  (2); T2  (2); T1  (2); T2  (2); T2  (2); T1  (2); T2  (2);
    T2  (2); T1  (2); T2  (2); T1  (2); T2  (2); T1  (2); T1  (2); T2  (2);
    T2  (2); T1  (2); T2  (2); T1  (2); T1  (2); T2  (2); [44] ;
U : U2- U3- (2); U1+ U4+ (2); U1- U4- (2); U2- U3- (2); U2+ U3+ (2); U1+ U4+ (2);
    U1+ U4+ (2); U2- U3- (2); U2- U3- (2); U1+ U4+ (2); U1- U4- (2); U2+ U3+ (2);
    U2- U3- (2); U1+ U4+ (2); U2+ U3+ (2); U2- U3- (2); U1+ U4+ (2); U1- U4- (2);
    U1+ U4+ (2); U2- U3- (2); U1+ U4+ (2); U2+ U3+ (2); [44] ;
X : X1  (2); X2  (2); X1  (2); X1  (2); X2  (2); X1  (2); X1  (2); X1  (2);
    X1  (2); X1  (2); X2  (2); X1  (2); X2  (2); X1  (2); X2  (2); X1  (2);
    X1  (2); X1  (2); X2  (2); X2  (2); X1  (2); X1  (2); [44] ;
Y : Y1  Y2  (4); Y1  (2); Y2  (2); Y1  (2); Y2  (2); Y2  (2); Y1  (2); Y1  (2);
    Y2  (2); Y2  (2); Y1  (2); Y2  (2); Y1  (2); Y1  Y2  (4); Y2  (2); Y1  (2);
    Y1  (2); Y2  (2); Y2  (2); Y1  (2); [44] ;
Z : Z1  (2); Z1  (2); Z2  (2); Z1  (2); Z2  (2); Z1  (2); Z1  (2); Z1  (2);
    Z1  (2); Z1  (2); Z2  (2); Z1  (2); Z2  (2); Z1  (2); Z2  (2); Z1  (2);
    Z1  (2); Z1  (2); Z2  (2); Z1  (2); Z2  (2); Z1  (2); [44];
\end{lstlisting}
\hyperref[tab:electride]{Back to the table}

\subsubsection*{636724 PrGeIr}
\label{sec:tqc636724}
\noindent Essential BR: $Ag@4a$ \\
\noindent RSI:
\begin{flalign*}
&\delta_{1}@4a\equiv -m(Ag)+m(Au) = -1,&
\end{flalign*}
\lstset{language=bash, keywordstyle=\color{blue!70}, basicstyle=\ttfamily, frame=shadowbox}

\hyperref[tab:electride]{Back to the table}

\subsubsection*{420415 ScCoSi}
\label{sec:tqc420415}
\noindent Essential BR: $Ag@4a$ \\
\noindent RSI:
\begin{flalign*}
&\delta_{1}@4a\equiv -m(Ag)+m(Au) = -1,&
\end{flalign*}
\lstset{language=bash, keywordstyle=\color{blue!70}, basicstyle=\ttfamily, frame=shadowbox}
\begin{lstlisting}
Computed bands:  1 - 32
GM: GM1+(1); GM4+(1); GM3-(1); GM2-(1); GM3-(1); GM4+(1); GM4+(1); GM2-(1);
    GM1+(1); GM3+(1); GM4-(1); GM1-(1); GM2-(1); GM3-(1); GM2+(1); GM3-(1);
    GM1+(1); GM2+(1); GM3+(1); GM2-(1); GM1+(1); GM1-(1); GM4+(1); GM3-(1);
    GM4+(1); GM1+(1); GM4+(1); GM4-(1); GM2+(1); GM2-(1); GM3+(1); GM1+(1);
    [32] ;
R : R1  R2  (4); R1  R2  (4); R1  R2  (4); R1  R2  (4); R1  R2  (4); R1  R2  (4);
    R1  R2  (4); R1  R2  (4); [32] ;
S : S1  S2  (4); S1  S2  (4); S1  S2  (4); S1  S2  (4); S1  S2  (4); S1  S2  (4);
    S1  S2  (4); S1  S2  (4); [32] ;
T : T1  (2); T2  (2); T2  (2); T1  (2); T2  (2); T1  (2); T2  (2); T1  (2);
    T1  (2); T2  (2); T2  (2); T1  (2); T2  (2); T1  (2); T2  (2); T1  (2);
    [32] ;
U : U1+ U4+ (2); U2- U3- (2); U1+ U4+ (2); U2- U3- (2); U1- U4- (2); U2+ U3+ (2);
    U2- U3- (2); U1+ U4+ (2); U2+ U3+ (2); U2- U3- (2); U1+ U4+ (2); U1- U4- (2);
    U1+ U4+ (2); U2- U3- (2); U1+ U4+ (2); U2+ U3+ (2); [32] ;
X : X1  (2); X1  (2); X1  (2); X1  (2); X2  (2); X1  (2); X1  (2); X2  (2);
    X2  (2); X1  (2); X1  (2); X1  (2); X2  (2); X1  (2); X2  (2); X1  (2);
    [32] ;
Y : Y2  (2); Y1  (2); Y1  (2); Y2  (2); Y2  (2); Y1  (2); Y2  (2); Y1  (2);
    Y1  (2); Y2  (2); Y2  (2); Y1  (2); Y1  (2); Y1  (2); Y2  (2); Y2  (2);
    [32] ;
Z : Z1  (2); Z1  (2); Z1  (2); Z1  (2); Z1  (2); Z2  (2); Z2  (2); Z1  (2);
    Z2  (2); Z1  (2); Z1  (2); Z1  (2); Z2  (2); Z1  (2); Z2  (2); Z1  (2); [32];
\end{lstlisting}
\hyperref[tab:electride]{Back to the table}

\subsubsection*{648028 PRuSe}
\label{sec:tqc648028}
\noindent Essential BR: $Ag@2d$ \\
\noindent RSI:
\begin{flalign*}
&\delta_{1}@2d\equiv -m(Ag)+m(Au) = -1,&
\end{flalign*}
\lstset{language=bash, keywordstyle=\color{blue!70}, basicstyle=\ttfamily, frame=shadowbox}

\hyperref[tab:electride]{Back to the table}

\subsubsection*{610996 NdAs$_{2}$}
\label{sec:tqc610996}
\noindent Essential BR: $Ag@2b$ \\
\noindent RSI:
\begin{flalign*}
&\delta_{1}@2b\equiv -m(Ag)+m(Au) = -1,&
\end{flalign*}
\lstset{language=bash, keywordstyle=\color{blue!70}, basicstyle=\ttfamily, frame=shadowbox}

\hyperref[tab:electride]{Back to the table}

\subsubsection*{648826 PdSe$_{2}$}
\label{sec:tqc648826}
\noindent Essential BR: $Ag@4b$ \\
\noindent RSI:
\begin{flalign*}
&\delta_{1}@4b\equiv -m(Ag)+m(Au) = -1,&
\end{flalign*}
\lstset{language=bash, keywordstyle=\color{blue!70}, basicstyle=\ttfamily, frame=shadowbox}
\begin{lstlisting}
Computed bands:  1 - 44
GM: GM1+(1); GM3+(1); GM2+(1); GM4+(1); GM4-(1); GM2-(1); GM3-(1); GM1-(1);
    GM2+(1); GM4+(1); GM3+(1); GM1+(1); GM4+(1); GM3+(1); GM1+(1); GM2+(1);
    GM3-(1); GM4-(1); GM2-(1); GM1-(1); GM2+(1); GM1+(1); GM1-(1); GM4+(1);
    GM3+(1); GM4-(1); GM4+(1); GM2+(1); GM1+(1); GM3+(1); GM2+GM3+(2); GM3-(1);
    GM4+(1); GM1+(1); GM2-(1); GM1+(1); GM4+(1); GM3+(1); GM2+(1); GM3+(1);
    GM2+(1); GM4+(1); GM1+(1); [44] ;
R : R1- R1- (4); R1+ R1+ (4); R1+ R1+ (4); R1+ R1+ (4); R1- R1- (4); R1+ R1+ (4);
    R1+ R1+ (4); R1+ R1+ (4); R1- R1- (4); R1+ R1+ (4); R1- R1- (4); [44] ;
S : S1  S2  (4); S1  S2  (4); S1  S2  (4); S1  S2  (4); S1  S2  (4); S1  S2  (4);
    S1  S2  (4); S1  S2  (4); S1  S2  (4); S1  S2  (4); S1  S2  (4); [44] ;
T : T1  T2  (4); T1  T2  (4); T1  T2  (4); T1  T2  (4); T1  T2  (4); T1  T2  (4);
    T1  T2  (4); T1  T2  (4); T1  T2  (4); T1  T2  (4); T1  T2  (4); [44] ;
U : U1  U2  (4); U1  U2  (4); U1  U2  (4); U1  U2  (4); U1  U2  (4); U1  U2  (4);
    U1  U2  (4); U1  U2  (4); U1  U2  (4); U1  U2  (4); U1  U2  (4); [44] ;
X : X1  (2); X2  (2); X2  (2); X1  (2); X1  (2); X2  (2); X2  (2); X1  (2);
    X2  (2); X1  (2); X1  (2); X2  (2); X1  (2); X2  (2); X1  (2); X2  (2);
    X2  (2); X1  (2); X2  (2); X1  (2); X2  (2); X1  (2); [44] ;
Y : Y2  (2); Y1  (2); Y2  (2); Y1  (2); Y1  (2); Y2  (2); Y1  (2); Y2  (2);
    Y2  (2); Y1  (2); Y2  (2); Y1  (2); Y2  (2); Y1  (2); Y1  (2); Y2  (2);
    Y1  (2); Y2  (2); Y2  (2); Y1  (2); Y2  (2); Y1  (2); [44] ;
Z : Z1  (2); Z2  (2); Z1  (2); Z2  (2); Z2  (2); Z1  (2); Z2  (2); Z1  (2);
    Z2  (2); Z1  (2); Z2  (2); Z1  (2); Z2  (2); Z1  (2); Z2  (2); Z1  (2);
    Z1  (2); Z2  (2); Z2  (2); Z1  (2); Z1  (2); Z2  (2); [44];
\end{lstlisting}
\hyperref[tab:electride]{Back to the table}

\subsubsection*{633093 FePSe}
\label{sec:tqc633093}
\noindent Essential BR: $Ag@2d$ \\
\noindent RSI:
\begin{flalign*}
&\delta_{1}@2d\equiv -m(Ag)+m(Au) = -1,&
\end{flalign*}
\lstset{language=bash, keywordstyle=\color{blue!70}, basicstyle=\ttfamily, frame=shadowbox}

\hyperref[tab:electride]{Back to the table}

\subsubsection*{38316 CoP$_{2}$}
\label{sec:tqc38316}
\noindent Essential BR: $Ag@2d$ \\
\noindent RSI:
\begin{flalign*}
&\delta_{1}@2d\equiv -m(Ag)+m(Au) = -1,&
\end{flalign*}
\lstset{language=bash, keywordstyle=\color{blue!70}, basicstyle=\ttfamily, frame=shadowbox}

\hyperref[tab:electride]{Back to the table}

\subsubsection*{421335 GdS$_{2}$}
\label{sec:tqc421335}
\noindent Essential BR: $Ag@2c$ \\
\noindent RSI:
\begin{flalign*}
&\delta_{1}@2c\equiv -m(Ag)+m(Au) = -1,&
\end{flalign*}
\lstset{language=bash, keywordstyle=\color{blue!70}, basicstyle=\ttfamily, frame=shadowbox}

\hyperref[tab:electride]{Back to the table}

\subsubsection*{409183 SmP$_{5}$}
\label{sec:tqc409183}
\lstset{language=bash, keywordstyle=\color{blue!70}, basicstyle=\ttfamily, frame=shadowbox}

\hyperref[tab:electride]{Back to the table}

\subsubsection*{82721 ReTeS}
\label{sec:tqc82721}
\lstset{language=bash, keywordstyle=\color{blue!70}, basicstyle=\ttfamily, frame=shadowbox}
\begin{lstlisting}
Computed bands:  1 - 38
GM: GM1 (1); GM4 (3); GM1 (1); GM4 (3); GM1 (1); GM4 (3); GM3 (2); GM1 (1);
    GM4 (3); GM5 (3); GM3 (2); GM1 (1); GM4 (3); GM3 (2); GM5 (3); GM4 (3);
    GM4 (3); [38] ;
X : X1  (1); X3  (1); X5  (2); X3  (1); X1  (1); X5  (2); X3  (1); X1  (1);
    X5  (2); X3  (1); X5  (2); X4  (1); X1  (1); X5  (2); X2  (1); X1  (1);
    X3  (1); X4  (1); X2  (1); X1  (1); X5  (2); X3  (1); X2  (1); X1  (1);
    X1  X5  (3); X3  (1); X5  (2); X5  (2); [38] ;
L : L1  (1); L1  (1); L3  (2); L1  (1); L1  (1); L3  (2); L1  (1); L3  (2);
    L1  (1); L3  (2); L1  (1); L3  (2); L1  (1); L3  (2); L2  (1); L3  (2);
    L1  (1); L3  (2); L1  (1); L3  (2); L1  (1); L3  (2); L2  (1); L3  (2);
    L1  (1); L3  (2); [38] ;
W : W1  (1); W3  (1); W4  (1); W2  (1); W4  (1); W2  (1); W3  (1); W1  (1);
    W3  (1); W1  (1); W2  (1); W4  (1); W3  (1); W4  (1); W2  (1); W1  (1);
    W3  (1); W2  (1); W4  (1); W1  (1); W2  (1); W3  (1); W4  (1); W1  (1);
    W4  (1); W1  (1); W3  (1); W2  (1); W1  (1); W2  (1); W1  (1); W4  (1);
    W2  (1); W3  (1); W2  (1); W4  (1); W1  (1); W3  (1); [38];
\end{lstlisting}
\hyperref[tab:electride]{Back to the table}

\subsubsection*{166387 SrGaSn}
\label{sec:tqc166387}
\noindent Essential BR: $A1'@2c$ \\
\noindent RSI:
\begin{flalign*}
&\delta_{1}@2c\equiv m(A1')+m(A2')-m(A2'')-m(A1'')-m(E')+m(E'') = 1,&
\end{flalign*}
\lstset{language=bash, keywordstyle=\color{blue!70}, basicstyle=\ttfamily, frame=shadowbox}
\begin{lstlisting}
Computed bands:  1 - 34
A : A1  (2); A1  (2); A1  (2); A3  (4); A1  (2); A3  (4); A1  (2); A1  (2);
    A1  (2); A1  (2); A3  (4); A3  (4); A1  (2); [34] ;
GM: GM1+GM3+(2); GM1+GM4-(2); GM2-(1); GM4-(1); GM5-GM6-(4); GM3+(1); GM2-(1);
    GM5+GM6-(4); GM1+(1); GM4-(1); GM3+(1); GM2-(1); GM1+(1); GM4-(1); GM3+(1);
    GM2-(1); GM5+(2); GM6-(2); GM1+(1); GM6+(2); GM5-(2); GM4-(1); [34] ;
H : H3  (2); H3  (2); H3  (2); H1  H2  (4); H3  (2); H1  (2); H2  (2); H2  (2);
    H1  (2); H1  (2); H2  (2); H3  (2); H3  (2); H1  (2); H2  (2); H1  (2);
    [34] ;
K : K1  K2  (2); K1  K4  (2); K3  K4  (2); K5  K6  (4); K2  K3  (2); K5  (2);
    K5  (2); K5  (2); K6  (2); K5  (2); K6  (2); K1  (1); K4  (1); K2  (1);
    K3  (1); K5  (2); K5  (2); K6  (2); [34] ;
L : L1  (2); L1  (2); L1  (2); L1  (2); L2  (2); L1  (2); L1  (2); L2  (2);
    L1  (2); L1  (2); L1  (2); L1  (2); L1  (2); L2  (2); L2  (2); L1  (2);
    L1  (2); [34] ;
M : M1+ M3+ (2); M1+ M4- (2); M2- M4- (2); M2- M4- (2); M1- M3- (2); M1+ M4- (2);
    M2- M3+ (2); M2+ M3- (2); M1+ (1); M3+ (1); M4- (1); M2- (1); M1+ (1);
    M4- (1); M3+ (1); M2- (1); M4- (1); M1+ (1); M4+ (1); M1- (1); M1+ (1);
    M2+ (1); M3- (1); M2- (1); M3+ (1); M4- (1); [34];
\end{lstlisting}
\hyperref[tab:electride]{Back to the table}

\subsubsection*{358 NdP$_{5}$}
\label{sec:tqc358}
\lstset{language=bash, keywordstyle=\color{blue!70}, basicstyle=\ttfamily, frame=shadowbox}

\hyperref[tab:electride]{Back to the table}

\subsubsection*{620416 CdSe$_{2}$}
\label{sec:tqc620416}
\noindent Essential BR: $A1g@4b$ \\
\noindent RSI:
\begin{flalign*}
&\delta_{1}@4b\equiv -m(2Eg)+m(2Eu) = 0,&
\\
&\delta_{2}@4b\equiv -m(A1g)+m(A1u) = -1,&
\end{flalign*}
\lstset{language=bash, keywordstyle=\color{blue!70}, basicstyle=\ttfamily, frame=shadowbox}
\begin{lstlisting}
Computed bands:  1 - 48
GM: GM1+(1); GM4+(3); GM4-(3); GM1-(1); GM4+(3); GM2+GM3+(2); GM4+(3); GM4+(3);
    GM2+GM3+(2); GM4+(3); GM1+(1); GM4+(3); GM4+(3); GM4-(3); GM1+(1); GM4+(3);
    GM2-GM3-(2); GM4-(3); GM4+(3); GM2+GM3+(2); [48] ;
R : R1- R3- (4); R1+ R3+ (4); R2+ R2+ (4); R1+ R3+ (4); R2+ R2+ (4); R1+ R3+ (4);
    R1+ R3+ (4); R1+ R3+ (4); R1- R3- (4); R2- R2- (4); R1- R3- (4); R2+ R2+ (4);
    [48] ;
M : M1  M2  (4); M1  M2  (4); M1  M2  (4); M1  M2  (4); M1  M2  (4); M1  M2  (4);
    M1  M2  (4); M1  M2  (4); M1  M2  (4); M1  M2  (4); M1  M2  (4); M1  M2  (4);
    [48] ;
X : X2  (2); X1  (2); X2  (2); X1  (2); X2  (2); X1  (2); X1  (2); X2  (2);
    X1  (2); X1  X2  (4); X2  (2); X2  (2); X1  (2); X1  (2); X2  (2); X1  (2);
    X2  (2); X1  (2); X2  (2); X2  (2); X1  (2); X1  (2); X2  (2); [48];
\end{lstlisting}
\hyperref[tab:electride]{Back to the table}

\subsubsection*{240755 IrN$_{2}$}
\label{sec:tqc240755}
\noindent Essential BR: $Ag@2a$ \\
\noindent RSI:
\begin{flalign*}
&\delta_{1}@2a\equiv -m(Ag)+m(Au) = -1,&
\end{flalign*}
\lstset{language=bash, keywordstyle=\color{blue!70}, basicstyle=\ttfamily, frame=shadowbox}

\hyperref[tab:electride]{Back to the table}

\subsubsection*{20664 KGa$_{3}$}
\label{sec:tqc20664}
\lstset{language=bash, keywordstyle=\color{blue!70}, basicstyle=\ttfamily, frame=shadowbox}
\begin{lstlisting}
Computed bands:  1 - 27
GM: GM1 (1); GM2 (1); GM1 (1); GM5 (2); GM1 (1); GM2 (1); GM5 (2); GM2 (1);
    GM5 (2); GM1 (1); GM2 (1); GM1 (1); GM5 (2); GM2 (1); GM1 (1); GM5 (2);
    GM1 (1); GM2 (1); GM4 (1); GM2 (1); GM5 (2); [27] ;
M : M1  (1); M2  (1); M2  (1); M5  (2); M1  (1); M1  (1); M2  (1); M5  (2);
    M5  (2); M1  (1); M2  (1); M2  (1); M5  (2); M1  (1); M2  (1); M5  (2);
    M1  (1); M1  (1); M5  (2); M2  (1); M4  (1); [27] ;
P : P2  (1); P1  (1); P3  (1); P2  (1); P4  (1); P1  (1); P3  (1); P4  (1);
    P3  (1); P2  (1); P1  (1); P4  (1); P1  (1); P4  (1); P2  (1); P3  (1);
    P4  (1); P1  (1); P3  (1); P2  (1); P4  (1); P2  (1); P3  (1); P1  (1);
    P2  (1); P4  (1); P3  (1); [27] ;
X : X1  (1); X2  (1); X3  (1); X1  (1); X4  (1); X2  (1); X3  (1); X3  (1);
    X4  (1); X1  (1); X2  (1); X4  (1); X1  (1); X3  (1); X2  (1); X4  (1);
    X3  (1); X1  (1); X4  (1); X3  (1); X1  (1); X2  (1); X3  (1); X2  (1);
    X4  (1); X1  (1); X4  (1); [27] ;
N : N1  (1); N1  (1); N1  (1); N1  (1); N2  (1); N1  (1); N1  (1); N1  (1);
    N2  (1); N1  (1); N1  (1); N2  (1); N1  (1); N1  (1); N1  (1); N1  (1);
    N2  (1); N1  (1); N1  (1); N2  (1); N1  (1); N1  (1); N2  (1); N1  (1);
    N2  (1); N1  (1); N1  (1); [27];
\end{lstlisting}
\hyperref[tab:electride]{Back to the table}

\subsubsection*{616892 BiOsSe}
\label{sec:tqc616892}
\noindent Essential BR: $Ag@2d$ \\
\noindent RSI:
\begin{flalign*}
&\delta_{1}@2d\equiv -m(Ag)+m(Au) = -1,&
\end{flalign*}
\lstset{language=bash, keywordstyle=\color{blue!70}, basicstyle=\ttfamily, frame=shadowbox}

\hyperref[tab:electride]{Back to the table}

\subsubsection*{409184 DyP$_{5}$}
\label{sec:tqc409184}
\lstset{language=bash, keywordstyle=\color{blue!70}, basicstyle=\ttfamily, frame=shadowbox}

\hyperref[tab:electride]{Back to the table}

\subsubsection*{633086 FePS}
\label{sec:tqc633086}
\noindent Essential BR: $Ag@2d$ \\
\noindent RSI:
\begin{flalign*}
&\delta_{1}@2d\equiv -m(Ag)+m(Au) = -1,&
\end{flalign*}
\lstset{language=bash, keywordstyle=\color{blue!70}, basicstyle=\ttfamily, frame=shadowbox}

\hyperref[tab:electride]{Back to the table}

\subsubsection*{652213 ZnSe$_{2}$}
\label{sec:tqc652213}
\noindent Essential BR: $A1g@4b$ \\
\noindent RSI:
\begin{flalign*}
&\delta_{1}@4b\equiv -m(2Eg)+m(2Eu) = 0,&
\\
&\delta_{2}@4b\equiv -m(A1g)+m(A1u) = -1,&
\end{flalign*}
\lstset{language=bash, keywordstyle=\color{blue!70}, basicstyle=\ttfamily, frame=shadowbox}
\begin{lstlisting}
Computed bands:  1 - 48
GM: GM1+(1); GM4+(3); GM4-(3); GM1-(1); GM4+(3); GM4+(3); GM2+GM3+(2); GM4+(3);
    GM2+GM3+(2); GM4+(3); GM1+(1); GM4+(3); GM4+(3); GM4-(3); GM1+(1);
    GM2-GM3-(2); GM4+(3); GM4-(3); GM4+(3); GM2+GM3+(2); [48] ;
R : R1- R3- (4); R1+ R3+ (4); R1+ R3+ (4); R2+ R2+ (4); R2+ R2+ (4); R1+ R3+ (4);
    R1+ R3+ (4); R1+ R3+ (4); R1- R3- (4); R2- R2- (4); R1- R3- (4); R2+ R2+ (4);
    [48] ;
M : M1  M2  (4); M1  M2  (4); M1  M2  (4); M1  M2  (4); M1  M2  (4); M1  M2  (4);
    M1  M2  (4); M1  M2  (4); M1  M2  (4); M1  M2  (4); M1  M2  (4); M1  M2  (4);
    [48] ;
X : X2  (2); X1  (2); X2  (2); X1  (2); X1  (2); X2  (2); X2  (2); X1  (2);
    X2  (2); X1  (2); X1  (2); X2  (2); X2  (2); X1  (2); X1  (2); X2  (2);
    X1  (2); X2  (2); X1  (2); X2  (2); X2  (2); X1  (2); X1  (2); X2  (2); [48];
\end{lstlisting}
\hyperref[tab:electride]{Back to the table}

\subsubsection*{409186 TmP$_{5}$}
\label{sec:tqc409186}
\lstset{language=bash, keywordstyle=\color{blue!70}, basicstyle=\ttfamily, frame=shadowbox}

\hyperref[tab:electride]{Back to the table}

\subsubsection*{34450 CuAsS}
\label{sec:tqc34450}
\noindent Essential BR: $Ag@4b$ \\
\noindent RSI:
\begin{flalign*}
&\delta_{1}@4b\equiv -m(Ag)+m(Au) = -1,&
\end{flalign*}
\lstset{language=bash, keywordstyle=\color{blue!70}, basicstyle=\ttfamily, frame=shadowbox}
\begin{lstlisting}
Computed bands:  1 - 44
GM: GM1+(1); GM4+(1); GM2-(1); GM3-(1); GM1+(1); GM4+(1); GM2-(1); GM3-(1);
    GM4+(1); GM3-(1); GM1+(1); GM3-(1); GM2-(1); GM1+(1); GM4+(1); GM3+(1);
    GM2+(1); GM4-(1); GM2-(1); GM1-(1); GM1+(1); GM4+(1); GM3+(1); GM2+(1);
    GM4+(1); GM3-(1); GM2-(1); GM1-(1); GM1+(1); GM2+(1); GM4-(1); GM3-(1);
    GM2-GM3+(2); GM1+(1); GM3-(1); GM4+(1); GM2-(1); GM1-(1); GM2+(1); GM4-(1);
    GM4+(1); GM1+(1); GM3+(1); [44] ;
R : R1  R2  (4); R1  R2  (4); R1  R2  (4); R1  R2  (4); R1  R2  (4); R1  R2  (4);
    R1  R2  (4); R1  R2  (4); R1  R2  (4); R1  R2  (4); R1  R2  (4); [44] ;
S : S1  S2  (4); S1  S2  (4); S1  S2  (4); S1  S2  (4); S1  S2  (4); S1  S2  (4);
    S1  S2  (4); S1  S2  (4); S1  S2  (4); S1  S2  (4); S1  S2  (4); [44] ;
T : T1  (2); T2  (2); T2  (2); T1  (2); T2  (2); T1  (2); T1  (2); T1  (2);
    T2  (2); T2  (2); T1  (2); T2  (2); T2  (2); T1  (2); T1  T2  (4); T1  (2);
    T2  (2); T1  (2); T2  (2); T2  (2); T1  (2); [44] ;
U : U2- U3- (2); U1+ U4+ (2); U2- U3- (2); U1+ U4+ (2); U1+ U4+ (2); U2- U3- (2);
    U1+ U4+ (2); U1- U4- (2); U2- U3- (2); U2+ U3+ (2); U2- U3- (2); U1- U4- (2);
    U1+ U4+ (2); U2- U3- (2); U2+ U3+ (2); U1- U4- (2); U1+ U4+ (2); U2- U3- (2);
    U1+ U4+ (2); U2+ U3+ (2); U2- U3- (2); U1- U4- (2); [44] ;
X : X1  (2); X1  (2); X1  (2); X1  (2); X1  (2); X1  (2); X1  (2); X1  X2  (4);
    X2  (2); X1  (2); X2  (2); X1  (2); X1  (2); X2  (2); X2  (2); X1  (2);
    X1  (2); X1  (2); X2  (2); X1  (2); X2  (2); [44] ;
Y : Y2  (2); Y1  (2); Y1  (2); Y2  (2); Y2  (2); Y1  (2); Y1  (2); Y2  (2);
    Y1  (2); Y2  (2); Y2  (2); Y1  (2); Y2  (2); Y1  (2); Y1  (2); Y2  (2);
    Y2  (2); Y1  (2); Y1  (2); Y2  (2); Y2  (2); Y1  (2); [44] ;
Z : Z1  (2); Z1  (2); Z1  (2); Z1  (2); Z1  (2); Z1  (2); Z1  (2); Z1  (2);
    Z2  (2); Z2  (2); Z1  (2); Z2  (2); Z1  (2); Z1  (2); Z2  (2); Z2  (2);
    Z1  (2); Z1  (2); Z1  (2); Z2  (2); Z1  (2); Z2  (2); [44];
\end{lstlisting}
\hyperref[tab:electride]{Back to the table}

\subsubsection*{41858 CoAsS}
\label{sec:tqc41858}
\lstset{language=bash, keywordstyle=\color{blue!70}, basicstyle=\ttfamily, frame=shadowbox}

\hyperref[tab:electride]{Back to the table}

\subsubsection*{422673 ErS$_{2}$}
\label{sec:tqc422673}
\noindent Essential BR: $Ag@2c$ \\
\noindent RSI:
\begin{flalign*}
&\delta_{1}@2c\equiv -m(Ag)+m(Au) = -1,&
\end{flalign*}
\lstset{language=bash, keywordstyle=\color{blue!70}, basicstyle=\ttfamily, frame=shadowbox}

\hyperref[tab:electride]{Back to the table}

\subsubsection*{414619 NdSe$_{2}$}
\label{sec:tqc414619}
\noindent Essential BR: $Ag@2c$ \\
\noindent RSI:
\begin{flalign*}
&\delta_{1}@2c\equiv -m(Ag)+m(Au) = -1,&
\end{flalign*}
\lstset{language=bash, keywordstyle=\color{blue!70}, basicstyle=\ttfamily, frame=shadowbox}

\hyperref[tab:electride]{Back to the table}

\subsubsection*{610526 FeAsSe}
\label{sec:tqc610526}
\noindent Essential BR: $Ag@2d$ \\
\noindent RSI:
\begin{flalign*}
&\delta_{1}@2d\equiv -m(Ag)+m(Au) = -1,&
\end{flalign*}
\lstset{language=bash, keywordstyle=\color{blue!70}, basicstyle=\ttfamily, frame=shadowbox}

\hyperref[tab:electride]{Back to the table}

\subsubsection*{422672 TbS$_{2}$}
\label{sec:tqc422672}
\noindent Essential BR: $Ag@2c$ \\
\noindent RSI:
\begin{flalign*}
&\delta_{1}@2c\equiv -m(Ag)+m(Au) = -1,&
\end{flalign*}
\lstset{language=bash, keywordstyle=\color{blue!70}, basicstyle=\ttfamily, frame=shadowbox}

\hyperref[tab:electride]{Back to the table}

\subsubsection*{15012 FeS$_{2}$}
\label{sec:tqc15012}
\noindent Essential BR: $A1g@4b$ \\
\noindent RSI:
\begin{flalign*}
&\delta_{1}@4b\equiv -m(2Eg)+m(2Eu) = 0,&
\\
&\delta_{2}@4b\equiv -m(A1g)+m(A1u) = -1,&
\end{flalign*}
\lstset{language=bash, keywordstyle=\color{blue!70}, basicstyle=\ttfamily, frame=shadowbox}
\begin{lstlisting}
Computed bands:  1 - 40
GM: GM1+(1); GM4+(3); GM4-(3); GM1-(1); GM4+(3); GM4-(3); GM2-GM3-(2); GM4+(3);
    GM1+(1); GM4+(3); GM2+GM3+(2); GM4-(3); GM4+(3); GM2+GM3+(2); GM4+(3);
    GM4+(3); GM1+(1); [40] ;
R : R1- R3- (4); R1+ R3+ (4); R1+ R3+ (4); R1- R3- (4); R2+ R2+ (4); R2- R2- (4);
    R1- R3- (4); R1+ R3+ (4); R2+ R2+ (4); R1+ R3+ (4); [40] ;
M : M1  M2  (4); M1  M2  (4); M1  M2  (4); M1  M2  (4); M1  M2  (4); M1  M2  (4);
    M1  M2  (4); M1  M2  (4); M1  M2  (4); M1  M2  (4); [40] ;
X : X2  (2); X1  (2); X2  (2); X1  (2); X1  (2); X2  (2); X1  (2); X1  (2);
    X2  (2); X2  (2); X1  (2); X2  (2); X2  (2); X1  (2); X1  (2); X1  (2);
    X2  (2); X2  (2); X2  (2); X1  (2); [40];
\end{lstlisting}
\hyperref[tab:electride]{Back to the table}

\subsubsection*{647944 PrP$_{2}$}
\label{sec:tqc647944}
\noindent Essential BR: $Ag@2b$ \\
\noindent RSI:
\begin{flalign*}
&\delta_{1}@2b\equiv -m(Ag)+m(Au) = -1,&
\end{flalign*}
\lstset{language=bash, keywordstyle=\color{blue!70}, basicstyle=\ttfamily, frame=shadowbox}

\hyperref[tab:electride]{Back to the table}

\subsubsection*{103943 RbGa$_{3}$}
\label{sec:tqc103943}
\lstset{language=bash, keywordstyle=\color{blue!70}, basicstyle=\ttfamily, frame=shadowbox}
\begin{lstlisting}
Computed bands:  1 - 27
GM: GM1 (1); GM2 (1); GM1 (1); GM5 (2); GM1 (1); GM2 (1); GM5 (2); GM2 (1);
    GM5 (2); GM1 (1); GM2 (1); GM1 (1); GM5 (2); GM2 (1); GM1 (1); GM5 (2);
    GM1 (1); GM2 (1); GM4 (1); GM2 (1); GM5 (2); [27] ;
M : M1  (1); M2  (1); M1  (1); M5  (2); M2  (1); M2  (1); M5  (2); M1  (1);
    M5  (2); M2  (1); M1  (1); M1  (1); M5  (2); M2  (1); M1  (1); M5  (2);
    M2  (1); M2  (1); M5  (2); M1  (1); M3  (1); [27] ;
P : P4  (1); P1  (1); P3  (1); P4  (1); P2  (1); P3  (1); P1  (1); P2  (1);
    P1  (1); P4  (1); P2  (1); P3  (1); P3  (1); P2  (1); P4  (1); P1  (1);
    P2  (1); P3  (1); P1  (1); P4  (1); P2  (1); P4  (1); P1  (1); P4  (1);
    P3  (1); P2  (1); P1  (1); [27] ;
X : X3  (1); X1  (1); X4  (1); X3  (1); X2  (1); X4  (1); X1  (1); X1  (1);
    X2  (1); X3  (1); X2  (1); X4  (1); X3  (1); X1  (1); X4  (1); X2  (1);
    X1  (1); X3  (1); X2  (1); X1  (1); X3  (1); X4  (1); X1  (1); X4  (1);
    X2  (1); X3  (1); X2  (1); [27] ;
N : N1  (1); N1  (1); N1  (1); N1  (1); N2  (1); N1  (1); N1  (1); N1  (1);
    N2  (1); N1  (1); N2  (1); N1  (1); N1  (1); N1  (1); N1  (1); N1  (1);
    N2  (1); N1  (1); N1  (1); N2  (1); N1  (1); N1  (1); N2  (1); N1  (1);
    N2  (1); N1  (1); N1  (1); [27];
\end{lstlisting}
\hyperref[tab:electride]{Back to the table}

\subsubsection*{409185 HoP$_{5}$}
\label{sec:tqc409185}
\lstset{language=bash, keywordstyle=\color{blue!70}, basicstyle=\ttfamily, frame=shadowbox}

\hyperref[tab:electride]{Back to the table}

\subsubsection*{152623 LaZnSn}
\label{sec:tqc152623}
\lstset{language=bash, keywordstyle=\color{blue!70}, basicstyle=\ttfamily, frame=shadowbox}
\begin{lstlisting}
Computed bands:  1 - 54
A : A1  (2); A1  (2); A1  (2); A3  (4); A1  (2); A3  (4); A1  (2); A1  (2);
    A3  (4); A3  (4); A3  (4); A1  (2); A3  (4); A1  (2); A1  (2); A1  (2);
    A3  (4); A3  (4); A1  (2); [54] ;
GM: GM1+GM3+(2); GM1+GM4-(2); GM2-(1); GM4-(1); GM5-GM6-(4); GM3+(1); GM2-(1);
    GM5+GM6-(4); GM1+(1); GM4-(1); GM3+(1); GM2-(1); GM6+(2); GM5-(2); GM5+(2);
    GM6-(2); GM5+(2); GM6-(2); GM1+(1); GM6+(2); GM5-(2); GM4-(1); GM3+(1);
    GM2-(1); GM1+(1); GM4-(1); GM3+(1); GM5+(2); GM6+(2); GM6-(2); GM1+(1);
    GM2-(1); GM5-(2); GM4-(1); [54] ;
H : H3  (2); H3  (2); H3  (2); H1  H2  (4); H1  H2  (4); H3  (2); H2  (2);
    H1  (2); H3  (2); H3  (2); H1  H3  (4); H2  (2); H1  (2); H3  (2); H2  (2);
    H1  (2); H2  (2); H1  (2); H2  (2); H3  (2); H3  (2); H2  (2); H1  (2);
    H1  (2); [54] ;
K : K1  K2  (2); K1  K4  (2); K3  (1); K5  (2); K6  (2); K4  (1); K5  (2);
    K2  K5  (3); K3  (1); K5  (2); K6  (2); K1  K4  (2); K1  (1); K2  K3  (2);
    K5  (2); K4  (1); K5  (2); K2  K6  (3); K3  (1); K6  (2); K6  (2); K5  (2);
    K5  (2); K6  (2); K1  (1); K4  (1); K2  (1); K3  (1); K5  (2); K6  (2);
    K5  (2); [54] ;
L : L1  (2); L1  (2); L1  (2); L1  (2); L2  (2); L1  (2); L1  (2); L2  (2);
    L1  (2); L1  (2); L2  (2); L2  (2); L1  (2); L1  (2); L1  L2  (4); L2  (2);
    L1  (2); L1  (2); L1  (2); L1  (2); L1  (2); L2  (2); L1  (2); L2  (2);
    L1  (2); L1  (2); [54] ;
M : M1+ M3+ (2); M1+ M4- (2); M4- (1); M2- (1); M2- (1); M4- (1); M1- M3- (2);
    M1+ (1); M4- (1); M3+ (1); M2- (1); M2+ M3- (2); M3+ (1); M1+ (1); M2- (1);
    M4- (1); M1- M4+ (2); M2+ M3- (2); M1+ (1); M4- (1); M1+ M3+ (2); M4- (1);
    M2+ (1); M2- M3- (2); M4+ (1); M1- (1); M3+ (1); M2- (1); M1+ (1); M3+ (1);
    M2- (1); M4- (1); M1+ (1); M4- (1); M3+ (1); M2- (1); M4+ (1); M1+ (1);
    M1- (1); M2+ (1); M1+ (1); M3+ (1); M4- (1); M3- (1); M2- (1); M4- (1); [54];
\end{lstlisting}
\hyperref[tab:electride]{Back to the table}

\subsubsection*{24774 HgO$_{2}$}
\label{sec:tqc24774}
\noindent Essential BR: $Ag@4b$ \\
\noindent RSI:
\begin{flalign*}
&\delta_{1}@4b\equiv -m(Ag)+m(Au) = -1,&
\end{flalign*}
\lstset{language=bash, keywordstyle=\color{blue!70}, basicstyle=\ttfamily, frame=shadowbox}
\begin{lstlisting}
Computed bands:  1 - 48
GM: GM1+(1); GM3+(1); GM2+(1); GM4+(1); GM3-(1); GM4-(1); GM2-(1); GM1-(1);
    GM4+(1); GM2+(1); GM3+(1); GM1+(1); GM1+(1); GM2+(1); GM4+(1); GM3+(1);
    GM3+(1); GM1+(1); GM2+(1); GM4+(1); GM2+(1); GM4-(1); GM3+(1); GM1+(1);
    GM2+(1); GM3-(1); GM3+(1); GM4+(1); GM4+(1); GM1+(1); GM2-(1); GM1-(1);
    GM1+(1); GM3-(1); GM4-(1); GM1-(1); GM3+(1); GM2-(1); GM2+(1); GM4+(1);
    GM4+(1); GM2+(1); GM3+(1); GM1+(1); GM1+(1); GM4+(1); GM2+(1); GM3+(1);
    [48] ;
R : R1- R1- (4); R1+ R1+ (4); R1+ R1+ (4); R1+ R1+ (4); R1- R1- (4); R1+ R1+ (4);
    R1+ R1+ (4); R1+ R1+ (4); R1+ R1+ (4); R1+ R1+ (4); R1- R1- (4); R1- R1- (4);
    [48] ;
S : S1  S2  (4); S1  S2  (4); S1  S2  (4); S1  S2  (4); S1  S2  (4); S1  S2  (4);
    S1  S2  (4); S1  S2  (4); S1  S2  (4); S1  S2  (4); S1  S2  (4); S1  S2  (4);
    [48] ;
T : T1  T2  (4); T1  T2  (4); T1  T2  (4); T1  T2  (4); T1  T2  (4); T1  T2  (4);
    T1  T2  (4); T1  T2  (4); T1  T2  (4); T1  T2  (4); T1  T2  (4); T1  T2  (4);
    [48] ;
U : U1  U2  (4); U1  U2  (4); U1  U2  (4); U1  U2  (4); U1  U2  (4); U1  U2  (4);
    U1  U2  (4); U1  U2  (4); U1  U2  (4); U1  U2  (4); U1  U2  (4); U1  U2  (4);
    [48] ;
X : X1  (2); X2  (2); X1  (2); X2  (2); X1  (2); X2  (2); X2  (2); X1  (2);
    X1  (2); X2  (2); X2  (2); X1  (2); X1  (2); X2  (2); X1  (2); X2  (2);
    X2  (2); X1  (2); X2  (2); X1  (2); X2  (2); X1  (2); X1  (2); X2  (2);
    [48] ;
Y : Y2  (2); Y1  (2); Y2  (2); Y1  (2); Y1  (2); Y2  (2); Y2  (2); Y1  (2);
    Y1  (2); Y2  (2); Y2  (2); Y1  (2); Y2  (2); Y1  (2); Y2  (2); Y2  (2);
    Y1  (2); Y1  (2); Y1  (2); Y2  (2); Y1  (2); Y2  (2); Y1  (2); Y2  (2);
    [48] ;
Z : Z1  (2); Z2  (2); Z2  (2); Z1  (2); Z2  (2); Z1  (2); Z1  (2); Z2  (2);
    Z1  (2); Z2  (2); Z2  (2); Z1  (2); Z2  (2); Z1  (2); Z1  (2); Z2  (2);
    Z1  (2); Z2  (2); Z2  (2); Z1  (2); Z2  (2); Z1  (2); Z2  (2); Z1  (2); [48];
\end{lstlisting}
\hyperref[tab:electride]{Back to the table}

\subsubsection*{191245 PdN$_{2}$}
\label{sec:tqc191245}
\lstset{language=bash, keywordstyle=\color{blue!70}, basicstyle=\ttfamily, frame=shadowbox}
\begin{lstlisting}
Computed bands:  1 - 40
GM: GM1+(1); GM4+(3); GM4-(3); GM1-(1); GM1+(1); GM4+(3); GM4-(3); GM2-GM3-(2);
    GM4-(3); GM4+(3); GM2+GM3+(2); GM4+(3); GM4+(3); GM2+GM3+(2); GM4+(3);
    GM1+(1); GM4+(3); [40] ;
R : R1- R3- (4); R1+ R3+ (4); R1+ R3+ (4); R2+ R2+ (4); R1- R3- (4); R2+ R2+ (4);
    R1+ R3+ (4); R1+ R3+ (4); R1+ R3+ (4); R2+ R2+ (4); [40] ;
M : M1  M2  (4); M1  M2  (4); M1  M2  (4); M1  M2  (4); M1  M2  (4); M1  M2  (4);
    M1  M2  (4); M1  M2  (4); M1  M2  (4); M1  M2  (4); [40] ;
X : X2  (2); X1  (2); X2  (2); X1  (2); X2  (2); X1  (2); X1  (2); X2  (2);
    X2  (2); X1  (2); X2  (2); X1  (2); X1  (2); X2  (2); X1  (2); X2  (2);
    X1  (2); X2  (2); X1  (2); X2  (2); [40];
\end{lstlisting}
\hyperref[tab:electride]{Back to the table}

\subsubsection*{160624 RhN$_{2}$}
\label{sec:tqc160624}
\noindent Essential BR: $Ag@2a$ \\
\noindent RSI:
\begin{flalign*}
&\delta_{1}@2a\equiv -m(Ag)+m(Au) = -1,&
\end{flalign*}
\lstset{language=bash, keywordstyle=\color{blue!70}, basicstyle=\ttfamily, frame=shadowbox}

\hyperref[tab:electride]{Back to the table}

\subsubsection*{600680 RuS$_{2}$}
\label{sec:tqc600680}
\noindent Essential BR: $A1g@4b$ \\
\noindent RSI:
\begin{flalign*}
&\delta_{1}@4b\equiv -m(2Eg)+m(2Eu) = 0,&
\\
&\delta_{2}@4b\equiv -m(A1g)+m(A1u) = -1,&
\end{flalign*}
\lstset{language=bash, keywordstyle=\color{blue!70}, basicstyle=\ttfamily, frame=shadowbox}
\begin{lstlisting}
Computed bands:  1 - 40
GM: GM1+(1); GM4+(3); GM4-(3); GM1-(1); GM4+(3); GM4-(3); GM4+(3); GM1+(1);
    GM2+GM3+(2); GM2-GM3-(2); GM4+(3); GM4-(3); GM4+(3); GM2+GM3+(2); GM4+(3);
    GM4+(3); GM1+(1); [40] ;
R : R1- R3- (4); R1+ R3+ (4); R1+ R3+ (4); R2+ R2+ (4); R1- R3- (4); R2- R2- (4);
    R1- R3- (4); R1+ R3+ (4); R2+ R2+ (4); R1+ R3+ (4); [40] ;
M : M1  M2  (4); M1  M2  (4); M1  M2  (4); M1  M2  (4); M1  M2  (4); M1  M2  (4);
    M1  M2  (4); M1  M2  (4); M1  M2  (4); M1  M2  (4); [40] ;
X : X2  (2); X1  (2); X2  (2); X1  (2); X1  (2); X2  (2); X1  (2); X2  (2);
    X1  (2); X2  (2); X1  (2); X2  (2); X2  (2); X1  (2); X1  (2); X1  (2);
    X2  (2); X2  (2); X2  (2); X1  (2); [40];
\end{lstlisting}
\hyperref[tab:electride]{Back to the table}

\subsubsection*{409182 PrP$_{5}$}
\label{sec:tqc409182}
\lstset{language=bash, keywordstyle=\color{blue!70}, basicstyle=\ttfamily, frame=shadowbox}

\hyperref[tab:electride]{Back to the table}

\subsubsection*{633475 FeSe$_{2}$}
\label{sec:tqc633475}
\noindent Essential BR: $A1g@4b$ \\
\noindent RSI:
\begin{flalign*}
&\delta_{1}@4b\equiv -m(2Eg)+m(2Eu) = 0,&
\\
&\delta_{2}@4b\equiv -m(A1g)+m(A1u) = -1,&
\end{flalign*}
\lstset{language=bash, keywordstyle=\color{blue!70}, basicstyle=\ttfamily, frame=shadowbox}
\begin{lstlisting}
Computed bands:  1 - 40
GM: GM1+(1); GM4+(3); GM4-(3); GM1-(1); GM4+(3); GM4-(3); GM4+(3); GM2-GM3-(2);
    GM1+(1); GM4+(3); GM2+GM3+(2); GM4-(3); GM4+(3); GM2+GM3+(2); GM4+(3);
    GM4+(3); GM1+(1); [40] ;
R : R1- R3- (4); R1+ R3+ (4); R1+ R3+ (4); R1- R3- (4); R2+ R2+ (4); R2- R2- (4);
    R1- R3- (4); R1+ R3+ (4); R2+ R2+ (4); R1+ R3+ (4); [40] ;
M : M1  M2  (4); M1  M2  (4); M1  M2  (4); M1  M2  (4); M1  M2  (4); M1  M2  (4);
    M1  M2  (4); M1  M2  (4); M1  M2  (4); M1  M2  (4); [40] ;
X : X2  (2); X1  (2); X2  (2); X1  (2); X1  (2); X2  (2); X1  (2); X1  (2);
    X2  (2); X2  (2); X2  (2); X1  (2); X2  (2); X1  (2); X1  (2); X1  X2  (4);
    X2  (2); X2  (2); X1  (2); [40];
\end{lstlisting}
\hyperref[tab:electride]{Back to the table}

\subsubsection*{102863 CsGa$_{3}$}
\label{sec:tqc102863}
\lstset{language=bash, keywordstyle=\color{blue!70}, basicstyle=\ttfamily, frame=shadowbox}
\begin{lstlisting}
Computed bands:  1 - 27
GM: GM1 (1); GM1 (1); GM2 (1); GM1 (1); GM5 (2); GM2 (1); GM1 (1); GM5 (2);
    GM2 (1); GM5 (2); GM2 (1); GM1 (1); GM5 (2); GM2 (1); GM1 (1); GM5 (2);
    GM1 (1); GM2 (1); GM2 (1); GM4 (1); GM5 (2); [27] ;
M : M1  (1); M2  (1); M1  (1); M2  (1); M5  (2); M2  (1); M5  (2); M1  (1);
    M5  (2); M2  (1); M1  (1); M1  (1); M5  (2); M2  (1); M1  (1); M5  (2);
    M2  (1); M2  (1); M1  (1); M5  (2); M3  (1); [27] ;
P : P1  (1); P3  (1); P4  (1); P3  (1); P2  (1); P4  (1); P1  (1); P2  (1);
    P1  (1); P3  (1); P2  (1); P4  (1); P4  (1); P2  P3  (2); P1  (1); P2  (1);
    P4  (1); P1  (1); P3  (1); P2  (1); P3  (1); P1  (1); P4  (1); P3  (1);
    P2  (1); P1  (1); [27] ;
X : X1  (1); X3  (1); X4  (1); X3  (1); X2  (1); X4  (1); X1  (1); X1  (1);
    X2  (1); X3  (1); X2  (1); X4  (1); X3  (1); X1  (1); X4  (1); X2  (1);
    X1  (1); X3  (1); X2  (1); X1  (1); X3  (1); X4  (1); X1  (1); X4  (1);
    X2  (1); X3  (1); X2  (1); [27] ;
N : N1  (1); N1  (1); N1  (1); N1  (1); N1  (1); N2  (1); N1  (1); N1  (1);
    N2  (1); N1  (1); N2  (1); N1  (1); N1  (1); N1  (1); N1  (1); N1  (1);
    N2  (1); N1  (1); N1  (1); N2  (1); N1  (1); N1  (1); N2  (1); N1  (1);
    N2  (1); N1  (1); N1  (1); [27];
\end{lstlisting}
\hyperref[tab:electride]{Back to the table}

\subsubsection*{633399 FeSbSe}
\label{sec:tqc633399}
\noindent Essential BR: $Ag@2d$ \\
\noindent RSI:
\begin{flalign*}
&\delta_{1}@2d\equiv -m(Ag)+m(Au) = -1,&
\end{flalign*}
\lstset{language=bash, keywordstyle=\color{blue!70}, basicstyle=\ttfamily, frame=shadowbox}

\hyperref[tab:electride]{Back to the table}

\subsubsection*{413527 PrSe$_{2}$}
\label{sec:tqc413527}
\noindent Essential BR: $Ag@2c$ \\
\noindent RSI:
\begin{flalign*}
&\delta_{1}@2c\equiv -m(Ag)+m(Au) = -1,&
\end{flalign*}
\lstset{language=bash, keywordstyle=\color{blue!70}, basicstyle=\ttfamily, frame=shadowbox}

\hyperref[tab:electride]{Back to the table}

\subsubsection*{22221 NiP$_{2}$}
\label{sec:tqc22221}
\noindent Essential BR: $A1g@4b$ \\
\noindent RSI:
\begin{flalign*}
&\delta_{1}@4b\equiv -m(2Eg)+m(2Eu) = 0,&
\\
&\delta_{2}@4b\equiv -m(A1g)+m(A1u) = -1,&
\end{flalign*}
\lstset{language=bash, keywordstyle=\color{blue!70}, basicstyle=\ttfamily, frame=shadowbox}
\begin{lstlisting}
Computed bands:  1 - 40
GM: GM1+(1); GM4+(3); GM4-(3); GM1-(1); GM4+(3); GM4-(3); GM4+(3); GM2-GM3-(2);
    GM4+(3); GM2+GM3+(2); GM1+(1); GM4+(3); GM2+GM3+(2); GM4+(3); GM1+(1);
    GM4+(3); GM4-(3); [40] ;
R : R1- R3- (4); R1+ R3+ (4); R1+ R3+ (4); R2+ R2+ (4); R1- R3- (4); R2+ R2+ (4);
    R1+ R3+ (4); R1+ R3+ (4); R2- R2- (4); R1- R3- (4); [40] ;
M : M1  M2  (4); M1  M2  (4); M1  M2  (4); M1  M2  (4); M1  M2  (4); M1  M2  (4);
    M1  M2  (4); M1  M2  (4); M1  M2  (4); M1  M2  (4); [40] ;
X : X2  (2); X1  (2); X2  (2); X1  (2); X1  (2); X2  (2); X1  (2); X1  (2);
    X2  (2); X2  (2); X1  (2); X2  (2); X2  (2); X1  (2); X1  (2); X2  (2);
    X1  (2); X2  (2); X2  (2); X1  (2); [40] ;
\end{lstlisting}
\hyperref[tab:electride]{Back to the table}

\subsubsection*{409187 LuP$_{5}$}
\label{sec:tqc409187}
\lstset{language=bash, keywordstyle=\color{blue!70}, basicstyle=\ttfamily, frame=shadowbox}

\hyperref[tab:electride]{Back to the table}

\subsubsection*{32530 LaSe$_{2}$}
\label{sec:tqc32530}
\noindent Essential BR: $Ag@2c$ \\
\noindent RSI:
\begin{flalign*}
&\delta_{1}@2c\equiv -m(Ag)+m(Au) = -1,&
\end{flalign*}
\lstset{language=bash, keywordstyle=\color{blue!70}, basicstyle=\ttfamily, frame=shadowbox}

\hyperref[tab:electride]{Back to the table}

\subsubsection*{409188 YP$_{5}$}
\label{sec:tqc409188}
\lstset{language=bash, keywordstyle=\color{blue!70}, basicstyle=\ttfamily, frame=shadowbox}

\hyperref[tab:electride]{Back to the table}

\subsubsection*{102867 CsIn$_{3}$}
\label{sec:tqc102867}
\lstset{language=bash, keywordstyle=\color{blue!70}, basicstyle=\ttfamily, frame=shadowbox}
\begin{lstlisting}
Computed bands:  1 - 27
GM: GM1 (1); GM2 (1); GM1 (1); GM5 (2); GM1 (1); GM2 (1); GM5 (2); GM2 (1);
    GM5 (2); GM1 (1); GM2 (1); GM1 (1); GM5 (2); GM2 (1); GM1 (1); GM5 (2);
    GM1 (1); GM2 (1); GM4 (1); GM2 (1); GM5 (2); [27] ;
M : M1  (1); M2  (1); M1  (1); M5  (2); M2  (1); M2  (1); M5  (2); M1  (1);
    M5  (2); M2  (1); M1  (1); M1  (1); M5  (2); M2  (1); M1  (1); M5  (2);
    M2  (1); M2  (1); M5  (2); M1  (1); M3  (1); [27] ;
P : P3  (1); P1  (1); P4  (1); P3  (1); P2  (1); P4  (1); P1  (1); P2  (1);
    P1  (1); P3  (1); P2  (1); P4  (1); P4  (1); P2  (1); P3  (1); P1  (1);
    P2  (1); P1  (1); P4  (1); P3  (1); P2  (1); P3  (1); P1  (1); P3  (1);
    P4  (1); P2  (1); P1  (1); [27] ;
X : X3  (1); X1  (1); X4  (1); X3  (1); X2  (1); X4  (1); X1  (1); X1  (1);
    X2  (1); X3  (1); X2  (1); X4  (1); X3  (1); X1  (1); X4  (1); X2  (1);
    X1  (1); X3  (1); X2  (1); X1  (1); X3  (1); X4  (1); X1  (1); X4  (1);
    X2  (1); X3  (1); X2  (1); [27] ;
N : N1  (1); N1  (1); N1  (1); N1  (1); N2  (1); N1  (1); N1  (1); N1  (1);
    N2  (1); N1  (1); N2  (1); N1  (1); N1  (1); N1  (1); N1  (1); N1  (1);
    N2  (1); N1  (1); N1  (1); N2  (1); N1  (1); N1  (1); N2  (1); N1  (1);
    N2  (1); N1  (1); N1  (1); [27];
\end{lstlisting}
\hyperref[tab:electride]{Back to the table}

\subsubsection*{92525 PrS$_{2}$}
\label{sec:tqc92525}
\noindent Essential BR: $Ag@2c$ \\
\noindent RSI:
\begin{flalign*}
&\delta_{1}@2c\equiv -m(Ag)+m(Au) = -1,&
\end{flalign*}
\lstset{language=bash, keywordstyle=\color{blue!70}, basicstyle=\ttfamily, frame=shadowbox}

\hyperref[tab:electride]{Back to the table}

\subsubsection*{615734 YbB$_{12}$}
\label{sec:tqc615734}
\lstset{language=bash, keywordstyle=\color{blue!70}, basicstyle=\ttfamily, frame=shadowbox}
\begin{lstlisting}
Computed bands:  1 - 22
GM: GM4-(3); GM1+(1); GM4-(3); GM5+(3); GM1+(1); GM2-(1); GM3+(2); GM4-(3);
    GM5+(3); GM3+(2); [22] ;
X : X3- (1); X5- (2); X1+ (1); X3- (1); X5- (2); X4+ (1); X1+ (1); X5+ (2);
    X2+ (1); X4- (1); X5- (2); X3- (1); X2- (1); X1+ (1); X4+ (1); X5- (2);
    X3- (1); [22] ;
L : L2- (1); L3- (2); L2- (1); L1+ (1); L3+ (2); L2- (1); L3- (2); L1+ (1);
    L3- (2); L3+ (2); L2- (1); L1+ (1); L1+ (1); L3+ (2); L3- (2); [22] ;
W : W5  (2); W2  (1); W2  (1); W5  (2); W1  (1); W5  (2); W1  (1); W2  (1);
    W3  (1); W2  (1); W5  (2); W4  (1); W1  (1); W5  (2); W1  (1); W5  (2); [22];
\end{lstlisting}
\hyperref[tab:electride]{Back to the table}

\subsubsection*{624284 Nb$_{7}$Co$_{6}$}
\label{sec:tqc624284}
\lstset{language=bash, keywordstyle=\color{blue!70}, basicstyle=\ttfamily, frame=shadowbox}

\hyperref[tab:electride]{Back to the table}

\subsubsection*{425778 Ba(P$_{2}$Au)$_{2}$}
\label{sec:tqc425778}
\noindent Essential BR: $Ag@16c$ \\
\noindent RSI:
\begin{flalign*}
&\delta_{1}@16c\equiv -m(Ag)+m(Au) = -1,&
\end{flalign*}
\lstset{language=bash, keywordstyle=\color{blue!70}, basicstyle=\ttfamily, frame=shadowbox}

\hyperref[tab:electride]{Back to the table}

\subsubsection*{262413 Sr$_{2}$Zn$_{2}$As$_{3}$}
\label{sec:tqc262413}
\noindent Essential BR: $Ag@2a$ \\
\noindent RSI:
\begin{flalign*}
&\delta_{1}@2a\equiv -m(Ag)+m(Au)-m(Bg)+m(Bu) = -1,&
\end{flalign*}
\lstset{language=bash, keywordstyle=\color{blue!70}, basicstyle=\ttfamily, frame=shadowbox}

\hyperref[tab:electride]{Back to the table}

\subsubsection*{406949 Bi$_{4}$RuI$_{2}$}
\label{sec:tqc406949}
\lstset{language=bash, keywordstyle=\color{blue!70}, basicstyle=\ttfamily, frame=shadowbox}
\begin{lstlisting}
Computed bands:  1 - 42
GM: GM1+(1); GM3-GM4-(2); GM2+(1); GM1+(1); GM3-GM4-(2); GM1+(1); GM3-GM4-(2);
    GM2+(1); GM2+(1); GM1-(1); GM2+(1); GM1+(1); GM3-GM4-(2); GM3+GM4+(2);
    GM2-(1); GM2+(1); GM1+(1); GM3-GM4-(2); GM1+GM2+(2); GM2-(1); GM3-GM4-(2);
    GM2+(1); GM1+(1); GM2-(1); GM1-(1); GM3+GM4+(2); GM1+(1); GM3+GM4+(2);
    GM3-GM4-(2); GM2+(1); GM1-(1); [42] ;
M : M1+ (1); M3- M4- (2); M2+ (1); M1+ (1); M3- M4- (2); M1+ (1); M3- M4- (2);
    M2+ (1); M2+ (1); M1- (1); M2+ (1); M1+ (1); M3- M4- (2); M2+ (1);
    M3+ M4+ (2); M3- M4- (2); M1+ (1); M2- (1); M1+ (1); M1- (1); M1+ M2- (2);
    M3- M4- (2); M2+ (1); M3+ M4+ (2); M2- (1); M2+ (1); M3- M4- (2);
    M3+ M4+ (2); M2+ (1); M1+ (1); M1- (1); [42] ;
P : P2  (1); P3  P4  (2); P1  (1); P1  (1); P2  (1); P3  P4  (2); P3  P4  (2);
    P1  (1); P2  (1); P1  (1); P2  (1); P3  P4  (2); P1  (1); P2  (1);
    P3  P4  (2); P3  P4  (2); P1  (1); P2  (1); P2  (1); P3  P4  (2); P1  (1);
    P2  (1); P1  (1); P2  (1); P1  (1); P3  P4  (2); P2  (1); P3  P4  (2);
    P1  (1); P3  P4  (2); P2  (1); P2  (1); [42] ;
X : X1+ (1); X2- (1); X2- (1); X1+ (1); X1+ (1); X2- (1); X1+ (1); X2- (1);
    X2- (1); X2- (1); X1+ (1); X1+ (1); X1- (1); X1+ (1); X1+ (1); X2- (1);
    X2- (1); X2+ (1); X1+ (1); X2- (1); X2- (1); X2+ (1); X1+ (1); X1- (1);
    X1- (1); X2- (1); X1+ (1); X2- (1); X1+ (1); X2+ (1); X1- (1); X1- (1);
    X1+ (1); X2+ (1); X1+ (1); X2- (1); X2+ (1); X1+ (1); X2+ (1); X2- (1);
    X1+ (1); X1- (1); [42] ;
N : N1- (1); N1+ (1); N1+ (1); N1- (1); N1+ (1); N1- (1); N1- (1); N1- (1);
    N1+ (1); N1+ (1); N1- (1); N1+ (1); N1- (1); N1+ (1); N1+ (1); N1- (1);
    N1+ (1); N1- (1); N1+ (1); N1+ (1); N1+ (1); N1- (1); N1- (1); N1- (1);
    N1+ (1); N1- (1); N1+ (1); N1- (1); N1+ (1); N1- (1); N1+ (1); N1+ (1);
    N1- (1); N1- (1); N1- (1); N1+ (1); N1+ (1); N1- (1); N1+ (1); N1- (1);
    N1- (1); N1+ (1); [42];
\end{lstlisting}
\hyperref[tab:electride]{Back to the table}

\subsubsection*{426082 Eu$_{2}$Zn$_{2}$P$_{3}$}
\label{sec:tqc426082}
\noindent Essential BR: $Ag@2a$ \\
\noindent RSI:
\begin{flalign*}
&\delta_{1}@2a\equiv -m(Ag)+m(Au)-m(Bg)+m(Bu) = -1,&
\end{flalign*}
\lstset{language=bash, keywordstyle=\color{blue!70}, basicstyle=\ttfamily, frame=shadowbox}

\hyperref[tab:electride]{Back to the table}

\subsubsection*{82533 Sr$_{3}$(GeN)$_{2}$}
\label{sec:tqc82533}
\noindent Essential BR: $Ag@2a$ \\
\noindent RSI:
\begin{flalign*}
&\delta_{1}@2a\equiv -m(Ag)+m(Au) = -1,&
\end{flalign*}
\lstset{language=bash, keywordstyle=\color{blue!70}, basicstyle=\ttfamily, frame=shadowbox}

\hyperref[tab:electride]{Back to the table}

\subsubsection*{247425 Sc$_{2}$Si$_{2}$Pt$_{3}$}
\label{sec:tqc247425}
\noindent Essential BR: $Ag@2d$ \\
\noindent RSI:
\begin{flalign*}
&\delta_{1}@2d\equiv -m(Ag)+m(Au)-m(Bg)+m(Bu) = -1,&
\end{flalign*}
\lstset{language=bash, keywordstyle=\color{blue!70}, basicstyle=\ttfamily, frame=shadowbox}

\hyperref[tab:electride]{Back to the table}

\subsubsection*{639447 Ho$_{3}$Ni$_{2}$}
\label{sec:tqc639447}
\noindent Essential BR: $A1g@3b$ \\
\noindent RSI:
\begin{flalign*}
&\delta_{1}@3b\equiv -m(2Eg)+m(2Eu) = 0,&
\\
&\delta_{2}@3b\equiv -m(A1g)+m(A1u) = -1,&
\end{flalign*}
\lstset{language=bash, keywordstyle=\color{blue!70}, basicstyle=\ttfamily, frame=shadowbox}

\hyperref[tab:electride]{Back to the table}

\subsubsection*{2150 Er$_{3}$Ni$_{2}$}
\label{sec:tqc2150}
\noindent Essential BR: $A1g@3a$ \\
\noindent RSI:
\begin{flalign*}
&\delta_{1}@3a\equiv -m(2Eg)+m(2Eu) = 0,&
\\
&\delta_{2}@3a\equiv -m(A1g)+m(A1u) = -1,&
\end{flalign*}
\lstset{language=bash, keywordstyle=\color{blue!70}, basicstyle=\ttfamily, frame=shadowbox}

\hyperref[tab:electride]{Back to the table}

\subsubsection*{281462 Ca$_{6}$Cr$_{2}$HN$_{6}$}
\label{sec:tqc281462}
\lstset{language=bash, keywordstyle=\color{blue!70}, basicstyle=\ttfamily, frame=shadowbox}
\begin{lstlisting}
Computed bands:  1 - 52
GM: GM1+(1); GM2-GM3-(2); GM2+GM3+(2); GM1-(1); GM2+GM3+(2); GM1-(1); GM1+(1);
    GM2-GM3-(2); GM2+GM3+(2); GM2-GM3-(2); GM2+GM3+(2); GM1+(1); GM1-(1);
    GM1+(1); GM1-(1); GM2-GM3-(2); GM1+(1); GM1-(1); GM2-GM3-(2); GM2+GM3+(2);
    GM1+(1); GM2+GM3+(2); GM2-GM3-(2); GM1-(1); GM1+(1); GM2-GM3-(2);
    GM2+GM3+(2); GM1+(1); GM2-GM3-(2); GM1-(1); GM2+GM3+(2); GM1+(1); GM1-(1);
    GM1+(1); GM2-GM3-(2); [52] ;
T : T1- (1); T2+ T3+ (2); T2- T3- (2); T1+ (1); T1+ (1); T2- T3- (2); T1- (1);
    T1+ (1); T2+ T3+ (2); T2- T3- (2); T1- (1); T1+ (1); T2+ T3+ (2);
    T2- T3- (2); T1- (1); T2+ T3+ (2); T1- (1); T2+ T3+ (2); T1+ (1);
    T2- T3- (2); T1- (1); T1+ (1); T2+ T3+ (2); T2- T3- (2); T2+ T3+ (2);
    T1- (1); T2- T3- (2); T1+ (1); T2+ T3+ (2); T1+ (1); T2- T3- (2); T1- (1);
    T1+ (1); T1- (1); T2+ T3+ (2); [52] ;
F : F1- (1); F1+ (1); F1- (1); F1+ (1); F1+ (1); F1- (1); F1- (1); F1+ (1);
    F1+ (1); F1- (1); F1+ (1); F1- (1); F1- (1); F1+ (1); F1- (1); F1+ (1);
    F1- (1); F1+ (1); F1- (1); F1+ (1); F1- (1); F1- (1); F1+ (1); F1+ (1);
    F1+ (1); F1- (1); F1+ (1); F1- (1); F1- (1); F1+ (1); F1+ (1); F1- (1);
    F1+ (1); F1+ (1); F1- (1); F1- (1); F1+ (1); F1- (1); F1- (1); F1+ (1);
    F1- (1); F1+ (1); F1- (1); F1- (1); F1+ (1); F1+ (1); F1+ (1); F1- (1);
    F1+ (1); F1+ (1); F1- (1); F1- (1); [52] ;
L : L1- (1); L1+ (1); L1+ (1); L1- (1); L1- (1); L1+ (1); L1+ (1); L1- (1);
    L1- (1); L1- (1); L1+ (1); L1- (1); L1+ (1); L1+ (1); L1- (1); L1+ (1);
    L1+ (1); L1- (1); L1+ (1); L1- (1); L1+ L1- (2); L1+ (1); L1- (1); L1- (1);
    L1+ (1); L1+ (1); L1- (1); L1- (1); L1+ (1); L1- (1); L1+ (1); L1+ (1);
    L1- (1); L1+ (1); L1- (1); L1+ (1); L1+ (1); L1- (1); L1+ (1); L1- (1);
    L1- (1); L1+ (1); L1+ (1); L1+ L1- (2); L1- (1); L1+ (1); L1- (1); L1- (1);
    L1+ (1); L1+ (1); [52];
\end{lstlisting}
\hyperref[tab:electride]{Back to the table}

\subsubsection*{10509 Mn$_{4}$Al$_{11}$}
\label{sec:tqc10509}
\lstset{language=bash, keywordstyle=\color{blue!70}, basicstyle=\ttfamily, frame=shadowbox}

\hyperref[tab:electride]{Back to the table}

\subsubsection*{173357 Ba$_{4}$Fe$_{2}$S$_{4}$I$_{5}$}
\label{sec:tqc173357}
\lstset{language=bash, keywordstyle=\color{blue!70}, basicstyle=\ttfamily, frame=shadowbox}

\hyperref[tab:electride]{Back to the table}

\subsubsection*{6031 Sb$_{2}$I$_{2}$F$_{11}$}
\label{sec:tqc6031}
\noindent Essential BR: $A@2a$ \\
\noindent RSI:
\begin{flalign*}
&\delta_{1}@2a\equiv -m(A)+m(B) = -1,&
\end{flalign*}
\lstset{language=bash, keywordstyle=\color{blue!70}, basicstyle=\ttfamily, frame=shadowbox}

\hyperref[tab:electride]{Back to the table}

\subsubsection*{66024 K$_{5}$Te$_{3}$}
\label{sec:tqc66024}
\noindent Essential BR: $Ag@2a$ \\
\noindent RSI:
\begin{flalign*}
&\delta_{1}@2a\equiv -m(Ag)+m(Au)-m(2Eg)+m(2Eu) = -1,&
\\
&\delta_{2}@2a\equiv m(Ag)-m(Au)-m(Bg)+m(Bu) = 1,&
\end{flalign*}
\lstset{language=bash, keywordstyle=\color{blue!70}, basicstyle=\ttfamily, frame=shadowbox}

\hyperref[tab:electride]{Back to the table}

\subsubsection*{174081 NaSi}
\label{sec:tqc174081}
\lstset{language=bash, keywordstyle=\color{blue!70}, basicstyle=\ttfamily, frame=shadowbox}
\begin{lstlisting}
Computed bands:  1 - 20
GM: GM1+(1); GM1-(1); GM2+(1); GM2-(1); GM2+(1); GM2-(1); GM1+(1); GM1-(1);
    GM1+(1); GM1-(1); GM2+(1); GM2-(1); GM1+(1); GM1+(1); GM2+(1); GM1-(1);
    GM1-(1); GM2-(1); GM1+(1); GM1-(1); [20] ;
Y : Y1+ (1); Y1- (1); Y2+ (1); Y2- (1); Y2+ (1); Y1+ (1); Y1- (1); Y2- (1);
    Y1+ (1); Y1- (1); Y1- (1); Y2+ (1); Y2- (1); Y1+ (1); Y1+ (1); Y1- (1);
    Y2+ (1); Y2- (1); Y1+ (1); Y1- (1); [20] ;
L : L1+ (1); L1- (1); L1- (1); L1+ (1); L1+ (1); L1- (1); L1+ (1); L1- (1);
    L1+ (1); L1- (1); L1- (1); L1+ (1); L1+ (1); L1- (1); L1+ (1); L1- (1);
    L1+ (1); L1+ (1); L1- (1); L1- (1); [20] ;
M : M1  (2); M1  (2); M1  (2); M1  (2); M1  (2); M1  (2); M1  (2); M1  (2);
    M1  (2); M1  (2); [20] ;
V : V1+ (1); V1- (1); V1- (1); V1+ (1); V1+ (1); V1- (1); V1- (1); V1+ (1);
    V1+ (1); V1- (1); V1- (1); V1+ (1); V1- (1); V1+ (1); V1- (1); V1+ (1);
    V1- (1); V1+ (1); V1- (1); V1+ (1); [20] ;
U : U1  U2  (2); U1  U2  (2); U1  U2  (2); U1  U2  (2); U1  U2  (2); U1  U2  (2);
    U1  U2  (2); U1  U2  (2); U1  U2  (2); U1  U2  (2); [20] ;
A : A1  (2); A1  (2); A1  (2); A1  (2); A1  (2); A1  (2); A1  (2); A1  (2);
    A1  (2); A1  (2); [20];
\end{lstlisting}
\hyperref[tab:electride]{Back to the table}

\subsubsection*{635229 Ga$_{3}$Ru}
\label{sec:tqc635229}
\noindent Essential BR: $Ag@2b$ \\
\noindent RSI:
\begin{flalign*}
&\delta_{1}@2b\equiv -m(Ag)+m(Au)-m(B1g)+m(B1u)-m(B3g)+m(B3u)-m(B2g)+m(B2u) = -1,&
\end{flalign*}
\lstset{language=bash, keywordstyle=\color{blue!70}, basicstyle=\ttfamily, frame=shadowbox}
\begin{lstlisting}
Computed bands:  1 - 34
A : A4  (2); A1  (2); A3  (2); A1  (2); A3  (2); A4  (2); A1  (2); A1  (2);
    A2  (2); A1  (2); A3  (2); A4  (2); A4  (2); A2  (2); A2  (2); A3  (2);
    A1  (2); [34] ;
GM: GM1+(1); GM4+(1); GM3+(1); GM2-(1); GM5-(2); GM4+(1); GM3-(1); GM5+(2);
    GM1+(1); GM2+(1); GM4-(1); GM2-(1); GM4+(1); GM5-(2); GM5-(2); GM4+(1);
    GM1+(1); GM5-(2); GM5+(2); GM1-(1); GM1+(1); GM5-(2); GM2+(1); GM3+(1);
    GM5+(2); GM3-(1); [34] ;
M : M5- (2); M1+ M4+ (2); M2- M3- (2); M5+ (2); M5- (2); M5- (2); M1+ M4+ (2);
    M1+ M4+ (2); M2+ M3+ (2); M2- M3- (2); M5- (2); M1- M4- (2); M5+ (2);
    M5- (2); M1+ M4+ (2); M5+ (2); M2+ M3+ (2); [34] ;
Z : Z1  (2); Z2  (2); Z3  (2); Z1  (2); Z4  (2); Z1  (2); Z4  (2); Z1  (2);
    Z3  (2); Z2  (2); Z1  (2); Z4  (2); Z4  (2); Z1  (2); Z3  (2); Z2  (2);
    Z3  (2); [34] ;
R : R1- (2); R1+ (2); R1+ (2); R1- (2); R1+ (2); R1- (2); R1- (2); R1+ (2);
    R1+ (2); R1+ (2); R1- (2); R1- (2); R1- (2); R1+ (2); R1- (2); R1+ (2);
    R1- (2); [34] ;
X : X2  (2); X1  (2); X2  (2); X2  (2); X2  (2); X1  (2); X2  (2); X1  (2);
    X2  (2); X1  (2); X2  (2); X2  (2); X2  (2); X2  (2); X1  (2); X2  (2);
    X1  (2); [34];
\end{lstlisting}
\hyperref[tab:electride]{Back to the table}

\subsubsection*{34048 CoSb$_{3}$}
\label{sec:tqc34048}
\lstset{language=bash, keywordstyle=\color{blue!70}, basicstyle=\ttfamily, frame=shadowbox}
\begin{lstlisting}
Computed bands:  1 - 48
GM: GM1+(1); GM2+GM3+(2); GM4-(3); GM4-(3); GM4+(3); GM1+(1); GM2+GM3+(2);
    GM4+(3); GM4-(3); GM1+(1); GM4+(3); GM2+GM3+(2); GM4-(3); GM4+(3); GM4+(3);
    GM4+(3); GM4-(3); GM2-GM3-(2); GM2+GM3+(2); GM1+(1); GM1-(1); [48] ;
H : H2+ H3+ (2); H4- (3); H1+ (1); H4- (3); H4+ (3); H1- (1); H1+ (1); H4- (3);
    H4- (3); H2+ H3+ (2); H4+ (3); H2- H3- (2); H1+ (1); H4- (3); H2+ H3+ (2);
    H4- (3); H4+ (3); H2- H3- (2); H4- (3); H1- (1); H4- (3); [48] ;
P : P4  (3); P1  (1); P4  (3); P2  P3  (2); P4  (3); P4  (3); P1  (1); P4  (3);
    P2  P3  (2); P4  (3); P4  (3); P4  (3); P2  P3  (2); P4  (3); P1  (1);
    P4  (3); P2  P3  (2); P4  (3); P4  (3); P1  (1); [48] ;
N : N2- (1); N1+ (1); N2- (1); N2+ (1); N1- (1); N1+ (1); N1+ (1); N2- (1);
    N1+ (1); N1- (1); N2- (1); N2+ (1); N2- (1); N1+ (1); N1- (1); N2+ (1);
    N2- (1); N1+ (1); N2- (1); N1+ (1); N2+ (1); N2- (1); N1+ (1); N1- (1);
    N2+ (1); N2- (1); N1- (1); N1+ (1); N1- (1); N2+ (1); N2- (1); N2+ (1);
    N1- (1); N1- N2+ (2); N1+ (1); N2- (1); N2+ (1); N1+ (1); N1- (1); N1+ (1);
    N2- (1); N2- (1); N1+ (1); N1- (1); N2+ (1); N1+ (1); N2- (1); [48];
\end{lstlisting}
\hyperref[tab:electride]{Back to the table}

\subsubsection*{55514 In$_{3}$Ru}
\label{sec:tqc55514}
\noindent Essential BR: $Ag@2b$ \\
\noindent RSI:
\begin{flalign*}
&\delta_{1}@2b\equiv -m(Ag)+m(Au)-m(B1g)+m(B1u)-m(B3g)+m(B3u)-m(B2g)+m(B2u) = -1,&
\end{flalign*}
\lstset{language=bash, keywordstyle=\color{blue!70}, basicstyle=\ttfamily, frame=shadowbox}
\begin{lstlisting}
Computed bands:  1 - 34
A : A4  (2); A1  (2); A3  (2); A1  (2); A3  (2); A4  (2); A1  (2); A1  (2);
    A2  (2); A1  (2); A3  (2); A4  (2); A4  (2); A2  (2); A2  (2); A1  (2);
    A3  (2); [34] ;
GM: GM1+(1); GM3+(1); GM4+(1); GM2-(1); GM5-(2); GM4+(1); GM3-(1); GM2+(1);
    GM5+(2); GM1+(1); GM4-(1); GM5-(2); GM4+(1); GM2-(1); GM5-(2); GM4+(1);
    GM1+(1); GM5-(2); GM5+(2); GM1-(1); GM1+(1); GM5-(2); GM2+(1); GM3+(1);
    GM3-(1); GM5+(2); [34] ;
M : M5- (2); M1+ M4+ (2); M2- M3- (2); M5+ (2); M5- (2); M1+ M4+ (2); M5- (2);
    M1+ M4+ (2); M2+ M3+ (2); M2- M3- (2); M5- (2); M5+ (2); M1- M4- (2);
    M1+ M4+ (2); M5- (2); M5+ (2); M2+ M3+ (2); [34] ;
Z : Z1  (2); Z2  (2); Z3  (2); Z1  (2); Z4  (2); Z1  (2); Z4  (2); Z1  (2);
    Z3  (2); Z1  (2); Z2  (2); Z4  (2); Z1  (2); Z4  (2); Z3  (2); Z3  (2);
    Z2  (2); [34] ;
R : R1- (2); R1+ (2); R1+ (2); R1- (2); R1+ (2); R1- (2); R1- (2); R1+ (2);
    R1+ (2); R1+ (2); R1- (2); R1- (2); R1- (2); R1+ (2); R1+ (2); R1- (2);
    R1- (2); [34] ;
X : X2  (2); X1  (2); X2  (2); X2  (2); X2  (2); X1  (2); X2  (2); X1  (2);
    X2  (2); X2  (2); X1  (2); X2  (2); X2  (2); X2  (2); X1  (2); X2  (2);
    X1  (2); [34];
\end{lstlisting}
\hyperref[tab:electride]{Back to the table}

\subsubsection*{35200 TcP$_{3}$}
\label{sec:tqc35200}
\noindent Essential BR: $Ag@4b$ \\
\noindent RSI:
\begin{flalign*}
&\delta_{1}@4b\equiv -m(Ag)+m(Au) = -1,&
\end{flalign*}
\lstset{language=bash, keywordstyle=\color{blue!70}, basicstyle=\ttfamily, frame=shadowbox}
\begin{lstlisting}
Computed bands:  1 - 44
GM: GM1+(1); GM4+(1); GM2-(1); GM3-(1); GM1+(1); GM4+(1); GM2-(1); GM3-(1);
    GM4+(1); GM1+(1); GM3-(1); GM2-(1); GM4+(1); GM3-(1); GM2-(1); GM3-(1);
    GM1+(1); GM4+(1); GM1+(1); GM2-(1); GM4+(1); GM2-(1); GM1+(1); GM3-(1);
    GM1+(1); GM4+(1); GM3+(1); GM2+(1); GM4-(1); GM3+(1); GM1-(1); GM2+(1);
    GM2-(1); GM3-(1); GM4+(1); GM1+(1); GM2-(1); GM1-(1); GM2+(1); GM3-(1);
    GM4-(1); GM1+(1); GM4+(1); GM3+(1); [44] ;
R : R1  R2  (4); R1  R2  (4); R1  R2  (4); R1  R2  (4); R1  R2  (4); R1  R2  (4);
    R1  R2  (4); R1  R2  (4); R1  R2  (4); R1  R2  (4); R1  R2  (4); [44] ;
S : S1  S2  (4); S1  S2  (4); S1  S2  (4); S1  S2  (4); S1  S2  (4); S1  S2  (4);
    S1  S2  (4); S1  S2  (4); S1  S2  (4); S1  S2  (4); S1  S2  (4); [44] ;
T : T1  (2); T2  (2); T2  (2); T1  (2); T2  (2); T1  (2); T1  (2); T2  (2);
    T1  (2); T2  (2); T1  (2); T2  (2); T1  (2); T2  (2); T2  (2); T2  (2);
    T1  (2); T1  (2); T1  (2); T2  (2); T1  (2); T2  (2); [44] ;
U : U2- U3- (2); U1+ U4+ (2); U2- U3- (2); U1+ U4+ (2); U2- U3- (2); U1+ U4+ (2);
    U1+ U4+ (2); U2- U3- (2); U2- U3- (2); U1+ U4+ (2); U2- U3- (2); U1+ U4+ (2);
    U1- U4- (2); U2+ U3+ (2); U2- U3- (2); U1+ U4+ (2); U1- U4- (2); U2- U3- (2);
    U1+ U4+ (2); U2- U3- (2); U2+ U3+ (2); U1- U4- (2); [44] ;
X : X1  (2); X1  (2); X1  (2); X1  (2); X1  (2); X1  (2); X1  (2); X1  (2);
    X1  (2); X1  (2); X1  (2); X1  (2); X2  (2); X1  (2); X2  (2); X2  (2);
    X1  (2); X1  (2); X1  (2); X2  (2); X2  (2); X1  (2); [44] ;
Y : Y2  (2); Y1  (2); Y1  (2); Y2  (2); Y2  (2); Y1  (2); Y2  (2); Y1  (2);
    Y2  (2); Y1  (2); Y2  (2); Y1  (2); Y1  (2); Y2  (2); Y2  (2); Y1  (2);
    Y1  (2); Y2  (2); Y1  (2); Y2  (2); Y2  (2); Y1  (2); [44] ;
Z : Z1  (2); Z1  (2); Z1  (2); Z1  (2); Z1  (2); Z1  (2); Z1  (2); Z1  (2);
    Z1  (2); Z1  (2); Z1  (2); Z2  (2); Z1  (2); Z2  (2); Z1  (2); Z1  (2);
    Z2  (2); Z1  (2); Z1  (2); Z1  (2); Z2  (2); Z2  (2); [44];
\end{lstlisting}
\hyperref[tab:electride]{Back to the table}

\subsubsection*{78364 LiSi}
\label{sec:tqc78364}
\noindent Essential BR: $Ag@8c$ \\
\noindent RSI:
\begin{flalign*}
&\delta_{1}@8c\equiv -m(Ag)+m(Au) = -1,&
\end{flalign*}
\lstset{language=bash, keywordstyle=\color{blue!70}, basicstyle=\ttfamily, frame=shadowbox}
\begin{lstlisting}
Computed bands:  1 - 20
GM: GM1+(1); GM1-(1); GM3+GM4+(2); GM3-GM4-(2); GM2+(1); GM2-(1); GM1+(1);
    GM2-(1); GM3+GM4+(2); GM1-(1); GM2+(1); GM3-GM4-(2); GM1+(1); GM3+GM4+(2);
    GM2+(1); [20] ;
M : M2  (2); M1  (2); M1  (2); M2  (2); M1  (2); M2  (2); M2  (2); M1  (2);
    M1  (2); M2  (2); [20] ;
P : P1  P4  (2); P2  P3  (2); P2  P3  (2); P1  P4  (2); P1  P4  (2); P2  P3  (2);
    P1  P4  (2); P2  P3  (2); P2  P3  (2); P1  P4  (2); [20] ;
X : X1  (2); X1  (2); X1  (2); X1  (2); X1  (2); X1  (2); X1  (2); X1  (2);
    X1  (2); X1  (2); [20] ;
N : N1+ (1); N1- (1); N1+ (1); N1- (1); N1+ (1); N1- (1); N1+ (1); N1- (1);
    N1+ (1); N1- (1); N1- (1); N1+ (1); N1+ (1); N1- (1); N1- (1); N1+ (1);
    N1+ (1); N1- (1); N1+ (1); N1+ (1); [20];
\end{lstlisting}
\hyperref[tab:electride]{Back to the table}

\subsubsection*{601137 ZnSb}
\label{sec:tqc601137}
\noindent Essential BR: $Ag@4a$ \\
\noindent RSI:
\begin{flalign*}
&\delta_{1}@4a\equiv -m(Ag)+m(Au) = -1,&
\end{flalign*}
\lstset{language=bash, keywordstyle=\color{blue!70}, basicstyle=\ttfamily, frame=shadowbox}

\hyperref[tab:electride]{Back to the table}

\subsubsection*{52831 CdSb}
\label{sec:tqc52831}
\noindent Essential BR: $Ag@4a$ \\
\noindent RSI:
\begin{flalign*}
&\delta_{1}@4a\equiv -m(Ag)+m(Au) = -1,&
\end{flalign*}
\lstset{language=bash, keywordstyle=\color{blue!70}, basicstyle=\ttfamily, frame=shadowbox}

\hyperref[tab:electride]{Back to the table}

\subsubsection*{34050 Sb$_{3}$Ir}
\label{sec:tqc34050}
\lstset{language=bash, keywordstyle=\color{blue!70}, basicstyle=\ttfamily, frame=shadowbox}
\begin{lstlisting}
Computed bands:  1 - 48
GM: GM1+(1); GM2+GM3+(2); GM4-(3); GM4-(3); GM4+(3); GM2+GM3+(2); GM1+(1);
    GM4+(3); GM4+(3); GM1+(1); GM4-(3); GM2+GM3+(2); GM4+(3); GM4+(3); GM4+(3);
    GM4-(3); GM4-(3); GM2+GM3+(2); GM2-GM3-(2); GM1+(1); GM1-(1); [48] ;
H : H2+ H3+ (2); H4- (3); H1+ (1); H4- (3); H4+ (3); H1- (1); H4- (3); H4- (3);
    H1+ (1); H2- H3- (2); H2+ H3+ (2); H4- (3); H4- (3); H4+ (3); H1+ (1);
    H2- H3- (2); H2+ H3+ (2); H1- (1); H4- (3); H4- (3); H4+ (3); [48] ;
P : P4  (3); P1  (1); P4  (3); P2  P3  (2); P4  (3); P4  (3); P1  (1); P4  (3);
    P2  P3  (2); P4  (3); P4  (3); P4  (3); P2  P3  (2); P4  (3); P1  (1);
    P4  (3); P4  (3); P2  P3  (2); P4  (3); P1  (1); [48] ;
N : N2- (1); N1+ (1); N2- (1); N1+ (1); N2+ (1); N1- (1); N2- (1); N1+ (1);
    N1+ (1); N1- (1); N2- (1); N2+ (1); N2- (1); N1+ (1); N2+ (1); N1- (1);
    N1+ (1); N2- (1); N2- (1); N1- (1); N1+ (1); N2+ (1); N1+ (1); N2- (1);
    N2- (1); N1- (1); N1+ (1); N2+ (1); N2+ (1); N2- (1); N1- (1); N2+ (1);
    N1- (1); N2+ (1); N1- (1); N1+ (1); N2- (1); N2+ (1); N1+ (1); N1+ (1);
    N1- (1); N2- (1); N2- (1); N1+ (1); N1+ (1); N1- (1); N2- (1); N2+ (1); [48];
\end{lstlisting}
\hyperref[tab:electride]{Back to the table}

\subsubsection*{23628 SrP$_{3}$}
\label{sec:tqc23628}
\lstset{language=bash, keywordstyle=\color{blue!70}, basicstyle=\ttfamily, frame=shadowbox}

\hyperref[tab:electride]{Back to the table}

\subsubsection*{162107 RhN$_{3}$}
\label{sec:tqc162107}
\lstset{language=bash, keywordstyle=\color{blue!70}, basicstyle=\ttfamily, frame=shadowbox}
\begin{lstlisting}
Computed bands:  1 - 48
GM: GM1+(1); GM2+GM3+(2); GM4-(3); GM4-(3); GM4+(3); GM1+(1); GM2+GM3+(2);
    GM1+(1); GM2+GM3+(2); GM4-(3); GM4+(3); GM4+(3); GM4-(3); GM4-(3); GM4+(3);
    GM2+GM3+(2); GM1+(1); GM4+(3); GM4+(3); GM2-GM3-(2); GM1-(1); [48] ;
H : H2+ H3+ (2); H1+ (1); H4- (3); H4- (3); H4+ (3); H1+ (1); H4- (3);
    H2+ H3+ (2); H2+ H3+ (2); H1+ (1); H4- (3); H4- (3); H2- H3- (2); H4+ (3);
    H4- (3); H1- (1); H4- (3); H4+ (3); H4- (3); H2- H3- (2); H1- (1); [48] ;
P : P4  (3); P1  (1); P2  P3  (2); P4  (3); P4  (3); P4  (3); P4  (3); P4  (3);
    P2  P3  (2); P1  (1); P4  (3); P4  (3); P2  P3  (2); P1  (1); P4  (3);
    P4  (3); P1  (1); P2  P3  (2); P4  (3); P4  (3); [48] ;
N : N2- (1); N1+ (1); N2- (1); N1+ (1); N1- (1); N1+ (1); N2+ (1); N2- (1);
    N1+ (1); N2- (1); N1- (1); N2+ (1); N1+ (1); N2- (1); N2- (1); N1+ (1);
    N2- (1); N2- (1); N2+ (1); N1+ (1); N2- (1); N1+ (1); N1- (1); N1- (1);
    N2+ (1); N1+ (1); N2- (1); N2+ (1); N1+ (1); N2- (1); N1- (1); N1+ (1);
    N1- (1); N1+ (1); N2+ (1); N2+ (1); N1- (1); N2- (1); N1+ N2- (2); N1- (1);
    N2+ (1); N2- (1); N1+ (1); N1- (1); N2+ (1); N1- (1); N2+ (1); [48];
\end{lstlisting}
\hyperref[tab:electride]{Back to the table}

\subsubsection*{83664 Al$_{2}$(FeSi)$_{3}$}
\label{sec:tqc83664}
\lstset{language=bash, keywordstyle=\color{blue!70}, basicstyle=\ttfamily, frame=shadowbox}

\hyperref[tab:electride]{Back to the table}

\subsubsection*{427612 ZnAs}
\label{sec:tqc427612}
\noindent Essential BR: $Ag@4a$ \\
\noindent RSI:
\begin{flalign*}
&\delta_{1}@4a\equiv -m(Ag)+m(Au) = -1,&
\end{flalign*}
\lstset{language=bash, keywordstyle=\color{blue!70}, basicstyle=\ttfamily, frame=shadowbox}

\hyperref[tab:electride]{Back to the table}

\subsubsection*{51976 Mg$_{3}$In}
\label{sec:tqc51976}
\lstset{language=bash, keywordstyle=\color{blue!70}, basicstyle=\ttfamily, frame=shadowbox}
\begin{lstlisting}
Computed bands:  1 - 18
GM: GM1+(1); GM1+(1); GM2-(1); GM2-(1); GM1+(1); GM2-(1); GM3+(2); GM1+(1);
    GM1+(1); GM3-(2); GM3+(2); GM3-(2); GM2-(1); GM1+(1); [18] ;
T : T2- (1); T1+ (1); T1+ (1); T2- (1); T1+ (1); T2- (1); T3- (2); T2- (1);
    T3+ (2); T3+ (2); T1+ (1); T1+ (1); T3- (2); T2- (1); [18] ;
F : F2- (1); F1+ (1); F2- (1); F1+ (1); F1+ (1); F1+ (1); F2- (1); F1- (1);
    F2- (1); F1+ (1); F2+ (1); F1- (1); F2- (1); F1+ (1); F2+ (1); F1+ F2- (2);
    F1+ (1); [18] ;
L : L1+ (1); L2- (1); L2- (1); L1+ (1); L1+ (1); L2- (1); L2- (1); L2+ (1);
    L1- (1); L1+ (1); L1+ (1); L2- (1); L1+ (1); L2- (1); L1- (1); L2+ (1);
    L1+ (1); L2- (1); [18];
\end{lstlisting}
\hyperref[tab:electride]{Back to the table}

\subsubsection*{647985 ReP$_{3}$}
\label{sec:tqc647985}
\noindent Essential BR: $Ag@4b$ \\
\noindent RSI:
\begin{flalign*}
&\delta_{1}@4b\equiv -m(Ag)+m(Au) = -1,&
\end{flalign*}
\lstset{language=bash, keywordstyle=\color{blue!70}, basicstyle=\ttfamily, frame=shadowbox}
\begin{lstlisting}
Computed bands:  1 - 44
GM: GM1+(1); GM4+(1); GM2-(1); GM3-(1); GM1+(1); GM4+(1); GM2-(1); GM3-(1);
    GM4+(1); GM1+(1); GM3-(1); GM2-(1); GM4+(1); GM3-(1); GM2-(1); GM1+(1);
    GM3-(1); GM4+(1); GM1+(1); GM2-(1); GM4+(1); GM2-(1); GM1+(1); GM3-(1);
    GM1+(1); GM4+(1); GM3+(1); GM2+(1); GM4-(1); GM3+(1); GM1-(1); GM2+(1);
    GM3-(1); GM2-(1); GM4+(1); GM1+(1); GM2-(1); GM1-(1); GM2+(1); GM3-(1);
    GM4-(1); GM1+(1); GM4+(1); GM3+(1); [44] ;
R : R1  R2  (4); R1  R2  (4); R1  R2  (4); R1  R2  (4); R1  R2  (4); R1  R2  (4);
    R1  R2  (4); R1  R2  (4); R1  R2  (4); R1  R2  (4); R1  R2  (4); [44] ;
S : S1  S2  (4); S1  S2  (4); S1  S2  (4); S1  S2  (4); S1  S2  (4); S1  S2  (4);
    S1  S2  (4); S1  S2  (4); S1  S2  (4); S1  S2  (4); S1  S2  (4); [44] ;
T : T1  (2); T2  (2); T2  (2); T1  (2); T1  (2); T2  (2); T1  (2); T2  (2);
    T1  (2); T2  (2); T1  (2); T2  (2); T2  (2); T1  (2); T2  (2); T2  (2);
    T1  (2); T1  (2); T1  (2); T2  (2); T1  (2); T2  (2); [44] ;
U : U2- U3- (2); U1+ U4+ (2); U2- U3- (2); U1+ U4+ (2); U2- U3- (2); U1+ U4+ (2);
    U1+ U4+ (2); U2- U3- (2); U2- U3- (2); U1+ U4+ (2); U2- U3- (2); U1+ U4+ (2);
    U1- U4- (2); U2+ U3+ (2); U2- U3- (2); U1+ U4+ (2); U1- U4- (2); U2- U3- (2);
    U1+ U4+ (2); U2- U3- (2); U2+ U3+ (2); U1- U4- (2); [44] ;
X : X1  (2); X1  (2); X1  (2); X1  (2); X1  (2); X1  (2); X1  (2); X1  (2);
    X1  (2); X1  (2); X1  (2); X1  (2); X2  (2); X1  (2); X2  (2); X2  (2);
    X1  (2); X1  (2); X2  (2); X1  (2); X2  (2); X1  (2); [44] ;
Y : Y2  (2); Y1  (2); Y1  (2); Y2  (2); Y2  (2); Y1  (2); Y2  (2); Y1  (2);
    Y2  (2); Y1  (2); Y1  (2); Y2  (2); Y1  (2); Y2  (2); Y2  (2); Y1  (2);
    Y1  (2); Y2  (2); Y1  (2); Y2  (2); Y2  (2); Y1  (2); [44] ;
Z : Z1  (2); Z1  (2); Z1  (2); Z1  (2); Z1  (2); Z1  (2); Z1  (2); Z1  (2);
    Z1  (2); Z1  (2); Z1  (2); Z2  (2); Z1  (2); Z2  (2); Z1  (2); Z1  (2);
    Z2  (2); Z1  (2); Z1  (2); Z1  (2); Z2  (2); Z2  (2); [44];
\end{lstlisting}
\hyperref[tab:electride]{Back to the table}

\subsubsection*{103448 Ga$_{3}$Fe}
\label{sec:tqc103448}
\noindent Essential BR: $Ag@2b$ \\
\noindent RSI:
\begin{flalign*}
&\delta_{1}@2b\equiv -m(Ag)+m(Au)-m(B1g)+m(B1u)-m(B3g)+m(B3u)-m(B2g)+m(B2u) = -1,&
\end{flalign*}
\lstset{language=bash, keywordstyle=\color{blue!70}, basicstyle=\ttfamily, frame=shadowbox}
\begin{lstlisting}
Computed bands:  1 - 34
A : A4  (2); A3  (2); A1  (2); A1  (2); A4  (2); A3  (2); A1  (2); A1  (2);
    A2  (2); A3  (2); A1  (2); A4  (2); A2  (2); A4  (2); A2  (2); A1  (2);
    A3  (2); [34] ;
GM: GM1+(1); GM3+(1); GM2-(1); GM4+(1); GM5-(2); GM4+(1); GM3-(1); GM2+(1);
    GM5+(2); GM4-(1); GM1+(1); GM5-(2); GM5-(2); GM4+(1); GM2-(1); GM1+(1);
    GM4+(1); GM5+(2); GM5-(2); GM5-(2); GM1+(1); GM1-(1); GM3-(1); GM5+(2);
    GM2+(1); GM3+(1); [34] ;
M : M5- (2); M1+ M4+ (2); M2- M3- (2); M5+ (2); M5- (2); M1+ M4+ (2); M5- (2);
    M2+ M3+ (2); M1+ M4+ (2); M2- M3- (2); M5- (2); M1- M4- (2); M5+ (2);
    M1+ M4+ (2); M5- (2); M5+ (2); M2+ M3+ (2); [34] ;
Z : Z1  (2); Z2  (2); Z3  (2); Z1  (2); Z4  (2); Z1  (2); Z4  (2); Z1  (2);
    Z3  (2); Z1  (2); Z2  (2); Z1  (2); Z3  (2); Z4  (2); Z4  (2); Z3  (2);
    Z2  (2); [34] ;
R : R1- (2); R1+ (2); R1+ (2); R1- (2); R1+ (2); R1- (2); R1- (2); R1+ (2);
    R1+ (2); R1- (2); R1+ (2); R1- (2); R1- (2); R1+ (2); R1+ (2); R1- (2);
    R1- (2); [34] ;
X : X2  (2); X1  (2); X2  (2); X2  (2); X2  (2); X1  (2); X2  (2); X1  (2);
    X2  (2); X2  (2); X1  (2); X2  (2); X2  (2); X1  (2); X2  (2); X2  (2);
    X1  (2); [34];
\end{lstlisting}
\hyperref[tab:electride]{Back to the table}

\subsubsection*{635023 Ga$_{3}$Os}
\label{sec:tqc635023}
\noindent Essential BR: $Ag@2a$ \\
\noindent RSI:
\begin{flalign*}
&\delta_{1}@2a\equiv -m(Ag)+m(Au)-m(B1g)+m(B1u)-m(B3g)+m(B3u)-m(B2g)+m(B2u) = -1,&
\end{flalign*}
\lstset{language=bash, keywordstyle=\color{blue!70}, basicstyle=\ttfamily, frame=shadowbox}
\begin{lstlisting}
Computed bands:  1 - 34
A : A1  (2); A3  (2); A4  (2); A1  (2); A4  (2); A1  (2); A3  (2); A1  (2);
    A2  (2); A1  (2); A4  (2); A3  (2); A3  (2); A2  (2); A1  (2); A2  (2);
    A4  (2); [34] ;
GM: GM1+(1); GM4+(1); GM2-(1); GM3+(1); GM5-(2); GM4+(1); GM3-(1); GM5+(2);
    GM1+(1); GM2+(1); GM4-(1); GM4+(1); GM2-(1); GM5-(2); GM5-(2); GM1+(1);
    GM4+(1); GM5-(2); GM5+(2); GM1+(1); GM1-(1); GM5-(2); GM3-(1); GM5+(2);
    GM2+(1); GM3+(1); [34] ;
M : M1+ M4+ (2); M5- (2); M2- M3- (2); M5+ (2); M5- (2); M1+ M4+ (2); M5- (2);
    M1+ M4+ (2); M2- M3- (2); M5- (2); M2+ M3+ (2); M1- M4- (2); M5+ (2);
    M1+ M4+ (2); M5- (2); M5+ (2); M2+ M3+ (2); [34] ;
Z : Z1  (2); Z1  (2); Z2  (2); Z3  (2); Z4  (2); Z1  (2); Z3  (2); Z1  (2);
    Z4  (2); Z1  (2); Z2  (2); Z3  (2); Z1  (2); Z3  (2); Z4  (2); Z4  (2);
    Z2  (2); [34] ;
R : R1+ (2); R1- (2); R1- (2); R1+ (2); R1- (2); R1+ (2); R1+ (2); R1- (2);
    R1- (2); R1- (2); R1+ (2); R1+ (2); R1+ (2); R1- (2); R1- (2); R1+ (2);
    R1+ (2); [34] ;
X : X2  (2); X1  (2); X2  (2); X2  (2); X2  (2); X1  (2); X2  (2); X2  (2);
    X1  (2); X2  (2); X1  (2); X2  (2); X2  (2); X2  (2); X1  (2); X2  (2);
    X1  (2); [34];
\end{lstlisting}
\hyperref[tab:electride]{Back to the table}

\subsubsection*{634441 Ga$_{3}$Ir}
\label{sec:tqc634441}
\lstset{language=bash, keywordstyle=\color{blue!70}, basicstyle=\ttfamily, frame=shadowbox}
\begin{lstlisting}
Computed bands:  1 - 36
A : A1  (2); A4  (2); A3  (2); A1  (2); A3  (2); A1  (2); A4  (2); A1  (2);
    A2  (2); A1  (2); A3  (2); A4  (2); A4  (2); A2  (2); A1  (2); A3  (2);
    A2  (2); A1  (2); [36] ;
GM: GM1+(1); GM4+(1); GM3+(1); GM2-(1); GM5-(2); GM4+(1); GM3-(1); GM1+(1);
    GM5+(2); GM2+(1); GM4-(1); GM4+(1); GM2-(1); GM5-(2); GM5-(2); GM1+(1);
    GM4+(1); GM5-(2); GM5+(2); GM1-(1); GM1+(1); GM5-(2); GM3+(1); GM2+(1);
    GM5+(2); GM3-(1); GM1+(1); GM4+(1); [36] ;
M : M1+ M4+ (2); M5- (2); M2- M3- (2); M5+ (2); M5- (2); M5- (2); M1+ M4+ (2);
    M1+ M4+ (2); M2+ M3+ (2); M5- (2); M2- M3- (2); M5+ (2); M1- M4- (2);
    M5- (2); M1+ M4+ (2); M5+ (2); M2+ M3+ (2); M1+ M4+ (2); [36] ;
Z : Z1  (2); Z1  (2); Z2  (2); Z4  (2); Z3  (2); Z1  (2); Z4  (2); Z1  (2);
    Z3  (2); Z1  (2); Z2  (2); Z4  (2); Z4  (2); Z1  (2); Z3  (2); Z2  (2);
    Z3  (2); Z1  (2); [36] ;
R : R1- (2); R1+ (2); R1+ (2); R1- (2); R1+ (2); R1- (2); R1- (2); R1+ (2);
    R1+ (2); R1- (2); R1+ (2); R1- (2); R1- (2); R1+ (2); R1+ (2); R1- (2);
    R1- (2); R1+ (2); [36] ;
X : X2  (2); X1  (2); X2  (2); X2  (2); X2  (2); X1  (2); X2  (2); X2  (2);
    X1  (2); X2  (2); X1  (2); X2  (2); X2  (2); X1  (2); X2  (2); X2  (2);
    X1  (2); X2  (2); [36];
\end{lstlisting}
\hyperref[tab:electride]{Back to the table}

\subsubsection*{648855 SiPd$_{3}$}
\label{sec:tqc648855}
\noindent Essential BR: $Ag@4b$ \\
\noindent RSI:
\begin{flalign*}
&\delta_{1}@4b\equiv -m(Ag)+m(Au) = -1,&
\end{flalign*}
\lstset{language=bash, keywordstyle=\color{blue!70}, basicstyle=\ttfamily, frame=shadowbox}

\hyperref[tab:electride]{Back to the table}

\subsubsection*{103448 Ga$_{3}$Fe}
\label{sec:tqc103448}
\noindent Essential BR: $Ag@2b$ \\
\noindent RSI:
\begin{flalign*}
&\delta_{1}@2b\equiv -m(Ag)+m(Au)-m(B1g)+m(B1u)-m(B3g)+m(B3u)-m(B2g)+m(B2u) = -1,&
\end{flalign*}
\lstset{language=bash, keywordstyle=\color{blue!70}, basicstyle=\ttfamily, frame=shadowbox}
\begin{lstlisting}
Computed bands:  1 - 34
A : A4  (2); A3  (2); A1  (2); A1  (2); A4  (2); A3  (2); A1  (2); A1  (2);
    A2  (2); A3  (2); A1  (2); A4  (2); A2  (2); A4  (2); A2  (2); A1  (2);
    A3  (2); [34] ;
GM: GM1+(1); GM3+(1); GM2-(1); GM4+(1); GM5-(2); GM4+(1); GM3-(1); GM2+(1);
    GM5+(2); GM4-(1); GM1+(1); GM5-(2); GM5-(2); GM4+(1); GM2-(1); GM1+(1);
    GM4+(1); GM5+(2); GM5-(2); GM5-(2); GM1+(1); GM1-(1); GM3-(1); GM5+(2);
    GM2+(1); GM3+(1); [34] ;
M : M5- (2); M1+ M4+ (2); M2- M3- (2); M5+ (2); M5- (2); M1+ M4+ (2); M5- (2);
    M2+ M3+ (2); M1+ M4+ (2); M2- M3- (2); M5- (2); M1- M4- (2); M5+ (2);
    M1+ M4+ (2); M5- (2); M5+ (2); M2+ M3+ (2); [34] ;
Z : Z1  (2); Z2  (2); Z3  (2); Z1  (2); Z4  (2); Z1  (2); Z4  (2); Z1  (2);
    Z3  (2); Z1  (2); Z2  (2); Z1  (2); Z3  (2); Z4  (2); Z4  (2); Z3  (2);
    Z2  (2); [34] ;
R : R1- (2); R1+ (2); R1+ (2); R1- (2); R1+ (2); R1- (2); R1- (2); R1+ (2);
    R1+ (2); R1- (2); R1+ (2); R1- (2); R1- (2); R1+ (2); R1+ (2); R1- (2);
    R1- (2); [34] ;
X : X2  (2); X1  (2); X2  (2); X2  (2); X2  (2); X1  (2); X2  (2); X1  (2);
    X2  (2); X2  (2); X1  (2); X2  (2); X2  (2); X1  (2); X2  (2); X2  (2);
    X1  (2); [34];
\end{lstlisting}
\hyperref[tab:electride]{Back to the table}

\subsubsection*{76500 Tl$_{2}$(CdSb)$_{3}$}
\label{sec:tqc76500}
\noindent Essential BR: $Ag@2c$ \\
\noindent RSI:
\begin{flalign*}
&\delta_{1}@2c\equiv -m(Ag)+m(Au)-m(Bg)+m(Bu) = -1,&
\end{flalign*}
\lstset{language=bash, keywordstyle=\color{blue!70}, basicstyle=\ttfamily, frame=shadowbox}

\hyperref[tab:electride]{Back to the table}

\subsubsection*{640899 P$_{3}$Ir}
\label{sec:tqc640899}
\lstset{language=bash, keywordstyle=\color{blue!70}, basicstyle=\ttfamily, frame=shadowbox}
\begin{lstlisting}
Computed bands:  1 - 48
GM: GM1+(1); GM2+GM3+(2); GM4-(3); GM4-(3); GM4+(3); GM2+GM3+(2); GM1+(1);
    GM4+(3); GM1+(1); GM4+(3); GM2+GM3+(2); GM4-(3); GM4+(3); GM4+(3); GM4-(3);
    GM4+(3); GM4-(3); GM2+GM3+(2); GM1+(1); GM2-GM3-(2); GM1-(1); [48] ;
H : H2+ H3+ (2); H1+ (1); H4- (3); H4- (3); H4+ (3); H4- (3); H4- (3); H1+ (1);
    H1- (1); H2- H3- (2); H2+ H3+ (2); H4- (3); H1+ (1); H2+ H3+ (2); H4- (3);
    H4+ (3); H4- (3); H2- H3- (2); H1- (1); H4- (3); H4+ (3); [48] ;
P : P4  (3); P1  (1); P2  P3  (2); P4  (3); P4  (3); P4  (3); P4  (3);
    P2  P3  (2); P1  (1); P4  (3); P4  (3); P4  (3); P2  P3  (2); P1  (1);
    P4  (3); P4  (3); P4  (3); P2  P3  (2); P4  (3); P1  (1); [48] ;
N : N2- (1); N1+ (1); N2- (1); N1+ (1); N1- (1); N2+ (1); N1+ (1); N2- (1);
    N1+ (1); N1- (1); N2- (1); N2+ (1); N2- (1); N1+ (1); N2+ (1); N2- (1);
    N1+ (1); N1- (1); N2- (1); N1+ (1); N1- (1); N2- (1); N2+ (1); N1+ (1);
    N2- (1); N1+ (1); N1- (1); N2+ (1); N2- (1); N1- (1); N2+ (1); N2+ (1);
    N1+ (1); N1- (1); N2+ (1); N1- (1); N2- (1); N1+ (1); N2- (1); N1+ (1);
    N2+ (1); N2- (1); N1+ (1); N1- (1); N1+ (1); N2- (1); N1- (1); N2+ (1); [48];
\end{lstlisting}
\hyperref[tab:electride]{Back to the table}

\subsubsection*{96509 LiSi}
\label{sec:tqc96509}
\noindent Essential BR: $Ag@8c$ \\
\noindent RSI:
\begin{flalign*}
&\delta_{1}@8c\equiv -m(Ag)+m(Au) = -1,&
\end{flalign*}
\lstset{language=bash, keywordstyle=\color{blue!70}, basicstyle=\ttfamily, frame=shadowbox}
\begin{lstlisting}
Computed bands:  1 - 20
GM: GM1+(1); GM1-(1); GM3+GM4+(2); GM3-GM4-(2); GM2+(1); GM2-(1); GM1+(1);
    GM2-(1); GM3+GM4+(2); GM1-(1); GM2+(1); GM3-GM4-(2); GM1+(1); GM3+GM4+(2);
    GM2+(1); [20] ;
M : M2  (2); M1  (2); M1  (2); M2  (2); M1  (2); M2  (2); M2  (2); M1  (2);
    M1  (2); M2  (2); [20] ;
P : P1  P4  (2); P2  P3  (2); P2  P3  (2); P1  P4  (2); P1  P4  (2); P2  P3  (2);
    P1  P4  (2); P2  P3  (2); P2  P3  (2); P1  P4  (2); [20] ;
X : X1  (2); X1  (2); X1  (2); X1  (2); X1  (2); X1  (2); X1  (2); X1  (2);
    X1  (2); X1  (2); [20] ;
N : N1+ (1); N1- (1); N1+ (1); N1- (1); N1+ (1); N1- (1); N1+ (1); N1- (1);
    N1+ (1); N1- (1); N1- (1); N1+ (1); N1+ (1); N1- (1); N1- (1); N1+ (1);
    N1+ (1); N1- (1); N1+ (1); N1+ (1); [20];
\end{lstlisting}
\hyperref[tab:electride]{Back to the table}

\subsubsection*{27159 NiP}
\label{sec:tqc27159}
\noindent Essential BR: $Ag@4b$ \\
\noindent RSI:
\begin{flalign*}
&\delta_{1}@4b\equiv -m(Ag)+m(Au) = -1,&
\end{flalign*}
\lstset{language=bash, keywordstyle=\color{blue!70}, basicstyle=\ttfamily, frame=shadowbox}

\hyperref[tab:electride]{Back to the table}

\subsubsection*{408324 Li$_{3}$CaMnN$_{3}$}
\label{sec:tqc408324}
\lstset{language=bash, keywordstyle=\color{blue!70}, basicstyle=\ttfamily, frame=shadowbox}
\begin{lstlisting}
Computed bands:  1 - 35
GM: GM1+(1); GM1-(1); GM2+GM3+(2); GM1-(1); GM1+(1); GM2-GM3-(2); GM1+(1);
    GM1-(1); GM2-GM3-(2); GM2+GM3+(2); GM1+(1); GM2+GM3+(2); GM1-(1); GM1+(1);
    GM2-GM3-(2); GM2-GM3-(2); GM2+GM3+(2); GM2-GM3-(2); GM1-(1); GM1+(1);
    GM1-(1); GM2+GM3+(2); GM1+(1); GM2-GM3-(2); [35] ;
T : T1+ (1); T1- (1); T2+ T3+ (2); T1- (1); T1+ (1); T2- T3- (2); T1+ (1);
    T1- (1); T2- T3- (2); T2+ T3+ (2); T1+ (1); T1- (1); T2- T3- (2);
    T2+ T3+ (2); T2- T3- (2); T1- (1); T1+ (1); T2+ T3+ (2); T2- T3- (2);
    T1- (1); T1+ (1); T2+ T3+ (2); T1+ (1); T2- T3- (2); [35] ;
F : F1- (1); F1+ (1); F1- (1); F1+ (1); F1+ (1); F1- (1); F1- (1); F1+ (1);
    F1+ (1); F1- (1); F1- (1); F1+ (1); F1+ (1); F1- (1); F1+ (1); F1- (1);
    F1+ (1); F1- (1); F1- (1); F1+ (1); F1+ (1); F1- (1); F1- (1); F1- (1);
    F1+ (1); F1+ (1); F1- (1); F1- (1); F1+ (1); F1+ (1); F1+ (1); F1- (1);
    F1+ (1); F1- (1); F1- (1); [35] ;
L : L1- (1); L1+ (1); L1- (1); L1+ (1); L1+ (1); L1- (1); L1- (1); L1+ (1);
    L1+ (1); L1- (1); L1- (1); L1+ (1); L1+ (1); L1- (1); L1+ (1); L1- (1);
    L1+ (1); L1- (1); L1+ (1); L1- (1); L1- (1); L1+ (1); L1- (1); L1+ (1);
    L1- (1); L1+ (1); L1- (1); L1+ (1); L1- (1); L1+ (1); L1+ (1); L1+ (1);
    L1- (1); L1- (1); L1- (1); [35];
\end{lstlisting}
\hyperref[tab:electride]{Back to the table}

\subsubsection*{638878 Hf$_{3}$Sb}
\label{sec:tqc638878}
\lstset{language=bash, keywordstyle=\color{blue!70}, basicstyle=\ttfamily, frame=shadowbox}
\begin{lstlisting}
Computed bands:  1 - 34
GM: GM1 (1); GM3 GM4 (2); GM2 (1); GM1 (1); GM2 (1); GM3 GM4 (2); GM3 GM4 (2);
    GM2 (1); GM2 (1); GM1 (1); GM1 (1); GM3 GM4 (2); GM1 (1); GM1 GM3 GM4 (3);
    GM2 (1); GM1 (1); GM2 (1); GM3 GM4 (2); GM1 (1); GM3 GM4 (2); GM2 (1);
    GM2 (1); GM1 (1); GM1 (1); GM3 GM4 (2); [ 34];
M : M2  (1); M3  M4  (2); M1  (1); M1  (1); M3  M4  (2); M2  (1); M2  (1);
    M3  M4  (2); M1  (1); M3  M4  (2); M2  (1); M1  (1); M1  (1); M2  (1);
    M3  M4  (2); M2  (1); M1  (1); M3  M4  (2); M1  (1); M3  M4  (2); M2  (1);
    M1  (1); M2  (1); M3  M4  (2); M1  (1); M2  (1); [ 34];
P : P3  (1); P2  (1); P4  (1); P1  (1); P1  (1); P4  (1); P3  (1); P2  (1);
    P4  (1); P1  (1); P2  (1); P3  (1); P2  (1); P3  (1); P1  (1); P4  (1);
    P1  (1); P2  (1); P3  (1); P4  (1); P1  (1); P4  (1); P2  (1); P3  (1);
    P2  (1); P4  (1); P3  (1); P1  (1); P2  (1); P4  (1); P3  (1); P1  (1);
    P4  (1); P2  (1); [ 34];
X : X2  (1); X1  (1); X1  (1); X2  (1); X2  (1); X1  (1); X1  (1); X2  (1);
    X1  (1); X2  (1); X1  (1); X2  (1); X2  (1); X1  (1); X2  (1); X1  (1);
    X2  (1); X1  (1); X2  (1); X1  (1); X2  (1); X2  (1); X1  (1); X2  (1);
    X1  (1); X2  (1); X1  (1); X1  (1); X1  (1); X2  (1); X1  (1); X1  (1);
    X2  (1); X2  (1); [ 34];
N : N1  (1); N1  (1); N1  (1); N1  (1); N1  (1); N1  (1); N1  (1); N1  (1);
    N1  (1); N1  (1); N1  (1); N1  (1); N1  (1); N1  (1); N1  (1); N1  (1);
    N1  (1); N1  (1); N1  (1); N1  (1); N1  (1); N1  (1); N1  (1); N1  (1);
    N1  (1); N1  (1); N1  (1); N1  (1); N1  (1); N1  (1); N1  (1); N1  (1);
    N1  (1); N1  (1); [ 34];
PA: PA3 (1); PA2 (1); PA4 (1); PA1 (1); PA1 (1); PA4 (1); PA3 (1); PA2 (1);
    PA4 (1); PA1 (1); PA2 (1); PA3 (1); PA2 (1); PA3 (1); PA1 (1); PA4 (1);
    PA1 (1); PA2 (1); PA3 (1); PA4 (1); PA1 (1); PA4 (1); PA2 (1); PA3 (1);
    PA2 (1); PA4 (1); PA3 (1); PA1 (1); PA2 (1); PA4 (1); PA3 (1); PA1 (1);
    PA4 (1); PA2 (1); [ 34];
\end{lstlisting}
\hyperref[tab:electride]{Back to the table}

\subsubsection*{610521 Pr(FeAs$_{3}$)$_{4}$}
\label{sec:tqc610521}
\lstset{language=bash, keywordstyle=\color{blue!70}, basicstyle=\ttfamily, frame=shadowbox}
\begin{lstlisting}
Computed bands:  1 - 52
GM: GM1+(1); GM4-(3); GM1+(1); GM2+GM3+(2); GM4-(3); GM4-(3); GM4+(3); GM1+(1);
    GM2+GM3+(2); GM4+(3); GM4-(3); GM1+(1); GM4+(3); GM2+GM3+(2); GM4-(3);
    GM4+(3); GM4-(3); GM4+(3); GM2-GM3-(2); GM4+(3); GM2+GM3+(2); GM1+(1);
    GM1-(1); [52] ;
H : H1+ (1); H4- (3); H2+ H3+ (2); H1+ (1); H4- (3); H4- (3); H4+ (3); H1+ (1);
    H1- (1); H4- (3); H4- (3); H2+ H3+ (2); H2+ H3+ (2); H1+ (1); H4+ (3);
    H2- H3- (2); H4- (3); H4+ (3); H4- (3); H4- (3); H1- (1); H2- H3- (2);
    H4- (3); [52] ;
P : P1  (1); P4  (3); P4  (3); P1  (1); P4  (3); P2  P3  (2); P4  (3); P4  (3);
    P1  (1); P4  (3); P2  P3  (2); P4  (3); P4  (3); P4  (3); P2  P3  (2);
    P1  (1); P4  (3); P4  (3); P2  P3  (2); P4  (3); P4  (3); P1  (1); [52] ;
N : N1+ (1); N2- (1); N1- N2- (2); N2- (1); N1+ (1); N2- (1); N1+ (1); N2+ (1);
    N1- (1); N1+ (1); N2- (1); N1+ (1); N1- (1); N2+ (1); N2- (1); N2- (1);
    N1+ (1); N2+ (1); N1+ (1); N1- (1); N2- (1); N1+ (1); N2- (1); N2+ (1);
    N2- (1); N1+ (1); N1- (1); N1+ (1); N2- (1); N2+ (1); N1- (1); N2- (1);
    N2+ (1); N1- (1); N2+ (1); N1+ (1); N1- (1); N2+ (1); N2- (1); N1+ (1);
    N1+ (1); N1- (1); N2- (1); N1- (1); N2+ (1); N1+ (1); N2- (1); N1+ (1);
    N2+ (1); N2- (1); N1- (1); [52];
\end{lstlisting}
\hyperref[tab:electride]{Back to the table}

\subsubsection*{621065 Ce(FeSb$_{3}$)$_{4}$}
\label{sec:tqc621065}
\lstset{language=bash, keywordstyle=\color{blue!70}, basicstyle=\ttfamily, frame=shadowbox}
\begin{lstlisting}
Computed bands:  1 - 52
GM: GM1+(1); GM4-(3); GM1+(1); GM2+GM3+(2); GM4-(3); GM4-(3); GM4+(3); GM1+(1);
    GM2+GM3+(2); GM4-(3); GM4+(3); GM1+(1); GM4+(3); GM2+GM3+(2); GM4-(3);
    GM4+(3); GM4-(3); GM4+(3); GM2-GM3-(2); GM4+(3); GM2+GM3+(2); GM1+(1);
    GM1-(1); [52] ;
H : H1+ (1); H4- (3); H2+ H3+ (2); H4- (3); H1+ (1); H4- (3); H4+ (3); H1- (1);
    H1+ (1); H4- (3); H4- (3); H2+ H3+ (2); H4+ (3); H2+ H3+ (2); H4+ (3);
    H1+ (1); H4- (3); H2- H3- (2); H4- (3); H4- (3); H1- (1); H2- H3- (2);
    H4- (3); [52] ;
P : P1  (1); P4  (3); P4  (3); P1  (1); P2  P3  (2); P4  (3); P4  (3); P4  (3);
    P1  (1); P4  (3); P2  P3  (2); P4  (3); P4  (3); P4  (3); P2  P3  (2);
    P1  (1); P4  (3); P4  (3); P2  P3  (2); P4  (3); P1  (1); P4  (3); [52] ;
N : N1+ (1); N1- N2- (2); N2- (1); N2- (1); N1+ (1); N2- (1); N1+ (1); N2+ (1);
    N1- (1); N1+ (1); N2- (1); N1+ (1); N1- (1); N2+ N2- (2); N2- (1); N1+ (1);
    N1- (1); N2+ (1); N2- (1); N1+ (1); N1+ (1); N2- (1); N2+ (1); N2- (1);
    N1+ (1); N1- (1); N1+ (1); N1- (1); N2- (1); N2+ (1); N1- (1); N2+ (1);
    N2- (1); N2+ (1); N1+ (1); N1- (1); N2+ (1); N1+ (1); N2- (1); N1- (1);
    N1- (1); N2+ (1); N2- (1); N1+ (1); N1+ (1); N1+ (1); N2- (1); N2+ (1);
    N2- (1); N1- (1); [52];
\end{lstlisting}
\hyperref[tab:electride]{Back to the table}

\subsubsection*{621988 Ce(Sb$_{3}$Ru)$_{4}$}
\label{sec:tqc621988}
\lstset{language=bash, keywordstyle=\color{blue!70}, basicstyle=\ttfamily, frame=shadowbox}
\begin{lstlisting}
Computed bands:  1 - 52
GM: GM1+(1); GM4-(3); GM1+(1); GM2+GM3+(2); GM4-(3); GM4-(3); GM4+(3); GM1+(1);
    GM2+GM3+(2); GM4+(3); GM4+(3); GM4-(3); GM1+(1); GM2+GM3+(2); GM4+(3);
    GM4-(3); GM4+(3); GM4+(3); GM4-(3); GM2+GM3+(2); GM2-GM3-(2); GM1+(1);
    GM1-(1); [52] ;
H : H1+ (1); H4- (3); H2+ H3+ (2); H4- (3); H1+ (1); H4- (3); H4+ (3); H1+ (1);
    H4- (3); H4- (3); H1- (1); H2+ H3+ (2); H2- H3- (2); H2+ H3+ (2); H1+ (1);
    H4- (3); H4+ (3); H4- (3); H4+ (3); H4- (3); H1- (1); H2- H3- (2); H4- (3);
    [52] ;
P : P1  (1); P4  (3); P4  (3); P1  (1); P4  (3); P2  P3  (2); P4  (3); P4  (3);
    P1  P4  (4); P2  P3  (2); P4  (3); P4  (3); P4  (3); P2  P3  (2); P1  (1);
    P4  (3); P4  (3); P2  P3  (2); P4  (3); P4  (3); P1  (1); [52] ;
N : N1+ (1); N1- N2- N2- (3); N2- (1); N1+ (1); N2- (1); N2+ (1); N1+ (1);
    N1- (1); N1+ (1); N2- (1); N1+ (1); N1- (1); N2+ (1); N2- (1); N2- (1);
    N1+ (1); N1+ (1); N2- (1); N2+ (1); N1- (1); N1+ N2- (2); N2+ (1); N1- (1);
    N1+ (1); N2- (1); N2- (1); N1+ (1); N2+ (1); N1- (1); N2+ (1); N1- (1);
    N2- (1); N2+ (1); N1+ (1); N1- (1); N2+ (1); N1- (1); N1+ (1); N2- (1);
    N2+ (1); N1+ (1); N1- (1); N2- (1); N2- (1); N1+ (1); N1+ (1); N2+ (1);
    N2- (1); N1- (1); [52];
\end{lstlisting}
\hyperref[tab:electride]{Back to the table}

\subsubsection*{610010 Ce(As$_{3}$Os)$_{4}$}
\label{sec:tqc610010}
\lstset{language=bash, keywordstyle=\color{blue!70}, basicstyle=\ttfamily, frame=shadowbox}
\begin{lstlisting}
Computed bands:  1 - 52
GM: GM1+(1); GM4-(3); GM1+(1); GM2+GM3+(2); GM4-(3); GM4-(3); GM4+(3);
    GM2+GM3+(2); GM1+(1); GM4+(3); GM4+(3); GM1+(1); GM4-(3); GM2+GM3+(2);
    GM4+(3); GM4-(3); GM4+(3); GM4-(3); GM4+(3); GM2-GM3-(2); GM2+GM3+(2);
    GM1+(1); GM1-(1); [52] ;
H : H1+ (1); H4- (3); H2+ H3+ (2); H4- (3); H1+ (1); H4- (3); H4+ (3); H4- (3);
    H1- (1); H4- (3); H1+ (1); H2+ H3+ (2); H2- H3- (2); H2+ H3+ (2); H4- (3);
    H1+ (1); H4+ (3); H4- (3); H4+ (3); H4- (3); H1- (1); H2- H3- (2); H4- (3);
    [52] ;
P : P1  (1); P4  (3); P4  (3); P1  (1); P4  (3); P2  P3  (2); P4  (3); P4  (3);
    P1  (1); P4  (3); P2  P3  (2); P4  (3); P4  (3); P4  (3); P2  P3  (2);
    P1  (1); P4  (3); P4  (3); P2  P3  (2); P4  (3); P4  (3); P1  (1); [52] ;
N : N1+ (1); N2- (1); N1- (1); N2- (1); N2- (1); N1+ (1); N2- (1); N1+ (1);
    N2+ (1); N1- (1); N2- (1); N1+ (1); N1+ (1); N1- (1); N2+ (1); N2- (1);
    N2- (1); N1+ (1); N2+ (1); N1- (1); N1+ (1); N2- (1); N1+ (1); N2- (1);
    N2+ (1); N1- (1); N1+ (1); N2- (1); N2- (1); N1+ (1); N2+ (1); N1- (1);
    N2+ (1); N2- (1); N1- (1); N2+ (1); N1- (1); N2+ (1); N1+ (1); N1- (1);
    N1+ (1); N2- (1); N2- (1); N1- (1); N1+ N2+ (2); N1+ (1); N2- (1); N1+ (1);
    N2+ (1); N2- (1); N1- (1); [52];
\end{lstlisting}
\hyperref[tab:electride]{Back to the table}

\subsubsection*{610776 La(As$_{3}$Os)$_{4}$}
\label{sec:tqc610776}
\lstset{language=bash, keywordstyle=\color{blue!70}, basicstyle=\ttfamily, frame=shadowbox}
\begin{lstlisting}
Computed bands:  1 - 52
GM: GM1+(1); GM4-(3); GM1+(1); GM2+GM3+(2); GM4-(3); GM4-(3); GM4+(3);
    GM2+GM3+(2); GM1+(1); GM4+(3); GM4+(3); GM1+(1); GM4-(3); GM2+GM3+(2);
    GM4+(3); GM4-(3); GM4+(3); GM4-(3); GM4+(3); GM2-GM3-(2); GM2+GM3+(2);
    GM1+(1); GM1-(1); [52] ;
H : H1+ (1); H4- (3); H2+ H3+ (2); H4- (3); H1+ (1); H4- (3); H4+ (3); H4- (3);
    H1- (1); H4- (3); H1+ (1); H2+ H3+ (2); H2- H3- (2); H2+ H3+ (2); H4- (3);
    H1+ (1); H4+ (3); H4- (3); H4+ (3); H4- (3); H1- (1); H2- H3- (2); H4- (3);
    [52] ;
P : P1  (1); P4  (3); P4  (3); P1  (1); P4  (3); P2  P3  (2); P4  (3); P4  (3);
    P1  (1); P4  (3); P2  P3  (2); P4  (3); P4  (3); P4  (3); P2  P3  (2);
    P1  (1); P4  (3); P4  (3); P4  (3); P2  P3  (2); P4  (3); P1  (1); [52] ;
N : N1+ (1); N2- (1); N1- (1); N2- (1); N2- (1); N1+ (1); N2- (1); N1+ (1);
    N2+ (1); N1- (1); N2- (1); N1+ (1); N1+ (1); N1- (1); N2+ (1); N2- (1);
    N2- (1); N1+ (1); N2+ (1); N1- (1); N1+ (1); N2- (1); N1+ (1); N2- (1);
    N2+ (1); N1- (1); N1+ (1); N2- (1); N2- (1); N1+ (1); N2+ (1); N1- (1);
    N2+ (1); N2- (1); N1- (1); N2+ (1); N1- (1); N2+ (1); N1+ (1); N1- (1);
    N1+ (1); N2- (1); N2- (1); N1- (1); N1+ (1); N2+ (1); N1+ (1); N2- (1);
    N1+ (1); N2+ (1); N2- (1); N1- (1); [52];
\end{lstlisting}
\hyperref[tab:electride]{Back to the table}

\subsubsection*{611007 Nd(As$_{3}$Os)$_{4}$}
\label{sec:tqc611007}
\lstset{language=bash, keywordstyle=\color{blue!70}, basicstyle=\ttfamily, frame=shadowbox}
\begin{lstlisting}
Computed bands:  1 - 52
GM: GM1+(1); GM4-(3); GM1+(1); GM2+GM3+(2); GM4-(3); GM4-(3); GM4+(3); GM1+(1);
    GM2+GM3+(2); GM4+(3); GM4+(3); GM1+(1); GM4-(3); GM2+GM3+(2); GM4+(3);
    GM4-(3); GM4+(3); GM4-(3); GM4+(3); GM2-GM3-(2); GM2+GM3+(2); GM1+(1);
    GM1-(1); [52] ;
H : H1+ (1); H4- (3); H2+ H3+ (2); H4- (3); H1+ (1); H4- (3); H4+ (3); H4- (3);
    H1- (1); H4- (3); H1+ (1); H2+ H3+ (2); H2- H3- (2); H2+ H3+ (2); H4- (3);
    H1+ (1); H4+ (3); H4- (3); H4+ (3); H4- (3); H1- (1); H2- H3- (2); H4- (3);
    [52] ;
P : P1  (1); P4  (3); P4  (3); P1  (1); P4  (3); P2  P3  (2); P4  (3); P4  (3);
    P1  (1); P4  (3); P2  P3  (2); P4  (3); P4  (3); P4  (3); P2  P3  (2);
    P1  (1); P4  (3); P4  (3); P2  P3  (2); P4  (3); P4  (3); P1  (1); [52] ;
N : N1+ (1); N2- (1); N1- (1); N2- (1); N2- (1); N1+ (1); N2- (1); N1+ (1);
    N2+ (1); N1- (1); N2- (1); N1+ (1); N1+ (1); N1- (1); N2+ (1); N2- (1);
    N2- (1); N1+ (1); N2+ (1); N1- (1); N1+ (1); N2- (1); N1+ (1); N2- (1);
    N2+ (1); N1- (1); N1+ (1); N2- (1); N2- (1); N1+ (1); N2+ (1); N1- (1);
    N2+ (1); N2- (1); N1- (1); N2+ (1); N1- (1); N2+ (1); N1+ (1); N1- (1);
    N1+ (1); N2- (1); N2- (1); N1- (1); N1+ N2+ (2); N1+ (1); N2- (1); N1+ (1);
    N2+ (1); N2- (1); N1- (1); [52];
\end{lstlisting}
\hyperref[tab:electride]{Back to the table}

\subsubsection*{23080 La(FeAs$_{3}$)$_{4}$}
\label{sec:tqc23080}
\lstset{language=bash, keywordstyle=\color{blue!70}, basicstyle=\ttfamily, frame=shadowbox}
\begin{lstlisting}
Computed bands:  1 - 52
GM: GM1+(1); GM4-(3); GM1+(1); GM2+GM3+(2); GM4-(3); GM4-(3); GM4+(3); GM1+(1);
    GM2+GM3+(2); GM4+(3); GM4-(3); GM1+(1); GM4+(3); GM2+GM3+(2); GM4-(3);
    GM4+(3); GM4-(3); GM4+(3); GM2-GM3-(2); GM4+(3); GM2+GM3+(2); GM1+(1);
    GM1-(1); [52] ;
H : H1+ (1); H4- (3); H2+ H3+ (2); H1+ (1); H4- (3); H4- (3); H4+ (3); H1+ (1);
    H1- (1); H4- (3); H4- (3); H2+ H3+ (2); H2+ H3+ (2); H1+ (1); H4+ (3);
    H2- H3- (2); H4- (3); H4+ (3); H4- (3); H4- (3); H1- (1); H2- H3- (2);
    H4- (3); [52] ;
P : P1  (1); P4  (3); P4  (3); P1  (1); P4  (3); P2  P3  (2); P4  (3); P4  (3);
    P1  (1); P4  (3); P2  P3  (2); P4  (3); P4  (3); P4  (3); P2  P3  (2);
    P1  (1); P4  (3); P4  (3); P2  P3  (2); P4  (3); P4  (3); P1  (1); [52] ;
N : N1+ (1); N2- (1); N1- (1); N2- (1); N2- (1); N1+ (1); N2- (1); N1+ (1);
    N2+ (1); N1- (1); N1+ (1); N2- (1); N1+ (1); N1- (1); N2+ (1); N2- (1);
    N2- (1); N1+ (1); N2+ (1); N1- (1); N1+ (1); N2- (1); N1+ (1); N2- (1);
    N2+ (1); N2- (1); N1+ (1); N1- (1); N1+ (1); N2- (1); N2+ (1); N1- (1);
    N2- (1); N2+ (1); N1- (1); N2+ (1); N1+ (1); N1- (1); N2- (1); N2+ (1);
    N1+ (1); N1- (1); N1+ (1); N2- (1); N1- (1); N2+ (1); N1+ (1); N2- (1);
    N1+ (1); N2+ (1); N2- (1); N1- (1); [52];
\end{lstlisting}
\hyperref[tab:electride]{Back to the table}

\subsubsection*{610003 Ce(FeAs$_{3}$)$_{4}$}
\label{sec:tqc610003}
\lstset{language=bash, keywordstyle=\color{blue!70}, basicstyle=\ttfamily, frame=shadowbox}
\begin{lstlisting}
Computed bands:  1 - 52
GM: GM1+(1); GM4-(3); GM1+(1); GM2+GM3+(2); GM4-(3); GM4-(3); GM4+(3); GM1+(1);
    GM2+GM3+(2); GM4+(3); GM4-(3); GM1+(1); GM4+(3); GM2+GM3+(2); GM4-(3);
    GM4+(3); GM4-(3); GM4+(3); GM2-GM3-(2); GM4+(3); GM2+GM3+(2); GM1+(1);
    GM1-(1); [52] ;
H : H1+ (1); H4- (3); H2+ H3+ (2); H1+ (1); H4- (3); H4- (3); H4+ (3); H1+ (1);
    H1- (1); H4- (3); H2+ H3+ H4- (5); H2+ H3+ (2); H1+ (1); H4+ (3);
    H2- H3- (2); H4- (3); H4+ (3); H4- (3); H4- (3); H1- (1); H2- H3- (2);
    H4- (3); [52] ;
P : P1  (1); P4  (3); P4  (3); P1  (1); P4  (3); P2  P3  (2); P4  (3); P4  (3);
    P1  (1); P4  (3); P2  P3  (2); P4  (3); P4  (3); P4  (3); P2  P3  (2);
    P1  (1); P4  (3); P4  (3); P2  P3  (2); P4  (3); P4  (3); P1  (1); [52] ;
N : N1+ (1); N2- (1); N1- (1); N2- (1); N2- (1); N1+ (1); N2- (1); N1+ (1);
    N2+ (1); N1- (1); N1+ (1); N2- (1); N1+ (1); N1- (1); N2+ (1); N2- (1);
    N2- (1); N1+ (1); N2+ (1); N1+ (1); N1- (1); N2- (1); N1+ (1); N2- (1);
    N2+ (1); N2- (1); N1+ (1); N1- (1); N1+ (1); N2- (1); N2+ (1); N1- (1);
    N2- (1); N2+ (1); N1- (1); N2+ (1); N1+ (1); N1- (1); N2+ (1); N2- (1);
    N1+ (1); N1+ (1); N1- (1); N2- (1); N1- (1); N2+ (1); N1+ (1); N2- (1);
    N1+ (1); N2+ (1); N2- (1); N1- (1); [52];
\end{lstlisting}
\hyperref[tab:electride]{Back to the table}

\subsubsection*{621737 Ce(Sb$_{3}$Os)$_{4}$}
\label{sec:tqc621737}
\lstset{language=bash, keywordstyle=\color{blue!70}, basicstyle=\ttfamily, frame=shadowbox}
\begin{lstlisting}
Computed bands:  1 - 52
GM: GM1+(1); GM4-(3); GM1+(1); GM2+GM3+(2); GM4-(3); GM4+(3); GM4-(3); GM1+(1);
    GM2+GM3+(2); GM4+(3); GM4+(3); GM1+(1); GM4-(3); GM4+(3); GM2+GM3+(2);
    GM4+(3); GM4-(3); GM4+(3); GM4-(3); GM2-GM3-(2); GM2+GM3+(2); GM1+(1);
    GM1-(1); [52] ;
H : H1+ (1); H4- (3); H2+ H3+ (2); H4- (3); H1+ (1); H4- (3); H4+ (3); H1- (1);
    H4- (3); H4- (3); H1+ (1); H2- H3- (2); H2+ H3+ (2); H4- (3); H2+ H3+ (2);
    H4+ (3); H4- (3); H1+ (1); H4+ (3); H1- (1); H4- (3); H2- H3- (2); H4- (3);
    [52] ;
P : P1  (1); P4  (3); P4  (3); P1  (1); P2  P3  (2); P4  (3); P4  (3); P4  (3);
    P1  (1); P4  (3); P2  P3  (2); P4  (3); P4  (3); P4  (3); P2  P3  (2);
    P1  (1); P4  (3); P4  (3); P2  P3  (2); P4  (3); P4  (3); P1  (1); [52] ;
N : N1+ (1); N1- N2- (2); N2- (1); N2- (1); N1+ (1); N2- (1); N2+ (1); N1+ (1);
    N1- (1); N1+ (1); N2- (1); N1+ (1); N1- (1); N2+ (1); N2- (1); N2- (1);
    N1+ (1); N1- (1); N2+ (1); N1+ N2- (2); N2- (1); N1+ (1); N2+ (1); N1- (1);
    N1+ (1); N2- (1); N2- (1); N1+ (1); N2+ (1); N1- (1); N2+ (1); N1- (1);
    N2+ (1); N2- (1); N1- (1); N2+ (1); N1+ (1); N1- (1); N1+ (1); N2+ (1);
    N1- (1); N2- (1); N1+ (1); N2- (1); N2- (1); N1+ (1); N1+ (1); N2+ (1);
    N2- (1); N1- (1); [52];
\end{lstlisting}
\hyperref[tab:electride]{Back to the table}

\subsubsection*{610013 Ce(As$_{3}$Ru)$_{4}$}
\label{sec:tqc610013}
\lstset{language=bash, keywordstyle=\color{blue!70}, basicstyle=\ttfamily, frame=shadowbox}
\begin{lstlisting}
Computed bands:  1 - 52
GM: GM1+(1); GM4-(3); GM1+(1); GM2+GM3+(2); GM4-(3); GM4-(3); GM4+(3);
    GM2+GM3+(2); GM1+(1); GM4+(3); GM4+(3); GM1+(1); GM4-(3); GM2+GM3+(2);
    GM4+(3); GM4-(3); GM4+(3); GM4-(3); GM4+(3); GM2+GM3+(2); GM2-GM3-(2);
    GM1+(1); GM1-(1); [52] ;
H : H1+ (1); H4- (3); H2+ H3+ (2); H4- (3); H1+ (1); H4- (3); H4+ (3); H4- (3);
    H1+ (1); H4- (3); H1- (1); H2+ H3+ (2); H2- H3- (2); H2+ H3+ (2); H1+ (1);
    H4- (3); H4+ (3); H4- (3); H4+ (3); H4- (3); H1- (1); H2- H3- (2); H4- (3);
    [52] ;
P : P1  (1); P4  (3); P4  (3); P1  (1); P4  (3); P2  P3  (2); P4  (3); P4  (3);
    P4  (3); P2  P3  (2); P1  (1); P4  (3); P4  (3); P4  (3); P2  P3  (2);
    P1  (1); P4  (3); P4  (3); P4  (3); P2  P3  (2); P4  (3); P1  (1); [52] ;
N : N1+ (1); N2- (1); N1- (1); N2- (1); N2- (1); N1+ (1); N2- (1); N1+ (1);
    N2+ (1); N1- (1); N2- (1); N1+ (1); N1+ (1); N1- (1); N2+ (1); N2- (1);
    N2- (1); N1+ (1); N1+ (1); N2- (1); N2+ (1); N1- (1); N1+ (1); N2+ (1);
    N2- (1); N1- (1); N1+ (1); N2- (1); N2- (1); N1+ (1); N2+ (1); N1- (1);
    N2- (1); N2+ (1); N1- (1); N2+ (1); N1+ N2+ (2); N1- (1); N1- (1); N1+ (1);
    N2- (1); N2- (1); N1+ (1); N2+ (1); N1- (1); N2- (1); N1+ (1); N1+ (1);
    N2+ (1); N2- (1); N1- (1); [52];
\end{lstlisting}
\hyperref[tab:electride]{Back to the table}

\subsubsection*{647712 Pr(P$_{3}$Os)$_{4}$}
\label{sec:tqc647712}
\lstset{language=bash, keywordstyle=\color{blue!70}, basicstyle=\ttfamily, frame=shadowbox}
\begin{lstlisting}
Computed bands:  1 - 52
GM: GM1+(1); GM4-(3); GM1+(1); GM2+GM3+(2); GM4-(3); GM4-(3); GM4+(3); GM1+(1);
    GM2+GM3+(2); GM1+(1); GM4+(3); GM4+(3); GM2+GM3+(2); GM4-(3); GM4-(3);
    GM4+(3); GM4-(3); GM4+(3); GM4+(3); GM2+GM3+(2); GM2-GM3-(2); GM1+(1);
    GM1-(1); [52] ;
H : H1+ (1); H4- (3); H2+ H3+ (2); H1+ (1); H4- (3); H4- (3); H4+ (3); H1+ (1);
    H4- (3); H4- (3); H2+ H3+ (2); H1- (1); H2+ H3+ (2); H2- H3- (2); H1+ (1);
    H4- (3); H4+ (3); H4- (3); H4- (3); H4+ (3); H1- (1); H4- (3); H2- H3- (2);
    [52] ;
P : P1  (1); P4  (3); P4  (3); P1  (1); P4  (3); P2  P3  (2); P4  (3); P4  (3);
    P4  (3); P1  (1); P2  P3  (2); P4  (3); P4  (3); P4  (3); P2  P3  (2);
    P1  (1); P4  (3); P4  (3); P4  (3); P2  P3  (2); P4  (3); P1  (1); [52] ;
N : N1+ (1); N2- (1); N1- (1); N2- (1); N2- (1); N1+ (1); N2- (1); N1+ (1);
    N2+ (1); N1- (1); N2- (1); N1+ (1); N1+ (1); N1- (1); N2+ (1); N2- (1);
    N2- (1); N1+ (1); N1+ (1); N2+ (1); N2- (1); N1- (1); N2- (1); N1+ (1);
    N2- (1); N2+ (1); N1- (1); N1+ (1); N2- (1); N1+ (1); N2- (1); N2+ (1);
    N1- (1); N2+ (1); N1- (1); N2+ (1); N1+ (1); N2- (1); N1+ (1); N1+ N2+ (2);
    N1- (1); N2- (1); N1- (1); N1+ (1); N1- (1); N2- (1); N2+ (1); N1+ (1);
    N2- (1); N2+ (1); N1- (1); [52];
\end{lstlisting}
\hyperref[tab:electride]{Back to the table}

\subsubsection*{183088 Nd(Sb$_{3}$Os)$_{4}$}
\label{sec:tqc183088}
\lstset{language=bash, keywordstyle=\color{blue!70}, basicstyle=\ttfamily, frame=shadowbox}
\begin{lstlisting}
Computed bands:  1 - 52
GM: GM1+(1); GM4-(3); GM1+(1); GM2+GM3+(2); GM4-(3); GM4-(3); GM4+(3); GM1+(1);
    GM2+GM3+(2); GM4+(3); GM4+(3); GM1+(1); GM4-(3); GM2+GM3+(2); GM4+(3);
    GM4+(3); GM4-(3); GM4+(3); GM4-(3); GM2-GM3-(2); GM2+GM3+(2); GM1+(1);
    GM1-(1); [52] ;
H : H1+ (1); H4- (3); H2+ H3+ (2); H4- (3); H1+ (1); H4- (3); H4+ (3); H1- (1);
    H4- (3); H1+ (1); H4- (3); H2+ H3+ (2); H2- H3- (2); H4- (3); H2+ H3+ (2);
    H4+ (3); H1+ (1); H4- (3); H4+ (3); H4- (3); H1- (1); H2- H3- (2); H4- (3);
    [52] ;
P : P1  (1); P4  (3); P4  (3); P1  (1); P4  (3); P2  P3  (2); P4  (3); P4  (3);
    P1  (1); P4  (3); P2  P3  (2); P4  (3); P4  (3); P4  (3); P2  P3  (2);
    P1  (1); P4  (3); P4  (3); P2  P3  (2); P4  (3); P4  (3); P1  (1); [52] ;
N : N1+ (1); N1- N2- N2- (3); N2- (1); N1+ (1); N2- (1); N1+ (1); N2+ (1);
    N1- (1); N1+ (1); N2- (1); N1+ (1); N1- (1); N2- (1); N2+ (1); N2- (1);
    N1+ (1); N1- (1); N2+ (1); N2- (1); N1+ (1); N1+ (1); N2- (1); N2+ (1);
    N1- (1); N1+ (1); N2- (1); N2- (1); N1+ (1); N2+ (1); N1- (1); N2+ (1);
    N1- (1); N2- (1); N2+ (1); N1- (1); N1+ N2+ (2); N1- (1); N1+ (1); N2- (1);
    N2+ (1); N1- (1); N1+ N2- (2); N1+ (1); N2- (1); N1+ (1); N2+ (1); N2- (1);
    N1- (1); [52];
\end{lstlisting}
\hyperref[tab:electride]{Back to the table}

\subsubsection*{611222 Pr(As$_{3}$Ru)$_{4}$}
\label{sec:tqc611222}
\lstset{language=bash, keywordstyle=\color{blue!70}, basicstyle=\ttfamily, frame=shadowbox}
\begin{lstlisting}
Computed bands:  1 - 52
GM: GM1+(1); GM4-(3); GM1+(1); GM2+GM3+(2); GM4-(3); GM4-(3); GM4+(3);
    GM2+GM3+(2); GM1+(1); GM4+(3); GM4+(3); GM1+(1); GM4-(3); GM2+GM3+(2);
    GM4+(3); GM4-(3); GM4+(3); GM4-(3); GM4+(3); GM2+GM3+(2); GM2-GM3-(2);
    GM1+(1); GM1-(1); [52] ;
H : H1+ (1); H4- (3); H2+ H3+ (2); H4- (3); H1+ (1); H4- (3); H4+ (3); H4- (3);
    H1+ (1); H4- (3); H1- (1); H2+ H3+ (2); H2- H3- (2); H2+ H3+ (2); H1+ (1);
    H4- (3); H4+ (3); H4- (3); H4+ (3); H4- (3); H1- (1); H2- H3- (2); H4- (3);
    [52] ;
P : P1  (1); P4  (3); P4  (3); P1  (1); P4  (3); P2  P3  (2); P4  (3); P4  (3);
    P4  (3); P2  P3  (2); P1  (1); P4  (3); P4  (3); P4  (3); P2  P3  (2);
    P1  (1); P4  (3); P4  (3); P4  (3); P2  P3  (2); P4  (3); P1  (1); [52] ;
N : N1+ (1); N2- (1); N1- (1); N2- (1); N2- (1); N1+ (1); N2- (1); N1+ (1);
    N2+ (1); N1- (1); N2- (1); N1+ (1); N1+ (1); N1- (1); N2+ (1); N2- (1);
    N2- (1); N1+ (1); N1+ (1); N2- (1); N2+ (1); N1- (1); N1+ (1); N2- (1);
    N2+ (1); N1- (1); N1+ (1); N2- (1); N2- (1); N1+ (1); N2+ (1); N1- (1);
    N2- (1); N2+ (1); N1- (1); N2+ (1); N1+ (1); N2+ (1); N1- (1); N1- (1);
    N1+ (1); N2- (1); N2- (1); N1+ (1); N2+ (1); N1- (1); N2- (1); N1+ (1);
    N1+ (1); N2+ (1); N2- (1); N1- (1); [52];
\end{lstlisting}
\hyperref[tab:electride]{Back to the table}

\subsubsection*{1286 La(FeP$_{3}$)$_{4}$}
\label{sec:tqc1286}
\lstset{language=bash, keywordstyle=\color{blue!70}, basicstyle=\ttfamily, frame=shadowbox}
\begin{lstlisting}
Computed bands:  1 - 52
GM: GM1+(1); GM4-(3); GM1+(1); GM2+GM3+(2); GM4-(3); GM4-(3); GM4+(3); GM1+(1);
    GM2+GM3+(2); GM1+(1); GM4-(3); GM4+(3); GM2+GM3+(2); GM4+(3); GM4-(3);
    GM4-(3); GM4+(3); GM4+(3); GM4+(3); GM2-GM3-(2); GM2+GM3+(2); GM1+(1);
    GM1-(1); [52] ;
H : H1+ (1); H4- (3); H2+ H3+ (2); H1+ (1); H4- (3); H4- (3); H4+ (3); H1+ (1);
    H4- (3); H2+ H3+ (2); H1- (1); H4- (3); H2+ H3+ (2); H1+ (1); H4+ (3);
    H4- (3); H2- H3- (2); H4+ (3); H4- (3); H4- (3); H1- (1); H2- H3- (2);
    H4- (3); [52] ;
P : P1  (1); P4  (3); P4  (3); P1  (1); P4  (3); P2  P3  (2); P4  (3); P4  (3);
    P4  (3); P1  (1); P2  P3  (2); P4  (3); P4  (3); P4  (3); P2  P3  (2);
    P4  (3); P1  (1); P4  (3); P2  P3  (2); P4  (3); P4  (3); P1  (1); [52] ;
N : N1+ (1); N2- (1); N2- (1); N1- (1); N2- (1); N1+ (1); N2- (1); N2+ (1);
    N1+ (1); N1+ (1); N2- (1); N1- (1); N1+ (1); N1- (1); N2+ (1); N2- (1);
    N1+ (1); N2- (1); N1+ (1); N2+ (1); N1- (1); N2- (1); N2- (1); N1+ (1);
    N2+ (1); N2- (1); N1+ (1); N1- (1); N2- (1); N1+ (1); N2- (1); N2+ (1);
    N1- (1); N2+ (1); N1- (1); N2+ (1); N1+ (1); N2- (1); N1+ (1); N1- (1);
    N2+ (1); N1+ (1); N2- (1); N1- (1); N1+ (1); N1- (1); N2+ (1); N2- (1);
    N1+ (1); N2+ (1); N2- (1); N1- (1); [52];
\end{lstlisting}
\hyperref[tab:electride]{Back to the table}

\subsubsection*{79927 Nd(FeSb$_{3}$)$_{4}$}
\label{sec:tqc79927}
\lstset{language=bash, keywordstyle=\color{blue!70}, basicstyle=\ttfamily, frame=shadowbox}
\begin{lstlisting}
Computed bands:  1 - 52
GM: GM1+(1); GM4-(3); GM1+(1); GM2+GM3+(2); GM4-(3); GM4-(3); GM4+(3); GM1+(1);
    GM2+GM3+(2); GM4-(3); GM4+(3); GM1+(1); GM4+(3); GM4-(3); GM2+GM3+(2);
    GM4+(3); GM4-(3); GM4+(3); GM2-GM3-(2); GM4+(3); GM2+GM3+(2); GM1+(1);
    GM1-(1); [52] ;
H : H1+ (1); H4- (3); H2+ H3+ (2); H4- (3); H1+ (1); H4- (3); H4+ (3); H1- (1);
    H1+ (1); H4- (3); H4- (3); H2+ H3+ (2); H4+ (3); H2+ H3+ (2); H4+ (3);
    H1+ (1); H2- H3- (2); H4- (3); H4- (3); H4- (3); H1- (1); H2- H3- (2);
    H4- (3); [52] ;
P : P1  (1); P4  (3); P4  (3); P1  (1); P2  P3  (2); P4  (3); P4  (3); P4  (3);
    P1  (1); P4  (3); P2  P3  (2); P4  (3); P4  (3); P4  (3); P1  (1);
    P2  P3  (2); P4  (3); P4  (3); P2  P3  (2); P4  (3); P1  (1); P4  (3); [52] ;
N : N1+ (1); N1- N2- N2- (3); N2- (1); N1+ (1); N2- (1); N1+ (1); N2+ (1);
    N1- (1); N1+ (1); N2- (1); N1+ (1); N1- (1); N2- (1); N2+ (1); N2- (1);
    N1+ (1); N1- (1); N2+ (1); N2- (1); N1+ (1); N1+ (1); N2- (1); N2+ (1);
    N2- (1); N1+ (1); N1- (1); N1+ (1); N1- (1); N2+ N2- (2); N1- (1); N2+ (1);
    N2+ N2- (2); N1+ (1); N1- (1); N2+ (1); N1+ (1); N1- (1); N2- (1); N1- (1);
    N2+ (1); N2- (1); N1+ (1); N1+ (1); N1+ N2- (2); N2+ (1); N2- (1); N1- (1);
    [52];
\end{lstlisting}
\hyperref[tab:electride]{Back to the table}

\subsubsection*{645809 Nd(Sb$_{3}$Ru)$_{4}$}
\label{sec:tqc645809}
\lstset{language=bash, keywordstyle=\color{blue!70}, basicstyle=\ttfamily, frame=shadowbox}
\begin{lstlisting}
Computed bands:  1 - 52
GM: GM1+(1); GM4-(3); GM1+(1); GM2+GM3+(2); GM4-(3); GM4-(3); GM4+(3); GM1+(1);
    GM2+GM3+(2); GM4+(3); GM4+(3); GM4-(3); GM1+(1); GM2+GM3+(2); GM4+(3);
    GM4-(3); GM4+(3); GM4+(3); GM4-(3); GM2+GM3+(2); GM2-GM3-(2); GM1+(1);
    GM1-(1); [52] ;
H : H1+ (1); H4- (3); H2+ H3+ (2); H4- (3); H1+ (1); H4- (3); H4+ (3); H1+ (1);
    H4- (3); H4- (3); H1- (1); H2+ H3+ (2); H2- H3- (2); H2+ H3+ (2); H4- (3);
    H1+ (1); H4+ (3); H4- (3); H4+ (3); H4- (3); H1- (1); H2- H3- (2); H4- (3);
    [52] ;
P : P1  (1); P4  (3); P4  (3); P1  (1); P4  (3); P2  P3  (2); P4  (3); P4  (3);
    P4  (3); P1  (1); P2  P3  (2); P4  (3); P4  (3); P4  (3); P2  P3  (2);
    P1  (1); P4  (3); P4  (3); P2  P3  (2); P4  (3); P4  (3); P1  (1); [52] ;
N : N1+ (1); N1- N2- N2- (3); N2- (1); N1+ (1); N2- (1); N1+ (1); N2+ (1);
    N1- (1); N1+ (1); N2- (1); N1+ (1); N1- (1); N2- (1); N2+ (1); N2- (1);
    N1+ (1); N2- (1); N1+ (1); N1- (1); N2+ (1); N1+ (1); N2- (1); N2+ (1);
    N1+ (1); N1- (1); N2- (1); N2- (1); N1+ (1); N2+ (1); N1- (1); N1- (1);
    N2+ (1); N2- (1); N2+ (1); N1+ (1); N1- (1); N2+ (1); N1- (1); N1+ (1);
    N2- (1); N2+ (1); N1- (1); N1+ (1); N2- (1); N2- (1); N1+ (1); N1+ (1);
    N2+ (1); N2- (1); N1- (1); [52];
\end{lstlisting}
\hyperref[tab:electride]{Back to the table}

\subsubsection*{155178 Pr(Sb$_{3}$Os)$_{4}$}
\label{sec:tqc155178}
\lstset{language=bash, keywordstyle=\color{blue!70}, basicstyle=\ttfamily, frame=shadowbox}
\begin{lstlisting}
Computed bands:  1 - 52
GM: GM1+(1); GM4-(3); GM1+(1); GM2+GM3+(2); GM4-(3); GM4-(3); GM4+(3); GM1+(1);
    GM2+GM3+(2); GM4+(3); GM4+(3); GM1+(1); GM4-(3); GM2+GM3+(2); GM4+(3);
    GM4+(3); GM4-(3); GM4+(3); GM4-(3); GM2-GM3-(2); GM2+GM3+(2); GM1+(1);
    GM1-(1); [52] ;
H : H1+ (1); H4- (3); H2+ H3+ (2); H4- (3); H1+ (1); H4- (3); H4+ (3); H1- (1);
    H4- (3); H1+ (1); H4- (3); H2+ H3+ (2); H2- H3- (2); H4- (3); H2+ H3+ (2);
    H4+ (3); H1+ (1); H4- (3); H4+ (3); H4- (3); H1- (1); H2- H3- (2); H4- (3);
    [52] ;
P : P1  (1); P4  (3); P4  (3); P1  (1); P4  (3); P2  P3  (2); P4  (3); P4  (3);
    P1  (1); P4  (3); P2  P3  (2); P4  (3); P4  (3); P4  (3); P2  P3  (2);
    P1  (1); P4  (3); P4  (3); P2  P3  (2); P4  (3); P4  (3); P1  (1); [52] ;
N : N1+ (1); N1- N2- N2- (3); N2- (1); N1+ (1); N2- (1); N1+ (1); N2+ (1);
    N1- (1); N1+ (1); N2- (1); N1+ (1); N1- (1); N2+ N2- (2); N2- (1); N1+ (1);
    N2+ (1); N1- (1); N2- (1); N1+ (1); N2- (1); N1+ (1); N2+ (1); N1- (1);
    N1+ (1); N2- (1); N2- (1); N1+ (1); N2+ (1); N1- (1); N2+ (1); N1- (1);
    N2- (1); N2+ (1); N1- (1); N2+ (1); N1+ (1); N1- (1); N1+ (1); N2- (1);
    N2+ (1); N1- (1); N1+ (1); N2- (1); N1+ (1); N2- (1); N1+ (1); N2+ (1);
    N2- (1); N1- (1); [52];
\end{lstlisting}
\hyperref[tab:electride]{Back to the table}

\subsubsection*{53490 La(FeSb$_{3}$)$_{4}$}
\label{sec:tqc53490}
\lstset{language=bash, keywordstyle=\color{blue!70}, basicstyle=\ttfamily, frame=shadowbox}
\begin{lstlisting}
Computed bands:  1 - 52
GM: GM1+(1); GM4-(3); GM1+(1); GM2+GM3+(2); GM4-(3); GM4-(3); GM4+(3); GM1+(1);
    GM2+GM3+(2); GM4-(3); GM4+(3); GM1+(1); GM4+(3); GM4-(3); GM2+GM3+(2);
    GM4+(3); GM4-(3); GM4+(3); GM2-GM3-(2); GM4+(3); GM2+GM3+(2); GM1+(1);
    GM1-(1); [52] ;
H : H1+ (1); H4- (3); H2+ H3+ (2); H4- (3); H1+ (1); H4- (3); H4+ (3); H1- (1);
    H1+ (1); H4- (3); H4- (3); H2+ H3+ (2); H4+ (3); H2+ H3+ (2); H4+ (3);
    H1+ (1); H2- H3- (2); H4- (3); H4- (3); H4- (3); H1- (1); H2- H3- (2);
    H4- (3); [52] ;
P : P1  (1); P4  (3); P4  (3); P1  (1); P4  (3); P2  P3  (2); P4  (3); P4  (3);
    P1  (1); P4  (3); P2  P3  (2); P4  (3); P4  (3); P4  (3); P2  P3  (2);
    P1  (1); P4  (3); P4  (3); P2  P3  (2); P4  (3); P4  (3); P1  (1); [52] ;
N : N1+ (1); N2- (1); N1- N2- (2); N2- (1); N1+ (1); N2- (1); N2+ (1); N1+ (1);
    N1- (1); N1+ (1); N2- (1); N1+ (1); N1- (1); N2+ (1); N2- (1); N2- (1);
    N1+ (1); N1- (1); N2+ (1); N2- (1); N1+ (1); N1+ (1); N2- (1); N2+ (1);
    N2- (1); N1+ (1); N1- (1); N1+ (1); N2- (1); N2+ (1); N1- (1); N1- (1);
    N2+ (1); N2- (1); N2+ (1); N1+ (1); N1- (1); N2+ (1); N2- (1); N1+ (1);
    N1- (1); N1- (1); N2+ (1); N2- (1); N1+ (1); N1+ (1); N2- (1); N1+ (1);
    N2+ (1); N2- (1); N1- (1); [52];
\end{lstlisting}
\hyperref[tab:electride]{Back to the table}

\subsubsection*{647760 Sm(Sb$_{3}$Os)$_{4}$}
\label{sec:tqc647760}
\lstset{language=bash, keywordstyle=\color{blue!70}, basicstyle=\ttfamily, frame=shadowbox}
\begin{lstlisting}
Computed bands:  1 - 52
GM: GM1+(1); GM4-(3); GM1+(1); GM2+GM3+(2); GM4-(3); GM4-(3); GM4+(3); GM1+(1);
    GM2+GM3+(2); GM4+(3); GM4+(3); GM1+(1); GM4-(3); GM2+GM3+(2); GM4+(3);
    GM4+(3); GM4-(3); GM4-(3); GM4+(3); GM2-GM3-(2); GM2+GM3+(2); GM1+(1);
    GM1-(1); [52] ;
H : H1+ (1); H4- (3); H2+ H3+ (2); H4- (3); H1+ (1); H4- (3); H4+ (3); H1- (1);
    H1+ (1); H4- (3); H4- (3); H2+ H3+ (2); H2- H3- (2); H4- (3); H2+ H3+ (2);
    H4+ (3); H1+ (1); H4- (3); H4+ (3); H4- (3); H1- (1); H2- H3- (2); H4- (3);
    [52] ;
P : P1  (1); P4  (3); P4  (3); P1  (1); P4  (3); P2  P3  (2); P4  (3); P4  (3);
    P1  (1); P4  (3); P2  P3  (2); P4  (3); P4  (3); P4  (3); P2  P3  (2);
    P1  (1); P4  (3); P4  (3); P2  P3  (2); P4  (3); P4  (3); P1  (1); [52] ;
N : N1+ (1); N1- N2- N2- (3); N2- (1); N1+ (1); N2- (1); N1+ (1); N2+ (1);
    N1- (1); N1+ (1); N2- (1); N1+ (1); N1- (1); N2- (1); N2+ (1); N2- (1);
    N1+ (1); N2+ (1); N1- (1); N2- (1); N1+ (1); N2- (1); N1+ (1); N2+ (1);
    N1- (1); N1+ (1); N2- (1); N2- (1); N1+ (1); N2+ (1); N1- (1); N2+ (1);
    N1- (1); N2- (1); N2+ (1); N1- (1); N1+ (1); N2+ (1); N1- (1); N1+ (1);
    N2- (1); N2+ (1); N1- (1); N2- (1); N1+ (1); N1+ (1); N2- (1); N1+ (1);
    N2+ (1); N2- (1); N1- (1); [52];
\end{lstlisting}
\hyperref[tab:electride]{Back to the table}

\subsubsection*{183085 La(Sb$_{3}$Os)$_{4}$}
\label{sec:tqc183085}
\lstset{language=bash, keywordstyle=\color{blue!70}, basicstyle=\ttfamily, frame=shadowbox}
\begin{lstlisting}
Computed bands:  1 - 52
GM: GM1+(1); GM4-(3); GM1+(1); GM2+GM3+(2); GM4-(3); GM4-(3); GM4+(3); GM1+(1);
    GM2+GM3+(2); GM4+(3); GM4+(3); GM1+(1); GM4-(3); GM2+GM3+(2); GM4+(3);
    GM4+(3); GM4-(3); GM4+(3); GM4-(3); GM2-GM3-(2); GM2+GM3+(2); GM1+(1);
    GM1-(1); [52] ;
H : H1+ (1); H4- (3); H2+ H3+ (2); H4- (3); H1+ (1); H4- (3); H4+ (3); H1- (1);
    H4- (3); H1+ (1); H4- (3); H2+ H3+ (2); H2- H3- (2); H4- (3); H2+ H3+ (2);
    H4+ (3); H1+ (1); H4- (3); H4+ (3); H4- (3); H1- (1); H2- H3- (2); H4- (3);
    [52] ;
P : P1  (1); P4  (3); P4  (3); P1  (1); P4  (3); P2  P3  (2); P4  (3); P4  (3);
    P1  (1); P4  (3); P2  P3  (2); P4  (3); P4  (3); P4  (3); P2  P3  (2);
    P1  (1); P4  (3); P4  (3); P2  P3  (2); P4  (3); P4  (3); P1  (1); [52] ;
N : N1+ (1); N2- (1); N1- N2- (2); N2- (1); N1+ (1); N2- (1); N1+ (1); N2+ (1);
    N1- (1); N1+ (1); N2- (1); N1+ (1); N1- (1); N2+ (1); N2- (1); N2- (1);
    N1+ (1); N2+ (1); N1- (1); N1+ (1); N2- (1); N1+ N2- (2); N2+ (1); N1- (1);
    N1+ (1); N2- (1); N2- (1); N1+ (1); N2+ (1); N1- (1); N2+ (1); N1- (1);
    N2- (1); N2+ (1); N1- (1); N2+ (1); N1+ (1); N1- (1); N1+ (1); N2- (1);
    N2+ (1); N1- (1); N2- (1); N1+ (1); N2- (1); N1+ (1); N1+ (1); N2+ (1);
    N2- (1); N1- (1); [52];
\end{lstlisting}
\hyperref[tab:electride]{Back to the table}

\subsubsection*{645670 Nd(P$_{3}$Os)$_{4}$}
\label{sec:tqc645670}
\lstset{language=bash, keywordstyle=\color{blue!70}, basicstyle=\ttfamily, frame=shadowbox}
\begin{lstlisting}
Computed bands:  1 - 52
GM: GM1+(1); GM4-(3); GM1+(1); GM2+GM3+(2); GM4-(3); GM4-(3); GM4+(3); GM1+(1);
    GM2+GM3+(2); GM4+(3); GM1+(1); GM4+(3); GM2+GM3+(2); GM4-(3); GM4-(3);
    GM4+(3); GM4-(3); GM4+(3); GM4+(3); GM2+GM3+(2); GM2-GM3-(2); GM1+(1);
    GM1-(1); [52] ;
H : H1+ (1); H4- (3); H2+ H3+ (2); H1+ (1); H4- (3); H4- (3); H4+ (3); H1+ (1);
    H4- (3); H4- (3); H1- (1); H2+ H3+ (2); H2+ H3+ (2); H2- H3- (2); H1+ (1);
    H4- (3); H4+ (3); H4- (3); H4- (3); H4+ (3); H1- (1); H4- (3); H2- H3- (2);
    [52] ;
P : P1  (1); P4  (3); P4  (3); P1  (1); P4  (3); P2  P3  (2); P4  (3); P4  (3);
    P4  (3); P1  (1); P2  P3  (2); P4  (3); P4  (3); P4  (3); P2  P3  (2);
    P1  (1); P4  (3); P4  (3); P4  (3); P2  P3  (2); P4  (3); P1  (1); [52] ;
N : N1+ (1); N2- (1); N1- (1); N2- (1); N2- (1); N1+ (1); N2- (1); N1+ (1);
    N2+ (1); N1- (1); N2- (1); N1+ (1); N1+ (1); N1- (1); N2+ (1); N2- (1);
    N2- (1); N1+ (1); N1+ (1); N2+ (1); N2- (1); N1- (1); N2- (1); N1+ (1);
    N2- (1); N2+ (1); N1- (1); N1+ (1); N2- (1); N1+ (1); N2- (1); N2+ (1);
    N1- (1); N2+ (1); N1- (1); N2+ (1); N1+ (1); N2- (1); N1+ (1); N2+ (1);
    N1+ (1); N1- (1); N2- (1); N1- (1); N1+ (1); N1- (1); N2- (1); N2+ (1);
    N1+ (1); N2- (1); N2+ (1); N1- (1); [52];
\end{lstlisting}
\hyperref[tab:electride]{Back to the table}

\subsubsection*{641615 La(P$_{3}$Os)$_{4}$}
\label{sec:tqc641615}
\lstset{language=bash, keywordstyle=\color{blue!70}, basicstyle=\ttfamily, frame=shadowbox}
\begin{lstlisting}
Computed bands:  1 - 52
GM: GM1+(1); GM4-(3); GM1+(1); GM2+GM3+(2); GM4-(3); GM4-(3); GM4+(3);
    GM2+GM3+(2); GM1+(1); GM4+(3); GM1+(1); GM4+(3); GM2+GM3+(2); GM4-(3);
    GM4-(3); GM4+(3); GM4-(3); GM4+(3); GM4+(3); GM2+GM3+(2); GM2-GM3-(2);
    GM1+(1); GM1-(1); [52] ;
H : H1+ (1); H4- (3); H2+ H3+ (2); H1+ (1); H4- (3); H4- (3); H4+ (3); H4- (3);
    H1+ (1); H4- (3); H1- (1); H2+ H3+ (2); H2+ H3+ (2); H2- H3- (2); H1+ (1);
    H4- (3); H4+ (3); H4- (3); H4- (3); H4+ (3); H1- (1); H4- (3); H2- H3- (2);
    [52] ;
P : P1  (1); P4  (3); P4  (3); P1  (1); P4  (3); P2  P3  (2); P4  (3); P4  (3);
    P4  (3); P1  (1); P2  P3  (2); P4  (3); P4  (3); P4  (3); P2  P3  (2);
    P1  (1); P4  (3); P4  (3); P4  (3); P2  P3  (2); P4  (3); P1  (1); [52] ;
N : N1+ (1); N2- (1); N1- (1); N2- (1); N2- (1); N1+ (1); N2- (1); N1+ (1);
    N2+ (1); N2- (1); N1- (1); N1+ (1); N1+ (1); N1- (1); N2+ (1); N2- (1);
    N2- (1); N1+ (1); N1+ (1); N2+ (1); N2- (1); N1- (1); N1+ (1); N2- (1);
    N2+ (1); N2- (1); N1- (1); N1+ (1); N2- (1); N1+ (1); N2- (1); N2+ (1);
    N1- (1); N2+ (1); N1- (1); N2+ (1); N1+ (1); N2- (1); N1+ (1); N2+ (1);
    N1+ N1- (2); N2- (1); N1- (1); N1+ (1); N1- (1); N2- (1); N2+ (1); N2- (1);
    N1+ (1); N2+ (1); N1- (1); [52];
\end{lstlisting}
\hyperref[tab:electride]{Back to the table}

\subsubsection*{280022 Ba$_{5}$Sb$_{4}$}
\label{sec:tqc280022}
\lstset{language=bash, keywordstyle=\color{blue!70}, basicstyle=\ttfamily, frame=shadowbox}

\hyperref[tab:electride]{Back to the table}

\subsubsection*{280143 Tm$_{5}$(ReO$_{6}$)$_{2}$}
\label{sec:tqc280143}
\lstset{language=bash, keywordstyle=\color{blue!70}, basicstyle=\ttfamily, frame=shadowbox}

\hyperref[tab:electride]{Back to the table}

\subsubsection*{36157 BeSO$_{8}$}
\label{sec:tqc36157}
\lstset{language=bash, keywordstyle=\color{blue!70}, basicstyle=\ttfamily, frame=shadowbox}

\hyperref[tab:electride]{Back to the table}

\subsubsection*{200210 Mg$_{3}$Nb$_{6}$O$_{11}$}
\label{sec:tqc200210}
\noindent Essential BR: $A1g@1a$ \\
\noindent RSI:
\begin{flalign*}
&\delta_{1}@1a\equiv -m(Eg)+m(Eu) = 0,&
\\
&\delta_{2}@1a\equiv -m(A1g)+m(A1u)-m(A2g)+m(A2u) = -1,&
\end{flalign*}
\lstset{language=bash, keywordstyle=\color{blue!70}, basicstyle=\ttfamily, frame=shadowbox}

\hyperref[tab:electride]{Back to the table}

\subsubsection*{246145 Bi$_{4}$I}
\label{sec:tqc246145}
\lstset{language=bash, keywordstyle=\color{blue!70}, basicstyle=\ttfamily, frame=shadowbox}

\hyperref[tab:electride]{Back to the table}

\subsubsection*{262307 Nd$_{2}$SbO$_{2}$}
\label{sec:tqc262307}
\lstset{language=bash, keywordstyle=\color{blue!70}, basicstyle=\ttfamily, frame=shadowbox}

\hyperref[tab:electride]{Back to the table}

\subsubsection*{422527 Sr(GaAs)$_{2}$}
\label{sec:tqc422527}
\lstset{language=bash, keywordstyle=\color{blue!70}, basicstyle=\ttfamily, frame=shadowbox}

\hyperref[tab:electride]{Back to the table}

\subsubsection*{26416 Ba(GeP)$_{2}$}
\label{sec:tqc26416}
\lstset{language=bash, keywordstyle=\color{blue!70}, basicstyle=\ttfamily, frame=shadowbox}

\hyperref[tab:electride]{Back to the table}

\subsubsection*{10032 AlSiP$_{3}$}
\label{sec:tqc10032}
\noindent Essential BR: $Ag@4b$ \\
\noindent RSI:
\begin{flalign*}
&\delta_{1}@4b\equiv -m(Ag)+m(Au) = -1,&
\end{flalign*}
\lstset{language=bash, keywordstyle=\color{blue!70}, basicstyle=\ttfamily, frame=shadowbox}
\begin{lstlisting}
Computed bands:  1 - 44
GM: GM1+(1); GM2+(1); GM4+(1); GM3-(1); GM3+(1); GM1+(1); GM2-(1); GM4+(1);
    GM2-(1); GM3-(1); GM4-(1); GM1-(1); GM2-(1); GM4+(1); GM1+(1); GM4-(1);
    GM3-(1); GM3+(1); GM4+(1); GM1+(1); GM1-(1); GM2+(1); GM3-(1); GM2-(1);
    GM4+(1); GM1+(1); GM4-(1); GM2-(1); GM1-(1); GM3+(1); GM4+(1); GM3-(1);
    GM2+(1); GM1-(1); GM2+(1); GM3+(1); GM1+(1); GM3-(1); GM2-(1); GM4-(1);
    GM4+(1); GM3+(1); GM1+(1); GM2+(1); [44] ;
R : R1  R2  (4); R1  R2  (4); R1  R2  (4); R1  R2  (4); R1  R2  (4); R1  R2  (4);
    R1  R2  (4); R1  R2  (4); R1  R2  (4); R1  R2  (4); R1  R2  (4); [44] ;
S : S1  S2  (4); S1  S2  (4); S1  S2  (4); S1  S2  (4); S1  S2  (4); S1  S2  (4);
    S1  S2  (4); S1  S2  (4); S1  S2  (4); S1  S2  (4); S1  S2  (4); [44] ;
T : T2  (2); T1  (2); T2  (2); T1  (2); T1  (2); T2  (2); T1  (2); T2  (2);
    T2  (2); T1  (2); T1  (2); T2  (2); T1  (2); T2  (2); T2  (2); T1  (2);
    T2  (2); T1  (2); T2  (2); T1  (2); T1  (2); T2  (2); [44] ;
U : U2- U3- (2); U1- U4- (2); U1+ U4+ (2); U2- U3- (2); U1+ U4+ (2); U2+ U3+ (2);
    U2- U3- (2); U1+ U4+ (2); U2+ U3+ (2); U1+ U4+ (2); U1- U4- (2); U2- U3- (2);
    U2- U3- (2); U2+ U3+ (2); U1- U4- (2); U1+ U4+ (2); U2- U3- (2); U1+ U4+ (2);
    U2- U3- (2); U1- U4- (2); U1- U4- (2); U2+ U3+ (2); [44] ;
X : X1  (2); X2  (2); X1  (2); X1  (2); X1  (2); X2  (2); X1  (2); X1  (2);
    X2  (2); X2  (2); X1  (2); X1  (2); X1  (2); X2  (2); X1  (2); X2  (2);
    X1  (2); X1  (2); X2  (2); X2  (2); X2  (2); X1  (2); [44] ;
Y : Y2  (2); Y1  (2); Y2  (2); Y1  (2); Y2  (2); Y1  (2); Y1  (2); Y2  (2);
    Y1  (2); Y2  (2); Y2  (2); Y1  (2); Y1  (2); Y2  (2); Y1  (2); Y1  (2);
    Y2  (2); Y2  (2); Y1  (2); Y2  (2); Y1  (2); Y2  (2); [44] ;
Z : Z1  (2); Z2  (2); Z1  (2); Z1  (2); Z1  (2); Z2  (2); Z1  (2); Z2  (2);
    Z1  (2); Z1  (2); Z2  (2); Z1  (2); Z1  (2); Z2  (2); Z1  (2); Z2  (2);
    Z1  (2); Z2  (2); Z1  (2); Z2  (2); Z1  (2); Z2  (2); [44];
\end{lstlisting}
\hyperref[tab:electride]{Back to the table}

\subsubsection*{280002 TiNb$_{3}$O$_{6}$}
\label{sec:tqc280002}
\noindent Essential BR: $A1g@3a$ \\
\noindent RSI:
\begin{flalign*}
&\delta_{1}@3a\equiv -m(2Eg)+m(2Eu) = 0,&
\\
&\delta_{2}@3a\equiv -m(A1g)+m(A1u) = -1,&
\end{flalign*}
\lstset{language=bash, keywordstyle=\color{blue!70}, basicstyle=\ttfamily, frame=shadowbox}

\hyperref[tab:electride]{Back to the table}

\subsubsection*{415950 Nb$_{2}$F$_{5}$}
\label{sec:tqc415950}
\lstset{language=bash, keywordstyle=\color{blue!70}, basicstyle=\ttfamily, frame=shadowbox}
\begin{lstlisting}
Computed bands:  1 - 86
GM: GM5+(3); GM4-(3); GM5-(3); GM1+(1); GM4+(3); GM4-(3); GM3+(2); GM1+(1);
    GM3+(2); GM5-(3); GM4-(3); GM5+(3); GM1+(1); GM3+(2); GM5-(3); GM1+(1);
    GM3+(2); GM4-(3); GM2-(1); GM5+(3); GM3+(2); GM2+(1); GM4-(3); GM4-(3);
    GM4+(3); GM5+(3); GM3-(2); GM5-(3); GM4-(3); GM4+(3); GM4-(3); GM5-(3);
    GM1+(1); GM5+(3); GM4-(3); GM2-(1); [86] ;
H : H5+ (3); H4- (3); H5- (3); H1+ (1); H4+ (3); H4- (3); H3+ (2); H3+ (2);
    H1+ (1); H4- (3); H5- (3); H5+ (3); H1+ (1); H3+ (2); H2- (1); H4- (3);
    H3+ (2); H2+ (1); H5- (3); H1+ (1); H5+ (3); H3+ H4- (5); H4+ (3); H5- (3);
    H4- (3); H4- (3); H5+ (3); H4+ (3); H3- (2); H5- (3); H4- (3); H1+ (1);
    H5+ (3); H4- (3); H2- (1); [86] ;
P : P4  (3); P4  (3); P5  (3); P1  (1); P5  (3); P4  (3); P3  (2); P1  (1);
    P3  (2); P4  (3); P5  (3); P4  (3); P4  (3); P1  (1); P5  (3); P4  (3);
    P3  (2); P1  (1); P2  (1); P3  (2); P4  (3); P1  (1); P5  (3); P4  (3);
    P3  (2); P4  (3); P5  (3); P3  (2); P4  (3); P5  (3); P4  (3); P5  (3);
    P1  (1); P4  (3); P4  (3); P1  (1); [86] ;
N : N3+ (1); N1+ (1); N4+ (1); N3- (1); N4- (1); N2- (1); N4- (1); N1- N3- (2);
    N1+ (1); N2+ N3+ N4+ (3); N3- (1); N4- (1); N2- (1); N1+ (1); N2+ (1);
    N1+ (1); N1+ (1); N2+ (1); N3- (1); N2- (1); N4- (1); N1- (1); N3- (1);
    N4- (1); N3+ (1); N1+ (1); N4+ (1); N3- N4- (2); N1+ (1); N2- (1); N4- (1);
    N3- (1); N1+ (1); N2+ (1); N1- (1); N3- (1); N1+ (1); N4- (1); N2- (1);
    N3+ (1); N2+ (1); N2+ (1); N1+ (1); N3+ (1); N1+ (1); N4+ (1); N2- (1);
    N1+ (1); N3- (1); N2+ (1); N2+ (1); N2- (1); N4+ (1); N3+ (1); N1+ N3- (2);
    N4- (1); N3- (1); N1- (1); N4+ (1); N4- (1); N1- (1); N2- (1); N2- (1);
    N4+ (1); N4- (1); N2+ (1); N3+ (1); N1+ (1); N4+ (1); N4- (1); N3- (1);
    N2+ (1); N3+ (1); N1+ (1); N3+ (1); N4+ (1); N3- (1); N4- (1); N1+ (1);
    N2- (1); N2- (1); [86];
\end{lstlisting}
\hyperref[tab:electride]{Back to the table}

\subsubsection*{62519 Ba(As$_{3}$Pt$_{2}$)$_{2}$}
\label{sec:tqc62519}
\lstset{language=bash, keywordstyle=\color{blue!70}, basicstyle=\ttfamily, frame=shadowbox}

\hyperref[tab:electride]{Back to the table}

\subsubsection*{657803 K$_{2}$Fe(PS$_{3}$)$_{2}$}
\label{sec:tqc657803}
\noindent Essential BR: $Ag@2d$ \\
\noindent RSI:
\begin{flalign*}
&\delta_{1}@2d\equiv -m(Ag)+m(Au) = -1,&
\end{flalign*}
\lstset{language=bash, keywordstyle=\color{blue!70}, basicstyle=\ttfamily, frame=shadowbox}

\hyperref[tab:electride]{Back to the table}

\subsubsection*{62517 Sr(P$_{3}$Pt$_{2}$)$_{2}$}
\label{sec:tqc62517}
\lstset{language=bash, keywordstyle=\color{blue!70}, basicstyle=\ttfamily, frame=shadowbox}

\hyperref[tab:electride]{Back to the table}

\subsubsection*{25766 Nb$_{3}$Br$_{8}$}
\label{sec:tqc25766}
\noindent Essential BR: $A1g@3b$ \\
\noindent RSI:
\begin{flalign*}
&\delta_{1}@3b\equiv -m(Eg)+m(Eu) = 0,&
\\
&\delta_{2}@3b\equiv -m(A1g)+m(A1u)-m(A2g)+m(A2u) = -1,&
\end{flalign*}
\lstset{language=bash, keywordstyle=\color{blue!70}, basicstyle=\ttfamily, frame=shadowbox}

\hyperref[tab:electride]{Back to the table}

\subsubsection*{202189 La$_{2}$Mo$_{2}$O$_{7}$}
\label{sec:tqc202189}
\lstset{language=bash, keywordstyle=\color{blue!70}, basicstyle=\ttfamily, frame=shadowbox}

\hyperref[tab:electride]{Back to the table}

\subsubsection*{82360 Ba$_{5}$CrN$_{5}$}
\label{sec:tqc82360}
\noindent Essential BR: $Ag@2c$ \\
\noindent RSI:
\begin{flalign*}
&\delta_{1}@2c\equiv -m(Ag)+m(Au)-m(Bg)+m(Bu) = -1,&
\end{flalign*}
\lstset{language=bash, keywordstyle=\color{blue!70}, basicstyle=\ttfamily, frame=shadowbox}

\hyperref[tab:electride]{Back to the table}

\subsubsection*{62518 Sr(As$_{3}$Pt$_{2}$)$_{2}$}
\label{sec:tqc62518}
\lstset{language=bash, keywordstyle=\color{blue!70}, basicstyle=\ttfamily, frame=shadowbox}

\hyperref[tab:electride]{Back to the table}

\subsubsection*{417514 Dy$_{4}$InRh}
\label{sec:tqc417514}
\lstset{language=bash, keywordstyle=\color{blue!70}, basicstyle=\ttfamily, frame=shadowbox}

\hyperref[tab:electride]{Back to the table}

\subsubsection*{417518 Tb$_{4}$InRh}
\label{sec:tqc417518}
\lstset{language=bash, keywordstyle=\color{blue!70}, basicstyle=\ttfamily, frame=shadowbox}

\hyperref[tab:electride]{Back to the table}

\subsubsection*{417517 Ho$_{4}$InRh}
\label{sec:tqc417517}
\lstset{language=bash, keywordstyle=\color{blue!70}, basicstyle=\ttfamily, frame=shadowbox}

\hyperref[tab:electride]{Back to the table}

\subsubsection*{417519 Tm$_{4}$InRh}
\label{sec:tqc417519}
\lstset{language=bash, keywordstyle=\color{blue!70}, basicstyle=\ttfamily, frame=shadowbox}

\hyperref[tab:electride]{Back to the table}

\subsubsection*{418265 Er$_{4}$InIr}
\label{sec:tqc418265}
\lstset{language=bash, keywordstyle=\color{blue!70}, basicstyle=\ttfamily, frame=shadowbox}

\hyperref[tab:electride]{Back to the table}

\subsubsection*{612757 CeCrB$_{4}$}
\label{sec:tqc612757}
\lstset{language=bash, keywordstyle=\color{blue!70}, basicstyle=\ttfamily, frame=shadowbox}

\hyperref[tab:electride]{Back to the table}

\subsubsection*{418270 Ho$_{4}$InIr}
\label{sec:tqc418270}
\lstset{language=bash, keywordstyle=\color{blue!70}, basicstyle=\ttfamily, frame=shadowbox}

\hyperref[tab:electride]{Back to the table}

\subsubsection*{418567 Y$_{4}$InIr}
\label{sec:tqc418567}
\lstset{language=bash, keywordstyle=\color{blue!70}, basicstyle=\ttfamily, frame=shadowbox}

\hyperref[tab:electride]{Back to the table}

\subsubsection*{417515 Er$_{4}$InRh}
\label{sec:tqc417515}
\lstset{language=bash, keywordstyle=\color{blue!70}, basicstyle=\ttfamily, frame=shadowbox}

\hyperref[tab:electride]{Back to the table}

\subsubsection*{20081 YB$_{4}$Mo}
\label{sec:tqc20081}
\lstset{language=bash, keywordstyle=\color{blue!70}, basicstyle=\ttfamily, frame=shadowbox}

\hyperref[tab:electride]{Back to the table}

\subsubsection*{616683 HfBi$_{2}$}
\label{sec:tqc616683}
\lstset{language=bash, keywordstyle=\color{blue!70}, basicstyle=\ttfamily, frame=shadowbox}

\hyperref[tab:electride]{Back to the table}

\subsubsection*{613559 TbCrB$_{4}$}
\label{sec:tqc613559}
\lstset{language=bash, keywordstyle=\color{blue!70}, basicstyle=\ttfamily, frame=shadowbox}

\hyperref[tab:electride]{Back to the table}

\subsubsection*{39429 TiCl$_{3}$}
\label{sec:tqc39429}
\lstset{language=bash, keywordstyle=\color{blue!70}, basicstyle=\ttfamily, frame=shadowbox}

\hyperref[tab:electride]{Back to the table}

\subsubsection*{613514 HoCrB$_{4}$}
\label{sec:tqc613514}
\lstset{language=bash, keywordstyle=\color{blue!70}, basicstyle=\ttfamily, frame=shadowbox}

\hyperref[tab:electride]{Back to the table}

\subsubsection*{648073 TmPS}
\label{sec:tqc648073}
\noindent Essential BR: $Ag@4b$ \\
\noindent RSI:
\begin{flalign*}
&\delta_{1}@4b\equiv -m(Ag)+m(Au) = -1,&
\end{flalign*}
\lstset{language=bash, keywordstyle=\color{blue!70}, basicstyle=\ttfamily, frame=shadowbox}

\hyperref[tab:electride]{Back to the table}

\subsubsection*{262063 KTl}
\label{sec:tqc262063}
\lstset{language=bash, keywordstyle=\color{blue!70}, basicstyle=\ttfamily, frame=shadowbox}

\hyperref[tab:electride]{Back to the table}

\subsubsection*{89380 FeS}
\label{sec:tqc89380}
\lstset{language=bash, keywordstyle=\color{blue!70}, basicstyle=\ttfamily, frame=shadowbox}

\hyperref[tab:electride]{Back to the table}

\subsubsection*{419780 CoGeTe}
\label{sec:tqc419780}
\noindent Essential BR: $Ag@4b$ \\
\noindent RSI:
\begin{flalign*}
&\delta_{1}@4b\equiv -m(Ag)+m(Au) = -1,&
\end{flalign*}
\lstset{language=bash, keywordstyle=\color{blue!70}, basicstyle=\ttfamily, frame=shadowbox}

\hyperref[tab:electride]{Back to the table}

\subsubsection*{648063 TbPS}
\label{sec:tqc648063}
\noindent Essential BR: $Ag@4b$ \\
\noindent RSI:
\begin{flalign*}
&\delta_{1}@4b\equiv -m(Ag)+m(Au) = -1,&
\end{flalign*}
\lstset{language=bash, keywordstyle=\color{blue!70}, basicstyle=\ttfamily, frame=shadowbox}

\hyperref[tab:electride]{Back to the table}

\subsubsection*{647960 PrPS}
\label{sec:tqc647960}
\noindent Essential BR: $Ag@4b$ \\
\noindent RSI:
\begin{flalign*}
&\delta_{1}@4b\equiv -m(Ag)+m(Au) = -1,&
\end{flalign*}
\lstset{language=bash, keywordstyle=\color{blue!70}, basicstyle=\ttfamily, frame=shadowbox}

\hyperref[tab:electride]{Back to the table}

\subsubsection*{630913 ErPS}
\label{sec:tqc630913}
\noindent Essential BR: $Ag@4b$ \\
\noindent RSI:
\begin{flalign*}
&\delta_{1}@4b\equiv -m(Ag)+m(Au) = -1,&
\end{flalign*}
\lstset{language=bash, keywordstyle=\color{blue!70}, basicstyle=\ttfamily, frame=shadowbox}

\hyperref[tab:electride]{Back to the table}

\subsubsection*{658658 DyCrB$_{4}$}
\label{sec:tqc658658}
\lstset{language=bash, keywordstyle=\color{blue!70}, basicstyle=\ttfamily, frame=shadowbox}

\hyperref[tab:electride]{Back to the table}

\subsubsection*{611500 TiAs$_{2}$}
\label{sec:tqc611500}
\lstset{language=bash, keywordstyle=\color{blue!70}, basicstyle=\ttfamily, frame=shadowbox}

\hyperref[tab:electride]{Back to the table}

\subsubsection*{16171 YCrB$_{4}$}
\label{sec:tqc16171}
\lstset{language=bash, keywordstyle=\color{blue!70}, basicstyle=\ttfamily, frame=shadowbox}

\hyperref[tab:electride]{Back to the table}

\subsubsection*{409820 YbNiB$_{4}$}
\label{sec:tqc409820}
\lstset{language=bash, keywordstyle=\color{blue!70}, basicstyle=\ttfamily, frame=shadowbox}

\hyperref[tab:electride]{Back to the table}

\subsubsection*{641637 LaPS}
\label{sec:tqc641637}
\noindent Essential BR: $Ag@4b$ \\
\noindent RSI:
\begin{flalign*}
&\delta_{1}@4b\equiv -m(Ag)+m(Au) = -1,&
\end{flalign*}
\lstset{language=bash, keywordstyle=\color{blue!70}, basicstyle=\ttfamily, frame=shadowbox}

\hyperref[tab:electride]{Back to the table}

\subsubsection*{35676 LiGeTe$_{2}$}
\label{sec:tqc35676}
\noindent Essential BR: $Ag@1g$ \\
\noindent RSI:
\begin{flalign*}
&\delta_{1}@1g\equiv -m(Ag)+m(Au) = -1,&
\end{flalign*}
\lstset{language=bash, keywordstyle=\color{blue!70}, basicstyle=\ttfamily, frame=shadowbox}

\hyperref[tab:electride]{Back to the table}

\subsubsection*{260373 GeTeRh}
\label{sec:tqc260373}
\noindent Essential BR: $Ag@4b$ \\
\noindent RSI:
\begin{flalign*}
&\delta_{1}@4b\equiv -m(Ag)+m(Au) = -1,&
\end{flalign*}
\lstset{language=bash, keywordstyle=\color{blue!70}, basicstyle=\ttfamily, frame=shadowbox}

\hyperref[tab:electride]{Back to the table}

\subsubsection*{280592 Ba$_{3}$Sb$_{2}$O}
\label{sec:tqc280592}
\lstset{language=bash, keywordstyle=\color{blue!70}, basicstyle=\ttfamily, frame=shadowbox}

\hyperref[tab:electride]{Back to the table}

\subsubsection*{645690 NdPS}
\label{sec:tqc645690}
\noindent Essential BR: $Ag@4b$ \\
\noindent RSI:
\begin{flalign*}
&\delta_{1}@4b\equiv -m(Ag)+m(Au) = -1,&
\end{flalign*}
\lstset{language=bash, keywordstyle=\color{blue!70}, basicstyle=\ttfamily, frame=shadowbox}

\hyperref[tab:electride]{Back to the table}

\subsubsection*{639545 HoPS}
\label{sec:tqc639545}
\noindent Essential BR: $Ag@4b$ \\
\noindent RSI:
\begin{flalign*}
&\delta_{1}@4b\equiv -m(Ag)+m(Au) = -1,&
\end{flalign*}
\lstset{language=bash, keywordstyle=\color{blue!70}, basicstyle=\ttfamily, frame=shadowbox}

\hyperref[tab:electride]{Back to the table}

\subsubsection*{613495 ErCrB$_{4}$}
\label{sec:tqc613495}
\lstset{language=bash, keywordstyle=\color{blue!70}, basicstyle=\ttfamily, frame=shadowbox}

\hyperref[tab:electride]{Back to the table}

\subsubsection*{630062 DyPS}
\label{sec:tqc630062}
\noindent Essential BR: $Ag@4b$ \\
\noindent RSI:
\begin{flalign*}
&\delta_{1}@4b\equiv -m(Ag)+m(Au) = -1,&
\end{flalign*}
\lstset{language=bash, keywordstyle=\color{blue!70}, basicstyle=\ttfamily, frame=shadowbox}

\hyperref[tab:electride]{Back to the table}

\subsubsection*{413194 SiSbPt}
\label{sec:tqc413194}
\noindent Essential BR: $Ag@4a$ \\
\noindent RSI:
\begin{flalign*}
&\delta_{1}@4a\equiv -m(Ag)+m(Au) = -1,&
\end{flalign*}
\lstset{language=bash, keywordstyle=\color{blue!70}, basicstyle=\ttfamily, frame=shadowbox}

\hyperref[tab:electride]{Back to the table}

\subsubsection*{404705 Ba$_{2}$LiGe$_{3}$}
\label{sec:tqc404705}
\lstset{language=bash, keywordstyle=\color{blue!70}, basicstyle=\ttfamily, frame=shadowbox}

\hyperref[tab:electride]{Back to the table}

\subsubsection*{648080 YPS}
\label{sec:tqc648080}
\noindent Essential BR: $Ag@4b$ \\
\noindent RSI:
\begin{flalign*}
&\delta_{1}@4b\equiv -m(Ag)+m(Au) = -1,&
\end{flalign*}
\lstset{language=bash, keywordstyle=\color{blue!70}, basicstyle=\ttfamily, frame=shadowbox}

\hyperref[tab:electride]{Back to the table}

\subsubsection*{300157 KAlSb$_{4}$}
\label{sec:tqc300157}
\noindent Essential BR: $Ag@4a$ \\
\noindent RSI:
\begin{flalign*}
&\delta_{1}@4a\equiv -m(Ag)+m(Au) = -1,&
\end{flalign*}
\lstset{language=bash, keywordstyle=\color{blue!70}, basicstyle=\ttfamily, frame=shadowbox}

\hyperref[tab:electride]{Back to the table}

\subsubsection*{410520 Ba$_{2}$Si$_{3}$Ag}
\label{sec:tqc410520}
\lstset{language=bash, keywordstyle=\color{blue!70}, basicstyle=\ttfamily, frame=shadowbox}

\hyperref[tab:electride]{Back to the table}

\subsubsection*{638875 HfSb$_{2}$}
\label{sec:tqc638875}
\lstset{language=bash, keywordstyle=\color{blue!70}, basicstyle=\ttfamily, frame=shadowbox}

\hyperref[tab:electride]{Back to the table}

\subsubsection*{409295 Sr$_{2}$LiSi$_{3}$}
\label{sec:tqc409295}
\lstset{language=bash, keywordstyle=\color{blue!70}, basicstyle=\ttfamily, frame=shadowbox}

\hyperref[tab:electride]{Back to the table}

\subsubsection*{648052 SmPS}
\label{sec:tqc648052}
\noindent Essential BR: $Ag@4b$ \\
\noindent RSI:
\begin{flalign*}
&\delta_{1}@4b\equiv -m(Ag)+m(Au) = -1,&
\end{flalign*}
\lstset{language=bash, keywordstyle=\color{blue!70}, basicstyle=\ttfamily, frame=shadowbox}

\hyperref[tab:electride]{Back to the table}

\subsubsection*{404707 Ba$_{2}$LiSi$_{3}$}
\label{sec:tqc404707}
\lstset{language=bash, keywordstyle=\color{blue!70}, basicstyle=\ttfamily, frame=shadowbox}

\hyperref[tab:electride]{Back to the table}

\subsubsection*{300158 KGaSb$_{4}$}
\label{sec:tqc300158}
\noindent Essential BR: $Ag@4a$ \\
\noindent RSI:
\begin{flalign*}
&\delta_{1}@4a\equiv -m(Ag)+m(Au) = -1,&
\end{flalign*}
\lstset{language=bash, keywordstyle=\color{blue!70}, basicstyle=\ttfamily, frame=shadowbox}

\hyperref[tab:electride]{Back to the table}

\subsubsection*{62520 BaP$_{3}$Pt$_{2}$}
\label{sec:tqc62520}
\noindent Essential BR: $Ag@2c$ \\
\noindent RSI:
\begin{flalign*}
&\delta_{1}@2c\equiv -m(Ag)+m(Au) = -1,&
\end{flalign*}
\lstset{language=bash, keywordstyle=\color{blue!70}, basicstyle=\ttfamily, frame=shadowbox}

\hyperref[tab:electride]{Back to the table}

\subsubsection*{66779 ZrSb$_{2}$}
\label{sec:tqc66779}
\lstset{language=bash, keywordstyle=\color{blue!70}, basicstyle=\ttfamily, frame=shadowbox}

\hyperref[tab:electride]{Back to the table}

\subsubsection*{613517 LuCrB$_{4}$}
\label{sec:tqc613517}
\lstset{language=bash, keywordstyle=\color{blue!70}, basicstyle=\ttfamily, frame=shadowbox}

\hyperref[tab:electride]{Back to the table}

\subsubsection*{42880 ZrBi$_{2}$}
\label{sec:tqc42880}
\lstset{language=bash, keywordstyle=\color{blue!70}, basicstyle=\ttfamily, frame=shadowbox}

\hyperref[tab:electride]{Back to the table}

\subsubsection*{1053 K(MoO$_{3}$)$_{3}$}
\label{sec:tqc1053}
\lstset{language=bash, keywordstyle=\color{blue!70}, basicstyle=\ttfamily, frame=shadowbox}

\hyperref[tab:electride]{Back to the table}

\subsubsection*{186915 Yb(Al$_{5}$Ru)$_{2}$}
\label{sec:tqc186915}
\lstset{language=bash, keywordstyle=\color{blue!70}, basicstyle=\ttfamily, frame=shadowbox}

\hyperref[tab:electride]{Back to the table}

\subsubsection*{151140 Yb(Al$_{5}$Fe)$_{2}$}
\label{sec:tqc151140}
\lstset{language=bash, keywordstyle=\color{blue!70}, basicstyle=\ttfamily, frame=shadowbox}

\hyperref[tab:electride]{Back to the table}

\subsubsection*{29261 Ba$_{3}$(Si$_{2}$P$_{3}$)$_{2}$}
\label{sec:tqc29261}
\noindent Essential BR: $Ag@2c$ \\
\noindent RSI:
\begin{flalign*}
&\delta_{1}@2c\equiv -m(Ag)+m(Au) = -1,&
\end{flalign*}
\lstset{language=bash, keywordstyle=\color{blue!70}, basicstyle=\ttfamily, frame=shadowbox}

\hyperref[tab:electride]{Back to the table}

\subsubsection*{62690 Tl(MoO$_{3}$)$_{3}$}
\label{sec:tqc62690}
\lstset{language=bash, keywordstyle=\color{blue!70}, basicstyle=\ttfamily, frame=shadowbox}

\hyperref[tab:electride]{Back to the table}

\subsubsection*{238040 Yb(Al$_{5}$Os)$_{2}$}
\label{sec:tqc238040}
\lstset{language=bash, keywordstyle=\color{blue!70}, basicstyle=\ttfamily, frame=shadowbox}

\hyperref[tab:electride]{Back to the table}

\subsubsection*{48168 Na$_{7}$Al$_{2}$Sb$_{5}$}
\label{sec:tqc48168}
\noindent Essential BR: $Ag@2d$ \\
\noindent RSI:
\begin{flalign*}
&\delta_{1}@2d\equiv -m(Ag)+m(Au) = -1,&
\end{flalign*}
\lstset{language=bash, keywordstyle=\color{blue!70}, basicstyle=\ttfamily, frame=shadowbox}

\hyperref[tab:electride]{Back to the table}

\subsubsection*{26565 Cs$_{3}$Zr$_{2}$I$_{9}$}
\label{sec:tqc26565}
\lstset{language=bash, keywordstyle=\color{blue!70}, basicstyle=\ttfamily, frame=shadowbox}

\hyperref[tab:electride]{Back to the table}

\subsubsection*{165596 CuAgPO$_{4}$}
\label{sec:tqc165596}
\noindent Essential BR: $Au@2b$ \\
\noindent RSI:
\begin{flalign*}
&\delta_{1}@2b\equiv -m(Ag)+m(Au) = 1,&
\end{flalign*}
\lstset{language=bash, keywordstyle=\color{blue!70}, basicstyle=\ttfamily, frame=shadowbox}

\hyperref[tab:electride]{Back to the table}

\subsubsection*{411801 La$_{3}$PI$_{3}$}
\label{sec:tqc411801}
\lstset{language=bash, keywordstyle=\color{blue!70}, basicstyle=\ttfamily, frame=shadowbox}

\hyperref[tab:electride]{Back to the table}

\subsubsection*{411803 La$_{3}$AsI$_{3}$}
\label{sec:tqc411803}
\lstset{language=bash, keywordstyle=\color{blue!70}, basicstyle=\ttfamily, frame=shadowbox}

\hyperref[tab:electride]{Back to the table}

\subsubsection*{418886 Ba$_{11}$(CdSb$_{2}$)$_{6}$}
\label{sec:tqc418886}
\noindent Essential BR: $Ag@2d$ \\
\noindent RSI:
\begin{flalign*}
&\delta_{1}@2d\equiv -m(Ag)+m(Au)-m(Bg)+m(Bu) = -1,&
\end{flalign*}
\lstset{language=bash, keywordstyle=\color{blue!70}, basicstyle=\ttfamily, frame=shadowbox}

\hyperref[tab:electride]{Back to the table}

\subsubsection*{413701 Sr$_{11}$(CdSb$_{2}$)$_{6}$}
\label{sec:tqc413701}
\noindent Essential BR: $Ag@2d$ \\
\noindent RSI:
\begin{flalign*}
&\delta_{1}@2d\equiv -m(Ag)+m(Au)-m(Bg)+m(Bu) = -1,&
\end{flalign*}
\lstset{language=bash, keywordstyle=\color{blue!70}, basicstyle=\ttfamily, frame=shadowbox}

\hyperref[tab:electride]{Back to the table}

\subsubsection*{107098 Ta$_{6}$Fe$_{16}$Si$_{7}$}
\label{sec:tqc107098}
\lstset{language=bash, keywordstyle=\color{blue!70}, basicstyle=\ttfamily, frame=shadowbox}

\hyperref[tab:electride]{Back to the table}

\subsubsection*{107097 Nb$_{6}$Fe$_{16}$Si$_{7}$}
\label{sec:tqc107097}
\lstset{language=bash, keywordstyle=\color{blue!70}, basicstyle=\ttfamily, frame=shadowbox}

\hyperref[tab:electride]{Back to the table}

\subsubsection*{427778 Ba$_{4}$Li$_{2}$(CdAs$_{2}$)$_{3}$}
\label{sec:tqc427778}
\noindent Essential BR: $Ag@4b$ \\
\noindent RSI:
\begin{flalign*}
&\delta_{1}@4b\equiv -m(Ag)+m(Au)-m(Bg)+m(Bu) = -1,&
\end{flalign*}
\lstset{language=bash, keywordstyle=\color{blue!70}, basicstyle=\ttfamily, frame=shadowbox}

\hyperref[tab:electride]{Back to the table}

\subsubsection*{68107 Mn$_{2}$MoP$_{12}$}
\label{sec:tqc68107}
\lstset{language=bash, keywordstyle=\color{blue!70}, basicstyle=\ttfamily, frame=shadowbox}

\hyperref[tab:electride]{Back to the table}

\subsubsection*{643287 Mn$_{2}$P$_{12}$W}
\label{sec:tqc643287}
\lstset{language=bash, keywordstyle=\color{blue!70}, basicstyle=\ttfamily, frame=shadowbox}

\hyperref[tab:electride]{Back to the table}

\subsubsection*{108870 Ta$_{3}$Ge}
\label{sec:tqc108870}
\lstset{language=bash, keywordstyle=\color{blue!70}, basicstyle=\ttfamily, frame=shadowbox}

\hyperref[tab:electride]{Back to the table}

\subsubsection*{165617 SrSn$_{3}$Sb$_{4}$}
\label{sec:tqc165617}
\noindent Essential BR: $Ag@4a$ \\
\noindent RSI:
\begin{flalign*}
&\delta_{1}@4a\equiv -m(Ag)+m(Au) = -1,&
\end{flalign*}
\lstset{language=bash, keywordstyle=\color{blue!70}, basicstyle=\ttfamily, frame=shadowbox}

\hyperref[tab:electride]{Back to the table}

\subsubsection*{636772 KGe}
\label{sec:tqc636772}
\lstset{language=bash, keywordstyle=\color{blue!70}, basicstyle=\ttfamily, frame=shadowbox}

\hyperref[tab:electride]{Back to the table}

\subsubsection*{409434 NaSn}
\label{sec:tqc409434}
\lstset{language=bash, keywordstyle=\color{blue!70}, basicstyle=\ttfamily, frame=shadowbox}
\begin{lstlisting}
Computed bands:  1 - 40
GM: GM1+(1); GM3+(1); GM2-(1); GM4-(1); GM5+(2); GM2+(1); GM4+(1); GM5+(2);
    GM5-(2); GM1-(1); GM5-(2); GM3-(1); GM1+(1); GM1-(1); GM4-(1); GM3+(1);
    GM5+(2); GM2-(1); GM5-(2); GM5+(2); GM1-(1); GM2+GM4+(2); GM4+(1); GM3-(1);
    GM2+(1); GM2-(1); GM4-(1); GM1+(1); GM5-(2); GM3+(1); GM3-(1); [40] ;
M : M2  (2); M1  (2); M4  (2); M3  (2); M2  (2); M1  (2); M4  (2); M3  (2);
    M1  (2); M4  (2); M2  (2); M1  (2); M3  (2); M2  (2); M4  (2); M2  (2);
    M1  (2); M2  (2); M3  (2); M1  (2); [40] ;
P : P1  P2  (4); P1  P2  (4); P1  P2  (4); P1  P2  (4); P1  P2  (4); P1  P2  (4);
    P1  P2  (4); P1  P2  (4); P1  P2  (4); P1  P2  (4); [40] ;
X : X1  (2); X2  (2); X1  (2); X2  (2); X1  (2); X2  (2); X2  (2); X1  (2);
    X2  (2); X1  (2); X1  (2); X2  (2); X1  (2); X1  (2); X2  (2); X2  (2);
    X1  (2); X2  (2); X2  (2); X1  (2); [40] ;
N : N1  (2); N1  (2); N1  (2); N1  (2); N1  (2); N1  (2); N1  (2); N1  (2);
    N1  (2); N1  (2); N1  (2); N1  (2); N1  (2); N1  (2); N1  (2); N1  (2);
    N1  (2); N1  (2); N1  (2); N1  (2); [40];
\end{lstlisting}
\hyperref[tab:electride]{Back to the table}

\subsubsection*{627104 CsSi}
\label{sec:tqc627104}
\lstset{language=bash, keywordstyle=\color{blue!70}, basicstyle=\ttfamily, frame=shadowbox}

\hyperref[tab:electride]{Back to the table}

\subsubsection*{409439 CsSn}
\label{sec:tqc409439}
\lstset{language=bash, keywordstyle=\color{blue!70}, basicstyle=\ttfamily, frame=shadowbox}

\hyperref[tab:electride]{Back to the table}

\subsubsection*{105156 NaPb}
\label{sec:tqc105156}
\lstset{language=bash, keywordstyle=\color{blue!70}, basicstyle=\ttfamily, frame=shadowbox}
\begin{lstlisting}
Computed bands:  1 - 40
GM: GM1+(1); GM3+(1); GM2-(1); GM4-(1); GM2+(1); GM5+(2); GM4+(1); GM5+(2);
    GM5-(2); GM1-(1); GM5-(2); GM3-(1); GM1+(1); GM1-(1); GM4-(1); GM3+(1);
    GM5+(2); GM5-(2); GM2-(1); GM1-(1); GM5+(2); GM4+(1); GM2+(1); GM4+(1);
    GM3-(1); GM2-(1); GM4-(1); GM5-(2); GM2+(1); GM1+(1); GM3+(1); GM3-(1);
    [40] ;
M : M2  (2); M1  (2); M4  (2); M3  (2); M1  (2); M2  (2); M4  (2); M3  (2);
    M1  (2); M4  (2); M2  (2); M1  (2); M3  (2); M2  (2); M4  (2); M2  (2);
    M1  (2); M2  (2); M3  (2); M1  (2); [40] ;
P : P1  P2  (4); P1  P2  (4); P1  P2  (4); P1  P2  (4); P1  P2  (4); P1  P2  (4);
    P1  P2  (4); P1  P2  (4); P1  P2  (4); P1  P2  (4); [40] ;
X : X1  (2); X2  (2); X1  (2); X2  (2); X1  (2); X2  (2); X2  (2); X1  (2);
    X2  (2); X1  (2); X1  (2); X1  (2); X2  (2); X1  (2); X2  (2); X2  (2);
    X1  (2); X2  (2); X2  (2); X1  (2); [40] ;
N : N1  (2); N1  (2); N1  (2); N1  (2); N1  (2); N1  (2); N1  (2); N1  (2);
    N1  (2); N1  (2); N1  (2); N1  (2); N1  (2); N1  (2); N1  (2); N1  (2);
    N1  (2); N1  (2); N1  (2); N1  (2); [40];
\end{lstlisting}
\hyperref[tab:electride]{Back to the table}

\subsubsection*{409435 KSn}
\label{sec:tqc409435}
\lstset{language=bash, keywordstyle=\color{blue!70}, basicstyle=\ttfamily, frame=shadowbox}
\begin{lstlisting}
Computed bands: 65 -104
GM: GM1+(1); GM3+(1); GM2-(1); GM4-(1); GM5+(2); GM2+(1); GM5+(2); GM4+(1);
    GM5-(2); GM1-(1); GM5-(2); GM3-(1); GM1+(1); GM3+(1); GM1-(1); GM4-(1);
    GM5+(2); GM2-(1); GM5-(2); GM5+(2); GM4+(1); GM4-(1); GM2+(1); GM1-(1);
    GM3-(1); GM4+(1); GM2+(1); GM1+(1); GM2-(1); GM5-(2); GM3+(1); GM3-(1);
    [40] ;
M : M2  (2); M1  (2); M4  (2); M3  (2); M2  (2); M1  (2); M4  (2); M3  (2);
    M1  (2); M2  (2); M4  (2); M1  (2); M3  (2); M2  (2); M4  (2); M1  (2);
    M2  (2); M2  (2); M3  (2); M1  (2); [40] ;
P : P1  P2  (4); P1  P2  (4); P1  P2  (4); P1  P2  (4); P1  P2  (4); P1  P2  (4);
    P1  P2  (4); P1  P2  (4); P1  P2  (4); P1  P2  (4); [40] ;
X : X1  (2); X2  (2); X1  (2); X1  (2); X2  (2); X2  (2); X2  (2); X1  (2);
    X2  (2); X1  (2); X1  (2); X2  (2); X1  (2); X1  (2); X2  (2); X2  (2);
    X1  (2); X2  (2); X2  (2); X1  (2); [40] ;
N : N1  (2); N1  (2); N1  (2); N1  (2); N1  (2); N1  (2); N1  (2); N1  (2);
    N1  (2); N1  (2); N1  (2); N1  (2); N1  (2); N1  (2); N1  (2); N1  (2);
    N1  (2); N1  (2); N1  (2); N1  (2); [40];
\end{lstlisting}
\hyperref[tab:electride]{Back to the table}

\subsubsection*{280615 Sb$_{2}$MoSe}
\label{sec:tqc280615}
\lstset{language=bash, keywordstyle=\color{blue!70}, basicstyle=\ttfamily, frame=shadowbox}

\hyperref[tab:electride]{Back to the table}

\subsubsection*{380397 RbNb$_{4}$Br$_{11}$}
\label{sec:tqc380397}
\lstset{language=bash, keywordstyle=\color{blue!70}, basicstyle=\ttfamily, frame=shadowbox}

\hyperref[tab:electride]{Back to the table}

\subsubsection*{79796 Ta$_{4}$FeTe$_{4}$}
\label{sec:tqc79796}
\lstset{language=bash, keywordstyle=\color{blue!70}, basicstyle=\ttfamily, frame=shadowbox}

\hyperref[tab:electride]{Back to the table}

\subsubsection*{88202 TlCu$_{5}$Se$_{3}$}
\label{sec:tqc88202}
\lstset{language=bash, keywordstyle=\color{blue!70}, basicstyle=\ttfamily, frame=shadowbox}

\hyperref[tab:electride]{Back to the table}

\subsubsection*{404695 K$_{6}$Na$_{14}$MgTl$_{18}$}
\label{sec:tqc404695}
\lstset{language=bash, keywordstyle=\color{blue!70}, basicstyle=\ttfamily, frame=shadowbox}
\begin{lstlisting}
Computed bands:  1 - 62
GM: GM1+GM2+GM3+(3); GM4-(3); GM1+GM2+GM3+(3); GM4-(3); GM4+(3); GM4-(3);
    GM4-(3); GM4+(3); GM1+(1); GM1+(1); GM4-(3); GM4-(3); GM2+GM3+(2); GM4+(3);
    GM2+GM3+(2); GM4-(3); GM1+(1); GM4-(3); GM1+(1); GM4+(3); GM1-(1); GM4-(3);
    GM2+GM3+(2); GM4+(3); GM4-(3); [62] ;
R : R4- (3); R4+ (3); R4- (3); R4+ (3); R1- R2- R3- (3); R1+ (1);
    R2+ R3+ R4+ (5); R4- (3); R1+ (1); R1- (1); R4- (3); R4+ (3); R4+ (3);
    R2+ R3+ (2); R2- R3- (2); R4- (3); R1+ (1); R4- (3); R1- (1); R4+ (3);
    R4- (3); R1- (1); R4+ (3); R4- (3); R2+ R3+ (2); [62] ;
M : M1+ M3- M4- (3); M2+ M3+ M3- (3); M1+ M3- M4- (3); M2+ M4+ M4- (3);
    M1- M1- M2+ (3); M1+ (1); M2- (1); M1+ M2+ M3+ M3- (4); M2- M4+ M4- (3);
    M1+ (1); M2+ (1); M3- (1); M4- (1); M2- (1); M4- (1); M3- (1); M1- (1);
    M2+ (1); M4+ (1); M1+ (1); M3+ (1); M1+ (1); M2+ (1); M2+ (1); M4- (1);
    M3- (1); M2- (1); M1+ (1); M4- (1); M3- (1); M2+ (1); M3- (1); M1- M3+ (2);
    M4+ (1); M4- (1); M1+ (1); M2- (1); M2+ (1); M4+ (1); M3- M4- (2); M1+ (1);
    M2- (1); M3+ (1); M1+ (1); M1- (1); [62] ;
X : X1+ X1+ X4- (3); X3+ X3- X4- (3); X1+ X1+ X4- (3); X2+ X2- X4- (3);
    X1- X2+ X3+ (3); X1+ X2- X3- (3); X3+ X3- X4- (3); X2+ X2- X4+ (3); X1+ (1);
    X4- (1); X4- (1); X3- (1); X2- (1); X1+ (1); X2+ X3+ (2); X1+ (1); X2+ (1);
    X3+ (1); X1+ (1); X4+ (1); X4- (1); X4- (1); X2- (1); X4- (1); X3- (1);
    X1+ (1); X4- (1); X3- (1); X1- (1); X2- (1); X4- (1); X1+ (1); X2+ (1);
    X2- (1); X4- (1); X3- (1); X1- (1); X2- (1); X3- (1); X1+ (1); X3+ (1);
    X4+ (1); X3+ (1); X2+ (1); X1+ (1); [62];
\end{lstlisting}
\hyperref[tab:electride]{Back to the table}

\subsubsection*{236348 K$_{6}$Na$_{14}$Tl$_{18}$Zn}
\label{sec:tqc236348}
\lstset{language=bash, keywordstyle=\color{blue!70}, basicstyle=\ttfamily, frame=shadowbox}
\begin{lstlisting}
Computed bands:  1 - 67
GM: GM1+GM2+GM3+(3); GM4-(3); GM1+GM2+GM3+(3); GM4-(3); GM4+(3); GM4-(3);
    GM4-(3); GM4+(3); GM1+(1); GM1+(1); GM4-(3); GM2+GM3+(2); GM4+(3); GM4-(3);
    GM2+GM3+(2); GM4+(3); GM2+GM3+(2); GM4-(3); GM1+(1); GM1+(1); GM4-(3);
    GM4+(3); GM1-(1); GM4-(3); GM2+GM3+(2); GM4+(3); GM4-(3); [67] ;
R : R4- (3); R4+ (3); R4- (3); R4+ (3); R1- R2- R3- (3); R1+ R4+ (4);
    R2+ R3+ (2); R4- (3); R1+ (1); R1- (1); R4- (3); R4+ (3); R2+ R3+ (2);
    R4+ (3); R4+ (3); R2+ R3+ (2); R2- R3- (2); R4- (3); R1+ (1); R4- (3);
    R1- (1); R4+ (3); R4- (3); R1- (1); R4+ (3); R4- (3); R2+ R3+ (2); [67] ;
M : M1+ M3- M4- (3); M2+ M4+ M4- (3); M1+ M3- M4- (3); M2+ M3+ M3- (3);
    M1- M1- M2+ (3); M1+ M2+ M4+ M4- (4); M2- (1); M1+ (1); M2- M3+ M3- (3);
    M1+ (1); M2+ (1); M4- (1); M3- (1); M2- (1); M1+ M2+ M3+ M4+ (4); M1+ (1);
    M3- (1); M4- (1); M1- (1); M2+ (1); M3+ (1); M1+ (1); M4+ (1); M1+ (1);
    M2+ (1); M2+ (1); M3- (1); M4- (1); M2- (1); M1+ (1); M3- (1); M4- (1);
    M2+ (1); M4- (1); M4+ (1); M1- (1); M3+ (1); M1+ (1); M3- (1); M2- (1);
    M2+ (1); M4- (1); M3- (1); M3+ (1); M2- (1); M1+ (1); M1+ (1); M4+ (1);
    M1- (1); [67] ;
X : X1+ X1+ X4- (3); X2+ X2- X4- (3); X1+ X1+ X4- (3); X3+ X3- X4- (3);
    X1- X2+ X3+ (3); X2+ X2- X4- (3); X1+ X2- X3- (3); X3+ X3- X4+ (3); X1+ (1);
    X4- (1); X4- (1); X2- (1); X3- (1); X1+ (1); X1+ X2+ X3+ X4+ (4); X1+ (1);
    X2+ X3+ (2); X1+ (1); X3+ (1); X2+ (1); X1+ (1); X4+ (1); X4- (1); X4- (1);
    X3- (1); X4- (1); X2- (1); X1+ (1); X4- (1); X2- (1); X1- (1); X3- (1);
    X4- (1); X1+ (1); X3+ (1); X3- (1); X4- (1); X2- (1); X1- (1); X3- (1);
    X2- (1); X1+ (1); X2+ (1); X4+ (1); X3+ (1); X2+ (1); X1+ (1); [67];
\end{lstlisting}
\hyperref[tab:electride]{Back to the table}

\subsubsection*{637466 Ge$_{3}$Os$_{2}$}
\label{sec:tqc637466}
\lstset{language=bash, keywordstyle=\color{blue!70}, basicstyle=\ttfamily, frame=shadowbox}

\hyperref[tab:electride]{Back to the table}

\subsubsection*{637743 Ge$_{3}$Ru$_{2}$}
\label{sec:tqc637743}
\lstset{language=bash, keywordstyle=\color{blue!70}, basicstyle=\ttfamily, frame=shadowbox}

\hyperref[tab:electride]{Back to the table}

\subsubsection*{647772 Si$_{3}$Os$_{2}$}
\label{sec:tqc647772}
\lstset{language=bash, keywordstyle=\color{blue!70}, basicstyle=\ttfamily, frame=shadowbox}

\hyperref[tab:electride]{Back to the table}

\subsubsection*{2344 Si$_{3}$Ru$_{2}$}
\label{sec:tqc2344}
\lstset{language=bash, keywordstyle=\color{blue!70}, basicstyle=\ttfamily, frame=shadowbox}

\hyperref[tab:electride]{Back to the table}

\subsubsection*{413518 Ba$_{11}$Sb$_{10}$}
\label{sec:tqc413518}
\lstset{language=bash, keywordstyle=\color{blue!70}, basicstyle=\ttfamily, frame=shadowbox}

\hyperref[tab:electride]{Back to the table}

\subsubsection*{410997 Tb$_{4}$BBr$_{6}$}
\label{sec:tqc410997}
\noindent Essential BR: $A@4e$ \\
\noindent RSI:
\begin{flalign*}
&\delta_{1}@4e\equiv -m(A)+m(B) = -1,&
\end{flalign*}
\lstset{language=bash, keywordstyle=\color{blue!70}, basicstyle=\ttfamily, frame=shadowbox}

\hyperref[tab:electride]{Back to the table}

\subsubsection*{82529 Ba$_{2}$(SnSb$_{2}$)$_{3}$}
\label{sec:tqc82529}
\noindent Essential BR: $Ag@4a$ \\
\noindent RSI:
\begin{flalign*}
&\delta_{1}@4a\equiv -m(Ag)+m(Au) = -1,&
\end{flalign*}
\lstset{language=bash, keywordstyle=\color{blue!70}, basicstyle=\ttfamily, frame=shadowbox}

\hyperref[tab:electride]{Back to the table}

\subsubsection*{39873 YB$_{7}$Mo$_{3}$}
\label{sec:tqc39873}
\noindent Essential BR: $Ag@4a$ \\
\noindent RSI:
\begin{flalign*}
&\delta_{1}@4a\equiv -m(Ag)+m(Au) = -1,&
\end{flalign*}
\lstset{language=bash, keywordstyle=\color{blue!70}, basicstyle=\ttfamily, frame=shadowbox}

\hyperref[tab:electride]{Back to the table}

\subsubsection*{402661 Y$_{4}$BBr$_{6}$}
\label{sec:tqc402661}
\noindent Essential BR: $A@4e$ \\
\noindent RSI:
\begin{flalign*}
&\delta_{1}@4e\equiv -m(A)+m(B) = -1,&
\end{flalign*}
\lstset{language=bash, keywordstyle=\color{blue!70}, basicstyle=\ttfamily, frame=shadowbox}

\hyperref[tab:electride]{Back to the table}

\subsubsection*{202674 SrNb$_{8}$O$_{14}$}
\label{sec:tqc202674}
\noindent Essential BR: $Ag@2b$ \\
\noindent RSI:
\begin{flalign*}
&\delta_{1}@2b\equiv -m(Ag)+m(Au)-m(Bg)+m(Bu) = -1,&
\end{flalign*}
\lstset{language=bash, keywordstyle=\color{blue!70}, basicstyle=\ttfamily, frame=shadowbox}

\hyperref[tab:electride]{Back to the table}

\subsubsection*{79976 BaNb$_{8}$O$_{14}$}
\label{sec:tqc79976}
\noindent Essential BR: $Ag@2b$ \\
\noindent RSI:
\begin{flalign*}
&\delta_{1}@2b\equiv -m(Ag)+m(Au)-m(Bg)+m(Bu) = -1,&
\end{flalign*}
\lstset{language=bash, keywordstyle=\color{blue!70}, basicstyle=\ttfamily, frame=shadowbox}

\hyperref[tab:electride]{Back to the table}

\subsubsection*{165180 K(NaIn$_{3}$)$_{3}$}
\label{sec:tqc165180}
\noindent Essential BR: $Au@4b$ \\
\noindent RSI:
\begin{flalign*}
&\delta_{1}@4b\equiv -m(Ag)+m(Au)-m(Bg)+m(Bu) = 1,&
\end{flalign*}
\lstset{language=bash, keywordstyle=\color{blue!70}, basicstyle=\ttfamily, frame=shadowbox}

\hyperref[tab:electride]{Back to the table}

\subsubsection*{72021 Nb$_{3}$Se$_{12}$I}
\label{sec:tqc72021}
\lstset{language=bash, keywordstyle=\color{blue!70}, basicstyle=\ttfamily, frame=shadowbox}

\hyperref[tab:electride]{Back to the table}

\subsubsection*{103709 K$_{3}$Ga$_{13}$}
\label{sec:tqc103709}
\lstset{language=bash, keywordstyle=\color{blue!70}, basicstyle=\ttfamily, frame=shadowbox}

\hyperref[tab:electride]{Back to the table}

\subsubsection*{165613 Na$_{5}$Sn$_{13}$}
\label{sec:tqc165613}
\noindent Essential BR: $Ag@4b$ \\
\noindent RSI:
\begin{flalign*}
&\delta_{1}@4b\equiv -m(Ag)+m(Au)-m(Bg)+m(Bu) = -1,&
\end{flalign*}
\lstset{language=bash, keywordstyle=\color{blue!70}, basicstyle=\ttfamily, frame=shadowbox}

\hyperref[tab:electride]{Back to the table}

\end{widetext}

\end{document}